\documentclass[letterpaper,english,10pt]{report}
\usepackage{babel}
\usepackage[T1]{fontenc}
\usepackage[latin9]{inputenc}
\usepackage{varioref}
\usepackage{amsmath}
\usepackage{makeidx}
\makeindex
\usepackage{graphicx}
\usepackage{amssymb}

\makeatletter

\providecommand{\LyX}{L\kern-.1667em\lower.25em\hbox{Y}\kern-.125emX\@}
\newcommand{\lyxline}[1][1pt]{%
  \par\noindent%
  \rule[.5ex]{\linewidth}{#1}\par}
\providecommand{\tabularnewline}{\\}

\usepackage{geometry} 
\newcounter{draft}

\usepackage{amssymb} 
\usepackage{ifthen}
\usepackage{calc} 
\usepackage{datetime}
\ifthenelse{\thedraft=0}{}{\usepackage[notref,notcite]{showkeys}}
\newcounter{alteFormelNr}\newcounter{alteFootnoteNr}
\newcommand{\rem}[1]{%
\ifthenelse{\thedraft=2}{%
\setcounter{alteFormelNr}{\value{equation}}
\setcounter{alteFootnoteNr}{\value{footnote}}
\marginpar[$\boldsymbol{==>}$]{$\boldsymbol{<==}$}
{\sc\tt << #1 >>}%
\setcounter{equation}{\value{alteFormelNr}}
\setcounter{footnote}{\value{alteFootnoteNr}}
}{}%
}
\newcommand{\frem}[1]{
\ifthenelse{\thedraft=2}{%
\setcounter{alteFormelNr}{\value{equation}}
{\sc\tt << #1 >>}%
\setcounter{equation}{\value{alteFormelNr}}
}{}%
}
\usepackage[hypertex,backref=page]{hyperref}
\usepackage{backrefx} 

\usepackage{multicol}

\usepackage{verbatim}
\ifthenelse{\thedraft=0}{
\geometry{verbose,a4paper,tmargin=15mm,bmargin=20mm,lmargin=20mm,rmargin=20mm}
}

\newcommand{\Ram}[2]{\centerline{\framebox[#1\textwidth]{\parbox{\textwidth}{#2}}}\vspace{.15cm}}
\newcommand{\Ramm}[2]{\framebox[#1\textwidth]{\parbox{\textwidth}{$#2$}}}
\newcommand{\vorRam}{\vspace{.1cm}\\ }
\newcommand{\vRam}[2]{\vorRam\Ram{#1}{#2}}

\newcommand{\bref}[1]{\def\theequation{\ref{#1}={\bf \thechapter}.\arabic{equation}}}
\newcommand{\eref}{\def\theequation{{\bf \thechapter}.\arabic{equation}}}

\newcounter{choice}
\newcommand{\nurTeil}{0} 
\newcommand{\Teil}[2]{
\ifthenelse{\equal{\nurTeil}{0}}{#2}{
\ifthenelse{\equal{\nurTeil}{#1}}{#2}{}
}
}

\newcommand{\lastmodified}{--not known--}
\newcommand{\nix}[1]{}

\newcommand{\inputTeil}[1]{
\newcommand{\title}[1]{}
\newcommand{\author}[1]{}
\newcommand{\date}[1]{\renewcommand{\lastmodified}{##1}}
\makeatletter\renewenvironment{abstract}{\collect@body \nix}{}\makeatother
\renewcommand{\nurTeil}{#1}
\renewcommand{\tableofcontents}{}
\renewcommand{\bibliography}[1]{}
\renewcommand{\appendix}{\rem{Achtung: Das Folgende war mal ein Appendix!!}}
\renewcommand{\printindex}{}
}

\newcommand{\remch}{\renewcommand{\rem}[1]{\frem{##1}}}

\newcounter{input}
\newtheorem{prop}{Proposition}
\newtheorem{thm}{Theorem}
\newtheorem{Def}{Definition}

{}

\padzeroes[3]
\newcommand{\thefoot}{\twodigit{\value{footnote}}}
\newcommand{\localappendix}{\setcounter{section}{0}
\ifthenelse{\theinput=1}{\renewcommand{\thesection}{\arabic{chapter}.\Alph{section}}}{\renewcommand{\thesection}{\Alph{section}}}}
\newcounter{localapp}

\renewcommand{\thefoot}{\twodigit{\value{chapter}}.\twodigit{\value{footnote}}@\arabic{footnote}}

\makeatother

\begin{document}
\newcommand{\bs}[1]{\boldsymbol{#1}}
\newcommand{\mf}[1]{\mathfrak{#1} }
\newcommand{\mc}[1]{\mathcal{#1}}
\newcommand{\norm}[1]{{\parallel#1 \parallel}}
\newcommand{\Norm}[1]{\left\Vert #1 \right\Vert }
\newcommand{\partiell}[2]{\frac{\partial#1 }{\partial#2 }}
\newcommand{\Partiell}[2]{\left( \frac{\partial#1 }{\partial#2 }\right) }
\newcommand{\ola}[1]{\overleftarrow{#1}}
\newcommand{\lpartial}{\overleftarrow{\partial}}
\newcommand{\partl}[1]{\frac{\partial}{\partial#1}}
\newcommand{\partr}[1]{\frac{\lpartial}{\partial#1}}
\newcommand{\funktional}[2]{\frac{\delta#1 }{\delta#2 }}
\newcommand{\funktl}[1]{\frac{\delta}{\delta#1}}
\newcommand{\funktr}[1]{\frac{\ola{\delta}}{\delta#1}}
\newcommand{\de}{{\bf d}\!}
\newcommand{\es}{{\bf s}\!}
\newcommand{\dew}{\bs{d}^{\textrm{w}}\!}
\newcommand{\Lie}{\bs{\mc{L}}}
\newcommand{\Liecov}{\Lie^{\textrm{(cov)}}}
\newcommand{\group}[1]{\mc{R}\!\left(#1\right)}
\newcommand{\Dorf}{\bs{\mc{D}}}
\newcommand{\pe}{\bs{\partial}}
\newcommand{\De}{\textrm{D}\!}
\newcommand{\total}[2]{\frac{\de#1 }{\de#2 }}
\newcommand{\Frac}[2]{\left( \frac{#1 }{#2 }\right) }
\newcommand{\To}{\rightarrow}
 \newcommand{\ket}[1]{|#1 >}
\newcommand{\bra}[1]{<#1 |}
\newcommand{\Ket}[1]{\left| #1 \right\rangle }
\newcommand{\Bra}[1]{\left\langle #1 \right| }
 \newcommand{\braket}[2]{<#1 |#2 >}
\newcommand{\Braket}[2]{\Bra{#1 }\left. #2 \right\rangle }
\newcommand{\kom}[2]{[#1 ,#2 ]}
\newcommand{\Kom}[2]{\left[ #1 ,#2 \right] }
\newcommand{\abs}[1]{\mid#1 \mid}
\newcommand{\Abs}[1]{\left| #1 \right| }
\newcommand{\erw}[1]{\langle#1\rangle}
\newcommand{\Erw}[1]{\left\langle #1 \right\rangle }
\newcommand{\bei}[2]{\left. #1 \right| _{#2 }}
\newcommand{\dann}{\Rightarrow}
\newcommand{\q}[1]{\underline{#1 }}
\newcommand{\hoch}[1]{^{#1 }}
\newcommand{\tief}[1]{_{#1 }}
\renewcommand{\hoch}[1]{{}^{#1}}\renewcommand{\tief}[1]{{}_{#1}}\newcommand{\lqn}[1]{\lefteqn{#1}}
\newcommand{\os}[2]{\overset{\lefteqn{{\scriptstyle #1}}}{#2}}
\newcommand{\us}[2]{\underset{\lqn{{\scriptstyle #2}}}{#1}}
\newcommand{\ous}[3]{\underset{\lefteqn{{\scriptstyle #3}}}{\os{#1}{#2}}}
\newcommand{\zwek}[2]{\begin{array}{c}
 #1\\
#2\end{array}}
\newcommand{\drek}[3]{\begin{array}{c}
 #1\\
#2\\
#3\end{array}}
\newcommand{\UB}[2]{\underbrace{#1 }_{\le{#2 }}}
\newcommand{\OB}[2]{\overbrace{#1 }^{\le{#2 }}}
\newcommand{\tr}{\textrm{tr}\,}
\newcommand{\Tr}{\textrm{Tr}\,}
\newcommand{\Det}{\textrm{Det}\,}
\newcommand{\diag}{\textrm{diag}\,}
\newcommand{\Diag}{\textrm{Diag}\,}
\newcommand{\sgn}{\textrm{sgn\,}}
\newcommand{\one}{1\!\!1}
\newcommand{\fussend}{\diamond}
\newcommand{\eps}{\varepsilon}
\newcommand{\dali}{\Box}
 \newcommand{\choice}[2]{\ifthenelse{\thechoice=1}{#1}{\ifthenelse{\thechoice=2}{#2}{\left\{  \begin{array}{c}
 #1\\
#2\end{array}\right\}  }}}
 \newcommand{\lchoice}[2]{\ifthenelse{\thechoice=1}{#1}{\ifthenelse{\thechoice=2}{#2}{\left\{  \begin{array}{c}
 #1\\
#2\end{array}\right.}}}
 \newcommand{\lcsign}{\ifthenelse{\thechoice=1}{+}{\ifthenelse{\thechoice=2}{-}{\pm}}}
 \newcommand{\lcmsign}{\ifthenelse{\thechoice=1}{-}{\ifthenelse{\thechoice=2}{+}{\mp}}}
\newcommand{\lcconst}{c}
 \newcommand{\weyl}{\alpha}
\newcommand{\greq}{=_{g}}
\newcommand{\grequiv}{\equiv_{g}}
\newcommand{\grdef}{:=_{g}}
\newcommand{\Greq}{=_{G}}
\newcommand{\Grequiv}{\equiv_{G}}
\newcommand{\Grdef}{:=_{G}}
\newcommand{\Ggreq}{=_{Gg}}
\newcommand{\Greqornot}{=_{(G)}}
\newcommand{\Grequivornot}{\equiv_{(G)}}
\newcommand{\Grdefornot}{:=_{(G)}}
\newcommand{\greqornot}{=_{(g)}}
\newcommand{\grequivornot}{\equiv_{(g)}}
\newcommand{\grdefornot}{:=_{(g)}}
\newcommand{\fatkomma}{\textrm{{\bf ,}}}
\newcommand{\basis}{\boldsymbol{\mf{t}}}
\newcommand{\ip}{\imath}
\newcommand{\Beta}{\textrm{\Large$\beta$}}
\newcommand{\sBeta}{\textrm{\large$\beta$}}
\newcommand{\be}{{\bs{b}}}
\newcommand{\ce}{{\bs{c}}}
\newcommand{\Q}{{\bs{Q}}}
\newcommand{\mm}{\bs{m}\ldots\bs{m}}
\newcommand{\nn}{\bs{n}\ldots\bs{n}}
\newcommand{\kk}{k\ldots k}
\newcommand{\OO}{\bs{\Omega}}
\newcommand{\oo}{\bs{o}}
\newcommand{\tet}{{\bs{\theta}}}
\newcommand{\Es}{{\bs{S}}}
\newcommand{\Ce}{{\bs{C}}}
\newcommand{\lam}{{\bs{\lambda}}}
\newcommand{\om}{{\bs{\omega}}}
\newcommand{\ro}{{\bs{\rho}}}
\newcommand{\cov}{\textrm{D}_{\tet}}
\newcommand{\cova}[1]{\mbox{D}_{\tet#1}}
\newcommand{\qua}[1]{\mbox{Q}_{\tet#1}}
\newcommand{\feps}{{\bs{\eps}}}
\newcommand{\qu}{\textrm{Q}_{\tet}}
\newcommand{\dimw}{d_{\textrm{w}}}
\newcommand{\mteta}{\mu(\tet)}
\newcommand{\msig}{d^{\lqn{\hoch{\dimw}}}\sigma}
\newcommand{\msigp}{d^{\hoch{{\scriptscriptstyle \dimw}\lqn{{\scriptscriptstyle -1}}}}\sigma}
\newcommand{\backtilde}{\!\!\tilde{}\,\,}
\newcommand{\lc}[1]{\us{#1}{\llcorner}}
\newcommand{\rc}[1]{\us{#1}{\lrcorner}}
\newcommand{\ix}{\bs{x}^{+}}
\newcommand{\Ph}{\bs{\Phi}^{+}}
\newcommand{\RR}{\mc{P}}
\newcommand{\rr}{\mf{p}}
\newcommand{\GB}{O}
\newcommand{\dP}{d}
\newcommand{\gem}[1]{\underline{#1}}
\newcommand{\gemOm}{\gem{\Omega}}
\newcommand{\gemT}{\gem{T}}
\newcommand{\gemR}{\gem{R}}
\newcommand{\gemnabla}{\gem{\nabla}}
\newcommand{\leftOm}{\os{\textsc{l}}{\Omega}}
\newcommand{\dil}{\Phi_{(ph)}}
\newcommand{\dilcomp}{\phi_{(ph)}}
\newcommand{\dilfunk}[1]{#1\Phi_{(ph)}}
\newcommand{\dilo}{\lambda}
\newcommand{\abstrakt}[1]{\lqn{\bigcirc}\,#1\,}
\newcommand{\av}[1]{\underleftrightarrow{#1}}
\newcommand{\avOm}{\av{\Omega}}
\newcommand{\xfull}{\,\os{{\scriptscriptstyle \twoheadrightarrow}}{x}\,}
\newcommand{\xboson}{\,\os{{\scriptscriptstyle \rightarrow}}{x}\,}
\newcommand{\xleft}{\,\os{\put(-2,0){${\scriptscriptstyle \rightarrow}$}\put(-1,0){${\scriptscriptstyle \rightharpoonup}$}}{x}\,}
\newcommand{\xright}{\,\os{\put(-2,0){${\scriptscriptstyle \rightarrow}$}\put(-1,0){${\scriptscriptstyle \rightharpoondown}$}}{x}\,}
\newcommand{\xteta}{\tet}
\newcommand{\xhatteta}{\hat{\tet}}
\newcommand{\xbothtetas}{\vec{\tet}}
\newcommand{\vecfull}[1]{\,\os{\multiput(-1.5,0)(-.2,.1){2}{${\scriptscriptstyle \bs{\twoheadrightarrow}}$}}{#1}\:}
\newcommand{\vecboson}[1]{\,\os{\multiput(-1,0)(-.2,0){2}{${\scriptscriptstyle \bs{\rightarrow}}$}}{#1}\,}
\newcommand{\formfull}[1]{\,\us{#1}{\multiput(-1,0)(-.2,0){2}{${\scriptscriptstyle \bs{\twoheadrightarrow}}$}}\,}
\newcommand{\formboson}[1]{\,\us{#1}{{\scriptscriptstyle \bs{\bs{\rightarrow}}}}\,}
\newcommand{\Qsp}{q}
\newcommand{\compensator}{\Phi}
\newcommand{\compcomp}{\phi}
\newcommand{\covPhi}[1]{\nabla_{#1}\compensator}
\newcommand{\hatcovPhi}[1]{\hat{\nabla}_{#1}\compensator}
\newcommand{\avcovPhi}[1]{\av{\nabla}_{#1}\compensator}
\newcommand{\checkcovPhi}[1]{\check{\nabla}_{#1}\compensator}
\newcommand{\gemcovPhi}[1]{\gem{\nabla}_{#1}\compensator}
\newcommand{\allfields}[1]{\phi_{\textrm{all}}^{\mc{#1}}}
\newcommand{\tildeallfields}[1]{\tilde{\phi}_{\textrm{all}}^{\mc{#1}}}
\newcommand{\bdry}[2]{(b\! c)_{#2}^{#1}}
\newcommand{\dshift}[1]{{\Xi^{(#1)}}}
\newcommand{\hatdshift}[1]{{\hat{\Xi}^{(#1)}}}
\newcommand{\brstfield}[1]{{\Upsilon^{(#1)}}}
\newcommand{\hatbrstfield}[1]{{\hat{\Upsilon}^{(#1)}}}
\newcommand{\tildebrstfield}[1]{{\tilde{\Upsilon}^{(#1)}}}
\newcommand{\tildehatbrstfield}[1]{{\tilde{\hat{\Upsilon}}^{(#1)}}}
\newcommand{\swedge}{\wedge_{\!\!\!\! s\,}}
\newcounter{dummyfatness}\newcommand{\thindummy}{\setcounter{dummyfatness}{1}}
\newcommand{\fatdummy}{\setcounter{dummyfatness}{0}}
\newcommand{\dumfat}[1]{{\ifthenelse{\thedummyfatness=0}{\bs{#1}}{#1}}}
\newcommand{\sdu}[1]{\ifthenelse{#1=0}{\dumfat{\varphi}}{\ifthenelse{#1=-1}{\dumfat{\phi}}{\ifthenelse{#1=-2}{\dumfat{\vartheta}}{\ifthenelse{#1=-3}{\dumfat{\theta}}{\dumfat{\varphi}_{#1}}}}}}
\newcommand{\hdu}[1]{\ifthenelse{#1=0}{\hat{\dumfat{\varphi}}}{\ifthenelse{#1=-1}{\hat{\dumfat{\phi}}}{\ifthenelse{#1=-2}{\hat{\dumfat{\vartheta}}}{\ifthenelse{#1=-3}{\hat{\dumfat{\theta}}}{\hat{\dumfat{\varphi}}_{#1}}}}}}
\newcommand{\bdu}[1]{\ifthenelse{#1=0}{f}{\ifthenelse{#1=-1}{g}{\ifthenelse{#1=-2}{h}{f_{#1}}}}}
\newcommand{\du}[1]{\ifthenelse{#1=0}{F}{\ifthenelse{#1=-1}{G}{\ifthenelse{#1=-2}{H}{F_{#1}}}}}

\renewcommand{\be}{\bs{\omega}}\renewcommand{\ce}{\bs{\lambda}}\setcounter{input}{1}

\title{~\vspace{-4cm}\\
\hfill \includegraphics{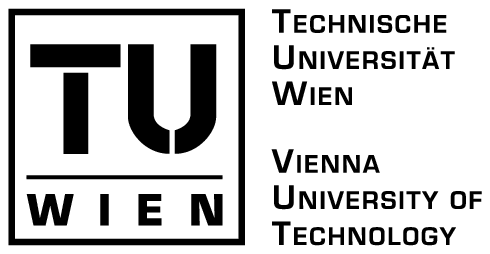}\vspace{2cm}\\
Dissertation:\\
Superstrings in General Backgrounds}

\author{ausgef\"uhrt zum Zwecke der Erlangung des akademischen Grades eines\\
Doktors der technischen Wissenschaften\vspace{.5cm} \\
unter der Leitung von\vspace{.5cm}\\
\textbf{A.o. Univ.-Prof. Dr. Maximilian Kreuzer}\\
Inst.Nr. 136\\
Institut f\"ur Theoretische Physik\vspace{.5cm}\\
eingereicht an der Technischen Universit\"at Wien,\\
Fakult\"at f\"ur Physik\vspace{.5cm}\\
von\vspace{.5cm}\\
\textbf{Sebastian Guttenberg}\\
Matrikelnummer 0427640\\
Jahngasse 14/21, A-1050 Wien \vspace{1cm}}

\date{Wien, am 16. August 2007\vspace{1cm}\\
\lyxline{\large}\parbox[b]{\textwidth}{\small The present copy of
the thesis is an improved version. Some sections have been added,
in particular to the first part about superspace conventions. They
contain mostly considerations from my PhD time that were not yet written
in a nice form at the submission date. Apart from that, the presentation
has been smoothed at some points and the supergravity transformation
of the gravitino is now reformulated in a way that it is comparable
to the literature. Also the appearance of the dilaton has been clarified
and a few additional constraints are extracted from the Bianchi identities.
The original version of the thesis can be found in the on-line dissertation
database of the TU Wien which is currently -- and hopefully in future
as well -- placed at http://www.ub.tuwien.ac.at/diss/AC05035309.pdf.
\\
$\quad$In spite of the changes, the character has remained that of
the original thesis and the text refers to the situation at the submission
date. Let me therefore acknowledge at this place the hospitality of
George Savvidy and the Institute of Nuclear Physics at the Demokritos
research institute in Athens, where part of the improvements were
implemented.\\
$\quad$Note finally that the address in Vienna given above is of
course out of date and was left untouched simply for sentimental reasons.{\normalsize \hspace{3cm}}Athens,
\textbf{July 31, 2008} \enlargethispage*{2cm}\vspace{.2cm} \\
A mistake in the argument for calculating the (correct) nilpotency
constraints (sections 5.9 and 5.10) has been corrected and the related
appendix section E.4 has been added in this new version (arXiv:0807.4968v2).
The presentation of the SUSY transformation for the dilatino has been
slightly improved. Page numbers may have changed with respect to the
previous arXiv version, but equation and section numbers remained
the same. Possible future corrections or comments will only be added
as errata or addendum. Thanks to the people of the theory department
at Turin university, in particular M. A. Rajabpour, for useful discussions.{\normalsize \hspace{2cm}}Torino,
\textbf{March 11, 2009}}}

\maketitle
\renewcommand{\thepage}{\roman{page}}\renewcommand{\abstractname}{Kurzfassung der Dissertation}\nopagebreak\thispagestyle{empty}

\begin{abstract}
\addcontentsline{toc}{chapter}{Kurzfassung der Dissertation}\index{Kurzfassung}\begin{tabular}{ll}
Kandidat: & Sebastian Guttenberg\tabularnewline
Erstprüfer / Betreuer: & Prof. Maximilian Kreuzer\tabularnewline
Zweitprüfer: & Prof. Ruben Minasian\tabularnewline
\end{tabular}\vspace{1cm}

\textbf{Superstrings in General Backgrounds}\vspace{0.5cm}

In der vorliegenden Arbeit werden einige Aspekte des Superstrings
im allgemeinen Hintergrund betrachtet. Die Arbeit besteht im Wesentlichen
aus drei Teilen: Der erste studiert die Vorraussetzungen, unter denen
man bosonische Strukturgleichungen in graduierte (z.B. im Superraum)
übertragen kann und formuliert diese in einem Satz. Auf diesen Betrachtungen
basierend werden Konventionen verwendet, die graduierungsabhängige
Vorzeichen absorbieren und die als Grundlage der Rechnungen des zweiten
Teils dienen.

Der zweite Teil beschreibt den Typ II Superstring mithilfe von Berkovits'
{}``pure spinor'' Formalismus. Die darin u.a. enthaltene Einbettung
in einen Target-Superraum ermöglicht im Gegensatz zum üblichen Ramond-Neveu-Schwarz
Formalismus eine direkte Kopplung des Strings an Ramond-Ramod-Felder.
Er eignet sich damit gut für ein Studium des Superstrings in allgemeinen
Hintergründen. In der Arbeit wird die Herleitung der {}``Supergravity
Constraints'' aus der klassischen BRST-Invarianz sorgfältig rekapituliert.
Die Herangehensweise unterscheidet sich dabei in einigen Punkten von
der ursprünglichen Herleitung von Berkovits und Howe. So bleibt die
Betrachtung im Unterschied zu deren Rechnung vollständig im Lagrange
Formalismus und zur besseren Strukturierung der Variationsrechung
wird ein kovariantes Variationsprinzip eingesetzt. Hinzu kommt die
Anwendung des im ersten Teil formulierten Satzes. Auch die Reihenfolge,
in der die Constraints erzielt werden, weicht von Berkovits und Howe
ab. Als neues Resultat werden die BRST Transformationen aller Weltflächen-Felder
hergeleitet, die bisher nur für den heterotischen Fall bekannt waren.
Ein entscheidender weiterer Schritt ist schließlich die Herleitung
der lokalen Supersymmetrie-Transformation der fermionischen Targetraum-Komponenten-Felder.

Dies liefert den Übergang zur sogenannten verallgemeinerten komplexen
Geometrie (GCG), die Bestandteil des letzten Teiles der Arbeit ist.
Die vierdimensionale effektive Supersymmetrie innerhalb einer zehndimensionalen
Typ-II Supergravitation bedingt eine {}``verallgemeinerte Calabi
Yau Mannigfaltigkeit'' als Kompaktifizierungsraum, welche wiederum
mit Methoden der GCG beschrieben werden kann. In der vorliegenden
Arbeit wird gezeigt, dass Poisson- oder Antiklammern in Sigmamodellen
auf natürliche Weise sogenannte {}``derived brackets'' im Targetraum
induzieren, darunter auch die Courant Klammer der GCG. Weiters wird
gezeigt, dass der verallgemeinerte Nijenhuis Tensor der GCG bis auf
einen de-Rham geschlossenen Term mit der {}``derived bracket'' der
verallgemeinerten Struktur mit sich selbst überein\-stimmt, und eine
neuartige Koordinatenform dieses Tensors wird präsentiert. Der Nutzen
der gewonnenen Erkenntnisse wird dann anhand von zwei Anwendungen
zur Integrabilität verallgemeinerter komplexer Strukturen demonstriert.

Der Anhang der Arbeit enthält eine Einführung in einige Aspekte von
GCG und {}``derived brackets''. Des\-weiteren werden u.a. das Noether
Theorem, Bianchi Identitäten, WZ-Eichung und $\Gamma$-Matrizen in
zehn Dimensionen besprochen. \end{abstract}\renewcommand{\abstractname}{Abstract}\begin{abstract}\addtocounter{page}{1}\addcontentsline{toc}{chapter}{\protect\hyperlink{Abstract}{Abstract}}\hypertarget{Abstract}{}\begin{tabular}{ll}
Candidate: & Sebastian Guttenberg\tabularnewline
First referee / supervisor: & Prof. Maximilian Kreuzer\tabularnewline
Second referee: & Prof. Ruben Minasian\tabularnewline
\end{tabular}\vspace{1cm}

\textbf{Superstrings in General Backgrounds}\vspace{0.5cm}

\index{abstract}In the present thesis, some aspects of superstrings
in general backgrounds are studied. The thesis divides into three
parts. The first is devoted to a careful study of very convenient
superspace conventions which are a basic tool for the second part.
We will formulate a theorem that gives a clear statement about when
the signs of a superspace calculation can be omitted. The second part
describes the type II superstring using Berkovits' pure spinor formalism.
Being effectively an embedding into superspace, target space supersymmetry
is manifest in the formulation and coupling to general backgrounds
(including Ramond-Ramond fields) is treatable. We will present a detailed
derivation of the supergravity constraints as it was given already
by Berkovits and Howe some years ago. The derivation will at several
points differ from the original one and will use new techniques like
a covariant variation principle. In addition, we will stay throughout
in the Lagrangian formalism in contrast to Berkovits and Howe. Also
the order in which we obtain the constraints and at some points the
logic will differ. As a new result we present the explicit form of
the BRST transformation of the worldsheet fields, which was before
given only for the heterotic case%
\footnote{These transformations were presented already in the original version
of August 16, 2007. In the meantime another paper \cite{D'Auria:2008ny}
independently presented BRST transformations for the type IIA string,
although in a very different setting, based on free differential algebras.
Note also another interesting paper on the pure spinor string in general
background \cite{Kluson:2008as} which has appeared in the meantime
and takes into account recent developments in Berkovits' formalism.$\quad\fussend$%
}. Having obtained all the constraints, we go one step further and
derive the form of local supersymmetry transformations of the fermionic
fields. This provides a contact point of the Berkovits string in general
background to those supergravity calculations which derive generalized
Calabi Yau conditions from effective four-dimensional supersymmetry.
The mathematical background for this setting is the so-called generalized
complex geometry (GCG) which is in turn the motivation for the last
part.

The third and last part is based on the author's paper on derived
brackets from sigma models which was motivated by the study of GCG.
It is shown in there, how derived brackets naturally arise in sigma-models
via Poisson- or antibrackets\rem{or by the quantum-commutator}, generalizing
an observation by Alekseev and Strobl. On the way to a precise formulation
of this relation, an explicit coordinate expression for the derived
bracket is obtained. The generalized Nijenhuis tensor of generalized
complex geometry is shown to coincide up to a de-Rham closed term
with the derived bracket of the structure with itself and a new coordinate
expression for this tensor is presented. The insight is applied to
two-dimensional sigma models in a background with generalized complex
structure. 

The appendix contains introductions to geometric brackets and to aspects
of generalized complex geometry. It further contains detailed reviews
on aspects of Noether's theorem, on the Bianchi identities (including
Dragon's theorem), on supergauge transformations and the WZ gauge
and on important relations for $\Gamma$-matrices (especially in ten
dimensions). A further appendix is devoted to the determination of
the (super)connection starting from different torsion- or invariance
constraints.
\end{abstract}
\addtocounter{page}{2}\newpage
\addcontentsline{toc}{chapter}{\protect\hyperlink{toc}{This table of Contents --- you are just reading it}\protect\hypertarget{toc}{}}

\tableofcontents{}\newpage

\chapter*{Some remarks in advance}

\index{remarks in advance}\addcontentsline{toc}{chapter}{Some remarks in advance} {

\begin{itemize}
\item The part about the superspace conventions is interesting in itself
and was a significant part of my research work. This is why it was
not put into the appendix. However, you can read the other parts without
this one. Only if you want to follow some calculations in detail,
you might miss some signs. Latest at this point you should study the
part about the superspace conventions before you assume that you have
found a mistake.
\item Capital indices $M$ in the part about derived brackets and generalized
geometry contain tangent and cotangent indices, while in the context
of superspace they contain bosonic and fermionic indices. In the latter
case we have $M=\{m,\bs{\mu},\hat{\bs{\mu}}\}$. The two fermionic
indices are sometimes collected in a capital curly index $\bs{\mc{M}}=\{\bs{\mu},\hat{\bs{\mu}}\}$.
\item The thesis-index at the end contains also a list of most of the used
symbols. So in case you start somewhere in the middle of the document
and would like to know, where some symbols or notations were introduced,
have a try to look at the index.
\item There are a couple of propositions contained in this thesis. They
simply contain more or less clear statements that one could have given
in the continuous text as well. In particular, their formulations
and proofs are mostly not of the same rigorousness as one would expect
it in mathematical literature. In addition, there is no clear rule
which statements are given as proposition and which are only given
in the text. The ones in propositions are important, but the ones
in the text can also be ...
\item Everything in this thesis has to be understood as graded. Graded antisymmetrization
will just be called 'antisymmetrization' and the square brackets $[\ldots]$
will be used to denote this, no matter if the graded antisymmetrized
objects are bosonic or fermionic. Likewise, the supervielbein will
often just be called 'vielbein'. Only at some points the terms 'graded'
or 'super' will be explicitly used.
\item It is a somewhat strange habit to desperately avoid the word {}``I''
in articles, in order to express ones own modesty. Writing instead
{}``the author'' seems unnecessary long and writing instead {}``we''
resembles the \emph{pluralis\index{pluralis} majestatis}, and I don't
see how this can possibly express modesty (although one then calls
it \emph{pluralis auctoris} or even \emph{pluralis modestiae}). In
spite of this, I got used myself to use frequently (and without thinking)
the word {}``we''. Understanding it as \emph{pluralis modestiae}
is probably only possible if one can replace {}``we'' with {}``the
reader and myself'', for example in {}``we will see in the following
...''. However, you, the reader, would probably loudly protest when
I write things like {}``we think ...'' or {}``we have no idea why...''
and claim that the reader is included. Nevertheless, I am afraid that
sentences like this will appear quite frequently and in order to avoid
inconsistencies, they have to be understood as the \emph{pluralis
majestatis ...}
\item The symbol \index{$\diamond$@$\fussend$|itext{end of footnote}}$\fussend$
marks the end of a footnote. If this mark is missing, it means that
the footnote is continued on the next page or that I simply forgot
to put it . (This remark was simply copied from my diploma thesis,
but at least I have changed the footnote symbol and the language)
\item This document was created with \LyX~\index{Lyx@\LyX} which is based
on \LaTeX\index{Latex@\LaTeX}.
\end{itemize}
}

\chapter*{Acknowledgements}

\addcontentsline{toc}{chapter}{Acknowledgements}{\inputTeil{0} \ifthenelse{\theinput=1}{}{}

\title{Acknowledgements}

\author{Sebastian Guttenberg}

\date{April 20, 2007}

\maketitle
\begin{abstract}
(DRacknowledge.lyx) Part of thesis
\end{abstract}
First of all I would like to thank my supervisor Maximilian Kreuzer
for all what he taught me, for his patience and for the chocolate
that he sometimes distributes in the coffee room. His keen mind and
challenging courses certainly had a big impact. Nevertheless I enjoyed
the freedom to work at whatever I was interested in. Only sometimes
I was told, not to waste my time on conventions (although inspired
by him) but to do some physics instead ;-) 

There were a lot of people accompanying my last few years at the Institute
for Theoretical Physics of the TU Vienna. In the beginning mainly
Emanuel Scheidegger, Ulrich Theis and Erwin Riegler had to suffer
under my numerous questions. With Erwin I shared the interest in many
recreational activities which were important to gather new energy
for our minds. I enjoyed this time very much.

Next I would like to thank Gernot Pauschenwein (who unfortunately
decided to change the group) and especially Johanna Knapp for collaboration
in the early phase of this work. Johanna is a truly pleasant colleague
and friend and I missed her collaboration when she moved on to CERN.

I had the pleasure to share an office in Vienna with Robert Wimmer,
Andreas Ipp, Andreas Gerhold, Paul Romatschke, Robert Schöfbeck, Urko
Reinosa and Christoph Mayrhofer. With all of them I had many discussions
and share several experiences. Andreas Ipp helped me with the algebraic
computer program Mathematica and wrote {}``decision programs'' for
me (climbing or working?), Andreas Gerhold for me was the shining
example concerning {}``efficiency''. I will never make it to this
level. Paul accompanied me to unfortunately only one ski-tour and
with the two Roberts I had numerous discussions on physics. Robert
W. was a frequent fellow at very-late-night-work and Christoph equipped
me with French songs and moral support when I left to Paris. There
is a special relationship to good old Urko, as he is probably still
grateful for the construction site which I had organized for him as
an {}``apartment''. ;-) Instead of being angry, he, and also Diego
Arbo and Maria-Jose Fernaud from {}``the other side'' of the tenth
floor, truely enriched my stay in Vienna. 

With Christoph, accompanied by Michal Michalcik, Rashid Ahmad, Maria
Schimpf, Viktor Atalla and finally Nils-Ole Walliser, the new generation
in our string group took over. Again all of them are more than just
colleagues and it is a pity to leave this group. In addition, Stefan
Stricker, Max Attems and recently arrived Aleksi Vuorinen in the group
of Anton Rebhan build an enriching counterpart with much overlap.
I would like to thank Max for help with the computer, Rashid and Michal
for checking some of my calculations and Viktor for being one of the
few people appreciating my conventions. 

With Radoslav Rashkov, I have by now a new pleasant office mate, and
I am happy to have access to his inexhaustible knowledge. I also should
not forget to mention Karl-Georg Schlesinger, who never hesitates
to share his deep mathematical insight. There were quite a lot of
other people at the ITP who made my stay enjoyable. I am grateful
to Herbert Balasin and Anton Rebhan for entertaining coffee-breaks,
discussions and lunch breaks and to Elfriede Mössmer and Franz Hochfellner
for helping with real-world-problems. 

During my PhD I also spent (in 2006) almost a year at the {}``Service
de Physique Th\'eorique'' of CEA in Saclay/Paris as a guest of the
string group consisting at that time of Mariana Gra\~na, Ruben Minasian,
Pierre Vanhove, Michael Chesterman, Kazuo Hazomichi and Yann Michel,
who warmly integrated me and took their time for all my questions
on pure spinors and generalized geometry. I am truly grateful to Ruben
who never hesitated to accept organizational efforts, in order to
enable that truly pleasant stay for me. He was rewarded by becoming
my second referee which implied yet additional work for him. Similarly,
Pierre offered any help and also did not hesitate to share his ideas
and thoughts. In particular I learned some gamma-matrix gymnastics
from him. Numerous other people deserve to be acknowledged, like all
the (not always) string-related PhD-students from Paris, who regularly
met for a very interesting and pedagogical seminar. In particular
I want to thank my former office-mate Marc Thormeier and also Michele
Frigerio for their cordiality during my stay. 

I also want to thank my long-time physics colleagues Marco Baumgartl,
Josef Dorfmeister, Steffen Metzger and Marcus Müller for keeping contact
and sharing experiences. Special thanks to Steffen for offering a
place to stay in Paris even in the final phase of his PhD thesis.
I am also grateful to Carlos Mafra for repeated encouragement concerning
my research work, even the one about conventions. A person who played
an important role in the early stage of my scientific career is Dierk
Schleicher, whom I want to thank for his believe in my abilities and
his enthusiasm about mathematics and physics. He enabled the stay
in Stony Brook during my diploma studies and was a reliable mentor
during that time. There are many other people who contributed at some
point, like Claus Jeschek, Frederik Witt, Stefan Antusch, ...

I also appreciate very much Peter van Nieuwenhuizen's help to understand
his and his collaborator's approach to the covariant quantization
at the early stage of my PhD studies. It is a pity that the work on
this subject came to an (intermediate?) end, just after the exchange
of some insight had started. 

Am Ende möchte ich meinen Eltern für ihren wertvollen Rückhalt und
ihre stete große Unterstützung in jeder Hinsicht danken. Ebenso meinem
Bruder, der sich in manch schwierigen Phasen wie kaum ein anderer
in mich hineinversetzen konnte. Auch aus der weiteren Verwandtschaft
kam viel Unterstützung und Zuspruch. Besonders bedanken möchte ich
mich in diesem Zusammenhang bei Brigitta Abele-Zöllner und Peter Abele.
Auch alle meine persönlichen Freunde, die sich regelmäßig nach dem
Fortgang meiner Arbeit erkundigt haben, und hier namentlich unerwähnt
bleiben, werden sich hoffentlich angesprochen fühlen. Zu guter Letzt
möchte ich Dir, Katharina, für all Deine Unterstützung, Geduld und
Verzicht in der Endphase dieser Arbeit danken. 
} \newpage\setcounter{page}{1}\renewcommand{\thepage}{\arabic{page}}\addcontentsline{toc}{part}{\protect\hyperlink{Intro}{Introduction}}

\part*{\hypertarget{Intro}{Introduction}}

{\inputTeil{0}\ifthenelse{\theinput=1}{}{}

\title{Introduction to Thesis}

\author{Sebastian Guttenberg}

\date{\today }

\maketitle
\begin{abstract}
This is supposed to be the Introduction to my phd-thesis...
\end{abstract}
This thesis is devoted to superstrings in general backgrounds, but
it will of course restrict to only some aspects, leaving out many
important areas. 

Apart from a few other simple cases, the quantized superstring is
well understood only in a flat background where the worldsheet fields
have basically free-field equations of motion. The physical spectrum
of a string in flat background, however, contains itself fluctuations
around this background. A huge number of strings therefore can sum
up to a non-vanishing mean background field, for example a curved
metric or even Ramond-Ramond bispinor-fields. The worldsheet dynamics
for the individual strings then has to be adjusted. In other words,
it is very natural to study the superstring in the most general background.
Consistency conditions from the worldsheet point of view implement
constraints and/or equations of motion on the background fields. On
the worldsheet level, the form of the consistency conditions depends
very much on the formalism one is using to describe the superstring.
In general, the gauge symmetries or alternatively BRST symmetries
of the action in flat background should be present in some form also
for the deformed action (string in general background), especially
after quantization. For the Ramond-Neveu-Schwarz (RNS) string, with
worldsheet fermions, this boils down to the quantum Weyl invariance
of the action, which also yields the critical dimension. For the Green
Schwarz (GS) string and for the Berkovits pure spinor string (to be
explained later), there are instead additional conditions. For the
Green Schwarz string, the so called $\kappa$ gauge symmetry has to
be preserved, while for the Berkovits pure spinor string one has to
guarantee the existence of a BRST operator which has the form $Q=\oint dz\,\ce^{\bs{\alpha}}d_{z\bs{\alpha}}$
in the flat case. In fact, in the latter two cases, the BRST symmetry
and the $\kappa$-symmetry are already strong enough to implement
the background field equations of motion at lowest order in $\alpha'$,
i.e. supergravity, such that quantum Weyl invariance does not give
additional constraints at this order.

There are of course backgrounds which are more interesting than others
for phenomenological reasons. First of all, as we are observing four
spacetime dimensions, we expect to live in a solution to the background
field equations where 6 of the 10 dimensions are compactified on a
small radius, such that they are effectively not visible. This compactification
has to be compatible with the supergravity equations, but without
restrictive boundary conditions there are infinitely many possibilities.
\rem{Even if one tries to classify them, the number of 'different'
solutions is terribly big.}\rem{praezisieren}For a long time, people
were hoping that there is a dynamical mechanism, preferring precisely
the compactification (or 'vacuum'\rem{, from the four dimensional
point of view}) that corresponds to our world. By now it seems more
and more likely that there is no such mechanism or at least not such
a strong one. Instead, the picture might be that we are simply sitting
in a huge 'landscape'\index{landscape} of possible vacua, where some
of them are more probable than others. As there is such a huge number
of effective four dimensional theories, it seems improbable that 'our
world' is not contained in them. Of course, being able to derive the
real world from string theory is a necessary requirement, if this
theory is supposed to be more than just interesting mathematics. By
now there exists a huge model building machinery. People are considering
orbi- and orientifolds and are putting intersecting D-branes into
the compactification manifold. The number of possibilities is huge.
Quite a lot of models come reasonably close to the standard model,
but none of them really matches. But even if there might be a lot
of justified criticism to string theory, this particular problem of
finding the real world is rather a matter of time. So far, only a
very tiny, mathematically treatable subset of solutions has been studied
and it would have been a lucky coincidence to find a suitable vacuum
in a simple setting. The bigger problem might show up only after finding
a vacuum which effectively reproduces the standard model: there might
be a still big number of different models which likewise reproduce
the standard model. Without knowing all of them and their common properties,
one cannot really make predictions about so far unknown physics. This
is, however, not an argument against string theory. If there is another
theory, unrelated to string theory, which also describes correctly
the standard model and gravity, then this model simply has to be added
to the set of all models which describe the so far observable physics
consistently. There is no reason to throw out the ones that might
have been obtained from string theory. Any approach that can consistently
describe the so far observable physics is of course admissible.

It is not the immediate aim of this thesis, however, to describe observable
physics, but to study the string in a general background in ten dimensions.
As argued above, one can be optimistic that someone will find real
physics within string theory. But sometimes it is easier to recognize
simplifying structures in the general setting and not in some particular
cases. Moreover, considerations like this should survive changes in
the communities opinion of what is an interesting model to look at.
This was the idea, but in the end, not everything in this thesis is
as general as it should be. First of all, mainly classical closed
strings in a type II background are considered. At some places we
keep boundary terms for later studies of open strings. Secondly a
whole part of the thesis is inspired by generalized complex geometry.
This in turn is related to a not very special but still special type
of compactifications. Let us recall this in the following lines:

Again for phenomenological reasons, in particular the hierarchy problem,
it is reasonable to expect that the four dimensional effective theory
resulting from compactification is $N=1$ supersymmetric. For that
reason, Candelas, Horowitz, Strominger and Witten introduced in 1985
\cite{Candelas:1985en} Calabi Yau manifolds into string theory. These
manifolds are Ricci flat and obey therefore the Einstein field equations
in vacuum. The supersymmetry constraint then corresponds to the existence
of a covariantly conserved (w.r.t. Levi Civita) $Spin(6)$-spinor.
Soon after, Strominger realized in \cite{Strominger:1986uh} that
a background B-field, in combination with a non-constant dilaton,
is also consistent with supersymmetric compactification. Nevertheless,
there has been very little activity on this more general case while
the Calabi-Yau case was intensively studied. This intensive study
lead to invaluable insights concerning dualities and the form of the
landscape in the Calabi-Yau case.\rem{Referenzen: Max und einige
Landscape...}

Only quite recently the importance of the general case including fluxes
was properly noticed. It was realized that the Calabi-Yau condition
gets replaced by a {}``generalized Calabi-Yau'' condition, which
brings the so-called generalized complex geometry into the game. See
the introduction to part \ref{part:Derived-Brackets-in} on page \pageref{part:Derived-Brackets-in}
for the relevant references. The derivation of this is mainly based
on supergravity calculations. Starting from ten dimensional type II
supergravity one demands effective $N=1$ supersymmetry in four dimensions
after compactification \cite{Grana:2004??,Grana:2004bg}. The results
could in general be modified by string corrections. In order to study
this, one has to set up the problem in the worldsheet language. In
other words, the superstring has to be placed into a general type
II background.

The first striking fact is that there is so far no treatable way to
couple the RNS string to Ramond-Ramond fields. Ramond-Ramond fields
can be either seen as bispinors (fields with two spinorial indices)
or equivalently (expanding in $\Gamma$-matrices) as a collection
of differential p-forms. Pullbacks of p-forms with p bigger than two
vanish on the worldsheet. Likewise we do not have elementary fields
with spacetime spinor indices in the RNS description. This is in short
the reason why coupling to the RR-fields is an open issue in the RNS
formalism. The natural alternative is the GS string which is basically
an embedding of the string into a target superspace. The fermionic
superspace coordinates or their momenta provide natural candidates
for the coupling to the RR-bispinor-fields. This formalism, however,
happens to have a fermionic gauge symmetry whose constraints are infinitely
reducible and would require an infinite tower of ghosts for ghosts
in the standard BRST covariant quantization procedure. It can be quantized
in flat space in the light cone gauge and shown to be equivalent to
RNS, but higher loop calculations are difficult because of the lack
of manifest covariance.

The problem of covariant quantization of the GS superstring was bothering
people for many painful years without real progress \rem{how many years??}
until Berkovits came up in 2000 with an alternative formalism \cite{Berkovits:2000fe},
based on commuting pure spinor ghost variables, which can be covariantly
quantized in the flat background. It is similar to the GS string in
that the target space is a supermanifold, but the origin of the pure
spinor ghost is still a bit mysterious. This ghost field and the corresponding
BRST operator are related to the $\kappa$-symmetry of the GS string,
but the relation is not very transparent. In addition, the pure spinor
condition is a quadratic constraint on the spinorial ghosts, which
seemed in the beginning not very attractive. For this reason there
were several attempts to get rid of this constraint or at least to
explain its occurrence. The beginning of my PhD research was devoted
to a promising approach by Grassi, Porrati, Policastro and van Nieuwenhuizen\cite{Nh:2001ug,Nh:2003cm,Nh:2003kq,Guttenberg:2004ht}
and I will give a few remarks about this at a later point. By now
the need for an alternative formalism has decreased, as Berkovits
managed to give a consistent multiloop picture in \cite{Berkovits:2004px}.
In any case the pure spinor formalism seems to provide the adequate
tool to study the superstring in curved background. On the classical
level this has already been done in \cite{Berkovits:2001ue}. It was
shown that classical BRST invariance of the pure spinor string in
general background already implies the supergravity constraints on
the background fields. 

One major subject of the thesis is to rederive this important result
with different techniques. All steps will be carefully motivated and
the calculations given in detail. Most importantly the calculation
given in this thesis can be seen as an independent check, as it is
done entirely in the Lagrangian formalism in contrast to \cite{Berkovits:2001ue}.
Moreover, a covariant variational principle will be established and
used to calculate the worldsheet equations of motion. Some results
are obtained in a different order but match in the end. One new result
is the explicit form for the BRST transformations of the worldsheet
fields of the type II string in general background, which were so
far only presented for the heterotic string in \cite{Chandia:2006ix}.
After the derivation of the constraints, we go one step further and
derive the supergravity transformations of the fermionic fields. The
transformations are in principle well known, but the idea is to obtain
them in the parametrization of the fields in which they enter the
pure spinor string. The supersymmetry transformations of the fermionic
fields are the starting point for the derivation of the generalized
complex Calabi-Yau conditions for supersymmetric compactifications.
Having a closed logical line from the pure spinor string to generalized
geometry hopefully opens the door for the study of quantum or string
corrections to this geometry. There is still a part missing in this
line from the Berkovits string to generalized complex geometry, as
we will end with the presentation of the supergravity transformations
and not proceed with the derivation of the generalized Calabi-Yau
conditions. Again, this calculation would not deliver new results
(following \cite{Grana:2004??,Grana:2004bg}), but it would be important
to have everything in the same setting and with the same conventions.
One might expect in addition that the superspace formulation will
give additional insight to the geometrical role of the RR-fields.
They are so far only spectators in generalized geometry. A bispinor
is from the superspace point of view just a part of a rank two tensor,
and it seems natural to include it into geometry by establishing some
version of generalized supergeometry. See also in the conclusions
for other possible extensions. 

Another new feature of the re-derivation of the supergravity constraints
from the pure spinor string is the rigorous (and in some sense very
unusual) application of some powerful superspace conventions. To be
more precise, we are going to use conventions where all the signs
which depend on the grading are absorbed via the use of a graded summation
convention and a graded equal sign. This a not a completely new idea
and northwest-southeast conventions (NW) or northeast-southwest conventions
(NE) already reflect this philosophy. \rem{lots of sugra authors,
e.g. PvN} Nevertheless most of the authors still write the signs
and take the rules of NW and NE only as a check. Only in \cite{Dragon:1978nf},
I have found an example where the signs were likewise absorbed. However,
a careful study, under which circumstances this is possible seemed
to be missing. This is the subject of part \ref{par:Superspace-Conventions}
on page \pageref{par:Superspace-Conventions}. This part is more than
just the declaration of the used conventions. The upshot is the formulation
of a theorem about when the grading dependent signs may be dropped.\rem{conjecture?}
The application to supermatrices shows that the underlying ideas lead
to slightly different definitions of e.g. supertraces or some matrix
operations. Using these definitions, all equations take exactly the
form they have for bosonic matrices. In particular the equation for
the superdeterminant reduces to an equation which holds in the very
same form for purely bosonic matrices.

Applying this philosophy to the Berkovits string calculation has some
strange effects. Most importantly, the commuting pure spinor ghosts
are treated as anticommuting objects. And likewise confusing, the
chiral blocks $\gamma_{\bs{\alpha\beta}}^{c}$ of the 10-dimensional
$\Gamma$-matrices are treated as antisymmetric objects although they
are in fact symmetric. This nevertheless makes perfect sense and the
confusion is not, because the conventions themselves are confusing,
but because of the difference to what one is used to. It is therefore
a very nice confirmation of the consistency of the conventions that
the quite lengthy calculation with the pure spinor string in general
background went through and led to the same results as the original
calculation. No single grading dependent sign had to be used. The
part about the superspace conventions -- although very interesting
in itself -- is not needed to understand the basic steps and ideas
of the other parts. Finally it should be mentioned that the appendix
about $\Gamma$-matrices in ten dimensions is written in ordinary
conventions for 'historical reasons'. It is, however, simple to translate
the equations to the other convention where needed.

There is finally part \vref{part:Derived-Brackets-in} of the thesis,
which is dealing basically with so called derived brackets and how
they arise in sigma models. This part is based on my paper \cite{Guttenberg:2006zi}.
The efforts to understand some aspects of the integrability of generalized
complex structures have led to the observation that super Poisson
brackets and super anti-brackets of worldsheet-supersymmetric or topological
sigma models induce quite naturally derived brackets in the target
space. A more detailed introduction and motivation for this part is
given at its beginning. 

The structure of the thesis is as follows: We start in part \vref{par:Superspace-Conventions}
with the discussion of the superspace conventions. In part \vref{par:PureSpinorString}
we will consider Berkovits pure spinor string. After a short motivation
for the formalism -- coming from the Green Schwarz string -- the derivation
of the supergravity constraints will be given and the supergravity
transformations of the fermionic fields will be derived. In part \vref{part:Derived-Brackets-in}
the appearance of derived brackets in sigma models and the relation
to integrability of generalized complex structures is discussed. All
parts contain their own small introduction. After the Conclusions
on page \pageref{par:Conclusion} there are a number of more or less
useful appendices. It starts with notations and conventions in appendix
\vref{cha:Notations-and-Conventions}. This appendix does of course
not contain the superspace conventions which are treated in part \ref{par:Superspace-Conventions}.
Note also that there is an index at the end of the thesis (page \pageref{index})
which should contain most of the used symbols. Appendices \vref{cha:Generalized-Complex-Geometry}
and \vref{cha:Derived-Brackets} give introductions to some aspects
of generalized complex geometry and derived brackets, respectively.
Appendix \vref{cha:Gamma-Matrices} summarizes some important facts
and equations for $\Gamma$-matrices with an emphasis on the ten-dimensional
case. In particular the explicit representation is given and the Fierz
identities for the chiral submatrices are derived. Appendix \vref{cha:Noether}
presents the Lagrangian version of the Noether theorem and the Noether
identities. Additional statements which are important for our BRST
invariance calculations of the pure spinor string are likewise given.
Appendix \vref{cha:BIs} recalls the general definitions of torsion,
curvature and H-field (valid as well in superspace) . It likewise
recalls the derivation of the Bianchi identities and gives the proof
for a slightly modified version of Dragon's theorem \cite{Dragon:1978nf}
about the relation of second and first Bianchi identities. Appendix
\vref{cha:ConnectionAppend} contains a general discussion on how
the connection is determined by invariance conditions and certain
constraints on torsion components. The simplest example is of course
the Levi Civita connection which is given by invariance of the metric
and vanishing torsion. In ten dimensional superspace there is no canonically
given superspace metric. In this appendix it will be discussed how
the connection is reconstructed from more general constraints, like
a given non-metricity or preserved structure constants. In addition
the Levi Civita connection will be extracted from a given general
superspace connection. And finally, in appendix \vref{cha:Supergauge-Transformations},
the Wess Zumino gauge will be reviewed in a general setting. This
gauge is useful and natural to eliminate auxiliary gauge degrees of
freedom. By fixing part of the superdiffeomorphism invariance, one
recovers ordinary diffeomorphism invariance and local supersymmetry.
This will be used in part \vref{par:PureSpinorString} to determine
the supergravity transformations of the fermionic background fields
of the pure spinor string. 
}

\part{Convenient Superspace Conventions}

\label{par:Superspace-Conventions}

{\inputTeil{0} \renewcommand{\be}{\bs{b}}\renewcommand{\ce}{\bs{c}}
\ifthenelse{\theinput=1}{}{}

\title{Powerful Conventions in Superspace}

\author{Sebastian Guttenberg}

\maketitle
\begin{abstract}
Part of thesis
\end{abstract}
\tableofcontents{}\newpage

\chapter{The general idea and setting}

Most bosonic definitions or equations have a natural generalization
to superspace. There are, however, always sign ambiguities in the
super-extensions of the definitions. For this reason, bosonic structural
equations only hold up to signs in the superspace or graded case.
The information that they hold up to signs is already a useful qualitative
statement, but it can be very cumbersome to determine the correct
signs. Rules like northwest-southeast or northeast-southwest were
introduced to fix the sign ambiguities. These rules in principle allow
to reconstruct the grading dependent signs from the structure of the
equation. It is then a natural step to drop all the signs during the
calculations and reintroduce them only at the very end. Or in other
words, simply take over the results from a bosonic calculation and
decorate it with the appropriate signs. But as usual, there exist
some subtle cases in which a strict application of the sign rules
compromises some other philosophy or is simply not possible. For this
reason a large majority of people working in that field prefer to
carry along all the signs and leave them away only in intermediate
steps where it is obvious that no problems will occur. A paper by
Dragon \cite{Dragon:1978nf} is the only example I know, where the
parity-dependent signs are left away completely. Nevertheless a precise
formulation of the conditions under which this is possible still seems
to be missing. Statements like {}``everything works basically the
same in the fermionic case, but one has to be careful with the signs''
are used frequently in talks. This is the reason, why we want to find
out the precise form of the above conditions. In addition, this idea
can probably be applied to many more situations than it was done so
far. In this first part of the thesis, we try to fill part of this
gap.

\section{Leading principle, graded Einstein summation convention}

The leading principle of our conventions is that every abstract calculation
looks formally exactly the same as in the bosonic case. All modifications
(signs etc) which are due to the fact that there are anticommuting
variables involved should be assigned only in the very end, to the
result of a purely bosonic calculation.

The conventions will be based on either northwest-southeast (NW for
short) or northeast-southwest (NE for short) conventions, which we
will explain a bit below. The NW convention is used for example in
some standard references as \cite{Wess:1992cp,VanNieuwenhuizen:1981ae}
while in B. DeWitt's book on supermanifolds \cite{DeWitt:1992cy}
the NE convention is used (although this is not immediately obvious,
due to his notation with some indices on the left). It is important,
however, that we will in the end have a formalism which looks exactly
the same for NW and NE. 

Our considerations will mainly treat objects with indices, for example
- but not necessarily - coordinates or tensor components. We assume
that there is an associative product among the objects being distributive
over a likewise present abelian group structure (the sum). Sometimes
we have even several of such products (tensor product or wedge product,
product of components,~...~), which all will be treated in the same
way. The described setting simply forms a general associative algebra.
But let us start with the motivating example.

Let $x^{M}$ be the coordinates in a local patch of a supermanifold.
Assume that the first components are bosonic and the following are
fermionic (anticommuting).\index{$x^M$|itext{coordinates of supermanifold}}\index{$x^{\bs{\mc{M}}}$|itext{fermionic coordinates}}\index{$x^m$|itext{bosonic coordinates}}\index{supermanifold!coordinates $x^M$ of a $\sim$}\begin{eqnarray}
x^{M} & \equiv & (x^{m},x^{\bs{\mc{M}}})\equiv(x^{m},\tet^{\mc{M}})\end{eqnarray}
The somewhat unusual choice of a curley capital letter for the fermionic
indices will be convenient for part \vref{par:PureSpinorString}.
There we have two different spinorial indices that we combine in the
capital curled one: $x^{\bs{\mc{M}}}\equiv(x^{\bs{\mu}},x^{\hat{\bs{\mu}}})$.
As usual, we assign a grading to the indices according to the split
into bosonic and fermionic variables. \begin{eqnarray}
\abs{x^{M}}\equiv\abs{M} & \equiv & \left\{ \begin{array}{c}
0\textrm{ for }M=m\\
1\textrm{ for }M=\bs{\mc{M}}\end{array}\right.\end{eqnarray}
For grading-dependent signs we use the shorthand notation\index{$(-)^{K(M+N)}$}\begin{eqnarray}
(-)^{M} & \equiv & (-1)^{\abs{M}}\\
(-)^{K(M+N)} & \equiv & (-1)^{\abs{K}\left(\abs{M}+\abs{N}\right)}\end{eqnarray}
A general object of interest is an object with $r_{u}$ upper and
$r_{l}$ lower indices (e.g. a rank $(r_{u},r_{l})$-tensor, but our
conventions should also extend to non-tensorial objects like connection-coefficients).
The overall grading of such an object is\begin{eqnarray}
\abs{T^{M_{1}\ldots M_{u}}\tief{N_{1}\ldots N_{l}}} & \equiv & \abs{T}+\abs{M_{1}}+\ldots+\abs{M_{u}}+\abs{N_{1}}+\ldots+\abs{N_{l}}\end{eqnarray}
where a nonvanishing grading $\abs{T}$ of the {}``body'' of the
object (let us call it the \textbf{rumpf}\index{rumpf|fett}, in order
not to mix it up with the body\index{body} of a supernumber) makes
sense when there are \emph{ghosts}\index{ghost} involved, i.e. objects,
with the same index-structure as the coordinates, but opposite grading.\index{$c^M$@$\ce^M$|itext{ghost}}\begin{eqnarray}
\abs{\ce^{M}}=\abs{\ce}+\abs{M} & \stackrel{\ce\textrm{ is a ghost}}{=} & 1+\abs{M}=\left\{ \begin{array}{c}
1\textrm{ for }M=m\\
0\textrm{ for }M=\mu\end{array}\right.\end{eqnarray}
 Also for differential forms we will have in general a grading that
differs from their index-grading. E.g. for the cotangent basis elements,
we will assign the grading $\abs{\de x^{M}}=\abs{\de}+\abs{M}=1+\abs{M}$.
\rem{We will discuss later whether it makes sense or not to use one
and the same grading for forms and indices, or whether it makes sense
to introduce an independent form-grading (both is consistent).}

Superspace coordinates $x^{M}$, the element $\de x^{M}$ of the exterior
algebra and the classical ghost field $\ce^{M}$ are examples of graded
commuting objects which are the main motivation for the following
discussion. Let us therefore give the definition:\index{graded commuting}\index{commuting!graded $\sim$}
\begin{equation}
a,b\mbox{ are \textbf{graded commuting }}:\iff ab=(-)^{ab}ba\end{equation}
For objects where part of the grading is assigned to the indices,
this simply becomes \begin{equation}
a^{M},b^{N}\mbox{ are \textbf{graded commuting }}:\iff a^{M}b^{N}=(-)^{(a+M)(b+N)}b^{N}a^{M}\end{equation}

\label{par:NWNE}Before we come to our conventions, let us quickly
remind the existing ones which already have the basic idea inherent.
The generalization of definitions from the commuting (bosonic) case
to the graded commuting case is not unique. A very simple example
is the interior product which has in local coordinates the form \index{$i_v$@$\ip_{v}\omega$}$\ip_{v}\omega=\sum_{m}v^{m}\omega_{m}=\sum_{m}\omega_{m}v^{m}$.
If one wants to extend this definition to vectors and forms that have
graded components as well, the order makes a difference. In the \textbf{northwest-southeast
convention} (\textbf{NW}\index{NW convention}\index{northwest-southeast|see{NW}}\index{convention!NW $\sim$}
for short) the extension is chosen in such a way that there is no
additional sign if the contraction of the indices is from the upper
left (northwest) to the lower right (southeast), i.e. $\ip_{v}\omega\equiv\sum_{M}v^{M}\omega_{M}=\sum_{M}(-)^{M}\omega_{M}v^{M}$.
Within the \textbf{northeast-southwest convention} (\textbf{NE}\index{NE convention}\index{northeast-southwest|see{NE}}\index{convention!NE $\sim$}
for short) instead, there is no sign when contracting from the lower
left to the upper right: $\ip_{v}\omega\equiv\sum_{M}\omega_{M}v^{M}=\sum_{M}(-)^{M}v^{M}\omega_{M}$.

It is also possible and sometimes very convenient to use a \textbf{mixed
convention}\index{mixed convention}\index{convention!mixed $\sim$}
with different summation conventions for different index subsets.
One could for example define $\ip_{v}\omega\equiv\sum_{m}\big(v^{m}\omega_{m}+v^{\bs{\mu}}\omega_{\bs{\mu}}+(-)^{\hat{\bs{\mu}}}v^{\hat{\bs{\mu}}}\omega_{\hat{\bs{\mu}}}\big)$.
We will come back to this below. \rem{This can be natural if $\hat{\bs{\mu}}$
upstairs is nothing but an index $\bs{\mu}$ downstairs. Another useful
example of mixed conventions will be discussed \vpageref{sub:A-simpler-way}.}

The above definitions are 'definitions by examples'. There will be
additional examples in what follows. In any case, the philosophy of
NW and NE is that for every new definition, possible ambiguities are
fixed by the contraction directions. This should give a unique way
of generalizing bosonic equations and already implies the possibility
that one can calculate in a purely bosonic manner and reconstruct
the signs at the very end, at least under certain conditions. 

In our convention, we will completely omit those signs which are encoded
in the structure of the terms. NW, NE or mixed conventions then formally
look the same, and there is no reason to decide a priori for one of
them. During the derivation and motivation we will always give the
signs for NW and only in important cases for NE. 

One of the main ingredients of our conventions will be what we call
the \textbf{graded Einstein summation convention}\index{graded summation convention|fett}\index{summation convention|fett}\index{Einstein!graded $\sim$ summation convention|fett}\index{convention!graded summation $\sim$}:
repeated indices in opposite positions (upper-lower) are summed over
their complete range, taking into account additional signs corresponding
to either NW, NE or mixed conventions.\begin{eqnarray}
a^{M}b_{M} & \equiv & \left\{ \begin{array}{c}
\sum_{M}(-)^{bM}a^{M}b_{M}\textrm{ for NW}\\
\sum_{M}(-)^{bM+M}a^{M}b_{M}\textrm{ for NE}\end{array}\right.\qquad b_{M}a^{M}\equiv\left\{ \begin{array}{c}
\sum_{M}(-)^{aM+M}b_{M}a^{M}\textrm{ for NW}\\
\sum_{M}(-)^{aM}b_{M}a^{M}\textrm{ for NE}\end{array}\right.\label{eq:gradedEinsteinSimple}\end{eqnarray}
The factor $(-)^{M}$ appears always in the {}``wrong'' contraction
direction (i.e. in a NE contraction in NW conventions and vice verse).
The factors $(-)^{aM}$ and $(-)^{bM}$ bring the contracted indices
next to each other. This definition of the graded summation convention
guarantees (in both cases, NW and NE) the following important properties: 

\begin{itemize}
\item All signs which depend on the grading of the dummy-indices, disappear
in the equation for graded commutativity. If $a^{M}$ and $b_{M}$
are graded commuting objects with $a^{M}b_{N}=(-)^{(a+M)(b+N)}b_{N}a^{M}$
then the definition (\ref{eq:gradedEinsteinSimple}) simply implies
for their contraction\begin{equation}
a^{M}b_{M}=(-)^{ab}b_{M}a^{M}\label{eq:grsum:comm}\end{equation}

\item In an associative algebra it is important that the definition of the
graded sum is compatible with associativity. Taking a third algebra
element $c$ (which may or may not have an index) and multiplying
from left, we have \begin{eqnarray}
c(a^{M}b_{M}) & = & (ca)^{M}b_{M}\label{eq:grsum:assoz}\end{eqnarray}
This is kind of trivial, because the grading of the first rumpf-symbol
in the sum in (\ref{eq:gradedEinsteinSimple}) does not enter the
definition. The other way round, however, we learn that the above
property forces the definition of the graded sum to avoid the grading
of the first element.
\end{itemize}
In fact one can see the above properties as the defining properties
of the graded summation convention. We could have made a more general
ansatz with a sign depending on the rumpfs $a,b$, the index $M$
and the contraction direction $\searrow$ or $\nearrow$: \begin{equation}
a^{M}b_{M}\equiv\sum_{M}(-)^{\phi(a,b,M,\searrow)}a^{M}b_{M},\quad b_{M}a^{M}\equiv\sum_{M}(-)^{\phi(b,a,M,\nearrow)}b_{M}a^{M}\end{equation}
Demanding the associativity property (\ref{eq:grsum:assoz}) implies
that $\phi(a,b,M,{\scriptstyle \searrow})=\phi(b,M,{\scriptstyle \searrow})$,
$\phi(b,a,M,{\scriptstyle \nearrow})=\phi(a,M,{\scriptstyle \nearrow})$.
The graded commutativity property (\ref{eq:grsum:comm}) then puts
an additional restriction \begin{equation}
(-)^{\phi(b,M,\searrow)+bM+M}=(-)^{\phi(a,M,\nearrow)+aM}\end{equation}
This fixes the $a$ and $b$ dependency of $(-)^{\phi}$ completely,
namely $(-)^{\phi(b,M,\searrow)}=(-)^{\phi_{0}(M,\searrow)+bM}$ and
$(-)^{\phi(a,M,\nearrow)}=(-)^{\phi_{0}(M,\nearrow)+aM}$. In addition
we have $(-)^{\phi_{0}(M,\searrow)}=(-)^{M}(-)^{\phi_{0}(M,\nearrow)}$
with some $\phi_{0}$. The most general definition of the graded summation
convention which has the above properties (\ref{eq:grsum:comm}) and
(\ref{eq:grsum:assoz}) therefore reads\index{$\phi_0(M)$|itext{sign $(-)^{\phi_0(M)}$ in graded summation}}%
\footnote{\index{footnote!\thefoot. distinct $\protect\mathbb{Z}_2$-gradings}Some
people prefer to have not one single $\mathbb{Z}_{2}$-grading which
governs the signs in a graded commutative algebra, but to have several
distinct $\mathbb{Z}_{2}$-gradings. For example one can distinguish
between the $\mathbb{Z}_{2}$ grading $\abs{\ldots}_{d}$ of differential
forms (even and odd) and the fermion grading $\abs{\ldots}_{f}$ (fermion
or boson). The graded summation convention can then be extended to
$a^{M}b_{M}\equiv\sum_{M}(-)^{\abs{b}_{d}\abs{M}_{d}}(-)^{\abs{b}_{f}\abs{M}_{f}}(-)^{\phi_{d}(M)}(-)^{\phi_{f}(M)}a^{M}b_{M}$.
One could even introduce a seperate grading for ghost fields $\abs{\ldots}_{g}$.
Although the present discussion uses only a single $\mathbb{Z}_{2}$
grading, basically everything works the same for distinct gradings.
As the summation convention swallows all the grading dependent signs
anyway, one can even decide only at the end, which picture one prefers.$\quad\fussend$%
}\begin{equation}
a^{M}b_{M}\equiv\sum_{M}(-)^{bM}(-)^{\phi_{0}(M)}a^{M}b_{M},\quad b_{M}a^{M}\equiv\sum_{M}(-)^{M+aM}(-)^{\phi_{0}(M)}b_{M}a^{M}\label{eq:gradedEinstein:mixed}\end{equation}
For $\phi_{0}(M)=0$, we arrive at NW-conventions, while for $\phi_{0}(M)=\abs{M}$
we are in NE. In general the function $\phi_{0}(M)$ may depend arbitrarily
on the index $M$. A natural condition is of course that for $M$
being a bosonic index, the summation should reduce to the ordinary
one, so that we require $\phi_{0}(M)=0$ for $\abs{M}=0$. For the
fermionic indices, we could in principle define the sign differently
for every single index. In superspace applications, however, the result
would then in general not be Lorentz invariant\rem{really?} and therefore
not very useful. But as mentioned already with the introductory example
of the interior product, it is consistent e.g. in extended superspace
to switch the sign between different subsets, each corresponding to
a representation of the Lorentz group. A mixed convention is also
useful in phase space considerations, where we combine configuration
space coordinates $x^{M}$ and momenta $p_{M}$ to Darboux coordinates
$z^{\q{M}}\equiv(x^{M},p_{M})$. The definition of the graded summation
convention for the combined indices $\q{M}$ will then change by $(-)^{M}$
when the index range goes from the coordinate index to the momentum
index.

By now we have defined in (\ref{eq:gradedEinsteinSimple}) or (\ref{eq:gradedEinstein:mixed})
only an index contraction between two graded commuting objects. The
first generalization is to allow $a^{M}$and $b_{M}$ to be not necessarily
graded commuting. The definitions (\ref{eq:gradedEinsteinSimple})
or (\ref{eq:gradedEinstein:mixed}) make still sense and (\ref{eq:grsum:assoz})
is still fulfilled, if $a^{M}$ and $b_{M}$ are elements of an associative
algebra. There is no good argument to modify the definition in this
more general case. Finally, we go one step further and assume that
$b$ in $a^{M}b_{M}$ is not necessarily an algebra element, but simply
a placeholder for either indices or rumpfs which can carry gradings.
Likewise $a$ will also be allowed to contain indices in addition
to one or more rumpfs. I.e., we could replace $b$ by an index $b\To\tief{N}$,
to get a definition for $a^{M}\tief{NM}$. We could even remove $b$
completely $b\To\{\}$ to obtain $a^{M}\tief{M}$, or replace both
by s.th. more complicated: $a\To A_{KL},\: b\To\hoch{PQ}\, B_{R}$
yields the definition for $A_{KL}\hoch{M}\tief{PQ}\, B_{RM}$. This
allows to define almost all possible contractions. Unfortunately,
we are in this way restricted to expressions which end with the dummy
index $M$. To close this gap we can introduce a third placeholder
and define $a^{M}b_{M}c\equiv\sum_{M}(-)^{bM}(-)^{\phi_{0}(M)}a^{M}b_{M}c$
and $b_{M}a^{M}c\equiv\sum_{M}(-)^{M+aM}(-)^{\phi_{0}(M)}b_{M}a^{M}c$.
Similar to $a$, $c$ is just a spectator and does not enter the signs
in the sums. We should now check that with this general definition
the graded sum is well defined, in particular when two index pairs
are contracted. 

\begin{itemize}
\item The graded summation for more than one index pair is \textbf{well-defined}
in the sense that the contraction-operations commute. 
\end{itemize}
In order to verify this statement, let $a,b,c,d$ and $e$ be placeholders
in the above sense. In the following two examples of index contractions
over $M$ and $N$ we will first start with the $M$-contraction followed
by the $N$-contraction and then reverse the order. The simple case
is when one contraction encloses the other:\begin{eqnarray}
a^{M}b_{N}c^{N}d_{M} & \stackrel{\mbox{\tiny first }M}{=} & \sum_{M}(-)^{\phi_{0}(M)}(-)^{M(b+N+c+N+d)}a^{M}b_{N}c^{N}d_{M}=\\
 & = & \sum_{M,N}(-)^{\phi_{0}(M)+\phi_{0}(N)}(-)^{M(b+N+c+N+d)}(-)^{Nc+N}a^{M}b_{N}c^{N}d_{M}\\
 & \stackrel{\mbox{\tiny first }N}{=} & \sum_{M}(-)^{\phi_{0}(N)}(-)^{cN+N}a^{M}b_{N}c^{N}d_{M}=\\
 & = & \sum_{M,N}(-)^{\phi_{0}(N)+\phi_{0}(M)}(-)^{Nc+N}(-)^{M(b+c+d)}a^{M}b_{N}c^{N}d_{M}\end{eqnarray}
There is certainly no problem with the above case. But also for the
case where the contractions intersect, everything goes fine if indices
which are already contracted are not taken into account in the second
contraction:\begin{eqnarray}
a_{N}b^{M}c^{N}d_{M} & \stackrel{\mbox{\tiny first }M}{=} & \sum_{M}(-)^{\phi_{0}(M)}(-)^{M(c+N+d)}a_{N}b^{M}c^{N}d_{M}=\\
 & = & \sum_{M,N}(-)^{\phi_{0}(M)+\phi_{0}(N)}(-)^{M(c+N+d)}(-)^{N(b+c)+N}a_{N}b^{M}c^{N}d_{M}\\
 & \stackrel{\mbox{\tiny first }N}{=} & \sum_{M}(-)^{\phi_{0}(N)}(-)^{N(b+M+c)+N}a_{N}b^{M}c^{N}d_{M}=\\
 & = & \sum_{M,N}(-)^{\phi_{0}(N)+\phi_{0}(M)}(-)^{N(b+M+c)+N}(-)^{M(c+d)}a_{N}b^{M}c^{N}d_{M}\end{eqnarray}
Let us give one last example in (NW) (upper line) and (NE) (lower
line)to clarify the general treatment:\begin{eqnarray}
\lqn{A^{M_{1}}\tief{KN_{1}N_{2}}\hoch{M_{2}}\tief{N_{3}}B^{N_{3}N_{1}}\tief{M_{1}M_{2}}\hoch{LN_{2}}\equiv}\label{eq:terribleSigns}\\
 &  & \hspace{-1.7cm}\equiv\left\{ \begin{array}{c}
\sum_{M_{1},M_{2},N_{1},N_{2},N_{3}}(-)^{M_{1}(K+N_{2}+M_{2}+B)+M_{2}(B+N_{1})+N_{1}(1+N_{2}+B)+N_{2}(1+B+L)+N_{3}(1+B)}A^{M_{1}}\tief{KN_{1}N_{2}}\hoch{M_{2}}\tief{N_{3}}B^{N_{3}N_{1}}\lqn{\tief{M_{1}M_{2}}\hoch{LN_{2}}}\\
\sum_{M_{1},M_{2},N_{1},N_{2},N_{3}}(-)^{M_{1}(1+K+N_{2}+M_{2}+B)+M_{2}(1+B+N_{1})+N_{1}(N_{2}+B)+N_{2}(B+L)+N_{3}B}A^{M_{1}}\tief{KN_{1}N_{2}}\hoch{M_{2}}\tief{N_{3}}B^{N_{3}N_{1}}\lqn{\tief{M_{1}M_{2}}\hoch{LN_{2}}}\end{array}\right.\nonumber \end{eqnarray}
The terrible\index{terrible signs} signs\index{signs!terrible $\sim$}
in the lower lines of (\ref{eq:terribleSigns}) are exactly those
which we want to omit during calculations. We thus will define every
calculational operation in such a way that it is consistent with this
graded summation convention, s.th. one can calculate only with expressions
as in the upper line of (\ref{eq:terribleSigns}) and assign the signs
only in the end of all the calculations.

By definition all the signs which depend on dummy indices are swallowed
by the definition of the graded summation. As mentioned, the equation
$a^{M}b_{N}=(-)^{(a+M)(b+N)}b_{N}a^{M}$ for graded commuting algebra
elements reduces in a sum to $a^{M}b_{M}=(-)^{ab}b_{M}a^{M}$. The
same simplification occurs for terms with several contracted indices,
like in (\ref{eq:terribleSigns}). Assuming that the objects there
are graded commuting as well, we get\begin{equation}
A^{M_{1}}\tief{KN_{1}N_{2}}\hoch{M_{2}}\tief{N_{3}}B^{N_{3}N_{1}}\tief{M_{1}M_{2}}\hoch{LN_{2}}=(-)^{(A+K)(B+L)}B^{N_{3}N_{1}}\tief{M_{1}M_{2}}\hoch{LN_{2}}A^{M_{1}}\tief{KN_{1}N_{2}}\hoch{M_{2}}\tief{N_{3}}\end{equation}
Although there are still signs depending on the naked indices, this
is far better than without the graded summation convention, where
we would have obtained instead the full sign factor \begin{equation}
(-)^{(A+M_{1}+K+N_{1}+N_{2}+M_{2}+N_{3})(B+N_{3}+N_{1}+M_{1}+M_{2}+L+N_{2})}\end{equation}

\section{Graded equal sign}

The graded summation convention takes care of all dummy indices. But
we can still be left with naked indices and/or graded rumpfs, which
likewise produce inconvenient signs. Also the summation convention
on its own might be dangerous. To show this, look at the following
example: Consider graded commutative variables $a^{M},b^{M},c^{M}$
and $d^{M}$ with bosonic rumpfs. Then the following equations, which
are obviously correct (using our graded summation convention) \begin{eqnarray}
a^{M}b^{N}c_{N}d_{M}-a^{M}b^{N}d_{M}c_{N} & = & 0\quad\forall\mbox{graded comm.}\: a^{M},b^{N},d_{M},c_{N}\label{eq:trivialExample}\\
\dann a^{M}b^{N}\left(c_{N}d_{M}-d_{M}c_{N}\right) & = & 0\quad\forall\mbox{graded\, comm.}\: a^{M},b^{N},d_{M},c_{N}\end{eqnarray}
could lead to the -- in general -- wrong assumption\begin{eqnarray}
c_{N}d_{M}-d_{M}c_{N} & = & 0\quad\forall\mbox{graded\, comm. }d_{M},c_{N}\textrm{ (not true in general!)}\end{eqnarray}
We therefore introduce a \textbf{graded equal\index{equal sign!graded $\sim$ $=_{g}$}
sign}\index{graded equal sign} $=_{g}$\index{$=_{g}$|itext{graded equal sign}},
which states that the equality holds if for each term a mismatch in
some common ordering of the indices is taken care of by an appropriate
sign factor:\begin{eqnarray}
c_{N}d_{M}-d_{M}c_{N}=_{g}0 & :\iff & c_{N}d_{M}-(-)^{MN}d_{M}c_{N}=0\label{eq:grequalSimple}\end{eqnarray}
If we imagine objects like in (\ref{eq:terribleSigns}), the graded
equal sign allows one to write down quickly correct equations without
bothering all the involved signs. And it will also lead as a guiding
line for all definitions of new objects, which should all be writable
in terms of the graded equal sign, in order to make them compatible
with the graded summation convention.

The idea of how to define the graded equal sign should be clear from
(\ref{eq:grequalSimple}), but in order to be able to write down a
definition for the general case, we have to be a little more careful.
For practical purposes it should be enough to have a look at the examples
following the general definition, to convince yourself that everything
is very natural and intuitive. 

Let us introduce the graded equal-sign for the most general case in
two steps. At first we look at equations with only bosonic rumpfs,
like in (\ref{eq:trivialExample}).

\subsubsection*{Graded equal sign for bosonic rumpfs}

Any term $T_{(i)}$ of the equation (which can be a product of a lot
of objects with indices) has some nonnegative integer number $k$
of naked indices (the vertical position of the indices does not play
a role for this definition, so we write them all upstairs, but the
very same definition holds for any position). We take the first term
in the equation, call it $T_{(1)}\hoch{M_{1}\ldots M_{k}}$, as reference
term. Any other term $T_{(i)}$ in the equation has to have the same
index set but perhaps with a different order or permutation\index{permutation}
$P_{(i)}$\index{$P_{(i)}$|itext{permutation}} of the indices. A
permutation of an index set $\left\{ M_{1},\ldots,M_{k}\right\} $
is defined via a permutation of the set $\left\{ 1,\ldots,k\right\} $\index{permutation}\index{$P_{(i)}$|itext{permutation}}\begin{eqnarray}
P_{(i)}(M_{1},\ldots,M_{k}) & \equiv & (M_{P_{(i)}(1)},\ldots,M_{P_{(i)}(k)}),\quad P_{(1)}\equiv\one\end{eqnarray}
In order to assign the appropriate signs to the terms, we introduce
for any of the $k$ indices $M_{i}$ an auxiliary graded commutative
object $o^{M_{i}}$\index{$o^{M_i}$} which carries the grading of
the index\begin{equation}
o^{M_{i}}o^{M_{j}}=(-)^{M_{i}M_{j}}o^{M_{j}}o^{M_{i}}\end{equation}
If $M_{i}$ are just supercoordinate-indices, then the supercoordinates
$x^{M}$ themselves can be taken instead of defining new variables
$o^{M}$. Let us now define something which we call a \textbf{grading\index{grading structure}
structure\index{structure!grading $\sim$}} for a given term, namely
a product of those objects $o$ with as many factors as the term has
naked indices: \begin{eqnarray}
\textrm{gs}(T_{(1)}^{M_{1}\ldots M_{k}}) & \equiv & o^{M_{1}}\cdots o^{M_{k}}\end{eqnarray}
In the grading structures of different terms, we can rearrange the
objects until all the naked indices have some common order. For example
for two terms with $3$ naked indices we have\index{$gs$@$\textrm{gs}(\ldots)$}\begin{eqnarray}
\hspace{-.5cm}\textrm{gs}(T_{(1)}^{M_{1}M_{2}M_{3}}) & = & o^{M_{1}}o^{M_{2}}o^{M_{3}}\\
\textrm{gs}(T_{(2)}^{M_{3}M_{2}M_{1}}) & = & o^{M_{3}}o^{M_{2}}o^{M_{1}}=(-)^{M_{1}(M_{2}+M_{3})+M_{2}M_{3}}o^{M_{1}}o^{M_{2}}o^{M_{3}}\end{eqnarray}
We call the resulting sign the \textbf{relative\index{relative sign of grading structures}
sign\index{sign!relative $\sim$ of grading structures} of the grading\index{grading structure!relative sign of $\sim$'s}
structures}\index{$sign$@$\textrm{sign}^g_{(\ldots)}(\ldots)$} \begin{eqnarray}
\textrm{gs}\bigl(T_{(i)}^{M_{P_{(i)}(1)}\ldots M_{P_{(i)}(k)}}\bigr) & \equiv & \textrm{sign}_{T_{(1)}^{M_{1}\ldots M_{k}}}^{g}\bigl(T_{(i)}^{M_{P_{(i)}(1)}\ldots M_{P_{(i)}(k)}}\bigr)\cdot\textrm{gs}\bigl(T_{(1)}^{M_{1}\ldots M_{k}}\bigr)\label{eq:relativeSignBos}\end{eqnarray}
As the rumpfs carry no grading so far, it is notationally more convenient
to replace $\textrm{sign}_{T_{(1)}^{M_{1}\ldots M_{k}}}^{g}\bigl(T_{(i)}^{M_{P_{(i)}(1)}\ldots M_{P_{(i)}(k)}}\bigr)$
by%
\footnote{\index{footnote!\thefoot. permutation signature}Note that this sign
does not in general coincide with the signature of a permutation.
The relative sign $\textrm{sign}_{M_{1}\ldots M_{k}}^{g}\left({\scriptstyle P_{(i)}(M_{1},\ldots,M_{k})}\right)$
coincides with the signature\index{signature of a permutation} of
the permutation $P_{(i)}$ (which is given by minus one to the number
of switches one needs to build the permutation) only if all indices
carry an odd grading.$\quad\fussend$%
} $\textrm{sign}_{M_{1}\ldots M_{k}}^{g}\bigl({\scriptstyle M_{P_{(i)}(1)}\ldots M_{P_{(i)}(k)}}\bigr)$.
For the above two terms with three naked indices we thus have \begin{eqnarray}
\textrm{sign}_{T_{(1)}^{M_{1}M_{2}M_{3}}}^{g}\left(T_{(2)}^{M_{3}M_{2}M_{1}}\right) & = & \textrm{sign}_{M_{1}M_{2}M_{3}}^{g}\left({\scriptstyle M_{3}M_{2}M_{1}}\right)=(-)^{M_{1}(M_{2}+M_{3})+M_{2}M_{3}}\end{eqnarray}
 Using this definition of the relative sign of grading structures,
we can now define the graded equal sign for an equation with general
terms (but still bosonic rumpfs) as\begin{equation}
\boxed{\sum_{i}T_{(i)}\hoch{M_{P_{(i)}(1)}\ldots M_{P_{(i)}(k)}}=_{g}0\quad:\iff\sum_{i}\textrm{sign}_{T_{(1)}^{M_{1}\ldots M_{k}}}^{g}(T_{(i)}\hoch{M_{P_{(i)}(1)}\ldots M_{P_{(i)}(k)}})\cdot T_{(i)}\hoch{M_{P_{(i)}(1)}\ldots M_{P_{(i)}(k)}}=0}\label{eq:greqBosonicRumpf}\end{equation}
This definition does not depend on the choice of the reference term
(above it is $T_{(1)}^{M_{1}\ldots M_{k}}$), because only the relative
sign is relevant. One can replace $\textrm{sign}_{T_{(1)}^{M_{1}\ldots M_{k}}}^{g}(T_{(i)}\hoch{M_{P_{(i)}(1)}\ldots M_{P_{(i)}(k)}})$
by $\textrm{sign}_{T_{(j)}^{M_{P_{(j)}(1)}\ldots M_{P_{(j)}(k)}}}^{g}(T_{(i)}\hoch{M_{P_{(i)}(1)}\ldots M_{P_{(i)}(k)}})$
for any $j$. As mentioned above we can also replace it by simply
$\textrm{sign}_{M_{1}\ldots M_{k}}^{g}({\scriptstyle M_{P_{(i)}(1)}\ldots M_{P_{(i)}(k)}})$.

In the following sections we will always give definitions and important
equations with the graded equal sign and with the ordinary one. The
somewhat long-winded definition of above should therefore become more
transparent in numerous examples later on. But let us first complete
our definition to the case involving graded rumpfs. One could get
rid of all graded rumpfs by shifting the grading to the indices (if
present), or create a new index with only one possible value. As this
would be notationally not very nice, we stay with graded rumpfs, but
we keep in mind that a graded rumpf is similar to a naked index. Problems
for including the rumpfs in the definition of the graded equal sign
appear, when the same rumpf appears several times in one term, which
is thus similar to to having coinciding naked indices:

\subsubsection*{Problem of coinciding\index{coinciding indices} indices: }

\label{sub:Problem-of-coinciding} The graded equal sign above (\ref{eq:greqBosonicRumpf})
is only well defined if all naked indices can be distinguished. In
general calculations one usually uses different letters for each index,
even if they are allowed to coincide, and then there is no problem.
What, however, if one looks at some special case with two coinciding
indices? Consider the following relations (which simply apply the
definition of the graded equal sign):\begin{eqnarray}
(a)\quad T_{(1)}\hoch{MN}=_{g}T_{(2)}\hoch{NM} & \iff & T_{(1)}\hoch{MN}=(-)^{NM}T_{(2)}\hoch{NM}\\
(b)\quad T_{(1)}\hoch{MN}=_{g}T_{(2)}\hoch{MN} & \iff & T_{(1)}\hoch{MN}=T_{(2)}\hoch{MN}\end{eqnarray}
For $M=N$ (no sum) this reads\index{coinciding indices}\begin{eqnarray}
(a)\quad T_{(1)}\hoch{MM}=_{g}T_{(2)}\hoch{MM} & \iff & T_{(1)}\hoch{MM}=(-)^{M}T_{(2)}\hoch{MM}\quad\textrm{no sum over }M\\
(b)\quad T_{(1)}\hoch{MM}=_{g}T_{(2)}\hoch{MM} & \iff & T_{(1)}\hoch{MM}=T_{(2)}\hoch{MM}\quad\textrm{no sum over }M\end{eqnarray}
Now (a) and (b) obviously contradict themselves and the graded equal
sign is therefore ill-defined\index{ill-defined!graded equal sign for coinciding indices}.
There are two options to solve this notational problem. The first
is to \emph{always rewrite the equation with an ordinary equal sign
before looking at any special case}. The second is to make apparent
the original name of the index in the following way (this is also
useful to suppress summation over repeated indices if it is not wanted)\begin{eqnarray}
(a)\quad T_{(1)}\hoch{M(N=M)}=_{g}T_{(2)}\hoch{(N=M)M} & \iff & T_{(1)}\hoch{M(N=M)}=(-)^{M}T_{(2)}\hoch{(N=M)M}\\
(b)\quad T_{(1)}\hoch{M(N=M)}=_{g}T_{(2)}\hoch{M(N=M)} & \iff & T_{(1)}\hoch{M(N=M)}=T_{(2)}\hoch{M(N=M)}\end{eqnarray}

\subsubsection*{Graded rumpfs}

A grading of a rumpf is like a naked index grading at the position
of the rumpf. The lesson from above is, that we can only include the
rumpfs completely into the definition of the graded equal sign, if
in each term every rumpf appears exactly once. As we can't rely that
this is the case in all equations of interest, we will include the
rumpfs only partially in the definition of the graded equal sign.
Namely, the graded equal sign will not compare the order of the rumpfs,
but the position of the indices with respect to the rumpfs. This is
again necessary to stay consistent with the graded summation convention.
Consider therefore the same trivial example as in (\ref{eq:trivialExample}),
however, now with graded rumpfs\begin{eqnarray}
a^{M}b^{N}c_{N}d_{M}-(-)^{cd}a^{M}b^{N}d_{M}c_{N} & = & 0\quad\forall\mbox{graded comm.}\: a^{M},b^{N},d_{M},c_{N}\label{eq:grTrivialExample}\\
\dann a^{M}b^{N}\left(c_{N}d_{M}-(-)^{cd}d_{M}c_{N}\right) & = & 0\quad\forall\mbox{graded comm.}\: a^{M},b^{N},d_{M},c_{N}\label{eq:grTrivialExampleAusgeklammert}\end{eqnarray}
We now want to simply read off\begin{eqnarray}
c_{N}d_{M}-(-)^{cd}d_{M}c_{N} & \greq & 0\quad\forall\mbox{graded comm.}\: d_{M},c_{N}\label{eq:grequalSimpleWithRumpf}\end{eqnarray}
In order for this to be correct, we have to extend the definition
of $\greq$ appropriately to the case of graded rumpfs. Let us therefore
write out the summation convention in (\ref{eq:grTrivialExampleAusgeklammert})
explicitely (in NW-conventions):\begin{eqnarray}
\sum_{M,N}a^{M}b^{N}\left((-)^{M(b+c+d)+Nc}c_{N}d_{M}-(-)^{M(b+N+d)+N(d+c)}(-)^{cd}d_{M}c_{N}\right) & = & 0\\
\dann(-)^{Mc}c_{N}d_{M}-(-)^{MN+Nd}(-)^{cd}d_{M}c_{N} & = & 0\\
\dann(-)^{Nd}c_{N}d_{M}-(-)^{MN+Mc}(-)^{cd}d_{M}c_{N} & = & 0\end{eqnarray}
Comparing the last line with (\ref{eq:grequalSimpleWithRumpf}) we
get\begin{eqnarray}
c_{N}d_{M}-(-)^{cd}d_{M}c_{N}\greq0 & :\iff & (-)^{Nd}c_{N}d_{M}-(-)^{MN+Mc}(-)^{cd}d_{M}c_{N}=0\label{eq:rumpfsbyhand}\end{eqnarray}
The graded equal sign therefore takes care of the order of the naked
indices via $(-)^{MN}$ and of the order of the naked indices with
respect to the rumpfs, i.e. it puts their grading to the very right
of all rumpfs via $(-)^{Nd}$ and $(-)^{Mc}$. Only the order of the
rumpfs among themselves is taken care of by hand via $(-)^{cd}$.
As stated before, the correct order of the rumpfs cannot a posteriori
be figured out, when some of them coincide. E.g. for $d=c$, the above
equivalence would become\begin{eqnarray}
c_{N}c_{M}-(-)^{c}c_{M}c_{N}\greq0 & \iff & (-)^{Nd}c_{N}c_{M}-(-)^{MN+Mc}(-)^{c}c_{M}c_{N}=0\label{eq:rummpfsbyhand:coinciding}\end{eqnarray}
There is no way to deduce the sign $(-)^{c}$ from the structure of
the equation itself, if one doesn't see it as a special case of (\ref{eq:rumpfsbyhand}).
The relative order of the rumpfs is not visible in (\ref{eq:rummpfsbyhand:coinciding}).
For that reason we did not a priori include the order of the rumpfs
into the definition of the graded equal sign, as it can be ill-defined
in such situations. Nevertheless we will make a suggestion a bit later,
how to include the rumpfs to some extent into a graded equal sign.
The nice observation so far is that we got rid of \textbf{all} index-dependent
signs! The use of the graded equal is in particular useful to define
\textbf{composite objects} of the form\begin{eqnarray}
A^{MN}\equiv_{g}B^{NK}C_{K}\hoch{M} & \iff & A^{MN}\equiv(-)^{CN+MN}B^{NK}C_{K}\hoch{M}\stackrel{NW}{=}(-)^{CN+MN}\sum_{K}(-)^{KC}B^{NK}C_{K}\hoch{M}\qquad\end{eqnarray}
This makes sure that the notation $A^{MN}$ is consistent with the
position of the gradings. This is again necessary to guarantee consistency
with the graded summation convention. I.e. for every $D_{MN}$ we
have (ordinary equal sign, all indices contracted)\begin{eqnarray}
A^{MN}D_{MN} & = & B^{NK}C_{K}\hoch{M}D_{MN}\end{eqnarray}
which would not be true for the definition $A^{MN}\equiv B^{NK}C_{K}\hoch{M}$
without the graded equal sign or the appropriate signs in front.

For a more general definition of the graded equal sign in the case
of graded rumpfs, we can again introduce auxiliary graded commuting
objects $o$ and extend our previous definition of the \textbf{grading\index{grading structure}
structure\index{structure!grading $\sim$}}, i.e. the product of these
objects $o$ with as many factors as there are naked indices and rumpfs
in a given term. For every rumpf which appears twice in a term we
have to introduce a second graded commuting object (call it $o'$),
because sticking to only one object would lead to $o^{c}o^{c}=0\textrm{ for }\abs{c}=1$.
Instead of giving a general definition, let us give two examples:\index{$gs$@$\textrm{gs}(\ldots)$}\begin{eqnarray}
\hspace{-.5cm}\textrm{gs}(c^{M}c^{N}T^{KL}x^{P}) & \equiv & o^{c}o^{M}o'^{c}o^{N}o^{T}o^{K}o^{L}o^{x}o^{P}\\
\textrm{gs}(x^{K}A^{MPN}c^{L}) & \equiv & o^{x}o^{K}o^{A}o^{M}o^{P}o^{N}o^{c}o^{L}\end{eqnarray}
In the grading structure, we can now rearrange the objects until all
the rumpfs are in the front (with unchanged relative position) and
the naked indices have some common order. E.g.\begin{eqnarray}
\hspace{-.5cm}\textrm{gs}(c^{M}c^{N}T^{KL}x^{P}) & = & (-)^{cM+T(M+N)+x(M+N+K+L)}o^{c}o'^{c}o^{T}o^{x}\cdot o^{M}o^{N}o^{K}o^{L}o^{P}\qquad\qquad\\
\textrm{gs}(x^{K}A^{MPN}c^{L}) & = & (-)^{AK+c(K+M+P+N)}o^{x}o^{A}o^{c}\cdot o^{K}o^{M}o^{P}o^{N}o^{L}=\\
 & = & (-)^{AK+c(K+M+P+N)}(-)^{MK+N(K+P)+LP}o^{x}o^{A}o^{c}\cdot o^{M}o^{N}o^{K}o^{L}o^{P}\end{eqnarray}
We call the resulting sign the \textbf{relative\index{relative sign of grading structures}
sign\index{sign!relative $\sim$ of grading structures} of the grading\index{grading structure!relative sign of $\sim$'s}
structures}\index{$sign$@$\textrm{sign}^g_{(\ldots)}(\ldots)$} \begin{equation}
\hspace{-.3cm}\,\textrm{sign}_{c^{M}c^{N}T^{KL}x^{P}}^{g}\left(x^{K}A^{MPN}c^{L}\right)=(-)^{cM+T(M+N)+x(M+N+K+L)}(-)^{AK+c(K+M+P+N)}(-)^{MK+N(K+P)+LP}\qquad\end{equation}
This definition of the relative sign reduces to (\ref{eq:relativeSignBos})
in the case of bosonic rumpfs. In order to write down the general
definition for the graded equal sign, let us replace the terms of
an equation (like $c^{M}c^{N}T^{KL}x^{P}$ and $x^{K}A^{MPN}c^{L}$
above) by placeholders $T_{(i)}$ (where $i$ just labels the different
terms). In the same way as for the bosonic rumpfs in (\ref{eq:greqBosonicRumpf})
we can finally give the definition for the graded equal sign in the
general case:\begin{Def}[graded equal sign '$\greq$'] \begin{equation}
\boxed{\sum_{i}T_{(i)}=_{g}0\quad:\iff\sum_{i}\textrm{sign}_{T_{(1)}}^{g}(T_{(i)})\cdot T_{(i)}=0}\label{eq:greqGradedRumpf}\end{equation}
Sometimes we call '$\,\greq$' also the {}``small\index{small graded equal sign}
graded equal sign''. \end{Def}In our example of above, this reads\begin{equation}
c^{M}c^{N}T^{KL}x^{P}-x^{K}A^{MPN}c^{L}=_{g}0\quad:\iff c^{M}c^{N}T^{KL}x^{P}-\textrm{sign}_{c^{M}c^{N}T^{KL}x^{P}}\left(x^{K}A^{MPN}c^{L}\right)\cdot x^{K}A^{MPN}c^{L}=0\end{equation}

\begin{prop}[Equivalence relation]\index{proposition!the graded equal sign is an equivalence relation}The
such defined graded equal sign obeys transitivity ($X\greq Y$, $Y\greq Z\dann X\greq Z$)
as well as reflexivity ($X\greq X$) and symmetry ($X\greq Y\dann Y\greq X)$
and is therefore an equivalence relation.\end{prop}

\noindent \emph{Proof: }Reflexivity: If the expression $X$ is a
sum of terms $T_{(i)}$, i.e. $X=\sum_{i}T_{(i)}$ then the claim
that $\sum_{i}T_{(i)}\greq\sum_{i}T_{(i)}$ is equivalent to $\sum_{i}\mbox{sign}_{T_{(1)}}^{g}(T_{(i)})\cdot T_{(i)}=\sum_{i}\mbox{sign}_{T_{(1)}}^{g}(T_{(i)})\cdot T_{(i)}$
which is obviously true. The symmetry is induced by the fact that
$\mbox{sign}_{T_{(i)}}T_{(j)}=\mbox{sign}_{T_{(j)}}T_{(i)}$. Transitivity
finally is seen as follows: Assume that we have $\sum_{i}T_{(i)}\greq\sum_{i}\tilde{T}_{(i)}$
(equivalent to $\sum_{i}\mbox{sign}_{T_{(1)}}^{g}(T_{(i)})\cdot T_{(i)}=\sum_{i}\mbox{sign}_{T_{(1)}}^{g}(\tilde{T}_{(i)})\cdot\tilde{T}_{(i)}$)
and $\sum_{i}\tilde{T}_{(i)}\greq\sum_{i}\tilde{\tilde{T}}_{(i)}$
(equivalent to $\sum_{i}\mbox{sign}_{\tilde{T}_{(1)}}^{g}(\tilde{T}_{(i)})\tilde{T}_{(i)}=\sum_{i}\mbox{sign}_{\tilde{T}_{(1)}}^{g}(\tilde{\tilde{T}}_{(i)})\tilde{\tilde{T}}_{(i)}$).
Then it follows (using transitivity of the ordinary equal sign) that
$\sum_{i}\mbox{sign}_{T_{(1)}}^{g}(T_{(i)})\cdot T_{(i)}=\mbox{sign}_{T_{(1)}}^{g}(\tilde{T}_{(1)})\sum_{i}\mbox{sign}_{\tilde{T}_{(1)}}^{g}(\tilde{\tilde{T}}_{(i)})\tilde{\tilde{T}}_{(i)}=\sum_{i}\mbox{sign}_{T_{(1)}}^{g}(\tilde{\tilde{T}}_{(i)})\tilde{\tilde{T}}_{(i)}$
which is in turn equivalent to $\sum_{i}T_{(i)}\greq\sum_{(i)}\tilde{\tilde{T}}_{(i)}.\qquad\square$

\paragraph{Remark:}

In part \ref{par:PureSpinorString}, beginning with chapter \vref{cha:curved-background},
we will throughout use the graded summation convention (based on NW)
and the graded equal sign $\greq$. The latter will then simply be
denoted with an ordinary equal sign $=$, in order to keep the notations
simple. \vspace{.3cm} 

Next we go one step further and define a \textbf{big\index{big graded equal sign}
graded\index{graded equal sign!big $\sim$} equal sign \index{$=_G$|itext{big graded equal sign}}$=_{G}$}
which also takes care of the order of as many rumpfs as possible.
 Let us give some simple examples\index{transposed!graded $\sim$}\index{complex conjugation!graded $\sim$}\index{hermitean conjugation!graded $\sim$}\index{graded!transposed}\index{graded!hermitean conjugation}\index{graded!complex conjugation}:\begin{eqnarray}
(AB)^{T}=_{G}B^{T}A^{T} & :\iff & (AB)^{T}=(-)^{AB}B^{T}A^{T}\\
(AB)^{\dagger}=_{G}B^{\dagger}A^{\dagger} & :\iff & (AB)^{\dagger}=(-)^{AB}B^{\dagger}A^{\dagger}\\
(ab)^{*}=_{G}a^{*}b^{*} & :\iff & (ab)^{*}=a^{*}b^{*}\\
D_{M}(AB)\Greq(D_{M}A)B+A(D_{M}B) & :\iff & D_{M}(AB)=(D_{M}A)B+(-)^{(D+M)A}A(D_{M}B)\label{eq:Beq-ex-IV}\end{eqnarray}
The above examples are well-designed. Every rumpf or naked index appears
in every term exactly once and a comparison of the order in each term
is possible. 

\begin{itemize}
\item In more general situations, the \textbf{big graded equal sign} $\Greq$
will be defined by first adding the signs corresponding to the use
of the small graded equal sign $\greq$ and then taking care of a
maximum of common (to all terms) and distinguishable (among themselves)
rumpf-symbols. For all remaining rumpf-symbols, a sign will be included
that assumes that their standard position is to the very left (not
changing their relative order).
\end{itemize}
Writing down a more formal definition of this idea in general would
probably be lengthy and not very illuminating, so let us again consider
some examples (which are not necessarily meaningful in real calculations):\begin{eqnarray}
ABAC\Greq\: CB & :\iff & (-)^{BA}ABAC=(-)^{CB}CB\end{eqnarray}
The maximum set of symbols common to each term is $\{B,C\}$. Their
relative order is different in the two terms, so that we get the factor
$(-)^{CB}$, while the factor $(-)^{BA}$ is the sign that compares
to the structure where all $A$'s (which do not belong to the common
set) are to the very left. Another example (with explanation right
afterwards):\begin{eqnarray}
0\Greq\quad A_{M}B_{N}A_{K}C_{L}\lqn{+B_{M}A_{N}A_{L}C_{K}+A_{N}B_{M}C_{K}A_{L}:\iff}\nonumber \\
0 & = & (-)^{BM+A(M+N)+C(M+N+K)}(-)^{BA}A_{M}B_{N}A_{K}C_{L}+\nonumber \\
 &  & +(-)^{AM+A(M+N)+C(M+N+L)}(-)^{LK}B_{M}A_{N}A_{L}C_{K}+\nonumber \\
 &  & +(-)^{BN+C(N+M)+A(N+M+K)}(-)^{NM}(-)^{(B+C)A}A_{N}B_{M}C_{K}A_{L}\end{eqnarray}
In a first step we have applied the small graded equal sign, which
includes moving all rumpf-gradings to the very left without changing
their relative order. This leads to the sign $(-)^{BM+A(M+N)+C(M+N+K)}$
for the first, $(-)^{AM+A(M+N)+C(M+N+L)}$ for the second and $(-)^{BN+C(N+M)+A(N+M+K)}$
for the third term. The small graded equal sign also takes care of
the relative order of the naked indices in all terms. If we take the
first term as reference term, this yields the factors $(-)^{LK}$
for the second and $(-)^{NM}$ for the third term. The additional
contribution from the big graded equal sign is obtained as follows:
This time the set of all rumpf-symbols $\{B,C,A\}$ is common to all
terms, but $A$ appears in two indistinguishable copies. The maximum
set of common (to all terms) and distinguishable (among themselves)
rumpf-symbols is thus again $\{B,C\}$. The gradings of the remaining
$A$'s are put to the very left, which yields a factor $(-)^{BA}$
for the first term, $(-)^{B(A+A)}=1$ for the second and $(-)^{(B+C)A}$
for the third term. Finally the relative order of $B$ and $C$ in
each term is compared which gives no extra factor in this example.

Note that the naked index in (\ref{eq:Beq-ex-IV}) was treated on
equal footing with the rumpfs. The big graded equal sign simply compared
the relative order of all involved symbols, no matter if they were
rumpf or naked index. In this case, where all rumpfs appear in each
term exactly once, this is equivalent to applying our more general
definition (given below (\ref{eq:Beq-ex-IV})), where we first apply
the small graded equal sign, which moves all the rumpf-gradings to
the very left. Indeed the example (\ref{eq:Beq-ex-IV}) can equivalently
be written as\begin{equation}
D_{M}(AB)\Greq(D_{M}A)B+A(D_{M}B):\iff(-)^{(A+B)M}D_{M}(AB)=(-)^{(A+B)M}(D_{M}A)B+(-)^{BM}(-)^{DA}A(D_{M}B)\end{equation}

There is a serious drawback of the so far given definition of the
big graded equal sign: it does not in general obey transitivity. We
will below modify the definition such that transitivity is guaranteed,
but let us first give examples where it is violated. If one defines
composite objects, like $A\Grequiv ba$, using the big graded equal
sign, it does not have any effect. The maximum set of symbols common
to all terms is empty. The symbol '$A$' on the lefthand side doesn't
appear in the term on the righthand side, and the symbols $'a'$ and
$'b'$ do not appear in the term on the lefthand side. The same reasoning
holds for $B\Grequiv ab$:\begin{equation}
A\Grequiv ba\iff A\equiv ba,\qquad B\Grequiv ab\iff B\equiv ab\end{equation}
Assume that we have $A\Greq B$ (which is equivalent to $A=B$, i.e.
$ba=ab$). Transitivity would then imply that $ba\Greq ab$ which
is equivalent to $ba=(-)^{ba}ab$ and does in general not agree with
the starting point $A=B$. A way out is to define the big graded equal
sign not for a single equation, but for the whole system of equations
under consideration. \begin{Def}[big graded equal sign '$\Greq$']

Given a system of equations, we first determine for each equation
$i$ the set $\mathbb{M}_{i}$ of rumpf-symbols which appear either
exactly once in each term or not at all in the given equation. Call
the intersection of these sets $\mathbb{M}\equiv\cap_{i}\,\mathbb{M}_{i}$.
The \textbf{big graded equal sign} '$\Greq$' in a system of equations
is now defined by first applying the sign rules corresponding to the
small graded equal sign '$\greq$' and then adding a sign that compares
the relative order of all rumpf-symbols which are in the set $\mathbb{M}$.
For all remaining rumpf-symbols, a sign will be included that assumes
that their standard position is to the very left (not changing their
relative order).

\end{Def}In the previous example this works as follows: The equations
under consideration are $A\Greq ba$, $B\Greq ab$ and $ab\Greq ba$.
The symbol '$A$' in the first equation appears once in the term on
the lefthand side, but not at all in the term on the righthand side.
It is thus not in the set $\mathbb{M}_{1}$. The same is true for
the rumpf symbols '$B$' in the second equation and for '$a$' and
'$b$' in the first and second equation. We thus have $\mathbb{M}_{1}=\{\}$,
$\mathbb{M}_{2}=\{\}$. Only for the last equation the rumpf symbols
'$a$' and '$b$' appear exactly once in each term so that $\mathbb{M}_{3}=\{a,b\}$.
The intersection, however, is still empty $\mathbb{M}=\mathbb{M}_{1}\cap\mathbb{M}_{2}\cap\mathbb{M}_{3}=\{\}$.
The big graded equal sign compares only the relative order of the
symbols in $\mathbb{M}$. In this case it therefore reduces to an
ordinary equal sign and transitivity is trivially preserved.

\begin{prop}[Transitivity]\index{proposition!transitivity of the big graded equal sign}In
addition to symmetry and reflexivity, the above defined big graded
equal sign $\Greq$ obeys transitivity within the given set of equations
that was used for its definition and is therefore an equivalence relation
within this set.\end{prop}

\noindent \emph{Proof: }Under the conditions of the definition (all
rumpf symbols appear for any given equation either exactly once in
each term or not at all in this equation) one can replace every rumpf
by a bosonic rumpf with an auxiliary naked index which carries the
grading. The big graded equal sign then reduces to the small graded
equal sign whose transitivity we have seen already.$\quad\square$

\section{Calculating with fermions as with bosons - a theorem}

Now we are equipped with the main tools that are necessary to turn
bosonic structural equations into graded structural equations. The
set $\mathbb{M}$ in the definition of the big graded equal sign contains
all symbols whose relative positions in a system of equations can
be uniquely determined. This is precisely the property that allows
to assign a grading to such a symbol and therefore deserves its own
definition.

\begin{Def}[Gradifiable] \label{Def:gradifiable}We call a naked
index or rumpf of an algebra element gradifiable\index{gradifiable}
in a given equation iff it either appears in every term of this equation
exactly once or it does not appear in the equation at all. We call
it gradifiable in a system of equations iff it is gradifiable in each
of them. In addition, every dummy index (one which appears in a single
term twice, once in upper and once in lower position) is also called
a gradifiable index.\end{Def}

\paragraph{Example}

In the equation $a^{M}b_{N}=b_{N}a^{M}$ all indices $\{M,N\}$ and
all rumpfs $\{a,b\}$ are gradifiable, because they appear in every
term exactly once. However in the set of equations $a^{M}b_{N}=b_{N}a^{M},\quad A^{M}\tief{N}=a^{M}b_{N}$
only the indices $\{M,N\}$ are gradifiable, while the rumpfs $\{A,a,b\}$
are not gradifiable any longer, as they all appear in the second equation,
but not exactly once in each term. The same set of equations, with
the second one written as $A(a,b)^{M}\tief{N}=a^{M}b_{N}$, however,
has gradifiable rumpf-symbols $a$ and $b$. The notion 'gradifiable'
therefore depends on the way how objects are denoted.

\begin{Def}[Gradification]\label{Def:gradification}The gradification\index{gradification}
of an index '$K$' or rumpf '$a$' assigns an undetermined parity
$\abs{K}$ or $\abs{a}$ to it, which will enter the graded summation
convention and the graded equal sign. The gradification of a given
set of algebraic equations is defined to be a new set of equations
with all gradifiable objects gradified, the equal sign replaced by
the big graded equal sign and the sum over dummy indices replaced
by the graded sum (using an arbitrary but well-defined sign rule like
NW or NE) over graded dummy indices. \end{Def}

\noindent More or less by definition, the following theorem holds:

\begin{thm}\label{thm:gradification}\index{theorem!gradification}If
a set of algebraic equations implies (perhaps via some intermediate
equations) a second set of algebraic equations, then the same holds
true for the gradification of the whole system. \end{thm}

\noindent \emph{Remark:}\textbf{ }According to the definition of
'gradifiable in a system of equations' only those indices and rumpfs
which are gradifiable in each equation (even the intermediate ones)
are gradifiable in the whole system.\vspace{.3cm} 

\noindent \emph{Comment on the proof: }All definitions were chosen
precisely with having in mind that the theorem should hold. Therefore
it seems that there is nothing to prove and the theorem just holds
by definition. Nevertheless, any attempts of mine to make this statement
more rigorous, failed so far. One might therefore insist on calling
the above theorem a 'conjecture' only. Calling it a conjecture, however,
would somehow implement that the proof is difficult. But as argued
above, I suspect that it is rather a triviality as soon as an appropriate
setting is used. A naive idea for a proof would be that the gradification
provides an isomorphism from one algebra to another. However, the
gradification map is not in general invertible. For example a commutative
but otherwise freely generated algebra is mapped to a graded commutative
(and otherwise freely generated) one. For odd generators, the square
is zero and therefore the gradification has less basis elements than
the original algebra, if the number of generators is the same. What
is mapped one to one is therefore not the algebra itself, but a certain
(sub)set of equations which characterize the algebra, namely the gradifiable
ones. ~$\square$

\paragraph*{Further remarks:}

\begin{itemize}
\item The example given after the definition of 'gradifiable' demonstrates
that the power of the theorem depends on how the original equations
are written. If one introduces auxiliary variables for composite objects
(like $A^{M}\tief{N}\equiv a^{M}b_{N}$), the number of gradifiable
objects may reduce, if the elementary variables are not denoted as
an argument (like in $A(a,b)^{M}\tief{N}$). The theorem gives no
statement about the best notation to use. It rather gives a statement
which holds for any notation, but the notation has an influence on
the number of gradifiable objects. Sometimes rumpf-symbols can be
turned gradifiable by a change of notation but sometimes this seems
impossible. It would be useful to characterize the 'best notation'
which makes as many symbols as possible gradifiable.
\item This theorem provides the possibility to use existing bosonic tensor
manipulation packages for Mathematica or other computer algebra systems
also for the graded case!
\item It is not excluded a priori that the original set of equations contains
fermionic variables which are then made bosonic (or are assigned an
undetermined grading). However, one has to make sure that equations
like \begin{eqnarray}
\tet\cdot\tet & = & 0\end{eqnarray}
are not contained in the set of equations that were needed to derive
something. In the above equation, $\tet$ obviously appears twice
in one term and is thus not gradifiable. This is also the reason why
anticommuting variables cannot be replaced completely by commuting
ones. In particular the sum of two nilpotent objects is not necessarily
nilpotent any longer in the commuting case. A recent paper \cite{Frydryszak:2006wk}
studies the properties of nilpotent\index{nilpotent commuting variables}
commuting\index{commuting nilpotent variables} variables where some
further differences (e.g. in the Leibniz rule) appear w.r.t. the anticommuting
case. 
\end{itemize}

\subsubsection*{Counterexample\index{gradification!counterexample}\index{counterexample!to the gradification theorem}s}

In the rest of this part of the thesis we will give a lot of examples
and applications of the theorem. There will, however, also be some
rather subtle examples which seem to be counterexamples at first sight.
One of those {}``counterexamples'' is the graded inverse of a matrix
with graded rumpf, treated in subsection \vref{sub:Graded-inverse}.
Another {}``counterexample'' is the derivative with respect to Grassmann
variables: the bosonic equation \begin{equation}
\partiell{}{x}x=1\end{equation}
suggests to define\index{derivative!for fermionic variables}\begin{equation}
\partiell{}{\theta}\theta\stackrel{?}{=}1\end{equation}
for fermionic variables. This definition makes perfect sense, but
results using this derivative cannot be derived via the theorem from
the bosonic case, as the rumpf theta does not appear excatly once
in every term. This problem can be omitted, if one introduces a new
index and puts the grading into the index. We discuss such derivatives
in subsection \vref{sub:Graded-rumpf-derivative}. 

Finally, a quite disturbing counterexample, which demonstrates that
intermediate equations have to be taken into account in the process
of gradification, is discussed on page \pageref{par:subtle-counterex}.

\chapter{Graded matrices (supermatrices) and graded matrix operations}

\index{graded matrix|see{supermatrix}}\index{supermatrix}Supermatrices
are the perfect objects to study the effects of our considerations.
We will drop the word 'super' or 'graded' in every definition, since
everything in this part has to be understood as graded. The equations
of this section will all be written in two ways: once in the left
column with the help of the (small) graded equal sign and the implicit
graded summation conventions and once on the righthand side with ordinary
equal sign, and the sum written out explicitely (in NW conventions),
in order to make the reader familiar with the new conventions.

Within this chapter, we will always consider four different kinds
of matrices, which differ in their index-positions:\index{$A^{MN},B^{M}\tief{N},C_{M}\hoch{N},D_{MN}$|itext{supermatrices}}\index{matrix!of type $A$,$B$,$C$ and $D$}\index{type $A$,$B$,$C$ and $D$ matrices}\begin{equation}
A^{MN},B^{M}\tief{N},C_{M}\hoch{N},D_{MN}\end{equation}

\paragraph{Remark:}

In case that we have several matrices of one type, e.g. type $B$,
we will denote them by $B_{1}$, $B_{2}$,$\ldots$ . It is important
to have in mind that we consider $B_{1}$ as a rumpf by itself and
not as a rumpf $B$ together with an index '1'.

\section{Transpose and hermitean conjugate}

Let us start with the definition of a transposed\index{transposed matrix}
matrix and a hermitean\index{hermitean conjugate matrix} conjugate
matrix in each of the four cases. The simple rule is to take the bosonic
definition and replace the equal sign by the big graded one (which
reduces to the small graded one in the below cases):\\
\begin{tabular}{cc}
\begin{minipage}[c][1\totalheight][t]{0.47\columnwidth}%
\begin{eqnarray*}
(A^{T})^{MN} & \grequiv & A^{NM}\\
(B^{T})_{M}\hoch{N} & \grequiv & B^{N}\tief{M}\\
(C^{T})^{M}\tief{N} & \grequiv & C_{N}\hoch{M}\\
(D^{T})_{MN} & \grequiv & D_{NM}\end{eqnarray*}
\end{minipage}%
 & %
\begin{minipage}[c][1\totalheight][t]{0.5\columnwidth}%
\begin{eqnarray}
(A^{T})^{MN} & \equiv & (-)^{MN}A^{NM}\\
(B^{T})_{M}\hoch{N} & \equiv & (-)^{MN}B^{N}\tief{M}\\
(C^{T})^{M}\tief{N} & \equiv & (-)^{MN}C_{N}\hoch{M}\\
(D^{T})_{MN} & \equiv & (-)^{MN}D_{NM}\end{eqnarray}
\end{minipage}%
\tabularnewline
\end{tabular}\\
\begin{tabular}{cc}
\begin{minipage}[c][1\totalheight][t]{0.47\columnwidth}%
\begin{eqnarray*}
(A^{\dag})^{MN} & \grequiv & (A^{NM})^{*}\\
(B^{\dag})_{M}\hoch{N} & \grequiv & (B^{N}\tief{M})^{*}\\
(C^{\dag})^{M}\tief{N} & \grequiv & (C_{N}\hoch{M})^{*}\\
(D^{\dag})_{MN} & \grequiv & (D_{NM})^{*}\end{eqnarray*}
\end{minipage}%
 & %
\begin{minipage}[c][1\totalheight][t]{0.5\columnwidth}%
\begin{eqnarray}
(A^{\dag})^{MN} & \equiv & (-)^{MN}(A^{NM})^{*}\\
(B^{\dag})_{M}\hoch{N} & \equiv & (-)^{MN}(B^{N}\tief{M})^{*}\\
(C^{\dag})^{M}\tief{N} & \equiv & (-)^{MN}(C_{N}\hoch{M})^{*}\\
(D^{\dag})_{MN} & \equiv & (-)^{MN}(D_{NM})^{*}\end{eqnarray}
\end{minipage}%
\tabularnewline
\end{tabular}\\
Clearly we have\begin{eqnarray}
(M^{T})^{T} & = & M\label{eq:doppeltranspose}\\
(M^{\dag})^{\dag} & = & M\label{eq:doppelhermit}\end{eqnarray}
for all matrices $M$, which is a first simple confirmation of the
theorem.

\section{Matrix multiplication}

We meet a first deviation from usual definitions when we consider
matrix\index{matrix multiplication!graded $\sim$} multiplications.%
\footnote{\index{footnote!\thefoot. matrix multiplication in B. DeWitt}Although
they seem to agree with the definitions in \cite{DeWitt:1992cy},
when one moves there all indices which are to the left of a rumpf
to the right with the corresponding sign according to that reference.$\qquad\fussend$%
} The definition of the matrix multiplication will depend on the index
structure of the matrix. Both, graded equal sign and the graded summation
convention have an influence now:

\begin{tabular}{cc}
\hspace{-1cm} %
\begin{minipage}[c][1\totalheight][t]{0.4\columnwidth}%
\begin{eqnarray*}
(AC)^{MN} & \grequiv & A^{MK}C_{K}\hoch{N}\\
 &  & \phantom{\sum_{K}}\\
(AD)^{M}\tief{N} & \grequiv & A^{MK}D_{KN}\\
 &  & \phantom{\sum_{K}}\\
(AB^{T})^{MN} & \grequiv & A^{MK}(B^{T})_{K}\hoch{N}\\
 & = & A^{MK}B^{N}\tief{K}\\
 &  & \phantom{\sum_{K}}\\
(BA)^{MN} & \grequiv & B^{M}\tief{K}A^{KN}\\
 &  & \phantom{\sum_{K}}\\
(B_{1}B_{2})^{M}\tief{N} & \grequiv & B_{1}^{\,\, M}\tief{K}B_{2}^{\,\, K}\tief{N}\\
 &  & \phantom{\sum_{K}}\\
 & \ddots\end{eqnarray*}
\end{minipage}%
 & %
\begin{minipage}[c][1\totalheight][t]{0.5\columnwidth}%
\begin{eqnarray}
(AC)^{MN} & \equiv & (-)^{MC}A^{MK}C_{K}\hoch{N}=\nonumber \\
 & \stackrel{NW}{=} & (-)^{MC}\sum_{K}(-)^{KC}A^{MK}C_{K}\hoch{N}\\
(AD)^{M}\tief{N} & \equiv & (-)^{MD}A^{MK}D_{KN}=\nonumber \\
 & \stackrel{NW}{=} & (-)^{MD}\sum_{K}(-)^{KD}A^{MK}D_{KN}\\
(AB^{T})^{MN} & \equiv & (-)^{MB}A^{MK}(B^{T})_{K}\hoch{N}=\nonumber \\
 & = & (-)^{MB}A^{MK}B^{N}\tief{K}=\nonumber \\
 & \stackrel{NW}{=} & (-)^{MB}\sum_{K}(-)^{K(B+N)}A^{MK}B^{N}\tief{K}\quad\\
(BA)^{MN} & \equiv & (-)^{MA}B^{M}\tief{K}A^{KN}=\nonumber \\
 & \stackrel{NW}{=} & (-)^{MA}\sum_{K}(-)^{K+KA}B^{M}\tief{K}A^{KN}\\
(B_{1}B_{2})^{M}\tief{N} & \equiv & (-)^{MB_{2}}B_{1}^{\,\, M}\tief{K}B_{2}^{\,\, K}\tief{N}=\nonumber \\
 & = & (-)^{MB_{2}}\sum_{K}(-)^{K+KB_{2}}B_{1}^{\,\, M}\tief{K}B_{2}^{\,\, K}\tief{N}\quad\quad\qquad\\
 & \ddots\nonumber \end{eqnarray}
\end{minipage}%
\tabularnewline
\end{tabular}

\subsection*{Associativity}

Up to now, we have used the graded equality and summation mainly for
definitions (apart from (\ref{eq:doppeltranspose}) and (\ref{eq:doppelhermit})).
Now we can apply our theorem by stating that the (graded) matrix multiplication
as defined above is associative\index{associativity!of graded matrix multiplication}\begin{eqnarray}
\left((B_{1}B_{2})B_{3}\right)^{M}\tief{N} & = & B_{1}(B_{2}B_{3})^{M}\tief{N}\\
\left((C_{1}C_{2})C_{3}\right)_{M}\hoch{N} & = & C_{1}(C_{2}C_{3})_{M}\hoch{N}\end{eqnarray}
The graded equal sign has no effect in these equation. Associativity
is guaranteed by theorem \ref{thm:gradification}. The full reasoning
in the $B$-case would be the following:

In the bosonic case we have\begin{equation}
(B_{1}B_{2})^{M}\tief{N}\equiv B_{1}^{\,\, M}\tief{K}B_{2}^{\,\, K}\tief{N}\dann\left((B_{1}B_{2})B_{3}\right)^{P}\tief{Q}=B_{1}(B_{2}B_{3})^{P}\tief{Q}\end{equation}
The dummy indices are by definition gradifiable. Each of the naked
indices $M$ and $N$ appears in every term of the first equation
exactly once and not at all in the second and is therefore gradifiable.
One could have written the second equation also with the same indices
$M$ and $N$ and they still would be gradifiable. The same reasoning
holds for $P$ and $Q$. Finally, $B_{1}$ and $B_{2}$ each appear
in every term of the first as well as of the second equation exactly
once, while $B_{3}$ does not appear in the first at all, but it appears
in the second in every term exactly once. All the rumpfs $B_{1},$$B_{2}$
and $B_{3}$ are thus gradifiable in this system of two equations.
The gradification of the whole system then reads \begin{equation}
(B_{1}B_{2})^{M}\tief{N}\Grequiv B_{1}^{\,\, M}\tief{K}B_{2}^{\,\, K}\tief{N}\dann\left((B_{1}B_{2})B_{3}\right)^{P}\tief{Q}\Greq B_{1}(B_{2}B_{3})^{P}\tief{Q}\end{equation}
where $B_{1}$, $B_{2}$ , $B_{3}$, $M$, $N$ , $P$ and $Q$ have
been assigned an undetermined grading, the sum over dummy indicies
now has to be understood as the graded sum and the equal signs were
replaced by the big graded equal sign (which reduces to the small
graded equal sign in the first equation and to the ordinary equal
sign in the second).

For this example it is still quite simple to check the validity of
the statement explicitly, e.g. in NW\begin{eqnarray}
\lefteqn{(-)^{MB_{3}}\sum_{L}(-)^{LB_{3}+L}\left((-)^{MB_{2}}\sum_{K}(-)^{KB_{2}+K}B_{1}\hoch{M}\tief{K}B_{2}\hoch{K}\tief{L}\right)B_{3}\hoch{L}\tief{N}=}\nonumber \\
 & = & (-)^{M(B_{2}+B_{3})}\sum_{K}(-)^{K(B_{2}+B_{3})+K}B_{1}\hoch{M}\tief{K}\left((-)^{KB_{3}}\sum_{L}(-)^{LB_{3}+L}B_{2}\hoch{K}\tief{L}B_{3}\hoch{L}\tief{N}\right)\end{eqnarray}
\rem{

\section{graded (anti)commutator }

\begin{tabular}{cc}
\begin{minipage}[c][1\totalheight][t]{0.47\columnwidth}%
\begin{eqnarray}
[A,B] & \equiv_{G} & AB-BA\end{eqnarray}
\end{minipage}%
 & %
\begin{minipage}[c][1\totalheight][t]{0.5\columnwidth}%
\begin{eqnarray}
[A,B] & \equiv & AB-(-)^{AB}BA\end{eqnarray}
\end{minipage}%
\tabularnewline
\end{tabular}}

\subsection*{Unit matrix}

The definition of the unit matrix is\index{unit matrix!graded $\sim$}\index{$1$@$\one$}\begin{eqnarray}
M\one & = & M\quad\forall M\end{eqnarray}
which implies via associativity (for the matrices of type $B$ and
$C$) that $M(\one N)=(M\one)N=MN\:\forall M,N$ and thus \begin{eqnarray}
\one N & = & N\quad\forall N\end{eqnarray}
For the different types of matricies $A,B,C$ and $D$, we have in
fact different types of unit matrices:\\
\begin{tabular}{cc}
\hspace{-1cm} %
\begin{minipage}[c][1\totalheight][t]{0.47\columnwidth}%
\begin{eqnarray*}
(A\one)^{MN}\equiv\phantom{\sum_{K}}A^{MK}\delta_{K}\hoch{N} & \stackrel{!}{=} & A^{MN}\\
(B\one)^{M}\tief{N}\equiv\phantom{\sum_{K}}B^{M}\tief{K}\delta^{K}\tief{N} & \stackrel{!}{=} & B^{M}\tief{N}\\
(C\one)_{M}\hoch{N}\equiv\phantom{\sum_{K}}C_{M}\hoch{K}\delta_{K}\hoch{N} & \stackrel{!}{=} & C_{M}\hoch{N}\\
(D\one)_{MN}\equiv\phantom{\sum_{K}}D_{MK}\delta^{K}\tief{N} & \stackrel{!}{=} & D_{MN}\end{eqnarray*}
\end{minipage}%
 & %
\begin{minipage}[c][1\totalheight][t]{0.54\columnwidth}%
\begin{eqnarray}
(A\one)^{MN}\stackrel{NW}{\equiv}\sum_{K}A^{MK}\delta_{K}\hoch{N} & \stackrel{!}{=} & A^{MN}\\
(B\one)^{M}\tief{N}\stackrel{NW}{\equiv}\sum_{K}(-)^{K}B^{M}\tief{K}\delta^{K}\tief{N} & \stackrel{!}{=} & B^{M}\tief{N}\\
(C\one)_{M}\hoch{N}\stackrel{NW}{\equiv}\sum_{K}C_{M}\hoch{K}\delta_{K}\hoch{N} & \stackrel{!}{=} & C_{M}\hoch{N}\qquad\\
(D\one)_{MN}\stackrel{NW}{\equiv}\sum_{K}(-)^{K}D_{MK}\delta^{K}\tief{N} & \stackrel{!}{=} & D_{MN}\qquad\end{eqnarray}
\end{minipage}%
\tabularnewline
\end{tabular} \\
From the righthand side we can see\index{$\delta^M\tief{N}$|itext{graded Kronecker}}\index{$\delta_M\hoch{N}$|itext{graded Kronecker}}\index{Kronecker delta!graded $\sim$}\index{graded!Kronecker delta}\begin{eqnarray}
\delta_{M}\hoch{N} & = & \left\{ \begin{array}{c}
\delta_{M}^{N}\textrm{ for NW}\\
(-)^{MN}\delta_{M}^{N}\textrm{ for NE}\end{array}\right.\end{eqnarray}
with \index{$\delta_M^N$|itext{numerical Kronecker delta}}$\delta_{M}^{N}$
being the numerical Kronecker delta, and\vspace{-.2cm}\\
\begin{tabular}{cc}
\begin{minipage}[c][1\totalheight][t]{0.47\columnwidth}%
\begin{eqnarray*}
\delta^{M}\tief{N} & \greq & \delta_{N}\hoch{M}\end{eqnarray*}
\end{minipage}%
 & %
\begin{minipage}[c][1\totalheight][t]{0.5\columnwidth}%
\begin{eqnarray}
\delta^{M}\tief{N} & = & (-)^{MN}\delta_{N}\hoch{M}\end{eqnarray}
\end{minipage}%
\tabularnewline
\end{tabular} \vspace{.2cm}\\
This graded Kronecker (the lefthand side shows that both versions
are graded equal anyway) of course also fullfils its task for vectors
and arbitrary rank tensors:%
\footnote{\index{Kronecker delta!for mixed conventions}\index{footnote!\thefoot. Kronecker for mixed conventions}If
the capital index combines two subsets of (small) indices with different
position, we might insist on NW (or any other convention) for the
small indices which leads to different definitions for the Kronecker
delta:\begin{eqnarray*}
a^{M} & = & (a^{m},a_{\mu})\\
a^{M}\delta_{M}\hoch{N} & = & a^{m}\delta_{m}\hoch{N}+a_{\mu}\delta^{\mu N}=\\
 & \stackrel{\textrm{mixed conv.}}{\equiv} & \sum_{m}a^{m}\delta_{m}\hoch{N}+\sum_{\mu}(-)^{\mu}a_{\mu}\delta^{\mu N}\stackrel{!}{=}a^{N}\\
\delta_{m}\hoch{N} & = & \delta_{m}^{N}\\
\delta^{\mu N} & = & (-)^{\mu}\delta^{\underset{\mu}{N}}\qquad\fussend\end{eqnarray*}
} \begin{eqnarray}
a^{M}\delta_{M}\hoch{N} & = & a^{N}\\
T_{M_{1}\ldots M_{r-1}K}\delta^{K}\tief{N} & = & T_{M_{1}\ldots M_{r-1}N}\end{eqnarray}

\section{Conjugations of matrix products -- hermitean scalar product}

Other simple applications of theorem \ref{thm:gradification} are
statements about the transpose\index{transpose!of matrix products}
and the hermitean\index{hermitean conjugate!of matrix products} conjugate
of a matrix product. Both, transposition and hermitean conjugation,
were defined as gradifications of the bosonic versions and thus the
equations for their action on matrix products will simply be the gradification
of the corresponding bosonic equation. We will start with the transposition.
The hermitean conjugation will follow a bit later after the discussion
of complex conjugation and hermitean scalar product.

\subsection{Transpose of matrix products}

The transpose of a matrix product in terms of the big graded equal
sign has the familiar bosonic behaviour.\\
\begin{tabular}{cc}
\begin{minipage}[c][1\totalheight][t]{0.47\columnwidth}%
\begin{eqnarray*}
\left((AC)^{T}\right)^{MN} & =_{G} & \left(C^{T}A^{T}\right)^{MN}\\
\left((AD)^{T}\right)^{M}\tief{N} & =_{G} & (D^{T}A^{T})^{M}\tief{N}\\
\left((BA)^{T}\right)^{MN} & =_{G} & (A^{T}B^{T})^{MN}\\
 & \ddots\end{eqnarray*}
\end{minipage}%
 & %
\begin{minipage}[c][1\totalheight][t]{0.47\columnwidth}%
\begin{eqnarray}
\left((AC)^{T}\right)^{MN} & = & (-)^{AC}\left(C^{T}A^{T}\right)^{MN}\\
\left((AD)^{T}\right)^{M}\tief{N} & = & (-)^{AD}(D^{T}A^{T})^{M}\tief{N}\\
\left((BA)^{T}\right)^{MN} & = & (-)^{AB}(A^{T}B^{T})^{MN}\\
 & \ddots\nonumber \end{eqnarray}
\end{minipage}%
\tabularnewline
\end{tabular}\\
Let us again verify explicitly that this is indeed true for e.g. the
first line (in NW conventions):\begin{eqnarray}
\left((AC)^{T}\right)^{MN} & = & (-)^{MN}(AC)^{NM}=\nonumber \\
 & = & (-)^{MN}(-)^{NC}\sum_{K}(-)^{CK}A^{NK}C_{K}\hoch{M}=\nonumber \\
 & = & (-)^{MN+NC}\sum_{K}(-)^{CK+(C+K+M)(A+N+K)}C_{K}\hoch{M}A^{NK}=\nonumber \\
 & = & \sum_{K}(-)^{CA+KA+KN+K+MA+MK}C_{K}\hoch{M}A^{NK}=\nonumber \\
 & = & (-)^{AC}(-)^{MA}\sum_{K}(-)^{KA+K}(C^{T})^{M}\tief{K}(A^{T})^{KN}=\nonumber \\
 & = & (-)^{AC}(-)^{MA}(C^{T})^{M}\tief{K}(A^{T})^{KN}=\nonumber \\
 & = & (-)^{AC}\left(C^{T}A^{T}\right)^{MN}\end{eqnarray}

\subsection{Complex conjugation of products of (graded) commuting variables}

Before we come to the discussion of hermitean scalar products and
hermitean conjugation of matrix products, we will have a short look
at complex\index{complex conjugation!of graded commuting variables|fett}
conjugation of graded commuting variables (we will often call them
graded numbers, or just numbers) and products of them. The reason
to do so, is that the complex conjugate of a product of two Grassmann
variables is often defined differently to our way, and we therefore
want to motivate it carefully.

Complex conjugation of usual complex numbers is just what it is. For
a (graded commuting) algebra based on a complex vector space one usually
defines some basis to be real, so that the complex conjugation acts
only on the expansion coefficients. Different definitions of the action
on the basis elements are possible and simply a matter of convenience.
However, the definition of the conjugation of the basis vectors should
at least obey the conjugation property $(\:)^{**}=(\:)$. For an algebra
whose vector-basis is generated by some generating set, the reality
properties of the composite objects are determined by the reality
properties of the generating set and the action of the complex conjugation
on the product of elements. It is natural to define $(ab)^{*}=a^{*}b^{*}$,
but using the opposite sign $(ab)^{*}=-a^{*}b^{*}$ for vectors $a,b$
would also be consistent. Indeed, in the case of an anticommuting
algebra this definition is very common because it can then be written
as $(ab)^{*}=b^{*}a^{*}$ and resembles the bosonic version of hermitean
conjugation where the order of objects is interchanged. Although there
is thus good reason to make this choice, we want to convince the reader
in the following that there is even better reason not to make this
choice. For a graded commuting algebra, where $a$ and $b$ are of
arbitrary grading, the choice \begin{equation}
(ab)^{*}\equiv a^{*}b^{*}\end{equation}
is certainly the one which fits into our philosophy, as it is the
gradification of the usual choice for (bosonic) commuting algebras.
This choice implies that the product of real objects is real again
and the real elements thus form a subalgebra. Indeed the above conjugation
rule can be derived from this reality condition. We could thus go
the other way round and define the complex conjugation simply by saying
that the product of two real products is always real. To derive the
above conjugation rule from that condition, consider the (graded)
commuting variable $a$ and decompose it into its real part \index{$R()$@$\Re(\ldots)$|itext{real part}}$\Re(a)$
and its imaginary part\index{$I()$@$\Im(\ldots)$|itext{imaginary part}}
$\Im(a)$, defined by (use of a graded equal sign makes no difference
here)\begin{eqnarray}
\Re(a) & \equiv & \frac{a+a^{*}}{2}\label{eq:Realteil}\\
\Im(a) & \equiv & \frac{a-a^{*}}{2i}\label{eq:Imaginaerteil}\end{eqnarray}
Both are real because $a**=a$\begin{eqnarray}
\Re(a)^{*} & = & \Re(a),\qquad\Im(a)^{*}=\Im(a)\end{eqnarray}
and we have\begin{eqnarray}
a & = & \Re(a)+i\Im(a)\\
a^{*} & = & \Re(a)-i\Im(a)\end{eqnarray}
We thus can seperate any number into a real and imaginary part, and
complex conjugation flips (as usual) the sign of the imaginary part.
Consider now the complex conjugation of the product of two graded
numbers\begin{eqnarray}
(ab)^{*} & = & \left[(\Re(a)\Re(b)-\Im(a)\Im(b))+i(\Re(a)\Im(b)+\Im(a)\Re(b))\right]^{*}=\nonumber \\
 & = & (\Re(a)\Re(b)-\Im(a)\Im(b))-i(\Re(a)\Im(b)+\Im(a)\Re(b))\\
a^{*}b^{*} & = & \left(\Re(a)-i\Im(b)\right)\left(\Re(a)-i\Im(b)\right)=\nonumber \\
 & = & (\Re(a)\Re(b)-\Im(a)\Im(b))-i(\Re(a)\Im(b)+\Im(a)\Re(b))\\
\dann(ab)^{*} & = & a^{*}b^{*}\label{eq:complexConjugation}\end{eqnarray}
From the first to the second line we have used that the product of
two real variables is real again. From our definitions of real and
imaginary part in (\ref{eq:Realteil}) and (\ref{eq:Imaginaerteil}),
which are just graded versions of the bosonic case, we could have
deduced (\ref{eq:complexConjugation}) as well via our theorem. We
just want to stress that in our context this is the only natural complex
conjugation. In order to allow a comparison with the 'usual' definition%
\footnote{\index{footnote!\thefoot. complex conjugation of Grassmann variables}It
seems that in the last decade, the definition $(ab)^{*}=a^{*}b^{*}$
has already become more popular (see for example \cite{Cartier:2002zp}),
while in \cite{DeWitt:1992cy} it was still defined with the opposite
order. Another discussion of complex conjugation can be found in \cite{Schmitt:1996hp}.$\qquad\fussend$%
}, let us for the moment denote the alternative version of complex
conjugation by $(\ldots)^{\tilde{*}}$.\begin{eqnarray}
(ab)^{\tilde{*}} & = & b^{\tilde{*}}a^{\tilde{*}}=(-)^{ab}a^{\tilde{*}}b^{\tilde{*}}\label{eq:complexConjugation:usual}\end{eqnarray}
As mentioned, this behaviour would not at all fit into our philosophy.
The same is true for the hermitean conjugation of the product of graded
matrices in the next but one subsection (as well as of graded operators
in the infinite dimensional case). How can we easily switch in applications
from one definition to the other? Instead of redefining the complex
conjugation itself, the switch of the behaviour from (\ref{eq:complexConjugation})
to (\ref{eq:complexConjugation:usual}) can also be achieved by redefining
the algebra product appropriately: \begin{eqnarray}
a\circ b & \equiv & i^{\epsilon_{a}\epsilon_{b}}a\cdot b\\
\dann(a\circ b)^{*} & = & (-i)^{\epsilon_{a}\epsilon_{b}}a^{*}b^{*}=(-)^{ab}a^{*}\circ b^{*}\end{eqnarray}
We used here the symbol $\epsilon_{a}$ to denote the parity, in order
to emphasize that the exponent of $'i'$ really should take only values
0 and 1, while for our usual prefactors $(-)^{ab}\equiv(-)^{\abs{a}\abs{b}}$,
the grading $\abs{a}$ does not need to be a $\mathbb{Z}_{2}$ grading.
The parity is given by $\epsilon_{a}\equiv\abs{a}\mod2$.

\subsection{Hermitean scalar product}

Using our above definition of complex conjugation also fixes the behaviour
of the graded version of a Hermitean scalar product. We use the index
notation $(v^{*})^{\bar{M}}\equiv(v^{M})^{*}$. The scalar product
(in a finite dimensional vector space for the beginning) then will
be defined as\vspace{-.5cm}\\
 \begin{tabular}{cc}
\begin{minipage}[t][1\totalheight]{0.47\columnwidth}%
\begin{eqnarray*}
\erw{v\mid w} & \Grequiv & \underbrace{(v^{*})^{\bar{M}}}_{(v^{M})^{*}}H_{\bar{M}N}w^{N}\\
 &  & \quad\mbox{with }\left(H_{\bar{M}N}\right)^{*}\Greq H_{\bar{N}M}\qquad\end{eqnarray*}
\end{minipage}%
 & %
\begin{minipage}[t][1\totalheight]{0.5\columnwidth}%
\begin{eqnarray}
\erw{v\mid w} & \stackrel{NW}{\equiv} & \sum_{\bar{M},N}(-)^{N+wN}\underbrace{(v^{*})^{\bar{M}}}_{(v^{M})^{*}}H_{\bar{M}N}w^{N}\nonumber \\
 &  & \quad\mbox{with }\left(H_{\bar{M}N}\right)^{*}=(-)^{MN}H_{\bar{N}M}\qquad\quad\end{eqnarray}
\end{minipage}%
\tabularnewline
\end{tabular}\vspace{.3cm}\\
where $H$ is a matrix of type '$D$' which is (graded) hermitean.
Strictly speaking, the rumpf $H$ appears only on the righthand side
and is therefore not gradifiable. However, if we identified on the
lefthand side the vertical line '$\mid$' as a placeholder for the
$H$-rumpf and also identify their grading, then it would be fine
to even gradify the rumpf $H$. For the following considerations we
will nevertheless stick to a bosonic rumpf $H$, i.e. $H_{\bar{M}N}$
should be considered as a bosonic supermatrix. The resulting scalar
product is (graded) sesquilinear in the sense\vspace{-.5cm} \\
\begin{tabular}{cc}
\begin{minipage}[t][1\totalheight]{0.47\columnwidth}%
\begin{eqnarray*}
\lqn{\erw{\lambda v_{1}+v_{2}\,\mid\,\mu w_{1}+w_{2}}\Greq}\\
 & \Greq & \lambda^{*}\mu\erw{v_{1}\,\mid\, w_{1}}+\lambda^{*}\erw{v_{1}\,\mid\, w_{2}}+\\
 &  & +\mu\erw{v_{2}\,\mid\, w_{1}}+\erw{v_{2}\,\mid\, w_{2}}\end{eqnarray*}
\end{minipage}%
 & %
\begin{minipage}[t][1\totalheight]{0.5\columnwidth}%
\begin{eqnarray}
\lqn{\erw{\lambda v_{1}+v_{2}\,\mid\,\mu w_{1}+w_{2}}=}\nonumber \\
 & = & (-)^{\mu v_{1}}\lambda^{*}\mu\erw{v_{1}\,\mid\, w_{1}}+\lambda^{*}\erw{v_{1}\,\mid\, w_{2}}+\nonumber \\
 &  & +(-)^{\mu v_{2}}\mu\erw{v_{2}\,\mid\, w_{1}}+\erw{v_{2}\,\mid\, w_{2}}\end{eqnarray}
\end{minipage}%
\tabularnewline
\end{tabular}\vspace{.3cm}\\
for $\lambda$ and $\mu$ being complex supernumbers. It is furthermore
(graded) hermitean, i.e.\vspace{-.5cm} \\
 \begin{tabular}{cc}
\begin{minipage}[t][1\totalheight]{0.47\columnwidth}%
\begin{eqnarray*}
\erw{v\mid w} & \Greq & \erw{w\mid v}^{*}\end{eqnarray*}
\end{minipage}%
 & %
\begin{minipage}[t][1\totalheight]{0.5\columnwidth}%
\begin{eqnarray}
\erw{v\mid w} & = & (-)^{vw}\erw{w\mid v}^{*}\end{eqnarray}
\end{minipage}%
\tabularnewline
\end{tabular}\vspace{.3cm}\\
The last equation implies that a scalar product of a vector with itself
obeys \begin{eqnarray}
\erw{v\mid v} & = & (-)^{v}\erw{v\mid v}^{*}\end{eqnarray}
and is therefore real only for even vectors and purely imaginary for
odd vectors. Note that a scalar product $\erw{\,\mid\,}_{0}$ which
obeys $\erw{v\mid w}_{0}=\erw{w\mid v}^{*}$ is obtained by either
replacing $*$ by $\tilde{*}$ of the previous subsection or by defining
$\erw{v\mid w}_{0}\equiv(-i)^{\epsilon_{v}\epsilon_{w}}\erw{v\mid w}$. 

The adjoint $B^{\dagger}$ of a matrix $B$ with respect to our scalar
product is defined as\vspace{-.5cm} \\
 \begin{tabular}{cc}
\begin{minipage}[t][1\totalheight]{0.47\columnwidth}%
\begin{eqnarray*}
\erw{v\mid Bw} & \Greq & \erw{B^{\dagger}v\mid w}\end{eqnarray*}
\end{minipage}%
 & %
\begin{minipage}[t][1\totalheight]{0.5\columnwidth}%
\begin{eqnarray}
\erw{v\mid Bw} & = & (-)^{Bv}\erw{B^{\dagger}v\mid w}\end{eqnarray}
\end{minipage}%
\tabularnewline
\end{tabular}\vspace{.3cm}\\
Assume that the hermitean matrix is non-degenerate in the sense that
it has an inverse \begin{eqnarray}
H_{\bar{M}K}H^{K\bar{N}} & = & \delta_{\bar{M}}\hoch{\bar{N}}\quad,\quad H^{M\bar{K}}H_{\bar{K}N}=\delta^{M}\tief{N}\end{eqnarray}
Although it is more common to use only the symmetric part of a scalar
product to pull indices up and down, we will in this section use $H_{\bar{M}N}$
and $H^{M\bar{N}}$ to pull indices. For a vector $v^{M}$ we thus
have the following additional possibilities of index-position and
form:\begin{eqnarray}
v_{\bar{M}} & \equiv & H_{\bar{M}N}v^{N}\\
(v^{*})_{M} & \equiv & (v_{\bar{M}})^{*}\\
(v^{*})^{\bar{M}} & \equiv & (v^{M})^{*}=(v^{*})_{N}H^{N\bar{M}}\quad\bigl(=(H_{\bar{N}K}v^{K})^{*}H^{N\bar{M}}\greq\,\underbrace{H_{\bar{K}N}H^{N\bar{M}}}_{\delta_{\bar{K}}\hoch{\bar{M}}}(v^{K})^{*}\bigr)\qquad\end{eqnarray}
\vspace{-.5cm}\\
Using the inverse matrix $H^{M\bar{N}}$, we can now give an explicit
expression for the adjoint matrix of $B$:\\
 $\erw{v\mid Bw}=(v^{*})^{\bar{M}}H_{\bar{M}N}(B^{N}\tief{K}w^{K})=(v^{*})^{\bar{M}}\underbrace{(H_{\bar{M}N}B^{N}\tief{L}H^{L\bar{P}})}_{\equiv B_{\bar{M}}\hoch{\bar{P}}}H_{\bar{P}K}w^{K}\Greq\,\bigl(\underbrace{(B^{*})_{M}\hoch{P}}_{\equiv(B_{\bar{M}}\hoch{\bar{P}})^{*}}v^{M}\bigr)^{*}H_{\bar{P}K}w^{K}$$\stackrel{!}{=}\erw{B^{\dagger}v\mid w}$.
From this calculation we can read off\begin{eqnarray}
(B^{\dagger})^{P}\tief{M} & \greq & (B_{\bar{M}}\hoch{\bar{P}})^{*}=\left(H_{\bar{M}N}B^{N}\tief{L}H^{L\bar{P}}\right)^{*}\greq H^{P\bar{L}}\underbrace{(B^{N}\tief{L})^{*}}_{(B^{\dagger})_{\bar{L}}\hoch{\bar{N}}}H_{\bar{N}M}\end{eqnarray}
Up to pulling indices with $H$ this agrees with our earlier definition
of the hermitean conjugate of a matrix $(B^{\dagger})_{\bar{L}}\hoch{\bar{N}}\greq\,(B^{N}\tief{L})^{*}$.

Having used indices all the time, we have implicitely chosen some
\textbf{basis}\begin{eqnarray}
\ket{e_{M}} & \equiv & \ket{\tief{M}}\end{eqnarray}
Every vector $\ket{v}$ of definite grading can be written as a linear
combination \begin{equation}
\ket{v}=v^{M}\ket{\tief{M}}\end{equation}
The complex conjugate basis is $\ket{\tief{\bar{M}}}\equiv\ket{\tief{M}}^{*}$,
so that $\ket{v^{*}}\equiv\ket{v}^{*}=(v^{*})^{\bar{M}}\ket{\tief{\bar{M}}}$.
Bra-vectors involve a complex conjugation. Because of $\bra{v^{M}e_{M}}=(v^{*})^{\bar{M}}\bra{e_{M}}$
it is convenient to denote \begin{eqnarray}
\bra{e_{M}} & \equiv & \bra{\tief{\bar{M}}}\end{eqnarray}
such that \begin{eqnarray}
\bra{v}=(v^{*})^{\bar{M}}\bra{\tief{\bar{M}}} & \mbox{ and } & \erw{\tief{\bar{M},N}}=H_{\bar{M}N}\end{eqnarray}
The \textbf{dual basis }will be denoted by $\bra{\hoch{M}}$ and it
is defined via \begin{eqnarray}
\erw{\hoch{M}\mid\tief{N}} & = & \delta^{M}\tief{N}\end{eqnarray}
After pulling down one index with $H$ one arrives at the equation
$\erw{\tief{\bar{M}}\mid\tief{N}}=H_{\bar{M}N}$ which we just had
before and which is in turn consistent with $\erw{v\mid w}=(v^{*})^{\bar{M}}H_{\bar{M}N}w^{N}$.
The dual basis $\bra{\hoch{M}}$ thus agrees with the {}``hermitean
conjugate'' $\bra{\tief{\bar{M}}}$ of $\ket{\tief{M}}$ up to raising
the index with $H^{M\bar{N}}$.

\paragraph{Clifford vacuum}

The above recall of some basic linear algebra will help us to understand
the graded version of creation and annihilation operators acting on
some Clifford vacuum. Let us denote just for this paragraph the index
of the creation operators by $k,l,m,\ldots$, although we used those
indices before for bosonic indices, while now we still assume them
to be graded and not purely bosonic. The creation operators generate
a complete basis from the Clifford vacuum, s.th. the indices $k,l,m,...$
are just a subset of the basis-indeces $M,N,\ldots$. Let us denote
the annihilation and creation operators by $a^{k}$ and $(a^{\dagger})_{k}$
respectively and the corresponding vectors or states by \begin{eqnarray}
\ket{\tief{k}} & \equiv & (a^{\dagger})_{k}\,\ket{0},\quad\ket{\tief{k_{1}k_{2}}}\equiv(a^{\dagger})_{k_{1}}(a^{\dagger})_{k_{2}}\,\ket{0},\quad\ket{\tief{k_{1}k_{2}k_{3}}}\equiv(a^{\dagger})_{k_{1}}(a^{\dagger})_{k_{2}}(a^{\dagger})_{k_{3}}\,\ket{0},\,\ldots\end{eqnarray}
The basis is then given by \begin{equation}
\ket{\tief{K}}\in\{\ket{0},\ket{\tief{k}},\ket{\tief{k_{1}k_{2}}},\ket{\tief{k_{1}k_{2}k_{3}}},\ldots\}\end{equation}
Finally we need the annihilation property of $a^{k}$ and their commutator
with the creation operators:\begin{equation}
a^{k}\ket{0}=0,\qquad\left[a^{k},(a^{\dagger})_{l}\right]\greq\;\delta^{k}\tief{l}\end{equation}
Assume that the Clifford vacuum is bosonic, so that we can normalize
it to one\begin{equation}
\erw{0\mid0}=1\end{equation}
This equation is not gradifiable, which is the reason why a bosonic
vacuum is preferrable. The dual basis is then given by the dual vacuum
$\bra{0}$ and its descendents \begin{eqnarray}
\bra{\hoch{k}} & \equiv & \bra{0}a^{k},\quad\tfrac{1}{2}\bra{\hoch{k_{1}k_{2}}}\equiv\tfrac{1}{2}\bra{0}a^{k_{1}}a^{k_{2}},\quad\tfrac{1}{3!}\bra{\hoch{k_{1}k_{2}k_{3}}}\equiv\tfrac{1}{3!}\bra{0}a^{k_{1}}a^{k_{2}}a^{k_{3}},\quad\ldots\\
\erw{\hoch{k}\mid\tief{l}} & = & \bra{0}a^{k}(a^{\dagger})_{l}\ket{0}=\delta^{k}\tief{l}\label{eq:dualNormalization}\\
\tfrac{1}{2}\erw{\hoch{k_{1}k_{2}}\mid\tief{l_{2}l_{1}}} & = & \!\!\!\tfrac{1}{2}\bra{\hoch{k_{1}}}a^{k_{2}}(a^{\dagger})_{l_{2}}\ket{\tief{l_{1}}}\greq\tfrac{1}{2}\bra{\hoch{k_{1}}}(a^{\dagger})_{l_{2}}a^{k_{2}}\ket{\tief{l_{1}}}+\tfrac{1}{2}\bra{\hoch{k_{1}}}\delta^{k_{2}}\tief{l_{2}}\ket{\tief{l_{1}}}\greq\delta^{k_{1}}\tief{(l_{2}}\delta^{k_{2}}\tief{l_{1})}\qquad\\
\tfrac{1}{3!}\erw{\hoch{k_{1}k_{2}k_{3}}\mid\tief{l_{3}l_{2}l_{1}}} & \greq & \delta^{k_{1}}\tief{(l_{1}}\delta^{k_{2}}\tief{l_{2}}\delta^{k_{3}}\tief{l_{3})}\\
 & \ddots\nonumber \\
\bra{\hoch{K}} & \in & \left\{ \bra{0},\bra{\hoch{k}},\tfrac{1}{2}\bra{\hoch{k_{1}k_{2}}},\tfrac{1}{3!}\bra{\hoch{k_{1}k_{2}k_{3}}},\ldots\right\} \end{eqnarray}
In the literature the indices of creation and annihilation operators
are usually put at the same vertical position, and the corresponding
states are normalized to be $\erw{\tief{k}\mid\tief{l}}=\delta_{kl}$.
The Kronecker delta on the righthand side corresponds to a special
choice of the scalar product and should in our context be replaced
by \begin{eqnarray}
\erw{\tief{\bar{k}}\mid\tief{l}} & = & H_{\bar{k}l}\end{eqnarray}
which agrees with (\ref{eq:dualNormalization}) after pulling one
index with $H$.

Note that the definition of a \textbf{norm\index{norm}} induced by
the scalar product will not be possible under the conditions of theorem
\ref{thm:gradification}. The bosonic definition $\norm{v}\equiv\erw{v,v}$
has the rumpf $v$ appearing twice on the righthand side which is
therefore not gradifiable. Still it makes sense to define a norm,
but it will not simply have gradified properties of the bosonic one.
In order to get a real norm, (while $\erw{v\mid v}$ is imaginary
for odd $v$), we have to include an imaginary factor in the fermionic
case and fix the arbitrary overall sign: E.g. \begin{equation}
\norm{v}\equiv\tfrac{1}{i^{\epsilon_{v}}}\erw{v,v}\end{equation}
Only at this point (choosing an appropriate $H_{\bar{M}N}$) we make
contact to the usual definitions in the literature. Physical observables
and probabilities should of course not depend on the conventions in
the end. In the same way as above, the definition of the probability
of some transition (which contains an absolute value square and is
therefore also not gradifiable) has to include an appropriate complex
factor. We are not going to work with Hilbert spaces in the second
part of this thesis anyway and therefore leave the details for further
studies. The leading thought was just to keep the idea of gradification
as long as possible and break it only in the last step, in the definition
of the norm and of probabilities.

\subsection{Hermitean conjugate of matrix products}

From our definition of a hermitean\index{hermitean conjugate!of matrix products}
conjugate and of complex conjugation of products of numbers, we get
via the theorem the natural rules for complex conjugation of (graded)
matrix products:\\
\begin{tabular}{cc}
\begin{minipage}[c][1\totalheight][t]{0.47\columnwidth}%
\begin{eqnarray*}
\left((AC)^{\dag}\right)^{MN} & =_{G} & \left(C^{\dag}A^{\dag}\right)^{MN}\\
\left((AD)^{\dag}\right)^{M}\tief{N} & =_{G} & (D^{\dag}A^{\dag})^{M}\tief{N}\\
\left((BA)^{\dag}\right)^{MN} & =_{G} & (A^{\dag}B^{\dag})^{MN}\\
 & \ddots\end{eqnarray*}
\end{minipage}%
 & %
\begin{minipage}[c][1\totalheight][t]{0.5\columnwidth}%
\begin{eqnarray}
\left((AC)^{\dag}\right)^{MN} & = & (-)^{AC}\left(C^{\dag}A^{\dag}\right)^{MN}\\
\left((AD)^{\dag}\right)^{M}\tief{N} & = & (-)^{AD}(D^{\dag}A^{\dag})^{M}\tief{N}\\
\left((BA)^{\dag}\right)^{MN} & = & (-)^{AB}(A^{\dag}B^{\dag})^{MN}\\
 & \ddots\nonumber \end{eqnarray}
\end{minipage}%
\tabularnewline
\end{tabular}\\
Similarly we expect for operators in the infinite dimensional case\\
\begin{tabular}{cc}
\begin{minipage}[c][1\totalheight][t]{0.47\columnwidth}%
\begin{eqnarray*}
(\hat{A}\hat{B})^{\dag} & =_{G} & \hat{B}^{\dag}\hat{A}^{\dag}\end{eqnarray*}
\end{minipage}%
 & %
\begin{minipage}[c][1\totalheight][t]{0.5\columnwidth}%
\begin{eqnarray}
(\hat{A}\hat{B})^{\dag} & = & (-)^{AB}\hat{B}^{\dag}\hat{A}^{\dag}\end{eqnarray}
\end{minipage}%
\tabularnewline
\end{tabular}\\
As mentioned in the context of complex conjugation, it is simply a
matter of redefining the operator product with a factor $(-i)^{\epsilon_{A}\epsilon_{B}}$
if one wants to make contact to the usual definition without sign.

\rem{

\subsection{Hermitean conjugation of operators in field theory}

The Hermitean conjugate of operators is defined to be the adjoint
operator w.r.t. a Hermitean scalar product. If a graded scalar product
is defined along the lines of theorem \ref{thm:gradification}, it
will be graded hermitean:\begin{eqnarray*}
\erw{\psi,\phi}=(-)^{\psi\phi}\erw{\phi,\psi} & \iff & \erw{\psi,\phi}=(-)^{\psi\phi}\erw{\phi,\psi}\end{eqnarray*}
As mentioned already for graded matricesApart from a few exceptions,
complex conjugation of Grassman-variables $\bs{a}$, $\bs{b}$ is
usually defined such that $(\bs{a}\bs{b})^{*}=\bs{b}^{*}\bs{a}^{*}=-\bs{a}^{*}\bs{b}^{*}$.
The argument for doing so is the consistency with $(\bs{A}\bs{B})^{\dagger}=\bs{A}^{\dagger}\bs{B}^{\dagger}$,
where $\bs{A},\bs{B}$ are fermionic operators. The latter behaviour
of operators in turn induced by a standard scalar product. We therefore
suggest the definition of a scalar product that respects the grading.
\begin{eqnarray}
\erw{\psi,\phi} & = & \int dx^{d}\,\psi*(x)\phi(x)\end{eqnarray}

}

\section{Graded inverse - a nice {}``counterexample'' to the theorem}

\label{sub:Graded-inverse}Consider for the beginning matrices with
even rumpf only\begin{equation}
\abs{A}=\abs{B}=\abs{C}=\abs{D}=0\end{equation}
We say $A$ is the \textbf{(graded) inverse}\index{graded inverse}\index{matrix inverse}
of $D$, $B_{2}$ the inverse of $B_{1}$ and $C^{2}$ the inverse
of $C^{1}$ iff\begin{eqnarray}
D_{MK}A^{KN} & = & \delta_{M}\hoch{N}\label{eq:grInverseI}\\
A^{MK}D_{KN} & = & \delta^{M}\tief{N}\\
B_{1}^{M}\tief{K}B_{2}^{K}\tief{N} & = & \delta^{M}\tief{N}\\
C_{M}^{1}\hoch{K}C_{K}^{2}\hoch{N} & = & \delta_{M}\hoch{N}\label{eq:grInverseIV}\end{eqnarray}
with \begin{eqnarray}
\delta_{M}\hoch{N} & = & (-)^{MN}\delta^{N}\tief{M}\end{eqnarray}
The so defined inverses in general do not coincide with the naive
inverses.%
\footnote{\index{footnote!\thefoot. inverse of a supermatrix}\index{supermatrix!inverse}\index{inverse of a supermatrix}To
verify this statement, write out the equations (\ref{eq:grInverseI})-(\ref{eq:grInverseIV})
in NW-conventions, using $\delta_{M}\hoch{N}=\delta_{M}^{N}$:\begin{eqnarray*}
\sum\quad D_{MK}(-)^{K}A^{KN} & = & \delta_{M}^{N}\\
\sum\quad A^{MK}D_{KN}(-)^{N} & = & \delta_{N}^{M}\\
\sum\quad B_{1}^{M}\tief{K}(-)^{K+N}B_{2}^{K}\tief{N} & = & \delta_{N}^{M}\\
\sum\quad C_{M}^{1}\hoch{K}C_{K}^{2}\hoch{N} & = & \delta_{M}^{N}\end{eqnarray*}
Only in the last case $C^{2}$ is the naive inverse of $C^{1}$.$\quad\fussend$%
}

From our theorem we can e.g. deduce that for matrices $M$ $N$ of
any type (with even rumpf) we have\begin{eqnarray}
(MN)^{-1} & =_{G} & (N^{-1}M^{-1})\\
\stackrel{\abs{M}=\abs{N}=0}{\dann}\quad(MN)^{-1} & = & (N^{-1}M^{-1})\end{eqnarray}
This is easily directly verified using associativity of our graded
matrix multiplication.

\subsubsection*{Counterexample}

\index{gradification!counterexample}\index{counterexample!to the gradification theorem}If
we take the rumpfs arbitrarily graded and still define an inverse
via $M^{-1}M=\one$, then we still have%
\footnote{\index{supermatrix!fermionic $\sim$}\index{fermionic supermatrix!inverse of $\sim$}\index{inverse of a fermionic supermatrix}\index{footnote!\thefoot. inverse of a fermionic supermatrix}Note
that although a Grassmann-variable has no inverse, a matrix with fermionic
rumpf can have an inverse. Take e.g. $x,y\neq0$ bosonic and $c$
fermionic, then we have\begin{eqnarray*}
\left(\begin{array}{cc}
c & x\\
y & 0\end{array}\right)\left(\begin{array}{cc}
0 & \frac{1}{y}\\
\frac{1}{x} & -\frac{c}{xy}\end{array}\right) & = & \left(\begin{array}{cc}
1 & 0\\
0 & 1\end{array}\right)\quad(\#)\end{eqnarray*}
The matrix multiplication above, however, is not according to our
graded matrix multiplication rules, which are\begin{eqnarray*}
\left(CC^{-1}\right)_{M}\hoch{N} & \equiv_{g} & C_{M}\hoch{K}(C^{-1})_{K}\hoch{N}=_{g}\delta_{M}\hoch{N}\\
\dann\left(CC^{-1}\right)_{M}\hoch{N} & \stackrel{NW}{=} & \sum_{K}(-)^{KA+MA}C_{M}\hoch{K}(C^{-1})_{K}\hoch{N}=\delta_{M}\hoch{N}\end{eqnarray*}
The following choice of matrices therefore correspond to the equation
(\#):\begin{eqnarray*}
C & = & \left(\begin{array}{cc}
c & -x\\
-y & 0\end{array}\right)\quad C^{-1}=\left(\begin{array}{cc}
0 & \frac{1}{y}\\
\frac{1}{x} & -\frac{c}{xy}\end{array}\right)\qquad\fussend\end{eqnarray*}
} \begin{eqnarray}
(MN)(N^{-1}M^{-1}) & \stackrel{\textrm{assoz}}{=} & M(NN^{-1})M^{-1}=\one\\
\dann\quad(MN)^{-1} & = & (N^{-1}M^{-1}),\quad\textrm{for any $\abs{M}$and }\abs{N}\end{eqnarray}
There is no expected prefactor $(-)^{MN}$ in the lower line! This
looks strange in terms of the big graded equal sign, which should
swallow the rumpf-dependend signs, but produces one here:\begin{eqnarray}
(MN)^{-1} & =_{G} & (-)^{MN}(N^{-1}M^{-1})\end{eqnarray}
The theorem thus is not applicable here! What went wrong? Our definition
of the inverse\begin{eqnarray}
(MM^{-1}) & = & \one\end{eqnarray}
is a non-valid gradification of the bosonic one: The theorem allows
us to assign a grading only to rumpfs which appear exactly once in
each term. The rumpf $M$ appears twice on the lefthand side and not
at all on the righthand side. Thus, the theorem does not allow to
give $M$ a grading. If we do so nevertheless, we can't derive known
rules from the bosonic case. The definition itself is of course ok,
but in order to stress that it is not simply a gradification of a
bosonic definition, we should better give it a new name, like \textbf{special
graded inverse}. 

The naked indices in (\ref{eq:grInverseI}) to (\ref{eq:grInverseIV})
appear excactly once in each term and can therefore be generalized
to graded indices.

\section{(Super) trace}

\index{supertrace|fett}\index{trace!graded matrix $\sim$}\index{supermatrix!trace}We
now come to another important deviation from usual supermatrix-definitions
which will enter an interesting result for superdeterminants. The
trace is the sum of the diagonal entries and makes sense for matrices
of type $C$ and $B$ only (matrices with one upper and one lower
index, i.e. endomorphisms)\begin{eqnarray}
\tr B & \equiv & B^{M}\tief{M}=\left\{ \begin{array}{c}
\sum_{M}B^{M}\tief{M}\quad NW\\
\sum_{M}(-)^{M}B^{M}\tief{M}\quad NE\end{array}\right.\\
\tr C & \equiv & C_{M}\hoch{M}=\left\{ \begin{array}{c}
\sum_{M}(-)^{M}C_{M}\hoch{M}\quad NW\\
\sum_{M}C_{M}\hoch{M}\quad NE\end{array}\right.\end{eqnarray}
The $(-)^{M}$ is familiar from usual definitions. We have it here,
however, either only for NW for matrices of type $C$ or for NE for
matrices of type $B$ while the other cases do not have the familiar
$(-)^{M}$ in the trace-definition. The reason is that e.g. for $B$-type
matrices in NW (where the trace has no sign factor) the $(-)^{M}$
instead is hidden in the matrix multiplication of two matrices. Thus,
either the matrix multpilication contains an extra $(-)^{M}$ and
the trace doesn't, or the other way round. In any case, the graded
cyclicity property of the trace holds:\begin{eqnarray}
\tr B_{1}B_{2} & = & B_{1}^{M}\tief{K}B_{2}\hoch{K}\tief{M}=(-)^{B_{2}B_{1}}\tr B_{2}B_{1}\label{eq:trCyclicity}\\
\iff\quad\tr\left[B_{1},B_{2}\right] & = & 0\label{eq:trOfCommutator}\end{eqnarray}
For matrices of type $A$ and $D$, we need a metric, in order to
define a meaningful trace:\begin{eqnarray}
\tr A & \equiv & A^{MN}G_{MN}\\
\tr D & \equiv & D_{MN}G^{MN}\end{eqnarray}

\section{(Super) determinant}

\index{superdeterminant}\index{determinant!super$\sim$|fett}\index{supermatrix!determinant}We
finally come to the so far most interesting demonstration of the use
of our conventions. Namely the definition of the superdeterminant.
As usual, we start from the definition via the exponential: \begin{eqnarray}
\det C & \equiv & e^{\tr\ln C},\quad\det B\equiv e^{\tr\ln B},\quad\label{eq:detUeberExp}\end{eqnarray}
Remember that in NW-conventions for a matrix of type B, the definition
of the trace matches the bosonic definition, while the definition
of the matrix product differs. For NE or for matrices of type C the
situation is just the other way round. In both cases the above definition
thus differs from the bosonic one, even if the matrix is purely bosonic
(but having two fermionic indices). Let us derive this in detail.

Consider the decomposition of $B$ in bosonic and fermionic blocks:\begin{eqnarray}
\left(B^{M}\tief{N}\right) & \equiv & \left(\begin{array}{cc}
B^{m}\tief{n} & B^{m}\tief{\nu}\\
B^{\mu}\tief{n} & B^{\mu}\tief{\nu}\end{array}\right)\equiv\left(\begin{array}{cc}
a^{m}\tief{n} & b^{m}\tief{\nu}\\
c^{\mu}\tief{n} & d^{\mu}\tief{\nu}\end{array}\right),\quad\abs{m}=0,\abs{\mu}=1\end{eqnarray}
Assuming that the matrix $(a)$ is invertible (which implies that
$a$ (and thus the rumpf of $B$) is bosonic, because a matrix with
purely fermionic entries cannot be inverted), one can seperate $C$
in a product of two block-triangular matrices\begin{eqnarray}
B & = & B_{1}B_{2}\\
B_{1} & = & \left(\begin{array}{cc}
a & 0\\
c & \one\end{array}\right)\left(\begin{array}{cc}
\one & (a^{-1}b)\\
0 & d-ca^{-1}b\end{array}\right)\end{eqnarray}
Now we will use two facts. One is that the trace of the logarithm
factorizes:\begin{eqnarray}
e^{F}e^{G} & \stackrel{BCH}{=} & e^{F+G+\frac{1}{2}\left[F,G\right]+\ldots}\\
B_{1}B_{2} & = & e^{\ln B_{1}+\ln B_{2}+\frac{1}{2}\left[\ln B_{1},\ln B_{2}\right]+\ldots}\\
\dann\tr\,\ln(B_{1}B_{2}) & \stackrel{(\ref{eq:trOfCommutator})}{=} & \tr\ln B_{1}+\tr\ln B_{2}\end{eqnarray}
And the other fact is that an arbitrary power of a block-triangular
matrix stays a blocktriangular matrix with the powers of the diagonal
blocks in the block diagonal:\begin{eqnarray}
\left(\begin{array}{cc}
a & 0\\
b & c\end{array}\right)^{n} & = & \left(\begin{array}{cc}
a^{n} & 0\\
* & c^{n}\end{array}\right)\\
\left(\begin{array}{cc}
a & b\\
0 & d\end{array}\right)^{n} & = & \left(\begin{array}{cc}
a^{n} & *\\
0 & d^{n}\end{array}\right)\quad\forall a,b,c,d\end{eqnarray}
In particular\begin{eqnarray}
(B_{1}-\one)^{n} & = & \left(\begin{array}{cc}
(a-\one)^{n} & 0\\
* & 0\end{array}\right)\\
(B_{2}-\one)^{n} & = & \left(\begin{array}{cc}
0 & 0\\
* & (d-ca^{-1}b-\one)^{n}\end{array}\right)\end{eqnarray}
Now we use the power series for the logarithm \begin{eqnarray}
\ln(1+x) & = & \sum_{n=1}^{\infty}\frac{1}{n!}\ln^{(n)}(1)x^{n}=\sum_{n=1}^{\infty}(-)^{n-1}\frac{x^{n}}{n}\\
\tr\ln(B_{1}) & = & \sum_{n=1}^{\infty}(-)^{n-1}\frac{\tr(B_{1}-\one)^{n}}{n}=\\
 & = & \sum_{n=1}^{\infty}\frac{(-)^{n-1}}{n}\tr\left(\begin{array}{cc}
(a-\one)^{n} & 0\\
* & 0\end{array}\right)=\\
 & = & \sum_{n=1}^{\infty}\frac{(-)^{n-1}}{n}\tr(a-\one)^{n}=\\
 & = & \tr\ln a\\
\tr\ln(B_{1}) & = & \tr\ln(d-ca^{-1}b)\end{eqnarray}
We thus get\index{determinant!super $\sim$}\index{superdeterminant}\begin{eqnarray}
\det B & = & \det B_{1}\cdot\det B_{2}=\\
 & = & \det a\cdot\det(d-ca^{-1}b)\end{eqnarray}
This result is true for every block-decomposition. $a,d$ do not necessarily
have to be bosonic as well as $b$ and $c$ do not have to be fermionic.
At first sight this seems to contradict the expression that one usually
finds in the literature, namely $\mbox{sdet}B=\det a/\det(d-ca^{-1}b)$. 

The reason for this mismatch lies simply in the graded definition
of the matrix multiplication (or the trace) and thus of the determinant
of a bosonic matrix with two fermionic indices. For NE-conventions,
the trace of the type-B submatrix $(d^{\mu}\tief{\nu})$ gives an
extra minus w.r.t. its naive bosonic trace. Its determinant defined
via the exponential and the graded trace is thus equal to $"1/\det(d)"$,
where now the determinant is the naive bosonic one, built with the
naive trace. The same is true, if we consider the corresponding submatrices
of a matrix of type $C$ in NW-conventions. For the determinant of
a matrix of type $B$ in NW (or likewise type $C$ in NE), however,
the comparison between our and the usual convention is a bit more
subtle. In the following we write terms in the usual convention in
quotation marks. At first, let us define the \textbf{dimension}\index{dimension!of a graded vector space|fett}\label{par:negativeDimension}
of a square matrix (or of the vector space it is acting on) as the
trace of the corresponding unit-matrix:\begin{eqnarray}
\dim(B) & \equiv & \delta^{M}\tief{M}="\dim(a)-\dim(d)"\\
\dim(d) & = & "-\dim(d)"\end{eqnarray}
I.e., fermionic dimensions are negative\index{negative dimension}
dimensions\index{dimension!negative $\sim$}!%
\footnote{\index{footnote!\thefoot. negative dimensions} The observation that
fermionic dimensions can be considered to be negative dimensions has
been made in literature at several places and with several arguments.
From the group theoretic point of view, this has been studied in \cite{Cvitanovic:1979qz,Cv2007:bt}.$\qquad\fussend$%
} The logarithm in the definition of the determinant has to be understood
as a power series, so that we first should look at simple powers of
the block $d$: \begin{eqnarray}
d^{2}\hoch{\mu}\tief{\nu} & = & d^{\mu}\tief{\lambda}d^{\lambda}\tief{\nu}=\\
 & \stackrel{NW}{=} & \sum_{\lambda}d^{\mu}\tief{\lambda}d^{\lambda}\tief{\nu}(-)^{\lambda}\\
\dann d^{n} & = & "(-1)^{n-1}d^{n}=-(-d)^{n}"\quad\textrm{naive matrix mult in quot}\label{eq:dHochn}\end{eqnarray}
Logarithm and determinant of $d^{\mu}\tief{\nu}$ can thus be written
as\begin{eqnarray}
\ln(d) & = & \sum_{n=1}^{\infty}\frac{(-)^{n-1}}{n}(d-\one)^{n}\underset{\textrm{and\,(\ref{eq:dHochn})}}{\stackrel{\one="-\one"}{=}}\\
 & \underset{\textrm{and\,(\ref{eq:dHochn})}}{\stackrel{\one="-\one"}{=}} & "-\sum_{n=1}^{\infty}\frac{(-)^{n-1}}{n}(-d-\one)^{n}"\\
 & = & "-\ln(-d)"\quad\textrm{naive matrix mult in quot}\\
\det(d) & = & \exp\tr\ln d="1/\det(-d)=(-1)^{\dim(d)}1/\det d\,"\end{eqnarray}
The sub-matrix $(d-ca^{-1}b)$ is of the same type as $d$, so that
we finally get\begin{eqnarray}
\det(d-ca^{-1}b) & \underset{ca^{-1}b="ca^{-1}b"}{\stackrel{a^{-1}="a^{-1}"}{=}} & "(-1)^{\dim(d)}1/\det(d-ca^{-1}b)"\\
\det B & = & "(-1)^{\dim(d)}\det a/\det(d-ca^{-1}b)"\quad\textrm{naive matrix mult in quot}\end{eqnarray}
For matrices of type $C$ in NW-convention, the situation is the same
as for matrices of type $B$ in NE-convention:$d^{n}="d^{n}",\:\one_{d}="\one_{d}",\:\ln d="\ln d",\:\tr\ln d="-\tr\ln d"$.
We thus get\begin{eqnarray}
 & \boxed{\det B=\det a\cdot\det(d-ca^{-1}b)=\left\{ \begin{array}{c}
"(-1)^{\dim(d)}\det a/\det(d-ca^{-1}b)"\quad\textrm{NW}\\
"\det a/\det(d-ca^{-1}b)"\quad\textrm{NE}\end{array}\right.}\\
 & \textrm{for }B^{M}\tief{N}={\left(\begin{array}{cc}
a & b\\
c & d\end{array}\right)^{M}}_{N}\end{eqnarray}
and\begin{eqnarray}
 & \boxed{\det C=\det a\cdot\det(d-ca^{-1}b)=\left\{ \begin{array}{c}
"\det a/\det(d-ca^{-1}b)"\quad\textrm{NW}\\
"(-1)^{\dim(d)}\det a/\det(d-ca^{-1}b)"\quad\textrm{NE}\end{array}\right.}\\
 & \textrm{for }C_{M}\hoch{N}={\left(\begin{array}{cc}
a & b\\
c & d\end{array}\right)_{M}}^{N}\end{eqnarray}
 As a check, let us take $C=B^{T}=\left(\begin{array}{cc}
a^{T} & c^{T}\\
b^{T} & d^{T}\end{array}\right)=\,"\left(\begin{array}{cc}
a^{T} & c^{T}\\
b^{T} & -d^{T}\end{array}\right)"$. Then we expect, following our theorem: \begin{equation}
\det B=\det B^{T}\end{equation}
Indeed, in NW-conventions this becomes in naive matrix-notations:
\begin{eqnarray}
"(-1)^{\dim(d)}\det(d-ca^{-1}b)" & \stackrel{!}{=} & "\det(-d^{T}-b^{T}(a^{-1})^{T}c^{T})"=\\
 & = & "\det\left(-d^{T}-(-)^{cb}ca^{-1}b\right)^{T}"=\\
 & = & "\det\left(-d+ca^{-1}b\right)"=\\
 & = & "(-1)^{\textrm{dim}(d)}\det(d-ca^{-1}b)"\quad\surd\end{eqnarray}

\rem{The alternative equation to (\ref{eq:detUeberExp}) for determinants
of bosonic matrices is\enlargethispage*{2cm} \begin{eqnarray}
\det A & = & \frac{1}{D!}\epsilon^{M_{1}\ldots M_{D}}\epsilon^{N_{1}\ldots N_{D}}A_{M_{1}N_{1}}\cdots A_{M_{D}N_{D}}\end{eqnarray}
but seems to be less useful for a generalization to the graded case.
(?)

\newpage Scalar-multiplication of a matrix\begin{eqnarray}
(\lambda\cdot C)^{M}\tief{N} & \equiv & \lambda C^{M}\tief{N}\\
(\lambda\cdot B)_{M}\hoch{N} & \equiv & \lambda B_{M}\hoch{N}\\
\tr\lambda M & = & \lambda\tr M\end{eqnarray}
Scaling-behaviour of the determinant:\begin{eqnarray}
\det(\lambda\cdot M) & = & \det(\lambda\cdot\one)\det M=\\
 & = & (\lambda)^{\dim(M)}\det M\\
\det(\lambda\cdot d) & = & \lambda^{\dim d}\det\lambda,\quad\dim d<0\end{eqnarray}
}\rem{

\section{Eigenvalues \& eigenvectors \& characteristic polynomial}

\index{characteristic polynomial}\index{eigenvector}\index{eigenvalue}\begin{eqnarray}
C^{M}\tief{N}v_{i}^{\: N} & = & \lambda_{i}v_{i}^{\: M}\end{eqnarray}
For $C_{d}$ diagonal and $\abs{v_{i}}=0$ in $NW$:\begin{eqnarray}
C_{d}^{M}\tief{N=M}v_{i}^{\: N=M}(-)^{N=M} & = & \lambda_{i}v_{i}^{\: M=N}=\lambda_{i=M}\delta^{M}\tief{N=M}v_{i=M}\hoch{N=M}(-)^{N=M}\\
\lambda_{i=M} & = & (-)^{M}C_{d}^{M}\tief{N=M}\\
C_{d}^{M}\tief{N=M} & = & \lambda_{i=M}\delta^{M}\tief{N=M}\\
v_{i} & = & (0,\ldots0,\underset{{\scriptstyle i}}{c},0\ldots0)\\
\tr C & = & C^{M}\tief{M}=\sum\lambda_{M}(-)^{M}=\tr(\lambda_{i=M}\delta^{M}\tief{N})\end{eqnarray}
Diagonal matrix of a Matrix (always bosonic)\begin{eqnarray}
\textrm{Diag }(C)^{M}\tief{N} & \equiv & C^{M}\tief{N=M}\delta^{M}\tief{N}\end{eqnarray}
Diagonal matrix from a vector\begin{eqnarray}
\textrm{Diag}(\lambda_{K})^{M}\tief{N} & \equiv & (\lambda_{K=M}\delta^{M}\tief{N})\end{eqnarray}
In that sense the trace is the sum of Eigenvalues. For the determinant
we have\begin{eqnarray}
\det\left(C_{d}\right) & = & \det\left(\lambda_{K=M}\delta^{M}\tief{N}\right)=\prod_{M}(\lambda_{M})^{(-)^{M}}\end{eqnarray}
}

\section{Graded gamma-matrices}

\label{sec:gradedGamma}\index{gamma matrix!graded $\sim$}\index{graded gamma matrix}Gamma
matrices and some of their properties are discussed in appendix \vref{app:gamma}.
Usually, they are considered to be ordinary bosonic matrices with
the anticommutator relation \begin{equation}
\left\{ \Gamma^{a},\Gamma^{b}\right\} =2\eta^{ab}\one\end{equation}
There are two ways how a grading can be introduced into the gamma-matrix
algebra. Either via the rumpf or via the indices. Let us start with
the rumpf. 

The anticommutator is for general matrices not a very natural object.
It does not automatically have derivative properties or a Jacobi identity
like the commutator. However, the gamma matrices can (in even dimensions)
be represented by \emph{off-diagonal} matrices. This offers the possibility
to regard them as fermionic supermatrices $\bs{\Gamma}^{a}$\index{$\Gamma$@$\protect\bs{\Gamma}^a$|itext{graded gamma matrix}}
whose fermionic diagonal blocks simply vanish. The anticommutator
above then simply becomes the graded commutator\begin{equation}
\left[\bs{\Gamma}^{a},\bs{\Gamma}^{b}\right]=2\eta^{ab}\one\end{equation}
Terms like $\bar{\psi}\bs{\Gamma}^{a}\partial_{a}\bs{\psi}$ in a
Lagrangian still stay bosonic, because $\bar{\psi}=\bs{\psi}^{\dagger}\bs{\Gamma}^{0}$
contains another odd gamma-matrix. This interpretation of a graded
algebra appears naturally in the RNS-string, where the spacetime spinors
are generated by acting with fermionic creation operators on a Clifford
vacuum. Linear combinations of these odd creation operators then correspond
to the (odd) gamma matrices.

It is interesting that in the graded picture the chirality matrix
plays a different role than the other gamma-matrices, because (as
a product of all gamma-matrices in even dimensions) it is an even
object $\Gamma^{\#}\propto\bs{\Gamma}^{0}\cdots\bs{\Gamma}^{d-1}$.
The anticommutation of it with the other matrices stays an anticommutation
even in the graded picture\begin{equation}
\{\Gamma^{\#},\bs{\Gamma}^{a}\}=0,\quad\{\Gamma^{\#},\Gamma^{\#}\}=2\one\end{equation}
This is actually also a hint that s.th. like the RNS string could
not work in the same way in odd (e.g. 11) dimensions, where one of
the gamma-matrices (and thus one of the generators acting on the clifford
vacuum) needs to be even.

The second possibility to re-distribute the grading, is to consider
the fermionic (Dirac) indices of $\bs{\Gamma}^{a\,\q{\bs{\alpha}}}\tief{\q{\bs{\beta}}}$
to carry an odd grading. (The underline simply shall distinguish the
Dirac-indices from Weyl indices, which are mainly used later on.)
As the fermionic indices come in pairs it does not change the overall
grading. We still assume the rumpf to be odd, too. The graded commutator
then becomes (in NW-conventions)\begin{eqnarray}
\left[\bs{\Gamma}^{a},\bs{\Gamma}^{b}\right]^{\q{\bs{\alpha}}}\tief{\q{\bs{\beta}}} & \equiv & \bs{\Gamma}^{a}\hoch{\q{\bs{\alpha}}}\tief{\q{\bs{\gamma}}}\bs{\Gamma}^{b}\hoch{\q{\bs{\gamma}}}\tief{\q{\bs{\beta}}}+\bs{\Gamma}^{b}\hoch{\q{\bs{\alpha}}}\tief{\q{\bs{\gamma}}}\bs{\Gamma}^{a}\hoch{\q{\bs{\gamma}}}\tief{\q{\bs{\beta}}}=\\
 & = & \sum_{\q{\bs{\gamma}}}\underbrace{(-)^{\q{\bs{\gamma}}+\q{\bs{\gamma}}\bs{\Gamma}}}_{1}\left(\bs{\Gamma}^{a}\hoch{\q{\bs{\alpha}}}\tief{\q{\bs{\gamma}}}\bs{\Gamma}^{b}\hoch{\q{\bs{\gamma}}}\tief{\q{\bs{\beta}}}+\bs{\Gamma}^{b}\hoch{\q{\bs{\alpha}}}\tief{\q{\bs{\gamma}}}\bs{\Gamma}^{a}\hoch{\q{\bs{\gamma}}}\tief{\q{\bs{\beta}}}\right)=\\
 & = & 2\eta^{ab}\underbrace{\left(-\delta^{\q{\bs{\alpha}}}\tief{\q{\bs{\beta}}}\right)}_{\delta_{\beta}^{\alpha}}\end{eqnarray}
The algebra thus changes the sign. It would not do so, however, if
we would grade only the indices and not the rumpfs. In any case, in
appendix \vref{app:gamma} we took the conventional point of view
of ordinary gamma-matrices with ungraded indices, because people are
more familiar with the equations in the conventional picture. For
our application to the Berkovits string in the second part of this
thesis, it is then necessary to make a grading-shift in the indices
to get the correct equations. However, for future applications in
superspace it might be more favourable to have all the equations in
the graded picture with graded rumpfs and indices. In this picture
it would also be more natural (though it was not done in this thesis)
to adjust the definition of the antisymmetrized products of gamma
matrices according to the graded summation. E.g. $\Gamma^{a_{1}a_{2}\,\q{\bs{\alpha}}}\tief{\q{\bs{\beta}}}\grequiv\bs{\Gamma}^{[a_{1}|\,\q{\bs{\alpha}}}\tief{\q{\bs{\gamma}}}\bs{\Gamma}^{|a_{2}]\,\q{\bs{\gamma}}}\tief{\q{\bs{\beta}}}$
with the graded summation convention and the graded equal sign instead
of the ordinary ones.

\chapter{Other Applications and Some Subtleties}

\section{Left and right derivative}

\subsection*{Bosonic rumpfs}

In the bosonic case we have for a variation of some function\begin{eqnarray}
\delta f(x) & = & \delta x^{m}\partiell{}{x^{m}}f=\underbrace{f\partr{x^{m}}}_{\partial f/\partial x^{m}}\delta x^{m}\end{eqnarray}
There is no difference between left and right derivative here, except
that we write it either on the left or on the right of the function.
\begin{eqnarray}
\partiell{}{x^{m}}f & = & \partial f/\partial x^{m}\end{eqnarray}
For the graded case with bosonic rumpfs, the situation is very similar.
We define (using graded summation; no need for graded equal in the
beginning, as there are no naked indices, but in the third equation
it is essential)\index{left derivative}\index{right derivative}\index{derivative!left- and right $\sim$}\index{$\partial_{M}(\ldots)\equiv\partiell{(\ldots)}{x^{M}}\equiv\partiell{}{x^{M}}(\ldots)$|itext{left derivative}}\index{$\partial$@$(\ldots)\partr{x^{K}}\equiv \partial(\ldots)/\partial{x^{K}}$|itext{right derivative}}\begin{eqnarray}
\delta f(x)\equiv_{g}\delta x^{M}\partiell{}{x^{M}}f & \equiv_{g} & \partial f/\partial x^{M}\delta x^{M}\\
\dann\quad0 & =_{g} & \delta x^{M}\left(\partiell{}{x^{M}}f-\partial f/\partial x^{M}\right)\\
\dann\quad\partiell{}{x^{M}}f & =_{g} & \partial f/\partial x^{M}\quad\iff\quad\partiell{}{x^{M}}f=(-)^{fM}\partial f/\partial x^{M}\end{eqnarray}
For $f=x^{M}$ we have\begin{eqnarray}
\delta x^{M} & = & \delta x^{K}\partiell{}{x^{K}}x^{M}=\partial x^{M}/\partial x^{K}\delta x^{K}\\
\dann\partiell{}{x^{K}}x^{M} & = & \delta_{K}\hoch{M}\\
\partial x^{M}/\partial x^{K} & = & \delta^{M}\tief{K}\end{eqnarray}
In the case of coordinates with bosonic rumpf, we will also use the
following symbols for derivatives\begin{eqnarray}
\partial_{M}f & \equiv & \frac{\quad\partial f}{\partial x^{M}\quad}\equiv\partiell{}{x^{M}}f\\
T_{MN,K} & \equiv & T_{MN}\partr{x^{K}}\equiv\partial T_{MN}/\partial x^{K}=(-)^{K(T+M+N)}\partial_{K}T_{MN}\end{eqnarray}
We will not use the notation $\partial_{M}$ for derivatives with
respect to ghosts or other objects with rumpf of odd or undetermined
grading, as the rumpf becomes invisible.

\subsection*{Graded rumpfs}

\label{sub:Graded-rumpf-derivative}For fermionic indices $\bs{\alpha}$
the above equations imply\begin{eqnarray}
\partiell{}{x^{\bs{\alpha}}}f & = & (-)^{f}\partial f/\partial x^{\bs{\alpha}}\label{eq:like-in}\\
\partiell{}{x^{\bs{\alpha}}}x^{\bs{\beta}} & = & -\partial x^{\bs{\beta}}/\partial x^{\bs{\alpha}}=\delta_{\bs{\alpha}}\hoch{\bs{\beta}}\end{eqnarray}
This would for fermionic objects $\ce$ without indices also suggest
to define left and right derivative such that\begin{eqnarray}
\partiell{}{\ce}\ce & \stackrel{?}{\equiv} & -\partial\ce/\partial\ce\end{eqnarray}
However, written without indices it is less intuitive and also not
common. We thus follow the literature and use the following definition
of \textbf{left derivative} and \textbf{right derivative} (now for
$c$ being of undetermined grading $\abs{c}$)\textbf{}\begin{eqnarray}
\delta F(c) & \equiv & \delta c\partiell{}{c}F(c)\equiv\partial F(c)/\partial c\,\delta c\label{eq:variation}\\
\partiell{}{c}F(c) & = & (-)^{c}(-)^{Fc}\partial F(c)/\partial c\label{eq:strange}\\
\partiell{}{c}c & = & \partial c/\partial c=1\label{eq:nice}\end{eqnarray}
Although (\ref{eq:variation}) and (\ref{eq:nice}) seem to be quite
intuitive, (\ref{eq:strange}) unfortunately is less intuitive. The
factor $(-)^{Fc}$ is expected, because we interchange the order of
$F$ and the derivative with respect to $c$. This factor could be
absorbed by using the big graded equal sign. The extra factor $(-)^{c}$,
however, stems from the fact that in (\ref{eq:variation}) the order
of $\partial/\partial c$ and $\delta c$ is exchanged, and the big
graded equal sign cannot figure that out, so that (\ref{eq:strange})
becomes $\partiell{}{c}F(c)=_{G}^{?}(-)^{c}\partial F(c)/\partial c$.
Thus for graded rumpfs, left and right derivative are simply not the
same operation (just written in a different order), but they differ
by a sign depending on the grading of the rumpf. The above definition
is thus not simply a gradifcation of a bosonic one. Indeed the rumpf
'$c$' was not gradifiable from the beginning. If one wants to use
statements derived via the theorem, one has to introduce an extra
index which carries the grading, like in (\ref{eq:like-in}). 

The generalization to the case with graded indices, however, is straight-forward
again\index{$\partiell{}{\ce^{M}}(\ldots)\equiv \partiell{(\ldots)}{\ce^{M}}$|itext{left derivative}}\index{$\partial$@$(\ldots)\partr{\ce^M}\equiv\partial(\ldots)/\partial{\ce^M}$|itext{right derivative}}:
\vspace{-.5cm}\\
 \begin{tabular}{cc}
\begin{minipage}[t][1\totalheight]{0.4\columnwidth}%
\begin{eqnarray*}
\hspace{-1cm}\partiell{}{c^{K}}F(c) & =_{g} & (-)^{c}(-)^{Fc}\partial F(c)/\partial c^{K}\\
\hspace{-1cm}\partiell{}{c^{M}}c^{N} & =_{g} & \delta_{M}\hoch{N}\\
\hspace{-1cm}\partial c^{M}/\partial c^{N} & =_{g} & \delta^{M}\tief{N}=_{g}\delta_{M}\hoch{N}\\
\hspace{-1cm}\partial c^{M}/\partial c^{N} & =_{g} & \partiell{}{c^{N}}c^{M}\end{eqnarray*}
\end{minipage}%
 & %
\begin{minipage}[t][1\totalheight]{0.56\columnwidth}%
\begin{eqnarray}
\partiell{}{c^{K}}F(c) & = & (-)^{FK}(-)^{c+cF}\partial F(c)/\partial c^{K}\qquad\qquad\\
(-)^{cM}\partiell{}{c^{M}}c^{N} & = & \delta_{M}\hoch{N}\quad\left(\stackrel{NW}{=}\delta_{M}^{N}\right)\\
(-)^{cM}\partial c^{M}/\partial c^{N} & = & \delta^{M}\tief{N}\\
(-)^{cM}\partial c^{M}/\partial c^{N} & = & (-)^{cN+NM}\partiell{}{c^{N}}c^{M}\end{eqnarray}
\end{minipage}%
\tabularnewline
\end{tabular}\vspace{.3cm}\\
This implies (using as always the graded summation convention) \begin{equation}
\delta F(c)=\delta c^{K}\partiell{}{c^{K}}F(c)=\partial F(c)/\partial c^{K}\,\delta c^{K}\end{equation}

\section{Tensor and wedge product }

Let us consider the wedge product \begin{equation}
\de x^{m}\de x^{n}\equiv\de x^{m}\wedge\de x^{n}\equiv\frac{1}{2}\left(\de x^{m}\otimes\de x^{n}-\de x^{n}\otimes\de x^{m}\right)\end{equation}
(The normalization $\tfrac{1}{2}$ implies that $p$-forms are written
as $\omega^{(p)}=\omega_{m_{1}\ldots m_{p}}\de x^{m_{1}}\cdots\de x^{m_{p}}$
without the usual prefactor $\tfrac{1}{p!}$.) The wedge product is
antisymmetric if $x^{m}$ are the coordinates of a bosonic manifold.
If one considers $\de x^{m}$ to be an odd object (w.r.t. the form
grading), the wedge product is a graded commuting product. As $x^{m}$
itself is even, the grading has to sit in '$\de\,$', and it is therefore
printed boldface. The form grading is a priori independent from the
Fermion grading but one can consistently combine them to have only
a single $\mathbb{Z}_{2}$ grading, where e.g. an odd differential
form which is at the same time Fermionic is considered to be even.
We will take exactly this point of view throughout the thesis, although
one should keep in mind that it is especially fitted to the exterior
algebra of forms. One can certainly define a symmetrized tensor product
as well, for which it would be more natural to consider $dx^{m}$
as an even object. However, it plays a less important role than the
wedge product. As argued already in the very beginning, it does not
really matter which point of view one takes, as the use of graded
equal sign and graded summation convention swallows all of the signs
anyway. One can therefore do all of the calculations without fixing
this issue and only in the end choose one or another version of graded
summation or graded equal sign.

Let us now consider some tensor of rank $(2,1)$:\begin{equation}
T^{(2,1)}=T_{mn}\hoch{k}\de x^{m}\otimes\de x^{n}\otimes\pe_{k}\end{equation}
Already before bringing any Fermion-grading into the game, we have
a graded equation which should match our philosophy of notations.
The grading on both sides is $\abs{T^{(2,1)}}=\abs{T}+2\abs{\de}+\abs{\pe}$.
It is therefore essential that we do not denote the tensor simply
by $T$, because then the tensor $T$ is odd while the rumpf $T$
is even which would lead to confusions. The superscript '(2,1)' therefore
should carry the grading $2\abs{\de}+\abs{\pe}$ of the basis elements.
Although we might not always write this superscript, it is always
understood that $\abs{T}$ is the grading of the rumpf and not of
the tensor.

All the indices in the above equation are dummy indices and are thus
gradifiable. The rumpf $T$ appears in every term exactly once (with
the above explanation) and is thus gradifiable as well. The rumpf
$x$, instead, is not gradifiable. The gradification of the tensor
definition reads \\
\begin{tabular}{cc}
\begin{minipage}[c][1\totalheight][t]{0.43\columnwidth}%
\[
\hspace{-.5cm}T^{(2,1)}\Greq T_{MN}\hoch{K}\de x^{M}\otimes\de x^{N}\otimes\pe_{K}\]
\end{minipage}%
 & %
\begin{minipage}[c][1\totalheight][t]{0.5\columnwidth}%
\begin{eqnarray}
T^{(2,1)} & \stackrel{NW}{=} & \sum_{M,N,K}(-)^{M+N}(-)^{M(N+K)+NK}(-)^{M\de+K\pe}\times\nonumber \\
 &  & \times T_{MN}\hoch{K}\de x^{M}\otimes\de x^{N}\otimes\pe_{K}\end{eqnarray}
\end{minipage}%
\tabularnewline
\end{tabular}\\
A two form e.g. takes the following form: \begin{eqnarray}
\omega^{(2)} & \equiv & \omega_{MN}\de x^{M}\wedge\de x^{N}\stackrel{NW}{=}\sum_{M,N}(-)^{MN+N}\omega_{MN}\de x^{M}\wedge\de x^{N}\end{eqnarray}
The grading of a p-form $\omega^{(p)}$ is $\abs{\omega^{(p)}}=\abs{\omega}+p$
and the graded Leibniz rule for the exterior derivative acting on
the wedge product $\omega^{(p)}\eta^{(q)}\equiv\omega^{(p)}\wedge\eta^{(q)}$
thus reads\\
\begin{tabular}{cc}
\begin{minipage}[c][1\totalheight][t]{0.46\columnwidth}%
\[
\zwek{}{\de(\omega^{(p)}\eta^{(q)})\Greq\de\omega^{(p)}\eta^{(q)}+\omega^{(p)}\de\eta^{(q)}}\]
\end{minipage}%
 & %
\begin{minipage}[c][1\totalheight][t]{0.5\columnwidth}%
\begin{equation}
\de(\omega^{(p)}\eta^{(q)})=\de\omega^{(p)}\eta^{(q)}+(-)^{\abs{\omega}+p}\omega^{(p)}\de\eta^{(q)}\end{equation}
\end{minipage}%
\tabularnewline
\end{tabular}

\paragraph{A subtle counterexample\index{counterexample} to the theorem}

\label{par:subtle-counterex}Gradification of the exterior algebra
is subtle, because we start with something anticommuting and turn
it in something commuting, which is less restrictive. One of the problems
one meets is the observation that there is no gradification of the
definition of the epsilon tensor, which provides the volume form in
the bosonic case. The more severe problem is the related to the nilpotency
of 1-forms:

We start from the gradifiable anticommutativity equation $\de x^{m_{1}}\de x^{m_{2}}=-\de x^{m_{1}}\de x^{m_{2}}$
(the indices are gradifiable) and the gradifiable definition of the
dimension $d\equiv\delta_{m}\hoch{m}$. In the bosonic case it follows
that $\de x^{m_{1}}\cdots\de x^{m_{d+1}}=0$. Also this last equation
is gradifiable in the indices but is wrong in the general graded case
and thus seems to contradict our theorem. But the theorem includes
also intermediate equations into the gradification. In the above case,
the reasoning goes from $\de x^{m_{1}}\de x^{m_{2}}=-\de x^{m_{1}}\de x^{m_{2}}$
via $\de x^{m}\de x^{m}=0$ (no sum) to the conclusion $\de x^{m_{1}}\cdots\de x^{m_{d+1}}=0$.
In the intermediate equation $\de x^{m}\de x^{m}=0$, the index $m$
is not gradifiable. 

Originally there was the hope that intermediate equations are irrelevant.
In particular, if all indices are fermionic, the dimension is negative.
The condition $\de x^{\mu_{1}}\cdots\de x^{\mu_{d+1}}=0$ then simply
would not be a restriction and everything is fine. For mixed fermionic
and bosonic variables, however, this mechanism breaks down.

It might be that including intermediate equations in the gradification
can be omitted by saying that an index is only gradifiable if the
number of copies in which it appears does not exceed the dimension.
We leave this for future studies.

\rem{

\section{Differential calculus}

\subsection{Tensor product}

The standard example for a Grassman algebra is the exterior algebra
of one-forms, where the basis-elements $\de x^{m}$ anticommute. However,
whether they anticommute or not depends not only on their form grading
but also on the product which we use. Actually we could use a commuting
product as well.

Products of degree $n$:

Graded algebra: tensor algebra (for the beginning of tensors with
upper indices, i.e. elements of $\bigoplus_{r=0}^{\infty}\otimes_{k=1}^{r}T^{*}$,
Grading: rank.\\
Consider e.g. variations of products of tensors, when we vary \begin{eqnarray}
\omega & \To & \omega+\delta\omega\\
\eta & \To & \eta+\delta\eta\end{eqnarray}
We allow for $\delta\omega$ and $\delta\eta$ to be of arbitrary
rank. For the tensor product we then have have \begin{eqnarray}
\delta(\omega\otimes\eta) & = & \left(\omega+\delta\omega\right)\otimes\left(\eta+\delta\eta\right)-\omega\otimes\eta=\\
 & = & \omega\otimes\delta\eta+\delta\omega\otimes\eta+\underbrace{\delta\omega\otimes\delta\eta}_{\To0}\\
\delta(\omega\otimes\eta) & = & \delta\omega\otimes\eta+\omega\otimes\delta\eta\end{eqnarray}
Or more general\begin{eqnarray}
\delta(\omega_{1}\otimes\ldots\otimes\omega_{k}) & = & \sum_{i=1}^{k}\omega_{1}\otimes\ldots\otimes\omega_{i-1}\otimes\delta\omega_{i}\otimes\omega_{i+1}\otimes\ldots\otimes\omega_{k}\end{eqnarray}
The tensor product thus doesn't care for the rank (grading) of the
variation. 

Consider also the covariant derivative in the direction $v$\begin{eqnarray}
D_{v}\omega & \equiv & v^{k}\nabla_{k}\omega_{m_{1}\ldots m_{k}}\de x^{m_{1}}\otimes\ldots\otimes\de x^{m_{p}}=\\
 & = & \de x^{k}(v)\nabla_{k}\omega_{m_{1}\ldots m_{k}}\de x^{m_{1}}\otimes\ldots\otimes\de x^{m_{p}}=\\
 & = & \nabla_{k}\omega_{m_{1}\ldots m_{k}}\de x^{k}\otimes\de x^{m_{1}}\otimes\ldots\otimes\de x^{m_{p}}(v,-,-,\ldots)\\
D\omega & \equiv & \nabla_{k}\omega_{m_{1}\ldots m_{p}}\de x^{k}\otimes\de x^{m_{1}}\otimes\ldots\otimes\de x^{m_{p}}\\
D\eta & = & \nabla_{l}\eta_{n_{1}\ldots n_{q}}\de x^{l}\otimes\de x^{n_{1}}\otimes\ldots\otimes\de x^{n_{q}}\\
D(\omega\otimes\eta) & = & \nabla_{m_{1}}\left(\omega_{m_{2}\ldots m_{p+1}}\eta_{m_{p+2}\ldots m_{p+q+1}}\right)\de x^{m_{1}}\otimes\ldots\otimes\de x^{m_{p+q+1}}\\
D\omega\otimes\eta+\omega\otimes D\eta & = & \nabla_{m_{1}}\omega_{m_{2}\ldots m_{p+1}}\eta_{m_{p+2}\ldots m_{p+q+1}}\de x^{m_{1}}\otimes\ldots\otimes\de x^{m_{p+q+1}}+\nonumber \\
 &  & +\omega_{m_{1}\ldots m_{p}}\nabla_{m_{p+1}}\eta_{m_{p+2}\ldots m_{p+q+1}}\de x^{m_{1}}\otimes\ldots\otimes\de x^{m_{p+q+1}}\\
D(\omega\otimes\eta) & \neq & D\omega\otimes\eta+\omega\otimes D\eta\end{eqnarray}
?Tensor product not invariant under coordinate trafos?$\omega'\otimes\eta'\neq\omega\otimes\eta?$\begin{eqnarray}
v^{m}\de x^{n} & = & \de x^{m}\otimes\de x^{n}(v,-)=\de x^{n}\otimes\de x^{m}(-,v)\end{eqnarray}

Now we introduce the antisymmetric wedge product\begin{eqnarray}
\de x^{m}\de x^{n}\equiv\de x^{m}\wedge\de x^{n} & \equiv & \de x^{m}\otimes\de x^{n}-\de x^{n}\otimes\de x^{m}\\
\de x^{m_{1}}\wedge\ldots\wedge\de x^{m_{r}} & \equiv & \sum_{P\in S_{r}}\sgn P\de x^{m_{P(1)}}\otimes\ldots\otimes\de x^{m_{P(r)}}\\
\omega & \equiv & \omega_{m_{1}\ldots m_{p}}\de x^{m_{1}}\otimes\ldots\otimes\de x^{m_{p}}=\frac{1}{p!}\omega_{m_{1}\ldots m_{p}}\de x^{m_{1}}\wedge\ldots\wedge\de x^{m_{p}}+\textrm{other stuff}\\
\eta & \equiv & \eta_{n_{1}\ldots n_{q}}\de x^{n_{1}}\otimes\ldots\otimes\de x^{n_{q}}\\
\omega\wedge\eta & = & \frac{1}{p!q!}\omega_{m_{1}\ldots m_{p}}\eta_{m_{p+1}\ldots m_{p+q}}\sum_{P\in S_{r}}\sgn P\de x^{m_{P(1)}}\otimes\ldots\otimes\de x^{m_{P(p+q)}}\\
\omega\wedge\eta & = & (-)^{pq}\eta\wedge\omega\end{eqnarray}
$\omega$ and $\eta$ do not necessarlily be forms (antisymmetric),
but their wedge-product will be! Consider now a variation $\delta$of
definite degree $n$ (general variations are sums of variations of
definite degree), i.e. \begin{eqnarray}
\deg(\delta\omega) & = & \deg(\omega)+n\\
\deg(\delta\eta) & = & \deg(\eta)+n\end{eqnarray}
\begin{eqnarray}
\delta\omega & = & (\delta\omega)_{m_{1}\ldots m_{p+n}}\de x^{m_{1}}\otimes\ldots\otimes\de x^{m_{p+n}}\\
\delta\eta & = & (\delta\eta)_{n_{1}\ldots n_{q+n}}\de x^{n_{1}}\otimes\ldots\otimes\de x^{n_{q+n}}\end{eqnarray}
Example for 1-forms:\begin{eqnarray*}
\omega & \equiv & \omega^{(1)}+\omega^{(2)}=\omega^{(1)}\\
\delta\omega & = & \delta\omega^{(2)}\\
\omega' & = & \omega^{(1)}+\delta\omega^{(2)}\\
\omega\wedge\eta & = & \omega^{(1)}\wedge\eta^{(1)}+\underbrace{\omega^{(2)}\wedge\eta^{(1)}+\omega^{(1)}\wedge\eta^{(2)}+\omega^{(2)}\wedge\eta^{(2)}}_{=0}\\
\delta\left(\omega\wedge\eta\right) & = & \delta\omega^{(2)}\wedge\eta^{(1)}+\omega^{(1)}\wedge\delta\eta^{(2)}\end{eqnarray*}
\begin{eqnarray}
\delta(\omega_{m}\de x^{m}) & = & \frac{1}{2}\alpha_{m_{1}m_{2}}\de x^{m_{1}}\wedge\de x^{m_{2}}\\
\omega' & = & \omega_{m}\de x^{m}+\frac{1}{2}\alpha_{m_{1}m_{2}}\left(\de x^{m_{1}}\otimes\de x^{m_{2}}-\de x^{m_{2}}\otimes\de x^{m_{1}}\right)\\
\eta' & = & \eta_{n}\de x^{n}+\frac{1}{2}\beta_{n_{1}n_{2}}\left(\de x^{n_{1}}\otimes\de x^{n_{2}}-\de x^{n_{2}}\otimes\de x^{n_{1}}\right)\end{eqnarray}
\begin{eqnarray*}
\end{eqnarray*}
\begin{eqnarray*}
\delta\left(\omega\wedge\eta\right) & = & \frac{1}{p!q!}\omega_{m_{1}\ldots m_{p}}\eta_{m_{p+1}\ldots m_{p+q}}\sum_{P\in S_{r}}\sgn P\de x^{m_{P(1)}}\otimes\ldots\otimes\de x^{m_{P(p+q)}}\end{eqnarray*}
\begin{eqnarray*}
\de x^{m}\swedge\de x^{n} & \equiv & \de x^{m}\otimes\de x^{n}+\de x^{n}\otimes\de x^{m}\end{eqnarray*}

\subsection{graded p-forms}

Graded forms are again a very subtle example, especially the wedge-product.
Take for example the bosonic variables $x^{m}$\begin{eqnarray}
\de x^{m}\de x^{n}\equiv\de x^{m}\wedge\de x^{n} & \equiv & \de x^{m}\otimes\de x^{n}-\de x^{n}\otimes\de x^{m}\label{eq:bosonicWedge}\end{eqnarray}
If we want to apply our theorem later on the definition of a graded
version of this equation, we have to make sure that it has a valid
gradification. The problem is, however, that already in the {}``bosonic
equation'' we have the graded rumpf $\de x$, and it appears twice
in every term which violates the conditions of the theorem for a valid
gradification.\rem{Why? Indices should still be gradifiable!?} One
solution is to treat de Ram grading and fermion grading seperately.
Then (\ref{eq:bosonicWedge}) has no graded rumpfs and we can make
the indices graded and use the graded equal sign to define $\de x^{M}\wedge\de x^{N}\equiv_{g}\de x^{M}\otimes\de x^{N}-\de x^{N}\otimes\de x^{M}$,
which reads explicitely $\de x^{M}\wedge\de x^{N}\equiv\de x^{M}\otimes\de x^{N}-(-)^{MN}\de x^{N}\otimes\de x^{M}$.
Actually, for a gradification of the above equation, the graded equal
sign lets us no other choice! 

Instead we can look for a better bosonic starting point, which still
admits to treat both gradings together. To this end we take two arbitrary
one-forms $\omega$ and $\eta$ (both odd in the bosonic case) and
take as starting point\begin{eqnarray}
\omega\wedge\eta & \equiv & \omega\otimes\eta-\eta\otimes\omega=\label{eq:bosonicWedgeII}\\
 & = & \omega\otimes\eta+(-)^{\eta\omega}\eta\otimes\omega\end{eqnarray}
to arrive at  \\
\begin{tabular}{cc}
\begin{minipage}[c][1\totalheight][t]{0.47\columnwidth}%
\begin{eqnarray}
\omega\wedge\eta & \equiv_{G} & \omega\otimes\eta+\eta\otimes\omega\end{eqnarray}
\end{minipage}%
 & %
\begin{minipage}[c][1\totalheight][t]{0.5\columnwidth}%
\begin{eqnarray}
\omega\wedge\eta & \equiv & \omega\otimes\eta+(-)^{\eta\omega}\eta\otimes\omega\end{eqnarray}
\end{minipage}%
\tabularnewline
\end{tabular}\\
which implies \begin{equation}
\de x^{M}\wedge\de x^{N}=(-)^{(N+1)(M+1)}\de x^{N}\wedge\de x^{M}\end{equation}
as a special case. Note that this last equation does not look nice
any longer with the big graded equal sign, as the rumpfs $\de x$
are not distinguished any longer. The reason is that is is a special
case and corresponds to the coinciding indices treated in section
\vref{sub:Problem-of-coinciding}. The somewhat disturbing point is
that the 'special case' is actually quite general, as it is the wedge
product among basis elements of the cotangent space. And also in the
formulation above with $\omega$ and $\eta$, the graded equal sign
(if rigorously applied) actually always looks at the ordering of the
elementary objects, which are again the $\de x$. This suggests that
we cannot apply our theorem, except we use the alternative to seperate
De-Ram and fermion grading...

In any case we have

\begin{eqnarray}
\omega & \equiv & \frac{1}{p!}\omega_{M_{1}\ldots M_{p}}\de x^{M_{1}}\ldots\de x^{M_{p}}\\
\de\omega & = & \frac{1}{p!}\de x^{M_{0}}\partial_{M_{0}}\omega_{M_{1}\ldots M_{p}}\de x^{M_{1}}\ldots\de x^{M_{p}}=\\
 & = & \frac{1}{p!}\partial_{[M_{1}}\omega_{M_{2}\ldots M_{p+1}]}\de x^{M_{1}}\ldots\de x^{M_{p+1}}\end{eqnarray}
}\rem{

\section{Other}

\subsection{graded OPEs}

see conventions.lyx

\subsection{Index-subsets with definite grading}

\subsection{Lie-Supergroups}

\subsection{De-Witt notation}

We do nowhere in the thesis (except here) use De Witt's notation of
writing some indices to the left, because a lot of people are kind
of afraid to use it. However, those who like it, can perfectly fit
it into our whole scheme. Just define e.g.\\
\begin{tabular}{cc}
\begin{minipage}[c][1\totalheight][t]{0.47\columnwidth}%
\begin{eqnarray}
_{A}b & \equiv_{G} & b_{A}\end{eqnarray}
\end{minipage}%
 & %
\begin{minipage}[c][1\totalheight][t]{0.47\columnwidth}%
\begin{eqnarray}
_{A}b & \equiv & (-)^{Ab}b_{A}\end{eqnarray}
\end{minipage}%
\tabularnewline
\end{tabular}\\
That's all.}

\section{Graded Poisson bracket}

\label{sec:graded-Poisson-bracket}\index{bracket!Poisson $\sim$}\index{graded Poisson bracket}\index{Poisson bracket!graded $\sim$}For
bosonic rumpfs '$q$' and '$p$' of the phase space variables $q^{M}$
and $p_{M}$, the bosonic Poisson bracket is easily generalized to
the graded case. The overall sign, i.e. whether one first takes the
derivative with respect to the momenta $p_{M}$ and then with respect
to the configuration space variables $q^{M}$ or the other way round
is already an ambiguity at the bosonic level and is only a matter
of taste. As it is just an overall sign, it is easily changed if preferred
differently. Our choice ($p_{M}$ first) was made in order to have
the Hamiltonian as the generator of time translations on the left
of the bracket. We always try to let generators or operators act from
the left. In any case the graded Poisson bracket is a simple gradification
of the bosonic one:\begin{eqnarray}
\left\{ F,G\right\}  & \equiv & \partial F/\partial p_{M}\partiell{}{q^{M}}G-\partial F/\partial q^{M}\partiell{}{p_{M}}G=\\
 & = & \partial F/\partial p_{M}\partiell{}{q^{M}}G-(-)^{FG}\partial G/\partial p_{M}\partiell{}{q^{M}}F=\\
 & = & \partiell{}{p_{M}}F\partiell{}{q^{M}}G-\partiell{}{q^{M}}F\partiell{}{p_{M}}G\\
\left\{ F,G\right\}  & = & -(-)^{FG}\left\{ G,F\right\} \\
\left\{ p_{M},q^{N}\right\}  & = & \delta_{M}\hoch{N}\stackrel{NW}{=}\delta_{M}^{N}\\
\left\{ q^{M},p_{N}\right\}  & = & -\delta^{M}\tief{N}\stackrel{NW}{=}-(-)^{M}\delta_{M}^{N}\end{eqnarray}
Like always, the sum over the index '$M$' has to be understood as
graded sum. The left and right-derivative with respect to variables
with bosonic rumpfs coincide (w.r.t. the graded equal sign) and the
generalization is therefore unique, as soon as the underlying summation
convention (NW or NE) is chosen. The sign $(-)^{FG}$ in the second
and fourth line of the above equation array would disappear upon the
use of the big graded equal sign. The rumpfs '$q$' and '$p$' are
a priori not gradifiable in these equations. 

Nevertheless the \emph{case of graded rumpfs} '$q$' and '$p$' can
be covered by just gradifying the indices. Assume for example that
we have in addition to $q^{M}$ and $p_{M}$ (with bosonic rumpfs)
also some ghost variables $\ce^{M}$ and $\be_{M}$ with the same
indices. In general, the indices of ghost variables would just cover
a subset of the index range of the original phase space, but this
subtlety does not matter for the present discussion. The rumpfs of
the ghost variables carry a grading and it is thus not uniquely fixed
how to extend the definition of the Poisson bracket to the ghost variables.
A natural way (having in mind the conditions for our theorem) is to
introduce some variables with two indices $z^{iM}$ containing $q^{M}$
as well as $\ce^{M}$ and the same for the momenta: \begin{eqnarray}
z^{iM}\equiv(q^{M},\ce^{M}),\quad z^{1M} & = & q^{M},\quad z^{\bs{2}M}=\ce^{M}\\
\pi_{iM}\equiv(p_{M},\be_{M}),\quad\pi_{1M} & = & p_{M},\quad\pi_{\bs{2}M}=\be_{M}\end{eqnarray}
The grading is now sitting in the additional index $i$, i.e. $\abs{i}=\left\{ \zwek{0\mbox{ for }i=1}{1\mbox{ for }i=2}\right.$.
One still has the freedom to decide whether this index should be upstairs
or downstairs for $z$ or equivalently whether we choose NW or NE
for the graded summation of this index. Choosing the position as above
and NW for the summation yields \begin{eqnarray}
z^{iM}\pi_{iM} & = & \sum_{i,M}(-)^{iM}z^{iM}\pi_{iM}=\sum_{i,M}\left(q^{M}p_{M}+(-)^{M}\ce^{M}\be_{M}\right)=q^{M}p_{M}+\ce^{M}\be_{M}\\
\pi_{iM}z^{iM} & = & \sum_{i,M}(-)^{iM+i+M}\pi_{iM}z^{iM}=\sum_{i,M}\left((-)^{M}p_{M}q^{M}-\be_{M}\ce^{M}\right)=p_{M}q^{M}-\be_{M}\ce^{M}\end{eqnarray}
Note the sign change of the last term from the first to the second
line. Now we can also write down the graded Poisson bracket for this
case, which looks in terms of the variables $(z^{iM},\pi_{iM})$ the
same as the one before in terms of ($q^{M},p_{M}$), but contains
an additional graded sum over the index $i$:\begin{eqnarray}
\left\{ F,G\right\}  & \equiv & \partial F/\partial\pi_{iM}\partiell{}{z^{iM}}G-\partial F/\partial z^{iM}\partiell{}{\pi_{iM}}G=\\
 & \stackrel{NW}{=} & \sum_{i,M}(-)^{iM}\partial F/\partial\pi_{iM}\partiell{}{z^{iM}}G-(-)^{iM+i+M}\partial F/\partial z^{iM}\partiell{}{\pi_{iM}}G\label{eq:PoissonInNWwithz}\end{eqnarray}
Before we rewrite this Poisson bracket in terms of $q^{M},p_{M},\ce^{M}$
and $\be_{M}$, let us recall the definition of left and right-derivative
of page \pageref{sub:Graded-rumpf-derivative}. With the graded equal
sign, left and right derivative w.r.t. $z^{iM}$ are simply given
by $\partl{z^{iM}}z^{jN}\greq\delta_{i}\hoch{j}\delta_{M}\hoch{N}\greq\partial z^{jN}/\partial z^{iM}$.
The same is true for the derivatives w.r.t. $\pi_{iM}$. Written with
the ordinary equal sign, this reads \begin{eqnarray}
\partl{z^{iM}}z^{jN} & = & (-)^{jM}\underbrace{\delta_{i}\hoch{j}}_{NW:\delta_{i}^{j}}\delta_{M}\hoch{N}=(-)^{(j+N)(i+M)}\partial z^{jN}/\partial z^{iM}\qquad\\
\partl{\pi_{iM}}\pi_{jN} & = & (-)^{jM}\underbrace{\delta^{i}\tief{j}}_{NW:-\delta_{j}^{i}}\delta^{M}\tief{N}=(-)^{(j+N)(i+M)}\partial\pi_{jN}/\partial\pi_{iM}\end{eqnarray}
For $i=j=1$ this agrees perfectly with the definition of left and
right derivative w.r.t. $q^{M}$ or $p_{M}$. For $i=j=2$ instead
(remember $z^{\bs{2}M}=\ce^{M}$ and $\pi_{\bs{2}M}=\be_{M}$), we
observe some mismatch (in NW for the right-derivative w.r.t. $\ce^{M}$
and for the left-derivative w.r.t. $\be_{M}$, in NE the other way
round)\begin{eqnarray}
\partl{\ce^{M}}\ce^{N}=\lqn{(-)^{M}\delta_{M}\hoch{N}=(-)^{M+N+MN}\partial\ce^{N}/\partial\ce^{M}\leftrightarrow}\nonumber \\
 & \leftrightarrow & \partl{z^{\bs{2}M}}z^{\bs{2}N}=(-)^{M}\underbrace{\delta_{\bs{2}}\hoch{\bs{2}}}_{NW:1}\delta_{M}\hoch{N}=-(-)^{N+M+NM}\partial z^{\bs{2}N}/\partial z^{\bs{2}M}\\
\partl{\be_{M}}\be_{N}=\lqn{(-)^{M}\delta^{M}\tief{N}=(-)^{M+N+MN}\partial\be_{N}/\partial\be_{M}\leftrightarrow}\nonumber \\
 & \leftrightarrow & \partl{\pi_{\bs{2}M}}\pi_{\bs{2}N}=(-)^{M}\underbrace{\delta^{\bs{2}}\tief{\bs{2}}}_{NW:-1}\delta^{M}\tief{N}=-(-)^{M+N+MN}\partial\pi_{\bs{2}N}/\partial\pi_{\bs{2}M}\end{eqnarray}
The definition of left and right derivative therefore depends on the
notation we use ($\ce^{M},\be_{M}$ or $z^{\bs{2}M},\pi_{\bs{2}M}$).
In NW-conventions (for the index $i$) we have\begin{eqnarray}
\partl{\ce^{M}} & \stackrel{NW}{=} & \partl{z^{\bs{2}M}},\quad\partr{\ce^{M}}\stackrel{NW}{=}-\partr{z^{\bs{2}M}}\\
\partl{\be_{M}} & \stackrel{NW}{=} & -\partl{\pi_{\bs{2}M}},\quad\partr{\be_{M}}\stackrel{NW}{=}\partr{\pi_{\bs{2}M}}\end{eqnarray}
In NE conventions (for the index $i$), we would have the opposite
signs. In the Poisson bracket, these signs always cancel (for NW and
for NE), because the left derivative w.r.t. $\be_{M}$ comes with
the right derivative w.r.t. $\ce^{M}$ and vice verse. Looking at
(\ref{eq:PoissonInNWwithz}) one can see that the only additional
sign which is not absorbed by the graded summation of the index $M$
is the $(-)^{i}$ in the second term due to the 'wrong' contraction
direction. This sign would come with the first term, if we had NE
conventions for the index $i$. The \textbf{Poisson bracket} given
before in terms of $z^{iM}$ and $\pi_{iM}$ can therefore be rewritten
(in graded summation conventions) as\vRam{1.02}{\begin{eqnarray}
\left\{ F,G\right\}  & = & \partial F/\partial p_{M}\partiell{}{q^{M}}G-\partial F/\partial q^{M}\partiell{}{p_{M}}G\pm\left(\partial F/\partial\be_{M}\partiell{}{\ce^{M}}G+\partial F/\partial\ce^{M}\partiell{}{\be_{M}}G\right)=\\
 & = & \partial F/\partial p_{M}\partiell{}{q^{M}}G-(-)^{FG}\partial G/\partial p_{M}\partiell{}{q^{M}}F\pm\left(\partial F/\partial\be_{M}\partiell{}{\ce^{M}}G-(-)^{FG}\partial G/\partial\be_{M}\partiell{}{\ce^{M}}F\right)\qquad\label{eq:PoissonBracketGradedcb}\end{eqnarray}
} The upper sign is for the choice of NW-conventions for the index
$i$ while the lower sign is for NE. This is in principle independent
of the summation convention for the index $M$. If one prefers overall
NE, where the minus in front of the bracket might be annoying, it
might be more natural to define the Poisson bracket with an overall
minus (or take NW only for the index $i$). If one wants to apply
the gradification theorem in order to derive true statements about
the graded Poisson bracket, it is in principle necessary to reintroduce
the extra index $i$ which carries the grading and rewrite the result
again in terms of the graded rumpfs after having applied the theorem.
In practice this is rarely necessary. For example, in order to show
the Jacobi identity for the graded Poisson bracket, it is enough to
know that one can write it as a gradification of a bosonic Poisson
bracket. The Jacobi identity itself does not explicitely contain the
variables $z^{iM}$ and therefore has the same form in terms of the
variables $q^{M}$ and $\ce^{M}$. The same is true for Leibniz rule
when acting on products of phase space functions:\begin{eqnarray}
\left\{ F,\left\{ G,H\right\} \right\}  & = & \left\{ \{F,G\},H\right\} +(-)^{FG}\left\{ G,\{F,H\}\right\} \\
\left\{ F,GH\right\}  & = & \left\{ F,G\right\} H+(-)^{FG}G\left\{ F,H\right\} \end{eqnarray}
The sign $(-)^{FG}$ would disappear when using the big graded equal
sign. Let us now fix the sign-ambiguity in (\ref{eq:PoissonBracketGradedcb}).
We will throughout use the more convenient \emph{upper sign} for the
definition of the Poisson bracket. This implies \begin{eqnarray}
\left\{ F,G\right\}  & = & -(-)^{FG}\left\{ G,F\right\} \\
\left\{ \be_{M},\ce^{N}\right\}  & =_{g} & \delta_{M}\hoch{N},\quad\left\{ p_{M},q^{M}\right\} =_{g}\delta_{M}\hoch{N}\\
\left\{ \ce^{M},\be_{N}\right\}  & =_{g} & \delta^{M}\tief{N},\quad\left\{ q^{M},p_{N}\right\} \greq-\delta^{M}\tief{N}\end{eqnarray}
Note again that this does not fix the summation convention for the
index $M$. We had only made a convenient choice for the auxiliary
index $i$ which is now absent anyway. The above equations further
imply\begin{eqnarray}
\left\{ \be_{M},\ldots\right\}  & = & \partiell{}{\ce^{M}}\left(\ldots\right),\quad\left\{ p_{M},\ldots\right\} =\partiell{}{q^{M}}\left(\ldots\right)\\
\left\{ \ldots,\be_{M}\right\}  & = & \partial\left(\ldots\right)/\partial\ce^{M},\quad\left\{ \ldots,p_{M}\right\} =-\partial\left(\ldots\right)/\partial q^{M}\end{eqnarray}
\begin{eqnarray}
\left\{ \ce^{M},\ldots\right\}  & = & \partiell{}{\be_{M}}\left(\ldots\right),\quad\left\{ q^{M},\ldots\right\} =-\partiell{}{p_{M}}\left(\ldots\right)\\
\left\{ \ldots,\ce^{M}\right\}  & = & \partial(\ldots)/\partial\be_{M},\quad\left\{ \ldots,q^{M}\right\} =\partial(\ldots)/\partial p_{M}\end{eqnarray}
\rem{Instead of introducing the additional index, we could replace
the index $M$ by an index which has the double range:\begin{eqnarray*}
z^{\q{M}} & \equiv & (q^{M},c^{\bs{M}})\equiv(q^{M},\ce^{M})\mbox{ with }\abs{\bs{M}}=\abs{M}+1\\
\pi_{\q{M}} & \equiv & (p_{M},b_{\bs{M}})=(p_{M},\be_{M})\end{eqnarray*}
\begin{eqnarray*}
z^{\q{M}}\pi_{\q{M}} & \stackrel{NE\mbox{ for 2nd}}{\equiv} & \sum_{M}\left(q^{M}p_{M}+(-)^{M+1}c^{\bs{M}}b_{\bs{M}}\right)\equiv q^{M}p_{M}-\ce^{M}\be_{M}\\
\pi_{\q{M}}z^{\q{M}} & \stackrel{NE\mbox{ for 2nd}}{\equiv} & \sum_{M}\left((-)^{M}p_{M}q^{M}+b_{\bs{M}}c^{\bs{M}}\right)=p_{M}q^{M}+\be_{M}\ce^{M}\end{eqnarray*}
\begin{eqnarray*}
\{F,G\} & = & \partial F/\partial\pi_{\q{M}}\partiell{}{z^{\q{M}}}G-\partial F/\partial z^{\q{M}}\partiell{}{\pi_{\q{M}}}G=\\
 & = & \partial F/\partial p_{M}\partiell{}{q^{M}}G-\partial F/\partial q^{M}\partiell{}{p_{M}}G+\partial F/\partial b_{\bs{M}}\partiell{}{c^{\bs{M}}}G-\partial F/\partial c^{\bs{M}}\partiell{}{c_{\bs{M}}}G=\\
 & = & \partial F/\partial p_{M}\partiell{}{q^{M}}G-\partial F/\partial q^{M}\partiell{}{p_{M}}G+\sum_{M}(-)^{M+1}\left(\partial F/\partial b_{\bs{M}}\partiell{}{c^{\bs{M}}}G-\partial F/\partial c^{\bs{M}}\partiell{}{b_{\bs{M}}}G\right)=\\
 & = & \partial F/\partial p_{M}\partiell{}{q^{M}}G-\partial F/\partial q^{M}\partiell{}{p_{M}}G-\partial F/\partial\be_{M}\partiell{}{\ce^{M}}G-\partial F/\partial\ce^{M}\partiell{}{\be_{M}}G\end{eqnarray*}
}

\paragraph{Antibracket}

\label{sub:Antibracket}A bracket which is closely related to the
Poisson bracket is the antibracket\index{antibracket}\index{bracket!anti $\sim$}.
It is defined in an extended configuration space with as many odd
variables (antifields) $\bs{q}_{M}^{+}$ as even variables $q^{M}$:\begin{eqnarray}
\left(F\bs{,}G\right) & = & \partial F/\partial\bs{q}_{M}^{+}\partiell{}{q^{M}}G-(-)^{(F+1)(G+1)}\partial G/\partial\bs{q}_{M}^{+}\partiell{}{q^{M}}F\end{eqnarray}
Note that this bracket is not simply a gradification of the Poisson
bracket. We had discussed before that the rumpfs '$p$' and '$q$'
in the Poisson bracket were not gradifiable but that this problem
can be removed by introducing an auxiliary index. However, this implies
that still $q$ and $p$ have the same parity, while here they have
opposite parity. On the other hand, the above equation can be seen
as the gradification of an antibracket defined for purely bosonic
rumpfs '$F$' and '$G$' and bosonic dummy index $M$. Rewriting it
in terms of the big graded equal sign $\Greq$, the sign $-(-)^{(F+1)(G+1)}$
would get replaced by a $+$ sign. Writing the antibracket without
the big graded equal sign better demonstrates its relation to the
Poisson bracket. In a sense, it behaves as if the gradings of '$F$'
and '$G$' were shifted by 1. The antibracket will be further discussed
at a later point (see e.g. footnote \ref{fn:The-antibracket-looks}
on page \pageref{fn:The-antibracket-looks} or footnote \ref{Lie-bracket-of-degree}
in the appendix on page \pageref{Lie-bracket-of-degree}). \rem{

\section{Grading shifts}

\subsection{Rumpf - index grading shifts}

\label{sec:RumpfIndexShift}Consider a supercommutative object with
a given rumpf-grading and a given index grading. For some reason it
could be interesting to shift the rumpf-grading to the indices or
vice versa. Other objects which are contracted with the variable under
consideration (as for example its conjugate momentum), have to be
shifted in a compensating way. Let's see, what happens:\begin{eqnarray}
\abs{c^{M}} & = & \abs{c}+\abs{M}\\
\abs{b_{M}} & = & \abs{b}+\abs{M}\\
\abs{\tilde{c}} & = & \abs{c}+1,\abs{\tilde{b}}=\abs{b}+1,\abs{\tilde{M}}=\abs{M}+1\\
c^{M}b_{M} & = & \sum_{M}(-)^{Mb}c^{M}b_{M}\\
b_{M}c^{M} & = & \sum_{M}(-)^{Mc+M}b_{M}c^{M}\\
\tilde{c}^{\tilde{M}}\tilde{b}_{\tilde{M}} & = & \sum_{\tilde{M}}(-)^{\tilde{M}\tilde{b}}\tilde{c}^{\tilde{M}}\tilde{b}_{\tilde{M}}=\sum_{\tilde{M}}(-)^{Mb}(-)^{M+b+1}\tilde{c}^{\tilde{M}}\tilde{b}_{\tilde{M}}\\
\tilde{b}_{\tilde{M}}\tilde{c}^{\tilde{M}} & = & \sum_{\tilde{M}}(-)^{\tilde{M}+\tilde{M}\tilde{c}}\tilde{b}_{\tilde{M}}\tilde{c}^{\tilde{M}}=\sum_{\tilde{M}}(-)^{M+1+Mc+M+c+1}\tilde{b}_{\tilde{M}}\tilde{c}^{\tilde{M}}=\sum_{\tilde{M}}(-)^{Mc+M}(-)^{M+c}\tilde{b}_{\tilde{M}}\tilde{c}^{\tilde{M}}\end{eqnarray}
If we define e.g.\begin{eqnarray}
\tilde{b}_{\tilde{M}} & \equiv & (-)^{M+b+1}b_{M}\\
\tilde{c}^{\tilde{M}} & \equiv & c^{M}\end{eqnarray}
we get\begin{eqnarray}
(-)^{cb}b_{M}c^{M}=c^{M}b_{M} & = & \tilde{c}^{\tilde{M}}\tilde{b}_{\tilde{M}}=(-)^{\tilde{b}\tilde{c}}\tilde{b}_{\tilde{M}}\tilde{c}^{\tilde{M}}\\
b_{M}c^{M} & = & (-)^{cb+\tilde{b}\tilde{c}}\tilde{b}_{\tilde{M}}\tilde{c}^{\tilde{M}}=(-)^{b+c+1}\tilde{b}_{\tilde{M}}\tilde{c}^{\tilde{M}}\end{eqnarray}
For the derivatives we get\begin{eqnarray}
\partiell{}{\tilde{c}^{\tilde{M}}}\tilde{c}^{\tilde{N}} & = & (-)^{\tilde{M}\tilde{c}}\delta_{\tilde{M}}\hoch{\tilde{N}}=(-)^{Mc}(-)^{M+c+1}\delta_{M}\hoch{N}\\
\partiell{}{\tilde{c}^{\tilde{M}}} & = & (-)^{M+c+1}\partiell{}{c^{M}}\\
\partial\tilde{c}^{\tilde{M}}/\partial\tilde{c}^{\tilde{N}} & = & (-)^{\tilde{M}\tilde{c}}\delta^{\tilde{M}}\tief{\tilde{N}}=(-)^{\tilde{M}\tilde{c}+\tilde{M}}\delta_{\tilde{N}}\hoch{\tilde{M}}=(-)^{Mc+c}\delta_{N}\hoch{M}=(-)^{Mc}(-)^{M+c}\delta^{M}\tief{N}\\
\partial/\partial\tilde{c}^{\tilde{M}} & \stackrel{!!}{=} & (-)^{M+c}\partial/\partial c^{M}\\
\partiell{}{\tilde{b}_{\tilde{M}}}\tilde{b}_{\tilde{N}} & = & (-)^{\tilde{M}\tilde{b}}\delta^{\tilde{M}}\tief{\tilde{N}}=(-)^{\tilde{M}\tilde{b}+\tilde{M}}\delta_{\tilde{N}}\hoch{\tilde{M}}=(-)^{Mb+b}\delta_{N}\hoch{M}=(-)^{Mb}(-)^{M+b}\delta^{M}\tief{N}\\
\partiell{}{\tilde{b}_{\tilde{M}}} & \stackrel{!!}{=} & -\partiell{}{b_{M}}\\
\partial\tilde{b}_{\tilde{M}}/\partial\tilde{b}_{\tilde{N}} & = & (-)^{\tilde{b}\tilde{M}}\delta_{\tilde{M}}\hoch{\tilde{N}}=(-)^{bM}(-)^{b+M+1}\delta_{M}\hoch{N}\\
\partial/\partial\tilde{b}_{\tilde{M}} & = & \partial/\partial b_{M}\end{eqnarray}
If $b$ and $c$ are conjugate (with $\abs{b}=\abs{c}$), the Poisson-bracket
(we will consider the Poisson-bracket in section \vref{sec:graded-Poisson-bracket})
reads:\begin{eqnarray}
(-)^{cM}\left\{ b_{M},c^{N}\right\}  & = & (-)^{cM}\partial b_{M}/\partial b_{K}\partiell{}{c^{K}}c^{N}-(-)^{cM}(-)^{bc}\partial b_{M}/\partial c^{K}\partiell{}{b_{K}}c^{N}\\
(-)^{cM}\left\{ b_{M},c^{N}\right\}  & = & \delta_{M}\hoch{K}\delta_{K}\hoch{N}=\delta_{M}\hoch{N}\quad(b_{M}\textrm{ corresponds to left derivative})\\
(-)^{bM}\left\{ c^{M},b_{N}\right\}  & = & -(-)^{bc}\delta^{M}\tief{K}\delta^{K}\tief{N}=-(-)^{bc}\delta^{M}\tief{N}\end{eqnarray}
Equivalently \begin{eqnarray}
(-)^{\tilde{M}\tilde{c}}\left\{ \tilde{b}_{\tilde{M}},\tilde{c}^{\tilde{N}}\right\}  & = & \delta_{\tilde{M}}\hoch{\tilde{N}}\\
\left\{ \tilde{b}_{\tilde{M}},\tilde{c}^{\tilde{N}}\right\}  & = & (-)^{M+c+1}\left\{ b_{M},c^{N}\right\} \\
(-)^{\tilde{M}\tilde{c}}\left\{ \tilde{c}^{\tilde{M}},\tilde{b}_{\tilde{N}}\right\}  & = & -(-)^{\tilde{b}\tilde{c}}\delta^{\tilde{M}}\tief{\tilde{N}}=-(-)^{bc+b+c+1}(-)^{\tilde{M}+M}\delta^{M}\tief{N}=-(-)^{bc}(-)^{b+c}\delta^{M}\tief{N}\\
\left\{ \tilde{c}^{\tilde{M}},\tilde{b}_{\tilde{N}}\right\}  & = & (-)^{M+b+1}\left\{ c^{M},b_{N}\right\} \end{eqnarray}
This is in perfect agreement with $\tilde{c}^{\tilde{M}}=c^{M}$ and
$\tilde{b}_{\tilde{M}}=(-)^{M+b+1}b_{M}$! (objects which don't carry
indices or rumpfs - like the bracket - should not depend on the variables
we use to describe them)

The antibracket (with $\abs{b}=\abs{c}+1$ and $\abs{,}=1$) reads
(we will consider the antibracket in more detail in section \vref{sub:Antibracket}):\begin{eqnarray}
(-)^{(c+1)M}\left(b_{M},c^{N}\right) & = & (-)^{(c+1)M}\partial b_{M}/\partial b_{K}\partiell{}{c^{K}}c^{N}-(-)^{(c+1)M}(-)^{bc}\partial b_{M}/\partial c^{K}\partiell{}{b_{K}}c^{N}\\
(-)^{(c+1)M}\left(b_{M},c^{N}\right) & = & \delta_{M}\hoch{K}\delta_{K}\hoch{N}=\delta_{M}\hoch{N}\quad(b_{M}\textrm{ corresponds to left derivative})\\
(-)^{cM}\left(c^{M},b_{N}\right) & = & (-)^{cM}\partial c^{M}/\partial b_{K}\partiell{}{c^{K}}b_{N}-(-)^{cM}(-)^{bc}\partial c^{M}/\partial c^{K}\partiell{}{b_{K}}b_{N}\\
(-)^{cM}\left(c^{M},b_{N}\right) & = & -\delta^{M}\tief{K}\delta^{K}\tief{N}=-\delta^{M}\tief{N}\end{eqnarray}
Equivalently \begin{eqnarray}
(-)^{\tilde{M}(\tilde{c}+1)}\left(\tilde{b}_{\tilde{M}},\tilde{c}^{\tilde{N}}\right) & = & \delta_{\tilde{M}}\hoch{\tilde{N}}\\
\left(\tilde{b}_{\tilde{M}},\tilde{c}^{\tilde{N}}\right) & = & (-)^{(M+1)c+M(c+1)}\left(b_{M},c^{N}\right)=(-)^{c+M}\left(b_{M},c^{N}\right)=(-)^{M+b+1}\left(b_{M},c^{N}\right)\\
(-)^{\tilde{c}\tilde{M}}\left(\tilde{c}^{\tilde{M}},\tilde{b}_{\tilde{N}}\right) & = & -\delta^{\tilde{M}}\tief{\tilde{N}}=-(-)^{\tilde{M}+M}\delta^{M}\tief{N}=\delta^{M}\tief{N}\\
\left(\tilde{c}^{\tilde{M}},\tilde{b}_{\tilde{N}}\right) & = & -(-)^{(c+1)(M+1)+cM}\left(c^{M},b_{N}\right)=(-)^{M+b+1}\left(c^{M},b_{N}\right)\end{eqnarray}

We could also have switched the summation convention to northeast
for the tilde indices:\begin{eqnarray}
\tilde{c}^{\tilde{M}}\tilde{b}_{\tilde{M}} & = & \sum_{\tilde{M}}(-)^{\tilde{M}+\tilde{M}\tilde{b}}\tilde{c}^{\tilde{M}}\tilde{b}_{\tilde{M}}=\sum_{\tilde{M}}(-)^{Mb+b}\tilde{c}^{\tilde{M}}\tilde{b}_{\tilde{M}}\\
\tilde{b}_{\tilde{M}}\tilde{c}^{\tilde{M}} & = & \sum_{\tilde{M}}(-)^{\tilde{M}\tilde{c}}\tilde{b}_{\tilde{M}}\tilde{c}^{\tilde{M}}=\sum_{\tilde{M}}(-)^{Mc+M+c+1}\tilde{b}_{\tilde{M}}\tilde{c}^{\tilde{M}}\end{eqnarray}
So if we define\begin{eqnarray}
\tilde{b}_{\tilde{M}} & \equiv & (-)^{b}b_{M}\\
\tilde{c}^{\tilde{M}} & \equiv & c^{M}\end{eqnarray}
we get\begin{eqnarray}
\tilde{c}^{\tilde{M}}\tilde{b}_{\tilde{M}} & = & \sum_{\tilde{M}}(-)^{Mb+b}\tilde{c}^{\tilde{M}}\tilde{b}_{\tilde{M}}=\sum_{M}(-)^{Mb}c^{M}b_{M}=c^{M}b_{M}\\
\tilde{b}_{\tilde{M}}\tilde{c}^{\tilde{M}} & = & \sum_{\tilde{M}}(-)^{Mc+M+c+1}\tilde{b}_{\tilde{M}}\tilde{c}^{\tilde{M}}=(-)^{b+c+1}b_{M}c^{M}\end{eqnarray}

\subsection{Creating a rumpf-index}

If a graded variable does not yet have any index, one has to create
one, in order to be able to shift the grading. Similarly, it is sometimes
not desireable to change the grading of an index, because the same
index is also carried by other objects, whose rumpf-grading we do
not want to change. Then it is also useful to introduce an additional
rumpf-index, which carries the grading of the rumpf. Let us first
consider the simple case of a graded object $c$ without indices together
with its conjugate momentum $b$. And introduce corresponding new
variables with index and a shifted grading:

\begin{eqnarray}
x^{i} & \equiv & c\quad i\in\left\{ 1\right\} \\
\abs{x} & = & \abs{c}+1,\quad\abs{i}=1\\
p_{i} & \equiv & -(-)^{b}b\quad i\in\left\{ 1\right\} \\
\abs{p} & = & \abs{b}+1,\quad\abs{i}=1\end{eqnarray}
There are now two ways to express $cb$ in the new variables:\begin{eqnarray}
p_{i}x^{i}=x^{i}p_{i} & = & -(-)^{p+b}cb=cb=(-)^{bc}bc\\
(-)^{ij+px+jx}p_{j}x^{i}=(-)^{ip}x^{i}p_{j} & = & cb=-bc\end{eqnarray}
(If $c$ and $b$ had grading $1$ and we had in addition objects
$x$ and $p$ with grading 0, then we can use a unifying notion, by
defining $x^{0}\equiv x,\quad p_{0}=p,\quad\abs{i}=i\in\left\{ 0,1\right\} $.)\begin{eqnarray}
(-)^{ix}\partiell{}{x^{i}}x^{j} & = & \delta_{i}\hoch{j}=1=\partiell{}{c}c\qquad i,j\in\left\{ 1\right\} \\
(-)^{ix}\partial x^{i}/\partial x^{j} & = & \delta^{i}\tief{j}=-1=-\partial c/\partial c\\
(-)^{ip}\partial p_{i}/\partial p_{j} & = & \delta_{i}\hoch{j}=1=\partial b/\partial b\\
(-)^{ip}\partiell{}{p_{i}}p_{j} & = & \delta^{i}\tief{j}=-1=-\partiell{}{b}b\end{eqnarray}
For $\abs{c}=\abs{b}=1$, we get for the Poisson bracket of ghosts:\begin{eqnarray}
\left\{ p_{i},x^{j}\right\}  & = & \partial p_{i}/\partial p_{k}\partiell{}{x^{k}}x^{j}=\delta_{i}\hoch{k}\delta_{k}\hoch{j}=\delta_{i}\hoch{j}=1=\partial b/\partial b\partiell{c}{c}=\left\{ b,c\right\} \\
\left\{ x^{i},p_{j}\right\}  & = & -\partial x^{i}/\partial x^{k}\partiell{}{p_{k}}p_{j}=-\delta^{i}\tief{k}\delta^{k}\tief{j}=-\delta^{i}\tief{j}=1=\partial c/\partial c\partiell{}{b}b=\left\{ c,b\right\} \end{eqnarray}

\section{Index-position-shifts}

Given variables $b_{M}$ and $c^{M}$ we want to introduce a new index
with the opposite position and replace \begin{equation}
b_{M}\To\bar{b}^{\bar{M}},\: c^{M}\To\bar{c}_{\bar{M}}\end{equation}
 not changing any gradings or summation conventions\begin{eqnarray}
\abs{\bar{b}} & = & \abs{b}\\
\abs{\bar{c}} & = & \abs{c}\\
\abs{M} & = & \abs{\bar{M}}\end{eqnarray}
\begin{eqnarray}
\bar{b}^{\bar{M}}\bar{c}_{\bar{M}} & = & \sum_{\bar{M}}(-)^{\bar{M}\bar{c}}\bar{b}^{\bar{M}}\bar{c}_{\bar{M}}=\sum_{M}(-)^{Mc}\bar{b}^{\bar{M}}\bar{c}_{\bar{M}}=(-)^{bc}\bar{c}_{\bar{M}}\bar{b}^{\bar{M}}\\
b_{M}c^{M} & = & \sum_{M}(-)^{M+Mc}b_{M}c^{M}=(-)^{bc}c^{M}b_{M}\end{eqnarray}
We can achieve \begin{eqnarray}
\bar{b}^{\bar{M}}\bar{c}_{\bar{M}} & = & b_{M}c^{M}\\
\bar{c}_{\bar{M}}\bar{b}^{\bar{M}} & = & c^{M}b_{M}\end{eqnarray}
by defining e.g.\begin{eqnarray}
\bar{b}^{\bar{M}} & = & (-)^{M}b_{M}\\
\bar{c}_{\bar{M}} & = & c^{M}\end{eqnarray}
or the other way round ($(-)^{M}$ in the definition of $\bar{c}^{\bar{M}}$).
For the derivatives we get:\begin{eqnarray}
\partiell{}{\bar{c}_{\bar{M}}}\bar{c}_{\bar{N}} & = & (-)^{Mc}\delta^{\bar{M}}\tief{\bar{N}}=(-)^{M+Mc}\delta_{M}\hoch{N}=(-)^{M}\partiell{}{c^{M}}c^{N}\\
\dann\partiell{}{\bar{c}_{\bar{M}}} & = & (-)^{M}\partiell{}{c^{M}}\\
\partial\bar{c}_{\bar{M}}/\partial\bar{c}_{\bar{N}} & = & (-)^{Mc}\delta_{\bar{M}}\hoch{\bar{N}}=(-)^{M+Mc}\delta^{M}\tief{N}=(-)^{M}\partial c^{M}/\partial c^{N}\\
\dann\partial/\partial\bar{c}_{\bar{N}} & = & (-)^{M}\partial/\partial c^{N}\end{eqnarray}
\begin{eqnarray}
\partiell{}{\bar{b}^{\bar{M}}}\bar{b}^{\bar{N}} & = & (-)^{bM}\delta_{\bar{M}}\hoch{\bar{N}}=(-)^{bM+M}\delta^{M}\tief{N}=(-)^{M}\partiell{}{b_{M}}b_{N}\\
\dann\partiell{}{\bar{b}^{\bar{M}}} & = & \partiell{}{b_{M}}\\
\partial\bar{b}^{\bar{M}}/\partial\bar{b}^{\bar{N}} & = & (-)^{bM}\delta^{\bar{M}}\tief{\bar{N}}=(-)^{bM+M}\delta_{M}\hoch{N}=(-)^{M}\partial b_{M}/\partial b_{N}\\
\dann\partial/\partial\bar{b}^{\bar{N}} & = & (-)^{M}\partial/\partial b_{N}\end{eqnarray}
}

\section{Lagrangian and Hamiltonian formalism}

The structural equations of the Lagrangian or Hamiltonian formalism
are good examples for the application of the gradification theorem.
Graded versions of the Lagrangian equations of motion will most probably
be very familiar to the reader. The intention here is only to carefully
demonstrate how at the one hand the choice of the summation convention
fixes all ambiguities and how on the other hand this choice need not
to be done a priori (apart from the choice for the auxiliary index
$i$ to be introduced again below).

Let us consider a Lagrangian $L(q,\ce,\dot{q},\dot{\ce})$ which depends
on variables $q^{M}$ with bosonic rumpf and ghost fields $\ce^{M}$with
fermionic rumpf and their time derivatives. The indices of $q$ and
$\ce$ will in general differ, but the assumption of the same index
simplifies the presentation. The variation of the action will contain
also derivatives w.r.t. $\ce^{M}$ and it is thus useful to introduce
again the variable $z^{iM}=(z^{1M},z^{\bs{2}M})=(q^{M},\ce^{M}).$\begin{eqnarray}
\delta S & = & \int dt\quad\delta z^{iM}\partl{z^{iM}}L+\delta\dot{z}^{iM}\partl{\dot{z}^{iM}}L=\\
 & = & \int dt\quad\delta z^{iM}\left(\partl{z^{iM}}L-\frac{d}{dt}(\partl{\dot{z}^{iM}}L)\right)+\mbox{bdry terms}\end{eqnarray}
The \textbf{equations of motion} thus have the form\begin{equation}
\partl{z^{iM}}L-\frac{d}{dt}(\partl{\dot{z}^{iM}}L)\greq0\end{equation}
where the graded equal sign has no effect here. As discussed earlier,
left and right derivative are graded equal and because $L$ is always
bosonic (at least in usual examples) they are in fact equal and there
is no arbitraryness of choosing left or right derivative. If we have
NW conventions for the auxiliary index $i$, the derivative w.r.t.
$z^{\bs{2}M}$ becomes the left derivative w.r.t. $\ce^{M}$ or minus
the right derivative w.r.t. $\ce^{M}$, although an overall minus
in the equations of motion is of course irrelevant.

In a similar way the definition of the \textbf{\index{conjugate momentum!graded definition}conjugate
\index{momentum!conjugate $\sim$, graded definition}momentum} is
already fixed by the choice of the summation convention. The definition
is simply \begin{equation}
\pi_{iM}\equiv\partiell{}{\dot{z}^{iM}}L=L\partr{\dot{z}^{iM}}\end{equation}
Again, left and right derivative coincide for bosonic rumpf $z$ (when
$L$ is bosonic) and their definition is fixed by the choice of the
summation convention. If we have NW conventions for the auxiliary
index $i$, this definition becomes \begin{eqnarray}
p_{M} & \equiv & \partl{\dot{q}^{M}}L=L\partr{\dot{q}^{M}}\\
\be_{M} & \equiv & \partl{\dot{\ce}^{M}}L=-L\partr{\dot{\ce}^{M}}\end{eqnarray}
For the choice of NE for the index $i$, the right derivative would
be without sign. Remember again that the choice of the summation convention
for the index $i$ does not fix the one for the index $M$.

The \textbf{Legendre\index{Legendre transformation!graded version}
transformation} to obtain the Hamiltonian is of course also fixed
by the summation convention\begin{eqnarray}
H(z,\pi) & \equiv & \int dt\quad\dot{z}^{iM}\pi_{iM}-L(z,\dot{z}(z,\pi))\end{eqnarray}
Although writing $\dot{z}^{iM}$ at the first position seems to fix
NW-conventions, this is not true. The signs are as usual hidden in
the summation. We thus have $\dot{z}^{iM}\pi_{iM}=\pi_{iM}\dot{z}^{iM}$
and are still free to decide in the end, which convention will enter
the actual summation. As before we have to make a choice for the summation
convention of the auxiliary index $i$, if we want to write this explicitely
in terms of $q^{M}$ and $\ce^{M}$ and its momenta:\begin{eqnarray}
H(q,\ce,p,\be) & \stackrel{\mbox{NW for }i}{\equiv} & \int dt\quad\dot{q}^{M}p_{M}+\dot{\ce}^{M}\be_{M}-L(q,\ce,\dot{q}(q,\ce,p,\be),\dot{\ce}(q,\ce,p,\be))\end{eqnarray}
 The same reasoning is applied for the second Legendre transformation
which yields the first order action $\tilde{L}(z,\pi,\dot{z},\dot{\pi})\equiv\int dt\quad\dot{z}^{iM}\pi_{iM}-H(z,\pi)$.

We had already mentioned that the summation convention for $i$ could
differ from the one for $M$ and that even within $M$ we could have
different summation conventions for different index-subsets. Applications
where the advantage of such \textbf{mixed\index{mixed summation conventions}\index{summation conventions!mixed $\sim$}
conventions }becomes obvious, are those where one joins several variable
with different index position to one variable, but wants to keep the
summation conventions of before. This is the case for example for
the introduction of Darboux\index{Darboux coordinates} coordinates
to parametrize the phase space. Let us forget for the moment about
the ghost variables. We can then define for example \begin{equation}
Z^{\q{M}}\equiv(q^{M},p_{M})\end{equation}
The Poisson bracket is then written with a mixed summation convention
for the index $\q{M}$ (based on NW for $M$) as \begin{eqnarray}
\left\{ F,G\right\}  & = & F\partr{Z^{\q{M}}}P^{\q{M}\q{N}}\partl{Z^{\q{N}}}G\stackrel{\mbox{mixed conv}}{\equiv}\\
 & \equiv & \sum_{M_{1},M_{2},N_{1},N_{2}}(-)^{M}F\partr{q^{M}}P^{MN}\partl{q^{N}}G+(-)^{M+N}F\partr{q^{M}}P^{M}\tief{N}\partl{p_{N}}G+\nonumber \\
 &  & +F\partr{p_{M}}P_{M}\hoch{N}\partl{q^{N}}G+(-)^{N}F\partr{p_{M}}P_{MN}\partl{p_{N}}G\end{eqnarray}
If we had NW conventions for the indices $\q{M}$ and $\q{N}$, the
definition of the graded summation would have a $(-)^{M}$ in front
of every of the four terms. For the special choice of coordinates
(with split in configuration space coordinates and momenta), the Poisson
bivector is simply \begin{equation}
P^{\q{M}\q{N}}=\left(\begin{array}{cc}
0 & -\delta^{M}\tief{N}\\
\delta_{M}\hoch{N} & 0\end{array}\right)\end{equation}
where the relation of the graded Kronecker deltas in NW-conventions
to the numerical $\delta_{M}^{N}$ is given by $\delta_{M}\hoch{N}=\delta_{M}^{N}=(-)^{MN}\delta^{N}\tief{M}$.
\rem{

\section{Conjugate momentum and Legendre transform}

Given a Lagrangian $L(q,c,\dot{q},\dot{c})$ including ghost fields
we define the conjugate momentum via the left derivative

\begin{eqnarray}
p_{M} & \equiv & \partiell{}{\dot{q}^{M}}L\\
b_{\mathcal{A}} & \equiv_{g} & \partiell{}{\dot{c}^{\mathcal{A}}}L\end{eqnarray}
Compatible with this definition, we define the Legendre transform
as\begin{eqnarray}
H(q,p,c,b) & \equiv & \dot{q}^{M}p_{M}+\dot{c}^{M}b_{M}-L(q,c,\dot{q}(q,p,c,b),\dot{c}(q,p,c,b))\end{eqnarray}
and the first order Lagrangian will be\begin{eqnarray}
\tilde{L}(q,p,c,b,\dot{q},\dot{c}) & \equiv & \dot{q}^{M}p_{M}+\dot{c}^{M}b_{M}-H(q,p,c,b)\\
\tilde{S} & \equiv & \int\de t\,\tilde{L}\end{eqnarray}
With these definitions we get the following variations of the first-order
action\begin{eqnarray}
\delta\tilde{S} & = & \int\de t\quad\delta\dot{q}^{M}p_{M}+\dot{q}^{M}\delta p_{M}+\delta\dot{c}^{M}b_{M}+\dot{c}^{M}\delta b_{M}+\nonumber \\
 &  & -\delta q^{M}\partial_{M}H-\partial H/\partial p_{M}\delta p_{M}-\delta c^{M}\partiell{}{c^{M}}H-\partial H/\partial b_{M}\delta b_{M}\stackrel{!}{=}0\\
\dot{p}_{M} & = & -\partial_{M}H=-\left\{ p_{M},H\right\} =\left\{ H,p_{M}\right\} \quad\left(=-\partial H/\partial q^{M}\right)\\
\dot{q}^{M} & = & \partial H/\partial p_{M}=\left\{ H,q^{M}\right\} \quad\left(=\partiell{}{p_{M}}H\right)\\
\dot{b}_{M} & = & -\partiell{}{c^{M}}H=-\left\{ b_{M},H\right\} =\left\{ H,b_{M}\right\} \quad\left(=\partial H/\partial c^{M}\right)\\
\dot{c}^{M} & = & \partial H/\partial b_{M}=\left\{ H,c^{M}\right\} \quad\left(=-\partiell{}{b_{M}}H\right)\end{eqnarray}
We get the familiar equations, if we use left derivatives for coordinates
and right derivatives for momenta.%
\footnote{We would always get the familiar equations, if we redefine $\partial/\partial c^{M}\To-\partial/\partial c^{M}$
and $\partiell{}{b_{M}}\To-\partiell{}{b_{M}}$, which would not affect
the definition of the Poisson-bracket. This would, however, imply
\begin{eqnarray*}
\delta f(c,b) & = & \delta c^{M}\partiell{}{c^{M}}f+\partial f/\partial b_{M}\delta b_{M}=-\partial f/c^{M}\delta c^{M}-\delta b_{M}\partiell{}{b_{M}}f\\
\partiell{}{c^{M}}c^{N} & \greq & \delta_{M}\hoch{N}\greq-\partial c^{N}/\partial c^{M}\\
\partial b_{M}/\partial b_{N} & \greq & \delta_{M}\hoch{N}\greq-\partiell{}{b_{N}}b_{M}\end{eqnarray*}
which becomes only intuitive when we assign an additional index $i\in\left\{ 1\right\} $
to $c$ and $b$, taking their grading ($\abs{b_{iM}}=\abs{i}+\abs{M}=1+\abs{M}$)\begin{eqnarray*}
\delta f(c,b) & = & \delta c^{iM}\partiell{}{c^{iM}}f+\partial f/\partial b_{iM}\delta b_{iM}=\partial f/c^{iM}\delta c^{iM}+\delta b_{iM}\partiell{}{b_{iM}}f\\
\partiell{}{c^{iM}}c^{jN} & \greq & \underbrace{\delta_{iM}\hoch{jN}}_{\greq\delta_{i}\hoch{j}\delta_{M}\hoch{N}}\greq\partial c^{jN}/\partial c^{iM}\\
\partial b_{iM}/\partial b_{jN} & \greq & \delta_{iM}\hoch{jN}\greq\partiell{}{b_{jN}}b_{iM}\qquad\fussend\end{eqnarray*}
}

\section{First order Action and symplectic 2-Form}

Given a first order action with physical fields $q^{M}$ and ghosts
$c^{\mathcal{A}}$ (where the index $\mathcal{A}$ usually is in some
subset of the index-range of $M$)\begin{eqnarray}
S[q,c] & = & \int\de t\quad\dot{q}^{M}P_{M}(q,c)+\dot{c}^{\mathcal{A}}B_{\mathcal{A}}(q,c)+V(q,c)=\label{eq:firstorderAction}\\
 & = & \int\quad\de q^{M}P_{M}(q,c)+\de c^{\mathcal{A}}B_{\mathcal{A}}(q,c)+\de t\, V(q,c)\label{eq:firstorderActionAsPullback}\end{eqnarray}
we get the following constraints on the conjugate momenta $p_{M}$
and $b_{\mathcal{A}}$ of $q^{M}$ and $c^{\mathcal{A}}$ respectively:\begin{eqnarray}
\phi_{M} & \equiv & p_{M}-P_{M}\stackrel{!}{=}0,\quad\abs{\phi}=0\\
\varphi_{\mathcal{A}} & \equiv & b_{\mathcal{A}}-B_{\mathcal{A}}\stackrel{!}{=}0,\quad\abs{\varphi}=1\end{eqnarray}
We regard them as target space one-forms\begin{eqnarray}
\phi_{M}\de q^{M}\equiv\phi & \equiv & p-P\equiv p_{M}\de q^{M}-P_{M}\de q^{M}\\
\varphi_{\mathcal{A}}\de c^{\mathcal{A}}\equiv\varphi & \equiv & b-B\equiv b_{\mathcal{A}}\de c^{\mathcal{A}}-B_{\mathcal{A}}\de c^{\mathcal{A}}\\
\Phi & \equiv & \phi+\varphi=(p+b)-(P+B)\end{eqnarray}
The constraint algebra reads\begin{eqnarray}
\left\{ \phi_{M},\phi_{N}\right\}  & = & -\left\{ p_{M},P_{N}\right\} -\left\{ P_{M},p_{N}\right\} \greq-\partial_{M}P_{N}+\partial P_{M}/\partial q^{N}=-2\partial_{[M}P_{N]}\\
\left\{ \varphi_{\mathcal{A}},\varphi_{\mathcal{B}}\right\}  & = & -\left\{ b_{\mathcal{A}},B_{\mathcal{B}}\right\} -\left\{ B_{\mathcal{A}},b_{\mathcal{B}}\right\} \greq-\partiell{}{c^{\mathcal{A}}}B_{\mathcal{B}}-\partial B_{\mathcal{A}}/\partial c^{\mathcal{B}}=-2\partiell{}{c^{(\mathcal{A}}}B_{\mathcal{B})}\,\,\left(\equiv-\partiell{}{c^{\mathcal{A}}}B_{\mathcal{B}}-(-)^{\mathcal{B}+\mathcal{A}+\mathcal{AB}}\partiell{}{c^{\mathcal{B}}}B_{\mathcal{A}}\right)\qquad\\
\left\{ \phi_{M},\varphi_{\mathcal{B}}\right\}  & = & -\left\{ p_{M},B_{\mathcal{B}}\right\} -\left\{ P_{M},b_{\mathcal{B}}\right\} \greq-\partial_{M}B_{\mathcal{B}}-\partial P_{M}/\partial c^{\mathcal{B}}\greq-2\cdot\frac{1}{2}\left(\partial_{M}B_{\mathcal{B}}-\partiell{}{c^{\mathcal{B}}}P_{M}\right)\\
\left\{ \varphi_{\mathcal{A}},\phi_{N}\right\}  & = & -\left\{ b_{\mathcal{A}},P_{N}\right\} -\left\{ B_{\mathcal{A}},p_{N}\right\} \greq-\partiell{}{c^{\mathcal{A}}}P_{N}+\partial B_{\mathcal{A}}/\partial q^{N}\greq-2\cdot\frac{1}{2}\left(\partiell{}{c^{\mathcal{A}}}P_{N}-\partial_{N}B_{\mathcal{A}}\right)\end{eqnarray}
Consider the \textbf{symplectic potential} \begin{eqnarray}
A & \equiv & \de q^{M}P_{M}+\de c^{\mathcal{A}}B_{\mathcal{A}}=\\
 & = & P_{N}\de q^{N}+B_{\mathcal{B}}\de c^{\mathcal{B}}\equiv P+B\end{eqnarray}
which serves as a potential for the \textbf{symplectic two-form\begin{eqnarray}
C & \equiv & -\de A=\\
 & = & -\de q^{M}\partial_{M}P_{N}\de q^{N}-\de c^{\mathcal{A}}\partiell{}{c^{\mathcal{A}}}P_{N}\de q^{N}-\de q^{M}\partial_{M}B_{\mathcal{B}}\de c^{\mathcal{B}}-\de c^{\mathcal{A}}\partiell{}{c^{\mathcal{A}}}B_{\mathcal{B}}\de c^{\mathcal{B}}=\label{eq:twoformcoeffsWithGhosts}\\
 & = & -\partial_{M}P_{N}\de q^{M}\de q^{N}-\partiell{}{c^{\mathcal{A}}}P_{N}\de c^{\mathcal{A}}\de q^{N}+\partial_{M}B_{\mathcal{B}}\de q^{M}\de c^{\mathcal{B}}-\partiell{}{c^{\mathcal{A}}}B_{\mathcal{B}}\de c^{\mathcal{A}}\de c^{\mathcal{B}}\equiv\\
 & \equiv & \de q^{M}C_{MN}\de q^{N}+\de c^{\mathcal{A}}\tilde{C}_{\mathcal{A}N}\de q^{N}-\de q^{M}\tilde{C}_{M\mathcal{B}}\de c^{\mathcal{B}}-\de c^{\mathcal{A}}\bar{C}_{\mathcal{AB}}\de c^{\mathcal{B}}\end{eqnarray}
}Using the one-form $\Phi$, the constraint algebra can be nicely
rewritten as\begin{eqnarray}
\frac{1}{2}\left\{ \Phi,\Phi\right\}  & = & C=-\de A=-\de(P+B)=\de\Phi-\de(p+b)\end{eqnarray}
where $\de$ is the target space differential (not the worldsheet-differential
as in the action (\ref{eq:firstorderActionAsPullback}) where we basically
have the pullback of the target-space one-forms on the worldline).
For $\de p=\de b=0$ ($\de$ being the configuration space differential),
we get the Maurer-Cartan-like equation\begin{equation}
\boxed{\frac{1}{2}\left\{ \Phi,\Phi\right\} =\de\Phi}\end{equation}
To build the Dirac bracket we need to invert the coefficient matrix
$\left(\begin{array}{cc}
\left\{ \phi_{M},\phi_{N}\right\}  & \left\{ \phi_{M},\varphi_{\mathcal{B}}\right\} \\
\left\{ \varphi_{\mathcal{A}},\phi_{N}\right\}  & \left\{ \varphi_{\mathcal{A}},\varphi_{\mathcal{B}}\right\} \end{array}\right)=_{g}\left(\begin{array}{cc}
C_{MN} & \tilde{C}_{M\mathcal{B}}\\
\tilde{C}_{\mathcal{A}N} & \bar{C}_{\mathcal{A\mathcal{B}}}\end{array}\right)$ where $\abs{C}=\abs{\bar{C}}=0$ and $\abs{\tilde{C}}=1$.\begin{eqnarray}
\left(\begin{array}{cc}
C_{MK} & \tilde{C}_{M\mathcal{C}}\\
\tilde{C}_{\mathcal{A}K} & \bar{C}_{\mathcal{A\mathcal{C}}}\end{array}\right)\left(\begin{array}{cc}
D^{KN} & \tilde{D}^{K\mathcal{B}}\\
\tilde{D}^{\mathcal{C}N} & \bar{D}^{\mathcal{CB}}\end{array}\right) & \stackrel{!}{\greq} & \left(\begin{array}{cc}
\delta_{M}\hoch{N} & 0\\
0 & \delta_{\mathcal{A}}\hoch{\mathcal{B}}\end{array}\right)\\
C_{MK}D^{KN}+\tilde{C}_{M\mathcal{C}}\tilde{D}^{\mathcal{C}N} & \stackrel{!}{\greq} & \delta_{M}\hoch{N}\label{eq:inverseeqnsWithGhostsI}\\
C_{MK}\tilde{D}^{K\mathcal{B}}+\tilde{C}_{M\mathcal{C}}\bar{D}^{\mathcal{CB}} & \stackrel{!}{\greq} & 0\\
\tilde{C}_{\mathcal{A}K}D^{KN}+\bar{C}_{\mathcal{AC}}\tilde{D}^{\mathcal{C}N} & \stackrel{!}{\greq} & 0\\
\tilde{C}_{\mathcal{A}K}\tilde{D}^{K\mathcal{B}}+\bar{C}_{\mathcal{AC}}\bar{D}^{\mathcal{CB}} & \stackrel{!}{\greq} & \delta_{\mathcal{A}}\hoch{\mathcal{C}}\label{eq:inverseeqnsWithGhostsIV}\end{eqnarray}
Using that our graded matrix multiplication is associative , this
also implies%
\footnote{\begin{eqnarray*}
(CD) & = & \one\\
(AC) & = & \one'\,\textrm{(different index structure)}\\
A & = & A(CD)=(AC)D=D\quad\fussend\end{eqnarray*}
}\begin{eqnarray}
\left(\begin{array}{cc}
D^{MK} & \tilde{D}^{M\mathcal{C}}\\
\tilde{D}^{\mathcal{A}K} & \bar{D}^{\mathcal{AC}}\end{array}\right)\left(\begin{array}{cc}
C_{KN} & \tilde{C}_{K\mathcal{B}}\\
\tilde{C}_{\mathcal{C}N} & \bar{C}_{\mathcal{CB}}\end{array}\right) & \greq & \left(\begin{array}{cc}
\delta^{M}\tief{N} & 0\\
0 & \delta^{\mathcal{A}}\tief{\mathcal{B}}\end{array}\right)\end{eqnarray}
The Dirac bracket then reads\begin{eqnarray}
\hspace{-1cm}\left\{ F,G\right\} _{D} & = & \left\{ F,G\right\} -\left\{ F,\phi_{M}\right\} D^{MN}\left\{ \phi_{N},G\right\} -\left\{ F,\phi_{M}\right\} \tilde{D}^{M\mathcal{B}}\left\{ \varphi_{\mathcal{B}},G\right\} -\left\{ F,\varphi_{\mathcal{A}}\right\} \tilde{D}^{\mathcal{A}N}\left\{ \phi_{N},G\right\} -\left\{ F,\varphi_{A}\right\} \bar{D}^{\mathcal{AB}}\left\{ \varphi_{\mathcal{B}},G\right\} =\qquad\\
 & = & \left\{ F,G\right\} -\left(\left\{ F,\phi_{M}\right\} ,\left\{ F,\varphi_{\mathcal{A}}\right\} \right)\left(\begin{array}{cc}
D^{MN} & \tilde{D}^{M\mathcal{B}}\\
\tilde{D}^{\mathcal{A}N} & \bar{D}^{\mathcal{AB}}\end{array}\right)\left(\begin{array}{c}
\left\{ \phi_{N},G\right\} \\
\left\{ \varphi_{\mathcal{B}},G\right\} \end{array}\right)\end{eqnarray}
which implies\begin{eqnarray}
\left(\begin{array}{cc}
\left\{ \phi_{K},\phi_{L}\right\} _{D} & \left\{ \phi_{K},\varphi_{\mathcal{D}}\right\} _{D}\\
\left\{ \varphi_{\mathcal{C}},\phi_{L}\right\} _{D} & \left\{ \varphi_{\mathcal{C}},\varphi_{\mathcal{D}}\right\} _{D}\end{array}\right) & =_{g} & \left(\begin{array}{cc}
C_{KL} & \tilde{C}_{K\mathcal{D}}\\
\tilde{C}_{\mathcal{C}L} & \bar{C}_{\mathcal{CD}}\end{array}\right)-\\
 &  & -\left(\begin{array}{cc}
\left\{ \phi_{K},\phi_{M}\right\}  & \left\{ \phi_{K},\varphi_{\mathcal{A}}\right\} \\
\left\{ \varphi_{\mathcal{C}},\phi_{M}\right\}  & \left\{ \varphi_{\mathcal{C}},\varphi_{\mathcal{A}}\right\} \end{array}\right)\left(\begin{array}{cc}
D^{MN} & \tilde{D}^{M\mathcal{B}}\\
\tilde{D}^{\mathcal{A}N} & \bar{D}^{\mathcal{AB}}\end{array}\right)\left(\begin{array}{cc}
\left\{ \phi_{N},\phi_{L}\right\}  & \left\{ \phi_{N},\varphi_{\mathcal{D}}\right\} \\
\left\{ \varphi_{\mathcal{B}},\phi_{L}\right\}  & \left\{ \varphi_{\mathcal{B}},\varphi_{\mathcal{D}}\right\} \end{array}\right)=\nonumber \\
 & = & \left(\begin{array}{cc}
0 & 0\\
0 & 0\end{array}\right)\end{eqnarray}

\subsection{Grading-shift in order to treat ghosts and physical fields together}

Let us repeat the same calculation while shifting the grading of $c$
and $b$ to their index as described in section \vref{sec:RumpfIndexShift}.

The grading of the ghosts sits partly in the rumpf: \begin{eqnarray*}
\abs{c^{\mathcal{A}}} & = & 1+\abs{\mathcal{A}}\\
\abs{b_{\mathcal{A}}} & = & 1+\abs{\mathcal{A}}\end{eqnarray*}
Define\begin{eqnarray}
p_{\tilde{\mathcal{A}}} & \equiv & (-)^{\mathcal{A}}b_{\mathcal{A}}\\
q^{\tilde{\mathcal{A}}} & \equiv & c^{\mathcal{A}}\end{eqnarray}
where the rumpfs are now even and the grading of the index gets shifted\begin{eqnarray}
\abs{q} & = & 0,\abs{p}=0\\
\abs{\tilde{\mathcal{A}}} & = & \abs{\mathcal{A}}+1\end{eqnarray}
The grading of the complete object remains of course the same\begin{eqnarray}
\abs{q^{\tilde{\mathcal{A}}}} & = & \abs{\tilde{A}}=1+\abs{\mathcal{A}}=\abs{c^{\mathcal{A}}}\\
\abs{p_{\tilde{\mathcal{A}}}} & = & \abs{\tilde{\mathcal{A}}}=1+\abs{\mathcal{A}}=\abs{b_{\mathcal{A}}}\end{eqnarray}
we get\begin{eqnarray}
-b_{\mathcal{A}}c^{\mathcal{A}}=c^{\mathcal{A}}b_{\mathcal{A}} & = & q^{\tilde{\mathcal{A}}}p_{\tilde{\mathcal{A}}}=p_{\tilde{\mathcal{A}}}q^{\tilde{\mathcal{A}}}\end{eqnarray}
For the derivatives we get\begin{eqnarray}
\partiell{}{q^{\tilde{\mathcal{A}}}} & \equiv & (-)^{\mathcal{A}}\partiell{}{c^{\mathcal{A}}},\quad\partial/\partial q^{\tilde{\mathcal{A}}}\stackrel{!!}{\equiv}-(-)^{\mathcal{A}}\partial/\partial c^{\mathcal{A}}\\
\dann\partiell{}{q^{\tilde{\mathcal{A}}}}q^{\tilde{\mathcal{B}}} & = & \delta_{\tilde{\mathcal{A}}}\hoch{\tilde{\mathcal{B}}}=\delta_{\mathcal{A}}\hoch{\mathcal{B}}\\
\partial q^{\tilde{\mathcal{A}}}/\partial q^{\tilde{\mathcal{B}}} & = & \delta^{\tilde{\mathcal{A}}}\tief{\tilde{\mathcal{B}}}=(-)^{\tilde{\mathcal{A}}}\delta_{\tilde{\mathcal{B}}}\hoch{\tilde{\mathcal{A}}}=(-)^{\mathcal{A}+1}\delta_{\mathcal{B}}\hoch{\mathcal{A}}=-\delta^{\mathcal{A}}\tief{\mathcal{B}}\\
\partiell{}{p_{\tilde{\mathcal{A}}}} & \stackrel{!!}{\equiv} & -\partiell{}{b_{\mathcal{A}}},\quad\partial/\partial p_{\tilde{\mathcal{A}}}\equiv\partial/\partial b_{\mathcal{A}}\\
\dann\partiell{}{p_{\tilde{\mathcal{A}}}}p_{\tilde{\mathcal{B}}} & = & \delta^{\tilde{\mathcal{A}}}\tief{\tilde{\mathcal{B}}}=(-)^{\tilde{\mathcal{A}}}\delta_{\tilde{\mathcal{B}}}\hoch{\tilde{\mathcal{A}}}=(-)^{\mathcal{A}+1}\delta_{\mathcal{B}}\hoch{\mathcal{A}}=-\delta^{\mathcal{A}}\tief{\mathcal{B}}\\
\partial p_{\tilde{\mathcal{A}}}/\partial p_{\tilde{\mathcal{B}}} & = & \delta_{\tilde{\mathcal{A}}}\hoch{\tilde{\mathcal{B}}}=\delta_{\mathcal{A}}\hoch{\mathcal{B}}\end{eqnarray}
We can now build combined variables with common even rumpf\begin{eqnarray}
q^{\q{M}} & \equiv & (q^{M},q^{\tilde{\mathcal{A}}})\\
p_{\q{M}} & \equiv & (p_{M},p_{\tilde{\mathcal{A}}})\\
\textrm{with }\abs{q}=\abs{p} & = & 0\end{eqnarray}
In those variables, the action (\ref{eq:firstorderAction}) takes
the simple form\begin{eqnarray}
S[q] & = & \int\de t\quad\dot{q}^{\q{M}}P_{\q{M}}(q)+V(q)=\\
 & = & \int\quad\de q^{\q{M}}P_{\q{M}}(q)+\de t\, V(q)\end{eqnarray}
with\begin{eqnarray}
P_{\q{M}}(q) & = & \left(P_{M}(q)\,,\, P_{\tilde{\mathcal{A}}}(q)\right)=\left(P_{M}(q,c)\,,\,(-)^{\mathcal{A}}B_{\mathcal{A}}(q,c)\right)\end{eqnarray}
We redefine the meaning of the one-form $P$ and $p$ formerly being
$P_{M}\de q^{M}$ and $p_{M}\de q^{M}$ as\begin{eqnarray}
P & \equiv & P_{\q{M}}\de q^{\q{M}}\quad(\textrm{formerly }P+B)\\
p & \equiv & p_{\q{m}}\de q^{\q{M}}\quad(\textrm{formerly }p+b)\end{eqnarray}
The constraints are\begin{eqnarray}
\Phi_{\q{M}} & \equiv & p_{\q{M}}-P_{\q{M}}=\left(p_{M}-P_{M},\, p_{\tilde{\mathcal{A}}}-P_{\tilde{\mathcal{A}}}\right)\\
\Phi & \equiv & \Phi_{\q{M}}\de q^{\q{M}}=\phi+\varphi=p-P\end{eqnarray}
with constraint algebra \begin{eqnarray}
\left\{ \Phi_{\q{M}},\Phi_{\q{N}}\right\}  & = & -\left\{ p_{\q{M}},P_{\q{N}}\right\} -\left\{ P_{\q{M}},p_{\q{N}}\right\} \greq-\partial_{\q{M}}P_{\q{N}}+\partial_{\q{N}}P_{\q{M}}\greq-2\partial_{[\q{M}}P_{\q{N}]}\equiv C_{\q{M}\q{N}}\end{eqnarray}
In terms of forms again - where the two-form $C$ is defined by\begin{eqnarray}
C & \equiv & \frac{1}{2}C_{\q{M}\q{N}}\de x^{\q{M}}\de x^{\q{N}}\end{eqnarray}
this reads again\begin{eqnarray}
\frac{1}{2}\left\{ \Phi,\Phi\right\}  & = & C=-\de P=\de\Phi\end{eqnarray}
\begin{equation}
\boxed{\frac{1}{2}\left\{ \Phi,\Phi\right\} =\de\Phi}\label{eq:MaurerCartanII}\end{equation}
 The grading shift was thus necessary to write common equations for
the components of the forms (like $C_{\q{M}\q{N}}=-2\partial_{[\q{M}}P_{\q{N}]}$)
which is not possible without the shift, as the notation indicates
that all the components have a common rumpf-grading. Coordinate independent
equations, however, like (\ref{eq:MaurerCartanII}) can be written
without the index-shift.

We can now finally build the Dirac-bracket with the graded inverse
of the component-matrix of the (symplectic) 2-form $C$, \begin{eqnarray}
\left\{ F,G\right\} _{D} & \equiv & \left\{ F,G\right\} -\left\{ F,\Phi_{\q{M}}\right\} (C^{-1})^{\q{M}\q{N}}\left\{ \Phi_{\q{N}},G\right\} =\\
 & = & \left\{ F,G\right\} +\partial F/\partial q^{\q{M}}\,(C^{-1})^{\q{M}\q{N}}\partiell{}{q^{\q{N}}}G\end{eqnarray}
On the constraint surface, where $F$ and $G$ depend on $q^{\q{M}}$
only (and not on $p_{\q{M}}$), this reduces to \begin{eqnarray}
\left\{ F(q),G(q)\right\} _{D} & \equiv & \partial F/\partial q^{\q{M}}\,(C^{-1})^{\q{M}\q{N}}\partiell{}{q^{\q{N}}}G\end{eqnarray}

\subsubsection*{Seperating the ghosts again (undo the grading shift)}

To compare the results with those we got above, let us undo the grading
shift\begin{eqnarray}
P_{\q{M}}(q) & = & \left(P_{M}(q),P_{\tilde{\mathcal{A}}}(q)\right)=\left(P_{M}(q,c),(-)^{\mathcal{A}}B_{\mathcal{A}}(q,c)\right)\end{eqnarray}
\begin{eqnarray}
C_{\q{M}\q{N}} & = & -2\partial_{[\q{M}}P_{\q{N}]}\end{eqnarray}

\begin{eqnarray*}
C & \equiv & \frac{1}{2}C_{\q{M}\q{N}}\de q^{\q{M}}\de q^{\q{N}}=\\
 & = & \frac{1}{2}\de q^{\q{M}}C_{\q{M}\q{N}}\de q^{\q{N}}\\
 & = & \frac{1}{2}\sum\quad\de q^{M}C_{MN}\de q^{N}+\de q^{M}C_{M\tilde{\mathcal{B}}}\de q^{\tilde{B}}+\de q^{\tilde{\mathcal{A}}}C_{\tilde{\mathcal{A}}N}\de q^{N}+\de q^{\tilde{\mathcal{A}}}C_{\tilde{\mathcal{A}}\tilde{\mathcal{B}}}\de q^{\tilde{\mathcal{B}}}=\\
 & \equiv & \frac{1}{2}\sum\quad\de q^{M}C_{MN}\de q^{N}+(-)^{\mathcal{B}+M}\de q^{M}\tilde{C}_{M\mathcal{B}}\de c^{\mathcal{B}}+(-)^{\mathcal{A}}\de c^{\mathcal{A}}\tilde{C}_{\mathcal{A}N}\de q^{N}+(-)^{\mathcal{B}}\de c^{\mathcal{A}}\bar{C}_{\mathcal{AB}}\de c^{\mathcal{B}}=\\
 & = & \frac{1}{2}\left(\de q^{M}C_{MN}\de q^{N}+\de q^{M}\tilde{C}_{M\mathcal{B}}\de c^{\mathcal{B}}+\de c^{\mathcal{A}}\tilde{C}_{\mathcal{A}N}\de q^{N}+\de c^{\mathcal{A}}\bar{C}_{\mathcal{AB}}\de c^{\mathcal{B}}\right)\end{eqnarray*}
with $\abs{C}=\abs{\bar{C}}=0$, $\abs{\tilde{C}}=1$, and\begin{eqnarray}
C_{MN} & \equiv & -2\partial_{[M}P_{N]}=-\partial_{M}P_{N}+(-)^{MN}\partial_{N}P_{M}\\
\tilde{C}_{M\mathcal{B}} & \equiv & (-)^{\mathcal{B}+M}C_{M\tilde{\mathcal{B}}}=-(-)^{\mathcal{B}+M}\partial_{M}P_{\tilde{\mathcal{B}}}+(-)^{M\tilde{\mathcal{B}}}(-)^{\mathcal{B}+M}\partial_{\tilde{\mathcal{B}}}P_{M}=\\
 & = & -(-)^{M}\partial_{M}B_{\mathcal{B}}+(-)^{M\mathcal{B}}\partiell{}{c^{\mathcal{B}}}P_{M}\\
\dann\tilde{C}_{M\mathcal{B}} & =_{g} & -\partial_{M}B_{\mathcal{B}}+\partiell{}{c^{\mathcal{B}}}P_{M}\\
\tilde{C}_{\mathcal{A}N} & \equiv & (-)^{\mathcal{A}}C_{\tilde{\mathcal{A}}N}=-(-)^{\mathcal{A}}\partial_{\tilde{\mathcal{A}}}P_{N}+(-)^{\mathcal{A}}(-)^{\tilde{\mathcal{A}}N}\partial_{N}P_{\tilde{\mathcal{A}}}=\\
 & = & -\partiell{}{c^{\mathcal{A}}}P_{N}+(-)^{\mathcal{A}N+N}\partial_{N}B_{\mathcal{A}}\\
\dann\tilde{C}_{\mathcal{A}N} & =_{g} & -\partiell{}{c^{\mathcal{A}}}P_{N}+\partial_{N}B_{\mathcal{A}}\\
\bar{C}_{\mathcal{AB}} & \equiv & (-)^{\mathcal{B}}C_{\tilde{\mathcal{A}}\tilde{\mathcal{B}}}=-(-)^{\mathcal{B}}\partial_{\tilde{\mathcal{A}}}P_{\tilde{\mathcal{B}}}+(-)^{\mathcal{B}}(-)^{\tilde{\mathcal{A}}\tilde{\mathcal{B}}}\partial_{\tilde{\mathcal{B}}}P_{\tilde{\mathcal{A}}}=\\
 & = & -(-)^{\mathcal{A}}\partiell{}{c^{\mathcal{A}}}B_{\mathcal{B}}-(-)^{\mathcal{AB}+\mathcal{B}}\partiell{}{c^{\mathcal{B}}}B_{\mathcal{A}}\\
\dann\bar{C}_{\mathcal{AB}} & =_{g} & -\partiell{}{c^{\mathcal{A}}}B_{\mathcal{B}}-\partiell{}{c^{\mathcal{B}}}B_{\mathcal{A}}\end{eqnarray}
We now recognize that $C_{MN},\tilde{C}_{M\mathcal{B}},\tilde{C}_{\mathcal{A}N}$
and $\bar{C}_{\mathcal{AB}}$ coincide with the coefficients in our
former calculation (\ref{eq:twoformcoeffsWithGhosts}) and everything
is consistent. 

For comparing the Dirac brackets we have to recall the definitions
of $C^{-1}$ or $D$ respectively. On the one hand we have\begin{eqnarray}
\sum\quad C_{\q{M}\q{K}}(C^{-1})^{\q{K}\q{N}}(-)^{\q{K}} & = & \delta_{\q{M}}\hoch{\q{N}}\\
\sum\quad(-)^{K}C_{MK}(C^{-1})^{KN}+(-)^{\tilde{\mathcal{C}}}C_{M\tilde{\mathcal{C}}}(C^{-1})^{\tilde{\mathcal{C}}N} & = & \delta_{M}\hoch{N}\\
\sum\quad(-)^{K}C_{MK}(C^{-1})^{K\tilde{\mathcal{B}}}+(-)^{\tilde{\mathcal{C}}}C_{M\tilde{\mathcal{C}}}(C^{-1})^{\tilde{\mathcal{C}}\tilde{\mathcal{B}}} & = & 0\\
\sum\quad(-)^{K}C_{\tilde{\mathcal{A}}K}(C^{-1})^{KN}+(-)^{\tilde{\mathcal{C}}}C_{\tilde{\mathcal{A}}\tilde{\mathcal{C}}}(C^{-1})^{\tilde{\mathcal{C}}N} & = & 0\\
\sum\quad(-)^{K}C_{\tilde{\mathcal{A}}K}(C^{-1})^{K\tilde{\mathcal{B}}}+(-)^{\tilde{\mathcal{C}}}C_{\tilde{\mathcal{A}}\tilde{\mathcal{C}}}(C^{-1})^{\tilde{\mathcal{C}}\tilde{\mathcal{B}}} & = & \delta_{\tilde{\mathcal{A}}}\hoch{\tilde{\mathcal{B}}}=\delta_{\mathcal{A}}\hoch{\mathcal{B}}\end{eqnarray}
which we can rewrite as\begin{eqnarray}
\sum\quad(-)^{K}C_{MK}(C^{-1})^{KN}-(-)^{M}\tilde{C}_{M\mathcal{C}}(C^{-1})^{\tilde{\mathcal{C}}N} & = & \delta_{M}\hoch{N}\\
\sum\quad(-)^{K}C_{MK}(C^{-1})^{K\tilde{\mathcal{B}}}-(-)^{M}\tilde{C}_{M\mathcal{C}}(C^{-1})^{\tilde{\mathcal{C}}\tilde{\mathcal{B}}} & = & 0\\
\sum\quad(-)^{K}(-)^{\mathcal{A}}\tilde{C}_{\mathcal{A}K}(C^{-1})^{KN}-\bar{C}_{\mathcal{A\mathcal{C}}}(C^{-1})^{\tilde{\mathcal{C}}N} & = & 0\\
\sum\quad(-)^{K}(-)^{\mathcal{A}}\tilde{C}_{\mathcal{A}K}(C^{-1})^{K\tilde{\mathcal{B}}}-\bar{C}_{\mathcal{A\mathcal{C}}}(C^{-1})^{\tilde{\mathcal{C}}\tilde{\mathcal{B}}} & = & \delta_{\tilde{\mathcal{A}}}\hoch{\tilde{\mathcal{B}}}=\delta_{\mathcal{A}}\hoch{\mathcal{B}}\end{eqnarray}
On the other hand we have (\ref{eq:inverseeqnsWithGhostsI})-(\ref{eq:inverseeqnsWithGhostsIV}),
where we now write out the summation convention explicitely:\begin{eqnarray}
\sum\quad(-)^{K}C_{MK}D^{KN}+(-)^{M}\tilde{C}_{M\mathcal{C}}\tilde{D}^{\mathcal{C}N} & = & \delta_{M}\hoch{N}\\
\sum\quad(-)^{M}C_{MK}\tilde{D}^{K\mathcal{B}}+(-)^{\mathcal{C}}\tilde{C}_{M\mathcal{C}}\bar{D}^{\mathcal{CB}} & = & 0\\
\sum\quad(-)^{K}\tilde{C}_{\mathcal{A}K}D^{KN}+(-)^{\mathcal{A}}\bar{C}_{\mathcal{AC}}\tilde{D}^{\mathcal{C}N} & = & 0\\
\sum\quad(-)^{\mathcal{A}}\tilde{C}_{\mathcal{A}K}\tilde{D}^{K\mathcal{B}}+(-)^{\mathcal{C}}\bar{C}_{\mathcal{AC}}\bar{D}^{\mathcal{CB}} & = & \delta_{\mathcal{A}}\hoch{\mathcal{C}}\end{eqnarray}
From this we can read off \begin{eqnarray}
\left(\begin{array}{cc}
D^{KN} & \tilde{D}^{K\mathcal{B}}\\
\tilde{D}^{\mathcal{C}N} & \bar{D}^{\mathcal{CB}}\end{array}\right) & = & \left(\begin{array}{cc}
(C^{-1})^{KN} & (-)^{K}(C^{-1})^{K\tilde{\mathcal{B}}}\\
-(C^{-1})^{\tilde{\mathcal{C}}N} & -(-)^{\mathcal{C}}(C^{-1})^{\tilde{\mathcal{C}}\tilde{\mathcal{B}}}\end{array}\right)\end{eqnarray}
Now we can verify\begin{eqnarray}
\sum(-)^{M}\left\{ F,\phi_{M}\right\} D^{MN}\left\{ \phi_{N},G\right\}  & = & \sum(-)^{M}\left\{ F,\Phi_{M}\right\} (C^{-1})^{MN}\left\{ \Phi_{N},G\right\} \\
\sum(-)^{\mathcal{B}}\left\{ F,\phi_{M}\right\} \tilde{D}^{M\mathcal{B}}\left\{ \varphi_{\mathcal{B}},G\right\}  & = & \sum(-)^{M}\left\{ F,\Phi_{M}\right\} (C^{-1})^{M\mathcal{\tilde{B}}}\left\{ \Phi_{\mathcal{\tilde{B}}},G\right\} \\
\sum\left\{ F,\varphi_{\mathcal{A}}\right\} \tilde{D}^{\mathcal{A}N}\left\{ \phi_{N},G\right\}  & = & \sum(-)^{\mathcal{\tilde{A}}}\left\{ F,\Phi_{\mathcal{\tilde{A}}}\right\} (C^{-1})^{\mathcal{\tilde{A}}N}\left\{ \Phi_{N},G\right\} \\
\sum(-)^{\mathcal{A}+\mathcal{B}}\left\{ F,\varphi_{\mathcal{A}}\right\} \bar{D}^{\mathcal{AB}}\left\{ \varphi_{\mathcal{B}},G\right\}  & = & \sum(-)^{\mathcal{\tilde{A}}}\left\{ F,\Phi_{\mathcal{\tilde{A}}}\right\} (C^{-1})^{\mathcal{\tilde{A}\tilde{B}}}\left\{ \Phi_{\mathcal{\tilde{B}}},G\right\} \end{eqnarray}
and have therefore convinced ourselves that also the Dirac-bracket
coincides and the two ways to calculate it are equivalent.

\subsection{First order formulation of a second order action}

If the action is just the first order formulation of an action which
had originally no constraints, then this Dirac bracket coincides with
the Poisson-bracket of the original action:\begin{eqnarray}
q^{\q{M}} & \equiv & (q^{\q{m}},q^{\tilde{\q{m}}})\equiv(q^{\q{m}},p_{\q{m}}),\quad\partial_{\q{M}}=(\partial_{\q{m}},\,\partial_{\tilde{\q{m}}})\equiv(\partial_{\q{m}},\,(-)^{\q{m}}\partiell{}{p_{\q{m}}})\equiv(\partial_{\q{m}},\,(-)^{\q{m}}\partial^{\q{m}})\\
S & = & \int\dot{q}^{\q{M}}P_{\q{M}}+V=\int\dot{q}^{\q{m}}p_{\q{m}}+V\\
\hspace{-1cm}\textrm{where }P_{\q{M}} & = & \left(P_{\q{m}},P_{\tilde{\q{m}}}\right)=\left(p_{\q{m}},0\right)\\
P & = & P_{\q{M}}\de q^{\q{M}}=p_{\q{m}}\de q^{\q{m}}=\de q^{\q{m}}p_{\q{m}}\\
C_{\q{M}\q{N}} & = & -2\partial_{[\q{M}}P_{\q{N}]}=-\left(\begin{array}{cc}
2\partial_{[\q{m}}P_{\q{n}]} & 2\partial_{[\q{m}}P_{\q{\tilde{n}}]}\\
2\partial_{[\q{\tilde{m}}}P_{\q{n}]} & 2\partial_{[\q{\tilde{m}}}P_{\q{\tilde{n}}]}\end{array}\right)=-\left(\begin{array}{cc}
0 & -(-)^{\q{m}\q{\tilde{n}}}\partial_{\q{\tilde{n}}}p_{\q{m}}\\
\partial_{\q{\tilde{m}}}p_{\q{n}} & 0\end{array}\right)=\left(\begin{array}{cc}
0 & \overbrace{\delta^{\q{n}}\tief{\q{m}}}^{(-)^{\q{m}}\delta_{\q{m}}\hoch{\q{n}}}\\
-(-)^{\q{m}}\delta^{\q{m}}\tief{\q{n}} & 0\end{array}\right)\qquad\\
C & = & \frac{1}{2}C_{\q{M}\q{N}}\de q^{\q{M}}\de q^{\q{N}}=\frac{1}{2}\sum(-)^{\q{N}\q{M}+\q{N}}C_{\q{M}\q{N}}\de q^{\q{M}}\de q^{\q{N}}=\\
 & = & \frac{1}{2}\sum(-)^{\q{n}\q{m}+\q{n}}\delta^{\q{n}}\tief{\q{m}}\de q^{\q{m}}\de p_{\q{n}}-(-)^{\q{n}\q{m}+\q{n}}(-)^{\q{m}}\delta^{\q{m}}\tief{\q{n}}\de p_{\q{m}}\de q^{\q{n}}=\\
 & = & \frac{1}{2}\sum(-)^{\q{n}}\de q^{\q{m}}\delta_{\q{m}}\hoch{\q{n}}\de p_{\q{n}}-\de p_{\q{m}}(-)^{\q{m}}\delta^{\q{m}}\tief{\q{n}}\de q^{\q{n}}=\\
 & = & \frac{1}{2}\de q^{\q{m}}\de p_{\q{m}}-\frac{1}{2}\de p_{\q{m}}\de q^{\q{m}}=\de q^{\q{m}}\de p_{\q{m}}\end{eqnarray}
\begin{eqnarray*}
(C^{-1})^{\q{M}\q{N}} & = & (-)^{\q{M}}\left(\begin{array}{cc}
0 & -(-)^{\q{m}}\delta^{\q{m}}\tief{\q{n}}\\
(-)^{\q{m}}\delta_{\q{m}}\hoch{\q{n}} & 0\end{array}\right)=\left(\begin{array}{cc}
0 & -\delta^{\q{m}}\tief{\q{n}}\\
\delta_{\q{m}}\hoch{\q{n}} & 0\end{array}\right)\end{eqnarray*}
\begin{eqnarray}
\left\{ F(q^{\q{M}}),G(q^{\q{M}})\right\}  & = & \partial F/\partial q^{\q{M}}(C^{-1})^{\q{M}\q{N}}\partiell{}{q^{\q{N}}}G=\\
 & = & \sum(-)^{\q{M}}\partial F/\partial q^{\q{M}}(C^{-1})^{\q{M}\q{N}}\partiell{}{q^{\q{N}}}G\\
 & = & \sum(-)^{\q{m}}(-)^{\q{m}}\partial F/\partial p_{\q{m}}\delta_{\q{m}}\hoch{\q{n}}\partiell{}{q^{\q{n}}}G-(-)^{\q{m}}\partial F/\partial q^{\q{m}}\delta^{\q{m}}\tief{\q{n}}(-)^{n}\partiell{}{p_{\q{n}}}G=\\
 & = & \partial F/\partial p_{\q{m}}\partiell{}{q^{\q{m}}}G-\partial F/\partial q^{\q{m}}\partiell{}{p_{\q{m}}}G\end{eqnarray}

\subsubsection*{A simpler way}

\label{sub:A-simpler-way}An alternative (simpler) way is to not insist
on NW for the capital indices, but only for the small ones:\begin{eqnarray}
q^{\q{M}} & \equiv & (q^{\q{m}},p_{\q{m}}),\quad\partial_{\q{M}}\equiv(\partial_{\q{m}},\,\partial^{\q{m}})\\
S & = & \int\dot{q}^{\q{M}}P_{\q{M}}+V=\int\dot{q}^{\q{m}}p_{\q{m}}+V\\
P_{\q{M}} & = & \left(P_{\q{m}},P^{\q{m}}\right)=\left(p_{\q{m}},0\right)\\
P & = & P_{\q{M}}\de q^{\q{M}}=p_{\q{m}}\de q^{\q{m}}\\
C_{\q{M}\q{N}} & = & -2\partial_{[\q{M}}P_{\q{N}]}=-\left(\begin{array}{cc}
2\partial_{[\q{m}}P_{\q{n}]} & 2\partial_{[\q{m}}P^{\q{n}]}\\
2\partial^{[\q{m}}P_{\q{n}]} & 2\partial^{[\q{m}}P^{\q{n}]}\end{array}\right)\greq\left(\begin{array}{cc}
0 & \partial^{\q{n}}p_{\q{m}}\\
-\partial^{\q{m}}p_{\q{n}} & 0\end{array}\right)\greq\left(\begin{array}{cc}
0 & \delta_{\q{m}}\hoch{\q{n}}\\
-\delta^{\q{m}}\tief{\q{n}} & 0\end{array}\right)\\
C & = & \frac{1}{2}C_{\q{M}\q{N}}\de q^{\q{M}}\de q^{\q{N}}=\\
 & = & -\frac{1}{2}\delta^{\q{m}}\tief{\q{n}}\de p_{\q{m}}\de q^{\q{n}}+\frac{1}{2}\delta_{\q{m}}\hoch{\q{n}}\de q^{\q{m}}\de p_{\q{n}}=\de q^{\q{m}}\de p_{\q{m}}\\
(C^{-1})^{\q{M}\q{N}} & = & \left(\begin{array}{cc}
0 & -\delta^{\q{m}}\tief{\q{n}}\\
\delta_{\q{m}}\hoch{\q{n}} & 0\end{array}\right)\end{eqnarray}
\begin{eqnarray}
\left\{ F(q^{\q{M}}),G(q^{\q{M}})\right\}  & = & \partial F/\partial q^{\q{M}}(C^{-1})^{\q{M}\q{N}}\partiell{}{q^{\q{N}}}G=\\
 & = & \partial F/\partial p_{\q{m}}\partiell{}{q^{\q{m}}}G-\partial F/\partial q^{\q{m}}\partiell{}{p_{\q{m}}}G\end{eqnarray}
}\rem{

\section{Equivalence of fermions to nilpotent bosons?}

(compare to \cite{Frydryszak:2006wk}) The formal equivalence of all
the graded equations to the bosonic equations naturally leads to the
question what differences remain between fermions and bosons. The
main difference is the nilpotency of fermions which was implemented
by anticommutativity but which could also be enforced by hand. However,
the nilpotency-equation is not gradifiable and differences in the
outcome are therfore unavoidable:

\begin{eqnarray*}
\tet\tet & = & 0\\
\eta\eta & = & 0\\
(\tet_{1}+\tet_{2})(\tet_{1}+\tet_{2}) & = & \tet_{1}^{2}+\tet_{1}\tet_{2}+\tet_{2}\tet_{1}+\tet_{2}^{2}=0\\
(\eta_{1}+\eta_{2})(\eta_{1}+\eta_{2}) & = & \eta_{1}\eta_{1}+\eta_{1}\eta_{2}+\eta_{2}\eta_{1}+\eta_{2}\eta_{2}=2\eta_{1}\eta_{2}\end{eqnarray*}
Related to that, graded antisymmetric matrices in the fermionic description
(so symmetric matricies) cannot be mapped one-to-one to antisymmetric
matrices in a bosonic description. This would only be possible if
the diagonals of the matrices are treated seperately.

A related problem is that for fermions $\theta$ the differential
$\de\theta$ is no longer nilpotent and we can have arbitrary high
powers of them. The question is, however, if higher powers play a
physical role, because integration always picks out the correct powers...

Argument for equivalence: assigning a seperate grading to every single
component of a fermion would correspond to describe it as a nilpotent
boson, as the component commutes with everything except with itself.
This, however, could break Lorentz-invariance...(?)

It is also not clear, how the integration of fermionic variables fits
into the whole picture. Probably, a nilpotent boson has to be integrated
in a similar way, in order to establish the equivalence.

}

\section{Lie-groups and -algebras}

\subsection{Gradifiable and not gradifiable group definitions}

\index{groups!super $\sim$}\index{supergroups}The positive experience
with the graded definition of matrix multiplication demands its application
to supergroups. The first question arising is, which supergroup definitions
have a natural gradification and which do not. Let us just give a
few examples to make the idea transparent.

The \textbf{general\index{general linear group!supergroup} linear
group}, i.e. the group of all invertible matrices $GL(n)$ is easily
gradifiable, because we know how to gradify the matrix multiplication
and we have (for bosonic supermatrices, i.e. matrices with bosonic
rumpf) a clear notion of invertability. If the index of the matrix
runs over $b$ bosonic and $f$ fermionic indices, the resulting group
is denoted by \index{$GL(b\mid f)$}$GL(b|f)$(see e.g. \cite[p.90]{Frappat:1996pb}).
Also the definition of the \textbf{special\index{special linear group!supergroup}
linear group} is gradifiable, because the definition of the determinant
is gradifiable as we discussed earlier, and the condition $\det M=1$
thus makes sense in the graded case as well. Because of $\det(M\cdot N)=\det M\cdot\det N$,
this condition defines a subgroup which is denoted as \index{$SL(b\mid f)$}$SL(b|f)$.

For bosonic matrices, the \textbf{unitary\index{unitary group} group}
is defined via \begin{equation}
U^{\dagger}U=\one\end{equation}
Or with indices \begin{eqnarray}
(U^{\dagger})_{m}\hoch{k}\delta_{kl}U^{l}\tief{n} & = & \delta_{mn}\end{eqnarray}
We have a well defined notion of graded hermitean conjugation and
also of a graded unity in the sense of a graded Kronecker delta with
one lower and one upper index. There is no natural gradification,
however, of a Kronecker delta with two indices at the same position.
It is strictly speaking a metric and not a unit operator. In even
dimensions we could use $\left(\begin{array}{cc}
0 & \one\\
-\one & 0\end{array}\right)$ as metric for the fermionic subspace, but this would be an ad-hoc
choice. The problem is that there is no characteristic property of
$\delta_{mn}$ which is gradifiable in our sense to uniquely give
its graded version. The characterization that it is a diagonal matrix
with only 1's in the diagonal is certainly not suitable for gradification,
because for fermionic dimensions the metric should still be graded
symmetric (i.e. antisymmetric) and is therefore necessarily off-diagonal.
There is thus at first sight no natural gradification of the definition
of the unitary group. Note that there exists nevertheless the notion
of a unitary supergroup \index{$U(b\mid f)$}$U(b|f)$ in the literature
(see e.g. \cite[p.90]{Frappat:1996pb}) . 

The practical meaning of the unitary group is that it leaves the canonical
scalar product $\delta_{\bar{m}n}$ in $\mathbb{C}^{d}$ invariant.
Suppose we have a more general scalar product $\erw{a,b}=(\bar{a})^{\bar{m}}g_{\bar{m}n}b^{n}$
and make a basis change. $a^{m}=U^{m}\tief{k}\tilde{a}^{k},\quad b^{n}=U^{n}\tief{l}\tilde{b}^{l}$.
Then we obtain $\erw{a,b}=(U^{m}\tief{k}\tilde{a}^{k})^{*}g_{\bar{m}n}U^{n}\tief{l}\tilde{b}^{l}\stackrel{!}{=}(\tilde{a}^{k})^{*}\tilde{g}_{\bar{k}l}\tilde{b}^{l}$.
The hermitean scalar product $g_{\bar{m}n}$ therefore transforms
like \begin{eqnarray}
\tilde{g}_{\bar{k}l} & = & (U^{m}\tief{k})^{*}g_{\bar{m}n}U^{n}\tief{l}=(U^{\dagger})_{\bar{k}}\hoch{\bar{m}}g_{\bar{m}n}U^{n}\tief{l}\end{eqnarray}
 We could define a matrix to be \textbf{unitary with respect to $g_{\bar{m}n}$}
iff \begin{equation}
(U^{\dagger})_{\bar{k}}\hoch{\bar{m}}g_{\bar{m}n}U^{n}\tief{l}=g_{\bar{m}n}\end{equation}
This is a gradifiable definition, because it is based on some generic
$g_{\bar{m}n}$ instead of the specific $\delta_{\bar{m}n}$. As discussed
above there is no defining property of $\delta_{\bar{m}n}$ which
is gradifiable.

The situation is the same for the \textbf{Lorentz group} (or likewise
for the orthorgonal group) with \begin{equation}
(\Lambda^{T})_{m}\hoch{k}\eta_{kl}\Lambda^{l}\tief{n}=\eta_{mn}\end{equation}
where we are again missing a gradification of the definition of $\eta_{mn}$.

The situation is a bit different for the \textbf{symplectic\index{symplectic group!supergroup}
group}, although its definition is very close to the above two. Symplectic
structures need even dimensional spaces. Assigning upper indices $\hoch{k}$
to the first $d$ dimensions and lower indices $\tief{k}$ to the
second $d$ dimensions and combine both into one index $\hoch{\q{k}}\equiv\left(\hoch{k},\tief{k}\right)$,
then the canonical symplectic form (being the matrix-inverse of the
canonical Poisson structure of the previous section) can be written
as\begin{equation}
B_{\q{k}\q{l}}=\left(\begin{array}{cc}
0 & \delta_{k}\hoch{l}\\
-\delta^{k}\tief{l} & 0\end{array}\right)\end{equation}
In contrast to the metrics of before, the symplectic form is gradifiable,
because it contains two unit operators in subspaces of which we know
the gradification. Elements $S$ of the symplectic group $SP(2d)$
are then given by\begin{eqnarray}
(S^{T})_{\q{m}}\hoch{\q{k}}B_{\q{k}\q{l}}S^{\q{l}}\tief{\q{n}} & = & B_{\q{k}\q{l}}\end{eqnarray}
Simply gradifying the indices yields the graded definition of the
symplectic group. The body $S$ of the symplectic matrix, however,
is not gradifiable, as it appears twice in the term on the left and
not at all on the right. If the index $k$ runs over $b$ bosonic
and $f$ fermionic indices, the resulting group could be denoted by
\index{$SP(2b\mid 2f)$}$SP(2b|2f)$, while in literature it is common
to introduce instead the notion of an orthosymplectic group which
differs, however, a bit from this group (see e.g. \cite[p.90]{Frappat:1996pb}).
The precise form of the group elements $S\in SP(2b|2f)$ depends on
the choice of either NW or NE for the definition of the matrix multiplication
and of the position of the indices at the matrix (first index up and
second down or vice verse).

Having seen the above example, it is obvious that gradification also
works for $O(d,d)$ or $SO(d,d)$ based on the metric $\eta_{\q{m}\q{n}}=\left(\begin{array}{cc}
0 & \delta_{m}\hoch{n}\\
\delta^{m}\tief{n} & 0\end{array}\right)$. If the indices $m,n$ take $d$ values, this metric has in the bosonic
case the signature $(d,d)$. Containing two off-diagonal Kronecker
deltas, the graded version of the metric looks just the same. If $d$
splits into $b$ bosonic and $f$ fermionic dimensions, the resulting
supergroups could be denoted as \index{$O(b,b\mid f,f)$}$O(b,b|f,f)$
and \index{$SO(b,b\mid f,f)$}$SO(b,b|f,f)$. For the fermionic subspace
we have $\delta^{\mu}\tief{\nu}=-\delta_{\nu}\hoch{\mu}$, and the
corresponding matrix block of the metric is numerically just the matrix
of a bosonic symplectic form. In this sense, $O(d,d)$ and $SP(2d)$
interchange their role in the bosonic and fermionic subspaces: \begin{equation}
O(d,d|0,0)\cong SP(0,0|d,d)\quad\mbox{and}\quad O(0,0|d,d)\cong SP(d,d|0,0)\label{eq:O-SP-relation}\end{equation}

Note finally that all supergroups which cannot be seen as a gradification
of a bosonic group, of course still make perfect sense. The message
is only that properties of those supergroups must be studied independently
and cannot be deduced from the corresponding bosonic groups via the
gradification theorem. The main example are groups of fermionic supermatrices.
The bosonic definition of a group requires the existence of an inverse
matrix. As we discussed already in the chapter on supermatrices, the
notion of an inverse matrix can only be gradified in the case of a
bosonic supermatrix, while the definition of a 'special graded inverse'
of a fermionic supermatrix cannot be used to take advantage of the
gradification theorem. 

In \cite{Cvitanovic:1979qz,Cv2007:bt} it was observed that $SO(d)$
can be seen as $SP(-d)$ (with $d$ fermionic, i.e. negative\index{negative dimension}
dimension\index{dimension!negative $\sim$}s -- see page \pageref{par:negativeDimension})
and that $SP(d)$ can be seen as $SO(-d)$. Understanding $SP(-d)\equiv SP(0|d)$
and $SO(-d)\equiv SO(0|d)$, this does almost but not completely match
with our above observation (\ref{eq:O-SP-relation}) which holds only
for split signature. This might be due to different definitions of
the supergroups and it would be interesting to make the comparison
in more detail.

\subsection{Graded Lie algebra}

\index{graded Lie algebra}\index{Lie algebra!graded $\sim$}In the
previous subsection we have just discussed a few examples for the
gradification of some Lie groups, although a more detailed study would
be a very interesting subject. Likewise we are not going to discuss
(graded) Lie algebras in any detail in this subsection, but instead
want to stress a few minor points, related to the summation convention.
In the previous subsection we were only discussing supergroups whose
elements are bosonic supermatrices, i.e. graded matrices with bosonic
rumpf, because only there we have a natural gradified version of an
inverse matrix. Nevertheless, even when the group matrices of a Lie
Group are all bosonic, its infinitesimal generators (when based on
the module of supernumbers) might well be expanded in a basis $T_{A}$
that contains fermionic matrices. Each of the $T_{A}$'s is a supermatrix,
and it depends on the index $A$, whether it is a fermionic or a bosonic
one:\begin{equation}
\abs{(T_{A})^{M}\tief{N}}=\abs{A}+\abs{M}+\abs{N}\end{equation}
Like in the bosonic case, group elements in the connected component
of the unity can be parametrized by%
\footnote{\label{fn:hermiticityOfGen}\index{footnote!\thefoot. hermiticity and unitarity and BCH for supergroups}Note
that due to our definition of complex conjugation and hermitean conjugation
$x^{A}T_{A}$ is hermitean if $T_{A}$ is hermitean and $x^{A}$ is
real: $(x^{A}T_{A})^{\dagger}=(x^{A})^{*}T_{A}^{\dagger}=x^{A}T_{A}$.
The group element $e^{ix^{A}T_{A}}$ thus would correspond to a unitary
group element. This would disagree with the statement before that
there is no natural gradification of unitary matrices. In fact, already
for the hermiticity we were too sloppy in the above reasoning: A graded
hermitean matrix is defined only when both indices are at the same
position. If one index is upstairs and the other is downstairs, one
needs a metric to define hermiticity, and this is again missing in
general in the graded case.

Note further that sometimes it is convenient to parametrize the group
element differently, namely by exponentiating seperately the bosonic
and the fermionic contributions:\[
g(x)=e^{ix^{A}T_{A}}\stackrel{!}{=}e^{iy^{a}T_{a}}e^{iy^{\bs{\alpha}}T_{\bs{\alpha}}}\equiv g(y)\]
The relation between $x$ and $y$ is obtained by using the graded
version of the Baker\index{Baker-Campbell-Hausdorff formula}-Campbell\index{Campbell!Baker-$\sim$-Hausdorff-formula}-Hausdorff\index{Hausdorff!Baker-Campbell-$\sim$-formula}
formula, which is simply the gradification of the bosonic one, i.e.
$e^{A}e^{B}=e^{A+B+\frac{1}{2}[A,B]+\frac{1}{12}[A,[A,B]]+\frac{1}{12}[[A,B],B]+\mc{O}([.,.]^{3})}$.$\qquad\fussend$%
} \begin{equation}
g(x)=e^{ix^{A}T_{A}}\end{equation}
where $x^{A}$ are some coordinates whose grading $\abs{A}$ is the
same as the one of the generators $T_{A}$, so that the group element
is a bosonic supermatrix. For example, for $g(x)$ to be in $GL$,
the exponent can be any (small) supermatrix, while for $g(x)$ to
be in $SL$, it has to be traceless ($\det e^{ix^{A}T_{A}}=\exp ix^{A}\tr T_{A}$).
One possible basis of the algebra of all supermatrices consists of
the matrices with one entry 1 and zero everywhere else. If the 1 is
in one of the diagonal blocks, the corresponding basis matrix $T_{a}$
is a bosonic one, while if the 1 is in one of the off-diagonal blocks,
$T_{\bs{\mc{A}}}$ is considered as a fermionic supermatrix (although
it has bosonic entries only). The fermionic supermatrices $T_{\bs{\mc{A}}}$
are contracted with a fermionic parameter $\tet^{\mc{A}}\equiv x^{\bs{\mc{A}}}$,
so that the resulting group element $g(x)$ is a bosonic supermatrix.

The algebra is determined by providing the \textbf{structure\index{structure constants!real $\sim$}
constants} for the (graded) commutator \begin{eqnarray}
[T_{A},T_{B}] & \greq & if_{AB}\hoch{C}T_{C}\label{eq:struc-const}\end{eqnarray}
The graded equal sign has no effect here again, because the naked
indices $A$ and $B$ are in the same order on both sides. If one
is dealing naively (see remark in footnote \ref{fn:hermiticityOfGen})
with (graded) hermitean matrices (or operators) $T_{A}^{\dagger}=T_{A}$,
then the commutator is always graded antihermitean $[T_{A},T_{B}]^{\dagger}\greq[T_{B}^{\dagger},T_{A}^{\dagger}]\greq[T_{B},T_{A}]\greq-[T_{A},T_{B}]$,
no matter whether the indices $A$ and $B$ are bosonic or fermionic.
Extracting the imaginary unit '$i$' then leads to real structure
constants. Note that in most of the literature, fermionic and bosonic
operators are treated differently in this issue, because of the different
definition of hermitean conjugation. An immediate application of the
gradification theorem is the \textbf{Jacobi\index{Jacobi identity!for the structure constants}
identity} in terms of the structure constants, which has of course
the same form as in the bosonic case, but with graded summation and
graded antisymmetrization:\begin{eqnarray}
f_{[AB|}\hoch{D}f_{D|C]}\hoch{E} & = & 0\end{eqnarray}
An \textbf{invariant metric }\begin{eqnarray}
\erw{T_{A},T_{B}} & \equiv & \mc{H}_{AB}\end{eqnarray}
is defined to obey \begin{eqnarray}
\erw{\left[T_{C},T_{A}\right],T_{B}}+\erw{T_{A},\left[T_{C},T_{B}\right]} & \greq & 0\end{eqnarray}
In terms of the structure constants (with $f_{ABC}\equiv f_{AB}\hoch{D}\mc{H}_{DC}$),
this reads \begin{eqnarray}
f_{CAB}+f_{CBA} & \greq & 0\end{eqnarray}
which means that the structure constants are also (graded) antisymmetric
in the last two indices and therefore in all indices. Indices are
pulled up again with the graded inverse of $\mc{H}_{AB}$ which is
defined by \begin{equation}
\mc{H}_{AC}\mc{H}^{CB}=\delta_{A}\hoch{B}\end{equation}
or equivalently $\mc{H}^{AC}\mc{H}_{CB}=\delta^{A}\tief{B}$. The
graded inverse $\mc{H}^{AB}$ differs from the naive (numerical) inverse
by a factor $(-)^{A}$ in NW and by a factor $(-)^{B}$ in NE. 

The defining equation for the structure constants (\ref{eq:struc-const})
seems to suggest that we already have fixed NW conventions, but it
can also be rewritten to enfavour NE. To this end we need the fact
that in the case of the existence of a group invariant metric to pull
up and down the indices $A,B$ and $C$, the structure constants with
all indices down are completely (graded) antisymmetric $f_{ABC}\greq f_{CAB}$.
The commutator (\ref{eq:struc-const}) then reads \begin{eqnarray}
[T_{A},T_{B}] & \greq & iT_{C}f^{C}\tief{AB}\end{eqnarray}
In both versions of the equation, the actual summation convention
has not yet been fixed. Let us finally write down the original form
(\ref{eq:struc-const}) of this commutator explicitely in NW-conventions,
including the matrix indices:\begin{equation}
\sum_{K}\left\{ (-)^{(M+K)B}(-)^{K}(T_{A})\hoch{M}\tief{K}(T_{B})^{K}\tief{N}-(-)^{AB}(-)^{(M+K)A}(-)^{K}(T_{B})\hoch{M}\tief{K}(T_{A})^{K}\tief{N}\right\} =\sum_{C}if_{AB}\hoch{C}(T_{C})^{M}\tief{N}\end{equation}
The position of the supermatrix indices (first one upstairs, second
downstairs) is more natural for NE conventions, where the sign $(-)^{K}$
would not appear in the terms on the lefthand side. 

Natural applications of the above considerations appear in the study
of WZNW-models\index{WZNW-model} based on graded Lie algebras (e.g.
in our study \cite{Guttenberg:2004ht} of a WZNW-like model \cite{Nh:2003kq},
where we however not yet rigorously applied the present conventions).

\section{Remark on the pure spinor ghosts}

In part \ref{par:PureSpinorString}, we will make frequent use of
the presented conventions. In particular, we will always use the graded
summation convention and the small graded equal sign without denoting
it explicitely! There are some effects that one needs to get used
to. The formalism contains among others the variables $x^{m}$, $\tet^{\mu}$,
$\hat{\tet}^{\mu}$ and a commuting ghost variable $\lambda^{\mu}$.
When we want to describe the first three as just components of a supercoodinate
$x^{M}$, we have to assign all the grading to the indices: $\tet^{\mu}\To\theta^{\bs{\mu}}\equiv x^{\bs{\mu}}$.
We call that a {}``rumpf-index\index{grading shift}\index{rumpf-index grading shift}
grading shift''. The fermionic variable $\tet^{\mu}=\theta^{\bs{\mu}}$
can be treated in both ways, either as odd rumpf with even index or
as even rumpf with odd index. The boldface notation should serve as
a reminder, which point of view we take. When we are considering the
combining object $x^{M}$, we have no choice, because all entries
share the same rumpf 'x'. Therefore we have to assign the grading
to the index and have to do the same for the ghost index, because
it simply is the same index:\begin{equation}
\lambda^{\mu}\To\lam^{\bs{\mu}}\end{equation}
When we leave away in calculations all index-dependent signs, the
pure spinor ghost will effectively be treated as an anticommuting
variable, because the rumpf is anticommuting! Another similar effect
is the switch of the symmetry properties of bispinors. E.g. the chiral
$\gamma$-matrices\index{$\gamma^c_{\bs{\alpha\beta}}$} \begin{equation}
\gamma_{(\alpha\beta)}^{c}\To\gamma_{[\bs{\alpha\beta}]}^{c}\end{equation}
which are symmetric before the grading shift, become effectively antisymmetric
afterwards. As an example, consider the following term\index{$\lam^{\bs{\alpha}}$|itext{pure spinor ghost}}\begin{eqnarray}
(\lambda\gamma^{c}\partial\lambda) & = & \lambda^{\alpha}\gamma_{(\alpha\beta)}^{c}\partial\lambda^{\beta}=\partial\lambda^{\alpha}\gamma_{(\alpha\beta)}^{c}\lambda^{\beta}=(\partial\lambda\gamma^{c}\lambda)\end{eqnarray}
The calculation goes through in the same way after the shift, because
the antisymmetry of the $\gamma$-matrix is compensated by the {}``anticommutativity''
of the ghosts. \begin{equation}
\lam\gamma^{c}\partial\lam\equiv\lam^{\bs{\alpha}}\gamma_{[\bs{\alpha}\bs{\beta}]}^{c}\partial\lam^{\bs{\beta}}=\partial\lam^{\bs{\alpha}}\gamma_{[\bs{\alpha}\bs{\beta}]}^{c}\lam^{\bs{\beta}}=\partial\lam\gamma^{c}\lam\end{equation}
As one of the summations is over a graded rumpf and another is in
the wrong direction, the contraction coincides with the one for ungraded
indices. This is not true for $\theta$, where we have a sign change
(for NW as well as for NE):\begin{eqnarray}
\lam\gamma^{c}\partial\lam & = & \lambda\gamma^{c}\partial\lambda\\
\theta\gamma^{c}\partial\theta & = & -\bs{\theta}\gamma^{c}\partial\bs{\theta}\end{eqnarray}
Note finally that the rumpf of $\gamma^{c}$ (the off-diagonal block
of $\Gamma^{c}$) stays bosonic, even when $\Gamma^{c}\To\bs{\Gamma}^{c}$
is reinterpreted as a fermionic supermatrix as suggested in section
\vref{sec:gradedGamma}. 

\bibliographystyle{fullsort}
\bibliography{phd,Proposal,/home/basti/ebooks}
\printindex{}
}

\part{Berkovits' Pure Spinor String in General Background}

\label{par:PureSpinorString}

\chapter{Motivation of the Pure Spinor String in Flat background}

\label{cha:Flat-background}\index{flat background}{\inputTeil{0}
\ifthenelse{\theinput=1}{}{}

\title{Pure Spinor String in Flat Background}

\author{Sebastian Guttenberg}

\date{August 12, '07}

\maketitle
\begin{abstract}
Part of thesis, 
\end{abstract}
\tableofcontents{}

\rem{To do:

\begin{itemize}
\item Don't panic
\end{itemize}
}

\section{From Green-Schwarz to Berkovits}

\index{string|see{pure spinor and Green Schwarz}}\index{Green Schwarz string}The
classical type II Green Schwarz (GS) superstring describes the embedding
of a string worldsheet into a target type II superspace with coordinates
$x^{M}\equiv(x^{m},\tet^{\mu},\hat{\tet}^{\hat{\mu}})$\index{$x^M$}\index{$x^m$}\index{$\tet^{\mu}$}\index{$\tet$@$\hat\tet^{\hat{\mu}}$}.
The bosonic coordinates $x^{m}$ locally parametrize the ten-dimensional
spacetime manifold, while the fermionic coordinates $\tet^{\mu}$
and $\hat{\tet}^{\hat{\mu}}$ have the dimension of Majorana Weyl
spinors and thus have each 16 real components. The Lorentz transformation
of spinors is from the supermanifold point of view a structure group
transformation in the tangent space of the supermanifold. In the flat
case, where one can identify the manifold with its tangent space,
the $\tet$'s are clearly spinors themselves. In the context of a
curved supermanifold that we will treat later on, this will not be
the case a priori. The $\tet$'s then only transform under super-diffeomorphisms
and not under structure group transformations. However, the supergravity
constraints will allow to choose a gauge (WZ-gauge) in which the two
transformations are coupled and the $\tet'$s likewise transform under
a structure group transformation. This is just a remark on the use
of the {}``curved index'' $\mu$. Objects that transform a priori
under the structure group carry the flat index $A$ or in particular
$\alpha$. 

The cases type IIA and IIB will be treated at the same time via the
choice $\hat{\tet}^{\hat{\mu}}\equiv\hat{\tet}_{\mu}$ for IIA and
$\hat{\tet}^{\hat{\mu}}\equiv\hat{\tet}^{\mu}$ for IIB. The supersymmetry
transformation in flat superspace reads\index{SUSY!in flat superspace}\index{superspace!flat}
\begin{eqnarray}
\delta\tet^{\mu} & = & \feps^{\mu},\qquad\delta\hat{\tet}^{\hat{\mu}}=\hat{\feps}^{\hat{\mu}}\\
\delta x^{m} & = & \feps\gamma^{m}\tet+\hat{\feps}\gamma^{m}\tet\end{eqnarray}
The small $\gamma$-matrices are discussed in the appendix \ref{cha:Gamma-Matrices}.
In order to build a supersymmetric theory, it is reasonable to consider
supersymmetric building blocks, in particular supersymmetric one-forms
(vielbeins)\index{flat superspace}\begin{equation}
E^{A}\equiv\de x^{M}E_{M}\hoch{A}=\big(\underbrace{\de x^{a}+\de\tet\gamma^{a}\tet+\de\hat{\tet}\gamma^{a}\hat{\tet}}_{\Pi^{a}}\quad,\quad\de\tet^{\alpha}\quad,\quad\de\hat{\tet}^{\hat{\alpha}}\big)\label{eq:vielbeinInFlatSuperspace}\end{equation}
\index{$E^A$!in flat superspace}\index{vielbein 1-form!in flat superspace}\index{supersymmetry-invariant 1-form}\index{invariant 1-form}Its
pullback to the worldsheet will be denoted by \begin{equation}
\Pi_{z/\bar{z}}^{A}\equiv\partial_{z/\bar{z}}x^{M}E_{M}\hoch{A}\end{equation}
We do not distinguish notationally between the coordinates of the
superspace and the embedding functions. The bosonic components $\Pi_{z}^{a}$\index{$\Pi_z^a$!in flat superspace}
are known as the supersymmetric momentum\begin{eqnarray}
\Pi_{z/\bar{z}}^{a} & = & \partial_{z/\bar{z}}x^{a}+\partial_{z/\bar{z}}\tet\gamma^{a}\tet+\partial_{z/\bar{z}}\hat{\tet}\gamma^{a}\hat{\tet}\end{eqnarray}

The introduction to the Green Schwarz string and the motivation for
the pure spinor formalism will be rather quick and sketchy. We will
be much more careful when we start to discuss the pure spinor string
in general background.

The classical Green Schwarz superstring in flat background consists
of the square of this momentum plus a Wess-Zumino term which establishes
a fermionic gauge symmetry. This gauge symmetry, called $\kappa$-symmetry\index{kappa-symmetry@$\kappa$-symmetry},
guarantees the matching of the physical fermionic and bosonic degrees
of freedom. The GS\index{Green Schwarz action} action has in conformal
gauge the following form:\index{$S_{GS}$|itext{Green Schwarz action}}\index{Wess-Zumino part of GS action}\index{$L_{WZ}$@$\mc{L}_{WZ}$|itext{Wess Zumino term}}
\begin{eqnarray}
S_{GS} & = & \int d^{2}z\quad\frac{1}{2}\Pi_{z}^{a}\eta_{ab}\Pi_{\bar{z}}^{b}+\mc{L}_{WZ}\\
\mc{L}_{WZ} & = & -\frac{1}{2}\Pi_{zm}\left(\tet\gamma^{m}\bar{\partial}\tet-\hat{\tet}\gamma^{m}\bar{\partial}\hat{\tet}\right)+\frac{1}{2}(\tet\gamma^{m}\partial\tet)(\hat{\tet}\gamma_{m}\bar{\partial}\hat{\tet})-(z\leftrightarrow\bar{z})\end{eqnarray}
It is covariant and almost manifestly spacetime supersymmetric. In
this last feature it differs from the RNS string, where space time
supersymmetry only comes in after GSO projection. The problem for
the Green Schwarz string on the other hand is that a covariant quantization
with the standard BRST procedure does not work. The reason for this
misery is a set of 16 mixed first and second class constraints $\bs{d}_{z\alpha}$
that cannot be split easily into first and second class type in a
covariant manner. The conjugate momentum $\bs{p}_{z\alpha}$\index{$p_{z\bs{\alpha}}$}
of $\tet^{\alpha}$ can be entirely expressed in terms of other phase
space variables and the corresponding fermionic phase space constraint
is just $\bs{d}_{z\alpha}$. It has the following explicit form (the
form of conjugate momentum to $x^{m}$ was already plugged in)\index{$d_{z\bs{\alpha}}$@$\dP_{z\bs{\alpha}}$}
\begin{eqnarray}
\bs{d}_{z\alpha} & \equiv & \bs{p}_{z\alpha}-(\gamma_{a}\tet)_{\alpha}\left(\partial x^{a}-\frac{1}{2}\tet\gamma^{a}\partial\tet-\frac{1}{2}\hat{\tet}\gamma^{a}\partial\hat{\tet}\right)\label{eq:d-definition-flach}\end{eqnarray}
Half of these constraints are first class and correspond to the above
mentioned fermionic $\kappa$ gauge symmetry. The fact that they have
a second-class part can be seen in a non-closure of the Poisson-algebra,
which has the following schematica form: \begin{eqnarray}
\left\{ \bs{d}_{z\alpha}(\sigma),\bs{d}_{z\beta}(\sigma')\right\}  & \propto & 2\gamma_{\alpha\beta}^{a}\Pi_{za}\delta(\sigma-\sigma')\end{eqnarray}
Siegel \cite{Siegel:1986xj} had the idea to make $\bs{d}_{z\alpha}$
part of a closed algebra by just adding the generators that arise
via the Poisson bracket, which leads to a (centrally extended), but
otherwise closed algebra\begin{eqnarray}
\left\{ \bs{d}_{z\alpha},\Pi_{za}\right\}  & \propto & 2\gamma_{a\,\alpha\beta}\partial\theta^{\beta}\delta(\sigma-\sigma')\\
\left\{ \Pi_{za},\Pi_{zb}\right\}  & \propto & \eta_{ab}\delta'(\sigma-\sigma')\\
\left\{ \bs{d}_{z\alpha},\partial\theta^{\beta}\right\}  & \propto & \delta_{\alpha}^{\beta}\delta'(\sigma-\sigma')\end{eqnarray}
The important observation is now that the same chiral algebra can
be obtained from a free-field Lagrangian, where the variable $\bs{p}_{z\alpha}$
is independent and cannot be integrated out:\index{$d_{zb}$@$\hat{d}_{\bar{z}\hat{\bs{\alpha}}}$}\index{$p_{zb}$@$\hat{p}_{\bar{z}\hat{\bs{\alpha}}}$}\begin{eqnarray}
S_{free} & = & \int d^{2}z\quad\frac{1}{2}\partial x^{m}\eta_{mn}\bar{\partial}x^{n}+\bar{\partial}\tet^{\alpha}\bs{p}_{z\alpha}+\partial\hat{\tet}^{\hat{\alpha}}\hat{\bs{p}}_{\bar{z}\hat{\alpha}}=\\
 & = & \int d^{2}z\quad\underbrace{\frac{1}{2}\Pi_{z}^{a}\eta_{ab}\Pi_{\bar{z}}^{b}+\mc{L}_{WZ}}_{\mc{L}_{GS}}+\bar{\partial}\tet^{\alpha}\bs{d}_{z\alpha}+\partial\hat{\tet}^{\hat{\alpha}}\hat{\bs{d}}_{\bar{z}\hat{\alpha}}\end{eqnarray}
In the second line we have used the original definition (\ref{eq:d-definition-flach})
for $\bs{d}_{z\alpha}$. Remarkably, this action coincides with the
Green Schwarz action for $\bs{d}_{\alpha}=\hat{\bs{d}}_{\hat{\alpha}}=0$.
In the above free theory, however, $\bs{d}_{z\alpha}$ is a priori
not a Hamiltonian constraint, but still a generator of a chiral (not
local) symmetry. In any case, the reformulation does not remove the
mixed first-second class property of $\bs{d}_{z\alpha}$, but it provides
a simple free-field Lagrangian. Berkovits \cite{Berkovits:2000fe}
had the idea to implement the constraints cohomologically with a BRST
operator disregarding its non-closure. The corresponding current ($\Q=\oint dz\:\bs{j}_{z}$)
for the left-moving and the right-moving sector take respectively
the simple form\index{$j$@$\bs{j}_{z}$}\index{$j$@$\hat{\bs{\jmath}}_{\bar{z}}$}
\begin{eqnarray}
\bs{j}_{z} & = & \lambda^{\alpha}\bs{d}_{z\alpha},\quad\bs{j}_{\bar{z}}=0\label{BerkovitsBRST}\\
\hat{\bs{\jmath}}_{\bar{z}} & = & \hat{\lambda}^{\alpha}\hat{\bs{d}}_{\bar{z}\hat{\alpha}},\quad\hat{\bs{\jmath}}_{z}=0\end{eqnarray}
where $\lambda^{\alpha}$ is a commuting ghost. For first class constraints
the BRST cohomology can be built, because the BRST operator is nilpotent
due to the closure of the algebra. For second class constraints, however,
the non-closure implies a lack of nilpotency of the BRST operator.
To overcome this problem, Berkovits put a constraint on the ghost
field $\lambda$ and $\hat{\lambda}$, the so called pure spinor constraint
\begin{equation}
\lambda\gamma^{c}\lambda=0,\qquad\hat{\lambda}\gamma^{c}\hat{\lambda}=0\end{equation}
This enforces nilpotency of the BRST operator and provides a well-defined
theory. The pure spinor constraint and the ghost kinetic term have
to be added to the original free action:\index{Berkovits string|see{pure spinor string}}\index{pure spinor string}\index{pure spinor string!in flat background}\index{$L_{gh}$@$\mc{L}_{gh}$|itext{ghost Lagrangian}}\begin{eqnarray}
S_{ps} & = & \int d^{2}z\quad\frac{1}{2}\partial x^{m}\eta_{mn}\bar{\partial}x^{n}+\bar{\partial}\tet^{\alpha}\bs{p}_{z\alpha}+\partial\hat{\tet}^{\hat{\alpha}}\hat{\bs{p}}_{\bar{z}\hat{\alpha}}+\mc{L}_{gh}\\
 & = & \int d^{2}z\quad\frac{1}{2}\Pi_{z}^{a}\eta_{ab}\Pi_{\bar{z}}^{b}+\mc{L}_{WZ}+\bar{\partial}\tet^{\alpha}\bs{d}_{z\alpha}+\partial\hat{\tet}^{\hat{\alpha}}\hat{\bs{d}}_{\bar{z}\hat{\alpha}}+\mc{L}_{gh}\\
\Pi_{z}^{a} & = & \partial x^{a}+\partial\tet\gamma^{a}\tet+\partial\hat{\tet}\gamma^{a}\hat{\tet}\\
\de_{z\alpha} & = & \bs{p}_{z\alpha}-(\gamma_{m}\tet)_{\alpha}\left(\partial x^{m}-\frac{1}{2}\tet\gamma^{m}\partial\tet-\frac{1}{2}\hat{\tet}\gamma^{m}\partial\hat{\tet}\right)\\
\mc{L}_{WZ} & = & -\frac{1}{2}\Pi_{zm}\left(\tet\gamma^{m}\bar{\partial}\tet-\hat{\tet}\gamma^{m}\bar{\partial}\hat{\tet}\right)+\frac{1}{2}(\tet\gamma^{m}\partial\tet)(\hat{\tet}\gamma_{m}\bar{\partial}\hat{\tet})-(z\leftrightarrow\bar{z})\label{eq:flat:WZ-term}\\
\mc{L}_{gh} & = & \bar{\partial}\lambda^{\beta}\omega_{z\beta}+\partial\hat{\lambda}^{\hat{\beta}}\hat{\omega}_{\bar{z}\hat{\beta}}+\frac{1}{2}L_{z\bar{z}a}(\lambda\gamma^{a}\lambda)+\frac{1}{2}\hat{L}_{z\bar{z}a}(\hat{\lambda}\gamma^{a}\hat{\lambda})\end{eqnarray}
\index{$L_{z\bar{z}a}$|itext{Lagrange multiplier}}\index{$L_{zz}$@$\hat{L}_{\bar{z}za}$}\index{$\lambda$@$\lam^{\bs{\alpha}}$|itext{pure spinor ghost}}\index{$\lambdab$@$\hat{\lam}^{\hat{\bs{\alpha}}}$|itext{right-moving pure spinor ghost}}\index{$\omega$@$\bs{\omega}_{z\bs{\alpha}}$|itext{antighost}}\index{$\omegab$@$\hat{\bs{\omega}}_{\bar{z}\hat{\bs{\alpha}}}$|itext{right-moving antighost}}The
pure spinor constraints seem like a replacement of one problem by
another. The constraints turn now out to be first class but infinitely
reducible. They generate antighost\index{antighost gauge symmetry}
gauge symmetries of the form \begin{equation}
\delta_{(\mu)}\omega_{z\alpha}=\mu_{za}(\gamma^{a}\lambda)_{\alpha},\qquad\delta_{(\mu)}\hat{\omega}_{\bar{z}\hat{\bs{\alpha}}}=\hat{\mu}_{\bar{z}a}(\gamma^{a}\hat{\lambda})_{\alpha}\end{equation}
accompanied by some transformation of the Lagrange multipliers. We
will discuss this in more detail in the general background-case. In
spite of this, the pure spinor constraint can be better handled than
the original constraint. One can solve the pure spinor constraint
explicitely in a U(5)-parametrization and calculate operator products.
Although the U(5) coordinates break manifest ten-dimensional Lorentz-covariance,
the resulting gauge-invariant OPE's all have a Lorentz covariant form
and the quantization is effectively Lorentz covariant. Berkovits showed
in the above cited papers the equivalence to the ordinary string.
In \cite{Berkovits:2004px} he presented a consistent description
for the calculation of higher loop amplitudes. There are still many
conceptual problems. The pure spinor formalism starts in the conformal
gauge and does not have worldsheet diffeomorphism invariance any longer.
Attempts to construct a composite b-ghost (as homotopy for the energy
momentum tensor) always involved inverse powers of the gost field.
In \cite{Berkovits:2005bt}, Berkovits recovered a $N=2$ algebra
by the introduction of additional worldsheet fields, which is now
known as {}``non-minimal formalism''. Multiloop calculations were
described or performed by Berkovits, Mafra, Nekrasov and Stahn in
\cite{Berkovits:2005ng,Berkovits:2006bk,Berkovits:2006vi,Stahn:2007uw}
(Since the last version of this thesis new results were obtained.
A recent detailed review is provided in \cite{Mafra:2009wq}\label{Carlos-Zitat}).
However, there is still a clear picture of the origin of the pure
spinor constraint missing. Attempts to relate the pure spinor string
to the Green Schwarz string via similarity transformations and redefinitions
were successful in \cite{Berkovits:2004tw}, but not very enlightening.
An additional task is the resolving of the tip-singularity of the
pure-spinor-cone. These questions were adressed in \cite{Nekrasov:2005wg}
and \cite{Berkovits:2006ik}. 

We should finally mention that the pure spinor approach of Berkovits
differs significantly from the hybrid formalism\cite{Berkovits:1994wr},
which was developped by the same author and shares only some of the
properties of the pure spinor approach. Two recent presentations of
this formalism including the numerous relevant references can be found
in \cite{Kappeli:2006fj}\cite{Linch:2006ig}.

\section{Efforts to remove or explain the pure spinor constraint}

\index{getting rid off the ps-constraint}\index{alternatives to pure spinor}There
were plenty of efforts to get rid of the pure spinor constraint in
the years after Berkovits presented his approach the first time. A
quite natural ansatz was followed by Chesterman\cite{Chesterman:2002ey,Chesterman:2004xt},
who implemented the first-class pure spinor constraint cohomologically,
via a second BRST operator. Due to the infinite\index{infinite reducible}
reducibility of this constraint, there arises an infinite number of
ghost for ghosts. Nevertheless he was able to extract the most important
information and avoided solving the pure spinor constraint explicitly.

Somehow related are the considerations of Aisaka and Kazama\cite{Aisaka:2002sd,Aisaka:2003mw,Aisaka:2004ga,Aisaka:2005vn,Aisaka:2006by}.
They were able to construct a BRST operator with five additional ghost
fields and no pure spinor constraint, using however U(5) parametrization
and breaking manifest Lorentz invariance. The relation to Chesterman's
approach can be established as follows: The infinitely reducible pure
spinor constraint can be replaced by an irreducible one in an U(5)
parametrization. This constraint can be implemented cohomologically
via a second BRST operator in a relative cohomology, and via homological
perturbation theory one can replace the two operators by a single
one. Within their 'doubled spinor formalism', they provided in \cite{Aisaka:2005vn}
a derivation of the pure spinor string from the Green Schwarz String
on the quantum level.

Another enlightening approach by Oda, Tonin et al.\cite{SorokinMatone:2002ft}
was the interpretation of the pure spinor formalism as a twisted and
gauge fixed version of the superembedding\index{superembedding formalism}
formalism. This led to a slightly modified version of the pure spinior
formalism, the Y-formalism\index{Y-formalism}, and to new insight
about the missing antighost b-field\cite{Oda:2004bg,Oda:2005sd,Oda:2005wu,Oda:2007ak}.

There was finally yet another approach by Grassi, Policastro, Porrati
and van Nieuwenhuizen, at that time most of them in Stony Brook, which
we will discuss shortly in a seperate section, as it was subject of
my early PhD studies.

\section{Some more words on the Stony-Brook-approach}

In a series of papers \cite{Nh:2001ug,Nh:2002tz,Nh:2002xf,Nh:2003cm,Nh:2003kq,Nh:2004nz,Nh:2004cz}
Grassi, Policastro, Porrati and van Nieuwenhuizen have removed the
pure spinor constraint by adding additional ghost variables. They
realized in \cite{Nh:2003kq} that their theory has the stucture of
a gauged WZNW\index{WZNW-model} model with the complete diagonal
subgroup gauged. It is based on the chiral algebra above. A current
can be set to zero by gauging the corresponding symmetry and thus
making it a first class constraint. However, $\bs{d}_{z\alpha}$ does
not form a subalgebra and thus cannot be gauged on its own. So if
one starts gauging $\bs{d}_{z\alpha}$ and tries to make the resulting
BRST-operator (\ref{BerkovitsBRST}) nilpotent by adding further ghosts,
one automatically arrives at a BRST operator that corresponds to a
theory where also $\Pi_{zm}$ and $\partial\theta^{\alpha}$ are gauged
(see e.g. \cite[p.7]{Nh:2003cm} or \cite[p.4]{Nh:2003kq}; this fact
was later also used to describe a topological model in \cite{Grassi:2004tv}).
In the gauged WZNW description this means that the complete diagonal
subgroup is gauged. Therefore a grading or filtration had to be introduced,
in order to obtain the correct cohomology. In \cite{Nh:2004cz} it
was argued that for any (simple) Lie algebra one can in general gauge
a coset (in our case the algebra that corresponds to $\bs{d}_{z\alpha}$,
modding out the subalgebra) by gauging the complete algebra and later
undo the gauging of the subalgebra by building the relative cohomology
with respect to a second BRST operator. This corresponds to the former
grading. Despite its elegance there are some puzzling points about
the WZNW action:

\begin{itemize}
\item For the heterotic string one starts with a chiral algebra and gets
from the WZNW model a chiral as well as an antichiral algebra. Somehow
one has to get rid of the antichiral one.
\item For the type II string one starts with a chiral and antichiral algebra.
Both of them double and the Jacobi identity forces one to mix those
algebras. Thus it has not been possible yet to produce a WZNW model
for the type II string.
\item The classical WZNW theory is not a free field theory which might cause
problems for calculating OPEs.
\end{itemize}
For those reasons, we avoided in \cite{Guttenberg:2004ht} the WZNW
action. Although the cited paper contains the work of the early stage
of my PhD, it will not be presented in this thesis in detail. The
reason is that it would open yet another field, whereas the presented
parts share some common aim. Let me therefore just sketch the results:
We started in \cite{Guttenberg:2004ht} with the free field action
of above, discussed its off-shell symmetry algebra generated by the
current $\bs{d}_{z\alpha}$ and gauged it, in order to turn $\bs{d}_{z\alpha}$
into a constraint. Before actually gauging the algebra via the Noether
procedure, we had to make it close off-shell. To this aim we introduced
auxiliary fields $P_{zm}$ and $P_{\bar{z}m}$. There still remained
double poles in the current algebra, which caused trouble in the gauging
procedure. They were be eliminated by doubling all fields as it was
done in \cite{Nh:2003kq}, in order to establish nilpotent BRST transformations.
Gauge fixing leads to the BRST-transformations as they are given in
\cite{Nh:2003kq}. 

Finally, we had a closer look at the final BRST operator proposed
in \cite{Nh:2003kq}, which includes diffeomorphism invariance by
adding a topological ghost quartet. We came to the conclusion that
this operator has to be modified via a second quartett of ghost fields
in order to become nilpotent. More details can be found in \cite{Guttenberg:2004ht}
and \cite{Knapp:2004Dipl}.

A last major progress was achieved in \cite{Nh:2004we} by establishing
an $N=4$ algebra in this formalism. There exist also independent
studies of WZNW models based on supergroups like for example on PSU(1,1|2)
in \cite{Gotz:2006qp} .

\bibliographystyle{fullsort}
\bibliography{phd,Proposal}
\printindex{}
}

\chapter{Closed Pure Spinor Superstring in general type II background}

\label{cha:curved-background}{\inputTeil{0}\ifthenelse{\theinput=1}{}{\renewcommand{\be}{\bs{\omega}}\renewcommand{\ce}{\bs{\lambda}}}

\title{Berkovits string in general background}

\author{Sebastian Guttenberg}

\date{last modified January 4, '08}

\maketitle
\tableofcontents{}\ifthenelse{\theinput=1}{}{\newpage}

\rem{To do:

\begin{itemize}
\item keep cool
\end{itemize}
}

The \rem{type II?}pure spinor string in general background was first
studied by Berkovits in \cite{Berkovits:2001ue}. The one-loop conformal
invariance of the heterotic version was studied in \cite{Chandia:2003hn}.
The classical worldsheet BRST transformations of the heterotic string
in general background were derived in \cite{Chandia:2006ix}. The
one-loop conformal invariance of the type II string finally was shown
in \cite{Bedoya:2006ic} where also the derivation of the supergravity
constraints was reviewed. Note also \cite{Kluson:2006wq,Bianchi:2006im,Kluson:2008as}
for another useful presentation of some aspects of the pure spinor
string in general or AdS5xS5 background. In the following we will
present again the derivation of the supergravity constraints as it
was done in \cite{Berkovits:2001ue},\cite{Bedoya:2006ic} but we
will explain in more detail several steps and also we will use a different
method to derive the constraints. In particular we will not go to
the Hamiltonian formalism in order to derive the BRST transformations
as generated via charge and Poisson bracket but we will stay in the
Lagrangian formalism and will use what we call {}``inverse Noether''.
In addition we will use a spacetime covariant variation in order to
derive the classical equations of motion in a spacetime covariant
manner and we will present the BRST transformations of all the worldsheet
fields for the type II string in general background. This has so far
been done only for the heterotic string in \cite{Chandia:2006ix}.
Having derived the supergravity constraints we will finally go to
the Wess Zumino gauge and derive the local supersymmetry transformations
of at least the fermionic fields in order to make contact to generalized
complex geometry.

Note that there was a carefull study in \cite{Grassi:2004ih} of how
to construct type II vertex operators in the pure spinor formalism.
This is at least for massless fields directly related to the deformations
of the action that we are going to study now. (After the first arXiv-version
of this thesis, another thesis by O. Bedoya\label{BedoyaZitat} \cite{Bedoya:2008yw}
studying and reviewing many aspects of the pure spinor string in general
background has appeared).

\section{Ansatz for action and BRST operators and some EOM's}

In the following we will consider the closed pure spinor string coupled
to general background fields. One can either add small perturbations
(integrated vertex operators) to the action or simply consider the
most general classically conformally invariant\rem{or better renormalizable, adding tachyons}
action with the given field content and the same antighost gauge symmetry
(generated by the pure spinor constraint). The action, however, is
not enough to specify the string completely. In addition, we need
two (one left-moving and one right-moving) BRST operators in the general
background. The existence of two such BRST operators which have to
be nilpotent and conserved (holomorphic and antiholomorphic respectively)
turns out to be equivalent to supergravity constraints on the background
fields. The important steps of this calculation will be carefully
motivated in the following.

The idea is to start from the most general renormalizable action with
the given field content. It is convenient to throw away immediately
the tachyon term which is allowed by renormalizability, but which
is not even BRST invariant for the undeformed BRST transformations,
at least for a non-constant tachyon field. The starting point then
reduces to the most general classically conformally invariant action.
In order to write down a classically conformally invariant action
(ghost number zero in each sector), we have to combine elementary
fields to terms with conformal\index{conformal weight} weight\index{weight!conformal $\sim$}
(1,1). There are no fields with negative conformal weight. The a priory
possible elementary building\index{building blocks!of ps action}
blocks of ghost number (0,0) are thus\begin{eqnarray*}
\mbox{weight (0,0)} &  & x^{M}\\
\mbox{weight (1,0)} &  & \partial x^{M},\dP_{z\bs{\alpha}},\ce^{\bs{\alpha}}\be_{z\bs{\beta}}\\
\mbox{weight (0,1)} &  & \bar{\partial}x^{M},\hat{\dP}_{\bar{z}\hat{\bs{\alpha}}},\hat{\ce}^{\hat{\bs{\alpha}}}\hat{\be}_{\bar{z}\hat{\bs{\beta}}}\\
\mbox{weight (1,1)} &  & \partial\bar{\partial}x^{M},\bar{\partial}\ce^{\bs{\alpha}}\be_{z\bs{\beta}},\partial\hat{\ce}^{\hat{\bs{\alpha}}}\hat{\be}_{\bar{z}\hat{\bs{\beta}}},\,\bar{\partial}\dP_{z\bs{\alpha}},\,\partial\hat{\dP}_{\bar{z}\hat{\bs{\alpha}}}\end{eqnarray*}
We now can combine an arbitrary function of $x^{M}$ (background field)
with either a (1,1)-building block or with one (1,0) combined with
one (0,1) building block. Via partial integration, a $\partial\bar{\partial}x^{M}$-term
with an arbitrary $x$-dependent coefficient can always be rewritten
as a $\partial x^{M}\bar{\partial}x^{N}$-term%
\footnote{\index{footnote!\thefoot. second x-derivative and bdry}This, however,
contributes to the surface term. In the case of open strings, adding
a $\partial\bar{\partial}x^{M}$-term is therefore equivalent to the
modification of the boundary part of the action.$\qquad\fussend$%
}. Before writing down the resulting action, let us note that we will
immediately absorb the $x$-dependent coefficient coming with $\bar{\partial}\ce^{\bs{\alpha}}\be_{z\bs{\beta}}$
in a reparametrization of $\be_{z\bs{\beta}}$ so that we simply get
the free ghost kinetic term $\bar{\partial}\ce^{\bs{\alpha}}\be_{z\bs{\alpha}}$.
Likewise for the hatted variables.

The most general classically conformally invariant (or renormalizable,
adding Tachyon\index{Tachyon} term\rem{noch dazufuegen!?}) action
with the same field content (including the pure spinor constraint
on the ghosts) with independently conserved left and right ghost number
now reads\vRam{1.01}{\begin{eqnarray}
\hspace{-.1cm}S & = & \int d^{2}z\quad\frac{1}{2}\partial x^{M}(\underbrace{G_{MN}(\xfull)+B_{MN}(\xfull)}_{\equiv\GB_{MN}(\xfull)})\bar{\partial}x^{N}+\bar{\partial}x^{M}E_{M}\hoch{\bs{\alpha}}(\xfull)\:\dP_{z\bs{\alpha}}+\partial x^{M}E_{M}\hoch{\hat{\bs{\alpha}}}(\xfull)\:\hat{\dP}_{\bar{z}\hat{\bs{\alpha}}}+\nonumber \\
 &  & +\dP_{z\bs{\alpha}}\RR^{\bs{\alpha}\hat{\bs{\beta}}}(\xfull)\:\hat{\dP}_{\bar{z}\hat{\bs{\beta}}}+\ce^{\bs{\alpha}}C_{\bs{\alpha}}\hoch{\bs{\beta}\hat{\bs{\gamma}}}(\xfull)\:\be_{z\bs{\beta}}\hat{\dP}_{\bar{z}\hat{\bs{\gamma}}}+\hat{\ce}^{\hat{\bs{\alpha}}}\hat{C}_{\hat{\bs{\alpha}}}\hoch{\hat{\bs{\beta}}\bs{\gamma}}(\xfull)\:\hat{\be}_{\bar{z}\hat{\bs{\beta}}}\dP_{z\bs{\gamma}}+\ce^{\bs{\alpha}}\hat{\ce}^{\hat{\bs{\alpha}}}S_{\bs{\alpha}\hat{\bs{\alpha}}}\hoch{\bs{\beta}\hat{\bs{\beta}}}(\xfull)\:\be_{z\bs{\beta}}\hat{\be}_{\bar{z}\hat{\bs{\beta}}}+\nonumber \\
 &  & +\underbrace{\left(\bar{\partial}\ce^{\bs{\beta}}+\ce^{\bs{\alpha}}\bar{\partial}x^{M}\Omega_{M\bs{\alpha}}\hoch{\bs{\beta}}(\xfull)\right)}_{\equiv\nabla_{\bar{z}}\ce^{\bs{\beta}}}\be_{z\bs{\beta}}+\underbrace{\left(\partial\hat{\ce}^{\hat{\bs{\beta}}}+\hat{\ce}^{\hat{\bs{\alpha}}}\partial x^{M}\hat{\Omega}_{M\hat{\bs{\alpha}}}\hoch{\hat{\bs{\beta}}}(\xfull)\right)}_{\equiv\hat{\nabla}_{z}\ce^{\hat{\bs{\beta}}}\frem{\equiv\nabla_{z}\ce^{\hat{\bs{\beta}}}+\ce^{\hat{\bs{\alpha}}}\partial x^{M}\Delta_{M\hat{\bs{\alpha}}}\hoch{\hat{\bs{\beta}}}}}\hat{\be}_{\bar{z}\hat{\bs{\beta}}}+\nonumber \\
 &  & +\frac{1}{2}L_{z\bar{z}a}(\ce\gamma^{a}\ce)+\frac{1}{2}\hat{L}_{\bar{z}z\hat{a}}(\hat{\ce}\gamma^{\hat{a}}\hat{\ce})\frem{+\alpha'\overbrace{\underbrace{\sqrt{h}}_{\frac{1}{2}e^{2\weyl}}\underbrace{R_{h}}_{-8e^{-2\weyl}\bar{\partial}\partial\weyl}}^{-4\partial\bar{\partial}\weyl}\dil(\xfull)}\qquad\label{eq:BiBaction}\end{eqnarray}
}\\
\index{action!in general background}\index{$S$|itext{action}}\rem{missing dilaton and tachyon}\index{$O$@$\GB_{MN}$}\index{$G_{MN}$}\index{$B_{MN}$}\index{$E_M\hoch{\bs\alpha}$}\index{$E_M\hoch{\hat{\bs\alpha}}$}\index{$P$@$\RR^{\bs{\alpha}\hat{\bs{\beta}}}$|itext{RR-field}}\index{$C_{\bs{\alpha}}\hoch{\bs{\beta}\hat{\bs{\gamma}}}$}\index{$C$@$\hat{C}_{\hat{\bs{\alpha}}}\hoch{\hat{\bs{\beta}}\bs{\gamma}}$}\index{$S_{\bs{\alpha}\hat{\bs{\alpha}}}\hoch{\bs{\beta}\hat{\bs{\beta}}}$}\index{$\Omega_{M\bs{\alpha}}\hoch{\bs{\beta}}$}\index{$\Omega$@$\hat{\Omega}_{M\hat{\bs{\alpha}}}\hoch{\hat{\bs{\beta}}}$}\index{$\nabla_{\bar{z}}\bs{\lambda}^{\bs{\beta}}$}\index{$\nabla$@$\hat{\nabla}_{z}\bs{\lambda}^{\hat{\bs{\beta}}}$}Note
that we denote with $\xfull$\index{$x$@$\xfull$} the complete set
$x^{M}$ of superspace coordinates, while $\xboson$\index{$x$@$\xboson$}
will only denote the bosonic subset $x^{m}$. As stated already above,
the kinetic ghost\index{ghost!kinetic term}\index{kinetic ghost term}
term $\bar{\partial}\ce^{\bs{\beta}}\be_{z\bs{\beta}}$ can always
be brought to this simple form by a redefinition of $\be$. We will
discuss this and other worldsheet reparametrizations below in detail.
The motivation for the definition of the covariant derivative $\nabla_{\bar{z}}\ce^{\bs{\beta}}$
will also be given at a later point. For the moment, $\Omega_{M\bs{\alpha}}\hoch{\bs{\beta}}(x)$
is just an arbitrary coefficient function or background field. Like
in the flat case, we implement the pure spinor constraints via two
Lagrange multipliers.

In order to complete the theory, we need two BRST operators which
reduce to the well known ones in the flat case. Their nilpotency and
(anti)holomorphicity will be checked later and lead to the supergravity
constraints. For the moment, let us just write down the most general
ansatz of their currents, which have to be of conformal weight (1,0)
and (0,1) and ghost number (1,0) and (0,1) respectively\index{BRST-current}\index{$j$@$\bs{j}_z$|itext{BRST current}}\index{$j$@$\hat{\bs{\jmath}}_{\bar z}$|itext{right-moving BRST current}}
\begin{eqnarray}
\bs{j}_{z} & = & \ce^{\bs{\alpha}}\left(\dP_{z\bs{\alpha}}+\brstfield{2}_{\bs{\alpha}M}(\xfull)\:\partial_{z}x^{M}+\ce^{\bs{\gamma}}\brstfield{3}_{\bs{\alpha}\bs{\gamma}}\hoch{\bs{\beta}}(\xfull)\be_{z\bs{\beta}}\frem{+\alpha'\brstfield{4}_{\bs{\alpha}}(\xfull)\partial_{z}\weyl}\right),\quad\bs{j}_{\bar{z}}=0\label{eq:BiBbrst}\\
\hat{\bs{\jmath}}_{\bar{z}} & = & \hat{\ce}^{\hat{\bs{\alpha}}}\left(\hat{\dP}_{\bar{z}\hat{\bs{\alpha}}}+\hatbrstfield{2}_{\hat{\bs{\alpha}}M}(\xfull)\:\partial_{\bar{z}}x^{M}+\hat{\ce}^{\hat{\bs{\gamma}}}\hatbrstfield{3}_{\hat{\bs{\alpha}}\hat{\bs{\gamma}}}\hoch{\hat{\bs{\beta}}}(\xfull)\hat{\be}_{\bar{z}\hat{\bs{\beta}}}\frem{+\alpha'\hatbrstfield{4}_{\hat{\bs{\alpha}}}(\xfull)\partial_{\bar{z}}\weyl}\right),\quad\hat{\bs{\jmath}}_{z}=0\label{eq:BiBbrstHat}\end{eqnarray}
Like for the ghost kinetic term, we have immediately absorbed any
$\xfull$-dependent coefficient $\brstfield{1}_{\bs{\alpha}}\hoch{\bs{\beta}}(\xfull)$
coming with $\ce^{\bs{\alpha}}\dP_{z\bs{\beta}}$ and its hatted version
in a redefinition of $\dP_{z\bs{\beta}}$ and $\hat{\dP}_{\bar{z}\hat{\bs{\beta}}}$.%
\footnote{\index{footnote!\thefoot. degenerate limit}If one wants to study
degenerate limits of the theory, one should remember and reintroduce
the coefficients $\brstfield{1}$, $\hatbrstfield{1}$ and the one
coming with the ghost kinetic terms.$\quad\fussend$%
} Of course one can further redefine $\dP_{z\bs{\alpha}}$ and $\hat{\dP}_{\bar{z}\hat{\bs{\alpha}}}$,
such that we arrive at the standard form $\bs{j}_{z}=\ce^{\bs{\alpha}}\dP_{z\bs{\alpha}}$
and $\hat{\bs{\jmath}}_{\bar{z}}=\hat{\ce}^{\hat{\bs{\alpha}}}\dP_{\bar{z}\hat{\bs{\alpha}}}$.
This does not change the general form of the action. We will discuss
the reparametrizations more carefully in the next section.

The following observation is important to reduce the computations
one has to do. Let us first define \begin{eqnarray}
\hat{\GB}_{MN} & \equiv & \GB_{NM},\quad\left(\hat{G}=G,\hat{B}=-B,\hat{H}=-H\right)\label{eq:GBhat}\\
\hat{\RR}^{\hat{\bs{\gamma}}\bs{\gamma}} & \equiv & \RR^{\bs{\gamma}\hat{\bs{\gamma}}}\\
\hat{S}_{\hat{\bs{\alpha}}\bs{\alpha}}\hoch{\hat{\bs{\beta}}\bs{\beta}} & \equiv & S_{\bs{\alpha}\hat{\bs{\alpha}}}\hoch{\bs{\beta}\hat{\bs{\beta}}}\end{eqnarray}
\index{$Sb$@$\hat{S}_{\hat{\bs{\alpha}}\bs{\alpha}}\hoch{\hat{\bs{\beta}}\bs{\beta}}$}\index{$O$@$\hat{\GB_{MN}}$}\index{$Pb$@$\hat{\RR}^{\hat{\bs{\gamma}}\bs{\gamma}}$}Then
-- rather obviously -- the following statement holds \begin{prop}[left-right symmetry]\index{symmetry!left-right}\index{left-right symmetry}\index{proposition!left-right symmetry}\label{prop:left-right-symmetry}The
complete theory (action +BRST operators) is invariant under the exchange
of hatted and unhatted objects if at the same time their indices are
flipped from hatted to unhatted and from $z$ to $\bar{z}$ and vice
verse, and $\partial$ is exchanged with $\bar{\partial}$: \begin{equation}
\begin{array}{c}
\dP\leftrightarrow\hat{\dP},\ce\leftrightarrow\hat{\ce},\be\leftrightarrow\hat{\be},L\leftrightarrow\hat{L},\GB\leftrightarrow\hat{\GB},\RR\leftrightarrow\hat{\RR},S\leftrightarrow\hat{S},C\leftrightarrow\hat{C},\Omega\leftrightarrow\hat{\Omega},\nabla\leftrightarrow\hat{\nabla},\brstfield{i}\leftrightarrow\hatbrstfield{i},\bs{j}\leftrightarrow\lqn{\hat{\bs{\jmath}}}\\
\partial\leftrightarrow\bar{\partial},\textrm{ indices: }\bs{\alpha}\leftrightarrow\hat{\bs{\alpha}},z\leftrightarrow\bar{z}\end{array}\label{eq:leftrightSymmetry}\end{equation}
In particular the replacement $\GB\leftrightarrow\hat{\GB}$ implies
due to (\ref{eq:GBhat}) that \begin{equation}
B\leftrightarrow-B,\qquad G\leftrightarrow G\end{equation}
\end{prop}

\paragraph{Simple eom's}

Before we close this section, let us quickly give the equations of
motion of those worldsheet variables (all but $x^{K}$) which can
be seen from the target superspace point of view as tangent or cotangent
vectors. This refers to the form of their reparametrizations that
will be discussed on page \pageref{par:local-strucgroup-transformations}.
Their equations of motion are comparatively simple:\begin{eqnarray}
\funktional{S}{\dP_{z\bs{\gamma}}} & = & \bar{\partial}x^{M}E_{M}\hoch{\bs{\gamma}}+\RR^{\bs{\gamma}\hat{\bs{\gamma}}}\hat{\dP}_{\bar{z}\hat{\bs{\gamma}}}+\hat{\ce}^{\hat{\bs{\alpha}}}\hat{C}_{\hat{\bs{\alpha}}}\hoch{\hat{\bs{\beta}}\bs{\gamma}}\hat{\be}_{\bar{z}\hat{\bs{\beta}}}\label{eq:eomI}\\
\funktional{S}{\hat{\dP}_{\bar{z}\hat{\bs{\gamma}}}} & = & \partial x^{M}E_{M}\hoch{\hat{\bs{\gamma}}}+\dP_{z\bs{\gamma}}\RR^{\bs{\gamma}\hat{\bs{\gamma}}}+\ce^{\bs{\alpha}}C_{\bs{\alpha}}\hoch{\bs{\beta}\hat{\bs{\gamma}}}\be_{z\bs{\beta}}\\
\funktional{S}{\be_{z\bs{\beta}}} & = & -\left(\nabla_{\bar{z}}\ce^{\bs{\beta}}+\ce^{\bs{\alpha}}\left(C_{\bs{\alpha}}\hoch{\bs{\beta}\hat{\bs{\gamma}}}\hat{\dP}_{\bar{z}\hat{\bs{\gamma}}}-\hat{\ce}^{\hat{\bs{\alpha}}}S_{\bs{\alpha}\hat{\bs{\alpha}}}\hoch{\bs{\beta}\hat{\bs{\beta}}}\hat{\be}_{\bar{z}\hat{\bs{\beta}}}\right)\right)\equiv-\mc{D}_{\bar{z}}\ce^{\bs{\beta}}\label{eq:yetanothercovDerI}\\
\funktional{S}{\hat{\be}_{\bar{z}\hat{\bs{\beta}}}} & = & -\left(\hat{\nabla}_{z}\hat{\ce}^{\hat{\bs{\beta}}}+\hat{\ce}^{\hat{\bs{\alpha}}}\left(\hat{C}_{\hat{\bs{\alpha}}}\hoch{\hat{\bs{\beta}}\bs{\gamma}}\dP_{z\bs{\gamma}}-\ce^{\bs{\alpha}}S_{\bs{\alpha}\hat{\bs{\alpha}}}\hoch{\bs{\beta}\hat{\bs{\beta}}}\be_{z\bs{\beta}}\right)\right)\equiv-\hat{\mc{D}}_{z}\hat{\ce}^{\hat{\bs{\beta}}}\\
\funktional{S}{\ce^{\bs{\alpha}}} & = & -\left(\nabla_{\bar{z}}\be_{z\bs{\alpha}}-\left(C_{\bs{\alpha}}\hoch{\bs{\beta}\hat{\bs{\gamma}}}\hat{\dP}_{\bar{z}\hat{\bs{\gamma}}}-\hat{\ce}^{\hat{\bs{\alpha}}}S_{\bs{\alpha}\hat{\bs{\alpha}}}\hoch{\bs{\beta}\hat{\bs{\beta}}}\hat{\be}_{\bar{z}\hat{\bs{\beta}}}\right)\be_{z\bs{\beta}}\right)+L_{z\bar{z}a}(\gamma^{a}\ce)_{\bs{\alpha}}\equiv-\mc{D}_{\bar{z}}\be_{z\bs{\alpha}}+L_{z\bar{z}a}(\gamma^{a}\ce)_{\bs{\alpha}}\\
\funktional{S}{\hat{\ce}^{\hat{\bs{\alpha}}}} & = & -\left(\hat{\nabla}_{z}\hat{\be}_{\bar{z}\hat{\bs{\alpha}}}-\left(\hat{C}_{\hat{\bs{\alpha}}}\hoch{\hat{\bs{\beta}}\bs{\gamma}}\dP_{z\bs{\gamma}}-\ce^{\bs{\alpha}}S_{\bs{\alpha}\hat{\bs{\alpha}}}\hoch{\bs{\beta}\hat{\bs{\beta}}}\be_{z\bs{\beta}}\right)\hat{\be}_{\bar{z}\hat{\bs{\beta}}}\right)+\hat{L}_{z\bar{z}a}(\gamma^{a}\hat{\ce})_{\hat{\bs{\alpha}}}\equiv-\hat{\mc{D}}_{z}\hat{\be}_{\bar{z}\hat{\bs{\alpha}}}+\hat{L}_{z\bar{z}a}(\gamma^{a}\hat{\ce})_{\hat{\bs{\alpha}}}\qquad\label{eq:yetanothercovDerIV}\\
\funktional{S}{L_{z\bar{z}a}} & = & \frac{1}{2}(\ce\gamma^{a}\ce),\qquad\funktional{S}{\hat{L}_{z\bar{z}a}}=\frac{1}{2}(\hat{\ce}\gamma^{a}\hat{\ce})\label{eq:eomVIII}\end{eqnarray}
\rem{\begin{equation}
\funktional{S}{\weyl}=-4\alpha'\gemnabla_{\bar{z}}\partial x^{M}\nabla_{M}\dil-4\alpha'\partial x^{M}\bar{\partial}x^{N}\gemnabla_{N}\gemnabla_{M}\dil\end{equation}
}In (\ref{eq:yetanothercovDerI})-(\ref{eq:yetanothercovDerIV})
we have introduced yet two other {}``covariant derivatives'' $\mc{D}_{\bar{z}}$
and $\hat{\mc{D}}_{z}$:\begin{eqnarray}
\mc{D}_{\bar{z}}\ce^{\bs{\beta}} & \equiv & \bar{\partial}\ce^{\bs{\beta}}+A_{\bar{z}\bs{\alpha}}\hoch{\bs{\beta}}\ce^{\bs{\alpha}},\qquad A_{\bar{z}\bs{\alpha}}\hoch{\bs{\beta}}\equiv\bar{\partial}x^{M}\Omega_{M\bs{\alpha}}\hoch{\bs{\beta}}+C_{\bs{\alpha}}\hoch{\bs{\beta}\hat{\bs{\gamma}}}\hat{\dP}_{\bar{z}\hat{\bs{\gamma}}}-\hat{\ce}^{\hat{\bs{\alpha}}}S_{\bs{\alpha}\hat{\bs{\alpha}}}\hoch{\bs{\beta}\hat{\bs{\beta}}}\hat{\be}_{\bar{z}\hat{\bs{\beta}}}\label{eq:AundMathcalD}\\
\hat{\mc{D}}_{z}\hat{\ce}^{\hat{\bs{\beta}}} & \equiv & \partial\hat{\ce}^{\hat{\bs{\beta}}}+\hat{A}_{z\hat{\bs{\alpha}}}\hoch{\hat{\bs{\beta}}}\hat{\ce}^{\hat{\bs{\alpha}}},\qquad\hat{A}_{z\hat{\bs{\alpha}}}\hoch{\hat{\bs{\beta}}}\equiv\partial x^{M}\hat{\Omega}_{M\hat{\bs{\alpha}}}\hoch{\hat{\bs{\beta}}}+\hat{C}_{\hat{\bs{\alpha}}}\hoch{\hat{\bs{\beta}}\bs{\gamma}}\dP_{z\bs{\gamma}}-\ce^{\bs{\alpha}}S_{\bs{\alpha}\hat{\bs{\alpha}}}\hoch{\bs{\beta}\hat{\bs{\beta}}}\be_{z\bs{\beta}}\label{eq:AhutUndMathcalDhut}\end{eqnarray}
These covariant derivatives are introduced simply for calculational
convenience and we do not give a geometric interpretation -- although
this might be interesting.\rem{Feldstaerke? siehe auch Diplarbeit!
Extension $A_{z\bs{\alpha}}\hoch{\bs{\beta}}\stackrel{?}{=}\partial x^{M}\Omega_{M\bs{\alpha}}\hoch{\bs{\beta}}$.}
For the covariant derivatives $\nabla_{\bar{z}}$ and $\hat{\nabla}_{z}$
defined in (\ref{eq:BiBaction}) instead, there exists a simple geometric
interpretation. They are pullbacks of the covariant target super tangent
space derivatives with connection coefficients $\Omega_{M\bs{\alpha}}\hoch{\bs{\beta}}$
and $\hat{\Omega}_{M\hat{\bs{\alpha}}}\hoch{\hat{\bs{\beta}}}$ to
the worldsheet. The reason why these two background fields can be
seen as connections will be given in the following.

Note that the derivation of the still missing variational derivative
with respect to $x^{K}$ is quite involved and will only be given
in section \ref{sec:Covariant-eoms} on page \pageref{sec:Covariant-eoms}
using a covariant variational principle.

\section{Vielbeins, worldsheet reparametrizations and target space symmetries}

\label{sec:vielbeins-and-reps}There are several ways to reparametrize
the worldsheet fields in the above action and the BRST currents. One
can use such reparametrizations to simplify the form of the action
(as we did already implicitly in order to get a simple ghost kinetic
term) or of the BRST currents. 

Before we come to the first convenient reparametrization, let us observe
the following: The two background fields $E_{M}\hoch{\bs{\alpha}}$
and $E_{M}\hoch{\hat{\bs{\alpha}}}$, combined to a $42\times32$
matrix $E_{M}\hoch{\bs{\mc{A}}}$,$\bs{\mc{A}}\in\{\bs{\alpha},\hat{\bs{\alpha}}\}$
have maximal rank 32 in a small perturbation around the string in
flat background. Or in other words, the quadratic block $E_{\bs{\mc{M}}}\hoch{\bs{\mc{A}}}$
is invertible%
\footnote{\index{footnote!\thefoot. degenerate limit}Again it might be interesting
to study also degenerate limits.$\qquad\fussend$%
}. It can thus be completed by some $E_{M}\hoch{a}$ to an invertible
$42\times42$ matrix which we can interpret as (super)vielbein. The
only requirement for $E_{M}\hoch{a}$ to be a valid completion is
that its bosonic sub-matrix $E_{m}\hoch{a}$ is invertible%
\footnote{\index{footnote!\thefoot. invertible bosonic supermatrix}The bosonic
supermatrix $\left(\begin{array}{cc}
E_{m}\hoch{a} & E_{m}\hoch{\bs{\mc{A}}}\\
E_{\bs{\mc{M}}}\hoch{a} & E_{\bs{\mc{M}}}\hoch{\bs{\mc{A}}}\end{array}\right)$ is invertible, iff its bosonic blocks $\left(E_{m}\hoch{a}\right)$
and $\left(E_{\bs{\mc{M}}}\hoch{\bs{\mc{A}}}\right)$ are invertible.$\qquad\fussend$%
}. The {}``background field'' $E_{M}\hoch{a}$ does not appear in
the action and nothing should depend on it. Let us from now on use
the completed vielbein $E_{M}\hoch{A}$ and its inverse $E_{A}\hoch{M}$
to switch from curved to flat indices and vice verse. In particular
we define\begin{eqnarray}
G_{AB} & \equiv & E_{A}\hoch{M}G_{MN}E_{B}\hoch{N}\label{eq:GAB}\end{eqnarray}
For later usage we denote the components of the pullback of the vielbein
$E^{A}$ to the worldsheet as\index{$\Pi_z^A$} \begin{eqnarray}
\Pi_{z}^{A} & \equiv & \partial x^{M}E_{M}\hoch{A}\label{eq:Pidefz}\\
\Pi_{\bar{z}}^{A} & \equiv & \bar{\partial}x^{M}E_{M}\hoch{A}\label{eq:Pidefbarz}\end{eqnarray}
In flat space, $\Pi_{z/\bar{z}}^{a}$ will just be the supersymmetric
momentum and the fermionic component will reduce to the worldsheet
derivative of the fermionic coordinates: $\Pi_{z/\bar{z}}^{\bs{\mc{A}}}\stackrel{\textrm{flat}}{\to}\partial_{z/\bar{z}}\tet^{\mc{A}}$.

Let us now study the possible reparametrizations of the worldsheet
variables systematically.

\paragraph{Possible reparametrizations }

We denote by $\allfields{I}$ the collection of all worldsheet fields.
If we make some reparametrization $\tildeallfields{I}=f[\allfields{I}]$,
the Jacobi matrix has to be invertible in order to lead to equivalent
equations of motion:\begin{equation}
\funktional{S}{\allfields{I}(\sigma)}=\int d^{2}\tilde{\sigma}\quad\funktional{\tildeallfields{J}(\tilde{\sigma})}{\allfields{I}(\sigma)}\funktional{S}{\tildeallfields{J}(\tilde{\sigma})}\end{equation}
The following reparametrizations are the most general ones which respect
the conformal weight as well as the left and right-moving ghost numbers
(note that the Lagrange multipliers have ghost number $(-2,0)$ and
$(0,-2)$ respectively): $\frem{\allfields{I}=(x^{M},\dP_{z\bs{\alpha}},\dP_{\bar{z}\hat{\bs{\alpha}}},\ce^{\bs{\alpha}},\be_{z\bs{\alpha}},\hat{\ce}^{\hat{\bs{\alpha}}},\hat{\be}_{\hat{\bs{\alpha}}},L_{z\bar{z}\bs{\alpha}},\hat{L}_{\bar{z}z\bs{\alpha}},\weyl}$\begin{eqnarray}
\tilde{x}^{M} & = & f^{M}(\xfull)\label{eq:repI}\\
\tilde{\ce}^{\bs{\alpha}} & = & \Lambda_{\bs{\beta}}\hoch{\bs{\alpha}}(\xfull)\ce^{\bs{\beta}},\qquad\tilde{\hat{\ce}}^{\hat{\bs{\alpha}}}=\hat{\Lambda}_{\hat{\bs{\beta}}}\hoch{\hat{\bs{\alpha}}}(\xfull)\hat{\ce}^{\hat{\bs{\beta}}}\label{eq:repII}\\
\tilde{\dP}_{z\bs{\alpha}} & = & \dshift{1}_{\bs{\alpha}}\hoch{\bs{\beta}}(\xfull)\dP_{z\bs{\beta}}+\dshift{2}_{\bs{\alpha}M}(\xfull)\partial x^{M}+\dshift{3}_{\bs{\alpha}\bs{\gamma}}\hoch{\bs{\delta}}(\xfull)\ce^{\bs{\gamma}}\be_{z\bs{\delta}}\\
\tilde{\hat{\dP}}_{\bar{z}\hat{\bs{\alpha}}} & = & \hatdshift{1}_{\hat{\bs{\alpha}}}\hoch{\hat{\bs{\beta}}}(\xfull)\hat{\dP}_{\bar{z}\hat{\bs{\beta}}}+\hatdshift{2}_{\hat{\bs{\alpha}}N}(\xfull)\bar{\partial}x^{N}+\hatdshift{3}_{\hat{\bs{\alpha}}\hat{\bs{\gamma}}}\hoch{\hat{\bs{\delta}}}(\xfull)\hat{\ce}^{\hat{\bs{\gamma}}}\hat{\be}_{\bar{z}\hat{\bs{\delta}}}\\
\tilde{\be}_{z\bs{\alpha}} & = & \dshift{4}_{\bs{\alpha}}\hoch{\bs{\beta}}(\xfull)\be_{z\bs{\beta}},\qquad\tilde{\hat{\be}}_{\bar{z}\hat{\bs{\alpha}}}=\hatdshift{4}_{\hat{\bs{\alpha}}}\hoch{\hat{\bs{\beta}}}(\xfull)\hat{\be}_{\bar{z}\hat{\bs{\beta}}}\\
\tilde{L}_{z\bar{z}a} & = & \dshift{5}_{a}\hoch{b}(\xfull)L_{z\bar{z}b},\qquad\tilde{\hat{L}}_{\bar{z}za}=\hatdshift{5}_{a}\hoch{b}(\xfull)\hat{L}_{\bar{z}zb}\label{eq:repVI}\end{eqnarray}
$f^{M}$ has to be an invertible function and $\Lambda$, $\dshift{1},\dshift{4},\dshift{5}$
and their hatted equivalents have to be invertible matrices. For a
general reparametrization, $\Lambda_{\bs{\alpha}}\hoch{\bs{\beta}}$
can be a general invertible matrix, but if we want to leave the form
of the action invariant, it has to be an element of the spin group
or a simple scaling. We will discuss that below. Note also, that we
have already used $\dshift{4}$ and $\dshift{1}$ and their hatted
versions to get a simple ghost-kinetic term in the action and a simple
first term of the BRST operator.

\paragraph{Shift reparametrization }

\label{par:Shift-reparametrization}Let us first study the effect
of the shift-reparametrizations \begin{eqnarray}
\dP_{z\bs{\alpha}} & = & \tilde{\dP}_{z\bs{\alpha}}-\dshift{2}_{\bs{\alpha}M}(\xfull)\partial x^{M}-\dshift{3}_{\bs{\alpha}\bs{\gamma}}\hoch{\bs{\delta}}(\xfull)\ce^{\bs{\gamma}}\be_{z\bs{\delta}},\qquad\dshift{1}_{\bs{\alpha}}\hoch{\bs{\beta}}=\delta_{\bs{\alpha}}\hoch{\bs{\beta}}\label{eq:shiftRep}\\
\hat{\dP}_{\bar{z}\hat{\bs{\alpha}}} & = & \tilde{\hat{\dP}}_{\bar{z}\hat{\bs{\alpha}}}-\hatdshift{2}_{\hat{\bs{\alpha}}N}(\xfull)\bar{\partial}x^{N}-\hatdshift{3}_{\hat{\bs{\alpha}}\hat{\bs{\gamma}}}\hoch{\hat{\bs{\delta}}}(\xfull)\hat{\ce}^{\hat{\bs{\gamma}}}\hat{\be}_{\bar{z}\hat{\bs{\delta}}},\qquad\hatdshift{1}_{\hat{\bs{\alpha}}}\hoch{\hat{\bs{\beta}}}=\delta_{\hat{\bs{\alpha}}}\hoch{\hat{\bs{\beta}}}\label{eq:shiftHatRep}\end{eqnarray}
 on the form of the action. Plugging the above reparametrization into
(\ref{eq:BiBaction})-(\ref{eq:BiBbrstHat}), the form of the action
and the BRST currents does not change if the background fields are
redefined accordingly. \rem{Im Detail sieht das so aus:\begin{eqnarray}
S & = & \int\quad\frac{1}{2}\partial x^{M}E_{M}\hoch{A}E_{N}\hoch{B}\bar{\partial}x^{N}\times\nonumber \\
 &  & \hspace{-2cm}\times\lqn{\left(\begin{array}{ccc}
\GB_{ab}+2\dshift{2}_{\bs{\gamma}a}\RR^{\bs{\gamma}\hat{\bs{\gamma}}}\hatdshift{2}_{\hat{\bs{\gamma}}b} & \GB_{a\bs{\beta}}-2\dshift{2}_{\bs{\beta}a}+2\dshift{2}_{\bs{\gamma}a}\RR^{\bs{\gamma}\hat{\bs{\gamma}}}\hatdshift{2}_{\hat{\bs{\gamma}}\bs{\beta}} & \GB_{a\hat{\bs{\beta}}}+2\dshift{2}_{\bs{\gamma}a}\RR^{\bs{\gamma}\hat{\bs{\gamma}}}\hatdshift{2}_{\hat{\bs{\gamma}}\hat{\bs{\beta}}}\\
\GB_{\bs{\alpha}b}+2\dshift{2}_{\bs{\gamma}\bs{\alpha}}\RR^{\bs{\gamma}\hat{\bs{\gamma}}}\hatdshift{2}_{\hat{\bs{\gamma}}b} & \GB_{\bs{\alpha}\bs{\beta}}-2\dshift{2}_{\bs{\beta}\bs{\alpha}}+2\dshift{2}_{\bs{\gamma}\bs{\alpha}}\RR^{\bs{\gamma}\hat{\bs{\gamma}}}\hatdshift{2}_{\hat{\bs{\gamma}}\bs{\beta}} & \GB_{\bs{\alpha}\hat{\bs{\beta}}}+2\dshift{2}_{\bs{\gamma}\bs{\alpha}}\RR^{\bs{\gamma}\hat{\bs{\gamma}}}\hatdshift{2}_{\hat{\bs{\gamma}}\hat{\bs{\beta}}}\\
\GB_{\hat{\bs{\alpha}}b}-2\hatdshift{2}_{\hat{\bs{\alpha}}b}+2\dshift{2}_{\bs{\gamma}\hat{\bs{\alpha}}}\RR^{\bs{\gamma}\hat{\bs{\gamma}}}\hatdshift{2}_{\hat{\bs{\gamma}}b} & \GB_{\hat{\bs{\alpha}}\bs{\beta}}-2\dshift{2}_{\bs{\beta}\hat{\bs{\alpha}}}-2\hatdshift{2}_{\hat{\bs{\alpha}}\bs{\beta}}+2\dshift{2}_{\bs{\gamma}\hat{\bs{\alpha}}}\RR^{\bs{\gamma}\hat{\bs{\gamma}}}\hatdshift{2}_{\hat{\bs{\gamma}}\bs{\beta}} & \GB_{\hat{\bs{\alpha}}\hat{\bs{\beta}}}-2\hatdshift{2}_{\hat{\bs{\alpha}}\hat{\bs{\beta}}}+2\dshift{2}_{\bs{\gamma}\hat{\bs{\alpha}}}\RR^{\bs{\gamma}\hat{\bs{\gamma}}}\hatdshift{2}_{\hat{\bs{\gamma}}\hat{\bs{\beta}}}\end{array}\right)_{AB}+}\nonumber \\
 &  & +\bar{\partial}x^{N}\left(E_{N}\hoch{\bs{\alpha}}-\RR^{\bs{\alpha}\hat{\bs{\alpha}}}\hatdshift{2}_{\hat{\bs{\alpha}}N}\right)\tilde{\dP}_{z\bs{\alpha}}+\partial x^{M}\left(E_{M}\hoch{\hat{\bs{\alpha}}}-\dshift{2}_{\bs{\alpha}M}\RR^{\bs{\alpha}\hat{\bs{\alpha}}}\right)\tilde{\hat{\dP}}_{\bar{z}\hat{\bs{\alpha}}}+\tilde{\dP}_{z\bs{\alpha}}\RR^{\bs{\alpha}\hat{\bs{\alpha}}}\tilde{\hat{\dP}}_{\bar{z}\hat{\bs{\alpha}}}+\nonumber \\
 &  & +\hat{\ce}^{\hat{\bs{\alpha}}}\left(\hat{C}_{\hat{\bs{\alpha}}}\hoch{\hat{\bs{\beta}}\bs{\alpha}}-\RR^{\bs{\alpha}\hat{\bs{\gamma}}}\hatdshift{3}_{\hat{\bs{\gamma}}\hat{\bs{\alpha}}}\hoch{\hat{\bs{\beta}}}\right)\hat{\be}_{\bar{z}\hat{\bs{\beta}}}\tilde{\dP}_{z\bs{\alpha}}+\nonumber \\
 &  & +\ce^{\bs{\alpha}}\left(C_{\bs{\alpha}}\hoch{\bs{\beta}\hat{\bs{\alpha}}}-\dshift{3}_{\bs{\gamma}\bs{\alpha}}\hoch{\bs{\beta}}\RR^{\bs{\gamma}\hat{\bs{\alpha}}}\right)\be_{z\bs{\beta}}\tilde{\hat{\dP}}_{\bar{z}\hat{\bs{\alpha}}}+\nonumber \\
 &  & +\ce^{\bs{\alpha}}\hat{\ce}^{\hat{\bs{\alpha}}}\left(S_{\bs{\alpha}\hat{\bs{\alpha}}}\hoch{\bs{\beta}\hat{\bs{\beta}}}+\hat{C}_{\hat{\bs{\alpha}}}\hoch{\hat{\bs{\beta}}\bs{\gamma}}\dshift{3}_{\bs{\gamma}\bs{\alpha}}\hoch{\bs{\beta}}+C_{\bs{\alpha}}\hoch{\bs{\beta}\hat{\bs{\gamma}}}\hatdshift{3}_{\hat{\bs{\gamma}}\hat{\bs{\alpha}}}\hoch{\hat{\bs{\beta}}}-\dshift{3}_{\bs{\gamma}\bs{\alpha}}\hoch{\bs{\beta}}\RR^{\bs{\gamma}\hat{\bs{\gamma}}}\hatdshift{3}_{\hat{\bs{\gamma}}\hat{\bs{\alpha}}}\hoch{\hat{\bs{\beta}}}\right)\be_{z\bs{\beta}}\hat{\be}_{\bar{z}\hat{\bs{\beta}}}+\nonumber \\
 &  & +\left(\bar{\partial}\ce^{\bs{\beta}}+\ce^{\bs{\alpha}}\bar{\partial}x^{M}\left(\Omega_{M\bs{\alpha}}\hoch{\bs{\beta}}-C_{\bs{\alpha}}\hoch{\bs{\beta}\hat{\bs{\alpha}}}\hatdshift{2}_{\hat{\bs{\alpha}}M}-E_{M}\hoch{\bs{\gamma}}\dshift{3}_{\bs{\gamma}\bs{\alpha}}\hoch{\bs{\beta}}+\dshift{3}_{\bs{\gamma}\bs{\alpha}}\hoch{\bs{\beta}}\RR^{\bs{\gamma}\hat{\bs{\alpha}}}\hatdshift{2}_{\hat{\bs{\alpha}}M}\right)\right)\be_{z\bs{\beta}}+\nonumber \\
 &  & +\left(\partial\hat{\ce}^{\hat{\bs{\beta}}}+\hat{\ce}^{\hat{\bs{\alpha}}}\partial x^{M}\left(\hat{\Omega}_{M\hat{\bs{\alpha}}}\hoch{\hat{\bs{\beta}}}-\hat{C}_{\hat{\bs{\alpha}}}\hoch{\hat{\bs{\beta}}\bs{\alpha}}\dshift{2}_{\bs{\alpha}M}-E_{M}\hoch{\hat{\bs{\gamma}}}\hatdshift{3}_{\hat{\bs{\gamma}}\hat{\bs{\alpha}}}\hoch{\hat{\bs{\beta}}}+\dshift{2}_{\bs{\alpha}M}\RR^{\bs{\alpha}\hat{\bs{\gamma}}}\hatdshift{3}_{\hat{\bs{\gamma}}\hat{\bs{\alpha}}}\hoch{\hat{\bs{\beta}}}\right)\right)\hat{\be}_{\bar{z}\hat{\bs{\beta}}}+\nonumber \\
 &  & +\frac{1}{2}L_{z\bar{z}a}(\ce\gamma^{a}\ce)+\frac{1}{2}\hat{L}_{z\bar{z}a}(\hat{\ce}\gamma^{a}\hat{\ce})\frem{-4\alpha'\partial\bar{\partial}\weyl\cdot\dil}\end{eqnarray}
and \begin{eqnarray}
\bs{j}_{z} & = & \ce^{\bs{\alpha}}\left(\tilde{\dP}_{z\bs{\alpha}}+\left(\brstfield{2}_{\bs{\alpha}M}-\dshift{2}_{\bs{\alpha}M}\right)\partial x^{M}+\ce^{\bs{\gamma}}\left(\brstfield{3}_{\bs{\alpha}\bs{\gamma}}\hoch{\bs{\beta}}-\dshift{3}_{\bs{\alpha}\bs{\gamma}}\hoch{\bs{\beta}}\right)\be_{z\bs{\beta}}\frem{+\alpha'\brstfield{4}_{\bs{\alpha}}\partial_{z}\weyl}\right)\\
\hat{\bs{\jmath}}_{\bar{z}} & = & \hat{\ce}^{\hat{\bs{\alpha}}}\left(\tilde{\hat{\dP}}_{\bar{z}\hat{\bs{\alpha}}}+\left(\hatbrstfield{2}_{\hat{\bs{\alpha}}N}-\hatdshift{2}_{\hat{\bs{\alpha}}N}\right)\bar{\partial}x^{N}+\left(\hatbrstfield{3}_{\hat{\bs{\alpha}}\hat{\bs{\gamma}}}\hoch{\hat{\bs{\beta}}}-\hatdshift{3}_{\hat{\bs{\alpha}}\hat{\bs{\gamma}}}\hoch{\hat{\bs{\beta}}}\right)\hat{\ce}^{\hat{\bs{\gamma}}}\hat{\be}_{\bar{z}\hat{\bs{\beta}}}\frem{+\alpha'\hatbrstfield{4}_{\hat{\bs{\alpha}}}\partial_{\bar{z}}\weyl}\right),\quad\hat{\bs{\jmath}}_{z}=0\end{eqnarray}
}The shift-reparametrization thus induces an effective transformation
of the background fields:\begin{eqnarray}
\tilde{E}_{N}\hoch{\bs{\gamma}} & = & E_{N}\hoch{\bs{\gamma}}-\RR^{\bs{\gamma}\hat{\bs{\alpha}}}\hatdshift{2}_{\hat{\bs{\alpha}}B}E_{N}\hoch{B},\qquad\tilde{E}_{M}\hoch{\hat{\bs{\gamma}}}=E_{M}\hoch{\hat{\bs{\gamma}}}-\dshift{2}_{\bs{\alpha}A}E_{M}\hoch{A}\RR^{\bs{\alpha}\hat{\bs{\gamma}}}\label{eq:vielbein-shift-transformation}\\
\tilde{\Omega}_{M\bs{\alpha}}\hoch{\bs{\beta}} & = & \Omega_{M\bs{\alpha}}\hoch{\bs{\beta}}-C_{\bs{\alpha}}\hoch{\bs{\beta}\hat{\bs{\alpha}}}\hatdshift{2}_{\hat{\bs{\alpha}}A}E_{M}\hoch{A}-E_{M}\hoch{\bs{\gamma}}\dshift{3}_{\bs{\gamma}\bs{\alpha}}\hoch{\bs{\beta}}+\dshift{3}_{\bs{\gamma}\bs{\alpha}}\hoch{\bs{\beta}}\RR^{\bs{\gamma}\hat{\bs{\alpha}}}\hatdshift{2}_{\hat{\bs{\alpha}}A}E_{M}\hoch{A}\\
\tilde{\hat{\Omega}}_{M\hat{\bs{\alpha}}}\hoch{\hat{\bs{\beta}}} & = & \hat{\Omega}_{M\hat{\bs{\alpha}}}\hoch{\hat{\bs{\beta}}}-\hat{C}_{\hat{\bs{\alpha}}}\hoch{\hat{\bs{\beta}}\bs{\alpha}}\dshift{2}_{\bs{\alpha}A}E_{M}\hoch{A}-E_{M}\hoch{\hat{\bs{\gamma}}}\hatdshift{3}_{\hat{\bs{\gamma}}\hat{\bs{\alpha}}}\hoch{\hat{\bs{\beta}}}+\dshift{2}_{\bs{\alpha}A}E_{M}\hoch{A}\RR^{\bs{\alpha}\hat{\bs{\gamma}}}\hatdshift{3}_{\hat{\bs{\gamma}}\hat{\bs{\alpha}}}\hoch{\hat{\bs{\beta}}}\\
\tilde{C}_{\bs{\alpha}}\hoch{\bs{\beta}\hat{\bs{\gamma}}} & = & C_{\bs{\alpha}}\hoch{\bs{\beta}\hat{\bs{\gamma}}}-\dshift{3}_{\bs{\gamma}\bs{\alpha}}\hoch{\bs{\beta}}\RR^{\bs{\gamma}\hat{\bs{\gamma}}},\qquad\tilde{\hat{C}}_{\hat{\bs{\alpha}}}\hoch{\hat{\bs{\beta}}\bs{\alpha}}=\hat{C}_{\hat{\bs{\alpha}}}\hoch{\hat{\bs{\beta}}\bs{\alpha}}-\RR^{\bs{\alpha}\hat{\bs{\gamma}}}\hatdshift{3}_{\hat{\bs{\gamma}}\hat{\bs{\alpha}}}\hoch{\hat{\bs{\beta}}}\\
\tilde{S}_{\bs{\alpha}\hat{\bs{\alpha}}}\hoch{\bs{\beta}\hat{\bs{\beta}}} & = & S_{\bs{\alpha}\hat{\bs{\alpha}}}\hoch{\bs{\beta}\hat{\bs{\beta}}}+\hat{C}_{\hat{\bs{\alpha}}}\hoch{\hat{\bs{\beta}}\bs{\gamma}}\dshift{3}_{\bs{\gamma}\bs{\alpha}}\hoch{\bs{\beta}}+C_{\bs{\alpha}}\hoch{\bs{\beta}\hat{\bs{\gamma}}}\hatdshift{3}_{\hat{\bs{\gamma}}\hat{\bs{\alpha}}}\hoch{\hat{\bs{\beta}}}-\dshift{3}_{\bs{\gamma}\bs{\alpha}}\hoch{\bs{\beta}}\RR^{\bs{\gamma}\hat{\bs{\gamma}}}\hatdshift{3}_{\hat{\bs{\gamma}}\hat{\bs{\alpha}}}\hoch{\hat{\bs{\beta}}}\\
\tildebrstfield{2}_{\bs{\alpha}M} & = & \brstfield{2}_{\bs{\alpha}M}-\dshift{2}_{\bs{\alpha}M},\qquad\tildehatbrstfield{2}_{\hat{\bs{\alpha}}N}=\hatbrstfield{2}_{\hat{\bs{\alpha}}N}-\hatdshift{2}_{\hat{\bs{\alpha}}N}\label{eq:BRST-shift-transformationII}\\
\tildebrstfield{3}_{\bs{\alpha}\bs{\gamma}}\hoch{\bs{\beta}} & = & \brstfield{3}_{\bs{\alpha}\bs{\gamma}}\hoch{\bs{\beta}}-\dshift{3}_{\bs{\alpha}\bs{\gamma}}\hoch{\bs{\beta}},\qquad\tildehatbrstfield{3}_{\hat{\bs{\alpha}}\hat{\bs{\gamma}}}\hoch{\hat{\bs{\beta}}}=\hatbrstfield{3}_{\hat{\bs{\alpha}}\hat{\bs{\gamma}}}\hoch{\hat{\bs{\beta}}}-\hatdshift{3}_{\hat{\bs{\alpha}}\hat{\bs{\gamma}}}\hoch{\hat{\bs{\beta}}}\label{eq:BRST-shift-transformationIII}\end{eqnarray}
Finally we have the transformation of $\GB_{MN}=G_{MN}+B_{MN}$ which
we split after the transformation again into its symmetric and antisymmetric
part:{\footnotesize \begin{eqnarray}
\tilde{G}_{MN} & = & E_{M}\hoch{A}E_{N}\hoch{B}\times\qquad\qquad\qquad\qquad\qquad\qquad\qquad\qquad\qquad\qquad\qquad\qquad\qquad\qquad\qquad\qquad\qquad\label{eq:shift-transformedG}\\
 &  & \hspace{-4cm}\lqn{\left(\begin{array}{ccc}
G_{ab}+2\dshift{2}_{\bs{\gamma}(a|}\RR^{\bs{\gamma}\hat{\bs{\gamma}}}\hatdshift{2}_{\hat{\bs{\gamma}}|b)} & G_{a\bs{\beta}}-\dshift{2}_{\bs{\beta}a}+2\dshift{2}_{\bs{\gamma}(a|}\RR^{\bs{\gamma}\hat{\bs{\gamma}}}\hatdshift{2}_{\hat{\bs{\gamma}}|\bs{\beta})} & G_{a\hat{\bs{\beta}}}-\hatdshift{2}_{\hat{\bs{\beta}}a}+2\dshift{2}_{\bs{\gamma}(a|}\RR^{\bs{\gamma}\hat{\bs{\gamma}}}\hatdshift{2}_{\hat{\bs{\gamma}}|\hat{\bs{\beta}})}\\
G_{\bs{\alpha}b}-\dshift{2}_{\bs{\alpha}b}+2\dshift{2}_{\bs{\gamma}(\bs{\alpha}|}\RR^{\bs{\gamma}\hat{\bs{\gamma}}}\hatdshift{2}_{\hat{\bs{\gamma}}|b)} & G_{\bs{\alpha}\bs{\beta}}-2\dshift{2}_{(\bs{\alpha}\bs{\beta})}+2\dshift{2}_{\bs{\gamma}(\bs{\alpha}|}\RR^{\bs{\gamma}\hat{\bs{\gamma}}}\hatdshift{2}_{\hat{\bs{\gamma}}|\bs{\beta})} & G_{\bs{\alpha}\hat{\bs{\beta}}}-\dshift{2}_{\bs{\alpha}\hat{\bs{\beta}}}-\hatdshift{2}_{\hat{\bs{\beta}}\bs{\alpha}}+2\dshift{2}_{\bs{\gamma}(\bs{\alpha}|}\RR^{\bs{\gamma}\hat{\bs{\gamma}}}\hatdshift{2}_{\hat{\bs{\gamma}}|\hat{\bs{\beta}})}\\
G_{\hat{\bs{\alpha}}b}-\hatdshift{2}_{\hat{\bs{\alpha}}b}+2\dshift{2}_{\bs{\gamma}(\hat{\bs{\alpha}}|}\RR^{\bs{\gamma}\hat{\bs{\gamma}}}\hatdshift{2}_{\hat{\bs{\gamma}}|b)} & G_{\hat{\bs{\alpha}}\bs{\beta}}-\dshift{2}_{\bs{\beta}\hat{\bs{\alpha}}}-\hatdshift{2}_{\hat{\bs{\alpha}}\bs{\beta}}+2\dshift{2}_{\bs{\gamma}(\hat{\bs{\alpha}}|}\RR^{\bs{\gamma}\hat{\bs{\gamma}}}\hatdshift{2}_{\hat{\bs{\gamma}}|\bs{\beta})} & G_{\hat{\bs{\alpha}}\hat{\bs{\beta}}}-2\hatdshift{2}_{(\hat{\bs{\alpha}}\hat{\bs{\beta}})}+2\dshift{2}_{\bs{\gamma}(\hat{\bs{\alpha}}|}\RR^{\bs{\gamma}\hat{\bs{\gamma}}}\hatdshift{2}_{\hat{\bs{\gamma}}|\hat{\bs{\beta}})}\end{array}\right)_{AB}}\nonumber \\
\nonumber \\\tilde{B}_{MN} & = & E_{M}\hoch{A}E_{N}\hoch{B}\times\qquad\qquad\qquad\qquad\qquad\qquad\qquad\qquad\qquad\qquad\qquad\qquad\qquad\qquad\qquad\qquad\qquad\label{eq:shift-transformedB}\\
 &  & \hspace{-4cm}\lqn{\left(\begin{array}{ccc}
B_{ab}+2\dshift{2}_{\bs{\gamma}[a|}\RR^{\bs{\gamma}\hat{\bs{\gamma}}}\hatdshift{2}_{\hat{\bs{\gamma}}|b]} & B_{a\bs{\beta}}-\dshift{2}_{\bs{\beta}a}+2\dshift{2}_{\bs{\gamma}[a|}\RR^{\bs{\gamma}\hat{\bs{\gamma}}}\hatdshift{2}_{\hat{\bs{\gamma}}|\bs{\beta}]} & B_{a\hat{\bs{\beta}}}+\hatdshift{2}_{\hat{\bs{\beta}}a}+2\dshift{2}_{\bs{\gamma}[a|}\RR^{\bs{\gamma}\hat{\bs{\gamma}}}\hatdshift{2}_{\hat{\bs{\gamma}}|\hat{\bs{\beta}}]}\\
B_{\bs{\alpha}b}+\dshift{2}_{\bs{\alpha}b}+2\dshift{2}_{\bs{\gamma}[\bs{\alpha}|}\RR^{\bs{\gamma}\hat{\bs{\gamma}}}\hatdshift{2}_{\hat{\bs{\gamma}}|b]} & B_{\bs{\alpha}\bs{\beta}}+2\dshift{2}_{[\bs{\alpha}\bs{\beta}]}+2\dshift{2}_{\bs{\gamma}[\bs{\alpha}|}\RR^{\bs{\gamma}\hat{\bs{\gamma}}}\hatdshift{2}_{\hat{\bs{\gamma}}|\bs{\beta}]} & B_{\bs{\alpha}\hat{\bs{\beta}}}+\dshift{2}_{\bs{\alpha}\hat{\bs{\beta}}}+\hatdshift{2}_{\hat{\bs{\beta}}\bs{\alpha}}+2\dshift{2}_{\bs{\gamma}[\bs{\alpha}|}\RR^{\bs{\gamma}\hat{\bs{\gamma}}}\hatdshift{2}_{\hat{\bs{\gamma}}|\hat{\bs{\beta}}]}\\
B_{\hat{\bs{\alpha}}b}-\hatdshift{2}_{\hat{\bs{\alpha}}b}+2\dshift{2}_{\bs{\gamma}[\hat{\bs{\alpha}}|}\RR^{\bs{\gamma}\hat{\bs{\gamma}}}\hatdshift{2}_{\hat{\bs{\gamma}}|b]} & B_{\hat{\bs{\alpha}}\bs{\beta}}-\dshift{2}_{\bs{\beta}\hat{\bs{\alpha}}}-\hatdshift{2}_{\hat{\bs{\alpha}}\bs{\beta}}+2\dshift{2}_{\bs{\gamma}[\hat{\bs{\alpha}}|}\RR^{\bs{\gamma}\hat{\bs{\gamma}}}\hatdshift{2}_{\hat{\bs{\gamma}}|\bs{\beta}]} & B_{\hat{\bs{\alpha}}\hat{\bs{\beta}}}-2\hatdshift{2}_{[\hat{\bs{\alpha}}\hat{\bs{\beta}}]}+2\dshift{2}_{\bs{\gamma}[\hat{\bs{\alpha}}|}\RR^{\bs{\gamma}\hat{\bs{\gamma}}}\hatdshift{2}_{\hat{\bs{\gamma}}|\hat{\bs{\beta}}]}\end{array}\right)_{AB}}\nonumber \end{eqnarray}
 }Interestingly, looking at (\ref{eq:shift-transformedG}), one
can bring $G_{AB}$ to the block diagonal form $G_{AB}=\diag(G_{ab},0,0)$
at least for vanishing $\RR^{\bs{\gamma}\hat{\bs{\gamma}}}$. For
general $\RR^{\bs{\gamma}\hat{\bs{\gamma}}}$, this is less clear
because the equations become at first sight quadratic%
\footnote{\index{footnote!\thefoot. bringing $G_{AB}$ to a simple form via rep's}Note
that the matrices in (\ref{eq:shift-transformedG}) and (\ref{eq:shift-transformedB})
do not yet correspond to $\tilde{G}_{AB}$ and $\tilde{B}_{AB}$ given
by $\tilde{G}_{MN}=\tilde{E}_{M}\hoch{A}\tilde{E}_{N}\hoch{B}\tilde{G}_{AB}$
and the equivalent equation for $\tilde{B}_{MN}$, as we have expressed
$\tilde{G}_{MN}$ and $\tilde{B}_{MN}$ in terms of the untransformed
vielbeins. Due to (\ref{eq:vielbein-shift-transformation}), the vielbeins
transformation has the form \begin{eqnarray*}
\tilde{E}_{M}\hoch{A} & = & \left(E_{M}\hoch{c},E_{M}\hoch{\bs{\gamma}},E_{M}\hoch{\hat{\bs{\gamma}}}\right)\left(\begin{array}{ccc}
\delta_{c}\hoch{a} & -\RR^{\bs{\alpha}\hat{\bs{\delta}}}\hatdshift{2}_{\hat{\bs{\delta}}c} & -\dshift{2}_{\bs{\delta}c}\RR^{\bs{\delta}\hat{\bs{\alpha}}}\\
0 & \delta_{\bs{\gamma}}\hoch{\bs{\alpha}}-\RR^{\bs{\alpha}\hat{\bs{\delta}}}\hatdshift{2}_{\hat{\bs{\delta}}\bs{\gamma}} & -\dshift{2}_{\bs{\delta}\bs{\gamma}}\RR^{\bs{\delta}\hat{\bs{\alpha}}}\\
0 & -\RR^{\bs{\alpha}\hat{\bs{\delta}}}\hatdshift{2}_{\hat{\bs{\delta}}\hat{\bs{\gamma}}} & \delta_{\hat{\bs{\gamma}}}\hoch{\hat{\bs{\alpha}}}-\dshift{2}_{\bs{\delta}\hat{\bs{\gamma}}}\RR^{\bs{\delta}\hat{\bs{\alpha}}}\end{array}\right)\end{eqnarray*}
For non-vanishing $\RR^{\bs{\gamma}\hat{\bs{\gamma}}}$, the inverse
of this matrix would enter the final form of $\tilde{G}_{AB}$ and
make the problem of finding a reparametrization with $\tilde{G}_{AB}=\diag(\tilde{G}_{ab},0,0)$
more complicated.$\qquad\fussend$%
} in the transformation parameters. It is thus more convenient to use
the shift reparametrization to bring the BRST-currents to their standard
form, i.e. simply shift $\brstfield{2},\,\brstfield{3}$, and their
hatted counterparts to zero. From now on we will thus use the simple
BRST-currents:\vRam{.5}{\begin{eqnarray}
\bs{j}_{z} & = & \ce^{\bs{\alpha}}\dP_{z\bs{\alpha}}\frem{+\alpha'\ce^{\bs{\alpha}}\brstfield{4}_{\bs{\alpha}}(\xfull)\partial_{z}\weyl},\quad\bs{j}_{\bar{z}}=0\label{eq:BiBbrstSimple}\\
\hat{\bs{\jmath}}_{\bar{z}} & = & \hat{\ce}^{\hat{\bs{\alpha}}}\hat{\dP}_{\bar{z}\hat{\bs{\alpha}}}\frem{+\alpha'\hat{\ce}^{\hat{\bs{\alpha}}}\hatbrstfield{4}_{\hat{\bs{\alpha}}}(\xfull)\partial_{\bar{z}}\weyl},\quad\hat{\bs{\jmath}}_{z}=0\label{eq:BiBbrstHatSimple}\end{eqnarray}
}\\
In \cite{Berkovits:2001ue} the authors start with both, the simple
form of the BRST currents as well as the above mentioned special form
of $G_{AB}$ and thus a reduced rank of $G_{MN}$. As we cannot reach
both at the same time with the shift reparametrizations, the simplified
form of the symmetric two-tensor has to be a result of BRST invariance
or likewise on-shell holomorphicity of the BRST-current. We will discover
this result soon. Only then we will use the freedom of the choice
of the auxiliary vielbein components $E_{M}\hoch{a}$ (which do not
appear in the action), in order to fix $G_{ab}$ to $\eta_{ab}$,
or at least proportional to it. For the moment, however, we do not
assume any restrictions on $G_{MN}$, $E_{M}\hoch{a}$ and $G_{AB}$
apart from the invertability of $E_{m}\hoch{a}$.

\paragraph{Local target space symmetries}

There are still many reparametrizations left and we could try to further
simplify the form of the action. It is, however, convenient not to
fix all freedom. As we do not want to destroy the form of action and
BRST currents that we have already obtained, the freedom consists
of 'stabilizing' reparametrizations. I.e. we have to restrict to those
reparametrizations out of (\ref{eq:repI})-(\ref{eq:repVI}) which
leave the form of the action (\ref{eq:BiBaction}) and the simple
BRST currents (\ref{eq:BiBbrstSimple}) and (\ref{eq:BiBbrstHatSimple})
invariant if one transforms the background fields accordingly. These
reparametrizations are in general not symmetries from the worldsheet
point of view as the compensating transformation of the background
fields corresponds to a change of the coupling constants. However,
as the action remains formally invariant, all the constraints on the
background fields which will be derived later will also remain formally
invariant. From the target space point of view the transformations
of the background fields (going along with the $\xfull$-dependent
reparametrizations) thus correspond to local symmetries of the target
space effective theory. What we have done so far by e.g. eliminating
the coefficient fields $\brstfield{i}$ in the BRST operator, corresponds
to a target space gauge fixing of auxiliary background fields.

\paragraph{Residual shift symmetry}

\label{par:Residual-shift-symmetry}Any further shift reparametrization
of $\dP_{z\bs{\alpha}}$ and $\hat{\dP}_{\bar{z}\hat{\bs{\alpha}}}$
changes off-shell the form of the BRST currents (\ref{eq:BiBbrstSimple})
and (\ref{eq:BiBbrstHatSimple}). But we may still allow changes of
the current up to the pure spinor constraint. The pure spinor constraint
generates a gauge transformation as we will see in the next section.
Any change of the BRST currents proportional to the pure spinor constraint
thus can be compensated by a gauge transformation. Under the reparametrizations
\begin{eqnarray}
\dP_{z\bs{\alpha}} & = & \tilde{\dP}_{z\bs{\alpha}}-\dshift{3}_{b}\hoch{\bs{\delta}}(\xfull)(\gamma^{b}\ce)_{\bs{\alpha}}\be_{z\bs{\delta}},\qquad\dann\dshift{3}_{\bs{\alpha}\bs{\gamma}}\hoch{\bs{\delta}}\equiv\gamma_{\bs{\alpha}\bs{\gamma}}^{b}\dshift{3}_{b}\hoch{\bs{\delta}}\label{eq:residualRep}\\
\hat{\dP}_{\bar{z}\hat{\bs{\alpha}}} & = & \tilde{\hat{\dP}}_{\bar{z}\hat{\bs{\alpha}}}-\hatdshift{3}_{b}\hoch{\hat{\bs{\delta}}}(\xfull)(\gamma^{b}\hat{\ce})_{\hat{\bs{\alpha}}}\hat{\be}_{\bar{z}\hat{\bs{\delta}}},\qquad\dann\hatdshift{3}_{\hat{\bs{\alpha}}\hat{\bs{\gamma}}}\hoch{\hat{\bs{\delta}}}\equiv\gamma_{\hat{\bs{\alpha}}\hat{\bs{\gamma}}}^{b}\hatdshift{3}_{b}\hoch{\hat{\bs{\delta}}}\label{eq:residualHatRep}\end{eqnarray}
the BRST currents change to \begin{eqnarray}
\bs{j}_{z} & = & \ce^{\bs{\alpha}}\tilde{\dP}_{z\bs{\alpha}}-\dshift{3}_{b}\hoch{\bs{\delta}}(\xfull)(\ce\gamma^{b}\ce)\be_{z\bs{\delta}}\frem{+\alpha'\ce^{\bs{\alpha}}\brstfield{4}_{\bs{\alpha}}(\xfull)\partial_{z}\weyl},\quad\bs{j}_{\bar{z}}=0\label{eq:residualBRSTchange}\\
\hat{\bs{\jmath}}_{\bar{z}} & = & \hat{\ce}^{\hat{\bs{\alpha}}}\tilde{\hat{\dP}}_{\bar{z}\hat{\bs{\alpha}}}-\hatdshift{3}_{b}\hoch{\hat{\bs{\delta}}}(\xfull)(\hat{\ce}\gamma^{b}\hat{\ce})\hat{\be}_{\bar{z}\hat{\bs{\delta}}}\frem{+\alpha'\hat{\ce}^{\hat{\bs{\alpha}}}\hatbrstfield{4}_{\hat{\bs{\alpha}}}(\xfull)\partial_{\bar{z}}\weyl},\quad\hat{\bs{\jmath}}_{z}=0\label{eq:residualHatBRSTchange}\end{eqnarray}
Global symmetries like the BRST transformation can always be redefined
by a gauge transformation without changing their physical meaning.
Doing this brings us back to the simple form of the BRST currents.
The transformation of the background fields under this reparametrization
is\rem{%
\footnote{Note the consistency of the given transformation of $C$ and the transformation
of $C$ derived via the relation to the covariant derivative of $\RR$
and the torsion! \begin{eqnarray*}
\tilde{C}_{\bs{\alpha}}\hoch{\bs{\beta}\hat{\bs{\gamma}}} & = & \tilde{\nabla}_{\bs{\alpha}}\RR^{\bs{\beta}\hat{\bs{\gamma}}}-2\tilde{T}_{\bs{\alpha\delta}}\hoch{\bs{\beta}}\RR^{\bs{\delta}\hat{\bs{\gamma}}}=\partial_{\bs{\alpha}}\RR^{\bs{\beta}\hat{\bs{\gamma}}}+\tilde{\Omega}_{\bs{\alpha}\bs{\delta}}\hoch{\bs{\beta}}\RR^{\bs{\delta}\hat{\bs{\gamma}}}+\tilde{\hat{\Omega}}_{\bs{\alpha}\hat{\bs{\delta}}}\hoch{\hat{\bs{\gamma}}}\RR^{\bs{\beta}\hat{\bs{\delta}}}-2\tilde{T}_{\bs{\alpha\delta}}\hoch{\bs{\beta}}\RR^{\bs{\delta}\hat{\bs{\gamma}}}=\\
 & = & \tilde{C}_{\bs{\alpha}}\hoch{\bs{\beta}\hat{\bs{\gamma}}}-\gamma_{\bs{\alpha}\bs{\delta}}^{b}\dshift{3}_{b}\hoch{\bs{\beta}}\RR^{\bs{\delta}\hat{\bs{\gamma}}}+2\gamma_{\bs{\alpha}\bs{\delta}}^{b}\dshift{3}_{b}\hoch{\bs{\beta}}\RR^{\bs{\delta}\hat{\bs{\gamma}}}=\\
 & = & \tilde{C}_{\bs{\alpha}}\hoch{\bs{\beta}\hat{\bs{\gamma}}}+\underbrace{\gamma_{\bs{\alpha}\bs{\delta}}^{b}}_{-\gamma_{\bs{\delta}\bs{\alpha}}^{b}}\dshift{3}_{b}\hoch{\bs{\beta}}\RR^{\bs{\delta}\hat{\bs{\gamma}}}\qquad\fussend\end{eqnarray*}
} }\begin{eqnarray}
\tilde{\Omega}_{M\bs{\alpha}}\hoch{\bs{\beta}} & = & \Omega_{M\bs{\alpha}}\hoch{\bs{\beta}}-E_{M}\hoch{\bs{\gamma}}\gamma_{\bs{\gamma}\bs{\alpha}}^{b}\dshift{3}_{b}\hoch{\bs{\beta}}\\
\tilde{\hat{\Omega}}_{M\hat{\bs{\alpha}}}\hoch{\hat{\bs{\beta}}} & = & \hat{\Omega}_{M\hat{\bs{\alpha}}}\hoch{\hat{\bs{\beta}}}-E_{M}\hoch{\hat{\bs{\gamma}}}\gamma_{\hat{\bs{\gamma}}\hat{\bs{\alpha}}}^{b}\hatdshift{3}_{b}\hoch{\hat{\bs{\beta}}}\\
\tilde{C}_{\bs{\alpha}}\hoch{\bs{\beta}\hat{\bs{\gamma}}} & = & C_{\bs{\alpha}}\hoch{\bs{\beta}\hat{\bs{\gamma}}}-\gamma_{\bs{\gamma}\bs{\alpha}}^{b}\dshift{3}_{b}\hoch{\bs{\beta}}\RR^{\bs{\gamma}\hat{\bs{\gamma}}},\qquad\tilde{\hat{C}}_{\hat{\bs{\alpha}}}\hoch{\hat{\bs{\beta}}\bs{\alpha}}=\hat{C}_{\hat{\bs{\alpha}}}\hoch{\hat{\bs{\beta}}\bs{\alpha}}-\RR^{\bs{\alpha}\hat{\bs{\gamma}}}\gamma_{\hat{\bs{\gamma}}\hat{\bs{\alpha}}}^{b}\hatdshift{3}_{b}\hoch{\hat{\bs{\beta}}}\\
\tilde{S}_{\bs{\alpha}\hat{\bs{\alpha}}}\hoch{\bs{\beta}\hat{\bs{\beta}}} & = & S_{\bs{\alpha}\hat{\bs{\alpha}}}\hoch{\bs{\beta}\hat{\bs{\beta}}}+\hat{C}_{\hat{\bs{\alpha}}}\hoch{\hat{\bs{\beta}}\bs{\gamma}}\gamma_{\bs{\gamma}\bs{\alpha}}^{b}\dshift{3}_{b}\hoch{\bs{\beta}}+C_{\bs{\alpha}}\hoch{\bs{\beta}\hat{\bs{\gamma}}}\gamma_{\hat{\bs{\gamma}}\hat{\bs{\alpha}}}^{b}\hatdshift{3}_{b}\hoch{\hat{\bs{\beta}}}-\gamma_{\bs{\gamma}\bs{\alpha}}^{a}\dshift{3}_{a}\hoch{\bs{\beta}}\RR^{\bs{\gamma}\hat{\bs{\gamma}}}\gamma_{\hat{\bs{\gamma}}\hat{\bs{\alpha}}}^{b}\hatdshift{3}_{b}\hoch{\hat{\bs{\beta}}}\end{eqnarray}
This target space gauge symmetry will be fixed at a later point in
section \ref{sec:Residual-shift-reparametrization} on page \pageref{sec:Residual-shift-reparametrization}.
\rem{It induces in particular a transformation of some torsion components
$T_{MN}\hoch{\bs{\beta}}=\partial_{[M}E_{N]}\hoch{\bs{\beta}}+\Omega_{[M|\bs{\alpha}}\hoch{\bs{\beta}}E_{|N]}\hoch{\bs{\alpha}}$:\begin{eqnarray}
\delta T_{MN}\hoch{\bs{\beta}} & = & -E_{[M|}\hoch{\bs{\gamma}}\gamma_{\bs{\gamma}\bs{\alpha}}^{b}\dshift{3}_{b}\hoch{\bs{\beta}}E_{|N]}\hoch{\bs{\alpha}}\\
\delta T_{\bs{\alpha}\bs{\beta}}\hoch{\bs{\delta}} & = & -\gamma_{\bs{\alpha}\bs{\beta}}^{b}\dshift{3}_{b}\hoch{\bs{\delta}}\end{eqnarray}
}

\paragraph{Superdiffeomorphisms }

\label{par:superdiffeomorphisms}Let us now consider the general reparametrizations
(\ref{eq:repI}) of the superspace-embedding functions $x^{M}$ which
correspond to target space super-diffeomorphisms. \begin{equation}
\tilde{x}^{M}=f^{M}(\xfull)\end{equation}
The worldsheet derivatives of the embedding functions transform like
target space vectors\begin{equation}
\bar{\partial}\tilde{x}^{M}=\partial\tilde{x}^{M}/\partial x^{N}\:\cdot\bar{\partial}x^{N}\end{equation}
For the action and the BRST-operators to remain form-invariant, the
background fields have to transform tensorial according to the appearance
of the curved index $M$, e.g. $\tilde{\Omega}_{M\bs{\alpha}}\hoch{\bs{\beta}}(\os{\tilde{{\scriptscriptstyle \twoheadrightarrow}}}{x})=\Omega_{N\bs{\alpha}}\hoch{\bs{\beta}}(\xfull)\,\partial x^{N}/\partial\tilde{x}^{M}$.
All objects with only flat indices or no indices have to transform
like scalars. In this way we observe that the resulting effective
equations for the background fields will be superdiffeomorphism invariant.

\paragraph{Gauge transformation of the B-field}

\label{par:B-field-gauge-trafo}\index{gauge transformation!of the B-field}\index{B-field@$B$-field!gauge transformation}One
of the gauge transformations of the background fields is a bit special,
as it is not related to a worldsheet reparametrization. It is the
shift $B\mapsto B+\de\Lambda$ with some one-form $\Lambda$. This
does not change the action at all, as the total derivative term simply
drops out (for closed strings). It is, however, again not a worldsheet
symmetry, as we do not transform the worldsheet fields but the coupling
constants. The background field-constraints will in the end be the
same for the transformed $B$ and we thus have again a gauge symmetry
from the target space point of view.

\paragraph{Local Lorentz transformations and local scale transformations }

\label{par:LocalStructureGroupTransformations}Next we consider reparametrizations
of the ghost $\ce^{\bs{\alpha}}$. An admissible reparametrizations
(\ref{eq:repII}) of $\ce^{\bs{\alpha}}$ turns the pure spinor term
$L_{z\bar{z}a}(\ce^{T}\gamma^{a}\ce)$ into $L_{z\bar{z}a}(\tilde{\ce}^{T}\Lambda^{-1}\gamma^{a}\Lambda^{T\,-1}\tilde{\ce})$.
In order to obtain the old pure spinor term also in the new variables,
the reparametrization of the ghosts has to be accompanied by an appropriate
reparametrization $L_{z\bar{z}b}=\Lambda_{b}\hoch{a}(\xfull)\cdot\tilde{L}_{z\bar{z}a}$\rem{
(\ref{eq:repVI})} of the Lagrange multiplier $L_{z\bar{z}a}$. The
condition for the invariance of the pure spinor term under the reparametrization
then reads%
\footnote{\index{footnote!\thefoot. reasoning for choice of structure group index positions}The
fact that we use the index structure $\Lambda_{\bs{\beta}}\hoch{\bs{\alpha}}$
instead of $\Lambda^{\bs{\alpha}}\tief{\bs{\beta}}$ is only for later
notational convenience. It is not necessarily related to using NW-conventions,
although $\tilde{\ce}^{\bs{\alpha}}=\ce^{\bs{\beta}}\Lambda_{\bs{\beta}}\hoch{\bs{\alpha}}$
contains a nice NW-contraction. For us the reason is simply that the
alternative index position would be very inconvenient for the associated
connection. The symbol $\Omega_{M\bs{\beta}}\hoch{\bs{\alpha}}$ is
just much simpler to type (and looks better) than $\Omega_{M}\hoch{\bs{\alpha}}\tief{\bs{\beta}}$.$\qquad\fussend$ %
} \begin{eqnarray}
\gamma_{\bs{\alpha\beta}}^{a} & \stackrel{!}{=} & \Lambda_{b}\hoch{a}(\Lambda^{-1})_{\bs{\alpha}}\hoch{\bs{\gamma}}\gamma_{\bs{\gamma}\bs{\delta}}^{b}(\Lambda^{-1})_{\bs{\beta}}\hoch{\bs{\delta}}\label{eq:LorentzScaleCondition}\end{eqnarray}
For infinitesimal reparametrizations we can rewrite it as\begin{eqnarray}
2L_{[\bs{\alpha}|}\hoch{\bs{\delta}}\gamma_{\bs{\delta}|\bs{\beta}]}^{a} & \stackrel{!}{=} & L_{b}\hoch{a}\gamma_{\bs{\alpha\beta}}^{b}\mbox{\qquad(\mbox{infini})}\label{eq:LorentzScaleConditionInfi}\\
\mbox{with }\Lambda_{\bs{\alpha}}\hoch{\bs{\beta}} & \equiv & \delta_{\bs{\alpha}}\hoch{\bs{\beta}}+L_{\bs{\alpha}}\hoch{\bs{\beta}},\qquad\Lambda_{a}\hoch{b}\equiv\delta_{a}^{b}+L_{a}\hoch{b}\end{eqnarray}
To obey this, both reparametrizations are restricted to local Lorentz
transformations and local scale transformations%
\footnote{\label{foot:LorentzScaleReason}\index{footnote!\thefoot. reason for restriction to Lorentz and scale trafos}The
$32\times32$ unity and the antisymmetrized $\Gamma$-matrices $\Gamma^{a_{1}\ldots a_{p}}$
(see appendix \ref{app:gamma} on page \pageref{app:gamma}ff) form
a basis of the vector space of all $32\times32$ matrices. The $16\times16$
sub-matrices $\delta_{\bs{\alpha}}\hoch{\bs{\delta}},\:\gamma^{a_{1}a_{2}}\tief{\bs{\alpha}}\hoch{\bs{\delta}},\:\ldots,\:\gamma^{a_{1}\ldots a_{10}}\tief{\bs{\alpha}}\hoch{\bs{\delta}}$
in the block-diagonal (they vanish for an odd number $p$ of bosonic
antisymmetrized indices, see (\ref{eq:antisymSmallGammasEven}) on
page \pageref{eq:antisymSmallGammasEven}) therefore span all the
$16\times16$ matrices. And due to the relations (\ref{eq:chiralHodgeI})-(\ref{eq:chiralHodgeIV})
on page \pageref{eq:chiralHodgeI}, i.e. $\gamma^{[p]}\propto\gamma^{[n-p]}$,
already the matrices $\delta_{\bs{\alpha}}\hoch{\bs{\delta}},\:\gamma^{a_{1}a_{2}}\tief{\bs{\alpha}}\hoch{\bs{\delta}}\mbox{ and }\gamma^{a_{1}\ldots a_{4}}\tief{\bs{\alpha}}\hoch{\bs{\delta}}$
form a complete basis of all $16\times16$-matrices\frem{$1+\frac{10\cdot9}{2}+\frac{10\cdot9\cdot8\cdot7}{4\cdot3\cdot2}=1+45+210=256=2^{8}=(2^{4})^{2}=16\times16$}.
We thus can expand the infinitesimal generator $L_{\bs{\alpha}}\hoch{\bs{\delta}}$
of the reparametrization matrix (i.e. $\Lambda_{\bs{\alpha}}\hoch{\bs{\delta}}=\delta_{\bs{\alpha}}\hoch{\bs{\delta}}+L_{\bs{\alpha}}\hoch{\bs{\delta}}$)
as follows:\begin{eqnarray*}
L_{\bs{\alpha}}\hoch{\bs{\delta}} & = & \frac{1}{2}L^{(D)}\delta_{\bs{\alpha}}\hoch{\bs{\delta}}+\frac{1}{4}L_{a_{1}a_{2}}^{(L)}\gamma^{a_{1}a_{2}}\tief{\bs{\alpha}}\hoch{\bs{\delta}}+L_{a_{1}\ldots a_{4}}\gamma^{a_{1}\ldots a_{4}}\tief{\bs{\alpha}}\hoch{\bs{\delta}}\end{eqnarray*}
Plugging this expansion into the condition (\ref{eq:LorentzScaleConditionInfi})
yields\begin{eqnarray*}
L_{b}\hoch{a}\gamma_{\bs{\alpha\beta}}^{b} & \stackrel{!}{=} & 2L_{[\bs{\alpha}|}\hoch{\bs{\delta}}\gamma_{\bs{\delta}|\bs{\beta}]}^{a}=L^{(D)}\gamma_{\bs{\alpha}\bs{\beta}}^{a}+\frac{1}{2}L_{a_{1}a_{2}}^{(L)}\underbrace{\gamma^{a_{1}a_{2}}\tief{[\bs{\alpha}|}\hoch{\bs{\delta}}\gamma_{\bs{\delta}|\bs{\beta}]}^{a}}_{\propto\gamma_{\bs{\alpha\beta}}^{[1]}+\underbrace{\gamma_{[\bs{\alpha}\bs{\beta}]}^{[3]}}_{0}}+2L_{a_{1}\ldots a_{4}}\underbrace{\gamma^{a_{1}\ldots a_{4}}\tief{[\bs{\alpha}|}\hoch{\bs{\delta}}\gamma_{\bs{\delta}|\bs{\beta}]}^{a}}_{\propto\underbrace{\gamma_{[\bs{\alpha}\bs{\beta}]}^{[3]}}_{0}+\gamma_{\bs{\alpha\beta}}^{[5]}}\qquad(*)\end{eqnarray*}
Below the curly bracket, we have indicated the schematic expansion
(\ref{eq:product-exp-schem-chiral}) of page \pageref{eq:product-exp-schem-chiral}.
Due to (\ref{eq:antisymSmallGammasOdd}), all the $\gamma^{[3]}$'s
vanish because of the graded antisymmetrization. We can thus concentrate
on the $\gamma^{[1]}$ and $\gamma^{[5]}$-part:\begin{eqnarray*}
\gamma^{a_{1}a_{2}}\tief{[\bs{\alpha}|}\hoch{\bs{\delta}}\gamma_{\bs{\delta}|\bs{\beta}]}^{a} & \stackrel{(\ref{eq:gammaIgammal})}{=} & 2\gamma^{[a_{1}}\tief{\bs{\alpha\beta}}\eta^{a_{2}]a}\\
\gamma^{a_{1}\ldots a_{4}}\tief{[\bs{\alpha}|}\hoch{\bs{\delta}}\gamma_{\bs{\delta}|\bs{\beta}]}^{a} & \stackrel{(\ref{eq:gammaIgammal})}{=} & \gamma^{a_{1}\ldots a_{4}a}\tief{\bs{\alpha\beta}}\end{eqnarray*}
The righthand side of ({*}) has to be a linear combination of $\gamma^{a}$'s
which is not true with a remaining $\gamma^{[5]}$-term $L_{a_{1}\ldots a_{4}}\gamma^{a_{1}\ldots a_{4}a}\tief{\bs{\alpha\beta}}$.
We thus have to demand\[
L_{a_{1}\ldots a_{4}}\stackrel{!}{=}0\]
With this condition, ({*}) and therefore (\ref{eq:LorentzScaleConditionInfi})
are fulfilled and the relation between the reparametrization of the
ghosts and of the Lagrange multipliers is given by \begin{eqnarray*}
L_{\bs{\alpha}}\hoch{\bs{\delta}} & = & \frac{1}{2}L^{(D)}\delta_{\bs{\alpha}}\hoch{\bs{\delta}}+\frac{1}{4}L_{a_{1}a_{2}}^{(L)}\gamma^{a_{1}a_{2}}\tief{\bs{\alpha}}\hoch{\bs{\delta}}\\
L_{b}\hoch{a} & = & L^{(D)}\delta_{b}^{a}+L_{bc}^{(L)}\eta^{ca}\qquad\fussend\end{eqnarray*}
\frem{The coefficients $\Omega_{M}^{(D)}$ and $\Omega_{M\, a_{1}a_{2}}^{(L)}$
can be extracted using $\delta_{\bs{\alpha}}\hoch{\bs{\alpha}}=-16$
and $\gamma^{a_{1}a_{2}}\tief{\bs{\alpha}}\hoch{\bs{\beta}}\gamma_{b_{2}b_{1}\,\bs{\beta}}\hoch{\bs{\alpha}}=-32\delta_{b_{1}b_{2}}^{a_{1}a_{2}}$
(graded version of (\ref{eq:gammapgammapSpurEven}) on page \pageref{eq:gammapgammapSpurEven})
\begin{eqnarray*}
\Omega_{M} & = & -\frac{1}{8}\Omega_{M\bs{\alpha}}\hoch{\bs{\alpha}}\\
\Omega_{Ma_{1}a_{2}} & = & -\frac{1}{8}\gamma_{a_{1}a_{2}\,\bs{\beta}}\hoch{\bs{\alpha}}\Omega_{M\bs{\alpha}}\hoch{\bs{\beta}}\qquad\fussend\end{eqnarray*}
}. %
}\rem{One could remove this restriction by introducing further background fields but there is no advantage in doing so}.
The infinitesimal generators thus have the following explicit form:
\begin{eqnarray}
L_{\bs{\alpha}}\hoch{\bs{\beta}} & = & L_{\:\bs{\alpha}}^{(D)}\hoch{\bs{\beta}}+L_{\:\bs{\alpha}}^{(L)}\hoch{\bs{\beta}},\qquad L_{a}\hoch{b}=L_{\; a}^{(D)}\hoch{b}+L_{\; a}^{(L)}\hoch{b}\\
L_{\:\bs{\alpha}}^{(D)}\hoch{\bs{\beta}} & \equiv & \frac{1}{2}L^{(D)}\delta_{\bs{\alpha}}\hoch{\bs{\beta}},\qquad L_{\:\bs{\alpha}}^{(L)}\hoch{\bs{\beta}}=\frac{1}{4}L_{ab}^{(L)}\gamma^{ab}\tief{\bs{\alpha}}\hoch{\bs{\beta}},\qquad L_{ab}^{(L)}=-L_{ba}^{(L)}\\
L_{\: a}^{(D)}\hoch{b} & \equiv & L^{(D)}\delta_{a}^{b},\qquad L_{\: a}^{(L)}\hoch{b}=L_{cd}^{(L)}\delta_{a}^{[c}\eta^{d]b},\qquad L_{cd}^{(L)}=-L_{dc}^{(L)}\end{eqnarray}
The reparametrization so far reads\label{par:local-strucgroup-transformations}\begin{eqnarray}
\tilde{\ce}^{\bs{\alpha}} & = & \Lambda_{\bs{\beta}}\hoch{\bs{\alpha}}\ce^{\bs{\beta}}\frem{=\left(\exp(\frac{1}{4}L_{ab}\gamma^{ab})\right)\tief{\bs{\beta}}\hoch{\bs{\alpha}}\ce^{\bs{\beta}}}\\
\tilde{L}_{z\bar{z}a} & = & \Lambda_{\: a}^{-1}\hoch{b}L_{z\bar{z}b}\frem{=(\exp(-L))_{a}\hoch{b}L_{z\bar{z}b}}\end{eqnarray}
Note that in our notation $\Lambda$ contains both, Lorentz transformations
and scale transformations (dilatations). 

In order to maintain the special form of the ghost kinetic term and
of the BRST-operator, we likewise have to transform\begin{eqnarray}
\tilde{\dP}_{z\bs{\alpha}} & = & (\Lambda^{-1})_{\bs{\alpha}}\hoch{\bs{\beta}}\dP_{z\bs{\beta}}\\
\tilde{\be}_{z\bs{\alpha}} & = & (\Lambda^{-1})_{\bs{\alpha}}\hoch{\bs{\beta}}\be_{z\bs{\beta}}\end{eqnarray}
with infinitesimally $(\Lambda^{-1})\tief{\bs{\alpha}}\hoch{\bs{\beta}}=\delta_{\bs{\alpha}}\hoch{\bs{\beta}}-L_{\bs{\alpha}}\hoch{\bs{\beta}}$.
The background fields can again be reparametrized in a way that the
complete action plus the BRST operators remain form-invariant: Just
transform every background field with unhatted spinorial indices accordingly.
E.g.\begin{eqnarray}
\tilde{C}_{\bs{\alpha}}\hoch{\bs{\beta}\hat{\bs{\gamma}}} & = & (\Lambda^{-1})_{\bs{\alpha}}\hoch{\bs{\gamma}}\Lambda_{\bs{\delta}}\hoch{\bs{\beta}}C_{\bs{\gamma}}\hoch{\bs{\delta}\hat{\bs{\gamma}}},\quad\ldots\end{eqnarray}
Only the field $\Omega_{M\bs{\alpha}}\hoch{\bs{\beta}}$ must not
transform like a tensor, but like a connection, in order to keep the
form-invariance of the action\rem{%
\footnote{The finite Lorentz transformation (in the connected component of the
unity) is $\Lambda=e^{L}$. For this we have \begin{align*}
\Lambda^{T}\eta\Lambda & =\eta,\qquad\Lambda^{-1}=\eta^{-1}\Lambda^{T}\eta\end{align*}
\begin{eqnarray*}
\tilde{\ce} & = & \Lambda\ce,\quad\tilde{\ce}^{T}=\ce^{T}\Lambda^{T}\\
\tilde{\be} & = & \left(\Lambda^{T}\right)^{-1}\be=\eta\Lambda\eta^{-1}\be\\
\tilde{\Omega}_{M} & = & -\left(\Lambda^{T}\right)^{-1}\partial_{M}\Lambda^{T}+\left(\Lambda^{T}\right)^{-1}\Omega_{M}\Lambda^{T}\end{eqnarray*}
Transformation of the connection:\begin{eqnarray*}
\nabla_{M}\ce & \To & \nabla_{M}(\Lambda\ce)=\partial_{M}(\Lambda\ce)+\tilde{\Omega}_{M}^{T}\Lambda\ce=\\
 &  & =\partial_{M}\Lambda\cdot\ce+\Lambda\left(\partial_{M}\ce+\Omega_{M}^{T}\ce\right)-\Lambda\Omega_{M}^{T}\ce+\tilde{\Omega}_{M}^{T}\Lambda\ce\\
\dann\tilde{\Omega}_{M}^{T}\Lambda & = & -\partial_{M}\Lambda+\Lambda\Omega_{M}^{T}\\
\tilde{\Omega}_{M} & = & -\left(\Lambda^{T}\right)^{-1}\partial_{M}\Lambda^{T}+\left(\Lambda^{T}\right)^{-1}\Omega_{M}\Lambda^{T}=\\
 & = & \Omega_{M}-\partial_{M}L^{T}-L^{T}\Omega_{M}+\Omega_{M}L^{T}+\mc{O}(L^{2})=\\
 & = & \Omega_{M}+\partial_{M}L+[L,\Omega_{M}]+\mc{O}(L^{2})\end{eqnarray*}
A quicker derivation in form-language:\begin{eqnarray*}
\bs{\nabla}v & = & \de v+\Omega v\\
\tilde{v} & = & Av\\
A(\nabla v) & \stackrel{!}{=} & \tilde{\nabla}\tilde{v}=\de(Av)+\tilde{\Omega}Av=\\
 & = & \de A\, v+A(\de v+\Omega v)-A\Omega v+\tilde{\Omega}Av\\
\dann\tilde{\Omega} & = & -\de A\, A^{-1}+A\Omega A^{-1}\stackrel{A=1+a}{=}\\
 & = & \Omega\underbrace{-\de a+[a,\Omega]}_{\delta\Omega}\qquad\fussend\end{eqnarray*}
}}\begin{eqnarray}
\tilde{\Omega}_{M\bs{\alpha}}\hoch{\bs{\beta}} & = & -\partial_{M}\Lambda_{\bs{\alpha}}\hoch{\bs{\beta}}+(\Lambda^{-1})_{\bs{\alpha}}\hoch{\bs{\gamma}}\Lambda_{\bs{\delta}}\hoch{\bs{\beta}}\Omega_{M\bs{\gamma}}\hoch{\bs{\delta}}\label{eq:connectionTransformsLikeConnection}\end{eqnarray}
\label{par:reasonForCovDer}This is exactly the reason why we have
combined it to a covariant derivative in the ghost kinetic term right
from the beginning. For the effective field equations all this means
that they will be invariant under a local Lorentz transformation and
dilatation acting on all the indices of the background fields which
are coupled to the ghosts, the ghost-momenta and the variables $\dP_{z\bs{\alpha}}$,
or in other words, acting on all unhatted flat spinorial indices.

We get an equivalent but in the beginning completely independent local
Lorentz transformation and scaling $\hat{\Lambda}_{\hat{\bs{\alpha}}}\hoch{\hat{\bs{\beta}}}$
acting on the hatted indices. In addition we may redefine the bosonic
vielbein $E^{a}=\de x^{M}E_{M}\hoch{a}$, which we introduced by hand.
Remember, it is related to $G_{AB}$ via $G_{MN}=E_{M}\hoch{A}G_{AB}E_{N}\hoch{B}$
and we did not yet restrict $G_{AB}$. The matrices $E_{M}\hoch{a}$
(of maximal rank 10) can thus be redefined by an arbitrary GL(10)
transformation on the index $a$, accompanied by a compensating transformation
of $G_{AB}$. At a later point, we will obtain a restriction on $G_{AB}$
which then allows only Lorentz and scale transformations $\check{\Lambda}_{a}\hoch{b}$
acting on the index $a$ of $E_{M}\hoch{a}$. This transformation,
acting on bosonic flat indices only, is again independent of the other
two local structure group transformations (acting on the spinorial
indices). The relation of the three transformations will in the end
be fixed (see page \pageref{Intermezzo:fixingTwoLorentz}) by a convenient
gauge fixing of some torsion components. In contrast to the fermionic
transformations, the bosonic local Lorentz transformation is not coupled
to a reparametrization of an elementary field (from the worldsheet
point of view), but only to the transformation of $G_{ab}$:\begin{eqnarray}
\tilde{E}_{M}\hoch{a} & = & \check{\Lambda}_{c}\hoch{a}E_{M}\hoch{c}\\
\tilde{G}_{ab} & = & (\check{\Lambda}^{-1})_{a}\hoch{c}G_{cd}(\check{\Lambda}^{-1})_{b}\hoch{d}\label{eq:structureGroupTrafoOfG}\end{eqnarray}
The transformation of the background fields is determined by their
flat indices. Combining the bosonic and fermionic flat indices to
$A\equiv(a,\bs{\alpha},\hat{\bs{\alpha}})$, we have a block diagonal
\textbf{structure group transformation} \index{structure group!Lorentz and scale|fett}acting
on the target super tangent space:\begin{eqnarray}
\gem{\Lambda}_{A}\hoch{B} & \equiv & \left(\begin{array}{ccc}
\check{\Lambda}_{a}\hoch{b} & 0 & 0\\
0 & \Lambda_{\bs{\alpha}}\hoch{\bs{\beta}} & 0\\
0 & 0 & \hat{\Lambda}_{\hat{\bs{\alpha}}}\hoch{\hat{\bs{\beta}}}\end{array}\right)\label{eq:BlockDiagLambda}\end{eqnarray}
All three blocks are independent. $\Lambda_{a}\hoch{b}$ instead,
which is acting on the Lagrange multiplier (but on no background field!),
was induced by $\Lambda_{\bs{\alpha}}\hoch{\bs{\beta}}$ via the invariance
of $\gamma_{\bs{\alpha\beta}}^{a}$. Also keep in mind that $\check{\Lambda}_{a}\hoch{b}$
is so far not restricted to Lorentz transformations or scalings. It
will be so at a later point.

\section{Connection}

We have seen in equation (\ref{eq:connectionTransformsLikeConnection})
\vpageref{eq:connectionTransformsLikeConnection} that $\Omega_{M\bs{\alpha}}\hoch{\bs{\beta}}$
and $\hat{\Omega}_{M\hat{\bs{\alpha}}}\hoch{\hat{\bs{\beta}}}$ transform
like connections under structure group transformations. Let us introduce
some auxiliary target space field $\check{\Omega}_{Ma}\hoch{b}$ which
transforms like a connection under the transformation $\check{\Lambda}_{a}\hoch{b}$
of the bosonic tangent space. As the field $\check{\Omega}_{Ma}\hoch{b}$
does not appear in the worldsheet action, nothing should depend on
it in the end. We can now combine the three objects to a structure
group connection on the target super tangent space (let's call it
the \textbf{mixed connection}\index{mixed connection}\index{connection!mixed}\index{$\Omega$@$\gemOm_{MA}\hoch{B}$}\index{$\Omega_{Ma}\hoch{b}$@$\check{\Omega}_{Ma}\hoch{b}$})
\begin{eqnarray}
\gemOm_{MA}\hoch{B} & \equiv & \left(\begin{array}{ccc}
\check{\Omega}_{Ma}\hoch{b} & 0 & 0\\
0 & \Omega_{M\bs{\alpha}}\hoch{\bs{\beta}} & 0\\
0 & 0 & \hat{\Omega}_{M\hat{\bs{\alpha}}}\hoch{\hat{\bs{\beta}}}\end{array}\right)\label{eq:mixedConnection}\end{eqnarray}
The underline will help us later to distinguish this connection from
alternative choices. This underline will decorate all objects referring
to this connection. The corresponding superspace connection coefficients
$\gem{\Gamma}_{MN}\hoch{K}$\index{$\Gamma$@$\gem{\Gamma}_{MN}\hoch{K}$}
are now given via \begin{eqnarray}
0 & \stackrel{!}{=} & \gem{\nabla}_{M}E_{N}\hoch{A}\equiv\partial_{M}E_{N}\hoch{A}-\gem{\Gamma}_{MN}\hoch{K}E_{K}\hoch{A}+\gem{\Omega}_{MB}\hoch{A}E_{N}\hoch{B}\end{eqnarray}
Due to the block-diagonal form of the connection, the curvature $\gemR_{A}\hoch{B}\equiv\de\gemOm_{A}\hoch{B}-\gemOm_{A}\hoch{C}\wedge\gemOm_{C}\hoch{B}$
is block diagonal as well\index{$R$@$\gemR_{A}\hoch{B}$}\begin{eqnarray}
\gemR_{A}\hoch{B} & = & \left(\begin{array}{ccc}
\check{R}_{a}\hoch{b} & 0 & 0\\
0 & R_{\bs{\alpha}}\hoch{\bs{\beta}} & 0\\
0 & 0 & \hat{R}_{\hat{\bs{\alpha}}}\hoch{\hat{\bs{\beta}}}\end{array}\right)\label{eq:mixedCurvature}\end{eqnarray}
and the upper index of the torsion $\gemT^{A}\equiv\de E^{A}-E^{C}\wedge\gemOm_{C}\hoch{A}$
tells us by which block of the connection it is determined:\index{$T$@$\gemT^{A}$}\begin{eqnarray}
\gemT^{A} & = & (\check{T}^{a},T^{\bs{\alpha}},\hat{T}^{\hat{\bs{\alpha}}})\label{eq:mixedTorsion}\end{eqnarray}

\paragraph{Remark}

Although the connection coefficients which act on the spinorial indices
have the correct transformation properties, we did not yet check that
they are Lie algebra valued, i.e. that the matrices $\Omega_{M\,\cdot}\hoch{\cdot}$
and $\hat{\Omega}_{M\,\cdot}\hoch{\cdot}$ are not general matrices,
but are restricted to the structure group algebra of Lorentz and scale
transformations. We will show this partwise below in section \ref{sec:Antighost-gauge-symmetry}
when we discuss the antighost gauge symmetry and will complete the
argument when we study the holomorphicity of the BRST current in section
\ref{sec:Holomorphic-BRST-current}. Let us already here give the
result for completeness:\index{$\Omega_{Ma_{1}a_{2}}^{(L)}$|itext{Lorentz connection}}\index{$\Omega_{M}^{(D)}$|itext{scale connection}}\index{Lorentz connection}\index{connection!Lorentz $\sim$}\index{scale connection}\index{connection!scale $\sim$}
\begin{eqnarray}
\Omega_{M\bs{\alpha}}\hoch{\bs{\beta}} & = & \frac{1}{2}\Omega_{M}^{(D)}\delta_{\bs{\alpha}}\hoch{\bs{\beta}}+\frac{1}{4}\Omega_{Ma_{1}a_{2}}^{(L)}\gamma^{a_{1}a_{2}}\tief{\bs{\alpha}}\hoch{\bs{\beta}},\qquad\hat{\Omega}_{M\hat{\bs{\alpha}}}\hoch{\hat{\bs{\beta}}}=\frac{1}{2}\hat{\Omega}_{M}^{(D)}\delta_{\hat{\bs{\alpha}}}\hoch{\hat{\bs{\beta}}}+\frac{1}{4}\hat{\Omega}_{Ma_{1}a_{2}}^{(L)}\gamma^{a_{1}a_{2}}\tief{\hat{\bs{\alpha}}}\hoch{\hat{\bs{\beta}}}\label{eq:specialFormOfOmega}\end{eqnarray}
The labels $(D)$ and $(L)$ distinguish the dilatation (or scaling)
part from the Lorentz part. 

This special form of the connection of course induces a special form
of the curvature (see (\ref{eq:mixedCurvature}) and (\ref{eq:R-Zerfall-bosonic}),(\ref{eq:R-Zerfall-ferm})
and (\ref{eq:R-Zerfall-ferm-hut}) on page \ref{eq:R-Zerfall-ferm}).
The curvature is blockdiagonal in the last two indices (\ref{eq:mixedCurvature})
and each block decays into a scale (or dilatation) part and a Lorentz
part:\index{$R_{ABC}$@$\gem{R}_{ABC}\hoch{D}$}\index{field strength!scale $\sim$}\index{scale field strength}\index{$R_{ABC}$@$\gem{R}^{(L)}_{ABC}\hoch{D}$}\index{$F^{(D)}_{MN}$|itext{scale field strength}}\begin{eqnarray}
\gem{R}_{MNC}\hoch{D} & = & \diag(\check{R}_{MNc}\hoch{d},R_{MN\bs{\gamma}}\hoch{\bs{\delta}},\hat{R}_{MN\hat{\bs{\gamma}}}\hoch{\hat{\bs{\delta}}})\label{eq:blockformOfR}\\
\check{R}_{MNc}\hoch{d} & = & \check{F}_{MN}^{(D)}\delta_{c}^{d}+\check{R}_{MN\: c}^{(L)}\hoch{d},\qquad\check{F}_{MN}^{(D)}=\frac{1}{10}\check{R}_{MNc}\hoch{c}\label{eq:decayofRcheck}\\
R_{MN\bs{\gamma}}\hoch{\bs{\delta}} & = & \frac{1}{2}F_{MN}^{(D)}\delta_{\bs{\gamma}}\hoch{\bs{\delta}}+\frac{1}{4}R_{MN}^{(L)}\tief{a_{1}}\hoch{b}\eta_{ba_{2}}\gamma^{a_{1}a_{2}}\tief{\bs{\gamma}}\hoch{\bs{\delta}},\qquad F_{MN}^{(D)}=-\frac{1}{8}R_{MN\bs{\gamma}}\hoch{\bs{\gamma}}\label{eq:decayofR}\\
\hat{R}_{MN\hat{\bs{\gamma}}}\hoch{\hat{\bs{\delta}}} & = & \frac{1}{2}\hat{F}_{MN}^{(D)}\delta_{\hat{\bs{\gamma}}}\hoch{\hat{\bs{\delta}}}+\frac{1}{4}\hat{R}_{MN}^{(L)}\tief{a_{1}}\hoch{b}\eta_{ba_{2}}\gamma^{a_{1}a_{2}}\tief{\hat{\bs{\gamma}}}\hoch{\hat{\bs{\delta}}},\qquad\hat{F}_{MN}^{(D)}=-\frac{1}{8}\hat{R}_{MN\hat{\bs{\gamma}}}\hoch{\hat{\bs{\gamma}}}\label{eq:decayofRhat}\end{eqnarray}
with the scale field strength \rem{%
\footnote{It is tempting to write $F^{(D)}\equiv\de\Omega^{(D)}\quad\iff\quad F_{AB}^{(D)}=\gemnabla_{[A}\Omega_{B]}+\gemT_{AB}\hoch{C}\Omega_{C}^{(D)}$
which is formally true if we act with the covariant derivative on
$\gemOm_{B}$ like on a tensorial object. We prefer the point of view
to act with the connection always in the same representations as the
local structure group transformation would do. There we have $\delta\Omega_{M}^{(D)}=-\partial_{M}\Lambda^{(D)}$.
The covariant derivative acting on $\Omega_{M}^{(D)}$ can thus be
defined as \[
\gemnabla_{M}\Omega_{N}^{(D)}\equiv\partial_{M}\Omega_{N}^{(D)}-\partial_{N}\gemOm_{M}^{(D)}??\qquad\fussend\]
} }\begin{equation}
\check{F}^{(D)}\equiv\de\check{\Omega}^{(D)},\qquad F^{(D)}\equiv\de\Omega^{(D)},\qquad\hat{F}^{(D)}\equiv\de\hat{\Omega}^{(D)}\label{eq:defF}\end{equation}

The major part of the covariant derivation of the last equation of
motion in section \ref{sec:Covariant-eoms}, where we have not yet
completed the argument that the mixed connection is structure group
valued, does not refer to this fact. Only the variation of the pure
spinor term will be affected and this will be discussed carefully.
\rem{Now in DRbianchi\_identities.lyx:We certainly want a spacetime
connection which leaves $\gamma_{\alpha\beta}^{a}$ invariant. The
choice of $\Omega_{M\bs{\alpha}}\hoch{\bs{\beta}}$ therefore determines
$\Omega_{Ma}\hoch{b}$ which in turn determines $\Omega_{M\hat{\alpha}}\hoch{\hat{\beta}}$.
Those three blocks combine to a complete \textbf{left mover structure
group connection} $\Omega_{MA}\hoch{B}$. The same argumentation holds
for $\hat{\Omega}_{M\hat{\bs{\alpha}}}\hoch{\hat{\bs{\beta}}}$ which
leads to a second (in general independent) \textbf{right mover structure
group} connection $\hat{\Omega}_{MA}\hoch{B}$. The difference of
the two connection coefficients \rem{%
\footnote{For a different vielbein $\tilde{E}^{A}=E^{B}J_{B}\hoch{A}$ with
$\tilde{\nabla}_{M}\tilde{E}^{A}=0$, we get\begin{eqnarray*}
\tilde{T}^{A} & = & \de\tilde{E}^{A}-\tilde{E}^{C}\wedge\Omega_{C}\hoch{A}=\\
 & = & \de E^{B}J_{B}\hoch{A}-E^{B}\wedge\de J_{B}\hoch{A}-E^{B}J_{B}\hoch{C}\wedge\Omega_{C}\hoch{A}=\\
 & = & T^{B}J_{B}\hoch{A}-E^{B}\wedge\bs{\nabla}J_{B}\hoch{A}=\\
 & = & T_{\bs{MM}}\hoch{B}J_{B}\hoch{A}+\nabla_{\bs{M}}J_{\bs{M}}\hoch{A}\\
\tilde{R}_{A}\hoch{B} & = & R_{A}\hoch{B}?or\tilde{\Omega}=J\Omega J^{-1}?\qquad\fussend\end{eqnarray*}
}}\begin{eqnarray}
\Delta_{MA}\hoch{B} & \equiv & \hat{\Omega}_{MA}\hoch{B}-\Omega_{MA}\hoch{B}\end{eqnarray}
is a tensor (transforms covariantly under local Lorentz transformations).
The torsion and curvature tensors change as follows under a change
of the connection:%
\footnote{\index{footnote!\thefoot. change of torsion and curvature for different connection}\begin{eqnarray*}
\hat{T}^{A} & = & \de E^{A}-E^{C}\wedge\underbrace{\hat{\Omega}_{C}\hoch{A}}_{\Delta+\Omega}=\\
 & = & T^{A}-E^{C}\wedge\Delta_{C}\hoch{A}\\
\hat{R}_{A}\hoch{B} & = & \de\hat{\Omega}_{A}\hoch{B}-\hat{\Omega}_{A}\hoch{C}\wedge\hat{\Omega}_{C}\hoch{B}=\\
 & = & R_{A}\hoch{B}+\de\Delta_{A}\hoch{B}-\Delta_{A}\hoch{C}\wedge\Omega_{C}\hoch{B}-\Omega_{A}\hoch{C}\wedge\Delta_{C}\hoch{B}-\Delta_{A}\hoch{C}\wedge\Delta_{C}\hoch{B}=\\
 & = & R_{A}\hoch{B}+\bs{\nabla}\Delta_{A}\hoch{B}+T^{K}\Delta_{KA}\hoch{B}-\Delta_{A}\hoch{C}\wedge\Delta_{C}\hoch{B}\qquad\fussend\end{eqnarray*}
}\begin{eqnarray}
\hat{T}_{MN}\hoch{A} & = & T_{MM}\hoch{A}+\Delta_{[MN]}\hoch{A}\\
\hat{R}_{MNA}\hoch{B} & = & R_{MNA}\hoch{B}+\nabla_{[M}\Delta_{N]A}\hoch{B}+T_{MN}\hoch{K}\Delta_{KA}\hoch{B}-\Delta_{[M|A}\hoch{K}\Delta_{|N]K}\hoch{B}\end{eqnarray}
}

\section{Antighost gauge symmetry}

\label{sec:Antighost-gauge-symmetry}The pure spinor constraints
$\ce\gamma^{a}\ce=\hat{\ce}\gamma^{a}\hat{\ce}=0$ are first class
constraints at least in the flat case and thus generate gauge symmetries.
The same should be true in the curved case. We can see this fact,
however, without referring to the Hamiltonian language, simply as
a consistency condition on the equations of motion.

For the ghost field we have two equations of motion which have to
be consistent in order to allow any solutions:\begin{eqnarray}
\funktional{S}{\be_{z\bs{\beta}}} & = & -\left(\bar{\partial}\ce^{\bs{\beta}}+\ce^{\bs{\alpha}}\left(\bar{\partial}x^{M}\Omega_{M\bs{\alpha}}\hoch{\bs{\beta}}+C_{\bs{\alpha}}\hoch{\bs{\beta}\hat{\bs{\gamma}}}\hat{\dP}_{\bar{z}\hat{\bs{\gamma}}}-\hat{\ce}^{\hat{\bs{\alpha}}}S_{\bs{\alpha}\hat{\bs{\alpha}}}\hoch{\bs{\beta}\hat{\bs{\beta}}}\hat{\be}_{\bar{z}\hat{\bs{\beta}}}\right)\right)\equiv-\mc{D}_{\bar{z}}\ce^{\bs{\beta}}\label{eq:ceom-antigauge}\\
\funktional{S}{L_{z\bar{z}a}} & = & \frac{1}{2}(\ce\gamma^{a}\ce)\end{eqnarray}
Every linear combination of the second line, $\frac{\mu_{a}}{2}(\ce\gamma^{a}\ce)$,
obviously is still on-shell zero for any set of local parameters $\mu_{a}$.
When we act with $\bar{\partial}$ on this expression, the result
still has to vanish on-shell. I.e. for any $\mu_{a}$, we need to
have:\rem{$-\delta_{(\mu)}\allfields{I}\funktional{S}{\allfields{I}}\:\ous{!}{=}{\hspace{-.2cm}\mbox{on-shell}}$}\begin{eqnarray}
0 & \ous{!}{=}{\hspace{-.2cm}\mbox{on-shell}} & \:\bar{\partial}\left(\frac{\mu_{a}}{2}\ce\gamma^{a}\ce\right)=\qquad\qquad\qquad\qquad\qquad\qquad\qquad\qquad\forall\mu_{a}(z,\bar{z})\nonumber \\
 & \stackrel{(\ref{eq:AundMathcalD})}{=} & \bar{\partial}\mu_{a}\cdot\underbrace{\frac{1}{2}(\ce\gamma^{a}\ce)}_{\funktional{S}{L_{z\bar{z}a}}}+\mu_{a}(\ce\gamma^{a})_{\bs{\beta}}\underbrace{\mc{D}_{\bar{z}}\ce^{\bs{\beta}}}_{-\funktional{S}{\be_{z\bs{\beta}}}}-\mu_{a}\ce^{\bs{\alpha}}\underbrace{\left(\Pi_{\bar{z}}^{C}\Omega_{C[\bs{\alpha}|}\hoch{\bs{\delta}}+C_{[\bs{\alpha}|}\hoch{\bs{\delta}\hat{\bs{\gamma}}}\hat{\dP}_{\bar{z}\hat{\bs{\gamma}}}-\hat{\ce}^{\hat{\bs{\alpha}}}S_{[\bs{\alpha}|\hat{\bs{\alpha}}}\hoch{\bs{\delta}\hat{\bs{\beta}}}\hat{\be}_{\bar{z}\hat{\bs{\beta}}}\right)}_{A_{\bar{z}[\bs{\alpha}|}\hoch{\bs{\delta}}}\gamma_{\bs{\delta}|\bs{\beta}]}^{a}\ce^{\bs{\beta}}\qquad\label{eq:on-shell-cond-antigauge}\end{eqnarray}
The first two terms in the last line vanish on-shell, so we may concentrate
on the rest. Following footnote \ref{foot:LorentzScaleReason} on
page \pageref{foot:LorentzScaleReason} (with $A_{\bar{z}[\bs{\alpha}|}\hoch{\bs{\delta}}$
taking the role of $L_{[\bs{\alpha}|}\hoch{\bs{\delta}}$) we can
expand $A_{\bar{z}[\bs{\alpha}|}\hoch{\bs{\delta}}$ in antisymmetrized
$\gamma$-matrices and obtain for the last term in \eqref{eq:on-shell-cond-antigauge}\begin{eqnarray}
-\mu_{a}\ce^{\bs{\alpha}}A_{\bar{z}[\bs{\alpha}|}\hoch{\bs{\delta}}\gamma_{\bs{\delta}|\bs{\beta}]}^{a}\ce^{\bs{\beta}} & = & -\mu_{a}\ce^{\bs{\alpha}}\left(\frac{1}{2}A_{\bar{z}}^{(D)}\gamma_{\bs{\alpha}\bs{\beta}}^{a}+\frac{1}{2}A_{\bar{z}\, a_{1}a_{2}}^{(L)}\gamma^{[a_{1}}\tief{\bs{\alpha\beta}}\eta^{a_{2}]a}+A_{\bar{z}\, a_{1}\ldots a_{4}}\gamma^{a_{1}\ldots a_{4}a}\tief{\bs{\alpha\beta}}\right)\ce^{\bs{\beta}}=\nonumber \\
 & = & -\underbrace{\left(A_{\bar{z}}^{(D)}\delta_{a}^{b}+A_{\bar{z}\, a}^{(L)}\hoch{b}\right)}_{\equiv A_{\bar{z}\, a}\hoch{b}}\mu_{b}\cdot\underbrace{\frac{1}{2}(\ce\gamma^{a}\ce)}_{\funktional{S}{L_{z\bar{z}a}}}-\mu_{a}A_{\bar{z}\, a_{1}\ldots a_{4}}(\ce\gamma^{a_{1}\ldots a_{4}a}\ce)\end{eqnarray}
It is natural to view $A_{\bar{z}\, a}\hoch{b}$ as the connection
coefficients corresponding to $\mc{D}_{\bar{z}}$ when acting on bosonic
indices. It is built from the expansion coefficients of $A_{\bar{z}\bs{\alpha}}\hoch{\bs{\beta}}$
which are in turn built from the expansion coefficients of $\Omega_{M\bs{\alpha}}\hoch{\bs{\beta}}$,
$C_{\bs{\alpha}}\hoch{\bs{\beta}\hat{\bs{\gamma}}}$ and $S_{\bs{\alpha}\hat{\bs{\alpha}}}\hoch{\bs{\beta}\hat{\bs{\beta}}}$
(all seen as matrices in $\bs{\alpha}$ and $\bs{\beta}$ -- compare
again to footnote \ref{foot:LorentzScaleReason} on page \pageref{foot:LorentzScaleReason})%
\footnote{\index{footnote!\thefoot. extracting dilatation and Lorentz part of connection}The
coefficients $\Omega_{M}^{(D)}$ and $\Omega_{M\, a_{1}a_{2}}^{(L)}$
can be extracted from the given $\Omega_{M\bs{\alpha}}\hoch{\bs{\beta}}$
using $\delta_{\bs{\alpha}}\hoch{\bs{\alpha}}=-16$ and $\gamma^{a_{1}a_{2}}\tief{\bs{\alpha}}\hoch{\bs{\beta}}\gamma_{b_{2}b_{1}\,\bs{\beta}}\hoch{\bs{\alpha}}=-32\delta_{b_{1}b_{2}}^{a_{1}a_{2}}$
(graded version of (\ref{eq:gammapgammapSpurEven}) on page \pageref{eq:gammapgammapSpurEven})
\begin{eqnarray*}
\Omega_{M} & = & -\frac{1}{8}\Omega_{M\bs{\alpha}}\hoch{\bs{\alpha}}\\
\Omega_{Ma_{1}a_{2}} & = & -\frac{1}{8}\gamma_{a_{1}a_{2}\,\bs{\beta}}\hoch{\bs{\alpha}}\Omega_{M\bs{\alpha}}\hoch{\bs{\beta}}\qquad\fussend\end{eqnarray*}
} \begin{eqnarray}
\mc{D}_{\bar{z}}\mu_{a} & \equiv & \bar{\partial}\mu_{a}-A_{\bar{z}a}\hoch{b}\mu_{b},\qquad A_{\bar{z}a}\hoch{b}\equiv\underbrace{\bar{\partial}x^{M}\Omega_{Ma}\hoch{b}}_{\Pi_{\bar{z}}^{C}\Omega_{Ca}\hoch{b}}+C_{a}\hoch{b\hat{\bs{\gamma}}}\hat{\dP}_{\bar{z}\hat{\bs{\gamma}}}-\hat{\ce}^{\hat{\bs{\alpha}}}S_{a\hat{\bs{\alpha}}}\hoch{b\hat{\bs{\beta}}}\hat{\be}_{\bar{z}\hat{\bs{\beta}}}\label{eq:AundMathcalDboson}\\
\hspace{-1cm}\mbox{with }\quad\Omega_{Ma}\hoch{b} & \equiv & \Omega_{M}^{(D)}\delta_{a}^{b}+\Omega_{Ma}^{(L)}\hoch{b}\qquad\Leftarrow\Omega_{M\bs{\alpha}}\hoch{\bs{\beta}}=\frac{1}{2}\Omega_{M}^{(D)}\delta_{\bs{\alpha}}\hoch{\bs{\beta}}+\frac{1}{4}\Omega_{Mab}^{(L)}\gamma^{ab}\tief{\bs{\alpha}}\hoch{\bs{\beta}}+\underbrace{\Omega_{Ma_{1}\ldots a_{4}}}_{=0\:{\rm (later)}}\gamma^{a_{1}\ldots a_{4}}\tief{\bs{\alpha}}\hoch{\bs{\beta}}\label{eq:OmegaMitBosonIndizes}\\
C_{a}\hoch{b\hat{\bs{\gamma}}} & \equiv & C^{\hat{\bs{\gamma}}}\delta_{a}^{b}+C^{\hat{\bs{\gamma}}}\tief{ac}\eta^{cb}\qquad\Leftarrow C_{\bs{\alpha}}\hoch{\bs{\beta}\hat{\bs{\gamma}}}=\frac{1}{2}C^{\hat{\bs{\gamma}}}\delta_{\bs{\alpha}}\hoch{\bs{\beta}}+\frac{1}{4}C^{\hat{\bs{\gamma}}}\tief{ab}\gamma^{ab}\tief{\bs{\alpha}}\hoch{\bs{\beta}}+\underbrace{C^{\hat{\bs{\gamma}}}\tief{a_{1}\ldots a_{4}}}_{=0\:{\rm (later)}}\gamma^{a_{1}\ldots a_{4}}\tief{\bs{\alpha}}\hoch{\bs{\beta}}\\
S_{a\hat{\bs{\alpha}}}\hoch{b\hat{\bs{\beta}}} & \equiv & S_{\hat{\bs{\alpha}}}\hoch{\hat{\bs{\beta}}}\delta_{a}^{b}+S_{\hat{\bs{\alpha}}}\hoch{\hat{\bs{\beta}}}\tief{ac}\eta^{cb}\qquad\Leftarrow S_{\bs{\alpha}\hat{\bs{\alpha}}}\hoch{\bs{\beta}\hat{\bs{\beta}}}=\frac{1}{2}S_{\hat{\bs{\alpha}}}\hoch{\hat{\bs{\beta}}}\delta_{\bs{\alpha}}\hoch{\bs{\beta}}+\frac{1}{4}S_{\hat{\bs{\alpha}}}\hoch{\hat{\bs{\beta}}}\tief{ac}\gamma^{ab}\tief{\bs{\alpha}}\hoch{\bs{\beta}}+\underbrace{S_{\hat{\bs{\alpha}}}\hoch{\hat{\bs{\beta}}}\tief{a_{1}\ldots a_{4}}}_{=0\:{\rm (later)}}\gamma^{a_{1}\ldots a_{4}}\tief{\bs{\alpha}}\hoch{\bs{\beta}}\label{eq:S:expand}\end{eqnarray}
The coefficient $\Omega_{Ma_{1}\ldots a_{4}}$ and the other $\gamma^{[4]}$-coefficients
do not enter the definitions of $\Omega_{Ma}\hoch{b}$, $C_{a}\hoch{b\hat{\bs{\gamma}}}$
and $S_{a\hat{\bs{\alpha}}}\hoch{b\hat{\bs{\beta}}}$. At a later
point we will find that the $\gamma^{[4]}$-coefficients actually
have to vanish, which then implies $\mc{D}_{\bar{z}}\gamma_{\bs{\alpha\beta}}^{a}=0$.
This is the actual motivation for this choice of bosonic connection.
It is tempting to argue that \begin{equation}
A_{\bar{z}\, a_{1}\ldots a_{4}}\equiv\Pi_{\bar{z}}^{C}\Omega_{Ca_{1}\ldots a_{4}}+\hat{\dP}_{\bar{z}\hat{\bs{\gamma}}}C^{\hat{\bs{\gamma}}}\tief{a_{1}\ldots a_{4}}+\hat{\ce}^{\hat{\bs{\alpha}}}S_{\hat{\bs{\alpha}}}\hoch{\hat{\bs{\beta}}}\tief{a_{1}\ldots a_{4}}\hat{\be}_{\bar{z}\hat{\bs{\beta}}}\label{eq:Afour}\end{equation}
 has to vanish already at this point, in order for all the terms in
\eqref{eq:on-shell-cond-antigauge} to vanish on-shell. But the condition
will be a bit weaker, as there is yet another equation of motion applicable%
\footnote{\label{foot:weakerThanBerk}\index{footnote!\thefoot. different antighost gauge symmetry}In
the original derivation of the supergravity constraints from Berkovits'
pure spinor string in \cite{Berkovits:2001ue} it is argued that the
action has to be invariant under the gauge transformation $\delta\omega_{\bs{\alpha}}=\mu_{a}(\gamma^{a}\ce)_{\bs{\alpha}}$
(the gauge symmetry generated by the pure spinor constraint in flat
space). In our notation this implies exactly $A_{\bar{z}\, a_{1}\ldots a_{4}}=0$.
However, there is no reason a priory, why the form of the gauge symmetry
should not be modified in curved space, as long as this modification
vanishes for the flat case. We will indeed discover such a modification
in the following, and with this modification the restriction on the
background fields is weaker. Nevertheless we will obtain the same
result in the end, as $A_{\bar{z}\, a_{1}\ldots a_{4}}=0$ will be
a consequence of BRST invariance later. $\qquad\fussend$%
}. We can replace $\Pi_{\bar{z}}^{\bs{\gamma}}$ (appearing in (\eqref{eq:Afour})
and (\ref{eq:AundMathcalDboson}), and defined in \eqref{eq:Pidefbarz})
with the equation of motion \eqref{eq:eomI}:$\Pi_{\bar{z}}^{\bs{\gamma}}=\funktional{S}{\dP_{z\bs{\gamma}}}-\RR^{\bs{\gamma}\hat{\bs{\gamma}}}\hat{\dP}_{\bar{z}\hat{\bs{\gamma}}}-\hat{\ce}^{\hat{\bs{\alpha}}}\hat{C}_{\hat{\bs{\alpha}}}\hoch{\hat{\bs{\beta}}\bs{\gamma}}\hat{\be}_{\bar{z}\hat{\bs{\beta}}}$
. Putting now all the last equations together, we arrive at\begin{eqnarray}
\bar{\partial}\left(\frac{\mu_{a}}{2}\ce\gamma^{a}\ce\right) & = & \mc{D}_{\bar{z}}\mu_{a}\cdot\funktional{S}{L_{z\bar{z}a}}-\mu_{a}(\ce\gamma^{a})_{\bs{\beta}}\funktional{S}{\be_{z\bs{\beta}}}-\mu_{a}\Omega_{\bs{\gamma}a_{1}\ldots a_{4}}(\ce\gamma^{a_{1}\ldots a_{4}a}\ce)\funktional{S}{\dP_{z\bs{\gamma}}}+\nonumber \\
 &  & -\mu_{a}\Big[\Pi_{\bar{z}}^{\{c,\hat{\bs{\gamma}}\}}\Omega_{\{c,\hat{\bs{\gamma}}\}a_{1}\ldots a_{4}}+\hat{\dP}_{\bar{z}\hat{\bs{\gamma}}}\left(C^{\hat{\bs{\gamma}}}\tief{a_{1}\ldots a_{4}}-\RR^{\bs{\gamma}\hat{\bs{\gamma}}}\Omega_{\bs{\gamma}a_{1}\ldots a_{4}}\right)+\nonumber \\
 &  & +\hat{\ce}^{\hat{\bs{\alpha}}}\left(S_{\hat{\bs{\alpha}}}\hoch{\hat{\bs{\beta}}}\tief{a_{1}\ldots a_{4}}-\hat{C}_{\hat{\bs{\alpha}}}\hoch{\hat{\bs{\beta}}\bs{\gamma}}\Omega_{\bs{\gamma}a_{1}\ldots a_{4}}\right)\hat{\be}_{\bar{z}\hat{\bs{\beta}}}\Big](\ce\gamma^{a_{1}\ldots a_{4}a}\ce)\qquad\label{eq:antigauge-divergence}\end{eqnarray}
The dummy indices in curly brackets $\{c,\hat{\bs{\gamma}}\}$ in
the second line simply should indicate a sum over $c$ and $\hat{\bs{\gamma}}$
only, and not over $\bs{\gamma}$. The first line on the righthand
side vanishes on-shell. The next two lines also have to vanish for
every $\mu_{a}$, because the left-hand side vanishes on-shell. At
this point we cannot make use of further equations of motion. In particular
the equation of motion for $x^{K}$, which we have not yet derived,
would be of conformal weight (1,1) (containing terms like $\partial\bar{\partial}x^{M}$)
and would therefore not be applicable. For consistency of the equations
of motion, we thus get the following restrictions on the background
fields\vRam{.4}{\begin{eqnarray}
\Omega_{c\, a_{1}\ldots a_{4}} & = & \Omega_{\hat{\bs{\gamma}}\, a_{1}\ldots a_{4}}=0\label{eq:OmegaCond}\\
C^{\hat{\bs{\gamma}}}\tief{a_{1}\ldots a_{4}} & = & \RR^{\bs{\gamma}\hat{\bs{\gamma}}}\Omega_{\bs{\gamma}a_{1}\ldots a_{4}}\label{eq:Ccond}\\
S_{\hat{\bs{\alpha}}}\hoch{\hat{\bs{\beta}}}\tief{a_{1}\ldots a_{4}} & = & \hat{C}_{\hat{\bs{\alpha}}}\hoch{\hat{\bs{\beta}}\bs{\gamma}}\Omega_{\bs{\gamma}a_{1}\ldots a_{4}}\label{eq:Scond}\end{eqnarray}
}\\ This condition is weaker as the one given in \cite{Berkovits:2001ue}
(see footnote \eqref{foot:weakerThanBerk}). It coincides exactly
iff we impose in addition $\Omega_{\bs{\gamma}\, a_{1}\ldots a_{4}}=0$
(see the remark at the end of this section). This additional restriction
will, however, only be a result of BRST invariance.

According to Noether, every symmetry transformation corresponds to
a divergence free current and vice verse. For a given current $j^{\zeta}$,
we can calculate the corresponding transformations by reading of the
coefficients of the variational derivatives of $S$ in the off-shell
divergence of the current (see (\ref{eq:noet:currentdivergence})):\begin{equation}
\partial_{\zeta}j_{(\rho)}^{\zeta}=-\delta_{(\rho)}\allfields{I}\funktional{S}{\allfields{I}}\end{equation}
If we take $j_{z}\equiv\frac{\mu_{za}}{2}(\ce\gamma^{a}\ce),\quad j_{\bar{z}}\equiv0$,
the condition (\ref{eq:on-shell-cond-antigauge}) tells that the current
is on-shell divergence free. We have chosen a parameter of weight
$(1,0)$, in order to get a current of correct weight. From (\ref{eq:antigauge-divergence})
we can now read off the corresponding symmetry transformations:\begin{eqnarray}
\delta_{(\mu)}\be_{z\bs{\alpha}} & = & \mu_{za}(\ce\gamma^{a})_{\bs{\alpha}}\label{eq:gaugetrafoAntighost}\\
\delta_{(\mu)}L_{z\bar{z}a} & = & -\mc{D}_{\bar{z}}\mu_{za}\label{eq:gaugetrafoLagrange}\\
\delta_{(\mu)}\dP_{z\bs{\gamma}} & = & \mu_{za}\Omega_{\bs{\gamma}a_{1}\ldots a_{4}}(\ce\gamma^{a_{1}\ldots a_{4}a}\ce)\label{eq:gaugetrafod}\end{eqnarray}
The current is divergence free for arbitrary (local) $\mu_{za}$ and
we therefore have a gauge symmetry. This is the \textbf{antighost
gauge symmetry} generated by the pure spinor constraint. For a flat
background we have $\Omega_{\bs{\gamma}a_{1}\ldots a_{4}}=0$ and
the transformation reduces to the usual form. As stated several times
already, we will obtain $\Omega_{\bs{\gamma}a_{1}\ldots a_{4}}=0$
also in the curved background, but only later as a result of BRST
invariance.\rem{here is a note hidden with a long old calculation}

With the same reasoning we get a gauge transformation corresponding
to the pure spinor constraint on the hatted ghost fields. This leads
to equivalent restrictions on the hatted connection $\hat{\Omega}_{M\hat{\bs{\alpha}}}\hoch{\hat{\bs{\beta}}}$
and also on $\hat{C}_{\hat{\bs{\alpha}}}\hoch{\hat{\bs{\beta}}\bs{\beta}}$
(seen as matrix in $\hat{\bs{\alpha}}$ and $\hat{\bs{\beta}}$).
The background field $S_{\bs{\alpha}\hat{\bs{\alpha}}}\hoch{\bs{\beta}\hat{\bs{\beta}}}$
is special, because the hatted version of the expansion (\ref{eq:S:expand})
together with the condition (\ref{eq:Scond}) is again a condition
on the expansion of $S$, now in its hatted indices. Once it is seen
as matrix in $\bs{\alpha}$ and $\bs{\beta}$ and once as matrix in
$\hat{\bs{\alpha}}$ and $\hat{\bs{\beta}}$. This is better treatable
in the special case considered in the following remark.

\paragraph{Remark on $\Omega_{\bs{\gamma}a_{1}\ldots a_{4}}=\hat{\Omega}_{\hat{\bs{\gamma}}a_{1}\ldots a_{4}}=0$:}

\label{remark:structureGroupValuedConnection}Although we will discover
these two additional constraints only later in (\ref{eq:endlichConnectionConstraint})
on page \pageref{eq:endlichConnectionConstraint}, it is nice to have
everything at one place. So let us continue the discussion of $S_{\bs{\alpha}\hat{\bs{\alpha}}}\hoch{\bs{\beta}\hat{\bs{\beta}}}$
in this case. As indicated above, we can expand it in two steps: \begin{eqnarray}
S_{\bs{\alpha}\hat{\bs{\alpha}}}\hoch{\bs{\beta}\hat{\bs{\beta}}} & = & \frac{1}{2}S_{\hat{\bs{\alpha}}}\hoch{\hat{\bs{\beta}}}\delta_{\bs{\alpha}}\hoch{\bs{\beta}}+\frac{1}{4}S_{\hat{\bs{\alpha}}}\hoch{\hat{\bs{\beta}}}\tief{a_{1}a_{2}}\gamma^{a_{1}a_{2}}\tief{\bs{\alpha}}\hoch{\bs{\beta}}=\nonumber \\
 & = & \frac{1}{2}\left(\frac{1}{2}S\delta_{\hat{\bs{\alpha}}}\hoch{\hat{\bs{\beta}}}+\frac{1}{4}S_{a_{1}a_{2}}\gamma^{a_{1}a_{2}}\tief{\hat{\bs{\alpha}}}\hoch{\hat{\bs{\beta}}}\right)\delta_{\bs{\alpha}}\hoch{\bs{\beta}}+\nonumber \\
 &  & +\frac{1}{4}\left(\frac{1}{2}\hat{S}_{a_{1}a_{2}}\delta_{\hat{\bs{\alpha}}}\hoch{\hat{\bs{\beta}}}+\frac{1}{4}S_{a_{1}a_{2}b_{1}b_{2}}\gamma^{b_{1}b_{2}}\tief{\hat{\bs{\alpha}}}\hoch{\hat{\bs{\beta}}}\right)\gamma^{a_{1}a_{2}}\tief{\bs{\alpha}}\hoch{\bs{\beta}}\end{eqnarray}
Let us summarize the result for all the involved fields:\vRam{0.85}{\begin{eqnarray}
\Omega_{M\bs{\alpha}}\hoch{\bs{\beta}} & = & \frac{1}{2}\Omega_{M}^{(D)}\delta_{\bs{\alpha}}\hoch{\bs{\beta}}+\frac{1}{4}\Omega_{Ma_{1}a_{2}}^{(L)}\gamma^{a_{1}a_{2}}\tief{\bs{\alpha}}\hoch{\bs{\beta}},\qquad\hat{\Omega}_{M\hat{\bs{\alpha}}}\hoch{\hat{\bs{\beta}}}=\frac{1}{2}\hat{\Omega}_{M}^{(D)}\delta_{\hat{\bs{\alpha}}}\hoch{\hat{\bs{\beta}}}+\frac{1}{4}\hat{\Omega}_{Ma_{1}a_{2}}^{(L)}\gamma^{a_{1}a_{2}}\tief{\hat{\bs{\alpha}}}\hoch{\hat{\bs{\beta}}}\label{eq:reducedConnectionForm}\\
C_{\bs{\alpha}}\hoch{\bs{\beta}\hat{\bs{\gamma}}} & = & \frac{1}{2}C^{\hat{\bs{\gamma}}}\delta_{\bs{\alpha}}\hoch{\bs{\beta}}+\frac{1}{4}C_{a_{1}a_{2}}^{\hat{\bs{\gamma}}}\gamma^{a_{1}a_{2}}\tief{\bs{\alpha}}\hoch{\bs{\beta}},\qquad\hat{C}_{\hat{\bs{\alpha}}}\hoch{\hat{\bs{\beta}}\bs{\gamma}}=\frac{1}{2}\hat{C}^{\bs{\gamma}}\delta_{\hat{\bs{\alpha}}}\hoch{\hat{\bs{\beta}}}+\frac{1}{4}\hat{C}_{a_{1}a_{2}}^{\bs{\gamma}}\gamma^{a_{1}a_{2}}\tief{\hat{\bs{\alpha}}}\hoch{\hat{\bs{\beta}}}\label{eq:reducedCform}\\
S_{\bs{\alpha}\hat{\bs{\alpha}}}\hoch{\bs{\beta}\hat{\bs{\beta}}} & = & \frac{1}{4}S\delta_{\bs{\alpha}}\hoch{\bs{\beta}}\delta_{\hat{\bs{\alpha}}}\hoch{\hat{\bs{\beta}}}+\frac{1}{8}S_{a_{1}a_{2}}\delta_{\bs{\alpha}}\hoch{\bs{\beta}}\gamma^{a_{1}a_{2}}\tief{\hat{\bs{\alpha}}}\hoch{\hat{\bs{\beta}}}+\nonumber \\
 &  & +\frac{1}{8}\hat{S}_{a_{1}a_{2}}\gamma^{a_{1}a_{2}}\tief{\bs{\alpha}}\hoch{\bs{\beta}}\delta_{\hat{\bs{\alpha}}}\hoch{\hat{\bs{\beta}}}+\frac{1}{16}S_{a_{1}a_{2}b_{1}b_{2}}\gamma^{a_{1}a_{2}}\tief{\bs{\alpha}}\hoch{\bs{\beta}}\gamma^{b_{1}b_{2}}\tief{\hat{\bs{\alpha}}}\hoch{\hat{\bs{\beta}}}\label{eq:reducedSform}\end{eqnarray}
}\\
Seen as a matrix in $\bs{\alpha}$ and $\bs{\beta}$ (or $\hat{\bs{\alpha}}$
and $\hat{\bs{\beta}}$ respectively), they are sums of generators
of Lorentz and scale transformations. Remembering the definition of
$\mc{D}_{\bar{z}}$ given in (\ref{eq:AundMathcalD}) and its extension
to bosonic indices in (\ref{eq:AundMathcalDboson}), it leaves invariant
the $\gamma$-matrices:%
\footnote{\index{footnote!\thefoot. covariant derivative on gamma}\begin{eqnarray*}
\mc{D}_{\bar{z}}\gamma_{\bs{\alpha\beta}}^{a} & = & \underbrace{\bar{\partial}\gamma_{\bs{\alpha\beta}}^{a}}_{=0}+\left(\bar{\partial}x^{M}\Omega_{Mb}\hoch{a}+C_{b}\hoch{a\hat{\bs{\gamma}}}\hat{\dP}_{\bar{z}\hat{\bs{\gamma}}}-\hat{\ce}^{\hat{\bs{\alpha}}}S_{b\hat{\bs{\alpha}}}\hoch{a\hat{\bs{\beta}}}\hat{\be}_{\bar{z}\hat{\bs{\beta}}}\right)\gamma_{\bs{\alpha\beta}}^{b}-2\left(\bar{\partial}x^{M}\Omega_{M[\bs{\alpha}|}\hoch{\bs{\delta}}+C_{[\bs{\alpha}|}\hoch{\bs{\delta}\hat{\bs{\gamma}}}\hat{\dP}_{\bar{z}\hat{\bs{\gamma}}}-\hat{\ce}^{\hat{\bs{\alpha}}}S_{[\bs{\alpha}|\hat{\bs{\alpha}}}\hoch{\bs{\delta}\hat{\bs{\beta}}}\hat{\be}_{\bar{z}\hat{\bs{\beta}}}\right)\gamma_{\bs{\delta}|\bs{\beta}]}^{a}\qquad\fussend\end{eqnarray*}
} \begin{eqnarray}
\mc{D}_{\bar{z}}\gamma_{\bs{\alpha\beta}}^{a} & = & 0,\qquad\hat{\mc{D}}_{z}\gamma_{\hat{\bs{\alpha}}\hat{\bs{\beta}}}^{a}=0\end{eqnarray}
\rem{In particular\begin{eqnarray}
\nabla_{M}\gamma_{\bs{\alpha\beta}}^{a} & = & 0,\qquad\hat{\nabla}_{M}\gamma_{\hat{\bs{\alpha}}\hat{\bs{\beta}}}^{a}=0\end{eqnarray}
}

The expressions $\ce^{\bs{\alpha}}\be_{z\bs{\alpha}}$ and $\ce^{\bs{\alpha}}\gamma^{a_{1}a_{2}}\tief{\bs{\alpha}}\hoch{\bs{\beta}}\be_{z\bs{\beta}}$
are the only gauge invariant quantities (on the constraint surface
$\ce\gamma^{a}\ce=0$) which are linear in ghost and antighost. The
reasoning is as follows: the most general combination is $\ce^{\bs{\alpha}}X_{\bs{\alpha}}\hoch{\bs{\beta}}\be_{z\bs{\beta}}$
with some general matrix $X_{\bs{\alpha}}\hoch{\bs{\beta}}$ which
can be expanded in $\gamma^{[0]},\,\gamma^{[2]}$ and $\gamma^{[4]}$.
Upon acting with a gauge transformation on this term, we get the products
$\gamma^{[0]}\gamma^{[1]}=\gamma^{[1]}$, $\gamma^{[2]}\gamma^{[1]}\propto\gamma^{[1]}+\gamma^{[3]}$,
and $\gamma^{[4]}\gamma^{[1]}\propto\gamma^{[3]}+\gamma^{[5]}$. As
$\gamma^{[5]}$ does not vanish when contracted with two ghosts, the
$\gamma^{[4]}$-part of the expansion has to vanish and we have shown
the above statement. The gauge invariant expression $\ce^{\bs{\alpha}}\be_{z\bs{\alpha}}$
is nothing but the ghost\index{ghost current!gauge invariant} current
(\ref{eq:ghostCurrent}), while $\ce^{\bs{\alpha}}\gamma^{a_{1}a_{2}}\tief{\bs{\alpha}}\hoch{\bs{\beta}}\be_{z\bs{\beta}}$
is part of the Lorentz current\index{Lorentz current}, which is discussed
in Berkovits' papers.

\section{Covariant variational principle \& EOM's}

\label{sec:Covariant-eoms}Remember the form of the action (\ref{eq:BiBaction}):\begin{eqnarray}
S & = & \int d^{2}z\quad\frac{1}{2}\Pi_{z}^{A}\underbrace{(G_{AB}+B_{AB})}_{\equiv\GB_{AB}}\Pi_{\bar{z}}^{B}+\Pi_{\bar{z}}^{\bs{\gamma}}\dP_{z\bs{\gamma}}+\Pi_{z}^{\hat{\bs{\gamma}}}\hat{\dP}_{\bar{z}\hat{\bs{\gamma}}}+\dP_{z\bs{\gamma}}\RR^{\bs{\gamma}\hat{\bs{\gamma}}}\hat{\dP}_{\bar{z}\hat{\bs{\gamma}}}+\nonumber \\
 &  & +\ce^{\bs{\alpha}}C_{\bs{\alpha}}\hoch{\bs{\beta}\hat{\bs{\gamma}}}\be_{z\bs{\beta}}\hat{\dP}_{\bar{z}\hat{\bs{\gamma}}}+\hat{\ce}^{\hat{\bs{\alpha}}}\hat{C}_{\hat{\bs{\alpha}}}\hoch{\hat{\bs{\beta}}\bs{\gamma}}\hat{\be}_{\bar{z}\hat{\bs{\beta}}}\dP_{z\bs{\gamma}}+\ce^{\bs{\alpha}}\hat{\ce}^{\hat{\bs{\alpha}}}S_{\bs{\alpha}\hat{\bs{\alpha}}}\hoch{\bs{\beta}\hat{\bs{\beta}}}\be_{z\bs{\beta}}\hat{\be}_{\bar{z}\hat{\bs{\beta}}}+\nonumber \\
 &  & +\nabla_{\bar{z}}\ce^{\bs{\beta}}\be_{z\bs{\beta}}+\hat{\nabla}\hat{\ce}^{\hat{\bs{\beta}}}\hat{\be}_{\bar{z}\hat{\bs{\beta}}}+\frac{1}{2}L_{z\bar{z}a}(\ce\gamma^{a}\ce)+\frac{1}{2}\hat{L}_{z\bar{z}a}(\hat{\ce}\gamma^{a}\hat{\ce})\frem{-4\alpha'\partial\bar{\partial}\weyl\cdot\dil}\label{eq:BiBagain}\end{eqnarray}
 In order to check if the BRST currents (\ref{eq:BiBbrstSimple})
and (\ref{eq:BiBbrstHatSimple}) are on-shell conserved (holomorphic
and antiholomorphic respectively), it is first of all necessary to
calculate the remaining classical equation of motion, the variation
with respect to $x^{K}$. Remember, the other equations of motion
were given already in (\ref{eq:eomI})-(\ref{eq:eomVIII}) on page
\pageref{eq:eomI}.

\paragraph{Covariant variation}

Deriving the variational derivative with respect to $x^{K}$ is quite
involved if we do not organize it properly. In the end we want to
have equations which transform covariantly under superdiffeomorphisms
and local structure group transformations. We therefore want to introduce
a method where we stay covariant right from the beginning, e.g. a
target space covariant variation of the action. In order to motivate
the following definitions, let us consider only the variation of one
simple term of the Lagrangian, e.g. the RR-term:\begin{eqnarray}
\lqn{\delta\left(\dP_{z\bs{\gamma}}\RR^{\bs{\gamma}\hat{\bs{\gamma}}}(\xfull)\hat{\dP}_{\bar{z}\hat{\bs{\gamma}}}\right)=}\nonumber \\
 & = & \delta\dP_{z\bs{\gamma}}\RR^{\bs{\gamma}\hat{\bs{\gamma}}}\hat{\dP}_{\bar{z}\hat{\bs{\gamma}}}+\dP_{z\bs{\gamma}}\delta x^{M}\partial_{M}\RR^{\bs{\gamma}\hat{\bs{\gamma}}}\hat{\dP}_{\bar{z}\hat{\bs{\gamma}}}+\dP_{z\bs{\gamma}}\RR^{\bs{\gamma}\hat{\bs{\gamma}}}\delta\hat{\dP}_{\bar{z}\hat{\bs{\gamma}}}=\\
 & = & \underbrace{\left(\delta\dP_{z\bs{\gamma}}-\delta x^{M}\Omega_{M\bs{\gamma}}\hoch{\bs{\beta}}\dP_{z\bs{\beta}}\right)}_{\equiv\delta_{cov}\dP_{z\bs{\gamma}}}\RR^{\bs{\gamma}\hat{\bs{\gamma}}}\hat{\dP}_{\bar{z}\hat{\bs{\gamma}}}+\dP_{z\bs{\gamma}}\underbrace{\delta x^{M}\gemnabla_{M}\RR^{\bs{\gamma}\hat{\bs{\gamma}}}}_{\equiv\delta_{\gem{cov}}\RR^{\bs{\gamma}\hat{\bs{\gamma}}}}\hat{\dP}_{\bar{z}\hat{\bs{\gamma}}}+\dP_{z\bs{\gamma}}\RR^{\bs{\gamma}\hat{\bs{\gamma}}}\underbrace{\left(\delta\hat{\dP}_{\bar{z}\hat{\bs{\gamma}}}-\delta x^{M}\hat{\Omega}_{M\hat{\bs{\gamma}}}\hoch{\hat{\bs{\alpha}}}\hat{\dP}_{\bar{z}\hat{\bs{\alpha}}}\right)}_{\equiv\delta_{\hat{cov}}\dP_{\bar{z}\hat{\bs{\gamma}}}}\qquad\label{eq:motivatingCovariantVariation}\end{eqnarray}
In order to arrive at the target space covariant expression $\gem{\nabla}_{M}\RR^{\bs{\gamma}\hat{\bs{\gamma}}}$,
it is thus convenient to group part of the $x^{K}$-variation to the
variation of $\dP_{z\bs{\gamma}}$ or $\hat{\dP}_{\bar{z}\hat{\bs{\gamma}}}$
as done above. Of course we could have chosen any connection for the
above rewriting, as long as we use for each contracted index pair
the same connection. For the variation of the complete action, however,
it is most convenient to choose the mixed connection, defined in (\ref{eq:mixedConnection}),
\begin{equation}
\gemOm_{MA}\hoch{B}\equiv\left(\begin{array}{ccc}
\check{\Omega}_{Ma}\hoch{b} & 0 & 0\\
0 & \Omega_{M\bs{\alpha}}\hoch{\bs{\beta}} & 0\\
0 & 0 & \hat{\Omega}_{M\hat{\bs{\alpha}}}\hoch{\hat{\bs{\beta}}}\end{array}\right)\end{equation}
\rem{As it contains already two of the background fields, the number
of terms will be reduced.%
\footnote{We have already written the ghost kinetic term as $\nabla_{\bar{z}}\ce^{\bs{\beta}}\be_{z\bs{\beta}}$.
In the covariant variation scheme that we are going to describe, we
can apply rules how to commute covariant derivatives with the covariant
variation. If we use a different connection than $\Omega_{M\bs{\alpha}}\hoch{\bs{\beta}}$,
we get an extra ghost term containing the difference tensor, which
has to be varied separately. %
}}Like for the structure group transformation, the connection $\Omega_{M\bs{\alpha}}\hoch{\bs{\beta}}$
acts on the unhatted fermionic indices and (!) on $L_{z\bar{z}a}$,
while $\hat{\Omega}_{M\hat{\bs{\alpha}}}\hoch{\hat{\bs{\beta}}}$
acts on the hatted indices and (!) on $L_{\bar{z}za}$. The third
independent block $\check{\Omega}_{Ma}\hoch{b}$ acts only on the
bosonic indices that appear via the bosonic vielbein and not on elementary
fields.\rem{besser erklaeren!Bezieht sich auch auf spezielle Form der Konnexion (Lorentz+Scale)}

Similar considerations as for the RR-term hold for the other terms
of the action. This motivates the definition of the \textbf{covariant
variation\index{variation!covariant $\sim$}\index{covariant variation}\index{$\delta_{cov}$}}
of the elementary fields in the above spirit: \begin{eqnarray}
\delta_{cov}\ce^{\bs{\alpha}} & \equiv & \delta\ce^{\bs{\alpha}}+\delta x^{M}\Omega_{M\bs{\beta}}\hoch{\bs{\alpha}}\ce^{\bs{\beta}},\qquad\delta_{cov}\be_{z\bs{\alpha}}\equiv\delta\be_{z\bs{\alpha}}-\delta x^{M}\Omega_{M\bs{\alpha}}\hoch{\bs{\beta}}\be_{z\bs{\beta}}\label{eq:covVarI}\\
\delta_{cov}\dP_{z\bs{\alpha}} & \equiv & \delta\dP_{z\bs{\alpha}}-\delta x^{M}\Omega_{M\bs{\alpha}}\hoch{\bs{\beta}}\dP_{z\bs{\beta}},\qquad\delta_{cov}L_{z\bar{z}a}\equiv\delta L_{z\bar{z}a}-\delta x^{M}\Omega_{Ma}\hoch{b}L_{z\bar{z}b}\label{eq:covVarII}\\
\delta_{\hat{cov}}\hat{\ce}^{\hat{\bs{\alpha}}} & \equiv & \delta\hat{\ce}^{\hat{\bs{\alpha}}}+\delta x^{M}\hat{\Omega}_{M\hat{\bs{\beta}}}\hoch{\hat{\bs{\alpha}}}\hat{\ce}^{\hat{\bs{\beta}}},\qquad\delta_{\hat{cov}}\hat{\be}_{\bar{z}\hat{\bs{\alpha}}}\equiv\delta\be_{\bar{z}\hat{\bs{\alpha}}}-\delta x^{M}\hat{\Omega}_{M\hat{\bs{\alpha}}}\hoch{\hat{\bs{\beta}}}\hat{\be}_{\bar{z}\hat{\bs{\beta}}}\label{eq:covVarIII}\\
\delta_{\hat{cov}}\hat{\dP}_{\bar{z}\hat{\bs{\alpha}}} & \equiv & \delta\hat{\dP}_{\bar{z}\hat{\bs{\alpha}}}-\delta x^{M}\hat{\Omega}_{M\hat{\bs{\beta}}}\hoch{\hat{\bs{\alpha}}}\hat{\dP}_{\bar{z}\hat{\bs{\alpha}}},\qquad\delta_{\hat{cov}}\hat{L}_{\bar{z}za}\equiv\delta\hat{L}_{\bar{z}za}-\delta x^{M}\hat{\Omega}_{Ma}\hoch{b}\hat{L}_{\bar{z}zb}\label{eq:covVarIV}\\
\delta_{\gem{cov}}x^{K} & \equiv & \delta x^{K}\label{eq:covVarV}\end{eqnarray}
Unfortunately this idea is not completely new. Similar versions of
covariant variations have been presented in \cite{Minkevich:1982a,Luckock:1989jr}
which in turn refer to \cite{Minkevich:1968a,Minkevich:1975a}.  As
already indicated in (\ref{eq:motivatingCovariantVariation}), we
understand the covariant variation acting on arbitrary background
tensor fields $T_{MA}^{NB}(\xfull)$ as \begin{eqnarray}
\delta_{\gem{cov}}T_{MA}^{NB}(\xfull) & \equiv & \delta x^{K}\gem{\nabla}_{K}T_{MA}^{NB}=\label{eq:covVarVI}\\
 & = & \delta T_{MA}^{NB}+\delta x^{K}\left(\gem{\Gamma}_{KL}\hoch{N}T_{MA}^{LB}+\gem{\Omega}_{KC}\hoch{B}T_{MA}^{NC}-\gem{\Gamma}_{KM}\hoch{L}T_{LA}^{NB}-\gem{\Omega}_{KA}\hoch{C}T_{MC}^{NB}\right)\qquad\label{eq:covVarVII}\end{eqnarray}
In the last line we discover that the covariant variation acts on
background fields in the same way as it acts on elementary fields
if the index structure is the same. Note that the covariant variation
cannot be understood as a variation (of e.g. $x^{K}$) in the ordinary
sense. The covariant variation is simply a derivation which only reduces
to an ordinary variation when acting on target space scalars, e.g.
on the Lagrangian. 

From the target space point of view, also objects like $\nabla_{\bar{z}}\ce^{\bs{\beta}}$
(target space covariant worldsheet derivatives of worldsheet variables)
transform tensorial under structure group transformations and diffeomorphisms.
The covariant variation is then simply defined according to their
target space transformation properties:\begin{eqnarray}
\delta_{cov}\nabla_{\bar{z}}\ce^{\bs{\beta}} & \equiv & \delta\nabla_{\bar{z}}\ce^{\bs{\beta}}+\delta x^{K}\Omega_{K\bs{\alpha}}\hoch{\bs{\beta}}\nabla_{\bar{z}}\ce^{\bs{\alpha}}\label{eq:covVarVIII}\\
\delta_{\hat{cov}}\hat{\nabla}_{z}\hat{\ce}^{\hat{\bs{\beta}}} & \equiv & \delta\hat{\nabla}_{z}\hat{\ce}^{\hat{\bs{\beta}}}+\delta x^{K}\hat{\Omega}_{K\hat{\bs{\alpha}}}\hoch{\hat{\bs{\beta}}}\hat{\nabla}_{z}\hat{\ce}^{\hat{\bs{\alpha}}}\label{eq:covVarIX}\end{eqnarray}
This is also the reason why the Lagrange multiplier is varied with
help of the connection $\Omega_{Ma}\hoch{b}$ (defined in (\ref{eq:OmegaMitBosonIndizes})
on page \pageref{eq:OmegaMitBosonIndizes}) which is induced by $\Omega_{M\bs{\alpha}}\hoch{\bs{\beta}}$,
and not with the independent $\check{\Omega}_{Ma}\hoch{b}$ that we
have introduced to act on the bosonic vielbein indices: In the reparametrization
corresponding to the structure group transformations, the transformation
of the Lagrange multiplier is directly coupled to the transformation
of the ghost. 

Next we define the \textbf{covariant variational derivative}s $\funktional{_{cov}S}{\allfields{I}}$
via\begin{eqnarray}
\delta S & \equiv & \int_{\Sigma}d^{2}z\quad\delta_{cov}\allfields{I}(z,\bar{z})\funktional{_{cov}S}{\allfields{I}(z,\bar{z})}\label{eq:covVarDerDef}\end{eqnarray}
We will soon give a statement about the relation to the ordinary variational
derivative. But let us first collect some tools to calculate it. In
order to arrive at the righthand side of (\ref{eq:covVarDerDef}),
we need to extract the covariant variations of the elementary fields.
In expressions like $\delta_{cov}\nabla_{\bar{z}}\ce^{\bs{\beta}}$
in (\ref{eq:covVarVIII}) this would require to commute e.g. the covariant
variation $\delta_{cov}$ with the covariant derivative $\nabla_{\bar{z}}$
and then do some partial integration. It was probably already noticed
by the reader that the covariant variation resembles very much the
target space covariant worldsheet derivative $\nabla_{z/\bar{z}}$
anyway. In fact the latter can be seen as a special case of it, namely
when we have $\delta\allfields{I}=\partial_{z/\bar{z}}\allfields{I}$.
Let us therefore consider the commutators of two arbitrary covariant
variations which will contain the desired commutator $[\delta_{cov},\nabla_{\bar{z}}]$
in the mentioned special case: \begin{eqnarray}
\left[\delta_{\gem{cov}}^{(1)},\delta_{\gem{cov}}^{(2)}\right]x^{K} & = & \left[\delta^{(1)},\delta^{(2)}\right]x^{K}+2\delta^{(1)}x^{M}\gem{T}_{MN}\hoch{K}\delta^{(2)}x^{N}\label{eq:commutatorOfCovVarI}\\
\left[\delta_{\gem{cov}}^{(1)},\delta_{\gem{cov}}^{(2)}\right]\varphi^{AM}\tief{B} & = & \left[\delta^{(1)},\delta^{(2)}\right]_{\gem{cov}}\varphi^{AM}\tief{B}+\nonumber \\
 &  & +2\delta^{(1)}x^{K}\delta^{(2)}x^{L}\left(\gem{R}_{KLC}\hoch{A}\varphi^{CM}\tief{B}+\gem{R}_{KLN}\hoch{M}\varphi^{AN}\tief{B}-\gem{R}_{KLB}\hoch{C}\varphi^{AM}\tief{C}\right)\qquad\label{eq:commutatorOfCovVarII}\end{eqnarray}
Here $\varphi^{AM}\tief{B}$ is just a representative example for
some elementary or composite field which transforms tensorial under
the target space transformations (super-diffeomorphisms and local
structure group transformations).

The covariant variation of the complete action coincides with the
ordinary one as all indices are contracted. However, the covariant
variational derivative defined in (\ref{eq:covVarDerDef}), differs
from the ordinary variational derivatives. The important thing is,
that nevertheless they define a set of equations of motion which is
equivalent the usual one -- and target space covariant. Let us see
the equivalence explicitly and reformulate the ordinary variation
into the covariant one:\begin{eqnarray}
\delta S & = & \int d^{2}z\quad\delta\dP_{z\bs{\gamma}}\funktional{S}{\dP_{z\bs{\gamma}}}+\delta\hat{\dP}_{\bar{z}\hat{\bs{\gamma}}}\funktional{S}{\hat{\dP}_{\bar{z}\hat{\bs{\gamma}}}}+\delta\ce^{\bs{\alpha}}\funktional{S}{\ce^{\bs{\alpha}}}+\delta\hat{\ce}^{\hat{\bs{\alpha}}}\funktional{S}{\hat{\ce}^{\hat{\bs{\alpha}}}}+\delta\be_{z\bs{\beta}}\funktional{S}{\be_{z\bs{\beta}}}+\delta\hat{\be}_{\bar{z}\hat{\bs{\beta}}}\funktional{S}{\hat{\be}_{\bar{z}\hat{\bs{\beta}}}}+\nonumber \\
 &  & +\delta L_{z\bar{z}a}\funktional{S}{L_{z\bar{z}a}}+\delta\hat{L}_{\bar{z}za}\funktional{S}{\hat{L}_{\bar{z}za}}+\delta x^{K}\funktional{S}{x^{K}}=\\
 & = & \int d^{2}z\quad\delta_{cov}\dP_{z\bs{\gamma}}\funktional{S}{\dP_{z\bs{\gamma}}}+\delta_{\hat{cov}}\hat{\dP}_{\bar{z}\hat{\bs{\gamma}}}\funktional{S}{\hat{\dP}_{\bar{z}\hat{\bs{\gamma}}}}+\delta_{cov}\ce^{\bs{\alpha}}\funktional{S}{\ce^{\bs{\alpha}}}+\delta_{\hat{cov}}\hat{\ce}^{\hat{\bs{\alpha}}}\funktional{S}{\hat{\ce}^{\hat{\bs{\alpha}}}}+\delta_{cov}\be_{z\bs{\beta}}\funktional{S}{\be_{z\bs{\beta}}}+\delta_{\hat{cov}}\be_{\bar{z}\hat{\bs{\beta}}}\funktional{S}{\hat{\be}_{\bar{z}\hat{\bs{\beta}}}}+\nonumber \\
 &  & +\delta_{cov}L_{z\bar{z}a}\funktional{S}{L_{z\bar{z}a}}+\delta_{\hat{cov}}\hat{L}_{z\bar{z}a}\funktional{S}{\hat{L}_{z\bar{z}a}}+\delta x^{K}\Big(\funktional{S}{x^{K}}+\Omega_{K\bs{\gamma}}\hoch{\bs{\delta}}\dP_{z\bs{\delta}}\funktional{S}{\dP_{z\bs{\gamma}}}+\hat{\Omega}_{K\hat{\bs{\gamma}}}\hoch{\hat{\bs{\delta}}}\hat{\dP}_{\bar{z}\hat{\bs{\delta}}}\funktional{S}{\hat{\dP}_{\bar{z}\hat{\bs{\gamma}}}}-\Omega_{K\bs{\beta}}\hoch{\bs{\alpha}}\ce^{\bs{\beta}}\funktional{S}{\ce^{\bs{\alpha}}}+\nonumber \\
 &  & -\hat{\Omega}_{K\hat{\bs{\beta}}}\hoch{\hat{\bs{\alpha}}}\hat{\ce}^{\hat{\bs{\beta}}}\funktional{S}{\hat{\ce}^{\hat{\bs{\alpha}}}}+\Omega_{K\bs{\beta}}\hoch{\bs{\alpha}}\be_{z\bs{\alpha}}\funktional{S}{\be_{z\bs{\beta}}}+\hat{\Omega}_{K\hat{\bs{\beta}}}\hoch{\hat{\bs{\alpha}}}\be_{\bar{z}\hat{\bs{\alpha}}}\funktional{S}{\hat{\be}_{\bar{z}\hat{\bs{\beta}}}}+\Omega_{Ka}\hoch{b}L_{z\bar{z}b}\funktional{S}{L_{z\bar{z}a}}+\hat{\Omega}_{Ka}\hoch{b}\hat{L}_{z\bar{z}b}\funktional{S}{\hat{L}_{z\bar{z}a}}\Big)\end{eqnarray}
We can now read off the \textbf{covariant variational derivative\index{covariant variational derivative}\index{variational derivative!covariant $\sim$}\index{$\frac{\delta_{\protect\underline{cov}}S}{\delta x^{K}}$}}
$\funktional{S_{\gem{cov}}}{x^{K}}$ w.r.t. $x^{K}$ as the coefficient
of $\delta x^{K}$:%
\footnote{\index{footnote!\thefoot. covariant derivative of a multivector valued form}Note
the analogy to the tangent space covariant derivative of some multivector
valued form\[
K(x,\bs{e},\tilde{\bs{e}})\equiv K_{a_{1}\ldots a_{k}}\hoch{b_{1}\ldots b_{k'}}(x)\cdot\bs{e}^{a_{1}}\cdots\bs{e}^{a_{k}}\tilde{\bs{e}}_{b_{1}}\cdots\tilde{\bs{e}}_{b_{k'}}\]
written in the following way\begin{eqnarray*}
\nabla_{m}K & = & \partial_{m}K(x,\bs{e},\tilde{\bs{e}})-\bs{e}^{a}\Omega_{ma}\hoch{b}\partl{\bs{e}^{b}}K+\tilde{\bs{e}}_{a}\Omega_{mb}\hoch{a}\partl{\tilde{\bs{e}}_{b}}K\qquad\fussend\end{eqnarray*}
}\begin{eqnarray}
\funktional{_{\gem{cov}}S}{x^{K}} & = & \funktional{S}{x^{K}}+\Omega_{K\bs{\gamma}}\hoch{\bs{\delta}}\dP_{z\bs{\delta}}\funktional{S}{\dP_{z\bs{\gamma}}}+\hat{\Omega}_{K\hat{\bs{\gamma}}}\hoch{\hat{\bs{\delta}}}\hat{\dP}_{\bar{z}\hat{\bs{\delta}}}\funktional{S}{\hat{\dP}_{\bar{z}\hat{\bs{\gamma}}}}-\Omega_{K\bs{\beta}}\hoch{\bs{\alpha}}\ce^{\bs{\beta}}\funktional{S}{\ce^{\bs{\alpha}}}-\hat{\Omega}_{K\hat{\bs{\beta}}}\hoch{\hat{\bs{\alpha}}}\hat{\ce}^{\hat{\bs{\beta}}}\funktional{S}{\hat{\ce}^{\hat{\bs{\alpha}}}}+\nonumber \\
 &  & +\Omega_{K\bs{\beta}}\hoch{\bs{\alpha}}\be_{z\bs{\alpha}}\funktional{S}{\be_{z\bs{\beta}}}+\hat{\Omega}_{K\hat{\bs{\beta}}}\hoch{\hat{\bs{\alpha}}}\be_{\bar{z}\hat{\bs{\alpha}}}\funktional{S}{\hat{\be}_{\bar{z}\hat{\bs{\beta}}}}+\Omega_{Ka}\hoch{b}L_{z\bar{z}b}\funktional{S}{L_{z\bar{z}a}}+\hat{\Omega}_{Ka}\hoch{b}\hat{L}_{z\bar{z}b}\funktional{S}{\hat{L}_{z\bar{z}a}}\label{eq:covVariationalx}\end{eqnarray}
All the other variational derivatives (\ref{eq:eomI})-(\ref{eq:eomVIII})
remain untouched:\begin{eqnarray}
\funktional{_{\gem{cov}}S}{\dP_{z\bs{\alpha}}} & = & \funktional{S}{\dP_{z\bs{\alpha}}},\quad\ldots,\,\funktional{_{\gem{cov}}S}{\hat{L}_{\bar{z}za}}=\funktional{S}{\hat{L}_{\bar{z}za}}\label{eq:covVariationalRest}\end{eqnarray}
 According to (\ref{eq:covVariationalx}), $\delta_{\gem{cov}}S/\delta x^{K}$
coincides with $\delta S/\delta x^{K}$ when all the other equations
of motion are fulfilled. This leads to the following obvious but important
statement: \begin{prop}Setting the covariant variational derivatives
defined via (\ref{eq:covVariationalx}) and (\ref{eq:covVariationalRest})
to zero, leads to a set of equations which is equivalent to the equations
of motion obtained by the ordinary variational derivatives: \begin{equation}
\funktional{_{\gem{cov}}S}{\left(x^{K},\dP_{z\bs{\alpha}},\ce^{\bs{\alpha}},\be_{z\bs{\alpha}},\hat{\dP}_{\bar{z}\hat{\bs{\alpha}}},\hat{\ce}^{\hat{\bs{\alpha}}},\hat{\be}_{\bar{z}\hat{\bs{\alpha}}},L_{z\bar{z}a},\hat{L}_{\bar{z}za}\right)}=0\iff\funktional{S}{\left(x^{K},\dP_{z\bs{\alpha}},\ce^{\bs{\alpha}},\be_{z\bs{\alpha}},\hat{\dP}_{\bar{z}\hat{\bs{\alpha}}},\hat{\ce}^{\hat{\bs{\alpha}}},\hat{\be}_{\bar{z}\hat{\bs{\alpha}}},L_{z\bar{z}a},\hat{L}_{\bar{z}za}\right)}=0\qquad\end{equation}
The covariant variational derivatives in turn are obtained by using
the covariant variation defined in (\ref{eq:covVarI})-(\ref{eq:covVarVIII})
and the commutators (\ref{eq:commutatorOfCovVarI}) and (\ref{eq:commutatorOfCovVarII}).\end{prop}\rem{Let
us now assume a Lagrangian of the form $\mc{L}(x,\partial x,P,\ce,\nabla\ce,\be)$,
where $x^{M}$ are coordinates of the target-space, $\partial_{\mu}x^{M}$
are treated as tangent space vectors (or basis elements of the cotangent
space). Likewise $P_{\mu A}$ and $\be_{\mu A}$ are cotangent space
vectors (with flat indices) and $\ce^{A}$~(and $\Pi_{\mu}^{A}=\partial_{\mu}x^{M}E_{M}\hoch{A}$)
are tangent space vectors with the covariant derivatives and variations
acting accordingly. The variation of the action then reads ($\delta_{cov}\mc{L}=\delta\mc{L}$)\begin{eqnarray}
\delta S & = & \int\quad\delta x^{M}\nabla_{M}\mc{L}+\underbrace{\delta_{cov}(\partial_{\mu}x^{K})}_{\nabla_{\mu}\delta x^{K}+2T_{cov,\mu}\hoch{K}}\partiell{\mc{L}}{(\partial_{\mu}x^{K})}+\nonumber \\
 &  & +\delta_{cov}P_{\mu A}\partl{P_{\mu A}}\mc{L}+\delta_{cov}\be_{\mu A}\partl{\be_{\mu A}}\mc{L}+\nonumber \\
 &  & +\delta_{cov}\ce^{A}\partl{\ce^{A}}\mc{L}+\underbrace{\delta_{cov}(\nabla_{\mu}\ce^{A})}_{\nabla_{\mu}\delta_{cov}\ce^{A}+2R_{cov,\mu B}\hoch{A}\ce^{B}}\partl{(\nabla_{\mu}\ce^{A})}\mc{L}\\
 & = & \int\quad\partial_{\mu}\left(\delta x^{M}\partiell{\mc{L}}{(\partial_{\mu}x^{M})}+\delta_{cov}\ce^{A}\partl{(\nabla_{\mu}\ce^{A})}\mc{L}\right)+\nonumber \\
 &  & +\delta x^{M}\left(\nabla_{M}\mc{L}-\nabla_{\mu}\partiell{\mc{L}}{(\partial_{\mu}x^{M})}+2\partial_{\mu}x^{N}T_{MN}\hoch{K}\partiell{\mc{L}}{(\partial_{\mu}x^{K})}+2\partial_{\mu}x^{N}R_{MN\, B}\hoch{A}\ce^{B}\partl{(\nabla_{\mu}\ce^{A})}\mc{L}\right)+\nonumber \\
 &  & +\delta_{cov}\ce^{A}\left(\partl{\ce^{A}}\mc{L}-\nabla_{\mu}\partl{(\nabla_{\mu}\ce^{A})}\mc{L}\right)+\nonumber \\
 &  & +\delta_{cov}P_{\mu A}\partl{P_{\mu A}}\mc{L}+\delta_{cov}\be_{\mu A}\partl{\be_{\mu A}}\mc{L}\end{eqnarray}
 }

\paragraph{The last equation of motion}

We are now ready to calculate the last equation of motion, the variation
with respect to $x^{K}$. Admittedly introducing a new tool like the
covariant variation for just one equation seems a bit of overkill.
However, in any case we would have been forced during the calculation
to organize the result into covariant expressions and the covariant
variation gives a general recipe how to do that. Although we described
the covariant variation for the Berkovits string, it is a tool which
is very useful in any other nonlinear sigma model. In addition it
should be noted that the above concept works for an arbitrary connection
and not only for the connection $\gemOm_{MA}\hoch{B}$ or the corresponding
$\gem{\Gamma}_{MN}\hoch{K}$. The calculation just simplifies at some
points, if one restricts to connections with special properties, or
to connections which are already present in the action. E.g. only
because we are choosing $\gemOm_{MA}\hoch{B}$, we can make use of
(\ref{eq:commutatorOfCovVarI}) and (\ref{eq:commutatorOfCovVarII})
in order to commute the covariant variation with the target space
covariant worldsheet derivative. In addition we will make use of the
fact that the covariant variation annihilates the vielbein:

\begin{equation}
\delta_{\gem{cov}}E_{M}\hoch{A}(\xfull)=0\frem{,\qquad\gemnabla_{K}G_{AB}=2\checkcovPhi{K}G_{AB}}\label{eq:deltaCovEisNix}\end{equation}
 Note also how the antisymmetrized covariant derivative of the $B$-field
can be written in terms of its exterior derivative $H$ and the torsion:
\begin{equation}
\bs{\gemnabla}B\equiv\gemnabla_{\bs{M}}B_{\bs{MM}}=\de B-\ip_{\gemT}B=H_{\bs{MMM}}-2\gemT_{\bs{MM}}\hoch{K}B_{K\bs{M}}\label{eq:covariantExtDerOfB}\end{equation}
The important contributions to the (covariant) variation of the action
come from the covariant variation of the (spacetime covariant) worldsheet
derivatives of the elementary fields, like $\delta_{cov}\nabla_{\bar{z}}\ce^{\bs{\alpha}}$
and $\delta_{\gem{cov}}\Pi_{z/\bar{z}}^{A}$. For the latter we have
(compare to the equation before (2.12) in \cite{Bedoya:2006ic})\begin{eqnarray}
\delta_{\gem{cov}}\Pi_{z/\bar{z}}^{A} & \stackrel{(\ref{eq:deltaCovEisNix})}{=} & \delta_{\gem{cov}}\partial_{z/\bar{z}}x^{K}\cdot E_{K}\hoch{A}=\\
 & \stackrel{(\ref{eq:commutatorOfCovVarI})}{=} & \gemnabla_{z/\bar{z}}\delta x^{K}\cdot E_{K}\hoch{A}+2\delta x^{M}\gem{T}_{MN}\hoch{A}\partial_{z/\bar{z}}x^{N}\label{eq:covVarOfPi}\end{eqnarray}
For the ghost terms we obtain curvature expressions instead of torsion
expressions:\begin{eqnarray}
\delta_{cov}\nabla_{\bar{z}}\ce^{\bs{\beta}} & \stackrel{(\ref{eq:commutatorOfCovVarII})}{=} & \nabla_{\bar{z}}\delta_{cov}\ce^{\bs{\beta}}+2\delta x^{K}\bar{\partial}x^{L}R_{KL\bs{\alpha}}\hoch{\bs{\beta}}\ce^{\bs{\alpha}}\\
\delta_{\hat{cov}}\hat{\nabla}_{z}\hat{\ce}^{\hat{\bs{\beta}}} & \stackrel{(\ref{eq:commutatorOfCovVarII})}{=} & \hat{\nabla}_{z}\delta_{\hat{cov}}\hat{\ce}^{\hat{\bs{\beta}}}+2\delta x^{K}\partial x^{L}\hat{R}_{KL\hat{\bs{\alpha}}}\hoch{\hat{\bs{\beta}}}\hat{\ce}^{\hat{\bs{\alpha}}}\end{eqnarray}
As a last ingredient, before we vary the action, we should note a
specialty of the pure spinor term. The covariant variation on the
Lagrange multiplier is chosen in such a way that the covariant variation
of $\gamma_{\bs{\alpha}\bs{\beta}}^{a}$ is almost zero. But as we
discussed at length in section \ref{sec:Antighost-gauge-symmetry}
on page \pageref{sec:Antighost-gauge-symmetry} the structure group
is not yet for all components of the connection reduced to Lorentz
plus scale transformations and we have in general a non-vanishing
$\gamma^{[4]}$-part $\Omega_{\bs{\gamma}a_{1}\ldots a_{4}}$. At
least formally we therefore obtain a non-vanishing covariant derivative
on $\gamma_{\bs{\alpha\beta}}^{a}$ (with $\Omega_{M\bs{\alpha}}\hoch{\bs{\beta}}$
acting on the spinorial indices and $\Omega_{Ma}\hoch{b}$ of (\ref{eq:OmegaMitBosonIndizes})
acting on the bosonic one):\begin{eqnarray}
\nabla_{M}\gamma_{\bs{\alpha}\bs{\beta}}^{a} & = & -2E_{M}\hoch{\bs{\gamma}}\Omega_{\bs{\gamma}\, a_{1}\ldots a_{4}}\gamma^{a_{1}\ldots a_{4}}\tief{[\bs{\alpha}|}\hoch{\bs{\delta}}\gamma_{\bs{\delta}|\bs{\beta}]}^{a}\stackrel{(\ref{eq:gammaIgammal})}{=}-2E_{M}\hoch{\bs{\gamma}}\Omega_{\bs{\gamma}\, a_{1}\ldots a_{4}}\gamma^{a_{1}\ldots a_{4}a}\tief{\bs{\alpha\beta}}\end{eqnarray}

Due to (\ref{eq:covVariationalx}) and (\ref{eq:covVariationalRest})
we know already that only the variational derivative with respect
to $x^{K}$ gets modified while the others remain untouched. We therefore
collect the terms which are proportional to the $x^{K}$-variation
only. In particular we do not need to consider the first term respectively
of the above two equations. For completeness, however, we keep the
total derivatives coming from the corresponding partial integration.
Apart from the variation of $\Pi_{z/\bar{z}}^{A}$, $\nabla_{\bar{z}}\ce^{\bs{\beta}}$
and $\hat{\nabla}_{z}\hat{\ce}^{\hat{\bs{\beta}}}$ we only have covariant
variations of the background fields. The (covariant) variation of
the action (\ref{eq:BiBagain}) thus takes the following form\begin{eqnarray}
\delta S & = & \int d^{2}z\quad\delta x^{K}\Big[\frac{1}{2}\Pi_{z}^{A}\gemnabla_{K}\GB_{AB}\Pi_{\bar{z}}^{B}+\dP_{z\bs{\gamma}}\gemnabla_{K}\RR^{\bs{\gamma}\hat{\bs{\gamma}}}\hat{\dP}_{\bar{z}\hat{\bs{\gamma}}}+\nonumber \\
 &  & +\ce^{\bs{\alpha}}\gemnabla_{K}C_{\bs{\alpha}}\hoch{\bs{\beta}\hat{\bs{\gamma}}}\be_{z\bs{\beta}}\hat{\dP}_{\bar{z}\hat{\bs{\gamma}}}+\hat{\ce}^{\hat{\bs{\alpha}}}\gemnabla_{K}\hat{C}_{\hat{\bs{\alpha}}}\hoch{\hat{\bs{\beta}}\bs{\gamma}}\hat{\be}_{\bar{z}\hat{\bs{\beta}}}\dP_{z\bs{\gamma}}+\ce^{\bs{\alpha}}\hat{\ce}^{\hat{\bs{\alpha}}}\gemnabla_{K}S_{\bs{\alpha}\hat{\bs{\alpha}}}\hoch{\bs{\beta}\hat{\bs{\beta}}}\be_{z\bs{\beta}}\hat{\be}_{\bar{z}\hat{\bs{\beta}}}\frem{-4\alpha'\partial\bar{\partial}\weyl\cdot\gemnabla_{K}\dil}\Big]+\nonumber \\
 &  & +\frac{1}{2}\underbrace{\left(\gemnabla_{z}\delta x^{K}\cdot E_{K}\hoch{A}+2\delta x^{M}\gem{T}_{MN}\hoch{A}\partial_{z}x^{N}\right)}_{\delta_{\gem{cov}}\Pi_{z}^{A}}\GB_{AB}\Pi_{\bar{z}}^{B}+\frac{1}{2}\Pi_{z}^{A}\GB_{AB}\underbrace{\left(\gemnabla_{\bar{z}}\delta x^{K}\cdot E_{K}\hoch{B}+2\delta x^{M}\gem{T}_{MN}\hoch{B}\partial_{\bar{z}}x^{N}\right)}_{\delta_{\gem{cov}}\Pi_{\bar{z}}^{B}}+\nonumber \\
 &  & +\underbrace{\left(\gemnabla_{\bar{z}}\delta x^{K}\cdot E_{K}\hoch{\bs{\gamma}}+2\delta x^{M}\gem{T}_{MN}\hoch{\bs{\gamma}}\partial_{\bar{z}}x^{N}\right)}_{\delta_{\gem{cov}}\Pi_{\bar{z}}^{\bs{\gamma}}}\dP_{z\bs{\gamma}}+\underbrace{\left(\gemnabla_{z}\delta x^{K}\cdot E_{K}\hoch{\hat{\bs{\gamma}}}+2\delta x^{M}\gem{T}_{MN}\hoch{\hat{\bs{\gamma}}}\partial_{z}x^{N}\right)}_{\delta_{\gem{cov}}\Pi_{z}^{\hat{\bs{\gamma}}}}\hat{\dP}_{\bar{z}\hat{\bs{\gamma}}}+\nonumber \\
 &  & +\underbrace{2\delta x^{K}\bar{\partial}x^{L}R_{KL\bs{\alpha}}\hoch{\bs{\beta}}\ce^{\bs{\alpha}}}_{\delta_{cov}\nabla_{\bar{z}}\ce^{\bs{\beta}}-\nabla_{\bar{z}}\delta_{cov}\ce^{\bs{\beta}}}\be_{z\bs{\beta}}+\underbrace{2\delta x^{K}\partial x^{L}\hat{R}_{KL\hat{\bs{\alpha}}}\hoch{\hat{\bs{\beta}}}\hat{\ce}^{\hat{\bs{\alpha}}}}_{\delta_{\hat{cov}}\hat{\nabla}_{z}\hat{\ce}^{\hat{\bs{\beta}}}-\hat{\nabla}_{z}\delta_{\hat{cov}}\hat{\ce}^{\hat{\bs{\beta}}}}\hat{\be}_{\bar{z}\hat{\bs{\beta}}}+\nonumber \\
 &  & -\delta x^{K}E_{K}\hoch{\bs{\gamma}}\Omega_{\bs{\gamma}\, a_{1}\ldots a_{4}}(\ce\gamma^{a_{1}\ldots a_{4}a}\ce)\cdot L_{z\bar{z}a}-\delta x^{K}E_{K}\hoch{\hat{\bs{\gamma}}}\hat{\Omega}_{\hat{\bs{\gamma}}\, a_{1}\ldots a_{4}}(\hat{\ce}\gamma^{a_{1}\ldots a_{4}a}\hat{\ce})\cdot\hat{L}_{\bar{z}za}+\nonumber \\
 &  & +\delta_{cov}\dP_{z\bs{\alpha}}\funktional{S}{\dP_{z\bs{\alpha}}}+\delta_{\hat{cov}}\hat{\dP}_{\bar{z}\hat{\bs{\alpha}}}\funktional{S}{\hat{\dP}_{\bar{z}\bs{\alpha}}}+\delta_{cov}\ce^{\bs{\alpha}}\funktional{S}{\ce^{\bs{\alpha}}}+\delta_{\hat{cov}}\hat{\ce}^{\hat{\bs{\alpha}}}\funktional{S}{\hat{\ce}^{\hat{\bs{\alpha}}}}+\delta_{cov}\be_{z\bs{\alpha}}\funktional{S}{\be_{z\bs{\alpha}}}+\delta_{\hat{cov}}\hat{\be}_{\bar{z}\hat{\bs{\alpha}}}\funktional{S}{\hat{\be}_{\bar{z}\hat{\bs{\alpha}}}}+\nonumber \\
 &  & +\delta_{cov}L_{z\bar{z}a}\funktional{S}{L_{z\bar{z}a}}+\delta_{\hat{cov}}\hat{L}_{\bar{z}za}\funktional{S}{\hat{L}_{\bar{z}za}}\frem{+\delta\weyl\funktional{S}{\weyl}-\partial\bar{\partial}\left(4\alpha'\weyl\cdot\dil\right)}+\partial_{\bar{z}}\left(\delta_{cov}\ce^{\bs{\beta}}\be_{z\bs{\beta}}\right)+\partial_{z}\left(\delta_{\hat{cov}}\hat{\ce}^{\hat{\bs{\beta}}}\hat{\be}_{\bar{z}\hat{\bs{\beta}}}\right)\qquad\qquad\end{eqnarray}
We finally make a partial integration for the terms in the third and
fourth line (keeping again the total derivatives as a reference for
future studies of the open string) and arrive at \begin{eqnarray}
\delta S & = & \int d^{2}z\quad\delta x^{K}E_{K}\hoch{C}\Big[-\frac{1}{2}\GB_{CB}\gemnabla_{z}\Pi_{\bar{z}}^{B}-\frac{1}{2}\gemnabla_{\bar{z}}\Pi_{z}^{A}\GB_{AC}+\nonumber \\
 &  & +\frac{1}{2}\Pi_{z}^{A}\left(\gemnabla_{C}\GB_{AB}-\gemnabla_{A}\GB_{CB}-\gemnabla_{B}\GB_{AC}+2\gem{T}_{CA}\hoch{D}\GB_{DB}+2\GB_{AD}\gem{T}_{CB}\hoch{D}\right)\Pi_{\bar{z}}^{B}+\nonumber \\
 &  & -\delta_{C}\hoch{\bs{\gamma}}\gemnabla_{\bar{z}}\dP_{z\bs{\gamma}}-\delta_{C}\hoch{\hat{\bs{\gamma}}}\gemnabla_{z}\hat{\dP}_{\bar{z}\hat{\bs{\gamma}}}+2\gem{T}_{CB}\hoch{\bs{\gamma}}\Pi_{\bar{z}}^{B}\dP_{z\bs{\gamma}}+2\gem{T}_{CA}\hoch{\hat{\bs{\gamma}}}\Pi_{z}^{A}\hat{\dP}_{\bar{z}\hat{\bs{\gamma}}}+\nonumber \\
 &  & +2\Pi_{\bar{z}}^{B}R_{CB\bs{\alpha}}\hoch{\bs{\beta}}\ce^{\bs{\alpha}}\be_{z\bs{\beta}}+2\Pi_{z}^{A}\hat{R}_{CA\hat{\bs{\alpha}}}\hoch{\hat{\bs{\beta}}}\hat{\ce}^{\hat{\bs{\alpha}}}\hat{\be}_{\bar{z}\hat{\bs{\beta}}}+\nonumber \\
 &  & +\dP_{z\bs{\gamma}}\gemnabla_{C}\RR^{\bs{\gamma}\hat{\bs{\gamma}}}\hat{\dP}_{\bar{z}\hat{\bs{\gamma}}}+\ce^{\bs{\alpha}}\gemnabla_{C}C_{\bs{\alpha}}\hoch{\bs{\beta}\hat{\bs{\gamma}}}\be_{z\bs{\beta}}\hat{\dP}_{\bar{z}\hat{\bs{\gamma}}}+\hat{\ce}^{\hat{\bs{\alpha}}}\gemnabla_{C}\hat{C}_{\hat{\bs{\alpha}}}\hoch{\hat{\bs{\beta}}\bs{\gamma}}\hat{\be}_{\bar{z}\hat{\bs{\beta}}}\dP_{z\bs{\gamma}}+\nonumber \\
 &  & +\ce^{\bs{\alpha}}\hat{\ce}^{\hat{\bs{\alpha}}}\gemnabla_{C}S_{\bs{\alpha}\hat{\bs{\alpha}}}\hoch{\bs{\beta}\hat{\bs{\beta}}}\be_{z\bs{\beta}}\hat{\be}_{\bar{z}\hat{\bs{\beta}}}-\delta_{C}\hoch{\bs{\gamma}}\Omega_{\bs{\gamma}\, a_{1}\ldots a_{4}}(\ce\gamma^{a_{1}\ldots a_{4}a}\ce)\cdot L_{z\bar{z}a}-\delta_{C}\hoch{\hat{\bs{\gamma}}}\hat{\Omega}_{\hat{\bs{\gamma}}\, a_{1}\ldots a_{4}}(\hat{\ce}\gamma^{a_{1}\ldots a_{4}a}\hat{\ce})\cdot\hat{L}_{\bar{z}za}\frem{-4\alpha'\partial\bar{\partial}\weyl\cdot\gemnabla_{C}\dil}\Big]+\nonumber \\
 &  & +\delta_{cov}\dP_{z\bs{\alpha}}\funktional{S}{\dP_{z\bs{\alpha}}}+\delta_{\hat{cov}}\hat{\dP}_{\bar{z}\hat{\bs{\alpha}}}\funktional{S}{\hat{\dP}_{\bar{z}\bs{\alpha}}}+\delta_{cov}\ce^{\bs{\alpha}}\funktional{S}{\ce^{\bs{\alpha}}}+\delta_{\hat{cov}}\hat{\ce}^{\hat{\bs{\alpha}}}\funktional{S}{\hat{\ce}^{\hat{\bs{\alpha}}}}+\delta_{cov}\be_{z\bs{\alpha}}\funktional{S}{\be_{z\bs{\alpha}}}+\delta_{\hat{cov}}\hat{\be}_{\bar{z}\hat{\bs{\alpha}}}\funktional{S}{\hat{\be}_{\bar{z}\hat{\bs{\alpha}}}}+\nonumber \\
 &  & +\delta_{cov}L_{z\bar{z}a}\funktional{S}{L_{z\bar{z}a}}+\delta_{\hat{cov}}\hat{L}_{\bar{z}za}\funktional{S}{\hat{L}_{\bar{z}za}}\frem{+\delta\weyl\funktional{S}{\weyl}-\partial\bar{\partial}\left(4\alpha'\weyl\cdot\dil\right)}+\nonumber \\
 &  & +\partial_{\bar{z}}\left(\delta_{cov}\ce^{\bs{\beta}}\be_{z\bs{\beta}}+\frac{1}{2}\Pi_{z}^{A}\GB_{AK}\delta x^{K}+\delta x^{K}\cdot E_{K}\hoch{\bs{\gamma}}\dP_{z\bs{\gamma}}\right)+\nonumber \\
 &  & +\partial_{z}\left(\delta_{\hat{cov}}\hat{\ce}^{\hat{\bs{\beta}}}\hat{\be}_{\bar{z}\hat{\bs{\beta}}}+\frac{1}{2}\delta x^{K}\GB_{KB}\Pi_{\bar{z}}^{B}+\delta x^{K}\cdot E_{K}\hoch{\hat{\bs{\gamma}}}\hat{\dP}_{\bar{z}\hat{\bs{\gamma}}}\right)\end{eqnarray}
Now we can read off the covariant variational derivative with respect
to $x^{K}$. But let us note two further relations first:\begin{eqnarray}
\lqn{\gem{\nabla}_{C}\GB_{AB}-\gem{\nabla}_{A}\GB_{CB}-\gem{\nabla}_{B}\GB_{AC}=}\nonumber \\
 & \stackrel{(\ref{eq:covariantExtDerOfB})}{=} & 3H_{CAB}-2\gemT_{AB}\hoch{D}B_{DC}-2\gemT_{CA}\hoch{D}B_{DB}-2\gemT_{BC}\hoch{D}B_{DA}+\gemnabla_{C}G_{AB}-\gemnabla_{A}G_{CB}-\gemnabla_{B}G_{AC}\end{eqnarray}
and \begin{eqnarray}
\gemnabla_{z}\Pi_{\bar{z}}^{D} & \stackrel{(\ref{eq:commutatorOfCovVarI})}{=} & \gemnabla_{\bar{z}}\Pi_{z}^{D}+2\Pi_{z}^{A}\Pi_{\bar{z}}^{B}\gemT_{AB}\hoch{D}\end{eqnarray}
In addition we define \index{$T$@$\gemT_{AB\mid C}$}\begin{eqnarray}
\gemT_{AB|C} & \equiv & \gemT_{AB}\hoch{D}G_{DC}\end{eqnarray}
\rem{which is different from zero only for the last index being bosonic:
$C=c$. } Note that we use the symmetric rank two tensor $G_{AB}$
only to pull indices down. Pulling them up again is in general not
possible as $G_{AB}$ might be degenerate. In fact we will learn soon
that it has to be degenerate. 

The \textbf{final result} of the variation now reads\begin{eqnarray}
\delta S & = & \int d^{2}z\quad\delta x^{K}\funktional{_{\gem{cov}}S}{x^{K}}+\delta_{cov}\dP_{z\bs{\alpha}}\funktional{S}{\dP_{z\bs{\alpha}}}+\delta_{\hat{cov}}\hat{\dP}_{\bar{z}\hat{\bs{\alpha}}}\funktional{S}{\hat{\dP}_{\bar{z}\bs{\alpha}}}+\nonumber \\
 &  & +\delta_{cov}\ce^{\bs{\alpha}}\funktional{S}{\ce^{\bs{\alpha}}}+\delta_{\hat{cov}}\hat{\ce}^{\hat{\bs{\alpha}}}\funktional{S}{\hat{\ce}^{\hat{\bs{\alpha}}}}+\delta_{cov}\be_{z\bs{\alpha}}\funktional{S}{\be_{z\bs{\alpha}}}+\delta_{\hat{cov}}\hat{\be}_{\bar{z}\hat{\bs{\alpha}}}\funktional{S}{\hat{\be}_{\bar{z}\hat{\bs{\alpha}}}}+\nonumber \\
 &  & +\delta_{cov}L_{z\bar{z}a}\funktional{S}{L_{z\bar{z}a}}+\delta_{\hat{cov}}\hat{L}_{\bar{z}za}\funktional{S}{\hat{L}_{\bar{z}za}}\frem{+\delta\weyl\funktional{S}{\weyl}-\partial\bar{\partial}\left(4\alpha'\weyl\cdot\dil\right)}+\nonumber \\
 &  & +\partial_{\bar{z}}\left(\delta_{cov}\ce^{\bs{\beta}}\be_{z\bs{\beta}}+\frac{1}{2}\Pi_{z}^{A}\GB_{AK}\delta x^{K}+\delta x^{K}\cdot E_{K}\hoch{\bs{\gamma}}\dP_{z\bs{\gamma}}\right)+\nonumber \\
 &  & +\partial_{z}\left(\delta_{\hat{cov}}\hat{\ce}^{\hat{\bs{\beta}}}\hat{\be}_{\bar{z}\hat{\bs{\beta}}}+\frac{1}{2}\delta x^{K}\GB_{KB}\Pi_{\bar{z}}^{B}+\delta x^{K}\cdot E_{K}\hoch{\hat{\bs{\gamma}}}\hat{\dP}_{\bar{z}\hat{\bs{\gamma}}}\right)\end{eqnarray}
with the following covariant variational derivatives or equations
of motion (remember (\ref{eq:eomI})-(\ref{eq:eomVIII})):\vRam{1.05}{\begin{eqnarray}
\funktional{_{\gem{cov}}S}{x^{K}} & = & E_{K}\hoch{C}\Big[\underbrace{-\gemnabla_{\bar{z}}\Pi_{z}^{D}}_{-\gemnabla_{z}\Pi_{\bar{z}}^{D}\lqn{{\scriptstyle +2\Pi_{z}^{A}\Pi_{\bar{z}}^{B}\gemT_{AB}\hoch{D}}}}G_{DC}+\Pi_{z}^{A}\left(\frac{3}{2}H_{CAB}-\gem{T}_{AB|C}+2\gem{T}_{C(A|B)}+\frac{1}{2}\gemnabla_{C}G_{AB}-\gemnabla_{(A}G_{B)C}\right)\Pi_{\bar{z}}^{B}+\nonumber \\
 &  & -\delta_{C}\hoch{\bs{\gamma}}\nabla_{\bar{z}}\dP_{z\bs{\gamma}}-\delta_{C}\hoch{\hat{\bs{\gamma}}}\hat{\nabla}_{z}\hat{\dP}_{\bar{z}\hat{\bs{\gamma}}}+2T_{CB}\hoch{\bs{\gamma}}\Pi_{\bar{z}}^{B}\dP_{z\bs{\gamma}}+2\hat{T}_{CA}\hoch{\hat{\bs{\gamma}}}\Pi_{z}^{A}\hat{\dP}_{\bar{z}\hat{\bs{\gamma}}}+\nonumber \\
 &  & +\dP_{z\bs{\gamma}}\gemnabla_{C}\RR^{\bs{\gamma}\hat{\bs{\gamma}}}\hat{\dP}_{\bar{z}\hat{\bs{\gamma}}}+\ce^{\bs{\alpha}}\gemnabla_{C}C_{\bs{\alpha}}\hoch{\bs{\beta}\hat{\bs{\gamma}}}\be_{z\bs{\beta}}\hat{\dP}_{\bar{z}\hat{\bs{\gamma}}}+\hat{\ce}^{\hat{\bs{\alpha}}}\gemnabla_{C}\hat{C}_{\hat{\bs{\alpha}}}\hoch{\hat{\bs{\beta}}\bs{\gamma}}\hat{\be}_{\bar{z}\hat{\bs{\beta}}}\dP_{z\bs{\gamma}}+\nonumber \\
 &  & +\ce^{\bs{\alpha}}\hat{\ce}^{\hat{\bs{\alpha}}}\gemnabla_{C}S_{\bs{\alpha}\hat{\bs{\alpha}}}\hoch{\bs{\beta}\hat{\bs{\beta}}}\be_{z\bs{\beta}}\hat{\be}_{\bar{z}\hat{\bs{\beta}}}-\delta_{C}\hoch{\bs{\gamma}}\Omega_{\bs{\gamma}\, a_{1}\ldots a_{4}}(\ce\gamma^{a_{1}\ldots a_{4}a}\ce)\cdot L_{z\bar{z}a}-\delta_{C}\hoch{\hat{\bs{\gamma}}}\hat{\Omega}_{\hat{\bs{\gamma}}\, a_{1}\ldots a_{4}}(\hat{\ce}\gamma^{a_{1}\ldots a_{4}a}\hat{\ce})\cdot\hat{L}_{\bar{z}za}\frem{-4\alpha'\partial\bar{\partial}\weyl\cdot\gemnabla_{C}\dil}+\nonumber \\
 &  & +2\Pi_{\bar{z}}^{B}R_{CB\bs{\alpha}}\hoch{\bs{\beta}}\ce^{\bs{\alpha}}\be_{z\bs{\beta}}+2\Pi_{z}^{A}\hat{R}_{CA\hat{\bs{\alpha}}}\hoch{\hat{\bs{\beta}}}\hat{\ce}^{\hat{\bs{\alpha}}}\hat{\be}_{\bar{z}\hat{\bs{\beta}}}\Big]\label{eq:eomx}\\
\funktional{S}{\dP_{z\bs{\gamma}}} & = & \Pi_{\bar{z}}^{\bs{\gamma}}+\RR^{\bs{\gamma}\hat{\bs{\gamma}}}\hat{\dP}_{\bar{z}\hat{\bs{\gamma}}}+\hat{\ce}^{\hat{\bs{\alpha}}}\hat{C}_{\hat{\bs{\alpha}}}\hoch{\hat{\bs{\beta}}\bs{\gamma}}\hat{\be}_{\bar{z}\hat{\bs{\beta}}}\label{eq:eomPibar}\\
\funktional{S}{\hat{\dP}_{\bar{z}\hat{\bs{\gamma}}}} & = & \Pi_{z}^{\hat{\bs{\gamma}}}+\dP_{z\bs{\gamma}}\RR^{\bs{\gamma}\hat{\bs{\gamma}}}+\ce^{\bs{\alpha}}C_{\bs{\alpha}}\hoch{\bs{\beta}\hat{\bs{\gamma}}}\be_{z\bs{\beta}}\label{eq:eomPi}\\
\funktional{S}{\be_{z\bs{\beta}}} & = & -\left(\nabla_{\bar{z}}\ce^{\bs{\beta}}+\ce^{\bs{\alpha}}\left(C_{\bs{\alpha}}\hoch{\bs{\beta}\hat{\bs{\gamma}}}\hat{\dP}_{\bar{z}\hat{\bs{\gamma}}}-\hat{\ce}^{\hat{\bs{\alpha}}}S_{\bs{\alpha}\hat{\bs{\alpha}}}\hoch{\bs{\beta}\hat{\bs{\beta}}}\hat{\be}_{\bar{z}\hat{\bs{\beta}}}\right)\right)\equiv-\mc{D}_{\bar{z}}\ce^{\bs{\beta}}\\
\funktional{S}{\hat{\be}_{\bar{z}\hat{\bs{\beta}}}} & = & -\left(\hat{\nabla}_{z}\hat{\ce}^{\hat{\bs{\beta}}}+\hat{\ce}^{\hat{\bs{\alpha}}}\left(\hat{C}_{\hat{\bs{\alpha}}}\hoch{\hat{\bs{\beta}}\bs{\gamma}}\dP_{z\bs{\gamma}}-\ce^{\bs{\alpha}}S_{\bs{\alpha}\hat{\bs{\alpha}}}\hoch{\bs{\beta}\hat{\bs{\beta}}}\be_{z\bs{\beta}}\right)\right)\equiv-\hat{\mc{D}}_{z}\hat{\ce}^{\hat{\bs{\beta}}}\\
\funktional{S}{\ce^{\bs{\alpha}}} & = & -\left(\nabla_{\bar{z}}\be_{z\bs{\alpha}}-\left(C_{\bs{\alpha}}\hoch{\bs{\beta}\hat{\bs{\gamma}}}\hat{\dP}_{\bar{z}\hat{\bs{\gamma}}}-\hat{\ce}^{\hat{\bs{\alpha}}}S_{\bs{\alpha}\hat{\bs{\alpha}}}\hoch{\bs{\beta}\hat{\bs{\beta}}}\hat{\be}_{\bar{z}\hat{\bs{\beta}}}\right)\be_{z\bs{\beta}}\right)+L_{z\bar{z}a}(\gamma^{a}\ce)_{\bs{\alpha}}\equiv-\mc{D}_{\bar{z}}\be_{z\bs{\alpha}}+L_{z\bar{z}a}(\gamma^{a}\ce)_{\bs{\alpha}}\\
\funktional{S}{\hat{\ce}^{\hat{\bs{\alpha}}}} & = & -\left(\hat{\nabla}_{z}\hat{\be}_{\bar{z}\hat{\bs{\alpha}}}-\left(\hat{C}_{\hat{\bs{\alpha}}}\hoch{\hat{\bs{\beta}}\bs{\gamma}}\dP_{z\bs{\gamma}}-\ce^{\bs{\alpha}}S_{\bs{\alpha}\hat{\bs{\alpha}}}\hoch{\bs{\beta}\hat{\bs{\beta}}}\be_{z\bs{\beta}}\right)\hat{\be}_{\bar{z}\hat{\bs{\beta}}}\right)+\hat{L}_{z\bar{z}a}(\gamma^{a}\hat{\ce})_{\hat{\bs{\alpha}}}\equiv-\hat{\mc{D}}_{z}\hat{\be}_{\bar{z}\hat{\bs{\alpha}}}+\hat{L}_{z\bar{z}a}(\gamma^{a}\hat{\ce})_{\hat{\bs{\alpha}}}\qquad\\
\funktional{S}{L_{z\bar{z}a}} & = & \frac{1}{2}(\ce\gamma^{a}\ce),\qquad\funktional{S}{\hat{L}_{z\bar{z}a}}=\frac{1}{2}(\hat{\ce}\gamma^{a}\hat{\ce})\label{eq:eomVIIIb}\end{eqnarray}
\frem{\begin{equation}
\funktional{S}{\weyl}=-4\alpha'\gemnabla_{\bar{z}}\Pi_{z}^{D}\nabla_{D}\dil-4\alpha'\Pi_{z}^{A}\Pi_{\bar{z}}^{B}\gemnabla_{B}\gemnabla_{A}\dil\end{equation}
}}
Note that we used for the covariant variation an independent connection
$\check{\Omega}_{Ma}\hoch{b}$ for the bosonic subspace. This connection
is a priory not a background field of the string metric. We are free
to choose it in a convenient way. \rem{The only restrictions that
we have put onto it so far is that it contains only a Lorentz part
and a scaling part which implies%
\footnote{\index{footnote!\thefoot. eqaution of motion independent of bosonic connection}This
implies that in particular the first line of (\ref{eq:eomx}) does
not depend on the choice of the connection $\check{\Omega}_{Ma}\hoch{b}$.
Let us see this explicitly and collect the terms depending on $\check{\Omega}_{Ma}\hoch{b}$:\begin{eqnarray*}
\lqn{-\gemnabla_{\bar{z}}\Pi_{z}^{D}G_{DC}+\frac{1}{2}\Pi_{z}^{A}\left(3H_{CAB}-2\gem{T}_{AB|C}+4\gem{T}_{C(A|B)}+2\checkcovPhi{C}G_{AB}-4\checkcovPhi{(A}G_{B)C}\right)\Pi_{\bar{z}}^{B}=}\\
 & = & \frac{1}{2}\Pi_{z}^{A}\big(-2\check{\Omega}_{Ba|c}\underbrace{-2\check{\Omega}_{[AB]|c}+2\check{\Omega}_{[CA]|b}+2\check{\Omega}_{[CB]|a}-2\check{\Omega}_{C}^{(D)}G_{AB}+4\check{\Omega}_{(A}^{(D)}G_{B)C}}_{-2\check{\Omega}_{Bc|a}+4\check{\Omega}_{B}^{(D)}G_{ac}=2\check{\Omega}_{Ba|c}}\big)\Pi_{\bar{z}}^{B}+\ldots\qquad\fussend\end{eqnarray*}
} $\gemnabla_{K}G_{AB}=2\checkcovPhi{K}G_{AB}$.}

\section{Ghost current}

\label{sec:Ghost-current}\index{ghost current|fett}Let us assign
ghost numbers $(1,0)$ and $(-1,0)$ to the fields $\ce^{\bs{\alpha}}$
and $\be_{z\bs{\alpha}}$. The corresponding transformation (with
some global transformation parameter $\rho$) is \begin{eqnarray}
\delta\ce^{\alpha} & = & \rho\ce^{\bs{\alpha}},\qquad\delta\be_{z\bs{\alpha}}=-\rho\be_{z\bs{\alpha}}\end{eqnarray}
For the action to remain unchanged, we also need to transform the
Lagrange multiplier \begin{equation}
\delta L_{z\bar{z}a}=-2\rho L_{z\bar{z}a}\end{equation}
which therefore has ghost number $-2$. Varying the action with a
local parameter, we arrive at \begin{eqnarray}
\delta S & = & \int_{\Sigma}d^{2}z\quad\bar{\partial}\rho\cdot(\ce^{\bs{\beta}}\be_{z\bs{\beta}})+\mbox{bdry-terms}\end{eqnarray}
According to (\ref{eq:noet:trick}) and footnote \ref{foot:noet:trick}
on page \pageref{foot:noet:trick}, we can read off the ghost current
as \begin{equation}
j^{gh}=\ce^{\bs{\alpha}}\be_{z\bs{\alpha}}\label{eq:ghostCurrent}\end{equation}
It has the same form as in flat space.

In section \ref{sec:Holomorphic-BRST-current}, we will derive the
BRST transformations of the worldsheet fields from the given BRST
current via {}``inverse Noether'' (see (\ref{eq:noet:currentdivergence})).
The idea is to calculate the divergence of the current and try to
express it in terms of the equations of motion. The transformations
of the worldsheet fields can then be read off as coefficients. This
avoids switching to the Hamiltonian formalism and using the Poisson
bracket to generate the transformations. It might be instructive to
see, how {}``inverse Noether'' works for the simple example of the
ghost current before we come to the BRST current later:\begin{eqnarray}
-\delta\allfields{I}\funktional{S}{\allfields{I}} & \stackrel{!}{=} & \bar{\partial}(\ce^{\bs{\alpha}}\be_{z\bs{\alpha}})=\nonumber \\
 & = & \mc{D}_{\bar{z}}\ce^{\bs{\alpha}}\cdot\be_{z\bs{\alpha}}+\ce^{\bs{\alpha}}\mc{D}_{\bar{z}}\be_{z\bs{\alpha}}=\nonumber \\
 & = & -\funktional{S}{\be_{z\bs{\alpha}}}\be_{z\bs{\alpha}}+\ce^{\bs{\alpha}}\left(-\funktional{S}{\ce^{\bs{\alpha}}}+L_{a}(\gamma^{a}\ce)_{\bs{\alpha}}\right)=\nonumber \\
 & = & \be_{z\bs{\alpha}}\funktional{S}{\be_{z\bs{\alpha}}}-\ce^{\bs{\alpha}}\funktional{S}{\ce^{\bs{\alpha}}}+2L_{z\bar{z}a}\funktional{S}{L_{z\bar{z}a}}\end{eqnarray}
From this one can read off the transformations with which we had begun. 

The ghost current and the corresponding transformations for the hatted
variables are obtained via proposition \ref{prop:left-right-symmetry}
on page \pageref{prop:left-right-symmetry}.

\section{Holomorphic BRST current}

\label{sec:Holomorphic-BRST-current}We now come to the main part
of the derivation of the supergravity constraints from the pure spinor
string. The pure spinor string in flat background had two (graded)
commuting and nilpotent BRST differentials which defined the physical
spectrum. Putting the string in a curved background is a matter of
consistent deformation. I.e., gauge symmetries and BRST symmetries
have to survive. They may be deformed, but the number of physical
degrees of worldsheet variables cannot simply change as soon as there
is a backreaction from the background that was produced by the strings
themselves. This is a similar consistency like the demand for vanishing
quantum anomalies. It is therefore legitimate to demand (apart from
the two antighost gauge symmetries) also two (graded) commuting BRST
symmetries. Remember, we already have simplified in (\ref{eq:BiBbrstSimple})
and (\ref{eq:BiBbrstHatSimple}) the general ansatz for the BRST currents
by reparametrizations to the simple form \begin{eqnarray}
\bs{j}_{z} & = & \ce^{\bs{\gamma}}\dP_{z\bs{\gamma}}\frem{+\alpha'\ce^{\bs{\gamma}}Y_{\bs{\gamma}}(\xfull)\partial_{z}\weyl},\quad\bs{j}_{\bar{z}}=0\\
\hat{\bs{\jmath}}_{\bar{z}} & = & \hat{\ce}^{\hat{\bs{\gamma}}}\hat{\dP}_{\bar{z}\hat{\bs{\gamma}}}\frem{+\alpha'\hat{\ce}^{\hat{\bs{\gamma}}}\hat{Y}_{\hat{\bs{\gamma}}}(\xfull)\partial_{\bar{z}}\weyl},\quad\hat{\bs{\jmath}}_{z}=0\end{eqnarray}
Instead of deriving the corresponding BRST transformations in the
Hamiltonian formalism using the Poisson bracket, we stay in the Lagrangian
formalism and apply Noether's theorem (see (\ref{eq:noet:NoethersTheorem}))
inversely in the sense that we try to express the divergence of the
given currents as linear combinations of the equations of motion in
order to derive the corresponding transformations:\begin{eqnarray}
\bar{\partial}\bs{j}_{z} & \stackrel{!}{=} & -\es\allfields{I}\funktional{S}{\allfields{I}}=-\es_{\gem{cov}}\allfields{I}\funktional{_{\gem{cov}}S}{\allfields{I}}\label{eq:threeInOneI}\\
\partial\hat{\bs{\jmath}}_{\bar{z}} & \stackrel{!}{=} & -\hat{\es}\allfields{I}\funktional{S}{\allfields{I}}=-\hat{\es}_{\gem{cov}}\allfields{I}\funktional{_{\gem{cov}}S}{\allfields{I}}\label{eq:threeInOneII}\end{eqnarray}
Here $\allfields{I}$ is the collection of all the worldsheet fields.
BRST invariance of the action is according to Noether equivalent to
having this special form of the divergences of the currents. These
two equations thus do three things at the same time: The possibility
to write the divergence of the currents as linear combinations of
the equations of motion fixes the precise form of the BRST current.
At the same time it puts constraints on the background fields\rem{conventional supergravity constraints?}:
all terms not proportional to equations of motion have to vanish.
And finally it determines the form of the (covariant) BRST transformations. 

After determining the BRST transformation, the nilpotency conditions
$\es^{2}=0,\left[\es,\hat{\es}\right]=0$ and $\hat{\es}^{2}=0$ put
further constraints on the background fields including the torsion.
Some torsion components can then be further simplified by using two
of the three local Lorentz transformations and scale transformations
which leads to only one remaining local Lorentz transformation and
one local scale transformation. Putting these restrictions on some
torsion components induces via the Bianchi identities further constraints
on other components. All the constraints on the torsion and other
functionals of the background fields combine finally to the target
space supergravity equations of motion. Note that our approach differs
from the one in \cite{Berkovits:2001ue} in two major points. First
of all we stay in the Lagrangian formalism throughout. Second, we
first check the holomorphicity and then the nilpotency. In fact, we
need to do so, because only in the first step we can determine the
BRST transformations of the worldsheet fields which we need in the
Lagrangian formalism to check nilpotency. The BRST transformations
have so far been given only for the heterotic string in \cite{Chandia:2006ix},
so that the transformations in the type II case are a new result.

Let us now perform in more detail the program sketched above:\begin{eqnarray}
\bar{\partial}\bs{j}_{z} & = & \mc{D}_{\bar{z}}\ce^{\bs{\gamma}}\dP_{z\bs{\gamma}}\frem{+\alpha'\mc{D}_{\bar{z}}\ce^{\bs{\gamma}}Y_{\bs{\gamma}}(\xfull)\partial_{z}\weyl}+\ce^{\bs{\gamma}}\mc{D}_{\bar{z}}\dP_{z\bs{\gamma}}\frem{+\alpha'\ce^{\bs{\gamma}}\mc{D}_{\bar{z}}(Y_{\bs{\gamma}}(\xfull)\partial_{z}\weyl)}=\qquad\\
 & = & -\dP_{z\bs{\gamma}}\funktional{S}{\be_{z\bs{\gamma}}}\frem{-\alpha'Y_{\bs{\gamma}}(\xfull)\partial_{z}\weyl\funktional{S}{\be_{z\bs{\gamma}}}}+\ce^{\bs{\gamma}}\mc{D}_{\bar{z}}\dP_{z\bs{\gamma}}\frem{+\alpha'\ce^{\bs{\gamma}}\mc{D}_{\bar{z}}Y_{\bs{\gamma}}(\xfull)\partial_{z}\weyl+\alpha'\ce^{\bs{\gamma}}Y_{\bs{\gamma}}(\xfull)\partial_{\bar{z}}\partial_{z}\weyl}\qquad\label{eq:firstBRSTdivergence}\end{eqnarray}
\rem{with \begin{eqnarray*}
\mc{D}_{\bar{z}}Y_{\bs{\gamma}} & = & \left(\Pi_{\bar{z}}^{B}\gemnabla_{B}Y_{\bs{\gamma}}-\left(C_{\bs{\gamma}}\hoch{\bs{\delta}\hat{\bs{\delta}}}\hat{\dP}_{\bar{z}\hat{\bs{\delta}}}-\hat{\ce}^{\hat{\bs{\alpha}}}S_{\bs{\gamma}\hat{\bs{\alpha}}}\hoch{\bs{\delta}\hat{\bs{\beta}}}\hat{\be}_{\bar{z}\hat{\bs{\beta}}}\right)Y_{\bs{\delta}}\right)\end{eqnarray*}
}In the following we will replace all occurrences of $\mc{D}_{\bar{z}}\dP_{z\bs{\gamma}}$,
$\Pi_{z}^{\hat{\bs{\gamma}}}$, $\Pi_{\bar{z}}^{\bs{\gamma}}$, $\mc{D}_{\bar{z}}\ce^{\bs{\alpha}}$,
$\hat{\mc{D}}_{z}\hat{\ce}^{\hat{\bs{\alpha}}}$, $\mc{D}_{\bar{z}}\be_{z\bs{\alpha}}$,
$\hat{\mc{D}}_{z}\hat{\be}_{\bar{z}\hat{\bs{\alpha}}}$, $\ce\gamma^{a}\ce$
and $\hat{\ce}\gamma^{a}\hat{\ce}$ by the equations of motion (\ref{eq:eomx})-(\ref{eq:eomVIIIb}).
In the end, all terms which are not proportional to the equations
of motion have to vanish which leads to some of the supergravity constraints
while the terms proportional to the equations of motion tell us the
BRST transformation of the elementary fields. In order to extract
$\mc{D}_{\bar{z}}\dP_{z\bs{\gamma}}$ from the $x^{K}$-equation of
motion (\ref{eq:eomx}), let us project (\ref{eq:eomx}) to a flat
spinorial index $\bs{\alpha}$ using some index relabeling: \begin{eqnarray}
\mc{D}_{\bar{z}}\dP_{z\bs{\alpha}} & = & -E_{\bs{\alpha}}\hoch{K}\funktional{_{\gem{cov}}S}{x^{K}}-\gemnabla_{\bar{z}}\Pi_{z}^{D}G_{D\bs{\alpha}}+\nonumber \\
 &  & +\Pi_{z}^{C}\left(\frac{3}{2}H_{\bs{\alpha}CD}-\gemT_{CD|\bs{\alpha}}+2\gem{T}_{\bs{\alpha}(C|D)}+\frac{1}{2}\gemnabla_{\bs{\alpha}}G_{CD}\frem{\leftarrow\checkcovPhi{\bs{\alpha}}G_{CD}}-\gemnabla_{(C}G_{D)\bs{\alpha}}\right)\Pi_{\bar{z}}^{D}+\nonumber \\
 &  & +2T_{\bs{\alpha}D}\hoch{\bs{\gamma}}\Pi_{\bar{z}}^{D}\dP_{z\bs{\gamma}}+2\hat{T}_{\bs{\alpha}C}\hoch{\hat{\bs{\gamma}}}\Pi_{z}^{C}\hat{\dP}_{\bar{z}\hat{\bs{\gamma}}}+\nonumber \\
 &  & +\dP_{z\bs{\gamma}}\left(\gemnabla_{\bs{\alpha}}\RR^{\bs{\gamma}\hat{\bs{\gamma}}}-C_{\bs{\alpha}}\hoch{\bs{\gamma}\hat{\bs{\gamma}}}\right)\hat{\dP}_{\bar{z}\hat{\bs{\gamma}}}+\ce^{\bs{\alpha}_{2}}\gemnabla_{\bs{\alpha}}C_{\bs{\alpha}_{2}}\hoch{\bs{\beta}\hat{\bs{\gamma}}}\be_{z\bs{\beta}}\hat{\dP}_{\bar{z}\hat{\bs{\gamma}}}+\hat{\ce}^{\hat{\bs{\alpha}}}\left(\gemnabla_{\bs{\alpha}}\hat{C}_{\hat{\bs{\alpha}}}\hoch{\hat{\bs{\beta}}\bs{\gamma}}+S_{\bs{\alpha}\hat{\bs{\alpha}}}\hoch{\bs{\gamma}\hat{\bs{\beta}}}\right)\hat{\be}_{\bar{z}\hat{\bs{\beta}}}\dP_{z\bs{\gamma}}+\nonumber \\
 &  & +\ce^{\bs{\alpha}_{2}}\hat{\ce}^{\hat{\bs{\alpha}}}\gemnabla_{\bs{\alpha}}S_{\bs{\alpha}_{2}\hat{\bs{\alpha}}}\hoch{\bs{\beta}\hat{\bs{\beta}}}\be_{z\bs{\beta}}\hat{\be}_{\bar{z}\hat{\bs{\beta}}}-\Omega_{\bs{\alpha}\, a_{1}\ldots a_{4}}(\ce\gamma^{a_{1}\ldots a_{4}a}\ce)\cdot L_{z\bar{z}a}\frem{-4\alpha'\partial\bar{\partial}\weyl\cdot\gemnabla_{\bs{\alpha}}\dil}+\nonumber \\
 &  & +2\Pi_{\bar{z}}^{D}R_{\bs{\alpha}D\bs{\alpha}_{2}}\hoch{\bs{\beta}}\ce^{\bs{\alpha}_{2}}\be_{z\bs{\beta}}+2\Pi_{z}^{C}\hat{R}_{\bs{\alpha}C\hat{\bs{\alpha}}}\hoch{\hat{\bs{\beta}}}\hat{\ce}^{\hat{\bs{\alpha}}}\hat{\be}_{\bar{z}\hat{\bs{\beta}}}\label{eq:projectedEomx}\end{eqnarray}
Already at this point we can determine some constraints on the background
fields. The divergence of the BRST current given in (\ref{eq:firstBRSTdivergence})
has to become a linear combination of the equations of motion. The
term $\gemnabla_{\bar{z}}\Pi_{z}^{D}G_{D\bs{\alpha}}$ in (\ref{eq:projectedEomx})
cannot be compensated by any other term and it also cannot be replaced
by a further equation of motion. The same is true for our beloved
$\Omega_{\bs{\alpha}\, a_{1}\ldots a_{4}}(\ce\gamma^{a_{1}\ldots a_{4}a}\ce)\cdot L_{z\bar{z}a}$.
Using in addition proposition \ref{prop:left-right-symmetry} for
the constraints from the antiholomorphicity of the right-mover BRST
current, we can demand\vRam{.5}{\begin{eqnarray}
G_{A\mc{B}} & \stackrel{!}{=} & 0\quad(\mbox{only }G_{ab}\neq0)\label{eq:endlichMetrikConstraint}\\
\Omega_{\bs{\alpha}\, a_{1}\ldots a_{4}} & \stackrel{!}{=} & 0,\qquad\hat{\Omega}_{\hat{\bs{\alpha}}\, a_{1}\ldots a_{4}}\stackrel{!}{=}0\label{eq:endlichConnectionConstraint}\end{eqnarray}
}\\
With (\ref{eq:endlichConnectionConstraint}) we have finally obtained
the missing ingredient for the reduction of the spinorial connection
coefficients to Lorentz plus scale transformations as it was summarized
already in the remark on page \pageref{remark:structureGroupValuedConnection}
at the end of the section \ref{sec:Antighost-gauge-symmetry} about
the antighost gauge symmetry.

Equation (\ref{eq:endlichMetrikConstraint}) allows us to choose a
frame where $G_{ab}=e^{2\Phi}\eta_{ab}$, such that we reduce also
the bosonic structure group to Lorentz plus scale transformations.
Let us discuss this in more detail in the following intermezzo.\vspace{.5cm}

\lyxline{\normalsize}\vspace{-.25cm}\lyxline{\normalsize}

\subsubsection*{Intermezzo\index{structure group!bosonic $\sim$}\index{intermezzo!reduced bosonic structure group}
about the reduced bosonic structure group}

\label{Intermezzo:bosonic structure group}\index{structure group!bosonic}\index{bosonic structure group!Lorentz plus scale}Due
to (\ref{eq:endlichMetrikConstraint}) we know that $G_{AB}$ is of
the block-diagonal form $G_{AB}=\diag(G_{ab},0,0)$. This means that
the symmetric rank two tensor is of the form\begin{equation}
G_{MN}=E_{M}\hoch{a}G_{ab}E_{N}\hoch{b}\label{eq:metricDegenerate}\end{equation}
In particular we have $G_{mn}=E_{m}\hoch{a}G_{ab}E_{n}\hoch{b}$.
As the $E_{M}\hoch{a}$ were introduced by hand, we may choose $E_{m}\hoch{a}$
orthonormal as usual, i.e. such that $G_{ab}$ becomes the Minkowski
metric. This is at least for the leading component $G_{mn}(\xboson)$
(i.e. $\xbothtetas=0$) a familiar thing to do, but it holds also
in the $\xbothtetas$-dependent case:\vspace{-.4cm} 

\begin{prop}\rem{\index{Sylvester for superspace}\index{proposition!Sylvester for superspace}}\index{orthonormal basis}\index{proposition!orthonormal basis}\label{prop:Superspace-Sylvester}\rem{Literatur?}For
all symmetric rank two tensor fields $G_{mn}(\overbrace{\xfull}^{\{\xboson\lqn{,\xbothtetas\}}})$
whose real body ($\xbothtetas=0$-part) has signature (1,9), there
exists locally a frame $E_{m}\hoch{a}(\xfull)$, such that \begin{equation}
G_{mn}(\underbrace{\xfull}_{\{\xboson\lqn{,\xbothtetas\}}})=E_{m}\hoch{a}(\xfull)\eta_{ab}E_{n}\hoch{b}(\xfull)\label{eq:superSylvester}\end{equation}
Note: In contrast to the ordinary bosonic version, the entries of
the matrices are supernumbers.\end{prop}

\paragraph{Proof }

Due to usual linear algebra, there is an orthonormal basis with respect
to the real symmetric matrix $G_{mn}(\xboson)$, i.e. we can always
find locally $E_{m}\hoch{a}(\xboson)$, s.t. (\ref{eq:superSylvester})
is fulfilled for $\xbothtetas=0$. In order to prove the same for
$\xbothtetas\neq0$, we will make a $\xbothtetas$-expansion of (\ref{eq:superSylvester})
and show that we can always construct a solution $E_{m}\hoch{a}(\xboson,\xbothtetas)$
for arbitrary $\xbothtetas$ from the bosonic solution $E_{m}\hoch{a}(\xboson)$.
Remember the notations $x^{\bs{\mc{M}}}\equiv\tet^{\mc{M}}$ and $\bei{G_{mn}}{}=\bei{G_{mn}}{\xbothtetas=0}$.
The $\xbothtetas$-expansion of (\ref{eq:superSylvester}) then reads
\begin{eqnarray}
\lqn{\sum_{n\geq0}\frac{1}{n!}x^{\bs{\mc{M}}_{1}}\cdots x^{\bs{\mc{M}}_{n}}\bei{(\partial_{\bs{\mc{M}}_{1}}\ldots\partial_{\bs{\mc{M}}_{n}}G_{mn})}{}\stackrel{!}{=}}\nonumber \\
 & \stackrel{!}{=} & \sum_{k,l\geq0}\frac{1}{k!}x^{\bs{\mc{K}}_{1}}\cdots x^{\bs{\mc{K}}_{k}}\bei{(\partial_{\bs{\mc{K}}_{1}}\ldots\partial_{\bs{\mc{K}}_{k}}E_{m}\hoch{a})}{}\eta_{ab}\frac{1}{l!}x^{\bs{\mc{L}}_{1}}\cdots x^{\bs{\mc{L}}_{l}}\bei{(\partial_{\bs{\mc{L}}_{1}}\ldots\partial_{\bs{\mc{L}}_{l}}E_{n}\hoch{b})}{}=\nonumber \\
 & = & \sum_{n\geq0}\frac{1}{n!}x^{\bs{\mc{M}}_{1}}\cdots x^{\bs{\mc{M}}_{n}}\sum_{m=0}^{n}\left(\zwek{n}{m}\right)\bei{(\partial_{\bs{\mc{M}}_{1}}\ldots\partial_{\bs{\mc{M}}_{m}}E_{m}\hoch{a})}{}\eta_{ab}\bei{(\partial_{\bs{\mc{M}}_{m+1}}\ldots\partial_{\bs{\mc{M}}_{n}}E_{n}\hoch{b})}{}\qquad\end{eqnarray}
At $n=0$ we have the solvable bosonic equation $G_{mn}(\xboson)\stackrel{!}{=}E_{m}\hoch{a}(\xboson)\eta_{ab}E_{n}\hoch{b}(\xboson)$
to start with. At higher orders $n$ we have\begin{eqnarray}
\lqn{\bei{(\partial_{\bs{\mc{M}}_{1}}\ldots\partial_{\bs{\mc{M}}_{n}}G_{mn})\stackrel{!}{=}}{}}\nonumber \\
 & \stackrel{!}{=} & \sum_{m=0}^{n}\left(\zwek{n}{m}\right)\bei{(\partial_{\bs{\mc{M}}_{1}}\ldots\partial_{\bs{\mc{M}}_{m}}E_{m}\hoch{a})}{}\eta_{ab}\bei{(\partial_{\bs{\mc{M}}_{m+1}}\ldots\partial_{\bs{\mc{M}}_{n}}E_{n}\hoch{b})}{}=\nonumber \\
 & = & 2\bei{E_{m}\hoch{a}}{}\eta_{ab}\bei{(\partial_{\bs{\mc{M}}_{1}}\ldots\partial_{\bs{\mc{M}}_{n}}E_{n}\hoch{b})}{}+\sum_{m=1}^{n-1}\left(\zwek{n}{m}\right)\bei{(\partial_{\bs{\mc{M}}_{1}}\ldots\partial_{\bs{\mc{M}}_{m}}E_{m}\hoch{a})}{}\eta_{ab}\bei{(\partial_{\bs{\mc{M}}_{m+1}}\ldots\partial_{\bs{\mc{M}}_{n}}E_{n}\hoch{b})}{}\qquad\end{eqnarray}
We thus have the iterative explicit expression for the n-th $\xbothtetas$-derivative
of the vielbein in terms of the $(n-1)$-th and all lower derivatives.
\begin{eqnarray}
\lqn{\bei{(\partial_{\bs{\mc{M}}_{1}}\ldots\partial_{\bs{\mc{M}}_{n}}E_{n}\hoch{d})=}{}}\\
 & = & \frac{1}{2}\eta^{cd}\bei{E_{c}\hoch{m}}{}\Big[\bei{(\partial_{\bs{\mc{M}}_{1}}\ldots\partial_{\bs{\mc{M}}_{n}}G_{mn})}{}-\sum_{m=1}^{n-1}\left(\zwek{n}{m}\right)\bei{(\partial_{\bs{\mc{M}}_{1}}\ldots\partial_{\bs{\mc{M}}_{m}}E_{m}\hoch{a})}{}\eta_{ab}\bei{(\partial_{\bs{\mc{M}}_{m+1}}\ldots\partial_{\bs{\mc{M}}_{n}}E_{n}\hoch{b})}{}\Big]\qquad\nonumber \end{eqnarray}
This completes the proof of the proposition.\hfill$\square$\vspace{.3cm}\enlargethispage*{1.8cm}

In spite of the above proposition, we will not fix $G_{ab}$ to $\eta_{ab}$,
but only up to a conformal factor. This is of course possible by a
redefinition of $E_{M}\hoch{a}$ with the square root of this conformal
factor. The reason for us to do this is the fact that we have for
the spinorial indices not only Lorentz-, but also scale transformations.
It seems natural to keep this scale invariance also for the bosonic
indices, as long as we do not fix the fermionic one (in particular
if we aim at structure group invariant $\gamma$-matrices $\gamma_{\bs{\alpha\beta}}^{a}$).
We thus introduce an auxiliary \textbf{compensator field}\index{compensator field $\Phi$}\index{$\Phi$|itext{compensator field}}
$\Phi(\xfull)$ and choose $E_{m}\hoch{a}$ such that \begin{equation}
\boxed{G_{ab}=e^{2\Phi}\eta_{ab}}\label{eq:metricWithCompensator}\end{equation}
As soon as $E_{m}\hoch{a}(\xfull)$ is chosen appropriately, the remaining
vielbein components $E_{\bs{\mc{M}}}\hoch{a}$ are uniquely determined
via: \begin{eqnarray}
G_{\bs{\mc{M}}n} & \stackrel{!}{=} & E_{\bs{\mc{M}}}\hoch{a}e^{2\Phi}\eta_{ab}E_{n}\hoch{b}\qquad\dann E_{\bs{\mc{M}}}\hoch{a}=G_{\bs{\mc{M}}n}E_{b}\hoch{n}e^{-2\Phi}\eta^{ba}\end{eqnarray}
In summary this means that there is locally always a choice for the
bosonic 1-form $E^{a}=\de x^{M}E_{M}\hoch{a}$, such that $G_{MN}=E_{M}\hoch{a}e^{2\Phi}\eta_{ab}E_{N}\hoch{b}$
or $G_{MN}=E_{M}\hoch{a}\eta_{ab}E_{N}\hoch{b}$, if one does not
introduce the compensator field. The latter form of $G_{MN}$ was
the starting point in \cite{Berkovits:2001ue}, probably motivated
by the integrated vertex operator of the flat space. 

With the compensator field included, the bosonic structure group with
infinitesimal generator $\check{L}_{a}\hoch{b}$ (compare to page
\pageref{eq:structureGroupTrafoOfG} with $\check{\Lambda}_{a}\hoch{b}=\delta_{a}^{b}+\check{L}_{a}\hoch{b}$)
is -- like the fermionic ones -- restricted to Lorentz plus scale
transformations. We should of course also restrict the auxiliary connection
accordingly. \begin{eqnarray}
\check{L}_{a}\hoch{b} & = & \check{L}^{(D)}\delta_{a}^{b}+\check{L}_{a}^{(L)}\hoch{b},\qquad\check{L}_{ab}\equiv\check{L}_{a}\hoch{c}\eta_{cb}=-\check{L}_{ba}\\
\check{\Omega}_{Ma}\hoch{b} & = & \check{\Omega}_{M}^{(D)}\delta_{a}^{b}+\check{\Omega}_{a}^{(L)}\hoch{b},\qquad\check{\Omega}_{Mab}\equiv\check{\Omega}_{Ma}\hoch{c}\eta_{cb}=-\check{\Omega}_{Mba}\end{eqnarray}
The compensator field is a scalar with respect to superdiffeomorphisms.
With respect to the structure group, however, it has to transform
in a special way, in order to make $G_{ab}$ transforming covariantly.
The infinitesimal transformation of $G_{ab}$ under structure group
transformations is $\delta G_{ab}=-2\check{L}_{(a|}\hoch{c}G_{c|b)}=-2\check{L}^{(D)}G_{ab}$
(see (\ref{eq:structureGroupTrafoOfG}) on page \pageref{eq:structureGroupTrafoOfG}).
This transformation results in a simple shift of the compensator field.
For the same reason, also the covariant derivative contains a shift
of $\Phi$:\index{$\nabla$@$\checkcovPhi{M}$}\begin{eqnarray}
\delta\Phi & = & -\check{L}^{(D)}\\
\checkcovPhi{M} & \equiv & \partial_{M}\Phi-\check{\Omega}_{M}^{(D)}\label{eq:checkcovPhi}\\
\gem{\nabla}_{M}G_{AB} & = & 2\checkcovPhi{M}G_{AB}\qquad(=\partial_{M}G_{AB}-2\gem{\Omega}_{M(A|}\hoch{C}G_{C|B)})\vspace{-.5cm}\,\label{eq:covDerG}\end{eqnarray}
\vspace{-.3cm} \lyxline{\normalsize}\vspace{-.25cm}\lyxline{\normalsize}

\vspace{.5cm} Let us return to the calculation of the divergence
of the BRST current and let us finally replace $\mc{D}_{\bar{z}}\dP_{z\bs{\alpha}}$in
(\ref{eq:firstBRSTdivergence}) by the $x^{K}$ equation of motion
given in (\ref{eq:projectedEomx}) (already using (\ref{eq:endlichMetrikConstraint})
and (\ref{eq:endlichConnectionConstraint}))%
\footnote{\label{foot:bosonic-d}\index{footnote!\thefoot. suggestion for bosonic $d_{za}$}The
comparison of the rewritten bosonic $x^{K}$-equation \begin{eqnarray*}
\lqn{\frac{1}{2}\gemnabla_{\bar{z}}(\Pi_{z}^{e}G_{ea})+\frac{1}{2}\gemnabla_{z}(\Pi_{\bar{z}}^{e}G_{ea})=}\\
 & = & -E_{a}\hoch{K}\funktional{_{\gem{cov}}S}{x^{K}}+\Pi_{z}^{C}\left(\frac{3}{2}H_{aCD}+2\gem{T}_{a(C|D)}+\checkcovPhi{a}G_{CD}\right)\Pi_{\bar{z}}^{B}+2T_{aD}\hoch{\bs{\gamma}}\Pi_{\bar{z}}^{D}\dP_{z\bs{\gamma}}+2\hat{T}_{aC}\hoch{\hat{\bs{\gamma}}}\Pi_{z}^{C}\hat{\dP}_{\bar{z}\hat{\bs{\gamma}}}+\\
 &  & +\dP_{z\bs{\gamma}}\gemnabla_{a}\RR^{\bs{\gamma}\hat{\bs{\gamma}}}\hat{\dP}_{\bar{z}\hat{\bs{\gamma}}}+\ce^{\bs{\alpha}}\gemnabla_{a}C_{\bs{\alpha}}\hoch{\bs{\beta}\hat{\bs{\gamma}}}\be_{z\bs{\beta}}\hat{\dP}_{\bar{z}\hat{\bs{\gamma}}}+\hat{\ce}^{\hat{\bs{\alpha}}}\gemnabla_{a}\hat{C}_{\hat{\bs{\alpha}}}\hoch{\hat{\bs{\beta}}\bs{\gamma}}\hat{\be}_{\bar{z}\hat{\bs{\beta}}}\dP_{z\bs{\gamma}}+\\
 &  & +\ce^{\bs{\alpha}}\hat{\ce}^{\hat{\bs{\alpha}}}\gemnabla_{a}S_{\bs{\alpha}\hat{\bs{\alpha}}}\hoch{\bs{\beta}\hat{\bs{\beta}}}\be_{z\bs{\beta}}\hat{\be}_{\bar{z}\hat{\bs{\beta}}}\frem{-4\alpha'\partial\bar{\partial}\weyl\cdot\gemnabla_{a}\dil}+2\Pi_{\bar{z}}^{D}R_{aD\bs{\alpha}}\hoch{\bs{\beta}}\ce^{\bs{\alpha}}\be_{z\bs{\beta}}+2\Pi_{z}^{C}\hat{R}_{aC\hat{\bs{\alpha}}}\hoch{\hat{\bs{\beta}}}\hat{\ce}^{\hat{\bs{\alpha}}}\hat{\be}_{\bar{z}\hat{\bs{\beta}}}\\
\hspace{-1.5cm}\textrm{with }\quad\nabla_{\bar{z}}\dP_{z\bs{\alpha}} & = & -E_{\bs{\alpha}}\hoch{K}\funktional{_{\gem{cov}}S}{x^{K}}+\Pi_{z}^{C}\left(\frac{3}{2}H_{\bs{\alpha}CD}+2\gem{T}_{\bs{\alpha}(C|D)}+\checkcovPhi{\bs{\alpha}}G_{CD}\right)\Pi_{\bar{z}}^{D}+2T_{\bs{\alpha}D}\hoch{\bs{\gamma}}\Pi_{\bar{z}}^{D}\dP_{z\bs{\gamma}}+2\hat{T}_{\bs{\alpha}C}\hoch{\hat{\bs{\gamma}}}\Pi_{z}^{C}\hat{\dP}_{\bar{z}\hat{\bs{\gamma}}}+\\
 &  & +\dP_{z\bs{\gamma}}\gemnabla_{\bs{\alpha}}\RR^{\bs{\gamma}\hat{\bs{\gamma}}}\hat{\dP}_{\bar{z}\hat{\bs{\gamma}}}+\ce^{\bs{\alpha}_{2}}\gemnabla_{\bs{\alpha}}C_{\bs{\alpha}_{2}}\hoch{\bs{\beta}\hat{\bs{\gamma}}}\be_{z\bs{\beta}}\hat{\dP}_{\bar{z}\hat{\bs{\gamma}}}+\hat{\ce}^{\hat{\bs{\alpha}}}\gemnabla_{\bs{\alpha}}\hat{C}_{\hat{\bs{\alpha}}}\hoch{\hat{\bs{\beta}}\bs{\gamma}}\hat{\be}_{\bar{z}\hat{\bs{\beta}}}\dP_{z\bs{\gamma}}+\\
 &  & +\ce^{\bs{\alpha}_{2}}\hat{\ce}^{\hat{\bs{\alpha}}}\gemnabla_{\bs{\alpha}}S_{\bs{\alpha}_{2}\hat{\bs{\alpha}}}\hoch{\bs{\beta}\hat{\bs{\beta}}}\be_{z\bs{\beta}}\hat{\be}_{\bar{z}\hat{\bs{\beta}}}\frem{-4\alpha'\partial\bar{\partial}\weyl\cdot\gemnabla_{\bs{\alpha}}\dil}+2\Pi_{\bar{z}}^{D}R_{\bs{\alpha}D\bs{\alpha}_{2}}\hoch{\bs{\beta}}\ce^{\bs{\alpha}_{2}}\be_{z\bs{\beta}}+2\Pi_{z}^{C}\hat{R}_{\bs{\alpha}C\hat{\bs{\alpha}}}\hoch{\hat{\bs{\beta}}}\hat{\ce}^{\hat{\bs{\alpha}}}\hat{\be}_{\bar{z}\hat{\bs{\beta}}}\end{eqnarray*}
and with $\hat{\nabla}_{z}\hat{\dP}_{\bar{z}\hat{\bs{\alpha}}}$\frem{\begin{eqnarray*}
\hat{\nabla}_{z}\hat{\dP}_{\bar{z}\hat{\bs{\alpha}}} & = & -E_{\hat{\bs{\alpha}}}\hoch{K}\funktional{_{\gem{cov}}S}{x^{K}}+\Pi_{\bar{z}}^{C}\left(-\frac{3}{2}H_{\hat{\bs{\alpha}}CD}+2\gem{T}_{\hat{\bs{\alpha}}(C|D)}+\checkcovPhi{\bs{\alpha}}G_{CD}\right)\Pi_{z}^{D}+2\hat{T}_{\hat{\bs{\alpha}}D}\hoch{\hat{\bs{\gamma}}}\Pi_{z}^{D}\hat{\dP}_{\bar{z}\hat{\bs{\gamma}}}+2T_{\hat{\bs{\alpha}}C}\hoch{\bs{\gamma}}\Pi_{\bar{z}}^{C}\dP_{z\bs{\gamma}}+\\
 &  & +\dP_{z\bs{\gamma}}\gemnabla_{\hat{\bs{\alpha}}}\RR^{\bs{\gamma}\hat{\bs{\gamma}}}\hat{\dP}_{\bar{z}\hat{\bs{\gamma}}}+\ce^{\bs{\alpha}}\gemnabla_{\hat{\bs{\alpha}}}C_{\bs{\alpha}}\hoch{\bs{\beta}\hat{\bs{\gamma}}}\be_{z\bs{\beta}}\hat{\dP}_{\bar{z}\hat{\bs{\gamma}}}+\hat{\ce}^{\hat{\bs{\alpha}}_{2}}\gemnabla_{\hat{\bs{\alpha}}}\hat{C}_{\hat{\bs{\alpha}}_{2}}\hoch{\hat{\bs{\beta}}\bs{\gamma}}\hat{\be}_{\bar{z}\hat{\bs{\beta}}}\dP_{z\bs{\gamma}}+\\
 &  & +\ce^{\bs{\alpha}}\hat{\ce}^{\hat{\bs{\alpha}}_{2}}\gemnabla_{\hat{\bs{\alpha}}}S_{\bs{\alpha}\hat{\bs{\alpha}}_{2}}\hoch{\bs{\beta}\hat{\bs{\beta}}}\be_{z\bs{\beta}}\hat{\be}_{\bar{z}\hat{\bs{\beta}}}\frem{-4\alpha'\partial\bar{\partial}\weyl\cdot\gemnabla_{\hat{\bs{\alpha}}}\dil}+2\Pi_{\bar{z}}^{D}R_{\hat{\bs{\alpha}}D\bs{\alpha}}\hoch{\bs{\beta}}\ce^{\bs{\alpha}}\be_{z\bs{\beta}}+2\Pi_{z}^{C}\hat{R}_{\hat{\bs{\alpha}}C\hat{\bs{\alpha}}_{2}}\hoch{\hat{\bs{\beta}}}\hat{\ce}^{\hat{\bs{\alpha}}_{2}}\hat{\be}_{\bar{z}\hat{\bs{\beta}}}\end{eqnarray*}
} suggests the introduction of \begin{eqnarray*}
\dP_{za} & \equiv & \frac{1}{2}\Pi_{z}^{e}G_{ea},\qquad\dP_{\bar{z}a}\equiv\frac{1}{2}\Pi_{\bar{z}}^{e}G_{ea}\qquad\fussend\end{eqnarray*}
\frem{Hier sind noch Rechnungen versteckt zur Liouville-Feld Bewegungsgleichung}\\
------%
}:\begin{eqnarray}
\bar{\partial}\bs{j}_{z} & = & -\dP_{z\bs{\gamma}}\funktional{S}{\be_{z\bs{\gamma}}}\frem{-\alpha'Y_{\bs{\gamma}}\cdot\partial_{z}\weyl\funktional{S}{\be_{z\bs{\gamma}}}}-\ce^{\bs{\alpha}}E_{\bs{\alpha}}\hoch{K}\funktional{_{\gem{cov}}S}{x^{K}}+\nonumber \\
 &  & +\ce^{\bs{\alpha}}\Pi_{z}^{C}\underbrace{\left(\frac{3}{2}H_{\bs{\alpha}CD}+2\gem{T}_{\bs{\alpha}(C|D)}+\checkcovPhi{\bs{\alpha}}G_{CD}\right)}_{\equiv Y_{\bs{\alpha}CD}}\Pi_{\bar{z}}^{D}+\nonumber \\
 &  & +2\ce^{\bs{\alpha}}T_{\bs{\alpha}D}\hoch{\bs{\gamma}}\Pi_{\bar{z}}^{D}\dP_{z\bs{\gamma}}+2\ce^{\bs{\alpha}}\hat{T}_{\bs{\alpha}C}\hoch{\hat{\bs{\gamma}}}\Pi_{z}^{C}\hat{\dP}_{\bar{z}\hat{\bs{\gamma}}}+\nonumber \\
 &  & +\ce^{\bs{\alpha}}\dP_{z\bs{\gamma}}\left(\gemnabla_{\bs{\alpha}}\RR^{\bs{\gamma}\hat{\bs{\gamma}}}-C_{\bs{\alpha}}\hoch{\bs{\gamma}\hat{\bs{\gamma}}}\right)\hat{\dP}_{\bar{z}\hat{\bs{\gamma}}}+\ce^{\bs{\alpha}}\ce^{\bs{\alpha}_{2}}\gemnabla_{\bs{\alpha}}C_{\bs{\alpha}_{2}}\hoch{\bs{\beta}\hat{\bs{\gamma}}}\be_{z\bs{\beta}}\hat{\dP}_{\bar{z}\hat{\bs{\gamma}}}+\ce^{\bs{\alpha}}\hat{\ce}^{\hat{\bs{\alpha}}}\left(\gemnabla_{\bs{\alpha}}\hat{C}_{\hat{\bs{\alpha}}}\hoch{\hat{\bs{\beta}}\bs{\gamma}}+S_{\bs{\alpha}\hat{\bs{\alpha}}}\hoch{\bs{\gamma}\hat{\bs{\beta}}}\right)\hat{\be}_{\bar{z}\hat{\bs{\beta}}}\dP_{z\bs{\gamma}}+\nonumber \\
 &  & +\ce^{\bs{\alpha}}\ce^{\bs{\alpha}_{2}}\hat{\ce}^{\hat{\bs{\alpha}}}\gemnabla_{\bs{\alpha}}S_{\bs{\alpha}_{2}\hat{\bs{\alpha}}}\hoch{\bs{\beta}\hat{\bs{\beta}}}\be_{z\bs{\beta}}\hat{\be}_{\bar{z}\hat{\bs{\beta}}}+\nonumber \\
 &  & +2\ce^{\bs{\alpha}}\Pi_{\bar{z}}^{D}R_{\bs{\alpha}D\bs{\alpha}_{2}}\hoch{\bs{\beta}}\ce^{\bs{\alpha}_{2}}\be_{z\bs{\beta}}+2\ce^{\bs{\alpha}}\Pi_{z}^{C}\hat{R}_{\bs{\alpha}C\hat{\bs{\alpha}}}\hoch{\hat{\bs{\beta}}}\hat{\ce}^{\hat{\bs{\alpha}}}\hat{\be}_{\bar{z}\hat{\bs{\beta}}}\label{eq:secondBRSTdivergence}\end{eqnarray}
\rem{\begin{eqnarray*}
 &  & +\alpha'\ce^{\bs{\gamma}}\left(\Pi_{\bar{z}}^{B}\gemnabla_{B}Y_{\bs{\gamma}}-\left(C_{\bs{\gamma}}\hoch{\bs{\delta}\hat{\bs{\delta}}}\hat{\dP}_{\bar{z}\hat{\bs{\delta}}}-\hat{\ce}^{\hat{\bs{\alpha}}}S_{\bs{\gamma}\hat{\bs{\alpha}}}\hoch{\bs{\delta}\hat{\bs{\beta}}}\hat{\be}_{\bar{z}\hat{\bs{\beta}}}\right)Y_{\bs{\delta}}\right)\partial_{z}\weyl+\\
 &  & +\alpha'\ce^{\bs{\gamma}}\left(Y_{\bs{\gamma}}-4\gemnabla_{\bs{\gamma}}\dil\right)\partial_{\bar{z}}\partial_{z}\weyl\end{eqnarray*}
}Before we plug in further equations of motion (replacing $\Pi_{\bar{z}}^{\bs{\delta}}$
and $\Pi_{z}^{\hat{\bs{\gamma}}}$) we should notice that we can already
read off some more constraints. Namely $Y_{\bs{\alpha}cd}=Y_{\bs{\alpha}c\hat{\bs{\delta}}}=Y_{\bs{\alpha}\bs{\gamma}d}=Y_{\bs{\alpha}\bs{\gamma}\hat{\bs{\delta}}}=0$\rem{and
$Y_{\bs{\gamma}}=4\alpha'\gemnabla_{\bs{\gamma}}\dil$}. The first
constraint $Y_{\bs{\alpha}cd}=0$ can be separated into symmetric
and antisymmetric part of the indices $c$ and $d$. In addition,
we already add everywhere the constraints coming from the right-moving
BRST current , using proposition \ref{prop:left-right-symmetry} on
page \pageref{prop:left-right-symmetry} ($H\To-H$, $\check{T}\To\check{T}$,
$\check{\nabla}\To\check{\nabla}$)%
\footnote{\index{footnote!\thefoot. independence of choice of bosonic connection $\check \Omega_{Ma}\hoch{b}$}At
first we should remember that $\gemT_{AC}\hoch{B}=\diag(\check{T}_{AC}\hoch{b},T_{AC}\hoch{\bs{\beta}},\hat{T}_{AC}\hoch{\hat{\bs{\beta}}})$.
As $G_{bd}$ are the only non-vanishing components of $G_{BD}$, the
contraction of the upper torsion index with $G_{BD}$ projects out
the first block-diagonal and we can write \[
\gem{T}_{AC|D}=\check{T}_{AC|D}\]
The next important observation is that the constraints are independent
of the choice of the auxiliary bosonic connection $\check{\Omega}_{Ma}\hoch{b}$,
as it should be . The only condition is that it obeys $\check{\Omega}_{M(a|b)}=\check{\Omega}_{M}^{(D)}G_{ab}$
which we used during the derivation by taking $\gemnabla_{M}G_{AB}=2\checkcovPhi{M}G_{AB}$
(see (\ref{eq:covDerG})). Remember also that $\checkcovPhi{\bs{\alpha}}=E_{\bs{\alpha}}\hoch{M}\partial_{M}\Phi-\check{\Omega}_{\bs{\alpha}}^{(D)}$
(\ref{eq:checkcovPhi}). $\check{\Omega}_{Ma}\hoch{b}$ enters the
terms $Y_{\bs{\alpha}CD}$ (defined in (\ref{eq:secondBRSTdivergence})
and containing the constraints) only in the combination $2\check{T}_{\bs{\alpha}(C|D)}-\check{\Omega}_{\bs{\alpha}}^{(D)}G_{CD}$,
where it completely cancels:\begin{eqnarray*}
2\check{T}_{\bs{\alpha}(C|D)}-\check{\Omega}_{\bs{\alpha}}^{(D)}G_{CD} & = & 2(\de E^{b})_{\bs{\alpha}(C|}G_{b|D)}+\check{\Omega}_{\bs{\alpha}(C|D)}-\underbrace{\check{\Omega}_{(C|\bs{\alpha}|D)}}_{=0}-\check{\Omega}_{\bs{\alpha}}^{(D)}G_{CD}=\\
 & = & 2E_{\bs{\alpha}}\hoch{M}E_{(C|}\hoch{N}\partial_{[M}E_{N]}\hoch{b}G_{b|D)}\end{eqnarray*}
In particular the connection does not enter at all the following torsion
component: \begin{eqnarray*}
\check{T}_{\bs{\alpha}\hat{\bs{\delta}}|c} & = & (\de E^{d})_{\bs{\alpha}\hat{\bs{\delta}}}G_{dc}\end{eqnarray*}
The constraints (\ref{eq:Y-constrII})-(\ref{eq:Y-constrIV}) are
therefore independent of the choice of $\check{\Omega}_{Ma}\hoch{b}$.
In particular, we can choose $\Omega_{Ma}\hoch{b}$ (defined by $\Omega_{M\bs{\alpha}}\hoch{\bs{\beta}}$
via $\nabla_{M}\gamma_{\bs{\alpha\beta}}^{c}=0$) or $\hat{\Omega}_{Ma}\hoch{b}$
(defined by $\hat{\Omega}_{M\hat{\bs{\alpha}}}\hoch{\hat{\bs{\beta}}}$
via $\hat{\nabla}_{M}\gamma_{\hat{\bs{\alpha}}\hat{\bs{\beta}}}^{c}=0$).$\qquad\fussend$%
}.\newpage{}\rem{\begin{eqnarray}
Y_{\bs{\alpha}CD} & = & \frac{3}{2}H_{\bs{\alpha}CD}+2\check{T}_{\bs{\alpha}(C|}\hoch{L}G_{L|D)}+\checkcovPhi{\bs{\alpha}}G_{CD}\label{eq:Y}\end{eqnarray}
}\Ram{.6}{\begin{eqnarray}
H_{\bs{\mc{A}}cd} & = & 0\label{eq:Y-constrI}\\
\check{T}_{\bs{\mc{A}}(c|d)} & = & -\frac{1}{2}\checkcovPhi{\bs{\mc{A}}}G_{cd}\label{eq:Y-constrII}\\
\zwek{\frac{3}{2}H_{\bs{\alpha}c\hat{\bs{\delta}}}+\check{T}_{\bs{\alpha}\hat{\bs{\delta}}|c}\lqn{\qquad\!=\quad0}}{-\frac{3}{2}H_{\hat{\bs{\alpha}}c\bs{\delta}}+\check{T}_{\hat{\bs{\alpha}}\bs{\delta}|c}\lqn{\quad\,\;=\quad0}} &  & \quad\quad\Bigg\}\qquad\dann\quad H_{\bs{\alpha}\hat{\bs{\delta}}c}=\check{T}_{\bs{\alpha}\hat{\bs{\delta}}|c}=0\label{eq:Y-constrIII}\\
\frac{3}{2}H_{\bs{\alpha}\bs{\gamma}d}+\check{T}_{\bs{\alpha}\bs{\gamma}|d} & = & 0,\qquad-\frac{3}{2}H_{\hat{\bs{\alpha}}\hat{\bs{\gamma}}d}+\check{T}_{\hat{\bs{\alpha}}\hat{\bs{\gamma}}|d}=0\label{eq:Y-constrIV}\\
H_{\bs{\alpha}\bs{\gamma}\hat{\bs{\delta}}} & = & 0,\qquad H_{\hat{\bs{\alpha}}\hat{\bs{\gamma}}\bs{\delta}}=0\label{eq:Y-constrV}\end{eqnarray}
}\rem{We can already rewrite some of the remaining components of
$Y$ \begin{eqnarray}
Y_{\bs{\alpha}CD} & = & \frac{3}{2}H_{\bs{\alpha}CD}+2\check{T}_{\bs{\alpha}(C|D)}+\checkcovPhi{\bs{\alpha}}G_{CD}\\
Y_{\bs{\alpha}c\bs{\delta}} & = & 3H_{\bs{\alpha}c\bs{\delta}}=2T_{\bs{\alpha}\bs{\delta}|c},\qquad\hat{Y}_{\hat{\bs{\alpha}}c\hat{\bs{\delta}}}=-3H_{\hat{\bs{\alpha}}c\hat{\bs{\delta}}}=2\hat{T}_{\hat{\bs{\alpha}}\hat{\bs{\delta}}|c}\\
Y_{\bs{\alpha}\bs{\gamma}\bs{\delta}} & = & \frac{3}{2}H_{\bs{\alpha}\bs{\gamma}\bs{\delta}},\qquad\hat{Y}_{\hat{\bs{\alpha}}\hat{\bs{\gamma}}\hat{\bs{\delta}}}=-\frac{3}{2}H_{\hat{\bs{\alpha}}\hat{\bs{\gamma}}\hat{\bs{\delta}}}\\
Y_{\bs{\alpha}\hat{\bs{\gamma}}\bs{\delta}} & = & \frac{3}{2}H_{\bs{\alpha}\hat{\bs{\gamma}}\bs{\delta}}=0,\qquad\ddots\\
Y_{\bs{\alpha}\hat{\bs{\gamma}}\hat{\bs{\delta}}} & = & \frac{3}{2}H_{\bs{\alpha}\hat{\bs{\gamma}}\hat{\bs{\delta}}}=0\\
Y_{\bs{\alpha}\hat{\bs{\gamma}}d} & = & 0\end{eqnarray}
}\rem{The divergence of the current can now be written as\begin{eqnarray}
\bar{\partial}j_{z} & = & -P_{z\bs{\gamma}}\funktional{S}{\be_{z\bs{\gamma}}}-\ce^{\bs{\alpha}}E_{\bs{\alpha}}\hoch{K}\funktional{_{\gem{cov}}}{x^{K}}S+\nonumber \\
 &  & +\ce^{\bs{\alpha}}\Pi_{z}^{c}\underbrace{3H_{\bs{\alpha}c\bs{\delta}}}_{2T_{\bs{\alpha}\bs{\delta}|c}}\Pi_{\bar{z}}^{\bs{\delta}}+\frac{3}{2}\ce^{\bs{\alpha}}\Pi_{z}^{\bs{\gamma}}H_{\bs{\alpha}\bs{\gamma}\bs{\delta}}\Pi_{\bar{z}}^{\bs{\delta}}+\nonumber \\
 &  & +2\Pi_{\bar{z}}^{\bs{\delta}}\ce^{\bs{\alpha}}T_{\bs{\alpha}\bs{\delta}}\hoch{\bs{\gamma}}P_{z\bs{\gamma}}+2\Pi_{\bar{z}}^{(d,\hat{\bs{\delta}})}\ce^{\bs{\alpha}}T_{\bs{\alpha}(d,\hat{\bs{\delta}})}\hoch{\bs{\gamma}}P_{z\bs{\gamma}}+\nonumber \\
 &  & +2\Pi_{z}^{\hat{\bs{\gamma}}}\ce^{\bs{\alpha}}\hat{T}_{\bs{\alpha}\hat{\bs{\gamma}}}\hoch{\hat{\bs{\delta}}}\hat{P}_{\bar{z}\hat{\bs{\delta}}}+2\Pi_{z}^{(c,\bs{\gamma})}\ce^{\bs{\alpha}}\hat{T}_{\bs{\alpha}(c,\bs{\gamma})}\hoch{\hat{\bs{\gamma}}}\hat{P}_{\bar{z}\hat{\bs{\gamma}}}+\nonumber \\
 &  & +2\ce^{\bs{\alpha}}\hat{\ce}^{\hat{\bs{\alpha}}}\Pi_{z}^{\hat{\bs{\gamma}}}\hat{R}_{\bs{\alpha}\hat{\bs{\gamma}}\hat{\bs{\alpha}}}\hoch{\hat{\bs{\beta}}}\hat{\be}_{\bar{z}\hat{\bs{\beta}}}+2\ce^{\bs{\alpha}}\hat{\ce}^{\hat{\bs{\alpha}}}\Pi_{z}^{(c,\bs{\gamma})}\hat{R}_{\bs{\alpha}(c,\bs{\gamma})\hat{\bs{\alpha}}}\hoch{\hat{\bs{\beta}}}\hat{\be}_{\bar{z}\hat{\bs{\beta}}}+\nonumber \\
 &  & +2\Pi_{\bar{z}}^{\bs{\delta}}\ce^{\bs{\alpha}_{2}}R_{\bs{\alpha}_{2}\bs{\delta}\bs{\alpha}}\hoch{\bs{\beta}}\ce^{\bs{\alpha}}\be_{z\bs{\beta}}+2\Pi_{\bar{z}}^{(d,\hat{\bs{\delta}})}\ce^{\bs{\alpha}_{2}}R_{\bs{\alpha}_{2}(d,\hat{\bs{\delta}})\bs{\alpha}}\hoch{\bs{\beta}}\ce^{\bs{\alpha}}\be_{z\bs{\beta}}+\nonumber \\
 &  & +P_{z\bs{\gamma}}\ce^{\bs{\alpha}}\left(\tilde{\nabla}_{\bs{\alpha}}\RR^{\bs{\gamma}\hat{\bs{\gamma}}}-C_{\bs{\alpha}}\hoch{\bs{\gamma}\hat{\bs{\gamma}}}\right)\hat{P}_{\bar{z}\hat{\bs{\gamma}}}+\nonumber \\
 &  & +\ce^{\bs{\alpha}_{2}}\ce^{\bs{\alpha}}\tilde{\nabla}_{\bs{\alpha}_{2}}C_{\bs{\alpha}}\hoch{\bs{\beta}\hat{\bs{\gamma}}}\be_{z\bs{\beta}}\hat{P}_{\bar{z}\hat{\bs{\gamma}}}+\ce^{\bs{\alpha}}\hat{\ce}^{\hat{\bs{\alpha}}}\left(\tilde{\nabla}_{\bs{\alpha}}\hat{C}_{\hat{\bs{\alpha}}}\hoch{\hat{\bs{\beta}}\bs{\gamma}}+S_{\bs{\alpha}\hat{\bs{\alpha}}}\hoch{\bs{\gamma}\hat{\bs{\beta}}}\right)\hat{\be}_{\bar{z}\hat{\bs{\beta}}}P_{z\bs{\gamma}}+\nonumber \\
 &  & +\ce^{\bs{\alpha}_{2}}\ce^{\bs{\alpha}}\hat{\ce}^{\hat{\bs{\alpha}}}\tilde{\nabla}_{\bs{\alpha}_{2}}S_{\bs{\alpha}\hat{\bs{\alpha}}}\hoch{\bs{\beta}\hat{\bs{\beta}}}\be_{z\bs{\beta}}\hat{\be}_{\bar{z}\hat{\bs{\beta}}}\end{eqnarray}
\rem{\begin{eqnarray*}
 &  & +4\alpha'\ce^{\bs{\gamma}}\left(\Pi_{\bar{z}}^{B}\gemnabla_{B}\gemnabla_{\bs{\gamma}}\dil-\left(C_{\bs{\gamma}}\hoch{\bs{\delta}\hat{\bs{\delta}}}\hat{\dP}_{\bar{z}\hat{\bs{\delta}}}-\hat{\ce}^{\hat{\bs{\alpha}}}S_{\bs{\gamma}\hat{\bs{\alpha}}}\hoch{\bs{\delta}\hat{\bs{\beta}}}\hat{\be}_{\bar{z}\hat{\bs{\beta}}}\right)\gemnabla_{\bs{\delta}}\dil\right)\partial_{z}\weyl\end{eqnarray*}
}}So far we have used only the equations of motion obtained by the
variational derivative with respect to the antighosts and with respect
to $x^{K}$. There still remain the ones with respect to the ghosts,
with respect to the Lagrange multipliers and with respect to $\dP_{z\bs{\alpha}}$
and $\hat{\dP}_{\bar{z}\hat{\bs{\alpha}}}$. The first ones simply
will not enter the calculation and the pure spinor constraints (coming
from the Lagrange multipliers) will be used at the very end. So let
us remind ourselves the variational derivatives with respect to $\dP_{z\bs{\alpha}}$
and $\hat{\dP}_{\bar{z}\hat{\bs{\alpha}}}$ ((\ref{eq:eomPi}) and
(\ref{eq:eomPibar})): \begin{eqnarray}
\Pi_{\bar{z}}^{\bs{\delta}} & = & \funktional{S}{\dP_{z\bs{\delta}}}-\RR^{\bs{\delta}\hat{\bs{\gamma}}}\hat{\dP}_{\bar{z}\hat{\bs{\gamma}}}-\hat{\ce}^{\hat{\bs{\alpha}}}\hat{C}_{\hat{\bs{\alpha}}}\hoch{\hat{\bs{\beta}}\bs{\delta}}\hat{\be}_{\bar{z}\hat{\bs{\beta}}},\qquad\Pi_{z}^{\hat{\bs{\gamma}}}=\funktional{S}{\hat{\dP}_{\bar{z}\hat{\bs{\gamma}}}}-\dP_{z\bs{\gamma}}\RR^{\bs{\gamma}\hat{\bs{\gamma}}}-\ce^{\bs{\alpha}}C_{\bs{\alpha}}\hoch{\bs{\beta}\hat{\bs{\gamma}}}\be_{z\bs{\beta}}\end{eqnarray}
Together with the new constraints (\ref{eq:Y-constrI})-(\ref{eq:Y-constrV})
we plug them into the divergence (\ref{eq:secondBRSTdivergence})
of the BRST current \rem{and arrive at \begin{eqnarray}
\bar{\partial}\bs{j}_{z} & = & -\dP_{z\bs{\gamma}}\funktional{S}{\be_{z\bs{\gamma}}}-\ce^{\bs{\alpha}}E_{\bs{\alpha}}\hoch{K}\funktional{_{\tilde{cov}}}{x^{K}}S+\nonumber \\
 &  & +\ce^{\bs{\alpha}}\Pi_{z}^{c}\underbrace{3H_{\bs{\alpha}c\bs{\delta}}}_{2T_{\bs{\alpha}\bs{\delta}|c}}\left(\funktional{S}{P_{z\bs{\delta}}}-\RR^{\bs{\delta}\hat{\bs{\gamma}}}\hat{P}_{\bar{z}\hat{\bs{\gamma}}}-\hat{\ce}^{\hat{\bs{\alpha}}}\hat{C}_{\hat{\bs{\alpha}}}\hoch{\hat{\bs{\beta}}\bs{\delta}}\hat{\be}_{\bar{z}\hat{\bs{\beta}}}\right)+\nonumber \\
 &  & +\frac{3}{2}\ce^{\bs{\alpha}}\Pi_{z}^{\bs{\gamma}}H_{\bs{\alpha}\bs{\gamma}\bs{\delta}}\left(\funktional{S}{P_{z\bs{\delta}}}-\RR^{\bs{\delta}\hat{\bs{\gamma}}}\hat{P}_{\bar{z}\hat{\bs{\gamma}}}-\hat{\ce}^{\hat{\bs{\alpha}}}\hat{C}_{\hat{\bs{\alpha}}}\hoch{\hat{\bs{\beta}}\bs{\delta}}\hat{\be}_{\bar{z}\hat{\bs{\beta}}}\right)+\nonumber \\
 &  & +2\left(\funktional{S}{P_{z\bs{\delta}}}-\RR^{\bs{\delta}\hat{\bs{\gamma}}}\hat{P}_{\bar{z}\hat{\bs{\gamma}}}-\hat{\ce}^{\hat{\bs{\alpha}}}\hat{C}_{\hat{\bs{\alpha}}}\hoch{\hat{\bs{\beta}}\bs{\delta}}\hat{\be}_{\bar{z}\hat{\bs{\beta}}}\right)\ce^{\bs{\alpha}}T_{\bs{\alpha}\bs{\delta}}\hoch{\bs{\gamma}}P_{z\bs{\gamma}}+2\Pi_{\bar{z}}^{(d,\hat{\bs{\delta}})}\ce^{\bs{\alpha}}T_{\bs{\alpha}(d,\hat{\bs{\delta}})}\hoch{\bs{\gamma}}P_{z\bs{\gamma}}+\nonumber \\
 &  & +2\left(\funktional{S}{\hat{P}_{\bar{z}\hat{\bs{\gamma}}}}-P_{z\bs{\gamma}}\RR^{\bs{\gamma}\hat{\bs{\gamma}}}-\ce^{\bs{\alpha}_{2}}C_{\bs{\alpha}_{2}}\hoch{\bs{\beta}\hat{\bs{\gamma}}}\be_{z\bs{\beta}}\right)\ce^{\bs{\alpha}}\hat{T}_{\bs{\alpha}\hat{\bs{\gamma}}}\hoch{\hat{\bs{\delta}}}\hat{P}_{\bar{z}\hat{\bs{\delta}}}+2\Pi_{z}^{(c,\bs{\gamma})}\ce^{\bs{\alpha}}\hat{T}_{\bs{\alpha}(c,\bs{\gamma})}\hoch{\hat{\bs{\gamma}}}\hat{P}_{\bar{z}\hat{\bs{\gamma}}}+\nonumber \\
 &  & +2\ce^{\bs{\alpha}}\hat{\ce}^{\hat{\bs{\alpha}}}\left(\funktional{S}{\hat{P}_{\bar{z}\hat{\bs{\gamma}}}}-P_{z\bs{\gamma}}\RR^{\bs{\gamma}\hat{\bs{\gamma}}}-\ce^{\bs{\alpha}_{2}}C_{\bs{\alpha}_{2}}\hoch{\bs{\beta}\hat{\bs{\gamma}}}\be_{z\bs{\beta}}\right)\hat{R}_{\bs{\alpha}\hat{\bs{\gamma}}\hat{\bs{\alpha}}}\hoch{\hat{\bs{\beta}}}\hat{\be}_{\bar{z}\hat{\bs{\beta}}}+2\ce^{\bs{\alpha}}\hat{\ce}^{\hat{\bs{\alpha}}}\Pi_{z}^{(c,\bs{\gamma})}\hat{R}_{\bs{\alpha}(c,\bs{\gamma})\hat{\bs{\alpha}}}\hoch{\hat{\bs{\beta}}}\hat{\be}_{\bar{z}\hat{\bs{\beta}}}+\nonumber \\
 &  & +2\left(\funktional{S}{P_{z\bs{\delta}}}-\RR^{\bs{\delta}\hat{\bs{\gamma}}}\hat{P}_{\bar{z}\hat{\bs{\gamma}}}-\hat{\ce}^{\hat{\bs{\alpha}}}\hat{C}_{\hat{\bs{\alpha}}}\hoch{\hat{\bs{\beta}}\bs{\delta}}\hat{\be}_{\bar{z}\hat{\bs{\beta}}}\right)\ce^{\bs{\alpha}_{2}}R_{\bs{\alpha}_{2}\bs{\delta}\bs{\alpha}}\hoch{\bs{\beta}}\ce^{\bs{\alpha}}\be_{z\bs{\beta}}+2\Pi_{\bar{z}}^{(d,\hat{\bs{\delta}})}\ce^{\bs{\alpha}_{2}}R_{\bs{\alpha}_{2}(d,\hat{\bs{\delta}})\bs{\alpha}}\hoch{\bs{\beta}}\ce^{\bs{\alpha}}\be_{z\bs{\beta}}+\nonumber \\
 &  & +P_{z\bs{\gamma}}\ce^{\bs{\alpha}}\left(\tilde{\nabla}_{\bs{\alpha}}\RR^{\bs{\gamma}\hat{\bs{\gamma}}}-C_{\bs{\alpha}}\hoch{\bs{\gamma}\hat{\bs{\gamma}}}\right)\hat{P}_{\bar{z}\hat{\bs{\gamma}}}+\nonumber \\
 &  & +\ce^{\bs{\alpha}_{2}}\ce^{\bs{\alpha}}\tilde{\nabla}_{\bs{\alpha}_{2}}C_{\bs{\alpha}}\hoch{\bs{\beta}\hat{\bs{\gamma}}}\be_{z\bs{\beta}}\hat{P}_{\bar{z}\hat{\bs{\gamma}}}+\ce^{\bs{\alpha}}\hat{\ce}^{\hat{\bs{\alpha}}}\left(\tilde{\nabla}_{\bs{\alpha}}\hat{C}_{\hat{\bs{\alpha}}}\hoch{\hat{\bs{\beta}}\bs{\gamma}}+S_{\bs{\alpha}\hat{\bs{\alpha}}}\hoch{\bs{\gamma}\hat{\bs{\beta}}}\right)\hat{\be}_{\bar{z}\hat{\bs{\beta}}}P_{z\bs{\gamma}}+\nonumber \\
 &  & +\ce^{\bs{\alpha}_{2}}\ce^{\bs{\alpha}}\hat{\ce}^{\hat{\bs{\alpha}}}\tilde{\nabla}_{\bs{\alpha}_{2}}S_{\bs{\alpha}\hat{\bs{\alpha}}}\hoch{\bs{\beta}\hat{\bs{\beta}}}\be_{z\bs{\beta}}\hat{\be}_{\bar{z}\hat{\bs{\beta}}}\end{eqnarray}
\rem{\begin{eqnarray*}
 &  & +\left(4\alpha'\partial_{z}\weyl\cdot\ce^{\bs{\alpha}}\gemnabla_{\bs{\beta}}\gemnabla_{\bs{\alpha}}\dil\right)\funktional{S}{\dP_{z\bs{\beta}}}+4\alpha'\ce^{\bs{\alpha}}\Pi_{\bar{z}}^{(b,\hat{\bs{\beta}})}\left(\gemnabla_{(b,\hat{\bs{\beta}})}\gemnabla_{\bs{\alpha}}\dil\right)\partial_{z}\weyl-4\alpha'\ce^{\bs{\alpha}}\left(C_{\bs{\alpha}}\hoch{\bs{\delta}\hat{\bs{\delta}}}\gemnabla_{\bs{\delta}}\dil+\RR^{\bs{\beta}\hat{\bs{\delta}}}\gemnabla_{\bs{\beta}}\gemnabla_{\bs{\alpha}}\dil\right)\partial_{z}\weyl\,\hat{\dP}_{\bar{z}\hat{\bs{\delta}}}+\\
 &  & +4\alpha'\ce^{\bs{\alpha}}\hat{\ce}^{\hat{\bs{\alpha}}}\left(S_{\bs{\alpha}\hat{\bs{\alpha}}}\hoch{\bs{\delta}\hat{\bs{\beta}}}\gemnabla_{\bs{\delta}}\dil-\hat{C}_{\hat{\bs{\alpha}}}\hoch{\hat{\bs{\beta}}\bs{\beta}}\gemnabla_{\bs{\beta}}\gemnabla_{\bs{\alpha}}\dil\right)\hat{\be}_{\bar{z}\hat{\bs{\beta}}}\partial_{z}\weyl\end{eqnarray*}
}}In a last effort we sort all the terms with respect to the appearance
of the elementary fields and finally arrive at\begin{eqnarray}
\bar{\partial}\bs{j}_{z} & = & -\dP_{z\bs{\gamma}}\funktional{S}{\be_{z\bs{\gamma}}}-\ce^{\bs{\alpha}}E_{\bs{\alpha}}\hoch{K}\funktional{_{\gem{cov}}}{x^{K}}S+\nonumber \\
 &  & +\ce^{\bs{\alpha}}\Big(\frac{3}{2}\Pi_{z}^{\bs{\gamma}}H_{\bs{\alpha}\bs{\gamma}\bs{\delta}}+2T_{\bs{\alpha}\bs{\delta}}\hoch{\bs{\gamma}}\dP_{z\bs{\gamma}}-2\ce^{\bs{\alpha}_{2}}R_{\bs{\alpha}_{2}\bs{\delta}\bs{\alpha}}\hoch{\bs{\beta}}\be_{z\bs{\beta}}+\Pi_{z}^{c}\underbrace{3H_{\bs{\alpha}c\bs{\delta}}}_{2\check{T}_{\bs{\alpha}\bs{\delta}|c}\,\lqn{{\scriptstyle (\ref{eq:Y-constrIV})}}}\Big)\funktional{S}{\dP_{z\bs{\delta}}}+\nonumber \\
 &  & +\ce^{\bs{\alpha}}\Big(2\hat{T}_{\bs{\alpha}\hat{\bs{\gamma}}}\hoch{\hat{\bs{\delta}}}\hat{\dP}_{\bar{z}\hat{\bs{\delta}}}+2\hat{\ce}^{\hat{\bs{\alpha}}}\hat{R}_{\bs{\alpha}\hat{\bs{\gamma}}\hat{\bs{\alpha}}}\hoch{\hat{\bs{\beta}}}\hat{\be}_{\bar{z}\hat{\bs{\beta}}}\Big)\funktional{S}{\hat{\dP}_{\bar{z}\hat{\bs{\gamma}}}}+\nonumber \\
\nonumber \\ &  & +\ce^{\bs{\alpha}}\Pi_{z}^{c}\Big(-\underbrace{3H_{\bs{\alpha}c\bs{\delta}}}_{2\check{T}_{\bs{\alpha}\bs{\delta}|c}\,\lqn{{\scriptstyle (\ref{eq:Y-constrIV})}}}\RR^{\bs{\delta}\hat{\bs{\gamma}}}+2\hat{T}_{\bs{\alpha}c}\hoch{\hat{\bs{\gamma}}}\Big)\hat{\dP}_{\bar{z}\hat{\bs{\gamma}}}+\ce^{\bs{\alpha}}\Pi_{z}^{\bs{\gamma}}\left(2\hat{T}_{\bs{\alpha}\bs{\gamma}}\hoch{\hat{\bs{\gamma}}}-\frac{3}{2}H_{\bs{\alpha}\bs{\gamma}\bs{\delta}}\RR^{\bs{\delta}\hat{\bs{\gamma}}}\right)\hat{\dP}_{\bar{z}\hat{\bs{\gamma}}}+\nonumber \\
 &  & +\ce^{\bs{\alpha}}\dP_{z\bs{\gamma}}\left(2T_{\bs{\alpha}d}\hoch{\bs{\gamma}}\right)\Pi_{\bar{z}}^{d}+2\ce^{\bs{\alpha}}\dP_{z\bs{\gamma}}\left(T_{\bs{\alpha}\hat{\bs{\delta}}}\hoch{\bs{\gamma}}\right)\Pi_{\bar{z}}^{\hat{\bs{\delta}}}+\nonumber \\
 &  & +\ce^{\bs{\alpha}}\dP_{z\bs{\gamma}}\left(\gemnabla_{\bs{\alpha}}\RR^{\bs{\gamma}\hat{\bs{\gamma}}}-C_{\bs{\alpha}}\hoch{\bs{\gamma}\hat{\bs{\gamma}}}-2T_{\bs{\alpha}\bs{\delta}}\hoch{\bs{\gamma}}\RR^{\bs{\delta}\hat{\bs{\gamma}}}-2\hat{T}_{\bs{\alpha}\hat{\bs{\delta}}}\hoch{\hat{\bs{\gamma}}}\RR^{\bs{\gamma}\hat{\bs{\delta}}}\right)\hat{\dP}_{\bar{z}\hat{\bs{\gamma}}}+\nonumber \\
 &  & +\ce^{\bs{\alpha}}\hat{\ce}^{\hat{\bs{\alpha}}}\Pi_{z}^{c}\Big(-\underbrace{3H_{\bs{\alpha}c\bs{\delta}}}_{2\check{T}_{\bs{\alpha}\bs{\delta}|c}\,\lqn{{\scriptstyle (\ref{eq:Y-constrIV})}}}\hat{C}_{\hat{\bs{\alpha}}}\hoch{\hat{\bs{\beta}}\bs{\delta}}+2\hat{R}_{\bs{\alpha}c\hat{\bs{\alpha}}}\hoch{\hat{\bs{\beta}}}\Big)\hat{\be}_{\bar{z}\hat{\bs{\beta}}}+\ce^{\bs{\alpha}}\hat{\ce}^{\hat{\bs{\alpha}}}\Pi_{z}^{\bs{\gamma}}\left(2\hat{R}_{\bs{\alpha}\bs{\gamma}\hat{\bs{\alpha}}}\hoch{\hat{\bs{\beta}}}-\frac{3}{2}H_{\bs{\alpha}\bs{\gamma}\bs{\delta}}\hat{C}_{\hat{\bs{\alpha}}}\hoch{\hat{\bs{\beta}}\bs{\delta}}\right)\hat{\be}_{\bar{z}\hat{\bs{\beta}}}+\nonumber \\
 &  & +\ce^{\bs{\alpha}}\hat{\ce}^{\hat{\bs{\alpha}}}\dP_{z\bs{\gamma}}\left(\gemnabla_{\bs{\alpha}}\hat{C}_{\hat{\bs{\alpha}}}\hoch{\hat{\bs{\beta}}\bs{\gamma}}+S_{\bs{\alpha}\hat{\bs{\alpha}}}\hoch{\bs{\gamma}\hat{\bs{\beta}}}-2T_{\bs{\alpha}\bs{\delta}}\hoch{\bs{\gamma}}\hat{C}_{\hat{\bs{\alpha}}}\hoch{\hat{\bs{\beta}}\bs{\delta}}-2\hat{R}_{\bs{\alpha}\hat{\bs{\gamma}}\hat{\bs{\alpha}}}\hoch{\hat{\bs{\beta}}}\RR^{\bs{\gamma}\hat{\bs{\gamma}}}\right)\hat{\be}_{\bar{z}\hat{\bs{\beta}}}+\ce^{\bs{\alpha}_{1}}\ce^{\bs{\alpha}_{2}}X_{\bs{\alpha}_{1}\bs{\alpha}_{2}}\qquad\label{eq:thirdBRSTdivergence}\end{eqnarray}
\rem{\begin{eqnarray*}
 &  & +\left(4\alpha'\partial_{z}\weyl\cdot\ce^{\bs{\alpha}}\gemnabla_{\bs{\beta}}\gemnabla_{\bs{\alpha}}\dil\right)\funktional{S}{\dP_{z\bs{\beta}}}+4\alpha'\ce^{\bs{\alpha}}\Pi_{\bar{z}}^{(b,\hat{\bs{\beta}})}\left(\gemnabla_{(b,\hat{\bs{\beta}})}\gemnabla_{\bs{\alpha}}\dil\right)\partial_{z}\weyl-4\alpha'\ce^{\bs{\alpha}}\left(C_{\bs{\alpha}}\hoch{\bs{\delta}\hat{\bs{\delta}}}\gemnabla_{\bs{\delta}}\dil+\RR^{\bs{\beta}\hat{\bs{\delta}}}\gemnabla_{\bs{\beta}}\gemnabla_{\bs{\alpha}}\dil\right)\partial_{z}\weyl\,\hat{\dP}_{\bar{z}\hat{\bs{\delta}}}+\\
 &  & +4\alpha'\ce^{\bs{\alpha}}\hat{\ce}^{\hat{\bs{\alpha}}}\left(S_{\bs{\alpha}\hat{\bs{\alpha}}}\hoch{\bs{\delta}\hat{\bs{\beta}}}\gemnabla_{\bs{\delta}}\dil-\hat{C}_{\hat{\bs{\alpha}}}\hoch{\hat{\bs{\beta}}\bs{\beta}}\gemnabla_{\bs{\beta}}\gemnabla_{\bs{\alpha}}\dil\right)\hat{\be}_{\bar{z}\hat{\bs{\beta}}}\partial_{z}\weyl\end{eqnarray*}
}where we defined an extra symbol for the terms coming quadratic
in the ghost $\ce^{\bs{\alpha}}$:\begin{eqnarray}
X_{\bs{\alpha}_{1}\bs{\alpha}_{2}} & \equiv & 2\left(R_{[\bs{\alpha}_{1}|d|\bs{\alpha}_{2}]}\hoch{\bs{\beta}}\right)\Pi_{\bar{z}}^{d}\be_{z\bs{\beta}}+2\Pi_{\bar{z}}^{\hat{\bs{\delta}}}\left(R_{[\bs{\alpha}_{1}|\hat{\bs{\delta}}|\bs{\alpha}_{2}]}\hoch{\bs{\beta}}\right)\be_{z\bs{\beta}}+\nonumber \\
 &  & +\left(\gemnabla_{[\bs{\alpha}_{1}}C_{\bs{\alpha}_{2}]}\hoch{\bs{\beta}\hat{\bs{\gamma}}}-2\hat{T}_{[\bs{\alpha}_{1}|\hat{\bs{\delta}}}\hoch{\hat{\bs{\gamma}}}C_{|\bs{\alpha}_{2}]}\hoch{\bs{\beta}\hat{\bs{\delta}}}-2R_{[\bs{\alpha}_{1}|\bs{\delta}|\bs{\alpha}_{2}]}\hoch{\bs{\beta}}\RR^{\bs{\delta}\hat{\bs{\gamma}}}\right)\hat{\dP}_{\bar{z}\hat{\bs{\gamma}}}\be_{z\bs{\beta}}+\nonumber \\
 &  & +\hat{\ce}^{\hat{\bs{\alpha}}}\left(\gemnabla_{[\bs{\alpha}_{1}}S_{\bs{\alpha}_{2}]\hat{\bs{\alpha}}}\hoch{\bs{\beta}\hat{\bs{\beta}}}+2\hat{R}_{[\bs{\alpha}_{1}|\hat{\bs{\gamma}}\hat{\bs{\alpha}}}\hoch{\hat{\bs{\beta}}}C_{|\bs{\alpha}_{2}]}\hoch{\bs{\beta}\hat{\bs{\gamma}}}+2R_{[\bs{\alpha}_{1}|\bs{\delta}|\bs{\alpha}_{2}]}\hoch{\bs{\beta}}\hat{C}_{\hat{\bs{\alpha}}}\hoch{\hat{\bs{\beta}}\bs{\delta}}\right)\be_{z\bs{\beta}}\hat{\be}_{\bar{z}\hat{\bs{\beta}}}\label{eq:Xbispinor}\end{eqnarray}
Summarizing, we observe that we managed -- with the help of the equations
of motion -- to turn the simple equation (\ref{eq:firstBRSTdivergence})
into a quite lengthy one ... We are not going to copy the whole long
equation again for the next step. The only equation of motion that
we may still apply, is the pure spinor constraint \begin{equation}
\funktional{S}{L_{z\bar{z}a}}=\frac{1}{2}(\ce\gamma^{a}\ce)\end{equation}
We therefore can concentrate on the term $\ce^{\bs{\alpha}_{1}}X_{\bs{\alpha}_{1}\bs{\alpha}_{2}}\ce^{\bs{\alpha}_{2}}$,
where the pure spinor combination $\ce\gamma^{a}\ce$ might appear.
As discussed in footnote \ref{foot:LorentzScaleReason} on page \pageref{foot:LorentzScaleReason}
(see also the appendix-subsection \ref{sub:Vanishing-of-gamma-traces}
on page \pageref{sub:Vanishing-of-gamma-traces}), all graded antisymmetric
$16\times16$ matrices can be expanded in $\gamma^{[1]}$ and $\gamma^{[5]}$:\begin{eqnarray}
X_{\bs{\alpha}_{1}\bs{\alpha}_{2}} & \equiv & X_{a}\gamma_{\bs{\alpha}_{1}\bs{\alpha}_{2}}^{a}+X_{a_{1}\ldots a_{5}}\gamma_{\bs{\alpha}_{1}\bs{\alpha}_{2}}^{a_{1}\ldots a_{5}}\label{eq:XbispinorExpansion}\\
X_{a} & \stackrel{(\ref{eq:gamma:gammaexpansionOddII})}{=} & \frem{\frac{1}{16}X_{\alpha_{1}\alpha_{2}}\gamma_{a}^{\alpha_{1}\alpha_{2}}=}\frac{1}{16}\gamma_{a}^{\bs{\alpha}_{2}\bs{\alpha}_{1}}X_{\bs{\alpha}_{1}\bs{\alpha}_{2}}\qquad\left(=-\frac{1}{16}\gamma_{a}^{\bs{\alpha}_{1}\bs{\alpha}_{2}}X_{\bs{\alpha}_{1}\bs{\alpha}_{2}}\right)\label{eq:XbispinorEinsform}\\
X_{a_{1}\ldots a_{5}} & \stackrel{(\ref{eq:gamma:gammaexpansionOddII})}{=} & \frac{1}{32\cdot5!}\gamma_{a_{5}\ldots a_{1}}^{\bs{\alpha}_{2}\bs{\alpha}_{1}}X_{\bs{\alpha}_{1}\bs{\alpha}_{2}}\label{eq:XbispinorFiveform}\end{eqnarray}
We can use this to rewrite the quadratic ghost term as follows:\begin{eqnarray}
\ce^{\bs{\alpha}_{1}}X_{\bs{\alpha}_{1}\bs{\alpha}_{2}}\ce^{\bs{\alpha}_{2}} & = & -\frac{1}{8}\gamma_{a}^{\bs{\alpha}_{1}\bs{\alpha}_{2}}X_{\bs{\alpha}_{1}\bs{\alpha}_{2}}\funktional{S}{L_{z\bar{z}a}}+\frac{1}{32\cdot5!}\gamma_{a_{1}\ldots a_{5}}^{\bs{\alpha}_{2}\bs{\alpha}_{1}}X_{\bs{\alpha}_{1}\bs{\alpha}_{2}}(\ce\gamma^{a_{1}\ldots a_{5}}\ce)\label{eq:thirdBRSTdivergenceSupplement}\end{eqnarray}
This was the last ingredient to determine all remaining constraints
on the background fields and also to be able to read off all BRST
transformations (including the one for the Lagrange multiplier). Let
us start with the constraints. In addition to (\ref{eq:Y-constrI})-(\ref{eq:Y-constrV}),
we get the following constraints on the background fields:\vRam{1.01}{\begin{eqnarray}
\hat{T}_{\bs{\alpha}c}\hoch{\hat{\bs{\gamma}}} & = & \underbrace{\check{T}_{\bs{\alpha}\bs{\delta}|c}}_{\frac{3}{2}H_{\bs{\alpha}c\bs{\delta}}\,\lqn{{\scriptstyle (\ref{eq:Y-constrIV})}}}\RR^{\bs{\delta}\hat{\bs{\gamma}}},\qquad T_{\hat{\bs{\alpha}}c}\hoch{\bs{\gamma}}=\underbrace{\check{T}_{\hat{\bs{\alpha}}\hat{\bs{\delta}}|c}}_{-\frac{3}{2}H_{\hat{\bs{\alpha}}c\hat{\bs{\delta}}}\,\lqn{{\scriptstyle (\ref{eq:Y-constrIV})}}}\RR^{\bs{\gamma}\hat{\bs{\delta}}}\label{eq:holConstrI}\\
\hat{T}_{\bs{\alpha}\bs{\gamma}}\hoch{\hat{\bs{\gamma}}} & = & \frac{3}{4}H_{\bs{\alpha}\bs{\gamma}\bs{\delta}}\RR^{\bs{\delta}\hat{\bs{\gamma}}},\qquad T_{\hat{\bs{\alpha}}\hat{\bs{\gamma}}}\hoch{\bs{\gamma}}=-\frac{3}{4}H_{\hat{\bs{\alpha}}\hat{\bs{\gamma}}\hat{\bs{\delta}}}\RR^{\bs{\gamma}\hat{\bs{\delta}}}\label{eq:holConstrII}\\
T_{\bs{\alpha}d}\hoch{\bs{\gamma}} & = & 0,\qquad\hat{T}_{\hat{\bs{\alpha}}d}\hoch{\hat{\bs{\gamma}}}=0\label{eq:holConstrIII}\\
T_{\bs{\alpha}\hat{\bs{\delta}}}\hoch{\bs{\gamma}} & = & 0,\qquad\hat{T}_{\hat{\bs{\alpha}}\bs{\delta}}\hoch{\hat{\bs{\gamma}}}=0,\quad\stackrel{(\ref{eq:Y-constrIII})}{\dann}\gemT_{\bs{\alpha}\hat{\bs{\alpha}}}\hoch{K}=0\label{eq:holConstrIV}\\
C_{\bs{\alpha}}\hoch{\bs{\gamma}\hat{\bs{\gamma}}} & = & \gemnabla_{\bs{\alpha}}\RR^{\bs{\gamma}\hat{\bs{\gamma}}}-2T_{\bs{\alpha}\bs{\delta}}\hoch{\bs{\gamma}}\RR^{\bs{\delta}\hat{\bs{\gamma}}}-2\underbrace{\hat{T}_{\bs{\alpha}\hat{\bs{\delta}}}\hoch{\hat{\bs{\gamma}}}}_{=0\,(\ref{eq:holConstrIV})}\RR^{\bs{\gamma}\hat{\bs{\delta}}}\label{eq:holConstrVa}\\
\hat{C}_{\hat{\bs{\alpha}}}\hoch{\hat{\bs{\gamma}}\bs{\gamma}} & = & \gemnabla_{\hat{\bs{\alpha}}}\RR^{\bs{\gamma}\hat{\bs{\gamma}}}-2\hat{T}_{\hat{\bs{\alpha}}\hat{\bs{\delta}}}\hoch{\hat{\bs{\gamma}}}\RR^{\bs{\gamma}\hat{\bs{\delta}}}\label{eq:holConstrVb}\\
\hat{R}_{\bs{\alpha}c\hat{\bs{\alpha}}}\hoch{\hat{\bs{\beta}}} & = & \underbrace{\frac{3}{2}H_{\bs{\alpha}c\bs{\delta}}}_{\check{T}_{\bs{\alpha}\bs{\delta}|c}\,\lqn{{\scriptstyle (\ref{eq:Y-constrIV})}}}\hat{C}_{\hat{\bs{\alpha}}}\hoch{\hat{\bs{\beta}}\bs{\delta}},\qquad R_{\hat{\bs{\alpha}}c\bs{\alpha}}\hoch{\bs{\beta}}=\underbrace{-\frac{3}{2}H_{\hat{\bs{\alpha}}c\hat{\bs{\delta}}}}_{\check{T}_{\hat{\bs{\alpha}}\hat{\bs{\delta}}|c}\,\lqn{{\scriptstyle (\ref{eq:Y-constrIV})}}}C_{\bs{\alpha}}\hoch{\bs{\beta}\hat{\bs{\delta}}}\label{eq:holConstrVI}\\
\hat{R}_{\bs{\alpha}\bs{\gamma}\hat{\bs{\alpha}}}\hoch{\hat{\bs{\beta}}} & = & \frac{3}{4}H_{\bs{\alpha}\bs{\gamma}\bs{\delta}}\hat{C}_{\hat{\bs{\alpha}}}\hoch{\hat{\bs{\beta}}\bs{\delta}},\qquad R_{\hat{\bs{\alpha}}\hat{\bs{\gamma}}\bs{\alpha}}\hoch{\bs{\beta}}=-\frac{3}{4}H_{\hat{\bs{\alpha}}\hat{\bs{\gamma}}\hat{\bs{\delta}}}C_{\bs{\alpha}}\hoch{\bs{\beta}\hat{\bs{\delta}}}\label{eq:holConstrVII}\\
S_{\bs{\alpha}\hat{\bs{\alpha}}}\hoch{\bs{\gamma}\hat{\bs{\beta}}} & = & -\gemnabla_{\bs{\alpha}}\underbrace{\hat{C}_{\hat{\bs{\alpha}}}\hoch{\hat{\bs{\beta}}\bs{\gamma}}}_{\gemnabla_{\hat{\bs{\alpha}}}\RR^{\bs{\gamma}\hat{\bs{\beta}}}-2\hat{T}_{\hat{\bs{\alpha}}\hat{\bs{\delta}}}\hoch{\hat{\bs{\beta}}}\RR^{\bs{\gamma}\hat{\bs{\delta}}}\lqn{{\scriptstyle \,(\ref{eq:holConstrVb})}}}+2T_{\bs{\alpha}\bs{\delta}}\hoch{\bs{\gamma}}\hat{C}_{\hat{\bs{\alpha}}}\hoch{\hat{\bs{\beta}}\bs{\delta}}+2\hat{R}_{\bs{\alpha}\hat{\bs{\gamma}}\hat{\bs{\alpha}}}\hoch{\hat{\bs{\beta}}}\RR^{\bs{\gamma}\hat{\bs{\gamma}}}\label{eq:holConstrVIIIa}\\
S_{\bs{\alpha}\hat{\bs{\alpha}}}\hoch{\bs{\beta}\hat{\bs{\gamma}}} & = & -\gemnabla_{\hat{\bs{\alpha}}}\underbrace{C_{\bs{\alpha}}\hoch{\bs{\beta}\hat{\bs{\gamma}}}}_{\gemnabla_{\bs{\alpha}}\RR^{\bs{\beta}\hat{\bs{\gamma}}}-2T_{\bs{\alpha}\bs{\delta}}\hoch{\bs{\beta}}\RR^{\bs{\delta}\hat{\bs{\gamma}}}\lqn{{\scriptstyle \,(\ref{eq:holConstrVa})}}}+2\hat{T}_{\hat{\bs{\alpha}}\hat{\bs{\delta}}}\hoch{\hat{\bs{\gamma}}}C_{\bs{\alpha}}\hoch{\bs{\beta}\hat{\bs{\delta}}}+2R_{\hat{\bs{\alpha}}\bs{\gamma}\bs{\alpha}}\hoch{\bs{\beta}}\RR^{\bs{\gamma}\hat{\bs{\gamma}}}\label{eq:holConstrVIIIb}\\
\gamma_{a_{1}\ldots a_{5}}^{\bs{\alpha}_{1}\bs{\alpha}_{2}}R_{d\bs{\alpha}_{1}\bs{\alpha}_{2}}\hoch{\bs{\beta}} & = & 0,\qquad\gamma_{a_{1}\ldots a_{5}}^{\hat{\bs{\alpha}}_{1}\hat{\bs{\alpha}}_{2}}\hat{R}_{d\hat{\bs{\alpha}}_{1}\hat{\bs{\alpha}}_{2}}\hoch{\hat{\bs{\beta}}}=0\label{eq:holConstrIX}\\
\gamma_{a_{1}\ldots a_{5}}^{\bs{\alpha}_{1}\bs{\alpha}_{2}}R_{\hat{\bs{\delta}}\bs{\alpha}_{1}\bs{\alpha}_{2}}\hoch{\bs{\beta}} & = & 0,\qquad\gamma_{a_{1}\ldots a_{5}}^{\hat{\bs{\alpha}}_{1}\hat{\bs{\alpha}}_{2}}\hat{R}_{\bs{\delta}\hat{\bs{\alpha}}_{1}\hat{\bs{\alpha}}_{2}}\hoch{\hat{\bs{\beta}}}=0\label{eq:holConstrX}\\
\gamma_{a_{1}\ldots a_{5}}^{\bs{\alpha}_{1}\bs{\alpha}_{2}}\left(\gemnabla_{\bs{\alpha}_{2}}C_{\bs{\alpha}_{1}}\hoch{\bs{\beta}\hat{\bs{\gamma}}}\right) & = & 2\gamma_{a_{1}\ldots a_{5}}^{\bs{\alpha}_{1}\bs{\alpha}_{2}}\Big(R_{\bs{\alpha}_{2}\bs{\delta}\bs{\alpha}_{1}}\hoch{\bs{\beta}}\RR^{\bs{\delta}\hat{\bs{\gamma}}}-\underbrace{\hat{T}_{\bs{\alpha}_{1}\hat{\bs{\delta}}}\hoch{\hat{\bs{\gamma}}}}_{=0\lqn{{\scriptstyle \,(\ref{eq:holConstrIV})}}}C_{\bs{\alpha}_{2}}\hoch{\bs{\beta}\hat{\bs{\delta}}}\Big),\qquad\mbox{plus hatted version }\ldots\label{eq:holConstrXIa}\\
\gamma_{a_{1}\ldots a_{5}}^{\bs{\alpha}_{1}\bs{\alpha}_{2}}\left(\gemnabla_{\bs{\alpha}_{2}}S_{\bs{\alpha}_{1}\hat{\bs{\alpha}}}\hoch{\bs{\beta}\hat{\bs{\beta}}}\right) & = & 2\gamma_{a_{1}\ldots a_{5}}^{\bs{\alpha}_{1}\bs{\alpha}_{2}}\left(\hat{R}_{\bs{\alpha}_{1}\hat{\bs{\gamma}}\hat{\bs{\alpha}}}\hoch{\hat{\bs{\beta}}}C_{\bs{\alpha}_{2}}\hoch{\bs{\beta}\hat{\bs{\gamma}}}-R_{\bs{\alpha}_{2}\bs{\delta}\bs{\alpha}_{1}}\hoch{\bs{\beta}}\hat{C}_{\hat{\bs{\alpha}}}\hoch{\hat{\bs{\beta}}\bs{\delta}}\right),\qquad\mbox{plus hatted version}\ldots\qquad\label{eq:holConstrXIIa}\end{eqnarray}
}\\
Note that on the constraint surface the condition $\gamma_{a_{1}\ldots a_{5}}^{\bs{\alpha}_{1}\bs{\alpha}_{2}}X_{\bs{\alpha}_{1}\bs{\alpha}_{2}}=0$
is equivalent to the vanishing of $X_{\bs{\alpha}_{1}\bs{\alpha}_{2}}$
when contracted with two ghost fields:\begin{equation}
\gamma_{a_{1}\ldots a_{5}}^{\bs{\alpha}_{1}\bs{\alpha}_{2}}X_{\bs{\alpha}_{1}\bs{\alpha}_{2}}=0\qquad\stackrel{(\ref{eq:XbispinorExpansion})-(\ref{eq:XbispinorFiveform})}{\iff}X_{[\bs{\alpha}_{1}\bs{\alpha}_{2}]}=\frac{1}{16}(\gamma_{a}^{\bs{\alpha}_{4}\bs{\alpha}_{3}}X_{\bs{\alpha}_{3}\bs{\alpha}_{4}})\gamma_{\bs{\alpha}_{1}\bs{\alpha}_{2}}^{a}\qquad\stackrel{(\ce\gamma^{a}\ce)=0}{\iff}\qquad\ce^{\bs{\alpha}_{1}}X_{\bs{\alpha}_{1}\bs{\alpha}_{2}}\ce^{\bs{\alpha}_{2}}=0\label{eq:bispinorConstraintInDifferentVersions}\end{equation}
The above equivalences hold for general bispinors, not only for the
one defined in (\ref{eq:Xbispinor}). It is not necessary to memorize
the constraints (\ref{eq:holConstrXIa}) and (\ref{eq:holConstrXIIa})
as they are a consequence of other constraints anyway. We will show
this fact at the end of section \vref{sec:Residual-shift-reparametrization}.

Let us now devote a new section to the BRST transformations that we
can likewise read off from (\ref{eq:thirdBRSTdivergence}).

\section{The covariant BRST transformations}

Remember that we started on page \pageref{eq:threeInOneI} with the
demand $\bar{\partial}\bs{j}_{z}\stackrel{!}{=}-\es_{\gem{cov}}\allfields{I}\funktional{_{\gem{cov}}S}{\allfields{I}}$.
The covariant BRST transformations $\es_{\gem{cov}}\allfields{I}$
have to be understood in the sense of the covariant variation defined
in (\ref{eq:covVarI})-(\ref{eq:covVarV}). We have for example $\es_{\gem{cov}}\hat{\ce}^{\hat{\bs{\alpha}}}=\es_{\hat{cov}}\hat{\ce}^{\hat{\bs{\alpha}}}=\es\hat{\ce}^{\hat{\bs{\alpha}}}+\es x^{M}\hat{\Omega}_{M\hat{\bs{\beta}}}\hoch{\hat{\bs{\alpha}}}\hat{\ce}^{\hat{\bs{\beta}}}$.
When the constraints of the end of last section are fulfilled, we
can read off the covariant BRST transformations $\es_{\gem{cov}}\allfields{I}$
from equation (\ref{eq:thirdBRSTdivergence}) together with (\ref{eq:thirdBRSTdivergenceSupplement}).
Again we give at the same time (using proposition \vref{prop:left-right-symmetry})
the results for the right-mover BRST-symmetry $\hat{\es}$, defined
via%
\footnote{\label{foot:BRSTofd}\index{footnote!\thefoot. BRST of d, mixed first-second order formalism}Another
way to write down the BRST transformations for $\dP_{z\bs{\delta}}$
and $\hat{\dP}_{\bar{z}\hat{\bs{\gamma}}}$ is the following\begin{eqnarray*}
\es_{cov}\dP_{z\bs{\delta}} & = & -\frac{3}{2}\ce^{\bs{\alpha}}\Pi_{z}^{\{c,\bs{\gamma}\}}H_{\bs{\alpha}\{c,\bs{\gamma}\}\bs{\delta}}-\ce^{\bs{\alpha}}\gemT_{\bs{\alpha}\bs{\delta}}\hoch{\{c,\bs{\gamma}\}}\{G_{cd}\Pi_{z}^{d}\,,\,2\dP_{z\bs{\gamma}}\}+2\ce^{\bs{\alpha}}\ce^{\bs{\alpha}_{2}}R_{\bs{\alpha}_{2}\bs{\delta}\bs{\alpha}}\hoch{\bs{\beta}}\be_{z\bs{\beta}}\\
\es_{\hat{cov}}\hat{\dP}_{\bar{z}\hat{\bs{\gamma}}} & = & -\frac{3}{2}\ce^{\bs{\alpha}}\Pi_{\bar{z}}^{\{d,\hat{\bs{\delta}}\}}\underbrace{H_{\bs{\alpha}\{d,\hat{\bs{\delta}}\}\hat{\bs{\gamma}}}}_{=0}-\ce^{\bs{\alpha}}\underbrace{\gemT_{\bs{\alpha}\hat{\bs{\gamma}}}\hoch{\{d,\hat{\bs{\delta}}\}}}_{=0}\{G_{dc}\Pi_{\bar{z}}^{c}\,,\,2\hat{\dP}_{\bar{z}\hat{\bs{\delta}}}\}-2\ce^{\bs{\alpha}}\hat{\ce}^{\hat{\bs{\alpha}}}\hat{R}_{\bs{\alpha}\hat{\bs{\gamma}}\hat{\bs{\alpha}}}\hoch{\hat{\bs{\beta}}}\hat{\be}_{\bar{z}\hat{\bs{\beta}}}\end{eqnarray*}
In the second line for the first two terms, we have just used a complicated
way to write zero. The reason was to bring it to a form similar to
the one in the first line. In any case, at least the first line suggests
again the introduction of the variables\begin{eqnarray*}
\dP_{zc} & \equiv & \frac{1}{2}G_{cd}\Pi_{z}^{d},\qquad\dP_{\bar{z}c}\equiv\frac{1}{2}G_{cd}\Pi_{\bar{z}}^{d}\end{eqnarray*}
that we already proposed in footnote \vref{foot:bosonic-d}. Indeed,
their BRST transformation takes the form\begin{eqnarray*}
\es_{\check{cov}}\dP_{z\, c} & = & -\frac{3}{2}\ce^{\bs{\alpha}}\Pi_{z}^{\bs{\beta}}H_{\bs{\alpha}\bs{\beta}c}-2\ce^{\bs{\alpha}}\check{T}_{\bs{\alpha}c}\hoch{d}\dP_{z\, d}\end{eqnarray*}
\frem{Steps inbetween are \begin{eqnarray*}
\es_{\check{cov}}\dP_{z/\bar{z}\, c} & = & \ce^{\bs{\alpha}}\underbrace{\checkcovPhi{\bs{\alpha}}G_{cd}}_{-2\check{T}_{\bs{\alpha}(c|d)}}\Pi_{z/\bar{z}}^{d}+\ce^{\bs{\alpha}}\Pi_{z/\bar{z}}^{B}\check{T}_{\bs{\alpha}B|c}=\\
 & = & -2\ce^{\bs{\alpha}}\check{T}_{\bs{\alpha}(c|d)}\Pi_{z/\bar{z}}^{d}+\ce^{\bs{\alpha}}\Pi_{z/\bar{z}}^{d}\check{T}_{\bs{\alpha}d|c}+\ce^{\bs{\alpha}}\Pi_{z/\bar{z}}^{\bs{\beta}}\underbrace{\check{T}_{\bs{\alpha}\bs{\beta}|c}}_{=-\frac{3}{2}H_{\bs{\alpha}\bs{\beta}c}}+\ce^{\bs{\alpha}}\Pi_{z/\bar{z}}^{\hat{\bs{\beta}}}\underbrace{\check{T}_{\bs{\alpha}\hat{\bs{\beta}}|c}}_{0}\end{eqnarray*}
}Using $H_{a\bs{\beta}c}=T_{\bs{\alpha}c}\hoch{\bs{\delta}}=0$ and
at (least for $\ce\gamma^{a}\ce=0$) $\ce^{\bs{\alpha}}\ce^{\bs{\alpha}_{2}}R_{\bs{\alpha}_{2}d\bs{\alpha}}\hoch{\bs{\beta}}=0$,
the transformation of $\dP_{z\, c}$ takes the same form as the one
of $\dP_{z\bs{\delta}}$ and we can write \begin{eqnarray*}
\es_{\gem{cov}}\dP_{z\{d,\bs{\delta}\}} & = & -\frac{3}{2}\ce^{\bs{\alpha}}\Pi_{z}^{\{c,\bs{\gamma}\}}H_{\bs{\alpha}\{c,\bs{\gamma}\}\{d,\bs{\delta}\}}-2\ce^{\bs{\alpha}}\gemT_{\bs{\alpha}\{d,\bs{\delta}\}}\hoch{\{c,\bs{\gamma}\}}\dP_{z\{c,\bs{\gamma}\}}-2\ce^{\bs{\alpha}_{1}}\ce^{\bs{\alpha}_{2}}R_{\{d,\bs{\delta}\}\bs{\alpha}_{2}\bs{\alpha}_{1}}\hoch{\bs{\beta}}\be_{z\bs{\beta}}\quad\mbox{for }(\ce\gamma^{a}\ce)=0\end{eqnarray*}
We suggest to introduce $\dP_{zd}$ as an independent variable into
the action, with an on-shell value $\dP_{zc}\equiv\frac{1}{2}G_{cd}\Pi_{z}^{d}$.
Doing this, one would arrive at a formalism where the $G_{MN}$ term
is replaced by a first order term, while the $B_{MN}$ term remains.
This would therefore be a mixed first-second order formalism which
would be suitable to couple it to e.g. the components of a generalized
complex structure. \frem{Guess:\begin{eqnarray*}
\es_{\tilde{cov}}P_{zD} & = & -\frac{3}{2}\ce^{\bs{\alpha}}\Pi_{z}^{C}H_{\bs{\alpha}CD}-2\ce^{\bs{\alpha}}T_{\bs{\alpha}D}\hoch{C}P_{zC}-2\ce^{\bs{\alpha}}R_{\bs{\alpha}D\bs{\alpha}_{2}}\hoch{\bs{\beta}}\ce^{\bs{\alpha}_{2}}\be_{z\bs{\beta}}\\
\es_{\tilde{cov}}\hat{P}_{\bar{z}D} & = & \frac{3}{2}\ce^{\bs{\alpha}}\Pi_{\bar{z}}^{C}H_{\bs{\alpha}CD}-2\ce^{\bs{\alpha}}\hat{T}_{\bs{\alpha}D}\hoch{C}\hat{P}_{\bar{z}C}-2\ce^{\bs{\alpha}}\hat{R}_{\bs{\alpha}D\hat{\bs{\alpha}}}\hoch{\hat{\bs{\beta}}}\hat{\ce}^{\hat{\bs{\alpha}}}\hat{\be}_{\bar{z}\hat{\bs{\beta}}}\end{eqnarray*}
}$\qquad\fussend$%
} $\partial\hat{\bs{\jmath}}_{\bar{z}}\stackrel{!}{=}-\hat{\es}_{\gem{cov}}\allfields{I}\funktional{_{\gem{cov}}S}{\allfields{I}}$:\vRam{1.05}{\begin{eqnarray}
\es x^{M} & = & \ce^{\bs{\alpha}}E_{\bs{\alpha}}\hoch{M},\qquad\hat{\es}x^{M}=\hat{\ce}^{\hat{\bs{\alpha}}}E_{\hat{\bs{\alpha}}}\hoch{M}\label{eq:covBRSTofx}\\
\es_{cov}\ce^{\bs{\alpha}} & = & 0=\hat{\es}_{cov}\ce^{\bs{\alpha}},\qquad\hat{\es}_{\hat{cov}}\hat{\ce}^{\hat{\bs{\alpha}}}=0=\es_{\hat{cov}}\hat{\ce}^{\hat{\bs{\alpha}}}\label{eq:covBRSTofc}\\
\es_{cov}\be_{z\bs{\alpha}} & = & \dP_{z\bs{\alpha}},\qquad\hat{\es}_{cov}\be_{z\bs{\alpha}}=0,\qquad\hat{\es}_{\hat{cov}}\hat{\be}_{\bar{z}\hat{\bs{\alpha}}}=\hat{\dP}_{\bar{z}\hat{\bs{\alpha}}},\qquad\es_{\hat{cov}}\hat{\be}_{\bar{z}\hat{\bs{\alpha}}}=0\label{eq:covBRSTofb}\\
\es_{cov}\dP_{z\bs{\delta}} & = & -\ce^{\bs{\alpha}}\Pi_{z}^{c}\underbrace{3H_{\bs{\alpha}c\bs{\delta}}}_{2\check{T}_{\bs{\alpha}\bs{\delta}|c}}-\frac{3}{2}\ce^{\bs{\alpha}}\Pi_{z}^{\bs{\gamma}}H_{\bs{\alpha}\bs{\gamma}\bs{\delta}}-2\ce^{\bs{\alpha}}T_{\bs{\alpha}\bs{\delta}}\hoch{\bs{\gamma}}\dP_{z\bs{\gamma}}+2\ce^{\bs{\alpha}}\ce^{\bs{\alpha}_{2}}R_{\bs{\alpha}_{2}\bs{\delta}\bs{\alpha}}\hoch{\bs{\beta}}\be_{z\bs{\beta}}\label{eq:covBRSTofd}\\
\hat{\es}_{\hat{cov}}\hat{\dP}_{\bar{z}\hat{\bs{\delta}}} & = & \hat{\ce}^{\hat{\bs{\alpha}}}\Pi_{\bar{z}}^{c}\underbrace{3H_{\hat{\bs{\alpha}}c\hat{\bs{\delta}}}}_{-2\check{T}_{\hat{\bs{\alpha}}\hat{\bs{\delta}}|c}}+\frac{3}{2}\hat{\ce}^{\hat{\bs{\alpha}}}\Pi_{\bar{z}}^{\hat{\bs{\gamma}}}H_{\hat{\bs{\alpha}}\hat{\bs{\gamma}}\hat{\bs{\delta}}}-2\hat{\ce}^{\hat{\bs{\alpha}}}\hat{T}_{\hat{\bs{\alpha}}\hat{\bs{\delta}}}\hoch{\hat{\bs{\gamma}}}\hat{\dP}_{\bar{z}\hat{\bs{\gamma}}}+2\hat{\ce}^{\hat{\bs{\alpha}}}\hat{\ce}^{\hat{\bs{\alpha}}_{2}}\hat{R}_{\hat{\bs{\alpha}}_{2}\hat{\bs{\delta}}\hat{\bs{\alpha}}}\hoch{\hat{\bs{\beta}}}\be_{\bar{z}\hat{\bs{\beta}}}\qquad\label{eq:covhatBRSTofhatd}\\
\es_{\hat{cov}}\hat{\dP}_{\bar{z}\hat{\bs{\gamma}}} & = & -2\ce^{\bs{\alpha}}\underbrace{\hat{T}_{\bs{\alpha}\hat{\bs{\gamma}}}\hoch{\hat{\bs{\delta}}}}_{=0}\hat{\dP}_{\bar{z}\hat{\bs{\delta}}}-2\ce^{\bs{\alpha}}\hat{\ce}^{\hat{\bs{\alpha}}}\hat{R}_{\bs{\alpha}\hat{\bs{\gamma}}\hat{\bs{\alpha}}}\hoch{\hat{\bs{\beta}}}\hat{\be}_{\bar{z}\hat{\bs{\beta}}}\label{eq:covBRSTofhatd}\\
\hat{\es}_{cov}\dP_{z\bs{\gamma}} & = & -2\hat{\ce}^{\hat{\bs{\alpha}}}\underbrace{T_{\hat{\bs{\alpha}}\bs{\gamma}}\hoch{\bs{\delta}}}_{=0}\dP_{z\bs{\delta}}-2\hat{\ce}^{\hat{\bs{\alpha}}}\ce^{\bs{\alpha}}R_{\hat{\bs{\alpha}}\bs{\gamma}\bs{\alpha}}\hoch{\bs{\beta}}\be_{z\bs{\beta}}\label{eq:covhatBRSTofd}\\
\es_{cov}L_{z\bar{z}a} & = & \frac{1}{8}\gamma_{a}^{\bs{\alpha}_{1}\bs{\alpha}_{2}}X_{\bs{\alpha}_{1}\bs{\alpha}_{2}},\qquad\hat{\es}_{cov}L_{z\bar{z}a}=0,\qquad\hat{\es}_{\hat{cov}}\hat{L}_{\bar{z}z\, a}=\frac{1}{8}\gamma_{a}^{\hat{\bs{\alpha}}_{1}\hat{\bs{\alpha}}_{2}}\hat{X}_{\hat{\bs{\alpha}}_{1}\hat{\bs{\alpha}}_{2}},\qquad\es_{\hat{cov}}\hat{L}_{\bar{z}z}=0\qquad\label{eq:covBRSTofL}\end{eqnarray}
} The composite object $X_{\bs{\alpha}_{1}\bs{\alpha}_{2}}$ is given
in (\ref{eq:Xbispinor}). Let us for completeness also give the BRST
transformation of the supersymmetric momentum \begin{eqnarray}
\es_{\gem{cov}}\Pi_{z/\bar{z}}^{A} & \stackrel{(\ref{eq:covVarOfPi}\,)}{=} & \nabla_{z/\bar{z}}\ce^{\bs{\alpha}}\delta_{\bs{\alpha}}\hoch{A}+2\ce^{\bs{\alpha}}\Pi_{z/\bar{z}}^{B}\gemT_{\bs{\alpha}B}\hoch{A}\label{eq:covBRSTofPi}\\
\hat{\es}_{\gem{cov}}\Pi_{z/\bar{z}}^{A} & \stackrel{(\ref{eq:covVarOfPi}\,)}{=} & \hat{\nabla}_{z/\bar{z}}\hat{\ce}^{\hat{\bs{\alpha}}}\delta_{\hat{\bs{\alpha}}}\hoch{A}+2\hat{\ce}^{\hat{\bs{\alpha}}}\Pi_{z/\bar{z}}^{B}\gemT_{\hat{\bs{\alpha}}B}\hoch{A}\label{eq:covHatBRSTofPi}\end{eqnarray}
All these BRST transformations are similar to those for the heterotic
string, given in \cite{Chandia:2006ix}. There it was also noted that
the BRST transformations always contain a Lorentz transformation (multiplication
with the connection). We have absorbed this term into the definition
of the covariant variation. The advantage is that we then have expressions
all the time that are covariant with respect to the target space structure
group. Although the ordinary BRST differential $\es$ is needed to
calculate the cohomology (as it squares to zero), the calculations
are simpler if they are performed with $\es_{\gem{cov}}$ and only
in the end transferred to $\es$. When acting on a target space scalar,
the two coincide anyway.

\section{Graded commutation of left- and right-moving BRST differential}

\label{sec:Graded-commutation}We have started in flat background
with two independent BRST symmetries, the left-moving and the right-moving
one, which both squared to zero and graded commuted. As they define
the physical spectrum and identify physically equivalent states, these
facts should not change in a consistent theory, at least on-shell.
This is similar to the fact that gauge symmetries should not be broken.
We have already derived the constraints coming from a vanishing divergence
of the BRST currents. The ansatz for the currents was such that this
corresponds to holomorphicity for $\bs{j}_{z}$ and antiholomorphicity
for $\hat{\bs{\jmath}}_{\bar{z}}$. Having on-shell a holomorphic
$\bs{j}_{z}$ and an antiholomorphic $\hat{\bs{\jmath}}_{\bar{z}}$
is in a conformal theory already enough to make the corresponding
symmetries commute\rem{true?}. For example on the level of operators,
the operator product between a holomorphic and an antiholomorphic
current always vanishes on-shell. The same is true for the charges
which generate the symmetry.  The on-shell vanishing of the commutators
is all that we can demand for consistency. Therefore we do not expect
any additional information from the graded commutation of left- and
right-moving BRST differential. Nevertheless it is instructive to
calculate the graded commutators and consider it as a further check.
In particular it is interesting to see the terms which prevent an
off-shell commutation of the differentials. The starting point is
the request that we have\begin{eqnarray}
[\hat{\es},\es]\allfields{I} & \stackrel{!}{=} & \delta_{(\mu)}\allfields{I}+\delta_{(\hat{\mu})}\allfields{I}+\delta_{\mbox{triv}}\allfields{I}\label{eq:commutationCondition}\end{eqnarray}
where $\delta_{\mbox{triv}}\allfields{I}$ is a trivial and thus
on-shell vanishing gauge transformation (see page \pageref{thm:trafoOnshellZero}
in the appendix) while $\delta_{(\mu)}$ and $\delta_{(\hat{\mu})}$
are the antighost gauge transformations. Spelled out in words, (\ref{eq:commutationCondition})
means that the graded commutator $[\hat{\es},\es]$ has to vanish
on shell up to antighost gauge transformations. There are at least
two ways to check this. Either we calculate the commutator of the
transformations on each worldsheet field or we calculate the transformations
of the Noether currents. This is directly related to calculating the
Poisson brackets of the generating charges in the Hamiltonian formalism.

\paragraph{Determining $[\es,\hat{\es}]$ via the transformation of the currents }

We start with the defining equations of the BRST currents: \begin{eqnarray}
\bar{\partial}\bs{j}_{z} & = & -\es\allfields{I}\funktional{S}{\allfields{I}},\quad\partial\hat{\bs{\jmath}}_{\bar{z}}=-\hat{\es}\allfields{I}\funktional{S}{\allfields{I}}\label{eq:BRSI}\end{eqnarray}
 \rem{\[
\bs{j}_{z}=\ce^{\bs{\alpha}}\dP_{z\bs{\alpha}}=-\es(\ce^{\bs{\alpha}}\be_{z\bs{\alpha}}),\qquad\hat{\bs{\jmath}}_{\bar{z}}=\hat{\ce}^{\hat{\bs{\alpha}}}\dP_{\bar{z}\hat{\bs{\alpha}}}=-\hat{\es}(\hat{\ce}^{\hat{\bs{\alpha}}}\hat{\be}_{\bar{z}\hat{\bs{\alpha}}})\]
}The current for the graded commutator $[\hat{\es},\es]$ is given
only on-shell by $\hat{\es}\bs{j}_{z}$ or $\es\hat{\bs{\jmath}}_{\bar{z}}$
(one would expect this from the Hamiltonian formalism). A correct
off-shell expression can be obtained by acting on (\ref{eq:BRSI})
with $\hat{\es}$ or $\es$ respectively. The derivation of the current
$j_{[\hat{\es},\es]}$ corresponding to $[\es,\hat{\es}]$ was too
simple and indeed not correct in the original version of this thesis,
so that by now I have moved a more careful and general derivation
into the appendix. From there we can adopt the result from equation
(\ref{eq:noet:jcommisdeltaj}) on page \pageref{eq:noet:jcommisdeltaj}:
\begin{eqnarray}
j_{[\hat{\es},\es]z} & = & \hat{\es}\bs{j}_{z}+\Bigl(\funktional{S}{\allfields{I}}\partiell{(\hat{\es}\allfields{I})}{(\partial_{\bar{z}}\allfields{K})}\Bigr)\cdot\es\allfields{K},\quad j_{[\hat{\es},\es]\bar{z}}=\Bigl(\funktional{S}{\allfields{I}}\partiell{(\hat{\es}\allfields{I})}{(\partial\allfields{K})}\Bigr)\cdot\es\allfields{K}\label{eq:currentForCommutator}\end{eqnarray}
or equivalently (interchanging the role of $\es$ and $\hat{\es}$)
\begin{eqnarray}
j_{[\hat{\es},\es]z} & = & \Bigl(\funktional{S}{\allfields{I}}\partiell{(\es\allfields{I})}{(\partial_{\bar{z}}\allfields{K})}\Bigr)\cdot\hat{\es}\allfields{K},\quad j_{[\hat{\es},\es]\bar{z}}=\es\hat{\bs{\jmath}}_{\bar{z}}+\Bigl(\funktional{S}{\allfields{I}}\partiell{(\es\allfields{I})}{(\partial\allfields{K})}\Bigr)\cdot\hat{\es}\allfields{K}\label{eq:currentForCommutatorII}\end{eqnarray}
For consistency we need only that $[\es,\hat{\es}]$ vanishes up to
trivial and other gauge transformations. It is thus enough to demand
that the corresponding current $j_{[\hat{\es},\es]}$ vanishes on-shell,
because on-shell vanishing currents correspond to gauge transformations
(see proposition \ref{prop:currentOnshellZero} on \pageref{prop:currentOnshellZero}
in the appendix). If we take the expression for $j_{[\hat{\es},\es]z}$
from (\ref{eq:currentForCommutatorII}) and the expression for $j_{[\hat{\es},\es]\bar{z}}$
from (\ref{eq:currentForCommutator}), we can observe that both components
of the current vanish on-shell without any extra conditions on the
background fields! As claimed at the beginning of this section this
happens due to the fact that left- and right-mover BRST currents $\bs{j}_{z}$
and $\hat{\bs{\jmath}}_{\bar{z}}$ are on-shell holomorphic and antiholomorphic
respectively. 

In principle we are already done with the commutator $[\es,\hat{\es}]$,
but it is a good check to see, whether we obtain the same result if
we do it the other way round and take the expression for $j_{[\hat{\es},\es]z}$
from (\ref{eq:currentForCommutator}) and the expression for $j_{[\hat{\es},\es]\bar{z}}$
from (\ref{eq:currentForCommutatorII}). This corresponds to demanding
$\hat{\es}\bs{j}_{z}\stackrel{{\rm on\, shell}}{=}0,\quad\es\hat{\bs{\jmath}}_{\bar{z}}\stackrel{{\rm on\, shell}}{=}0$.
In order to calculate $\hat{\es}\bs{j}_{z}$, remember the form of
the BRST current $\bs{j}_{z}=\ce^{\bs{\alpha}}\dP_{z\bs{\alpha}}$
(\ref{eq:BiBbrstSimple}) and also note that it is a target space
scalar. The BRST differential can thus be replaced by the covariant
one: \begin{eqnarray}
\hat{\es}\bs{j}_{z} & = & \frem{\hat{\es}_{cov}(\ce^{\bs{\gamma}}\dP_{z\bs{\gamma}})=}-\ce^{\bs{\gamma}}\hat{\es}_{cov}\dP_{z\bs{\gamma}}=-2\hat{\ce}^{\hat{\bs{\alpha}}}\ce^{\bs{\gamma}}\ce^{\bs{\alpha}}R_{\hat{\bs{\alpha}}\bs{\gamma}\bs{\alpha}}\hoch{\bs{\beta}}\be_{z\bs{\beta}}\quad\ous{(\ref{eq:holConstrX})}{=}{(\ref{eq:bispinorConstraintInDifferentVersions})}\quad-\frac{1}{8}\hat{\ce}^{\hat{\bs{\alpha}}}\gamma_{a}^{\bs{\alpha}\bs{\gamma}}R_{\hat{\bs{\alpha}}\bs{\gamma}\bs{\alpha}}\hoch{\bs{\beta}}\be_{z\bs{\beta}}\underbrace{(\ce\gamma^{a}\ce)}_{2\funktional{S}{L_{z\bar{z}}}}\label{eq:above-on-shell-vanishing-term}\end{eqnarray}
Using the left-right-symmetry of proposition \vref{prop:left-right-symmetry}
we get the corresponding expression for $\es\hat{\bs{\jmath}}_{\bar{z}}$.\rem{\[
\es\hat{\bs{\jmath}}_{\bar{z}}=\es(\hat{\ce}^{\hat{\bs{\alpha}}}\hat{\dP}_{\bar{z}\hat{\bs{\alpha}}})=-\hat{\ce}^{\hat{\bs{\gamma}}}\es_{\hat{cov}}\hat{\dP}_{\bar{z}\hat{\bs{\gamma}}}=2\hat{\ce}^{\hat{\bs{\gamma}}}\ce^{\bs{\alpha}}\hat{\ce}^{\hat{\bs{\alpha}}}\hat{R}_{\bs{\alpha}\hat{\bs{\gamma}}\hat{\bs{\alpha}}}\hoch{\hat{\bs{\beta}}}\hat{\be}_{\bar{z}\hat{\bs{\beta}}}\stackrel{(\ref{eq:holConstrX})}{=}0\]
} Both vanish on the pure spinor constraint surface $(\ce\gamma^{a}\ce)=(\hat{\ce}\gamma^{a}\hat{\ce})=0$
so that indeed the Noether current belonging to $[\hat{\es},\es]$
vanishes on-shell and thus $[\es,\hat{\es}]$ will vanish on-shell
up to gauge transformations. 

If we wanted to know also the non-trivial gauge transformations that
appear in the commutator, we would have to calculate also the additional
on-shell vanishing terms that are added to $\hat{\es}\bs{j}_{z}$
in the expression of $j_{[\es,\hat{\es}]z}$ in (\ref{eq:currentForCommutator}).
It turns out that only $\Bigl(\funktional{S}{\hat{\dP}_{\bar{z}\hat{\bs{\delta}}}}\partiell{(\hat{\es}\hat{\dP}_{\bar{z}\hat{\bs{\delta}}})}{(\bar{\partial}x^{K})}\Bigr)\cdot sx^{K}=\tfrac{3}{2}\funktional{S}{\hat{\dP}_{\bar{z}\hat{\bs{\delta}}}}\hat{\lambda}^{\hat{\bs{\alpha}}}\hat{\lambda}^{\hat{\bs{\gamma}}}H_{\hat{\bs{\alpha}}\hat{\bs{\gamma}}\hat{\bs{\delta}}}$
is contributing a priori. However, we will see later that $H_{\hat{\bs{\alpha}}\hat{\bs{\gamma}}\hat{\bs{\delta}}}$
is required to vanish from the nilpotency demand of the BRST transformation
as well as from the Bianchi identities.

The (non-trivial) gauge transformations that will appear in the commutator
$[\es,\hat{\es}]$ are thus given precisely by the above off-shell
non-vanishing term (\ref{eq:above-on-shell-vanishing-term}). Namely
if we take $\mu_{za}\equiv-\frac{1}{4}\hat{\ce}^{\hat{\bs{\alpha}}}\gamma_{a}^{\bs{\alpha}\bs{\gamma}}R_{\hat{\bs{\alpha}}\bs{\gamma}\bs{\alpha}}\hoch{\bs{\beta}}\be_{z\bs{\beta}}$
we obtain \begin{eqnarray}
\hat{\es}\bs{j}_{z} & = & \tfrac{1}{2}\mu_{za}(\ce\gamma^{a}\ce)\end{eqnarray}
 which is precisely the current of the antighost gauge transformation
given on the lefthand side of (\ref{eq:on-shell-cond-antigauge})
with corresponding antighost gauge transformations $\delta_{(\mu)}\be_{z\bs{\alpha}}=\mu_{za}(\ce\gamma^{a})_{\bs{\alpha}}$
(\ref{eq:gaugetrafoAntighost}) and $\delta_{(\mu)}L_{z\bar{z}a}=-\mc{D}_{\bar{z}}\mu_{za}$
(\ref{eq:gaugetrafoLagrange}).\rem{Let us see this explicitely,
and plug the expressions for $\hat{\es}\bs{j}_{z}$ and $\es\hat{\bs{\jmath}}_{\bar{z}}$
into the lefthand side of (\ref{eq:currentForCommutator}) using the
covariant derivatives $\mc{D}_{\bar{z}}$ and $\hat{\mc{D}}_{z}$
of (\ref{eq:AundMathcalD}), (\ref{eq:AhutUndMathcalDhut}) and (\ref{eq:AundMathcalDboson}):\begin{eqnarray*}
\lqn{\bar{\partial}(\hat{\es}\bs{j}_{z})+\partial(\es\hat{\bs{\jmath}}_{\bar{z}})=}\\
 & = & -\frac{1}{8}\mc{D}_{\bar{z}}\left(\hat{\ce}^{\hat{\bs{\alpha}}}\gamma_{a}^{\bs{\alpha}\bs{\gamma}}R_{\hat{\bs{\alpha}}\bs{\gamma}\bs{\alpha}}\hoch{\bs{\beta}}\be_{z\bs{\beta}}\right)\underbrace{(\ce\gamma^{a}\ce)}_{2\funktional{S}{L_{z\bar{z}}}}-\frac{1}{4}\hat{\ce}^{\hat{\bs{\alpha}}}\gamma_{a}^{\bs{\alpha}\bs{\gamma}}R_{\hat{\bs{\alpha}}\bs{\gamma}\bs{\alpha}}\hoch{\bs{\beta}}\be_{z\bs{\beta}}(\ce\gamma^{a}\underbrace{\mc{D}_{\bar{z}}\ce}_{-\funktional{S}{\be_{z}}})+\mbox{hatted version}\end{eqnarray*}
} Having a current that coincides with the one of a gauge transformation,
the form of $[\es,\hat{\es}]$ can only differ by a trivial gauge
transformation. In any case we have obtained the result that the commutator
vanishes up to gauge transformations. A safe way to figure out potentially
appearing trivial gauge transformations in the commutator is to calculate
it on each single worldsheet field separately.

\paragraph{Acting on each field separately}

Although this method would lead to the precise off-shell form of all
the commutators, we are for now satisfied with the result we already
obtained and give the explicit commutator only for the most simple
cases. Starting with the covariant BRST transformations of the elementary
fields (given in (\ref{eq:covBRSTofx})-(\ref{eq:covBRSTofL}) on
page \pageref{eq:covBRSTofx}), we will first calculate the commutator
$[\hat{\es}_{\gem{cov}},\es_{\gem{cov}}]$ and only after that determine
the ordinary commutator via the relations (\ref{eq:commutatorOfCovVarI})
and (\ref{eq:commutatorOfCovVarII}). For the embedding functions
$x^{K}$, the ghosts $\ce^{\bs{\alpha}},\hat{\ce}^{\hat{\bs{\alpha}}}$
and the antighosts $\be_{z\bs{\alpha}}$ and $\hat{\be}_{\bar{z}\hat{\bs{\alpha}}}$
the calculation is very simple and we immediately obtain\begin{eqnarray}
\left[\hat{\es}_{\gem{cov}},\es_{\gem{cov}}\right]x^{K} & = & 0\\
\left[\hat{\es}_{\gem{cov}},\es_{\gem{cov}}\right]\ce^{\bs{\gamma}} & = & 0,\qquad\left[\es_{\gem{cov}},\hat{\es}_{\gem{cov}}\right]\hat{\ce}^{\hat{\bs{\gamma}}}=0\\
\left[\hat{\es}_{\gem{cov}},\es_{\gem{cov}}\right]\be_{z\bs{\gamma}} & = & \hat{\es}_{cov}\dP_{z\bs{\gamma}}=-2\hat{\ce}^{\hat{\bs{\alpha}}}\ce^{\bs{\alpha}}R_{\hat{\bs{\alpha}}\bs{\gamma}\bs{\alpha}}\hoch{\bs{\beta}}\be_{z\bs{\beta}},\qquad\left[\es_{\gem{cov}},\hat{\es}_{\gem{cov}}\right]\hat{\be}_{\bar{z}\hat{\bs{\gamma}}}=-2{\ce^{\bs{\alpha}}\hat{\ce}}^{\hat{\bs{\alpha}}}\hat{R}_{\bs{\alpha}\bs{\gamma}\hat{\bs{\alpha}}}\hoch{\hat{\bs{\beta}}}\hat{\be}_{\bar{z}\hat{\bs{\beta}}}\end{eqnarray}
 The transformations of the remaining fields are much more complicated
and we prefer not to study them. Let us now derive the ordinary commutators:\begin{eqnarray}
[\hat{\es},\es]x^{K} & \stackrel{(\ref{eq:commutatorOfCovVarI})}{=} & \underbrace{\left[\hat{\es}_{\gem{cov}},\es_{\gem{cov}}\right]x^{K}}_{=0}-2\hat{\ce}^{\hat{\bs{\alpha}}}\underbrace{\gem{T}_{\hat{\bs{\alpha}}\bs{\alpha}}\hoch{K}}_{=0\,(\ref{eq:holConstrIV})}\ce^{\bs{\alpha}}=0\label{eq:commute:interesting}\\
\left[\es,\hat{\es}\right]_{cov}\ce^{\bs{\gamma}} & \stackrel{(\ref{eq:commutatorOfCovVarII})}{=} & \underbrace{\left[\es_{\gem{cov}},\hat{\es}_{\gem{cov}}\right]\ce^{\bs{\gamma}}}_{=0}-2\underbrace{\ce^{\bs{\alpha}}\hat{\ce}^{\hat{\bs{\alpha}}}R_{\bs{\alpha}\hat{\bs{\alpha}}\bs{\beta}}\hoch{\bs{\gamma}}\ce^{\bs{\beta}}}_{=0\,(\ref{eq:holConstrX})}=0\\
\left[\es,\hat{\es}\right]_{cov}\be_{z\bs{\gamma}} & \stackrel{(\ref{eq:commutatorOfCovVarII})}{=} & \underbrace{\left[\hat{\es}_{\gem{cov}},\es_{\gem{cov}}\right]\be_{z\bs{\gamma}}}_{=-2\hat{\ce}^{\hat{\bs{\alpha}}}\ce^{\bs{\alpha}}R_{\hat{\bs{\alpha}}\bs{\gamma}\bs{\alpha}}\hoch{\bs{\beta}}\be_{z\bs{\beta}}}+2\ce^{\bs{\alpha}}\hat{\ce}^{\hat{\bs{\alpha}}}R_{\bs{\alpha}\hat{\bs{\alpha}}\bs{\gamma}}\hoch{\bs{\beta}}\be_{z\bs{\beta}}=\nonumber \\
 & = & 4\hat{\ce}^{\hat{\bs{\alpha}}}\ce^{\bs{\alpha}}R_{\hat{\bs{\alpha}}[\bs{\alpha}\bs{\gamma}]}\hoch{\bs{\beta}}\be_{z\bs{\beta}}\end{eqnarray}
Again we get the corresponding equations for $\hat{\ce}^{\hat{\bs{\alpha}}}$
and $\hat{\be}_{\bar{z}\hat{\bs{\gamma}}}$. The last line corresponds
excactly to the gauge transformation with gauge parameter $\mu_{za}=-\frac{1}{4}\hat{\ce}^{\hat{\bs{\alpha}}}\gamma_{a}^{\bs{\alpha}\bs{\gamma}}R_{\hat{\bs{\alpha}}\bs{\gamma}\bs{\alpha}}\hoch{\bs{\beta}}\be_{z\bs{\beta}}$
that we found already above. This is strictly speaking true only if
$H_{\hat{\bs{\alpha}}\hat{\bs{\beta}}\hat{\bs{\gamma}}}=0$ (remember
the off-shell terms that were mentioned after (\ref{eq:above-on-shell-vanishing-term})),
a constraint that we will obtain only in the next section from nilpotency.
The explanation is that the different ways of calculating the same
quantity $[\es,\hat{\es}]$ certainly assume the validity of the Bianchi
identities which already at this point would imply the above extra
constraint. However, we will do a careful analysis of the Bianchi
identities only in the end, after having obtained the additional constraints
from nilpotency. It is further interesting to see in (\ref{eq:commute:interesting}),
that some holomorphicity constraints like $\gem{T}_{\hat{\bs{\alpha}}\bs{\alpha}}\hoch{K}=0$
are needed for the commutation. In fact, in \cite{Bedoya:2006ic}
this constraint was derived by demanding a vanishing Poisson bracket
between the two generators of the BRST symmetries. The constraint
$\gem{T}_{\hat{\bs{\alpha}}\bs{\alpha}}\hoch{K}=0$ did not appear
in our derivation via the currents above. The reason is that we already
started the derivation in (\ref{eq:BRSI}) from an equation which
assumes on-shell holomorphicity.

\section{Nilpotency of the BRST differentials}

\label{sec:Nilpotency}\index{nilpotency|fett}While the last section
was rather a check than bringing much new information, the nilpotency
of the BRST differentials will give us additional constraints on the
background fields. The nilpotency is essential to define the physical
spectrum as in the flat case via the cohomology. It would be inconsistent
if this prescription breakes down, as soon as a nonvanishing background
is generated by the strings. Demanding nilpotency at least on-shell
and up to gauge transformations is thus legitimate.

\paragraph{Nilpotency constraints from the BRST transformation of the current}

In the same way as in the previous section, we can examine the BRST-transformation
of the BRST-current instead of studying nilpotency on every single
worldsheet field. Start from the defining equation of the BRST current
\begin{eqnarray}
\bar{\partial}\bs{j}_{z} & = & -\es\allfields{I}\funktional{S}{\allfields{I}}\end{eqnarray}
Again the current for the graded commutator $[\es,\es]=2\es^{2}$
is given only on-shell by $\es\bs{j}_{z}$ (what one would expect
from the Hamiltonian formalism). To obtain the off-shell expression
one can act with $\es$ for a second time on the above equation.
From the appendix we can adopt the result from equation (\ref{eq:noet:jcommisdeltaj})
on page \pageref{eq:noet:jcommisdeltaj}: \begin{eqnarray}
j_{[\es,\es]z} & = & \es\bs{j}_{z}+\Bigl(\funktional{S}{\allfields{I}}\partiell{(\es\allfields{I})}{(\partial_{\bar{z}}\allfields{K})}\Bigr)\cdot\es\allfields{K},\quad j_{[\es,\es]\bar{z}}=\Bigl(\funktional{S}{\allfields{I}}\partiell{(\es\allfields{I})}{(\partial\allfields{K})}\Bigr)\cdot\es\allfields{K}\end{eqnarray}
The BRST transformation of the BRST current $\es\bs{j}_{z}$ is therefore
at least on-shell the Noether current for the transformation $2\es^{2}$.
For consistency we need only that $\es^{2}$ vanishes up to trivial
and other gauge transformations. Due to proposition \ref{prop:currentOnshellZero}
on page \pageref{prop:currentOnshellZero} in the appendix, every
gauge transformation has (up to trivially conserved terms) an on-shell
vanishing Noether current. Demanding that $\es\bs{j}_{z}$ vanishes
on-shell is therefore a necessary condition.%
\footnote{\index{trivially conserved current}\index{footnote!\thefoot. no trivially conserved part}There
are no trivially conserved parts in $\es\bs{j}_{z}$. A trivially
conserved part is of the form $\partial_{\zeta}S^{[\zeta\xi]}$ for
some rank two tensor $S^{\zeta\xi}$. In the conformal gauge this
would take the form $\partial_{z}S_{[\bar{z}z]}$ which is of conformal
weight (2,1). Such a term is certainly not present in our current.$\quad\fussend$%
} Also due to proposition \ref{prop:currentOnshellZero} it is a sufficient
condition, as we know already that $\es\bs{j}_{z}$ is a Noether current
for a symmetry transformation and if this current vanishes on-shell,
the transformation can be extended to a local one, i.e. it is a gauge
transformation.

As the BRST current is a target space scalar, we can replace the BRST
differential with the covariant one when calculating $\es\bs{j}_{z}$:
\begin{eqnarray}
\es\bs{j}_{z} & = & \es_{\gem{cov}}\left(\ce^{\bs{\delta}}\dP_{z\bs{\delta}}\right)=-\ce^{\bs{\delta}}\es_{\gem{cov}}\dP_{z\bs{\delta}}=\nonumber \\
 & \stackrel{(\ref{eq:covBRSTofd})}{=} & -\ce^{\bs{\delta}}\ce^{\bs{\alpha}}\underbrace{3H_{\bs{\alpha}c\bs{\delta}}}_{2\check{T}_{\bs{\alpha}\bs{\delta}|c}}\Pi_{z}^{c}-\frac{3}{2}\ce^{\bs{\delta}}\ce^{\bs{\alpha}}H_{\bs{\alpha}\bs{\gamma}\bs{\delta}}\Pi_{z}^{\bs{\gamma}}-2\ce^{\bs{\delta}}\ce^{\bs{\alpha}}T_{\bs{\alpha}\bs{\delta}}\hoch{\bs{\gamma}}\dP_{z\bs{\gamma}}+2\ce^{\bs{\delta}}\ce^{\bs{\alpha}}\ce^{\bs{\alpha}_{2}}R_{\bs{\alpha}_{2}\bs{\delta}\bs{\alpha}}\hoch{\bs{\beta}}\be_{z\bs{\beta}}\qquad\end{eqnarray}
\rem{Note that $\es\bs{j}_{z}=\es^{2}(\ce^{\bs{\alpha}}\be_{z\bs{\alpha}})$.}
The only equations of motion, which can make $\es\bs{j}_{z}$ vanish
on-shell are the pure spinor constraints $\ce\gamma^{a}\ce=0$. We
therefore get the following conditions on the background fields

\begin{eqnarray}
\dann\ce^{\bs{\delta}}H_{\bs{\alpha}C\bs{\delta}}\ce^{\bs{\alpha}} & = & 0,\qquad\ce^{\bs{\delta}}\ce^{\bs{\alpha}}T_{\bs{\alpha}\bs{\delta}}\hoch{\bs{\gamma}}=0,\qquad\ce^{\bs{\delta}}\ce^{\bs{\alpha}_{1}}\ce^{\bs{\alpha}_{2}}R_{\bs{\alpha}_{2}\bs{\delta}\bs{\alpha}_{1}}\hoch{\bs{\beta}}=0,\qquad\mbox{(on shell)}\label{eq:nilpotency-constraints}\end{eqnarray}
 Remembering that we have the constraints $\check{T}_{\bs{\alpha}\bs{\delta}|c}=\frac{3}{2}H_{\bs{\alpha}c\bs{\delta}}$
(\ref{eq:Y-constrIV}) and $\hat{T}_{\bs{\alpha\delta}}\hoch{\hat{\bs{\gamma}}}=\frac{3}{4}H_{\bs{\alpha\delta\beta}}\RR^{\bs{\beta}\hat{\bs{\gamma}}}$,
we can extend the above condition on the torsion on all indices\begin{equation}
\ce^{\bs{\delta}}\ce^{\bs{\alpha}}\gemT_{\bs{\alpha}\bs{\delta}}\hoch{C}=0\qquad\mbox{(on-shell)}\label{eq:nilpotency-constraint-torsion}\end{equation}
All these on-shell conditions can be formulated in an off-shell version
with the help of $\gamma$-matrices by using (\ref{eq:bispinorConstraintInDifferentVersions})
on page \pageref{eq:bispinorConstraintInDifferentVersions}. Either
we write that the terms are linear combinations of $\gamma^{[1]}$'s,
or equivalently we can write that the $\gamma^{[5]}$-part vanishes.
We thus can rewrite the constraints on torsion and $H$-field as\index{$f_d\hoch{C}$}
\begin{eqnarray}
\gemT_{\bs{\alpha\beta}}\hoch{C} & = & \gamma_{\bs{\alpha\beta}}^{d}f_{d}\hoch{C}\quad\mbox{with }f_{d}\hoch{C}\equiv\frac{1}{16}\gamma_{d}^{\bs{\eps\delta}}\gemT_{\bs{\delta\eps}}\hoch{C}\label{eq:TproptoGamma}\\
H_{C\bs{\alpha}\bs{\beta}} & = & H_{Ca}\gamma_{\bs{\alpha\beta}}^{a}\quad\mbox{with }H_{Ca}\equiv\frac{1}{16}H_{C\bs{\delta\eps}}\gamma_{a}^{\bs{\eps\delta}}\label{eq:HproptoGamma}\end{eqnarray}
In particular for $C=\bs{\gamma}$, due to the (graded) total antisymmetry
of $H_{\bs{\gamma}\bs{\alpha\beta}}$, this should at the same time
be proportional to $\gamma_{\bs{\gamma\alpha}}^{a}$ and $\gamma_{\bs{\beta\gamma}}^{a}$:
\begin{eqnarray}
H_{\bs{\gamma}\bs{\alpha\beta}} & \stackrel{(\ref{eq:HproptoGamma})}{=} & H_{[\bs{\gamma}|a}\gamma_{|\bs{\alpha\beta}]}^{a}\stackrel{(\ref{eq:HproptoGamma})}{=}\frac{1}{16}H_{[\bs{\gamma}|\bs{\delta\eps}}\gamma_{a}^{\bs{\eps\delta}}\gamma_{|\bs{\alpha\beta}]}^{a}\stackrel{(\ref{eq:HproptoGamma})}{=}\frac{1}{16}H_{\bs{\eps}b}\gamma_{[\bs{\gamma}|\bs{\delta}}^{b}\gamma_{a}^{\bs{\eps\delta}}\gamma_{|\bs{\alpha\beta}]}^{a}\ous{(\ref{eq:smallClifford})}{=\qquad}{(\ref{eq:LittleFierz})}\frac{1}{8}H_{[\bs{\gamma}|a}\gamma_{|\bs{\alpha\beta}]}^{a}\label{eq:HisZeroArgument}\end{eqnarray}
In the last step we used the Clifford algebra (\ref{eq:smallClifford})
for the first two $\gamma$'s and then the Fierz identity (\ref{eq:LittleFierz})
to throw away one of the resulting terms. Remember that the appendix
about $\Gamma$-matrices doesn't use the graded summation convention.
For the Fierz identity we thus have a (graded) antisymmetrization,
instead of the symmetrization and for the Clifford algebra we get
an extra minus sign because of the NW-definition of the Kronecker-delta. 

The second and the last term of the above equation (\ref{eq:HisZeroArgument})
contradict each other if they do not vanish and thus $H_{\bs{\eps}\bs{\alpha}\bs{\beta}}$
has to vanish. The components $H_{\hat{\bs{\eps}}\bs{\alpha}\bs{\beta}}$
were constraint to be zero already before. Of the components in (\ref{eq:HproptoGamma}),
we thus have only $H_{c\bs{\alpha\beta}}$ nonvanishing. Because of
$\check{T}_{\bs{\alpha}\bs{\beta}|c}=-\frac{3}{2}H_{c\bs{\alpha}\bs{\beta}}$
(\ref{eq:Y-constrIV}) and $\hat{T}_{\bs{\alpha\beta}}\hoch{\hat{\bs{\gamma}}}\stackrel{(\ref{eq:holConstrII})}{=}\frac{3}{4}H_{\bs{\alpha\beta\delta}}\RR^{\bs{\delta}\hat{\bs{\gamma}}}=0$
, we have in addition \begin{eqnarray}
f_{dc} & = & -\frac{3}{2}H_{cd},\quad f_{d}\hoch{\hat{\bs{\gamma}}}=0\label{eq:BeziehungHuf}\end{eqnarray}
 The new constraints on $H$ and on the torsion thus read (the constraints
in brackets follow from the other ones in combination with (\ref{eq:Y-constrIV})
and (\ref{eq:holConstrII}) and are thus redundant):\vRam{.6}{\begin{eqnarray}
H_{\bs{\eps}\bs{\alpha}\bs{\beta}} & = & 0,\qquad H_{c\bs{\alpha\beta}}=-\frac{2}{3}\gamma_{\bs{\alpha\beta}}^{a}f_{ac}\label{eq:nilpotency-constraint-onH}\\
T_{\bs{\alpha\beta}}\hoch{\bs{\gamma}} & = & \gamma_{\bs{\alpha\beta}}^{d}f_{d}\hoch{\bs{\gamma}},\qquad(\check{T}_{\bs{\alpha\beta}}\hoch{c}=\gamma_{\bs{\alpha\beta}}^{d}f_{d}\hoch{c},\qquad\hat{T}_{\bs{\alpha\beta}}\hoch{\hat{\bs{\gamma}}}=0)\label{eq:nilpotency-constraint-onTfinal}\end{eqnarray}
}\\
As a remark let us note that the action in flat superspace with the
ordinary WZ-term of the GS-string corresponds to $H_{c\bs{\alpha\beta}}=-\frac{2}{3}\gamma_{c\,\bs{\alpha\beta}}$
and thus to $f_{ac}=\eta_{ac}$.  We can now analyze in a similar
way the constraint on the curvature in (\ref{eq:nilpotency-constraints}).
As the pure spinor constraint is quadratic it can be equivalently
written as $\ce^{\bs{\alpha}_{1}}\ce^{\bs{\alpha}_{2}}R_{[\bs{\alpha}_{2}\bs{\delta}\bs{\alpha}_{1}]}\hoch{\bs{\beta}}=0$
(on-shell). For this expression, one can do the same reasoning as
above with $H_{\bs{\eps\alpha\beta}}$ and arrives at\begin{equation}
\boxed{R_{[\bs{\gamma}\bs{\delta}\bs{\alpha}]}\hoch{\bs{\beta}}=0}\label{eq:nilpotency-constraint-onR}\end{equation}
We will get the same constraint from the Bianchi identities later
in (\ref{eq:(delta|0,3,0)}) in case one feels uncomfortable with
that line of arguments.

Of course we get all the correponding constraints also in the hatted
version from the right-mover BRST current according to the left-right
symmetry of page \pageref{prop:left-right-symmetry}:\vRam{.6}{\begin{eqnarray}
H_{\hat{\bs{\eps}}\hat{\bs{\alpha}}\hat{\bs{\beta}}} & = & 0,\qquad H_{c\hat{\bs{\alpha}}\hat{\bs{\beta}}}=\frac{2}{3}\gamma_{\hat{\bs{\alpha}}\hat{\bs{\beta}}}^{a}\hat{f}_{ac}\label{eq:nilpotency-constraint-onH-hat}\\
\hat{T}_{\hat{\bs{\alpha}}\hat{\bs{\beta}}}\hoch{\hat{\bs{\gamma}}} & = & \gamma_{\hat{\bs{\alpha}}\hat{\bs{\beta}}}^{d}\hat{f}_{d}\hoch{\hat{\bs{\gamma}}},\qquad(\check{T}_{\hat{\bs{\alpha}}\hat{\bs{\beta}}}\hoch{c}=\gamma_{\hat{\bs{\alpha}}\hat{\bs{\beta}}}^{d}\hat{f}_{d}\hoch{c},\qquad T_{\hat{\bs{\alpha}}\hat{\bs{\beta}}}\hoch{\bs{\gamma}}=0)\label{eq:nilpotency-constraint-onTfinal-hat}\\
\hat{R}_{[\hat{\bs{\gamma}}\hat{\bs{\delta}}\hat{\bs{\alpha}}]}\hoch{\hat{\bs{\beta}}} & = & 0\label{eq:nilpotency-constraint-onR-hat}\end{eqnarray}
} Remember that the curvature is structure group valued in the last
two indices and decays into Lorentz and scale part (see (\ref{eq:R-Zerfall-ferm})
in the appendix on page \pageref{eq:R-Zerfall-ferm}): $R_{\bs{\gamma\delta\alpha}}\hoch{\bs{\beta}}=\frac{1}{2}F_{\bs{\gamma\delta}}^{(D)}\delta_{\bs{\alpha}}\hoch{\bs{\beta}}+R_{\bs{\gamma\delta\alpha}}^{(L)}\hoch{\bs{\beta}}$
with $R_{\bs{\gamma\delta\alpha}}^{(L)}\hoch{\bs{\beta}}=\frac{1}{4}R^{(L)}\tief{\bs{\gamma\delta}a_{1}a_{2}}\gamma^{a_{1}a_{2}}\tief{\bs{\alpha}}\hoch{\bs{\beta}}$.
The constraint (\ref{eq:nilpotency-constraint-onR}), i.e. $0=R_{[\bs{\gamma}\bs{\delta}\bs{\alpha}]}\hoch{\bs{\beta}}\propto R_{\bs{\gamma}\bs{\delta}\bs{\alpha}}\hoch{\bs{\beta}}+2R_{\bs{\alpha}[\bs{\gamma}\bs{\delta}]}\hoch{\bs{\beta}}$,
therefore implies that $R_{\bs{\alpha}[\bs{\gamma}\bs{\delta}]}\hoch{\bs{\beta}}$
is Lie algebra valued in $\bs{\alpha}$ and $\bs{\beta}$ as well.
This means in particular that $R_{\bs{\alpha}[\bs{\gamma}\bs{\delta}]}\hoch{\bs{\beta}}\gamma^{a_{1}\ldots a_{4}}\tief{\bs{\beta}}\hoch{\bs{\alpha}}=0$.\rem{\begin{eqnarray*}
R_{\bs{\alpha}[\bs{\gamma}\bs{\delta}]}\hoch{\bs{\beta}} & = & -\frac{1}{4}F_{\bs{\gamma\delta}}^{(D)}\delta_{\bs{\alpha}}\hoch{\bs{\beta}}-\frac{1}{8}R^{(L)}\tief{\bs{\gamma\delta}a_{1}a_{2}}\gamma^{a_{1}a_{2}}\tief{\bs{\alpha}}\hoch{\bs{\beta}}\end{eqnarray*}
} Let us finally give the trace (in $\bs{\alpha}$ and $\bs{\beta}$)
of (\ref{eq:nilpotency-constraint-onR}) and its hatted equivalent
(\ref{eq:nilpotency-constraint-onR-hat}): \begin{eqnarray}
0 & = & R_{[\bs{\gamma\delta\alpha}]}\hoch{\bs{\alpha}}\\
 & = & \frac{1}{2}F_{[\bs{\gamma\delta}}^{(D)}\delta_{\bs{\alpha}]}\hoch{\bs{\alpha}}+R_{[\bs{\gamma\delta\alpha}]}^{(L)}\hoch{\bs{\alpha}}=\\
 & = & -\frac{9}{3}F_{\bs{\gamma\delta}}^{(D)}+\frac{2}{3}R_{\bs{\alpha}[\bs{\gamma\delta}]}^{(L)}\hoch{\bs{\alpha}}\end{eqnarray}
The scale curvature can be expressed in terms of the Lorentz curvature
as\begin{equation}
F_{\bs{\gamma\delta}}^{(D)}=\frac{2}{9}R_{\bs{\alpha}[\bs{\gamma\delta}]}^{(L)}\hoch{\bs{\alpha}}\:,\quad\hat{F}_{\hat{\bs{\gamma}}\hat{\bs{\delta}}}^{(D)}=\frac{2}{9}\hat{R}_{\hat{\bs{\alpha}}[\hat{\bs{\gamma}}\hat{\bs{\delta}}]}^{(L)}\hoch{\hat{\bs{\alpha}}}\label{eq:nilpotency:Falphbet}\end{equation}

\paragraph{Nilpotency on the single fields}

Just to get a flavour of how the calculation would work if we act
on each field twice with the BRST differential, we perform this for
the simplest cases. One discovers immediately that acting on $x^{K}$
and $\ce^{\bs{\alpha}}$ twice with the covariant BRST transformation
yields zero. The reformulation of $\es_{\gem{cov}}^{2}$ in terms
of the square of the ordinary differential $\es^{2}$ gives a torsion
or a curvature term respectively. These terms have to vanish on-shell
in order to have an on-shell vanishing $\es^{2}$: \begin{eqnarray}
0 & = & \es_{\gem{cov}}^{2}x^{K}=\underbrace{\es^{2}x^{K}}_{\stackrel{!}{=}0\:\lqn{({\rm on-shell})}}+2\ce^{\bs{\alpha}}\gemT_{\bs{\alpha}\bs{\beta}}\hoch{K}\ce^{\bs{\beta}}\qquad\dann\ce^{\bs{\alpha}}\gemT_{\bs{\alpha}\bs{\beta}}\hoch{K}\ce^{\bs{\beta}}\stackrel{!}{=}0\quad({\rm on-shell})\end{eqnarray}
\begin{eqnarray}
0 & = & \es_{cov}^{2}\ce^{\bs{\alpha}}=\underbrace{(\es^{2})_{cov}\ce^{\bs{\alpha}}}_{\stackrel{!}{=}0\:\lqn{({\rm on-shell})}}+2\ce^{\bs{\gamma}}\ce^{\bs{\delta}}R_{\bs{\gamma}\bs{\delta}\bs{\beta}}\hoch{\bs{\alpha}}\ce^{\bs{\beta}}\qquad\dann\ce^{\bs{\gamma}}\ce^{\bs{\delta}}R_{\bs{\gamma}\bs{\delta}\bs{\beta}}\hoch{\bs{\alpha}}\ce^{\bs{\beta}}\stackrel{!}{=}0\quad({\rm on-shell})\end{eqnarray}
On the antighosts we have $\es_{cov}^{2}\be_{z\bs{\alpha}}=\es_{cov}\dP_{z\bs{\alpha}}$
which will not vanish, but which will correspond to a gauge transformation.
The same should be true for $L_{z\bar{z}a}$. The calculation of $\es^{2}\dP_{z\bs{\gamma}}$
is quite involved to calculate and will probably contain also constraints
that follow from the earlier ones via Bianchi identities. We will
calculate the identities anyway in sections \vref{sec:Bianchi-identities-forH}
and \vref{sec:Bianchi-identitiesForT}.

\section{Residual shift-reparametrization}

\label{sec:Residual-shift-reparametrization}\index{residual shift-reparametrization}\index{shift-reparametrization!residual}Before
we are going to collect all the constraints on the background fields
which we have obtained so far, let us eventually make use of the residual
shift-symmetry discussed in the paragraph on page \pageref{par:Residual-shift-symmetry}
(which in turn refers to the paragraph about shift-reparametrization
on page \pageref{par:Shift-reparametrization}). It is a target space
symmetry that is based on a residual shift reparametrization of the
fermionic momenta: \begin{eqnarray}
\dP_{z\bs{\alpha}} & = & \tilde{\dP}_{z\bs{\alpha}}-\dshift{3}_{b}\hoch{\bs{\delta}}(\xfull)(\gamma^{b}\ce)_{\bs{\alpha}}\be_{z\bs{\delta}}\end{eqnarray}
The BRST current gets changed under this reparametrization by a linear
combination of the pure spinor constraints (\ref{eq:residualBRSTchange}),
but this change can be undone by a redefinition of the BRST transformations
with the corresponding antighost gauge transformations. This does
of course not change the on-shell holomorphicity of the BRST current,
as the pure spinor term vanishes on-shell. 

Apart from the change of the BRST current, we have the following induced
transformations of the background fields coming along with this reparametrization:\begin{eqnarray}
\tilde{\Omega}_{M\bs{\alpha}}\hoch{\bs{\beta}} & = & \Omega_{M\bs{\alpha}}\hoch{\bs{\beta}}-E_{M}\hoch{\bs{\gamma}}\gamma_{\bs{\gamma}\bs{\alpha}}^{b}\dshift{3}_{b}\hoch{\bs{\beta}}\\
\tilde{C}_{\bs{\alpha}}\hoch{\bs{\beta}\hat{\bs{\gamma}}} & = & C_{\bs{\alpha}}\hoch{\bs{\beta}\hat{\bs{\gamma}}}-\gamma_{\bs{\gamma}\bs{\alpha}}^{b}\dshift{3}_{b}\hoch{\bs{\beta}}\RR^{\bs{\gamma}\hat{\bs{\gamma}}}\\
\tilde{S}_{\bs{\alpha}\hat{\bs{\alpha}}}\hoch{\bs{\beta}\hat{\bs{\beta}}} & = & S_{\bs{\alpha}\hat{\bs{\alpha}}}\hoch{\bs{\beta}\hat{\bs{\beta}}}+\hat{C}_{\hat{\bs{\alpha}}}\hoch{\hat{\bs{\beta}}\bs{\gamma}}\gamma_{\bs{\gamma}\bs{\alpha}}^{b}\dshift{3}_{b}\hoch{\bs{\beta}}\end{eqnarray}
Note that the transformations of $C_{\bs{\alpha}}\hoch{\bs{\beta}\hat{\bs{\gamma}}}$
and $S_{\bs{\alpha}\hat{\bs{\alpha}}}\hoch{\bs{\beta}\hat{\bs{\beta}}}$
are in agreement with the holomorphicity constraints\rem{%
\footnote{\index{footnote!\thefoot. consistency of deltaC and deltaOmega}Note
the consistency of the given transformation of $C$ and the transformation
of $C$ derived via the relation to the covariant derivative of $\RR$
and the torsion! \begin{eqnarray*}
\tilde{C}_{\bs{\alpha}}\hoch{\bs{\beta}\hat{\bs{\gamma}}} & = & \tilde{\nabla}_{\bs{\alpha}}\RR^{\bs{\beta}\hat{\bs{\gamma}}}-2\tilde{T}_{\bs{\alpha\delta}}\hoch{\bs{\beta}}\RR^{\bs{\delta}\hat{\bs{\gamma}}}=\partial_{\bs{\alpha}}\RR^{\bs{\beta}\hat{\bs{\gamma}}}+\tilde{\Omega}_{\bs{\alpha}\bs{\delta}}\hoch{\bs{\beta}}\RR^{\bs{\delta}\hat{\bs{\gamma}}}+\tilde{\hat{\Omega}}_{\bs{\alpha}\hat{\bs{\delta}}}\hoch{\hat{\bs{\gamma}}}\RR^{\bs{\beta}\hat{\bs{\delta}}}-2\tilde{T}_{\bs{\alpha\delta}}\hoch{\bs{\beta}}\RR^{\bs{\delta}\hat{\bs{\gamma}}}=\\
 & = & \tilde{C}_{\bs{\alpha}}\hoch{\bs{\beta}\hat{\bs{\gamma}}}-\gamma_{\bs{\alpha}\bs{\delta}}^{b}\dshift{3}_{b}\hoch{\bs{\beta}}\RR^{\bs{\delta}\hat{\bs{\gamma}}}+2\gamma_{\bs{\alpha}\bs{\delta}}^{b}\dshift{3}_{b}\hoch{\bs{\beta}}\RR^{\bs{\delta}\hat{\bs{\gamma}}}=\\
 & = & \tilde{C}_{\bs{\alpha}}\hoch{\bs{\beta}\hat{\bs{\gamma}}}+\underbrace{\gamma_{\bs{\alpha}\bs{\delta}}^{b}}_{-\gamma_{\bs{\delta}\bs{\alpha}}^{b}}\dshift{3}_{b}\hoch{\bs{\beta}}\RR^{\bs{\delta}\hat{\bs{\gamma}}}\qquad\fussend\end{eqnarray*}
} } (\ref{eq:holConstrVa}) and (\ref{eq:holConstrVIIIb}), relating
them to $\Omega_{M\bs{\alpha}}\hoch{\bs{\beta}}$. It is thus enough
to memorize the transformation of the connection $\Omega_{M\bs{\alpha}}\hoch{\bs{\beta}}$.
Remember now the definition of the torsion as $\gem{T}^{A}=\de E^{A}-E^{B}\wedge\gem{\Omega}_{B}\hoch{A}$.
This implies the following transformation of the corresponding torsion
component (see also (\ref{eq:TwithNewConn}) in the appendix on page
\pageref{eq:TwithNewConn}): \begin{eqnarray}
\tilde{T}_{\bs{\alpha}_{1}\bs{\alpha}_{2}}\hoch{\bs{\beta}} & = & T_{\bs{\alpha}_{1}\bs{\alpha}_{2}}\hoch{\bs{\beta}}-\gamma_{\bs{\alpha}_{1}\bs{\alpha}_{2}}^{b}\dshift{3}_{b}\hoch{\bs{\beta}}\end{eqnarray}
Due to the nilpotency constraints we have $T_{\bs{\alpha}_{1}\bs{\alpha}_{2}}\hoch{\bs{\beta}}\propto\gamma_{\bs{\alpha}_{1}\bs{\alpha}_{2}}^{b}$.
In addition, the left-right symmetry of proposition \vref{prop:left-right-symmetry}
induces the same statements for $\hat{T}_{\hat{\bs{\alpha}}_{1}\hat{\bs{\alpha}}_{2}}\hoch{\hat{\bs{\beta}}}$
and the second residual shift symmetry related to the reparametrization
of $\hat{\dP}_{\hat{\bs{\gamma}}}$. We can therefore completely fix
the two residual gauge symmetries by choosing the (obviously accessible)
gauge\rem{alternativ:$T_{\bs{\alpha\beta}}\hoch{\bs{\gamma}}\To-\hat{T}_{\bs{\alpha\beta}}\hoch{\bs{\gamma}},\quad T_{\hat{\bs{\alpha}}\hat{\bs{\beta}}}\hoch{\hat{\bs{\gamma}}}\To-\hat{T}_{\hat{\bs{\alpha}}\hat{\bs{\beta}}}\hoch{\hat{\bs{\gamma}}}$?}
\begin{equation}
\boxed{T_{\bs{\alpha\beta}}\hoch{\bs{\gamma}}=0,\qquad\hat{T}_{\hat{\bs{\alpha}}\hat{\bs{\beta}}}\hoch{\hat{\bs{\gamma}}}=0}\label{eq:convConstr}\end{equation}
We can now immediately take advantage of this additional (conventional)
constraint and check the validity of the constraints (\ref{eq:holConstrXIa})
and (\ref{eq:holConstrXIIa}) on page \pageref{eq:holConstrXIa}.

\section{Further discussion of some selected constraints}

\label{sec:Further-discussion}There are some constraints which deserve
further examination, before we move on to study the Bianchi identities.
First, the four constraints (\ref{eq:holConstrXIa}), (\ref{eq:holConstrXIIa})
and their hatted versions on page \pageref{eq:holConstrXIIa} do not
look very useful as they stand. We will show that they are actually
consequences of other constraints. Second, with (\ref{eq:holConstrVIIIa})
and (\ref{eq:holConstrVIIIb}) we have two equations for $S_{\bs{\alpha}\hat{\bs{\alpha}}}\hoch{\bs{\beta}\hat{\bs{\beta}}}$
and it is interesting to know whether they are equivalent or not.
Let us start with this last problem:

\paragraph{Consistency of (\ref{eq:holConstrVIIIa}) and (\ref{eq:holConstrVIIIb})}

In the following we will (actually just for convenience) frequently
use the new conventional constraint $T_{\bs{\alpha\beta}}\hoch{\bs{\gamma}}=0=\hat{T}_{\hat{\bs{\alpha}}\hat{\bs{\beta}}}\hoch{\hat{\bs{\gamma}}}$
(\ref{eq:convConstr}). Starting with (\ref{eq:holConstrVIIIa}),
the tensor of interest is given by \begin{eqnarray}
S_{\bs{\alpha}\hat{\bs{\alpha}}}\hoch{\bs{\beta}\hat{\bs{\beta}}} & \ous{(\ref{eq:holConstrVIIIa})}{=}{(\ref{eq:holConstrVa})} & -\gemnabla_{\bs{\alpha}}\gemnabla_{\hat{\bs{\alpha}}}\RR^{\bs{\beta}\hat{\bs{\beta}}}\frem{+2T_{\bs{\alpha}\bs{\delta}}\hoch{\bs{\beta}}\gemnabla_{\hat{\bs{\alpha}}}\RR^{\bs{\delta}\hat{\bs{\beta}}}}+2\hat{R}_{\bs{\alpha}\hat{\bs{\gamma}}\hat{\bs{\alpha}}}\hoch{\hat{\bs{\beta}}}\RR^{\bs{\beta}\hat{\bs{\gamma}}}=\nonumber \\
 & \stackrel{(\ref{eq:generalCommutatorOfCovDer})}{=} & -\gemnabla_{\hat{\bs{\alpha}}}\gemnabla_{\bs{\alpha}}\RR^{\bs{\beta}\hat{\bs{\beta}}}+2\underbrace{\gemT_{\bs{\alpha}\hat{\bs{\alpha}}}\hoch{D}}_{=0\,\lqn{(\ref{eq:holConstrIV})}}\gemnabla_{D}\RR^{\bs{\beta}\hat{\bs{\delta}}}-2R_{\bs{\alpha}\hat{\bs{\alpha}}\bs{\delta}}\hoch{\bs{\beta}}\RR^{\bs{\delta}\hat{\bs{\beta}}}-2\hat{R}_{\bs{\alpha}\hat{\bs{\alpha}}\hat{\bs{\delta}}}\hoch{\hat{\bs{\beta}}}\RR^{\bs{\beta}\hat{\bs{\delta}}}+2\hat{R}_{\bs{\alpha}\hat{\bs{\gamma}}\hat{\bs{\alpha}}}\hoch{\hat{\bs{\beta}}}\RR^{\bs{\beta}\hat{\bs{\gamma}}}\qquad\end{eqnarray}
In order for this to be compatible with (\ref{eq:holConstrVIIIb}),
i.e. with \begin{eqnarray}
S_{\bs{\alpha}\hat{\bs{\alpha}}}\hoch{\bs{\beta}\hat{\bs{\beta}}} & \ous{(\ref{eq:holConstrVIIIb})}{=}{(\ref{eq:holConstrVb})} & -\gemnabla_{\hat{\bs{\alpha}}}\gemnabla_{\bs{\alpha}}\RR^{\bs{\beta}\hat{\bs{\beta}}}\frem{+2\hat{T}_{\hat{\bs{\alpha}}\hat{\bs{\delta}}}\hoch{\hat{\bs{\beta}}}\gemnabla_{\bs{\alpha}}\RR^{\bs{\beta}\hat{\bs{\delta}}}}+2R_{\hat{\bs{\alpha}}\bs{\gamma}\bs{\alpha}}\hoch{\bs{\beta}}\RR^{\bs{\gamma}\hat{\bs{\beta}}}\end{eqnarray}
the curvature has to obey\begin{equation}
R_{\hat{\bs{\alpha}}[\bs{\alpha}\bs{\delta}]}\hoch{\bs{\beta}}\RR^{\bs{\delta}\hat{\bs{\beta}}}-\hat{R}_{\bs{\alpha}[\hat{\bs{\alpha}}\hat{\bs{\delta}}]}\hoch{\hat{\bs{\beta}}}\RR^{\bs{\beta}\hat{\bs{\delta}}}=0\label{eq:consistencyOfSconstr}\end{equation}
In fact, this condition will be a simple consequence of the torsion
Bianchi identities that we will obtain in (\ref{eq:(delta|0,2,1)})
and (\ref{eq:(hdelta|0,1,2)}).

\paragraph{Check of (\ref{eq:holConstrXIa})}

The constraint (\ref{eq:holConstrXIa}) contains the covariant derivative
of $C_{\bs{\alpha}}\hoch{\bs{\beta}\hat{\bs{\gamma}}}$ for which
we can use in turn the constraint (\ref{eq:holConstrVa}) together
with our new constraint (\ref{eq:convConstr}).\begin{eqnarray}
\lqn{\gemnabla_{[\bs{\alpha}_{2}}C_{\bs{\alpha}_{1}]}\hoch{\bs{\beta}\hat{\bs{\gamma}}}-2R_{[\bs{\alpha}_{2}|\bs{\delta}|\bs{\alpha}_{1}]}\hoch{\bs{\beta}}\RR^{\bs{\delta}\hat{\bs{\gamma}}}=}\nonumber \\
 & \stackrel{(\ref{eq:holConstrVa})}{=} & \gemnabla_{[\bs{\alpha}_{2}}\gemnabla_{\bs{\alpha}_{1}]}\RR^{\bs{\beta}\hat{\bs{\gamma}}}-2R_{[\bs{\alpha}_{2}|\bs{\delta}|\bs{\alpha}_{1}]}\hoch{\bs{\beta}}\RR^{\bs{\delta}\hat{\bs{\gamma}}}=\nonumber \\
 & \stackrel{(\ref{eq:generalCommutatorOfCovDer})}{=} & -\gem{T}_{\bs{\alpha}_{2}\bs{\alpha}_{1}}\hoch{D}\gemnabla_{D}\RR^{\bs{\beta}\hat{\bs{\gamma}}}+3\underbrace{R_{[\bs{\alpha}_{2}\bs{\alpha}_{1}\bs{\delta}]}\hoch{\bs{\beta}}}_{=0\,\lqn{{\scriptstyle (\ref{eq:nilpotency-constraint-onR})}}}\RR^{\bs{\delta}\hat{\bs{\gamma}}}+\underbrace{\hat{R}_{\bs{\alpha}_{2}\bs{\alpha}_{1}\hat{\bs{\delta}}}\hoch{\hat{\bs{\gamma}}}}_{=0\,\lqn{{\scriptstyle (\ref{eq:holConstrVII}),(\ref{eq:nilpotency-constraint-onH})}}}\RR^{\bs{\beta}\hat{\bs{\delta}}}\end{eqnarray}
Only the first term remains, but recalling the nilpotency constraint
(\ref{eq:nilpotency-constraint-torsion}) in combination with (\ref{eq:bispinorConstraintInDifferentVersions}),
we observe that also this term vanishes, when contracted with $\gamma_{a_{1}\ldots a_{5}}^{\bs{\alpha}_{1}\bs{\alpha}_{2}}$.
The constraint (\ref{eq:holConstrXIa}) therefore does not give new
information and will be omitted in future listings. The same is true
of course for its hatted version due to the left-right symmetry.

\paragraph{Relating (\ref{eq:holConstrXIIa}) to a Bianchi identity}

For the constraint (\ref{eq:holConstrXIIa}) we have to consider the
following combination\begin{eqnarray}
\lqn{\gemnabla_{[\bs{\alpha}_{2}}S_{\bs{\alpha}_{1}]\hat{\bs{\alpha}}}\hoch{\bs{\beta}\hat{\bs{\beta}}}-2\hat{R}_{[\bs{\alpha}_{1}|\hat{\bs{\gamma}}\hat{\bs{\alpha}}}\hoch{\hat{\bs{\beta}}}C_{{|\bs{\alpha}}_{2}]}\hoch{\bs{\beta}\hat{\bs{\gamma}}}+2R_{[\bs{\alpha}_{2}|\bs{\delta}|\bs{\alpha}_{1}]}\hoch{\bs{\beta}}\hat{C}_{\hat{\bs{\alpha}}}\hoch{\hat{\bs{\beta}}\bs{\delta}}=}\nonumber \\
 & \ous{(\ref{eq:holConstrVIIIa})}{=}{(\ref{eq:holConstrVa})} & -\gemnabla_{[\bs{\alpha}_{2}|}\left(\gemnabla_{|\bs{\alpha}_{1}]}\gemnabla_{\hat{\bs{\alpha}}}\RR^{\bs{\beta}\hat{\bs{\beta}}}-2\hat{R}_{|\bs{\alpha}_{1}]\hat{\bs{\gamma}}\hat{\bs{\alpha}}}\hoch{\hat{\bs{\beta}}}\RR^{\bs{\beta}\hat{\bs{\gamma}}}\right)-2\hat{R}_{[\bs{\alpha}_{1}|\hat{\bs{\gamma}}\hat{\bs{\alpha}}}\hoch{\hat{\bs{\beta}}}\gemnabla_{{|\bs{\alpha}}_{2}]}\RR^{\bs{\beta}\hat{\bs{\gamma}}}+2R_{[\bs{\alpha}_{2}|\bs{\delta}|\bs{\alpha}_{1}]}\hoch{\bs{\beta}}\gemnabla_{\hat{\bs{\alpha}}}\RR^{\hat{\bs{\beta}}\bs{\delta}}=\nonumber \\
 & \stackrel{(\ref{eq:generalCommutatorOfCovDer})}{=} & \gemT_{\bs{\alpha}_{2}\bs{\alpha}_{1}}\hoch{C}\gemnabla_{C}\gemnabla_{\hat{\bs{\alpha}}}\RR^{\bs{\beta}\hat{\bs{\beta}}}+\underbrace{\hat{R}_{\bs{\alpha}_{2}\bs{\alpha}_{1}\hat{\bs{\alpha}}}\hoch{\hat{\bs{\gamma}}}}_{=0\,\lqn{{\scriptstyle (\ref{eq:holConstrVII}),(\ref{eq:nilpotency-constraint-onH})}}}\gemnabla_{\hat{\bs{\gamma}}}\RR^{\bs{\beta}\hat{\bs{\beta}}}-\underbrace{\hat{R}_{\bs{\alpha}_{2}\bs{\alpha}_{1}\hat{\bs{\gamma}}}\hoch{\hat{\bs{\beta}}}}_{=0\,\lqn{{\scriptstyle (\ref{eq:holConstrVII}),(\ref{eq:nilpotency-constraint-onH})}}}\gemnabla_{\hat{\bs{\alpha}}}\RR^{\bs{\beta}\hat{\bs{\gamma}}}+\nonumber \\
 &  & +2\gemnabla_{[\bs{\alpha}_{2}|}\hat{R}_{|\bs{\alpha}_{1}]\hat{\bs{\gamma}}\hat{\bs{\alpha}}}\hoch{\hat{\bs{\beta}}}\RR^{\bs{\beta}\hat{\bs{\gamma}}}+\underbrace{2R_{[\bs{\alpha}_{2}\bs{\delta}\bs{\alpha}_{1}]}\hoch{\bs{\beta}}}_{=0\,\lqn{{\scriptstyle (\ref{eq:nilpotency-constraint-onR})}}}\gemnabla_{\hat{\bs{\alpha}}}\RR^{\bs{\delta}\hat{\bs{\beta}}}=\nonumber \\
 & = & \gemT_{\bs{\alpha}_{2}\bs{\alpha}_{1}}\hoch{C}\gemnabla_{C}\gemnabla_{\hat{\bs{\alpha}}}\RR^{\bs{\beta}\hat{\bs{\beta}}}+2\gemnabla_{[\bs{\alpha}_{2}|}\hat{R}_{|\bs{\alpha}_{1}]\hat{\bs{\gamma}}\hat{\bs{\alpha}}}\hoch{\hat{\bs{\beta}}}\RR^{\bs{\beta}\hat{\bs{\gamma}}}\end{eqnarray}
The first term vanishes again when contracted with $\gamma_{a_{1}\ldots a_{5}}^{\bs{\alpha}_{1}\bs{\alpha}_{2}}$
((\ref{eq:nilpotency-constraint-torsion}) and (\ref{eq:bispinorConstraintInDifferentVersions}))
and the constraint (\ref{eq:holConstrXIIa}) reduces to\begin{equation}
\gamma_{a_{1}\ldots a_{5}}^{\bs{\alpha}_{1}\bs{\alpha}_{2}}\gemnabla_{[\bs{\alpha}_{2}|}\hat{R}_{|\bs{\alpha}_{1}]\hat{\bs{\gamma}}\hat{\bs{\alpha}}}\hoch{\hat{\bs{\beta}}}\RR^{\bs{\beta}\hat{\bs{\gamma}}}=0\label{eq:RestVomLangenConstraint}\end{equation}
We will see in a second that this equation is automatically fulfilled
when the Bianchi identity for the curvature is fulfilled. We will
study the Bianchi identities at a later point, but not all of those
for the curvature, because we intend to make use of Dragon's theorem,
relating second to first Bianchi identity. Let us therefore write
down at this point the Bianchi\index{second Bianchi identity|see curvature Bianchi identity|see{curvature BI}}\index{curvature!Bianchi identity}\index{Bianchi identity!for curvature}
identity that we have in mind (see (\ref{eq:BIforRcov}) on page \pageref{eq:BIforRcov}):\begin{eqnarray}
0 & \stackrel{!}{=} & \gemnabla_{[\bs{\alpha}_{2}|}\hat{R}_{|\bs{\alpha}_{1}\hat{\bs{\gamma}}]\hat{\bs{\alpha}}}\hoch{\hat{\bs{\beta}}}+2\gemT_{[\bs{\alpha}_{2}\bs{\alpha}_{1}|}\hoch{D}\hat{R}_{D|\hat{\bs{\gamma}}]\hat{\bs{\alpha}}}\hoch{\hat{\bs{\beta}}}=\nonumber \\
 & = & \frac{2}{3}\gemnabla_{[\bs{\alpha}_{2}|}\hat{R}_{|\bs{\alpha}_{1}]\hat{\bs{\gamma}}\hat{\bs{\alpha}}}\hoch{\hat{\bs{\beta}}}+\frac{1}{3}\gemnabla_{\hat{\bs{\gamma}}}\underbrace{\hat{R}_{\bs{\alpha}_{2}\bs{\alpha}_{1}\hat{\bs{\alpha}}}\hoch{\hat{\bs{\beta}}}}_{\hspace{-1cm}=0\,\lqn{{\scriptstyle (\ref{eq:holConstrVII}),(\ref{eq:nilpotency-constraint-onH})}}}+\frac{4}{3}\underbrace{\gemT_{\hat{\bs{\gamma}}[\bs{\alpha}_{2}|}\hoch{D}}_{\hspace{.5cm}=0\,\lqn{(\ref{eq:holConstrIV})}}\hat{R}_{D|\bs{\alpha}_{1}]\hat{\bs{\alpha}}}\hoch{\hat{\bs{\beta}}}+\frac{2}{3}\gemT_{\bs{\alpha}_{2}\bs{\alpha}_{1}}\hoch{D}\hat{R}_{D\hat{\bs{\gamma}}\hat{\bs{\alpha}}}\hoch{\hat{\bs{\beta}}}\end{eqnarray}
Once again the last torsion term vanishes when contracted with $\gamma_{a_{1}\ldots a_{5}}^{\bs{\alpha}_{1}\bs{\alpha}_{2}}$,
so that the above Bianchi identity implies \begin{equation}
\gamma_{a_{1}\ldots a_{5}}^{\bs{\alpha}_{1}\bs{\alpha}_{2}}\gemnabla_{[\bs{\alpha}_{2}|}\hat{R}_{|\bs{\alpha}_{1}]\hat{\bs{\gamma}}\hat{\bs{\alpha}}}\hoch{\hat{\bs{\beta}}}=0\label{eq:staerkerAlsRestVomLangen}\end{equation}
which is even stronger than (\ref{eq:RestVomLangenConstraint}). 
Of course we also get a hatted version of this constraint.

\section{BI's \& Collected constraints}

\label{sec:Collected-constraints}The next step ist to study all the
Bianchi\index{Bianchi identitiy} identities. The logic is as follows:
We have obtained certain constraints on the $H$-field, on the torsion
and on the curvature. As these objects are defined in terms of $B$-field,
vielbein and connection via $H=\de B$, $T^{A}=\de E^{A}-E^{B}\wedge\Omega_{B}\hoch{A}$
and $R_{A}\hoch{B}=\de\Omega_{A}\hoch{B}-\Omega_{A}\hoch{C}\wedge\Omega_{C}\hoch{A}$,
the constraints can be seen as differential equations for the elementary
fields. If one solved these equations and calculated again $H$-field,
torsion and curvature, one would observe additional constraints that
one had not seen in the beginning. Solving the differential equations
is a very hard problem, but the additional constraints on the derived
objects ($H$-field, torsion and curvature) can be obtained by the
Bianchi identites, without knowing the explicit solutions for the
elementary fields. Indeed the Bianchi identities can help to derive
the solutions. Depending on the point of view, the identities are
a direct consequence of  either the nilpotency of the de Rham differential
$\de^{2}=0$ (see appendix \vref{cha:BIs}) or of the Jacobi identity
for the commutator. Their explicit form, using the schematic index
notation of \pageref{par:Schematic-index-notation}, reads:\begin{eqnarray}
\gemnabla_{\bs{A}}H_{\bs{AAA}}+3\gemT_{\bs{AA}}\hoch{C}H_{C\bs{AA}} & \stackrel{!}{=} & 0\label{eq:BI-H-main}\\
\gemnabla_{\bs{A}}\gemT_{\bs{AA}}\hoch{D}+2\gemT_{\bs{AA}}\hoch{C}\gemT_{C\bs{A}}\hoch{D} & \stackrel{!}{=} & \gemR_{\bs{AAA}}\hoch{D}\label{eq:BI-T-main}\\
\gemnabla_{\bs{A}}\gemR_{\bs{AA}B}\hoch{C}+2\gemT_{\bs{AA}}\hoch{D}\gemR_{D\bs{A}B}\hoch{C} & \stackrel{!}{=} & 0\label{eq:BI-R-main}\end{eqnarray}
Repeated bold indices at the same altitude are simply antisymmetrized
ones. Dragon's theorem (see page \pageref{thm:Dragon}) tells us that
-- when the torsion Bianchi identity is fulfilled -- we can replace
the curvature Bianchi identity by the weaker condition\begin{eqnarray}
\lqn{\gemR_{\bs{CC}B}\hoch{A}\gemT_{\bs{CC}}\hoch{B}=}\nonumber \\
 & = & \gem{\nabla}_{\bs{C}}\gem{\nabla}_{\bs{C}}\gem{T}_{\bs{CC}}\hoch{A}+\gem{T}_{\bs{CC}}\hoch{D}\gemnabla_{D}\gemT_{\bs{CC}}\hoch{A}+2\left(\gemnabla_{\bs{C}}\gemT_{\bs{CC}}\hoch{B}+2\gemT_{\bs{CC}}\hoch{D}\gemT_{D\bs{C}}\hoch{B}\right)\gemT_{B\bs{C}}\hoch{A}\label{eq:replaceBI-R-main}\end{eqnarray}
We will anyway concentrate on the Bianchi identities for $H$-field
and torsion, because they provide most directly useful new algebraic
constraints. 

Note that all constraints so far were obtained for objects based on
$\gemOm_{MA}\hoch{B}=\diag(\check{\Omega}_{Ma}\hoch{b},\Omega_{M\bs{\alpha}}\hoch{\bs{\beta}},\hat{\Omega}_{M\hat{\bs{\alpha}}}\hoch{\hat{\bs{\beta}}})$,
the mixed connection defined in (\ref{eq:mixedConnection}) on page
\pageref{eq:mixedConnection}. It contains three a priori independent
blocks which all decay further in a Lorentz and a scale connection.
One of the important results from the study of the Bianchi identities
is that the torsion components $\check{T}_{\bs{\alpha\beta}}\hoch{c}$
and $\check{T}_{\hat{\bs{\alpha}}\hat{\bs{\beta}}}\hoch{c}$ are related
to $\gamma_{\bs{\alpha\beta}}^{c}$ and $\gamma_{\hat{\bs{\alpha}}\hat{\bs{\beta}}}^{c}$
respectively by a Lorentz plus scale transformation. It is discussed
in an intermezzo on page \pageref{Intermezzo:fixingTwoLorentz} (and
was also used in Berkovit's and Howe's original work \cite{Berkovits:2001ue})
that this can be used to fix two of the three independent blocks.
One is thus left with one independent copy of Lorentz plus scale which
should leave invariant $\gamma_{\bs{\alpha\beta}}^{c}$ and $\gamma_{\hat{\bs{\alpha}}\hat{\bs{\beta}}}^{c}$.
\index{fixing two of three Lorentz trafos}After this partial gauge
fixing, the mixed connection is not an appropriate choice any longer,
as it does not in general respect the gauge. We therefore introduce
three alternative connections, namely the \textbf{left-mover connection}\index{connection!left mover $\sim$}\index{left mover connection}\index{$\Omega_{MA}\hoch{B}$|itext{left mover connection}}
(defined by $\Omega_{M\bs{\alpha}}\hoch{\bs{\beta}}$ and invariance
of the gamma-matrices), the \textbf{right-mover connection} \index{connection!right mover $\sim$}\index{right mover connection}\index{$\Omega$@$\hat{\Omega}_{MA}\hoch{B}$|itext{right mover connection}}
(defined by $\hat{\Omega}_{M\hat{\bs{\alpha}}}\hoch{\hat{\bs{\beta}}}$
and invariance of the gamma-matrices) and the \textbf{average connection}\index{$\Omega$@$\avOm_{MA}\hoch{B}$|itext{average connection}}\index{average connection $\avOm_{MA}\hoch{B}$}\index{connection!average $\sim$ $\avOm_{MA}\hoch{B}$}
(see beginning of appendix \vref{cha:ConnectionAppend} for more details)\textbf{
}\begin{eqnarray}
\Omega_{MA}\hoch{B} & \equiv & \diag(\Omega_{Ma}\hoch{b},\Omega_{M\bs{\alpha}}\hoch{\bs{\beta}},\Omega_{M\hat{\bs{\alpha}}}\hoch{\hat{\bs{\beta}}}),\quad\nabla_{M}\gamma_{\bs{\alpha\beta}}^{c}=\nabla_{M}\gamma_{\hat{\bs{\alpha}}\hat{\bs{\beta}}}^{c}=0\\
\hat{\Omega}_{MA}\hoch{B} & \equiv & \diag(\hat{\Omega}_{Ma}\hoch{b},\hat{\Omega}_{M\bs{\alpha}}\hoch{\bs{\beta}},\hat{\Omega}_{M\hat{\bs{\alpha}}}\hoch{\hat{\bs{\beta}}}),\quad\hat{\nabla}_{M}\gamma_{\bs{\alpha\beta}}^{c}=\hat{\nabla}_{M}\gamma_{\hat{\bs{\alpha}}\hat{\bs{\beta}}}^{c}=0\\
\avOm_{MA}\hoch{B} & \equiv & \frac{1}{2}\left(\Omega_{MA}\hoch{B}+\hat{\Omega}_{MA}\hoch{B}\right)\end{eqnarray}
In addition we define the \textbf{difference\index{difference tensor}
tensor }\index{$\Delta_{MA}\hoch{B}$} \begin{eqnarray}
\Delta_{MA}\hoch{B} & \equiv & \hat{\Omega}_{MA}\hoch{B}-\Omega_{MA}\hoch{B}=\diag(\Delta_{Ma}\hoch{b},\Delta_{M\bs{\alpha}}\hoch{\bs{\beta}},\Delta_{M\hat{\bs{\alpha}}}\hoch{\hat{\bs{\beta}}})\label{eq:diff-tensor-in-collection}\end{eqnarray}
The Bianchi identities (\ref{eq:BI-H-main})-(\ref{eq:BI-R-main})
should of course also hold when all objects are based on the above
newly defined connections. This does not put restrictions on $\Delta_{MA}\hoch{B}$.
All different versions (based on different connections) of the Bianchi
identities will lead to equivalent information (see proposition \vref{prop:BI-with-shifted-connection}).
As they are most conveniently written down in terms of the mixed connection,
we will follow this path. Only the bosonic block $\check{\Omega}_{Ma}\hoch{b}$
will, depending on possible simplifications, be chosen to coincide
with either the left-mover connection $\Omega_{Ma}\hoch{b}$ or the
right-mover connection $\hat{\Omega}_{Ma}\hoch{b}$. The corresponding
calculations are lengthy and mostly not very elluminating, so we put
them into the local appendices at the end of this part of the thesis.
There we first start with collecting all constraints that we have
derived so far in appendix \vref{sec:constraintsBeforeBI} and then
discuss the Bianchi identities in detail starting from page \pageref{sec:Bianchi-identities-forH}.
Some conceptually more interesting discussions within these appendicies
are seperated in intermezzi. The first intermezzo on page \pageref{Intermezzo:fixingTwoLorentz}
is, as already mentioned, about the fixing of two of the three copies
of Lorentz plus scale transformations. The next on page \pageref{Intermezzo:differenceTensor}
is about how to determine the complete difference tensor from the
obtained constraints. There is finally a third intermezzo on page
\pageref{intermezzo:RR} which discusses the relation between constraints
on the RR-bispinors and constraints (or equations of motion) for the
corresponding p-forms.

After all this work in the local appendices, we will now collect all
the constraints\index{constraints!collected $\sim$ on the background fields}\index{collected constraints}
on the background fields that we have obtained, including the ones
from the Bianchi identities. If we later, within the derivation of
the supergravity transformations of some component fields, make use
of some explicit form of components of torsion, curvature or other
background fields without giving the explicit equation number, the
corresponding equation should be among the following ones. 

Not all equations that we are going to write are independent. It
is sometimes convenient to have them in different versions and grouped
in different ways. In particular we will give for later convenience
the explicit form of the torsion components based on left-mover, right-mover
and average connection, although this contains a lot of redundancy.

\paragraph{Restricted structure group constraints}

The first set of constraints is related to the restriction\index{restriction of the structure group to Lorentz and scale}
of the structure\index{structure group!restriction to Lorentz and scale}
group (of the supermanifold) to a a block diagonal form with three
copies of Lorentz and scale transformations. \rem{$(\check{SO}(1,9)\times\check{U}(1))\times\left(Spin(1,9)\times U(1)\right)\times\left(\hat{Spin}(1,9)\times\hat{U}(1)\right)$}This
was discussed in a paragraph on pages \pageref{eq:BlockDiagLambda}-\pageref{par:LocalStructureGroupTransformations},
in the remark on page \pageref{remark:structureGroupValuedConnection}
and in the intermezzo on page \pageref{Intermezzo:bosonic structure group}.
The following equations are taken from (\ref{eq:reducedConnectionForm})-(\ref{eq:reducedSform}),
(\ref{eq:endlichMetrikConstraint}) or (\ref{eq:metricDegenerate})
and (\ref{eq:metricWithCompensator})\begin{eqnarray}
\Omega_{M\bs{\alpha}}\hoch{\bs{\beta}} & = & \frac{1}{2}\Omega_{M}^{(D)}\delta_{\bs{\alpha}}\hoch{\bs{\beta}}+\frac{1}{4}\Omega_{Ma_{1}a_{2}}^{(L)}\gamma^{a_{1}a_{2}}\tief{\bs{\alpha}}\hoch{\bs{\beta}},\qquad\hat{\Omega}_{M\hat{\bs{\alpha}}}\hoch{\hat{\bs{\beta}}}=\frac{1}{2}\hat{\Omega}_{M}^{(D)}\delta_{\hat{\bs{\alpha}}}\hoch{\hat{\bs{\beta}}}+\frac{1}{4}\hat{\Omega}_{Ma_{1}a_{2}}^{(L)}\gamma^{a_{1}a_{2}}\tief{\hat{\bs{\alpha}}}\hoch{\hat{\bs{\beta}}}\\
C_{\bs{\alpha}}\hoch{\bs{\beta}\hat{\bs{\gamma}}} & = & \frac{1}{2}C^{\hat{\bs{\gamma}}}\delta_{\bs{\alpha}}\hoch{\bs{\beta}}+\frac{1}{4}C_{a_{1}a_{2}}^{\hat{\bs{\gamma}}}\gamma^{a_{1}a_{2}}\tief{\bs{\alpha}}\hoch{\bs{\beta}},\qquad\hat{C}_{\hat{\bs{\alpha}}}\hoch{\hat{\bs{\beta}}\bs{\gamma}}=\frac{1}{2}\hat{C}^{\bs{\gamma}}\delta_{\hat{\bs{\alpha}}}\hoch{\hat{\bs{\beta}}}+\frac{1}{4}\hat{C}_{a_{1}a_{2}}^{\bs{\gamma}}\gamma^{a_{1}a_{2}}\tief{\hat{\bs{\alpha}}}\hoch{\hat{\bs{\beta}}}\label{eq:coll:C:strgr-restr}\\
S_{\bs{\alpha}\hat{\bs{\alpha}}}\hoch{\bs{\beta}\hat{\bs{\beta}}} & = & \frac{1}{4}S\delta_{\bs{\alpha}}\hoch{\bs{\beta}}\delta_{\hat{\bs{\alpha}}}\hoch{\hat{\bs{\beta}}}+\frac{1}{8}S_{a_{1}a_{2}}\delta_{\bs{\alpha}}\hoch{\bs{\beta}}\gamma^{a_{1}a_{2}}\tief{\hat{\bs{\alpha}}}\hoch{\hat{\bs{\beta}}}+\nonumber \\
 &  & +\frac{1}{8}\hat{S}_{a_{1}a_{2}}\gamma^{a_{1}a_{2}}\tief{\bs{\alpha}}\hoch{\bs{\beta}}\delta_{\hat{\bs{\alpha}}}\hoch{\hat{\bs{\beta}}}+\frac{1}{16}S_{a_{1}a_{2}b_{1}b_{2}}\gamma^{a_{1}a_{2}}\tief{\bs{\alpha}}\hoch{\bs{\beta}}\gamma^{b_{1}b_{2}}\tief{\hat{\bs{\alpha}}}\hoch{\hat{\bs{\beta}}}\\
G_{MN} & = & E_{M}\hoch{a}G_{ab}E_{N}\hoch{b},\qquad G_{ab}=e^{2\Phi}\eta_{ab}\end{eqnarray}

\index{$G_{MN}$}\index{$B_{MN}$}\index{$E_M\hoch{\bs\alpha}$}\index{$E_M\hoch{\hat{\bs\alpha}}$}\index{$P$@$\RR^{\bs{\alpha}\hat{\bs{\beta}}}$|itext{RR-field}}\index{$C_{\bs{\alpha}}\hoch{\bs{\beta}\hat{\bs{\gamma}}}$}\index{$C$@$\hat{C}_{\hat{\bs{\alpha}}}\hoch{\hat{\bs{\beta}}\bs{\gamma}}$}\index{$S_{\bs{\alpha}\hat{\bs{\alpha}}}\hoch{\bs{\beta}\hat{\bs{\beta}}}$}\index{$\Omega_{M\bs{\alpha}}\hoch{\bs{\beta}}$}\index{$\Omega$@$\hat{\Omega}_{M\hat{\bs{\alpha}}}\hoch{\hat{\bs{\beta}}}$}The
above equations (without the last one) are equivalent to\begin{eqnarray}
\gamma^{a_{1}\ldots a_{4}}\tief{\bs{\beta}}\hoch{\bs{\alpha}}\Omega_{M\bs{\alpha}}\hoch{\bs{\beta}} & = & \gamma^{a_{1}\ldots a_{4}}\tief{\hat{\bs{\beta}}}\hoch{\hat{\bs{\alpha}}}\hat{\Omega}_{M\hat{\bs{\alpha}}}\hoch{\hat{\bs{\beta}}}=0\\
\gamma^{a_{1}\ldots a_{4}}\tief{\bs{\beta}}\hoch{\bs{\alpha}}C_{\bs{\alpha}}\hoch{\bs{\beta}\hat{\bs{\gamma}}} & = & \gamma^{a_{1}\ldots a_{4}}\tief{\hat{\bs{\beta}}}\hoch{\hat{\bs{\alpha}}}\hat{C}_{\hat{\bs{\alpha}}}\hoch{\hat{\bs{\beta}}\bs{\gamma}}=0\\
\gamma^{a_{1}\ldots a_{4}}\tief{\bs{\beta}}\hoch{\bs{\alpha}}S_{\bs{\alpha}\hat{\bs{\alpha}}}\hoch{\bs{\beta}\hat{\bs{\beta}}} & = & \gamma^{a_{1}\ldots a_{4}}\tief{\hat{\bs{\beta}}}\hoch{\hat{\bs{\alpha}}}S_{\bs{\alpha}\hat{\bs{\alpha}}}\hoch{\bs{\beta}\hat{\bs{\beta}}}=0\end{eqnarray}

\paragraph{Further constraints on $C$ and $S$ and indirectly on $\RR$}

The constraints (\ref{eq:holConstrVa}) and (\ref{eq:holConstrVb})
on $C$ and (\ref{eq:holConstrVIIIa}) and (\ref{eq:holConstrVIIIb})
on $S$ (all on page \pageref{eq:holConstrVIIIa}) can be regarded
as defining equations. We have already shown in section \ref{sec:Further-discussion}
that the two equations for $S$ are equivalent up to Bianchi identities.
\begin{eqnarray}
C_{\bs{\alpha}}\hoch{\bs{\gamma}\hat{\bs{\gamma}}} & = & \gemnabla_{\bs{\alpha}}\RR^{\bs{\gamma}\hat{\bs{\gamma}}}\label{eq:coll:C}\\
\hat{C}_{\hat{\bs{\alpha}}}\hoch{\hat{\bs{\gamma}}\bs{\gamma}} & = & \gemnabla_{\hat{\bs{\alpha}}}\RR^{\bs{\gamma}\hat{\bs{\gamma}}}\label{eq:coll:Chat}\\
S_{\bs{\alpha}\hat{\bs{\alpha}}}\hoch{\bs{\gamma}\hat{\bs{\beta}}} & = & -\gemnabla_{\bs{\alpha}}\underbrace{\hat{C}_{\hat{\bs{\alpha}}}\hoch{\hat{\bs{\beta}}\bs{\gamma}}}_{\gemnabla_{\hat{\bs{\alpha}}}\RR^{\bs{\gamma}\hat{\bs{\beta}}}}+2\hat{R}_{\bs{\alpha}\hat{\bs{\gamma}}\hat{\bs{\alpha}}}\hoch{\hat{\bs{\beta}}}\RR^{\bs{\gamma}\hat{\bs{\gamma}}}\\
S_{\bs{\alpha}\hat{\bs{\alpha}}}\hoch{\bs{\beta}\hat{\bs{\gamma}}} & = & -\gemnabla_{\hat{\bs{\alpha}}}\underbrace{C_{\bs{\alpha}}\hoch{\bs{\beta}\hat{\bs{\gamma}}}}_{\gemnabla_{\bs{\alpha}}\RR^{\bs{\beta}\hat{\bs{\gamma}}}}+2R_{\hat{\bs{\alpha}}\bs{\gamma}\bs{\alpha}}\hoch{\bs{\beta}}\RR^{\bs{\gamma}\hat{\bs{\gamma}}}\end{eqnarray}
In addition we have from the Bianchi identities the equations (\ref{eq:nablaP:hat})
and (\ref{eq:nablaP}): \begin{eqnarray}
\gemnabla_{\hat{\bs{\alpha}}}\RR^{\bs{\alpha}\hat{\bs{\beta}}} & = & -\frac{1}{2}\RR^{\bs{\alpha}\hat{\bs{\gamma}}}\hatcovPhi{\hat{\bs{\gamma}}}\cdot\delta_{\hat{\bs{\alpha}}}\hoch{\hat{\bs{\beta}}}+\left(T_{cd}\hoch{\bs{\alpha}}-\frac{1}{2}\hatcovPhi{\bs{\gamma}}\RR^{\bs{\alpha}\bs{\delta}}\tilde{\gamma}_{cd\,\bs{\delta}}\hoch{\bs{\gamma}}\right)\tilde{\gamma}^{cd}\tief{\hat{\bs{\alpha}}}\hoch{\hat{\bs{\beta}}}\\
\gemnabla_{\bs{\alpha}}\RR^{\bs{\beta}\hat{\bs{\alpha}}} & = & -\frac{1}{2}\RR^{\bs{\gamma}\hat{\bs{\alpha}}}\covPhi{\bs{\gamma}}\cdot\delta_{\bs{\alpha}}\hoch{\bs{\beta}}+\left(\hat{T}_{cd}\hoch{\hat{\bs{\alpha}}}-\frac{1}{2}\covPhi{\bs{\gamma}}\RR^{\bs{\delta}\hat{\bs{\alpha}}}\tilde{\gamma}_{cd\,\bs{\delta}}\hoch{\bs{\gamma}}\right)\tilde{\gamma}^{cd}\tief{\bs{\alpha}}\hoch{\bs{\beta}}\end{eqnarray}
In the intermezzo on page \pageref{intermezzo:RR} we give a qualitative
discussion how these equations are related to field equations for
the corresponding RR-p-form-field-strengths. The above expressions
for the spinorial derivatives of the RR-bispinors (which coincide
with $C$ and $\hat{C}$ according to (\ref{eq:coll:C}) and (\ref{eq:coll:Chat}))
already take into account the restricted structure group according
to (\ref{eq:coll:C:strgr-restr}). In addition they imply upon taking
the trace that \begin{eqnarray}
\gemnabla_{\hat{\bs{\alpha}}}\RR^{\bs{\alpha}\hat{\bs{\alpha}}} & = & 8\RR^{\bs{\alpha}\hat{\bs{\alpha}}}\hatcovPhi{\hat{\bs{\alpha}}}\quad\mbox{or}\quad\gemnabla_{\hat{\bs{\alpha}}}(e^{-8\Phi}\RR^{\bs{\alpha}\hat{\bs{\alpha}}})=0\label{eq:coll:nablahalphaPtrace}\\
\gemnabla_{\bs{\alpha}}\RR^{\bs{\alpha}\hat{\bs{\alpha}}} & = & 8\RR^{\bs{\alpha}\hat{\bs{\alpha}}}\covPhi{\bs{\alpha}}\quad\mbox{or}\quad\gemnabla_{\bs{\alpha}}(e^{-8\Phi}\RR^{\bs{\alpha}\hat{\bs{\alpha}}})=0\label{eq:coll:nablaalphaPtrace}\end{eqnarray}

\paragraph{Constraints on $H$}

Due to (\ref{eq:Y-constrI})-(\ref{eq:Y-constrV}), (\ref{eq:nilpotency-constraint-onH}),
(\ref{eq:nilpotency-constraint-onH-hat}), (\ref{eq:(0,4,0)''}),
(\ref{eq:(0,0,4)''}) and the total antisymmetry of $H$, its only
nonvanishing components are\index{$H_{ABC}$}\begin{eqnarray}
H_{abc} & \neq & 0\qquad(\mbox{in general})\\
H_{\bs{\alpha}\bs{\beta}c} & = & -\frac{2}{3}\tilde{\gamma}_{c\,\bs{\alpha\beta}}\equiv-\frac{2}{3}e^{2\Phi}\eta_{cd}\gamma_{\bs{\alpha\beta}}^{d}\\
H_{\hat{\bs{\alpha}}\hat{\bs{\beta}}c} & = & \frac{2}{3}\tilde{\gamma}_{c\,\hat{\bs{\alpha}}\hat{\bs{\beta}}}\equiv\frac{2}{3}e^{2\Phi}\eta_{cd}\gamma_{\hat{\bs{\alpha}}\hat{\bs{\beta}}}^{d}\end{eqnarray}
The vanishing components are thus (written a bit redundantly)\begin{eqnarray}
H_{ab\bs{\mc{C}}} & = & H_{\bs{\alpha}\hat{\bs{\beta}}C}=H_{\bs{\mc{A}\mc{B}\mc{C}}}=0\end{eqnarray}
Note that the constraints for $H_{\bs{\alpha}\bs{\beta}c}$ and $H_{\hat{\bs{\alpha}}\hat{\bs{\beta}}c}$
(coming from (\ref{eq:(0,4,0)''}) and (\ref{eq:(0,0,4)''})) are
related to the torsion constraints for $\check{T}_{\bs{\alpha\beta}}\hoch{c}$
and $\check{T}_{\hat{\bs{\alpha}}\hat{\bs{\beta}}}\hoch{c}$ and thus
(as mentioned in the beginning of this section) contain the gauge
fixing of two of the three initially independent Lorentz and scale
transformations (\ref{eq:BlockDiagLambda}). This is explained in
detail at page \pageref{Intermezzo:fixingTwoLorentz}. 

Further conditions on $H$, coming from the Bianchi identities (\ref{eq:(3,1,0)}),
(\ref{eq:(3,0,1)}) and (\ref{eq:(4,0,0)}), are\begin{eqnarray}
\nabla_{\hat{\bs{\delta}}}H_{abc} & = & -4\hat{T}_{[ab|}\hoch{\hat{\bs{\eps}}}\tilde{\gamma}_{|c]\hat{\bs{\eps}}\hat{\bs{\delta}}}\\
\hat{\nabla}_{\bs{\delta}}H_{abc} & = & 4T_{[ab|}\hoch{\bs{\eps}}\tilde{\gamma}_{|c]\bs{\eps}\bs{\delta}}\\
\nabla_{[a}H_{bcd]} & = & -\frac{9}{2}H_{[ab|}\hoch{e}H_{e|cd]}\end{eqnarray}
More information on the torsion components $T_{ab}\hoch{\eps}$ and
$\hat{T}_{ab}\hoch{\hat{\eps}}$ will be given in the corresponding
paragraph below.\rem{In matrix notation, $H$ reads:\begin{eqnarray}
H_{ABc} & = & \left(\begin{array}{ccc}
H_{abc} & H_{a\bs{\beta}c}=0 & H_{a\hat{\bs{\beta}}c}=0\\
H_{\bs{\alpha}bc}=0 & H_{\bs{\alpha\beta}c}=-\frac{2}{3}\tilde{\gamma}_{c\,\bs{\alpha\beta}} & H_{\bs{\alpha}\hat{\bs{\beta}}c}=0\\
H_{\hat{\bs{\alpha}}bc}=0 & H_{\hat{\bs{\alpha}}\bs{\beta}c}=0 & H_{\hat{\bs{\alpha}}\hat{\bs{\beta}}c}=\frac{2}{3}\tilde{\gamma}_{c\,\hat{\bs{\alpha}}\hat{\bs{\beta}}}\end{array}\right)_{AB}\\
H_{AB\bs{\gamma}} & = & \left(\begin{array}{ccc}
H_{ab\bs{\gamma}}=0 & H_{a\bs{\beta}\bs{\gamma}}=-\frac{2}{3}\tilde{\gamma}_{a\,\bs{\beta\gamma}} & H_{a\hat{\bs{\beta}}\bs{\gamma}}=0\\
H_{\bs{\alpha}b\bs{\gamma}}=\frac{2}{3}\tilde{\gamma}_{b\,\bs{\alpha\gamma}} & H_{\bs{\alpha\beta}\bs{\gamma}}=0 & H_{\bs{\alpha}\hat{\bs{\beta}}\bs{\gamma}}=0\\
H_{\hat{\bs{\alpha}}b\bs{\gamma}}=0 & H_{\hat{\bs{\alpha}}\bs{\beta}\bs{\gamma}}=0 & H_{\hat{\bs{\alpha}}\hat{\bs{\beta}}\bs{\gamma}}=0\end{array}\right)_{AB}\\
H_{AB\hat{\bs{\gamma}}} & = & \left(\begin{array}{ccc}
H_{ab\hat{\bs{\gamma}}}=0 & H_{a\bs{\beta}\hat{\bs{\gamma}}}=0 & H_{a\hat{\bs{\beta}}\hat{\bs{\gamma}}}=\frac{2}{3}\tilde{\gamma}_{a\,\hat{\bs{\beta}}\hat{\bs{\gamma}}}\\
H_{\bs{\alpha}b\hat{\bs{\gamma}}}=0 & H_{\bs{\alpha\beta}\hat{\bs{\gamma}}}=0 & H_{\bs{\alpha}\hat{\bs{\beta}}\hat{\bs{\gamma}}}=0\\
H_{\hat{\bs{\alpha}}b\hat{\bs{\gamma}}}=-\frac{2}{3}\tilde{\gamma}_{b\,\hat{\bs{\alpha}}\hat{\bs{\gamma}}} & H_{\hat{\bs{\alpha}}\bs{\beta}\hat{\bs{\gamma}}}=0 & H_{\hat{\bs{\alpha}}\hat{\bs{\beta}}\hat{\bs{\gamma}}}=0\end{array}\right)_{AB}\end{eqnarray}
}

\paragraph{Constraints on the torsion}

Let us now collect the information of the constraints (\ref{eq:Y-constrII})-(\ref{eq:Y-constrIV}),
(\ref{eq:holConstrI})-(\ref{eq:holConstrIV}), (\ref{eq:nilpotency-constraint-onTfinal}),
(\ref{eq:nilpotency-constraint-onTfinal-hat}), (\ref{eq:convConstr})
and the Bianchi identities (\ref{eq:(0,4,0)'}), (\ref{eq:(0,0,4)'}),
(\ref{eq:Tbetaca}), (\ref{eq:Thatbetaca}), (\ref{eq:T^{c}undOmega}),
(\ref{eq:hatT^{c}und-hatOmega}), (\ref{eq:(2,2,0)}), (\ref{eq:(2,0,2)}),
(\ref{eq:(delta|1,0,2)b}) and (\ref{eq:(hdelta|1,2,0)b}). The only
(a priori) nonvanishing components of the torsion $\gem{T}_{AB}\hoch{C}$
are\rem{%
\footnote{For (\ref{eq:T^{c}-sym}) take into account that \begin{eqnarray*}
\Delta_{[\bs{\mc{A}}c]}\hoch{d} & = & \frac{1}{2}\Delta_{\bs{\mc{A}}c}\hoch{d}=\frac{1}{2}\Delta_{\bs{\mc{A}}}\delta_{c}\hoch{d}+\frac{1}{2}\Delta_{\bs{\mc{A}}\, c}^{(L)}\hoch{d}\\
\hat{T}_{\bs{\mc{A}}(c|d)} & = & T_{\bs{\mc{A}}(c|d)}+\frac{1}{2}\Delta_{\bs{\mc{A}}(c|d)}=\\
 & = & T_{\bs{\mc{A}}(c|d)}+\frac{1}{2}\Delta_{\bs{\mc{A}}}G_{cd}\end{eqnarray*}
Remember also that the torsion constraints on $T_{\bs{\alpha}(c|d)}$
are constraints on the vielbein only, not on the connection: \begin{eqnarray*}
\underbrace{2T_{\bs{\mc{A}}(c|d)}-\Omega_{\bs{\mc{A}}}G_{cd}}_{2E_{\bs{\mc{A}}}\hoch{M}E_{(c|}\hoch{N}\left(\partial_{[M}E_{N]}\hoch{e}\right)G_{e|d)}} & = & -\nabla_{\bs{\mc{A}}}\Phi\, G_{cd}\qquad\fussend\end{eqnarray*}
}}\index{$T_{AB}$@$\gem{T}_{AB}\hoch{C}$} \begin{eqnarray}
\check{T}_{\bs{\alpha}\bs{\beta}}\hoch{c} & = & \gamma_{\bs{\alpha\beta}}^{c},\qquad\check{T}_{\hat{\bs{\alpha}}\hat{\bs{\beta}}}\hoch{c}=\gamma_{\hat{\bs{\alpha}}\hat{\bs{\beta}}}^{c}\\
T_{\bs{\alpha}b}\hoch{c} & = & -\frac{1}{2}\covPhi{\bs{\alpha}}\delta_{b}^{c}-\frac{1}{2}\gamma_{b}\hoch{c}\tief{\bs{\alpha}}\hoch{\bs{\beta}}\covPhi{\bs{\beta}},\qquad\hat{T}_{\hat{\bs{\alpha}}b}\hoch{c}=-\frac{1}{2}\hatcovPhi{\hat{\bs{\alpha}}}\delta_{b}^{c}-\frac{1}{2}\gamma_{b}\hoch{c}\tief{\hat{\bs{\alpha}}}\hoch{\hat{\bs{\beta}}}\hatcovPhi{\hat{\bs{\beta}}}\frem{\dann\check{T}_{\bs{\mc{A}}c}\hoch{c}=-5\checkcovPhi{\bs{\mc{A}}}}\\
T_{ab}\hoch{c} & = & \frac{3}{2}H_{ab}\hoch{c},\qquad\hat{T}_{ab}\hoch{c}=-\frac{3}{2}H_{ab}\hoch{c}\\
T_{\hat{\bs{\alpha}}c}\hoch{\bs{\gamma}} & = & \tilde{\gamma}_{c\,\hat{\bs{\alpha}}\hat{\bs{\delta}}}\RR^{\bs{\gamma}\hat{\bs{\delta}}},\qquad\hat{T}_{\bs{\alpha}c}\hoch{\hat{\bs{\gamma}}}=\tilde{\gamma}_{c\,\bs{\alpha\delta}}\RR^{\bs{\delta}\hat{\bs{\gamma}}}\\
T_{ab}\hoch{\bs{\gamma}} & = & \frac{1}{16}\left(\gemnabla_{\hat{\bs{\gamma}}}\RR^{\bs{\gamma}\hat{\bs{\delta}}}+8\hatcovPhi{\hat{\bs{\gamma}}}\RR^{\bs{\gamma}\hat{\bs{\delta}}}\right)\tilde{\gamma}_{ab\,\hat{\bs{\delta}}}\hoch{\hat{\bs{\gamma}}}\frem{vgl\,\nabla H},\qquad\hat{T}_{ab}\hoch{\hat{\bs{\gamma}}}=\frac{1}{16}\left(\gemnabla_{\bs{\gamma}}\RR^{\bs{\delta}\hat{\bs{\gamma}}}+8\covPhi{\bs{\gamma}}\RR^{\bs{\delta}\hat{\bs{\gamma}}}\right)\tilde{\gamma}_{ab\,\bs{\delta}}\hoch{\bs{\gamma}}\end{eqnarray}
The remaining components vanish, which can be written (again a bit
redundantly) as\begin{eqnarray}
\gem{T}_{\bs{\mc{A}\mc{B}}}\hoch{\bs{\mc{C}}} & = & \gem{T}_{\bs{\alpha}\hat{\bs{\alpha}}}\hoch{C}=T_{\bs{\alpha}d}\hoch{\bs{\gamma}}=\hat{T}_{\hat{\bs{\alpha}}d}\hoch{\hat{\bs{\gamma}}}=T_{\hat{\bs{\alpha}}b}\hoch{c}=\hat{T}_{\bs{\alpha}b}\hoch{c}=0\end{eqnarray}
We obtain some additional constraints from the Bianchi identities
(\ref{eq:(delta|2,0,1)}), (\ref{eq:(delta|2,1,0)}), (\ref{eq:(delta|3,0,0)})
and (\ref{eq:(hdelta|3,0,0)}): \begin{eqnarray}
\nabla_{\hat{\bs{\alpha}}}T_{bc}\hoch{\bs{\delta}} & = & -2\tilde{\gamma}_{[b|\,\hat{\bs{\alpha}}\hat{\bs{\delta}}}\gemnabla_{|c]}\RR^{\bs{\delta}\hat{\bs{\delta}}}-3H_{bce}\gamma_{\hat{\bs{\alpha}}\hat{\bs{\delta}}}^{e}\RR^{\bs{\delta}\hat{\bs{\delta}}}\\
\hat{\nabla}_{\bs{\alpha}}\hat{T}_{bc}\hoch{\hat{\bs{\delta}}} & = & -2\tilde{\gamma}_{[b|\,\bs{\alpha}\bs{\delta}}\gemnabla_{|c]}\RR^{\bs{\delta}\hat{\bs{\delta}}}+3H_{bce}\gamma_{\bs{\alpha}\bs{\delta}}^{e}\RR^{\bs{\delta}\hat{\bs{\delta}}}\end{eqnarray}
\begin{eqnarray}
\nabla_{[a}T_{bc]}\hoch{\bs{\delta}} & = & -3H_{[ab|}\hoch{e}T_{e|c]}\hoch{\bs{\delta}}-2\hat{T}_{[ab|}\hoch{\hat{\bs{\eps}}}\tilde{\gamma}_{|c]\,\hat{\bs{\eps}}\hat{\bs{\delta}}}\RR^{\bs{\delta}\hat{\bs{\delta}}}\\
\hat{\nabla}_{[a}\hat{T}_{bc]}\hoch{\hat{\bs{\delta}}} & = & 3H_{[ab|}\hoch{e}\hat{T}_{e|c]}\hoch{\hat{\bs{\delta}}}-2T_{[ab|}\hoch{\bs{\eps}}\tilde{\gamma}_{|c]\,\bs{\eps}\bs{\delta}}\RR^{\bs{\delta}\hat{\bs{\delta}}}\end{eqnarray}

\paragraph{Difference tensor}

With the help of the constraints obtained from the Bianchi identities
the explicit form (\ref{eq:DiffTenI})-(\ref{eq:DiffTenVI}) of the
difference tensor is derived in the intermezzo on page \pageref{Intermezzo:differenceTensor}.
The components with bosonic structure group indices are given by\begin{eqnarray}
\Delta_{Ab}\hoch{c}:\qquad\Delta_{ab|c} & = & -3H_{abc}\label{eq:coll:DiffTenI}\\
\Delta_{\bs{\alpha}b|c} & = & \covPhi{\bs{\alpha}}G_{bc}+\tilde{\gamma}_{bc}\tief{\bs{\alpha}}\hoch{\bs{\delta}}\covPhi{\bs{\delta}}\label{eq:coll:DiffTenII}\\
\Delta_{\hat{\bs{\alpha}}b|c} & = & -\hatcovPhi{\hat{\bs{\alpha}}}G_{bc}-\tilde{\gamma}_{bc}\tief{\hat{\bs{\alpha}}}\hoch{\hat{\bs{\delta}}}\hatcovPhi{\hat{\bs{\delta}}}\label{eq:coll:DiffTenIII}\end{eqnarray}
They determine the components with fermionic structure group indices
to be of the form\begin{eqnarray}
\Delta_{A\bs{\mc{B}}}\hoch{\bs{\mc{A}}}:\quad\Delta_{a\bs{\beta}}\hoch{\bs{\gamma}} & = & -\frac{3}{4}H_{abc}\tilde{\gamma}^{bc}\tief{\bs{\beta}}\hoch{\bs{\gamma}}\qquad,\qquad\qquad\qquad\quad\Delta_{a\hat{\bs{\beta}}}\hoch{\hat{\bs{\gamma}}}=-\frac{3}{4}H_{abc}\tilde{\gamma}^{bc}\tief{\hat{\bs{\beta}}}\hoch{\hat{\bs{\gamma}}}\label{eq:coll:DiffTenIV}\\
\Delta_{\bs{\alpha}\bs{\beta}}\hoch{\bs{\gamma}} & = & \frac{1}{2}\covPhi{\bs{\alpha}}\delta_{\bs{\beta}}\hoch{\bs{\gamma}}+\frac{1}{4}\gamma_{bc}\tief{\bs{\alpha}}\hoch{\bs{\delta}}\covPhi{\bs{\delta}}\gamma^{bc}\tief{\bs{\beta}}\hoch{\bs{\gamma}}\:,\quad\,\Delta_{\hat{\bs{\alpha}}\hat{\bs{\beta}}}\hoch{\hat{\bs{\gamma}}}=-\frac{1}{2}\hatcovPhi{\hat{\bs{\alpha}}}\delta_{\hat{\bs{\beta}}}\hoch{\hat{\bs{\gamma}}}-\frac{1}{4}\gamma_{bc}\tief{\hat{\bs{\alpha}}}\hoch{\hat{\bs{\delta}}}\hatcovPhi{\hat{\bs{\delta}}}\gamma^{bc}\tief{\hat{\bs{\beta}}}\hoch{\hat{\bs{\gamma}}}\qquad\label{eq:coll:DiffTenV}\\
\Delta_{\hat{\bs{\alpha}}\bs{\beta}}\hoch{\bs{\gamma}} & = & -\frac{1}{2}\hatcovPhi{\hat{\bs{\alpha}}}\delta_{\bs{\beta}}\hoch{\bs{\gamma}}-\frac{1}{4}\gamma_{bc}\tief{\hat{\bs{\alpha}}}\hoch{\hat{\bs{\delta}}}\hatcovPhi{\hat{\bs{\delta}}}\gamma^{bc}\tief{\bs{\beta}}\hoch{\bs{\gamma}}\:,\;\Delta_{\bs{\alpha}\hat{\bs{\beta}}}\hoch{\hat{\bs{\gamma}}}=\frac{1}{2}\covPhi{\bs{\alpha}}\delta_{\hat{\bs{\beta}}}\hoch{\hat{\bs{\gamma}}}+\frac{1}{4}\gamma_{bc}\tief{\bs{\alpha}}\hoch{\bs{\delta}}\covPhi{\bs{\delta}}\gamma^{bc}\tief{\hat{\bs{\beta}}}\hoch{\hat{\bs{\gamma}}}\qquad\label{eq:coll:DiffTenVI}\end{eqnarray}
The above equations imply in particular for the scale part (via taking
the trace)\begin{eqnarray}
\dann\Delta_{a}^{(D)} & = & 0\label{eq:DeltaDil-I}\\
\Delta_{\bs{\alpha}}^{(D)} & = & \covPhi{\bs{\alpha}}\label{eq:DeltaDil-II}\\
\Delta_{\hat{\bs{\alpha}}}^{(D)} & = & -\hatcovPhi{\hat{\bs{\alpha}}}\label{eq:DeltaDil-III}\end{eqnarray}
As we meet here the covariant derivatives of the compensator field,
it is useful to add at this place also the constraints (\ref{eq:T^{c}undOmega}),(\ref{eq:hatT^{c}und-hatOmega})
and (\ref{eq:Omegaa}) on the covariant derivative of the compensator
field coming from the Bianchi identities: \begin{equation}
\covPhi{\hat{\bs{\alpha}}}=\hatcovPhi{\bs{\alpha}}=\covPhi{a}=\hatcovPhi{a}=0\end{equation}
Remember that the covariant derivative of the compensator field is
given by $\covPhi{A}=E_{A}\hoch{M}(\partial_{M}\Phi-\Omega_{M}^{(D)})$.

\paragraph{Torsion constraints rewritten in various ways }

Due to the explicit knowledge of the difference tensor, we can write
down all components of $T_{AB}\hoch{C}$, $\hat{T}_{AB}\hoch{C}$
and $\av{T}_{AB}\hoch{C}$ (using e.g. $T_{AB}\hoch{\hat{\bs{\gamma}}}=\hat{T}_{AB}\hoch{\hat{\bs{\gamma}}}-\Delta_{[AB]}\hoch{\hat{\bs{\gamma}}}$).
They will be needed to derive the supersymmetry transformations in
the corresponding gauge. Before we start, let us stress once more
that the scale transformations (or dilatations) are still part of
our superspace structure group. If one prefers to fix the compensator
field $\compensator$ to zero immediately (which would correspond
to \cite{Berkovits:2001ue}), one needs to restrict to the Lorentz
part $\Omega_{MA}^{(L)}\hoch{B}$, $\hat{\Omega}_{MA}^{(L)}\hoch{B}$
or $\av{\Omega}_{MA}^{(L)}\hoch{B}$ of the corresponding connection.
The Lorentz part of the torsion can be obtained via \begin{eqnarray}
T_{AB}^{(L)}\hoch{C} & = & T_{AB}\hoch{C}-\Omega_{[AB]}^{(D)}\hoch{C}\quad\mbox{with }\Omega_{Mb}^{(D)}\hoch{c}=\Omega_{M}^{(D)}\delta_{b}^{c}\mbox{ and }\Omega_{M\bs{\mc{B}}}^{(D)}\hoch{\bs{\mc{C}}}=\tfrac{1}{2}\Omega_{M}^{(D)}\delta_{\bs{\mc{B}}}\hoch{\bs{\mc{C}}}\label{eq:LorentzPartOfTorsion}\end{eqnarray}
This will be made more explicit below for each case. 

Let us now start with the \emph{left-mover torsion}, whose components
$T_{AB}\hoch{C}$ are\begin{eqnarray}
T_{AB}\hoch{c} & \equiv & \left(\begin{array}{ccc}
T_{ab}\hoch{c} & T_{a\bs{\beta}}\hoch{c} & T_{a\hat{\bs{\beta}}}\hoch{c}\\
T_{\bs{\alpha}b}\hoch{c} & T_{\bs{\alpha}\bs{\beta}}\hoch{c} & T_{\bs{\alpha}\hat{\bs{\beta}}}\hoch{c}\\
T_{\hat{\bs{\alpha}}b}\hoch{c} & T_{\hat{\bs{\alpha}}\bs{\beta}}\hoch{c} & T_{\hat{\bs{\alpha}}\hat{\bs{\beta}}}\hoch{c}\end{array}\right)=\left(\begin{array}{ccc}
\frac{3}{2}H_{ab}\hoch{c} & \frac{1}{2}\covPhi{\bs{\beta}}\delta_{a}^{c}+\frac{1}{2}\gamma_{a}\hoch{c}\tief{\bs{\beta}}\hoch{\bs{\delta}}\covPhi{\bs{\delta}} & 0\\
-\frac{1}{2}\covPhi{\bs{\alpha}}\delta_{b}^{c}-\frac{1}{2}\gamma_{b}\hoch{c}\tief{\bs{\alpha}}\hoch{\bs{\delta}}\covPhi{\bs{\delta}} & \gamma_{\bs{\alpha\beta}}^{c} & 0\\
0 & 0 & \gamma_{\hat{\bs{\alpha}}\hat{\bs{\beta}}}^{c}\end{array}\right)\label{eq:coll:leftTorsionI}\\
T_{AB}\hoch{\bs{\gamma}} & \equiv & \left(\begin{array}{ccc}
T_{ab}\hoch{\bs{\gamma}} & T_{a\bs{\beta}}\hoch{\bs{\gamma}} & T_{a\hat{\bs{\beta}}}\hoch{\bs{\gamma}}\\
T_{\bs{\alpha}b}\hoch{\bs{\gamma}} & T_{\bs{\alpha}\bs{\beta}}\hoch{\bs{\gamma}} & T_{\bs{\alpha}\hat{\bs{\beta}}}\hoch{\bs{\gamma}}\\
T_{\hat{\bs{\alpha}}b}\hoch{\bs{\gamma}} & T_{\hat{\bs{\alpha}}\bs{\beta}}\hoch{\bs{\gamma}} & T_{\hat{\bs{\alpha}}\hat{\bs{\beta}}}\hoch{\bs{\gamma}}\end{array}\right)=\left(\begin{array}{ccc}
\frac{1}{16}\left(\gemnabla_{\hat{\bs{\eps}}}\RR^{\bs{\gamma}\hat{\bs{\delta}}}+8\hatcovPhi{\hat{\bs{\eps}}}\RR^{\bs{\gamma}\hat{\bs{\delta}}}\right)\tilde{\gamma}_{ab\,\hat{\bs{\delta}}}\hoch{\hat{\bs{\eps}}} & 0 & -\tilde{\gamma}_{a\,\hat{\bs{\beta}}\hat{\bs{\delta}}}\RR^{\bs{\gamma}\hat{\bs{\delta}}}\\
0 & 0 & 0\\
\tilde{\gamma}_{b\,\hat{\bs{\alpha}}\hat{\bs{\delta}}}\RR^{\bs{\gamma}\hat{\bs{\delta}}} & 0 & 0\end{array}\right)\label{eq:coll:leftTorsionII}\\
T_{AB}\hoch{\hat{\bs{\gamma}}} & \equiv & \left(\begin{array}{ccc}
T_{ab}\hoch{\hat{\bs{\gamma}}} & T_{a\bs{\beta}}\hoch{\hat{\bs{\gamma}}} & T_{a\hat{\bs{\beta}}}\hoch{\hat{\bs{\gamma}}}\\
T_{\bs{\alpha}b}\hoch{\hat{\bs{\gamma}}} & T_{\bs{\alpha}\bs{\beta}}\hoch{\hat{\bs{\gamma}}} & T_{\bs{\alpha}\hat{\bs{\beta}}}\hoch{\hat{\bs{\gamma}}}\\
T_{\hat{\bs{\alpha}}b}\hoch{\hat{\bs{\gamma}}} & T_{\hat{\bs{\alpha}}\bs{\beta}}\hoch{\hat{\bs{\gamma}}} & T_{\hat{\bs{\alpha}}\hat{\bs{\beta}}}\hoch{\hat{\bs{\gamma}}}\end{array}\right)=\label{eq:coll:leftTorsionIII}\\
 &  & \hspace{-2cm}\left(\begin{array}{ccc}
\frac{1}{16}\left(\gemnabla_{\bs{\gamma}}\RR^{\bs{\delta}\hat{\bs{\gamma}}}+8\covPhi{\bs{\gamma}}\RR^{\bs{\delta}\hat{\bs{\gamma}}}\right)\tilde{\gamma}_{ab\,\bs{\delta}}\hoch{\bs{\gamma}} & -\tilde{\gamma}_{a\,\bs{\beta\delta}}\RR^{\bs{\delta}\hat{\bs{\gamma}}} & \frac{3}{8}H_{ade}\tilde{\gamma}^{de}\tief{\hat{\bs{\beta}}}\hoch{\hat{\bs{\gamma}}}\\
\tilde{\gamma}_{b\,\bs{\alpha\delta}}\RR^{\bs{\delta}\hat{\bs{\gamma}}} & 0 & (-\frac{1}{8}\gamma_{de}\tief{\bs{\alpha}}\hoch{\bs{\delta}}\gamma^{de}\tief{\hat{\bs{\beta}}}\hoch{\hat{\bs{\gamma}}}\covPhi{\bs{\delta}}-\frac{1}{4}\covPhi{\bs{\alpha}}\delta_{\hat{\bs{\beta}}}\hoch{\hat{\bs{\gamma}}})\\
-\frac{3}{8}H_{bde}\tilde{\gamma}^{de}\tief{\hat{\bs{\alpha}}}\hoch{\hat{\bs{\gamma}}} & (\frac{1}{8}\gamma_{de}\tief{\bs{\beta}}\hoch{\bs{\delta}}\gamma^{de}\tief{\hat{\bs{\alpha}}}\hoch{\hat{\bs{\gamma}}}\covPhi{\bs{\delta}}+\frac{1}{4}\covPhi{\bs{\beta}}\delta_{\hat{\bs{\alpha}}}\hoch{\hat{\bs{\gamma}}}) & (\frac{1}{4}\gamma_{de\,[\hat{\bs{\alpha}}}\hoch{\hat{\bs{\delta}}}\gamma^{de}\tief{\hat{\bs{\beta}}]}\hoch{\hat{\bs{\gamma}}}\hatcovPhi{\hat{\bs{\delta}}}+\frac{1}{2}\hatcovPhi{[\hat{\bs{\alpha}}}\delta_{\hat{\bs{\beta}}]}\hoch{\hat{\bs{\gamma}}})\end{array}\right)\nonumber \end{eqnarray}
If we want to extract the Lorentz part, only a few of the components
change. Remember $\covPhi{a}=0$ and $\covPhi{\hat{\bs{\alpha}}}=0$
and assume only for this step that $\compensator$ was fixed to zero,
which implies $\nabla_{M}\Phi\To-\Omega_{M}^{(D)}$ and thus $\Omega_{a}^{(D)}=0$
and $\Omega_{\hat{\bs{\alpha}}}^{(D)}=0$. According to (\ref{eq:LorentzPartOfTorsion})
we then have \begin{eqnarray}
T_{\bs{\alpha}b}^{(L)}\hoch{c} & \stackrel{\compensator=0}{=} & T_{\bs{\alpha}b}\hoch{c}-\tfrac{1}{2}\Omega_{\bs{\alpha}}^{(D)}\delta_{b}\hoch{c}=\tfrac{1}{2}\gamma_{b}\hoch{c}\tief{\bs{\alpha}}\hoch{\bs{\delta}}\Omega_{\bs{\delta}}^{(D)}\label{eq:leftTorsionLorentzI}\\
T_{\bs{\alpha\beta}}^{(L)}\hoch{\bs{\gamma}} & = & T_{\bs{\alpha\beta}}\hoch{\bs{\gamma}}-\tfrac{1}{2}\Omega_{[\bs{\alpha}}^{(D)}\delta_{\bs{\beta}]}\hoch{\bs{\gamma}}=-\tfrac{1}{2}\Omega_{[\bs{\alpha}}^{(D)}\delta_{\bs{\beta}]}\hoch{\bs{\gamma}}\\
T_{\bs{\alpha}\hat{\bs{\beta}}}^{(L)}\hoch{\hat{\bs{\gamma}}} & \stackrel{\compensator=0}{=} & T_{\bs{\alpha}\hat{\bs{\beta}}}\hoch{\hat{\bs{\gamma}}}-\tfrac{1}{4}\Omega_{\bs{\alpha}}^{(D)}\delta_{\hat{\bs{\beta}}}\hoch{\hat{\bs{\gamma}}}=\tfrac{1}{8}\gamma_{de}\tief{\bs{\alpha}}\hoch{\bs{\delta}}\gamma^{de}\tief{\hat{\bs{\beta}}}\hoch{\hat{\bs{\gamma}}}\Omega_{\bs{\delta}}^{(D)}\label{eq:leftTorsionLorentzIII}\end{eqnarray}
All other components of $T^{(L)}$ coincide with $T$ for $\compensator=0$
(and $\nabla_{M}\Phi\To-\Omega_{M}^{(D)}$). 

The \emph{right-mover torsion} components $\hat{T}_{AB}\hoch{C}$
are\begin{eqnarray}
\hat{T}_{AB}\hoch{c} & \equiv & \left(\begin{array}{ccc}
\hat{T}_{ab}\hoch{c} & \hat{T}_{a\bs{\beta}}\hoch{c} & \hat{T}_{a\hat{\bs{\beta}}}\hoch{c}\\
\hat{T}_{\bs{\alpha}b}\hoch{c} & \hat{T}_{\bs{\alpha}\bs{\beta}}\hoch{c} & \hat{T}_{\bs{\alpha}\hat{\bs{\beta}}}\hoch{c}\\
\hat{T}_{\hat{\bs{\alpha}}b}\hoch{c} & \hat{T}_{\hat{\bs{\alpha}}\bs{\beta}}\hoch{c} & \hat{T}_{\hat{\bs{\alpha}}\hat{\bs{\beta}}}\hoch{c}\end{array}\right)=\left(\begin{array}{ccc}
-\frac{3}{2}H_{ab}\hoch{c} & 0 & \frac{1}{2}\hatcovPhi{\hat{\bs{\beta}}}\delta_{a}^{c}+\frac{1}{2}\gamma_{a}\hoch{c}\tief{\hat{\bs{\beta}}}\hoch{\hat{\bs{\delta}}}\hatcovPhi{\hat{\bs{\delta}}}\\
0 & \gamma_{\bs{\alpha\beta}}^{c} & 0\\
-\frac{1}{2}\hatcovPhi{\hat{\bs{\alpha}}}\delta_{b}^{c}-\frac{1}{2}\gamma_{b}\hoch{c}\tief{\hat{\bs{\alpha}}}\hoch{\hat{\bs{\delta}}}\hatcovPhi{\hat{\bs{\delta}}} & 0 & \gamma_{\hat{\bs{\alpha}}\hat{\bs{\beta}}}^{c}\end{array}\right)\\
\hat{T}_{AB}\hoch{\bs{\gamma}} & \equiv & \left(\begin{array}{ccc}
\hat{T}_{ab}\hoch{\bs{\gamma}} & \hat{T}_{a\bs{\beta}}\hoch{\bs{\gamma}} & \hat{T}_{a\hat{\bs{\beta}}}\hoch{\bs{\gamma}}\\
\hat{T}_{\bs{\alpha}b}\hoch{\bs{\gamma}} & \hat{T}_{\bs{\alpha}\bs{\beta}}\hoch{\bs{\gamma}} & \hat{T}_{\bs{\alpha}\hat{\bs{\beta}}}\hoch{\bs{\gamma}}\\
\hat{T}_{\hat{\bs{\alpha}}b}\hoch{\bs{\gamma}} & \hat{T}_{\hat{\bs{\alpha}}\bs{\beta}}\hoch{\bs{\gamma}} & \hat{T}_{\hat{\bs{\alpha}}\hat{\bs{\beta}}}\hoch{\bs{\gamma}}\end{array}\right)=\\
 &  & \hspace{-2cm}\left(\begin{array}{ccc}
\frac{1}{16}\left(\gemnabla_{\hat{\bs{\gamma}}}\RR^{\bs{\gamma}\hat{\bs{\delta}}}+8\hatcovPhi{\hat{\bs{\gamma}}}\RR^{\bs{\gamma}\hat{\bs{\delta}}}\right)\tilde{\gamma}_{ab\,\hat{\bs{\delta}}}\hoch{\hat{\bs{\gamma}}} & -\frac{3}{8}H_{ade}\tilde{\gamma}^{de}\tief{\bs{\beta}}\hoch{\bs{\gamma}} & -\tilde{\gamma}_{a\,\hat{\bs{\beta}}\hat{\bs{\delta}}}\RR^{\bs{\gamma}\hat{\bs{\delta}}}\\
\frac{3}{8}H_{bde}\tilde{\gamma}^{de}\tief{\bs{\alpha}}\hoch{\bs{\gamma}} & (\frac{1}{4}\gamma_{de}\tief{[\bs{\alpha}}\hoch{\bs{\delta}}\gamma^{de}\tief{\bs{\beta}]}\hoch{\bs{\gamma}}\covPhi{\bs{\delta}}+\frac{1}{2}\covPhi{[\bs{\alpha}}\delta_{\bs{\beta}]}\hoch{\bs{\gamma}}) & (\frac{1}{8}\gamma_{de}\tief{\hat{\bs{\beta}}}\hoch{\hat{\bs{\delta}}}\gamma^{de}\tief{\bs{\alpha}}\hoch{\bs{\gamma}}\hatcovPhi{\hat{\bs{\delta}}}+\frac{1}{4}\hatcovPhi{\hat{\bs{\beta}}}\delta_{\bs{\alpha}}\hoch{\bs{\gamma}})\\
\tilde{\gamma}_{b\,\hat{\bs{\alpha}}\hat{\bs{\delta}}}\RR^{\bs{\gamma}\hat{\bs{\delta}}} & (-\frac{1}{8}\gamma_{de}\tief{\hat{\bs{\alpha}}}\hoch{\hat{\bs{\delta}}}\gamma^{de}\tief{\bs{\beta}}\hoch{\bs{\gamma}}\hatcovPhi{\hat{\bs{\delta}}}-\frac{1}{4}\hatcovPhi{\hat{\bs{\alpha}}}\delta_{\bs{\beta}}\hoch{\bs{\gamma}}) & 0\end{array}\right)\nonumber \\
\hat{T}_{AB}\hoch{\hat{\bs{\gamma}}} & \equiv & \left(\begin{array}{ccc}
\hat{T}_{ab}\hoch{\hat{\bs{\gamma}}} & \hat{T}_{a\bs{\beta}}\hoch{\hat{\bs{\gamma}}} & \hat{T}_{a\hat{\bs{\beta}}}\hoch{\hat{\bs{\gamma}}}\\
\hat{T}_{\bs{\alpha}b}\hoch{\hat{\bs{\gamma}}} & \hat{T}_{\bs{\alpha}\bs{\beta}}\hoch{\hat{\bs{\gamma}}} & \hat{T}_{\bs{\alpha}\hat{\bs{\beta}}}\hoch{\hat{\bs{\gamma}}}\\
\hat{T}_{\hat{\bs{\alpha}}b}\hoch{\hat{\bs{\gamma}}} & \hat{T}_{\hat{\bs{\alpha}}\bs{\beta}}\hoch{\hat{\bs{\gamma}}} & \hat{T}_{\hat{\bs{\alpha}}\hat{\bs{\beta}}}\hoch{\hat{\bs{\gamma}}}\end{array}\right)=\left(\begin{array}{ccc}
\frac{1}{16}\left(\gemnabla_{\bs{\eps}}\RR^{\bs{\delta}\hat{\bs{\gamma}}}+8\covPhi{\bs{\eps}}\RR^{\bs{\delta}\hat{\bs{\gamma}}}\right)\tilde{\gamma}_{ab\,\bs{\delta}}\hoch{\bs{\eps}} & -\tilde{\gamma}_{a\,\bs{\beta\delta}}\RR^{\bs{\delta}\hat{\bs{\gamma}}} & 0\\
\tilde{\gamma}_{b\,\bs{\alpha\delta}}\RR^{\bs{\delta}\hat{\bs{\gamma}}} & 0 & 0\\
0 & 0 & 0\end{array}\right)\end{eqnarray}
In order to extract the Lorentz part, remember $\hatcovPhi{a}=0$
and $\hatcovPhi{\bs{\alpha}}=0$. For $\compensator=0$ $(\nabla_{M}\Phi\To-\Omega_{M}^{(D)})$
this implies $\hat{\Omega}_{a}^{(D)}=0$ and $\hat{\Omega}_{\bs{\alpha}}^{(D)}=0$.
According to (\ref{eq:LorentzPartOfTorsion}) we then have \begin{eqnarray}
\hat{T}_{\hat{\bs{\alpha}}b}^{(L)}\hoch{c} & \stackrel{\compensator=0}{=} & \hat{T}_{\hat{\bs{\alpha}}b}\hoch{c}-\tfrac{1}{2}\hat{\Omega}_{\hat{\bs{\alpha}}}^{(D)}\delta_{b}\hoch{c}=\tfrac{1}{2}\gamma_{b}\hoch{c}\tief{\hat{\bs{\alpha}}}\hoch{\hat{\bs{\delta}}}\hat{\Omega}_{\hat{\bs{\delta}}}^{(D)}\label{eq:rightTorsionLorentzI}\\
\hat{T}_{\hat{\bs{\alpha}}\hat{\bs{\beta}}}^{(L)}\hoch{\hat{\bs{\gamma}}} & = & \hat{T}_{\hat{\bs{\alpha}}\hat{\bs{\beta}}}\hoch{\hat{\bs{\gamma}}}-\tfrac{1}{2}\hat{\Omega}_{[\hat{\bs{\alpha}}}^{(D)}\delta_{\hat{\bs{\beta}}]}\hoch{\hat{\bs{\gamma}}}=-\tfrac{1}{2}\hat{\Omega}_{[\hat{\bs{\alpha}}}^{(D)}\delta_{\hat{\bs{\beta}}]}\hoch{\hat{\bs{\gamma}}}\\
\hat{T}_{\hat{\bs{\alpha}}\bs{\beta}}^{(L)}\hoch{\bs{\gamma}} & \stackrel{\compensator=0}{=} & \hat{T}_{\hat{\bs{\alpha}}\bs{\beta}}\hoch{\bs{\gamma}}-\tfrac{1}{4}\hat{\Omega}_{\hat{\bs{\alpha}}}^{(D)}\delta_{\bs{\beta}}\hoch{\bs{\gamma}}=\tfrac{1}{8}\gamma_{de}\tief{\hat{\bs{\alpha}}}\hoch{\hat{\bs{\delta}}}\gamma^{de}\tief{\bs{\beta}}\hoch{\bs{\gamma}}\hat{\Omega}_{\hat{\bs{\delta}}}^{(D)}\label{eq:rightTorsionLorentzIII}\end{eqnarray}
All other components of $\hat{T}^{(L)}$ coincide with $\hat{T}$
for $\compensator=0$ (and $\nabla_{M}\Phi\To-\Omega_{M}^{(D)}$). 

Finally we give the components of the \emph{average torsion} $\av{T}_{AB}\hoch{C}\equiv\frac{1}{2}\left(T_{AB}\hoch{C}+\hat{T}_{AB}\hoch{C}\right)$:\begin{eqnarray}
\av{T}_{AB}\hoch{c} & \equiv & \left(\begin{array}{ccc}
\av{T}_{ab}\hoch{c} & \av{T}_{a\bs{\beta}}\hoch{c} & \av{T}_{a\hat{\bs{\beta}}}\hoch{c}\\
\av{T}_{\bs{\alpha}b}\hoch{c} & \av{T}_{\bs{\alpha}\bs{\beta}}\hoch{c} & \av{T}_{\bs{\alpha}\hat{\bs{\beta}}}\hoch{c}\\
\av{T}_{\hat{\bs{\alpha}}b}\hoch{c} & \av{T}_{\hat{\bs{\alpha}}\bs{\beta}}\hoch{c} & \av{T}_{\hat{\bs{\alpha}}\hat{\bs{\beta}}}\hoch{c}\end{array}\right)=\nonumber \\
 & = & \left(\begin{array}{ccc}
0 & \frac{1}{4}\covPhi{\bs{\beta}}\delta_{a}^{c}+\frac{1}{4}\gamma_{a}\hoch{c}\tief{\bs{\beta}}\hoch{\bs{\delta}}\covPhi{\bs{\delta}} & \frac{1}{4}\hatcovPhi{\hat{\bs{\beta}}}\delta_{a}^{c}+\frac{1}{4}\gamma_{a}\hoch{c}\tief{\hat{\bs{\beta}}}\hoch{\hat{\bs{\delta}}}\hatcovPhi{\hat{\bs{\delta}}}\\
-\frac{1}{4}\covPhi{\bs{\alpha}}\delta_{b}^{c}-\frac{1}{4}\gamma_{b}\hoch{c}\tief{\bs{\alpha}}\hoch{\bs{\delta}}\covPhi{\bs{\delta}} & \gamma_{\bs{\alpha\beta}}^{c} & 0\\
-\frac{1}{4}\hatcovPhi{\hat{\bs{\alpha}}}\delta_{b}^{c}-\frac{1}{4}\gamma_{b}\hoch{c}\tief{\hat{\bs{\alpha}}}\hoch{\hat{\bs{\delta}}}\hatcovPhi{\hat{\bs{\delta}}} & 0 & \gamma_{\hat{\bs{\alpha}}\hat{\bs{\beta}}}^{c}\end{array}\right)\label{eq:coll:avTc}\\
\av{T}_{AB}\hoch{\bs{\gamma}} & \equiv & \left(\begin{array}{ccc}
\av{T}_{ab}\hoch{\bs{\gamma}} & \av{T}_{a\bs{\beta}}\hoch{\bs{\gamma}} & \av{T}_{a\hat{\bs{\beta}}}\hoch{\bs{\gamma}}\\
\av{T}_{\bs{\alpha}b}\hoch{\bs{\gamma}} & \av{T}_{\bs{\alpha}\bs{\beta}}\hoch{\bs{\gamma}} & \av{T}_{\bs{\alpha}\hat{\bs{\beta}}}\hoch{\bs{\gamma}}\\
\av{T}_{\hat{\bs{\alpha}}b}\hoch{\bs{\gamma}} & \av{T}_{\hat{\bs{\alpha}}\bs{\beta}}\hoch{\bs{\gamma}} & \av{T}_{\hat{\bs{\alpha}}\hat{\bs{\beta}}}\hoch{\bs{\gamma}}\end{array}\right)=\label{eq:coll:avTgam}\\
 &  & \hspace{-2cm}\left(\begin{array}{ccc}
\frac{1}{16}\left(\gemnabla_{\hat{\bs{\eps}}}\RR^{\bs{\gamma}\hat{\bs{\delta}}}+8\hatcovPhi{\hat{\bs{\eps}}}\RR^{\bs{\gamma}\hat{\bs{\delta}}}\right)\tilde{\gamma}_{ab\,\hat{\bs{\delta}}}\hoch{\hat{\bs{\eps}}} & -\frac{3}{16}H_{ade}\tilde{\gamma}^{de}\tief{\bs{\beta}}\hoch{\bs{\gamma}} & -\tilde{\gamma}_{a\,\hat{\bs{\beta}}\hat{\bs{\delta}}}\RR^{\bs{\gamma}\hat{\bs{\delta}}}\\
\frac{3}{16}H_{bde}\tilde{\gamma}^{de}\tief{\bs{\alpha}}\hoch{\bs{\gamma}} & (\frac{1}{8}\gamma_{de}\tief{[\bs{\alpha}}\hoch{\bs{\delta}}\gamma^{de}\tief{\bs{\beta}]}\hoch{\bs{\gamma}}\covPhi{\bs{\delta}}+\frac{1}{4}\covPhi{[\bs{\alpha}}\delta_{\bs{\beta}]}\hoch{\bs{\gamma}}) & (\frac{1}{16}\gamma_{de}\tief{\hat{\bs{\beta}}}\hoch{\hat{\bs{\delta}}}\gamma^{de}\tief{\bs{\alpha}}\hoch{\bs{\gamma}}\hatcovPhi{\hat{\bs{\delta}}}+\frac{1}{8}\hatcovPhi{\hat{\bs{\beta}}}\delta_{\bs{\alpha}}\hoch{\bs{\gamma}})\\
\tilde{\gamma}_{b\,\hat{\bs{\alpha}}\hat{\bs{\delta}}}\RR^{\bs{\gamma}\hat{\bs{\delta}}} & (-\frac{1}{16}\gamma_{de}\tief{\hat{\bs{\alpha}}}\hoch{\hat{\bs{\delta}}}\gamma^{de}\tief{\bs{\beta}}\hoch{\bs{\gamma}}\hatcovPhi{\hat{\bs{\delta}}}-\frac{1}{8}\hatcovPhi{\hat{\bs{\alpha}}}\delta_{\bs{\beta}}\hoch{\bs{\gamma}}) & 0\end{array}\right)\nonumber \\
\av{T}_{AB}\hoch{\hat{\bs{\gamma}}} & \equiv & \left(\begin{array}{ccc}
\av{T}_{ab}\hoch{\hat{\bs{\gamma}}} & \av{T}_{a\bs{\beta}}\hoch{\hat{\bs{\gamma}}} & \av{T}_{a\hat{\bs{\beta}}}\hoch{\hat{\bs{\gamma}}}\\
\av{T}_{\bs{\alpha}b}\hoch{\hat{\bs{\gamma}}} & \av{T}_{\bs{\alpha}\bs{\beta}}\hoch{\hat{\bs{\gamma}}} & \av{T}_{\bs{\alpha}\hat{\bs{\beta}}}\hoch{\hat{\bs{\gamma}}}\\
\av{T}_{\hat{\bs{\alpha}}b}\hoch{\hat{\bs{\gamma}}} & \av{T}_{\hat{\bs{\alpha}}\bs{\beta}}\hoch{\hat{\bs{\gamma}}} & \av{T}_{\hat{\bs{\alpha}}\hat{\bs{\beta}}}\hoch{\hat{\bs{\gamma}}}\end{array}\right)=\label{eq:coll:avThgam}\\
 &  & \hspace{-2cm}\left(\begin{array}{ccc}
\frac{1}{16}\left(\gemnabla_{\bs{\eps}}\RR^{\bs{\delta}\hat{\bs{\gamma}}}+8\covPhi{\bs{\eps}}\RR^{\bs{\delta}\hat{\bs{\gamma}}}\right)\tilde{\gamma}_{ab\,\bs{\delta}}\hoch{\bs{\eps}} & -\tilde{\gamma}_{a\,\bs{\beta\delta}}\RR^{\bs{\delta}\hat{\bs{\gamma}}} & \frac{3}{16}H_{ade}\tilde{\gamma}^{de}\tief{\hat{\bs{\beta}}}\hoch{\hat{\bs{\gamma}}}\\
\tilde{\gamma}_{b\,\bs{\alpha\delta}}\RR^{\bs{\delta}\hat{\bs{\gamma}}} & 0 & (-\frac{1}{16}\gamma_{de}\tief{\bs{\alpha}}\hoch{\bs{\delta}}\gamma^{de}\tief{\hat{\bs{\beta}}}\hoch{\hat{\bs{\gamma}}}\covPhi{\bs{\delta}}-\frac{1}{8}\covPhi{\bs{\alpha}}\delta_{\hat{\bs{\beta}}}\hoch{\hat{\bs{\gamma}}})\\
-\frac{3}{16}H_{bde}\tilde{\gamma}^{de}\tief{\hat{\bs{\alpha}}}\hoch{\hat{\bs{\gamma}}} & (\frac{1}{16}\gamma_{de}\tief{\bs{\beta}}\hoch{\bs{\delta}}\gamma^{de}\tief{\hat{\bs{\alpha}}}\hoch{\hat{\bs{\gamma}}}\covPhi{\bs{\delta}}+\frac{1}{8}\covPhi{\bs{\beta}}\delta_{\hat{\bs{\alpha}}}\hoch{\hat{\bs{\gamma}}}) & (\frac{1}{8}\gamma_{de\,[\hat{\bs{\alpha}}}\hoch{\hat{\bs{\delta}}}\gamma^{de}\tief{\hat{\bs{\beta}}]}\hoch{\hat{\bs{\gamma}}}\hatcovPhi{\hat{\bs{\delta}}}+\frac{1}{4}\hatcovPhi{[\hat{\bs{\alpha}}}\delta_{\hat{\bs{\beta}}]}\hoch{\hat{\bs{\gamma}}})\end{array}\right)\nonumber \end{eqnarray}
The unfortunate situation that neither $\av{T}_{\bs{\alpha\beta}}\hoch{\bs{\gamma}}$
nor $\av{T}_{\hat{\bs{\alpha}}\hat{\bs{\beta}}}\hoch{\hat{\bs{\gamma}}}$
vanish raises the question whether the conventional constraints $T_{\bs{\alpha\beta}}\hoch{\bs{\gamma}}=\hat{T}_{\hat{\bs{\alpha}}\hat{\bs{\beta}}}\hoch{\hat{\bs{\gamma}}}=0$
were a clever choice or better should be replaced by a constraint
on the average torsion.

Once more, in order to extract the Lorentz part, we need (for $\compensator=0$)
the constraints $\av{\Omega}_{a}^{(D)}=0$, $\av{\Omega}_{\bs{\alpha}}^{(D)}=\tfrac{1}{2}\Omega_{\bs{\alpha}}^{(D)}$
and $\av{\Omega}_{\hat{\bs{\alpha}}}^{(D)}=\tfrac{1}{2}\hat{\Omega}_{\hat{\bs{\alpha}}}^{(D)}$.
According to (\ref{eq:LorentzPartOfTorsion}) we then have \begin{eqnarray}
\av{T}_{\bs{\alpha}b}^{(L)}\hoch{c} & \stackrel{\compensator=0}{=} & \av{T}_{\bs{\alpha}b}\hoch{c}-\tfrac{1}{4}\Omega_{\bs{\alpha}}^{(D)}\delta_{b}\hoch{c}=\tfrac{1}{4}\gamma_{b}\hoch{c}\tief{\bs{\alpha}}\hoch{\bs{\delta}}\Omega_{\bs{\delta}}^{(D)}\label{eq:avTorsionLorentzI}\\
\av{T}_{\bs{\alpha\beta}}^{(L)}\hoch{\bs{\gamma}} & \stackrel{\compensator=0}{=} & \av{T}_{\bs{\alpha\beta}}\hoch{\bs{\gamma}}-\tfrac{1}{4}\Omega_{[\bs{\alpha}}^{(D)}\delta_{\bs{\beta}]}\hoch{\bs{\gamma}}=-\tfrac{1}{8}\gamma_{de}\tief{[\bs{\alpha}}\hoch{\bs{\delta}}\gamma^{de}\tief{\bs{\beta}]}\hoch{\bs{\gamma}}\Omega_{\bs{\delta}}^{(D)}-\tfrac{1}{2}\Omega_{[\bs{\alpha}}^{(D)}\delta_{\bs{\beta}]}\hoch{\bs{\gamma}}\\
\av{T}_{\bs{\alpha}\hat{\bs{\beta}}}^{(L)}\hoch{\hat{\bs{\gamma}}} & \stackrel{\compensator=0}{=} & \av{T}_{\bs{\alpha}\hat{\bs{\beta}}}\hoch{\hat{\bs{\gamma}}}-\tfrac{1}{8}\Omega_{\bs{\alpha}}^{(D)}\delta_{\hat{\bs{\beta}}}\hoch{\hat{\bs{\gamma}}}=\tfrac{1}{16}\gamma_{de}\tief{\bs{\alpha}}\hoch{\bs{\delta}}\gamma^{de}\tief{\hat{\bs{\beta}}}\hoch{\hat{\bs{\gamma}}}\Omega_{\bs{\delta}}^{(D)}\label{eq:avTorsionLorentzIII}\end{eqnarray}
\begin{eqnarray}
\av{T}_{\hat{\bs{\alpha}}b}^{(L)}\hoch{c} & \stackrel{\compensator=0}{=} & \av{T}_{\hat{\bs{\alpha}}b}\hoch{c}-\tfrac{1}{4}\hat{\Omega}_{\hat{\bs{\alpha}}}^{(D)}\delta_{b}\hoch{c}=\tfrac{1}{4}\gamma_{b}\hoch{c}\tief{\hat{\bs{\alpha}}}\hoch{\hat{\bs{\delta}}}\hat{\Omega}_{\hat{\bs{\delta}}}^{(D)}\label{eq:avTorsionLorentzIV}\\
\av{T}_{\hat{\bs{\alpha}}\hat{\bs{\beta}}}^{(L)}\hoch{\hat{\bs{\gamma}}} & \stackrel{\compensator=0}{=} & \av{T}_{\hat{\bs{\alpha}}\hat{\bs{\beta}}}\hoch{\hat{\bs{\gamma}}}-\tfrac{1}{4}\hat{\Omega}_{[\hat{\bs{\alpha}}}^{(D)}\delta_{\hat{\bs{\beta}}]}\hoch{\hat{\bs{\gamma}}}=-\tfrac{1}{8}\gamma_{de\,[\hat{\bs{\alpha}}}\hoch{\hat{\bs{\delta}}}\gamma^{de}\tief{\hat{\bs{\beta}}]}\hoch{\hat{\bs{\gamma}}}\hat{\Omega}_{\hat{\bs{\delta}}}^{(D)}-\tfrac{1}{2}\hat{\Omega}_{[\hat{\bs{\alpha}}}^{(D)}\delta_{\hat{\bs{\beta}}]}\hoch{\hat{\bs{\gamma}}}\\
\av{T}_{\hat{\bs{\alpha}}\bs{\beta}}^{(L)}\hoch{\bs{\gamma}} & \stackrel{\compensator=0}{=} & \av{T}_{\hat{\bs{\alpha}}\bs{\beta}}\hoch{\bs{\gamma}}-\tfrac{1}{8}\hat{\Omega}_{\hat{\bs{\alpha}}}^{(D)}\delta_{\bs{\beta}}\hoch{\bs{\gamma}}=\tfrac{1}{16}\gamma_{de}\tief{\hat{\bs{\alpha}}}\hoch{\hat{\bs{\delta}}}\gamma^{de}\tief{\bs{\beta}}\hoch{\bs{\gamma}}\hat{\Omega}_{\hat{\bs{\delta}}}^{(D)}\label{eq:avTorsionLorentzVI}\end{eqnarray}
The remaining components of $\av{T}^{(L)}$ coincide with $\av{T}$
for $\compensator=0$ (and $\nabla_{M}\Phi\To-\Omega_{M}^{(D)}$).

\paragraph{Constraints on the curvature}

Induced by the restricted structure group constraints on the connection,
we have such constraints likewise for the curvature (see (\ref{eq:mixedCurvature})
on page \pageref{eq:mixedCurvature} and (\ref{eq:R-Zerfall-bosonic}),(\ref{eq:R-Zerfall-ferm})
and (\ref{eq:R-Zerfall-ferm-hut}) on page \ref{eq:R-Zerfall-ferm}.
The curvature is blockdiagonal and each part decays into a scale part
and a Lorentz part:\index{$R_{ABC}$@$\gem{R}_{ABC}\hoch{D}$}\begin{eqnarray}
\gem{R}_{ABC}\hoch{D} & = & \diag(\check{R}_{ABc}\hoch{d},R_{AB\bs{\gamma}}\hoch{\bs{\delta}},\hat{R}_{AB\hat{\bs{\gamma}}}\hoch{\hat{\bs{\delta}}})\\
\check{R}_{ABc}\hoch{d} & = & \check{F}_{AB}^{(D)}\delta_{c}^{d}+\check{R}_{AB\: c}^{(L)}\hoch{d},\qquad\check{F}_{AB}^{(D)}=\frac{1}{10}\check{R}_{ABc}\hoch{c}\\
R_{AB\bs{\gamma}}\hoch{\bs{\delta}} & = & \frac{1}{2}F_{AB}^{(D)}\delta_{\bs{\gamma}}\hoch{\bs{\delta}}+\frac{1}{4}R_{AB}^{(L)}\tief{a_{1}}\hoch{b}\eta_{ba_{2}}\gamma^{a_{1}a_{2}}\tief{\bs{\gamma}}\hoch{\bs{\delta}},\qquad F_{AB}^{(D)}=-\frac{1}{8}R_{AB\bs{\gamma}}\hoch{\bs{\gamma}}\\
\hat{R}_{AB\hat{\bs{\alpha}}}\hoch{\hat{\bs{\beta}}} & = & \frac{1}{2}\hat{F}^{(D)}\delta_{\hat{\bs{\alpha}}}\hoch{\hat{\bs{\beta}}}+\frac{1}{4}\hat{R}_{AB}^{(L)}\tief{a_{1}}\hoch{b}\eta_{ba_{2}}\gamma^{a_{1}a_{2}}\tief{\hat{\bs{\alpha}}}\hoch{\hat{\bs{\beta}}},\qquad\hat{F}_{AB}^{(D)}=-\frac{1}{8}\hat{R}_{AB\hat{\bs{\gamma}}}\hoch{\hat{\bs{\gamma}}}\end{eqnarray}
with the scale field strength \rem{%
\footnote{It is tempting to write $F^{(D)}\equiv\de\Omega^{(D)}\quad\iff\quad F_{AB}^{(D)}=\gemnabla_{[A}\Omega_{B]}+\gemT_{AB}\hoch{C}\Omega_{C}^{(D)}$
which is formally true if we act with the covariant derivative on
$\gemOm_{B}$ like on a tensorial object. We prefer the point of view
to act with the connection always in the same representations as the
local structure group transformation would do. There we have $\delta\Omega_{M}^{(D)}=-\partial_{M}\Lambda^{(D)}$.
The covariant derivative acting on $\Omega_{M}^{(D)}$ can thus be
defined as \[
\gemnabla_{M}\Omega_{N}^{(D)}\equiv\partial_{M}\Omega_{N}^{(D)}-\partial_{N}\gemOm_{M}^{(D)}??\qquad\fussend\]
} }\begin{equation}
\check{F}^{(D)}\equiv\de\check{\Omega}^{(D)},\qquad F^{(D)}\equiv\de\Omega^{(D)},\qquad\hat{F}^{(D)}\equiv\de\hat{\Omega}^{(D)}\end{equation}
The bosonic field strength is also obtained via the commutator of
covariant derivatives acting on the compensator field $\Phi$. Only
the bosonic block $\check{\Omega}_{Ma}\hoch{b}$ of the mixed connection
$\gemOm_{MA}\hoch{B}$ acts on $\Phi$, because $\Phi$ is a compensator
for the transformation of $G_{ab}=e^{2\Phi}\eta_{ab}$ (with bosonic
indices only). But as the different blocks of the structure group
got related by partial gauge fixing, we may as well act with the left-
or right-mover connection on it:\begin{eqnarray}
\check{F}_{MN}^{(D)} & = & -\gemnabla_{[M}\checkcovPhi{N]}-\gemT_{MN}\hoch{K}\checkcovPhi{K}\\
F_{MN}^{(D)} & = & -\nabla_{[M}\covPhi{N]}-T_{MN}\hoch{K}\covPhi{K}\\
\hat{F}_{MN}^{(D)} & = & -\hat{\nabla}_{[M}\hatcovPhi{N]}-\hat{T}_{MN}\hoch{K}\hatcovPhi{K}\end{eqnarray}

Finallly we collect the holomorphicity (\ref{eq:holConstrVI}),(\ref{eq:holConstrVII}),(\ref{eq:holConstrIX}),(\ref{eq:holConstrX})
and nilpotency constraints (\ref{eq:nilpotency-constraint-onR}),(\ref{eq:nilpotency-constraint-onR-hat})
on the curvature, together with the Bianchi identities (\ref{eq:(delta|0,3,0)}),
(\ref{eq:(hdelta|0,0,3)}), (\ref{eq:(delta|0,2,1)}), (\ref{eq:(hdelta|0,1,2)}),
(\ref{eq:(delta|1,2,0)}), (\ref{eq:(hdelta|1,0,2)}), (\ref{eq:(delta|2,1,0)})
and (\ref{eq:(hdelta|2,0,1)}):\begin{eqnarray}
R_{\hat{\bs{\alpha}}c\bs{\alpha}}\hoch{\bs{\beta}} & = & \tilde{\gamma}_{c\,\hat{\bs{\alpha}}\hat{\bs{\delta}}}\gemnabla_{\bs{\alpha}}\RR^{\bs{\beta}\hat{\bs{\delta}}},\qquad\hat{R}_{\bs{\alpha}c\hat{\bs{\alpha}}}\hoch{\hat{\bs{\beta}}}=\tilde{\gamma}_{c\,\bs{\alpha\delta}}\gemnabla_{\hat{\bs{\alpha}}}\RR^{\bs{\delta}\hat{\bs{\beta}}}\label{eq:coll:Rhalphab:ferm}\\
R_{\hat{\bs{\alpha}}\hat{\bs{\gamma}}\bs{\alpha}}\hoch{\bs{\beta}} & = & 0,\qquad\hat{R}_{\bs{\alpha}\bs{\gamma}\hat{\bs{\alpha}}}\hoch{\hat{\bs{\beta}}}=0\\
R_{c[\bs{\alpha}\bs{\beta}]}\hoch{\bs{\gamma}} & = & \gamma_{\bs{\alpha\beta}}^{d}T_{dc}\hoch{\bs{\gamma}},\qquad\hat{R}_{c[\hat{\bs{\alpha}}\hat{\bs{\beta}}]}\hoch{\hat{\bs{\gamma}}}=\gamma_{\hat{\bs{\alpha}}\hat{\bs{\beta}}}^{d}\hat{T}_{dc}\hoch{\hat{\bs{\gamma}}}\\
R_{\hat{\bs{\gamma}}[\bs{\alpha}\bs{\beta}]}\hoch{\bs{\delta}} & = & -\gamma_{\bs{\alpha\beta}}^{e}\tilde{\gamma}_{e\,\hat{\bs{\gamma}}\hat{\bs{\delta}}}\RR^{\bs{\delta}\hat{\bs{\delta}}},\qquad\hat{R}_{\bs{\gamma}[\hat{\bs{\alpha}}\hat{\bs{\beta}}]}\hoch{\hat{\bs{\delta}}}=-\gamma_{\hat{\bs{\alpha}}\hat{\bs{\beta}}}^{e}\tilde{\gamma}_{e\,\bs{\gamma}\bs{\delta}}\RR^{\bs{\delta}\hat{\bs{\delta}}}\\
\hspace{-.5cm}R_{[\bs{\alpha}_{1}\bs{\alpha}_{2}\bs{\alpha}_{3}]}\hoch{\bs{\beta}} & = & 0,\qquad\hat{R}_{[\hat{\bs{\alpha}}_{1}\hat{\bs{\alpha}}_{2}\hat{\bs{\alpha}}_{3}]}\hoch{\hat{\bs{\beta}}}=0\\
R_{bc\bs{\alpha}}\hoch{\bs{\delta}} & = & \bei{\gemnabla_{\bs{\alpha}}T_{bc}\hoch{\bs{\delta}}}{\check{\Omega}=\hat{\Omega}}+4\tilde{\gamma}_{[b|\,\bs{\alpha\gamma}}\RR^{\bs{\gamma}\hat{\bs{\eps}}}\tilde{\gamma}_{|c]\,\hat{\bs{\eps}}\hat{\bs{\delta}}}\RR^{\bs{\delta}\hat{\bs{\delta}}},\quad\hat{R}_{bc\hat{\bs{\alpha}}}\hoch{\hat{\bs{\delta}}}=\bei{\gemnabla_{\hat{\bs{\alpha}}}\hat{T}_{bc}\hoch{\hat{\bs{\delta}}}}{\check{\Omega}=\Omega}+4\tilde{\gamma}_{[b|\,\hat{\bs{\alpha}}\hat{\bs{\gamma}}}\RR^{\bs{\eps}\hat{\bs{\gamma}}}\tilde{\gamma}_{|c]\,\bs{\eps\delta}}\RR^{\bs{\delta}\hat{\bs{\delta}}}\qquad\end{eqnarray}
Taking the trace of the first two curvature constraints (using (\ref{eq:coll:nablaalphaPtrace})
and (\ref{eq:coll:nablahalphaPtrace})) gives further informations
on the Dilatation-Field-strength (and thus indirectly also on the
Lorentz curvature)\rem{%
\footnote{\label{foot:A-little-later} A little later we will find the constraint
$\hat{\Omega}_{\bs{\gamma}}=\nabla_{\bs{\gamma}}\Phi$ and $\Omega_{\hat{\bs{\gamma}}}=\nabla_{\hat{\bs{\gamma}}}\Phi$
and will gauge fix $T_{\bs{\alpha\gamma}}\hoch{c}=\gamma_{\bs{\alpha\gamma}}^{c}$.
We can use this to calculate $\hat{\Omega}_{a}$ and $\Omega_{a}$
from (\ref{eq:F-Dil-constrII}):\begin{eqnarray*}
\hat{F}_{\bs{\alpha\gamma}}^{(D)}=\underbrace{\nabla_{[\bs{\alpha}}\nabla_{\bs{\gamma}]}\Phi}_{-T_{\bs{\alpha\gamma}}\hoch{c}\nabla_{c}\Phi}+T_{\bs{\alpha\gamma}}^{c}\hat{\Omega}_{c} & = & 0,\qquad F_{\hat{\bs{\alpha}}\hat{\bs{\gamma}}}^{(D)}=\underbrace{\hat{\nabla}_{[\hat{\bs{\alpha}}}\hat{\nabla}_{\hat{\bs{\gamma}}]}\Phi}_{-T_{\hat{\bs{\alpha}}\hat{\bs{\gamma}}}\hoch{c}\hat{\nabla}_{c}\Phi}+T_{\hat{\bs{\alpha}}\hat{\bs{\gamma}}}\hoch{c}\Omega_{c}=0\end{eqnarray*}
\[
\boxed{\hat{\Omega}_{c}=\Omega_{c}=\nabla_{c}\Phi}\]
From (\ref{eq:F-Dil-constrI}) instead, we get\begin{eqnarray*}
\hat{F}_{\bs{\alpha}c}^{(D)} & = & \tilde{\nabla}_{[\bs{\alpha}}\hat{\Omega}_{c]}+\tilde{T}_{\bs{\alpha}c}\hoch{D}\hat{\Omega}_{D}=\\
 & = & \tilde{\nabla}_{[\bs{\alpha}}\tilde{\nabla}_{c]}\Phi+\tilde{T}_{\bs{\alpha}c}\hoch{D}\hat{\Omega}_{D}=\\
 & = & \tilde{T}_{\bs{\alpha}c}\hoch{D}\left(-\tilde{\nabla}_{D}\Phi+\hat{\Omega}_{D}\right)=\\
 & = & \hat{T}_{\bs{\alpha}c}\hoch{\hat{\bs{\delta}}}\left(-\tilde{\nabla}_{\hat{\bs{\delta}}}\Phi+\hat{\Omega}_{\hat{\bs{\delta}}}\right)=\\
 & = & \gamma_{c\,\bs{\alpha\gamma}}\RR^{\bs{\gamma}\hat{\bs{\delta}}}\left(-\tilde{\nabla}_{\hat{\bs{\delta}}}\Phi+\hat{\Omega}_{\hat{\bs{\delta}}}\right)=\\
 & \stackrel{!}{=} & -\frac{1}{8}\gamma_{c\,\bs{\alpha\delta}}\tilde{\nabla}_{\hat{\bs{\alpha}}}\RR^{\bs{\delta}\hat{\bs{\alpha}}}\end{eqnarray*}
 \[
\boxed{\tilde{\nabla}_{\hat{\bs{\alpha}}}\RR^{\bs{\delta}\hat{\bs{\alpha}}}=8\RR^{\bs{\delta}\hat{\bs{\delta}}}\left(\tilde{\nabla}_{\hat{\bs{\delta}}}\Phi-\hat{\Omega}_{\hat{\bs{\delta}}}\right)}\qquad\fussend\]
}}\begin{eqnarray}
\hat{F}_{c\bs{\alpha}}^{(D)} & = & \tilde{\gamma}_{c\,\bs{\alpha}\bs{\delta}}\RR^{\bs{\delta}\hat{\bs{\alpha}}}\hatcovPhi{\hat{\bs{\alpha}}},\qquad F_{c\hat{\bs{\alpha}}}^{(D)}=\tilde{\gamma}_{c\,\hat{\bs{\alpha}}\hat{\bs{\delta}}}\RR^{\bs{\alpha}\hat{\bs{\delta}}}\covPhi{\bs{\alpha}}\label{eq:F-Dil-constrI}\\
\hat{F}_{\bs{\alpha\gamma}}^{(D)} & = & 0,\qquad F_{\hat{\bs{\alpha}}\hat{\bs{\gamma}}}^{(D)}=0\label{eq:F-Dil-constrII}\end{eqnarray}
\rem{\begin{eqnarray*}
\hat{R}_{\bs{\alpha}ca}\hoch{b} & = & \hat{F}_{\bs{\alpha}c}^{(D)}\delta_{a}\hoch{b}+\frac{1}{8}\hat{R}_{\bs{\alpha}c\hat{\bs{\alpha}}}^{(L)}\hoch{\hat{\bs{\beta}}}\gamma_{a}^{\: b}\tief{\hat{\bs{\beta}}}\hoch{\hat{\bs{\alpha}}}\\
\hat{R}_{\bs{\alpha}\bs{\gamma}a}\hoch{b} & = & 0,\qquad R_{\hat{\bs{\alpha}}\hat{\bs{\gamma}}a}\hoch{b}=0\end{eqnarray*}
}

\paragraph{Remaining BI's}

Finally we get a couple of constraints on curvature components where
the structure group indices are bosonic. They are related to the above
ones as we shall discuss after presenting them:

\begin{eqnarray}
R_{\bs{\alpha\beta}c}\hoch{d} & \stackrel{{\rm \tiny(\ref{eq:(d|1,2,0):expl})}}{=} & -\nabla_{[\bs{\alpha}}\covPhi{\bs{\beta}]}\delta_{c}\hoch{d}+\gamma_{c}\hoch{d}\tief{[\bs{\alpha}}\hoch{\bs{\delta}}\nabla_{\bs{\beta}]}\covPhi{\bs{\delta}}+3\gamma_{\bs{\alpha\beta}}^{e}H_{ec}\hoch{d}+\gamma_{c}\hoch{e}\tief{[\bs{\alpha}|}\hoch{\bs{\gamma}}\covPhi{\bs{\gamma}}\gamma_{e}\hoch{d}\tief{|\bs{\beta}]}\hoch{\bs{\delta}}\covPhi{\bs{\delta}}\qquad\label{eq:coll:Ralphabeta:bos}\\
\hat{R}_{\hat{\bs{\alpha}}\hat{\bs{\beta}}c}\hoch{d} & \stackrel{{\rm \tiny(\ref{eq:(d|1,0,2):expl})}}{=} & -\hat{\nabla}_{[\hat{\bs{\alpha}}}\hatcovPhi{\hat{\bs{\beta}}]}\delta_{c}\hoch{d}+\gamma_{c}\hoch{d}\tief{[\hat{\bs{\alpha}}}\hoch{\hat{\bs{\delta}}}\hat{\nabla}_{\hat{\bs{\beta}}]}\hatcovPhi{\hat{\bs{\delta}}}-3\gamma_{\hat{\bs{\alpha}}\hat{\bs{\beta}}}^{e}H_{ec}\hoch{d}+\gamma_{c}\hoch{e}\tief{[\hat{\bs{\alpha}}|}\hoch{\hat{\bs{\gamma}}}\hatcovPhi{\hat{\bs{\gamma}}}\gamma_{e}\hoch{d}\tief{|\hat{\bs{\beta}}]}\hoch{\hat{\bs{\delta}}}\hatcovPhi{\hat{\bs{\delta}}}\qquad\label{eq:coll:hRhalphahbeta:bos}\end{eqnarray}
\begin{eqnarray}
R_{\bs{\alpha}\hat{\bs{\beta}}c}\hoch{d} & \stackrel{{\rm \tiny(\ref{eq:(d|1,1,1)'})}}{=} & \frac{1}{2}\nabla_{\hat{\bs{\beta}}}\covPhi{\bs{\alpha}}\delta_{c}^{d}+\frac{1}{2}\gamma_{c}\hoch{d}\tief{\bs{\alpha}}\hoch{\bs{\gamma}}\nabla_{\hat{\bs{\beta}}}\covPhi{\bs{\gamma}}-2\tilde{\gamma}_{c\,\bs{\alpha}\bs{\beta}}\RR^{\bs{\beta}\hat{\bs{\eps}}}\gamma_{\hat{\bs{\eps}}\hat{\bs{\beta}}}^{d}+2\tilde{\gamma}_{c\,\hat{\bs{\beta}}\hat{\bs{\delta}}}\RR^{\bs{\eps}\hat{\bs{\delta}}}\gamma_{\bs{\eps}\bs{\alpha}}^{d}\label{eq:coll:Ralphahbeta:bos}\\
\hat{R}_{\hat{\bs{\alpha}}\bs{\beta}c}\hoch{d} & \stackrel{{\rm \tiny(\ref{eq:(d|1,1,1):equiv'})}}{=} & \frac{1}{2}\hat{\nabla}_{\bs{\beta}}\hatcovPhi{\hat{\bs{\alpha}}}\delta_{c}^{d}+\frac{1}{2}\gamma_{c}\hoch{d}\tief{\hat{\bs{\alpha}}}\hoch{\hat{\bs{\gamma}}}\hat{\nabla}_{\bs{\beta}}\hatcovPhi{\hat{\bs{\gamma}}}-2\tilde{\gamma}_{c\,\hat{\bs{\alpha}}\hat{\bs{\beta}}}\RR^{\bs{\eps}\hat{\bs{\beta}}}\gamma_{\bs{\eps}\bs{\beta}}^{d}+2\tilde{\gamma}_{c\,\bs{\beta}\bs{\delta}}\RR^{\bs{\delta}\hat{\bs{\eps}}}\gamma_{\hat{\bs{\eps}}\hat{\bs{\alpha}}}^{d}\label{eq:coll:hRhalphabeta:bos}\end{eqnarray}
\begin{eqnarray}
R_{\hat{\bs{\alpha}}[bc]d} & \stackrel{{\rm \tiny(\ref{eq:(d|2,0,1)'})}}{=} & -\frac{1}{8}\gemnabla_{\bs{\gamma}}\RR^{\bs{\delta}\hat{\bs{\eps}}}\tilde{\gamma}_{d[b|\,\bs{\delta}}\hoch{\bs{\gamma}}\gamma_{|c]\hat{\bs{\eps}}\hat{\bs{\alpha}}}+G_{d[b|}\tilde{\gamma}_{|c]\,\hat{\bs{\alpha}}\hat{\bs{\delta}}}\RR^{\bs{\eps}\hat{\bs{\delta}}}\covPhi{\bs{\eps}}\label{eq:coll:Rhalphab:bos}\\
\hat{R}_{\bs{\alpha}[bc]d} & \stackrel{{\rm \tiny(\ref{eq:(d|2,1,0)'})}}{=} & -\frac{1}{8}\gemnabla_{\hat{\bs{\gamma}}}\RR^{\bs{\eps}\hat{\bs{\delta}}}\tilde{\gamma}_{d[b|\,\hat{\bs{\delta}}}\hoch{\hat{\bs{\gamma}}}\gamma_{|c]\bs{\eps}\bs{\alpha}}+G_{d[b|}\tilde{\gamma}_{|c]\,\bs{\alpha\delta}}\RR^{\bs{\delta}\hat{\bs{\eps}}}\hatcovPhi{\hat{\bs{\eps}}}\label{eq:coll:hRalphab:bos}\end{eqnarray}
\begin{eqnarray}
R_{[abc]}\hoch{d} & \stackrel{{\rm \tiny(\ref{eq:(d|3,0,0)})}}{=} & \frac{3}{2}\nabla_{[a}H_{bc]}\hoch{d}+\frac{9}{2}H_{[ab|}\hoch{e}H_{e|c]}\hoch{d}+2T_{[ab|}\hoch{\bs{\eps}}T_{\bs{\eps}|c]}\hoch{d}\label{eq:coll:Rab:bos}\\
\hat{R}_{[abc]}\hoch{d} & \stackrel{{\rm \tiny(\ref{eq:(dhat|3,0,0)})}}{=} & -\frac{3}{2}\hat{\nabla}_{[a}H_{bc]}\hoch{d}+\frac{9}{2}H_{[ab|}\hoch{e}H_{e|c]}\hoch{d}+2\hat{T}_{[ab|}\hoch{\hat{\bs{\eps}}}\hat{T}_{\hat{\bs{\eps}}|c]}\hoch{d}\label{eq:coll:hRab:bos}\end{eqnarray}
 From the structure group constraints on the curvature, we know
that the components split into Lorentz and scale part $R_{ABc}\hoch{d}=F_{AB}^{(D)}\delta_{c}^{d}+R_{ABc}^{(L)}\hoch{d}$.
The same is true for the componets with fermionic structure group
indices, where we had the split $R_{AB\bs{\gamma}}\hoch{\bs{\delta}}=\frac{1}{2}F_{AB}^{(D)}\delta_{\bs{\gamma}}\hoch{\bs{\delta}}+\frac{1}{4}R_{ABc}^{(L)}\hoch{d}\gamma^{c}\tief{d\,\bs{\gamma}}\hoch{\bs{\delta}}$.
The coefficients $F_{AB}^{(D)}$ and $R_{ABc}^{(L)}\hoch{d}$ are
the same, when the bosonic block of $\check{\Omega}_{Ma}\hoch{b}$
was chosen to coincide with the left-mover connection. They can be
extracted from $R_{ABc}\hoch{d}$ just as $\tfrac{1}{10}$ of the
trace part and as the antisymmetric part respectively. To extract
the coefficients instead from $R_{AB\bs{\gamma}}\hoch{\bs{\delta}}$,
we need the fermionic trace $\delta_{\bs{\gamma}}\hoch{\bs{\gamma}}=-16$
which yields $F_{AB}^{(D)}=-\frac{1}{8}R_{AB\bs{\gamma}}\hoch{\bs{\gamma}}$
and the identity $\gamma_{ab}\tief{\bs{\delta}}\hoch{\bs{\gamma}}\gamma^{cd}\tief{\bs{\gamma}}\hoch{\bs{\delta}}=32\delta_{ab}^{cd}$
that allows to extract the Lorentz part as $R_{ABc}^{(L)}\hoch{d}=\frac{1}{8}\gamma_{c}\hoch{d}\tief{\bs{\delta}}\hoch{\bs{\gamma}}R_{AB\bs{\gamma}}\hoch{\bs{\delta}}$.
Then we can relate both curvature blocks directly in the following
way:\begin{eqnarray}
R_{ABc}\hoch{d} & = & -\frac{1}{8}R_{AB\bs{\gamma}}\hoch{\bs{\gamma}}\delta_{c}^{d}+\frac{1}{8}\gamma_{c}\hoch{d}\tief{\bs{\delta}}\hoch{\bs{\gamma}}R_{AB\bs{\gamma}}\hoch{\bs{\delta}}\\
R_{AB\bs{\gamma}}\hoch{\bs{\delta}} & = & \frac{1}{20}R_{ABc}\hoch{c}\delta_{\bs{\gamma}}\hoch{\bs{\delta}}+\frac{1}{4}R_{ABc}\hoch{d}\gamma^{c}\tief{d\,\bs{\gamma}}\hoch{\bs{\delta}}\end{eqnarray}
In the same way we can relate $\hat{R}_{ABc}\hoch{d}$ and $\hat{R}_{AB\hat{\bs{\gamma}}}\hoch{\hat{\bs{\delta}}}$
and compare their constraints which should reveal additional information.
This was used for example in footnote \ref{fn:dilaton-constraint}
on page \pageref{fn:dilaton-constraint} to derive the constraint\begin{equation}
\nabla_{\hat{\bs{\beta}}}\covPhi{\bs{\alpha}}=-\tilde{\gamma}_{d\,\bs{\alpha\rho}}\RR^{\bs{\rho}\hat{\bs{\eps}}}\gamma_{\hat{\bs{\eps}}\hat{\bs{\beta}}}^{d}\end{equation}
 on the compensator superfield.

\section{The dilaton superfield}

While we have found the covariant derivatives $\covPhi{a}=\hatcovPhi{a}=\covPhi{\hat{\bs{\alpha}}}=\hatcovPhi{\bs{\alpha}}$
of the compensator field $\compensator$ to be forced to vanish, the
remaining components $\covPhi{\bs{\alpha}}=E_{\bs{\alpha}}\hoch{M}(\partial_{M}\compensator-\Omega_{M}^{(D)})$
and $\hatcovPhi{\hat{\bs{\alpha}}}=E_{\hat{\bs{\alpha}}}\hoch{M}(\partial_{M}\compensator-\hat{\Omega}_{M}^{(D)})$
seem to contain physical fermionic degrees of freedom. Indeed, the
leading components of the scale connections $\Omega_{\bs{\alpha}}^{(D)}$
and $\hat{\Omega}_{\hat{\bs{\alpha}}}^{(D)}$ were identified in \cite{Berkovits:2001ue}
up to a constant factor with the dilatinos. As we have not yet fixed
the local scale invariance (guaranteed by the compensator field $\compensator$),
those connections are not covariant and we take instead the just mentioned
covariant derivatives of the compensator field. That is, we define
the \textbf{dilatinos} as\index{dilatino|itext{$\dilo_{\bs{\mu}}$}}\index{$\lambda_\mu$@$\dilo_{\bs{\mu}}$|itext{dilatino}}\index{$\lambda_{\mu}$@$\hat{\dilo}_{\hat{\bs{\mu}}}$|itext{dilatino}}\begin{equation}
\dilo_{\bs{\alpha}}\equiv\bei{\covPhi{\bs{\alpha}}}{\xbothtetas=0},\quad\hat{\dilo}_{\hat{\bs{\alpha}}}\equiv\bei{\hatcovPhi{\hat{\bs{\alpha}}}}{\xbothtetas=0}\label{eq:dilatinoDefI}\end{equation}
We are still completely missing the dilaton itself, whose appearance
is a bit hidden. It does not show up explicitely in the action. Although
we did not manually include it via the Fradkin Tseytlin term, its
physical degrees of freedom  should already be present in this setting.%
\footnote{\index{footnote!\thefoot. remark on the dilaton}Thanks to N. Berkovits
for clarifying this issue. In \cite{Berkovits:2001ue,Bedoya:2006ic}
the dilaton was added as an extra field via the Fradkin\index{Fradkin-Tseytlin term}-Tseytlin\index{Tseytlin!Fradkin-$\sim$-term}
term $S_{FT}=\int\quad\alpha'r\dil$ (with $r$ being the worldsheet
curvature) and then related to the already present field content via
a quantum consistency argument. Their result was $E_{\bs{\alpha}}\hoch{M}\partial_{M}\dil=4\Omega_{\bs{\alpha}}$
and $E_{\hat{\bs{\alpha}}}\hoch{M}\partial_{M}\dil=4\hat{\Omega}_{\hat{\bs{\alpha}}}$.
Because of the introduction of our compensator field $\Phi$, their
relations would modify in our case to \begin{eqnarray*}
E_{\bs{\alpha}}\hoch{M}\partial_{M}(\dil+4\Phi) & = & 4\Omega_{\bs{\alpha}}\qquad\iff\qquad-4\covPhi{\bs{\alpha}}=\nabla_{\bs{\alpha}}\dil\\
E_{\hat{\bs{\alpha}}}\hoch{M}\partial_{M}(\dil+4\Phi) & = & 4\hat{\Omega}_{\hat{\bs{\alpha}}}\qquad\iff\qquad-4\hatcovPhi{\hat{\bs{\alpha}}}=\hat{\nabla}_{\hat{\bs{\alpha}}}\dil\end{eqnarray*}
Our definition (\ref{eq:DilatonDef}) of the dilaton is thus consistent
with this result, although the definitions differ by a factor $-4$.$\quad\fussend$%
} Usually one would suspect the dilatinos to be components at first
order in $\xbothtetas$ of a scalar dilaton superfield instead of
being the component of a (non-covariantly transforming) compensator
field. The idea to recover such a scalar superfield is to equate its
spinorial derivative with the covariant spinorial derivatives of the
compensator field and let the algebra fix the missing bosonic derivative.
So let us simply {}``define'' the scalar \textbf{dilaton\index{dilaton-superfield}\index{dilaton}
superfield} $\dil$\index{$\Phi_{(ph)}$|itext{dilaton superfield}}
via\begin{equation}
\nabla_{\bs{\alpha}}\dil\equiv\covPhi{\bs{\alpha}},\quad\hat{\nabla}_{\hat{\bs{\alpha}}}\dil\equiv\hatcovPhi{\hat{\bs{\alpha}}}\label{eq:DilatonDef}\end{equation}
The different behaviour of the fields under scale transformations
is reflected in the different action of the covariant derivative.
While for the dilaton it acts like a partial derivative $\nabla_{\bs{\alpha}}\dil=E_{\bs{\alpha}}\hoch{M}\partial_{M}\dil$,
the action on the compensator field -- as mentioned already above
-- includes a shift $\nabla_{\bs{\alpha}}\compensator=E_{\bs{\alpha}}\hoch{M}(\partial_{M}\dil-\Omega_{M}^{(D)})$.
Of course we have to make sure that this definition does not put additional
restrictions on the already present field content, in particular on
the scale field strength. As $\dil$ is supposed to be a scalar field
(where the commutator of covariant derivatives does not contain any
curvature terms), while $\compensator$ is a compensator field (where
the commutator of covariant derivatives contains the scale field strength),
it is instructive to compare the derivative commutators acting on
them:\begin{eqnarray}
\bei{\gemnabla_{[\bs{\alpha}}\gemnabla_{\bs{\beta}]}\dil}{\check{\Omega}=\Omega} & = & -\bei{\gem{T}_{\bs{\alpha}\bs{\beta}}\hoch{C}}{\check{\Omega}=\Omega}\nabla_{C}\dil=-\gamma_{\bs{\alpha\beta}}^{c}\nabla_{c}\dil\nonumber \\
\bei{\gemnabla_{[\bs{\alpha}}\gemcovPhi{\bs{\beta}]}}{\check{\Omega}=\Omega} & = & -\bei{\gem{T}_{\bs{\alpha}\bs{\beta}}\hoch{C}}{\check{\Omega}=\Omega}\covPhi{C}-F_{\bs{\alpha}\bs{\beta}}^{(D)}=-F_{\bs{\alpha\beta}}^{(D)}\end{eqnarray}
Similar equations hold for the hatted indices. Consistency then requires
\begin{equation}
\gamma_{\bs{\alpha\beta}}^{c}\nabla_{c}\dil=F_{\bs{\alpha\beta}}^{(D)},\quad\gamma_{\hat{\bs{\alpha}}\hat{\bs{\beta}}}^{c}\nabla_{c}\dil=\hat{F}_{\hat{\bs{\alpha}}\hat{\bs{\beta}}}^{(D)}\label{eq:nablacDil}\end{equation}
In contrast to $\covPhi{c}$ and $\hatcovPhi{c}$, the bosonic derivative
$\nabla_{c}\dil$ of the dilaton superfield is in general nonzero.
For the validity of the above 'definition' it is important to observe
that because of the constraints (\ref{eq:FalphbetForDilaton}) the
equations (\ref{eq:nablacDil}) do not put an additional artificial
restriction on $F_{\bs{\alpha}\bs{\beta}}^{(D)}$ and $\hat{F}_{\hat{\bs{\alpha}}\hat{\bs{\beta}}}^{(D)}$.
Instead  (\ref{eq:nablacDil}) consistently completes (\ref{eq:DilatonDef})
to a complete superspace derivative of the superfield and we can use
the supervielbein to switch to curved coordinates where the covariant
derivative $\nabla_{M}\dil$ on the scalar field coincides with the
partial derivative $\partial_{M}\dil$. Integrating it, we are just
missing a constant, the dilaton zero mode (responsible for the string-coupling
in the loop-expansion). The dilaton superfield is thus well-defined
by (\ref{eq:DilatonDef}) up to an integration constant.\rem{

We can define a dilaton-superfield implicitely via \begin{eqnarray}
\nabla_{\bs{\alpha}}\dil & \equiv & \covPhi{\bs{\alpha}}=\bei{\gemcovPhi{\bs{\alpha}}}{\check{\Omega}=\Omega}\\
\nabla_{\hat{\bs{\alpha}}}\dil & \equiv & \hatcovPhi{\hat{\bs{\alpha}}}=\bei{\gemcovPhi{\hat{\bs{\alpha}}}}{\check{\Omega}=\hat{\Omega}}\end{eqnarray}
There is no connection present in $\nabla_{A}\dil$, so we can act
with any covariant derivative which is convenient, while for the compensator
it is important which connection is used. For the above implicit definition
to be consistent, the algebra of covariant derivatives acting on $\dil$
has to agree with the one obtained when acting on the compensator
$\Phi$. \begin{eqnarray}
\nabla_{[\bs{\alpha}}\nabla_{\bs{\beta}]}\dil=-\gamma_{\bs{\alpha\beta}}\hoch{c}\nabla_{c}\dil & \stackrel{!}{=} & \nabla_{[\bs{\alpha}},\nabla_{\bs{\beta}]}\Phi=-\gamma_{\bs{\alpha\beta}}\hoch{c}\underbrace{\covPhi{c}}_{0}-F_{\bs{\alpha\beta}}^{(D)}\end{eqnarray}
From this equation and its hatted equivalent we see that \begin{equation}
\gamma_{\bs{\alpha\beta}}\hoch{c}\nabla_{c}\dil=F_{\bs{\alpha\beta}}^{(D)},\qquad\gamma_{\hat{\bs{\alpha}}\hat{\bs{\beta}}}\hoch{c}\nabla_{c}\dil=\hat{F}_{\hat{\bs{\alpha}}\hat{\bs{\beta}}}^{(D)}\end{equation}
Next we have \begin{eqnarray}
[\gem{\nabla}_{\hat{\bs{\alpha}}},\gemnabla_{\bs{\beta}}]\dil=0 & \stackrel{!}{=} & \nabla_{\hat{\bs{\alpha}}}\covPhi{\bs{\beta}}-\hat{\nabla}_{\bs{\beta}}\hatcovPhi{\hat{\bs{\alpha}}}=\\
 & = & [\gemnabla_{\hat{\bs{\alpha}}},\gemnabla_{\bs{\beta}}]\bei{\Phi}{\check{\Omega}=\Omega}+[\gemnabla_{\hat{\bs{\alpha}}},\gemnabla_{\bs{\beta}}]\bei{\Phi}{\check{\Omega}=\hat{\Omega}}\end{eqnarray}
so that we need \begin{equation}
F_{\hat{\bs{\alpha}}\bs{\beta}}+\hat{F}_{\hat{\bs{\alpha}}\bs{\beta}}\stackrel{!}{=}0\end{equation}
What remains is the commutator between bosonic and fermionic derivative:
\begin{eqnarray}
[\gem{\nabla}_{a},\gemnabla_{\bs{\beta}}]\dil & = & -2\gemT_{a\bs{\beta}}\hoch{\hat{\bs{\gamma}}}\nabla_{\hat{\bs{\gamma}}}\dil-2\gemT_{a\bs{\beta}}\hoch{c}\nabla_{c}\dil\end{eqnarray}

\paragraph{Fradkin-Tseytlin}

So far our worldsheet action does not contain any dilaton\index{dilaton field}
field. In order to include the dilaton, however, we would like to
couple it to the worldsheet curvature. But the pure spinor string
is special in the sense that it does not contain any independent worldsheet
metric. So how can we couple to it?

A possible interpretation is the following: Even in the bosonic and
in the RNS string the conformal factor of the metric drops out of
the classical action completely due to the Weyl invariance. It is
pure gauge (at classical level). However, the worldsheet curvature
scalar depends only on this conformal factor. We thus have the same
situation as for the Berkovits string. We can, however, always add
the conformal factor as independent field and the corresponding ghost
will excatly cancel its contribution to the central charge. We thus
add just as in \cite{Berkovits:2001ue} and in \cite{Bedoya:2006ic}
the Fradkin\index{Fradkin Tseytlin term} Tseytlin term\begin{eqnarray}
S_{FT} & = & \int\quad\alpha'r\dil\end{eqnarray}
to the action in general background...}

\section{Local SUSY-transformation of the fermionic fields}

\label{sec:Local-SUSY-transformation}In order to make contact to
generalized complex geometry, we are interested in the local\index{local SUSY!of the fermionic fields}
supersymmetry\index{SUSY!local $\sim$ of the fermionic fields} transformations
of the fermionic fields, i.e. the gravitino and the dilatino. Note
that the superdiffeomorphisms and the local structure group transformations
contain a huge number of auxiliary gauge degrees of freedom in the
$\xbothtetas$-expansion of the transformation parameters. The physical
fields are recovered by choosing a gauge, in particular the so-called
WZ-gauge. Remaining bosonic diffeomorphisms, local structure group
transformations of the bosonic manifold and local supersymmetry are
then part of the stabilizer transformations of the chosen gauge. In
the appendix \vref{chapter:susy}, this procedure is carefully explained
and the supergravity transformations are derived for a general setting,
following roughly \cite{Wess:1992cp}.

\subsection{Connection to choose}

\index{connection}As mentioned above, in the appendix \vref{chapter:susy}
we describe the ususal procedure of choosing the Wess Zumino gauge
$\bei{E_{\bs{\mc{M}}}\hoch{A}}{}=\delta_{\bs{\mc{M}}}\hoch{A}$ and
$\bei{\Omega_{\bs{\mc{M}}A}\hoch{B}}{}=0$ (see (\ref{eq:WZ-gauge})
and (\ref{eq:WZ-gauge-connection})). This gauge fixing is possible
with any connection as long as it takes the same values (in the Lie
algebra) as the gauge transformations (Remember, a connection is a
Lie algebra valued one form). However, the present case is a bit special
in the following sense: We have derived the supergravity constraints
using the connection\index{connection!mixed $\sim$}\index{mixed connection}
\begin{eqnarray}
\gemOm_{MA}\hoch{B} & \equiv & \left(\begin{array}{ccc}
\check{\Omega}_{Ma}\hoch{b} & 0 & 0\\
0 & \Omega_{M\bs{\alpha}}\hoch{\bs{\beta}} & 0\\
0 & 0 & \hat{\Omega}_{M\hat{\bs{\alpha}}}\hoch{\hat{\bs{\beta}}}\end{array}\right)\end{eqnarray}
After that we have coupled the independent structure group transformations
of the three blocks by a gauge fixing s.t. $T_{\bs{\alpha\beta}}\hoch{c}=\gamma_{\bs{\alpha}\bs{\beta}}^{c}$
and $T_{\hat{\bs{\alpha}}\hat{\bs{\beta}}}\hoch{c}=\gamma_{\hat{\bs{\alpha}}\hat{\bs{\beta}}}^{c}$.
The remaining gauge symmetry has to leave this gauge fixing invariant
which reduces the structure group to only one copy of the Lorentz
group plus one scale group. The above connection however does not
leave the gauge fixing invariant (the covariant derivatives do not
vanish in general). In order to be consistent, we thus have to reformulate
the equations in terms of a connection which leaves $\gamma_{\bs{\alpha}\bs{\beta}}^{c}$
and $\gamma_{\hat{\bs{\alpha}}\hat{\bs{\beta}}}^{c}$ invariant. Possible
choices are either the left mover connection $\Omega_{MA}\hoch{B}$
(defined by $\Omega_{M\bs{\alpha}}\hoch{\bs{\beta}}$ and $\nabla_{M}\gamma_{\bs{\alpha}\bs{\beta}}^{c}=\nabla_{M}\gamma_{\hat{\bs{\alpha}}\hat{\bs{\beta}}}^{c}=0$)
or the right-mover connection $\hat{\Omega}_{MA}\hoch{B}$ (defined
by $\hat{\Omega}_{M\hat{\bs{\alpha}}}\hoch{\hat{\bs{\beta}}}$) or
the \textbf{average\index{average connection} connection}\index{connection!average $\sim$}\index{$\Omega$@$\avOm_{MA}\hoch{B}$|itext{average connection}}\begin{eqnarray}
\avOm_{MA}\hoch{B} & \equiv & \frac{1}{2}\left(\Omega_{MA}\hoch{B}+\hat{\Omega}_{MA}\hoch{B}\right)=\Omega_{MA}\hoch{B}+\frac{1}{2}\Delta_{MA}\hoch{B}\end{eqnarray}
We will study the choices $\Omega_{MA}\hoch{B}$ and $\avOm_{MA}\hoch{B}$.
The first has the advantage that at least the left mover equations
stay simple while the second has the advantage that the symmetry between
left and right movers is preserved. Corresponding to the the first
choice the connection part of the WZ gauge simply reads\index{gauge I}\begin{equation}
\boxed{\bei{\Omega_{\bs{\mc{M}}\, A}\hoch{B}}{}=0}\quad\textrm{(gauge I)}\label{eq:WZ-gauge-connectionIa}\end{equation}
In this gauge all the equations derived in appendix \vref{chapter:susy}
hold literally. The average connection becomes in this gauge $\bei{\avOm_{\bs{\mc{M}}\, A}\hoch{B}}{}=\frac{1}{2}\bei{\Delta_{\bs{\mc{M}}A}\hoch{B}}{}$,
while the mixed connection can be written as $\bei{\gemOm_{\bs{\mc{M}}\, A}\hoch{B}}{\check{\Omega}=\Omega,\tet=0}=\diag(0,0,\bei{\Delta_{\bs{\mc{M}}\,\hat{\bs{\alpha}}}\hoch{\hat{\bs{\beta}}}}{})$.
Alternatively to gauge-I we could put $\bei{\hat{\Omega}_{\bs{\mc{M}}\, A}\hoch{B}}{}=0$
or equivalently $\bei{\avOm_{\bs{\mc{M}}\, A}\hoch{B}}{}=-\frac{1}{2}\bei{\Delta_{\bs{\mc{M}}A}\hoch{B}}{}$
which would be the same type of gauge with simply the role of hatted
and unhatted variables interchanged. 

However, a qualitatively different but likewise natural gauge fixing
(preserving the symmetry in hatted and unhatted variables) is\index{gauge II}\begin{equation}
\boxed{\bei{\avOm_{\bs{\mc{M}}\, A}\hoch{B}}{}=0}\quad\textrm{(gauge II)}\label{eq:WZ-gauge-connectionIIa}\end{equation}
In this gauge we have to replace in all equations of appendix \vref{chapter:susy}
the objects $\Omega_{MA}\hoch{B}$, $\nabla_{M}$, $T_{MN}\hoch{A}$
and $R_{MNA}\hoch{B}$ with $\avOm_{MA}\hoch{B}$, $\av{\nabla}_{M}$,
$\av{T}_{MN}\hoch{A}$ and $\av{R}_{MNA}\hoch{B}$ respectively. The
mixed connection in this gauge becomes $\bei{\gemOm_{\bs{\mc{M}}A}\hoch{B}}{\check{\Omega}=\Omega,\tet=0}=\bei{\diag(-\frac{1}{2}\Delta_{\bs{\mc{M}}a}\hoch{b},-\frac{1}{2}\Delta_{\bs{\mc{M}}\bs{\alpha}}\hoch{\bs{\beta}},\frac{1}{2}\Delta_{\bs{\mc{M}}\hat{\bs{\alpha}}}\hoch{\hat{\bs{\beta}}})}{}$.

\subsection{Denoting the physical component fields}

\label{sub:notationComponentFields}We will try (where possible) to
use a small letter to denote the leading component of a superfield.
One should keep in mind that the notation for the component fields
is a bit subtle, because the bosonic vielbein offers a second useful
possibility to change from flat to curved indices. We will also make
use of this possibility for the component fields, but one has to be
careful. Defining for example $h_{mnk}\equiv\bei{H_{mnk}}{}$ \index{H-field!bosonic $\sim$ $h_{mnk}$}\index{$h_{mnk}$|itext{bosonic H-field}}and
then changing to flat indices with the bosonic vielbein\index{vielbein!bosonic $\sim\quad e_m\hoch{a}$}\index{$e_m\hoch{a}$|itext{bosonic vielbein}},
is different from first changing to flat indices with the supervielbein
and then taking the leading component: $h_{abc}\neq\bei{H_{abc}}{}$.
In the following we will provide the definitions of the component
fields. If the same component field is given later with changed indices
(flat to curved or vice verse), then this is done via the bosonic
vielbein. \begin{eqnarray}
\bei{E_{M}\hoch{A}}{} & \equiv & \left(\begin{array}{cc}
e_{m}\hoch{a} & \psi_{m}\hoch{\bs{\mc{A}}}\\
0 & \delta_{\bs{\mc{M}}}\hoch{\bs{\mc{A}}}\end{array}\right)\label{eq:DefBosVielbein}\\
\bei{\Omega_{mA}\hoch{B}}{} & \equiv & \omega_{mA}\hoch{B},\quad(\bei{\Omega_{\bs{\mc{M}}A}\hoch{B}}{}=0)\\
\bei{\compensator}{} & \equiv & \compcomp,\qquad\bei{\dil(\xfull)}{}\equiv\dilcomp(\xboson)\label{eq:dilaton}\\
\bei{G_{mn}}{} & \equiv & e^{2\compcomp}g_{mn}=e_{m}\hoch{a}e_{n}\hoch{b}e^{2\compcomp}\eta_{ab}\label{eq:bosonicMetricWithComp}\\
\bei{B_{mn}}{} & \equiv & b_{mn},\qquad\bei{H_{mnk}}{}\equiv h_{mnk}\quad\dann\quad h_{mnk}=\partial_{[m}b_{nk]}\label{eq:DefBosHandB}\end{eqnarray}
 The second line which defines the bosonic connection certainly has
to be adjusted according to the superconnection on which the WZ-gauge
is based. For gauge II the definition of the bosonic connection would
thus change to $\bei{\avOm_{mA}\hoch{B}}{}\equiv\av{\omega}_{mA}\hoch{B}$,
$(\bei{\avOm_{\bs{\mc{M}}A}\hoch{B}}{}=0)$.  \index{metric!bosonic $\sim$ $g_{mn}$}\index{$g_{mn}$|itext{bosonic metric}}\index{antisymmetric tensor field!bosonic $\sim$ $b_{mn}$}\index{$b_{mn}$|itext{antisymmetric tensor field}}In
the fourth line we see that we can use the bosonic compensator\index{compensator field!bosonic}
field $\compcomp$\index{$\phi$|itext{bosonic compensator}} to switch
from string\index{string frame} frame\index{frame!Einstein- and string $\sim$}
(vanishing $\compcomp$) to the Einstein\index{Einstein frame} frame
where $\compcomp$ should be gauge fixed to be proportional to the
dilaton. In the third line we have defined the bosonic \textbf{dilaton}
\index{dilaton}\index{$\phi_{ph}$|itext{dilaton}} $\dilcomp$ as
the leading component of the dilaton superfield. In contrast to the
compensator field, it contains a physical degree of freedom which
cannot be gauged away.%
\footnote{\index{footnote!\thefoot. bosonic local scale invariance and bosonic covariant derivative}There
are some more words to say about the remaining scale invariance. The
fact that the definition of the bosonic metric includes the compensator
field leads to a loss of the correspondance between scaling behaviour
and flat index. Define alternatively \begin{eqnarray*}
\bei{G_{mn}}{} & \equiv & \tilde{g}_{mn}=e_{m}\hoch{a}e_{n}\hoch{b}\tilde{g}_{ab}\quad(=e_{m}\hoch{a}e_{n}\hoch{b}e^{2\compcomp}\eta_{ab}),\qquad\tilde{g}^{mn}=e_{a}\hoch{m}e_{b}\hoch{n}g^{ab}\quad(\equiv e_{a}\hoch{m}e_{b}\hoch{n}e^{-2\compcomp}\eta^{ab})\end{eqnarray*}
For a scale transformation $\delta\compcomp=-\varphi$, we have the
following transformations of the other fields:\begin{eqnarray*}
\delta e_{m}\hoch{a} & = & \varphi e_{m}\hoch{a},\quad\delta e_{a}\hoch{m}=-\varphi e_{a}\hoch{m}\\
\delta\tilde{g}_{ab} & = & -2\varphi\,\tilde{g}_{ab},\quad\delta\tilde{g}^{ab}=2\varphi\tilde{g}^{ab}\qquad\leftrightarrow\quad\delta\eta_{ab}=\delta\eta^{ab}=0\\
\delta\tilde{g}_{mn} & = & \delta\tilde{g}^{mn}=0\qquad\qquad\leftrightarrow\quad\delta g_{mn}=2\varphi g_{mn},\quad\delta g^{mn}\\
\delta b_{mn} & = & \delta h_{mnk}=\delta\dilcomp=0\\
\delta\rr^{\bs{\alpha}\hat{\bs{\beta}}} & = & \varphi\rr^{\bs{\alpha}\hat{\bs{\beta}}}\\
\delta\psi_{m}\hoch{\bs{\mc{A}}} & = & \tfrac{1}{2}\varphi\psi_{m}\hoch{\bs{\mc{A}}}\\
\delta\dilo_{\bs{\mc{A}}} & = & -\tfrac{1}{2}\varphi\dilo_{\bs{\mc{A}}}\end{eqnarray*}
While for the use of $\tilde{g}_{mn}$ and $\tilde{g}_{ab}$ the scaling
behaviour is coupled to the flat indices, this is not the case for
$g_{mn}$ and $\eta_{ab}$. Before the scale invariance is not fixed,
we thus should not use $g_{mn}$ or $\eta_{ab}$ to lower or raise
indices. 

Similar considerations hold for the covariant derivative. Denote for
the moment the bosonic spacetime-connection with $\gamma_{mk}\hoch{l}$.
We will use it only in this footnote and should not mix it up with
an antisymmetrized product of three $\gamma$-matrices. This spacetime
connection will not be defined as the leading component of $\Gamma_{mk}\hoch{l}$,
but via \[
\nabla_{m}e_{k}\hoch{a}=0\mbox{ with }\nabla_{m}\equiv\partial_{m}\pm\gamma_{ml}\hoch{k}\pm\omega_{mA}\hoch{B}\]
which implies $\gamma_{mk}\hoch{n}e_{n}\hoch{a}=\bei{\Gamma_{mk}\hoch{N}}{}\bei{E_{N}\hoch{a}}{}$.
The scaling part of the so defined bosonic covariant derivative acts
on $\tilde{g}_{mn}$ and $\tilde{g}_{ab}$ according to their indices
but not on $g_{mn}$ and $\eta_{ab}.$$\qquad\fussend$ %
}

For the definition of the leading component of the RR-bispinor $\RR^{\bs{\alpha}\hat{\bs{\beta}}}$
we first need a motivating observation. Because of the definition
of the dilaton superfield in (\ref{eq:DilatonDef}) via the spinorial
covariant derivative of the compensator field, the latter can be replaced
in (\ref{eq:coll:nablaalphaPtrace}),(\ref{eq:coll:nablahalphaPtrace})
by the spinorial derivative of the dilaton superfield and those equations
can be rewritten as\begin{eqnarray}
\gemnabla_{\hat{\bs{\alpha}}}(e^{-8\dil}\RR^{\bs{\delta}\hat{\bs{\alpha}}})=0 & = & ,\quad\gemnabla_{\bs{\alpha}}(e^{-8\dil}\RR^{\bs{\alpha}\hat{\bs{\delta}}})=0\end{eqnarray}
This is the motivation to define the RR-fields as \begin{eqnarray}
\rr^{\bs{\alpha}\hat{\bs{\beta}}} & \equiv & e^{-8\dilcomp}\bei{\RR^{\bs{\alpha}\hat{\bs{\delta}}}}{}\label{eq:DefRR}\end{eqnarray}
We had defined the dilatino\index{dilatino} already in the previous
section in (\ref{eq:dilatinoDefI}). Having now the scalar dilaton
superfield at hand, it is convenient to use (\ref{eq:DilatonDef})
in order to write them as components of this superfield: 

\begin{equation}
\dilo_{\bs{\mc{A}}}\equiv\bei{\nabla_{\bs{\mc{A}}}\dil}{}\label{eq:DilatinoDefII}\end{equation}

The subtleties of having bosonic and superspace vielbein at the same
time were mentioned already in the beginning of this subsection. An
example for the issues is provided by the inverse vielbein whose leading
components are given by \begin{eqnarray}
\bei{E_{A}\hoch{M}}{} & = & \left(\begin{array}{cc}
e_{a}\hoch{m} & -\psi_{a}\hoch{\bs{\mc{M}}}\\
0 & \delta_{\bs{\mc{A}}}\hoch{\bs{\mc{M}}}\end{array}\right)\end{eqnarray}
where $e_{a}\hoch{m}$ is the inverse of $e_{m}\hoch{a}$ and the
indices of the gravitino were converted via bosonic vielbein and fermionic
Kronecker delta respectively:\begin{eqnarray}
e_{a}\hoch{m}e_{m}\hoch{b} & = & \delta_{a}^{b}\\
\psi_{a}\hoch{\bs{\mc{M}}} & \equiv & e_{a}\hoch{m}\psi_{m}\hoch{\bs{\mc{A}}}\delta_{\bs{\mc{A}}}\hoch{\bs{\mc{M}}}\end{eqnarray}
In the same way we define\begin{eqnarray}
b_{ab} & \equiv & e_{a}\hoch{m}e_{b}\hoch{n}b_{mn}\\
h_{abc} & \equiv & e_{a}\hoch{m}e_{b}\hoch{n}e_{c}\hoch{n}h_{mnk}\\
g_{ab} & \equiv & e_{a}\hoch{m}e_{b}\hoch{n}g_{mn}=\eta_{ab}\end{eqnarray}
As mentioned above, these expressions do in general not coincide with
the leading components of the corresponding superfields\begin{eqnarray}
\bei{G_{ab}}{} & = & e^{2\phi}\eta_{ab}-2e_{[a}\hoch{m}\psi_{b]}\hoch{\bs{\mc{N}}}\bei{G_{m\bs{\mc{N}}}}{}+\psi_{a}\hoch{\bs{\mc{M}}}\psi_{b}\hoch{\bs{\mc{N}}}\bei{G_{\bs{\mc{MN}}}}{}=\\
 & = & e^{2\phi}\eta_{ab}-2\psi_{[b}\hoch{\bs{\mc{B}}}\bei{G_{a]\bs{\mc{B}}}}{}-\psi_{a}\hoch{\bs{\mc{A}}}\psi_{b}\hoch{\bs{\mc{B}}}\bei{G_{\bs{\mc{AB}}}}{}\\
\bei{B_{ab}}{} & = & b_{ab}-2e_{[a}\hoch{m}\psi_{b]}\hoch{\bs{\mc{N}}}\bei{B_{m\bs{\mc{N}}}}{}+\psi_{a}\hoch{\bs{\mc{M}}}\psi_{b}\hoch{\bs{\mc{N}}}\bei{B_{\bs{\mc{MN}}}}{}=\\
 & = & b_{ab}-2\psi_{[b}\hoch{\bs{\mc{B}}}\bei{B_{a]\bs{\mc{B}}}}{}-\psi_{a}\hoch{\bs{\mc{A}}}\psi_{b}\hoch{\bs{\mc{B}}}\bei{B_{\bs{\mc{AB}}}}{}\\
\bei{H_{abc}}{} & = & h_{abc}-3e_{[a}\hoch{m}e_{b}\hoch{n}\psi_{c]}\hoch{\bs{\mc{K}}}\bei{H_{mn\bs{\mc{K}}}}{}+3\psi_{a}\hoch{\bs{\mc{M}}}\psi_{b}\hoch{\bs{\mc{N}}}e_{c}\hoch{k}\bei{H_{\bs{\mc{MN}}k}}{}-\psi_{a}\hoch{\bs{\mc{M}}}\psi_{b}\hoch{\bs{\mc{N}}}\psi_{c}\hoch{\bs{\mc{K}}}\bei{H_{\bs{\mc{MN}}\bs{\mc{K}}}}{}=\\
 & = & h_{abc}-3\psi_{[c}\hoch{\bs{\mc{C}}}\bei{H_{ab]\bs{\mc{C}}}}{}-3\psi_{[a}\hoch{\bs{\mc{A}}}\psi_{b|}\hoch{\bs{\mc{B}}}\bei{H_{\bs{\mc{AB}}|c]}}{}-\psi_{a}\hoch{\bs{\mc{A}}}\psi_{b}\hoch{\bs{\mc{B}}}\psi_{c}\hoch{\bs{\mc{C}}}\bei{H_{\bs{\mc{ABC}}}}{}\end{eqnarray}
Note that for vanishing gravitino $\psi_{m}\hoch{\bs{\mc{A}}}$ there
is no difference between the usage of bosonic vielbein or supervielbein
to change from flat to curved indices. For non-vanishing gravitino
the expressions already simplify significantly, if we take into account
the WZ-like gauge $\bei{B_{\bs{\mc{MN}}}}{}=\bei{B_{m\bs{\mc{N}}}}{}=0$
for the B-field and the supergravity constraints of $H$-field and
rank-two tensor $G_{AB}$. The latter has $G_{ab}$ as only nonvanishing
component.\begin{eqnarray}
\bei{G}{ab} & = & e^{2\compcomp}\eta_{ab}\\
\bei{B_{ab}}{} & = & b_{ab}\\
\bei{H_{abc}}{} & = & h_{abc}+2e^{2\compcomp}e_{[a}\hoch{m}e_{b}\hoch{n}\gamma_{c]\,\bs{\alpha\beta}}\psi_{m}\hoch{\bs{\alpha}}\psi_{n}\hoch{\bs{\beta}}-2e^{2\compcomp}e_{[a}\hoch{m}e_{b}\hoch{n}\gamma_{c]\,\hat{\bs{\alpha}}\hat{\bs{\beta}}}\hat{\psi}_{m}\hoch{\hat{\bs{\alpha}}}\hat{\psi}_{n}\hoch{\hat{\bs{\beta}}}\label{eq:HabcLeading}\end{eqnarray}
Let us eventually see how the \textbf{bosonic\index{bosonic torsion}
torsion\index{torsion!bosonic $\sim$}} \begin{equation}
t^{a}\equiv\de e^{a}-e^{c}\wedge\omega_{c}\hoch{a}\end{equation}
is related to the leading component of the superspace torsion:\begin{eqnarray}
\bei{T_{mn}\hoch{a}}{} & = & \partial_{[m}\bei{E_{n]}\hoch{a}}{}+\bei{E_{[n}\hoch{c}}{}\bei{\Omega_{m]c}\hoch{a}}{}+\bei{E_{[n}\hoch{\bs{\mc{C}}}}{}\bei{\Omega_{m]\bs{\mc{C}}}\hoch{a}}{}=\\
 & = & \partial_{[m}e_{n]}\hoch{a}+e_{[n}\hoch{c}\omega_{m]c}\hoch{a}=t_{mn}\hoch{a}\end{eqnarray}
Rewriting the superspace connection in terms of components with flat
indices yields\begin{eqnarray}
t_{mn}\hoch{a} & = & e_{m}\hoch{c}e_{n}\hoch{d}\bei{T_{cd}\hoch{a}}{}+2e_{[m}\hoch{c}\psi_{n]}\hoch{\bs{\mc{D}}}\bei{T_{c\bs{\mc{D}}}\hoch{a}}{}+\psi_{m}\hoch{\bs{\mc{C}}}\psi_{n}\hoch{\bs{\mc{D}}}\bei{T_{\bs{\mc{CD}}}\hoch{a}}{}\end{eqnarray}
which implies \begin{eqnarray}
t_{cd}\hoch{a} & = & \bei{T_{cd}\hoch{a}}{}+2e_{[c|}\hoch{m}\psi_{m}\hoch{\bs{\mc{C}}}\bei{T_{\bs{\mc{C}}|d]}\hoch{a}}{}+e_{c}\hoch{m}e_{d}\hoch{n}\psi_{m}\hoch{\bs{\mc{C}}}\psi_{n}\hoch{\bs{\mc{D}}}\bei{T_{\bs{\mc{CD}}}\hoch{a}}{}\end{eqnarray}
Similarly we have for the \textbf{bosonic\index{bosonic curvature}
curvature\index{curvature!bosonic $\sim$}}\begin{equation}
r_{a}\hoch{b}\equiv\de\omega_{a}\hoch{b}-\omega_{a}\hoch{c}\wedge\omega_{c}\hoch{b}\end{equation}
the following relations to the superspace curvature:\begin{eqnarray}
\bei{R_{mna}\hoch{b}}{} & = & r_{mna}\hoch{b}\\
r_{cda}\hoch{b} & = & \bei{R_{cda}\hoch{b}}{}+2e_{[c|}\hoch{m}\psi_{m}\hoch{\bs{\mc{C}}}\bei{R_{\bs{\mc{C}}|d]a}\hoch{b}}{}+e_{c}\hoch{m}e_{d}\hoch{n}\psi_{m}\hoch{\bs{\mc{C}}}\psi_{n}\hoch{\bs{\mc{D}}}\bei{R_{\bs{\mc{CD}}a}\hoch{b}}{}\end{eqnarray}
For gauge II the above expressions again have to be understood in
terms of the average connection. As we have not yet plugged any torsion
or curvature constraints into the equations, they are still valid
for both gauges.

\subsection{The gravitino transformation}

\subsubsection{General form}

In the appendix, the general form of the gravitino transformation
is given in equation (\ref{eq:generalGravitinoSusyTrafo}), which
we repeat here for convenience:

\begin{eqnarray}
\delta_{\eps}\psi_{m}\hoch{\bs{\mc{A}}} & = & \underbrace{\partial_{m}\eps^{\bs{\mc{A}}}+\omega_{m\bs{\mc{C}}}\hoch{\bs{\mc{A}}}\eps^{\bs{\mc{C}}}}_{\nabla_{m}\eps^{\bs{\mc{A}}}}+2\eps^{\bs{\mc{C}}}e_{m}\hoch{b}\bei{T_{\bs{\mc{C}}b}\hoch{\bs{\mc{A}}}}{}+2\eps^{\bs{\mc{C}}}\psi_{m}\hoch{\bs{\mc{B}}}\bei{T_{\bs{\mc{C}}\bs{\mc{B}}}\hoch{\bs{\mc{A}}}}{}\label{eq:bib:generalGravitinoSusyTrafo}\end{eqnarray}
where $\omega_{m\bs{\mc{A}}}\hoch{\bs{\mc{B}}}\equiv\bei{\Omega_{m\bs{\mc{A}}}\hoch{\bs{\mc{B}}}}{}$.\index{$\omega_m$@$\omega_{m\bs{\mc{A}}}\hoch{\bs{\mc{B}}}$}
The connection appearing explicitely and implicitely (in the torsion)
in this transformation has to be the same connection as the one on
which the WZ gauge fixing condition was put. The above equation can
thus be understood literally if we choose gauge I (based on the left-mover
connection $\Omega_{MA}\hoch{B}$) while for gauge II (based on the
average connection $\av{\Omega}_{MA}\hoch{B}$) every implicit or
explicit appearance of $\Omega_{MA}\hoch{B}$ has to be replaced by
$\av{\Omega}_{MA}\hoch{B}$. We can continue the considerations for
a while without deciding, whether we are in gauge I or gauge II, although
the notation will suggest that we are in gauge I (with connection
$\Omega_{MA}\hoch{B}$).

For the transformation of the gravitino(s) given above, we still need
additional information about the connection $\omega_{m\bs{\mc{C}}}\hoch{\bs{\mc{A}}}$,
which does not necessarily coincide with the Levi Civita connection.
In bosonic manifolds, the connection is completely determined by torsion
and (non)metricity, if a metric is given. If no metric is given, one
can likewise demand the preservation of other structures or structure
constants. In particular in 10-dimensional superspace we do not have
a non-degenerate superspace-metric. Only the bosonic block $G_{ab}$
of the symmetric rank two tensor $G_{AB}$ has full rank. In order
to determine the full superspace connection, one thus needs more than
the information about the covariant derivative of the symmetric rank
two tensor. A natural candidate is the covariant derivative of the
gamma-matrices, the structure constants of the supersymmetry algebra.
This logic is carefully described in appendix \ref{cha:ConnectionAppend}.

The derivation of (\ref{eq:bib:generalGravitinoSusyTrafo}) in the
appendix did not assume any restrictions on the structure group, apart
from being blockdiagonal w.r.t. bosonic and fermionic indices. Right
now, we make use of the fact that we have (for gauge I as well as
for gauge II) a connection with \begin{eqnarray}
\nabla_{M}\gamma_{\bs{\alpha\beta}}^{c} & \stackrel{!}{=} & \nabla_{M}\gamma_{\hat{\bs{\alpha}}\hat{\bs{\beta}}}^{c}\stackrel{!}{=}0\end{eqnarray}
which relates the three blocks of $\Omega_{MA}\hoch{B}$ and restricts
the structure group to local Lorentz and local scale transformations.
It is convenient to write \begin{eqnarray}
\gamma_{\bs{\mc{AB}}}^{c} & \equiv & \left(\begin{array}{cc}
\gamma_{\bs{\alpha\beta}}^{c} & 0\\
0 & \gamma_{\hat{\bs{\alpha}}\hat{\bs{\beta}}}^{c}\end{array}\right)\label{eq:combinedGamma}\end{eqnarray}
Only in type IIA this matrix coincides with $A\Gamma^{c}$ (where
$A$ is the intertwiner responsible for the Dirac-conjugate: $\bar{\Psi}=\Psi^{\dagger}A$).

We can then make use of equation (\ref{eq:LCinSuperspaceConnectionFermionic})
of appendix \ref{cha:ConnectionAppend}, which relates the leading
components of the superspace connection, in particular the ones with
fermionic structure group indices \begin{eqnarray}
\omega_{m\bs{\mc{A}}}\hoch{\bs{\mc{B}}} & \equiv & \bei{\Omega_{m\bs{\mc{A}}}\hoch{\bs{\mc{B}}}}{}\:,\end{eqnarray}
to the Levi Civita connection and a somewhat lengthy rest:\begin{eqnarray}
\omega_{m\bs{\mc{B}}}\hoch{\bs{\mc{A}}} & = & \omega_{m\bs{\mc{B}}}^{(LC)}\hoch{\bs{\mc{A}}}+\nonumber \\
 &  & +\frac{1}{4}e_{m}\hoch{a}\biggl\{2e^{-2\compcomp}\bei{T_{a[b|c]}}{}-e^{-2\compcomp}\bei{T_{bc|a}}{}-2\bigl(\bei{\covPhi{[b|}}{}-e_{[b|}\hoch{k}\partial_{k}\compcomp\bigr)\eta_{|c]a}-2e_{[b}\hoch{n}\eta_{c]a}\psi_{n}\hoch{\bs{\mc{C}}}\bei{(\covPhi{\bs{\mc{C}}})}{}+\nonumber \\
 &  & +\left(2e_{a}\hoch{k}e_{[b}\hoch{n}\eta_{c]d}-e_{b}\hoch{k}e_{c}\hoch{n}\eta_{ad}\right)\psi_{k}\hoch{\bs{\mc{C}}}\psi_{n}\hoch{\bs{\mc{D}}}\bei{T_{\bs{\mc{C}}\bs{\mc{D}}}\hoch{d}}{}+\nonumber \\
 &  & +e^{-2\compcomp}\Bigl(2e_{a}\hoch{n}\psi_{n}\hoch{\bs{\mc{C}}}\bei{T_{\bs{\mc{C}}[b|c]}}{}-2e_{b}\hoch{n}\psi_{n}\hoch{\bs{\mc{C}}}\bei{T_{\bs{\mc{C}}(a|c)}}{}+2e_{c}\hoch{n}\psi_{n}\hoch{\bs{\mc{C}}}\bei{T_{\bs{\mc{C}}(a|b)}}{}\Bigr)\biggr\}\gamma^{bc}\tief{\bs{\mc{B}}}\hoch{\bs{\mc{A}}}\nonumber \\
 &  & -\frac{1}{2}\left(\psi_{m}\hoch{\bs{\mc{C}}}\bei{\covPhi{\bs{\mc{C}}}}{}+e_{m}\hoch{a}\bei{\covPhi{a}}{}-\partial_{m}\compcomp\right)\delta_{\bs{\mc{B}}}\hoch{\bs{\mc{A}}}\label{eq:susy:connectionleading:both}\end{eqnarray}
where the Levi Civita connection $\omega_{m\bs{\mc{B}}}^{(LC)}\hoch{\bs{\mc{A}}}$
is the one with respect to the metric $g_{mn}=e_{m}\hoch{a}\eta_{ab}e_{n}\hoch{b}$.
We should note that the Levi Civita connection is not a suitable connection
for scale transformations, because it is only Lorentz group valued.
The terms $\partial_{k}\compcomp$ with the partial derivative of
the compensator field do not transform covariantly under scale transformations
and are the minimal extension of the Levi Civita connection to make
it a structure group valued connection. On the other hand, if one
decides to simply fix $\compcomp$ to zero and thus ending up only
with Lorentz transformations, these terms disappear. The last line
which is dilatation-valued can then not any longer be seen as part
of the connection. 

Together with (\ref{eq:bib:generalGravitinoSusyTrafo}) the above
expression for the connection determines the supergravity transformation
of the gravitino. In order to plug in the explicit constraints for
the torsion, we have to decide in which gauge we work.

\subsubsection{In gauge I}

In gauge I, we can take the above equations literally and plug in
the corresponding torsion constraints (\ref{eq:coll:leftTorsionI})-(\ref{eq:coll:leftTorsionIII}).
We will need in addition that according to (\ref{eq:HabcLeading})
the leading component of the H-field with flat coordinates is related
to the bosonic h-field via $\bei{H_{abc}}{}=h_{abc}+2e^{2\compcomp}e_{[a}\hoch{m}e_{b}\hoch{n}\gamma_{c]\,\bs{\alpha\beta}}\psi_{m}\hoch{\bs{\alpha}}\psi_{n}\hoch{\bs{\beta}}-2e^{2\compcomp}e_{[a}\hoch{m}e_{b}\hoch{n}\gamma_{c]\,\hat{\bs{\alpha}}\hat{\bs{\beta}}}\hat{\psi}_{m}\hoch{\hat{\bs{\alpha}}}\hat{\psi}_{n}\hoch{\hat{\bs{\beta}}}$.
The connection becomes \begin{eqnarray}
\omega_{m\bs{\mc{B}}}\hoch{\bs{\mc{A}}} & = & \omega_{m\bs{\mc{B}}}^{(LC)}\hoch{\bs{\mc{A}}}+\frac{1}{4}e_{m}\hoch{a}\biggl\{\tfrac{3}{2}h_{abc}e^{-2\compcomp}+2e_{[b|}\hoch{k}\partial_{k}\compcomp\eta_{|c]a}+4e_{a}\hoch{k}e_{[b}\hoch{n}\eta_{c]d}\psi_{k}\hoch{\bs{\gamma}}\psi_{n}\hoch{\bs{\delta}}\gamma_{\bs{\gamma\delta}}^{d}+\nonumber \\
 &  & -2e_{b}\hoch{k}e_{c}\hoch{n}\eta_{ad}\hat{\psi}_{k}\hoch{\hat{\bs{\gamma}}}\hat{\psi}_{n}\hoch{\hat{\bs{\delta}}}\gamma_{\hat{\bs{\gamma}}\hat{\bs{\delta}}}^{d}-e_{a}\hoch{n}\psi_{n}\hoch{\bs{\gamma}}\gamma_{bc\,\bs{\gamma}}\hoch{\bs{\delta}}\dilo_{\bs{\delta}}\biggr\}\gamma^{bc}\tief{\bs{\mc{B}}}\hoch{\bs{\mc{A}}}+\nonumber \\
 &  & +\tfrac{1}{2}\left(\partial_{m}\compcomp-\psi_{m}\hoch{\bs{\gamma}}\dilo_{\bs{\gamma}}\right)\delta_{\bs{\mc{B}}}\hoch{\bs{\mc{A}}}\label{eq:bosonic-leftmoverconn-final}\end{eqnarray}
The constraints needed for the left-mover version of the transformation
(\ref{eq:bib:generalGravitinoSusyTrafo}) are rather simple. In particular
all the components $T_{\bs{\mc{CB}}}\hoch{\bs{\alpha}}$ vanish. The
local supersymmetry transformation of the left-mover gravitino turns
into\begin{equation}
\boxed{\delta_{\eps}\psi_{m}\hoch{\bs{\alpha}}=\underbrace{\partial_{m}\eps^{\bs{\alpha}}+\omega_{m\bs{\gamma}}\hoch{\bs{\alpha}}\eps^{\bs{\gamma}}}_{\nabla_{m}\eps^{\bs{\alpha}}}+2\eps^{\hat{\bs{\gamma}}}e_{m}\hoch{b}e^{2\compcomp+8\dilcomp}\gamma_{b\,\hat{\bs{\gamma}}\hat{\bs{\delta}}}\rr^{\bs{\alpha}\hat{\bs{\delta}}}}\label{eq:finalGravitinoTrafoLeftAlpha}\end{equation}
If we want to fix the local scale invariance by setting the compensator
field to zero, this gauge has to be respected by the supersymmetry
transformation which then has to be redefined according to (\ref{eq:additionalScaleStabilizer})
with a dilatation with parameter $\eps^{\bs{\gamma}}\dilo_{\bs{\gamma}}$
, which would add a term $\tfrac{1}{2}(\eps^{\bs{\gamma}}\dilo_{\bs{\gamma}})\psi_{m}\hoch{\bs{\alpha}}$
to the above transformation.

For the right-mover transformation, the torsion constraints are more
involved and we arrive at \begin{eqnarray}
\delta_{\eps}\psi_{m}\hoch{\hat{\bs{\alpha}}} & = & \underbrace{\partial_{m}\hat{\eps}^{\hat{\bs{\alpha}}}+\omega_{m\hat{\bs{\gamma}}}\hoch{\hat{\bs{\alpha}}}\hat{\eps}^{\hat{\bs{\gamma}}}}_{\nabla_{m}\eps^{\hat{\bs{\alpha}}}}+\nonumber \\
 &  & +\tfrac{1}{4}e_{m}\hoch{a}\Bigl(-3e^{-2\compcomp}h_{abc}-6e_{[a}\hoch{m}e_{b}\hoch{n}\gamma_{c]\,\bs{\alpha\beta}}\psi_{m}\hoch{\bs{\alpha}}\psi_{n}\hoch{\bs{\beta}}+6e_{[a}\hoch{m}e_{b}\hoch{n}\gamma_{c]\,\hat{\bs{\alpha}}\hat{\bs{\beta}}}\hat{\psi}_{m}\hoch{\hat{\bs{\alpha}}}\hat{\psi}_{n}\hoch{\hat{\bs{\beta}}}+\nonumber \\
 &  & +e_{a}\hoch{n}\psi_{n}\hoch{\bs{\beta}}\gamma_{bc}\tief{\bs{\beta}}\hoch{\bs{\delta}}\dilo_{\bs{\delta}}-e_{a}\hoch{n}\hat{\psi}_{n}\hoch{\hat{\bs{\beta}}}\gamma_{bc\,\hat{\bs{\beta}}}\hoch{\hat{\bs{\delta}}}\hat{\dilo}_{\hat{\bs{\delta}}}\Bigr)\hat{\eps}^{\hat{\bs{\gamma}}}\gamma^{bc}\tief{\hat{\bs{\gamma}}}\hoch{\hat{\bs{\alpha}}}+\nonumber \\
 &  & +\tfrac{1}{2}(\psi_{m}\hoch{\bs{\beta}}\dilo_{\bs{\beta}}-\psi_{m}\hoch{\hat{\bs{\beta}}}\hat{\dilo}_{\hat{\bs{\beta}}})\hat{\eps}^{\hat{\bs{\alpha}}}+\nonumber \\
 &  & +2\eps^{\bs{\gamma}}e_{m}\hoch{b}e^{2\compcomp+8\dilcomp}\gamma_{b\,\bs{\gamma\delta}}\rr^{\bs{\delta}\hat{\bs{\alpha}}}+\nonumber \\
 &  & +\underbrace{\tfrac{1}{4}(\hat{\eps}^{\hat{\bs{\gamma}}}\gamma_{de\,\hat{\bs{\gamma}}}\hoch{\hat{\bs{\delta}}}\hat{\dilo}_{\hat{\bs{\delta}}}-\eps^{\bs{\gamma}}\gamma_{de}\tief{\bs{\gamma}}\hoch{\bs{\delta}}\dilo_{\bs{\delta}})\hat{\psi}_{m}\hoch{\hat{\bs{\beta}}}\gamma^{de}\tief{\hat{\bs{\beta}}}\hoch{\hat{\bs{\alpha}}}}_{{\rm Lorentz\, trafo}}+\underbrace{\tfrac{1}{2}(\hat{\eps}^{\hat{\bs{\gamma}}}\hat{\dilo}_{\hat{\bs{\gamma}}}-\eps^{\bs{\gamma}}\dilo_{\bs{\gamma}})\psi_{m}\hoch{\hat{\bs{\alpha}}}}_{{\rm dilatation}}\end{eqnarray}
It is obvious that {}``gauge I'' prefers the left-movers and destroys
the left-right symmetry. The last two terms correspond to a Lorentz
and a scale transformation of the gravitino with gauge parameters
$(\hat{\eps}^{\hat{\bs{\gamma}}}\gamma_{de\,\hat{\bs{\gamma}}}\hoch{\hat{\bs{\delta}}}\hat{\dilo}_{\hat{\bs{\delta}}}-\eps^{\bs{\gamma}}\gamma_{de}\tief{\bs{\gamma}}\hoch{\bs{\delta}}\dilo_{\bs{\delta}})$
and $(\hat{\eps}^{\hat{\bs{\gamma}}}\hat{\dilo}_{\hat{\bs{\gamma}}}-\eps^{\bs{\gamma}}\dilo_{\bs{\gamma}})$
respectively. They could be removed by redefining the local supersymmetry
transformation, but then they would show up for the left-mover. For
the right-mover one can combine some terms, if we plug the explicit
expression of the connection into the above equation:\vRam{1.02}{\begin{eqnarray}
\delta_{\eps}\psi_{m}\hoch{\hat{\bs{\alpha}}} & = & \nabla_{m}^{(LC)}\hat{\eps}^{\hat{\bs{\alpha}}}+\tfrac{1}{4}e_{m}\hoch{a}\Bigl(-\tfrac{3}{2}h_{abc}e^{-2\compcomp}+2e_{[b|}\hoch{k}\partial_{k}\compcomp\eta_{|c]a}+\nonumber \\
 &  & -2e_{b}\hoch{m}e_{c}\hoch{n}\gamma_{a\,\bs{\alpha\beta}}\psi_{m}\hoch{\bs{\alpha}}\psi_{n}\hoch{\bs{\beta}}+4e_{a}\hoch{m}e_{[b}\hoch{n}\gamma_{c]\,\hat{\bs{\alpha}}\hat{\bs{\beta}}}\hat{\psi}_{m}\hoch{\hat{\bs{\alpha}}}\hat{\psi}_{n}\hoch{\hat{\bs{\beta}}}-e_{a}\hoch{n}\hat{\psi}_{n}\hoch{\hat{\bs{\beta}}}\gamma_{bc\,\hat{\bs{\beta}}}\hoch{\hat{\bs{\delta}}}\hat{\dilo}_{\hat{\bs{\delta}}}\Bigr)\hat{\eps}^{\hat{\bs{\gamma}}}\gamma^{bc}\tief{\hat{\bs{\gamma}}}\hoch{\hat{\bs{\alpha}}}+\nonumber \\
 &  & +\tfrac{1}{2}(\partial_{m}\compcomp-\psi_{m}\hoch{\hat{\bs{\beta}}}\hat{\dilo}_{\hat{\bs{\beta}}})\hat{\eps}^{\hat{\bs{\alpha}}}+\nonumber \\
 &  & +2\eps^{\bs{\gamma}}e_{m}\hoch{b}e^{2\compcomp+8\dilcomp}\gamma_{b\,\bs{\gamma\delta}}\rr^{\bs{\delta}\hat{\bs{\alpha}}}+\nonumber \\
 &  & +\underbrace{\tfrac{1}{4}(\hat{\eps}^{\hat{\bs{\gamma}}}\gamma_{de\,\hat{\bs{\gamma}}}\hoch{\hat{\bs{\delta}}}\hat{\dilo}_{\hat{\bs{\delta}}}-\eps^{\bs{\gamma}}\gamma_{de}\tief{\bs{\gamma}}\hoch{\bs{\delta}}\dilo_{\bs{\delta}})\hat{\psi}_{m}\hoch{\hat{\bs{\beta}}}\gamma^{de}\tief{\hat{\bs{\beta}}}\hoch{\hat{\bs{\alpha}}}}_{{\rm Lorentz\, trafo}}+\underbrace{\tfrac{1}{2}(\hat{\eps}^{\hat{\bs{\gamma}}}\hat{\dilo}_{\hat{\bs{\gamma}}}-\eps^{\bs{\gamma}}\dilo_{\bs{\gamma}})\psi_{m}\hoch{\hat{\bs{\alpha}}}}_{{\rm dilatation}}\label{eq:finalGravitinoTrafoLefthalpha}\end{eqnarray}
}  Comparing the first three lines with the left-mover connection
(\ref{eq:bosonic-leftmoverconn-final}), we recognize its hatted version,
i.e. the right-mover connection. The first three lines thus combine
to $\hat{\nabla}_{m}\hat{\eps}^{\hat{\bs{\alpha}}}$. We would have
obtained the same result without the last line if we had started with
the right-mover super-connection instead of the left-mover one. Using
a different gauge thus corresponds to redefining the supersymmetry
transformation by a local Lorentz and scale transformation. Also this
transformation needs to be modified in the case that $\phi$ is fixed
to zero. The stabilizing dilatation with parameter $\eps^{\bs{\gamma}}\dilo_{\bs{\gamma}}$
would add the term $\tfrac{1}{2}(\eps^{\bs{\gamma}}\dilo_{\bs{\gamma}})\psi_{m}\hoch{\hat{\bs{\alpha}}}$
and thus cancel the last term.

\subsubsection{In gauge II}

For gauge II, we need to replace the connection $\Omega_{MA}\hoch{B}$
everywhere in the gravitino transformation (\ref{eq:susy:connectionleading:both})
and (\ref{eq:bib:generalGravitinoSusyTrafo}) by the average connection
$\avOm_{MA}\hoch{B}$ (with $\av{\omega}_{m\bs{\mc{A}}}\hoch{\bs{\mc{B}}}\equiv\bei{\av{\Omega}_{m\bs{\mc{A}}}\hoch{\bs{\mc{B}}}}{}$)\index{$\omega_m$@$\av{\omega}_{m\bs{\mc{A}}}\hoch{\bs{\mc{B}}}$}.
This implies that we also have to replace the torsion components $T_{AB}\hoch{C}$
by $\av{T}_{AB}\hoch{C}$. The constraints on the corresponding torsion
$\av{T}_{AB}\hoch{C}$ are collected in (\ref{eq:coll:avTc})-(\ref{eq:coll:avThgam}).
The explicit form of the transformation becomes quite lengthy if we
split the fermionic index $\bs{\mc{A}}$ into the left and right-mover
spinorial indices $\bs{\alpha}$ and $\hat{\bs{\alpha}}$. For that
reason it is advantageous to try to rewrite the constraints (\ref{eq:coll:avTc})-(\ref{eq:coll:avThgam})
with the combined fermionic indices. To this end we define \begin{eqnarray}
\epsilon_{(\bs{\mc{A}})} & = & \left\{ \zwek{1\quad\bs{\mc{A}}=\bs{\alpha}}{-1\quad\mbox{for }\bs{\mc{A}}=\hat{\bs{\alpha}}}\right.\\
\RR^{\bs{\mc{C}}\bs{\mc{D}}} & \equiv & \left(\begin{array}{cc}
0 & \RR^{\bs{\gamma}\hat{\bs{\delta}}}\\
\RR^{\bs{\delta}\hat{\bs{\gamma}}} & 0\end{array}\right)\end{eqnarray}
In order to keep the left-right symmetry we should think of $\hat{\epsilon}_{(\bs{\mc{A}})}\equiv-\epsilon_{(\bs{\mc{A}})}$.
Remembering also the definition of $\gamma_{\bs{\mc{AB}}}^{c}$ in
(\ref{eq:combinedGamma}) and the relation of the spinorial derivative
$\nabla_{\bs{\mc{A}}}\dil$ of the dilaton superfield to the one of
the compensator (\ref{eq:DilatonDef}), the torsion constraints (\ref{eq:coll:avTc})-(\ref{eq:coll:avThgam})
can be written as\begin{eqnarray}
\av{T}_{AB}\hoch{c} & \equiv & \left(\begin{array}{cc}
\av{T}_{ab}\hoch{c} & \av{T}_{a\bs{\mc{B}}}\hoch{c}\\
\av{T}_{\bs{\mc{A}}b}\hoch{c} & \av{T}_{\bs{\mc{AB}}}\hoch{c}\end{array}\right)=\nonumber \\
 & = & \left(\begin{array}{cc}
0 & \frac{1}{4}\nabla_{\bs{\mc{B}}}\dil\delta_{a}^{c}+\frac{1}{4}\gamma_{a}\hoch{c}\tief{\bs{\mc{B}}}\hoch{\bs{\mc{D}}}\nabla_{\bs{\mc{D}}}\dil\\
-\frac{1}{4}\nabla_{\bs{\mc{A}}}\dil\delta_{b}^{c}-\frac{1}{4}\gamma_{b}\hoch{c}\tief{\bs{\mc{A}}}\hoch{\bs{\mc{D}}}\nabla_{\bs{\mc{D}}}\dil & \gamma_{\bs{\mc{AB}}}^{c}\end{array}\right)\label{eq:avTorsionCombined-c}\\
\av{T}_{AB}\hoch{\bs{\mc{C}}} & \equiv & \left(\begin{array}{cc}
\av{T}_{ab}\hoch{\bs{\mc{C}}} & \av{T}_{a\bs{\mc{B}}}\hoch{\bs{\mc{C}}}\\
\av{T}_{\bs{\mc{A}}b}\hoch{\bs{\mc{C}}} & \av{T}_{\bs{\mc{AB}}}\hoch{\bs{\mc{C}}}\end{array}\right)=\nonumber \\
 & \!\!\!\!\!\!\!\!= & \!\!\!\!\!\!\!\!\left(\begin{array}{cc}
\!\!\!\!\left(\frac{1}{16}\gemnabla_{\bs{\mc{E}}}\RR^{\bs{\mc{C}}\bs{\mc{D}}}+\tfrac{1}{2}\nabla_{\bs{\mc{E}}}\dil\RR^{\bs{\mc{C}}\bs{\mc{D}}}\right)\tilde{\gamma}_{ab\,\bs{\mc{D}}}\hoch{\bs{\mc{E}}} & -\epsilon_{(\bs{\mc{C}})}\frac{3}{16}H_{ade}\tilde{\gamma}^{de}\tief{\bs{\mc{B}}}\hoch{\bs{\mc{C}}}-\tilde{\gamma}_{a\,\bs{\mc{B}}\bs{\mc{D}}}\RR^{\bs{\mc{C}}\bs{\mc{D}}}\\
\epsilon_{(\bs{\mc{C}})}\frac{3}{16}H_{bde}\tilde{\gamma}^{de}\tief{\bs{\mc{A}}}\hoch{\bs{\mc{C}}}+\tilde{\gamma}_{b\,\bs{\mc{AD}}}\RR^{\bs{\mc{C}}\bs{\mc{D}}} & \!\!\!\!\frac{\epsilon_{(\bs{\mc{A}})}\epsilon_{(\bs{\mc{B}})}}{8}(\gamma_{de}\tief{[\bs{\mc{A}}}\hoch{\bs{\mc{D}}}\gamma^{de}\tief{\bs{\mc{B}}]}\hoch{\bs{\mc{C}}}\nabla_{\bs{\mc{D}}}\dil+2\nabla_{[\bs{\mc{A}}}\dil\delta_{\bs{\mc{B}}]}\hoch{\bs{\mc{C}}})\!\!\!\!\end{array}\right)\qquad\label{eq:avTorsionCombinedFerm}\end{eqnarray}
In case that one has fixed the compensator superfield $\compensator$
already to zero, the Lorentz part of the above torsion differs according
to (\ref{eq:avTorsionLorentzI})-(\ref{eq:avTorsionLorentzVI}) only
in the following components: \begin{eqnarray}
\av{T}_{\bs{\mc{A}}b}^{(L)}\hoch{c} & \stackrel{\compensator=0}{=} & \av{T}_{\bs{\mc{A}}b}\hoch{c}+\tfrac{1}{4}\nabla_{\bs{\mc{A}}}\dil\delta_{b}\hoch{c}=-\tfrac{1}{4}\gamma_{b}\hoch{c}\tief{\bs{\mc{A}}}\hoch{\bs{\mc{D}}}\nabla_{\bs{\mc{D}}}\dil\\
\av{T}_{\bs{\mc{AB}}}^{(L)}\hoch{\bs{\mc{C}}} & \stackrel{\compensator=0}{=} & \av{T}_{\bs{\mc{AB}}}\hoch{\bs{\mc{C}}}+\tfrac{1}{4}\nabla_{[\bs{\mc{A}}|}\dil\delta_{|\bs{\mc{B}}]}\hoch{\bs{\mc{C}}}=\nonumber \\
 & = & \tfrac{1}{8}\epsilon_{(\bs{\mc{A}})}\epsilon_{(\bs{\mc{B}})}\gamma_{de}\tief{[\bs{\mc{A}}}\hoch{\bs{\mc{D}}}\gamma^{de}\tief{\bs{\mc{B}}]}\hoch{\bs{\mc{C}}}\nabla_{\bs{\mc{D}}}\dil+\tfrac{\epsilon_{(\bs{\mc{A}})}\epsilon_{(\bs{\mc{B}})}+1}{4}\nabla_{[\bs{\mc{A}}|}\dil\delta_{|\bs{\mc{B}}]}\hoch{\bs{\mc{C}}}\qquad\end{eqnarray}
For the components $\av{T}_{\bs{\mc{AB}}}\hoch{\bs{\mc{C}}}$ at $\xbothtetas=0$
(appearing in (\ref{eq:bib:generalGravitinoSusyTrafo})) we need to
remember the dilatino-definition $\bei{\nabla_{\bs{\mc{A}}}\dil}{}=\dilo_{\bs{\mc{A}}}$
(\ref{eq:DilatinoDefII}) and for $\av{T}_{\bs{\mc{A}}b}\hoch{\bs{\mc{C}}}$
at $\xbothtetas=0$ we need $\bei{H_{abc}}{}=h_{abc}+2e^{2\compcomp}e_{[a}\hoch{m}e_{b}\hoch{n}\gamma_{c]\,\bs{\mc{AB}}}\epsilon_{(\bs{\mc{A}})}\psi_{m}\hoch{\bs{\mc{A}}}\psi_{n}\hoch{\bs{\mc{B}}}$
(\ref{eq:HabcLeading}), $\bei{\RR^{\bs{\mc{A}}\bs{\mc{B}}}}{}=e^{8\dilcomp}\rr^{\bs{\mc{A}}\bs{\mc{B}}}$
(\ref{eq:DefRR}), $\tilde{\gamma}^{de}\tief{\bs{\mc{A}}}\hoch{\bs{\mc{C}}}=e^{-2\compensator}\gamma^{de}\tief{\bs{\mc{A}}}\hoch{\bs{\mc{C}}}$,
$\tilde{\gamma}_{b\,\bs{\mc{AD}}}=e^{2\compensator}\tilde{\gamma}_{b\,\bs{\mc{AD}}}$
and $\bei{\compensator}{}=\compcomp$. Now we can plug the constraints
into (\ref{eq:bib:generalGravitinoSusyTrafo})  to arrive at:\begin{eqnarray}
\delta_{\eps}\psi_{m}\hoch{\bs{\mc{A}}} & = & \underbrace{\partial_{m}\eps^{\bs{\mc{A}}}+\av{\omega}_{m\bs{\mc{C}}}\hoch{\bs{\mc{A}}}\eps^{\bs{\mc{C}}}}_{\av{\nabla}_{m}\eps^{\bs{\mc{A}}}}+\tfrac{1}{4}\eps^{\bs{\mc{C}}}\epsilon_{(\bs{\mc{C}})}\gamma_{de}\tief{[\bs{\mc{C}}|}\hoch{\bs{\mc{D}}}\dilo_{\bs{\mc{D}}}\psi_{m}\hoch{\bs{\mc{B}}}\epsilon_{(\bs{\mc{B}})}\gamma^{de}\tief{|\bs{\mc{B}}]}\hoch{\bs{\mc{A}}}+\tfrac{1}{2}\eps^{\bs{\mc{C}}}\psi_{m}\hoch{\bs{\mc{B}}}\epsilon_{(\bs{\mc{C}})}\epsilon_{(\bs{\mc{B}})}\dilo_{[\bs{\mc{C}}}\delta_{\bs{\mc{B}}]}\hoch{\bs{\mc{A}}}+\nonumber \\
 &  & +\eps^{\bs{\mc{C}}}e_{m}\hoch{b}\left(\epsilon_{(\bs{\mc{C}})}\frac{3}{8}\left(h_{bde}e^{-2\compcomp}+2e_{[b}\hoch{k}e_{d}\hoch{l}\gamma_{e]\,\bs{\mc{DB}}}\epsilon_{(\bs{\mc{D}})}\psi_{k}\hoch{\bs{\mc{D}}}\psi_{l}\hoch{\bs{\mc{B}}}\right)\gamma^{de}\tief{\bs{\mc{C}}}\hoch{\bs{\mc{A}}}+2e^{2\compcomp+8\dilcomp}\gamma_{b\,\bs{\mc{CD}}}\rr^{\bs{\mc{A}}\bs{\mc{D}}}\right)\qquad\label{eq:interGraviSugra}\end{eqnarray}
If we instead have $\compensator=0$ and restrict to the Lorentz-part
of the torsion, the last term in the first line has to be replaced
by $\tfrac{1}{2}\eps^{\bs{\mc{C}}}\psi_{m}\hoch{\bs{\mc{B}}}(\epsilon_{(\bs{\mc{C}})}\epsilon_{(\bs{\mc{B}})}+1)\dilo_{[\bs{\mc{C}}}\delta_{\bs{\mc{B}}]}\hoch{\bs{\mc{A}}}$
and the bosonic connection $\av{\omega}_{m\bs{\mc{C}}}\hoch{\bs{\mc{A}}}$
by its Lorentz part $\av{\omega}_{m\bs{\mc{C}}}^{(L)}\hoch{\bs{\mc{A}}}$.
In order to determine the connection from (\ref{eq:susy:connectionleading:both})
we make use of further torsion constraints from (\ref{eq:avTorsionCombined-c})
and (\ref{eq:avTorsionCombinedFerm}) and the constraint $\avcovPhi{a}=0$.
The result is\begin{eqnarray}
\lqn{\omega_{m\bs{\mc{B}}}\hoch{\bs{\mc{A}}}=}\nonumber \\
 & = & \omega_{m\bs{\mc{B}}}^{(LC)}\hoch{\bs{\mc{A}}}+\tfrac{1}{4}e_{m}\hoch{a}\biggl\{2e_{[b|}\hoch{k}\partial_{k}\compcomp\eta_{|c]a}+\left(2e_{a}\hoch{k}e_{[b}\hoch{n}\eta_{c]d}-e_{b}\hoch{k}e_{c}\hoch{n}\eta_{ad}\right)\psi_{k}\hoch{\bs{\mc{C}}}\psi_{n}\hoch{\bs{\mc{D}}}\gamma_{\bs{\mc{C}}\bs{\mc{D}}}^{d}-\tfrac{1}{2}e_{a}\hoch{n}\psi_{n}\hoch{\bs{\mc{C}}}\gamma_{bc}\tief{\bs{\mc{C}}}\hoch{\bs{\mc{D}}}\dilo_{\bs{\mc{D}}}\biggr\}\gamma^{bc}\tief{\bs{\mc{B}}}\hoch{\bs{\mc{A}}}\nonumber \\
 &  & -\tfrac{1}{4}\left(\psi_{m}\hoch{\bs{\mc{C}}}\dilo_{\bs{\mc{C}}}-2\partial_{m}\compcomp\right)\delta_{\bs{\mc{B}}}\hoch{\bs{\mc{A}}}\end{eqnarray}
where the second line is the Lorentz part $\omega_{m\bs{\mc{B}}}^{(L)}\hoch{\bs{\mc{A}}}$
of the connection. Some terms in the gravitino transformation can
be further combined if we plug back this explicit expression for the
connection into (\ref{eq:interGraviSugra}):\vRam{1.02}{\begin{eqnarray}
\delta_{\eps}\psi_{m}\hoch{\bs{\mc{A}}} & = & \nabla_{m}^{(LC)}\eps^{\bs{\mc{A}}}+2e^{2\compcomp+8\dilcomp}\eps^{\bs{\mc{B}}}e_{m}\hoch{b}\gamma_{b\,\bs{\mc{BD}}}\rr^{\bs{\mc{A}}\bs{\mc{D}}}+\nonumber \\
 &  & +\tfrac{1}{4}e_{m}\hoch{a}\Bigl\{2e_{[b|}\hoch{k}\partial_{k}\compcomp\eta_{|c]a}+\psi_{k}\hoch{\bs{\mc{C}}}\gamma_{\bs{\mc{C}}\bs{\mc{D}}}^{d}\left(2e_{a}\hoch{k}e_{[b}\hoch{n}\eta_{c]d}-e_{b}\hoch{k}e_{c}\hoch{n}\eta_{ad}+3\epsilon_{(\bs{\mc{A}})}\epsilon_{(\bs{\mc{D}})}e_{[a}\hoch{k}e_{b}\hoch{n}\eta_{c]d}\right)\psi_{n}\hoch{\bs{\mc{D}}}+\nonumber \\
 &  & -e_{a}\hoch{n}\psi_{n}\hoch{\bs{\mc{C}}}\tfrac{1}{2}\left(1+\epsilon_{(\bs{\mc{A}})}\epsilon_{(\bs{\mc{C}})}\right)\gamma_{bc}\tief{\bs{\mc{C}}}\hoch{\bs{\mc{D}}}\dilo_{\bs{\mc{D}}}+\tfrac{3}{2}\epsilon_{(\bs{\mc{A}})}h_{abc}e^{-2\compcomp}\Bigr\}\gamma^{bc}\tief{\bs{\mc{B}}}\hoch{\bs{\mc{A}}}\eps^{\bs{\mc{B}}}+\nonumber \\
 &  & -\tfrac{1}{2}\Bigl\{\tfrac{1}{2}\left(1+\epsilon_{(\bs{\mc{A}})}\epsilon_{(\bs{\mc{C}})}\right)\psi_{m}\hoch{\bs{\mc{C}}}\dilo_{\bs{\mc{C}}}-\partial_{m}\compcomp\Bigr\}\eps^{\bs{\mc{A}}}\nonumber \\
 &  & +\underbrace{\tfrac{1}{8}\epsilon_{(\bs{\mc{A}})}(\eps^{\bs{\mc{B}}}\epsilon_{(\bs{\mc{B}})}\gamma_{de}\tief{\bs{\mc{B}}}\hoch{\bs{\mc{D}}}\dilo_{\bs{\mc{D}}})\psi_{m}\hoch{\bs{\mc{C}}}\gamma^{de}\tief{\bs{\mc{C}}}\hoch{\bs{\mc{A}}}}_{{\rm Lorentz\, trafo}}+\underbrace{\tfrac{1}{4}\epsilon_{(\bs{\mc{A}})}(\eps^{\bs{\mc{B}}}\epsilon_{(\bs{\mc{B}})}\dilo_{\bs{\mc{B}}})\psi_{m}\hoch{\bs{\mc{A}}}}_{{\rm dilatation}}\label{eq:finalGravitinoTrafoAv}\end{eqnarray}
} Note that we still have local structure group invariance, so that
we can change the last terms by simply redefining the supersymmetry
transformation with a Lorentz transformation and a dilatation\index{scale transformation!contribution to SUSY}\index{dilatation!contribution to SUSY}.
However, we cannot remove the terms for left- and rightmovers at the
same time, because the corresponding gauge parameter differs in sign
due to the factor $\epsilon({\scriptstyle \bs{\mc{A}}})$ which is
$+1$ for ${\scriptstyle \bs{\alpha}}$ and $-1$ for ${\scriptstyle \hat{\bs{\alpha}}}$.
Note also that if the compensator superfield $\compensator$ was fixed
to zero already in the beginning, the dilatation part changes to $\tfrac{1+\epsilon_{(\bs{\mc{A}})}\epsilon_{(\bs{\mc{B}})}}{4}\eps^{\bs{\mc{B}}}\dilo_{\bs{\mc{B}}}\psi_{m}\hoch{\bs{\mc{A}}}$
 and thus corresponds to a redefinition of the supersymmetry transformation
by a dilatation with parameter $\tfrac{1}{2}\eps^{\bs{\mc{B}}}\dilo_{\bs{\mc{B}}}$.
This is the same minimal modification which is necessary when we only
fix the leading component $\phi$ to zero in the end and need to stabilize
it with a compensating dilatation according to according to (\ref{eq:additionalScaleStabilizer}).
The above transformation can be seen as the final result, but it is
at this point instructive to introduce eventually the split of the
collective fermionic index into left and right-mover:\index{SUSY!gravitino}\index{gravitino!local SUSY}\index{local SUSY!gravitino}\begin{eqnarray}
\delta_{\eps}\psi_{m}\hoch{\bs{\alpha}} & = & \nabla_{m}^{(LC)}\eps^{\bs{\alpha}}+2e^{2\compcomp+8\dilcomp}\eps^{\hat{\bs{\beta}}}e_{m}\hoch{b}\gamma_{b\,\hat{\bs{\beta}}\hat{\bs{\delta}}}\rr^{\bs{\alpha}\hat{\bs{\delta}}}+\nonumber \\
 &  & +\tfrac{1}{4}e_{m}\hoch{a}\Bigl\{2e_{[b|}\hoch{k}\partial_{k}\compcomp\eta_{|c]a}+4e_{a}\hoch{k}e_{[b}\hoch{n}\eta_{c]d}\left(\psi_{n}\hoch{\bs{\mc{\delta}}}\psi_{k}\hoch{\bs{\gamma}}\gamma_{\bs{\gamma\delta}}^{d}\right)-2e_{b}\hoch{k}e_{c}\hoch{n}\eta_{ad}\left(\hat{\psi}_{n}\hoch{\hat{\bs{\delta}}}\hat{\psi}_{k}\hoch{\hat{\bs{\gamma}}}\gamma_{\hat{\bs{\gamma}}\hat{\bs{\delta}}}^{d}\right)+\nonumber \\
 &  & -e_{a}\hoch{n}\psi_{n}\hoch{\bs{\gamma}}\gamma_{bc}\tief{\bs{\gamma}}\hoch{\bs{\delta}}\dilo_{\bs{\delta}}+\tfrac{3}{2}h_{abc}e^{-2\compcomp}\Bigr\}\gamma^{bc}\tief{\bs{\beta}}\hoch{\bs{\alpha}}\eps^{\bs{\beta}}-\tfrac{1}{2}\left(\psi_{m}\hoch{\bs{\gamma}}\dilo_{\bs{\gamma}}-\partial_{m}\compcomp\right)\eps^{\bs{\alpha}}+\nonumber \\
 &  & \underbrace{+\tfrac{1}{8}(\eps^{\bs{\beta}}\gamma_{de}\tief{\bs{\beta}}\hoch{\bs{\delta}}\dilo_{\bs{\delta}}-\hat{\eps}^{\hat{\bs{\beta}}}\gamma_{de}\tief{\hat{\bs{\beta}}}\hoch{\hat{\bs{\delta}}}\hat{\dilo}_{\hat{\bs{\delta}}})\psi_{m}\hoch{\bs{\gamma}}\gamma^{de}\tief{\bs{\gamma}}\hoch{\bs{\alpha}}}_{{\rm Lorentz\, trafo}}+\underbrace{\tfrac{1}{4}(\eps^{\bs{\beta}}\dilo_{\bs{\beta}}-\hat{\eps}^{\hat{\bs{\beta}}}\hat{\dilo}_{\hat{\bs{\beta}}})\psi_{m}\hoch{\bs{\alpha}}}_{{\rm dilatation}}\\
\delta_{\eps}\hat{\psi}_{m}\hoch{\hat{\bs{\alpha}}} & = & \nabla_{m}^{(LC)}\hat{\eps}^{\hat{\bs{\alpha}}}+2e^{2\compcomp+8\dilcomp}\eps^{\bs{\beta}}e_{m}\hoch{b}\gamma_{b\,\bs{\beta\delta}}\rr^{\bs{\delta}\hat{\bs{\alpha}}}+\nonumber \\
 &  & +\tfrac{1}{4}e_{m}\hoch{a}\Bigl\{2e_{[b|}\hoch{k}\partial_{k}\compcomp\eta_{|c]a}+4e_{a}\hoch{k}e_{[b}\hoch{n}\eta_{c]d}\left(\hat{\psi}_{n}\hoch{\hat{\bs{\delta}}}\hat{\psi}_{k}\hoch{\hat{\bs{\gamma}}}\gamma_{\hat{\bs{\gamma}}\hat{\bs{\delta}}}^{d}\right)-2e_{b}\hoch{k}e_{c}\hoch{n}\eta_{ad}\left(\psi_{n}\hoch{\bs{\mc{\delta}}}\psi_{k}\hoch{\bs{\gamma}}\gamma_{\bs{\gamma\delta}}^{d}\right)+\nonumber \\
 &  & -e_{a}\hoch{n}\hat{\psi}_{n}\hoch{\hat{\bs{\gamma}}}\gamma_{bc}\tief{\hat{\bs{\gamma}}}\hoch{\hat{\bs{\delta}}}\hat{\dilo}_{\hat{\bs{\delta}}}-\tfrac{3}{2}h_{abc}e^{-2\compcomp}\Bigr\}\gamma^{bc}\tief{\hat{\bs{\beta}}}\hoch{\hat{\bs{\alpha}}}\hat{\eps}^{\hat{\bs{\beta}}}-\tfrac{1}{2}\Bigl\{\hat{\psi}_{m}\hoch{\hat{\bs{\gamma}}}\hat{\dilo}_{\hat{\bs{\gamma}}}-\partial_{m}\compcomp\Bigr\}\hat{\eps}^{\hat{\bs{\alpha}}}+\nonumber \\
 &  & +\underbrace{\tfrac{1}{8}(\hat{\eps}^{\hat{\bs{\beta}}}\gamma_{de}\tief{\hat{\bs{\beta}}}\hoch{\hat{\bs{\delta}}}\hat{\dilo}_{\hat{\bs{\delta}}}-\eps^{\bs{\beta}}\gamma_{de}\tief{\bs{\beta}}\hoch{\bs{\delta}}\dilo_{\bs{\delta}})\hat{\psi}_{m}\hoch{\hat{\bs{\gamma}}}\gamma^{de}\tief{\hat{\bs{\gamma}}}\hoch{\hat{\bs{\alpha}}}}_{{\rm Lorentz\, trafo}}+\underbrace{\tfrac{1}{4}(\hat{\eps}^{\hat{\bs{\beta}}}\hat{\dilo}_{\hat{\bs{\beta}}}-\eps^{\bs{\beta}}\dilo_{\bs{\beta}})\hat{\psi}_{m}\hoch{\hat{\bs{\alpha}}}}_{{\rm dilatation}}\end{eqnarray}
Comparing these results with the ones obtained in {}``gauge I'',
i.e. with (\ref{eq:finalGravitinoTrafoLefthalpha}) for $\delta_{\eps}\hat{\psi}_{m}\hoch{\hat{\bs{\alpha}}}$
and with (\ref{eq:finalGravitinoTrafoLeftAlpha}) together with the
left-mover connection (\ref{eq:bosonic-leftmoverconn-final}) for
$\delta_{\eps}\psi_{m}\hoch{\bs{\alpha}}$, we recognize that they
again differ just in the last lines and are related by a local Lorentz
and scale transformation.

One can rewrite the result a bit using (\ref{eq:FromLittleFierz})
whose graded version reads\begin{eqnarray}
\gamma_{ab}\tief{\bs{\beta}}\hoch{\bs{\delta}}\gamma^{ab}\tief{\bs{\gamma}}\hoch{\bs{\alpha}} & = & \gamma_{ab}\tief{\bs{\gamma}}\hoch{\bs{\delta}}\gamma^{ab}\tief{\bs{\beta}}\hoch{\bs{\alpha}}+8\gamma_{\bs{\beta}\bs{\gamma}}^{a}\gamma_{a}^{\bs{\alpha}\bs{\delta}}+20\delta_{[\bs{\beta}}\hoch{\bs{\alpha}}\delta_{\bs{\gamma}]}\hoch{\bs{\delta}}\end{eqnarray}
This leads to \begin{eqnarray}
\tfrac{1}{8}(\eps^{\bs{\beta}}\gamma_{ab}\tief{\bs{\beta}}\hoch{\bs{\delta}}\dilo_{\bs{\delta}})\psi_{m}\hoch{\bs{\gamma}}\gamma^{ab}\tief{\bs{\gamma}}\hoch{\bs{\alpha}} & = & \tfrac{1}{8}(\psi_{m}\hoch{\bs{\gamma}}\gamma_{ab}\tief{\bs{\gamma}}\hoch{\bs{\delta}}\dilo_{\bs{\delta}})\gamma^{ab}\tief{\bs{\beta}}\hoch{\bs{\alpha}}\eps^{\bs{\beta}}+(\eps^{\bs{\beta}}\gamma_{\bs{\beta}\bs{\gamma}}^{a}\psi_{m}\hoch{\bs{\gamma}})\gamma_{a}^{\bs{\alpha}\bs{\delta}}\dilo_{\bs{\delta}}+\nonumber \\
 &  & +\tfrac{5}{4}(\psi_{m}\hoch{\bs{\gamma}}\dilo_{\bs{\gamma}})\eps^{\bs{\alpha}}-\tfrac{5}{4}(\eps^{\bs{\beta}}\dilo_{\bs{\beta}})\psi_{m}\hoch{\bs{\alpha}}\end{eqnarray}
but is of no real advantage. However, the above gravitino transformation
simplifies significantly, if we consider it at $\psi_{m}\hoch{\bs{\mc{A}}}=\lambda_{\bs{\mc{A}}}=0$
which is of special interest when we want to consider a string vacuum
with vanishing vacuum expectation value of the fermionic fields. In
addition we finally fix the bosonic compensator field $\compcomp$
to zero and arrive at\vRam{.8}{\begin{eqnarray}
\bei{\delta_{\eps}\psi_{m}\hoch{\bs{\mc{A}}}}{\psi=\lambda=0} & = & \nabla_{m}^{(LC)}\eps^{\bs{\mc{A}}}+\tfrac{3}{8}\epsilon_{(\bs{\mc{A}})}e_{m}\hoch{a}h_{abc}\eps^{\bs{\mc{B}}}\gamma^{bc}\tief{\bs{\mc{B}}}\hoch{\bs{\mc{A}}}+2e^{8\dilcomp}\eps^{\bs{\mc{B}}}e_{m}\hoch{b}\gamma_{b\,\bs{\mc{BD}}}\rr^{\bs{\mc{A}}\bs{\mc{D}}}\label{eq:finalGravitinoTrafoSimple}\end{eqnarray}
}  For convenience of the reader we present the result again with
the split of the fermionic index:\begin{eqnarray}
\bei{\delta_{\eps}\psi_{m}\hoch{\bs{\alpha}}}{\psi=\lambda=0} & = & \nabla_{m}^{(LC)}\eps^{\bs{\alpha}}+\tfrac{3}{8}e_{m}\hoch{a}h_{abc}\eps^{\bs{\beta}}\gamma^{bc}\tief{\bs{\beta}}\hoch{\bs{\alpha}}+2e^{8\dilcomp}\hat{\eps}^{\hat{\bs{\beta}}}e_{m}\hoch{b}\gamma_{b\,\hat{\bs{\beta}}\hat{\bs{\delta}}}\rr^{\bs{\alpha}\hat{\bs{\delta}}}\\
\bei{\delta_{\eps}\hat{\psi}_{m}\hoch{\hat{\bs{\alpha}}}}{\psi=\lambda=0} & = & \nabla_{m}^{(LC)}\hat{\eps}^{\hat{\bs{\alpha}}}-\tfrac{3}{8}e_{m}\hoch{a}h_{abc}\eps^{\hat{\bs{\beta}}}\gamma^{bc}\tief{\hat{\bs{\beta}}}\hoch{\hat{\bs{\alpha}}}+2e^{8\dilcomp}\eps^{\bs{\beta}}e_{m}\hoch{b}\gamma_{b\,\bs{\beta\delta}}\rr^{\bs{\delta}\hat{\bs{\alpha}}}\end{eqnarray}
This differs from the form that one can find in the literature (e.g.
\cite{Bergshoeff:2001pv}) by a redefinition $8\dilcomp\To\dilcomp$,
$\rr^{\bs{\alpha}\hat{\bs{\delta}}}\To\tfrac{1}{32}\rr^{\bs{\alpha}\hat{\bs{\delta}}}$
and by a redefinition $3H_{mnk}\To H_{mnk}$ where the latter discrepancy
was simply due to our different definition of the wedge product.

\subsection{The dilatino transformation}

According to (\ref{eq:DilatinoDefII}), the dilatinos are related
to the dilaton superfield via \begin{eqnarray}
\dilo_{\bs{\mc{A}}} & = & \bei{\nabla_{\bs{\mc{A}}}\dil}{}=\av{\nabla}_{\bs{\mc{A}}}\dil\end{eqnarray}
Note that for the dilaton $\dil$ (in contrast to the compensator
field $\compensator$) it does not make a difference with which connection
we act, because it is a scalar field. As described in the appendix,
the covariant derivative of the scalar field transforms like a vector
under supergauge transformations which leads to the following simple
local supersymmetry transformation of the dilatino (see in the appendix
on page \pageref{eq:generalDilatinoSusyTrafo:old}): \begin{equation}
\delta_{\eps}\dilo_{\bs{\mc{A}}}=\eps^{\bs{\mc{C}}}\bei{\nabla_{\bs{\mc{C}}}\nabla_{\bs{\mc{A}}}\dil}{}\label{eq:bib:diloTrafoboth}\end{equation}
For the second action of the covariant derivative the connection of
course plays a role and $\nabla_{\bs{\mc{C}}}$ has to be replaced
by $\av{\nabla}_{\bs{\mc{C}}}$ in gauge II. The transformation can
be rewritten in terms of the $\xbothtetas^{2}$ component of the dilaton
superfeld according to \ref{eq:generalDilatinoSusyTrafo:long} on
page \pageref{eq:generalDilatinoSusyTrafo:old} as\begin{eqnarray}
\delta_{\eps}\dilo_{\bs{\mc{A}}} & = & -\eps^{\bs{\mc{C}}}\bei{T_{\bs{\mc{C}}\bs{\mc{A}}}\hoch{b}}{}e_{b}\hoch{k}\partial_{k}\dilcomp+\eps^{\bs{\mc{C}}}\bei{T_{\bs{\mc{C}}\bs{\mc{A}}}\hoch{b}}{}\psi_{b}\hoch{\bs{\mc{K}}}\dilo_{\bs{\mc{K}}}-\eps^{\bs{\mc{C}}}\bei{T_{\bs{\mc{C}}\bs{\mc{A}}}\hoch{\bs{\mc{B}}}}{}\dilo_{\bs{\mc{B}}}+\nonumber \\
 &  & +\eps^{\bs{\mc{C}}}\delta_{\bs{\mc{C}}}\hoch{\bs{\mc{M}}}\delta_{\bs{\mc{A}}}\hoch{\bs{\mc{K}}}\bei{\partial_{\bs{\mc{M}}}\partial_{\bs{\mc{K}}}\dil}{}\label{eq:bib:diloTrafoboth:long}\end{eqnarray}
In any case we need more information about constraints on the dilaton
superfield, in order to write down the explicit transformation. In
footnote \ref{fn:dilaton-constraint} on page \pageref{fn:dilaton-constraint}
we have derived a constraint on $\nabla_{\hat{\bs{\beta}}}\covPhi{\bs{\alpha}}=\nabla_{\hat{\bs{\beta}}}\nabla_{\bs{\alpha}}\dil$,
and in a similar way it should be possible to extract more information
on $\nabla_{\bs{\beta}}\nabla_{\bs{\alpha}}\dil$. Without such constraints
it is therefore not yet very useful to write down the transformation
in both gauges. An interesting difference of the two gauges, however,
is the location of the dilatinos in the compensator superfield, which
we will quickly discuss:

\paragraph{Gauge I}

In gauge I we have in particular $\bei{\Omega_{\bs{\mc{A}}}^{(D)}}{}=0$.
The constraint $\covPhi{\hat{\bs{\alpha}}}=0$ and the relation $\covPhi{\bs{\alpha}}=\nabla_{\bs{\alpha}}\dil$
thus imply \begin{eqnarray}
\bei{\partial_{\hat{\bs{\mu}}}\compensator}{} & = & 0\\
\bei{\partial_{\bs{\mu}}\compensator}{} & = & \bei{\partial_{\bs{\mu}}\dil}{}=\dilo_{\bs{\mu}}\end{eqnarray}
The relation $\hatcovPhi{\hat{\bs{\alpha}}}=\nabla_{\bs{\alpha}}\dil$
(together with the above $\bei{\partial_{\hat{\bs{\mu}}}\compensator}{}=0$)
and the constraint $\hatcovPhi{\bs{\alpha}}=0$ (together with the
above $\bei{\partial_{\bs{\mu}}\compensator}{}=\bei{\partial_{\bs{\mu}}\dil}{}$)
on the other hand imply \begin{eqnarray}
\bei{\hat{\Omega}_{\hat{\bs{\mu}}}^{(D)}}{} & = & -\bei{\partial_{\hat{\bs{\mu}}}\dil}{}=-\hat{\dilo}_{\hat{\bs{\mu}}}\\
\bei{\hat{\Omega}_{\bs{\mu}}^{(D)}}{} & = & \bei{\partial_{\bs{\mu}}\dil}{}=\dilo_{\bs{\mu}}\end{eqnarray}
Only one of the dilatinos is thus part of the compensator field, while
both are contained in $\hat{\Omega}_{\bs{\mc{M}}}^{(D)}$ which should
in this gauge not be seen as scale part of the connection but as scale
part of the difference tensor\begin{equation}
\bei{\Delta_{\bs{\mc{M}}}^{(D)}}{}=\bei{\hat{\Omega}_{\bs{\mc{M}}}^{(D)}}{}-\bei{\Omega_{\bs{\mc{M}}}^{(D)}}{}=\bei{\hat{\Omega}_{\bs{\mc{M}}}^{(D)}}{}\end{equation}
Let me add one more step in this new version of the document. With
the information that we already had in the first arXiv version (namely
the constraint $\nabla_{\hat{\bs{\beta}}}\covPhi{\bs{\alpha}}=-\tilde{\gamma}_{d\,\bs{\alpha\rho}}\RR^{\bs{\rho}\hat{\bs{\eps}}}\gamma_{\hat{\bs{\eps}}\hat{\bs{\beta}}}^{d}$
of footnote \ref{fn:dilaton-constraint}), we can actually write down
explicitely at least half of the supersymmetry transformation of the
dilatino. Simply start with (\ref{eq:bib:diloTrafoboth}) and plug
in everything we know\begin{eqnarray*}
\delta_{\eps}\dilo_{\bs{\mc{\alpha}}} & = & \eps^{\bs{\gamma}}\bei{\nabla_{\bs{\gamma}}\nabla_{\bs{\mc{\alpha}}}\dil}{}+\hat{\eps}^{\hat{\bs{\gamma}}}\bei{\nabla_{\hat{\bs{\gamma}}}\nabla_{\bs{\mc{\alpha}}}\dil}{}=\\
 & = & \eps^{\bs{\gamma}}\bei{\nabla_{\bs{\gamma}}\nabla_{\bs{\mc{\alpha}}}\dil}{}-\hat{\eps}^{\hat{\bs{\gamma}}}e^{2\compcomp}\gamma_{d\,\bs{\alpha\rho}}e^{8\dilcomp}\rr^{\bs{\rho}\hat{\bs{\eps}}}\gamma_{\hat{\bs{\eps}}\hat{\bs{\gamma}}}^{d}\\
\delta_{\eps}\hat{\dilo}_{\hat{\bs{\alpha}}} & = & \eps^{\bs{\gamma}}\bei{\nabla_{\bs{\gamma}}\nabla_{\hat{\bs{\alpha}}}\dil}{}+\hat{\eps}^{\hat{\bs{\gamma}}}\bei{\nabla_{\hat{\bs{\gamma}}}\nabla_{\hat{\bs{\alpha}}}\dil}{}=\\
 & = & \eps^{\bs{\gamma}}\bei{\nabla_{\hat{\bs{\alpha}}}\nabla_{\bs{\gamma}}\dil}{}-2\eps^{\bs{\gamma}}\bei{T_{\bs{\gamma}\hat{\bs{\alpha}}}\hoch{C}\nabla_{C}\dil}{}+\hat{\eps}^{\hat{\bs{\gamma}}}\bei{\nabla_{\hat{\bs{\gamma}}}\nabla_{\hat{\bs{\alpha}}}\dil}{}=\\
 & = & -\eps^{\bs{\gamma}}e^{2\compcomp}\gamma_{d\,\bs{\gamma\rho}}e^{8\dilcomp}\rr^{\bs{\rho}\hat{\bs{\eps}}}\gamma_{\hat{\bs{\eps}}\hat{\bs{\alpha}}}^{d}+\eps^{\bs{\gamma}}\left(\tfrac{1}{4}\gamma_{de}\tief{\bs{\gamma}}\hoch{\bs{\delta}}\gamma^{de}\tief{\hat{\bs{\alpha}}}\hoch{\hat{\bs{\gamma}}}\dilo_{\bs{\delta}}+\tfrac{1}{2}\dilo_{\bs{\gamma}}\delta_{\hat{\bs{\alpha}}}\hoch{\hat{\bs{\gamma}}}\right)\hat{\dilo}_{\hat{\bs{\delta}}}+\hat{\eps}^{\hat{\bs{\gamma}}}\bei{\nabla_{\hat{\bs{\gamma}}}\nabla_{\hat{\bs{\alpha}}}\dil}{}\end{eqnarray*}
The second term of the last line would vanish for $\lambda=\psi=0$.
As mentioned before, we need some additional constraints on $\nabla_{\bs{\gamma}}\nabla_{\bs{\mc{\alpha}}}\dil$
and $\nabla_{\hat{\bs{\gamma}}}\nabla_{\hat{\bs{\alpha}}}\dil$ to
determine the missing second half of the transformations respectively.

\paragraph{Gauge II}

In gauge II the situation is fortunately more symmetric and we have
$\bei{\av{\Omega}_{\bs{\mc{A}}}^{(D)}}{}=0$ and $\avcovPhi{\mc{\bs{A}}}=\tfrac{1}{2}\left(\covPhi{\bs{\mc{A}}}+\hatcovPhi{\bs{\mc{A}}}\right)=\tfrac{1}{2}\nabla_{\bs{\mc{A}}}\dil$.
This (together with $\hatcovPhi{\bs{\alpha}}=\covPhi{\hat{\bs{\alpha}}}=0$)
implies \begin{eqnarray}
\bei{\partial_{\bs{\mc{M}}}\compensator}{} & = & \tfrac{1}{2}\dilo_{\bs{\mc{M}}}\\
\bei{\Omega_{\hat{\bs{\mu}}}^{(D)}}{} & = & \tfrac{1}{2}\hat{\dilo}_{\hat{\bs{\mu}}}=-\bei{\hat{\Omega}_{\hat{\bs{\mu}}}^{(D)}}{}\quad\dann\bei{\Delta_{\hat{\bs{\mu}}}^{(D)}}{}=-\hat{\dilo}_{\hat{\bs{\mu}}}\\
\bei{\hat{\Omega}_{\bs{\mu}}^{(D)}}{} & = & \tfrac{1}{2}\dilo_{\bs{\mu}}=-\bei{\Omega_{\bs{\mu}}^{(D)}}{}\quad\dann\bei{\Delta_{\bs{\mu}}^{(D)}}{}=\dilo_{\bs{\mu}}\end{eqnarray}
According to the first line both dilatinos are contained in the compensator
superfield in this gauge. Their local supersymmetry transformation
could thus also be determined by the transformation of the compensator
superfield which is, however, of no advantage and gives the same result.

Again we add one more step with respect to the 1st arXiv version of
this document, in order to obtain at least half of the SUSY transformation
in an explicit form. In the gauge II, (\ref{eq:bib:diloTrafoboth})
becomes for ${\scriptstyle \bs{\mc{A}}}={\scriptstyle \bs{\alpha}}$
\begin{eqnarray*}
\delta_{\eps}\dilo_{\bs{\alpha}} & = & \eps^{\bs{\gamma}}\bei{\av{\nabla}_{\bs{\gamma}}\nabla_{\bs{\alpha}}\dil}{}+\hat{\eps}^{\hat{\bs{\gamma}}}\bei{\av{\nabla}_{\hat{\bs{\gamma}}}\nabla_{\bs{\alpha}}\dil}{}=\\
 & = & \eps^{\bs{\gamma}}\bei{\nabla_{\bs{\gamma}}\nabla_{\bs{\alpha}}\dil}{}-\tfrac{1}{2}\eps^{\bs{\gamma}}\bei{\Delta_{\bs{\gamma\alpha}}\hoch{\bs{\delta}}\nabla_{\bs{\delta}}\dil}{}+\hat{\eps}^{\hat{\bs{\gamma}}}\bei{\nabla_{\hat{\bs{\gamma}}}\nabla_{\bs{\alpha}}\dil}{}-\tfrac{1}{2}\hat{\eps}^{\hat{\bs{\gamma}}}\bei{\Delta_{\hat{\bs{\gamma}}\bs{\alpha}}\hoch{\bs{\delta}}\nabla_{\bs{\delta}}\dil}{}=\\
 & = & \eps^{\bs{\gamma}}\bei{\nabla_{\bs{\gamma}}\nabla_{\bs{\alpha}}\dil}{}-\left(\tfrac{1}{4}(\eps^{\bs{\gamma}}\dilo_{\bs{\gamma}})\dilo_{\bs{\alpha}}+\tfrac{1}{8}(\eps^{\bs{\gamma}}\gamma_{bc\,\bs{\gamma}}\hoch{\bs{\eps}}\dilo_{\bs{\eps}})\gamma^{bc}\tief{\bs{\alpha}}\hoch{\bs{\delta}}\dilo_{\bs{\delta}}\right)+\\
 &  & -\hat{\eps}^{\hat{\bs{\gamma}}}e^{2\compcomp}\gamma_{d\,\bs{\alpha\rho}}e^{8\dilcomp}\rr^{\bs{\rho}\hat{\bs{\eps}}}\gamma_{\hat{\bs{\eps}}\hat{\bs{\gamma}}}^{d}+\left(\tfrac{1}{4}(\hat{\eps}^{\hat{\bs{\gamma}}}\hat{\dilo}_{\hat{\bs{\gamma}}})\dilo_{\bs{\alpha}}+\tfrac{1}{8}(\hat{\eps}^{\hat{\bs{\gamma}}}\gamma_{bc\,\hat{\bs{\gamma}}}\hoch{\hat{\bs{\eps}}}\hat{\dilo}_{\hat{\bs{\eps}}})\gamma^{bc}\tief{\bs{\alpha}}\hoch{\bs{\delta}}\dilo_{\bs{\delta}}\right)\end{eqnarray*}
For $\lambda=\psi=0$ the terms in the brackets disappear and we end
up with the same expression as in gauge I. In gauge II the transformation
of $\hat{\dilo}_{\hat{\bs{\alpha}}}$ can be simply obtained by the
unbroken left-right symmetry.

The transformation of the remaining fields in a general form (constraints
not yet plugged into the equations) can be found in the appendix after
page \pageref{sub:The-supersymmetry-transformation}.

\appendix
\ifthenelse{\theinput=1}{\renewcommand{\thesection}{\arabic{chapter}.\Alph{localapp}}}{}\addtocounter{localapp}{1}

\section{Constraints before the BI's}

\label{sec:constraintsBeforeBI}

\paragraph{Reduced structure group constraints}

The following equations are taken from (\ref{eq:reducedConnectionForm})-(\ref{eq:reducedSform}),
(\ref{eq:endlichMetrikConstraint}) or (\ref{eq:metricDegenerate})
and (\ref{eq:metricWithCompensator})\begin{eqnarray}
\Omega_{M\bs{\alpha}}\hoch{\bs{\beta}} & = & \frac{1}{2}\Omega_{M}^{(D)}\delta_{\bs{\alpha}}\hoch{\bs{\beta}}+\frac{1}{4}\Omega_{Ma_{1}a_{2}}^{(L)}\gamma^{a_{1}a_{2}}\tief{\bs{\alpha}}\hoch{\bs{\beta}},\qquad\hat{\Omega}_{M\hat{\bs{\alpha}}}\hoch{\hat{\bs{\beta}}}=\frac{1}{2}\hat{\Omega}_{M}^{(D)}\delta_{\hat{\bs{\alpha}}}\hoch{\hat{\bs{\beta}}}+\frac{1}{4}\hat{\Omega}_{Ma_{1}a_{2}}^{(L)}\gamma^{a_{1}a_{2}}\tief{\hat{\bs{\alpha}}}\hoch{\hat{\bs{\beta}}}\\
C_{\bs{\alpha}}\hoch{\bs{\beta}\hat{\bs{\gamma}}} & = & \frac{1}{2}C^{\hat{\bs{\gamma}}}\delta_{\bs{\alpha}}\hoch{\bs{\beta}}+\frac{1}{4}C_{a_{1}a_{2}}^{\hat{\bs{\gamma}}}\gamma^{a_{1}a_{2}}\tief{\bs{\alpha}}\hoch{\bs{\beta}},\qquad\hat{C}_{\hat{\bs{\alpha}}}\hoch{\hat{\bs{\beta}}\bs{\gamma}}=\frac{1}{2}\hat{C}^{\bs{\gamma}}\delta_{\hat{\bs{\alpha}}}\hoch{\hat{\bs{\beta}}}+\frac{1}{4}\hat{C}_{a_{1}a_{2}}^{\bs{\gamma}}\gamma^{a_{1}a_{2}}\tief{\hat{\bs{\alpha}}}\hoch{\hat{\bs{\beta}}}\\
S_{\bs{\alpha}\hat{\bs{\alpha}}}\hoch{\bs{\beta}\hat{\bs{\beta}}} & = & \frac{1}{4}S\delta_{\bs{\alpha}}\hoch{\bs{\beta}}\delta_{\hat{\bs{\alpha}}}\hoch{\hat{\bs{\beta}}}+\frac{1}{8}S_{a_{1}a_{2}}\delta_{\bs{\alpha}}\hoch{\bs{\beta}}\gamma^{a_{1}a_{2}}\tief{\hat{\bs{\alpha}}}\hoch{\hat{\bs{\beta}}}+\nonumber \\
 &  & +\frac{1}{8}\hat{S}_{a_{1}a_{2}}\gamma^{a_{1}a_{2}}\tief{\bs{\alpha}}\hoch{\bs{\beta}}\delta_{\hat{\bs{\alpha}}}\hoch{\hat{\bs{\beta}}}+\frac{1}{16}S_{a_{1}a_{2}b_{1}b_{2}}\gamma^{a_{1}a_{2}}\tief{\bs{\alpha}}\hoch{\bs{\beta}}\gamma^{b_{1}b_{2}}\tief{\hat{\bs{\alpha}}}\hoch{\hat{\bs{\beta}}}\\
G_{MN} & = & E_{M}\hoch{a}G_{ab}E_{N}\hoch{b},\qquad G_{ab}=e^{2\Phi}\eta_{ab}\end{eqnarray}
The above equations (without the last one) are equivalent to \begin{eqnarray}
\gamma^{a_{1}\ldots a_{4}}\tief{\bs{\beta}}\hoch{\bs{\alpha}}\Omega_{M\bs{\alpha}}\hoch{\bs{\beta}} & = & \gamma^{a_{1}\ldots a_{4}}\tief{\hat{\bs{\beta}}}\hoch{\hat{\bs{\alpha}}}\hat{\Omega}_{M\hat{\bs{\alpha}}}\hoch{\hat{\bs{\beta}}}=0\label{eq:OmegaCond2}\\
\gamma^{a_{1}\ldots a_{4}}\tief{\bs{\beta}}\hoch{\bs{\alpha}}C_{\bs{\alpha}}\hoch{\bs{\beta}\hat{\bs{\gamma}}} & = & \gamma^{a_{1}\ldots a_{4}}\tief{\hat{\bs{\beta}}}\hoch{\hat{\bs{\alpha}}}\hat{C}_{\hat{\bs{\alpha}}}\hoch{\hat{\bs{\beta}}\bs{\gamma}}=0\label{eq:Ccond2}\\
\gamma^{a_{1}\ldots a_{4}}\tief{\bs{\beta}}\hoch{\bs{\alpha}}S_{\bs{\alpha}\hat{\bs{\alpha}}}\hoch{\bs{\beta}\hat{\bs{\beta}}} & = & \gamma^{a_{1}\ldots a_{4}}\tief{\hat{\bs{\beta}}}\hoch{\hat{\bs{\alpha}}}S_{\bs{\alpha}\hat{\bs{\alpha}}}\hoch{\bs{\beta}\hat{\bs{\beta}}}=0\label{eq:Scond2}\end{eqnarray}
As discussed in the appendix \vref{cha:ConnectionAppend}, the spinorial
left-mover connection $\Omega_{M\bs{\alpha}}\hoch{\bs{\beta}}$ induces
via invariance of the small gamma-matrices a whole superspace left-mover
connection $\Omega_{MA}\hoch{B}$. Likewise the spinorial rightmover-connection
$\hat{\Omega}_{M\hat{\bs{\alpha}}}\hoch{\hat{\bs{\beta}}}$ induces
a superspace right-mover connection $\hat{\Omega}_{MA}\hoch{B}$.
The constraints (\ref{eq:OmegaCond2}) then apply in the same way
for $\hat{\Omega}_{M\bs{\alpha}}\hoch{\bs{\beta}}$ and $\Omega_{M\hat{\bs{\alpha}}}\hoch{\hat{\bs{\beta}}}$:\begin{eqnarray}
\gamma^{a_{1}\ldots a_{4}}\tief{\bs{\beta}}\hoch{\bs{\alpha}}\check{\Omega}_{M\bs{\alpha}}\hoch{\bs{\beta}} & = & \gamma^{a_{1}\ldots a_{4}}\tief{\hat{\bs{\beta}}}\hoch{\hat{\bs{\alpha}}}\check{\Omega}_{M\hat{\bs{\alpha}}}\hoch{\hat{\bs{\beta}}}=0\qquad\mbox{for any }\check{\Omega}\mbox{ which is Lorentz plus scale}\end{eqnarray}
Let us denote the difference\index{difference tensor} one-form between
the left-mover and the rightmover connection by\index{$\Delta_{MA}\hoch{B}$}\begin{eqnarray}
\Delta_{MA}\hoch{B} & \equiv & \hat{\Omega}_{MA}\hoch{B}-\Omega_{MA}\hoch{B}=\left(\begin{array}{ccc}
\Delta_{Ma}\hoch{b} & 0 & 0\\
0 & \Delta_{M\bs{\alpha}}\hoch{\bs{\beta}} & 0\\
0 & 0 & \Delta_{M\hat{\bs{\alpha}}}\hoch{\hat{\bs{\beta}}}\end{array}\right)\label{eq:BI:DifferenceTensor}\end{eqnarray}
The above restrictions on the spinorial connections induces the same
restrictions on the difference tensor \begin{eqnarray}
\gamma^{a_{1}\ldots a_{4}}\tief{\bs{\beta}}\hoch{\bs{\alpha}}\Delta_{C\bs{\alpha}}\hoch{\bs{\beta}}=\gamma^{a_{1}\ldots a_{4}}\tief{\hat{\bs{\beta}}}\hoch{\hat{\bs{\alpha}}}\Delta_{C\hat{\bs{\alpha}}}\hoch{\hat{\bs{\beta}}} & = & 0\label{eq:DeltaWithReducedStrGr}\end{eqnarray}

\paragraph{Further constraints on $C$ and $S$ and indirectly on $\mc{P}$}

The constraints (\ref{eq:holConstrVa}) and (\ref{eq:holConstrVb})
on $C$ and (\ref{eq:holConstrVIIIa}) and (\ref{eq:holConstrVIIIb})
on $S$ (all on page \pageref{eq:holConstrVIIIa}) can be regarded
as defining equations. We have already shown in section \ref{sec:Further-discussion}
that the two equations for $S$ are equivalent up to Bianchi identities.\begin{eqnarray}
C_{\bs{\alpha}}\hoch{\bs{\gamma}\hat{\bs{\gamma}}} & = & \gemnabla_{\bs{\alpha}}\RR^{\bs{\gamma}\hat{\bs{\gamma}}}\\
\hat{C}_{\hat{\bs{\alpha}}}\hoch{\hat{\bs{\gamma}}\bs{\gamma}} & = & \gemnabla_{\hat{\bs{\alpha}}}\RR^{\bs{\gamma}\hat{\bs{\gamma}}}\\
S_{\bs{\alpha}\hat{\bs{\alpha}}}\hoch{\bs{\gamma}\hat{\bs{\beta}}} & = & -\gemnabla_{\bs{\alpha}}\underbrace{\hat{C}_{\hat{\bs{\alpha}}}\hoch{\hat{\bs{\beta}}\bs{\gamma}}}_{\gemnabla_{\hat{\bs{\alpha}}}\RR^{\bs{\gamma}\hat{\bs{\beta}}}}+2\hat{R}_{\bs{\alpha}\hat{\bs{\gamma}}\hat{\bs{\alpha}}}\hoch{\hat{\bs{\beta}}}\RR^{\bs{\gamma}\hat{\bs{\gamma}}}\\
S_{\bs{\alpha}\hat{\bs{\alpha}}}\hoch{\bs{\beta}\hat{\bs{\gamma}}} & = & -\gemnabla_{\hat{\bs{\alpha}}}\underbrace{C_{\bs{\alpha}}\hoch{\bs{\beta}\hat{\bs{\gamma}}}}_{\gemnabla_{\bs{\alpha}}\RR^{\bs{\beta}\hat{\bs{\gamma}}}}+2R_{\hat{\bs{\alpha}}\bs{\gamma}\bs{\alpha}}\hoch{\bs{\beta}}\RR^{\bs{\gamma}\hat{\bs{\gamma}}}\end{eqnarray}
Combining them with the reduced structure group constraints (\ref{eq:OmegaCond2}),(\ref{eq:Ccond2})
and (\ref{eq:Scond2}), we obtain:\begin{eqnarray}
\gamma^{a_{1}\ldots a_{4}}\tief{\bs{\beta}}\hoch{\bs{\alpha}}\gemnabla_{\bs{\alpha}}\RR^{\bs{\beta}\hat{\bs{\gamma}}} & = & 0,\qquad\gamma^{a_{1}\ldots a_{4}}\tief{\hat{\bs{\beta}}}\hoch{\hat{\bs{\alpha}}}\gemnabla_{\hat{\bs{\alpha}}}\RR^{\bs{\gamma}\hat{\bs{\beta}}}=0\label{eq:RRconstrFromReducedStruct}\end{eqnarray}
The reduced structure group of $S$ instead doesn't provide additional
information. It is induced%
\footnote{\index{footnote!\thefoot. comment on the reduced structure group of $S_{\alpha\hat\alpha}\hoch{\beta\hat\beta}$}We
have $\nabla_{M}\gamma^{a_{1}a_{2}}\tief{a_{3}a_{4}\,\bs{\alpha}}\hoch{\bs{\beta}}=\hat{\nabla}_{M}\gamma^{a_{1}a_{2}}\tief{a_{3}a_{4}\,\bs{\alpha}}\hoch{\bs{\beta}}=\check{\nabla}_{M}\gamma^{a_{1}a_{2}}\tief{a_{3}a_{4}\,\bs{\alpha}}\hoch{\bs{\beta}}=0$
by definition, because $\nabla_{M}$ and the others are defined via
$\nabla_{M}\gamma_{\bs{\alpha\beta}}^{a}=0$. Therefore we have for
the mixed connection \begin{eqnarray*}
\gemnabla_{M}\tilde{\gamma}^{a_{1}a_{2}a_{3}a_{4}}\tief{\bs{\alpha}}\hoch{\bs{\beta}} & = & -4\check{\nabla}_{M}\Phi\cdot\tilde{\gamma}^{a_{1}a_{2}a_{3}a_{4}}\tief{\bs{\alpha}}\hoch{\bs{\beta}}+4(\check{\Omega}-\Omega)_{Mc}\hoch{[a_{1}|}\tilde{\gamma}^{c|a_{2}a_{3}a_{4}]}\tief{\bs{\alpha}}\hoch{\bs{\beta}}\end{eqnarray*}
All terms on the righthand side are proportional to $\gamma^{[4]}$,
and therefore we have schematically \[
\gamma^{[4]}S\propto\gamma^{[4]}\left(-\gemnabla C+2R\RR\right)\propto-\gemnabla(\underbrace{\gamma^{[4]}C}_{=0})+\underbrace{(\gemnabla\gamma^{[4]})C}_{\propto\gamma^{[4]}C=0}+2\underbrace{\gamma^{[4]}R}_{=0}\RR=0\]
The reduced structure group condition $\gamma^{[4]}S=0$ is thus a
consequence of $\gamma^{[4]}C=0$ and $\gamma^{[4]}R=0$.$\quad\fussend$%
} by the reduced structure group property of $C$ and of the curvature
$R$. \rem{\begin{eqnarray}
\mbox{or likewise }\gemnabla_{M}\gamma^{a_{1}a_{2}}\tief{a_{3}a_{4}\,\bs{\alpha}}\hoch{\bs{\beta}} & = & 2(\check{\Omega}-\Omega)_{Mc}\hoch{[a_{1}|}\gamma^{c|a_{2}]}\tief{a_{3}a_{4}\,\bs{\alpha}}\hoch{\bs{\beta}}-2(\check{\Omega}-\Omega)_{M[a_{3}|}\hoch{c}\gamma^{a_{1}a_{2}}\tief{c|a_{4}]\,\bs{\alpha}}\hoch{\bs{\beta}}=\\
 & = & 2(\check{\Omega}-\Omega)_{Mc}\hoch{[a_{1}|}\gamma^{c|a_{2}]}\tief{a_{3}a_{4}\,\bs{\alpha}}\hoch{\bs{\beta}}-2(\check{\Omega}-\Omega)_{M[a_{3}|}\hoch{c}\gamma^{a_{1}a_{2}}\tief{c|a_{4}]\,\bs{\alpha}}\hoch{\bs{\beta}}\end{eqnarray}
In addition, we could write:\begin{eqnarray*}
\gamma^{a_{1}\ldots a_{4}}\tief{\bs{\beta}}\hoch{\bs{\alpha}}T_{\{c,\hat{\bs{\alpha}}\}\bs{\alpha}}\hoch{\bs{\beta}} & = & \gamma^{a_{1}\ldots a_{4}}\tief{\bs{\beta}}\hoch{\bs{\alpha}}(\de E^{\bs{\beta}})_{\{c,\hat{\bs{\alpha}}\}\bs{\alpha}}=\gamma^{a_{1}\ldots a_{4}}\tief{\bs{\beta}}\hoch{\bs{\alpha}}\hat{T}_{\{c,\hat{\bs{\alpha}}\}\bs{\alpha}}\hoch{\bs{\beta}}\\
\gamma^{a_{1}\ldots a_{4}}\tief{\bs{\beta}}\hoch{\bs{\alpha}}\gemnabla_{\hat{\bs{\alpha}}}\gemnabla_{\bs{\alpha}}\RR^{\bs{\beta}\hat{\bs{\beta}}} & = & 0\quad\dann\quad\left(\gemnabla_{\hat{\bs{\alpha}}}\tilde{\gamma}^{a_{1}\ldots a_{4}}\tief{\bs{\beta}}\hoch{\bs{\alpha}}\right)\gemnabla_{\bs{\alpha}}\RR^{\bs{\beta}\hat{\bs{\beta}}}=0\end{eqnarray*}
The first doesn't tell too much, the second nothing new. What about
the following:\begin{eqnarray}
X^{\bs{\alpha}\hat{\bs{\alpha}}a_{1}\ldots a_{4}b_{1}\ldots b_{4}} & \equiv & \gamma^{a_{1}\ldots a_{4}}\tief{\bs{\beta}}\hoch{\bs{\alpha}}\RR^{\bs{\beta}\hat{\bs{\gamma}}}\gamma^{b_{1}\ldots b_{4}}\tief{\hat{\bs{\gamma}}}\hoch{\hat{\bs{\alpha}}}\\
0 & = & \gemnabla_{\bs{\alpha}}X^{\bs{\alpha}\hat{\bs{\alpha}}a_{1}\ldots a_{4}b_{1}\ldots b_{4}}=\gemnabla_{\hat{\bs{\alpha}}}X^{\bs{\alpha}\hat{\bs{\alpha}}a_{1}\ldots a_{4}b_{1}\ldots b_{4}}\end{eqnarray}
}

\paragraph{Constraints on $H$}

Due to (\ref{eq:Y-constrI})-(\ref{eq:Y-constrV}), (\ref{eq:nilpotency-constraint-onH}),
(\ref{eq:nilpotency-constraint-onH-hat}) and the total antisymmetry
of $H$, its only nonvanishing components are\index{$H_{ABC}$}\begin{eqnarray}
H_{abc} & \neq & 0\qquad(\mbox{in general})\\
H_{\bs{\alpha}\bs{\beta}c} & = & \frem{\gamma_{\bs{\alpha\beta}}^{a}\frac{1}{16}H_{\bs{\gamma\delta}c}\gamma_{a}^{\bs{\gamma\delta}}=}-\frac{2}{3}\check{T}_{\bs{\alpha}\bs{\beta}|c}\equiv-\frac{2}{3}\gamma_{\bs{\alpha\beta}}^{a}f_{ac}\\
H_{\hat{\bs{\alpha}}\hat{\bs{\beta}}c} & = & \frem{\gamma_{\hat{\bs{\alpha}}\hat{\bs{\beta}}}^{a}\frac{1}{16}H_{\hat{\bs{\gamma}}\hat{\bs{\delta}}c}\gamma_{a}^{\hat{\bs{\gamma}}\hat{\bs{\delta}}}=}\frac{2}{3}\check{T}_{\hat{\bs{\alpha}}\hat{\bs{\beta}}|c}\equiv\frac{2}{3}\gamma_{\hat{\bs{\alpha}}\hat{\bs{\beta}}}^{a}\hat{f}_{ac}\end{eqnarray}
The vanishing components are thus (written a bit redundantly)\begin{eqnarray}
H_{ab\bs{\mc{C}}} & = & H_{\bs{\alpha}\hat{\bs{\beta}}C}=H_{\bs{\mc{A}\mc{B}\mc{C}}}=0\end{eqnarray}
\rem{In matrix notation, $H$ reads:\begin{eqnarray}
H_{ABc} & = & \left(\begin{array}{ccc}
H_{abc} & H_{a\bs{\beta}c}=0 & H_{a\hat{\bs{\beta}}c}=0\\
H_{\bs{\alpha}bc}=0 & H_{\bs{\alpha\beta}c}=-\frac{2}{3}\gamma_{\bs{\alpha\beta}}^{a}f_{ac} & H_{\bs{\alpha}\hat{\bs{\beta}}c}=0\\
H_{\hat{\bs{\alpha}}bc}=0 & H_{\hat{\bs{\alpha}}\bs{\beta}c}=0 & H_{\hat{\bs{\alpha}}\hat{\bs{\beta}}c}=\frac{2}{3}\gamma_{\hat{\bs{\alpha}}\hat{\bs{\beta}}}^{a}\hat{f}_{ac}\end{array}\right)_{AB}\\
H_{AB\bs{\gamma}} & = & \left(\begin{array}{ccc}
H_{ab\bs{\gamma}}=0 & H_{a\bs{\beta}\bs{\gamma}}=-\frac{2}{3}\gamma_{\bs{\beta\gamma}}^{c}f_{ca} & H_{a\hat{\bs{\beta}}\bs{\gamma}}=0\\
H_{\bs{\alpha}b\bs{\gamma}}=\frac{2}{3}\gamma_{\bs{\alpha\gamma}}^{c}f_{cb} & H_{\bs{\alpha\beta}\bs{\gamma}}=0 & H_{\bs{\alpha}\hat{\bs{\beta}}\bs{\gamma}}=0\\
H_{\hat{\bs{\alpha}}b\bs{\gamma}}=0 & H_{\hat{\bs{\alpha}}\bs{\beta}\bs{\gamma}}=0 & H_{\hat{\bs{\alpha}}\hat{\bs{\beta}}\bs{\gamma}}=0\end{array}\right)_{AB}\\
H_{AB\hat{\bs{\gamma}}} & = & \left(\begin{array}{ccc}
H_{ab\hat{\bs{\gamma}}}=0 & H_{a\bs{\beta}\hat{\bs{\gamma}}}=0 & H_{a\hat{\bs{\beta}}\hat{\bs{\gamma}}}=\frac{2}{3}\gamma_{\hat{\bs{\beta}}\hat{\bs{\gamma}}}^{c}\hat{f}_{ca}\\
H_{\bs{\alpha}b\hat{\bs{\gamma}}}=0 & H_{\bs{\alpha\beta}\hat{\bs{\gamma}}}=0 & H_{\bs{\alpha}\hat{\bs{\beta}}\hat{\bs{\gamma}}}=0\\
H_{\hat{\bs{\alpha}}b\hat{\bs{\gamma}}}=-\frac{2}{3}\gamma_{\hat{\bs{\alpha}}\hat{\bs{\gamma}}}^{c}\hat{f}_{cb} & H_{\hat{\bs{\alpha}}\bs{\beta}\hat{\bs{\gamma}}}=0 & H_{\hat{\bs{\alpha}}\hat{\bs{\beta}}\hat{\bs{\gamma}}}=0\end{array}\right)_{AB}\end{eqnarray}
}

\paragraph{Constraints on the torsion}

Let us now collect the information of the constraints (\ref{eq:Y-constrII})-(\ref{eq:Y-constrIV}),
(\ref{eq:holConstrI})-(\ref{eq:holConstrIV}) and (\ref{eq:nilpotency-constraint-onTfinal}),(\ref{eq:nilpotency-constraint-onTfinal-hat}),(\ref{eq:convConstr}).
The only (a priori) nonvanishing components of the torsion $\gem{T}_{AB}\hoch{C}$
are\rem{%
\footnote{For (\ref{eq:T^{c}-sym}) take into account that \begin{eqnarray*}
\Delta_{[\bs{\mc{A}}c]}\hoch{d} & = & \frac{1}{2}\Delta_{\bs{\mc{A}}c}\hoch{d}=\frac{1}{2}\Delta_{\bs{\mc{A}}}\delta_{c}\hoch{d}+\frac{1}{2}\Delta_{\bs{\mc{A}}\, c}^{(L)}\hoch{d}\\
\hat{T}_{\bs{\mc{A}}(c|d)} & = & T_{\bs{\mc{A}}(c|d)}+\frac{1}{2}\Delta_{\bs{\mc{A}}(c|d)}=\\
 & = & T_{\bs{\mc{A}}(c|d)}+\frac{1}{2}\Delta_{\bs{\mc{A}}}G_{cd}\end{eqnarray*}
Remember also that the torsion constraints on $T_{\bs{\alpha}(c|d)}$
are constraints on the vielbein only, not on the connection: \begin{eqnarray*}
\underbrace{2T_{\bs{\mc{A}}(c|d)}-\Omega_{\bs{\mc{A}}}G_{cd}}_{2E_{\bs{\mc{A}}}\hoch{M}E_{(c|}\hoch{N}\left(\partial_{[M}E_{N]}\hoch{e}\right)G_{e|d)}} & = & -\nabla_{\bs{\mc{A}}}\Phi\, G_{cd}\qquad\fussend\end{eqnarray*}
}}\index{$T_{AB}$@$\gem{T}_{AB}\hoch{C}$} \begin{eqnarray}
\check{T}_{\bs{\mc{A}}(c|d)} & = & -\frac{1}{2}\checkcovPhi{\bs{\mc{A}}}G_{cd}\quad\frem{\dann\check{T}_{\bs{\mc{A}}c}\hoch{c}=-5\checkcovPhi{\bs{\mc{A}}}}\label{eq:T^{c}-sym'}\\
\check{T}_{\bs{\alpha}\bs{\beta}|c} & = & -\frac{3}{2}H_{\bs{\alpha}\bs{\beta}c}=\gamma_{\bs{\alpha\beta}}^{d}f_{dc},\qquad\check{T}_{\hat{\bs{\alpha}}\hat{\bs{\beta}}|c}=\frac{3}{2}H_{\hat{\bs{\alpha}}\hat{\bs{\beta}}c}=\gamma_{\hat{\bs{\alpha}}\hat{\bs{\beta}}}^{d}\hat{f}_{dc}\\
\hat{T}_{\bs{\alpha}c}\hoch{\hat{\bs{\gamma}}} & = & \check{T}_{\bs{\alpha}\bs{\delta}|c}\RR^{\bs{\delta}\hat{\bs{\gamma}}}=\gamma_{\bs{\alpha\delta}}^{d}f_{dc}\RR^{\bs{\delta}\hat{\bs{\gamma}}},\qquad T_{\hat{\bs{\alpha}}c}\hoch{\bs{\gamma}}=\check{T}_{\hat{\bs{\alpha}}\hat{\bs{\delta}}|c}\RR^{\bs{\gamma}\hat{\bs{\delta}}}=\gamma_{\hat{\bs{\alpha}}\hat{\bs{\delta}}}^{d}\hat{f}_{dc}\RR^{\bs{\gamma}\hat{\bs{\delta}}}\\
\gemT_{ab}\hoch{C} & \neq & 0\qquad(\mbox{in general})\end{eqnarray}
The remaining components all vanish, which can be written (again
a bit redundantly) as\begin{eqnarray}
\gem{T}_{\bs{\mc{A}\mc{B}}}\hoch{\bs{\mc{C}}} & = & \gem{T}_{\bs{\alpha}\hat{\bs{\alpha}}}\hoch{C}=T_{\bs{\alpha}d}\hoch{\bs{\gamma}}=\hat{T}_{\hat{\bs{\alpha}}d}\hoch{\hat{\bs{\gamma}}}=0\end{eqnarray}
 \rem{In matrix notation, we have \begin{eqnarray}
T_{AB}\hoch{c} & = & \left(\begin{array}{ccc}
T_{ab}\hoch{c} & T_{a\bs{\beta}}\hoch{c} & T_{a\hat{\bs{\beta}}}\hoch{c}\\
T_{\bs{\alpha}b}\hoch{c} & T_{\bs{\alpha}\bs{\beta}}\hoch{c}=\gamma_{\bs{\alpha\beta}}^{d}f_{dc} & T_{\bs{\alpha}\hat{\bs{\beta}}}\hoch{c}=0\\
T_{\hat{\bs{\alpha}}b}\hoch{c} & T_{\hat{\bs{\alpha}}\bs{\beta}}\hoch{c}=0 & T_{\hat{\bs{\alpha}}\hat{\bs{\beta}}}\hoch{c}=\gamma_{\hat{\bs{\alpha}}\hat{\bs{\beta}}}^{d}\hat{f}_{dc}\end{array}\right)=\hat{T}_{AB}\hoch{c}-\left(\begin{array}{ccc}
\Delta_{[ab]}\hoch{c} & -\frac{1}{2}\Delta_{\bs{\beta}a}\hoch{c} & -\frac{1}{2}\Delta_{\hat{\bs{\beta}}a}\hoch{c}\\
\frac{1}{2}\Delta_{\bs{\alpha}b}\hoch{c} & 0 & 0\\
\frac{1}{2}\Delta_{\hat{\bs{\alpha}}b}\hoch{c} & 0 & 0\end{array}\right)\\
T_{AB}\hoch{\bs{\gamma}} & = & \left(\begin{array}{ccc}
T_{ab}\hoch{\bs{\gamma}} & T_{a\bs{\beta}}\hoch{\bs{\gamma}}=0 & T_{a\hat{\bs{\beta}}}\hoch{\bs{\gamma}}=-\gamma_{\hat{\bs{\beta}}\hat{\bs{\delta}}}^{d}\hat{f}_{da}\RR^{\bs{\gamma}\hat{\bs{\delta}}}\\
T_{\bs{\alpha}b}\hoch{\bs{\gamma}}=0 & T_{\bs{\alpha}\bs{\beta}}\hoch{\bs{\gamma}}=0 & T_{\bs{\alpha}\hat{\bs{\beta}}}\hoch{\bs{\gamma}}=0\\
T_{\hat{\bs{\alpha}}b}\hoch{\bs{\gamma}}=\gamma_{\hat{\bs{\alpha}}\hat{\bs{\delta}}}^{d}\hat{f}_{db}\RR^{\bs{\gamma}\hat{\bs{\delta}}} & T_{\hat{\bs{\alpha}}\bs{\beta}}\hoch{\bs{\gamma}}=0 & T_{\hat{\bs{\alpha}}\hat{\bs{\beta}}}\hoch{\bs{\gamma}}=0\end{array}\right)\\
\hat{T}_{AB}\hoch{\hat{\bs{\gamma}}} & = & \left(\begin{array}{ccc}
\hat{T}_{ab}\hoch{\hat{\bs{\gamma}}} & \hat{T}_{a\bs{\beta}}\hoch{\hat{\bs{\gamma}}}=-\gamma_{\bs{\beta\delta}}^{d}f_{da}\RR^{\bs{\delta}\hat{\bs{\gamma}}} & \hat{T}_{a\hat{\bs{\beta}}}\hoch{\hat{\bs{\gamma}}}=0\\
\hat{T}_{\bs{\alpha}b}\hoch{\hat{\bs{\gamma}}}=\gamma_{\bs{\alpha\delta}}^{d}f_{db}\RR^{\bs{\delta}\hat{\bs{\gamma}}} & \hat{T}_{\bs{\alpha}\bs{\beta}}\hoch{\hat{\bs{\gamma}}}=0 & \hat{T}_{\bs{\alpha}\hat{\bs{\beta}}}\hoch{\hat{\bs{\gamma}}}=0\\
\hat{T}_{\hat{\bs{\alpha}}b}\hoch{\hat{\bs{\gamma}}}=0 & \hat{T}_{\hat{\bs{\alpha}}\bs{\beta}}\hoch{\hat{\bs{\gamma}}}=0 & \hat{T}_{\hat{\bs{\alpha}}\hat{\bs{\beta}}}\hoch{\hat{\bs{\gamma}}}=0\end{array}\right)=T_{AB}\hoch{\hat{\bs{\gamma}}}+\left(\begin{array}{ccc}
0 & 0 & \frac{1}{2}\Delta_{a\hat{\bs{\beta}}}\hoch{\hat{\bs{\gamma}}}\\
0 & 0 & \frac{1}{2}\Delta_{\bs{\alpha}\hat{\bs{\beta}}}\hoch{\hat{\bs{\gamma}}}\\
-\frac{1}{2}\Delta_{b\hat{\bs{\alpha}}}\hoch{\hat{\bs{\gamma}}} & -\frac{1}{2}\Delta_{\bs{\beta}\hat{\bs{\alpha}}}\hoch{\hat{\bs{\gamma}}} & \Delta_{[\hat{\bs{\alpha}}\hat{\bs{\beta}}]}\hoch{\hat{\bs{\gamma}}}\end{array}\right)\end{eqnarray}
}The above constraints are constraints on the torsion $\gemT_{AB}\hoch{C}=(\check{T}_{AB}\hoch{c},T_{AB}\hoch{\bs{\gamma}},\hat{T}_{AB}\hoch{\hat{\bs{\gamma}}})$,
which is based on the mixed connection $\gemOm_{AB}\hoch{C}$ defined
in (\ref{eq:mixedConnection}) on page \pageref{eq:mixedConnection}.
When solving the Bianchi identities in the next local appendix, the
bosonic block $\check{\Omega}_{Ma}\hoch{b}$ of the connection will
be chosen for convenience to sometimes coincide with the left-mover
connection $\Omega_{Ma}\hoch{b}$ (induced by $\Omega_{M\bs{\alpha}}\hoch{\bs{\beta}}$)
or with the right mover connection (induced by $\hat{\Omega}_{M\hat{\bs{\alpha}}}\hoch{\bs{\beta}}$;
see appendix \vref{cha:ConnectionAppend}). Not only for the bosonic
block, but also for the fermionic blocks, information on torsion based
on left-or right-mover connection, instead of the mixed connection
will be important later. This information is in principle given by
the difference-tensor $\Delta_{MA}\hoch{B}$, introduced above in
(\ref{eq:BI:DifferenceTensor}). Complete knowledge of the difference
tensor, allows to calculate the corresponding torsion components via\begin{eqnarray}
\hat{T}_{AB}\hoch{C}-T_{AB}\hoch{C} & = & \Delta_{[AB]}\hoch{C}\label{eq:DifferenceOfTandHatT}\end{eqnarray}
Due to the block diagonality of the connection and the difference
tensor, some of these torsion components do not contain the connection
at all. If we denote by $\check{\Omega}_{MA}\hoch{B}$ the connection
which is induced by the bosonic block of the mixed connection (i.e.
it is block diagonal and Lorentz plus scale, but otherwise arbitrary),
then we have \begin{eqnarray}
\hat{T}_{\bs{\mc{AB}}}\hoch{c} & = & T_{\bs{\mc{AB}}}\hoch{c}=\check{T}_{\bs{\mc{AB}}}\hoch{c}=(\de E^{c})_{\bs{\mc{AB}}}\label{eq:checkTbosonic}\\
\hat{T}_{\{a,\hat{\bs{\alpha}}\}\{b,\hat{\bs{\beta}}\}}\hoch{\bs{\gamma}} & = & T_{\{a,\hat{\bs{\alpha}}\}\{b,\hat{\bs{\beta}}\}}\hoch{\bs{\gamma}}=\check{T}_{\{a,\hat{\bs{\alpha}}\}\{b,\hat{\bs{\beta}}\}}\hoch{\bs{\gamma}}=(\de E^{\bs{\gamma}})_{\{a,\hat{\bs{\alpha}}\}\{b,\hat{\bs{\beta}}\}}\label{eq:checkTfermionic}\\
\hat{T}_{\{a,\bs{\alpha}\}\{b,\bs{\beta}\}}\hoch{\hat{\bs{\gamma}}} & = & T_{\{a,\bs{\alpha}\}\{b,\bs{\beta}\}}\hoch{\hat{\bs{\gamma}}}=\check{T}_{\{a,\bs{\alpha}\}\{b,\bs{\beta}\}}\hoch{\hat{\bs{\gamma}}}=(\de E^{\hat{\bs{\gamma}}})_{\{a,\bs{\alpha}\}\{b,\bs{\beta}\}}\label{eq:checkTfermionicHat}\end{eqnarray}
The brackets $\{a,\bs{\alpha}\}\{b,\bs{\beta}\}$ shall denote that
the equation holds if the index $A$ is either $a$ or $\bs{\alpha}$
(but not $\hat{\bs{\alpha}}$), while the index $B$ is either $b$
or $\bs{\beta}$ (but not $\hat{\bs{\beta}}$).

\paragraph{Constraints on the curvature}

Induced by the restricted structure group constraints on the connection,
we have such constraints likewise for the curvature (see (\ref{eq:mixedCurvature})
on page \pageref{eq:mixedCurvature} and (\ref{eq:R-Zerfall-bosonic}),(\ref{eq:R-Zerfall-ferm})
and (\ref{eq:R-Zerfall-ferm-hut}) on page \ref{eq:R-Zerfall-ferm}.
The curvature is blockdiagonal and each part decays into a scale part
and a Lorentz part:\index{$R_{ABC}$@$\gem{R}_{ABC}\hoch{D}$}\begin{eqnarray}
\gem{R}_{ABC}\hoch{D} & = & \diag(\check{R}_{ABc}\hoch{d},R_{AB\bs{\gamma}}\hoch{\bs{\delta}},\hat{R}_{AB\hat{\bs{\gamma}}}\hoch{\hat{\bs{\delta}}})\\
\check{R}_{ABc}\hoch{d} & = & \check{F}_{AB}^{(D)}\delta_{c}^{d}+\check{R}_{AB\: c}^{(L)}\hoch{d},\qquad\check{F}_{AB}^{(D)}=\frac{1}{10}\check{R}_{ABc}\hoch{c}\\
R_{AB\bs{\gamma}}\hoch{\bs{\delta}} & = & \frac{1}{2}F_{AB}^{(D)}\delta_{\bs{\gamma}}\hoch{\bs{\delta}}+\frac{1}{4}R_{AB}^{(L)}\tief{a_{1}}\hoch{b}\eta_{ba_{2}}\gamma^{a_{1}a_{2}}\tief{\bs{\gamma}}\hoch{\bs{\delta}},\qquad F_{AB}^{(D)}=-\frac{1}{8}R_{AB\bs{\gamma}}\hoch{\bs{\gamma}}\\
\hat{R}_{AB\hat{\bs{\gamma}}}\hoch{\hat{\bs{\gamma}}} & = & \frac{1}{2}\hat{F}^{(D)}\delta_{\hat{\bs{\alpha}}}\hoch{\hat{\bs{\beta}}}+\frac{1}{4}\hat{R}_{AB}^{(L)}\tief{a_{1}}\hoch{b}\eta_{ba_{2}}\gamma^{a_{1}a_{2}}\tief{\hat{\bs{\alpha}}}\hoch{\hat{\bs{\beta}}},\qquad\hat{F}_{AB}^{(D)}=-\frac{1}{8}\hat{R}_{AB\hat{\bs{\gamma}}}\hoch{\hat{\bs{\gamma}}}\end{eqnarray}
with the scale field strength \rem{%
\footnote{\label{fn:covariant-der-of-connection}It is tempting to write $F^{(D)}\equiv\de\Omega^{(D)}\quad\iff\quad F_{AB}^{(D)}=\gemnabla_{[A}\Omega_{B]}+\gemT_{AB}\hoch{C}\Omega_{C}^{(D)}$
which is formally true if we act with the covariant derivative on
$\gemOm_{B}$ like on a tensorial object. We prefer the point of view
to act with the connection always in the same representations as the
infinitesimal local structure group transformation would do. There
we have $\delta\Omega_{M}^{(D)}=-\partial_{M}L^{(D)}$. The covariant
derivative acting on $\Omega_{M}^{(D)}$ can thus be defined as $\gemnabla_{M}\Omega_{N}^{(D)}\equiv\partial_{M}\Omega_{N}^{(D)}-\gem{\Gamma}_{MN}\hoch{K}\Omega_{K}^{(D)}-\partial_{N}\gemOm_{M}^{(D)}=2F_{MN}^{(D)}-\gem{\Gamma}_{MN}\hoch{K}\Omega_{K}^{(D)}$.
On a general connection $\Omega_{MA}\hoch{B}$, the infinitesimal
structure group transformation reads $\delta\Omega_{MA}\hoch{B}=-\partial_{M}L_{A}\hoch{B}-[L,\Omega_{M}]_{A}\hoch{B}\equiv\mc{R}(L_{\,\cdot}\hoch{\cdot})\Omega_{MA}\hoch{B}$.
One can thus define \begin{eqnarray*}
\nabla_{K}\Omega_{MA}\hoch{B} & \equiv & \partial_{K}\Omega_{MA}\hoch{B}-\Gamma_{KM}\hoch{L}\Omega_{LA}\hoch{B}+\mc{R}(\Omega_{K\,\cdot}\hoch{\cdot})\Omega_{MA}\hoch{B}=\\
 & = & -\Gamma_{KM}\hoch{L}\Omega_{LA}\hoch{B}+\partial_{K}\Omega_{MA}\hoch{B}-\partial_{M}\Omega_{KA}\hoch{B}-[\Omega_{K},\Omega_{M}]_{A}\hoch{B}=\\
 & = & -\Gamma_{KM}\hoch{L}\Omega_{LA}\hoch{B}+2R_{KMA}\hoch{B}\end{eqnarray*}
\frem{not space-time covariant any longer! Better base on supergauge trafo?}A
partial derivative of a covariant object like a vector $v^{A}$ or
a covector $w_{A}$ does likewise not transform covariantly but with
an inhomogenous term with the partial derivative of the gauge parameter,
i.e. $\delta\partial_{M}w=-\partial_{M}(Lw)=-\partial_{M}L\cdot w-L\cdot\partial_{M}w\equiv\mc{R}(L)\partial_{M}w$.
The covariant derivative on this non-covariant object will then be
defined as\begin{eqnarray*}
\nabla_{K}\partial_{M}w & \equiv & \partial_{K}\partial_{M}w-\Gamma_{KM}\hoch{L}\partial_{L}w+\mc{R}(\Omega_{K})\partial_{M}w=\\
 & = & \partial_{K}\partial_{M}w-\Gamma_{KM}\hoch{L}\partial_{L}w-\Omega_{K}\cdot\partial_{M}w-\partial_{M}\Omega_{K}\cdot w\end{eqnarray*}
To see that the definitions are consistent, we can convince ourselves
that the expressions combine correctly, if we add the non-covariant
objects $\partial_{M}w$ and $\Omega_{M}w$ to the covariant object
$\nabla_{M}w=\partial_{M}w-\Omega_{M}w$:\begin{eqnarray*}
\nabla_{K}\left(\partial_{M}w-\Omega_{M}w\right) & = & \nabla_{K}\partial_{M}w-\nabla_{K}\Omega_{M}\cdot w-\Omega_{M}\nabla_{K}w=\\
 & = & \partial_{K}\partial_{M}w-\Gamma_{KM}\hoch{L}\partial_{L}w-\Omega_{K}\cdot\partial_{M}w-\partial_{M}\Omega_{K}\cdot w+\Gamma_{KM}\hoch{L}\Omega_{L}w+\\
 &  & -\partial_{K}\Omega_{M}w+\partial_{M}\Omega_{K}w+[\Omega_{K},\Omega_{M}]w-\Omega_{M}\partial_{K}w+\Omega_{M}\Omega_{K}w=\\
 & = & \partial_{K}\left(\partial_{M}w-\Omega_{M}w\right)-\Gamma_{KM}\hoch{L}\left(\partial_{L}w-\Omega_{L}w\right)-\Omega_{K}\cdot\left(\partial_{M}w-\Omega_{M}w\right)\quad\surd\qquad\fussend\end{eqnarray*}
} }\begin{eqnarray}
\check{F}^{(D)} & \equiv & \de\check{\Omega}^{(D)},\qquad F^{(D)}\equiv\de\Omega^{(D)},\qquad\hat{F}^{(D)}\equiv\de\hat{\Omega}^{(D)}\end{eqnarray}
The bosonic field strength is also obtained via the commutator of
covariant derivatives acting on the compensator field $\Phi$. Only
the bosonic block $\check{\Omega}_{Ma}\hoch{b}$ of the mixed connection
$\gemOm_{MA}\hoch{B}$ acts on $\Phi$, because $\Phi$ is a compensator
for the transformation of $G_{ab}$ (with bosonic indices):\begin{eqnarray}
\check{F}_{MN}^{(D)} & = & -\gemnabla_{[M}\checkcovPhi{N]}-\gemT_{MN}\hoch{K}\checkcovPhi{K}\end{eqnarray}
 \rem{only after the different blocks of the structure group are related by partial gauge fixing, we may write:\begin{eqnarray*}
F_{MN}^{(D)} & = & -\nabla_{[M}\covPhi{N]}-T_{MN}\hoch{K}\covPhi{K}\\
\hat{F}_{MN}^{(D)} & = & -\hat{\nabla}_{[M}\hatcovPhi{N]}-\hat{T}_{MN}\hoch{K}\hatcovPhi{K}\\
\check{F}_{MN}^{(D)} & = & -\check{\nabla}_{[M}\checkcovPhi{N]}-\check{T}_{MN}\hoch{K}\checkcovPhi{K}\end{eqnarray*}
}Finallly we had a couple of holomorphicity (\ref{eq:holConstrVI}),(\ref{eq:holConstrVII}),(\ref{eq:holConstrIX}),(\ref{eq:holConstrX})
and nilpotency constraints (\ref{eq:nilpotency-constraint-onR}),(\ref{eq:nilpotency-constraint-onR-hat})
on the curvature:\begin{eqnarray}
\hat{R}_{\bs{\alpha}c\hat{\bs{\alpha}}}\hoch{\hat{\bs{\beta}}} & = & \underbrace{\check{T}_{\bs{\alpha}\bs{\delta}|c}}_{\gamma_{\bs{\alpha\delta}}^{d}f_{dc}}\underbrace{\hat{C}_{\hat{\bs{\alpha}}}\hoch{\hat{\bs{\beta}}\bs{\delta}}}_{\gemnabla_{\hat{\bs{\alpha}}}\RR^{\bs{\delta}\hat{\bs{\beta}}}}\frem{\hat{R}_{\bs{\alpha}bc}\hoch{d},\hat{R}_{\bs{\alpha}b\bs{\gamma}}\hoch{\bs{\delta}}},\qquad R_{\hat{\bs{\alpha}}c\bs{\alpha}}\hoch{\bs{\beta}}=\underbrace{\check{T}_{\hat{\bs{\alpha}}\hat{\bs{\delta}}|c}}_{\gamma_{\hat{\bs{\alpha}}\hat{\bs{\delta}}}^{d}\hat{f}_{dc}}\underbrace{C_{\bs{\alpha}}\hoch{\bs{\beta}\hat{\bs{\delta}}}}_{\gemnabla_{\bs{\alpha}}\RR^{\bs{\beta}\hat{\bs{\delta}}}}\frem{R_{\hat{\bs{\alpha}}bc}\hoch{d},R_{\hat{\bs{\alpha}}b\hat{\bs{\gamma}}}\hoch{\hat{\bs{\delta}}}}\\
\hat{R}_{\bs{\alpha}\bs{\gamma}\hat{\bs{\alpha}}}\hoch{\hat{\bs{\beta}}} & = & 0,\frem{\hat{R}_{\bs{\alpha\beta\gamma}}\hoch{\bs{\delta}}=\hat{R}_{\bs{\alpha\beta}c}\hoch{d}=0}\qquad R_{\hat{\bs{\alpha}}\hat{\bs{\gamma}}\bs{\alpha}}\hoch{\bs{\beta}}=0\frem{R_{\hat{\bs{\alpha}}\hat{\bs{\beta}}\hat{\bs{\gamma}}}\hoch{\hat{\bs{\delta}}}=R_{\hat{\bs{\alpha}}\hat{\bs{\beta}}c}\hoch{d}=0}\\
\gamma_{a_{1}\ldots a_{5}}^{\bs{\alpha}_{1}\bs{\alpha}_{2}}R_{d\bs{\alpha}_{1}\bs{\alpha}_{2}}\hoch{\bs{\beta}} & = & 0,\qquad\gamma_{a_{1}\ldots a_{5}}^{\hat{\bs{\alpha}}_{1}\hat{\bs{\alpha}}_{2}}\hat{R}_{d\hat{\bs{\alpha}}_{1}\hat{\bs{\alpha}}_{2}}\hoch{\hat{\bs{\beta}}}=0\\
\gamma_{a_{1}\ldots a_{5}}^{\bs{\alpha}_{1}\bs{\alpha}_{2}}R_{\hat{\bs{\delta}}\bs{\alpha}_{1}\bs{\alpha}_{2}}\hoch{\bs{\beta}} & = & 0,\qquad\gamma_{a_{1}\ldots a_{5}}^{\hat{\bs{\alpha}}_{1}\hat{\bs{\alpha}}_{2}}\hat{R}_{\bs{\delta}\hat{\bs{\alpha}}_{1}\hat{\bs{\alpha}}_{2}}\hoch{\hat{\bs{\beta}}}=0\\
R_{[\bs{\alpha}_{1}\bs{\alpha}_{2}\bs{\alpha}_{3}]}\hoch{\bs{\beta}} & = & 0,\frem{R_{\bs{\alpha\beta}c}\hoch{d}}\qquad\hat{R}_{[\hat{\bs{\alpha}}_{1}\hat{\bs{\alpha}}_{2}\hat{\bs{\alpha}}_{3}]}\hoch{\hat{\bs{\beta}}}=0\frem{\hat{R}_{\hat{\bs{\alpha}}\hat{\bs{\beta}}c}\hoch{d}}\end{eqnarray}
Taking the trace of the first two curvature constraints gives further
informations on dilatation-Field-strength and Lorentz curvature\begin{eqnarray}
\hat{F}_{\bs{\alpha}c}^{(D)} & = & -\frac{1}{8}\check{T}_{\bs{\alpha}\bs{\delta}|c}\gemnabla_{\hat{\bs{\alpha}}}\RR^{\bs{\delta}\hat{\bs{\alpha}}},\qquad F_{\hat{\bs{\alpha}}c}^{(D)}=-\frac{1}{8}\check{T}_{\hat{\bs{\alpha}}\hat{\bs{\delta}}|c}\gemnabla_{\bs{\alpha}}\RR^{\bs{\alpha}\hat{\bs{\delta}}}\label{eq:F-Dil-constrI'}\\
\hat{F}_{\bs{\alpha\gamma}}^{(D)} & = & 0,\qquad F_{\hat{\bs{\alpha}}\hat{\bs{\gamma}}}^{(D)}=0\label{eq:F-Dil-constrII'}\end{eqnarray}
The trace of the last curvature constraint we had provided already
in (\ref{eq:nilpotency:Falphbet}):\begin{equation}
F_{\bs{\gamma\delta}}^{(D)}=\frac{2}{9}R_{\bs{\alpha}[\bs{\gamma\delta}]}^{(L)}\hoch{\bs{\alpha}}\:,\quad\hat{F}_{\hat{\bs{\gamma}}\hat{\bs{\delta}}}^{(D)}=\frac{2}{9}\hat{R}_{\hat{\bs{\alpha}}[\hat{\bs{\gamma}}\hat{\bs{\delta}}]}^{(L)}\hoch{\hat{\bs{\alpha}}}\end{equation}
\rem{\begin{eqnarray*}
\hat{R}_{\bs{\alpha}ca}\hoch{b} & = & \hat{F}_{\bs{\alpha}c}^{(D)}\delta_{a}\hoch{b}+\frac{1}{8}\hat{R}_{\bs{\alpha}c\hat{\bs{\alpha}}}^{(L)}\hoch{\hat{\bs{\beta}}}\gamma_{a}^{\: b}\tief{\hat{\bs{\beta}}}\hoch{\hat{\bs{\alpha}}}\\
\hat{R}_{\bs{\alpha}\bs{\gamma}a}\hoch{b} & = & 0,\qquad R_{\hat{\bs{\alpha}}\hat{\bs{\gamma}}a}\hoch{b}=0\end{eqnarray*}
}\addtocounter{localapp}{1}

\section{Bianchi identities for H}

\label{sec:Bianchi-identities-forH} \index{Bianchi identity!for $H$}In
this local appendix we will study explicitly all the Bianchi identities
for the $H$-field. They are of the form

\begin{eqnarray}
0 & \stackrel{!}{=} & \gemnabla_{\bs{A}}H_{\bs{AAA}}+3\gemT_{\bs{AA}}\hoch{C}H_{C\bs{AA}}\label{eq:BI:H-BI}\end{eqnarray}
This is equivalent to $\de H=0$ and is independent of the connection,
in particular independent of the precise form of $\check{\Omega}$.
Sometimes it is thus convenient to calculate with the left-mover connection
$\check{\Omega}_{a}\hoch{b}=\Omega_{a}\hoch{b}$ (the latter defined
via $\nabla_{M}\gamma_{\bs{\alpha}\bs{\beta}}^{a}=0$, see appendix
\vref{cha:ConnectionAppend}) and sometimes we set $\check{\Omega}_{a}\hoch{b}=\hat{\Omega}_{a}\hoch{b}$
(defined via $\hat{\nabla}_{M}\gamma_{\hat{\bs{\alpha}}\hat{\bs{\beta}}}^{a}=0$). 

...

Let us now go back to the Bianchi identity (\ref{eq:BI:H-BI}), where
we make use of $\gemT_{AB}\hoch{C}$ instead of $T_{AB}\hoch{C}$
or $\hat{T}_{AB}\hoch{C}$. What we have just discussed is thus for
the moment only relevant for the the third index being bosonic $C=c$,
as we might choose $\gemT_{AB}\hoch{c}\equiv\check{T}_{AB}\hoch{c}$
to be either $T_{AB}\hoch{c}$ or $\hat{T}_{AB}\hoch{c}$. 

Every index $A$ of the Bianchi identity (\ref{eq:BI:H-BI}) can be
either $a$, $\bs{\alpha}$ or $\hat{\bs{\alpha}}$. As all indices
are antisymmetrized, we can distinguish the cases by specifying how
often each type of index appears. We denote in brackets first the
number of bosonic indices, then the number of unhatted fermionic indices
and finally the number of hatted fermionic indices:(\#$a$,\#$\bs{\alpha}$,\#$\hat{\bs{\alpha}}$).
The sum has to add up to four: \#$a$+\#$\bs{\alpha}$+\#$\hat{\bs{\alpha}}=4$.
Each number is in $\{0,\ldots,4\}$ which has five elements. If $\#a$
is $0$ there are five possibilities left for \#$\bs{\alpha}$ which
fixes \#$\hat{\bs{\alpha}}=$4-\#$\hat{\bs{\alpha}}$. If \#$a$ is
$1$, there are four possibilities left for \#$\bs{\alpha}$, and
so on. Altogether there are $5+4+3+2+1=15$ distinct cases. However,
some of them are related by the symmetry between hatted and unhatted
indices: (\#$a$,\#$\bs{\alpha}$,\#$\hat{\bs{\alpha}}$)$\leftrightarrow$(\#$a$,\#$\hat{\bs{\alpha}}$,\#$\bs{\alpha}$).
This map has {}``fixed points'' only for (\#$\hat{\bs{\alpha}}$,\#$\bs{\alpha}$)$\in\left\{ (0,0),(1,1),(2,2)\right\} $.
The effective number of equations we have to calculate is thus $\frac{15-3}{2}+3=9$.
In the following we go through all these cases. \vspace{.3cm}\\
$\bullet\quad$\underbar{(0,4,0)$\bs{\alpha\beta\gamma\delta}\leftrightarrow$((0,0,4)$\hat{\bs{\alpha}}\hat{\bs{\beta}}\hat{\bs{\gamma}}\hat{\bs{\delta}}$):}%
\footnote{\index{footnote!\thefoot. about the torsion in the H-BI}It might
be confusing that we obtain in (\ref{eq:BI-intermediate}) a constraint
not only on some components of $H_{ABC}$, but on a bilinear combination
of $H_{ABC}$ and $\gem{T}_{AB}\hoch{C}$. At first sight this seems
to contradict the equivalence to $\de H=0$ which is clearly only
a constraint on $H$. However, $H_{ABC}$ depends on $H$ (with components
$H_{MNK}$) AND the vielbein. And the torsion component $\gemT_{\bs{\alpha\beta}}\hoch{c}=(\de E^{c})_{\bs{\alpha\beta}}+\gemOm_{\bs{\alpha\beta}}\hoch{c}=(\de E^{c})_{\bs{\alpha\beta}}$
happens to depend only on the vielbein. The bilinear constraint thus
boils down to $(\de H)_{\bs{\alpha\beta\gamma\delta}}=0$, as it should
be. $\qquad\fussend$%
}\begin{eqnarray}
0 & \stackrel{!}{=} & \gemnabla_{[\bs{\alpha}}\underbrace{H_{\bs{\beta\gamma\delta}]}}_{=0\,(\ref{eq:nilpotency-constraint-onH})}+3\gemT_{[\bs{\alpha\beta}|}\hoch{C}H_{C|\bs{\gamma\delta}]}=\\
 & \ous{(\ref{eq:nilpotency-constraint-onH})}{=}{(\ref{eq:nilpotency-constraint-onTfinal})} & \quad3\gemT_{[\bs{\alpha\beta}|}\hoch{c}H_{c|\bs{\gamma\delta}]}=\label{eq:BI-intermediate}\\
 & = & -2\gamma_{[\bs{\alpha\beta}|}^{d}f_{d}\hoch{c}\gamma_{|\bs{\gamma}]\bs{\delta}}^{e}f_{ec}\label{eq:BI-fquadrat}\end{eqnarray}
The last line can only reduce to the Fierz identity $\gamma_{[\bs{\alpha\beta}|}^{d}\gamma_{d\,|\bs{\gamma}]\bs{\delta}}=0$
for%
\footnote{\index{footnote!\thefoot. torsion differs from $\gamma^c_{\bs{\alpha\beta}}$ only by Lorentz plus scale trafo}Let
us make this somewhat fishy argument more precise and contract (\ref{eq:BI-fquadrat})
with two chiral gamma matrices. In order to be able to apply some
equations of appendix \ref{cha:Gamma-Matrices} we will switch for
a moment to ungraded summation conventions (or equivalently perform
a grading shift of the fermionic index). We also multiply the whole
equation by $-\frac{3}{2}$ for convenience: \begin{eqnarray*}
0 & \stackrel{!}{=} & 3\gamma_{a}^{\beta\alpha}\gamma_{b}^{\delta\gamma}\gamma_{(\alpha\beta|}^{d}f_{d}\hoch{c}\gamma_{|\gamma)\delta}^{e}f_{ec}=\\
 & = & \gamma_{a}^{\beta\alpha}\gamma_{\alpha\beta}^{d}\gamma_{b}^{\delta\gamma}\gamma_{\gamma\delta}^{e}f_{d}\hoch{c}f_{ec}+\gamma_{a}^{\beta\alpha}\gamma_{\gamma\alpha}^{d}\gamma_{b}^{\delta\gamma}\gamma_{\beta\delta}^{e}f_{d}\hoch{c}f_{ec}+\gamma_{a}^{\beta\alpha}\gamma_{\beta\gamma}^{d}\gamma_{b}^{\delta\gamma}\gamma_{\alpha\delta}^{e}f_{d}\hoch{c}f_{ec}=\\
 & \ous{(\ref{eq:gammagammaSpur})}{=}{(\ref{eq:smallClifford})}= & \quad(16)^{2}f_{a}\hoch{c}f_{bc}+2\cdot\left(\delta_{a}^{d}\delta_{\gamma}^{\beta}+\gamma_{a}\hoch{d\,\beta}\tief{\gamma}\right)\left(\delta_{b}^{e}\delta_{\beta}^{\gamma}+\gamma_{b}\hoch{e\,\gamma}\tief{\beta}\right)f_{d}\hoch{c}f_{ec}=\\
 & \ous{(\ref{eq:gammatracelessness})}{=}{(\ref{eq:gammapgammapSpurEven})}\quad & (16)^{2}f_{a}\hoch{c}f_{bc}+32\delta_{a}^{d}\delta_{b}^{e}f_{d}\hoch{c}f_{ec}+2\cdot32G_{af}\delta_{eb}^{fd}f_{d}\hoch{c}f_{\: c}^{e}=\\
 & = & 16\cdot18f_{a}\hoch{c}f_{bc}-32G_{ab}f_{e}\hoch{c}f_{\: c}^{e}+32f_{b}\hoch{c}f_{ac}=\\
 & = & 16\cdot20\cdot f_{a}\hoch{c}f_{bc}-32G_{ab}f_{e}\hoch{c}f_{\: c}^{e}\end{eqnarray*}
We can now read off $f_{a}\hoch{c}f_{bc}=(\frac{1}{10}f_{e}\hoch{c}f_{\: c}^{e})G_{ab}$
or $f\eta f^{T}=\frac{1}{10}\tr(f\eta f^{T})\cdot\eta$, which means
simply that $f_{a}\hoch{b}$ is proportional to a Lorentz transformation.$\qquad\fussend$%
} \begin{equation}
f_{d}\hoch{c}G_{cb}f_{e}\hoch{b}=(f\cdot G\cdot f^{T})_{de}\stackrel{!}{\propto}G_{de}\propto\eta_{de}\label{eq:(0,4,0)}\end{equation}
The same is true for $\hat{f}$:\begin{equation}
(\hat{f}\cdot G\cdot\hat{f}^{T})_{ab}\propto G_{ab}\label{eq:(0,0,4)}\end{equation}
That means, $f$ and $\hat{f}$ are proportional to a Lorentz transformation.
In other words, \emph{If nonzero, $f$ and $\hat{f}$ are a composition
of a Lorentz transformation and a scaling}. 

\vspace{.5cm}

\lyxline{\normalsize}\vspace{-.25cm}\lyxline{\normalsize}

\subsubsection*{Intermezzo\index{intermezzo!fixing two of three Lorentz-plus-scale transformations}
on the fixing of two blocks of the structure group}

\label{Intermezzo:fixingTwoLorentz}\index{structure group!fixing two of three blocks}\index{gauge fixing!of two Lorentz-plus-scale transformations}\index{Lorentz transformation!fixing two of three $\sim$'s}

The above result provides a possibility to relate the three (a priori
independent) blocks of the structure group on the tangent space of
the supermanifold. We can thus use the local Lorentz transformation
(acting only on the unhatted spinor indices) and the local scale transformation
(likewise acting only on the unhatted spinor indices) to fix $f$
to unity and likewise use the hatted transformations to fix $\hat{f}$
to unity as it was done in \cite{Berkovits:2001ue}. We will do the
same, although -- regarding the subtleties discussed below -- one
should keep in mind that other kinds of gauge fixing might also have
their advantages. The gauge fixing leads to the following constraints:
\vRam{0.6}{\begin{eqnarray}
\check{T}_{\bs{\alpha}\bs{\beta}}\hoch{c} & = & \gamma_{\bs{\alpha\beta}}^{c}\frem{\stackrel{\Delta_{[\bs{\alpha\beta}]}\hoch{c}=0}{=}\hat{T}_{\bs{\alpha\beta}}\hoch{c}},\qquad(f_{a}\hoch{b}=\delta_{a}^{b})\label{eq:(0,4,0)'}\\
\check{T}_{\hat{\bs{\alpha}}\hat{\bs{\beta}}}\hoch{c} & = & \gamma_{\hat{\bs{\alpha}}\hat{\bs{\beta}}}^{c}\frem{=T_{\hat{\bs{\alpha}}\hat{\bs{\beta}}}\hoch{c}},\qquad(\hat{f}_{a}\hoch{b}=\delta_{a}^{b})\label{eq:(0,0,4)'}\\
\dann H_{\bs{\alpha}\bs{\beta}c} & = & -\frac{2}{3}\gamma_{\bs{\alpha\beta}}^{d}G_{dc}=-\frac{2}{3}e^{2\Phi}\gamma_{\bs{\alpha\beta}}^{d}\eta_{dc}\equiv-\frac{2}{3}\tilde{\gamma}_{c\,\bs{\alpha\beta}}\label{eq:(0,4,0)''}\\
H_{\hat{\bs{\alpha}}\hat{\bs{\beta}}c} & = & \frac{2}{3}\gamma_{\hat{\bs{\alpha}}\hat{\bs{\beta}}}^{d}G_{dc}=\frac{2}{3}e^{2\Phi}\gamma_{\hat{\bs{\alpha}}\hat{\bs{\beta}}}^{d}\eta_{dc}\equiv\frac{2}{3}\tilde{\gamma}_{c\,\hat{\bs{\alpha}}\hat{\bs{\beta}}}\label{eq:(0,0,4)''}\end{eqnarray}
\index{$\gamma_{c\,\bs{\alpha\beta}}$@$\tilde{\gamma}_{c\,\bs{\alpha\beta}}$}\index{$\gamma_{c\,\hat{\bs{\alpha}}\hat{\bs{\beta}}}$@$\tilde{\gamma}_{c\,\hat{\bs{\alpha}}\hat{\bs{\beta}}}$}}
The constraints (\ref{eq:(0,4,0)'}) and (\ref{eq:(0,0,4)'}) have
to be valid for any bosonic connection-block $\check{\Omega}_{Ma}\hoch{b}$,
in particular for the left and right-mover connections: $T_{\bs{\alpha\beta}}\hoch{c}=\hat{T}_{\bs{\alpha\beta}}\hoch{c}=\gamma_{\bs{\alpha\beta}}^{c}$.
Due to $\Delta_{[\bs{\alpha\beta}]}\hoch{c}=\Delta_{[\hat{\bs{\alpha}}\hat{\bs{\beta}}]}\hoch{c}=0$,
the constraints for $\check{T}_{\bs{\alpha}\bs{\beta}}\hoch{c}$ and
$\check{T}_{\hat{\bs{\alpha}}\hat{\bs{\beta}}}\hoch{c}$ are constraints
on the vielbein only. Having fixed the torsion components to the chiral
gamma matrices, the latter should remain invariant under the reduced
structure group. If we act with an infinitesimal transformation \begin{equation}
L_{a}\hoch{b}=L^{(D)}\delta_{a}^{b}+L_{a}^{(L)\, b},\qquad\mbox{with }L_{ab}^{(L)}=-L_{ba}^{(L)}\label{eq:coupledGroupTrafosI}\end{equation}
 on the bosonic index, it has to be compensated by the appropriate
actions on the fermionic indices (compare to footnote \vref{foot:LorentzScaleReason}
for a derivation):\begin{eqnarray}
L_{\bs{\alpha}}\hoch{\bs{\beta}} & = & \frac{1}{2}L^{(D)}\delta_{\bs{\alpha}}\hoch{\bs{\beta}}+\frac{1}{4}L_{ab}^{(L)}\gamma^{ab}\tief{\bs{\alpha}}\hoch{\bs{\beta}}\label{eq:coupledGroupTrafosII}\\
L_{\hat{\bs{\alpha}}}\hoch{\hat{\bs{\beta}}} & = & \frac{1}{2}L^{(D)}\delta_{\hat{\bs{\alpha}}}\hoch{\hat{\bs{\beta}}}+\frac{1}{4}L_{ab}^{(L)}\gamma^{ab}\tief{\hat{\bs{\alpha}}}\hoch{\hat{\bs{\beta}}}\label{eq:coupledGroupTrafosIII}\end{eqnarray}
This guarantuees\rem{%
\footnote{\index{footnote!\thefoot. IIA <-> IIB}\index{type IIA / IIB}For
type IIB the chiralities of $\bs{\alpha}$ and $\hat{\bs{\alpha}}$
are the same, so that the matrices $\gamma_{\bs{\alpha\beta}}^{a}$
and $\gamma_{\hat{\bs{\alpha}}\hat{\bs{\beta}}}^{a}$ coincide. This
means of course that also the structure group transformations of the
indices $L_{\bs{\alpha}}\hoch{\bs{\beta}}$ and $L_{\hat{\bs{\alpha}}}\hoch{\hat{\bs{\beta}}}$
coincide. For type IIA the situation is different. The matrix $\gamma_{\hat{\bs{\alpha}}\hat{\bs{\beta}}}^{a}$
then coincides with $\gamma^{a\,\bs{\alpha}\bs{\beta}}$. Although
numerically equal, $L_{\hat{\bs{\alpha}}}\hoch{\hat{\bs{\beta}}}$
has to act differently on $\gamma_{\hat{\bs{\alpha}}\hat{\bs{\beta}}}^{a}$
than $L_{\bs{\alpha}}\hoch{\bs{\beta}}$ acts on $\gamma^{a\,\bs{\alpha}\bs{\beta}}$,
otherwise it would be impossible for $\gamma_{\hat{\bs{\alpha}}\hat{\bs{\beta}}}^{a}$
and $\gamma_{\bs{\alpha\beta}}^{a}$ to be invariant under scale transformations
at the same time. One solution would be to fix $T_{\hat{\bs{\alpha}}\hat{\bs{\beta}}}\hoch{c}=e^{2\Phi}\gamma_{\hat{\bs{\alpha}}\hat{\bs{\beta}}}\hoch{c}$
for type IIA. We prefer, however, to treat IIA and IIB in the same
way. Another solution is then simply to fix in the end the (auxiliary)
scale transformations, at least when considering IIA.$\qquad\fussend$%
}} \begin{eqnarray}
\delta_{(L)}\gamma_{\bs{\alpha\beta}}^{a} & \equiv & L_{c}\hoch{a}\gamma_{\bs{\alpha\beta}}^{c}-2L_{[\bs{\alpha}}\hoch{\bs{\gamma}}\gamma_{\bs{\beta}]\bs{\gamma}}^{a}=0\\
\delta_{(L)}\gamma_{\hat{\bs{\alpha}}\hat{\bs{\beta}}}^{a} & \equiv & L_{c}\hoch{a}\gamma_{\hat{\bs{\alpha}}\hat{\bs{\beta}}}^{c}-2L_{[\hat{\bs{\alpha}}}\hoch{\hat{\bs{\gamma}}}\gamma_{\hat{\bs{\beta}}]\hat{\bs{\gamma}}}^{a}=0\end{eqnarray}
It is important to realize that $\gamma_{\bs{\alpha}\bs{\beta}}^{a}$
and $\gamma_{\hat{\bs{\alpha}}\hat{\bs{\beta}}}^{a}$ are not covariantly
constant with respect to the mixed connection $\gemOm_{MA}\hoch{B}$
that we have used so far. For the choice $\check{\Omega}_{Ma}\hoch{b}=\Omega_{Ma}\hoch{b}$
we get $\gemnabla_{M}\gamma_{\hat{\bs{\alpha}}\hat{\bs{\beta}}}^{a}\neq0$,
for $\check{\Omega}_{Ma}\hoch{b}=\hat{\Omega}_{Ma}\hoch{b}$ we get
$\gemnabla_{M}\gamma_{\bs{\alpha\beta}}^{a}\neq0$ and for any other
choice of $\check{\Omega}_{Ma}\hoch{b}$ none of the $\gamma$-matrices
will be covariantly conserved in general. Although all the equations
written in terms of $\gemOm_{MA}\hoch{B}$ remain of course formally
valid, it is geometrically not a suitable connection any longer. Parallel
transport would destroy our gauge. As mentioned at the beginning of
section \vref{cha:ConnectionAppend}, there are at least three natural
choices for connections which leave the gamma matrices invariant,
for example $\Omega_{MA}\hoch{B}$ (defined by the left-mover connection),
$\hat{\Omega}_{MA}\hoch{B}$ (defined by the rightmover connection)
and the average $\avOm_{MA}\hoch{B}\equiv\frac{1}{2}\left(\Omega_{MA}\hoch{B}+\hat{\Omega}_{MA}\hoch{B}\right)$.
These will be in particular relevant for the discussion of the WZ-gauge.
For the further discussion of the Bianchi identities after this intermezzo,
however, we stick formally to $\gemOm_{MA}\hoch{B}$.

\paragraph{Type IIA/IIB}

\label{typedistinction}\index{type IIA/IIB distinction}\index{two!type IIA/IIB distinction}\index{hatted index!distinction IIA/IIB}Let
us also give an important remark about the differences of type IIA
and type IIB which become important only at this point. In type IIB,
the hatted index $\hoch{\hat{\bs{\alpha}}}$ should be of the same
chirality, while in type IIA, $\hoch{\hat{\bs{\alpha}}}$ should be
of opposite chirality as $\hoch{\bs{\alpha}}$. This statement makes
only sense, when the Lorentz-transformations of hatted and unhatted
indeces are coupled, which was done only in the last steps above.
Before, the distinction between IIA and IIB was merely deciding whether
$\gamma_{\hat{\bs{\alpha}}\hat{\bs{\beta}}}^{c}$ is numerically equal
to $\gamma_{\bs{\alpha}\bs{\beta}}^{c}$ (IIB) or to $\gamma^{c\,\bs{\alpha}\bs{\beta}}$
(IIA). 

The transcription from the general equations (with hatted indices)
to the case of \textbf{type IIB} \index{IIB}is quite simple and direct,
as the index positions do not change. The conditions $\nabla_{M}\gamma_{\bs{\alpha\beta}}^{c}=0$
and $\nabla_{M}\gamma_{\hat{\bs{\alpha}}\hat{\bs{\beta}}}^{c}=0$
become numerically the same and imply that $\Omega_{M\hat{\bs{\alpha}}}\hoch{\hat{\bs{\beta}}}=\Omega_{M\bs{\alpha}}\hoch{\bs{\beta}}$
(same for the average connection). The hatted indices thus indeed
transform with the same chirality (w.r.t. Lorentz) and in addition
with the same representation of the scale transformation and the hats
of the indices can simply be dropped.

For \textbf{type IIA} \index{IIA}the situation is a bit more involved
and requires some familiarity with the graded summation convention
discussed around page \pageref{eq:gradedEinsteinSimple} in the first
part of the thesis. A downstairs hatted index $\tief{\hat{\bs{\alpha}}}$
should in IIA in the end correspond to an upstairs unhatted index
and vice verse. In a first step, we will still distinguish it from
the unhatted index and write it (just for this paragraph) as a \textbf{tilded}
\textbf{index} $\hoch{\tilde{\bs{\alpha}}}$ at opposite vertical
position. NW conventions for the hatted indices would then correspond
to NE conventions for the tilded index. We could stick to such mixed
conventions (NW for the unhatted indices and NE for the tilded indices),
but in order to make a comparison of the tilded with the undecorated
index, it is better to switch back to NW for the tilded index as well.
In principle this works as follows: spell out the NW summation conventions
for the hatted indices explicitely, replace the hatted by the tilded
in opposite vertical position and write it again in terms of the graded
summation convention based on NW. We call this an \textbf{index-position-shift}\index{index-position-shift}.
For example for the action of the covariant derivative on a spinor
with upper hatted index, this yields\begin{eqnarray}
\nabla_{M}\psi^{\hat{\bs{\alpha}}} & = & \partial_{M}\psi^{\hat{\bs{\alpha}}}+\Omega_{M\hat{\bs{\gamma}}}\hoch{\hat{\bs{\alpha}}}\psi^{\hat{\bs{\gamma}}}=\\
 & = & \partial_{M}\psi^{\hat{\bs{\alpha}}}+\sum_{\hat{\bs{\gamma}}}\underbrace{(-)^{\hat{\bs{\gamma}}\hat{\bs{\alpha}}+\hat{\bs{\gamma}}}}_{1}\Omega_{M\hat{\bs{\gamma}}}\hoch{\hat{\bs{\alpha}}}\psi^{\hat{\bs{\gamma}}}=\\
 & = & \partial_{M}\psi_{\tilde{\bs{\alpha}}}-\sum_{\tilde{\bs{\gamma}}}(-)^{\tilde{\bs{\gamma}}\tilde{\bs{\alpha}}}\Omega_{M}\hoch{\tilde{\bs{\gamma}}}\tief{\tilde{\bs{\alpha}}}\psi_{\tilde{\bs{\gamma}}}=\\
 & = & \partial_{M}\psi_{\tilde{\bs{\alpha}}}-\Omega_{M}\hoch{\tilde{\bs{\gamma}}}\tief{\tilde{\bs{\alpha}}}\psi_{\tilde{\bs{\gamma}}}\end{eqnarray}
 In order to get back our usual index position for the connection
(first fermionic index down, second up), we finally define \begin{eqnarray}
\Omega_{M\tilde{\bs{\gamma}}}\hoch{\tilde{\bs{\alpha}}} & \equiv & \Omega_{M}\hoch{\tilde{\bs{\alpha}}}\tief{\tilde{\bs{\gamma}}}\quad(=\Omega_{M\hat{\bs{\gamma}}}\hoch{\hat{\bs{\alpha}}})\label{eq:identificationOfconnections}\end{eqnarray}
where the equalities should be understood as graded equalities in
the sense of (\ref{eq:grequalSimple}) on page \pageref{eq:grequalSimple}.
Upon this identification, the action of the covariant derivative on
a lower tilded index takes the usual form $\nabla_{M}\psi_{\tilde{\bs{\alpha}}}=\partial_{M}\psi_{\tilde{\bs{\alpha}}}-\Omega_{M\tilde{\bs{\alpha}}}\hoch{\tilde{\bs{\gamma}}}\psi_{\tilde{\bs{\gamma}}}$.
Equation (\ref{eq:identificationOfconnections}) also guarantees that
the action of a covariant derivative on a lower hatted index becomes
the correct action on the corresponding upper tilded index, i.e. $\nabla_{M}\psi_{\hat{\bs{\alpha}}}=\partial_{M}\psi_{\hat{\bs{\alpha}}}-\Omega_{M\hat{\bs{\alpha}}}\hoch{\hat{\bs{\gamma}}}\psi_{\hat{\bs{\gamma}}}=\partial_{M}\psi^{\tilde{\bs{\alpha}}}+\Omega_{M\tilde{\bs{\gamma}}}\hoch{\tilde{\bs{\alpha}}}\psi^{\tilde{\bs{\gamma}}}=\nabla_{M}\psi^{\tilde{\bs{\alpha}}}$.
Now we are finally able to compare the connections $\Omega_{M\tilde{\bs{\alpha}}}\hoch{\tilde{\bs{\beta}}}$
and $\Omega_{M\bs{\alpha}}\hoch{\bs{\beta}}$ and see whether we can
identify them like in type IIB. First note that like for the symmetry
algebra generators (\ref{eq:coupledGroupTrafosII}) and (\ref{eq:coupledGroupTrafosIII})
themselves, the invariance conditions $\nabla_{M}\gamma_{\bs{\alpha\beta}}^{c}=0$
and $\nabla_{M}\gamma_{\hat{\bs{\alpha}}\hat{\bs{\beta}}}^{c}=0$
determine the spinorial connections to be of the form (see again footnote
\vref{foot:LorentzScaleReason} for a derivation)\begin{eqnarray}
\Omega_{M\bs{\alpha}}\hoch{\bs{\beta}} & = & \tfrac{1}{2}\Omega_{M}^{(D)}\delta_{\bs{\alpha}}\hoch{\bs{\beta}}+\tfrac{1}{4}\Omega_{Mab}^{(L)}\gamma^{ab}\tief{\bs{\alpha}}\hoch{\bs{\beta}}\\
\Omega_{M\hat{\bs{\alpha}}}\hoch{\hat{\bs{\beta}}} & = & \tfrac{1}{2}\Omega_{M}^{(D)}\delta_{\hat{\bs{\alpha}}}\hoch{\hat{\bs{\beta}}}+\tfrac{1}{4}\Omega_{Mab}^{(L)}\gamma^{ab}\tief{\hat{\bs{\alpha}}}\hoch{\hat{\bs{\beta}}}\end{eqnarray}
The Kronecker delta in the second line will be rewritten upon the
index-position shift as $\delta_{\hat{\bs{\alpha}}}\hoch{\hat{\bs{\beta}}}=\delta_{\hat{\bs{\alpha}}}^{\hat{\bs{\beta}}}=\delta_{\tilde{\bs{\beta}}}^{\tilde{\bs{\alpha}}}=-\delta^{\tilde{\bs{\alpha}}}\tief{\tilde{\bs{\beta}}}$.
Finally we make use of the facts that $\gamma^{ab\,\tilde{\bs{\alpha}}}\tief{\tilde{\bs{\beta}}}$
is graded equal to $\gamma^{ab\,}\tief{\tilde{\bs{\beta}}}\hoch{\tilde{\bs{\alpha}}}$
(according to (\ref{eq:antisymSmallGammasEven}) in the appendix),
$\delta^{\tilde{\bs{\alpha}}}\tief{\tilde{\bs{\beta}}}$ is graded
equal to $\delta\tief{\tilde{\bs{\beta}}}\hoch{\tilde{\bs{\alpha}}}$
and of the identification (\ref{eq:identificationOfconnections})
to arrive at \begin{eqnarray}
\Omega_{M\tilde{\bs{\alpha}}}\hoch{\tilde{\bs{\beta}}} & = & -\tfrac{1}{2}\Omega_{M}^{(D)}\delta_{\tilde{\bs{\alpha}}}\hoch{\tilde{\bs{\beta}}}+\tfrac{1}{4}\Omega_{Mab}^{(L)}\gamma^{ab}\tief{\tilde{\bs{\alpha}}}\hoch{\tilde{\bs{\beta}}}\end{eqnarray}
Therefore the tilded indices transform in the same way under Lorentz,
but with opposite sign under scale transformations as the untilded
indices. Only when the scale transformations are fixed, tilded and
untilded indices can be identified. This can be seen differently,
by simply doing the identification and imposing $\nabla_{M}\gamma_{\bs{\alpha\beta}}^{c}=\nabla_{M}\gamma^{c\,\bs{\alpha}\bs{\beta}}=0$
which implies via the Clifford algebra ($\gamma^{(a|\,\bs{\alpha\gamma}}\gamma_{\bs{\gamma\beta}}^{|b)}=-\eta^{ab}\delta^{\bs{\alpha}}\tief{\bs{\beta}}$,
the graded version of (\ref{eq:smallClifford}) of page \pageref{eq:smallClifford})
that $\nabla_{M}\eta^{ab}=0$. But scale transformation do not leave
invariant the Minkowski metric. In summary, keeping the (anyway auxiliary)
scale transformations unfixed seems a bit artificial in type IIA and
is more natural in type IIB.

Let us now proceed with the discussion of the Bianchi identities for
the $H$-field. \lyxline{\normalsize}\vspace{-.25cm}\lyxline{\normalsize}

\vspace{.5cm} \hspace{-.6cm} $\bullet\quad$\underbar{(0,3,1)$\bs{\alpha\beta\gamma}\hat{\bs{\delta}}\leftrightarrow$((0,1,3)$\hat{\bs{\alpha}}\hat{\bs{\beta}}\hat{\bs{\gamma}}\bs{\delta}$):}\begin{eqnarray}
0 & \stackrel{!}{=} & \gemnabla_{[\bs{\alpha}}H_{\bs{\beta\gamma}\hat{\bs{\delta}}]}+3\gemT_{[\bs{\alpha\beta}|}\hoch{C}H_{C|\bs{\gamma}\hat{\bs{\delta}}]}=0\quad(\mbox{due to }(\ref{eq:Y-constrIII}),(\ref{eq:Y-constrV}),\mbox{ and }(\ref{eq:nilpotency-constraint-onH}))\end{eqnarray}
No new constraints from this one.

\paragraph{Remark:}

As in the above equation we will make use of all the constraints that
we have derived from the BRST invariance and nilpotency. As it is
cumbersome to specify each time explicitely which constraint we have
used, we will not do it everywhere. Every constraint that we use without
referring to its equation number will be taken from (\ref{eq:Y-constrI})-(\ref{eq:Y-constrV})
(page \pageref{eq:Y-constrI}), (\ref{eq:holConstrI})-(\ref{eq:holConstrXIIa})
(page \pageref{eq:holConstrI}), (\ref{eq:nilpotency-constraint-onH})-(\ref{eq:nilpotency-constraint-onR-hat})
(page \pageref{eq:nilpotency-constraint-onH}) and (\ref{eq:convConstr})
on page \pageref{eq:convConstr}. These are all the framed equations.
However, to the newly gained constraints within this local appendix
(which will be framed as well) we will refer explicitely.\vspace{.3cm}\\
$\bullet\quad$\underbar{(0,2,2)$\bs{\alpha\beta}\hat{\bs{\gamma}}\hat{\bs{\delta}}$:}\begin{eqnarray}
0 & \stackrel{!}{=} & \gemnabla_{[\bs{\alpha}}H_{\bs{\beta}\hat{\bs{\gamma}}\hat{\bs{\delta}}]}+3\gemT_{[\bs{\alpha\beta}|}\hoch{C}H_{C|\hat{\bs{\gamma}}\hat{\bs{\delta}}]}=\\
 & \propto & \gemT_{\bs{\alpha\beta}}\hoch{c}H_{c\hat{\bs{\gamma}}\hat{\bs{\delta}}}+\gemT_{\hat{\bs{\gamma}}\hat{\bs{\delta}}}\hoch{c}H_{c\bs{\alpha\beta}}=\\
 & \propto & \gamma_{\bs{\alpha\beta}}^{a}f_{a}\hoch{c}\gamma_{\hat{\bs{\gamma}}\hat{\bs{\delta}}}^{b}\hat{f}_{bc}-\gamma_{\hat{\bs{\gamma}}\hat{\bs{\delta}}}^{b}\hat{f}_{b}\hoch{c}\gamma_{\bs{\alpha\beta}}^{a}f_{ac}=\\
 & = & \gamma_{\bs{\alpha\beta}}^{a}\gamma_{\hat{\bs{\gamma}}\hat{\bs{\delta}}}^{b}\left(f_{a}\hoch{c}\hat{f}_{bc}-\hat{f}_{b}\hoch{c}f_{ac}\right)=0\end{eqnarray}
$\bullet\quad$\underbar{(1,3,0)$\bs{\alpha\beta\gamma}d\leftrightarrow$((1,0,3)$\hat{\bs{\alpha}}\hat{\bs{\beta}}\hat{\bs{\gamma}}d$):}%
\footnote{\index{footnote!\thefoot. about $T_{\bs{\alpha}(c"|"d)}$}Remember
$\check{T}_{\bs{\alpha}(c|d)}=-\frac{1}{2}\checkcovPhi{\bs{\alpha}}G_{cd}=\frac{1}{2}E_{\bs{\alpha}}\hoch{M}(\check{\Omega}_{M}^{(D)}-\partial_{M}\Phi)G_{cd}$.
This can be reformulated as a condition on the vielbein only: \begin{eqnarray*}
\check{T}_{\bs{\alpha}c|d} & = & (\de E^{e})_{\bs{\alpha}c}G_{ed}+\underbrace{\check{\Omega}_{[\bs{\alpha}c]}\hoch{e}G_{ed}}_{\equiv\check{\Omega}_{[\bs{\alpha}c]|d}}\\
\check{T}_{\bs{\alpha}(c|d)} & = & (\de E^{e})_{\bs{\alpha}(c}G_{d)e}+\frac{1}{2}\check{\Omega}_{\bs{\alpha}(c|d)}=\\
 & = & (\de E^{e})_{\bs{\alpha}(c}G_{d)e}+\frac{1}{2}\check{\Omega}_{\bs{\alpha}}^{(D)}G_{cd}\\
\dann(\de E)_{\bs{\alpha}(c|d)}\equiv(\de E^{e})_{\bs{\alpha}(c}G_{d)e} & = & -\frac{1}{2}E_{\bs{\alpha}}\hoch{M}\partial_{M}\Phi\, G_{cd}\end{eqnarray*}
Reparametrizing $\tilde{E}_{M}\hoch{A}\equiv e^{\Phi}E_{M}\hoch{A}$,
this can be rewritten as \begin{eqnarray*}
(\de\tilde{E})_{\bs{\alpha}(c|d)} & = & \left(E_{[\bs{\alpha}|}\hoch{M}\partial_{M}\Phi\cdot e^{\Phi}G_{|c]d}-\frac{1}{2}e^{\Phi}E_{\bs{\alpha}}\hoch{M}\partial_{M}\Phi\, G_{cd}\right)=0\quad\mbox{or }\tilde{\check{T}}_{\bs{\alpha}(c|d)}=0\end{eqnarray*}
in accordance with \cite{Berkovits:2001ue}.$\qquad\fussend$ %
}\begin{eqnarray}
0 & \stackrel{!}{=} & \gemnabla_{[\bs{\alpha}}H_{\bs{\beta\gamma}d]}+3\gemT_{[\bs{\alpha\beta}|}\hoch{C}H_{C|\bs{\gamma}d]}=\\
 & = & \frac{3}{4}\gemnabla_{[\bs{\alpha}}H_{\bs{\beta\gamma}]d}+\frac{3}{2}\check{T}_{[\bs{\beta}|d}\hoch{c}H_{c|\bs{\gamma\alpha}]}=\\
 & = & -\frac{1}{2}\gemnabla_{[\bs{\alpha}}(\gamma_{\bs{\beta\gamma}]}^{c}G_{cd})-\check{T}_{[\bs{\beta}|d|c}\gamma_{\:|\bs{\gamma\alpha}]}^{c}=\\
 & \stackrel{\check{\Omega}_{Ma}\hoch{b}=\Omega_{Ma}\hoch{b}}{=} & -\gamma_{[\bs{\beta\gamma}}^{c}\Big(\covPhi{\bs{\alpha}]}\cdot G_{cd}+\underbrace{T_{\bs{\alpha}]d|c}}_{T_{\bs{\alpha}][d|c]}-\frac{1}{2}\covPhi{\bs{\alpha}}\lqn{{\scriptstyle \cdot G_{dc}=-T_{\bs{\alpha}]c|d}-\covPhi{\bs{\alpha}}\cdot G_{dc}}}}\Big)=\\
 & = & \gamma_{[\bs{\beta\gamma}}^{c}T_{\bs{\alpha}]c|d}\label{eq:(1,3,0)}\end{eqnarray}
In the fourth line we made the choice of $\check{\Omega}_{Ma}\hoch{b}$
in such a way that $\gemnabla_{\bs{\alpha}}\gamma_{\bs{\beta}\bs{\gamma}}^{c}=0$.
In the following calculations we will use a lot of gamma-matrix identities
from appendix \ref{cha:Gamma-Matrices} where we did not use graded
conventions. We will therefore temporarily switch to non-graded conventions
(or equivalently perform a grading shift of the fermionic indices). 

As a first step to solve the constraint (\ref{eq:(1,3,0)}), let us
contract it with $\gamma_{a}^{\alpha\beta}$:\begin{eqnarray}
0 & \stackrel{!}{=} & \gamma_{a}^{\alpha\beta}\gamma_{\alpha\beta}^{c}T_{\gamma c|d}+\gamma_{a}^{\alpha\beta}\gamma_{\gamma\alpha}^{c}T_{\beta c|d}+\gamma_{a}^{\alpha\beta}\gamma_{\beta\gamma}^{c}T_{\alpha c|d}=\\
 & \stackrel{(\ref{eq:smallClifford}),(\ref{eq:antisymSmallGammasEven})}{=} & 16T_{\gamma a|d}+2\left(\delta_{a}^{c}\delta_{\gamma}^{\beta}+\gamma^{c}\tief{a\,\gamma}\hoch{\beta}\right)T_{\beta c|d}=\\
\dann9T_{\gamma a|d} & = & \gamma_{a}\hoch{c}\tief{\gamma}\hoch{\beta}T_{\beta c|d}\label{eq:puh}\end{eqnarray}
Although the contraction with $\gamma_{a}^{\alpha\beta}$ looks like
a projection, the new equation (\ref{eq:puh}) still contains all
the information of (\ref{eq:(1,3,0)}) (in the nongraded version,
the graded antisymmetrization becomes an ordinary symmetrization):\begin{eqnarray}
\gamma_{(\beta\gamma}^{c}T_{\alpha)c|d} & \stackrel{(\ref{eq:puh})}{=} & \frac{1}{9}\gamma_{(\beta\gamma|}^{c}\gamma_{c}\hoch{e}\tief{|\alpha)}\hoch{\delta}T_{\delta e|d}=\\
 & \stackrel{(\ref{eq:smallClifford}),(\ref{eq:antisymSmallGammasEven})}{=} & \frac{1}{9}\gamma_{(\beta\gamma|}^{c}\left(\gamma_{c|\alpha)}\hoch{\eps}\gamma^{e}\tief{\eps}\hoch{\delta}-\delta_{c}^{e}\delta_{|\alpha)}^{\delta}\right)T_{\delta e|d}=\\
 & \stackrel{(\ref{eq:LittleFierz})}{=} & -\frac{1}{9}\gamma_{(\beta\gamma|}^{c}T_{|\alpha)c|d}\end{eqnarray}
Comparing the first and the last line leads back to (\ref{eq:(1,3,0)}).
This was just to argue that we can forget now about (\ref{eq:(1,3,0)}),
and take (\ref{eq:puh}) as new starting point. Remember that we have
already a constraint for the symmetrized part (in $c$ and $d$) of
$T_{\alpha c|d}$ and let let us in addition introduce a temporary
notation for the yet unknown antisymmetrized part: \begin{equation}
T_{\alpha(c|d)}=-\frac{1}{2}\covPhi{\alpha}G_{cd},\qquad T_{\alpha[c|d]}\equiv\dot{T}_{\alpha cd}\end{equation}
Now we split (\ref{eq:puh}) into its symmetric and its antisymmetric
part in $a$ and $d$ (the symmetric part is multiplied by (-2) for
convenience):%
\footnote{\index{$\gamma^{ab}$@$\tilde\gamma^{ab}\tief\alpha\hoch\beta$}\index{footnote!\thefoot. scaling weight; $\tilde\gamma_{a\alpha\beta}$}The
tilde on gamma matrices or antisymmetriced products between them just
takes into account the correct scaling weight: $\gamma_{\alpha\beta}^{a}$
is invariant under scale transformations, if the transformations acting
on bosonic and fermionic indices are coupled as in (\ref{eq:coupledGroupTrafosI})-(\ref{eq:coupledGroupTrafosIII}),
i.e. if the fermionic scale transformation has an extra factor $\frac{1}{2}$.
The bosonic metric $G_{ab}\equiv e^{2\Phi}\eta_{ab}$ and its inverse
$G^{ab}\equiv e^{-2\Phi}\eta^{ab}$, used to lower and raise bosonic
flat indices, however, are not scale invariant. Lowering an index
of the gamma-matrix yields $\tilde{\gamma}_{a\,\alpha\beta}\equiv G_{ab}\gamma_{\alpha\beta}^{b}=e^{2\Phi}\gamma_{a\,\alpha\beta}$.
The reason for the tilde is thus only to indicate that the gamma matrix
is not the numerical one but has a Weyl factor in it which corresponds
to the weight indicated by the index structure. Similarly we have
\[
\tilde{\gamma}^{ab}\tief{\alpha}\hoch{\beta}\equiv e^{-2\Phi}\gamma^{ab}\tief{\alpha}\hoch{\beta}\qquad\fussend\]
}\begin{eqnarray}
9\covPhi{\gamma}G_{ad} & = & 2\tilde{\gamma}_{c(a|}\tief{\gamma}\hoch{\beta}\dot{T}_{\beta}\hoch{c}\tief{|d)}=\tilde{\gamma}_{ca}\tief{\gamma}\hoch{\beta}\dot{T}_{\beta}\hoch{c}\tief{d}+\tilde{\gamma}_{cd}\tief{\gamma}\hoch{\beta}\dot{T}_{\beta}\hoch{c}\tief{a}\frem{\dann45\covPhi{\delta}=\tilde{\gamma}^{ad}\tief{\delta}\hoch{\gamma}\dot{T}_{\gamma}\tief{ad}}\label{eq:puhI}\\
9\dot{T}_{\gamma ad} & = & \tilde{\gamma}_{[a|c}\tief{\gamma}\hoch{\beta}\dot{T}_{\beta}\hoch{c}\tief{|d]}-\frac{1}{2}\tilde{\gamma}_{ad}\tief{\gamma}\hoch{\beta}\covPhi{\beta}\label{eq:puhII}\end{eqnarray}
In order to solve this kind of equations, it always helps to take
traces (we will use the trace of (\ref{eq:puhI}) soon) and to contract
with several combinations of $\gamma$-matrices. Here it turns out
to be useful to contract (\ref{eq:puhI}) with $\tilde{\gamma}^{ab}\tief{\alpha}\hoch{\gamma}$.
The antisymmetrization in the bosonic indices of the result will produce
a term similar to the one in (\ref{eq:puhII}), s.th. the equations
can then be combined. But let us first perform the contraction. We
will use the following gamma-matrix identities (see (\ref{eq:gammaIIgammaII})
on page \pageref{eq:gammaIIgammaII}):\begin{eqnarray}
\tilde{\gamma}^{ab}\tilde{\gamma}_{ca} & = & \delta_{a}^{a}\gamma^{b}\tief{c}-\delta_{a}^{b}\gamma^{a}\tief{c}-\delta_{c}^{a}\gamma^{b}\tief{a}+\delta_{a}^{a}\delta_{c}^{b}\one-\delta_{c}^{a}\delta_{a}^{b}\one=8\gamma^{b}\tief{c}+9\delta_{c}^{b}\one\label{eq:gammaabca}\\
\tilde{\gamma}^{ab}\tilde{\gamma}_{cd} & = & \gamma^{ab}\tief{cd}+\delta_{c}^{b}\gamma^{a}\tief{d}+\delta_{d}^{a}\gamma^{b}\tief{c}-\delta_{d}^{b}\gamma^{a}\tief{c}-\delta_{c}^{a}\gamma^{b}\tief{d}+\delta_{d}^{a}\delta_{c}^{b}\one-\delta_{c}^{a}\delta_{d}^{b}\one\end{eqnarray}
The $\gamma^{[4]}$ part in the second equation could be removed
by taking a symmetrization. This, however, would in the end only lead
back to (\ref{eq:puhI}). Instead, note that the same $\gamma^{[4]}$
is produced in the product $\gamma_{d}\hoch{b}\gamma_{c}\hoch{a}$.
And this combination is more useful, as we can then apply $\gamma_{c}\hoch{a}\tief{\alpha}\hoch{\beta}T_{\beta}\hoch{c}\tief{a}\stackrel{(\ref{eq:puhI})}{=}45\covPhi{\alpha}$:\begin{eqnarray}
-\gamma_{d}\hoch{b}\gamma_{c}\hoch{a} & = & \gamma^{ab}\tief{cd}+\delta_{c}^{b}\gamma^{a}\tief{d}-\delta_{d}^{a}\gamma^{b}\tief{c}+G^{ba}\tilde{\gamma}_{dc}+G_{cd}\tilde{\gamma}^{ba}-\delta_{d}^{a}\delta_{c}^{b}\one+G_{cd}G^{ba}\one\\
\dann\tilde{\gamma}^{ab}\tilde{\gamma}_{cd} & = & -\gamma_{d}\hoch{b}\gamma_{c}\hoch{a}+2\delta_{d}^{a}\gamma^{b}\tief{c}-\delta_{d}^{b}\gamma^{a}\tief{c}-\delta_{c}^{a}\gamma^{b}\tief{d}-G^{ba}\tilde{\gamma}_{dc}-G_{cd}\tilde{\gamma}^{ba}+\left(2\delta_{d}^{a}\delta_{c}^{b}-\delta_{c}^{a}\delta_{d}^{b}-G_{cd}G^{ba}\right)\one\label{eq:gammaabcd}\end{eqnarray}
The contraction of (\ref{eq:puhI}) with $\tilde{\gamma}^{ab}\tief{\alpha}\hoch{\gamma}$
then yields (using (\ref{eq:gammaabca}), (\ref{eq:gammaabcd}) and
$\gamma_{c}\hoch{a}\tief{\alpha}\hoch{\beta}T_{\beta}\hoch{c}\tief{a}\stackrel{(\ref{eq:puhI})}{=}45\covPhi{\alpha}$):\begin{eqnarray}
\lqn{9\tilde{\gamma}_{d}\hoch{b}\tief{\alpha}\hoch{\gamma}\covPhi{\gamma}\stackrel{!}{=}}\nonumber \\
 & \stackrel{!}{=} & \left(8\gamma^{b}\tief{c\alpha}\hoch{\beta}+9\delta_{c}^{b}\delta_{\alpha}^{\beta}\right)\dot{T}_{\beta}\hoch{c}\tief{d}+\\
 &  & +\left(-\gamma_{d}\hoch{b}\gamma_{c}\hoch{a}+2\delta_{d}^{a}\gamma^{b}\tief{c}-\delta_{d}^{b}\gamma^{a}\tief{c}-\delta_{c}^{a}\gamma^{b}\tief{d}-G^{ba}\tilde{\gamma}_{dc}-G_{cd}\tilde{\gamma}^{ba}+\left(2\delta_{d}^{a}\delta_{c}^{b}-\delta_{c}^{a}\delta_{d}^{b}-G_{cd}G^{ba}\right)\one\right)_{\alpha}\hoch{\beta}\dot{T}_{\beta}\hoch{c}\tief{a}=\qquad\\
 & = & 8\gamma^{b}\tief{c\alpha}\hoch{\beta}\dot{T}_{\beta}\hoch{c}\tief{d}+9\dot{T}_{\alpha}\hoch{b}\tief{d}+\nonumber \\
 &  & -45\gamma_{d}\hoch{b}\tief{\alpha}\hoch{\beta}\covPhi{\beta}+2\gamma^{b}\tief{c\alpha}\hoch{\beta}\dot{T}_{\beta}\hoch{c}\tief{d}+45\delta_{d}^{b}\covPhi{\alpha}-\gamma^{b}\tief{d}\underbrace{\dot{T}\hoch{a}\tief{a}}_{=0}+\nonumber \\
 &  & -\tilde{\gamma}_{dc\,\alpha}\hoch{\beta}\dot{T}_{\beta}\hoch{cb}-\tilde{\gamma}^{ba}\tief{\alpha}\hoch{\beta}\dot{T}_{\beta da}+2\dot{T}_{\alpha}\hoch{b}\tief{d}-\delta_{d}^{b}\underbrace{\dot{T}_{\alpha}\hoch{c}\tief{c}}_{=0}-\dot{T}_{\alpha d}\hoch{b}=\\
 & = & 45\gamma^{b}\tief{d\,\alpha}\hoch{\gamma}\covPhi{\gamma}+45\delta_{d}^{b}\covPhi{\alpha}+12\dot{T}_{\alpha}\hoch{b}\tief{d}+\nonumber \\
 &  & +10\gamma^{b}\tief{c\alpha}\hoch{\beta}\dot{T}_{\beta}\hoch{c}\tief{d}+\tilde{\gamma}^{bc}\tief{\alpha}\hoch{\beta}\dot{T}_{\beta cd}-\tilde{\gamma}_{dc\,\alpha}\hoch{\beta}\dot{T}_{\beta}\hoch{cb}\end{eqnarray}
Putting everything on one side and taking the antisymmetric part (in
b,d) of this equation leads to\begin{eqnarray}
0 & \stackrel{!}{=} & 54\tilde{\gamma}_{bd}\tief{\alpha}\hoch{\gamma}\covPhi{\gamma}+12\dot{T}_{\alpha bd}+12\tilde{\gamma}_{[b|c\,\alpha}\hoch{\beta}\dot{T}_{\beta}\hoch{c}\tief{|d]}=\\
 & \stackrel{(\ref{eq:puhII})}{=} & 54\tilde{\gamma}_{bd}\tief{\alpha}\hoch{\gamma}\covPhi{\gamma}+12\dot{T}_{\alpha bd}+12\left(9\dot{T}_{\alpha bd}+\frac{1}{2}\tilde{\gamma}_{bd\alpha}\hoch{\beta}\covPhi{\beta}\right)\\
\dann\dot{T}_{\alpha bd} & = & -\frac{1}{2}\tilde{\gamma}_{bd}\tief{\alpha}\hoch{\gamma}\covPhi{\gamma}\end{eqnarray}
Let us switch back to the graded conventions. After this somewhat
tedious calculation, we only need to combine this antisymmetric part
($T_{\bs{\alpha}[b|d]}\equiv\dot{T}_{\bs{\alpha}bd}$) with the symmetric
one $T_{\bs{\alpha}(b|d)}=-\frac{1}{2}\covPhi{\bs{\alpha}}G_{bd}$,
in order to end up with the final result for the Bianchi identity
(\ref{eq:(1,3,0)})\begin{equation}
\boxed{T_{\bs{\beta}c}\hoch{a}=-\frac{1}{2}\covPhi{\bs{\beta}}\delta_{c}\hoch{a}-\frac{1}{2}\gamma_{c}\hoch{a}\tief{\bs{\beta}}\hoch{\bs{\gamma}}\covPhi{\bs{\gamma}}}\label{eq:Tbetaca}\end{equation}
Via the left-right symmetry, we get correspondingly \begin{equation}
\boxed{\hat{T}_{\hat{\bs{\beta}}c}\hoch{a}=-\frac{1}{2}\hatcovPhi{\hat{\bs{\beta}}}\delta_{c}\hoch{a}-\frac{1}{2}\gamma_{c}\hoch{a}\tief{\hat{\bs{\beta}}}\hoch{\hat{\bs{\gamma}}}\hatcovPhi{\hat{\bs{\gamma}}}}\label{eq:Thatbetaca}\end{equation}
\\
$\bullet\quad$\underbar{(1,2,1)$\bs{\alpha\beta}\hat{\bs{\gamma}}d\leftrightarrow$((1,1,2)$\hat{\bs{\alpha}}\hat{\bs{\beta}}\bs{\gamma}d$):}\begin{eqnarray}
0 & \stackrel{!}{=} & \gemnabla_{[\bs{\alpha}}H_{\bs{\beta}\hat{\bs{\gamma}}d]}+3\gemT_{[\bs{\alpha}\bs{\beta}|}\hoch{E}H_{E|\hat{\bs{\gamma}}d]}=\\
 & = & \frac{1}{4}\gemnabla_{\hat{\bs{\gamma}}}H_{\bs{\alpha}\bs{\beta}d}+\frac{1}{2}\hat{T}_{\bs{\alpha}\bs{\beta}}\hoch{\hat{\bs{\eps}}}H_{\hat{\bs{\eps}}\hat{\bs{\gamma}}d}+\frac{1}{2}\check{T}_{\hat{\bs{\gamma}}d}\hoch{e}H_{e\bs{\alpha}\bs{\beta}}=\\
 & = & -\frac{1}{6}\gemnabla_{\hat{\bs{\gamma}}}(\gamma_{\bs{\alpha\beta}}^{c}f_{cd})-\frac{1}{3}\check{T}_{\hat{\bs{\gamma}}d}\hoch{e}\gamma_{\bs{\alpha}\bs{\beta}}^{c}f_{ce}=\\
 & \ous{f_{ce}=G_{ce}\,(\ref{eq:(0,4,0)'})}{=}{\check{\Omega}=\Omega} & \qquad-\frac{1}{3}\gamma_{\bs{\alpha\beta}}^{c}\underbrace{\left(\covPhi{\hat{\bs{\gamma}}}G_{cd}+T_{\hat{\bs{\gamma}}d|c}\right)}_{-T_{\hat{\bs{\gamma}}c|d}\,(\ref{eq:Y-constrII})}\label{eq:(1,2,1)}\end{eqnarray}
\begin{equation}
\stackrel{(\ref{eq:T^{c}-sym'})}{\dann}\boxed{T_{\hat{\bs{\gamma}}c}\hoch{d}=0,\quad\covPhi{\hat{\bs{\gamma}}}=0}\label{eq:T^{c}undOmega}\end{equation}
Likewise we have \begin{equation}
\boxed{\hat{T}_{\bs{\alpha}b}\hoch{c}=0,\qquad\hatcovPhi{\bs{\gamma}}=0}\label{eq:hatT^{c}und-hatOmega}\end{equation}
These results can also be used to determine $\covPhi{a}$:\begin{eqnarray*}
\hat{\nabla}_{[\bs{\alpha}}\underbrace{\hatcovPhi{\bs{\beta}]}}_{=0} & = & -\hat{T}_{\bs{\alpha\beta}}\hoch{C}\hatcovPhi{C}-\hat{F}_{\bs{\alpha\beta}}^{(D)}=\\
 & = & -\underbrace{\hat{T}_{\bs{\alpha\beta}}\hoch{c}}_{=\gamma_{\bs{\alpha\beta}}^{c}}\hatcovPhi{c}-\hat{T}_{\bs{\alpha\beta}}\hoch{\bs{\gamma}}\underbrace{\hatcovPhi{\bs{\gamma}}}_{=0}-\underbrace{\hat{T}_{\bs{\alpha\beta}}\hoch{\hat{\bs{\gamma}}}}_{=0}\hatcovPhi{\hat{\bs{\gamma}}}-\underbrace{\hat{F}_{\bs{\alpha\beta}}^{(D)}}_{=0\:(\ref{eq:F-Dil-constrII'})}\end{eqnarray*}
The above equation and its hatted counterpart imply \begin{equation}
\boxed{\hatcovPhi{c}=\covPhi{c}=0}\label{eq:Omegaa}\end{equation}
We can play this game once more and consider the commutator\rem{%
\footnote{From (\ref{eq:F-Dil-constrI}) instead, we get\begin{eqnarray*}
\hat{F}_{\bs{\alpha}c}^{(D)} & = & \tilde{\nabla}_{[\bs{\alpha}}\hat{\Omega}_{c]}+\tilde{T}_{\bs{\alpha}c}\hoch{D}\hat{\Omega}_{D}=\\
 & = & \tilde{\nabla}_{[\bs{\alpha}}\tilde{\nabla}_{c]}\Phi+\tilde{T}_{\bs{\alpha}c}\hoch{D}\hat{\Omega}_{D}=\\
 & = & \tilde{T}_{\bs{\alpha}c}\hoch{D}\left(-\tilde{\nabla}_{D}\Phi+\hat{\Omega}_{D}\right)=\\
 & = & \hat{T}_{\bs{\alpha}c}\hoch{\hat{\bs{\delta}}}\left(-\tilde{\nabla}_{\hat{\bs{\delta}}}\Phi+\hat{\Omega}_{\hat{\bs{\delta}}}\right)=\\
 & = & \gamma_{c\,\bs{\alpha\gamma}}\RR^{\bs{\gamma}\hat{\bs{\delta}}}\left(-\tilde{\nabla}_{\hat{\bs{\delta}}}\Phi+\hat{\Omega}_{\hat{\bs{\delta}}}\right)=\\
 & \stackrel{!}{=} & -\frac{1}{8}\gamma_{c\,\bs{\alpha\delta}}\tilde{\nabla}_{\hat{\bs{\alpha}}}\RR^{\bs{\delta}\hat{\bs{\alpha}}}\end{eqnarray*}
 \[
\boxed{\tilde{\nabla}_{\hat{\bs{\alpha}}}\RR^{\bs{\delta}\hat{\bs{\alpha}}}=8\RR^{\bs{\delta}\hat{\bs{\delta}}}\left(\tilde{\nabla}_{\hat{\bs{\delta}}}\Phi-\hat{\Omega}_{\hat{\bs{\delta}}}\right)}\qquad\fussend\]
}} \begin{eqnarray}
\underbrace{\hat{\nabla}_{[\bs{\alpha}}\hatcovPhi{b]}}_{=0} & = & -\hat{T}_{\bs{\alpha}b}\hoch{C}\hatcovPhi{C}-\hat{F}_{\bs{\alpha}b}^{(D)}=\\
 & = & -\hat{T}_{\bs{\alpha}b}\hoch{c}\underbrace{\hatcovPhi{c}}_{=0}-\hat{T}_{\bs{\alpha}b}\hoch{\bs{\gamma}}\underbrace{\hatcovPhi{\bs{\gamma}}}_{=0}-\underbrace{\hat{T}_{\bs{\alpha}b}\hoch{\hat{\bs{\gamma}}}}_{=\tilde{\gamma}_{b\bs{\alpha\delta}}\RR^{\bs{\delta}\hat{\bs{\gamma}}}}\hatcovPhi{\hat{\bs{\gamma}}}-\hat{F}_{\bs{\alpha}b}^{(D)}\end{eqnarray}
Due to (\ref{eq:F-Dil-constrI'}) we have $\hat{F}_{\bs{\alpha}b}^{(D)}=-\frac{1}{8}\tilde{\gamma}_{b\,\bs{\alpha\delta}}\gemnabla_{\hat{\bs{\gamma}}}\RR^{\bs{\delta}\hat{\bs{\gamma}}}$
and therefore \begin{equation}
\boxed{\gemnabla_{\hat{\bs{\alpha}}}\RR^{\bs{\delta}\hat{\bs{\alpha}}}=8\RR^{\bs{\delta}\hat{\bs{\beta}}}\hatcovPhi{\hat{\bs{\beta}}}}\label{eq:RR-eoms}\end{equation}
The hatted version of this equation reads\begin{equation}
\boxed{\gemnabla_{\bs{\alpha}}\RR^{\bs{\alpha}\hat{\bs{\delta}}}=8\RR^{\bs{\beta}\hat{\bs{\delta}}}\covPhi{\bs{\beta}}}\label{eq:RR-eomsII}\end{equation}
\\
$\bullet\quad$\underbar{(2,2,0)$ab\bs{\alpha}\bs{\beta}\leftrightarrow$((2,0,2)$ab\hat{\bs{\alpha}}\hat{\bs{\beta}}$):}%
\footnote{\index{footnote!\thefoot. combinatorical remark}Combinatorically
$[ab][\bs{\alpha}\bs{\beta}]$ arises 4 times in all 24 possibilities$\dann\frac{4}{24}=\frac{1}{6}\qquad\fussend$%
}\begin{eqnarray}
0 & \stackrel{!}{=} & \gemnabla_{[a}H_{b\bs{\alpha\beta}]}+3\gemT_{[ab|}\hoch{C}H_{C|\bs{\alpha\beta}]}=\\
 & = & \frac{1}{2}\gemnabla_{[a}H_{b]\bs{\alpha\beta}}+\frac{1}{2}\check{T}_{ab}\hoch{c}H_{c\bs{\alpha\beta}}+\frac{1}{2}\check{T}_{\bs{\alpha\beta}}\hoch{c}H_{cab}=\\
 & = & \frac{1}{2}\gemnabla_{[a|}(-\frac{2}{3}\gamma_{\bs{\alpha\beta}}^{c}f_{c|b]})-\frac{1}{2}\gamma_{\bs{\alpha\beta}}^{d}\left(\frac{2}{3}\check{T}_{ab}\hoch{c}f_{dc}-f_{d}\hoch{c}H_{cab}\right)=\\
 & \ous{f_{cb}=G_{cb}}{=}{\check{\Omega}=\Omega} & \qquad-\frac{1}{3}\gamma_{\bs{\alpha\beta}}^{d}\big(T_{ab|d}-\frac{3}{2}H_{dab}+2\underbrace{\covPhi{[a}}_{0\,(\ref{eq:Omegaa})}G_{b]d}\big)\end{eqnarray}
Using $\frac{1}{16}\gamma_{\bs{\alpha\beta}}^{d}\gamma_{c}^{\bs{\alpha\beta}}=\delta_{c}^{d}$
we get\begin{equation}
\boxed{T_{ab|d}=\frac{3}{2}H_{abd}}\label{eq:(2,2,0)}\end{equation}
Likewise we have%
\footnote{\index{footnote!\thefoot. some consistency check}As a consitency
check, we compute the BI's for the index-combination $ab\hat{\bs{\alpha}}\hat{\bs{\beta}}$
explicitely with $T$ (not $\hat{T}$):\begin{eqnarray*}
0 & \stackrel{!}{=} & \nabla_{[a}H_{b\hat{\bs{\alpha}}\hat{\bs{\beta}}]}+3T_{[ab|}\hoch{C}H_{C|\hat{\bs{\alpha}}\hat{\bs{\beta}}]}=\\
 & = & \frac{1}{2}\nabla_{[a}H_{b]\hat{\bs{\alpha}}\hat{\bs{\beta}}}+\frac{1}{2}T_{ab}\hoch{c}H_{c\hat{\bs{\alpha}}\hat{\bs{\beta}}}+\frac{1}{2}T_{\hat{\bs{\alpha}}\hat{\bs{\beta}}}\hoch{c}H_{cab}=\\
 & = & \frac{1}{2}\nabla_{[a|}(\frac{2}{3}\gamma_{\hat{\bs{\alpha}}\hat{\bs{\beta}}}^{c}\hat{f}_{c|b]})+\frac{1}{2}\gamma_{\hat{\bs{\alpha}}\hat{\bs{\beta}}}^{d}\left(\frac{2}{3}T_{ab}\hoch{c}\hat{f}_{dc}+\hat{f}_{d}\hoch{c}H_{cab}\right)=\\
 & \stackrel{\hat{f}_{cb}=G_{cb}}{=} & \frac{1}{3}\nabla_{[a|}(\gamma_{\hat{\bs{\alpha}}\hat{\bs{\beta}}}^{c}G_{c|b]})+\frac{1}{3}\gamma_{\hat{\bs{\alpha}}\hat{\bs{\beta}}}^{d}\left(T_{ab|d}+\frac{3}{2}H_{dab}\right)=\\
 & = & \frac{1}{3}\nabla_{[a|}(\gamma_{\hat{\bs{\alpha}}\hat{\bs{\beta}}}^{c})G_{c|b]}+\frac{1}{3}\gamma_{\hat{\bs{\alpha}}\hat{\bs{\beta}}}^{d}\left(T_{ab|d}+\frac{3}{2}H_{dab}+2\covPhi{a}G_{b]d}\right)=\\
 & = & \frac{1}{3}\gamma_{\hat{\bs{\alpha}}\hat{\bs{\beta}}}^{d}\Big(T_{ab|d}+\frac{3}{2}H_{dab}\underbrace{-\Delta_{[a|d||b]}}_{+\Delta_{[ab]|d}-2\Delta_{[a}G_{b]d}}\Big)=\\
 & = & \frac{1}{3}\gamma_{\hat{\bs{\alpha}}\hat{\bs{\beta}}}^{d}\Big(\hat{T}_{ab|d}+\frac{3}{2}H_{dab}\Big)\qquad\fussend\end{eqnarray*}
}\begin{equation}
\boxed{\hat{T}_{ab|d}=-\frac{3}{2}H_{abd}}\label{eq:(2,0,2)}\end{equation}

\vspace{.5cm}

\lyxline{\normalsize}\vspace{-.25cm}\lyxline{\normalsize}

\subsubsection*{Intermezzo\index{intermezzo!difference tensor} on the difference
tensor}

\label{Intermezzo:differenceTensor}\index{difference tensor!intermezzo on $\sim$}We
have finally obtained the last ingredient to calculate the explicit
form of the difference tensor (\ref{eq:BI:DifferenceTensor}) between
the connections $\hat{\Omega}$ and $\Omega$. The difference tensor
is block-diagonal like the connections and we have in particular $\Delta_{[\bs{\mc{A}}\bs{\mc{B}}]}\hoch{c}=0$.
Using $\Delta_{[AB]}\hoch{c}=\hat{T}_{AB}\hoch{c}-T_{AB}\hoch{c}$
with $\hat{T}_{ab|c}=\frac{3}{2}H_{abc}$, $T_{ab|c}=-\frac{3}{2}H_{abc}$
and $\hat{T}_{\bs{\alpha}b|c}=T_{\hat{\bs{\alpha}}b|c}=0$, we can
give a simple expression for $\Delta_{[AB]}\hoch{c}$. At the same
time we have information about the difference tensor when it is symmetrized
in its last two (bosonic) indices: $\Delta_{A(b|c)}=\bigl(\hat{\Omega}_{A}^{(D)}-\Omega_{A}^{(D)}\bigr)G_{bc}=\bigl(\covPhi{A}-\hatcovPhi{A}\bigr)G_{bc}$
with $\covPhi{\hat{\bs{\alpha}}}=\hatcovPhi{\bs{\alpha}}=\covPhi{a}=\hatcovPhi{a}=0$.
We can thus write down explicitely the antisymmetrized (in the first
two indices) and the symmetrized (in the last two indices) difference
tensor between left and right-mover connection\index{$\Delta_{AB}\hoch{C}$}\begin{eqnarray}
\Delta_{[AB]}\hoch{c} & = & \left(\begin{array}{ccc}
-3H_{ab}\hoch{c} & -T_{a\bs{\beta}}\hoch{c} & \hat{T}_{a\hat{\bs{\beta}}}\hoch{c}\\
-T_{\bs{\alpha}b}\hoch{c} & 0 & 0\\
\hat{T}_{\hat{\bs{\alpha}}b}\hoch{c} & 0 & 0\end{array}\right)\\
\Delta_{a(b|c)} & = & 0,\quad\Delta_{\bs{\alpha}(b|c)}=\covPhi{\bs{\alpha}}G_{bc},\quad\Delta_{\hat{\bs{\alpha}}(b|c)}=-\hatcovPhi{\hat{\bs{\alpha}}}G_{bc}\label{eq:DeltaSymInLast}\end{eqnarray}
As $\Delta_{AB}\hoch{C}$ is block diagonal in the last two indices,
we know that $\Delta_{\bs{\mc{A}}b}\hoch{c}=2\Delta_{[\bs{\mc{A}}b]}\hoch{c}.$For
$\Delta_{ab}\hoch{c}$ we can use (see (\ref{eq:diffTensorInTermsOfSymAndAsym}))
\begin{eqnarray}
\Delta_{ab|c} & = & \Delta_{[ab]|c}+\Delta_{[ca]|b}-\Delta_{[bc]|a}+\Delta_{a(c|b)}+\Delta_{b(c|a)}-\Delta_{c(b|a)}\end{eqnarray}
The difference tensor with bosonic structure group indices is thus
completely determined to be\vRam{.6}{ \begin{eqnarray}
\Delta_{Ab}\hoch{c}:\qquad\Delta_{ab|c} & = & -3H_{abc}\label{eq:DiffTenI}\\
\Delta_{\bs{\alpha}b|c} & = & -2T_{\bs{\alpha}b|c}\stackrel{(\ref{eq:Tbetaca})}{=}\covPhi{\bs{\alpha}}G_{bc}+\tilde{\gamma}_{bc}\tief{\bs{\alpha}}\hoch{\bs{\delta}}\covPhi{\bs{\delta}}\label{eq:DiffTenII}\\
\Delta_{\hat{\bs{\alpha}}b|c} & = & 2\hat{T}_{\hat{\bs{\alpha}}b|c}\stackrel{(\ref{eq:Thatbetaca})}{=}-\hatcovPhi{\hat{\bs{\alpha}}}G_{bc}-\tilde{\gamma}_{bc}\tief{\hat{\bs{\alpha}}}\hoch{\hat{\bs{\delta}}}\hatcovPhi{\hat{\bs{\delta}}}\label{eq:DiffTenIII}\end{eqnarray}
} This is consistent with (\ref{eq:DeltaSymInLast}) as well as
with the left-right symmetry, if one defines $\hat{\Delta}\equiv-\Delta$.
The components of the difference tensor with fermionic group indices
are induced by the ones with bosonic group indices via \begin{equation}
\Delta_{A\bs{\beta}}\hoch{\bs{\gamma}}=\frac{1}{2}\Delta_{A}^{(D)}\delta_{\bs{\beta}}\hoch{\bs{\gamma}}+\frac{1}{4}\Delta_{A[b|c]}\tilde{\gamma}^{bc}\tief{\bs{\beta}}\hoch{\bs{\gamma}},\quad\Delta_{A\hat{\bs{\beta}}}\hoch{\hat{\bs{\gamma}}}=\frac{1}{2}\Delta_{A}^{(D)}\delta_{\hat{\bs{\beta}}}\hoch{\hat{\bs{\gamma}}}+\frac{1}{4}\Delta_{A[b|c]}\tilde{\gamma}^{bc}\tief{\hat{\bs{\beta}}}\hoch{\hat{\bs{\gamma}}}\end{equation}
Remember that this is due to the fact that both connections $\Omega_{MA}\hoch{B}$
and $\hat{\Omega}_{MA}\hoch{B}$ are defined to leave the chiral $\gamma$-matrices
invariant. The components with fermionic group indices are accordingly\vRam{1.02}{
\begin{eqnarray}
\Delta_{A\bs{\mc{B}}}\hoch{\bs{\mc{A}}}:\qquad\Delta_{a\bs{\beta}}\hoch{\bs{\gamma}} & = & -\frac{3}{4}H_{abc}\tilde{\gamma}^{bc}\tief{\bs{\beta}}\hoch{\bs{\gamma}}\qquad,\qquad\qquad\qquad\quad\Delta_{a\hat{\bs{\beta}}}\hoch{\hat{\bs{\gamma}}}=-\frac{3}{4}H_{abc}\tilde{\gamma}^{bc}\tief{\hat{\bs{\beta}}}\hoch{\hat{\bs{\gamma}}}\label{eq:DiffTenIV}\\
\Delta_{\bs{\alpha}\bs{\beta}}\hoch{\bs{\gamma}} & = & \frac{1}{2}\covPhi{\bs{\alpha}}\delta_{\bs{\beta}}\hoch{\bs{\gamma}}+\frac{1}{4}\gamma_{bc}\tief{\bs{\alpha}}\hoch{\bs{\delta}}\covPhi{\bs{\delta}}\gamma^{bc}\tief{\bs{\beta}}\hoch{\bs{\gamma}}\:,\quad\,\Delta_{\hat{\bs{\alpha}}\hat{\bs{\beta}}}\hoch{\hat{\bs{\gamma}}}=-\frac{1}{2}\hatcovPhi{\hat{\bs{\alpha}}}\delta_{\hat{\bs{\beta}}}\hoch{\hat{\bs{\gamma}}}-\frac{1}{4}\gamma_{bc}\tief{\hat{\bs{\alpha}}}\hoch{\hat{\bs{\delta}}}\hatcovPhi{\hat{\bs{\delta}}}\gamma^{bc}\tief{\hat{\bs{\beta}}}\hoch{\hat{\bs{\gamma}}}\label{eq:DiffTenV}\\
\Delta_{\hat{\bs{\alpha}}\bs{\beta}}\hoch{\bs{\gamma}} & = & -\frac{1}{2}\hatcovPhi{\hat{\bs{\alpha}}}\delta_{\bs{\beta}}\hoch{\bs{\gamma}}-\frac{1}{4}\gamma_{bc}\tief{\hat{\bs{\alpha}}}\hoch{\hat{\bs{\delta}}}\hatcovPhi{\hat{\bs{\delta}}}\gamma^{bc}\tief{\bs{\beta}}\hoch{\bs{\gamma}}\:,\;\Delta_{\bs{\alpha}\hat{\bs{\beta}}}\hoch{\hat{\bs{\gamma}}}=\frac{1}{2}\covPhi{\bs{\alpha}}\delta_{\hat{\bs{\beta}}}\hoch{\hat{\bs{\gamma}}}+\frac{1}{4}\gamma_{bc}\tief{\bs{\alpha}}\hoch{\bs{\delta}}\covPhi{\bs{\delta}}\gamma^{bc}\tief{\hat{\bs{\beta}}}\hoch{\hat{\bs{\gamma}}}\qquad\label{eq:DiffTenVI}\end{eqnarray}
}  \rem{Because of the block-diagonality in the last two indices
of $\Delta_{AB}\hoch{C}$, we have \begin{eqnarray}
\Delta_{[AB]}\hoch{c} & = & \left(\begin{array}{ccc}
\Delta_{[ab]}\hoch{c} & -\frac{1}{2}\Delta_{\bs{\beta}a}\hoch{c} & -\frac{1}{2}\Delta_{\hat{\bs{\beta}}a}\hoch{c}\\
\frac{1}{2}\Delta_{\bs{\alpha}b}\hoch{c} & 0 & 0\\
\frac{1}{2}\Delta_{\hat{\bs{\alpha}}b}\hoch{c} & 0 & 0\end{array}\right)\nonumber \\
\Delta_{[AB]}\hoch{\bs{\gamma}} & = & \left(\begin{array}{ccc}
0 & \frac{1}{2}\Delta_{a\bs{\beta}}\hoch{\bs{\gamma}} & 0\\
-\frac{1}{2}\Delta_{b\bs{\alpha}}\hoch{\bs{\gamma}} & \Delta_{[\bs{\alpha\beta}]}\hoch{\bs{\gamma}} & -\frac{1}{2}\Delta_{\hat{\bs{\beta}}\bs{\alpha}}\hoch{\bs{\gamma}}\\
0 & \frac{1}{2}\Delta_{\hat{\bs{\alpha}}\bs{\beta}}\hoch{\bs{\gamma}} & 0\end{array}\right),\quad\Delta_{[AB]}\hoch{\hat{\bs{\gamma}}}=\left(\begin{array}{ccc}
0 & 0 & \frac{1}{2}\Delta_{a\hat{\bs{\beta}}}\hoch{\hat{\bs{\gamma}}}\\
0 & 0 & \frac{1}{2}\Delta_{\bs{\alpha}\hat{\bs{\beta}}}\hoch{\hat{\bs{\gamma}}}\\
-\frac{1}{2}\Delta_{b\hat{\bs{\alpha}}}\hoch{\hat{\bs{\gamma}}} & -\frac{1}{2}\Delta_{\bs{\beta}\hat{\bs{\alpha}}}\hoch{\hat{\bs{\gamma}}} & \Delta_{[\hat{\bs{\alpha}}\hat{\bs{\beta}}]}\hoch{\hat{\bs{\gamma}}}\end{array}\right)\qquad\end{eqnarray}
 \begin{eqnarray}
\Delta_{[AB]}\hoch{\bs{\gamma}} & = & \left(\begin{array}{ccc}
0 & -\frac{3}{8}H_{ade}\tilde{\gamma}^{de}\tief{\bs{\beta}}\hoch{\bs{\gamma}} & 0\\
\frac{3}{8}H_{bde}\tilde{\gamma}^{de}\tief{\bs{\alpha}}\hoch{\bs{\gamma}} & (\frac{1}{4}\gamma_{de}\tief{[\bs{\alpha}}\hoch{\bs{\delta}}\gamma^{de}\tief{\bs{\beta}]}\hoch{\bs{\gamma}}\covPhi{\bs{\delta}}+\frac{1}{2}\covPhi{[\bs{\alpha}}\delta_{\bs{\beta}]}\hoch{\bs{\gamma}}) & (\frac{1}{8}\gamma_{de}\tief{\hat{\bs{\beta}}}\hoch{\hat{\bs{\delta}}}\gamma^{de}\tief{\bs{\alpha}}\hoch{\bs{\gamma}}\hatcovPhi{\hat{\bs{\delta}}}+\frac{1}{4}\hatcovPhi{\hat{\bs{\beta}}}\delta_{\bs{\alpha}}\hoch{\bs{\gamma}})\\
0 & (-\frac{1}{8}\gamma_{de}\tief{\hat{\bs{\alpha}}}\hoch{\hat{\bs{\delta}}}\gamma^{de}\tief{\bs{\beta}}\hoch{\bs{\gamma}}\hatcovPhi{\hat{\bs{\delta}}}-\frac{1}{4}\hatcovPhi{\hat{\bs{\alpha}}}\delta_{\bs{\beta}}\hoch{\bs{\gamma}}) & 0\end{array}\right)\qquad\end{eqnarray}
\begin{eqnarray}
\Delta_{[AB]}\hoch{\hat{\bs{\gamma}}} & = & \left(\begin{array}{ccc}
0 & 0 & -\frac{3}{8}H_{ade}\tilde{\gamma}^{de}\tief{\hat{\bs{\beta}}}\hoch{\hat{\bs{\gamma}}}\\
0 & 0 & (\frac{1}{8}\gamma_{de}\tief{\bs{\alpha}}\hoch{\bs{\delta}}\gamma^{de}\tief{\hat{\bs{\beta}}}\hoch{\hat{\bs{\gamma}}}\covPhi{\bs{\delta}}+\frac{1}{4}\covPhi{\bs{\alpha}}\delta_{\hat{\bs{\beta}}}\hoch{\hat{\bs{\gamma}}})\\
\frac{3}{8}H_{bde}\tilde{\gamma}^{de}\tief{\hat{\bs{\alpha}}}\hoch{\hat{\bs{\gamma}}} & (-\frac{1}{8}\gamma_{de}\tief{\bs{\beta}}\hoch{\bs{\delta}}\gamma^{de}\tief{\hat{\bs{\alpha}}}\hoch{\hat{\bs{\gamma}}}\covPhi{\bs{\delta}}-\frac{1}{4}\covPhi{\bs{\beta}}\delta_{\hat{\bs{\alpha}}}\hoch{\hat{\bs{\gamma}}}) & (-\frac{1}{4}\gamma_{de\,[\hat{\bs{\alpha}}}\hoch{\hat{\bs{\delta}}}\gamma^{de}\tief{\hat{\bs{\beta}}]}\hoch{\hat{\bs{\gamma}}}\hatcovPhi{\hat{\bs{\delta}}}-\frac{1}{2}\hatcovPhi{[\hat{\bs{\alpha}}}\delta_{\hat{\bs{\beta}}]}\hoch{\hat{\bs{\gamma}}})\end{array}\right)\qquad\end{eqnarray}
} We will use this difference tensor from now on frequently to change
from one connection to another. \rem{In particular, the constraints
(\ref{eq:RR-eoms}) and (\ref{eq:RR-eomsII}) can be rewritten as
\begin{eqnarray}
\hat{\nabla}_{\hat{\bs{\alpha}}}\RR^{\bs{\delta}\hat{\bs{\alpha}}} & = & \frac{15}{2}\RR^{\bs{\delta}\hat{\bs{\beta}}}\hatcovPhi{\hat{\bs{\beta}}}-\frac{1}{4}\gamma_{bc}\tief{\hat{\bs{\alpha}}}\hoch{\hat{\bs{\delta}}}\hatcovPhi{\hat{\bs{\delta}}}\gamma^{bc}\tief{\bs{\beta}}\hoch{\bs{\delta}}\RR^{\bs{\beta}\hat{\bs{\alpha}}}\\
\nabla_{\hat{\bs{\alpha}}}\RR^{\bs{\delta}\hat{\bs{\alpha}}} & = & \frac{17}{2}\RR^{\bs{\delta}\hat{\bs{\beta}}}\hatcovPhi{\hat{\bs{\beta}}}+\frac{1}{4}\gamma_{bc}\tief{\hat{\bs{\alpha}}}\hoch{\hat{\bs{\delta}}}\hatcovPhi{\hat{\bs{\delta}}}\gamma^{bc}\tief{\hat{\bs{\beta}}}\hoch{\hat{\bs{\alpha}}}\RR^{\bs{\delta}\hat{\bs{\beta}}}\\
\nabla_{\bs{\alpha}}\RR^{\bs{\alpha}\hat{\bs{\delta}}} & = & \frac{15}{2}\RR^{\bs{\beta}\hat{\bs{\delta}}}\covPhi{\bs{\beta}}-\frac{1}{4}\gamma_{bc}\tief{\bs{\alpha}}\hoch{\bs{\delta}}\covPhi{\bs{\delta}}\gamma^{bc}\tief{\hat{\bs{\beta}}}\hoch{\hat{\bs{\delta}}}\RR^{\bs{\alpha}\hat{\bs{\beta}}}\\
\hat{\nabla}_{\bs{\alpha}}\RR^{\bs{\alpha}\hat{\bs{\delta}}} & = & \frac{17}{2}\RR^{\bs{\beta}\hat{\bs{\delta}}}\covPhi{\bs{\beta}}+\frac{1}{4}\gamma_{bc}\tief{\bs{\alpha}}\hoch{\bs{\delta}}\covPhi{\bs{\delta}}\gamma^{bc}\tief{\bs{\beta}}\hoch{\bs{\alpha}}\RR^{\bs{\beta}\hat{\bs{\delta}}}\end{eqnarray}
}Let us take immediate advantage of the difference tensor to rewrite
some constraints on the curvature with the help of equation (\pageref{eq:RwithNewConnII})
of the appendix. \begin{eqnarray}
\hat{R}_{\hat{\bs{\gamma}}\hat{\bs{\gamma}}\hat{\bs{\alpha}}}\hoch{\hat{\bs{\beta}}} & = & \underbrace{R_{\hat{\bs{\gamma}}\hat{\bs{\gamma}}\hat{\bs{\alpha}}}\hoch{\hat{\bs{\beta}}}}_{=0}+\hat{\nabla}_{\hat{\bs{\gamma}}}\Delta_{\hat{\bs{\gamma}}\hat{\bs{\alpha}}}\hoch{\hat{\bs{\beta}}}+\gamma_{\hat{\bs{\gamma}}\hat{\bs{\gamma}}}\hoch{c}\Delta_{c\hat{\bs{\alpha}}}\hoch{\hat{\bs{\beta}}}+\Delta_{\hat{\bs{\gamma}}\hat{\bs{\alpha}}}\hoch{\hat{\bs{\delta}}}\Delta_{\hat{\bs{\gamma}}\hat{\bs{\delta}}}\hoch{\hat{\bs{\beta}}}=\\
 & = & -\frac{1}{2}\hat{\nabla}_{\hat{\bs{\gamma}}}\hatcovPhi{\hat{\bs{\gamma}}}\delta_{\hat{\bs{\alpha}}}\hoch{\hat{\bs{\beta}}}+\frac{1}{4}\gamma_{ab}\tief{\hat{\bs{\gamma}}}\hoch{\hat{\bs{\delta}}}\hat{\nabla}_{\hat{\bs{\gamma}}}\hatcovPhi{\hat{\bs{\delta}}}\gamma^{ab}\tief{\hat{\bs{\alpha}}}\hoch{\hat{\bs{\beta}}}-\frac{3}{4}\gamma_{\hat{\bs{\gamma}}\hat{\bs{\gamma}}}\hoch{c}H_{cab}\tilde{\gamma}^{ab}\tief{\hat{\bs{\alpha}}}\hoch{\hat{\bs{\beta}}}+\nonumber \\
 &  & +\left(-\frac{1}{2}\hatcovPhi{\hat{\bs{\gamma}}}\delta_{\hat{\bs{\alpha}}}\hoch{\hat{\bs{\delta}}}-\frac{1}{4}\gamma_{ab}\tief{\hat{\bs{\gamma}}}\hoch{\hat{\bs{\eps}}}\hatcovPhi{\hat{\bs{\eps}}}\gamma^{ab}\tief{\hat{\bs{\alpha}}}\hoch{\hat{\bs{\delta}}}\right)\left(-\frac{1}{2}\hatcovPhi{\hat{\bs{\gamma}}}\delta_{\hat{\bs{\delta}}}\hoch{\hat{\bs{\beta}}}-\frac{1}{4}\gamma_{cd}\tief{\hat{\bs{\gamma}}}\hoch{\hat{\bs{\varphi}}}\hatcovPhi{\hat{\bs{\varphi}}}\gamma^{cd}\tief{\hat{\bs{\delta}}}\hoch{\hat{\bs{\beta}}}\right)=\\
 & = & -\frac{1}{2}\hat{\nabla}_{\hat{\bs{\gamma}}}\hatcovPhi{\hat{\bs{\gamma}}}\delta_{\hat{\bs{\alpha}}}\hoch{\hat{\bs{\beta}}}+\frac{1}{4}\gamma_{ab}\tief{\hat{\bs{\gamma}}}\hoch{\hat{\bs{\delta}}}\hat{\nabla}_{\hat{\bs{\gamma}}}\hatcovPhi{\hat{\bs{\delta}}}\gamma^{ab}\tief{\hat{\bs{\alpha}}}\hoch{\hat{\bs{\beta}}}-\frac{3}{4}\gamma_{\hat{\bs{\gamma}}\hat{\bs{\gamma}}}\hoch{c}H_{cab}\tilde{\gamma}^{ab}\tief{\hat{\bs{\alpha}}}\hoch{\hat{\bs{\beta}}}+\nonumber \\
 &  & +\frac{1}{16}(\gamma_{ab}\tief{\hat{\bs{\gamma}}}\hoch{\hat{\bs{\eps}}}\hatcovPhi{\hat{\bs{\eps}}})(\gamma_{cd}\tief{\hat{\bs{\gamma}}}\hoch{\hat{\bs{\varphi}}}\hatcovPhi{\hat{\bs{\varphi}}})\gamma^{ab}\tief{\hat{\bs{\alpha}}}\hoch{\hat{\bs{\delta}}}\gamma^{cd}\tief{\hat{\bs{\delta}}}\hoch{\hat{\bs{\beta}}}\end{eqnarray}
In order to simplify the last term, let us suppress the fermionic
indices for a moment. The last line then reads $\frac{1}{32}\left((\gamma_{ab}\hat{\nabla}\Phi)(\gamma_{cd}\hat{\nabla}\Phi)-(\gamma_{cd}\hat{\nabla}\Phi)(\gamma_{ab}\hat{\nabla}\Phi)\right)\gamma^{ab}\gamma^{cd}$.
Now we can use \begin{eqnarray}
\gamma^{ab}\gamma^{cd} & = & \gamma^{abcd}+\eta^{bc}\gamma^{ad}+\eta^{ad}\gamma^{bc}-\eta^{ac}\gamma^{bd}-\eta^{bd}\gamma^{ac}+\eta^{bc}\eta^{ad}-\eta^{ac}\eta^{bd}\end{eqnarray}
Due to the contraction with $(\gamma_{cd}\hat{\nabla}\Phi)(\gamma_{ab}\hat{\nabla}\Phi)-(ab\leftrightarrow cd)$,
the $\gamma^{[4]}$-term and the $\gamma^{[0]}$-term ($\eta^{b[c}\eta^{d]a}$)
disappear. We are left with \begin{equation}
\frac{1}{32}\left((\gamma_{ab}\hat{\nabla}\Phi)(\gamma_{cd}\hat{\nabla}\Phi)-(ab\leftrightarrow cd)\right)\gamma^{ab}\gamma^{cd}=\frac{1}{4}(\gamma_{ab}\hat{\nabla}\Phi)\eta^{bc}(\gamma_{cd}\hat{\nabla}\Phi)\gamma^{ad}\end{equation}
The curvature component in question and its hatted version thus become\vRam{.83}{\begin{eqnarray}
\hat{R}_{\hat{\bs{\gamma}}\hat{\bs{\gamma}}\hat{\bs{\alpha}}}\hoch{\hat{\bs{\beta}}} & = & -\frac{1}{2}\hat{\nabla}_{\hat{\bs{\gamma}}}\hatcovPhi{\hat{\bs{\gamma}}}\delta_{\hat{\bs{\alpha}}}\hoch{\hat{\bs{\beta}}}+\nonumber \\
 &  & +\frac{1}{4}\left(\gamma_{ad}\tief{\hat{\bs{\gamma}}}\hoch{\hat{\bs{\delta}}}\hat{\nabla}_{\hat{\bs{\gamma}}}\hatcovPhi{\hat{\bs{\delta}}}+(\gamma_{ab}\tief{\hat{\bs{\gamma}}}\hoch{\hat{\bs{\eps}}}\hatcovPhi{\hat{\bs{\eps}}})\eta^{bc}(\gamma_{cd}\tief{\hat{\bs{\gamma}}}\hoch{\hat{\bs{\varphi}}}\hatcovPhi{\hat{\bs{\varphi}}})-3\gamma_{\hat{\bs{\gamma}}\hat{\bs{\gamma}}}\hoch{c}H_{cad}e^{-2\Phi}\right)\gamma^{ad}\tief{\hat{\bs{\alpha}}}\hoch{\hat{\bs{\beta}}}\label{eq:Rgamgamalphbet:hat}\\
R_{\bs{\gamma}\bs{\gamma}\bs{\alpha}}\hoch{\bs{\beta}} & = & -\frac{1}{2}\nabla_{\bs{\gamma}}\covPhi{\bs{\gamma}}\delta_{\bs{\alpha}}\hoch{\bs{\beta}}+\nonumber \\
 &  & +\frac{1}{4}\left(\gamma_{ad}\tief{\bs{\gamma}}\hoch{\bs{\delta}}\nabla_{\bs{\gamma}}\covPhi{\bs{\delta}}+(\gamma_{ab}\tief{\bs{\gamma}}\hoch{\bs{\eps}}\covPhi{\bs{\eps}})\eta^{bc}(\gamma_{cd}\tief{\bs{\gamma}}\hoch{\bs{\varphi}}\covPhi{\bs{\varphi}})+3\gamma_{\bs{\gamma}\bs{\gamma}}\hoch{c}H_{cad}e^{-2\Phi}\right)\gamma^{ad}\tief{\bs{\alpha}}\hoch{\bs{\beta}}\label{eq:Rgamgamalphbet}\end{eqnarray}
}  \rem{compare to \begin{eqnarray}
R_{\bs{\alpha\beta}c}\hoch{d} & = & -\nabla_{[\bs{\alpha}}\covPhi{\bs{\beta}]}\delta_{c}\hoch{d}+\gamma_{c}\hoch{d}\tief{[\bs{\alpha}}\hoch{\bs{\delta}}\nabla_{\bs{\beta}]}\covPhi{\bs{\delta}}+3\gamma_{\bs{\alpha\beta}}^{e}H_{ec}\hoch{d}+\nonumber \\
 &  & +\gamma_{c}\hoch{e}\tief{[\bs{\alpha}|}\hoch{\bs{\gamma}}\covPhi{\bs{\gamma}}\gamma_{e}\hoch{d}\tief{|\bs{\beta}]}\hoch{\bs{\delta}}\covPhi{\bs{\delta}}\end{eqnarray}
obtained later}We can compare this result to the nilpotency constraint
$R_{[\bs{\gamma\delta\alpha}]}\hoch{\bs{\beta}}=0$ or at least to
its trace $F_{\bs{\alpha\beta}}^{(D)}=\frac{2}{9}R_{\bs{\gamma}[\bs{\alpha}\bs{\beta}]}^{(L)}\hoch{\bs{\gamma}}$:~
Scaling and Lorentz component of (\ref{eq:Rgamgamalphbet}) are

\begin{eqnarray}
F_{\bs{\gamma\gamma}}^{(D)} & = & -\nabla_{\bs{\gamma}}\covPhi{\bs{\gamma}}\\
R_{\bs{\gamma}\bs{\gamma}\bs{\alpha}}^{(L)}\hoch{\bs{\beta}} & = & \frac{1}{4}\left(\gamma_{ad}\tief{\bs{\gamma}}\hoch{\bs{\delta}}\nabla_{\bs{\gamma}}\covPhi{\bs{\delta}}+(\gamma_{ab}\tief{\bs{\gamma}}\hoch{\bs{\eps}}\covPhi{\bs{\eps}})\eta^{bc}(\gamma_{cd}\tief{\bs{\gamma}}\hoch{\bs{\varphi}}\covPhi{\bs{\varphi}})+3\gamma_{\bs{\gamma}\bs{\gamma}}^{c}H_{cad}e^{-2\Phi}\right)\gamma^{ad}\tief{\bs{\alpha}}\hoch{\bs{\beta}}\end{eqnarray}
with trace \begin{eqnarray}
R_{\bs{\beta}\bs{\gamma}\bs{\alpha}}^{(L)}\hoch{\bs{\beta}} & = & \frac{1}{8}\gamma^{ad}\tief{\bs{\alpha}}\hoch{\bs{\beta}}\gamma_{ad}\tief{\bs{\beta}}\hoch{\bs{\delta}}\nabla_{\bs{\gamma}}\covPhi{\bs{\delta}}-\frac{1}{8}\gamma^{ad}\tief{\bs{\alpha}}\hoch{\bs{\beta}}\gamma_{ad}\tief{\bs{\gamma}}\hoch{\bs{\delta}}\nabla_{\bs{\beta}}\covPhi{\bs{\delta}}+\nonumber \\
 &  & +\frac{1}{4}(\gamma_{ab}\tief{\bs{\gamma}}\hoch{\bs{\eps}}\covPhi{\bs{\eps}})\eta^{bc}\gamma^{ad}\tief{\bs{\alpha}}\hoch{\bs{\beta}}(\gamma_{dc}\tief{\bs{\beta}}\hoch{\bs{\varphi}}\covPhi{\bs{\varphi}})+\frac{3}{4}\gamma^{ad}\tief{\bs{\alpha}}\hoch{\bs{\beta}}\gamma_{\bs{\beta}\bs{\gamma}}^{c}H_{cad}e^{-2\Phi}\end{eqnarray}
Now we use $\gamma^{ad}\tief{\bs{\alpha}}\hoch{\bs{\gamma}}\gamma_{dc}\tief{\bs{\gamma}}\hoch{\bs{\beta}}=8\gamma^{a}\tief{c\bs{\alpha}}\hoch{\bs{\beta}}+9\delta_{c}^{a}\delta_{\bs{\alpha}}\hoch{\bs{\beta}}$
(\ref{eq:gammaIIgammaIIContr}) and $\gamma^{ad}\tief{\bs{\alpha}}\hoch{\bs{\gamma}}\gamma_{ad}\tief{\bs{\gamma}}\hoch{\bs{\beta}}=-90\delta_{\bs{\alpha}}\hoch{\bs{\beta}}$
(\ref{eq:gammaIIgammaIIContrII}) to arrive at\begin{eqnarray}
R_{\bs{\beta}\bs{\gamma}\bs{\alpha}}^{(L)}\hoch{\bs{\beta}} & = & \frac{-90}{8}\nabla_{\bs{\gamma}}\covPhi{\bs{\alpha}}-\frac{1}{8}\gamma^{ad}\tief{\bs{\alpha}}\hoch{\bs{\beta}}\gamma_{ad}\tief{\bs{\gamma}}\hoch{\bs{\delta}}\nabla_{\bs{\beta}}\covPhi{\bs{\delta}}+\nonumber \\
 &  & +2\gamma^{ac}\tief{\bs{\gamma}}\hoch{\bs{\eps}}\covPhi{\bs{\eps}}\gamma\tief{ac\bs{\alpha}}\hoch{\bs{\varphi}}\covPhi{\bs{\varphi}}+\nonumber \\
 &  & +\frac{3}{4}\gamma^{cad}\tief{\bs{\alpha\gamma}}H_{cad}e^{-2\Phi}\end{eqnarray}
The antisymmetric part (in $\bs{\alpha},\bs{\gamma}$) is \begin{eqnarray}
R_{\bs{\beta}[\bs{\gamma}\bs{\alpha}]}^{(L)}\hoch{\bs{\beta}} & = & \frac{-90}{8}\nabla_{[\bs{\gamma}}\covPhi{\bs{\alpha}]}-\frac{1}{8}\gamma^{ad}\tief{[\bs{\alpha}|}\hoch{\bs{\beta}}\gamma_{ad}\tief{|\bs{\gamma}]}\hoch{\bs{\delta}}\nabla_{[\bs{\beta}}\covPhi{\bs{\delta}]}=\nonumber \\
 & = & \frac{45}{4}F_{\bs{\gamma\alpha}}^{(D)}-\frac{1}{8}\gamma^{ad}\tief{[\bs{\alpha}|}\hoch{\bs{\beta}}\gamma_{ad}\tief{|\bs{\gamma}]}\hoch{\bs{\delta}}F_{\bs{\delta\beta}}^{(D)}\quad\label{eq:star}\end{eqnarray}
Now we expand the scaling curvature in $\gamma$-matrices. Because
of the graded antisymmetry, only $\gamma^{[1]}$ and $\gamma^{[5]}$
appear: $F_{\bs{\delta\beta}}^{(D)}=F_{c}^{(D)}\gamma_{\bs{\delta\beta}}^{c}+F_{c_{1}\ldots c_{5}}^{(D)}\gamma_{\bs{\delta\beta}}^{c_{1}\ldots c_{5}}$.
In $\gamma^{ad}\tief{\bs{\gamma}}\hoch{\bs{\delta}}F_{\bs{\delta\beta}}^{(D)}$
we then need the following multiplications of $\gamma$-matrices (\ref{eq:gammaIIgammalohneIndex}):\begin{eqnarray}
\gamma^{ad}\tief{\bs{\gamma}}\hoch{\bs{\delta}}\gamma_{\bs{\delta\beta}}^{c} & = & \gamma_{\bs{\gamma}\bs{\beta}}^{adc}+\eta^{dc}\gamma_{\bs{\gamma}\bs{\beta}}^{a}-\eta^{ac}\gamma_{\bs{\gamma}\bs{\beta}}^{d}=\\
 & = & -\gamma_{\bs{\beta}\bs{\gamma}}^{dac}+2\eta^{c[a}\gamma_{\bs{\beta}\bs{\gamma}}^{d]}\\
\gamma^{ad}\tief{\bs{\gamma}}\hoch{\bs{\delta}}\gamma_{\bs{\delta\beta}}^{c_{1}\ldots c_{5}} & = & \gamma_{\bs{\gamma\beta}}^{adc_{1}\ldots c_{5}}+5\eta^{d[c_{1}|}\gamma_{\bs{\gamma\beta}}^{a|c_{2}\ldots c_{5}]}-5\eta^{a[c_{1}|}\gamma_{\bs{\gamma\beta}}^{d|c_{2}\ldots c_{5}]}-20\eta^{a[c_{1}|}\eta^{d|c_{2}}\gamma_{\bs{\gamma\beta}}^{c_{3}\ldots c_{5}]}=\\
 & = & \gamma_{\bs{\beta\gamma}}^{dac_{1}\ldots c_{5}}-5\eta^{d[c_{1}|}\gamma_{\bs{\beta\gamma}}^{a|c_{2}\ldots c_{5}]}+5\eta^{a[c_{1}|}\gamma_{\bs{\beta\gamma}}^{d|c_{2}\ldots c_{5}]}-20\eta^{a[c_{1}|}\eta^{d|c_{2}}\gamma_{\bs{\beta\gamma}}^{c_{3}\ldots c_{5}]}\end{eqnarray}
For the expression $\gamma_{ad}\tief{[\bs{\alpha}|}\hoch{\bs{\beta}}\gamma^{ad}\tief{|\bs{\gamma}]}\hoch{\bs{\delta}}F_{\bs{\delta\beta}}^{(D)}$
in (\ref{eq:star}), we can make use of (\ref{eq:gammaIgammalContr})-(\ref{eq:gammaIIgammalContrII})
and of the fact that $\gamma_{[\bs{\alpha\gamma}]}^{[3]}=\gamma_{[\bs{\alpha\gamma}]}^{[7]}=0$:\begin{eqnarray}
\gamma_{ad}\tief{[\bs{\alpha}|}\hoch{\bs{\beta}}\gamma_{\bs{\beta}|\bs{\gamma}]}^{dac} & = & 72\gamma_{\bs{\alpha\gamma}}^{c},\quad\gamma_{ad}\tief{[\bs{\alpha}|}\hoch{\bs{\beta}}\eta^{c[a}\gamma_{\bs{\beta}|\bs{\gamma}]}^{d]}=9\gamma_{\bs{\alpha\gamma}}^{c}\\
\gamma_{ad}\tief{[\bs{\alpha}|}\hoch{\bs{\beta}}\gamma_{\bs{\beta}|\bs{\gamma}]}^{dac_{1}\ldots c_{5}} & = & 20\gamma_{\bs{\alpha\gamma}}^{c_{1}\ldots c_{5}},\quad\gamma_{ad}\tief{[\bs{\alpha}|}\hoch{\bs{\beta}}\eta^{a[c_{1}|}\gamma_{\bs{\beta}|\bs{\gamma}]}^{d|c_{2}\ldots c_{5}]}=5\gamma_{\bs{\alpha\gamma}}^{c_{1}\ldots c_{5}}\\
\gamma_{ad}\tief{[\bs{\alpha}|}\hoch{\bs{\beta}}\eta^{a[c_{1}|}\eta^{d|c_{2}}\gamma_{\bs{\beta}|\bs{\gamma}]}^{c_{3}\ldots c_{5}]} & = & 0\end{eqnarray}

The equation ({*}) thus becomes\begin{eqnarray}
R_{\bs{\beta}[\bs{\gamma}\bs{\alpha}]}^{(L)}\hoch{\bs{\beta}} & = & \frac{45}{4}F_{\bs{\gamma\alpha}}^{(D)}+\frac{54}{8}F_{c}^{(D)}\gamma_{\bs{\alpha\gamma}}^{c}-\frac{70}{8}F_{c_{1}\ldots c_{5}}^{(D)}\gamma_{\bs{\alpha\gamma}}^{c_{1}\ldots c_{5}}=\\
 & = & \frac{9}{2}F_{\bs{\gamma\alpha}}^{(D)}+\frac{31}{2}F_{c_{1}\ldots c_{5}}^{(D)}\gamma_{\bs{\gamma\alpha}}^{c_{1}\ldots c_{5}}\end{eqnarray}
From our nilpotency constraint (\ref{eq:nilpotency:Falphbet}) we
can now deduce that $F_{c_{1}\ldots c_{5}}^{(D)}=0$ or equivalently
that \begin{equation}
\boxed{\gamma_{a_{1}\ldots a_{5}}^{\bs{\alpha\beta}}F_{\bs{\alpha\beta}}^{(D)}=0,\quad\gamma_{a_{1}\ldots a_{5}}^{\hat{\bs{\alpha}}\hat{\bs{\beta}}}\hat{F}_{\hat{\bs{\alpha}}\hat{\bs{\beta}}}^{(D)}=0}\label{eq:FalphbetForDilaton}\end{equation}
\rem{Let us use the difference tensor again to learn more from the
curvature constraint $R_{\hat{\bs{\gamma}}d\bs{\alpha}}\hoch{\bs{\beta}}=\tilde{\gamma}_{d\,\hat{\bs{\gamma}}\hat{\bs{\delta}}}\gemnabla_{\bs{\alpha}}\RR^{\bs{\beta}\hat{\bs{\delta}}}$:
\begin{eqnarray}
\hat{R}_{\hat{\bs{\gamma}}d\bs{\alpha}}\hoch{\bs{\beta}} & = & \underbrace{R_{\hat{\bs{\gamma}}d\bs{\alpha}}\hoch{\bs{\beta}}}_{=\tilde{\gamma}_{d\,\hat{\bs{\gamma}}\hat{\bs{\delta}}}\gemnabla_{\bs{\alpha}}\RR^{\bs{\beta}\hat{\bs{\delta}}}}+\nabla_{[\hat{\bs{\gamma}}}\Delta_{d]\bs{\alpha}}\hoch{\bs{\beta}}+T_{\hat{\bs{\gamma}}d}\hoch{C}\Delta_{C\bs{\alpha}}\hoch{\bs{\beta}}-\Delta_{[\hat{\bs{\gamma}}|\bs{\alpha}}\hoch{\bs{\delta}}\Delta_{|d]\bs{\delta}}\hoch{\bs{\beta}}=\\
 & \stackrel{\check{\Omega}=\Omega}{=} & \tilde{\gamma}_{d\,\hat{\bs{\gamma}}\hat{\bs{\delta}}}\gemnabla_{\bs{\alpha}}\RR^{\bs{\beta}\hat{\bs{\delta}}}+\gemnabla_{[\hat{\bs{\gamma}}}\Delta_{d]\bs{\alpha}}\hoch{\bs{\beta}}-\frac{1}{2}\Delta_{d\hat{\bs{\gamma}}}\hoch{\hat{\bs{\delta}}}\Delta_{\hat{\bs{\delta}}\bs{\alpha}}\hoch{\bs{\beta}}+\gemT_{\hat{\bs{\gamma}}d}\hoch{C}\Delta_{C\bs{\alpha}}\hoch{\bs{\beta}}-\Delta_{[\hat{\bs{\gamma}}d]}\hoch{\hat{\bs{\gamma}}}\Delta_{\hat{\bs{\gamma}}\bs{\alpha}}\hoch{\bs{\beta}}-\Delta_{[\hat{\bs{\gamma}}|\bs{\alpha}}\hoch{\bs{\delta}}\Delta_{|d]\bs{\delta}}\hoch{\bs{\beta}}\end{eqnarray}
}

\lyxline{\normalsize}\vspace{-.25cm}\lyxline{\normalsize}

\vspace{.5cm} \hspace{-.6cm}$\bullet\quad$\underbar{(2,1,1)$ab\bs{\alpha}\hat{\bs{\beta}}$:}\begin{eqnarray}
0 & \stackrel{!}{=} & \gemnabla_{[a}H_{b\bs{\alpha}\hat{\bs{\beta}}]}+3\gemT_{[ab|}\hoch{C}H_{C|\bs{\alpha}\hat{\bs{\beta}}]}=\\
 & = & -\gemT_{[a|\bs{\alpha}}\hoch{C}H_{C|b]\hat{\bs{\beta}}}-\gemT_{[b|\hat{\bs{\beta}}}\hoch{C}H_{C|a]\bs{\alpha}}=\\
 & \stackrel{f_{ac}=G_{ac}}{=} & -\frac{2}{3}\tilde{\gamma}_{[a|\bs{\alpha\delta}}\RR^{\bs{\delta}\hat{\bs{\gamma}}}\tilde{\gamma}_{|b]\,\hat{\bs{\gamma}}\hat{\bs{\beta}}}+\frac{2}{3}\tilde{\gamma}_{[b|\hat{\bs{\beta}}\hat{\bs{\delta}}}\RR^{\bs{\gamma}\hat{\bs{\delta}}}\tilde{\gamma}_{|a]\,\bs{\gamma}\bs{\alpha}}=\\
 & = & \frac{2}{3}\tilde{\gamma}_{[a|\bs{\alpha\delta}}\tilde{\gamma}_{|b]\,\hat{\bs{\delta}}\hat{\bs{\beta}}}\left(-\RR^{\bs{\delta}\hat{\bs{\delta}}}+\RR^{\bs{\delta}\hat{\bs{\delta}}}\right)=0\end{eqnarray}
$\bullet\quad$\underbar{(3,1,0)$abc\bs{\delta}\leftrightarrow$((3,0,1)$abc\hat{\bs{\delta}})$:}\begin{eqnarray}
0 & \stackrel{!}{=} & \gemnabla_{[a}H_{bc\hat{\bs{\delta}}]}+3\gemT_{[ab|}\hoch{E}H_{E|c\hat{\bs{\delta}}]}=\\
 & = & -\frac{1}{4}\check{\nabla}_{\hat{\bs{\delta}}}H_{abc}+\frac{3}{2}\gemT_{[ab|}\hoch{E}H_{E|c]\hat{\bs{\delta}}}-\frac{3}{2}\gemT_{\hat{\bs{\delta}}[a|}\hoch{E}H_{E|bc]}=\\
 & \stackrel{\check{\Omega}=\Omega}{=} & -\frac{1}{4}\nabla_{\hat{\bs{\delta}}}H_{abc}-\frac{3}{2}\hat{T}_{[ab|}\hoch{\hat{\bs{\eps}}}H_{|c]\hat{\bs{\eps}}\hat{\bs{\delta}}}-\frac{3}{2}T_{\hat{\bs{\delta}}[a|}\hoch{e}H_{e|bc]}=\\
 & \stackrel{f_{ab}=G_{ab}}{=} & -\frac{1}{4}\nabla_{\hat{\bs{\delta}}}H_{abc}-\hat{T}_{[ab|}\hoch{\hat{\bs{\eps}}}\tilde{\gamma}_{|c]\hat{\bs{\eps}}\hat{\bs{\delta}}}-\frac{3}{2}\underbrace{T_{\hat{\bs{\delta}}[a|}\hoch{e}}_{=0\,(\ref{eq:T^{c}undOmega})}H_{e|bc]}\end{eqnarray}
\Ram{0.6}{\begin{eqnarray}
\nabla_{\hat{\bs{\delta}}}H_{abc} & = & -4\hat{T}_{[ab|}\hoch{\hat{\bs{\eps}}}\tilde{\gamma}_{|c]\hat{\bs{\eps}}\hat{\bs{\delta}}}\label{eq:(3,1,0)}\\
\mbox{likewise }\hat{\nabla}_{\bs{\delta}}H_{abc} & = & 4T_{[ab|}\hoch{\bs{\eps}}\tilde{\gamma}_{|c]\bs{\eps}\bs{\delta}}\label{eq:(3,0,1)}\end{eqnarray}
} \rem{\\
Contracting with $\gamma^{d\,\hat{\delta}\hat{\alpha}}$ yields\begin{eqnarray}
\nabla_{\hat{\delta}}H_{abc}\gamma^{d\,\hat{\delta}\hat{\alpha}} & = & -\frac{4}{3}\hat{T}_{ab}\hoch{\hat{\eps}}\gamma_{c\hat{\eps}\hat{\delta}}\gamma^{d\,\hat{\delta}\hat{\alpha}}-\frac{4}{3}\hat{T}_{ca}\hoch{\hat{\eps}}\gamma_{b\hat{\eps}\hat{\delta}}\gamma^{d\,\hat{\delta}\hat{\alpha}}-\frac{4}{3}\hat{T}_{bc}\hoch{\hat{\eps}}\gamma_{a\hat{\eps}\hat{\delta}}\gamma^{d\,\hat{\delta}\hat{\alpha}}=\\
 & = & -\frac{4}{3}\hat{T}_{ab}\hoch{\hat{\eps}}\left(\delta_{c}^{d}\delta_{\hat{\eps}}^{\hat{\alpha}}+\gamma_{c}\hoch{d}\tief{\hat{\eps}}\hoch{\hat{\alpha}}\right)-\frac{4}{3}\hat{T}_{ca}\hoch{\hat{\bs{\eps}}}\left(\delta_{b}^{d}\delta_{\hat{\eps}}^{\hat{\alpha}}+\gamma_{b}\hoch{d}\tief{\hat{\eps}}\hoch{\hat{\alpha}}\right)-\\
 &  & -\frac{4}{3}\hat{T}_{bc}\hoch{\hat{\eps}}\left(\delta_{a}^{d}\delta_{\hat{\eps}}^{\hat{\alpha}}+\gamma_{a}\hoch{d}\tief{\hat{\eps}}\hoch{\hat{\alpha}}\right)\\
c=d:\quad\nabla_{\hat{\delta}}H_{abc}\gamma^{c\,\hat{\delta}\hat{\alpha}} & = & -\frac{32}{3}\hat{T}_{ab}\hoch{\hat{\alpha}}-\frac{4}{3}\hat{T}_{ca}\hoch{\hat{\eps}}\gamma_{b}\hoch{c}\tief{\hat{\eps}}\hoch{\hat{\alpha}}-\frac{4}{3}\hat{T}_{bc}\hoch{\hat{\eps}}\gamma_{a}\hoch{c}\tief{\hat{\eps}}\hoch{\hat{\alpha}}\quad??\end{eqnarray}
}\\
$\bullet\quad$\underbar{(4,0,0)$abcd:$}\begin{equation}
\boxed{0\stackrel{!}{=}\check{\nabla}_{[a}H_{bcd]}+3\check{T}_{[ab|}\hoch{e}H_{e|cd]}}\label{eq:(4,0,0)}\end{equation}
\rem{Define the bosonic vielbein as \begin{equation}
e_{m}\hoch{a}\equiv E_{m}\hoch{a}\end{equation}
and its inverse as \begin{eqnarray}
E_{m}\hoch{a}e_{a}\hoch{n} & = & \delta_{m}^{n},\qquad e_{a}\hoch{n}\neq E_{a}\hoch{n}\\
\mbox{compare to }E_{m}\hoch{a}E_{a}\hoch{n}+E_{m}\hoch{\bs{\mc{A}}}E_{\bs{\mc{A}}}\hoch{n} & = & \delta_{m}^{n}\end{eqnarray}
Acting with the bosonic vielbeins on the above BI leads to the fact
that \begin{eqnarray}
\de H' & = & 0\\
H'_{\bs{mmm}} & \equiv & E_{\bs{m}}\hoch{a_{1}}E_{\bs{m}}\hoch{a_{2}}E_{\bs{m}}\hoch{a_{3}}H_{a_{1}a_{2}a_{3}}=\\
 & = & E_{\bs{m}}\hoch{a_{1}}E_{\bs{m}}\hoch{a_{2}}E_{\bs{m}}\hoch{a_{3}}E_{a_{1}}\hoch{N_{1}}E_{a_{2}}\hoch{N_{2}}E_{a_{3}}\hoch{N_{3}}H_{N_{1}N_{2}N_{3}}\end{eqnarray}
}\addtocounter{localapp}{1}

\section{The Bianchi identities for the torsion}

\label{sec:Bianchi-identitiesForT}The Bianchi\index{Bianchi identity!for the torsion}
identity for the torsion reads

\begin{equation}
0\stackrel{!}{=}\gemnabla_{\bs{A}}\gemT_{\bs{AA}}\hoch{D}+2\gemT_{\bs{AA}}\hoch{C}\gemT_{C\bs{A}}\hoch{D}-\gemR_{\bs{AAA}}\hoch{D}\frem{\equiv\gem{I}_{\bs{AAA}}\hoch{D}}\end{equation}
Again, depending on what is more convenient, the bosonic part of the
connection $\check{\Omega}_{a}\hoch{b}$ will be chosen to be either
$\Omega_{a}\hoch{b}$ or $\hat{\Omega}_{a}\hoch{b}$. Due to proposition
\ref{prop:BI-with-shifted-connection} on page \pageref{prop:BI-with-shifted-connection},
both are equivalent. The index $A$ can again be either $a$, $\bs{\alpha}$
or $\hat{\bs{\alpha}}$. For fixed upper index the numbers of their
appearance as lower index are \#$a$, \#$\bs{\alpha}$, \#$\hat{\bs{\alpha}}\in\{0,1,2,3\}$.
In analogy to the Bianchi identities for $H$, we have for each fixed
upper index $4+3+2+1=10$ possibilities and thus altogether $30$
possibilities. The symmetry between hatted and unhatted indices relates
the 10 with upper index $\hat{\bs{\delta}}$ to the ten with upper
index $\bs{\delta}$. The remaining 10 have again an internal symmetry
with fixed points (\#$\bs{\alpha}$, \#$\hat{\bs{\alpha}})\in\{(0,0),(1,1)\}$,
so that there remain effectively $\frac{10-2}{2}+2=6$ of those 10.
Altogether we have thus effectively 16 equations to study.

$\bullet\quad$\underbar{(delta|0,3,0)$_{\bs{\alpha\beta\gamma}}\hoch{\bs{\delta}}\leftrightarrow$((hdelta|0,0,3)$_{\hat{\bs{\alpha}}\hat{\bs{\beta}}\hat{\bs{\gamma}}}\hoch{\hat{\bs{\delta}}}$),dim1:}\begin{eqnarray}
0 & \stackrel{!}{=} & \nabla_{[\bs{\alpha}}T_{\bs{\beta\gamma}]}\hoch{\bs{\delta}}+2\gem{T}_{[\bs{\alpha\beta}|}\hoch{E}T_{E|\bs{\gamma}]}\hoch{\bs{\delta}}-R_{[\bs{\alpha\beta\gamma}]}\hoch{\bs{\delta}}=\\
 & = & 2\check{T}_{[\bs{\alpha\beta}|}\hoch{e}\underbrace{T_{e|\bs{\gamma}]}\hoch{\bs{\delta}}}_{=0}-R_{[\bs{\alpha\beta\gamma}]}\hoch{\bs{\delta}}\end{eqnarray}
\Ram{0.4}{\begin{eqnarray}
R_{[\bs{\alpha\beta\gamma}]}\hoch{\bs{\delta}} & = & 0\label{eq:(delta|0,3,0)}\\
\hat{R}_{[\hat{\bs{\alpha}}\hat{\bs{\beta}}\hat{\bs{\gamma}}]}\hoch{\hat{\bs{\delta}}} & = & 0\label{eq:(hdelta|0,0,3)}\end{eqnarray}
} This is a confirmation of the nilpotency constraint (\ref{eq:nilpotency-constraint-onR})
that we had derived earlier. Taking the trace yields\begin{eqnarray}
0 & \stackrel{!}{=} & R_{\bs{\alpha\beta\gamma}}\hoch{\bs{\gamma}}+2R_{\bs{\gamma}[\bs{\alpha\beta}]}\hoch{\bs{\gamma}}=\\
 & = & -9F_{\bs{\alpha\beta}}^{(D)}+2R_{\bs{\gamma}[\bs{\alpha\beta}]}^{(L)}\hoch{\bs{\gamma}}\end{eqnarray}
\begin{equation}
\boxed{F_{\bs{\alpha\beta}}^{(D)}\frem{=-\nabla_{[\bs{\alpha}}\covPhi{\bs{\beta}]}-\gamma_{\bs{\alpha\beta}}^{c}\covPhi{c}}\stackrel{!}{=}\frac{2}{9}R_{\bs{\gamma}[\bs{\alpha\beta}]}^{(L)}\hoch{\bs{\gamma}}}\label{eq:FDilalphbet}\end{equation}
and \begin{equation}
\boxed{\hat{F}_{\hat{\bs{\alpha}}\hat{\bs{\beta}}}^{(D)}\frem{=-\hat{\nabla}_{[\hat{\bs{\alpha}}}\hatcovPhi{\hat{\bs{\beta}]}}-\gamma_{\hat{\bs{\alpha}}\hat{\bs{\beta}}}^{c}\hatcovPhi{c}}\stackrel{!}{=}\frac{2}{9}\hat{R}_{\hat{\bs{\gamma}}[\hat{\bs{\alpha}}\hat{\bs{\beta}}]}^{(L)}\hoch{\hat{\bs{\gamma}}}}\label{eq:FhDilhalphhbet}\end{equation}
\\
$\bullet\quad$\underbar{(delta|0,2,1)$_{\bs{\alpha}\bs{\beta}\hat{\bs{\gamma}}}\hoch{\bs{\delta}}\leftrightarrow$((hdelta|0,1,2)$_{\hat{\bs{\alpha}}\hat{\bs{\beta}}\bs{\gamma}}\hoch{\hat{\bs{\delta}}}$)dim1:}\begin{eqnarray}
0 & \stackrel{!}{=} & \gem{\nabla}_{[\bs{\alpha}}T_{\bs{\beta}\hat{\bs{\gamma}}]}\hoch{\bs{\delta}}+2T_{[\bs{\alpha\beta}|}\hoch{\gem{E}}T_{\gem{E}|\hat{\bs{\gamma}}]}\hoch{\bs{\delta}}-R_{[\bs{\alpha\beta}\hat{\bs{\gamma}}]}\hoch{\bs{\delta}}=\\
 & = & \frac{2}{3}T_{\bs{\alpha\beta}}\hoch{e}T_{e\hat{\bs{\gamma}}}\hoch{\bs{\delta}}-\frac{2}{3}R_{\hat{\bs{\gamma}}[\bs{\alpha\beta}]}\hoch{\bs{\delta}}=\\
 & \stackrel{f_{c}\hoch{e}=\delta_{c}^{e}}{=} & -\frac{2}{3}\gamma_{\bs{\alpha\beta}}\hoch{e}\tilde{\gamma}_{e\,\hat{\bs{\gamma}}\hat{\bs{\delta}}}\RR^{\bs{\delta}\hat{\bs{\delta}}}-\frac{2}{3}R_{\hat{\bs{\gamma}}[\bs{\alpha\beta}]}\hoch{\bs{\delta}}\end{eqnarray}
\Ram{0.4}{ \begin{eqnarray}
R_{\hat{\bs{\gamma}}[\bs{\alpha\beta}]}\hoch{\bs{\delta}} & = & -\gamma_{\bs{\alpha\beta}}\hoch{e}\tilde{\gamma}_{e\,\hat{\bs{\gamma}}\hat{\bs{\delta}}}\RR^{\bs{\delta}\hat{\bs{\delta}}}\label{eq:(delta|0,2,1)}\\
\hat{R}_{\bs{\gamma}[\hat{\bs{\alpha}}\hat{\bs{\beta}}]}\hoch{\hat{\bs{\delta}}} & = & -\gamma_{\hat{\bs{\alpha}}\hat{\bs{\beta}}}\hoch{e}\tilde{\gamma}_{e\,\bs{\gamma}\bs{\delta}}\RR^{\bs{\delta}\hat{\bs{\delta}}}\label{eq:(hdelta|0,1,2)}\end{eqnarray}
}  Again taking the trace gives additional information on the Dilatation
part\begin{eqnarray}
R_{\hat{\bs{\gamma}}\bs{\alpha\delta}}\hoch{\bs{\delta}}-R_{\hat{\bs{\gamma}}\bs{\delta\alpha}}\hoch{\bs{\delta}} & = & 2\gamma_{\bs{\alpha\delta}}\hoch{e}\RR^{\bs{\delta}\hat{\bs{\delta}}}\tilde{\gamma}_{e\,\hat{\bs{\delta}}\hat{\bs{\gamma}}}\\
-8F_{\hat{\bs{\gamma}}\bs{\alpha}}^{(D)}-\frac{1}{2}F_{\hat{\bs{\gamma}}\bs{\alpha}}^{(D)}-R_{\hat{\bs{\gamma}}\bs{\delta\alpha}}^{(L)}\hoch{\bs{\delta}} & = & 2\gamma_{\bs{\alpha\delta}}\hoch{e}\RR^{\bs{\delta}\hat{\bs{\delta}}}\tilde{\gamma}_{e\,\hat{\bs{\delta}}\hat{\bs{\gamma}}}\end{eqnarray}
\begin{equation}
\boxed{F_{\hat{\bs{\gamma}}\bs{\alpha}}^{(D)}=-\frac{4}{17}\gamma_{\bs{\alpha\delta}}^{e}\RR^{\bs{\delta}\hat{\bs{\delta}}}\tilde{\gamma}_{e\,\hat{\bs{\delta}}\hat{\bs{\gamma}}}-\frac{2}{17}R_{\hat{\bs{\gamma}}\bs{\delta\alpha}}^{(L)}\hoch{\bs{\delta}}}\label{eq:FDilhutgamalph}\end{equation}
\begin{equation}
\boxed{\hat{F}_{\bs{\gamma}\hat{\bs{\alpha}}}^{(D)}=-\frac{4}{17}\gamma_{\hat{\bs{\alpha}}\hat{\bs{\delta}}}^{e}\RR^{\bs{\delta}\hat{\bs{\delta}}}\tilde{\gamma}_{e\,\bs{\delta}\bs{\gamma}}-\frac{2}{17}\hat{R}_{\bs{\gamma}\hat{\bs{\delta}}\hat{\bs{\alpha}}}^{(L)}\hoch{\hat{\bs{\delta}}}}\label{eq:FhutDilgamhutalph}\end{equation}
$\bullet\quad$\underbar{(delta|0,1,2)$_{\bs{\alpha}\hat{\bs{\beta}}\hat{\bs{\gamma}}}\hoch{\bs{\delta}}\leftrightarrow$((hdelta|0,2,1)$_{\hat{\bs{\alpha}}\bs{\beta}\bs{\gamma}}\hoch{\hat{\bs{\delta}}}$)dim1:}\begin{eqnarray}
0 & \stackrel{!}{=} & \gem{\nabla}_{[\bs{\alpha}}T_{\hat{\bs{\beta}}\hat{\bs{\gamma}}]}\hoch{\bs{\delta}}+2\gem{T}_{[\bs{\alpha}\hat{\bs{\beta}}|}\hoch{E}T_{E|\hat{\bs{\gamma}}]}\hoch{\bs{\delta}}-R_{[\bs{\alpha}\hat{\bs{\beta}}\hat{\bs{\gamma}}]}\hoch{\bs{\delta}}=\\
 & = & \frac{2}{3}T_{\hat{\bs{\beta}}\hat{\bs{\gamma}}}\hoch{e}\underbrace{T_{e\bs{\alpha}}\hoch{\bs{\delta}}}_{=0}-\frac{1}{2}\underbrace{R_{\hat{\bs{\beta}}\hat{\bs{\gamma}}\bs{\alpha}}\hoch{\bs{\delta}}}_{=0}=0\end{eqnarray}
$\bullet\quad$\underbar{(delta|0,0,3)$_{\hat{\bs{\alpha}}\hat{\bs{\beta}}\hat{\bs{\gamma}}}\hoch{\bs{\delta}}\leftrightarrow$((hdelta|0,3,0)$_{\bs{\alpha}\bs{\beta}\bs{\gamma}}\hoch{\hat{\bs{\delta}}}$)dim1:}\begin{eqnarray}
0 & \stackrel{!}{=} & \gem{\nabla}_{[\hat{\bs{\alpha}}}T_{\hat{\bs{\beta}}\hat{\bs{\gamma}}]}\hoch{\bs{\delta}}+2\gem{T}_{[\hat{\bs{\alpha}}\hat{\bs{\beta}}|}\hoch{E}T_{E|\hat{\bs{\gamma}}]}\hoch{\bs{\delta}}-\underbrace{R_{[\hat{\bs{\alpha}}\hat{\bs{\beta}}\hat{\bs{\gamma}}]}\hoch{\bs{\delta}}}_{=0}=\\
 & = & 2T_{[\hat{\bs{\alpha}}\hat{\bs{\beta}}|}\hoch{e}T_{e|\hat{\bs{\gamma}}]}\hoch{\bs{\delta}}=\\
 & = & -2\gamma_{[\hat{\bs{\alpha}}\hat{\bs{\beta}}|}\hoch{e}\tilde{\gamma}_{e\,|\hat{\bs{\gamma}}]\hat{\bs{\delta}}}\RR^{\bs{\delta}\hat{\bs{\delta}}}\stackrel{\mbox{Fierz }}{=}0\end{eqnarray}
$\bullet\quad$\underbar{(delta|1,2,0)$_{\bs{\alpha\beta}c}\hoch{\bs{\delta}}\leftrightarrow$((hdelta|1,0,2)$_{\hat{\bs{\alpha}}\hat{\bs{\beta}}c}\hoch{\hat{\bs{\delta}}}$)dim$\frac{3}{2}$:}\begin{eqnarray}
0 & \stackrel{!}{=} & \gem{\nabla}_{[\bs{\alpha}}T_{\bs{\beta}c]}\hoch{\bs{\delta}}+2\gem{T}_{[\bs{\alpha\beta}|}\hoch{E}T_{E|c]}\hoch{\bs{\delta}}-R_{[\bs{\alpha\beta}c]}\hoch{\bs{\delta}}=\\
 & = & \frac{2}{3}\gem{T}_{\bs{\alpha\beta}}\hoch{E}T_{Ec}\hoch{\bs{\delta}}+\frac{4}{3}\gem{T}_{c[\bs{\alpha}|}\hoch{E}\underbrace{T_{E|\bs{\beta}]}\hoch{\bs{\delta}}}_{=0}-\frac{2}{3}R_{c[\bs{\alpha\beta}]}\hoch{\bs{\delta}}=\\
 & = & \frac{2}{3}\gamma_{\bs{\alpha\beta}}\hoch{e}T_{ec}\hoch{\bs{\delta}}-\frac{2}{3}R_{c[\bs{\alpha\beta}]}\hoch{\bs{\delta}}\end{eqnarray}
\Ram{0.4}{\begin{eqnarray}
R_{c[\bs{\alpha\beta}]}\hoch{\bs{\delta}} & = & \gamma_{\bs{\alpha\beta}}\hoch{e}T_{ec}\hoch{\bs{\delta}}\label{eq:(delta|1,2,0)}\\
\hat{R}_{c[\hat{\bs{\alpha}}\hat{\bs{\beta}}]}\hoch{\hat{\bs{\delta}}} & = & \gamma_{\hat{\bs{\alpha}}\hat{\bs{\beta}}}\hoch{e}\hat{T}_{ec}\hoch{\hat{\bs{\delta}}}\label{eq:(hdelta|1,0,2)}\end{eqnarray}
}  Taking the trace yields\begin{eqnarray}
0 & = & R_{c\bs{\alpha\delta}}\hoch{\bs{\delta}}-R_{c\bs{\delta\alpha}}\hoch{\bs{\delta}}-2\gamma_{\bs{\alpha\delta}}\hoch{e}T_{ec}\hoch{\bs{\delta}}=\\
 & = & -\frac{17}{2}F_{c\bs{\alpha}}^{(D)}-R_{c\bs{\delta\alpha}}^{(L)}\hoch{\bs{\delta}}-2\gamma_{\bs{\alpha\delta}}\hoch{e}T_{ec}\hoch{\bs{\delta}}\end{eqnarray}
\begin{equation}
\boxed{F_{c\bs{\alpha}}^{(D)}=-\frac{2}{17}R_{c\bs{\delta\alpha}}^{(L)}\hoch{\bs{\delta}}-\frac{4}{17}\gamma_{\bs{\alpha\delta}}^{e}T_{ec}\hoch{\bs{\delta}}}\label{eq:Fcalph}\end{equation}
\begin{equation}
\boxed{\hat{F}_{c\hat{\bs{\alpha}}}^{(D)}=-\frac{2}{17}\hat{R}_{c\hat{\bs{\delta}}\hat{\bs{\alpha}}}^{(L)}\hoch{\hat{\bs{\delta}}}-\frac{4}{17}\gamma_{\hat{\bs{\alpha}}\hat{\bs{\delta}}}^{e}\hat{T}_{ec}\hoch{\hat{\bs{\delta}}}}\label{eq:Fhatcalphhat}\end{equation}
$\bullet\quad$\underbar{(delta|1,1,1)$_{\bs{\alpha}\hat{\bs{\beta}}c}\hoch{\bs{\delta}}\leftrightarrow$((hdelta|1,1,1)$_{\hat{\bs{\alpha}}\bs{\beta}c}\hoch{\hat{\bs{\delta}}}$)dim$\frac{3}{2}$:}%
\footnote{\index{footnote!\thefoot. another calculational remark}\begin{eqnarray*}
\bei{\gem{\nabla}_{M}\tilde{\gamma}_{c\,\bs{\alpha}\bs{\beta}}}{\check{\Omega}=\Omega} & = & 2\gamma_{\bs{\alpha}\bs{\beta}}^{d}\covPhi{M}G_{dc}=2\tilde{\gamma}_{c\,\bs{\alpha}\bs{\beta}}\covPhi{M}\\
\bei{\gem{\nabla}_{M}\tilde{\gamma}_{c\,\bs{\alpha}\bs{\beta}}}{\check{\Omega}=\hat{\Omega}} & = & 2\tilde{\gamma}_{c\,\bs{\alpha}\bs{\beta}}\covPhi{M}-\Delta_{Mc}\hoch{d}\tilde{\gamma}_{d\,\bs{\alpha}\bs{\beta}}=\\
 & = & \gamma_{\bs{\alpha}\bs{\beta}}^{d}\left[2\covPhi{M}G_{dc}-\Delta_{Mc|d}\right]=\\
 & = & \gamma_{\bs{\alpha}\bs{\beta}}^{d}\left[\left(\covPhi{M}+\hatcovPhi{M}\right)G_{dc}-\Delta_{Mcd}^{(L)}\right]\end{eqnarray*}
And equivalently \begin{eqnarray*}
\bei{\gem{\nabla}_{M}\tilde{\gamma}_{c\,\hat{\bs{\alpha}}\hat{\bs{\beta}}}}{\check{\Omega}=\hat{\Omega}} & = & 2\tilde{\gamma}_{c\,\hat{\bs{\alpha}}\hat{\bs{\beta}}}\hatcovPhi{M}\\
\bei{\gem{\nabla}_{M}\tilde{\gamma}_{c\,\hat{\bs{\alpha}}\hat{\bs{\beta}}}}{\check{\Omega}=\Omega} & = & \gamma_{\hat{\bs{\alpha}}\hat{\bs{\beta}}}^{d}\left[\left(\covPhi{M}+\hatcovPhi{M}\right)G_{dc}+\Delta_{Mcd}^{(L)}\right]\qquad\fussend\end{eqnarray*}
}\begin{eqnarray}
0 & \stackrel{!}{=} & \gem{\nabla}_{[\bs{\alpha}}T_{\hat{\bs{\beta}}c]}\hoch{\bs{\delta}}+2\gem{T}_{[\bs{\alpha}\hat{\bs{\beta}}|}\hoch{E}T_{E|c]}\hoch{\bs{\delta}}-R_{[\bs{\alpha}\hat{\bs{\beta}}c]}\hoch{\bs{\delta}}=\\
 & \stackrel{\check{\Omega}=\hat{\Omega}}{=} & \frac{1}{3}\gem{\nabla}_{\bs{\alpha}}\underbrace{T_{\hat{\bs{\beta}}c}\hoch{\bs{\delta}}}_{\tilde{\gamma}_{c\,\hat{\bs{\beta}}\hat{\bs{\delta}}}\RR^{\bs{\delta}\hat{\bs{\delta}}}}+\frac{2}{3}\hat{T}_{c\bs{\alpha}}\hoch{\hat{\bs{\eps}}}\underbrace{T_{\hat{\bs{\eps}}\hat{\bs{\beta}}}\hoch{\bs{\delta}}}_{=0}+\frac{2}{3}\underbrace{\hat{T}_{c\bs{\alpha}}\hoch{e}}_{=0}T_{e\hat{\bs{\beta}}}\hoch{\bs{\delta}}-\frac{1}{3}\underbrace{R_{\hat{\bs{\beta}}c\bs{\alpha}}\hoch{\bs{\delta}}}_{\tilde{\gamma}_{c\,\hat{\bs{\beta}}\hat{\bs{\delta}}}C_{\bs{\alpha}}\hoch{\bs{\delta}\hat{\bs{\delta}}}}=\\
 & = & \frac{1}{3}\gem{\nabla}_{\bs{\alpha}}\left(\tilde{\gamma}_{c\,\hat{\bs{\beta}}\hat{\bs{\delta}}}\RR^{\bs{\delta}\hat{\bs{\delta}}}\right)-\frac{1}{3}\tilde{\gamma}_{c\,\hat{\bs{\beta}}\hat{\bs{\delta}}}\gem{\nabla}_{\bs{\alpha}}P^{\bs{\delta}\hat{\bs{\delta}}}=\\
 & = & \frac{1}{3}\gem{\nabla}_{\bs{\alpha}}\left(\tilde{\gamma}_{c\hat{\bs{\beta}}\hat{\bs{\delta}}}\right)\,\RR^{\bs{\delta}\hat{\bs{\delta}}}=\\
 & = & \tfrac{2}{3}\tilde{\gamma}_{c\,\hat{\bs{\beta}}\hat{\bs{\delta}}}\hatcovPhi{\bs{\alpha}}\RR^{\bs{\delta}\hat{\bs{\delta}}}=0\end{eqnarray}
\\
$\bullet\quad$\underbar{(delta|1,0,2)$_{\hat{\bs{\alpha}}\hat{\bs{\beta}}c}\hoch{\bs{\delta}}\leftrightarrow$((hdelta|1,2,0)$_{\bs{\alpha}\bs{\beta}c}\hoch{\hat{\bs{\delta}}}$)dim$\frac{3}{2}$:}\begin{eqnarray}
0 & \stackrel{!}{=} & \gem{\nabla}_{[\hat{\bs{\alpha}}}T_{\hat{\bs{\beta}}c]}\hoch{\bs{\delta}}+2\gem{T}_{[\hat{\bs{\alpha}}\hat{\bs{\beta}}|}\hoch{E}T_{E|c]}\hoch{\bs{\delta}}-R_{[\hat{\bs{\alpha}}\hat{\bs{\beta}}c]}\hoch{\bs{\delta}}=\\
 & = & \frac{2}{3}\gem{\nabla}_{[\hat{\bs{\alpha}}}\underbrace{T_{\hat{\bs{\beta}}]c}\hoch{\bs{\delta}}}_{\tilde{\gamma}_{c\,\hat{\bs{\beta}}]\hat{\bs{\gamma}}}\RR^{\bs{\delta}\hat{\bs{\gamma}}}}+\frac{2}{3}\check{T}_{\hat{\bs{\alpha}}\hat{\bs{\beta}}}\hoch{e}T_{ec}\hoch{\bs{\delta}}+\frac{4}{3}\check{T}_{c[\hat{\bs{\alpha}}|}\hoch{e}T_{e|\hat{\bs{\beta}}]}\hoch{\bs{\delta}}=\\
 & \stackrel{\check{\Omega}=\hat{\Omega}}{=} & \frac{4}{3}\hatcovPhi{[\hat{\bs{\alpha}|}}\tilde{\gamma}_{c\,|\hat{\bs{\beta}}]\hat{\bs{\gamma}}}\RR^{\bs{\delta}\hat{\bs{\gamma}}}+\frac{2}{3}\gem{\nabla}_{[\hat{\bs{\alpha}}}\RR^{\bs{\delta}\hat{\bs{\gamma}}}\tilde{\gamma}_{c\,\hat{\bs{\beta}}]\hat{\bs{\gamma}}}+\frac{2}{3}\gamma_{\hat{\bs{\alpha}}\hat{\bs{\beta}}}^{e}T_{ec}\hoch{\bs{\delta}}+\frac{4}{3}\hat{T}_{[\hat{\bs{\alpha}}|c}\hoch{e}\tilde{\gamma}_{e\,|\hat{\bs{\beta}}]\hat{\bs{\delta}}}\RR^{\bs{\delta}\hat{\bs{\delta}}}=\\
 & = & \left(\frac{4}{3}\left(\hatcovPhi{[\hat{\bs{\alpha}|}}\delta_{c}^{e}+\hat{T}_{[\hat{\bs{\alpha}}|c}\hoch{e}\right)\RR^{\bs{\delta}\hat{\bs{\gamma}}}+\frac{2}{3}\gem{\nabla}_{[\hat{\bs{\alpha}}|}\RR^{\bs{\delta}\hat{\bs{\gamma}}}\delta_{c}^{e}\right)\tilde{\gamma}_{e\,|\hat{\bs{\beta}}]\hat{\bs{\gamma}}}+\frac{2}{3}\gamma_{\hat{\bs{\alpha}}\hat{\bs{\beta}}}^{e}T_{ec}\hoch{\bs{\delta}}=\\
 & = & \frac{2}{3}\big(-\underbrace{2\hat{T}_{[\hat{\bs{\alpha}}|e|c}}_{\Delta_{[\hat{\bs{\alpha}}|e|c}}\RR^{\bs{\delta}\hat{\bs{\gamma}}}+\gem{\nabla}_{[\hat{\bs{\alpha}}|}\RR^{\bs{\delta}\hat{\bs{\gamma}}}G_{ec}\big)\gamma_{\,|\hat{\bs{\beta}}]\hat{\bs{\gamma}}}^{e}+\frac{2}{3}\gamma_{\hat{\bs{\alpha}}\hat{\bs{\beta}}}^{e}T_{ec}\hoch{\bs{\delta}}\end{eqnarray}
Contracting the above with $\gamma_{e}^{\hat{\bs{\alpha}}\hat{\bs{\beta}}}$
(using $\gamma_{e}^{\hat{\bs{\alpha}}\hat{\bs{\beta}}}\gamma_{\hat{\bs{\alpha}}\hat{\bs{\beta}}}^{f}=-\gamma_{e}^{\hat{\bs{\alpha}}\hat{\bs{\beta}}}\gamma_{\hat{\bs{\beta}}\hat{\bs{\alpha}}}^{f}=-\gamma_{e}^{\hat{\alpha}\hat{\beta}}\gamma_{\hat{\beta}\hat{\alpha}}^{f}=-16\delta_{e}^{f}$),
we get\begin{eqnarray}
T_{ec}\hoch{\bs{\delta}} & = & \frac{1}{16}\left(\gem{\nabla}_{[\hat{\bs{\alpha}}|}\RR^{\bs{\delta}\hat{\bs{\delta}}}G_{cd}-2\hat{T}_{[\hat{\bs{\alpha}}|d:c}\RR^{\bs{\delta}\hat{\bs{\delta}}}\right)\gamma_{|\hat{\bs{\beta}}]\hat{\bs{\delta}}}^{d}\gamma_{e}^{\hat{\bs{\alpha}}\hat{\bs{\beta}}}=\\
 & = & \frac{1}{16}\left(2\hat{T}_{\hat{\bs{\alpha}}d|c}\RR^{\bs{\delta}\hat{\bs{\delta}}}-\gem{\nabla}_{\hat{\bs{\alpha}}}\RR^{\bs{\delta}\hat{\bs{\delta}}}G_{cd}\right)\gamma_{\hat{\bs{\delta}}\hat{\bs{\beta}}}^{d}\gamma_{e}^{\hat{\bs{\alpha}}\hat{\bs{\beta}}}\end{eqnarray}
\\
\Ram{0.6}{\begin{eqnarray}
T_{ec}\hoch{\bs{\delta}} & = & \frac{1}{16}\left(2\hat{T}_{\hat{\bs{\alpha}}d|c}\RR^{\bs{\delta}\hat{\bs{\delta}}}-\gem{\nabla}_{\hat{\bs{\alpha}}}\RR^{\bs{\delta}\hat{\bs{\delta}}}G_{cd}\right)\gamma_{\hat{\bs{\delta}}\hat{\bs{\beta}}}^{d}\gamma_{e}^{\hat{\bs{\alpha}}\hat{\bs{\beta}}}\label{eq:(delta|1,0,2)}\\
\hat{T}_{ec}\hoch{\hat{\bs{\delta}}} & = & \frac{1}{16}\left(2T_{\bs{\alpha}d|c}\RR^{\bs{\delta}\hat{\bs{\delta}}}-\gem{\nabla}_{\bs{\alpha}}\RR^{\bs{\delta}\hat{\bs{\delta}}}G_{cd}\right)\gamma_{\bs{\delta}\bs{\beta}}^{d}\gamma_{e}^{\bs{\alpha}\bs{\beta}}\label{eq:(hdelta|1,2,0)}\end{eqnarray}
}  \\
The product of $\gamma$-matrices can be further expanded. \begin{eqnarray}
T_{ec}\hoch{\bs{\delta}} & = & \frac{1}{16}\left(2\hat{T}_{\hat{\bs{\alpha}}d|c}\RR^{\bs{\delta}\hat{\bs{\delta}}}-\gem{\nabla}_{\hat{\bs{\alpha}}}\RR^{\bs{\delta}\hat{\bs{\delta}}}G_{cd}\right)\left(\delta_{e}^{d}\delta_{\hat{\bs{\delta}}}\hoch{\hat{\bs{\alpha}}}+\gamma^{d}\tief{e\,\hat{\bs{\delta}}}\hoch{\hat{\bs{\alpha}}}\right)=\\
 & = & \frac{1}{16}\big(2\hat{T}_{\hat{\bs{\alpha}}e|c}\RR^{\bs{\delta}\hat{\bs{\alpha}}}-\gem{\nabla}_{\hat{\bs{\alpha}}}\RR^{\bs{\delta}\hat{\bs{\alpha}}}G_{ce}+\underbrace{2\hat{T}_{\hat{\bs{\alpha}}d|c}\gamma^{d}\tief{e\,\hat{\bs{\delta}}}\hoch{\hat{\bs{\alpha}}}}_{-18\hat{T}_{\hat{\bs{\delta}}e|c}\:(\ref{eq:puh})}\RR^{\bs{\delta}\hat{\bs{\delta}}}-\gem{\nabla}_{\hat{\bs{\alpha}}}\RR^{\bs{\delta}\hat{\bs{\delta}}}\tilde{\gamma}_{ce\,\hat{\bs{\delta}}}\hoch{\hat{\bs{\alpha}}}\big)\end{eqnarray}
 The result should be antisymmetric in $e$ and $c$. Remember now\begin{equation}
\gem{\nabla}_{\hat{\bs{\alpha}}}\RR^{\bs{\delta}\hat{\bs{\alpha}}}G_{ce}=8\RR^{\bs{\delta}\hat{\bs{\delta}}}\hatcovPhi{\hat{\bs{\delta}}}G_{ce}=-16\RR^{\bs{\delta}\hat{\bs{\delta}}}\hat{T}_{\hat{\bs{\delta}}(c|e)}\end{equation}
and we get\begin{eqnarray}
T_{ec}\hoch{\bs{\delta}} & = & \frac{1}{16}\left(-16\hat{T}_{\hat{\bs{\delta}}e|c}\RR^{\bs{\delta}\hat{\bs{\delta}}}+16\RR^{\bs{\delta}\hat{\bs{\delta}}}\hat{T}_{\hat{\bs{\delta}}(c|e)}-\gem{\nabla}_{\hat{\bs{\alpha}}}\RR^{\bs{\delta}\hat{\bs{\delta}}}\tilde{\gamma}_{ce\,\hat{\bs{\delta}}}\hoch{\hat{\bs{\alpha}}}\right)=\\
 & = & \frac{1}{16}\left(-16\hat{T}_{\hat{\bs{\delta}}[e|c]}\RR^{\bs{\delta}\hat{\bs{\delta}}}-\gem{\nabla}_{\hat{\bs{\alpha}}}\RR^{\bs{\delta}\hat{\bs{\delta}}}\tilde{\gamma}_{ce\,\hat{\bs{\delta}}}\hoch{\hat{\bs{\alpha}}}\right)\end{eqnarray}
Using $\hat{T}_{\hat{\bs{\delta}}[e|c]}=-\frac{1}{2}\gamma_{ec}\tief{\hat{\bs{\delta}}}\hoch{\hat{\bs{\gamma}}}\hatcovPhi{\hat{\bs{\gamma}}}$
leads to \begin{equation}
\boxed{T_{ec}\hoch{\bs{\delta}}=\frac{1}{16}\left(\gem{\nabla}_{\hat{\bs{\gamma}}}\RR^{\bs{\delta}\hat{\bs{\delta}}}+8\hatcovPhi{\hat{\bs{\gamma}}}\RR^{\bs{\delta}\hat{\bs{\delta}}}\right)\tilde{\gamma}_{ec\,\hat{\bs{\delta}}}\hoch{\hat{\bs{\gamma}}}}\label{eq:(delta|1,0,2)b}\end{equation}
\begin{equation}
\boxed{\hat{T}_{ec}\hoch{\hat{\bs{\delta}}}=\frac{1}{16}\left(\gem{\nabla}_{\bs{\gamma}}\RR^{\bs{\delta}\hat{\bs{\delta}}}+8\covPhi{\bs{\gamma}}\RR^{\bs{\delta}\hat{\bs{\delta}}}\right)\tilde{\gamma}_{ec\,\bs{\delta}}\hoch{\bs{\gamma}}}\label{eq:(hdelta|1,2,0)b}\end{equation}
\rem{In the special case $\gem{\nabla}_{\hat{\bs{\gamma}}}\RR^{\bs{\delta}\hat{\bs{\delta}}}=8\hatcovPhi{\hat{\bs{\gamma}}}\RR^{\bs{\delta}\hat{\bs{\delta}}}$,
we have $T_{ec}\hoch{\bs{\delta}}=\hatcovPhi{\hat{\bs{\gamma}}}\RR^{\bs{\delta}\hat{\bs{\delta}}}\tilde{\gamma}_{ec\,\hat{\bs{\delta}}}\hoch{\hat{\bs{\gamma}}}$
and not $0$! Is the sign of $\nabla\RR$ really correct?}Instead
of solving for the torsion component, we can also solve for the covariant
derivative of the RR-field:\begin{eqnarray}
\frac{1}{16}\gemnabla_{\hat{\bs{\gamma}}}\RR^{\bs{\delta}\hat{\bs{\delta}}}\tilde{\gamma}_{ec\,\hat{\bs{\delta}}}\hoch{\hat{\bs{\gamma}}} & = & T_{ec}\hoch{\bs{\delta}}-\frac{1}{2}\hatcovPhi{\hat{\bs{\gamma}}}\RR^{\bs{\delta}\hat{\bs{\delta}}}\tilde{\gamma}_{ec\,\hat{\bs{\delta}}}\hoch{\hat{\bs{\gamma}}}\end{eqnarray}
Together with (\ref{eq:RR-eoms}) and the fact that $C_{\bs{\alpha}}\hoch{\bs{\beta}\hat{\bs{\gamma}}}=\gemnabla_{\bs{\alpha}}\RR\hoch{\bs{\beta}\hat{\bs{\gamma}}}$
is structure group valued in $\bs{\alpha}$ and $\bs{\beta}$ (as
well as $\hat{C}$), we get\vRam{.8}{\begin{eqnarray}
\gemnabla_{\hat{\bs{\alpha}}}\RR^{\bs{\alpha}\hat{\bs{\beta}}} & = & -\frac{1}{2}\RR^{\bs{\alpha}\hdu{0}}\hatcovPhi{\hdu{0}}\cdot\delta_{\hat{\bs{\alpha}}}\hoch{\hat{\bs{\beta}}}+\left(T_{\bdu{0}\bdu{-1}}\hoch{\bs{\alpha}}-\frac{1}{2}\hatcovPhi{\hdu{0}}\RR^{\bs{\alpha}\hdu{-1}}\tilde{\gamma}_{\bdu{0}\bdu{-1}\,\hdu{-1}}\hoch{\hdu{0}}\right)\tilde{\gamma}^{\bdu{0}\bdu{-1}}\tief{\hat{\bs{\alpha}}}\hoch{\hat{\bs{\beta}}}\label{eq:nablaP:hat}\\
\gemnabla_{\bs{\alpha}}\RR^{\bs{\beta}\hat{\bs{\alpha}}} & = & -\frac{1}{2}\RR^{\sdu{0}\hat{\bs{\alpha}}}\covPhi{\sdu{0}}\cdot\delta_{\bs{\alpha}}\hoch{\bs{\beta}}+\left(\hat{T}_{\bdu{0}\bdu{-1}}\hoch{\hat{\bs{\alpha}}}-\frac{1}{2}\covPhi{\sdu{0}}\RR^{\sdu{-1}\hat{\bs{\alpha}}}\tilde{\gamma}_{\bdu{0}\bdu{-1}\,\sdu{-1}}\hoch{\sdu{0}}\right)\tilde{\gamma}^{\bdu{0}\bdu{-1}}\tief{\bs{\alpha}}\hoch{\bs{\beta}}\label{eq:nablaP}\end{eqnarray}
} Due to the algebra of covariant derivatives, the above equations
also contain informations on the spacetime derivative of $\RR^{\bs{\alpha}\hat{\bs{\beta}}}$.
It is thus of interest to study the commutator $\gemnabla_{[\bs{\gamma}}\gemnabla_{\bs{\alpha}]}\RR^{\bs{\beta}\hat{\bs{\alpha}}}$:
\begin{eqnarray}
\lqn{-\gamma_{\bs{\gamma\alpha}}^{d}\gemnabla_{d}\RR^{\bs{\beta}\hat{\bs{\alpha}}}+\gemR_{\bs{\gamma\alpha\delta}}\hoch{\bs{\beta}}\RR^{\bs{\delta}\hat{\bs{\alpha}}}+\underbrace{\gemR_{\bs{\gamma\alpha}\hat{\bs{\delta}}}\hoch{\hat{\bs{\alpha}}}}_{=0}\RR^{\bs{\beta}\hat{\bs{\delta}}}=\gemnabla_{[\bs{\gamma}}\gemnabla_{\bs{\alpha}]}\RR^{\bs{\beta}\hat{\bs{\alpha}}}=}\nonumber \\
 & = & -\frac{1}{2}\gemnabla_{[\bs{\gamma}|}\RR^{\bs{\delta}\hat{\bs{\alpha}}}\covPhi{\bs{\delta}}\cdot\delta_{|\bs{\alpha}]}\hoch{\bs{\beta}}-\frac{1}{2}\RR^{\bs{\delta}\hat{\bs{\alpha}}}\nabla_{[\bs{\gamma}|}\covPhi{\bs{\delta}}\cdot\delta_{|\bs{\alpha}]}\hoch{\bs{\beta}}+\nonumber \\
 &  & +\gemnabla_{[\bs{\gamma}|}\left(\hat{T}_{bc}\hoch{\hat{\bs{\alpha}}}-\frac{1}{2}\covPhi{\bs{\delta}}\RR^{\bs{\eps}\hat{\bs{\alpha}}}\tilde{\gamma}_{bc\,\bs{\eps}}\hoch{\bs{\delta}}\right)\tilde{\gamma}^{bc}\tief{|\bs{\alpha}]}\hoch{\bs{\beta}}+\left(\hat{T}_{bc}\hoch{\hat{\bs{\alpha}}}-\frac{1}{2}\covPhi{\bs{\delta}}\RR^{\bs{\eps}\hat{\bs{\alpha}}}\tilde{\gamma}_{bc\,\bs{\eps}}\hoch{\bs{\delta}}\right)\bei{\gemnabla_{[\bs{\gamma}|}\tilde{\gamma}^{bc}\tief{|\bs{\alpha}]}\hoch{\bs{\beta}}}{\check{\Omega}=\Omega}\end{eqnarray}
In particular, we obtain a Dirac-like operator acting on the first
index of $\RR^{\bs{\alpha}\hat{\bs{\beta}}}$ if we contract the indices
$\bs{\alpha}$ and $\bs{\beta}$: \begin{eqnarray}
\lqn{-\gamma_{\bs{\gamma\alpha}}^{d}\gemnabla_{d}\RR^{\bs{\alpha}\hat{\bs{\alpha}}}+\underbrace{R_{\bs{\gamma\alpha\delta}}\hoch{\bs{\alpha}}}_{-4F_{\bs{\gamma\delta}}^{(D)}=\lqn{{\scriptstyle 4\nabla_{[\bs{\gamma}}\covPhi{\bs{\delta}]}+4\gamma_{\bs{\gamma\delta}}^{c}\covPhi{c}}}}\RR^{\bs{\delta}\hat{\bs{\alpha}}}=}\nonumber \\
 & = & \frac{17}{2}\gemnabla_{\bs{\gamma}}\RR^{\bs{\delta}\hat{\bs{\alpha}}}\covPhi{\bs{\delta}}+\frac{17}{2}\RR^{\bs{\delta}\hat{\bs{\alpha}}}\nabla_{\bs{\gamma}}\covPhi{\bs{\delta}}-\frac{1}{2}\gemnabla_{\bs{\alpha}}\left(\hat{T}_{bc}\hoch{\hat{\bs{\alpha}}}-\frac{1}{2}\covPhi{\bs{\delta}}\RR^{\bs{\eps}\hat{\bs{\alpha}}}\tilde{\gamma}_{bc\,\bs{\eps}}\hoch{\bs{\delta}}\right)\tilde{\gamma}^{bc}\tief{\bs{\gamma}}\hoch{\bs{\alpha}}+\nonumber \\
 &  & -\frac{1}{2}\left(\hat{T}_{bc}\hoch{\hat{\bs{\alpha}}}-\frac{1}{2}\covPhi{\bs{\delta}}\RR^{\bs{\eps}\hat{\bs{\alpha}}}\tilde{\gamma}_{bc\,\bs{\eps}}\hoch{\bs{\delta}}\right)\gemnabla_{\bs{\alpha}}\tilde{\gamma}^{bc}\tief{\bs{\gamma}}\hoch{\bs{\alpha}}\qquad\label{eq:start-point-for-qualitativeRRdisc}\end{eqnarray}
In the same way we can obtain an equation for Dirac-like operator
acting on the second index of $\RR^{\bs{\alpha}\hat{\bs{\beta}}}$,
if we consider the hatted version of the above equation. 

Plugging further torsion constraints into these equations yields rather
lengthy expressions and we thus restrict ourselves to a qualitative
discussion of the further steps which would lead to field equations
for the RR-p-forms, to be presented in the following intermezzo. \vspace{.5cm}

\lyxline{\normalsize}\vspace{-.25cm}\lyxline{\normalsize}

\subsubsection*{Intermezzo\index{intermezzo!RR-field equations} on the RR-field-equations}

\label{intermezzo:RR}As just mentioned above, the equation (\ref{eq:start-point-for-qualitativeRRdisc})
and its hatted equivalent together with some other torsion constraints
of before determine the equations of motion of the RR-field strengths.
We will make a qualitive discussion and assume that the fermionic
fields vanish so that the equations in WZ gauge basically reduce to
$\gamma_{\bs{\gamma\alpha}}^{d}\gemnabla_{d}\rr^{\bs{\alpha}\hat{\bs{\alpha}}}=0$
and $\gamma_{\hat{\bs{\gamma}}\hat{\bs{\alpha}}}^{d}\gemnabla_{d}\rr^{\bs{\alpha}\hat{\bs{\alpha}}}=0$
where $\rr^{\bs{\alpha}\hat{\bs{\alpha}}}$ is the leading component
of $\RR^{\bs{\alpha}\hat{\bs{\alpha}}}$ in the $\xbothtetas$-expansion
(see page \pageref{sub:notationComponentFields}). 

In order to see that this corresponds to reasonable equations for
the RR\index{RR-p-form}-p-forms, let us first recall the translation
of field equations on the bispinor fields $\rr^{\bs{\alpha}\hat{\bs{\beta}}}$
into the equations on the level of differential forms in the flat
case. On the form level one expects for the RR-field strength's $g^{(p)}$
s.th. like $\de g^{(p)}=0$ and\index{$\star$|itext{Hodge star}}
$\star\de\star g^{(p)}=0$. As it is discussed in the appendix on
page \pageref{eq:actionOfDirac-op} and in the following, this corresponds
on the bispinor level precisely to two Dirac equations, one acting
on the first index and one on the second, i.e. $\not\partial_{\bs{\gamma}\bs{\alpha}}\rr^{\bs{\alpha}\hat{\bs{\beta}}}=\not\not\partial_{\hat{\bs{\gamma}}\hat{\bs{\beta}}}\rr^{\bs{\alpha}\hat{\bs{\beta}}}=0$
with $\not\partial_{\bs{\alpha}\bs{\beta}}=\gamma_{\bs{\alpha\beta}}^{m}\partial_{m}$.
Of course the equations are not yet the full truth, as they do not
reflect the curved background. In order to show the above correspondence,
we need to distinguish between \index{two!type IIA}\index{type IIA}IIA
(where ${\scriptstyle \bs{\alpha}}$ and ${\scriptstyle \hat{\bs{\alpha}}}$
are of opposite chirality) and type IIB\index{type IIB}\index{two!type IIB}
(where ${\scriptstyle \bs{\alpha}}$ and ${\scriptstyle \hat{\bs{\alpha}}}$
are of the same chirality). We will frequently use equations from
the appendix \vref{app:gamma} where we did not use the graded conventions.
We will therefore consider in this intermezzo the spinorial indices
ungraded in the summations (this refers to the graded summation convention
discussed in the first part of this thesis; if you have not read that
part, you can safely ignore the comment).

Assume we are in \textbf{type IIA} where we can expand the RR-bispinor
in even antisymmetrized products of $\gamma$-matrices: \begin{eqnarray}
\rr^{\alpha}\tief{\beta} & = & 2g^{(0)}\underbrace{\delta_{\beta}^{\alpha}}_{\gamma^{[0]}}+2g_{a_{1}a_{2}}^{(2)}\underbrace{\gamma^{a_{1}a_{2}\,\alpha}\tief{\beta}}_{\gamma^{[2]}}+2g_{a_{1}a_{2}a_{3}a_{4}}^{(4)}\underbrace{\gamma^{a_{1}a_{2}a_{3}a_{4}\alpha}\tief{\beta}}_{\gamma^{[4]}}\label{eq:RR-exp:IIA}\\
2g^{(0)} & = & \frac{1}{16}\rr^{\alpha}\tief{\beta}\delta_{\alpha}^{\beta}\\
2g_{a_{1}a_{2}}^{(2)} & = & \frac{1}{32}\rr^{\alpha}\tief{\beta}\gamma_{a_{2}a_{1}}\hoch{\beta}\tief{\alpha}\\
2g_{a_{1}a_{2}a_{3}a_{4}}^{(4)} & = & \frac{1}{16\cdot4!}\rr^{\alpha}\tief{\beta}\gamma_{a_{4}a_{3}a_{2}a_{1}}\hoch{\beta}\tief{\alpha}\end{eqnarray}
Usually the coeficients $g_{a_{1}\ldots a_{p}}^{(p)}$ which correspond
to p-forms (or better p-form field strengths) are denoted with a capital
$G$, but we want to keep the capital letters reserved for superfields.
The matrices $\gamma^{[0]},\gamma^{[2]}$ and $\gamma^{[4]}$ are
the chiral blocks of the antisymmetrized products of the Dirac gamma
matrices $\Gamma^{[2k]}$ which is block diagonal. Similarly, $\Gamma^{[2k+1]}$
is block off-diagonal and defines the chiral blocks $\gamma^{[2k+1]}$:
\begin{equation}
\Gamma^{[2k]\q{\alpha}}\tief{\q{\beta}}=\left(\begin{array}{cc}
\gamma^{[2k]\alpha}\tief{\beta} & 0\\
0 & \gamma^{[2k]}\tief{\alpha}\hoch{\beta}\end{array}\right),\quad\Gamma^{[2k+1]\q{\alpha}}\tief{\q{\beta}}=\left(\begin{array}{cc}
0 & \gamma^{[2k]\alpha\beta}\\
\gamma_{\alpha\beta}^{[2k+1]} & 0\end{array}\right)\end{equation}
The chiral blocks can be extracted via the chirality matrix $\Gamma^{\#\,\q{\alpha}}\tief{\q{\beta}}=\left(\begin{array}{cc}
\gamma^{\#\alpha}\tief{\beta} & 0\\
0 & \gamma_{\alpha}^{\#\,\beta}\end{array}\right)=\left(\begin{array}{cc}
\delta_{\beta}^{\alpha} & 0\\
0 & -\delta_{\alpha}^{\beta}\end{array}\right)$ which acts (when multiplied from the right) on the first coloumn
as the identity and on the second one as minus the identity:\begin{eqnarray}
\frac{1}{2}\left(\Gamma^{[2k]}(\one+\Gamma^{\#})\right)^{\q{\alpha}}\tief{\q{\beta}} & = & \left(\begin{array}{cc}
\gamma^{[2k]\alpha}\tief{\beta} & 0\\
0 & 0\end{array}\right),\quad\frac{1}{2}\left(\Gamma^{[2k]}(\one-\Gamma^{\#})\right)^{\q{\alpha}}\tief{\q{\beta}}=\left(\begin{array}{cc}
0 & 0\\
0 & \gamma^{[2k]}\tief{\alpha}\hoch{\beta}\end{array}\right)\qquad\label{eq:chiralBlocksEven}\\
\frac{1}{2}\left(\Gamma^{[2k+1]}(\one+\Gamma^{\#})\right)^{\q{\alpha}}\tief{\q{\beta}} & = & \left(\begin{array}{cc}
0 & 0\\
\gamma_{\alpha\beta}^{[2k+1]} & 0\end{array}\right),\quad\frac{1}{2}\left(\Gamma^{[2k+1]}(\one-\Gamma^{\#})\right)^{\q{\alpha}}\tief{\q{\beta}}=\left(\begin{array}{cc}
0 & \gamma^{[2k+1]\alpha\beta}\\
0 & 0\end{array}\right)\qquad\label{eq:chiralBlocksOdd}\end{eqnarray}
Via the clifford map, the $\Gamma^{[2k]}$ get mapped to even forms.
In addition we define the Hodge star operator such that it corresponds
via this mapping to the multiplication of the chirality matrix from
the right (see page \pageref{eq:chirality-on-pformII}). The chiral
blocks thus get mapped as follows\begin{eqnarray}
\gamma^{[2k]\alpha}\tief{\beta} & \stackrel{\slash^{-1}}{\mapsto} & \tfrac{1}{2}(1+\star)e^{a_{1}}\wedge\ldots\wedge e^{a_{2k}},\qquad\gamma^{[2k]}\tief{\alpha}\hoch{\beta}\stackrel{\slash^{-1}}{\mapsto}\tfrac{1}{2}(1-\star)e^{a_{1}}\wedge\ldots\wedge e^{a_{2k}}\label{eq:chiralBlocksClMapEven}\\
\gamma_{\alpha\beta}^{[2k+1]} & \stackrel{\slash^{-1}}{\mapsto} & \tfrac{1}{2}(1+\star)e^{a_{1}}\wedge\ldots\wedge e^{a_{2k+1}},\qquad\gamma^{[2k+1]\alpha\beta}\stackrel{\slash^{-1}}{\mapsto}\tfrac{1}{2}(1-\star)e^{a_{1}}\wedge\ldots\wedge e^{a_{2k+1}}\label{eq:chiralBlocksClMapOdd}\end{eqnarray}
and the bispinor field $\rr^{\alpha}\tief{\beta}$ therefore corresponds
to an even self-dual formal sum of differential forms:\begin{eqnarray}
\rr^{\alpha}\tief{\beta}=\not\! g^{\alpha}\tief{\beta} & \stackrel{\slash^{-1}}{\mapsto} & g\equiv g^{(0)}\Bigl(1+\underbrace{\frac{1}{10!}\epsilon\tief{b_{1}\ldots b_{10}}e^{b_{1}}\wedge\ldots\wedge e^{b_{10}}}_{\star1}\Bigr)+g_{a_{1}a_{2}}^{(2)}\Bigl(e^{a_{1}}\wedge e^{a_{2}}\underbrace{-\frac{1}{8!}\epsilon^{a_{1}a_{2}}\tief{b_{1}\ldots b_{8}}e^{b_{1}}\cdots e^{b_{8}}}_{+\star(e^{a_{1}}\wedge e^{a_{2}})}\Bigr)+\nonumber \\
 &  & +g_{a_{1}a_{2}a_{3}a_{4}}^{(4)}\Bigl(e^{a_{1}}\cdots e^{a_{4}}+\underbrace{\frac{1}{6!}\epsilon^{a_{1}\ldots a_{4}}\tief{b_{1}\ldots b_{6}}e^{b_{1}}\cdots e^{b_{6}}}_{\star(e^{a_{1}}\wedge\ldots\wedge e^{a_{4}})}\Bigr)\label{eq:RR-Clifford-map}\end{eqnarray}
According to (\ref{eq:actionOfDirac-op}) and (\ref{eq:actionOfDirac-op-right})
in the appendix, the action of the Dirac operator $\gamma_{\gamma\alpha}^{c}\nabla_{c}$
on the first or $\gamma^{c\,\gamma\beta}\nabla_{c}$ on the second
index (with a covariant derivative that leaves the gamma-matrices
invariant) yields\begin{eqnarray}
\gamma_{\gamma\alpha}^{c}\nabla_{c}\rr^{\alpha}\tief{\beta} & \stackrel{\slash^{-1}}{\mapsto} & \bs{\nabla}g+\star\bs{\nabla}\underbrace{\star g}_{g}\label{eq:expected}\\
\nabla_{c}\rr^{\alpha}\tief{\beta}\cdot\gamma^{c\,\beta\gamma} & \stackrel{\slash^{-1}}{\mapsto} & \bs{\nabla}g-\star\bs{\nabla}\underbrace{\star g}_{g}\label{eq:below-expected}\end{eqnarray}
When $\omega_{ab}\hoch{c}=\omega_{ab}^{(LC)}\hoch{c}+\frac{3}{2}h_{ab}\hoch{c}$
one might expect to get something like the $h$-twisted differential
on the righthand side, but this is not true for a connection that
respects the gamma-matrices as we assumed in the two equations above.
The expression in (\ref{eq:expected}) does not coincide with the
$h$-twisted differential for this choice of connection. It is important
therefore that we act with our {}``mixed'' connection which acts
on the first fermionic index with $\omega_{a\beta}\hoch{\gamma}=\tfrac{1}{4}\left(\omega_{ab}^{(LC)}\hoch{c}+\frac{3}{2}h_{ab}\hoch{c}\right)\gamma^{b}\tief{c\,\beta}\hoch{\gamma}$
and on the second with $\hat{\omega}_{a}\hoch{\beta}\tief{\gamma}=\tfrac{1}{4}\left(\omega_{ab}^{(LC)}\hoch{c}-\frac{3}{2}h_{ab}\hoch{c}\right)\gamma^{b}\tief{c}\hoch{\beta}\tief{\gamma}$.
This mixed connection does not leave both gamma-matrix blocks $\gamma_{\alpha\beta}^{c}$
and $\gamma^{c\,\alpha\beta}$ invariant at the same time. Depending
on the sign we choose for the action on the bosonic index, it either
leaves invariant only the first or only the second. The calculation
of above therefore does not go through in the same way and gets modified
as follows: 

Let us act with the left-mover connection $\omega_{ab}\hoch{c}=\omega_{ab}^{(LC)}\hoch{c}+\frac{3}{2}h_{ab}\hoch{c}$
on the bosonic indices and rewrite $\hat{\omega}_{a}\hoch{\delta}\tief{\beta}=\omega_{a}\hoch{\delta}\tief{\beta}+\Delta_{a}\hoch{\delta}\tief{\beta}=\omega_{a}\hoch{\delta}\tief{\beta}-\tfrac{3}{4}h_{ab}\hoch{c}\gamma^{b}\tief{c}\hoch{\delta}\tief{\beta}$.
We then have%
\footnote{\index{footnote!\thefoot. example for grading shift}In order to better
understand the sign in (\ref{eq:sign-to-be-understood}), note that
the action of the connection on the fermionic indices was defined
via graded conventions according to the first part of the thesis and
that the second (lower) index of the RR-bispinor used to be an upper
hatted index $\rr^{\bs{\alpha}\hat{\bs{\beta}}}=\rr^{\bs{\alpha}}\tief{\bs{\beta}}$.
The action of the covariant derivative is thus \begin{eqnarray*}
\gemnabla_{m}\rr^{\bs{\alpha}\hat{\bs{\beta}}} & \greq & \partial_{m}\rr^{\bs{\alpha}\hat{\bs{\beta}}}+\omega_{m\bs{\delta}}\hoch{\bs{\alpha}}\rr^{\bs{\delta}\hat{\bs{\beta}}}+\hat{\omega}_{m\hat{\bs{\delta}}}\hoch{\hat{\bs{\beta}}}\rr^{\bs{\alpha}\hat{\bs{\delta}}}\end{eqnarray*}
In this second part of the thesis we ususally did not denote the
graded equal sign explicitely. It had to be understood as such, whenever
graded indices appeared. For this explicit comparison, however, it
is useful to make a distinction. In terms of ordinary equal sign and
explicitely written summation (NW-conventions), this becomes: \begin{eqnarray*}
\gemnabla_{m}\rr^{\bs{\alpha}\hat{\bs{\beta}}} & = & \partial_{m}\rr^{\bs{\alpha}\hat{\bs{\beta}}}+\sum_{\bs{\delta}}\underbrace{(-)^{\bs{\delta}+\bs{\delta\alpha}}}_{1}\omega_{m\bs{\delta}}\hoch{\bs{\alpha}}\rr^{\bs{\delta}\hat{\bs{\beta}}}+\underbrace{(-)^{\bs{\alpha}\hat{\bs{\beta}}}}_{-1}\sum_{\hat{\bs{\delta}}}\underbrace{(-)^{\hat{\bs{\delta}}+\hat{\bs{\delta}}(\hat{\bs{\beta}}+\bs{\alpha})}}_{-1}\hat{\omega}_{m\hat{\bs{\delta}}}\hoch{\hat{\bs{\beta}}}\rr^{\bs{\alpha}\hat{\bs{\delta}}}\end{eqnarray*}
In other words, if we consider the indices to carry no grading, we
have\begin{eqnarray*}
\gemnabla_{m}\rr^{\alpha\hat{\beta}} & = & \partial_{m}\rr^{\alpha\hat{\beta}}+\omega_{m\delta}\hoch{\alpha}\rr^{\delta\hat{\beta}}+\hat{\omega}_{m\hat{\delta}}\hoch{\hat{\beta}}\rr^{\alpha\hat{\delta}}\\
\mbox{or }\gemnabla_{m}\rr^{\alpha}\tief{\beta} & = & \partial_{m}\rr^{\alpha}\tief{\beta}+\omega_{m\delta}\hoch{\alpha}\rr^{\delta}\tief{\beta}+\hat{\omega}_{m}\hoch{\delta}\tief{\beta}\rr^{\alpha}\tief{\delta}\quad\fussend\end{eqnarray*}
} \begin{eqnarray}
\gamma_{\gamma\alpha}^{c}\bei{\gemnabla_{c}\rr^{\alpha}\tief{\beta}}{\check{\omega}=\omega} & = & \gamma_{\gamma\alpha}^{c}\nabla_{c}\rr^{\alpha}\tief{\beta}-\tfrac{3}{4}\gamma_{\gamma\alpha}^{c}h_{ca}\hoch{b}\gamma^{a}\tief{b}\hoch{\delta}\tief{\beta}\rr^{\alpha}\tief{\delta}=\label{eq:sign-to-be-understood}\\
 & = & \gamma_{\gamma\alpha}^{c}\nabla_{c}\rr^{\alpha}\tief{\beta}-\tfrac{3}{4}h_{cab}\gamma_{\gamma\alpha}^{[c|}\rr^{\alpha}\tief{\delta}\gamma^{|ab]}\hoch{\delta}\tief{\beta}\label{eq:last-term-of}\end{eqnarray}
In the last term, we have two matrix multiplications between three
matrices (in the spinorial indices), which corresponds on the form
side to two Clifford-multiplications. According to (\ref{eq:chiralBlocksClMapOdd}),
the chiral gamma matrix $\gamma_{\gamma\alpha}^{c}$ can be seen as
the Clifford map of the self-dual projection of the vielbein $\tfrac{1}{2}(1+\star)e^{c}$.
The even form $g$, corresponding to $\rr^{\alpha}\tief{\delta}$,
is given in (\ref{eq:RR-Clifford-map}) and $\gamma^{ab}\hoch{\delta}\tief{\beta}$
corresponds according to (\ref{eq:chiralBlocksClMapEven}) to $\tfrac{1}{2}(1+\star)e^{a}\wedge e^{b}$.
Now we need the explicit expression for the Clifford multiplication
on the form-side and the fact that the Clifford multiplication of
two self-dually projected forms yields either zero or the self dual
projection of their Clifford multiplication (see equation (\ref{eq:Clifford-multI})
and below in the appendix): \begin{eqnarray}
\not\!\omega\not\!\rho & \stackrel{\slash^{-1}}{\mapsto} & \sum_{k\geq0}\frac{1}{k!}\omega\partr{e^{a_{1}}}\cdots\partr{e^{a_{k}}}\eta^{a_{1}b_{1}}\cdots\eta^{a_{k}b_{k}}\partl{e^{b_{k}}}\cdots\partl{e^{b_{1}}}\rho\label{eq:CliffordMultExpl}\\
\not\!\omega^{(p)}\tfrac{1}{2}(\one+\Gamma^{\#})\not\!\rho^{(r)}\tfrac{1}{2}(\one+\Gamma^{\#}) & = & \left\{ \zwek{\not\!\omega^{(p)}\not\!\rho^{(r)}\tfrac{1}{2}(\one+\Gamma^{\#})\quad\mbox{for }r\mbox{ even}}{0\quad\mbox{for }r\mbox{ }\mbox{odd}}\right.\end{eqnarray}
The differential forms $g$ and $e^{a}\wedge e^{b}$ both are even
so that now we can write down (using also (\ref{eq:expected})) the
inverse Clifford map of (\ref{eq:last-term-of}) \begin{eqnarray}
\gamma_{\gamma\alpha}^{c}\bei{\gemnabla_{c}\rr^{\alpha}\tief{\beta}}{\check{\omega}=\omega} & \stackrel{\slash^{-1}}{\mapsto} & \bs{\nabla}g+\star\bs{\nabla}\underbrace{\star g}_{g}+\nonumber \\
 &  & -\tfrac{3}{4}h_{cab}\tfrac{1}{2}(1+\star)\Bigl\{\sum_{l\geq0}\frac{1}{l!}\left(\sum_{k\geq0}\frac{1}{k!}e^{[c|}\partr{e^{a_{1}}}\cdots\partr{e^{a_{k}}}\eta^{a_{1}b_{1}}\cdots\eta^{a_{k}b_{k}}\partl{e^{b_{k}}}\cdots\partl{e^{b_{1}}}g\right)\partr{e^{c_{1}}}\cdots\partr{e^{c_{l}}}\times\nonumber \\
 &  & \times\eta^{c_{1}d_{1}}\cdots\eta^{c_{l}d_{l}}\partl{e^{d_{l}}}\cdots\partl{e^{d_{1}}}\left(e^{|a}\wedge e^{b]}\right)\Bigr\}\\
 & = & (1+\star)\bs{\nabla}g+\nonumber \\
 &  & -\tfrac{3}{4}h_{cab}\tfrac{1}{2}(1+\star)\Bigl\{\sum_{l\geq0}\frac{1}{l!}\left(e^{[c|}\wedge g+\eta^{[c|b_{1}}\partl{e^{b_{1}}}g\right)\partr{e^{c_{1}}}\cdots\partr{e^{c_{l}}}\times\nonumber \\
 &  & \times\eta^{c_{1}d_{1}}\cdots\eta^{c_{l}d_{l}}\partl{e^{d_{l}}}\cdots\partl{e^{d_{1}}}\left(e^{|a}\wedge e^{b]}\right)\Bigr\}=\\
 & = & (1+\star)\bs{\nabla}g+\nonumber \\
 &  & -\tfrac{3}{8}(1+\star)\bigl(\underbrace{h\wedge g}_{\ip_{h}g}-\underbrace{e^{a}e^{b}h_{ab}\hoch{c}\partl{e^{c}}g}_{\frac{2}{3}\ip_{t}g}-\underbrace{e^{a}h_{a}\hoch{bc}\partl{e^{b}}\partl{e^{c}}g}_{\tfrac{2}{3}\ip_{\tilde{t}}g=\tfrac{2}{3}\ip_{\tilde{t}}\star g}+\underbrace{h\hoch{abc}\partl{e^{a}}\partl{e^{b}}\partl{e^{c}}g}_{\ip_{\tilde{h}}g=\ip_{\tilde{h}}\star g}\bigr)\end{eqnarray}
In the last line below the underbraces we have considered the $h$-field
$h\equiv h_{abc}e^{a}e^{b}e^{c}$ as a 3-form, the corresponding torsion
$t=\tfrac{3}{2}h_{ab}\hoch{c}e^{a}e^{b}\otimes e_{c}$ as a vector-valued
2-form, $\tilde{t}\equiv\tfrac{3}{2}h_{a}\hoch{bc}e^{a}\otimes e_{b}e_{c}$
as a two-vector valued 1-form and $\tilde{h}\equiv h^{abc}e_{a}e_{b}e_{c}$
as a three-vector and have used the generalized definition of an interior
product with respect to a multivector valued form, given in (\ref{eq:bc-interior-product-II}).
Now we can use the result given in the appendix in equation (\ref{eq:stariKeta})
on page \pageref{eq:stariKeta} and in the discussion below, which
implies that \begin{eqnarray}
\star\ip_{\tilde{t}}\star g & = & \ip_{t}g,\qquad\star\ip_{\tilde{h}}\star g=\ip_{h}g=h\wedge g\end{eqnarray}
Remembering that $\bs{\nabla}=\de-\ip_{t}$, we thus get the final
result\begin{equation}
\boxed{\gamma_{\gamma\alpha}^{c}\bei{\gemnabla_{c}\rr^{\alpha}\tief{\beta}}{\check{\omega}=\omega}\stackrel{\slash^{-1}}{\mapsto}(1+\star)\Bigl\{\left(\de-\tfrac{3}{4}h\wedge\right)g-\tfrac{1}{2}\underbrace{\ip_{t}g}_{\mbox{or }\lqn{{\scriptstyle \star\ip_{t}\star g}}}\Bigr\}\quad\quad}\label{eq:DIracLeftFinalI}\end{equation}
with $\ip_{T}g=\tfrac{3}{2}e^{a}e^{b}h_{ab}\hoch{c}\partl{e^{c}}g$
and $\star\ip_{t}\star g=\ip_{\tilde{t}}g=\tfrac{3}{2}e^{a}h_{a}\hoch{bc}\partl{e^{b}}\partl{e^{c}}g$. 

Let's do the same analysis for the Dirac-operator acting on the second
index, which turns out to be a bit simpler, with only one Clifford
multiplication:\begin{eqnarray}
\bei{\gemnabla_{c}\rr^{\alpha}\tief{\beta}}{\check{\omega}=\omega}\gamma^{c\,\beta\gamma} & = & \nabla_{c}\rr^{\alpha}\tief{\beta}\cdot\gamma^{c\,\beta\gamma}-\tfrac{3}{4}\rr^{\alpha}\tief{\delta}\underbrace{h_{abc}\gamma^{ab}\hoch{\delta}\tief{\beta}\gamma^{c\,\beta\gamma}}_{h_{abc}\gamma^{abc\,\delta\gamma}}\end{eqnarray}
According to (\ref{eq:chiralBlocksOdd}), $h_{abc}\gamma^{abc\,\delta\gamma}=\left(\tfrac{1}{2}\not h(\one-\Gamma^{\#})\right)^{\delta\gamma}$.
Using (\ref{eq:below-expected}) and the explicit expression (\ref{eq:CliffordMultExpl})
for the Clifford multiplication on the form-side, the above derivative
operator is mapped to the following:\begin{eqnarray}
\lqn{\bei{\gemnabla_{c}\rr^{\alpha}\tief{\beta}}{\check{\omega}=\omega}\gamma^{c\,\beta\gamma}\stackrel{/^{-1}}{\mapsto}}\\
 & \stackrel{/^{-1}}{\mapsto} & \bigl(\bs{\nabla}g-\star\bs{\nabla}\underbrace{\star g}_{g}\bigr)+\nonumber \\
 &  & -\tfrac{3}{8}(1-\star)\sum_{k\geq0}\frac{1}{k!}g\partr{e^{a_{1}}}\cdots\partr{e^{a_{k}}}\eta^{a_{1}b_{1}}\cdots\eta^{a_{k}b_{k}}\partl{e^{b_{k}}}\cdots\partl{e^{b_{1}}}h\\
 & = & (1-\star)\bs{\nabla}g+\nonumber \\
 &  & -\tfrac{3}{8}(1-\star)\Bigl\{\underbrace{h\wedge g}_{\ip_{h}g}-\underbrace{3e^{a}e^{b}h_{ab}\hoch{c}\partl{e^{c}}g}_{2\ip_{t}g}+\underbrace{3e^{a}h_{a}\hoch{bc}\partl{e^{b}}\partl{e^{c}}g}_{2\ip_{\tilde{t}}\star g}-\underbrace{h^{abc}\,\partl{e^{a}}\partl{e^{b}}\partl{e^{c}}g}_{\ip_{\tilde{h}}\star g}\Bigr\}\end{eqnarray}
Using again that $\star\ip_{\tilde{t}}\star g=\ip_{t}g$, $\star\ip_{\tilde{h}}\star g=\ip_{h}g=h\wedge g$,
and $\bs{\nabla}g=\de g-\ip_{t}g$ we end up with \begin{equation}
\boxed{\gamma^{c\,\gamma\beta}\bei{\gemnabla_{c}\rr^{\alpha}\tief{\beta}}{\check{\omega}=\omega}\stackrel{/^{-1}}{\mapsto}(1-\star)\Bigl\{\left(\de-\tfrac{3}{4}h\wedge\right)g+\tfrac{1}{2}\underbrace{\ip_{t}g}_{\mbox{or }\lqn{{\scriptstyle -\star\ip_{t}\star g}}}\Bigr\}\qquad}\label{eq:DIracLeftFinalII}\end{equation}
with $\ip_{t}g=\tfrac{3}{2}e^{a}e^{b}h_{ab}\hoch{c}\partl{e^{c}}g$
and $\star\ip_{t}\star g=\ip_{\tilde{t}}g=\tfrac{3}{2}e^{a}h_{a}\hoch{bc}\partl{e^{b}}\partl{e^{c}}g$.
If both actions of the Dirac operator vanish, we thus get the following
condition on the form side (adding and subtracting (\ref{eq:DIracLeftFinalI})
and (\ref{eq:DIracLeftFinalII}) lead to equivalent equations)%
\footnote{\index{footnote!\thefoot. comment on the twisted differential}We
could try to absorb the somewhat disturbing contribution of $\star\ip_{t}\star g$
or $\ip_{t}g$ by reintroducing $\bs{\nabla}g$ via $\ip_{t}g=-\bs{\nabla}g+\de g$.
The result, however, looks even less natural and the twisted differential
gets modified at intermediate steps. The equations (\ref{eq:DIracLeftFinalI}),
(\ref{eq:DIracLeftFinalII}) and (\ref{eq:finalFormEom}) take the
following form\begin{eqnarray*}
\gamma_{\gamma\alpha}^{c}\bei{\gemnabla_{c}\rr^{\alpha}\tief{\beta}}{\check{\omega}=\omega} & \stackrel{\slash^{-1}}{\mapsto} & \tfrac{1}{2}(1+\star)\Bigl\{\bs{\nabla}g+\left(\de-\tfrac{3}{2}h\wedge\right)g\Bigr\}\quad\quad\\
\bei{\gemnabla_{c}\rr^{\alpha}\tief{\beta}}{\check{\omega}=\omega}\gamma^{c\,\beta\gamma} & \stackrel{/^{-1}}{\mapsto} & \tfrac{1}{2}(1-\star)\Bigl\{-\bs{\nabla}g+3\left(\de-\tfrac{1}{2}h\wedge\right)g\Bigr\}\qquad\\
\gamma_{\gamma\alpha}^{c}\bei{\gemnabla_{c}\rr^{\alpha}\tief{\beta}}{\check{\omega}=\omega}=\bei{\gemnabla_{c}\rr^{\alpha}\tief{\beta}}{\check{\omega}=\omega}\gamma^{c\,\beta\gamma}=0 & \iff & 2\left(\de-\tfrac{3}{4}h\wedge\right)g+\star\bs{\nabla}\star g-\star\de\star g=0\quad\fussend\end{eqnarray*}
}\begin{eqnarray}
\gamma_{\gamma\alpha}^{c}\bei{\gemnabla_{c}\rr^{\alpha}\tief{\beta}}{\check{\omega}=\omega}=\gamma^{c\,\gamma\beta}\bei{\gemnabla_{c}\rr^{\alpha}\tief{\beta}}{\check{\omega}=\omega}=0 & \iff & \left(\de-\tfrac{3}{4}h\wedge\right)g-\tfrac{1}{2}\star\ip_{t}\star g=0\label{eq:finalFormEom}\end{eqnarray}
Next we consider the \textbf{type IIB} case where we can expand the
RR-bispinor in odd antisymmetrized products of $\gamma$-matrices
(see (\ref{eq:gamma:gammaexpansionOddI}) on page \pageref{eq:gamma:gammaexpansionOddI}):
\begin{eqnarray}
\rr^{\alpha\beta} & = & 2g_{a}^{(1)}\underbrace{\gamma^{a\,\alpha\beta}}_{\gamma^{[1]}}+2g_{a_{1}a_{2}a_{3}}^{(3)}\underbrace{\gamma^{a_{1}a_{2}a_{3}\,\alpha\beta}}_{\gamma^{[3]}}+g_{a_{1}a_{2}a_{3}a_{4}a_{5}}^{(5)}\underbrace{\gamma^{a_{1}a_{2}a_{3}a_{4}a_{5}\alpha\beta}}_{\gamma^{[5]}}\label{eq:RR-exp:IIB}\\
2g_{a}^{(1)} & = & \frac{1}{16}\rr^{\alpha\beta}\gamma_{a\,\beta\alpha}\\
2g_{a_{1}a_{2}a_{3}}^{(3)} & = & \frac{1}{16\cdot3!}\rr^{\alpha\beta}\gamma_{a_{1}a_{2}a_{3}\,\beta\alpha}\\
g_{a_{1}a_{2}a_{3}a_{4}a_{5}}^{(5)} & = & \frac{1}{32\cdot5!}\rr^{\alpha\beta}\gamma_{a_{5}a_{4}a_{3}a_{2}a_{1}\,\beta\alpha}\end{eqnarray}
This is mapped to an odd anti self-dual form on the form-side \begin{eqnarray}
\rr^{\alpha\beta} & \stackrel{\slash^{-1}}{\mapsto} & (1-\star)\left(g_{a}^{(1)}e^{a}+g_{a_{1}a_{2}a_{3}}^{(3)}e^{a_{1}}\wedge e^{a_{2}}\wedge e^{a_{3}}+\frac{1}{2}g_{a_{1}\ldots a_{5}}^{(5)}e^{a_{1}}\wedge\ldots\wedge e^{a_{5}}\right)\equiv g\end{eqnarray}
on the form-side. According to (\ref{eq:actionOfDirac-op}) and (\ref{eq:actionOfDirac-op-right})
in the appendix, the action of the Dirac operator $\gamma_{\gamma\alpha}^{c}\nabla_{c}$
on the first or on the second index (with a covariant derivative that
leaves the gamma-matrices invariant) yields for an antiselfdual and
odd $g$\begin{eqnarray}
\gamma_{\gamma\alpha}^{c}\nabla_{c}\rr^{\alpha\beta} & \stackrel{\slash^{-1}}{\mapsto} & (1-\star)\bs{\nabla}g\\
\nabla_{c}\rr^{\alpha\beta}\cdot\gamma_{\beta\gamma}^{c} & \stackrel{\slash^{-1}}{\mapsto} & -(1+\star)\bs{\nabla}g\end{eqnarray}
Instead of a connection that leaves the gamma-matrices invariant,
we have again the mixed connection acting differently on left- and
right-movers. We thus act on the first fermionic index of $\rr^{\alpha\beta}$
with $\omega_{a\beta}\hoch{\gamma}=\tfrac{1}{4}(\omega_{ab}^{(LC)}\hoch{c}+\frac{3}{2}h_{ab}\hoch{c})\gamma^{b}\tief{c\,\beta}\hoch{\gamma}$
and on the second with $\hat{\omega}_{a\beta}\hoch{\gamma}=\tfrac{1}{4}(\omega_{ab}^{(LC)}\hoch{c}-\frac{3}{2}h_{ab}\hoch{c})\gamma^{b}\tief{c\,\beta}\hoch{\gamma}$.
Again we decide to act on the bosonic indices with the left mover
connection $\omega_{ab}\hoch{c}=\omega_{ab}^{(LC)}\hoch{c}+\frac{3}{2}h_{ab}\hoch{c}$
and rewrite $\hat{\omega}_{c\delta}\hoch{\beta}=\omega_{c\delta}\hoch{\beta}+\Delta_{c\delta}\hoch{\beta}=\omega_{c\delta}\hoch{\beta}-\tfrac{3}{4}h_{ca}\hoch{b}\gamma^{a}\tief{b\,\delta}\hoch{\beta}$.
We then have for the action of the Dirac operator (based on the mixed
connection) on the second index \begin{eqnarray}
\lqn{\bei{\gemnabla_{c}\rr^{\alpha\beta}}{\check{\omega}=\omega}\cdot\gamma_{\beta\gamma}^{c}=}\nonumber \\
 & = & \nabla_{c}\rr^{\alpha\beta}\cdot\gamma_{\beta\gamma}^{c}-\tfrac{3}{4}\rr^{\alpha\delta}h_{abc}\gamma^{abc}\tief{\delta\gamma}\\
 & = & \nabla_{c}\rr^{\alpha\beta}\cdot\gamma_{\beta\gamma}^{c}-\tfrac{3}{4}(\not g\,\not h\,)^{\alpha}\tief{\gamma}\stackrel{/^{-1}}{\mapsto}\\
 & \stackrel{/^{-1}}{\mapsto} & -(1+\star)\bs{\nabla}g-\tfrac{3}{8}(1+\star)\sum_{k\geq0}\frac{1}{k!}g\partr{e^{a_{1}}}\cdots\partr{e^{a_{k}}}\eta^{a_{1}b_{1}}\cdots\eta^{a_{k}b_{k}}\partl{e^{b_{k}}}\cdots\partl{e^{b_{1}}}h=\\
 & = & -(1+\star)(\de-\ip_{t})g+\tfrac{3}{8}(1+\star)\underbrace{h\wedge g}_{\ip_{h}g}-\tfrac{3}{8}(1+\star)\underbrace{3e^{a}e^{b}h_{ab}\hoch{c}\partl{e^{c}}g}_{2\ip_{t}g}+\nonumber \\
 &  & +\tfrac{3}{8}(1+\star)\underbrace{3e^{a}h_{a}\hoch{bc}\partl{e^{b}}\partl{e^{c}}g}_{2\ip_{\tilde{t}}g=2\star\ip_{t}\star g}-\tfrac{3}{8}(1+\star)\underbrace{h^{abc}\partl{e^{a}}\partl{e^{b}}\partl{e^{c}}g}_{\ip_{\tilde{h}}g=\star\ip_{h}\star g}\end{eqnarray}
After collecting all the terms, we arrive at \begin{equation}
\boxed{\bei{\gemnabla_{c}\rr^{\alpha\beta}}{\check{\omega}=\omega}\cdot\gamma_{\beta\gamma}^{c}\stackrel{/^{-1}}{\mapsto}-(1+\star)\Bigl\{\left(\de-\tfrac{3}{4}h\wedge\right)g+\tfrac{1}{2}\underbrace{\ip_{t}g}_{\mbox{or }\lqn{{\scriptstyle -\star\ip_{t}\star}g}}\Bigr\}\quad\quad}\end{equation}
For the action of the Dirac operator on the first index, finally,
we have\begin{eqnarray}
\lqn{\gamma_{\gamma\alpha}^{c}\bei{\gemnabla_{c}\rr^{\alpha\beta}}{\check{\omega}=\omega}=}\nonumber \\
 & = & \gamma_{\gamma\alpha}^{c}\nabla_{c}\rr^{\alpha\beta}-\tfrac{3}{4}h_{abc}\gamma_{\gamma\alpha}^{c}\rr^{\alpha\delta}\gamma^{ab}\tief{\delta}\hoch{\beta}\stackrel{/^{-1}}{\mapsto}\\
 & \stackrel{/^{-1}}{\mapsto} & (1-\star)\bs{\nabla}g+\nonumber \\
 &  & -\tfrac{3}{8}h_{abc}(1-\star)\sum_{k\geq0}\frac{1}{k!}\left(e^{c}\wedge g+\eta^{cd}\partl{e^{d}}g\right)\partr{e^{a_{1}}}\cdots\partr{e^{a_{k}}}\eta^{a_{1}b_{1}}\cdots\eta^{a_{k}b_{k}}\partl{e^{b_{k}}}\cdots\partl{e^{b_{1}}}\left(e^{a}\wedge e^{b}\right)\qquad\\
 & = & (1-\star)\underbrace{\bs{\nabla}g}_{\de g-\ip_{t}g}+\nonumber \\
 &  & -\tfrac{3}{8}(1-\star)\Bigl\{\underbrace{h\wedge g}_{\ip_{h}g}-\underbrace{e^{a}e^{b}h_{ab}\hoch{c}\partl{e^{c}}g}_{\frac{2}{3}\ip_{t}g}-\underbrace{e^{a}h_{a}\hoch{bc}\partl{e^{b}}\partl{e^{c}}g}_{\frac{2}{3}\ip_{\tilde{t}}g=\frac{2}{3}\star\ip_{t}\star g}+\underbrace{h^{abc}\partl{e^{a}}\partl{e^{b}}\partl{e^{c}}g}_{\ip_{\tilde{h}}g=\star\ip_{h}\star g}\Bigr\}\qquad\end{eqnarray}
The terms then combine to \begin{equation}
\boxed{\gamma_{\gamma\alpha}^{c}\bei{\gemnabla_{c}\rr^{\alpha\beta}}{\check{\omega}=\omega}\stackrel{/^{-1}}{\mapsto}(1-\star)\Bigl\{\left(\de-\tfrac{3}{4}h\wedge\right)g\underbrace{-\tfrac{1}{2}\ip_{t}g}_{\mbox{or }-\tfrac{1}{2}\lqn{\star\ip_{t}\star g}}\Bigr\}\quad\quad}\end{equation}
The equations on the form side thus look the same as for type IIA.
In particular we have \begin{eqnarray}
\gamma_{\gamma\alpha}^{c}\bei{\gemnabla_{c}\rr^{\alpha\beta}}{\check{\omega}=\omega}=\bei{\gemnabla_{c}\rr^{\alpha\beta}}{\check{\omega}=\omega}\cdot\gamma_{\beta\gamma}^{c}=0 & \iff & \left(\de-\tfrac{3}{4}h\wedge\right)g-\tfrac{1}{2}\star\ip_{t}\star g=0\end{eqnarray}

\lyxline{\normalsize}\vspace{-.25cm}\lyxline{\normalsize}\vspace{.5cm} $\bullet\quad$\underbar{(delta|2,1,0)$_{\bs{\alpha}bc}\hoch{\bs{\delta}}\leftrightarrow$((hdelta|2,0,1))$_{\hat{\bs{\alpha}}bc}\hoch{\hat{\bs{\delta}}}$)dim$\frac{4}{2}$:}\begin{eqnarray}
0 & \stackrel{!}{=} & \gem{\nabla}_{[\bs{\alpha}}T_{bc]}\hoch{\bs{\delta}}+2\gem{T}_{[\bs{\alpha}b|}\hoch{E}T_{E|c]}\hoch{\bs{\delta}}-R_{[\bs{\alpha}bc]}\hoch{\bs{\delta}}=\\
 & = & \frac{1}{3}\gem{\nabla}_{\bs{\alpha}}T_{bc}\hoch{\bs{\delta}}+\frac{4}{3}\gem{T}_{\bs{\alpha}[b|}\hoch{E}T_{E|c]}\hoch{\bs{\delta}}-\frac{1}{3}R_{bc\bs{\alpha}}\hoch{\bs{\delta}}=\\
 & = & \frac{1}{3}\gem{\nabla}_{\bs{\alpha}}T_{bc}\hoch{\bs{\delta}}+\frac{4}{3}\check{T}_{\bs{\alpha}[b|}\hoch{e}T_{e|c]}\hoch{\bs{\delta}}+\frac{4}{3}\hat{T}_{\bs{\alpha}[b|}\hoch{\hat{\bs{\eps}}}T_{\hat{\bs{\eps}}|c]}\hoch{\bs{\delta}}-\frac{1}{3}R_{bc\bs{\alpha}}\hoch{\bs{\delta}}=\\
 & = & \frac{1}{3}\gem{\nabla}_{\bs{\alpha}}T_{bc}\hoch{\bs{\delta}}+\frac{4}{3}\underbrace{\hatcovPhi{a}\check{T}_{\bs{\alpha}[b|}\hoch{e}}_{=0\,\mbox{for }\check{\Omega}=\hat{\Omega}}T_{e|c]}\hoch{\bs{\delta}}+\frac{4}{3}\tilde{\gamma}_{[b|\,\bs{\alpha\gamma}}\RR^{\bs{\gamma}\hat{\bs{\eps}}}\tilde{\gamma}_{|c]\,\hat{\bs{\eps}}\hat{\bs{\delta}}}\RR^{\bs{\delta}\hat{\bs{\delta}}}-\frac{1}{3}R_{bc\bs{\alpha}}\hoch{\bs{\delta}}\end{eqnarray}
\Ram{0.6}{\begin{eqnarray}
R_{bc\bs{\alpha}}\hoch{\bs{\delta}} & = & \bei{\gem{\nabla}_{\bs{\alpha}}T_{bc}\hoch{\bs{\delta}}}{\check{\Omega}=\hat{\Omega}}+4\tilde{\gamma}_{[b|\,\bs{\alpha\gamma}}\RR^{\bs{\gamma}\hat{\bs{\eps}}}\tilde{\gamma}_{|c]\,\hat{\bs{\eps}}\hat{\bs{\delta}}}\RR^{\bs{\delta}\hat{\bs{\delta}}}\label{eq:(delta|2,1,0)}\\
\hat{R}_{bc\hat{\bs{\alpha}}}\hoch{\hat{\bs{\delta}}} & = & \bei{\gem{\nabla}_{\hat{\bs{\alpha}}}\hat{T}_{bc}\hoch{\hat{\bs{\delta}}}}{\check{\Omega}=\Omega}+4\tilde{\gamma}_{[b|\,\hat{\bs{\alpha}}\hat{\bs{\gamma}}}\RR^{\bs{\eps}\hat{\bs{\gamma}}}\tilde{\gamma}_{|c]\,\bs{\eps\delta}}\RR^{\bs{\delta}\hat{\bs{\delta}}}\label{eq:(hdelta|2,0,1)}\end{eqnarray}
} \\
\rem{%
\footnote{\index{footnote!\thefoot. calculational remark}Taking at this point
the trace  leads to \begin{eqnarray*}
0 & \stackrel{!}{=} & 8F_{bc}^{(D)}+\bei{\gem{\nabla}_{\bs{\delta}}T_{bc}\hoch{\bs{\delta}}}{\check{\Omega}=\hat{\Omega}}+4\tilde{\gamma}_{[b|\,\bs{\delta\gamma}}\RR^{\bs{\gamma}\hat{\bs{\eps}}}\tilde{\gamma}_{|c]\,\hat{\bs{\eps}}\hat{\bs{\delta}}}\RR^{\bs{\delta}\hat{\bs{\delta}}}=\\
 & \stackrel{\textrm{if }\Omega_{\bs{\alpha}}=\partial_{\bs{\alpha}}\Phi}{=} & \frac{1}{16}\gem{\nabla}_{\bs{\delta}}\gem{\nabla}_{\hat{\bs{\gamma}}}\RR^{\bs{\delta}\hat{\bs{\delta}}}\tilde{\gamma}_{bc}\hoch{\,\hat{\bs{\gamma}}}\tief{\hat{\bs{\delta}}}+4\tilde{\gamma}_{[b|\,\bs{\delta\gamma}}\RR^{\bs{\gamma}\hat{\bs{\eps}}}\tilde{\gamma}_{|c]\,\hat{\bs{\eps}}\hat{\bs{\delta}}}\RR^{\bs{\delta}\hat{\bs{\delta}}}=\\
 & = & \frac{1}{16}\gem{\nabla}_{\hat{\bs{\gamma}}}\gem{\nabla}_{\bs{\delta}}\RR^{\bs{\delta}\hat{\bs{\delta}}}\tilde{\gamma}_{bc}\hoch{\,\hat{\bs{\gamma}}}\tief{\hat{\bs{\delta}}}+\frac{1}{8}R_{\bs{\delta}\hat{\bs{\gamma}}\bs{\eps}}\hoch{\bs{\delta}}\RR^{\bs{\eps}\hat{\bs{\delta}}}\tilde{\gamma}_{bc}\hoch{\,\hat{\bs{\gamma}}}\tief{\hat{\bs{\delta}}}+\frac{1}{8}\hat{R}_{\bs{\delta}\hat{\bs{\gamma}}\hat{\bs{\eps}}}\hoch{\hat{\bs{\delta}}}\RR^{\bs{\delta}\hat{\bs{\eps}}}\tilde{\gamma}_{bc}\hoch{\,\hat{\bs{\gamma}}}\tief{\hat{\bs{\delta}}}+4\tilde{\gamma}_{[b|\,\bs{\delta\gamma}}\RR^{\bs{\gamma}\hat{\bs{\eps}}}\tilde{\gamma}_{|c]\,\hat{\bs{\eps}}\hat{\bs{\delta}}}\RR^{\bs{\delta}\hat{\bs{\delta}}}\end{eqnarray*}
Remember now $\hat{R}_{\bs{\gamma}[\hat{\bs{\alpha}}\hat{\bs{\beta}}]}\hoch{\hat{\bs{\delta}}}=-\gamma_{\hat{\bs{\alpha}}\hat{\bs{\beta}}}^{e}\tilde{\gamma}_{e\,\bs{\gamma}\bs{\delta}}\RR^{\bs{\delta}\hat{\bs{\delta}}}$
and $R_{\hat{\bs{\gamma}}[\bs{\alpha}\bs{\beta}]}\hoch{\bs{\delta}}=-\gamma_{\bs{\alpha\beta}}^{e}\tilde{\gamma}_{e\,\hat{\bs{\gamma}}\hat{\bs{\delta}}}\RR^{\bs{\delta}\hat{\bs{\delta}}}$
and $\gem{\nabla}_{\bs{\alpha}}\RR^{\bs{\alpha}\hat{\bs{\delta}}}=8\RR^{\bs{\alpha}\hat{\bs{\delta}}}(\partial_{\bs{\alpha}}\Phi-\Omega_{\bs{\alpha}})\stackrel{\textrm{if }\Omega_{\bs{\alpha}}=\partial_{\bs{\alpha}}\Phi}{=}0$.\begin{eqnarray*}
0 & \underset{\textrm{if }\Omega_{\bs{\alpha}}=\partial_{\bs{\alpha}}\Phi}{\stackrel{!}{=}} & -\frac{1}{8}R_{\hat{\bs{\gamma}}\bs{\delta}\bs{\eps}}\hoch{\bs{\delta}}\RR^{\bs{\eps}\hat{\bs{\delta}}}\tilde{\gamma}_{bc}\hoch{\,\hat{\bs{\gamma}}}\tief{\hat{\bs{\delta}}}+\frac{1}{8}\hat{R}_{\bs{\delta}\hat{\bs{\gamma}}\hat{\bs{\eps}}}\hoch{\hat{\bs{\delta}}}\RR^{\bs{\delta}\hat{\bs{\eps}}}\tilde{\gamma}_{bc}\hoch{\,\hat{\bs{\gamma}}}\tief{\hat{\bs{\delta}}}+4\tilde{\gamma}_{[b|\,\bs{\delta\gamma}}\RR^{\bs{\gamma}\hat{\bs{\eps}}}\tilde{\gamma}_{|c]\,\hat{\bs{\eps}}\hat{\bs{\delta}}}\RR^{\bs{\delta}\hat{\bs{\delta}}}=\\
 & = & -\frac{1}{8}R_{\hat{\bs{\gamma}}\bs{\eps}\bs{\delta}}\hoch{\bs{\delta}}\RR^{\bs{\eps}\hat{\bs{\delta}}}\tilde{\gamma}_{bc}\hoch{\,\hat{\bs{\gamma}}}\tief{\hat{\bs{\delta}}}-\frac{1}{4}R_{\hat{\bs{\gamma}}[\bs{\delta}\bs{\eps}]}\hoch{\bs{\delta}}\RR^{\bs{\eps}\hat{\bs{\delta}}}\tilde{\gamma}_{bc}\hoch{\,\hat{\bs{\gamma}}}\tief{\hat{\bs{\delta}}}+\frac{1}{8}\hat{R}_{\bs{\delta}\hat{\bs{\gamma}}\hat{\bs{\eps}}}\hoch{\hat{\bs{\delta}}}\RR^{\bs{\delta}\hat{\bs{\eps}}}\tilde{\gamma}_{bc}\hoch{\,\hat{\bs{\gamma}}}\tief{\hat{\bs{\delta}}}+4\tilde{\gamma}_{[b|\,\bs{\delta\gamma}}\RR^{\bs{\gamma}\hat{\bs{\eps}}}\tilde{\gamma}_{|c]\,\hat{\bs{\eps}}\hat{\bs{\delta}}}\RR^{\bs{\delta}\hat{\bs{\delta}}}=\\
 & = & \underbrace{F_{\hat{\bs{\gamma}}\bs{\eps}}^{(D)}}_{\frac{1}{2}\nabla_{\hat{\bs{\gamma}}}(\Omega_{\bs{\eps}}-\partial_{\bs{\eps}}\Phi)}\RR^{\bs{\eps}\hat{\bs{\delta}}}\tilde{\gamma}_{bc}\hoch{\,\hat{\bs{\gamma}}}\tief{\hat{\bs{\delta}}}+\frac{1}{4}\gamma_{\bs{\delta}\bs{\eps}}^{e}\tilde{\gamma}_{e\,\hat{\bs{\gamma}}\hat{\bs{\eps}}}\RR^{\bs{\delta}\hat{\bs{\eps}}}\RR^{\bs{\eps}\hat{\bs{\delta}}}\tilde{\gamma}_{bc}\hoch{\,\hat{\bs{\gamma}}}\tief{\hat{\bs{\delta}}}+\frac{1}{8}\hat{R}_{\bs{\delta}\hat{\bs{\gamma}}\hat{\bs{\eps}}}\hoch{\hat{\bs{\delta}}}\RR^{\bs{\delta}\hat{\bs{\eps}}}\tilde{\gamma}_{bc}\hoch{\,\hat{\bs{\gamma}}}\tief{\hat{\bs{\delta}}}+4\tilde{\gamma}_{[b|\,\bs{\delta\gamma}}\RR^{\bs{\gamma}\hat{\bs{\eps}}}\tilde{\gamma}_{|c]\,\hat{\bs{\eps}}\hat{\bs{\delta}}}\RR^{\bs{\delta}\hat{\bs{\delta}}}=\\
 & \stackrel{\textrm{if }\Omega_{\bs{\alpha}}=\partial_{\bs{\alpha}}\Phi}{=} & \frac{1}{4}\gamma_{\bs{\delta}\bs{\eps}}^{e}\RR^{\bs{\delta}\hat{\bs{\eps}}}\RR^{\bs{\eps}\hat{\bs{\delta}}}\underbrace{\tilde{\gamma}_{bc}\tief{\hat{\bs{\delta}}}\hoch{\,\hat{\bs{\gamma}}}\tilde{\gamma}_{e\,\hat{\bs{\gamma}}\hat{\bs{\eps}}}}_{\gamma_{bce}+G_{ce}\gamma_{b}-G_{be}\gamma_{c}}+\frac{1}{8}\hat{R}_{\bs{\delta}\hat{\bs{\gamma}}\hat{\bs{\eps}}}\hoch{\hat{\bs{\delta}}}\RR^{\bs{\delta}\hat{\bs{\eps}}}\tilde{\gamma}_{bc}\hoch{\,\hat{\bs{\gamma}}}\tief{\hat{\bs{\delta}}}+4\tilde{\gamma}_{[b|\,\bs{\delta\gamma}}\RR^{\bs{\gamma}\hat{\bs{\eps}}}\tilde{\gamma}_{|c]\,\hat{\bs{\eps}}\hat{\bs{\delta}}}\RR^{\bs{\delta}\hat{\bs{\delta}}}=\\
 & = & \frac{1}{4}\underbrace{\gamma_{\bs{\delta}\bs{\eps}}^{e}\RR^{\bs{\eps}\hat{\bs{\delta}}}\overbrace{\gamma_{bce\,\hat{\bs{\delta}}\hat{\bs{\eps}}}}^{\textrm{gr.sym}}\RR^{\bs{\delta}\hat{\bs{\eps}}}}_{=0}+\frac{1}{2}\gamma_{[c|\bs{\delta}\bs{\eps}}\RR^{\bs{\eps}\hat{\bs{\delta}}}\gamma_{|b]\hat{\bs{\delta}}\hat{\bs{\eps}}}\RR^{\bs{\delta}\hat{\bs{\eps}}}+\frac{1}{8}\hat{R}_{\bs{\delta}\hat{\bs{\gamma}}\hat{\bs{\eps}}}\hoch{\hat{\bs{\delta}}}\RR^{\bs{\delta}\hat{\bs{\eps}}}\tilde{\gamma}_{bc}\hoch{\,\hat{\bs{\gamma}}}\tief{\hat{\bs{\delta}}}+4\tilde{\gamma}_{[b|\,\bs{\delta\gamma}}\RR^{\bs{\gamma}\hat{\bs{\eps}}}\tilde{\gamma}_{|c]\,\hat{\bs{\eps}}\hat{\bs{\delta}}}\RR^{\bs{\delta}\hat{\bs{\delta}}}\quad?\quad\fussend\end{eqnarray*}
}}Plugging in $T_{bc}\hoch{\bs{\delta}}=\frac{1}{16}\left(\gem{\nabla}_{\hat{\bs{\gamma}}}\RR^{\bs{\delta}\hat{\bs{\delta}}}+8\hat{\nabla}_{\hat{\bs{\gamma}}}\Phi\RR^{\bs{\delta}\hat{\bs{\delta}}}\right)\tilde{\gamma}_{bc}\hoch{\,\hat{\bs{\gamma}}}\tief{\hat{\bs{\delta}}}$yields\begin{eqnarray}
R_{bc\bs{\alpha}}\hoch{\bs{\delta}} & = & \frac{1}{16}\gem{\nabla}_{\bs{\alpha}}\left(\gem{\nabla}_{\hat{\bs{\gamma}}}\RR^{\bs{\delta}\hat{\bs{\delta}}}+8\hat{\nabla}_{\hat{\bs{\gamma}}}\Phi\RR^{\bs{\delta}\hat{\bs{\delta}}}\right)\cdot\tilde{\gamma}_{bc}\hoch{\,\hat{\bs{\gamma}}}\tief{\hat{\bs{\delta}}}+\nonumber \\
 &  & +\frac{1}{16}\left(\gem{\nabla}_{\hat{\bs{\gamma}}}\RR^{\bs{\delta}\hat{\bs{\delta}}}+8\hat{\nabla}_{\hat{\bs{\gamma}}}\Phi\RR^{\bs{\delta}\hat{\bs{\delta}}}\right)2\underbrace{\hatcovPhi{\bs{\alpha}}}_{=0}\tilde{\gamma}_{bc}\hoch{\,\hat{\bs{\gamma}}}\tief{\hat{\bs{\delta}}}+\nonumber \\
 &  & +4\tilde{\gamma}_{[b|\,\bs{\alpha\gamma}}\RR^{\bs{\gamma}\hat{\bs{\eps}}}\tilde{\gamma}_{|c]\,\hat{\bs{\eps}}\hat{\bs{\delta}}}\RR^{\bs{\delta}\hat{\bs{\delta}}}=\\
 & = & \left(\frac{1}{16}\gem{\nabla}_{\bs{\alpha}}\gem{\nabla}_{\hat{\bs{\gamma}}}\RR^{\bs{\delta}\hat{\bs{\delta}}}+\frac{8}{16}\gem{\nabla}_{\bs{\alpha}}\hat{\nabla}_{\hat{\bs{\gamma}}}\Phi\RR^{\bs{\delta}\hat{\bs{\delta}}}+\frac{8}{16}\hat{\nabla}_{\hat{\bs{\gamma}}}\Phi\gem{\nabla}_{\bs{\alpha}}\RR^{\bs{\delta}\hat{\bs{\delta}}}\right)\cdot\tilde{\gamma}_{bc}\hoch{\,\hat{\bs{\gamma}}}\tief{\hat{\bs{\delta}}}+\nonumber \\
 &  & +4\tilde{\gamma}_{[b|\,\bs{\alpha\gamma}}\RR^{\bs{\gamma}\hat{\bs{\eps}}}\tilde{\gamma}_{|c]\,\hat{\bs{\eps}}\hat{\bs{\delta}}}\RR^{\bs{\delta}\hat{\bs{\delta}}}=\\
 & = & \Big(\frac{1}{16}\gem{\nabla}_{\hat{\bs{\gamma}}}\gem{\nabla}_{\bs{\alpha}}\RR^{\bs{\delta}\hat{\bs{\delta}}}-\frac{1}{8}R_{\hat{\bs{\gamma}}\bs{\alpha}\bs{\eps}}\hoch{\bs{\delta}}\RR^{\bs{\eps}\hat{\bs{\delta}}}+\frac{1}{8}R_{\bs{\alpha}\hat{\bs{\gamma}}\hat{\eps}}\hoch{\hat{\bs{\delta}}}\RR^{\bs{\delta}\hat{\bs{\eps}}}+\nonumber \\
 &  & +\hat{F}_{\hat{\bs{\gamma}}\bs{\alpha}}^{(D)}\RR^{\bs{\delta}\hat{\bs{\delta}}}+\frac{1}{2}\hat{\nabla}_{\hat{\bs{\gamma}}}\Phi\gem{\nabla}_{\bs{\alpha}}\RR^{\bs{\delta}\hat{\bs{\delta}}}\Big)\cdot\tilde{\gamma}_{bc}\hoch{\,\hat{\bs{\gamma}}}\tief{\hat{\bs{\delta}}}+\nonumber \\
 &  & +4\tilde{\gamma}_{[b|\,\bs{\alpha\gamma}}\RR^{\bs{\gamma}\hat{\bs{\eps}}}\tilde{\gamma}_{|c]\,\hat{\bs{\eps}}\hat{\bs{\delta}}}\RR^{\bs{\delta}\hat{\bs{\delta}}}\end{eqnarray}
Taking the trace yields\begin{eqnarray}
-8F_{bc}^{(D)} & = & \Big(\frac{1}{16}\gem{\nabla}_{\hat{\bs{\gamma}}}\underbrace{\gem{\nabla}_{\bs{\alpha}}\RR^{\bs{\alpha}\hat{\bs{\delta}}}}_{8\RR^{\bs{\delta}\hat{\bs{\delta}}}\covPhi{\bs{\delta}}}-\frac{1}{8}R_{\hat{\bs{\gamma}}\bs{\alpha}\bs{\eps}}\hoch{\bs{\alpha}}\RR^{\bs{\eps}\hat{\bs{\delta}}}+\frac{1}{8}R_{\bs{\alpha}\hat{\bs{\gamma}}\hat{\eps}}\hoch{\hat{\bs{\delta}}}\RR^{\bs{\alpha}\hat{\bs{\eps}}}+\nonumber \\
 &  & +\hat{F}_{\hat{\bs{\gamma}}\bs{\alpha}}^{(D)}\RR^{\bs{\alpha}\hat{\bs{\delta}}}+\frac{1}{2}\hat{\nabla}_{\hat{\bs{\gamma}}}\Phi\underbrace{\gem{\nabla}_{\bs{\alpha}}\RR^{\bs{\alpha}\hat{\bs{\delta}}}}_{8\RR^{\bs{\delta}\hat{\bs{\delta}}}\covPhi{\bs{\delta}}}\Big)\cdot\tilde{\gamma}_{bc}\hoch{\,\hat{\bs{\gamma}}}\tief{\hat{\bs{\delta}}}+\nonumber \\
 &  & +4\tilde{\gamma}_{[b|\,\bs{\alpha\gamma}}\RR^{\bs{\gamma}\hat{\bs{\eps}}}\tilde{\gamma}_{|c]\,\hat{\bs{\eps}}\hat{\bs{\delta}}}\RR^{\bs{\alpha}\hat{\bs{\delta}}}\end{eqnarray}
\rem{\begin{eqnarray*}
 & = & \Big(\frac{1}{2}\gem{\nabla}_{\hat{\bs{\gamma}}}\RR^{\bs{\delta}\hat{\bs{\delta}}}\covPhi{\bs{\delta}}-\RR^{\bs{\delta}\hat{\bs{\delta}}}F_{\hat{\bs{\gamma}}\bs{\delta}}^{(D)}+F_{\hat{\bs{\gamma}}\bs{\eps}}^{(D)}\RR^{\bs{\eps}\hat{\bs{\delta}}}+\frac{1}{4}\RR^{\bs{\eps}\hat{\bs{\delta}}}\gamma_{\bs{\eps\delta}}^{e}\RR^{\bs{\delta}\hat{\bs{\eps}}}\tilde{\gamma}_{e\,\hat{\bs{\eps}}\hat{\bs{\gamma}}}+\frac{1}{8}R_{\bs{\alpha}\hat{\bs{\gamma}}\hat{\eps}}\hoch{\hat{\bs{\delta}}}\RR^{\bs{\alpha}\hat{\bs{\eps}}}+\\
 &  & +\hat{F}_{\hat{\bs{\gamma}}\bs{\alpha}}^{(D)}\RR^{\bs{\alpha}\hat{\bs{\delta}}}+4\hat{\nabla}_{\hat{\bs{\gamma}}}\Phi\RR^{\bs{\delta}\hat{\bs{\delta}}}\covPhi{\bs{\delta}}\Big)\cdot\tilde{\gamma}_{bc}\hoch{\,\hat{\bs{\gamma}}}\tief{\hat{\bs{\delta}}}+\\
 &  & +4\tilde{\gamma}_{[b|\,\bs{\alpha\gamma}}\RR^{\bs{\gamma}\hat{\bs{\eps}}}\tilde{\gamma}_{|c]\,\hat{\bs{\eps}}\hat{\bs{\delta}}}\RR^{\bs{\alpha}\hat{\bs{\delta}}}=...?\end{eqnarray*}
}$\bullet\quad$\underbar{(delta|2,0,1)$_{\hat{\bs{\alpha}}bc}\hoch{\bs{\delta}}\leftrightarrow$(hdelta|2,1,0)$_{\bs{\alpha}bc}\hoch{\hat{\bs{\delta}}}$,dim$\frac{4}{2}$}:\begin{eqnarray}
0 & \stackrel{!}{=} & \gem{\nabla}_{[\hat{\bs{\alpha}}}T_{bc]}\hoch{\bs{\delta}}+2\gem{T}_{[\hat{\bs{\alpha}}b|}\hoch{E}T_{E|c]}\hoch{\bs{\delta}}-R_{[\hat{\bs{\alpha}}bc]}\hoch{\bs{\delta}}=\\
 & = & \frac{1}{3}\gem{\nabla}_{\hat{\bs{\alpha}}}T_{bc}\hoch{\bs{\delta}}+\frac{2}{3}\gem{\nabla}_{[b}T_{c]\hat{\bs{\alpha}}}\hoch{\bs{\delta}}+\frac{4}{3}\gem{T}_{\hat{\bs{\alpha}}[b|}\hoch{E}T_{E|c]}\hoch{\bs{\delta}}+\frac{2}{3}\gem{T}_{bc}\hoch{E}T_{E\hat{\bs{\alpha}}}\hoch{\bs{\delta}}=\\
 & = & \frac{1}{3}\gem{\nabla}_{\hat{\bs{\alpha}}}T_{bc}\hoch{\bs{\delta}}-\frac{2}{3}\gem{\nabla}_{[b}\left(\tilde{\gamma}_{c]\,\hat{\bs{\alpha}}\hat{\bs{\delta}}}\RR^{\bs{\delta}\hat{\bs{\delta}}}\right)+\frac{4}{3}\check{T}_{\hat{\bs{\alpha}}[b|}\hoch{e}T_{e|c]}\hoch{\bs{\delta}}+\frac{2}{3}\check{T}_{bc}\hoch{e}T_{e\hat{\bs{\alpha}}}\hoch{\bs{\delta}}=\\
 & \stackrel{\check{\Omega}=\hat{\Omega}}{=} & \frac{1}{3}\bei{\gem{\nabla}_{\hat{\bs{\alpha}}}T_{bc}\hoch{\bs{\delta}}}{\check{\Omega}=\hat{\Omega}}+\frac{2}{3}\tilde{\gamma}_{[b\,\hat{\bs{\alpha}}\hat{\bs{\delta}}}\gem{\nabla}_{c]}\RR^{\bs{\delta}\hat{\bs{\delta}}}+\frac{4}{3}\hat{T}_{\hat{\bs{\alpha}}[b|}\hoch{e}T_{e|c]}\hoch{\bs{\delta}}+H_{bc}\hoch{e}\tilde{\gamma}_{e\hat{\bs{\alpha}}\hat{\bs{\beta}}}\RR^{\bs{\delta}\hat{\bs{\beta}}}\end{eqnarray}
or\begin{eqnarray}
 & \stackrel{\check{\Omega}=\Omega}{=} & \frac{1}{3}\bei{\gem{\nabla}_{\hat{\bs{\alpha}}}T_{bc}\hoch{\bs{\delta}}}{\check{\Omega}=\Omega}-\frac{2}{3}\gamma_{\hat{\bs{\alpha}}\hat{\bs{\delta}}}^{d}\Big[\big(\underbrace{\nabla_{[b|}\Phi}_{=0\,(\ref{eq:Omegaa})}+\underbrace{\hat{\nabla}_{[b|}\Phi}_{=0\,(\ref{eq:Omegaa})}\big)G_{d|c]}+\underbrace{\Delta_{[bc]d}}_{-3H_{bcd}}\Big]\RR^{\bs{\delta}\hat{\bs{\delta}}}+\nonumber \\
 &  & +\frac{2}{3}\tilde{\gamma}_{[b|\,\hat{\bs{\alpha}}\hat{\bs{\delta}}}\gem{\nabla}_{|c]}\RR^{\bs{\delta}\hat{\bs{\delta}}}-H_{bce}\gamma_{\hat{\bs{\alpha}}\hat{\bs{\delta}}}^{e}\RR^{\bs{\delta}\hat{\bs{\delta}}}=\\
 & = & \frac{1}{3}\nabla_{\hat{\bs{\alpha}}}T_{bc}\hoch{\bs{\delta}}+\frac{2}{3}\tilde{\gamma}_{[b|\,\hat{\bs{\alpha}}\hat{\bs{\delta}}}\gem{\nabla}_{|c]}\RR^{\bs{\delta}\hat{\bs{\delta}}}+H_{bce}\gamma_{\hat{\bs{\alpha}}\hat{\bs{\delta}}}^{e}\RR^{\bs{\delta}\hat{\bs{\delta}}}\end{eqnarray}
\Ram{0.6}{\begin{eqnarray}
\nabla_{\hat{\bs{\alpha}}}T_{bc}\hoch{\bs{\delta}} & = & -2\tilde{\gamma}_{[b|\,\hat{\bs{\alpha}}\hat{\bs{\delta}}}\gemnabla_{|c]}\RR^{\bs{\delta}\hat{\bs{\delta}}}-3H_{bce}\gamma_{\hat{\bs{\alpha}}\hat{\bs{\delta}}}^{e}\RR^{\bs{\delta}\hat{\bs{\delta}}}\label{eq:(delta|2,0,1)}\\
\hat{\nabla}_{\bs{\alpha}}\hat{T}_{bc}\hoch{\hat{\bs{\delta}}} & = & -2\tilde{\gamma}_{[b|\,\bs{\alpha}\bs{\delta}}\gemnabla_{|c]}\RR^{\bs{\delta}\hat{\bs{\delta}}}+3H_{bce}\gamma_{\bs{\alpha}\bs{\delta}}^{e}\RR^{\bs{\delta}\hat{\bs{\delta}}}\label{eq:(hdelta|2,1,0)}\end{eqnarray}
}\\
$\bullet\quad$\underbar{(delta|3,0,0)$_{abc}\hoch{\bs{\delta}}\leftrightarrow$((hdelta|3,0,0)$_{abc}\hoch{\hat{\bs{\delta}}}$)dim$\frac{5}{2}$:}\begin{eqnarray}
0 & \stackrel{!}{=} & \gem{\nabla}_{[a}T_{bc]}\hoch{\bs{\delta}}+2T_{[ab|}\hoch{E}T_{E|c]}\hoch{\bs{\delta}}-R_{[abc]}\hoch{\bs{\delta}}=\\
 & = & \gem{\nabla}_{[a}T_{bc]}\hoch{\bs{\delta}}+2\check{T}_{[ab|}\hoch{e}T_{e|c]}\hoch{\bs{\delta}}+2\hat{T}_{[ab|}\hoch{\hat{\bs{\eps}}}\tilde{\gamma}_{|c]\,\hat{\bs{\eps}}\hat{\bs{\delta}}}\RR^{\bs{\delta}\hat{\bs{\delta}}}\end{eqnarray}
\Ram{0.6}{\begin{eqnarray}
\nabla_{[a}T_{bc]}\hoch{\bs{\delta}} & = & -3H_{[ab|}\hoch{e}T_{e|c]}\hoch{\bs{\delta}}-2\hat{T}_{[ab|}\hoch{\hat{\bs{\eps}}}\tilde{\gamma}_{|c]\,\hat{\bs{\eps}}\hat{\bs{\delta}}}\RR^{\bs{\delta}\hat{\bs{\delta}}}\label{eq:(delta|3,0,0)}\\
\hat{\nabla}_{[a}\hat{T}_{bc]}\hoch{\hat{\bs{\delta}}} & = & 3H_{[ab|}\hoch{e}\hat{T}_{e|c]}\hoch{\hat{\bs{\delta}}}-2T_{[ab|}\hoch{\bs{\eps}}\tilde{\gamma}_{|c]\,\bs{\eps}\bs{\delta}}\RR^{\bs{\delta}\hat{\bs{\delta}}}\label{eq:(hdelta|3,0,0)}\end{eqnarray}
}\\
$\bullet\quad$\underbar{(d|0,3,0)$_{\bs{\alpha\beta\gamma}}\hoch{d}\leftrightarrow$((d|0,0,3)$_{\hat{\bs{\alpha}}\hat{\bs{\beta}}\hat{\bs{\gamma}}}\hoch{d}$)dim$\frac{1}{2}$:}\begin{eqnarray}
0 & \stackrel{!}{=} & \gem{\nabla}_{[\bs{\alpha}}\check{T}_{\bs{\beta\gamma}]}\hoch{d}+2\check{T}_{[\bs{\alpha\beta}|}\hoch{c}\check{T}_{c|\bs{\gamma}]}\hoch{d}-\underbrace{\check{R}_{[\bs{\alpha\beta\gamma}]}\hoch{d}}_{=0}=\\
 & = & \gem{\nabla}_{[\bs{\alpha}}\left(\gamma_{\bs{\beta\gamma}]}\hoch{c}f_{c}\hoch{d}\right)+2\gamma_{[\bs{\alpha\beta}|}^{e}f_{e}\hoch{c}\check{T}_{c|\bs{\gamma}]}\hoch{d}=\\
 & \stackrel{f_{c}\hoch{d}=\delta_{c}^{d}}{=} & \underbrace{\nabla_{[\bs{\alpha}}\left(\gamma_{\bs{\beta\gamma}]}^{d}\right)}_{=0}-2\underbrace{\gamma_{[\bs{\alpha\beta}}^{c}T_{\bs{\gamma}]c}\hoch{d}}_{=0\,(\ref{eq:(1,3,0)})}\end{eqnarray}
\\
$\bullet\quad$\underbar{(d|0,1,2)$_{\bs{\alpha}\hat{\bs{\beta}}\hat{\bs{\gamma}}}\hoch{a}\leftrightarrow$((d|0,2,1)$_{\hat{\bs{\alpha}}\bs{\beta}\bs{\gamma}}\hoch{a}$)dim$\frac{1}{2}$:}
\begin{eqnarray}
0 & \stackrel{!}{=} & \gem{\nabla}_{[\bs{\alpha}}\check{T}_{\hat{\bs{\beta}}\hat{\bs{\gamma}}]}\hoch{d}+2\gem{T}_{[\bs{\alpha}\hat{\bs{\beta}}|}\hoch{C}\check{T}_{C|\hat{\bs{\gamma}}]}\hoch{d}-\check{R}_{[\bs{\alpha}\hat{\bs{\beta}}\hat{\bs{\gamma}}]}\hoch{d}=\\
 & = & \frac{1}{3}\gem{\nabla}_{\bs{\alpha}}\check{T}_{\hat{\bs{\beta}}\hat{\bs{\gamma}}}\hoch{d}+\frac{2}{3}\check{T}_{\hat{\bs{\beta}}\hat{\bs{\gamma}}}\hoch{c}\check{T}_{c\bs{\alpha}}\hoch{d}=\\
 & = & \frac{2}{3}\gamma_{\hat{\bs{\beta}}\hat{\bs{\gamma}}}\hoch{c}\hat{T}_{c\bs{\alpha}}\hoch{d}=0\end{eqnarray}
 \\
$\bullet\quad$\underbar{(d|1,2,0)$_{\bs{\alpha\beta}c}\hoch{d}\leftrightarrow$((d|1,0,2)$_{\hat{\bs{\alpha}}\hat{\bs{\beta}}c}\hoch{d}$)dim1:}\\
\begin{eqnarray}
0 & \stackrel{!}{=} & \gem{\nabla}_{[\bs{\alpha}}\check{T}_{\bs{\beta}c]}\hoch{d}+2\gem{T}_{[\bs{\alpha\beta}|}\hoch{E}T_{E|c]}\hoch{d}-\check{R}_{[\bs{\alpha\beta}c]}\hoch{d}=\\
 & = & \frac{2}{3}\gem{\nabla}_{[\bs{\alpha}}\check{T}_{\bs{\beta}]c}\hoch{d}+\frac{1}{3}\gem{\nabla}_{c}T_{\bs{\alpha}\bs{\beta}}\hoch{d}+\frac{2}{3}\gem{T}_{\bs{\alpha\beta}}\hoch{E}\check{T}_{Ec}\hoch{d}+\frac{4}{3}\gem{T}_{c[\bs{\alpha}|}\hoch{E}\check{T}_{E|\bs{\beta}]}\hoch{d}-\frac{1}{3}\check{R}_{\bs{\alpha\beta}c}\hoch{d}-\frac{2}{3}\underbrace{R_{c[\bs{\alpha\beta}]}\hoch{d}}_{=0}=\\
 & \us{\stackrel{f_{e}\hoch{d}=\delta_{e}^{d}}{=}}{\check{\Omega}=\Omega} & \frac{2}{3}\nabla_{[\bs{\alpha}}T_{\bs{\beta}]c}\hoch{d}+\frac{1}{3}\underbrace{\nabla_{c}\gamma_{\bs{\alpha}\bs{\beta}}^{d}}_{=0}+\frac{2}{3}\gamma_{\bs{\alpha\beta}}^{e}\underbrace{T_{ec}\hoch{d}}_{\frac{3}{2}H_{ec}\hoch{d}}+\frac{4}{3}T_{[\bs{\alpha}|c}\hoch{e}T_{|\bs{\beta}]e}\hoch{d}+\frac{4}{3}\underbrace{T_{c[\bs{\alpha}|}\hoch{\bs{\eps}}}_{=0}\gamma_{\bs{\eps}|\bs{\beta}]}^{d}-\frac{1}{3}R_{\bs{\alpha\beta}c}\hoch{d}\end{eqnarray}
\Ram{0.6}{\begin{eqnarray}
R_{\bs{\alpha\beta}c}\hoch{d} & \stackrel{!}{=} & 2\nabla_{[\bs{\alpha}}T_{\bs{\beta}]c}\hoch{d}+3\gamma_{\bs{\alpha\beta}}^{e}H_{ec}\hoch{d}+4T_{[\bs{\alpha}|c}\hoch{e}T_{|\bs{\beta}]e}\hoch{d}\label{eq:(d|1,2,0)}\\
\hat{R}_{\hat{\bs{\alpha}}\hat{\bs{\beta}}c}\hoch{d} & \stackrel{!}{=} & 2\hat{\nabla}_{[\hat{\bs{\alpha}}}\hat{T}_{\hat{\bs{\beta}}]c}\hoch{d}-3\gamma_{\hat{\bs{\alpha}}\hat{\bs{\beta}}}^{e}H_{ec}\hoch{d}+4\hat{T}_{[\hat{\bs{\alpha}}|c}\hoch{e}\hat{T}_{|\hat{\bs{\beta}}]e}\hoch{d}\label{eq:(d|1,0,2)}\end{eqnarray}
} Taking the trace (using $R_{\bs{MM}a}\hoch{b}=F^{(D)}\tief{\bs{MM}}\delta_{a}^{b}+R_{\bs{MM}a}^{(L)}\hoch{b}$)
yields\begin{eqnarray}
10F_{\bs{\alpha\beta}}^{(D)} & \stackrel{!}{=} & -10\nabla_{[\bs{\alpha}}\covPhi{\bs{\beta}]},\qquad true\end{eqnarray}
Plugging in the torsion constraints yields\vRam{.7}{\begin{eqnarray}
R_{\bs{\alpha\beta}c}\hoch{d} & = & -\nabla_{[\bs{\alpha}}\covPhi{\bs{\beta}]}\delta_{c}\hoch{d}+\gamma_{c}\hoch{d}\tief{[\bs{\alpha}}\hoch{\bs{\delta}}\nabla_{\bs{\beta}]}\covPhi{\bs{\delta}}+3\gamma_{\bs{\alpha\beta}}^{e}H_{ec}\hoch{d}+\nonumber \\
 &  & +\gamma_{c}\hoch{e}\tief{[\bs{\alpha}|}\hoch{\bs{\gamma}}\covPhi{\bs{\gamma}}\gamma_{e}\hoch{d}\tief{|\bs{\beta}]}\hoch{\bs{\delta}}\covPhi{\bs{\delta}}\label{eq:(d|1,2,0):expl}\\
\hat{R}_{\hat{\bs{\alpha}}\hat{\bs{\beta}}c}\hoch{d} & = & -\hat{\nabla}_{[\hat{\bs{\alpha}}}\hatcovPhi{\hat{\bs{\beta}}]}\delta_{c}\hoch{d}+\gamma_{c}\hoch{d}\tief{[\hat{\bs{\alpha}}}\hoch{\hat{\bs{\delta}}}\hat{\nabla}_{\hat{\bs{\beta}}]}\hatcovPhi{\hat{\bs{\delta}}}-3\gamma_{\hat{\bs{\alpha}}\hat{\bs{\beta}}}^{e}H_{ec}\hoch{d}+\nonumber \\
 &  & +\gamma_{c}\hoch{e}\tief{[\hat{\bs{\alpha}}|}\hoch{\hat{\bs{\gamma}}}\hatcovPhi{\hat{\bs{\gamma}}}\gamma_{e}\hoch{d}\tief{|\hat{\bs{\beta}}]}\hoch{\hat{\bs{\delta}}}\hatcovPhi{\hat{\bs{\delta}}}\label{eq:(d|1,0,2):expl}\end{eqnarray}
} This agrees with (\ref{eq:Rgamgamalphbet}) and (\ref{eq:Rgamgamalphbet:hat}).\\
$\bullet\quad$\underbar{(d|1,1,1)$_{\bs{\alpha}\hat{\bs{\beta}}c}\hoch{d}$dim1:}\begin{eqnarray}
0 & \stackrel{!}{=} & \gem{\nabla}_{[\bs{\alpha}}\check{T}_{\hat{\bs{\beta}}c]}\hoch{d}+2\gem{T}_{[\bs{\alpha}\hat{\bs{\beta}}|}\hoch{E}\check{T}_{E|c]}\hoch{d}-\check{R}_{[\bs{\alpha}\hat{\bs{\beta}}c]}\hoch{d}=\\
 & = & \frac{1}{3}\gem{\nabla}_{\bs{\alpha}}\check{T}_{\hat{\bs{\beta}}c}\hoch{d}+\frac{1}{3}\gem{\nabla}_{\hat{\bs{\beta}}}\check{T}_{c\bs{\alpha}}\hoch{d}+\frac{2}{3}\gem{T}_{c\bs{\alpha}}\hoch{E}\check{T}_{E\hat{\bs{\beta}}}\hoch{d}+\frac{2}{3}\gem{T}_{\hat{\bs{\beta}}c}\hoch{E}\check{T}_{E\bs{\alpha}}\hoch{d}-\frac{1}{3}\check{R}_{\bs{\alpha}\hat{\bs{\beta}}c}\hoch{d}=\\
 & \stackrel{\check{\Omega}=\Omega}{=} & \frac{1}{3}\nabla_{\hat{\bs{\beta}}}T_{c\bs{\alpha}}\hoch{d}+\frac{2}{3}\hat{T}_{c\bs{\alpha}}\hoch{\hat{\bs{\eps}}}T_{\hat{\bs{\eps}}\hat{\bs{\beta}}}\hoch{d}+\frac{2}{3}T_{\hat{\bs{\beta}}c}\hoch{e}T_{e\bs{\alpha}}\hoch{d}+\frac{2}{3}T_{\hat{\bs{\beta}}c}\hoch{\bs{\eps}}T_{\bs{\eps}\bs{\alpha}}\hoch{d}-\frac{1}{3}R_{\bs{\alpha}\hat{\bs{\beta}}c}\hoch{d}=\\
 & = & \frac{1}{3}\nabla_{\hat{\bs{\beta}}}T_{c\bs{\alpha}}\hoch{d}-\frac{2}{3}\tilde{\gamma}_{c\,\bs{\alpha}\bs{\beta}}\RR^{\bs{\beta}\hat{\bs{\eps}}}\gamma_{\hat{\bs{\eps}}\hat{\bs{\beta}}}^{d}+\frac{2}{3}\tilde{\gamma}_{c\,\hat{\bs{\beta}}\hat{\bs{\alpha}}}\RR^{\bs{\eps}\hat{\bs{\alpha}}}\gamma_{\bs{\eps}\bs{\alpha}}^{d}-\frac{1}{3}R_{\bs{\alpha}\hat{\bs{\beta}}c}\hoch{d}\end{eqnarray}
\Ram{0.7}{\begin{eqnarray}
R_{\bs{\alpha}\hat{\bs{\beta}}c}\hoch{d} & = & \nabla_{\hat{\bs{\beta}}}T_{c\bs{\alpha}}\hoch{d}-2\tilde{\gamma}_{c\,\bs{\alpha}\bs{\beta}}\RR^{\bs{\beta}\hat{\bs{\eps}}}\gamma_{\hat{\bs{\eps}}\hat{\bs{\beta}}}^{d}+2\tilde{\gamma}_{c\,\hat{\bs{\beta}}\hat{\bs{\delta}}}\RR^{\bs{\eps}\hat{\bs{\delta}}}\gamma_{\bs{\eps}\bs{\alpha}}^{d}\label{eq:(d|1,1,1)}\\
\hat{R}_{\hat{\bs{\alpha}}\bs{\beta}c}\hoch{d} & = & \hat{\nabla}_{\bs{\beta}}\hat{T}_{c\hat{\bs{\alpha}}}\hoch{d}-2\tilde{\gamma}_{c\,\hat{\bs{\alpha}}\hat{\bs{\beta}}}\RR^{\bs{\eps}\hat{\bs{\beta}}}\gamma_{\bs{\eps}\bs{\beta}}^{d}+2\tilde{\gamma}_{c\,\bs{\beta}\bs{\delta}}\RR^{\bs{\delta}\hat{\bs{\eps}}}\gamma_{\hat{\bs{\eps}}\hat{\bs{\alpha}}}^{d}\quad\mbox{(equivalent)}\label{eq:(d|1,1,1):equiv}\end{eqnarray}
} Plugging the explicit expression for $T_{c\bs{\alpha}}\hoch{d}$
and $\hat{T}_{c\hat{\bs{\alpha}}}\hoch{d}$ into (\ref{eq:(d|1,1,1)})
and (\ref{eq:(d|1,1,1):equiv}) yields%
\footnote{\label{fn:dilaton-constraint}\index{footnote!\thefoot. constraint on dilaton from comparing different constraints on curvature}From
this constraint on $R_{\bs{\alpha}\hat{\bs{\beta}}c}\hoch{d}$ we
can also derive a further constraint on some spinorial components.
Remember that we have $R_{\bs{\alpha}\hat{\bs{\beta}}\bs{\gamma}}\hoch{\bs{\delta}}=\frac{1}{2}F_{\bs{\alpha}\hat{\bs{\beta}}}^{(D)}\delta_{\bs{\gamma}}\hoch{\bs{\delta}}+\frac{1}{4}R_{\bs{\alpha}\hat{\bs{\beta}}c}^{(L)}\hoch{d}\gamma^{c}\tief{d\,\bs{\gamma}}\hoch{\bs{\delta}}$
and therefore \begin{eqnarray*}
R_{\bs{\alpha}\hat{\bs{\beta}}\bs{\gamma}}\hoch{\bs{\delta}} & = & \tfrac{1}{4}\nabla_{\hat{\bs{\beta}}}\covPhi{\bs{\alpha}}\delta_{\bs{\gamma}}\hoch{\bs{\delta}}+\tfrac{1}{4}\left(\tfrac{1}{2}\gamma_{c}\hoch{d}\tief{\bs{\alpha}}\hoch{\bs{\eps}}\nabla_{\hat{\bs{\beta}}}\covPhi{\bs{\eps}}-2\tilde{\gamma}_{c\,\bs{\alpha}\bs{\beta}}\RR^{\bs{\beta}\hat{\bs{\eps}}}\gamma_{\hat{\bs{\eps}}\hat{\bs{\beta}}}^{d}+2\tilde{\gamma}_{c\,\hat{\bs{\beta}}\hat{\bs{\delta}}}\RR^{\bs{\eps}\hat{\bs{\delta}}}\gamma_{\bs{\eps}\bs{\alpha}}^{d}\right)\gamma^{c}\tief{d\,\bs{\gamma}}\hoch{\bs{\delta}}\end{eqnarray*}
The last terms can be combined and we arrive at\[
\boxed{R_{\bs{\alpha}\hat{\bs{\beta}}\bs{\gamma}}\hoch{\bs{\delta}}=\tfrac{1}{4}\nabla_{\hat{\bs{\beta}}}\covPhi{\bs{\alpha}}\delta_{\bs{\gamma}}\hoch{\bs{\delta}}+\tfrac{1}{8}\gamma_{cd}\tief{\bs{\alpha}}\hoch{\bs{\eps}}\gamma^{cd}\tief{\,\bs{\gamma}}\hoch{\bs{\delta}}\nabla_{\hat{\bs{\beta}}}\covPhi{\bs{\eps}}-\gamma_{\bs{\alpha}\bs{\eps}}^{c}\RR^{\bs{\eps}\hat{\bs{\eps}}}\gamma_{\hat{\bs{\eps}}\hat{\bs{\beta}}}^{d}\gamma\tief{cd\,\bs{\gamma}}\hoch{\bs{\delta}}}\]
Next we can compare whether this is consistent with our earlier constraint
$R_{\hat{\bs{\beta}}[\bs{\alpha\gamma}]}\hoch{\bs{\delta}}=-\gamma_{\bs{\alpha\gamma}}\hoch{e}\tilde{\gamma}_{e\,\hat{\bs{\beta}}\hat{\bs{\delta}}}\RR^{\bs{\delta}\hat{\bs{\delta}}}$
: \begin{eqnarray*}
R_{\hat{\bs{\beta}}[\bs{\alpha}\bs{\gamma}]}\hoch{\bs{\delta}} & = & \tfrac{1}{4}\nabla_{\hat{\bs{\beta}}}\covPhi{[\bs{\alpha}}\delta_{\bs{\gamma}]}\hoch{\bs{\delta}}+\tfrac{1}{8}\gamma_{cd}\tief{[\bs{\alpha}}\hoch{\bs{\eps}}\gamma^{cd}\tief{\bs{\gamma}]}\hoch{\bs{\delta}}\nabla_{\hat{\bs{\beta}}}\covPhi{\bs{\eps}}+\gamma_{\hat{\bs{\beta}}\hat{\bs{\eps}}}^{d}\RR^{\bs{\eps}\hat{\bs{\eps}}}\gamma_{\bs{\eps}[\bs{\alpha}|}^{c}\tilde{\gamma}\tief{cd\,|\bs{\gamma}]}\hoch{\bs{\delta}}\end{eqnarray*}
Being graded antisymmetric in ${\scriptstyle \bs{\alpha}}$ and ${\scriptstyle \bs{\gamma}}$,
it can be expanded in $\gamma_{\bs{\alpha\gamma}}^{a}$ and $\gamma_{\bs{\alpha\gamma}}^{a_{1}\ldots a_{5}}$,
where the coefficient for $\gamma^{[5]}$ should vanish and the other
coincide with the old expression. Before projecting the coefficients
by brute force one can do a first step in this direction by using
the identities $\gamma^{ab}\tief{[\bs{\alpha}|}\hoch{\bs{\eps}}\gamma_{ab}\tief{|\bs{\gamma}]}\hoch{\bs{\delta}}=-4\gamma_{a}^{\bs{\eps\delta}}\gamma_{\bs{\alpha\gamma}}^{a}-10\delta_{[\bs{\alpha}}\hoch{\bs{\eps}}\delta_{\bs{\gamma}]}\hoch{\bs{\delta}}$
(graded version of (\ref{eq:FromLittleFierz})) and $\gamma_{\bs{\eps}[\bs{\alpha}|}^{c}\tilde{\gamma}\tief{cd\,|\bs{\gamma}]}\hoch{\bs{\delta}}=-\tfrac{1}{2}\gamma_{\bs{\alpha}\bs{\gamma}}^{c}\tilde{\gamma}_{cd\,\bs{\eps}}\hoch{\bs{\delta}}-\tfrac{1}{2}\tilde{\gamma}_{d\,\bs{\alpha}\bs{\gamma}}\delta_{\bs{\eps}}\hoch{\bs{\delta}}-\tilde{\gamma}_{d\,\bs{\eps}[\bs{\alpha}|}\delta_{|\bs{\gamma}]}\hoch{\bs{\delta}}$,
which are both immediate consequences of the Fierz identity $\gamma_{[\bs{\alpha\beta}|}^{c}\gamma_{c\,|\bs{\gamma}]\bs{\delta}}=0$
.\begin{eqnarray*}
R_{\hat{\bs{\beta}}[\bs{\alpha}\bs{\gamma}]}\hoch{\bs{\delta}} & = & -\nabla_{\hat{\bs{\beta}}}\covPhi{[\bs{\alpha}}\delta_{\bs{\gamma}]}\hoch{\bs{\delta}}-\tfrac{1}{2}\gamma_{a}^{\bs{\eps\delta}}\gamma_{\bs{\alpha\gamma}}^{a}\nabla_{\hat{\bs{\beta}}}\covPhi{\bs{\eps}}+\\
 &  & -\tfrac{1}{2}\gamma_{\bs{\alpha}\bs{\gamma}}^{c}\tilde{\gamma}_{cd\,\bs{\eps}}\hoch{\bs{\delta}}\gamma_{\hat{\bs{\beta}}\hat{\bs{\eps}}}^{d}\RR^{\bs{\eps}\hat{\bs{\eps}}}-\tfrac{1}{2}\tilde{\gamma}_{d\,\bs{\alpha}\bs{\gamma}}\gamma_{\hat{\bs{\beta}}\hat{\bs{\eps}}}^{d}\RR^{\bs{\delta}\hat{\bs{\eps}}}-\tilde{\gamma}_{d\,\bs{\eps}[\bs{\alpha}|}\delta_{|\bs{\gamma}]}\hoch{\bs{\delta}}\gamma_{\hat{\bs{\beta}}\hat{\bs{\eps}}}^{d}\RR^{\bs{\eps}\hat{\bs{\eps}}}\end{eqnarray*}
Now let us write the expansion in $\gamma^{[1]}$ and $\gamma^{[5]}$
as $R_{\hat{\bs{\beta}}[\bs{\alpha}\bs{\gamma}]}\hoch{\bs{\delta}}=R_{\hat{\bs{\beta}}}^{[1]}\hoch{\bs{\delta}}\tief{a}\gamma_{\bs{\alpha}\bs{\gamma}}^{a}+R_{\hat{\bs{\beta}}}^{[5]}\hoch{\bs{\delta}}\tief{a_{1}\ldots a_{5}}\gamma_{\bs{\alpha}\bs{\gamma}}^{a_{1}\ldots a_{5}}$.
The second term has to vanish, so that the first condition is (projecting
with $\gamma_{a_{1}\ldots a_{5}}^{\bs{\alpha\gamma}}$): \[
\boxed{\gamma_{a_{1}\ldots a_{5}}^{\bs{\delta\alpha}}\left(\nabla_{\hat{\bs{\beta}}}\covPhi{\bs{\alpha}}+\tilde{\gamma}_{d\,\bs{\alpha}\bs{\eps}}\RR^{\bs{\eps}\hat{\bs{\eps}}}\gamma_{\hat{\bs{\eps}}\hat{\bs{\beta}}}^{d}\right)=0}\]
The other coefficient can be projected with $\gamma_{a}^{\bs{\gamma\alpha}}$
via $R_{\hat{\bs{\beta}}}^{[1]}\hoch{\bs{\delta}}\tief{a}=\tfrac{1}{16}\gamma_{a}^{\bs{\gamma\alpha}}R_{\hat{\bs{\beta}}[\bs{\alpha}\bs{\gamma}]}\hoch{\bs{\delta}}$,
which should coincide with $-\tilde{\gamma}_{a\,\hat{\bs{\beta}}\hat{\bs{\delta}}}\RR^{\bs{\delta}\hat{\bs{\delta}}}$.
We thus obtain as second condition\begin{eqnarray*}
-\tilde{\gamma}_{a\,\hat{\bs{\beta}}\hat{\bs{\delta}}}\RR^{\bs{\delta}\hat{\bs{\delta}}} & = & -\tfrac{1}{2}\gamma_{a}^{\bs{\eps\delta}}\nabla_{\hat{\bs{\beta}}}\covPhi{\bs{\eps}}-\tfrac{1}{2}\tilde{\gamma}_{ad\,\bs{\eps}}\hoch{\bs{\delta}}\gamma_{\hat{\bs{\beta}}\hat{\bs{\eps}}}^{d}\RR^{\bs{\eps}\hat{\bs{\eps}}}-\tfrac{1}{2}\tilde{\gamma}_{a\,\hat{\bs{\beta}}\hat{\bs{\eps}}}\RR^{\bs{\delta}\hat{\bs{\eps}}}\\
 &  & -\tfrac{1}{16}\gamma_{a}^{\bs{\delta\alpha}}\nabla_{\hat{\bs{\beta}}}\covPhi{\bs{\alpha}}-\tfrac{1}{16}\gamma_{a}^{\bs{\delta\alpha}}\tilde{\gamma}_{d\,\bs{\eps}\bs{\alpha}}\gamma_{\hat{\bs{\beta}}\hat{\bs{\eps}}}^{d}\RR^{\bs{\eps}\hat{\bs{\eps}}}\end{eqnarray*}
which can be further simplified to \[
\boxed{\gamma_{a}^{\bs{\delta\alpha}}\left(\nabla_{\hat{\bs{\beta}}}\covPhi{\bs{\alpha}}+\tilde{\gamma}_{d\,\bs{\alpha\rho}}\RR^{\bs{\rho}\hat{\bs{\eps}}}\gamma_{\hat{\bs{\eps}}\hat{\bs{\beta}}}^{d}\right)=0}\]
For this last equation we can finally use that $\gamma_{\bs{\beta\delta}}^{a}\gamma_{a}^{\bs{\delta\alpha}}=-10\delta_{\bs{\beta}}\hoch{\bs{\alpha}}$
which implies that already the bracket itself has to vanish and we
get the following constraint on the compensator superfield (and likewise
on the dilaton superfield): \[
\boxed{\nabla_{\hat{\bs{\beta}}}\covPhi{\bs{\alpha}}=-\tilde{\gamma}_{d\,\bs{\alpha\rho}}\RR^{\bs{\rho}\hat{\bs{\eps}}}\gamma_{\hat{\bs{\eps}}\hat{\bs{\beta}}}^{d}}\qquad\fussend\]
}\vRam{0.75}{ \begin{eqnarray}
R_{\bs{\alpha}\hat{\bs{\beta}}c}\hoch{d} & = & \frac{1}{2}\nabla_{\hat{\bs{\beta}}}\covPhi{\bs{\alpha}}\delta_{c}^{d}+\frac{1}{2}\gamma_{c}\hoch{d}\tief{\bs{\alpha}}\hoch{\bs{\gamma}}\nabla_{\hat{\bs{\beta}}}\covPhi{\bs{\gamma}}-2\tilde{\gamma}_{c\,\bs{\alpha}\bs{\beta}}\RR^{\bs{\beta}\hat{\bs{\eps}}}\gamma_{\hat{\bs{\eps}}\hat{\bs{\beta}}}^{d}+2\tilde{\gamma}_{c\,\hat{\bs{\beta}}\hat{\bs{\delta}}}\RR^{\bs{\eps}\hat{\bs{\delta}}}\gamma_{\bs{\eps}\bs{\alpha}}^{d}\label{eq:(d|1,1,1)'}\\
\hat{R}_{\hat{\bs{\alpha}}\bs{\beta}c}\hoch{d} & = & \frac{1}{2}\hat{\nabla}_{\bs{\beta}}\hatcovPhi{\hat{\bs{\alpha}}}\delta_{c}^{d}+\frac{1}{2}\gamma_{c}\hoch{d}\tief{\hat{\bs{\alpha}}}\hoch{\hat{\bs{\gamma}}}\hat{\nabla}_{\bs{\beta}}\hatcovPhi{\hat{\bs{\gamma}}}-2\tilde{\gamma}_{c\,\hat{\bs{\alpha}}\hat{\bs{\beta}}}\RR^{\bs{\eps}\hat{\bs{\beta}}}\gamma_{\bs{\eps}\bs{\beta}}^{d}+2\tilde{\gamma}_{c\,\bs{\beta}\bs{\delta}}\RR^{\bs{\delta}\hat{\bs{\eps}}}\gamma_{\hat{\bs{\eps}}\hat{\bs{\alpha}}}^{d}\label{eq:(d|1,1,1):equiv'}\end{eqnarray}
} Taking the trace of (\ref{eq:(d|1,1,1)'}) yields\begin{eqnarray}
10F_{\bs{\alpha}\hat{\bs{\beta}}}^{(D)} & = & 5\nabla_{\hat{\bs{\beta}}}\nabla_{\bs{\alpha}}\Phi-2\tilde{\gamma}_{c\,\bs{\alpha}\bs{\beta}}\RR^{\bs{\beta}\hat{\bs{\eps}}}\gamma_{\hat{\bs{\eps}}\hat{\bs{\beta}}}^{c}+2\tilde{\gamma}_{c\,\hat{\bs{\beta}}\hat{\bs{\delta}}}\RR^{\bs{\eps}\hat{\bs{\delta}}}\gamma_{\bs{\eps}\bs{\alpha}}^{c}\\
\dann\quad F_{\bs{\alpha}\hat{\bs{\beta}}}^{(D)} & = & \frac{1}{2}\nabla_{\hat{\bs{\beta}}}\nabla_{\bs{\alpha}}\Phi\end{eqnarray}
This does not give new information as it follows from $\gemT_{\bs{\alpha}\hat{\bs{\beta}}}\hoch{C}=0$,
$\covPhi{\hat{\bs{\beta}}}=0$ and the algebra $\bei{\gemnabla_{[\bs{\alpha}}\gemnabla_{\hat{\bs{\beta}}]}\Phi}{\check{\Omega}=\Omega}=-\bei{\gemT_{\bs{\alpha}\hat{\bs{\beta}}}\hoch{C}}{\check{\Omega}=\Omega}\nabla_{C}\Phi-F_{\bs{\alpha}\hat{\bs{\beta}}}^{(D)}$.
\\
$\bullet\quad$\underbar{(d|2,1,0)$_{\bs{\alpha}bc}\hoch{d}\leftrightarrow$((d|2,0,1)$_{\hat{\bs{\alpha}}bc}\hoch{d}$)dim$\frac{3}{2}$:}\begin{eqnarray}
0 & \stackrel{!}{=} & \gem{\nabla}_{[\bs{\alpha}}\check{T}_{bc]}\hoch{d}+2\gem{T}_{[\bs{\alpha}b|}\hoch{E}\check{T}_{E|c]}\hoch{d}-\check{R}_{[\bs{\alpha}bc]}\hoch{d}=\\
 & = & \frac{1}{3}\gem{\nabla}_{\bs{\alpha}}\check{T}_{bc}\hoch{d}+\frac{2}{3}\gem{\nabla}_{[b}\underbrace{\check{T}_{c]\bs{\alpha}}\hoch{d}}_{=0\,\mbox{for }\check{\Omega}=\hat{\Omega}}+\frac{4}{3}\gem{T}_{\bs{\alpha}[b|}\hoch{E}\check{T}_{E|c]}\hoch{d}+\frac{2}{3}\gem{T}_{bc}\hoch{E}\check{T}_{E\bs{\alpha}}\hoch{d}-\frac{2}{3}\check{R}_{\bs{\alpha}[bc]}\hoch{d}=\\
 & \stackrel{\check{\Omega}=\hat{\Omega}}{=} & -\frac{1}{2}\hat{\nabla}_{\bs{\alpha}}H_{bc}\hoch{d}+\frac{4}{3}\hat{T}_{\bs{\alpha}[b|}\hoch{\hat{\bs{\eps}}}\hat{T}_{\hat{\bs{\eps}}|c]}\hoch{d}+\frac{2}{3}T_{bc}\hoch{\bs{\eps}}\gamma_{\bs{\eps}\bs{\alpha}}^{d}-\frac{2}{3}\hat{R}_{\bs{\alpha}[bc]}\hoch{d}\end{eqnarray}
\Ram{0.6}{\begin{eqnarray}
\hat{R}_{\bs{\alpha}[bc]}\hoch{d} & = & -\frac{3}{4}\hat{\nabla}_{\bs{\alpha}}H_{bc}\hoch{d}+2\tilde{\gamma}_{[b|\,\bs{\alpha\delta}}\RR^{\bs{\delta}\hat{\bs{\eps}}}\hat{T}_{\hat{\bs{\eps}}|c]}\hoch{d}+T_{bc}\hoch{\bs{\eps}}\gamma_{\bs{\eps}\bs{\alpha}}^{d}\label{eq:(d|2,1,0)}\\
R_{\hat{\bs{\alpha}}[bc]}\hoch{d} & = & \frac{3}{4}\nabla_{\hat{\bs{\alpha}}}H_{bc}\hoch{d}+2\tilde{\gamma}_{[b|\,\hat{\bs{\alpha}}\hat{\bs{\delta}}}\RR^{\bs{\eps}\hat{\bs{\delta}}}T_{\bs{\eps}|c]}\hoch{d}+\hat{T}_{bc}\hoch{\hat{\bs{\eps}}}\gamma_{\hat{\bs{\eps}}\hat{\bs{\alpha}}}^{d}\label{eq:(d|2,0,1)}\end{eqnarray}
} At this point it is convenient to plug the constraints (\ref{eq:(3,1,0)})
and (\ref{eq:(3,0,1)}) into the above equations to obtain slightly
simplified expressions\begin{eqnarray}
\hat{R}_{\bs{\alpha}[bc]d} & = & -2T_{d[b|}\hoch{\bs{\eps}}\gamma_{|c]\bs{\eps}\bs{\alpha}}+2\tilde{\gamma}_{[b|\,\bs{\alpha\delta}}\RR^{\bs{\delta}\hat{\bs{\eps}}}\hat{T}_{\hat{\bs{\eps}}|c]d}\\
R_{\hat{\bs{\alpha}}[bc]d} & = & -2\hat{T}_{d[b|}\hoch{\hat{\bs{\eps}}}\gamma_{|c]\hat{\bs{\eps}}\hat{\bs{\alpha}}}+2\tilde{\gamma}_{[b|\,\hat{\bs{\alpha}}\hat{\bs{\delta}}}\RR^{\bs{\eps}\hat{\bs{\delta}}}T_{\bs{\eps}|c]d}\end{eqnarray}
Let us plug the explicit expressions for the torsion components into
the first equation:\begin{eqnarray}
\hat{R}_{\bs{\alpha}[bc]d} & = & -\frac{1}{8}\left(\gemnabla_{\hat{\bs{\gamma}}}\RR^{\bs{\eps}\hat{\bs{\delta}}}+8\hatcovPhi{\hat{\bs{\gamma}}}\RR^{\bs{\eps}\hat{\bs{\delta}}}\right)\tilde{\gamma}_{d[b|\,\hat{\bs{\delta}}}\hoch{\hat{\bs{\gamma}}}\gamma_{|c]\bs{\eps}\bs{\alpha}}+\nonumber \\
 &  & -\tilde{\gamma}_{[b|\,\bs{\alpha\delta}}\RR^{\bs{\delta}\hat{\bs{\eps}}}\left(\hatcovPhi{\hat{\bs{\eps}}}G_{|c]d}+\tilde{\gamma}_{|c]d\,\hat{\bs{\eps}}}\hoch{\hat{\bs{\delta}}}\hatcovPhi{\hat{\bs{\delta}}}\right)=\\
 & = & -\frac{1}{8}\gemnabla_{\hat{\bs{\gamma}}}\RR^{\bs{\eps}\hat{\bs{\delta}}}\tilde{\gamma}_{d[b|\,\hat{\bs{\delta}}}\hoch{\hat{\bs{\gamma}}}\gamma_{|c]\bs{\eps}\bs{\alpha}}+G_{d[b|}\tilde{\gamma}_{|c]\,\bs{\alpha\delta}}\RR^{\bs{\delta}\hat{\bs{\eps}}}\hatcovPhi{\hat{\bs{\eps}}}\end{eqnarray}
Including the hatted version, we thus get in summary\vRam{0.6}{\begin{eqnarray}
\hat{R}_{\bs{\alpha}[bc]d} & = & -\frac{1}{8}\gemnabla_{\hat{\bs{\gamma}}}\RR^{\bs{\eps}\hat{\bs{\delta}}}\tilde{\gamma}_{d[b|\,\hat{\bs{\delta}}}\hoch{\hat{\bs{\gamma}}}\gamma_{|c]\bs{\eps}\bs{\alpha}}+G_{d[b|}\tilde{\gamma}_{|c]\,\bs{\alpha\delta}}\RR^{\bs{\delta}\hat{\bs{\eps}}}\hatcovPhi{\hat{\bs{\eps}}}\label{eq:(d|2,1,0)'}\\
R_{\hat{\bs{\alpha}}[bc]d} & = & -\frac{1}{8}\gemnabla_{\bs{\gamma}}\RR^{\bs{\delta}\hat{\bs{\eps}}}\tilde{\gamma}_{d[b|\,\bs{\delta}}\hoch{\bs{\gamma}}\gamma_{|c]\hat{\bs{\eps}}\hat{\bs{\alpha}}}+G_{d[b|}\tilde{\gamma}_{|c]\,\hat{\bs{\alpha}}\hat{\bs{\delta}}}\RR^{\bs{\eps}\hat{\bs{\delta}}}\covPhi{\bs{\eps}}\label{eq:(d|2,0,1)'}\end{eqnarray}
} Finally we take the trace of the first equation in the indices
$c$ and $d$ \begin{eqnarray}
\frac{9}{2}\hat{F}_{\bs{\alpha}b}^{(D)}-\frac{1}{2}\hat{R}_{\bs{\alpha}db}^{(L)}\hoch{d} & = & -\frac{1}{16}\gemnabla_{\hat{\bs{\gamma}}}\RR^{\bs{\eps}\hat{\bs{\delta}}}\tilde{\gamma}_{db\,\hat{\bs{\delta}}}\hoch{\hat{\bs{\gamma}}}\gamma_{\bs{\eps}\bs{\alpha}}^{d}-\frac{9}{2}\tilde{\gamma}_{b\,\bs{\alpha\delta}}\RR^{\bs{\delta}\hat{\bs{\eps}}}\hatcovPhi{\hat{\bs{\eps}}}\end{eqnarray}
with $\hat{F}_{\bs{\alpha}b}^{(D)}=-\gemnabla_{[\bs{\alpha}}\hatcovPhi{b]}-\gem{T}_{\bs{\alpha}b}\hoch{C}\hatcovPhi{C}=-\tilde{\gamma}_{b\,\bs{\alpha\beta}}\RR^{\bs{\beta}\hat{\bs{\gamma}}}\hatcovPhi{\hat{\bs{\gamma}}}$
or eventually: \begin{equation}
\boxed{\hat{R}_{d\bs{\alpha}b}^{(L)}\hoch{d}=\frac{1}{8}\gem{\nabla}_{\hat{\bs{\gamma}}}\RR^{\bs{\eps}\hat{\bs{\eps}}}\tilde{\gamma}_{bc\,\hat{\bs{\eps}}}\hoch{\hat{\bs{\gamma}}}\gamma_{\bs{\eps}\bs{\alpha}}^{c}}\end{equation}
\begin{equation}
\boxed{R_{d\hat{\bs{\alpha}}b}^{(L)}\hoch{d}=\frac{1}{8}\gem{\nabla}_{\bs{\gamma}}\RR^{\bs{\eps}\hat{\bs{\eps}}}\tilde{\gamma}_{bc\,\bs{\eps}}\hoch{\bs{\gamma}}\gamma_{\hat{\bs{\eps}}\hat{\bs{\alpha}}}^{c}}\end{equation}
\\
$\bullet\quad$\underbar{(d|3,0,0)$_{abc}\hoch{d}$dim2:}\begin{eqnarray}
0 & \stackrel{!}{=} & \gem{\nabla}_{[a}\check{T}_{bc]}\hoch{d}+2\gem{T}_{[ab|}\hoch{E}\check{T}_{E|c]}\hoch{d}-\check{R}_{[abc]}\hoch{d}=\\
 & \stackrel{\check{\Omega}=\Omega}{=} & \nabla_{[a}T_{bc]}\hoch{d}+2T_{[ab|}\hoch{e}T_{e|c]}\hoch{d}+2T_{[ab|}\hoch{\bs{\eps}}T_{\bs{\eps}|c]}\hoch{d}-R_{[abc]}\hoch{d}=\\
 & = & \frac{3}{2}\nabla_{[a}H_{bc]}\hoch{d}+\frac{9}{2}H_{[ab|}\hoch{e}H_{e|c]}\hoch{d}+2T_{[ab|}\hoch{\bs{\eps}}T_{\bs{\eps}|c]}\hoch{d}-R_{[abc]}\hoch{d}\end{eqnarray}
\Ram{0.6}{\begin{eqnarray}
R_{[abc]}\hoch{d} & = & \frac{3}{2}\nabla_{[a}H_{bc]}\hoch{d}+\frac{9}{2}H_{[ab|}\hoch{e}H_{e|c]}\hoch{d}+2T_{[ab|}\hoch{\bs{\eps}}T_{\bs{\eps}|c]}\hoch{d}\label{eq:(d|3,0,0)}\\
\hat{R}_{[abc]}\hoch{d} & = & -\frac{3}{2}\hat{\nabla}_{[a}H_{bc]}\hoch{d}+\frac{9}{2}H_{[ab|}\hoch{e}H_{e|c]}\hoch{d}+2\hat{T}_{[ab|}\hoch{\hat{\bs{\eps}}}\hat{T}_{\hat{\bs{\eps}}|c]}\hoch{d}\label{eq:(dhat|3,0,0)}\end{eqnarray}
} \\
Taking the trace yields\begin{eqnarray}
0 & \stackrel{!}{=} & \frac{1}{2}\nabla_{d}H_{ab}\hoch{d}+3\underbrace{H_{d[a|}\hoch{e}H_{e|b]}\hoch{d}}_{=0}+\frac{2}{3}T_{ab}\hoch{\bs{\eps}}T_{\bs{\eps}d}\hoch{d}+\nonumber \\
 &  & +\frac{4}{3}T_{d[a|}\hoch{\bs{\eps}}T_{\bs{\eps}|b]}\hoch{d}-\frac{8}{3}F_{ab}^{(D)}+\frac{2}{3}R_{d[ab]}^{(L)}\hoch{d}=\\
 & = & \frac{1}{2}\nabla_{d}H_{ab}\hoch{d}-\frac{10}{3}T_{ab}\hoch{\bs{\eps}}\covPhi{\bs{\eps}}+\frac{4}{3}T_{d[a|}\hoch{\bs{\eps}}T_{\bs{\eps}|b]}\hoch{d}-\frac{8}{3}F_{ab}^{(D)}+\frac{2}{3}R_{d[ab]}^{(L)}\hoch{d}\end{eqnarray}
with $F_{ab}^{(D)}=-\bei{\gemnabla_{[a}\gemnabla_{b]}\Phi}{\check{\Omega}=\Omega}-\bei{\gemT_{ab}\hoch{C}}{\check{\Omega}=\Omega}\covPhi{C}=-T_{ab}\hoch{\bs{\gamma}}\covPhi{\bs{\gamma}}$.
We thus get the following trace constraint on the bosonic left-moving
and right-moving (via the left-right-symmetry) Lorentz curvature:
\begin{equation}
\boxed{-R_{d[ab]}^{(L)}\hoch{d}=\frac{3}{4}\nabla_{d}H_{ab}\hoch{d}-T_{ab}\hoch{\bs{\gamma}}\covPhi{\bs{\gamma}}+2T_{d[a|}\hoch{\bs{\eps}}T_{\bs{\eps}|b]}\hoch{d}}\end{equation}
\begin{equation}
\boxed{-\hat{R}_{d[ab]}^{(L)}\hoch{d}=-\frac{3}{4}\hat{\nabla}_{d}H_{ab}\hoch{d}-\hat{T}_{ab}\hoch{\hat{\bs{\gamma}}}\hatcovPhi{\hat{\bs{\gamma}}}+2\hat{T}_{d[a|}\hoch{\hat{\bs{\eps}}}\hat{T}_{\hat{\bs{\eps}}|b]}\hoch{d}}\end{equation}
\addtocounter{localapp}{1}

\section{Identities for the scaling field strength}

Instead of extracting in a clumsy way the information about the dilaton
field strength, we can obtain the information in a more direct way.
At some points this should also serve as a check of equations that
we have already obtained. From the torsion Bianchi identity (\ref{eq:BI-T-main})
we cannot easily extract the dilatation part, because the group indices
are antisymmetrized. Instead, we will study the algebra of covariant
derivatives acting on the compensator field. \rem{

\subsection{From the derivative algebra on the compensator}

}We start with the constraints \begin{equation}
\covPhi{\hat{\bs{\alpha}}}=\hatcovPhi{\bs{\alpha}}=\covPhi{a}=\hatcovPhi{a}=0\end{equation}
Remember, that on the compensator field the commutator of covariant
derivatives reads\begin{equation}
\gemnabla_{[A}\checkcovPhi{B]}=-\check{T}_{AB}\hoch{C}\checkcovPhi{C}-\check{F}_{AB}^{(D)}\end{equation}
Now we can plug in various indices:\\
$\bullet\quad$\underbar{$(a,b):$}\begin{eqnarray}
\underbrace{\gemnabla_{[a}\gemnabla_{b]}\Phi}_{0} & = & -\check{T}_{ab}\hoch{c}\checkcovPhi{c}-T_{ab}\hoch{\bs{\gamma}}\checkcovPhi{\bs{\gamma}}-\hat{T}_{ab}\hoch{\hat{\bs{\gamma}}}\checkcovPhi{\hat{\bs{\gamma}}}-\check{F}_{ab}^{(D)}=\\
 & \stackrel{\check{\Omega}=\Omega}{=} & -T_{ab}\hoch{\bs{\gamma}}\covPhi{\bs{\gamma}}-F_{ab}^{(D)}=\\
 & = & -\frac{1}{16}\bigl(\gemnabla_{\hat{\bs{\gamma}}}\RR^{\bs{\gamma}\hat{\bs{\delta}}}+8\hatcovPhi{\hat{\bs{\gamma}}}\RR^{\bs{\gamma}\hat{\bs{\delta}}}\bigr)\tilde{\gamma}_{ab\hat{\bs{\delta}}}\hoch{\hat{\bs{\gamma}}}\covPhi{\bs{\gamma}}-F_{ab}^{(D)}\end{eqnarray}
\begin{equation}
\boxed{F_{ab}=-\frac{1}{16}\bigl(\gemnabla_{\hat{\bs{\gamma}}}\RR^{\bs{\gamma}\hat{\bs{\delta}}}+8\hatcovPhi{\hat{\bs{\gamma}}}\RR^{\bs{\gamma}\hat{\bs{\delta}}}\bigr)\tilde{\gamma}_{ab\hat{\bs{\delta}}}\hoch{\hat{\bs{\gamma}}}\covPhi{\bs{\gamma}}}\label{eq:D:Fab}\end{equation}
\begin{equation}
\boxed{\hat{F}_{ab}=-\frac{1}{16}\bigl(\gemnabla_{\bs{\gamma}}\RR^{\bs{\delta}\hat{\bs{\gamma}}}+8\covPhi{\bs{\gamma}}\RR^{\bs{\delta}\hat{\bs{\gamma}}}\bigr)\tilde{\gamma}_{ab\bs{\delta}}\hoch{\bs{\gamma}}\hatcovPhi{\hat{\bs{\gamma}}}}\label{eq:D:Fab-hat}\end{equation}
$\bullet\quad$\underbar{$(a,\bs{\beta})\leftrightarrow(a,\hat{\bs{\beta}}):$}\begin{eqnarray}
\underbrace{\gemnabla_{[a}\checkcovPhi{\bs{\beta}]}}_{0\,\mbox{for }\check{\Omega}=\hat{\Omega}} & = & -\check{T}_{a\bs{\beta}}\hoch{c}\checkcovPhi{c}-T_{a\bs{\beta}}\hoch{\bs{\gamma}}\checkcovPhi{\bs{\gamma}}-\hat{T}_{a\bs{\beta}}\hoch{\hat{\bs{\gamma}}}\checkcovPhi{\hat{\bs{\gamma}}}-\check{F}_{a\bs{\beta}}^{(D)}=\\
 & \stackrel{\check{\Omega}=\hat{\Omega}}{=} & -\hat{T}_{a\bs{\beta}}\hoch{\hat{\bs{\gamma}}}\hatcovPhi{\hat{\bs{\gamma}}}-\hat{F}_{a\bs{\beta}}^{(D)}=\\
 & = & \tilde{\gamma}_{a\,\bs{\beta\delta}}\RR^{\bs{\delta}\hat{\bs{\gamma}}}\hatcovPhi{\hat{\bs{\gamma}}}-\hat{F}_{a\bs{\beta}}^{(D)}\end{eqnarray}
\begin{equation}
\boxed{\hat{F}_{a\bs{\beta}}^{(D)}=\tilde{\gamma}_{a\,\bs{\beta\delta}}\RR^{\bs{\delta}\hat{\bs{\gamma}}}\hatcovPhi{\hat{\bs{\gamma}}},\qquad F_{a\hat{\bs{\beta}}}^{(D)}=\tilde{\gamma}_{a\,\hat{\bs{\beta}}\hat{\bs{\delta}}}\RR^{\bs{\gamma}\hat{\bs{\delta}}}\covPhi{\bs{\gamma}}}\label{eq:D:Fhatabeta-and-hat}\end{equation}
For $\hat{\Omega}=\Omega$ instead, we obtain \begin{eqnarray}
\underbrace{\gemnabla_{[a}\checkcovPhi{\bs{\beta}]}}_{\frac{1}{2}\nabla_{a}\covPhi{\bs{\beta}}\,\mbox{for }\check{\Omega}=\Omega} & \stackrel{\check{\Omega}=\Omega}{=} & -F_{a\bs{\beta}}^{(D)}\end{eqnarray}
\begin{equation}
\boxed{F_{a\bs{\beta}}^{(D)}=\frac{1}{2}\nabla_{a}\covPhi{\bs{\beta}},\quad\hat{F}_{a\hat{\bs{\beta}}}^{(D)}=\frac{1}{2}\hat{\nabla}_{a}\hatcovPhi{\hat{\bs{\beta}}}}\label{eq:D:Fabeta-and-hat}\end{equation}
$\bullet\quad$\underbar{$(\bs{\alpha},\bs{\beta})\leftrightarrow(\hat{\bs{\alpha}},\hat{\bs{\beta}}):$}\begin{eqnarray}
\underbrace{\gemnabla_{[\bs{\alpha}}\checkcovPhi{\bs{\beta}]}}_{0\mbox{ for }\check{\Omega}=\hat{\Omega}} & = & -\check{T}_{\bs{\alpha\beta}}\hoch{c}\checkcovPhi{c}-T_{\bs{\alpha\beta}}\hoch{\bs{\gamma}}\checkcovPhi{\bs{\gamma}}-\hat{T}_{\bs{\alpha\beta}}\hoch{\hat{\bs{\gamma}}}\checkcovPhi{\hat{\bs{\gamma}}}-\check{F}_{\bs{\alpha\beta}}^{(D)}=\\
 & \stackrel{\check{\Omega}=\hat{\Omega}}{=} & -\hat{F}_{\bs{\alpha\beta}}^{(D)}\end{eqnarray}
\begin{equation}
\boxed{\hat{F}_{\bs{\alpha\beta}}=F_{\hat{\bs{\alpha}}\hat{\bs{\beta}}}=0}\label{eq:D:Fhatalphabeta-and-hat}\end{equation}
$\bullet\quad$\underbar{$(\bs{\alpha},\hat{\bs{\beta}}):$}\begin{eqnarray}
\underbrace{\gemnabla_{[\bs{\alpha}}\checkcovPhi{\hat{\bs{\beta}}]}}_{\frac{1}{2}\hat{\nabla}_{\bs{\alpha}}\hatcovPhi{\hat{\bs{\beta}}}\mbox{ for }\check{\Omega}=\hat{\Omega}} & = & -\check{T}_{\bs{\alpha}\hat{\bs{\beta}}}\hoch{c}\checkcovPhi{c}-T_{\bs{\alpha}\hat{\bs{\beta}}}\hoch{\bs{\gamma}}\checkcovPhi{\bs{\gamma}}-\hat{T}_{\bs{\alpha}\hat{\bs{\beta}}}\hoch{\hat{\bs{\gamma}}}\checkcovPhi{\hat{\bs{\gamma}}}-\check{F}_{\bs{\alpha}\hat{\bs{\beta}}}^{(D)}=\\
 & \stackrel{\check{\Omega}=\hat{\Omega}}{=} & -\hat{F}_{\bs{\alpha}\hat{\bs{\beta}}}^{(D)}\end{eqnarray}
\begin{equation}
\boxed{\hat{F}_{\bs{\alpha}\hat{\bs{\beta}}}^{(D)}=-\frac{1}{2}\hat{\nabla}_{\bs{\alpha}}\hatcovPhi{\hat{\bs{\beta}}},\quad F_{\hat{\bs{\alpha}}\bs{\beta}}^{(D)}=-\frac{1}{2}\nabla_{\hat{\bs{\alpha}}}\covPhi{\bs{\beta}}}\label{eq:D:Fhatalphahbeta-and-hat}\end{equation}
\rem{

\subsection{Bianchi identities for the field strength}

The curvature Bianchi identity (\ref{eq:BI-R-main}) is in its complete
form hard to analyse, because one has to distinguish so many cases.
That was the reason for replacing it by the somewhat simpler (\ref{eq:replaceBI-R-main})
via Dragon's theorem. However, in contrast to the torsion Bianchi
identity (\ref{eq:BI-T-main}), the one for the curvature decays into
the identity for Lorentz and scale part. The scale part of the curvature
Bianchi identity is much easier treatable than the full one:\begin{eqnarray}
\gemnabla_{\bs{A}}\check{F}_{\bs{AA}}+2\gemT_{\bs{AA}}\hoch{D}\check{F}_{D\bs{A}} & \stackrel{!}{=} & 0\end{eqnarray}
\rem{Strictly speaking, we have this equation for each block, i.e.
for $\check{F}$, $F$ and $\hat{F}$!\\
}$\bullet\quad$\underbar{$\bs{\alpha\alpha\alpha}:$}\begin{eqnarray}
0 & \stackrel{!}{=} & \nabla_{\bs{\alpha}}\check{F}_{\bs{\alpha\alpha}}+2\gamma_{\bs{\alpha\alpha}}\hoch{c}\check{F}_{c\bs{\alpha}}=\\
 & \stackrel{\check{\Omega}=\hat{\Omega}}{=} & 2\gamma_{\bs{\alpha\alpha}}\hoch{c}\hat{F}_{c\bs{\alpha}}\end{eqnarray}
This is nothing new, because it follows from (\ref{eq:D:Fhatabeta-and-hat}).
If we set $\check{\Omega}=\Omega$ instead, we get\begin{eqnarray}
0 & \stackrel{\check{\Omega}=\Omega}{=} & \nabla_{\bs{\alpha}}F_{\bs{\alpha\alpha}}+2\gamma_{\bs{\alpha\alpha}}\hoch{c}F_{c\bs{\alpha}}\end{eqnarray}
\begin{equation}
\boxed{\nabla_{[\bs{\alpha}}F_{\bs{\beta\gamma}]}=-\gamma_{[\bs{\alpha\beta}|}\hoch{c}\nabla_{c}\covPhi{|\bs{\gamma}]},\quad\hat{\nabla}_{[\hat{\bs{\alpha}}}\hat{F}_{\hat{\bs{\beta}}\hat{\bs{\gamma}}]}=-\gamma_{[\hat{\bs{\alpha}}\hat{\bs{\beta}}|}\hoch{c}\hat{\nabla}_{c}\hatcovPhi{|\hat{\bs{\gamma}}]}}\end{equation}
\rem{This is consistent with $F_{\bs{\beta\gamma}}=\gamma_{\bs{\beta\gamma}}^{c}\nabla_{c}\dil$:\begin{eqnarray}
\nabla_{[\bs{\alpha}}F_{\bs{\beta\gamma}]} & = & \gamma_{[\bs{\beta\gamma}}^{c}\nabla_{\bs{\alpha}]}\nabla_{c}\dil=\nonumber \\
 & = & \gamma_{[\bs{\beta\gamma}|}^{c}\nabla_{c}\nabla_{|\bs{\alpha}]}\dil+2\gamma_{[\bs{\beta\gamma}|}^{c}\bei{\gemT_{c|\bs{\alpha}]}\hoch{D}}{\check{\Omega}=\Omega}\nabla_{D}\dil=\nonumber \\
 & = & \gamma_{[\bs{\beta\gamma}|}^{c}\nabla_{c}\nabla_{|\bs{\alpha}]}\dil+\gamma_{[\bs{\beta\gamma}|}^{c}\left(\covPhi{\bs{\alpha}]}\delta_{c}^{d}+\gamma_{c}\hoch{d}\tief{|\bs{\alpha}]}\hoch{\bs{\gamma}}\covPhi{\bs{\gamma}}\right)\nabla_{d}\dil+\nonumber \\
 &  & -2\underbrace{\gamma_{[\bs{\beta\gamma}|}^{c}\tilde{\gamma}_{c\,|\bs{\alpha}]\bs{\delta}}}_{0}\RR^{\bs{\delta}\hat{\bs{\delta}}}\nabla_{\hat{\bs{\delta}}}\dil=\nonumber \\
 & = & \gamma_{[\bs{\beta\gamma}|}^{c}\nabla_{c}\nabla_{|\bs{\alpha}]}\dil+\gamma_{[\bs{\beta\gamma}|}^{c}\left(\covPhi{\bs{\alpha}]}\delta_{c}^{d}-\delta_{c}\hoch{d}\delta\tief{|\bs{\alpha}]}\hoch{\bs{\gamma}}\covPhi{\bs{\gamma}}\right)\nabla_{d}\dil\quad\surd\end{eqnarray}
\\
}$\bullet\quad$$\bs{\alpha\alpha}\hat{\bs{\alpha}}$\begin{eqnarray}
0 & \stackrel{!}{=} & \gemnabla_{[\bs{\alpha}}\check{F}_{\bs{\beta}\hat{\bs{\gamma}}]}+2\gemT_{[\bs{\alpha\beta}|}\hoch{D}\check{F}_{D|\hat{\bs{\gamma}}]}=\\
 & = & \frac{2}{3}\gemnabla_{[\bs{\alpha}}\check{F}_{\bs{\beta}]\hat{\bs{\gamma}}}+\frac{1}{3}\gemnabla_{\hat{\bs{\gamma}}}\check{F}_{\bs{\alpha}\bs{\beta}}+\frac{2}{3}\gamma_{\bs{\alpha\beta}}\hoch{d}\check{F}_{d\hat{\bs{\gamma}}}=\\
\mbox{either} & \stackrel{\check{\Omega}=\hat{\Omega}}{=} & \left(-\frac{1}{3}\gemnabla_{[\bs{\alpha}}\hat{\nabla}_{\bs{\beta}]}+\frac{1}{3}\gamma_{\bs{\alpha\beta}}\hoch{d}\hat{\nabla}_{d}\right)\hatcovPhi{\hat{\bs{\gamma}}}\\
\dann\hat{R}_{\bs{\alpha\beta}\hat{\bs{\gamma}}}\hoch{\hat{\bs{\delta}}}\hatcovPhi{\hat{\bs{\delta}}} & = & 0\quad\\
\mbox{or} & \stackrel{\check{\Omega}=\Omega}{=} & \frac{2}{3}\gemnabla_{[\bs{\alpha}|}\nabla_{\hat{\bs{\gamma}}}\covPhi{|\bs{\beta}]}+\frac{1}{3}\nabla_{\hat{\bs{\gamma}}}F_{\bs{\alpha}\bs{\beta}}+\frac{2}{3}\gamma_{\bs{\alpha\beta}}\hoch{d}\tilde{\gamma}_{d\,\hat{\bs{\gamma}}\hat{\bs{\delta}}}\RR^{\bs{\gamma}\hat{\bs{\delta}}}\covPhi{\bs{\gamma}}\end{eqnarray}
\begin{eqnarray}
0 & \stackrel{!}{=} & \gemnabla_{[a}\check{F}_{bc]}+2\gemT_{[ab|}\hoch{d}\check{F}_{d|c]}=\\
 & \stackrel{\check{\Omega}=\Omega}{=} & \gemnabla_{[a}F_{bc]}+3H_{[ab|}\hoch{d}F_{d|c]}\end{eqnarray}
\begin{eqnarray}
0 & \stackrel{!}{=} & \frac{1}{3}\gemnabla_{a}\check{F}_{\bs{\beta\gamma}}+\frac{2}{3}\gemnabla_{[\bs{\beta}}\check{F}_{\bs{\gamma}]a}+\frac{2}{3}\gemT_{[\bs{\beta\gamma}|}\hoch{D}\check{F}_{D|a]}+\frac{4}{3}\gemT_{a[\bs{\beta}|}\hoch{D}\check{F}_{D|\bs{\gamma}]}=\\
 & \stackrel{\check{\Omega}=\hat{\Omega}}{=} & -\frac{2}{3}\gemnabla_{[\bs{\beta}}\tilde{\gamma}_{a|\bs{\gamma}]\bs{\delta}}\RR^{\bs{\delta}\hat{\bs{\delta}}}\hatcovPhi{\hat{\bs{\delta}}}+\frac{2}{3}\gamma_{[\bs{\beta\gamma}|}\hoch{d}\hat{F}_{d|a]}-\frac{4}{3}\tilde{\gamma}_{a[\bs{\beta}|\bs{\delta}}\RR\hoch{\bs{\delta}\hat{\bs{\delta}}}\hat{F}_{\hat{\bs{\delta}}|\bs{\gamma}]}\end{eqnarray}
$\bullet\quad$\underbar{$(a,a,a):$}\begin{eqnarray}
0 & \stackrel{!}{=} & \check{\nabla}_{\bs{a}}\check{F}_{\bs{aa}}+2\check{T}_{\bs{aa}}\hoch{d}\check{F}_{d\bs{a}}+2T_{\bs{aa}}\hoch{\bs{\delta}}\check{F}_{\bs{\delta}\bs{a}}+2\hat{T}_{\bs{aa}}\hoch{\hat{\bs{\delta}}}\check{F}_{\bs{\delta}\bs{a}}\end{eqnarray}
$\bullet\quad$\underbar{$(\bs{\alpha},a,a):$}\begin{eqnarray}
0 & \stackrel{!}{=} & \frac{1}{3}\check{\nabla}_{\bs{\alpha}}\check{F}_{\bs{aa}}+\frac{2}{3}\check{\nabla}_{\bs{a}}\check{F}_{\bs{a\alpha}}+\frac{2}{3}\check{T}_{\bs{aa}}\hoch{d}\check{F}_{d\bs{\alpha}}+\frac{2}{3}T_{\bs{aa}}\hoch{\bs{\delta}}\check{F}_{\bs{\delta}\bs{\alpha}}+\frac{2}{3}\hat{T}_{\bs{aa}}\hoch{\hat{\bs{\delta}}}\check{F}_{\hat{\bs{\delta}}\bs{\alpha}}+\frac{4}{3}\hat{T}_{\bs{\alpha a}}\hoch{\hat{\bs{\delta}}}\check{F}_{\hat{\bs{\delta}}\bs{a}}\end{eqnarray}

\begin{eqnarray}
0 & \stackrel{!}{=} & \gemnabla_{\bs{A}}\check{F}_{\bs{AA}}+2\gemT_{\bs{AA}}\hoch{D}\check{F}_{D\bs{A}}\end{eqnarray}
}\rem{\addtocounter{localapp}{1}

\section{Dragon's identity}

We have already analysed the Bianchi identities for the $H$-field
and for the torsion. Due to Dragon, the curvature Bianchi identity
(\ref{eq:BI-R-main}) can be replaced by the somewhat weaker (\ref{eq:replaceBI-R-main}):\begin{eqnarray}
0 & = & -\gemR_{\bs{CC}B}\hoch{A}\gemT_{\bs{CC}}\hoch{B}+\gemnabla_{\bs{C}}\gemnabla_{\bs{C}}\gemT_{\bs{CC}}\hoch{A}+\gemT_{\bs{CC}}\hoch{B}\gemnabla_{B}\gemT_{\bs{CC}}\hoch{A}+2\left(\gemnabla_{\bs{C}}\gemT_{\bs{CC}}\hoch{D}+2\gemT_{\bs{CC}}\hoch{B}\gemT_{B\bs{C}}\hoch{D}\right)\gemT_{D\bs{C}}\hoch{A}\qquad\end{eqnarray}
We are not going through all cases but just through some selected
examples:\\
$\bullet\quad$\underbar{$(_{\bs{\gamma\gamma\gamma\gamma}}\hoch{a})\&(_{\hat{\bs{\gamma}}\hat{\bs{\gamma}}\hat{\bs{\gamma}}\hat{\bs{\gamma}}}\hoch{a}):$}\begin{eqnarray}
0 & = & -\gemR_{\bs{\gamma\gamma}b}\hoch{a}\underbrace{\gemT_{\bs{\gamma\gamma}}\hoch{b}}_{\gamma_{\bs{\gamma\gamma}}^{b}}+\gemnabla_{\bs{\gamma}}\gemnabla_{\bs{\gamma}}\underbrace{\gemT_{\bs{\gamma\gamma}}\hoch{a}}_{\gamma_{\bs{\gamma\gamma}}^{a}}+\gemT_{\bs{\gamma\gamma}}\hoch{B}\gemnabla_{B}\underbrace{\gemT_{\bs{\gamma\gamma}}\hoch{a}}_{\gamma_{\bs{\gamma\gamma}}^{a}}+2\Bigl(\gemnabla_{\bs{\gamma}}\gemT_{\bs{\gamma\gamma}}\hoch{D}+2\gemT_{\bs{\gamma\gamma}}\hoch{B}\gemT_{B\bs{\gamma}}\hoch{D}\Bigr)\gemT_{D\bs{\gamma}}\hoch{a}=\\
 & \stackrel{\check{\Omega}=\Omega}{=} & \left(-R_{\bs{\gamma\gamma}b}\hoch{a}+4T_{\bs{\gamma}b}\hoch{d}T_{\bs{\gamma}d}\hoch{a}\right)\gamma_{\bs{\gamma\gamma}}^{b}=\\
 & = & \left(-R_{\bs{\gamma\gamma}b}\hoch{a}+\left(\covPhi{\bs{\gamma}}\delta_{b}^{d}+\gamma_{b}\hoch{d}\tief{\bs{\gamma}}\hoch{\bs{\delta}}\covPhi{\bs{\delta}}\right)\left(\covPhi{\bs{\gamma}}\delta_{d}^{a}+\gamma_{d}\hoch{a}\tief{\bs{\gamma}}\hoch{\bs{\eps}}\covPhi{\bs{\eps}}\right)\right)\gamma_{\bs{\gamma\gamma}}^{b}=\\
 & = & \Bigl(-R_{\bs{\gamma\gamma}b}\hoch{a}+\underbrace{\gamma_{b}\hoch{d}\tief{\bs{\gamma}}\hoch{\bs{\delta}}}_{\gamma_{b\,\bs{\eps\gamma}}\gamma^{d\,\bs{\eps\delta}}-\delta_{b}^{d}\delta_{\bs{\gamma}}\hoch{\bs{\delta}}}\covPhi{\bs{\delta}}\gamma_{d}\hoch{a}\tief{\bs{\gamma}}\hoch{\bs{\eps}}\covPhi{\bs{\eps}}\Bigr)\gamma_{\bs{\gamma\gamma}}^{b}=\\
 & \stackrel{\gamma_{b\,\bs{\eps\gamma}}\gamma_{\bs{\gamma\gamma}}^{b}=0}{=} & \Bigl(-R_{\bs{\gamma\gamma}b}\hoch{a}-\covPhi{\bs{\gamma}}\underbrace{\gamma_{b}\hoch{a}\tief{\bs{\gamma}}\hoch{\bs{\eps}}}_{\gamma_{b\,\bs{\delta\gamma}}\gamma^{a\,\bs{\delta\eps}}-\delta_{b}^{a}\delta_{\bs{\gamma}}\hoch{\bs{\eps}}}\covPhi{\bs{\eps}}\Bigr)\gamma_{\bs{\gamma\gamma}}^{b}=\\
 & = & \Bigl(-R_{\bs{\gamma\gamma}b}\hoch{a}+\covPhi{\bs{\gamma}}\covPhi{\bs{\gamma}}\delta_{b}^{a}\Bigr)\gamma_{\bs{\gamma\gamma}}^{b}\end{eqnarray}
\begin{equation}
\boxed{R_{\bs{\gamma\gamma}b}\hoch{a}\gamma_{\bs{\gamma\gamma}}^{b}=0,\quad\hat{R}_{\hat{\bs{\gamma}}\hat{\bs{\gamma}}b}\hoch{a}\gamma_{\hat{\bs{\gamma}}\hat{\bs{\gamma}}}^{b}=0}\end{equation}
$\bullet\quad$\underbar{$(_{\bs{\gamma\gamma\gamma\gamma}}\hoch{\bs{\alpha}})\&(_{\hat{\bs{\gamma}}\hat{\bs{\gamma}}\hat{\bs{\gamma}}\hat{\bs{\gamma}}}\hoch{\hat{\bs{\alpha}}}):$}\begin{eqnarray}
0 & = & -\gemR_{\bs{\gamma\gamma\beta}}\hoch{\bs{\alpha}}\underbrace{\gemT_{\bs{\gamma\gamma}}\hoch{\bs{\beta}}}_{=0}+\gemnabla_{\bs{\gamma}}\gemnabla_{\bs{\gamma}}\underbrace{\gemT_{\bs{\gamma\gamma}}\hoch{\bs{\alpha}}}_{=0}+\gemT_{\bs{\gamma\gamma}}\hoch{B}\gemnabla_{B}\underbrace{\gemT_{\bs{\gamma\gamma}}\hoch{\bs{\alpha}}}_{=0}+2\Bigl(\gemnabla_{\bs{\gamma}}\gemT_{\bs{\gamma\gamma}}\hoch{D}+2\gemT_{\bs{\gamma\gamma}}\hoch{B}\gemT_{B\bs{\gamma}}\hoch{D}\Bigr)\underbrace{\gemT_{D\bs{\gamma}}\hoch{\bs{\alpha}}}_{=0}=\\
 & = & 0\end{eqnarray}
$\bullet\quad$\underbar{$(_{\bs{\gamma\gamma\gamma}\hat{\bs{\gamma}}}\hoch{\bs{\alpha}})\&(_{\hat{\bs{\gamma}}\hat{\bs{\gamma}}\hat{\bs{\gamma}}\bs{\gamma}}\hoch{\hat{\bs{\alpha}}}):$}\begin{eqnarray}
0 & = & -\gemR_{[\bs{\gamma\gamma}|\bs{\beta}}\hoch{\bs{\alpha}}\underbrace{\gemT_{|\bs{\gamma}\hat{\bs{\gamma}}]}\hoch{\bs{\beta}}}_{=0}+\gemnabla_{[\bs{\gamma}}\gemnabla_{\bs{\gamma}}\underbrace{\gemT_{\bs{\gamma}\hat{\bs{\gamma}}]}\hoch{\bs{\alpha}}}_{=0}+\gemT_{[\bs{\gamma\gamma}|}\hoch{B}\gemnabla_{B}\underbrace{\gemT_{|\bs{\gamma}\hat{\bs{\gamma}}]}\hoch{\bs{\alpha}}}_{=0}+2\Bigl(\gemnabla_{[\bs{\gamma}}\gemT_{\bs{\gamma\gamma}|}\hoch{D}+2\gemT_{[\bs{\gamma\gamma}|}\hoch{B}\gemT_{B|\bs{\gamma}|}\hoch{D}\Bigr)\gemT_{D|\hat{\bs{\gamma}}]}\hoch{\bs{\alpha}}=\\
 & = & \frac{1}{2}\Bigl(\gemnabla_{\bs{\gamma}}\gamma_{\bs{\gamma\gamma}}^{d}+2\gamma_{\bs{\gamma\gamma}}^{b}\gemT_{b\bs{\gamma}}\hoch{d}\Bigr)\gemT_{d\hat{\bs{\gamma}}}\hoch{\bs{\alpha}}=\\
 & \stackrel{\check{\Omega}=\Omega}{=} & \gamma_{\bs{\gamma\gamma}}^{b}\gemT_{b\bs{\gamma}}\hoch{d}\gemT_{d\hat{\bs{\gamma}}}\hoch{\bs{\alpha}}=\\
 & = & -\frac{1}{2}\gamma_{\bs{\gamma\gamma}}^{b}\Bigl(\covPhi{\bs{\gamma}}\delta_{b}^{d}+\underbrace{\gamma_{b}\hoch{d}\tief{\bs{\gamma}}\hoch{\bs{\delta}}}_{\gamma_{b\,\bs{\gamma}\bs{\eps}}\gamma^{d\,\bs{\delta}\bs{\eps}}-\delta_{b}^{d}\delta_{\bs{\gamma}}\hoch{\bs{\delta}}}\covPhi{\bs{\delta}}\Bigr)\tilde{\gamma}_{d\,\hat{\bs{\gamma}}\hat{\bs{\delta}}}\RR^{\bs{\alpha}\hat{\bs{\delta}}}=\\
 & \stackrel{\gamma_{\bs{\gamma\gamma}}^{b}\gamma_{b\,\bs{\gamma}\bs{\eps}}=0}{=} & 0\end{eqnarray}
$\bullet\quad$\underbar{$(_{\bs{\gamma\gamma}\hat{\bs{\gamma}}\hat{\bs{\gamma}}}\hoch{\bs{\alpha}})$\&$(_{\hat{\bs{\gamma}}\hat{\bs{\gamma}}\bs{\gamma\gamma}}\hoch{\hat{\bs{\alpha}}}):$}\begin{eqnarray}
0 & = & -\gemR_{[\bs{\gamma\gamma}|\bs{\beta}}\hoch{\bs{\alpha}}\underbrace{\gemT_{|\hat{\bs{\gamma}}\hat{\bs{\gamma}}]}\hoch{\bs{\beta}}}_{=0}+\gemnabla_{[\bs{\gamma}}\gemnabla_{\bs{\gamma}}\underbrace{\gemT_{\hat{\bs{\gamma}}\hat{\bs{\gamma}}]}\hoch{\bs{\alpha}}}_{=0}+\gemT_{[\bs{\gamma\gamma}|}\hoch{B}\gemnabla_{B}\underbrace{\gemT_{|\hat{\bs{\gamma}}\hat{\bs{\gamma}}]}\hoch{\bs{\alpha}}}_{=0}+2\Bigl(\gemnabla_{[\bs{\gamma}}\gemT_{\bs{\gamma}\hat{\bs{\gamma}}|}\hoch{D}+2\gemT_{[\bs{\gamma\gamma}|}\hoch{B}\gemT_{B|\hat{\bs{\gamma}}|}\hoch{D}\Bigr)\gemT_{D|\hat{\bs{\gamma}}]}\hoch{\bs{\alpha}}=\\
 & \propto & \gemnabla_{\hat{\bs{\gamma}}}\gamma_{\bs{\gamma}\bs{\gamma}}^{d}\gemT_{d\hat{\bs{\gamma}}}\hoch{\bs{\alpha}}+\gemnabla_{\bs{\gamma}}\gamma_{\hat{\bs{\gamma}}\hat{\bs{\gamma}}}^{d}\underbrace{\gemT_{d\bs{\gamma}}\hoch{\bs{\alpha}}}_{0}+2\gamma_{\bs{\gamma\gamma}}\hoch{b}\gemT_{b\hat{\bs{\gamma}}}\hoch{D}\gemT_{D\hat{\bs{\gamma}}}\hoch{\bs{\alpha}}+2\gamma_{\hat{\bs{\gamma}}\hat{\bs{\gamma}}}^{b}\gemT_{b\bs{\gamma}}\hoch{D}\underbrace{\gemT_{D\bs{\gamma}}\hoch{\bs{\alpha}}}_{0}=\\
 & \stackrel{\check{\Omega}=\Omega}{=} & 2\gamma_{\bs{\gamma\gamma}}\hoch{b}\gemT_{b\hat{\bs{\gamma}}}\hoch{d}\gemT_{d\hat{\bs{\gamma}}}\hoch{\bs{\alpha}}=\\
 & \propto & \gamma_{\bs{\gamma\gamma}}\hoch{b}\left(\hatcovPhi{\hat{\bs{\gamma}}}\delta_{b}^{d}+\gamma_{b}\hoch{d}\tief{\hat{\bs{\gamma}}}\hoch{\hat{\bs{\delta}}}\hatcovPhi{\hat{\bs{\delta}}}\right)\tilde{\gamma}_{d\,\hat{\bs{\gamma}}\hat{\bs{\delta}}}\RR^{\bs{\alpha}\hat{\bs{\delta}}}\end{eqnarray}
$\bullet\quad$\underbar{$(_{\bs{\gamma\gamma\gamma}\hat{\bs{\gamma}}}\hoch{\hat{\bs{\alpha}}})$\&$(_{\hat{\bs{\gamma}}\hat{\bs{\gamma}}\hat{\bs{\gamma}}\bs{\gamma}}\hoch{\bs{\alpha}}):$}\begin{eqnarray}
0 & = & -\gemR_{[\bs{\gamma\gamma}|\hat{\bs{\beta}}}\hoch{\hat{\bs{\alpha}}}\underbrace{\gemT_{|\bs{\gamma}\hat{\bs{\gamma}}]}\hoch{\hat{\bs{\beta}}}}_{=0}+\gemnabla_{[\bs{\gamma}}\gemnabla_{\bs{\gamma}}\underbrace{\gemT_{\bs{\gamma}\hat{\bs{\gamma}}]}\hoch{\hat{\bs{\alpha}}}}_{=0}+\gemT_{[\bs{\gamma\gamma}|}\hoch{B}\gemnabla_{B}\underbrace{\gemT_{|\bs{\gamma}\hat{\bs{\gamma}}]}\hoch{\hat{\bs{\alpha}}}}_{=0}+2\Bigl(\gemnabla_{[\bs{\gamma}}\gemT_{\bs{\gamma\gamma}|}\hoch{D}+2\gemT_{[\bs{\gamma\gamma}|}\hoch{B}\gemT_{B|\bs{\gamma}|}\hoch{D}\Bigr)\gemT_{D|\hat{\bs{\gamma}}]}\hoch{\hat{\bs{\alpha}}}=\\
 & = & -\frac{1}{2}\Bigl(\gemnabla_{\hat{\bs{\gamma}}}\gemT_{\bs{\gamma}\bs{\gamma}}\hoch{D}+2\gemT_{\bs{\gamma\gamma}}\hoch{B}\gemT_{B\hat{\bs{\gamma}}}\hoch{D}\Bigr)\gemT_{D\bs{\gamma}}\hoch{\hat{\bs{\alpha}}}=\\
 & \stackrel{\check{\Omega}=\Omega}{=} & -\gamma_{\bs{\gamma\gamma}}^{b}\gemT_{b\hat{\bs{\gamma}}}\hoch{d}\gemT_{d\bs{\gamma}}\hoch{\hat{\bs{\alpha}}}=\\
 & = & \frac{1}{2}\gamma_{\bs{\gamma\gamma}}^{b}\Bigl(\hatcovPhi{\hat{\bs{\gamma}}}\delta_{b}^{d}+\gamma_{b}\hoch{d}\tief{\hat{\bs{\gamma}}}\hoch{\hat{\bs{\delta}}}\hatcovPhi{\hat{\bs{\delta}}}\Bigr)\tilde{\gamma}_{d\,\bs{\gamma\delta}}\RR^{\bs{\delta}\hat{\bs{\alpha}}}=\\
 & = & \frac{1}{2}\gamma_{\bs{\gamma\gamma}}^{b}\gamma_{b}\hoch{d}\tief{\hat{\bs{\gamma}}}\hoch{\hat{\bs{\delta}}}\hatcovPhi{\hat{\bs{\delta}}}\tilde{\gamma}_{d\,\bs{\gamma\delta}}\RR^{\bs{\delta}\hat{\bs{\alpha}}}\end{eqnarray}
\begin{equation}
\boxed{\gamma_{\bs{\gamma\gamma}}^{b}\gamma_{\bs{\gamma\delta}}^{d}\gamma_{bd}\tief{\hat{\bs{\gamma}}}\hoch{\hat{\bs{\delta}}}\hatcovPhi{\hat{\bs{\delta}}}\RR^{\bs{\delta}\hat{\bs{\alpha}}}=0}\end{equation}
\begin{equation}
\boxed{\gamma_{\hat{\bs{\gamma}}\hat{\bs{\gamma}}}^{b}\gamma_{\hat{\bs{\gamma}}\hat{\bs{\delta}}}^{d}\gamma_{bd}\tief{\bs{\gamma}}\hoch{\bs{\delta}}\covPhi{\bs{\delta}}\RR^{\bs{\alpha}\hat{\bs{\delta}}}=0}\end{equation}
\underbar{$\bullet\quad$$(_{\bs{\gamma\gamma\gamma\gamma}}\hoch{\hat{\bs{\alpha}}})$\&$(_{\hat{\bs{\gamma}}\hat{\bs{\gamma}}\hat{\bs{\gamma}}\hat{\bs{\gamma}}}\hoch{\bs{\alpha}}):$}\begin{eqnarray}
0 & = & -\gemR_{\bs{\gamma\gamma}\hat{\bs{\beta}}}\hoch{\hat{\bs{\alpha}}}\underbrace{\gemT_{\bs{\gamma\gamma}}\hoch{\hat{\bs{\beta}}}}_{=0}+\gemnabla_{\bs{\gamma}}\gemnabla_{\bs{\gamma}}\underbrace{\gemT_{\bs{\gamma\gamma}}\hoch{\hat{\bs{\alpha}}}}_{=0}+\gemT_{\bs{\gamma\gamma}}\hoch{B}\gemnabla_{B}\underbrace{\gemT_{\bs{\gamma\gamma}}\hoch{\hat{\bs{\alpha}}}}_{=0}+2\Bigl(\gemnabla_{\bs{\gamma}}\gemT_{\bs{\gamma\gamma}}\hoch{D}+2\gemT_{\bs{\gamma\gamma}}\hoch{B}\gemT_{B\bs{\gamma}}\hoch{D}\Bigr)\gemT_{D\bs{\gamma}}\hoch{\hat{\bs{\alpha}}}=\\
 & = & 2\Bigl(\gemnabla_{\bs{\gamma}}\gamma_{\bs{\gamma\gamma}}\hoch{d}+2\gamma_{\bs{\gamma\gamma}}\hoch{b}\gemT_{b\bs{\gamma}}\hoch{d}\Bigr)\gemT_{d\bs{\gamma}}\hoch{\hat{\bs{\alpha}}}=\\
 & \stackrel{\check{\Omega}=\Omega}{=} & -2\gamma_{\bs{\gamma\gamma}}\hoch{b}\Bigl(\covPhi{\bs{\gamma}}\delta_{b}^{d}+\underbrace{\gamma_{b}\hoch{d}\tief{\bs{\gamma}}\hoch{\bs{\delta}}}_{\gamma_{b\,\bs{\eps\gamma}}\gamma^{d\,\bs{\eps\delta}}-\delta_{b}^{d}\delta_{\bs{\gamma}}\hoch{\bs{\delta}}}\covPhi{\bs{\delta}}\Bigr)\tilde{\gamma}_{d\,\bs{\gamma\delta}}\RR^{\bs{\delta}\hat{\bs{\alpha}}}=\\
 & = & 0\end{eqnarray}
\begin{eqnarray}
0 & = & -\gemR_{\bs{CC}B}\hoch{A}\gemT_{\bs{CC}}\hoch{B}+\gemnabla_{\bs{C}}\gemnabla_{\bs{C}}\gemT_{\bs{CC}}\hoch{A}+\gemT_{\bs{CC}}\hoch{B}\gemnabla_{B}\gemT_{\bs{CC}}\hoch{A}+2\left(\gemnabla_{\bs{C}}\gemT_{\bs{CC}}\hoch{D}+2\gemT_{\bs{CC}}\hoch{B}\gemT_{B\bs{C}}\hoch{D}\right)\gemT_{D\bs{C}}\hoch{A}\end{eqnarray}

}\rem{\newpage\addtocounter{localapp}{1}

\section{Conjugate momenta}

We have done all of the calculations in the Lagrangian formalism.
Nevertheless, knowing about the form of the conjugate momenta can
be a valuable additional check. Let us therefore give their form here
for completeness. To maintain the manifest chiral split, we define
two conjugate momenta, which correspond to choosing $\sigma^{+}$
or $\sigma^{-}$ as canonical time respectively%
\footnote{The relation to the {}``ordinary'' momentum, as given in \cite[p.14]{Berkovits:2001ue}
is the following\begin{eqnarray*}
p_{K}\equiv p_{zK}+p_{\bar{z}K} & = & \frac{1}{2}\left(\partial x^{M}+\bar{\partial}x^{M}\right)G_{MK}+\frac{1}{2}\left(\partial x^{M}-\bar{\partial}x^{M}\right)B_{MK}+E_{K}\hoch{\bs{\alpha}}\dP_{z\bs{\alpha}}+E_{K}\hoch{\hat{\bs{\alpha}}}\hat{\dP}_{\bar{z}\hat{\bs{\alpha}}}+\ce^{\bs{\alpha}}\Omega_{K\bs{\alpha}}\hoch{\bs{\beta}}(\xfull)\be_{z\bs{\beta}}+\hat{\ce}^{\hat{\bs{\alpha}}}\hat{\Omega}_{K\hat{\bs{\alpha}}}\hoch{\hat{\bs{\beta}}}(\xfull)\hat{\be}_{\bar{z}\hat{\bs{\beta}}}\end{eqnarray*}
This implies \begin{eqnarray*}
\dP_{z\bs{\gamma}} & = & E_{\bs{\gamma}}\hoch{K}\left(p_{K}-\frac{1}{2}\left(\partial x^{M}+\bar{\partial}x^{M}\right)G_{MK}-\frac{1}{2}\left(\partial x^{M}-\bar{\partial}x^{M}\right)B_{MK}-\ce^{\bs{\alpha}}\Omega_{K\bs{\alpha}}\hoch{\bs{\beta}}(\xfull)\be_{z\bs{\beta}}-\hat{\ce}^{\hat{\bs{\alpha}}}\hat{\Omega}_{K\hat{\bs{\alpha}}}\hoch{\hat{\bs{\beta}}}(\xfull)\hat{\be}_{\bar{z}\hat{\bs{\beta}}}\right)\qquad\fussend\end{eqnarray*}
} \begin{eqnarray}
p_{z/\bar{z}K} & \equiv & \partl{(\partial_{\bar{z}/z}x^{K})}\mc{L}\end{eqnarray}
\begin{eqnarray}
p_{zK} & = & \frac{1}{2}\partial x^{M}\GB_{MK}+E_{K}\hoch{\bs{\alpha}}\dP_{z\bs{\alpha}}+\ce^{\bs{\alpha}}\Omega_{K\bs{\alpha}}\hoch{\bs{\beta}}(\xfull)\be_{z\bs{\beta}}\\
p_{\bar{z}K} & = & \frac{1}{2}\GB_{KN}\bar{\partial}x^{N}+E_{K}\hoch{\hat{\bs{\alpha}}}\hat{\dP}_{\bar{z}\hat{\bs{\alpha}}}+\hat{\ce}^{\hat{\bs{\alpha}}}\hat{\Omega}_{K\hat{\bs{\alpha}}}\hoch{\hat{\bs{\beta}}}(\xfull)\hat{\be}_{\bar{z}\hat{\bs{\beta}}}\end{eqnarray}
The momenta $p_{zK}$ and $p_{\bar{z}K}$ are related via the symmetry
of (\ref{eq:leftrightSymmetry}). We can thus concentrate on the first
one only. The left moving (we will impose holomorphicity on it later)
BRST operator is given by\begin{eqnarray}
\bs{j}_{z} & = & \ce^{\bs{\alpha}}\left(\dP_{z\bs{\alpha}}+\brstfield{1}_{\bs{\alpha}M}(\xfull)\:\partial_{z}x^{M}+\ce^{\bs{\gamma}}\brstfield{2}_{\bs{\alpha}\bs{\gamma}}\hoch{\bs{\beta}}(\xfull)\be_{z\bs{\beta}}\frem{+\alpha'\brstfield{3}_{\bs{\alpha}}(\xfull)\partial_{z}\weyl}\right)=\\
 & = & \ce^{\bs{\alpha}}E_{\bs{\alpha}}\hoch{K}\left(p_{zK}+\left(-\frac{1}{2}\GB_{MK}+E_{K}\hoch{\bs{\beta}}\brstfield{1}_{\bs{\beta}M}\right)\partial_{z}x^{M}+\ce^{\bs{\gamma}}\left(\brstfield{2}_{K\bs{\gamma}}\hoch{\bs{\beta}}-\Omega_{K\bs{\gamma}}\hoch{\bs{\beta}}\right)\be_{z\bs{\beta}}\frem{+\alpha'\brstfield{3}_{\bs{\alpha}}(\xfull)\partial_{z}\weyl}\right)\end{eqnarray}
}

\section{Recovering flat-space action / comment on linearized SUGRA}

\index{flat superspace!as a solution of the pure spinor string in general background}If
all curvature components vanish, all higher components (in the $\xbothtetas$-expansion)
vanish in the extended WZ-gauge due to (\ref{eq:connectionInTermsOfRII})
and (\ref{eq:bosConnectionInTermsOfRII}) and the remaining bosonic
local Lorentz and scale transformations can be used to fix $\bei{\gemOm_{mA}\hoch{B}}{}=0$
such that all connection components vanish. The only torsion components
which are forced to be nonzero are $T_{\bs{\alpha\beta}}\hoch{c}=(\de E^{c})_{\bs{\alpha\beta}}=\gamma_{\bs{\alpha\beta}}^{c}$
and $T_{\hat{\bs{\alpha}}\hat{\bs{\beta}}}\hoch{c}=(\de E^{c})_{\hat{\bs{\alpha}}\hat{\bs{\beta}}}=\gamma_{\hat{\bs{\alpha}}\hat{\bs{\beta}}}^{c}$.
A solution which is compatible with the extended WZ-gauge (\ref{eq:vielbeinInTermsOfTandRII}),
(\ref{eq:bosVielbeinInTermsOfTandOmegaII}) and (\ref{eq:bosVielbeinInTermsOfTandRII}),
and which fixes also the remaining bosonic diffeomorphism invariance
is given by \begin{eqnarray}
E_{M}\hoch{A} & = & \left(\begin{array}{ccc}
\delta_{m}^{a} & 0 & 0\\
(\theta^{\bs{\beta}}\gamma_{\bs{\beta\mu}}^{a}) & \delta_{\bs{\mu}}\hoch{\bs{\alpha}} & 0\\
(\hat{\theta}^{\hat{\bs{\beta}}}\gamma_{\hat{\bs{\beta}}\hat{\bs{\mu}}}^{a}) & 0 & \delta_{\hat{\bs{\mu}}}\hoch{\hat{\bs{\alpha}}}\end{array}\right),\quad E_{A}\hoch{M}=\left(\begin{array}{ccc}
\delta_{a}^{m} & 0 & 0\\
-(\theta^{\bs{\beta}}\gamma_{\bs{\beta\alpha}}^{m}) & \delta_{\bs{\alpha}}\hoch{\bs{\mu}} & 0\\
-(\hat{\theta}^{\hat{\bs{\beta}}}\gamma_{\hat{\bs{\beta}}\hat{\bs{\alpha}}}^{m}) & 0 & \delta_{\hat{\bs{\alpha}}}\hoch{\hat{\bs{\mu}}}\end{array}\right)\end{eqnarray}
The supersymmetric invariant one -forms thus read \begin{eqnarray}
E^{A} & = & \de x^{M}E_{M}\hoch{A}=\left(\de x^{a}+\de\theta^{\bs{\mu}}\theta^{\bs{\beta}}\gamma_{\bs{\beta\mu}}^{a}+\de\hat{\theta}^{\hat{\bs{\mu}}}\hat{\theta}^{\hat{\bs{\beta}}}\gamma_{\hat{\bs{\beta}}\hat{\bs{\mu}}}^{a}\,,\,\de\theta^{\bs{\alpha}}\,,\,\de\hat{\theta}^{\hat{\bs{\alpha}}}\right)\end{eqnarray}
which agrees with (\ref{eq:vielbeinInFlatSuperspace}). 

The reasoning is similar for the B-field and its field-strength $H$.
The only components of $H$ which are forced to be nonzero are $H_{c\bs{\alpha\beta}}=-\frac{2}{3}\gamma_{c\bs{\alpha\beta}}$
and $H_{c\hat{\bs{\alpha}}\hat{\bs{\beta}}}=\frac{2}{3}\gamma_{c\hat{\bs{\alpha}}\hat{\bs{\beta}}}$.
A simple solution for $H_{CAB}=(\de B)_{CAB}=\nabla_{[C}B_{AB]}+2T_{[CA|}\hoch{D}B_{D|B]}$
(which is compatible with the WZ-like gauge (\ref{eq:BgaugefixI}),
(\ref{eq:BgaugefixII}) has the form \begin{equation}
B_{AB}=\left(\begin{array}{ccc}
0 & x^{\bs{\gamma}}\gamma_{a\,\bs{\gamma}\bs{\beta}} & -x^{\hat{\bs{\gamma}}}\gamma_{a\,\hat{\bs{\gamma}}\hat{\bs{\beta}}}\\
-x^{\bs{\gamma}}\gamma_{b\,\bs{\gamma}\bs{\alpha}} & 0 & -\left(\gamma_{\bs{\alpha}\bs{\gamma}}^{c}x^{\bs{\gamma}}\right)\left(x^{\hat{\bs{\gamma}}}\gamma_{c\,\hat{\bs{\gamma}}\hat{\bs{\beta}}}\right)\\
x^{\hat{\bs{\gamma}}}\gamma_{b\,\hat{\bs{\gamma}}\hat{\bs{\alpha}}} & \left(\gamma_{\bs{\beta}\bs{\gamma}}^{c}x^{\bs{\gamma}}\right)\left(x^{\hat{\bs{\gamma}}}\gamma_{c\,\hat{\bs{\gamma}}\hat{\bs{\alpha}}}\right) & 0\end{array}\right)\end{equation}
All other fields which appear in the Lagrangian can be chosen to vanish.
The curved-background action (\ref{eq:BiBagain}) thus reduces to\begin{eqnarray}
S_{0} & = & \int d^{2}z\quad\frac{1}{2}\Pi_{z}^{a}\eta_{ab}\Pi_{\bar{z}}^{b}+\frac{1}{2}\Pi_{z}^{A}B_{AB}\Pi_{\bar{z}}^{B}+\bar{\partial}\theta^{\bs{\gamma}}\dP_{z\bs{\gamma}}+\partial\hat{\theta}^{\hat{\bs{\gamma}}}\hat{\dP}_{\bar{z}\hat{\bs{\gamma}}}+\nonumber \\
 &  & +\bar{\partial}\ce^{\bs{\beta}}\be_{z\bs{\beta}}+\partial\hat{\ce}^{\hat{\bs{\beta}}}\hat{\be}_{\bar{z}\hat{\bs{\beta}}}+\frac{1}{2}L_{z\bar{z}a}(\ce\gamma^{a}\ce)+\frac{1}{2}\hat{L}_{z\bar{z}a}(\hat{\ce}\gamma^{a}\hat{\ce})\end{eqnarray}
The $B$-field term takes the explicit form \begin{eqnarray}
\frac{1}{2}\Pi_{z}^{A}B_{AB}\Pi_{\bar{z}}^{B} & = & \frac{1}{2}\Pi_{z}^{a}\left(B_{a\bs{\beta}}\Pi_{\bar{z}}^{\bs{\beta}}+B_{a\hat{\bs{\beta}}}\Pi_{\bar{z}}^{\hat{\bs{\beta}}}\right)+\frac{1}{2}\Pi_{z}^{\bs{\alpha}}B_{\bs{\alpha}\hat{\bs{\beta}}}\Pi_{\bar{z}}^{\hat{\bs{\beta}}}-(z\leftrightarrow\bar{z})=\\
 & = & \frac{1}{2}\Pi_{z}^{a}\left(\theta^{\bs{\gamma}}\gamma_{a\,\bs{\gamma}\bs{\beta}}\Pi_{\bar{z}}^{\bs{\beta}}-\theta^{\hat{\bs{\gamma}}}\gamma_{a\,\hat{\bs{\gamma}}\hat{\bs{\beta}}}\Pi_{\bar{z}}^{\hat{\bs{\beta}}}\right)-\frac{1}{2}\left(\Pi_{z}^{\bs{\alpha}}\gamma_{\bs{\alpha}\bs{\gamma}}^{c}\theta^{\bs{\gamma}}\right)\left(\theta^{\hat{\bs{\gamma}}}\gamma_{c\,\hat{\bs{\gamma}}\hat{\bs{\beta}}}\Pi_{\bar{z}}^{\hat{\bs{\beta}}}\right)-(z\leftrightarrow\bar{z})=\\
 & = & \frac{1}{2}\Pi_{z}^{a}\left(\Pi_{\bar{z}}^{\bs{\beta}}\theta^{\bs{\gamma}}\gamma_{a\,\bs{\gamma}\bs{\beta}}-\Pi_{\bar{z}}^{\hat{\bs{\beta}}}\theta^{\hat{\bs{\gamma}}}\gamma_{a\,\hat{\bs{\gamma}}\hat{\bs{\beta}}}\right)-\frac{1}{2}\left(\theta^{\bs{\gamma}}\Pi_{z}^{\bs{\alpha}}\gamma_{\bs{\alpha}\bs{\gamma}}^{c}\right)\left(\Pi_{\bar{z}}^{\hat{\bs{\beta}}}\theta^{\hat{\bs{\gamma}}}\gamma_{c\,\hat{\bs{\gamma}}\hat{\bs{\beta}}}\right)-(z\leftrightarrow\bar{z})\end{eqnarray}
Upon a shift of the grading from the fermionic indices to the rumpfs,
this coincides precisely with the form of the WZ-term given in (\ref{eq:flat:WZ-term}).
Only the antighost field has to be redefined with a minus sign, in
order to match the flat-space Lagrangian.

The BRST\index{BRST!in flat superspace} transformations (\ref{eq:covBRSTofx})-(\ref{eq:covBRSTofL})
reduce in flat space\index{flat superspace!BRST transformations}
to \begin{eqnarray}
\es_{0}x^{m} & = & \ce^{\bs{\alpha}}\gamma_{\bs{\alpha\beta}}^{m}\theta^{\bs{\beta}},\qquad\es_{0}\theta^{\bs{\mu}}=\ce^{\bs{\alpha}}\delta_{\bs{\alpha}}\hoch{\bs{\mu}}\\
\es_{0}\be_{z\bs{\alpha}} & = & \dP_{z\bs{\alpha}}\\
\es_{0}\dP_{z\bs{\delta}} & = & -2\ce^{\bs{\alpha}}\Pi_{z}^{c}\gamma_{c\,\bs{\alpha\delta}}\end{eqnarray}
The corresponding hatted equations are obtained for the hatted fields.
All other transformations vanish. In particular, the Lagrange multiplier
doesn't transform (the complicated $X_{\bs{\alpha\beta}}$ vanishes
in flat space). The pure spinor constraint guarantees the nilpotency
of $\es_{0}$ when acting twice on $\dP_{z\bs{\delta}}$. The BRST
transformation of the supersymmetric objects reduce to \begin{eqnarray}
\es_{0}\Pi_{z}^{a} & = & 2\ce^{\bs{\alpha}}\partial x^{\bs{\beta}}\gamma_{\bs{\alpha\beta}}^{a},\qquad\es_{0}\partial x^{\bs{\alpha}}=\partial\ce^{\bs{\alpha}}\end{eqnarray}
We can see the BRST transformation $\es$ of curved background as
a perturbation around the one in flat background\begin{eqnarray}
\es & \equiv & (\es_{0}+\bs{u})\end{eqnarray}
From the point of view of the string in flat background with action
$S_{0}$, the difference $U\equiv S-S_{0}$ to the action in curved
background is simply a vertex operator which should be BRST-invariant.
The condition of a conserved BRST current (which enforced the supergravity
constraints) corresponds to the invariance $\es S=0$ of the action,
or written as a perturbation: \begin{eqnarray}
0 & = & (\es_{0}+\bs{u})(S_{0}+U)=\\
 & = & \underbrace{\es_{0}S_{0}}_{0}+\bs{u}S_{0}+\es_{0}U+\bs{u}U\end{eqnarray}
At linearized level, we thus have\index{linearized SUGRA}\index{supergravity!linearized}
\begin{equation}
\bs{u}S_{0}+\es_{0}U=0\end{equation}
In the antifield formalism (which we did not really discuss in this
context), the BRST transformations are generated by the actions themselves
(enlarged with an antifield content) via the antibracket. The above
equation then reads\begin{equation}
(U\fatkomma S_{0})+(S_{0}\fatkomma U)=2(S_{0}\fatkomma U)=2\es_{0}U\stackrel{!}{=}0\end{equation}
This explains the well-known fact that the vertex operators of the
flat space pure spinor string have to obey linearized supergravity
constraints.

\printindex{}

\bibliographystyle{fullsort}
\bibliography{Proposal,phd}

}

\part[Derived Brackets in Sigma-Models]{Derived Brackets in Sigma-Models\vspace{3cm}\\ {\large "Don't make a break, make a bracket" (Kathi S.)\index{break!Don't make a $\sim$, make a bracket}\index{Don't make a break, make a bracket}\index{bracket!Don't make a break, make a $\sim$}}}\label{part:Derived-Brackets-in}

{\remch\inputTeil{0}\numberwithin{equation}{chapter}\renewcommand{\be}{{\bs{b}}}\renewcommand{\ce}{{\bs{c}}}\ifthenelse{\theinput=1}{}{\remch }

\title{Brackets, Sigma Models and Integrability of Generalized Complex
Structures}

\author{\textbf{Sebastian Guttenberg}}

\date{Last modified August 08, 2007}

\maketitle
\begin{abstract}
This text is part of the thesis and is not excatly the version which
was published as paper. The abstract is still the same: It is shown
how derived brackets naturally arise in sigma-models via Poisson-
or antibracket\rem{or by the quantum-commutator}, generalizing a
recent observation by Alekseev and Strobl. On the way to a precise
formulation of this relation, an explicit coordinate expression for
the derived bracket is obtained. The generalized Nijenhuis tensor
of generalized complex geometry is shown to coincide up to a de-Rham
closed term with the derived bracket of the structure with itself
and a new coordinate expression for this tensor is presented. The
insight is applied to two known two-dimensional sigma models in a
background with generalized complex structure. Introductions to geometric
brackets on the one hand and to generalized complex geometry on the
other hand are given in the appendix.
\end{abstract}
\rem{To do:\\
Zitate: Michael?\\
 H-twisting\\
Quantum-(J,J)-bracket\\
r-wedge <-> $\ip^{(p,q)}$\\
(pure) Spinors\\
Zucc: missing last condition\\
Hull: foot:dual-coord}\newpage 

\tableofcontents{}\newpage\eref

\section*{Introduction to the Bracket Part}

\addcontentsline{toc}{chapter}{Introduction to the Bracket Part}\setcounter{footnote}{0}\renewcommand{\thefoot}{\twodigit{\value{chapter}}.b\twodigit{\value{footnote}}@\arabic{footnote}}This
part of the thesis is based on the author's paper \cite{Guttenberg:2006zi}.
See also \cite{Guttenberg:2007ha} for a short article which contains
some of the basic ideas. In the meantime a paper by Klaus Bering \cite{Bering:2006eb}
was brought to my attention. Although it follows a different aim,
its geometrical setting, especially in its section 5, is very close
to the one presented here. Moreover, the geometrical meaning of the
variables is nicely presented there, e.g. in its table 7, and can
thus serve as a useful supplement to the reading of the present part
of the thesis.

There are quite a lot of different geometric brackets floating around
in the literature, like Schouten bracket, Nijenhuis bracket or in
generalized complex geometry the Dorfman bracket and Courant bracket,
to list just some of them. They are often related to integrability
conditions for some structures on manifolds. The vanishing of the
Nijenhuis bracket of a complex structure with itself, for example,
is equivalent to its integrability. The same is true for the Schouten
bracket and a Poisson structure. The above brackets can be unified
with the concept of derived brackets \cite{Kosmann-Schwarzbach:2003en}.
Within this concept, they are all just natural extensions of the Lie-bracket
of vector fields to higher rank tensor fields. 

It is well known that the antibracket appearing in the Lagrangian
formalism for sigma models is closely related to the Schouten-bracket
in target space. In addition it was recently observed by Alekseev
and Strobl that the Dorfman bracket for sums of vectors and one-forms
appears naturally in two dimensional sigma models%
\footnote{\index{footnote!\thefoot. Courant and Dorfman bracket}In \cite{Alekseev:2004np},
the non-symmetric bracket is called 'Courant bracket'. Following e.g.
Gualtieri \cite{Gualtieri:007} or \cite{Kosmann-Schwarzbach:2003en},
it will be called 'Dorfman bracket' in this thesis, while 'Courant
bracket' is reserved for its antisymmetrization (see (\ref{eq:Dorfman-bracket})
and (\ref{eq:Courant-bracket})).$\quad\fussend$%
} \cite{Alekseev:2004np}. This was generalized by Bonelli and Zabzine
\cite{Bonelli:2005ti} to a derived bracket for sums of vectors and
$p$-forms on a $p$-brane%
\footnote{\index{footnote!\thefoot. Vinogradov bracket}\index{Vinogradov bracket}\index{bracket!Vinogradov}The
\emph{Vinogradov bracket} appearing in \cite{Bonelli:2005ti} is just
the antisymmetrization of a derived bracket (see footnote\rem{spaeter vielleicht nimmer}
\ref{Vinogradov-bracket} on page \pageref{Vinogradov-bracket}).$\qquad\fussend$%
}. These observations lead to the natural question whether there is
a general relation between the sigma-model Poisson bracket or antibracket
and derived brackets in target space. Working out the precise relation
for sigma models with a special field content but undetermined dimension
and dynamics, is the major subject of the present part of the thesis.

One of the motivations for this part of the thesis was the application
to generalized complex geometry. The importance of the latter in string
theory is due to the observation that effective spacetime supersymmetry
after compactification requires the compactification manifold to be
a generalized Calabi-Yau manifold \cite{Hitchin:2004ut,Gualtieri:007,Grana:2004bg,Grana:2004??,Grana:2005ny,Grana:2005jc}.
Deviations from an ordinary Calabi Yau manifold are due to fluxes
and also the concept of mirror symmetry can be generalized in this
context. There are numerous other important articles on the subject,
like e.g. \cite{Kapustin:2004gv,Pestun:2005rp,Pestun:2006rj,Jeschek:2004je,Jeschek:2005ek,Cassani:2007pq,Grange:2004ah,Tomasiello:2007zq,Ikeda:2006pd,Ikeda:2007rn}
and many more. A more complete list of references can be found in
\cite{Grana:2005jc}. A major part of the considerations so far was
done from the supergravity point of view. Target space supersymmetry
is, however, related to an $N=2$ supersymmetry on the worldsheet.
For this reason the relation between an extended worldsheet supersymmetry
and the presence of an integrable generalized complex structure (GCS)
was studied in \cite{Lindstrom:2004iw} (the reviews \cite{Zabzine:2006uz,Lindstrom:2006ee}
on generalized complex geometry have this relation in mind). Zabzine
clarified in \cite{Zabzine:2005qf} the relation in a model independent
way in a Hamiltonian description and showed that the existence of
a second non-manifest worldsheet supersymmetry $\Q_{2}$ in an $N=1$
sigma-model is equivalent to the existence of an integrable GCS $\mc{J}$.
It is the observation that the integrability of the GCS $\mc{J}$
can be written as the vanishing of a generalized bracket $\left[\mc{J}\bs{,}\mc{J}\right]_{B}=0$
which leads to the natural question, whether there is a direct mapping
between $\left[\mc{J},\mc{J}\right]_{B}=0\,\&\,\mc{J}^{2}=-1$ on
the one side and $\{\Q_{2},\Q_{2}\}=2P$ on the other side. This will
be a natural application in subsection \ref{sub:Zabzine} of the more
general preceding considerations about the relation between (super-)Poisson
brackets in sigma models with special field content and derived brackets
in the target space. 

A second interesting application is Zucchini's Hitchin-sigma-model
\cite{Zucchini:2004ta}. There are up to now three more papers on
that subject \cite{Zucchini:2005rh,Zucchini:2005cq,Zucchini:2007ie},
but the present discussion refers only to the first one. Zucchini's
model is a two dimensional sigma-model in a target space with a generalized
complex structure (GCS). The sigma-model is topological when the GCS
is integrable, while the inverse does not hold. The condition for
the sigma model to be topological is the master equation $(S\bs{,}S)=0$.
Again we might wonder whether there is a direct mapping between the
antibracket and $S$ on the one hand and the geometric bracket and
$\mc{J}$ on the other hand and it will be shown in subsection \ref{sub:Zucchini}
how this mapping works as an application of the considerations in
subsection \ref{sub:antibracket}. In order to understand more about
geometric brackets in general, however, it was necessary to dive into
\rem{??} Kosmann-Schwarzbach's review on derived brackets \cite{Kosmann-Schwarzbach:2003en}
which led to observations that go beyond the application to the integrability
of a GCS . 

The structure of this part of the thesis is as follows: The general
relation between sigma models and derived brackets in target space
will be studied in the next section. The necessary geometric setup
will be established in \ref{sub:bc-phase-space}. Although there are
no new deep insights in \ref{sub:bc-phase-space}, the unconventional
idea to extend the exterior derivative on forms to multivector valued
forms (see (\ref{eq:dK-coord}) and (\ref{eq:d-auf-partial})) will
provide a tool to write down a coordinate expression for the general
derived bracket between multivector valued forms (\ref{eq:bc-derived-bracket-coord})
which to my knowledge does not yet exist in literature. The main results
in section \ref{sec:sigma-model-induced}, however, are the propositions
1 on page \pageref{eq:Proposition1} and 1b on page \pageref{eq:Proposition1b}
for the relation between the Poisson-bracket in a sigma-model with
special field content and the derived bracket in the target space,
and the proposition 3b on page \pageref{eq:PropositionIIIb} for the
relation between the antibracket in a sigma-model and the derived
bracket in target space. Proposition 2 on page \pageref{eq:Proposition2}
is just a short quantum consideration which only works for the particle
case. In section \ref{sec:Applications-in-string} the propositions
1b and 3b are finally applied to the two examples which were mentioned
above. 

Another result is the relation between the generalized Nijenhuis tensor
and the derived bracket of $\mc{J}$ with itself, given in (\ref{eq:relation-between-derived-and-Nij-Tens-in-main-part}).
The derivation of this can be found in the appendix on page \pageref{sub:Derivation-via-derived-bracket}.
In addition to this, there is a new coordinate form of the generalized
Nijenhuis tensor presented in (\ref{eq:generalized-integrabilityIII})
on page \pageref{eq:generalized-integrabilityIII}, which might be
easier to memorize than the known ones. There is also a short comment
in footnote \ref{foot:dual-coord} on page \pageref{foot:dual-coord}
on a possible relation to Hull's doubled geometry. 

This part of the thesis makes use of only three of the appendices.
Appendix \ref{sec:Conventions} on page \pageref{sec:Conventions}
summarizes the used conventions, while appendix \ref{sec:bracket-review}
on page \pageref{sec:bracket-review} is an introduction to geometric
brackets. Finally, appendix \ref{sec:Generalized-complex-geometry}
on page \pageref{sec:Generalized-complex-geometry} provides some
aspects of generalized complex geometry which might be necessary to
understand the two applications of above. \numberwithin{equation}{chapter}

\rem{Hier waere noch eine section ueber normale komplexe Struktur als Motivation. Probleme mit der Leibnizregel...}

\chapter{Sigma-model-induced brackets}

\label{sec:sigma-model-induced}

\section{Geometric brackets in phase space formulation}

\label{sub:bc-phase-space} In the following some basic geometric
ingredients which are necessary to formulate derived brackets will
be given. Although there is no sigma model and no physics explicitly
involved in this first subsection, the presentation and the techniques
will be very suggestive, s.th. there is visually no big change when
we proceed after that with considerations on sigma-models.

\subsection{Algebraic brackets}

\label{sub:Algebraic-brackets} Consider a real differentiable manifold
$M$. The interior\index{interior product} product\index{product!interior $\sim$|see{interior product}}
with a vector field\index{$v$|itext{general vector field}} $v=v^{k}\pe_{k}$
(in a local coordinate basis) acting on a differential form $\rho$
is a differential operator in the sense that it differentiates with
respect to the basis elements of the cotangent space:%
\footnote{\index{footnote!\thefoot. prefactor in forms}Note, that a convention
is used, were the prefactor $\frac{1}{r!}$ which usually comes along
with an $r$-form is absorbed into the definition of the wedge-product.
The common conventions can for all equations easily be recovered by
redefining all coefficients appropriately, e.g. $\rho_{m_{1}\ldots m_{r}}\To\frac{1}{r!}\rho_{m_{1}\ldots m_{r}}.\qquad\fussend$%
}\index{$\rho^{(r)}$|itext{r-form, $\rho$}}\index{$i_v$@$\ip_v \rho$|itext{interior product}}\begin{eqnarray}
\ip_{v}\rho^{(r)} & = & r\cdot v^{k}\rho_{km_{1}\ldots m_{r-1}}^{(r)}(x)\:\de x^{m_{1}}\cdots\de x^{m_{r-1}}=v^{k}\partl{(\de x^{k})}\left(\rho_{m_{1}\ldots m_{r}}\de x^{m_{1}}\cdots\de x^{m_{r}}\right)\end{eqnarray}
Let us rename%
\footnote{\index{ghost!as form}\index{footnote!\thefoot. ghosts and forms}The
similarity with ghosts is of course no accident. It is well known
(see e.g. \cite{Henneaux:1992ig}) that ghosts in a gauge theory can
be seen as 1-forms dual to the gauge-vector fields and the BRST differential
as the sum of the Koszul-Tate differential (whose homology implements
the restriction to the constraint surface) and the longitudinal exterior
derivative along the constraint surface. In that sense the present
description corresponds to a topological theory, where all degrees
of freedom are gauged away. But we will not necessarily always view
$\bs{c}^{m}$ as ghosts in the following. So let us in the beginning
see $\bs{c}^{m}$ just as another name for $\de x^{m}$. We do not
yet assume an underlying sigma-model, i.e. $\bs{b}_{m}$ and $\bs{c}^{m}$
do not necessarily depend on a worldsheet variable.$\quad\fussend$%
} \index{$c^m$@$\ce^m$|itext{$\equiv\de x^m$}}\index{$b_m$@$\be_m$|itext{$\equiv\pe_m$}}\index{$dx^m$@$\de x^m$}\index{$\partial$@$\pe_m$|itext{coordinate basis element of $TM$}}\begin{eqnarray}
\ce^{m} & \equiv & \de x^{m}\\
\be_{m} & \equiv & \pe_{m}\end{eqnarray}
The vector $v$ takes locally the form $v=v^{m}\be_{m}$ and when
we introduce a canonical graded Poisson bracket between $\ce^{m}$
and $\be_{m}$ via $\left\{ \be_{m},\ce^{n}\right\} =\delta_{m}^{n}$
, we get \begin{eqnarray}
\ip_{v}\rho & = & \left\{ v,\rho\right\} \end{eqnarray}
\rem{Es gilt auch\index{$i_v$@$\ip_v K$} $\ip_{v}K=\left\{ v,K\right\} $,
aber i.a.\index{$i_v$@$\ip_v T$|itext{$\neq \{v,T\}$}} $\ip_{v}T\neq\left\{ v,T\right\} $,
wenn $T$ $p$ enthaelt. Ferner gilt fuer 1-Formen $\left\{ \omega,\rho\right\} =\ip_{\omega}^{(1)}\rho=0\neq\ip_{\omega}^{(0)}\rho=\ip_{\omega}\rho$.
Entsprechend gilt fuer generalized vectors $\left\{ \mf{a},\rho\right\} =\ip_{\mf{a}}^{(1)}\rho=\ip_{a}\rho\neq\ip_{\mf{a}}\rho$
und \index{$(\{\})$@$\{\protect\mf{a},K\}$}$\left\{ \mf{a},K\right\} =\ip_{\mf{a}}^{(1)}K-(-)^{k-k'}\ip_{K}^{(1)}\mf{a}=\ip_{a}^{(1)}K-(-)^{k-k'}\ip_{K}^{(1)}\alpha$} 
Extending also the local $x$-coordinate-space to a phase space by
introducing the conjugate\index{conjugate momentum|itext{$p_m$}}
momentum\index{momentum!conjugate $\sim$ $p_m$} $p_{m}$\index{$p_m$|itext{$\hat=\partial_m$}}
(whose geometric interpretation we will discover soon), we have altogether
the (graded\index{graded!Poisson bracket|see{Poisson bracket}}) Poisson\index{Poisson bracket!in $T\oplus T^*$}
bracket\index{bracket!Poisson $\sim$!in $T\oplus T^*$}\index{$(\{\})$@$\{\ldots,\ldots\}$|itext{Poisson bracket}}
\begin{eqnarray}
\left\{ \be_{m},\ce^{n}\right\}  & = & \delta_{m}^{n}=\left\{ \ce^{n},\be_{m}\right\} \label{eq:Poisson-bracket-bc}\\
\left\{ p_{m},x^{n}\right\}  & = & \delta_{m}^{n}=-\left\{ x^{n},p_{m}\right\} \label{eq:Poisson-bracket-xp}\\
\left\{ A,B\right\}  & = & A\partr{\be_{k}}\partl{\ce^{k}}B+A\partr{p_{k}}\partl{x^{k}}B-(-)^{AB}\left(B\partr{\be_{k}}\partl{\ce^{k}}A+B\partr{p_{k}}\partl{x^{k}}A\right)\label{eq:Poisson-bracket}\end{eqnarray}
and can write the exterior derivative acting on forms as generated
via the Poisson-bracket by an odd phase-space function $\oo(\ce,p)$\index{generator!for exterior derivative $\de=\{\oo,\ldots\}$}\index{$o$@$\oo$|itext{generator for exterior derivative}}\index{$d$@$\de$|itext{exterior derivative}}\index{exterior derivative|itext{$\de$}}\index{$(\{\})$@$\{\ldots,\ldots\}$!$\{\oo,\rho^{(r)}\}=\de\rho^{(r)}$}
\begin{eqnarray}
\oo & \equiv & \oo(\ce,p)\equiv\ce^{k}p_{k}\label{eq:BRST-op}\\
\left\{ \oo,\rho^{(r)}\right\}  & = & \ce^{k}\left\{ p_{k},\rho_{m_{1}\ldots m_{r}}(x)\right\} \ce^{m_{1}}\cdots\ce^{m_{r}}=\de\rho^{(r)}\label{eq:exterior-derivative-via-BRST}\end{eqnarray}
The variables $\ce^{m}$,$\be_{m}$,$x^{m}$ and $p_{m}$ can be seen
as coordinates of $T^{*}(\Pi TM)$, \index{$T^*(\Pi TM)$}\index{$\Pi TM$}the
cotangent bundle of the tangent bundle with parity\index{parity inversed fiber}
inversed fiber. \rem{$\textrm{Fun}\left(\Pi TM\right)=\Gamma\left(\bigwedge^{\bullet}T^{*}M\right)$,
$\textrm{Fun}\left(T^{*}(\Pi TM)\right)\stackrel{?}{=}\Gamma\left(\textrm{End}(\Omega^{\bullet}(M))\right)$.
In the present thesis, however, we will stick to the interpretation
of $\ce^{m}$ and $\be_{m}$ as coordinate basis elements of $T^{*}M$
and $TM$ and we will assign an interpretation of $p_{m}$ in \ref{sub:Extended-exterior-derivative}
only via an embedding into the space of differential operators acting
on forms.}

\subsubsection*{Interior product and {}``quantization''}

Given a multivector valued form $K^{(k,k')}$\index{$K^{(k,k')}$|itext{multivector valued form}}\index{multivector valued form}
of form\index{form degree|itext{$k$}} degree $k$ and multivector\index{multivector degree|itext{$k'$}}
degree $k'$, it reads in the local coordinate patch with the new
symbols\index{$K_{\mm}\hoch{\nn}$|itext{schematic index notation of $K^{(k,k')}$}}
\begin{eqnarray}
K^{(k,k')}\equiv K^{(k,k')}(x,\ce,\be) & \equiv & K_{m_{1}\ldots m_{k}}\hoch{n_{1}\ldots n_{k'}}(x)\,\bs{c}^{m_{1}}\cdots\ce^{m_{k}}\be_{n_{1}}\cdots\be_{n_{k'}}\equiv K_{\mm}\hoch{\nn}\label{eq:multivector-valued-form-K}\end{eqnarray}
The notation $K(x,\ce,\be)$ should stress, that $K$ is locally a
(smooth on a $C^{\infty}$ manifold) function of the phase space variables
which will later be used for analytic continuation ($x$ will be allowed
to take c-number values of a superfunction). The last expression in
the above equation introduces a \textbf{schematic\index{schematic index notation}
index\index{index!schematic $\sim$ notation} notation\index{notation!schematic index $\sim$}}
which is useful to write down the explicit coordinate form for lengthy
expressions. See in the appendix \ref{fat-index} at page \pageref{fat-index}
for a more detailed description of its definition. It should, however,
be self-explanatory enough for a first reading of the thesis

One can define a natural generalization of the interior product with
a vector $\ip_{v}$ to an \textbf{interior\index{interior product!w.r.t. multivector valued form|itext{$\ip_K$}}
product} with a multivector valued form $\ip_{K}$ acting on some
$r$-form (in fact, it is more like a combination of an interior and
an exterior product -- see footnote \ref{foot:interior-product} on
page \pageref{foot:interior-product}\rem{Anhang} --, but we will
stick to this name)\index{$i_K$@$\ip_K\rho$}\begin{eqnarray}
\ip_{K^{(k,k')}}\rho^{(r)} & \equiv & (k')!\left(\zwek{r}{k'}\right)K_{\bs{m}\ldots\bs{m}}\hoch{l_{1}\ldots l_{k'}}\rho_{\underbrace{{\scriptstyle l_{k'}\ldots l_{1}\bs{m}\ldots\bs{m}}}_{r}}=\label{eq:bc-interior-product}\\
 & = & K_{m_{1}\ldots m_{k}}\hoch{n_{1}\ldots n_{k'}}\bs{c}^{m_{1}}\cdots\ce^{m_{k}}\left\{ \be_{n_{1}},\left\{ \cdots,\left\{ \be_{n_{k'}},\rho^{(r)}\right\} \right\} \right\} \label{eq:bc-interior-product-I}\\
 & = & K_{m_{1}\ldots m_{k}}\hoch{n_{1}\ldots n_{k'}}\bs{c}^{m_{1}}\cdots\ce^{m_{k}}\partl{\ce^{n_{1}}}\cdots\partl{\ce^{n_{k'}}}\rho^{(r)}\label{eq:bc-interior-product-II}\end{eqnarray}
It is a derivative of order $k'$ and thus not a derivative in the
usual sense like $\ip_{v}$. The third line shows the reason for the
normalization of the first line, while the second line is added for
later convenience. The interior product is commonly used as an \textbf{embedding\index{embedding!of multivector valued forms in operator space}}
of the multivector valued forms in the space of differential operators
acting on forms, i.e. $K\To\ip_{K}$, s.th. structures of the latter
can be induced on the space of multivector valued forms. In (\ref{eq:bc-interior-product-II})
the interior product $\ip_{K}$ can be seen, up to a factor of $\hbar/i$,
as the quantum operator corresponding to $K$, where the form $\rho$
plays the role of a wave function. The natural ordering\index{ordering}
is here to put the conjugate momenta to the right. We can therefore
fix the following {}``\textbf{quantization}\index{quantization!of a multivector valued form}''
rule (corresponding to $\hat{\be}=\frac{\hbar}{i}\partl{\ce}$)\index{$bb$@$\hat\be_m$|itext{quantized $\be$}}\index{$Kb$@$\hat K^{(k,k')}$|itext{$\propto \ip_K$}}\begin{eqnarray}
\hat{K}^{(k,k')} & \equiv & \left(\frac{\hbar}{i}\right)^{k'}\ip_{K^{(k,k')}}\label{eq:quantization}\\
\textrm{with }\ip_{K^{(k,k')}} & = & K_{\mm}\hoch{n_{1}\ldots n_{k'}}\frac{\partial^{k'}}{\partial\ce^{n_{1}}\cdots\partial\ce^{n_{k'}}}\label{eq:bc-interior-product-III}\end{eqnarray}
 The (graded) commutator of two interior products induces an algebraic\index{algebraic bracket}
bracket\index{bracket!algebraic $\sim$|itext{$[K,L]^\Delta$}} due
to Buttin \cite{Buttin:1974}\rem{(see also in the appendix, in subsection
\ref{sub:multivector-form-brackets})}, which is defined via \index{$([])$@$[\ldots,\ldots]$|itext{commutator}}\index{$([])$@$[\ldots,\ldots]$!$[\ip_K,\ip_L]$}\index{$([])$@$[\ldots,\ldots]^\Delta$|itext{algebraic bracket}}\index{$([])$@$[\ldots,\ldots]^\Delta$!$[K,L]^\Delta$}\index{bracket!commutator|itext{$[\ldots,\ldots]$}}\index{commutator}\begin{eqnarray}
\left[\ip_{K^{(k,k')}},\ip_{L^{(l,,l')}}\right] & \equiv & \ip_{\left[K,L\right]^{\Delta}}\label{eq:bc-algebraic-bracket}\end{eqnarray}
 A short calculation, using the obvious generalization of $\partial_{x}^{n}(f(x)g(x))=\sum_{p=0}^{n}\left(\zwek{n}{p}\right)\partial_{x}^{p}f(x)\partial_{x}^{n-p}g(x)$
leads to\index{product!of interior products}\index{product!star $\sim$}\index{star product}\index{Moyal product}
\begin{eqnarray}
\ip_{K}\ip_{L} & = & \sum_{p\geq0}\ip_{\ip_{K}^{(p)}L}=\ip_{K\wedge L}+\sum_{p\geq1}\ip_{\ip_{K}^{(p)}L}\label{eq:bc-product-of-interior-products}\end{eqnarray}
where we introduced the \textbf{interior product\index{interior product!of order $p$|itext{$\ip_K^{(p)}$}}
of order $p$} \rem{(see (\ref{eq:interior-productIII}))}\index{$i_K$@$\ip_{K^{(k,k')}}^{(p)}$}\begin{eqnarray}
\ip_{K^{(k,k')}}^{(p)} & \equiv & \left(\zwek{k'}{p}\right)K_{\bs{m}\ldots\bs{m}}\hoch{\bs{n}\ldots\bs{n}l_{1}\ldots l_{p}}\frac{\partial^{p}}{\partial\ce^{n_{1}}\cdots\partial\ce^{n_{p}}}=\label{eq:bc-interior-pproduct-I}\\
 & = & \frac{1}{p!}K\frac{\lpartial^{p}}{\partial\be_{n_{p}}\cdots\partial\be_{n_{1}}}\frac{\partial^{p}}{\partial\ce^{n_{1}}\cdots\partial\ce^{n_{p}}}\label{eq:bc-interior-pproductII}\\
\dann\,\ip_{K^{(k,k')}}^{(p)}L^{(l,l')} & = & (-)^{(k'-p)(l-p)}p!\left(\zwek{k'}{p}\right)\left(\zwek{l}{p}\right)K_{\bs{m}\ldots\bs{m}}\hoch{\bs{n}\ldots\bs{n}l_{1}\ldots l_{p}}L_{l_{p}\ldots l_{1}\bs{m}\ldots\bs{m}}\hoch{\bs{n}\ldots\bs{n}}\label{eq:bc-interior-pproduct-on-L}\end{eqnarray}
which contracts only $p$ of all $k'$ upper indices and therefore
coincides with the interior product of above when acting on forms
for $p=k'$ and with the wedge product for $p=0$.\index{$()$@$\wedge$|itext{wedge}}\index{$K\wedge L$}\begin{equation}
\ip_{K^{(k,k')}}^{(k')}\rho=\ip_{K^{(k,k')}}\rho,\qquad\ip_{K}^{(0)}L=K\wedge L\end{equation}
\rem{\[
K\wedge L=(-)^{k'l}K_{\bs{m}\ldots\bs{m}}\hoch{\bs{n}\ldots\bs{n}}L_{\bs{m}\ldots\bs{m}}\hoch{\bs{n}\ldots\bs{n}}\]
}\rem{Kann man noch weiter verallgemeinern (haengt mit r-wedge zusammen) zu
(\ref{eq:interior-productIV})\index{$i_Kb$@$\ip_{K^{(k,k')}}^{(p,q)}$}\begin{eqnarray*}
\ip_{K^{(k,k')}}^{(p,q)} & \equiv & \frac{1}{p!q!}K\frac{\lpartial^{p}}{\partial\be_{n_{p}}\cdots\partial\be_{n_{1}}}\frac{\lpartial^{q}}{\partial\ce^{k_{q}}\cdots\partial\ce^{k_{1}}}\frac{\partial^{q}}{\partial\be_{k_{1}}\cdots\partial\be_{k_{q}}}\frac{\partial^{p}}{\partial\ce^{n_{1}}\cdots\partial\ce^{n_{p}}}\end{eqnarray*}
} Using (\ref{eq:bc-product-of-interior-products}) the \textbf{algebraic\index{algebraic bracket|fett}
bracket}\index{bracket!algebraic $\sim$|fett} $[\:,\:]^{\Delta}$\index{$([])$@$[\ldots,\ldots]^\Delta$!$[K,L]^\Delta$|fett}
defined in (\ref{eq:bc-algebraic-bracket}) can thus be written as
\begin{eqnarray}
[K^{(k,k')},L^{(l,l')}]^{\Delta} & = & \sum_{p\geq1}\underbrace{\ip_{K}^{(p)}L-(-)^{(k-k')(l-l')}\ip_{L}^{(p)}K}_{\equiv[K,L]_{(p)}^{\Delta}}\label{eq:bc-algebraic-bracketI}\end{eqnarray}
 (\ref{eq:bc-interior-pproduct-on-L}) provides the explicit coordinate
form of this algebraic bracket. From (\ref{eq:bc-interior-pproductII})
we recover the known fact that the $p=1$ term of the algebraic bracket
is induced by the Poisson-bracket and therefore is itself an algebraic
bracket, called the \textbf{big\index{big bracket} bracket}\index{bracket!big $\sim$}
\cite{Kosmann-Schwarzbach:2003en} or \textbf{Buttin\index{Buttin's algebraic bracket}'s
algebraic\index{algebraic bracket!Buttin's $\sim$} bracket\index{bracket!Buttin's algebraic $\sim$}}\index{$([])$@$[\ldots,\ldots]_{(1)}^\Delta$|iText{big bracket}}\index{$([])$@$[\ldots,\ldots]_{(1)}^\Delta$!$[K,L]_{(1)}^\Delta$}
\cite{Buttin:1974}\rem{ (\ref{eq:bigbracket})}\index{$(\{\})$@$\{\ldots,\ldots\}$!$\{K,L\}\leftrightarrow[K,L]_{(1)}^\Delta$}\begin{eqnarray}
\hspace{-0.5cm}\lqn{\Ramm{.59}{\Big.}}\quad[K,L]_{(1)}^{\Delta} & = & \ip_{K}^{(1)}L-(-)^{(k-k')(l-l')}\ip_{L}^{(1)}K\quad\stackrel{(\ref{eq:bc-interior-pproductII})}{=}\left\{ K,L\right\} \quad=\label{eq:bc-big-bracket}\\
 & \stackrel{(\ref{eq:bc-interior-pproduct-on-L})}{=} & (-)^{(k'-1)(l-1)}k'l\, K_{\bs{m}\ldots\bs{m}}\hoch{\bs{n}\ldots\bs{n}l_{1}}L_{l_{1}\bs{m}\ldots\bs{m}}\hoch{\bs{n}\ldots\bs{n}}+\label{eq:bc-big-bracket-coord}\\
 &  & -(-)^{(k-k')(l-l')}(-)^{(l'-1)(k-1)}l'k\, L_{\bs{m}\ldots\bs{m}}\hoch{\bs{n}\ldots\bs{n}l_{1}}K_{l_{1}\bs{m}\ldots\bs{m}}\hoch{\bs{n}\ldots\bs{n}}\qquad\nonumber \end{eqnarray}
For $k'=l'=1$ it reduces to the Richardson\index{Richardson-Nijenhuis bracket}-Nijenhuis\index{Nijenhuis!Richardson-$\sim$ bracket}
bracket\index{bracket!Richardson-Nijenhuis $\sim$} (\ref{eq:Richardson-Nijenhuis-bracket-coord})
for vector valued forms. In \cite{Kosmann-Schwarzbach:2003en} the
big bracket is described as the canonical Poisson structure on $\bigwedge^{\bullet}(T\oplus T^{*})$
which matches with the observation in (\ref{eq:bc-big-bracket}).
The bracket takes an especially pleasant coordinate form for generalized\index{generalized form}\index{generalized multivector}
multivectors\index{form!generalized $\sim$}\index{multivector!generalized $\sim$}
as is presented in equation (\ref{eq:multvec-bigbrack}) on page \pageref{eq:multvec-bigbrack}.

The multivector-degree of the $p$-th term of the complete algebraic
bracket (\ref{eq:bc-algebraic-bracketI}) is $(k'+l'-p)$, so that
we can rewrite (\ref{eq:bc-algebraic-bracket}) in terms of {}``quantum''-operators
(\ref{eq:quantization}) in the following way:\index{$([])$@$[\ldots,\ldots]$!$[\hat{K},\hat{L}]$}
\begin{eqnarray}
\left[\hat{K}^{(k,k')},\hat{L}^{(l,l')}\right] & = & \sum_{p\geq1}\left(\frac{\hbar}{i}\right)^{p}\widehat{\left[K,L\right]_{(p)}^{\Delta}}=\label{eq:quantum-commutator1}\\
 & = & \left(\frac{\hbar}{i}\right)\widehat{\left\{ K,L\right\} }+\sum_{p\geq2}\left(\frac{\hbar}{i}\right)^{p}\widehat{\left[K,L\right]_{(p)}^{\Delta}}\label{eq:quantum-commutatorII}\end{eqnarray}
 The Poisson bracket is, as it should be, the leading order of the
quantum bracket.

\subsection{Extended exterior derivative and the derived bracket of the commutator}

\label{sub:Extended-exterior-derivative} In the previous subsection
the commutator of differential operators induced (via the interior
product as embedding) an algebraic bracket on the embedded tensors.
Also other structures from the operator space can be induced on the
tensors. Having the commutator at hand, one can build the \textbf{derived\index{derived bracket}
bracket\index{bracket!derived $\sim$}} (see footnote \ref{foot-derived-bracket}
on page \pageref{foot-derived-bracket}\rem{(\ref{eq:derived-bracketI})})
of the commutator by additionally commuting the first argument with
the exterior derivative. Being interested in the induced structure
on multivector valued forms, we consider as arguments only interior
products with those multivector valued forms\index{$([])$@$\left[\ip_{K},_{\de}\ip_{L}\right]$|itext{derived bracket of the commutator by $\de$}}\begin{eqnarray}
\left[\ip_{K},_{\de}\ip_{L}\right] & \equiv & \left[\left[\ip_{K},\de\,\right],\ip_{L}\right]\label{eq:bc-derived-bracket}\end{eqnarray}
One can likewise use other differentials to build a derived bracket,
e.g. the twisted differential $\left[\de+H,\ldots\right]$ with an
odd closed form $H$, which leads to so called twisted brackets. Let
us restrict to $\de$ for the moment.\rem{Kommt da noch was? Querverweis?}
The derived bracket is in general not skew-symmetric but it obeys
a graded Jacobi-identity \rem{(Leibniz rule, when acting on itself from the left)}
and is therefore what one calls a Loday bracket. When looking for
new brackets, the Jacobi identity is the property which is hardest
to check. A mechanism like above, which automatically provides it
is therefore very powerful. The above derived bracket will induce
brackets like the Schouten bracket or even the Dorfman bracket of
generalized complex geometry on the tensors. In general, however,
the interior products are not closed under its action, i.e. the result
of the bracket cannot necessarily be written as $\ip_{\tilde{K}}$
for some $\tilde{K}$. An expression for a general bracket on the
tensor level, which reduces in the corresponding cases to the well
known brackets therefore does not exist. Instead one normally has
to derive the brackets in the special cases separately. In the following,
however, a natural approach is discussed including the new variable
$p_{m}$, introduced in (\ref{eq:Poisson-bracket-xp}), which leads
to an explicit coordinate expression for the general bracket. This
expression is of course tensorial only in the mentioned special cases,
that is when terms with $p_{m}$ vanish. This is not an artificial
procedure, as the conjugate variable $p_{m}$ to $x^{m}$ is always
present in sigma-models, and it will in turn explain the geometric
meaning of $p_{m}$.

The exterior derivative $\de$ acting on forms is usually not defined
acting on multivector valued forms (otherwise we could build the derived
bracket of the algebraic bracket (\ref{eq:bc-algebraic-bracketI})
by $\de\,$ without lifting everything to operators via the interior
product). But via $\{\oo,K^{(k,k')}\}$ we can, at least formally,
define a differential on multivector valued forms. The result, however,
contains the variable $p_{k}$ which we have not yet interpreted geometrically.
After extending the definition of the interior product to objects
containing $p_{m}$, we will get the relation $\left[\de\,,\ip_{K}\right]=\ip_{\left\{ \oo,K\right\} }$,
i.e. $\{\oo,\ldots\}$ can be seen as an induced differential from
the space of operators. For forms $\omega^{(q)}$, this simply reads
$\left[\de\,,\ip_{\omega}\right]=\ip_{\de\omega}$. The definition
$\de K\equiv\left\{ \oo,K\right\} $ thus seems to be a reasonable
extension of the exterior derivative to multivector valued forms.
Let us first provide the necessary definitions.

Consider a phase space function, which is of arbitrary order in the
variable $p_{k}$\index{$T^{(t,t',t'')}(x,\ce,\be,p)$}\index{$T_{m_{1}\ldots m_{t}}\hoch{n_{1}\ldots n_{t'}k_{1}\ldots k_{t''}}(x)$}\begin{eqnarray}
T^{(t,t',t'')}(x,\ce,\be,p) & \equiv & T_{m_{1}\ldots m_{t}}\hoch{n_{1}\ldots n_{t'}k_{1}\ldots k_{t''}}(x)\,\ce^{m_{1}}\cdots\ce^{m_{t}}\be_{m_{1}}\cdots\be_{m_{t'}}p_{k_{1}}\cdots p_{k_{t''}}\label{eq:Tcbp}\end{eqnarray}
$T$ is symmetrized in $k_{1}\ldots k_{t''}\,$,while it is antisymmetrized
in the remaining indices. Using the usual quantization\index{quantization rules}
rules $\be\To\frac{\hbar}{i}\partl{\ce}$ and $p\To\frac{\hbar}{i}\partl{x}$
with the indicated ordering (conjugate momenta to the right) while
still insisting on (\ref{eq:quantization}) as the relation between
quantum operator and interior product, we get an extended definition
of the \textbf{interior\index{interior product!extended definition $\ip_{T^{(t,t',t'')}}$}
product}\index{product!interior $\sim$!extended}\index{$i_T$@$\ip_{T^{(t,t',t'')}}$}
(\ref{eq:bc-interior-product-I},\ref{eq:bc-interior-product-II}):\index{$T^{(t)}$@$\hat{T}^{(t,t',t'')}$}\begin{eqnarray}
\hspace{-0.8cm}\ip_{T^{(t,t',t'')}} & \equiv & \left(\frac{i}{\hbar}\right)^{t'+t''}\hat{T}^{(t,t',t'')}\equiv\label{eq:quantizationII}\\
 & \equiv & T_{m_{1}\ldots m_{t}}\hoch{n_{1}\ldots n_{t'}k_{1}\ldots k_{t''}}\ce^{m_{1}}\cdots\ce^{m_{t}}\frac{\partial^{t'}}{\partial\ce^{n_{1}}\cdots\partial\ce^{n_{t'}}}\frac{\partial^{t''}}{\partial x^{k_{1}}\cdots\partial x^{k_{t''}}}=\label{eq:quantizationIII}\\
\hspace{-0.6cm}\ip_{T^{(t,t',t'')}}\rho^{(r)} & = & T_{m_{1}\ldots m_{t}}\hoch{n_{1}\ldots n_{t'}k_{1}\ldots k_{t''}}\ce^{m_{1}}\cdots\ce^{m_{t}}\left\{ \be_{n_{1}},\left\{ \cdots,\left\{ \be_{n_{t'}},\left\{ p_{k_{1}},\left\{ \cdots,\left\{ p_{k_{t''}},\rho^{(r)}\right\} \right\} \right\} \right\} \right\} \right\} =\qquad\label{eq:i-T}\\
 & = & (t')!\left(\zwek{r}{t'}\right)T_{\mm}\hoch{n_{1}\ldots n_{t'}k_{1}\ldots k_{t''}}\frac{\partial^{t''}}{\partial x^{k_{1}}\cdots\partial x^{k_{t''}}}\rho_{n_{t'}\ldots n_{1}\mm}^{(r)}\label{eq:i-TII}\end{eqnarray}
\rem{\[
\ip_{T^{(t,t',t'')}}=\frac{1}{(t'+t'')!}\left(\zwek{t'+t''}{t'}\right)T\frac{\lpartial^{t'+t''}}{\partial p_{i_{t'+t''}}\ldots\partial p_{i_{t'+1}}\partial\be_{i_{t'}}\ldots\partial\be_{i_{1}}}\frac{\partial^{t'+t''}}{\partial\bs{c}^{i_{1}}\ldots\partial\bs{c}^{i_{t'}}\partial x^{i_{t'+1}}\ldots\partial x^{i_{t'+t''}}}\]
}The operator $\ip_{T}$ will serve us as an embedding of $T$ (a
phase space function, which lies in the extension of the space of
multivector valued forms by the basis element $p_{k}$) into the space
of differential operators acting on forms. Because of the partial
derivatives with respect to $x$, the last line is not a tensor and
$T$ in that sense not a well defined geometric object. Nevertheless
it can be a building block of a geometrically well defined object,
for example in the definition of the \textbf{exterior\index{exterior derivative!on multivector valued forms}
derivative}\index{derivative!extended exterior $\sim$} on multivector
valued forms which we suggested above. Namely, if we define%
\footnote{\index{footnote!\thefoot. exterior derivative versus BRST differential}\index{BRST differential!exterior derivative as $\sim$}This
can of course be seen as a BRST differential, which is well known
to be the sum of the longitudinal exterior derivate plus the Koszul
Tate differential. However, as the constraint surface in our case
corresponds to the configuration space ($p_{k}$ would be the first
class constraint generating the BRST-transformation), it is reasonable
to regard the BRST differential as a natural extension of the exterior
derivative of the configuration space.$\quad\fussend$%
}\index{$dK$@$\de K^{(k,k')}$} \rem{Relation zu Zucchini's Beobachtung?Klaus Bering;}\begin{eqnarray}
\de K^{(k,k')} & \equiv & \left\{ \oo,K^{(k,k')}\right\} =\label{eq:dK}\\
 & = & \partial_{\bs{m}}K_{\mm}\hoch{\nn}-(-)^{k-k'}k'\cdot K_{\mm}\hoch{\nn k}p_{k}\label{eq:dK-coord}\end{eqnarray}
We get via our extended embedding (\ref{eq:i-TII}) the nice relation
\index{$([])$@$[\ldots,\ldots]$!$[\de,\ip_K]$|itext{$\protect\Lie_K$}}\index{$L_K$@$\protect\Lie_K$}%
\footnote{\index{footnote!\thefoot. $[\de,\ip_K]\rho=\ip_{\de K}\rho$ }The
exterior derivative on forms has already earlier (\ref{eq:exterior-derivative-via-BRST})
been seen to coincide with the Poisson bracket with $\oo$, which
can be used to demonstrate (\ref{eq:dK-und-Lie}):\begin{eqnarray*}
\left[\de\,,\ip_{K}\right]\rho & = & \de\,(\ip_{K}\rho)-(-)^{\abs{K}}\ip_{K}(\de\rho)=\\
 & = & \left\{ \oo,\ip_{K}\rho\right\} -(-)^{\abs{K}}\ip_{K}\left\{ \oo,\rho\right\} =\\
 & \stackrel{(\ref{eq:bc-interior-product-I})}{=} & \partial_{m_{1}}K_{m_{2}\ldots m_{k+1}}\hoch{n_{1}\ldots n_{k'}}\bs{c}^{m_{1}}\cdots\ce^{m_{k+1}}\Big\{\be_{n_{1}},\left\{ \be_{n_{2}},\left\{ \cdots,\left\{ \be_{n_{k'}},\rho^{(r)}\right\} \right\} \right\} +\\
 &  & +(-)^{k}k'\cdot K_{m_{1}\ldots m_{k}}\hoch{n_{1}\ldots n_{k'}}\bs{c}^{m_{1}}\cdots\ce^{m_{k}}\Big\{\underbrace{\left\{ \oo,\be_{n_{1}}\right\} }_{p_{n_{1}}},\left\{ \be_{n_{2}},\left\{ \cdots,\left\{ \be_{n_{k'}},\rho^{(r)}\right\} \right\} \right\} \Big\}\us{\stackrel{(\ref{eq:i-T})}{=}}{(\ref{eq:dK-coord})}\ip_{\de K}\rho\qquad\fussend\end{eqnarray*}
}\begin{eqnarray}
\ip_{\de K}\rho & = & \left[\de\,,\ip_{K}\right]\rho\stackrel{(\ref{eq:Lie-derivativeI})}{=}-(-)^{k-k'}\Lie_{K}\rho\label{eq:dK-und-Lie}\\
\textrm{with }\quad\Lie_{K}\rho & = & (k')!\left(\zwek{r}{k'-1}\right)K_{\bs{m}\ldots\bs{m}}\hoch{l_{1}\ldots l_{k'}}\partial_{l_{k'}}\rho_{l_{k'-1}\ldots l_{1}\bs{m}\ldots\bs{m}}+\nonumber \\
 &  & -(-)^{k-k'}(k')!\left(\zwek{r}{k'}\right)\partial_{\bs{m}}K_{\bs{m}\ldots\bs{m}}\hoch{l_{1}\ldots l_{k'}}\rho_{l_{k'}\ldots l_{1}\bs{m}\ldots\bs{m}}\end{eqnarray}
 \rem{nur wahr, wenn man auf Formen wirkt!! $[\de,\ip_v]w\neq \Lie_v w$ ...
Lie Ableitung bisher nur im Anhang definiert! Dort vielleicht eine kurze subsection mit $\ip$ und Cartan-Gleichungen, aber ohne $\ip^{(p)}$!?}$\Lie_{K}\rho$ is the natural generalization of the Lie derivative
with respect to vectors acting on forms, which is given similarly
$\Lie_{v}\rho=[\ip_{v},\de\,]\rho$. \index{$L_v$@$\protect\Lie_v$|itext{Lie derivative}}\index{$([])$@$[\ldots,\ldots]$!$[\de,\ip_v]$|itext{$\protect\Lie_v$}}As
$\ip_{K}$ is a higher order derivative, also $\Lie_{K}$ is a higher
order derivative. Nevertheless, it will be called \textbf{Lie\index{Lie derivative}\index{Lie derivative!with respect to multivector valued form}
derivative\index{derivative!Lie $\sim$} with respect to} $K$ in
this thesis. Let us again recall this fact: if $p_{k}$ appears in
a combination like $\de K$, there is a well defined geometric meaning
and $\de K$ is up to a sign nothing else than the Lie derivative
with respect to $K$, when embedded in the space of differential operators
on forms. The commutator with the exterior derivative is a natural
differential in the space of differential operators acting on forms,
and via the embedding it induces the differential $\de$ on $K$.
It should perhaps be stressed that the above definition of $\de K$
corresponds to an extended action of the exterior derivative which
acts also on the basis elements of the tangent space\index{conjugate momentum}\index{$p_m$}\index{$d$@$\de(\pe_m)$}\begin{eqnarray}
\de(\pe_{m}) & = & p_{m}\label{eq:d-auf-partial}\end{eqnarray}
This approach will enable us to give explicit coordinate expressions
for the derived bracket of multivector valued forms even in the general
case where the result is not a tensor: In the space of differential
operators on forms, we have the commutator $[\ip_{K},\ip_{L}]$ and
its derived bracket (\ref{eq:derived-bracketI}) $[\ip_{K},_{\de}\ip_{L}]\equiv[[\ip_{K},\de\,],\ip_{L}]$,
while on the space of multivector valued forms we have the algebraic
bracket $\left[K,L\right]^{\Delta}$ and want to define its derived
bracket up to a sign as $[\de K,L]^{\Delta}$. To this end we also
have to extend the definition (\ref{eq:bc-interior-pproduct-I},\ref{eq:bc-interior-pproductII})
of $\ip^{(p)}$, which appears in the explicit expression of the algebraic
bracket in (\ref{eq:bc-algebraic-bracketI}) to objects that contain
$p_{k}$. This is done in a way that the old equations for the algebraic
bracket remain formally the same. So let us define\index{$i_T$@$\ip_{T^{(t,t',t'')}}^{(p)}$}%
\footnote{\index{footnote!\thefoot. combinatorical remark}Note that $\sum_{q=0}^{p}\left(\zwek{t'}{q}\right)\left(\zwek{t''}{p-q}\right)=\left(\zwek{t'+t''}{p}\right)$
\rem{and $\sum_{q=0}^{p}\left(\zwek{p}{q}\right)=2^{p}$}$\qquad\fussend$%
}\begin{eqnarray}
\hspace{-.5cm}\lqn{\ip_{T^{(t,t',t'')}}^{(p)}\equiv}\nonumber \\
 &  & \hspace{-.5cm}\equiv\sum_{q=0}^{p}\left(\zwek{t'}{q}\right)\left(\zwek{t''}{p-q}\right)T_{\bs{m}\ldots\bs{m}}\hoch{\bs{n}\ldots\bs{n}i_{1}\ldots i_{q}\,,\, i_{q+1}\ldots i_{p}k_{1}\ldots k_{t''-p+q}}p_{k_{1}}\cdots p_{k_{t''-p+q}}\frac{\partial^{p}}{\partial\bs{c}^{i_{1}}\ldots\partial\bs{c}^{i_{q}}\partial x^{i_{q+1}}\ldots\partial x^{i_{p}}}\qquad\quad\\
 &  & \hspace{-.5cm}=\frac{1}{p!}\sum_{q=0}^{p}\left(\zwek{p}{q}\right)T\frac{\lpartial^{p}}{\partial p_{i_{p}}\ldots\partial p_{i_{q+1}}\partial\be_{i_{q}}\ldots\partial\be_{i_{1}}}\frac{\partial^{p}}{\partial\bs{c}^{i_{1}}\ldots\partial\bs{c}^{i_{q}}\partial x^{i_{q+1}}\ldots\partial x^{i_{p}}}\end{eqnarray}
\rem{\begin{eqnarray*}
\ip_{T^{(t,t',t'')}}^{(p)}\tilde{T}^{(\tilde{t},\tilde{t}',\tilde{t}'')} & = & \sum_{q=0}^{p}(-)^{(t'-q)(\tilde{t}-q)}q!\left(\zwek{\tilde{t}}{q}\right)\left(\zwek{t'}{q}\right)\left(\zwek{t''}{p-q}\right)\times\\
 &  & \times T_{\bs{m}\ldots\bs{m}}\hoch{\bs{n}\ldots\bs{n}i_{1}\ldots i_{q}\,,\, i_{q+1}\ldots i_{p}k\ldots k}\frac{\partial^{p-q}}{\partial x^{i_{q+1}}\ldots\partial x^{i_{p}}}\tilde{T}_{i_{q}\ldots i_{1}\bs{m}\ldots\bs{m}}\hoch{\bs{n}\ldots\bs{n}\,,\, k\ldots k}\end{eqnarray*}
} For $p=t'+t''$ it coincides with the full interior product (\ref{eq:i-TII}):
$\ip_{T^{(t,t',t'')}}^{(t'+t'')}=\ip_{T^{(t,t',t'')}}$. In addition
we have with this definition (after some calculation) $\ip_{\de T}^{(p)}=[\de\,,\ip_{T}^{(p)}]$
and in particular \begin{equation}
\ip_{\de K}^{(p)}=[\de\,,\ip_{K}^{(p)}]\end{equation}
and the equations for the algebraic bracket (\ref{eq:bc-algebraic-bracket})-(\ref{eq:bc-algebraic-bracketI}))
indeed remain formally the same for objects containing $p_{m}$\index{$([])$@$[\ldots,\ldots]^\Delta$!$[T^{(t,t',t'')},\tilde{T}^{(\tilde{t},\tilde{t}',\tilde{t}'')}]^{\Delta}$}\index{$([])$@$[\ldots,\ldots]_{(1)}^\Delta$!$[T,\tilde{T}]_{(1)}^\Delta$}\index{$([])$@$[\ldots,\ldots]$!$[\ip_{T^{(t,t',t'')}},\ip_{\tilde{T}^{(\tilde{t},\tilde{t}',\tilde{t}'')}}]$}
\begin{eqnarray}
[\ip_{T^{(t,t',t'')}},\ip_{\tilde{T}^{(\tilde{t},\tilde{t}',\tilde{t}'')}}] & \equiv & \ip_{\left[T,\tilde{T}\right]^{\Delta}}\\
\ip_{T}\ip_{\tilde{T}} & = & \sum_{p\geq0}\ip_{\ip_{T}^{(p)}\tilde{T}}\\
{}[T^{(t,t',t'')},\tilde{T}^{(\tilde{t},\tilde{t}',\tilde{t}'')}]^{\Delta} & \equiv & \sum_{p\geq1}\underbrace{\ip_{T}^{(p)}\tilde{T}-(-)^{(t-t')(\tilde{t}-\tilde{t}')}\ip_{\tilde{T}}^{(p)}T}_{\equiv[T,\tilde{T}]_{(p)}^{\Delta}}\\
{}[T,\tilde{T}]_{(1)}^{\Delta} & = & \left\{ T,\tilde{T}\right\} \label{eq:bc-big-br-TtildeT}\end{eqnarray}
 which we can again rewrite in terms of {}``quantum''-operators
(\ref{eq:quantization}) as\index{$([])$@$[\ldots,\ldots]$!$[\hat{T}^{(k,k')},\hat{\tilde{T}}^{(l,l')}]$}\begin{eqnarray}
\left[\hat{T}^{(k,k')},\hat{\tilde{T}}^{(l,l')}\right] & = & \sum_{p\geq1}\left(\frac{\hbar}{i}\right)^{p}\widehat{\left[T,\tilde{T}\right]_{(p)}^{\Delta}}=\label{eq:quantum-commutator-T-Ttilde}\\
 & = & \left(\frac{\hbar}{i}\right)\widehat{\left\{ T,\tilde{T}\right\} }+\sum_{p\geq2}\left(\frac{\hbar}{i}\right)^{p}\widehat{\left[T,\tilde{T}\right]_{(p)}^{\Delta}}\end{eqnarray}
It should be stressed that -- although very useful -- $\ip^{(p)}$
is unfortunately NOT a geometric operation any longer in general,
in the sense that $\ip_{\de K}^{(p)}L$ and also $\ip_{L}^{(p)}\de K$
do not have a well defined geometric meaning, although $\de K$ and
$L$ have. $\ip_{\de K}\rho$ and $\ip_{K}^{(p)}L$ are in contrast
well defined. $\ip_{\de K}^{(p)}L$, for example, should rather be
understood as a building block of a coordinate calculation which combines
only in certain combinations (e.g. the bracket $[\,,\,]^{\Delta}$)
to s.th. geometrically meaningful. 

We are now ready to define the \textbf{derived\index{derived bracket!of the algebraic bracket}
bracket}\index{bracket!derived $\sim$ of the algebraic bracket} of
the algebraic bracket for multivector valued forms (see footnote \ref{foot-derived-bracket}
on page \pageref{foot-derived-bracket}) \index{$([])$@$[\ldots\bs{,}\ldots]$!$[K^{(k,k')}\bs{,}L^{(l,l')}]$}\index{$([])$@$[\ldots\bs{,}\ldots]$|iText{derived bracket of $[\ldots,\ldots]^\Delta$ by $\de$ }}\index{$([])$@$[\ldots,_\de\ldots]^\Delta$|see{$[\ldots\bs{,}\ldots]$}}\begin{eqnarray}
\hspace{-1cm}\left[K^{(k,k')}\bs{,}L^{(l,l')}\right] & \equiv & \left[K,_{\de\,}L\right]^{\Delta}\equiv-(-)^{k-k'}\left[\de K,L\right]^{\Delta}=\label{eq:bc-derived-bracketI}\\
 & = & \sum_{p\geq1}-(-)^{k-k'}\ip_{\de K}^{(p)}L+(-)^{(k+1-k')(l-l')+k-k'}\ip_{L}^{(p)}\de K=\label{eq:bc-derived-bracketII}\\
 & = & \sum_{p\geq1}-(-)^{k-k'}\ip_{\de K}^{(p)}L+(-)^{(k-k'+1)(l-l'+1)}(-)^{l-l'}\ip_{\de L}^{(p)}K+(-)^{(k-k')(l-l')+k-k'}\de(\ip_{L}^{(p)}K)\qquad\label{eq:bc-derived-bracketIII}\end{eqnarray}
 The result is geometrical in the sense that after embedding via the
interior product it is a well defined operator acting on forms. This
is the case, because due to our extended definitions we have for \textbf{all}
multivector valued forms the relation \begin{eqnarray}
\left[[\ip_{K},\de],\ip_{L}\right] & = & \ip_{\left[K^{(k,k')}\bs{,}L^{(l,l')}\right]}\end{eqnarray}
 and the lefthand side is certainly a well defined geometric object.\rem{Achtung! Lie-Derivative und interior Product nur geom wohldef, wenn sie auf Form wirken!! betrachte z.B. die Lie-Ableitung von einem Vektor nach einem Vektor.
(\ref{eq:dK-und-Lie}): Lie-derivative of order $p$ via\index{$L_K$@$\protect\Lie_K^{(p)}$}\begin{eqnarray*}
\Lie_{K}^{(p)} & \equiv & [\ip_{K}^{(p)},\de\,]=-(-)^{k-k'}\ip_{\de K}^{(p)}\end{eqnarray*}
}  A considerable effort went into getting a correct coordinate form
for the general derived bracket and for that reason, let us quickly
have a glance at the final result, although it is kind of ugly:%
\footnote{\index{footnote!\thefoot. building blocks of $[K\bs{,}L]$}The building
blocks are \begin{eqnarray*}
\ip_{\de K}^{(p)}L & = & (-)^{(k'-p)(l-p)}p!\left(\zwek{k'}{p}\right)\left(\zwek{l}{p}\right)\partial_{\bs{m}}K_{\bs{m}\ldots\bs{m}}\hoch{\bs{n}\ldots\bs{n}i_{1}\ldots i_{p}}L_{i_{p}\ldots i_{1}\bs{m}\ldots\bs{m}}\hoch{\bs{n}\ldots\bs{n}}+\\
 &  & -(-)^{k-k'}(-)^{(k'-1-p)(l-p)}(p+1)!\left(\zwek{k'}{p+1}\right)\left(\zwek{l}{p}\right)K_{\mm}\hoch{\nn i_{1}\ldots i_{p}k}L_{i_{p}\ldots i_{1}\bs{m}\ldots\bs{m}}\hoch{\bs{n}\ldots\bs{n}}p_{k}+\\
 &  & -(-)^{k-k'}(-)^{(k'-p)(l-p+1)}p!\left(\zwek{k'}{p}\right)\left(\zwek{l}{p-1}\right)K_{\mm}\hoch{\nn i_{1}\ldots i_{p-1}i_{p}}\partial_{i_{p}}L_{i_{p-1}\ldots i_{1}\bs{m}\ldots\bs{m}}\hoch{\bs{n}\ldots\bs{n}}\end{eqnarray*}
\begin{eqnarray*}
\ip_{L}^{(p)}\de K & = & (-)^{(l'-p)(k+1-p)+p}p!\left(\zwek{k}{p}\right)\left(\zwek{l'}{p}\right)L_{\bs{m}\ldots\bs{m}}\hoch{\bs{n}\ldots\bs{n}k_{1}\ldots k_{p}}\partial_{\bs{m}}K_{k_{p}\ldots k_{1}\bs{m}\ldots\bs{m}}\hoch{\bs{n}\ldots\bs{n}}+\\
 &  & +(-)^{(l'-p)(k+1-p)}p!\left(\zwek{k}{p-1}\right)\left(\zwek{l'}{p}\right)L_{\bs{m}\ldots\bs{m}}\hoch{\bs{n}\ldots\bs{n}k_{1}\ldots k_{p-1}l}\partial_{l}K_{k_{p-1}\ldots k_{1}\bs{m}\ldots\bs{m}}\hoch{\bs{n}\ldots\bs{n}}+\\
 &  & -(-)^{k-k'}(-)^{(l'-p)(k-p)}k'\cdot p!\left(\zwek{k}{p}\right)\left(\zwek{l'}{p}\right)L_{\bs{m}\ldots\bs{m}}\hoch{\bs{n}\ldots\bs{n}k_{1}\ldots k_{p}}K_{k_{p}\ldots k_{1}\bs{m}\ldots\bs{m}}\hoch{\bs{n}\ldots\bs{n}k}p_{k}\qquad\fussend\end{eqnarray*}
}\index{$([])$@$[\ldots\bs{,}\ldots]$!$[Kb]$@$[K\bs{,}L]$|itext{coordinate expression}}
\begin{eqnarray}
\left[K\bs{,}L\right] & = & \sum_{p\geq1}-(-)^{k-k'}(-)^{(k'-p)(l-p)}p!\left(\zwek{l}{p}\right)\left(\zwek{k'}{p}\right)\partial_{\bs{m}}K_{\bs{m}\ldots\bs{m}}\hoch{\bs{n}\ldots\bs{n}l_{1}\ldots l_{p}}L_{l_{p}\ldots l_{1}\bs{m}\ldots\bs{m}}\hoch{\bs{n}\ldots\bs{n}}+\nonumber \\
 &  & +(-)^{k+k'l+k'+p+pl+pk'}p!\left(\zwek{k}{p}\right)\left(\zwek{l'}{p}\right)\partial_{\bs{m}}K_{\bs{m}\ldots\bs{m}k_{p}\ldots k_{1}}\hoch{\bs{n}\ldots\bs{n}}L_{\bs{m}\ldots\bs{m}}\hoch{k_{1}\ldots k_{p}\bs{n}\ldots\bs{n}}+\nonumber \\
 &  & -(-)^{k'l+k'+pl+pk'}p!\left(\zwek{k}{p-1}\right)\left(\zwek{l'}{p}\right)\partial_{l}K_{\bs{m}\ldots\bs{m}k_{p-1}\ldots k_{1}}\hoch{\bs{n}\ldots\bs{n}}L_{\bs{m}\ldots\bs{m}}\hoch{k_{1}\ldots k_{p-1}l\bs{n}\ldots\bs{n}}+\nonumber \\
 &  & +(-)^{(k'-p)(l-p+1)}p!\left(\zwek{l}{p-1}\right)\left(\zwek{k'}{p}\right)K_{\bs{m}\ldots\bs{m}}\hoch{\bs{n}\ldots\bs{n}l_{1}\ldots l_{p-1}k}\partial_{k}L_{l_{p-1}\ldots l_{1}\bs{m}\ldots\bs{m}}\hoch{\bs{n}\ldots\bs{n}}+\nonumber \\
 &  & +(-)^{(k'-1-p)(l-p)}p!(k'-p)\left(\zwek{l}{p}\right)\left(\zwek{k'}{p}\right)K_{\bs{m}\ldots\bs{m}}\hoch{\bs{n}\ldots\bs{n}l_{1}\ldots l_{p}k}L_{l_{p}\ldots l_{1}\bs{m}\ldots\bs{m}}\hoch{\bs{n}\ldots\bs{n}}p_{k}+\nonumber \\
 &  & -(-)^{k'l+l+pk'+lp}k'\cdot p!\left(\zwek{k}{p}\right)\left(\zwek{l'}{p}\right)K_{\bs{m}\ldots\bs{m}k_{p}\ldots k_{1}}\hoch{\bs{n}\ldots\bs{n}k}L_{\bs{m}\ldots\bs{m}}\hoch{k_{1}\ldots k_{p}\bs{n}\ldots\bs{n}}p_{k}\label{eq:bc-derived-bracket-coord}\end{eqnarray}
The result is only a tensor, when both terms with $p_{k}$ on the
righthand side vanish, although the complete expression is in general
geometrically well-defined when considered to be a differential operator
acting on forms via $\ip_{\left[K\bs{,}L\right]}$ as this equals
per definition the well-defined $[[\ip_{K},\de],\ip_{L}]$. The above
coordinate form reduces in the appropriate cases to vector Lie-bracket\index{Lie bracket!of vector fields}\index{bracket!vector Lie $\sim$},
Schouten\index{Schouten bracket}-bracket\index{bracket!Schouten},
and (up to a total derivative) to the (Fr\"ohlicher\index{Fr\"ohlicher Nijenhuis bracket})-Nijenhuis\index{Nijenhuis bracket}-bracket\index{bracket!Fröhlicher Nijenhuis $\sim$}.
If one allows as well sums of a vector and a 1-form, we get the Dorfman\index{Dorfman bracket}
bracket\index{bracket!Dorfman $\sim$}, and also the sum of a vector
and a general form gives a result without $p$. \rem{ In the above
equation we have to hide $p_{k}$ in a total derivative in order to
have a chance to get a tensorial bracket! But let us first start again
with (\ref{eq:bc-derived-bracketIII})\begin{eqnarray*}
\left[K\bs{,}L\right] & = & \sum_{p=1}^{\infty}-(-)^{k-k'}(-)^{(k'-p)(l-p)}p!\left(\zwek{l}{p}\right)\left(\zwek{k'}{p}\right)\partial_{\bs{m}}K_{\bs{m}\ldots\bs{m}}\hoch{\bs{n}\ldots\bs{n}l_{1}\ldots l_{p}}L_{l_{p}\ldots l_{1}\bs{m}\ldots\bs{m}}\hoch{\bs{n}\ldots\bs{n}}+\\
 &  & +(-)^{(k'-p)(l-p+1)}p!\left(\zwek{l}{p-1}\right)\left(\zwek{k'}{p}\right)K_{\bs{m}\ldots\bs{m}}\hoch{\bs{n}\ldots\bs{n}l_{1}\ldots l_{p-1}k}\partial_{k}L_{l_{p-1}\ldots l_{1}\bs{m}\ldots\bs{m}}\hoch{\bs{n}\ldots\bs{n}}+\\
 &  & +(-)^{(k'-1-p)(l-p)}p!(k'-p)\left(\zwek{l}{p}\right)\left(\zwek{k'}{p}\right)K_{\bs{m}\ldots\bs{m}}\hoch{\bs{n}\ldots\bs{n}l_{1}\ldots l_{p}k}L_{l_{p}\ldots l_{1}\bs{m}\ldots\bs{m}}\hoch{\bs{n}\ldots\bs{n}}p_{k}+\\
 &  & +(-)^{(k-k'+1)(l-l'+1)}(-)^{l-l'}(-)^{(l'-p)(k-p)}p!\left(\zwek{k}{p}\right)\left(\zwek{l'}{p}\right)\partial_{\bs{m}}L_{\bs{m}\ldots\bs{m}}\hoch{\bs{n}\ldots\bs{n}l_{1}\ldots l_{p}}K_{l_{p}\ldots l_{1}\bs{m}\ldots\bs{m}}\hoch{\bs{n}\ldots\bs{n}}+\\
 &  & -(-)^{(k-k'+1)(l-l'+1)}(-)^{(l'-p)(k-p+1)}p!\left(\zwek{k}{p-1}\right)\left(\zwek{l'}{p}\right)L_{\bs{m}\ldots\bs{m}}\hoch{\bs{n}\ldots\bs{n}l_{1}\ldots l_{p-1}k}\partial_{k}K_{l_{p-1}\ldots l_{1}\bs{m}\ldots\bs{m}}\hoch{\bs{n}\ldots\bs{n}}+\\
 &  & -(-)^{(k-k'+1)(l-l'+1)}(-)^{(l'-1-p)(k-p)}p!(l'-p)\left(\zwek{k}{p}\right)\left(\zwek{l'}{p}\right)L_{\bs{m}\ldots\bs{m}}\hoch{\bs{n}\ldots\bs{n}l_{1}\ldots l_{p}k}K_{l_{p}\ldots l_{1}\bs{m}\ldots\bs{m}}\hoch{\bs{n}\ldots\bs{n}}p_{k}+\\
 &  & +(-)^{(k-k')(l-l'+1)}(-)^{(l'-p)(k-p)}p!\left(\zwek{k}{p}\right)\left(\zwek{l'}{p}\right)\de\,\left(L_{\bs{m}\ldots\bs{m}}\hoch{\bs{n}\ldots\bs{n}k_{1}\ldots k_{p}}K_{k_{p}\ldots k_{1}\bs{m}\ldots\bs{m}}\hoch{\bs{n}\ldots\bs{n}}\right)\end{eqnarray*}
} 

Due to our extended definition of the exterior derivative, we can
also define the \textbf{derived\index{derived bracket!of the Poisson bracket}\index{derived bracket!of the big bracket}
bracket\index{bracket!derived $\sim$ of the big $\sim$} of the big\index{big bracket!derived bracket of the $\sim$}
bracket} (the Poisson\index{Poisson bracket!derived bracket of the $\sim$}
bracket) via\index{$([])$@$[\ldots,_\de\ldots]_{(1)}^\Delta$|itext{derived bracket of the big bracket by $\de$}}
\begin{eqnarray}
\left[K^{(k,k')},_{\de\,}L^{(l,l')}\right]_{(1)}^{\Delta} & \equiv & -(-)^{k-k'}\left[\de K,L\right]_{(1)}^{\Delta}=\label{eq:bc-derived-of-bigbracket}\\
 & = & -(-)^{k-k'}\left\{ \de K,L\right\} \end{eqnarray}
which is just the $p=1$ term of the full derived bracket with the
explicit coordinate expression\begin{eqnarray}
\left[K,_{\de\,}L\right]_{(1)}^{\Delta} & = & -(-)^{k-k'}(-)^{(k'-1)(l-1)}lk'\partial_{\bs{m}}K_{\bs{m}\ldots\bs{m}}\hoch{\bs{n}\ldots\bs{n}l_{1}}L_{l_{1}\bs{m}\ldots\bs{m}}\hoch{\bs{n}\ldots\bs{n}}+\nonumber \\
 &  & -(-)^{k+k'l+l}kl'\partial_{\bs{m}}K_{\bs{m}\ldots\bs{m}k_{1}}\hoch{\bs{n}\ldots\bs{n}}L_{\bs{m}\ldots\bs{m}}\hoch{k_{1}\bs{n}\ldots\bs{n}}+\nonumber \\
 &  & -(-)^{k'l+l}l'\partial_{l}K_{\bs{m}\ldots\bs{m}}\hoch{\bs{n}\ldots\bs{n}}L_{\bs{m}\ldots\bs{m}}\hoch{l\bs{n}\ldots\bs{n}}+\nonumber \\
 &  & +(-)^{(k'-1)l}k'K_{\bs{m}\ldots\bs{m}}\hoch{\bs{n}\ldots\bs{n}k}\partial_{k}L_{\bs{m}\ldots\bs{m}}\hoch{\bs{n}\ldots\bs{n}}+\nonumber \\
 &  & +(-)^{k'(l-1)}(k'-1)lk'K_{\bs{m}\ldots\bs{m}}\hoch{\bs{n}\ldots\bs{n}l_{1}k}L_{l_{1}\bs{m}\ldots\bs{m}}\hoch{\bs{n}\ldots\bs{n}}p_{k}+\nonumber \\
 &  & -(-)^{k'l+k'}k'kl'K_{\bs{m}\ldots\bs{m}k_{1}}\hoch{\bs{n}\ldots\bs{n}k}L_{\bs{m}\ldots\bs{m}}\hoch{k_{1}\bs{n}\ldots\bs{n}}p_{k}\label{eq:bc-derived-of-bigbracket-coord}\end{eqnarray}
\begin{eqnarray}
\left[K\bs{,}L\right] & = & \left[K,_{\de\,}L\right]_{(1)}^{\Delta}-(-)^{k-k'}\sum_{p\geq2}\left[\de K,L\right]_{(p)}^{\Delta}\label{eq:derived-complete-vs-big}\end{eqnarray}
Like the big bracket itself, also its derived bracket takes a very
pleasant coordinate form for generalized multivectors (see (\ref{eq:derived-of-big-generalized})
on page \pageref{eq:derived-of-big-generalized}). \rem{ or\begin{eqnarray*}
\left[K\bs{,}_{\de\,}L\right]_{(1)}^{\Delta} & = & -(-)^{k-k'}(-)^{(k'-1)(l-1)}lk'\partial_{\bs{m}}K_{\bs{m}\ldots\bs{m}}\hoch{\bs{n}\ldots\bs{n}j}L_{j\bs{m}\ldots\bs{m}}\hoch{\bs{n}\ldots\bs{n}}+\\
 &  & +(-)^{(k'-1)l}k'K_{\bs{m}\ldots\bs{m}}\hoch{\bs{n}\ldots\bs{n}j}\partial_{j}L_{\bs{m}\ldots\bs{m}}\hoch{\bs{n}\ldots\bs{n}}+\\
 &  & +(-)^{k'(l-1)}lk'(k'-1)K_{\bs{m}\ldots\bs{m}}\hoch{\bs{n}\ldots\bs{n}ji}L_{j\bs{m}\ldots\bs{m}}\hoch{\bs{n}\ldots\bs{n}}p_{i}+\\
 &  & +(-)^{(k-k'+1)(l-l'+1)}(-)^{l-l'}(-)^{(l'-1)(k-1)}kl'\partial_{\bs{m}}L_{\bs{m}\ldots\bs{m}}\hoch{\bs{n}\ldots\bs{n}j}K_{j\bs{m}\ldots\bs{m}}\hoch{\bs{n}\ldots\bs{n}}+\\
 &  & -(-)^{(k-k'+1)(l-l'+1)+(l'-1)k}l'L_{\bs{m}\ldots\bs{m}}\hoch{\bs{n}\ldots\bs{n}j}\partial_{j}K_{\bs{m}\ldots\bs{m}}\hoch{\bs{n}\ldots\bs{n}}+\\
 &  & -(-)^{(k-k'+1)(l-l'+1)}(-)^{l'(k-1)}kl'(l'-1)L_{\bs{m}\ldots\bs{m}}\hoch{\bs{n}\ldots\bs{n}ji}K_{j\bs{m}\ldots\bs{m}}\hoch{\bs{n}\ldots\bs{n}}p_{i}+\\
 &  & +(-)^{(k-k')(l-l'+1)}(-)^{(l'-1)(k-1)}kl'\de\,\left(L_{\bs{m}\ldots\bs{m}}\hoch{\bs{n}\ldots\bs{n}j}K_{j\bs{m}\ldots\bs{m}}\hoch{\bs{n}\ldots\bs{n}}\right)\end{eqnarray*}
} In contrast to the full derived bracket, we have no guarantee for
this derived bracket to be geometrical itself. \rem{coordinate form for generalized vectors?Dorfmann-Schouten-bracket}

\rem{Gegenbeispiel: Consider the case $K=K^{(1,1)}=K_{\bs{m}}\hoch{\bs{n}}$
and $L=v^{(2)}=v^{\bs{m}\bs{m}}$ with \begin{eqnarray*}
\de K & = & \partial_{\bs{m}}K_{\bs{m}}\hoch{\bs{n}}-K_{\bs{m}}\hoch{i}p_{i}\end{eqnarray*}
 \begin{eqnarray*}
\left[K^{(1,1)},_{\de}v^{(2)}\right] & = & \ip_{v^{(2)}}^{(1)}\de K^{(1,1)}+\ip_{v^{(2)}}^{(2)}\de K^{(1,1)}-\ip_{\de K}^{(1)}v\end{eqnarray*}
It turns out that for $\ip_{\ip_{v^{(2)}}^{(1)}\de K^{(1,1)}+\ip_{v^{(2)}}^{(2)}\de K^{(1,1)}-\ip_{\de K}^{(1)}v}\rho^{(r)}$
the Christoffel symbols cancel only in the complete combination, including
$\ip^{(2)}$! (Detailliere Rechnung in Note) } \rem{Remembering
that $\oo=\ce^{k}p_{k}$, which implies $\ip_{\oo}=\ce^{k}\partl{x^{k}}$,
we should note that there are now several ways to express the exterior
derivative\begin{eqnarray*}
\oo & \equiv & \ce^{k}p_{k},\qquad\ip_{Q}=\ce^{k}\partl{x^{k}}=\frac{i}{\hbar}\hat{\oo}\\
\ip_{\oo}\rho & = & \frac{i}{\hbar}\hat{\oo}\rho=\de\rho\qquad\ip_{\oo}K\neq\de K\quad!\\
{}[\ip_{\oo},\ip_{K}] & = & \ip_{[\oo,K]^{\Delta}}=\ip_{\{\oo,K\}}=\ip_{\de K}\\
{}[\hat{\oo},\hat{K}] & = & \left(\frac{\hbar}{i}\right)\widehat{\de K}\\
\left[[\ip_{K},\de\,],\ip_{L}\right] & = & -(-)^{k-k'}\left[\ip_{\de K},\ip_{L}\right]=\left[\left[\ip_{K},\ip_{\oo}\right],\ip_{L}\right]=\ip_{\left[\left[K,\oo\right]^{\Delta},L\right]^{\Delta}}\end{eqnarray*}
} 

Let us eventually note how one can easily adjust the extended\index{extended exterior derivative!twisted}
exterior\index{exterior derivative!twisted $\sim$} derivative to
the twisted\index{twisted!exterior derivative} case:\index{$d_H$@$\de_H$|itext{twisted exterior derivative}}\index{H@$H$-twist}\index{$H$|itext{3-form}}\begin{eqnarray}
[\de+H\wedge\,,\,\ip_{K}] & \equiv & \ip_{\de_{H}K}\\
\de_{H}K & = & \de K+\left[H,K\right]^{\Delta}=\de K-(-)^{k-k'}\sum_{p\geq1}\ip_{K}^{(p)}H\end{eqnarray}
with $H$ being an odd closed differential form. It should be stressed
that $\de+H\wedge$ is not a differential, but on the operator level
its commutator $[\de+H\wedge,\ldots]$ is a differential and thus
the above defined $\de_{H}$ is a differential as well.

\section{Sigma-Models}

\label{sub:Sigma-Models} A sigma\index{sigma-model} model is a field
theory whose fields are embedding functions from a world\index{world-volume|itext{$\Sigma$}}-volume
$\Sigma$\index{$\Sigma$|itext{world-volume}} into a target\index{target space|itext{$M$}}
space $M$\index{$M$|itext{target space}}, like in string theory.
So far there was no sigma-model explicitly involved into our considerations.
One can understand the previous subsection simply as a convenient
way to formulate some geometry. The phase space introduced there,
however, is like the phase space of a (point particle) sigma model
with only one world-volume dimension -- the time -- which is not showing
up in the off-shell phase-space. Let us now naively consider the same
setting like before as a sigma model with the coordinates\index{coordinates!target space $\sim$ $x^m$}
$x^{m}$\index{$x^m$|itext{target space coordinates}} depending on
some worldsheet coordinates%
\footnote{\index{footnote!\thefoot. worldvolume index}The index $\mu$ will
not include the worldvolume time, when considering the phase space,
but it will contain the time in the Lagrangian formalism. As this
should be clear from the context, there will be no notational distinction.
$\qquad\fussend$%
} $\sigma^{\mu}$\index{$\sigma^\mu$|itext{worldvolume coordinates $\sigma^\mu$}}\index{coordinates!worldvolume $\sim$}.
The resulting model has a very special field content, because its
anticommuting fields $\ce^{m}(\sigma)$ have the same index structure
as the embedding coordinate $x^{m}(\sigma)$. In one and two worldvolume-dimensions,
$\ce^{m}$ can be regarded as worldvolume-fermions, and this will
be used in the stringy application in \ref{sub:Zabzine}. In general
worldvolume dimensions, $\ce^{m}$ could be seen as ghosts, leading
to a topological theory. In any case the dimension of the worldvolume
will not yet be fixed, as the described mechanism does not depend
on it. 

A multivector valued form on a $C^{\infty}$-manifold $M$ can locally
be regarded as an analytic function of $x^{m},\de x^{m}\equiv\ce^{m}$
and $\pe_{m}\equiv\be_{m}$\index{$K^{(k,k')}(x,\ce,\be)$}\begin{eqnarray}
K^{(k,k')}(x,\de x,\pe) & = & K_{m_{1}\ldots m_{k}}\hoch{n_{1}\ldots n_{k'}}(x)\de x^{m_{1}}\wedge\cdots\wedge\de x^{m_{k}}\wedge\pe_{n_{1}}\wedge\cdots\wedge\pe_{n_{k'}}=\label{eq:K-als-analytische-FunktionI}\\
 & \equiv & K_{m_{1}\ldots m_{k}}\hoch{n_{1}\ldots n_{k'}}(x)\ce^{m_{1}}\cdots\ce^{m_{k}}\be_{n_{1}}\cdots\be_{n_{k'}}=K^{(k,k')}(x,\ce,\be)\label{eq:K-als-analytische-FunktionII}\end{eqnarray}
 For sigma models, $x^{m}\To x^{m}(\sigma),p_{m}\To p_{m}(\sigma),\ce^{m}\To\ce^{m}(\sigma)$
and $\be_{m}\To\be_{m}(\sigma)$ become dependent on the worldvolume
variables $\sigma^{\mu}$. They are, however, for every $\sigma$
valid arguments of the function $K$. Frequently only the worldvolume
coordinate $\sigma$ will then be denoted as new argument of $K$,
which has to be understood in the following sense\index{$K^{(k,k')}(\sigma)$}\begin{eqnarray}
\hspace{-1cm}K^{(k,k')}(\sigma)\equiv K^{(k,k')}\left(x(\sigma),\ce(\sigma),\be(\sigma)\right) & = & K_{m_{1}\ldots m_{k}}\hoch{n_{1}\ldots n_{k'}}\left(x(\sigma)\right)\cdot\ce^{m_{1}}(\sigma)\cdots\ce^{m_{k}}(\sigma)\be_{n_{1}}(\sigma)\cdots\be_{n_{k'}}(\sigma)\label{eq:K-of-sigma}\end{eqnarray}
Also functions depending on $p_{m}$, like $\de K(x,\ce,\be,p)$ in
(\ref{eq:dK-coord}), or more general a function $T^{(t,t',t'')}(x,\ce,\be,p)$
as in (\ref{eq:Tcbp}) are denoted in this way\index{$T^{(t,t',t'')}(\sigma)$}\index{$o$@$\oo(\sigma)$}\index{$dK$@$\de K(\sigma)$}\begin{eqnarray}
T^{(t,t',t'')}(\sigma) & \equiv & T^{(t,t',t'')}\left(x(\sigma),\ce(\sigma),\be(\sigma),p(\sigma)\right)\quad(\textrm{see }(\ref{eq:Tcbp}))\label{eq:T-of-sigma}\\
\textrm{e.g. }\de K(\sigma) & \equiv & \de K\left(x(\sigma),\ce(\sigma),\be(\sigma),p(\sigma)\right)\quad(\textrm{see }(\ref{eq:dK-coord}))\label{eq:dK-of-sigma}\\
\textrm{or }\oo(\sigma) & \equiv & \oo\left(\ce(\sigma),p(\sigma)\right)=\ce^{m}(\sigma)p_{m}(\sigma)\quad(\textrm{see }(\ref{eq:BRST-op}))\label{eq:o-of-sigma}\end{eqnarray}
 The expression $\de K(\sigma)$ should \textbf{NOT} be mixed up with
the world-volume\index{world-volume exterior derivative $\dew$} exterior\index{exterior derivative!world-volume $\sim$ $\dew$}
derivative of $K$ which will be denoted by\index{$d^w$@$\dew$|itext{world-volume exterior derivative}}
$\dew K(\sigma)$.%
\footnote{\index{footnote!\thefoot. confusion about $\dew$} It is much better
to mix it up with a BRST transformation or with something similar
to a worldsheet supersymmetry transformation. We will come to that
later in subsection \ref{sub:Zabzine}. To make confusion perfect,
it should be added that in contrast it is not completely wrong in
subsection \ref{sub:antibracket} to mix up the target space exterior
derivative with the worldsheet exterior derivative...$\qquad\fussend$%
} Every operation of the previous section, like $\ip_{K}^{(p)}L$ or
the algebraic or derived brackets leads again to functions of $x,\ce,\be$
and sometimes $p$. Let us use for all of them the notation as above,
e.g. for the derived bracket of the big bracket (\ref{eq:bc-derived-of-bigbracket},\ref{eq:bc-derived-of-bigbracket-coord})\begin{eqnarray}
\left[K^{(k,k')},_{\de\,}L^{(l,l')}\right]_{(1)}^{\Delta}(\sigma) & \equiv & \left[K^{(k,k')}\bs{,}L^{(l,l')}\right]_{(1)}^{(\Delta)}\left(x(\sigma),\ce(\sigma),\be(\sigma),p(\sigma)\right)\label{eq:bracket-of-sigma}\end{eqnarray}
And even $\de x^{m}=\ce^{m}$ and $\de\be_{m}=p_{m}$ will be seen
as a function (identity) of $\ce^{m}$ or $\be_{m}$, s.th. we denote
\begin{eqnarray}
\de x^{m}(\sigma) & \equiv & \ce^{m}(\sigma)\label{eq:dx-of-sigma}\\
\de\be_{m}(\sigma) & \equiv & p_{m}(\sigma)\label{eq:db-of-sigma}\end{eqnarray}
Although $\de$ acts only in the target space on $x,\be,\ce$ and
$p$, the above obviously suggests to introduce a differential --
say $\es$\index{$s$@$\es(\ldots)$|itext{BRST differential}} -- in
the new phase space, which is compatible with the target space differential
in the sense \rem{(\ref{eq:d-auf-partial})}\begin{eqnarray}
\es\left(x^{m}(\sigma)\right) & = & \de x^{m}(\sigma)\equiv\ce^{m}(\sigma)\\
\es\left(\be_{m}(\sigma)\right) & = & \de\be_{m}(\sigma)\equiv p_{m}(\sigma)\end{eqnarray}
 We can generate $\es$ with the Poisson bracket in almost the same
way as $\de$ before in (\ref{eq:BRST-op}):\index{$\Omega$@$\OO$|itext{BRST operator}}\index{$d_{\rm w}$|itext{worldvolume dimension}}\begin{eqnarray}
\OO & \equiv & \int_{\Sigma}\msigp\quad\oo(\sigma)=\int\msigp\quad\ce^{m}(\sigma)p_{m}(\sigma),\qquad\textrm{\es}\,(\ldots)=\left\{ \OO,\ldots\right\} \label{eq:Omega}\end{eqnarray}
The Poisson\index{Poisson bracket} bracket between the conjugate
fields gets of course an additional delta function compared to (\ref{eq:Poisson-bracket-bc},\ref{eq:Poisson-bracket-xp}).\begin{eqnarray}
\left\{ p_{m}(\sigma'),x^{n}(\sigma)\right\}  & = & \delta_{m}^{n}\delta^{d_{{\rm w}}-1}(\sigma'-\sigma)\\
\left\{ \be_{m}(\sigma'),\ce^{n}(\sigma)\right\}  & = & \delta_{m}^{n}\delta^{d_{{\rm w}}-1}(\sigma'-\sigma)\end{eqnarray}
 The first important (but rather trivial) observation is then that
for $K(\sigma)$ being a function of $x(\sigma),\ce(\sigma),\be(\sigma)$
as in (\ref{eq:K-of-sigma}) (and not a functional, which could contain
derivatives on or integrations over $\sigma$) we have \begin{eqnarray}
\es\,(K(\sigma)) & = & \left(\ce^{m}(\sigma)\partl{(x^{m}(\sigma))}+p_{m}(\sigma)\partl{(\be_{m}(\sigma))}\right)K\left(x(\sigma),\ce(\sigma),\be(\sigma)\right)=\de K(\sigma)\label{eq:sK-gleich-dK}\end{eqnarray}
The same is true for more general objects of the form of $T$ in (\ref{eq:T-of-sigma}).
Because of this fact the distinction between $\de$ and $\es$ is
not very essential, but in subsection \ref{sub:antibracket} the replacement
of the arguments as in (\ref{eq:T-of-sigma}) will be different and
the distinction very essential in order not to get confused.

The relation between Poisson bracket and big bracket (\ref{eq:bc-big-bracket},\ref{eq:bc-big-br-TtildeT})
gets obviously modified by a delta function\begin{eqnarray}
\left\{ K^{(k,k')}(\sigma'),L^{(l,l')}(\sigma)\right\}  & = & \left[K^{(k,k')},L^{(l,l')}\right]_{(1)}^{\Delta}(\sigma)\,\delta^{d_{{\rm w}}-1}(\sigma'-\sigma)\label{eq:Poisson-big-br}\\
\textrm{or more general }\left\{ T^{(t,t',t'')}(\sigma'),\tilde{T}^{(\tilde{t},\tilde{t}',\tilde{t}'')}(\sigma)\right\}  & = & \left[T^{(t,t',t'')},\tilde{T}^{(\tilde{t},\tilde{t}',\tilde{t}'')}\right]_{(1)}^{\Delta}(\sigma)\,\delta^{d_{{\rm w}}-1}(\sigma'-\sigma)\label{eq:Poisson-big-br-T}\end{eqnarray}
The relation between the derived bracket (using $\es\,$) on the lefthand
side and the derived bracket (using $\de\,$) on the righthand side
is (omitting the overall sign in the definition of the derived bracket)\begin{eqnarray}
\left\{ \es K^{(k,k')}(\sigma'),L^{(l,l')}(\sigma)\right\}  & \stackrel{(\ref{eq:sK-gleich-dK})}{=} & \left\{ \de K^{(k,k')}(\sigma'),L^{(l,l')}(\sigma)\right\} \stackrel{(\ref{eq:Poisson-big-br-T})}{=}\left[\de K^{(k,k')},L^{(l,l')}\right]_{(1)}^{\Delta}(\sigma)\,\delta^{d_{{\rm w}}-1}(\sigma'-\sigma)\qquad\end{eqnarray}
The worldvolume coordinates $\sigma$ remain so far more or less only
spectators. In the subsection \ref{sub:antibracket}, the world-volume
coordinates play a more active part and already in the following subsection
a similar role is taken by an anticommuting extension of the worldsheet. 

Before we proceed, it should be stressed that the replacement of $x,\ce,\be$
and $p$ by $x(\sigma),\ce(\sigma),\be(\sigma)$ and $p(\sigma)$
was just the most naive replacement to do, and it will be a bit extended
in the following section until it can serve as a useful tool in an
application in \ref{sub:Zabzine}. But in principle, one can replace
those variables by any fields with matching index structure and parity
(even composite ones) and study the resulting relations between Poisson
bracket on the one side and geometric bracket on the other side. Also
the differential $\es\,$ can be replaced for example by the twisted
differential or by more general BRST-like transformations. In this
way it should be possible to implement other derived brackets, for
example those built with the Poisson-Lichnerowicz-differential (see
\cite{Kosmann-Schwarzbach:2003en}), in a sigma-model description.
In \ref{sub:antibracket}, a different (but also quite canonical)
replacement is performed and we will see that the different replacement
corresponds to a change of the role of $\sigma$ and an anticommuting
worldvolume coordinate $\tet$ which will be introduced in the following.\vspace{-.1cm}

\section{Natural appearance of derived brackets in Poisson brackets of superfields}\vspace{-.3cm} 

\label{sub:Natural-appearance} In the application to worldsheet theories
in section \ref{sec:Applications-in-string}, there appear superfields,
either in the sense of worldsheet supersymmetry or in the sense of
de-Rham superfields (see e.g. \cite{Cattaneo:1999fm,Zucchini:2004ta}).
Let us view a superfield in general as a method to implement a fermionic
transformation of the fields via a shift in a fermionic parameter
$\tet$ which can be regarded as fermionic extension of the worldvolume.
In our case the fermionic transformation is just the spacetime de-Rham-differential
$\de$, or more precisely $\es\,$, and is not necessarily connected
to worldvolume supersymmetry. In fact, in worldvolumes of dimension
higher than two, supersymmetry requires more than one fermionic parameter
while a single $\tet$ is enough for our purpose to implement $\es$.
In two dimensions, however, this single theta can really be seen as
a worldsheet fermion (see \ref{sub:Zabzine}). But let us neglect
this knowledge for a while, in order to clearly see the mechanism,
which will be a bit hidden again, when applied to the supersymmetric
case in \ref{sub:Zabzine}.

As just said above, we want to implement with superfields the fermionic
transformation $\es$ and not yet a supersymmetry. So let us define
in this section a \textbf{superfield\index{superfield}} as a function
of the phase space fields with additional dependence on $\tet$\index{$\theta$@$\tet$},
$Y=Y(x(\sigma),p(\sigma),\ce(\sigma),\be(\sigma),\tet)$, which obeys
\footnote{\index{footnote!\thefoot. about the superfield definition}If this
seems unfamiliar, compare with the case of worldsheet supersymmetry,
where one introduces a differential operator $\qu\equiv\partial_{\tet}+\tet\partial_{\sigma}$
and the definition of a superfield is, in contrast to here, $\delta_{\feps}Y\stackrel{!}{=}\feps\qu Y$,
where $\delta_{\feps}$ is the supersymmetry transformation of the
component fields (compare \ref{sub:Zabzine}).$\qquad\fussend$%
}\enlargethispage*{2cm}\begin{eqnarray}
\es Y(x(\sigma),p(\sigma),\ce(\sigma),\be(\sigma),\tet) & \stackrel{!}{=} & \partial_{\tet}Y(x(\sigma),p(\sigma),\ce(\sigma),\be(\sigma),\tet)\label{eq:superfield-definition}\\
\textrm{with } &  & \es x^{m}(\sigma)=\ce^{m}(\sigma),\es\be_{m}(\sigma)=p_{m}(\sigma)\quad(\es\tet=0)\end{eqnarray}
 With our given field content it is possible to define two basic conjugate%
\footnote{\index{footnote!\thefoot. super-Poisson bracket}\label{foot:super-Poisson-bracket}The
superfields $\Phi$ and $\Es$ are conjugate with respect to the following
\textbf{super\index{super-Poisson bracket}-Poisson\index{Poisson bracket!super-$\sim$}-bracket}\index{bracket!super-Poisson $\sim$}
\begin{eqnarray*}
\left\{ F(\sigma',\tet'),G(\sigma,\tet)\right\}  & \equiv & \int\msigp\backtilde\int d\tilde{\tet}\qquad\big(\delta F(\sigma',\tet')/\delta\Es_{k}(\tilde{\sigma},\tilde{\tet})\funktional{}{\Phi^{k}(\tilde{\sigma},\tilde{\tet})}G(\sigma,\tet)-\delta F(\sigma',\tet')/\delta\Phi^{k}(\tilde{\sigma},\tilde{\tet})\funktional{}{\Es_{k}(\tilde{\sigma},\tilde{\tet})}G(\sigma,\tet)\big)=\\
 & = & \int\msigp\backtilde\int d\tilde{\tet}\qquad\big(\delta F(\sigma',\tet')/\delta\Es_{k}(\tilde{\sigma},\tilde{\tet})\funktional{}{\Phi^{k}(\tilde{\sigma},\tilde{\tet})}G(\sigma,\tet)-(-)^{FG}\delta G(\sigma',\tet')/\delta\Es_{k}(\tilde{\sigma},\tilde{\tet})\funktional{}{\Phi^{k}(\tilde{\sigma},\tilde{\tet})}F(\sigma,\tet)\big)\end{eqnarray*}
which, however, boils down to taking the ordinary graded Poisson bracket
between the component fields (as can be seen in (\ref{eq:conjugate-superfields})).
The \textbf{functional\index{functional derivative} derivative\index{derivative!functional $\sim$}s}
from the left and from the right are defined as usual via\[
\delta_{S}A\equiv\int\msigp\backtilde\int d\tilde{\theta}\quad\delta A/\delta S_{k}(\tilde{\sigma},\tilde{\theta})\cdot\delta S_{k}(\tilde{\sigma},\tilde{\theta})\equiv\int\msigp\backtilde\int d\tilde{\theta}\quad\delta S_{k}(\tilde{\sigma},\tilde{\theta})\cdot\funktl{S_{k}(\tilde{\sigma},\tilde{\theta})}A\]
and similarly for $\Phi$, which leads to \begin{eqnarray*}
\funktl{\Es_{m}(\tilde{\sigma},\tilde{\tet})}\Es_{n}(\sigma,\tet) & = & \delta_{n}^{m}(\tet-\tilde{\tet})\delta^{d_{{\rm w}}-1}(\sigma-\tilde{\sigma})=-\delta\Es_{n}(\sigma,\tet)/\Es_{m}(\tilde{\sigma},\tilde{\tet})\\
\funktl{\Phi^{m}(\tilde{\sigma},\tilde{\tet})}\Phi^{n}(\sigma,\tet) & = & \delta_{m}^{n}(\tilde{\tet}-\tet)\delta^{d_{{\rm w}}-1}(\sigma-\tilde{\sigma})=\delta\Phi^{n}(\sigma,\tet)/\delta\Phi^{m}(\tilde{\sigma},\tilde{\tet})\end{eqnarray*}
The functional derivatives can also be split in those with respect
to the component fields\[
\funktl{\Es_{m}(\tilde{\sigma},\tilde{\tet})}=\funktl{p_{m}(\tilde{\sigma})}-\tilde{\tet}\funktl{\be_{m}(\tilde{\sigma})},\qquad\funktl{\Phi^{m}(\tilde{\sigma},\tilde{\tet})}=\funktl{\ce^{m}(\tilde{\sigma})}+\tilde{\tet}\funktl{x^{m}(\tilde{\sigma})}\qquad\fussend\]
} superfields $\Phi^{m}$ and $\Es_{m}$ \\
which build up a super-phase-space\enlargethispage*{1cm}%
\footnote{\index{footnote!\thefoot. delta function for Grassmann variables}\index{delta function!for Grassmann variables}For
Grassmann\index{Grassmann!delta function} variables $\delta(\tet'-\tet)=\tet'-\tet$
in the following sense\begin{eqnarray*}
\int\de\tet'(\tet'-\tet)F(\tet') & = & \int\de\tet'(\tet'-\tet)\left(F(\tet)+(\tet'-\tet)\partial_{\tet}F(\tet)\right)=\\
 & = & \int\de\tet'\quad\tet'F(\tet)-\tet'\tet\partial_{\tet}F(\tet)-\tet\tet'\partial_{\tet}F(\tet)=\\
 & = & F(\tet)\end{eqnarray*}
We have as usual\begin{eqnarray*}
\tet\delta(\tet'-\tet) & = & \tet(\tet'-\tet)=\tet\tet'=\tet'(\tet'-\tet)=\\
 & = & \tet'\delta(\tet'-\tet)\end{eqnarray*}
Pay attention to the antisymmetry\begin{eqnarray*}
\delta(\tet'-\tet) & = & -\delta(\tet-\tet')\qquad\fussend\end{eqnarray*}
} \index{$\Phi^{m}(\sigma,\tet)$}\index{$S_m$@$\Es_{m}(\sigma,\tet)$}\begin{eqnarray}
\Phi^{m}(\sigma,\tet) & \equiv & x^{m}(\sigma)+\bs{\theta}\ce^{m}(\sigma)=x^{m}(\sigma)+\bs{\theta}\es x^{m}(\sigma)\\
\Es_{m}(\sigma,\tet) & \equiv & \be_{m}(\sigma)+\tet p_{m}(\sigma)=\be_{m}(\sigma)+\tet\es\be_{m}(\sigma)\\
\left\{ \Es_{m}(\sigma,\tet),\Phi^{n}(\sigma',\tet')\right\}  & = & \left\{ \be_{m}(\sigma),\tet'\ce^{n}(\sigma')\right\} +\tet\left\{ p_{m}(\sigma),x^{n}(\sigma')\right\} =\label{eq:conjugate-superfields}\\
 & = & \underbrace{(\tet-\tet')}_{\equiv\delta(\tet-\tet')}\delta(\sigma-\sigma')\delta_{m}^{n}\end{eqnarray}
$\Phi$ and $\Es$ are obviously superfields in the above sense\begin{eqnarray}
\partial_{\tet}\Phi^{m}(\sigma,\tet) & = & \underbrace{\es x^{m}(\sigma)}_{\ce^{m}(\sigma)}\underbrace{+\tet\es\ce^{m}(\sigma)}_{=0}=\es\Phi^{m}(\sigma,\tet)\label{eq:theta-derivative-eq-exterior-derivative}\\
\partial_{\tet}\Es_{m} & = & \underbrace{\es\be_{m}(\sigma)}_{p_{m}(\sigma)}\underbrace{+\tet\es p_{m}(\sigma)}_{0}=\es\Es_{m}(\sigma,\tet)\label{eq:theta-derivative-eq-exterior-derivativeS}\end{eqnarray}
as well as $\es\Phi(\sigma,\tet)=\ce(\sigma)$ and $\es\Es(\sigma,\tet)=p(\sigma)$
are superfields, and every analytic function of those fields will
be a superfield again. \rem{and $\partial_\sigma\Phi$, aber hier wurscht...}

We will convince ourselves in this subsection that in the Poisson
brackets of general superfields, the derived brackets come with the
complete $\delta$-function (of $\sigma$ and $\tet$) while the corresponding
algebraic brackets come with a derivative of the delta-function. The
introduction of worldsheet coordinates $\sigma$ was not yet really
necessary for this discussion, but it will be useful for the comparison
with the subsequent subsection. Indeed, we do not specify the dimension
$d_{{\rm w}}$ of the worldsheet yet. An argument sigma is representing
several worldsheet coordinates $\sigma^{\mu}$. It should be stressed
again that the differential $\de$ should \textbf{NOT} be mixed up
with the worldsheet exterior derivative $\dew$, which does not show
up in this subsection. 

Similar as in \ref{sub:Sigma-Models}, equations (\ref{eq:K-of-sigma})-(\ref{eq:db-of-sigma}),we
will view all geometric objects as functions of local coordinates
and replace the arguments not by phase space fields but by the just
defined super-phase fields which reduces for $\tet=0$ to the previous
case. \index{$T^{(t,t',t'')}(\sigma,\tet)$}\begin{eqnarray}
T^{(t,t',t'')}(\sigma,\tet) & \equiv & T^{(t,t',t'')}\left(\Phi(\sigma,\tet),\es\Phi(\sigma,\tet),\Es(\sigma,\tet),\es\Es(\sigma,\tet)\right)\stackrel{\tet=0}{=}T^{(t,t',t'')}(\sigma)\quad(\textrm{see }(\ref{eq:T-of-sigma}))\label{eq:T-sig-tet}\end{eqnarray}
For example for a multivector valued form we write\index{$K^{(k,k')}(\sigma,\tet)$}
\begin{eqnarray}
K^{(k,k')}(\sigma,\tet) & \equiv & K^{(k,k')}\big(\Phi^{m}(\sigma,\tet),\underbrace{\es\Phi^{m}(\sigma,\tet)}_{\ce^{m}(\sigma)},\Es_{m}(\sigma,\tet)\big)=\label{eq:K-sig-tet-def}\\
 &  & \hspace{-1cm}=K_{m_{1}\ldots m_{k}}\hoch{n_{1}\ldots n_{k'}}\left(\Phi(\sigma,\tet)\right)\,\underbrace{\es\Phi^{m_{1}}(\sigma,\tet)}_{\ce^{m_{1}}(\sigma)}\ldots\es\Phi^{m_{k}}(\sigma,\tet)\Es_{n_{1}}(\sigma,\tet)\ldots\Es_{n_{k'}}(\sigma,\tet)\us{\stackrel{\tet=0}{=}}{(\ref{eq:K-of-sigma})}K^{(k,k')}(\sigma)\qquad\end{eqnarray}
Likewise for all the other examples of \ref{sub:Sigma-Models}:\index{$dK$@$\de K(\sigma,\tet)$}\index{$o$@$\oo(\sigma,\tet)$}\begin{eqnarray}
\textrm{e.g. }\de K(\sigma,\tet) & \equiv & \de K\left(\Phi(\sigma,\tet),\es\Phi(\sigma,\tet),\Es(\sigma,\tet),\es\Es(\sigma,\tet)\right)\label{eq:dK-of-sig-tet}\\
\textrm{or }\oo(\sigma,\tet) & \equiv & \oo\left(\es\Phi(\sigma,\tet),\es\Es(\sigma,\tet)\right)=\ce^{m}(\sigma)p_{m}(\sigma)=\oo(\sigma)\label{eq:o-of-sig-tet}\\
\hspace{-1.2cm}\left[K^{(k,k')},_{\de\,}L^{(l,l')}\right]_{(1)}^{\Delta}(\sigma,\tet) & \equiv & \left[K^{(k,k')}\bs{,}L^{(l,l')}\right]_{(1)}^{(\Delta)}\left(\Phi(\sigma,\tet),\es\Phi(\sigma,\tet),\Es(\sigma,\tet),\es\Es(\sigma,\tet)\right)\us{\stackrel{\tet=0}{=}}{(\ref{eq:bracket-of-sigma})}\left[K^{(k,k')}\bs{,}L^{(l,l')}\right]_{(1)}^{(\Delta)}(\sigma)\qquad\quad\\
\de x^{m}(\sigma,\tet) & \equiv & \es\Phi^{m}(\sigma,\tet)=\ce^{m}(\sigma)\\
\de\be_{m}(\sigma,\tet) & \equiv & \es\Es_{m}(\sigma,\tet)=p_{m}(\sigma)\label{eq:db-sig-tet}\end{eqnarray}
For functions of the type $T^{(t,t',t'')}(\sigma,\tet)$ we clearly
have \begin{eqnarray}
\de T^{(t,t',t'')}(\sigma,\tet) & = & \es\left(T^{(t,t',t'')}(\sigma,\tet)\right)\label{eq:dT-gleich-sT}\\
\textrm{in particular }\de K^{(k,k')}(\sigma,\tet) & = & \es\left(K^{(k,k')}(\sigma,\tet)\right)\end{eqnarray}
As all those analytic functions of the basic superfields are superfields
(in the sense of \ref{eq:superfield-definition}) themselves, $\partial_{\tet}$
can be replaced by $\es\,$ in a $\tet$-expansion, so that we get
the important relation \begin{eqnarray}
T^{(t,t',t'')}(\sigma,\tet) & = & T^{(t,t',t'')}(\sigma)+\tet\de T^{(t,t',t'')}(\sigma)\label{eq:wichtig}\\
K^{(k,k')}(\sigma,\tet) & = & K^{(k,k')}(\sigma)+\tet\de K^{(k,k')}(\sigma)\label{eq:wichtigII}\end{eqnarray}
This also implies that $\de T(\sigma,\tet)$ and in particular $\de K(\sigma,\tet)$
do actually not depend on $\tet$:\begin{equation}
\de K^{(k,k')}(\sigma,\tet)=\de K^{(k,k')}(\sigma)\label{eq:wichtigIII}\end{equation}
 Now comes the nice part:

\paragraph{Proposition\index{proposition!super-Poisson bracket of multivector valued forms}
1 }

\emph{For all multivector valued forms $K^{(k,k')},L^{(l,l')}$ on
the target space manifold, in a local coordinate patch seen as functions
of $x^{m}$,$\de x^{m}$ and $\pe_{m}$ as in (\ref{eq:multivector-valued-form-K}),
the following equation holds for the corresponding superfields (\ref{eq:K-sig-tet-def})
\begin{equation}
\hspace{-.4cm}\boxed{\{K^{(k,k')}(\sigma',\tet'),L^{(l,l')}(\sigma,\tet)\}=\delta(\tet'-\tet)\delta(\sigma-\sigma')\cdot\underbrace{[\de K,L]_{(1)}^{\Delta}}_{\lqn{-(-)^{k-k'}\left[K,_{\de}L\right]_{(1)}^{\Delta}}}(\sigma,\tet)+\underbrace{\partial_{\tet}\delta(\tet-\tet')}_{=1}\delta(\sigma-\sigma')[K,L]_{(1)}^{\Delta}(\sigma,\tet)}\!\!\!\label{eq:Proposition1}\end{equation}
 where $[K,L]_{(1)}^{\Delta}$ is the big bracket (\ref{eq:bc-big-bracket})
(Buttin's algebraic bracket, which was previously just the Poisson
bracket, being true now up to a $\delta(\sigma-\sigma')$ only after
setting $\tet=\tet'$) and $\left[K,_{\de}L\right]_{(1)}^{\Delta}$
is the derived bracket of the big bracket (\ref{eq:bc-derived-of-bigbracket}).}\vspace{.5cm}\rem{aehnliche Prop waere fuer die Ersetzung wie im Antifeld-Fall denkbar}

\emph{Proof}$\quad$ Using (\ref{eq:wichtigII}), we can simply plug
$K(\sigma',\tet')=K(\sigma')+\tet'\de K(\sigma')$ and $L(\sigma,\tet)=L(\sigma)+\tet\de L(\sigma)$
into the lefthand side:\begin{eqnarray}
\lqn{\left\{ K(\sigma',\tet'),L(\sigma,\tet)\right\} =}\nonumber \\
 & = & \left\{ K(\sigma'),L(\sigma)\right\} +\tet'\left\{ \de K(\sigma'),L(\sigma)\right\} +(-)^{k-k'}\tet\left\{ K(\sigma'),\de L(\sigma)\right\} +(-)^{k-k'}\tet\tet'\left\{ \de K(\sigma'),\de L(\sigma)\right\} =\qquad\\
 & = & \left\{ K(\sigma'),L(\sigma)\right\} +(\tet'-\tet)\left\{ \de K(\sigma'),L(\sigma)\right\} +\tet\de\left\{ K(\sigma'),L(\sigma)\right\} -\tet\tet'\de\left\{ \de K(\sigma'),L(\sigma)\right\} =\\
 & \stackrel{(\ref{eq:bc-big-bracket})}{=} & \delta(\sigma-\sigma')\left(\left[K,L\right]_{(1)}^{\Delta}(\sigma)+\tet\de\left[K,L\right]_{(1)}^{\Delta}(\sigma)\right)+(\tet'-\tet)\delta(\sigma-\sigma')\left(\left[\de K,L\right]_{(1)}^{\Delta}(\sigma)+\tet\de\left[\de K,L\right]_{(1)}^{\Delta}(\sigma)\right)=\qquad\quad\\
 & \stackrel{(\ref{eq:wichtig})}{=} & \delta(\sigma-\sigma')\left[K,L\right]_{(1)}^{\Delta}(\sigma,\tet)+(\tet'-\tet)\delta(\sigma-\sigma')\left[\de K,L\right]_{(1)}^{\Delta}(\sigma,\tet)\qquad\square\end{eqnarray}

There is yet another way to see that the bracket at the plain delta
functions is the derived bracket of the one at the derivative of the
delta-function, which will be useful later: Denote the coefficients
in front of the delta-functions by $A(K,L)$ and $B(K,L)$: \begin{equation}
\left\{ K(\sigma',\tet'),L(\sigma,\tet)\right\} =A(K,L)\cdot\delta(\tet'-\tet)\delta(\sigma-\sigma')+B(K,L)(\sigma,\tet)\underbrace{\partial_{\tet}\delta(\tet-\tet')}_{=1}\delta(\sigma-\sigma')\end{equation}
In order to hit the delta-functions, it is enough to integrate over
a patch $U(\sigma)$ containing the point parametrized by $\sigma$.
We can thus extract $A$ and $B$ via\allowdisplaybreaks \begin{eqnarray}
A(K,L)(\sigma,\tet) & = & \int\de\tet'\int_{U(\sigma)}\msigp'\left\{ K(\sigma',\tet'),L(\sigma,\tet)\right\} =\\
 & = & \int\de\tet'\int\msigp'\left\{ K(\sigma')+\tet'\de K(\sigma'),L(\sigma,\tet)\right\} =\\
 & = & \int\msigp'\{\underbrace{\de K(\sigma')}_{\stackrel{(\ref{eq:wichtigIII})}{=}\de K(\sigma',\tet)},L(\sigma,\tet)\}\\
B(K,L)(\sigma,\tet) & = & \int\de\tet'\int_{U(\sigma)}\msigp'(\tet'-\tet)\left\{ K(\sigma',\tet'),L(\sigma,\tet)\right\} =\\
 & = & \int\msigp'\left\{ K(\sigma',\tet'),L(\sigma,\tet)\right\} \mid_{\tet'=\tet}\\
\dann A(K,L) & = & B(\de K,L)\end{eqnarray}
It is thus enough to collect in a direct calculation the terms at
the derivative of the delta-function and verify that it leads to the
big bracket.$\qquad\square$

\section{Comment on the quantum case}

\label{sub:Canonical-commutator} In (\ref{eq:quantization}) the
embedding via the interior product into the space of operators acting
on forms was interpreted as quantization . In the presence of world-volume
dimensions, the partial derivative as Schroedinger\index{Schroedinger representation}
representation for conjugate momenta is no longer appropriate and
one has to switch to the functional derivative. Remember\begin{eqnarray}
\Phi^{m}(\sigma,\tet) & = & x^{m}(\sigma)+\tet\ce^{m}(\sigma),\qquad\de\Phi^{m}(\sigma,\tet)=\ce^{m}(\sigma)=\de\Phi(\sigma)\\
\Es_{m}(\sigma,\tet) & = & \be_{m}(\sigma)+\tet p_{m}(\sigma),\qquad\de\Es_{m}(\sigma,\tet)=p_{m}(\sigma)=\de\Es(\sigma)\end{eqnarray}
The quantization of the superfields in the Schroedinger representation
(conjugate momenta as super functional derivatives) is consistent
with the quantization of the component fields (see also footnote \ref{foot:super-Poisson-bracket})\begin{eqnarray}
\hat{\Es}_{m}(\sigma,\tet) & \equiv & \frac{\hbar}{i}\funktional{}{\Phi^{m}(\sigma,\tet)}=\frac{\hbar}{i}\funktional{}{\ce^{m}(\sigma)}+\tet\frac{\hbar}{i}\funktional{}{x^{m}(\sigma)}\\
\dann\left[\hat{\Es}_{m}(\sigma,\tet),\hat{\Phi}^{n}(\sigma',\tet')\right] & = & \frac{\hbar}{i}\left(\funktional{}{\ce^{m}(\sigma)}+\tet\funktional{}{x^{m}(\sigma)}\right)\left(x^{n}(\sigma')+\tet'\ce^{n}(\sigma')\right)=\\
 & = & \frac{\hbar}{i}\delta_{m}^{n}\left(\tet-\tet'\right)\delta(\sigma-\sigma')\end{eqnarray}
The quantization of a multivector valued form, containing several
operators $\hat{\Es}$ at the same worldvolume-point, however, leads
to powers of delta functions with the same argument when acting on
some wave functional. This is the usual problem in quantum field theory
and requires a model dependent regularization and renormalization.
We will stay model independent here and therefore will not treat the
quantum case for a present worldvolume coordinate $\sigma$.  Nevertheless
it is instructive to study it for absent $\sigma$, but keeping $\tet$
and considering {}``worldline-superfields'' of the form \begin{eqnarray}
\Phi^{m}(\tet) & = & x^{m}+\tet\ce^{m},\qquad\de\Phi^{m}(\tet)=\ce^{m}\\
\Es_{m}(\tet) & = & \be_{m}+\tet p_{m},\qquad\de\Es_{m}(\tet)=p_{m}\end{eqnarray}
Quantum operator and commutator simplify to\index{$S_m$@$\hat\Es_m(\tet)$}\begin{eqnarray}
\hat{\Es}_{m}(\tet) & \equiv & \frac{\hbar}{i}\funktional{}{\Phi^{m}(\tet)}=\frac{\hbar}{i}\partl{\ce^{m}}+\tet\frac{\hbar}{i}\partl{x^{m}}\\
\dann\left[\hat{\Es}_{m}(\tet),\hat{\Phi}^{n}(\tet')\right] & = & \frac{\hbar}{i}\delta_{m}^{n}\left(\tet-\tet'\right)\\
\left[\hat{\Es}_{m}(\tet),\widehat{\de\Phi}^{n}(\tet')\right] & = & \frac{\hbar}{i}\delta_{m}^{n}\end{eqnarray}
\rem{$\de\hat{\Phi}=\widehat{\de\Phi}$ in the sense that $\left[\de\,,\ip_{K}\right]=\ip_{\de K}$.
}In contrast to $\sigma$, products of $\tet$-delta functions are
no problem. 

The important relation $K(\tet)=K+\tet\de K$ (\ref{eq:wichtigII})
can be extended to the quantum case as seen when acting on some $r$-form.
\begin{eqnarray}
\ip_{K^{(k,k')}}\rho^{(r)}(\tet) & \stackrel{(\ref{eq:wichtig})}{=} & \ip_{K}\rho+\tet\de(\ip_{K}\rho)=\\
 & \stackrel{(\ref{eq:dK-und-Lie})}{=} & \ip_{K}\rho+\tet\left(\ip_{\de K}\rho+(-)^{k-k')}\ip_{K}\de\rho\right)=\\
 & = & \ip_{K}(\tet)\left(\rho(\tet)\right)\\
\textrm{with }\ip_{K}(\tet) & \equiv & \ip_{K}+\tet\left[\de\,,\ip_{K}\right]\label{eq:iK-of-tet}\end{eqnarray}
In that sense we have (remember $\hat{K}=\left(\frac{\hbar}{i}\right)^{k'}\ip_{K}$)\index{$K^{(k,k')}$@$\hat{K}^{(k,k')}(\tet)$}\begin{eqnarray}
\hat{K}^{(k,k')}(\tet) & = & \hat{K}^{(k,k')}+\tet\widehat{\de K}\label{eq:wichtig-quant}\\
\textrm{with }\widehat{\de K} & \stackrel{(\ref{eq:dK-und-Lie})}{=} & \left[\de\,,\hat{K}\right]\label{eq:dK-und-Lie-quant}\end{eqnarray}
where the explicit form of this quantized multivector valued form
reads \begin{eqnarray}
\hat{K}^{(k,k')}(\tet) & \equiv & \left(\frac{\hbar}{i}\right)^{k'}K_{m_{1}\ldots m_{k}}\hoch{n_{1}\ldots n_{k'}}\left(\Phi(\tet)\right)\,\underbrace{\de\Phi^{m_{1}}(\tet)}_{\ce^{m_{1}}}\ldots\de\Phi^{m_{k}}(\tet)\funktional{}{\Phi^{n_{1}}(\tet)}\ldots\funktional{}{\Phi^{n_{k'}}(\tet)}\label{eq:K-quant}\end{eqnarray}
In the derivation of (\ref{eq:iK-of-tet}), $\ip_{K}$ and $\rho$
both were evaluated at the same $\tet$. Let us eventually consider
the general case: \begin{eqnarray}
\hat{K}^{(k,k')}(\tet')\rho^{(r)}(\tet) & = & \left(\hat{K}+\tet'\widehat{\de K}\right)\left(\rho+\tet\de\rho\right)=\\
 & = & \hat{K}\rho+\tet'\widehat{\de K}\rho+(-)^{k-k'}\tet\hat{K}\de\rho+(-)^{k-k'}\tet\tet'\widehat{\de K}\de\rho=\\
 & = & \hat{K}\rho+\tet\de\left(\widehat{K}\rho\right)+(\tet'-\tet)\left(\widehat{\de K}\rho+\tet\de\left(\widehat{\de K}\rho\right)\right)\end{eqnarray}
The relation between quantum operators acting on forms and the interior
product therefore becomes modified in comparison to (\ref{eq:quantization})
and reads\begin{equation}
\boxed{\hat{K}^{(k,k')}(\tet')\rho^{(r)}(\tet)=\left(\frac{\hbar}{i}\right)^{k'}\Big(\ip_{K}\rho(\tet)+(\tet'-\tet)\underbrace{\ip_{\de K}\rho(\tet)}_{(-)^{k-k'}\Lie_{K}\rho}\Big)}\label{eq:quantization-for-superfields}\end{equation}

\paragraph{Proposition 2 }

\emph{\index{proposition!commutator of quantized multivector valued forms}For
all multivector valued forms $K^{(k,k')},L^{(l,l')}$ on the target
space manifold, in a local coordinate patch seen as functions of $x^{m}$,$\de x^{m}$
and $\pe_{m}$ as in (\ref{eq:multivector-valued-form-K}), the following
equations holds for the corresponding quantized worldline-superfields
(\ref{eq:K-quant}) $\hat{K}(\tet)$ and $\hat{L}(\tet)$:}\\
\emph{\Ram{1}{\begin{eqnarray}
[\hat{K}^{(k,k')}(\tet'),\hat{L}^{(l,l')}(\tet)] & = & \sum_{p\geq1}\left(\frac{\hbar}{i}\right)^{p}\Big(\underbrace{\partial_{\tet}\delta(\tet-\tet')}_{=1}\widehat{[K,L]_{(p)}^{\Delta}}(\tet)+\delta(\tet'-\tet)\widehat{[\de K,L]_{(p)}^{\Delta}}(\tet)\Big)\label{eq:Proposition2}\\
{}[\hat{K}^{(k,k')}(\tet'),\hat{L}^{(l,l')}(\tet)]\lqn{\rho(\tilde{\tet})=}\nonumber \\
 &  & \hspace{-2cm}=\left(\frac{\hbar}{i}\right)^{k'+l'}\Big(\ip_{\left[K,L\right]^{\Delta}}\rho^{(r)}(\tilde{\tet})+\delta(\tet-\tilde{\tet})\ip_{\de\left[K,L\right]^{\Delta}}\rho^{(r)}(\tilde{\tet})+\nonumber \\
 &  & \hspace{-2cm}\qquad\qquad\qquad\qquad+\delta(\tet'-\tet)\left(\ip_{\left[\de K,L\right]^{\Delta}}\rho^{(r)}(\tilde{\tet})+\delta(\tet-\tilde{\tet})\ip_{\de\left[\de K,L\right]^{\Delta}}\rho^{(r)}(\tilde{\tet})\right)\Big)\end{eqnarray}
} Again the algebraic bracket (\ref{eq:algebraic-bracketI}) comes
with the derivative of the delta function while the derived bracket
(\ref{eq:bc-derived-bracketI}) comes with the plain delta functions.
But this time the algebraic bracket is not only the big bracket $\left[\,,\,\right]_{(1)}^{\Delta}$,
but the full one.} \vspace{.5cm}

\emph{Proof}$\quad$Let us just plug in (\ref{eq:wichtig-quant})
into the lefthand side:\begin{eqnarray}
[\hat{K}(\tet'),\hat{L}(\tet)] & = & [\hat{K}+\tet'\widehat{\de K}\,,\,\hat{L}+\tet\widehat{\de L}]=\\
 & = & [\hat{K},\hat{L}]+\tet'[\widehat{\de K}\,,\,\hat{L}]+(-)^{k-k'}\tet[\hat{K}\,,\,\widehat{\de L}]-(-)^{k-k'}\tet'\tet[\widehat{\de K}\,,\,\widehat{\de L}]=\\
 & \stackrel{(\ref{eq:dK-und-Lie-quant})}{=} & [\hat{K},\hat{L}]+\tet\left[\de\,,[\hat{K}\,,\,\hat{L}]\right]+(\tet'-\tet)\left([\widehat{\de K}\,,\,\hat{L}]+\tet\left[\de\,,[\widehat{\de K}\,,\,\hat{L}]\right]\right)=\\
 & = & [\hat{K},\hat{L}](\tet)+(\tet'-\tet)[\widehat{\de K}\,,\,\hat{L}]\label{eq:lila-Pause}\end{eqnarray}
Remember now the algebraic bracket (\ref{eq:algebraic-bracket})\begin{eqnarray}
[\ip_{K^{(k,k')}},\ip_{L^{(l,,l')}}] & = & \ip_{[K,L]^{\Delta}}=\sum_{p\geq1}\ip_{[K,L]_{(p)}^{\Delta}}\\
\textrm{with }\left[K,L\right]_{(p)}^{\Delta} & \equiv & \ip_{K}^{(p)}L-(-)^{(k-k')(l-l')}\ip_{L}^{(p)}K\end{eqnarray}
or likewise written in terms of $\hat{K}$ and $\hat{L}$ \bref{eq:quantum-commutator1}\begin{equation}
[\hat{K}^{(k,k')},\hat{L}^{(l,l')}]=\sum_{p\geq1}\left(\frac{\hbar}{i}\right)^{p}\widehat{\left[K,L\right]_{(p)}^{\Delta}}\end{equation}
\eref Due to (\ref{eq:quantum-commutator-T-Ttilde}) we have exactly
the same equation for $[\widehat{\de K}\,,\,\hat{L}]$. Plugging this
back into (\ref{eq:lila-Pause}) completes the proof of (\ref{eq:Proposition2}).
The second equation in the proposition is just a simple rewriting,
when acting on a form, which enables to combine the $p$-th terms
of algebraic and derived bracket to the complete ones.$\qquad\square$

\section{Analogy for the antibracket}

\label{sub:antibracket}In the previous subsection the target space
exterior derivative $\de$ (realized in the $\sigma$-model phase-space
by $\es$) was induced by the the derivative $\partial_{\tet}$ with
respect to the anticommuting coordinate. But thinking of the pullback
of forms in the target space to worldvolume-forms, $\de$ can of course
also be induced to some extend by the derivative with respect to the
bosonic worldvolume coordinates $\sigma^{\mu}$ (including the time,
because we are in the Lagrangian formalism now) or better by the worldvolume\index{world-volume exterior derivative $\dew$}
exterior derivative $\dew$. To this end, however, we have to make
a different identification of the basis elements in tangent- and cotangent-space
of the target space with the fields on the worldvolume than before,
namely%
\footnote{\index{footnote!\thefoot. comparison with \cite{Alekseev:2004np} and \cite{Bonelli:2005ti}}\label{foot:Strobl}This
identification resembles the one in \cite{Alekseev:2004np} with $\pe_{m}\To p_{m}(z)$
and $\de x^{m}\To\partial x^{m}(z)$, or $\de x^{m_{1}}\cdots\de x^{m_{p}}\To\epsilon^{\mu_{1}\ldots\mu_{p}}\partial_{\mu_{1}}x^{m_{1}}(\sigma)\cdots\partial_{\mu_{p}}x^{m_{p}}(\sigma)$
in \cite{Bonelli:2005ti}. It is observed in \cite{Alekseev:2004np}
that the Poisson bracket induces the Dorfman bracket between sums
of vectors and 1-forms (in generalized geometry) and in \cite{Bonelli:2005ti}
more generally that the Poisson-bracket for the $p$-brane induces
the corresponding bracket between sums of vectors and $p$-forms (which
is called, Vinogradov\index{Vinogradov bracket} bracket in \cite{Bonelli:2005ti}).
As $\partial x^{m}$ and $p_{m}$ are commuting phase space variables,
higher rank tensors would automatically be symmetrized (only volume
forms, i.e. p-forms on a p-brane, can be implemented, using the epsilon-tensor).
Symmetrized tensors and brackets inbetween (e.g. the Schouten bracket
for symmetric multivectors) make sense and one could transfer the
present analysis to this setting, but in general a natural exterior
derivative is missing. Therefore the analysis for the above identifications
is done in the antifield-formalism. The appearing derived brackets
will also contain the Dorfman bracket and the corresponding bracket
for sums of vectors and p-forms and in that sense the present approach
is a generalization of the observations above.$\qquad\fussend$%
} \begin{eqnarray}
\de x^{m} & \To & \dew x^{m}(\sigma)=\dew\sigma^{\mu}\partial_{\mu}x^{m}(\sigma),\qquad\qquad\pe_{m}\To\ix_{m}(\sigma)\label{eq:indentification-of-basis-elements}\end{eqnarray}
where $\ix_{m}$\index{$x_m$@$\ix\tief{m}$|itext{antifield}} is the
antifield\index{antifield} of $x^{m}$, i.e. the conjugate field
to $x^{m}$ with respect to the antibracket\index{bracket!anti-$\sim$|itext{$(\ldots\bs{,}\ldots)$}}\index{antibracket}%
\footnote{\label{fn:The-antibracket-looks}\index{footnote!\thefoot. antibracket}The
antibracket looks similar to the Poisson-bracket, but their conjugate
fields have opposite parity, which leads to a different symmetry (namely
that of a Lie-bracket of degree +1 (or -1), i.e. the one in a Gerstenhaber\index{Gerstenhaber algebra}
algebra or Schouten\index{Schouten-algebra}-algebra, see footnote
\ref{Lie-bracket-of-degree} of Appendix C) \rem{evtl spaeter nimmer footnote}\begin{eqnarray*}
\left(A\bs{,}B\right) & \equiv & \int\msig\backtilde\quad\big(\delta A/\ix_{k}(\tilde{\sigma})\funktional{}{x^{k}(\tilde{\sigma})}B-\delta A/\delta x^{k}(\tilde{\sigma})\funktional{}{\ix_{k}(\tilde{\sigma})}B\big)=\\
 & = & \int\msig\backtilde\quad\big(\delta A/\ix_{k}(\tilde{\sigma})\funktional{}{x^{k}(\tilde{\sigma})}B-(-)^{(A+1)(B+1)}\delta B/\ix_{k}(\tilde{\sigma})\funktional{}{x^{k}(\tilde{\sigma})}A\big)\\
\left(A\bs{,}B\right) & = & -(-)^{(A+1)(B+1)}\left(B\bs{,}A\right)\\
\left(\ix_{m}(\sigma)\bs{,}B\right) & = & \funktional{}{x^{m}(\sigma)}B=-\left(B\bs{,}\ix_{m}(\sigma)\right)\\
\left(x^{m}(\sigma)\bs{,}B\right) & = & -\funktional{}{\ix_{m}(\sigma)}B=(-)^{B}\left(B\bs{,}x^{m}(\sigma)\right)\qquad\fussend\end{eqnarray*}
}. Let us rename\index{$\tet^{\mu}$} \begin{eqnarray}
\tet^{\mu} & \equiv & \dew\sigma^{\mu}\end{eqnarray}
For a target space $r$-form \begin{eqnarray}
\rho^{(r)}(x^{m},\de x^{m}) & \equiv & \rho_{m_{1}\ldots m_{r}}(x)\de x^{m_{1}}\cdots\de x^{m_{r}}\end{eqnarray}
we define (in analogy to (\ref{eq:K-sig-tet-def}), but indicating
that we allow in the beginning only a variation in $\sigma$)\index{$\rho_{\tet}^{(r)}(\sigma)$}\begin{eqnarray}
\rho_{\tet}^{(r)}(\sigma) & \equiv & \rho^{(r)}(x^{m}(\sigma),\dew x^{m}(\sigma))=\rho_{m_{1}\ldots m_{r}}(x(\sigma))\dew x^{m_{1}}(\sigma)\cdots\dew x^{m_{r}}(\sigma)\end{eqnarray}
\textbf{Attention:} this vanishes identically for $r>d_{\textrm{w}}$
(worldvolume dimension). 

The worldvolume exterior derivative then induces the target space
exterior derivative in the following sense\begin{eqnarray}
\dew\rho_{\tet}^{(r)}(\sigma) & = & (\de\rho^{(r)})_{\tet}(\sigma)\end{eqnarray}
Again both sides vanish identically for now $r+1>d_{\textrm{w}}$,
which means that in this way one can calculate with target space fields
of form degree not bigger than the worldvolume dimension. If we want
to have the same relation for $K_{\tet}^{(k,k')}(\sigma)$\index{$K^{(k,k')}$@$K_{\tet}^{(k,k')}(\sigma)$}
(defined in the analogous way), we have to extend the identification
in (\ref{eq:indentification-of-basis-elements}) by \begin{eqnarray}
p_{m} & \To & \dew\ix_{m}(\sigma)\label{eq:identification-of-basis-elements-p}\end{eqnarray}
and get \begin{eqnarray}
\dew K_{\tet}^{(k,k')}(\sigma) & = & (\de K^{(k,k')})_{\tet}(\sigma)\end{eqnarray}
with \begin{eqnarray}
K_{\tet}^{(k,k')}(\sigma) & \equiv & K^{(k,k')}\left(x^{m}(\sigma),\dew x^{m}(\sigma),\ix_{m}(\sigma)\right)\label{eq:sigma-model-realizationI}\\
(\de K^{(k,k')})_{\tet}(\sigma) & \equiv & \de K^{(k,k')}\left(x^{m}(\sigma),\dew x^{m}(\sigma),\ix_{m}(\sigma),\dew\ix_{m}(\sigma)\right)\label{eq:sigma-model-realization-II}\end{eqnarray}
The analysis is thus very similar to that of the previous section.
\rem{hier ist tildeOmega usw versteckt}

\paragraph{Proposition 3a}

\emph{\index{proposition!antibracket of multivector valued forms (3a)}For
all multivector valued forms $K^{(k,k')},L^{(l,l')}$ on the target
space manifold, in a local coordinate patch seen as functions of $x^{m}$,$\de x^{m}$
and $\pe_{m}$, the following equation holds for the corresponding
sigma-model realizations (\ref{eq:sigma-model-realizationI},\ref{eq:sigma-model-realization-II})\begin{equation}
\boxed{(K_{\tet}(\sigma')\bs{,}L_{\tet}(\sigma))=\big(\underbrace{[K,_{\de}L]_{(1)}^{\Delta}}_{\lqn{-(-)^{k-k'}\left[\de K,L\right]_{(1)}^{\Delta}}}\big)_{\tet}(\sigma)\delta^{d_{\textrm{w}}}(\sigma-\sigma')-(-)^{k-k'}\tet^{\mu}\partial_{\mu}\delta^{d_{\textrm{w}}}(\sigma-\sigma')\big(\left[K,L\right]_{(1)}^{\Delta}\big)_{\tet}(\sigma)}\label{eq:PropositionIII}\end{equation}
}

\emph{Proof}$\quad$The proof is very similar to that one of proposition
3b (\ref{eq:PropositionIIIb}) and is therefore omitted at this place.$\quad\square$

\paragraph{Conjugate Superfields}

With $\tet^{\mu}=\dew\sigma^{\mu}$ we have introduced anticommuting
coordinates and it would be nice to extend the anti-bracket of the
fields $x^{m}$ and $\ix_{m}$ to a super-antibracket of conjugate
superfields. Remember, in the previous subsection we had the superfields
$\Phi^{m}=x^{m}+\tet\ce^{m}$ and its conjugate $\Es_{m}$. There
we had one $\tet$ and two component fields. In general the number
of component fields has to exceed the worldvolume dimension $d_{\textrm{w}}$\index{$d_{\rm w}$}
(the number of $\tet$'s) by one, s.th. we have to introduce a lot
of new fields to realize conjugate superfields. But before, let us
define the fermionic integration\index{integration measure $\mu(\tet)$}
measure\index{measure $\mu(\tet)$} \index{$\mu(\tet)$|itext{fermionic integration measure}}$\mu(\tet)$
via \begin{equation}
\int\mu(\tet)f(\tet)=\partl{\tet^{d_{\textrm{w}}}}\cdots\partl{\tet^{1}}f(\tet)=\frac{1}{d_{\textrm{w}}!}\epsilon^{\mu_{1}\ldots\mu_{d_{\textrm{w}}}}\partl{\tet^{\mu_{1}}}\cdots\partl{\tet^{\mu_{d_{\textrm{w}}}}}f(\tet)\end{equation}
The corresponding $d_{\textrm{w}}$-dimensional $\delta$-function
is\begin{eqnarray}
\delta^{d_{\textrm{w}}}(\tet'-\tet) & \equiv & (\tet'^{1}-\tet^{1})\cdots(\tet'^{d_{\textrm{w}}}-\tet^{d_{\textrm{w}}})=\\
 & = & \frac{1}{d_{\textrm{w}}!}\epsilon_{\mu_{1}\ldots\mu_{d_{\textrm{w}}}}(\tet'^{\mu_{1}}-\tet^{\mu_{1}})\cdots(\tet'^{\mu_{d_{\textrm{w}}}}-\tet^{\mu_{d_{\textrm{w}}}})=\\
 & = & \sum_{k=0}^{d_{\textrm{w}}}\frac{1}{k!(d_{\textrm{w}}-k)!}\epsilon_{\mu_{1}\ldots\mu_{d_{\textrm{w}}}}\tet'^{\mu_{1}}\cdots\tet'^{\mu_{k}}\tet^{\mu_{k+1}}\cdots\tet^{\mu_{d_{\textrm{w}}}}\\
\int\mu(\tet')\delta^{d_{\textrm{w}}}(\tet'-\tet)f(\tet') & = & f(\tet)\\
\delta^{d_{\textrm{w}}}(\tet'-\tet) & = & (-)^{d_{\textrm{w}}}\delta^{d_{\textrm{w}}}(\tet-\tet')\end{eqnarray}
For the two conjugate superfields, call them $\Phi^{m}$ and $\Ph_{m}$,
we want to have the canonical super anti bracket\begin{equation}
\left(\Ph_{m}(\sigma',\tet')\bs{,}\Phi^{n}(\sigma,\tet)\right)=\delta_{m}^{n}\delta^{d_{\textrm{w}}}(\sigma'-\sigma)\delta^{d_{\textrm{w}}}(\tet'-\tet)=-\left(\Phi^{n}(\sigma,\tet),\Ph_{m}(\sigma',\tet')\right)\label{eq:super-antibracket}\end{equation}
From the above considerations about the fermionic delta function it
is now clear, how these superfields can be defined (they are known
as \textbf{de Rham\index{de Rham superfield} superfield\index{superfield!de Rham $\sim$}s},
because of the interpretation of $\tet^{\mu}$ as $\dew\sigma^{\mu}$;
see e.g. \cite{Cattaneo:1999fm,Zucchini:2004ta}):\index{$\Phi^{m}(\sigma,\tet)$}\index{$\Ph_{m}(\sigma',\tet')$|itext{anti-superfield}}\begin{eqnarray}
\Phi^{m}(\sigma,\tet) & \equiv & x^{m}(\sigma)+\bs{x}_{\mu_{d_{\textrm{w}}}}^{m}(\sigma)\tet^{\mu_{d_{\textrm{w}}}}+x_{\mu_{d_{\textrm{w}}-1}\mu_{d_{\textrm{w}}}}^{m}(\sigma)\tet^{\mu_{d_{\textrm{w}}-1}}\tet^{\mu_{d_{\textrm{w}}}}+\ldots+x_{\mu_{1}\ldots\mu_{d_{\textrm{w}}}}^{m}(\sigma)\tet^{\mu_{1}}\cdots\tet^{\mu_{d_{\textrm{w}}}}\label{eq:deRhamSuperfieldPhi}\\
\Ph_{m}(\sigma',\tet') & \equiv & \frac{1}{d_{\textrm{w}}!}\epsilon_{\mu_{1}\ldots\mu_{d_{\textrm{w}}}}\tet'^{\mu_{1}}\cdots\tet'^{\mu_{d_{\textrm{w}}}}\ix_{m}(\sigma')+\frac{1}{(d_{\textrm{w}}-1)!1!}\epsilon_{\mu_{1}\ldots\mu_{d_{\textrm{w}}}}\tet'^{\mu_{1}}\cdots\tet'^{\mu_{d_{\textrm{w}}-1}}x_{m}^{+}\hoch{\mu_{d_{\textrm{w}}}}(\sigma')+\nonumber \\
 &  & \hspace{-1.5cm}+\frac{1}{(d_{\textrm{w}}-2)!2!}\epsilon_{\mu_{1}\ldots\mu_{d_{\textrm{w}}}}\tet'^{\mu_{1}}\cdots\tet'^{\mu_{d_{\textrm{w}}-2}}\ix_{m}\hoch{\mu_{d_{\textrm{w}}-1}\mu_{d_{\textrm{w}}}}(\sigma')+\ldots+\frac{1}{d_{\textrm{w}}!}\epsilon_{\mu_{1}\ldots\mu_{d_{\textrm{w}}}}\ix_{m}\hoch{\mu_{1}\ldots\mu_{d_{\textrm{w}}}}(\sigma')\qquad\label{eq:deRhamSuperfieldPhipl}\end{eqnarray}
The component fields with the matching number of worldsheet indices
are conjugate to each other, e.g.\begin{eqnarray}
\left(\ix_{m}\hoch{\mu_{1}\mu_{2}}(\sigma')\bs{,}x_{\nu_{1}\nu_{2}}^{n}(\sigma)\right) & = & \delta_{m}^{n}\delta_{\nu_{1}\nu_{2}}^{\mu_{1}\mu_{2}}\delta^{d_{\textrm{w}}}(\sigma-\sigma')\end{eqnarray}
For the notation with boldface symbols for anticommuting variables,
the worldvolume was assumed to be even-dimensional. In this case,
one can analytically continue the coordinate form of multivector-valued
forms of the form \begin{equation}
K^{(k,k')}(x,\de x,\pe)\equiv K_{m_{1}\ldots m_{k}}\hoch{n_{1}\ldots n_{k'}}\de x^{m_{1}}\wedge\cdots\wedge\de x^{m_{k}}\wedge\pe_{n_{1}}\wedge\cdots\wedge\pe_{n_{k'}}\end{equation}
to functions of superfields (in odd worldvolume dimension one would
get a symmetrization of the multivector-indices) and redefine $K(\sigma,\tet)$
of (\ref{eq:K-sig-tet-def}) to\index{$K^{(k,k')}(\sigma,\tet)$}
\begin{eqnarray}
K^{(k,k')}(\sigma,\tet) & \equiv & K^{(k,k')}\left(\Phi(\sigma,\tet),\dew\Phi(\sigma,\tet),\Ph(\sigma,\tet)\right)=\label{eq:K-sig-tet-def-II}\\
 & = & K_{m_{1}\ldots m_{k}}\hoch{n_{1}\ldots n_{k'}}(\Phi)\dew\Phi^{m_{1}}\cdots\dew\Phi^{m_{k}}\Ph_{n_{1}}\cdots\Ph_{n_{k'}}\end{eqnarray}
All other geometric quantities have to be understood in this new sense
now:\index{$T^{(t,t',t'')}(\sigma,\tet)$}\begin{eqnarray}
T^{(t,t',t'')}(\sigma,\tet) & \equiv & T^{(t,t',t'')}\left(\Phi(\sigma,\tet),\es\Phi(\sigma,\tet),\Ph(\sigma,\tet),\dew\Ph(\sigma,\tet)\right)\quad(\textrm{see }(\ref{eq:Tcbp}))\label{eq:T-sig-tet-II}\end{eqnarray}
 To stay with the examples used in (\ref{eq:T-sig-tet})-(\ref{eq:db-sig-tet}):\index{$dK$@$\de K(\sigma,\tet)$}\begin{eqnarray}
\textrm{e.g. }\de K(\sigma,\tet) & \equiv & \de K\left(\Phi(\sigma,\tet),\dew\Phi(\sigma,\tet),\Ph(\sigma,\tet),\dew\Ph(\sigma,\tet)\right)\qquad(\textrm{compare }(\ref{eq:dK-coord}))\label{eq:dK-of-sig-tet-II}\\
\textrm{or }\oo(\sigma,\tet) & \equiv & \oo\left(\dew\Phi(\sigma,\tet),\dew\Ph(\sigma,\tet)\right)=\dew\Phi^{m}(\sigma,\tet)\dew\Ph_{m}(\sigma,\tet)\quad(\textrm{compare }\oo=\ce^{m}p_{m})\qquad\quad\label{eq:o-of-sig-tet-II}\\
\hspace{-1.2cm}\left[K^{(k,k')},_{\de\,}L^{(l,l')}\right]_{(1)}^{\Delta}(\sigma,\tet) & \equiv & \left[K^{(k,k')}\bs{,}L^{(l,l')}\right]_{(1)}^{(\Delta)}\left(\Phi(\sigma,\tet),\dew\Phi(\sigma,\tet),\Ph(\sigma,\tet),\dew\Ph(\sigma,\tet)\right)\\
\de x^{m}(\sigma,\tet) & \equiv & \dew\Phi^{m}(\sigma,\tet)\\
(\de\pe_{m})(\sigma,\tet)\equiv(\de\be_{m})(\sigma,\tet) & \equiv & \dew\Ph_{m}(\sigma,\tet)\label{eq:db-sig-tet-II}\end{eqnarray}
Note that the former relation $K(\sigma,\tet)=K(\sigma)+\tet\de K(\sigma)$
does NOT hold any longer with those new definitions! Nevertheless
we get a very similar statement as compared to propositions 2 on page
\pageref{eq:Proposition1}:\rem{stattdessen $K(\sigma,\tet)=K(\tet)+\sigma^{\mu}\partial_{\mu}K(\tet)+\ldots$
oder (?) $K(\sigma+\tet)=K(\sigma)+\tet^{\mu}\partial_{\mu}K(\sigma)$}

\paragraph{Proposition 3b}

\emph{\index{proposition!antibracket of multivector valued forms (3b)}For
all multivector valued forms $K^{(k,k')},L^{(l,l')}$ on the target
space manifold, in a local coordinate patch seen as functions of $x^{m}$,$\de x^{m}$
and $\pe_{m}$, the following equation holds for even worldvolume-dimension
$d_{\textrm{w}}$ for the corresponding superfields (\ref{eq:K-sig-tet-def-II}):\begin{equation}
\boxed{(K(\sigma',\tet')\bs{,}L(\sigma,\tet))=\delta^{d_{\textrm{w}}}(\sigma'-\sigma)\delta^{d_{\textrm{w}}}(\tet'-\tet)\underbrace{\left[K,_{\de}L\right]_{(1)}^{\Delta}}_{\lqn{-(-)^{k-k'}\left[\de K,L\right]_{(1)}^{\Delta}}}(\sigma,\tet)-(-)^{k-k'}\tet^{\mu}\partial_{\mu}\delta^{d_{\textrm{w}}}(\sigma-\sigma')\delta^{d_{\textrm{w}}}(\tet'-\tet)\left[K,L\right]_{(1)}^{\Delta}(\sigma,\tet)}\label{eq:PropositionIIIb}\end{equation}
where $[K,L]_{(1)}^{\Delta}$ is the big bracket (\ref{eq:bc-big-bracket})
and $\left[K,_{\de}L\right]_{(1)}^{\Delta}$ is the derived bracket
of the big bracket (\ref{eq:bc-derived-of-bigbracket}).}

\emph{Note that $\sigma$ and $\tet$ have switched their roles compared
to the previous subsection (\ref{eq:Proposition1}), where the algebraic
bracket came together with the derivative with respect to $\tet$
of the delta-functions, while now it comes along with $\partial_{\mu}$
of the delta-functions.\vspace{.5cm}}

\emph{Proof}$\quad$Let us use again the second idea in the proof
of proposition 2, i.e. first collect the terms with derivatives of
the delta function, only to show that one gets the algebraic bracket,
and after that argue that the term with plain delta functions is its
derived bracket. In doing this, however, we will need to prove an
extension of the above proposition to objects like $\de K$ (or more
general an object $T^{(t,t',t'')}$ as in (\ref{eq:Tcbp})) that contain
the basis element $p_{m}$, which is then replaced by $\dew\Ph_{m}$
as e.g. in (\ref{eq:dK-of-sig-tet-II}).\\
(i) The antibracket between two such objects $T$ and $\tilde{T}$
gets contributions to the derivative of the delta-function only from
the antibrackets between $\dew\Phi^{m}$ and $\Ph_{m}$ and between
$\Phi^{m}$ and $\dew\Ph_{m}$ (compare (\ref{eq:super-antibracket}))\begin{eqnarray}
\left(\Ph_{m}(\sigma',\tet')\bs{,}\dew\Phi^{n}(\sigma,\tet)\right) & = & \delta_{m}^{n}\tet^{\mu}\partial_{\mu}\delta^{d_{\textrm{w}}}(\sigma'-\sigma)\delta^{d_{\textrm{w}}}(\tet'-\tet)\\
\left(\dew\Phi^{n}(\sigma',\tet')\bs{,}\Ph_{m}(\sigma,\tet)\right) & = & \delta_{m}^{n}\tet^{\mu}\partial_{\mu}\delta^{d_{\textrm{w}}}(\sigma'-\sigma)\delta^{d_{\textrm{w}}}(\tet'-\tet)\\
\left(\dew\Ph_{m}(\sigma',\tet')\bs{,}\Phi^{n}(\sigma,\tet)\right) & = & -\delta_{m}^{n}\tet^{\mu}\partial_{\mu}\delta^{d_{\textrm{w}}}(\sigma'-\sigma)\delta^{d_{\textrm{w}}}(\tet'-\tet)\\
\left(\Phi^{n}(\sigma',\tet')\bs{,}\dew\Ph_{m}(\sigma,\tet)\right) & = & -\tet^{\mu}\left(\Phi^{n}(\sigma',\tet')\bs{,}\partial_{\mu}\Ph_{m}(\sigma,\tet)\right)=\delta_{m}^{n}\tet^{\mu}\partial_{\mu}\delta^{d_{\textrm{w}}}(\sigma'-\sigma)\delta^{d_{\textrm{w}}}(\tet'-\tet)\end{eqnarray}
The last case is the only one where we had to take care of an extra
sign stemming from $\tet$ jumping over the graded comma. Comparing
this to (\ref{eq:Poisson-bracket-bc}), where we had \begin{eqnarray}
\left\{ \be_{m},\ce^{n}\right\}  & = & \delta_{m}^{n}\\
\left\{ \ce^{n},\be_{m}\right\}  & = & \delta_{m}^{n}\\
\left\{ p_{m},x^{n}\right\}  & = & \delta_{m}^{n}\\
\left\{ x^{n},p_{m}\right\}  & = & -\delta_{m}^{n}\end{eqnarray}
one recognizes that the only difference is an overall odd factor $\tet^{\mu}\partial_{\mu}\delta^{d_{\textrm{w}}}(\sigma'-\sigma)\delta^{d_{\textrm{w}}}(\tet'-\tet)$
(the delta-function for $\tet$ is an even object for even worldvolume
dimension $d_{\textrm{w}}$) and an additional minus sign for the
lower two lines, but the corresponding indices just get contracted
like for the Poisson bracket. After such a bracket of basis elements
has been calculated (which happens just between the remaining factors
of $T$ (at $\sigma'$) on the left and the remaining factors of $\tilde{T}$
(at $\sigma$) on the right) this overall odd factor has to be pulled
to the very left which gives an additional factor of $(-)^{t-t'}$
(in the notation of (\ref{eq:Tcbp})) plus an additional minus sign
for the upper two lines which compensates the relative minus sign
of before and we get just an overall factor of $-(-)^{t-t'}\tet^{\mu}\partial_{\mu}\delta^{d_{\textrm{w}}}(\sigma'-\sigma)\delta^{d_{\textrm{w}}}(\tet'-\tet)$
in all cases at the very left as compared to the Poisson-bracket.
The remaining terms are still partly at $\sigma$ and partly at $\sigma'$,
but using\begin{equation}
A(\sigma)B(\sigma')\partial_{\mu}\delta(\sigma-\sigma')=A(\sigma)\partial_{\mu}B(\sigma)\delta(\sigma-\sigma')+A(\sigma)B(\sigma)\partial_{\mu}\delta(\sigma-\sigma')\quad\forall A,B\label{eq:hint}\end{equation}
we can take all remaining factors in $T(\sigma',\tet')$ at $\sigma$,
while $\tet'$ is set to $\tet$ anyway by the $\delta$-function.
We have thus verified one of the coefficients of the complete antibracket:\begin{eqnarray}
(T(\sigma',\tet'),\tilde{T}(\sigma,\tet)) & = & -(-)^{t-t'}\tet^{\mu}\partial_{\mu}\delta^{d_{\textrm{w}}}(\sigma-\sigma')\delta^{d_{\textrm{w}}}(\tet'-\tet)\left[T,\tilde{T}\right]_{(1)}^{\Delta}(\sigma,\tet)+\nonumber \\
 &  & +\delta^{d_{\textrm{w}}}(\sigma-\sigma')\delta^{d_{\textrm{w}}}(\tet'-\tet)A(\sigma,\tet)\label{eq:gruenerPunkt}\end{eqnarray}
with $A(\sigma,\tet)$ yet to be determined.\\
(ii) It remains to show that $A(\sigma,\tet)$ is a derived expression
of $\left[T,\tilde{T}\right]_{(1)}^{\Delta}$. A hint to this fact
is already given in (\ref{eq:hint}), but this is not enough, as there
is also a contribution from the $(\Phi^{m},\Ph_{n})$-brackets. In
order to get a precise relation between $A(\sigma,\tet)$ and $\left[T,\tilde{T}\right]_{(1)}^{\Delta}(\sigma,\tet)$,
let us see how one can extract them from the complete antibracket.
In order to hit the delta functions with the integration, it is enough
to integrate over the patch $U(\sigma)$ containing the point which
is parametrized by $\sigma^{\mu}$. The last term in (\ref{eq:gruenerPunkt})
is the only one contributing when integrating over $\sigma'$ and
$\tet$\begin{eqnarray}
A(\sigma,\tet) & = & \int_{U(\sigma)}\de^{d_{\textrm{w}}}\sigma'\int\mu(\tet')\quad(T(\sigma',\tet'),\tilde{T}(\sigma,\tet))\end{eqnarray}
That the first term on the righthand side of (\ref{eq:gruenerPunkt})
does not contribute is not obvious as $U(\sigma)$ might have a boundary.
However, for this term one ends up integrating a $d_{\textrm{w}}$-dimensional
delta-function over a boundary of dimension not higher than $d_{\textrm{w}}-1$,
so that one is left with an at least one-dimensional delta-function
on the boundary which vanishes as the boundary of the open neighbourhood
$U(\sigma)$ of $\sigma$ of course nowhere hits $\sigma$.

Extracting the algebraic bracket $\left[T,\tilde{T}\right]_{(1)}^{\Delta}$
is a bit more tricky. One can do it via\begin{eqnarray}
\hspace{-.7cm}\zwek{{\scriptstyle \textrm{for any fixed}}}{{\scriptstyle \textrm{index }\lambda}}:\,\left[T,\tilde{T}\right]_{(1)}^{\Delta}(\sigma,\tet) & = & -(-)^{t-t'}\int_{U(\sigma)}\de^{d_{\textrm{w}}}\sigma'\int\mu(\tet')\quad\left(\frac{e^{\sigma'^{\lambda}}}{e^{\sigma^{\lambda}}}-1\right)\partl{\tet^{\lambda}}(T(\sigma',\tet')\bs{,}\tilde{T}(\sigma,\tet))\quad\label{eq:rosa-Pause}\end{eqnarray}
The boundary term proportional to $\left(\frac{e^{\sigma'^{\lambda}}}{e^{\sigma^{\lambda}}}-1\right)\delta^{d_{\textrm{w}}}(\sigma-\sigma')$
appearing above on the righthand side after partial integration vanishes
as $\sigma'$ in the prefactor is set to $\sigma$ via the delta function.
\\
The claim is now that $A(\sigma,\tet)=-(-)^{t-t'}\left[\de T,\tilde{T}\right]_{(1)}^{\Delta}(\sigma,\tet)$.
So let us calculate the righthand side via (\ref{eq:rosa-Pause}):\begin{eqnarray}
\left[\de T,\tilde{T}\right]_{(1)}^{\Delta}(\sigma,\tet) & = & -(-)^{t+1-t'}\int_{U(\sigma)}\de^{d_{\textrm{w}}}\sigma'\int\mu(\tet')\quad\left(\frac{e^{\sigma'^{\lambda}}}{e^{\sigma^{\lambda}}}-1\right)\partl{\tet^{\lambda}}(\de T(\sigma',\tet')\bs{,}\tilde{T}(\sigma,\tet))=\\
 & = & -(-)^{t+1-t'}\int\de^{d_{\textrm{w}}}\sigma'\int\mu(\tet')\quad\left(\frac{e^{\sigma'^{\lambda}}}{e^{\sigma^{\lambda}}}-1\right)\partl{\tet^{\lambda}}\tet'^{\mu}\partial_{\mu}'(T(\sigma',\tet')\bs{,}\tilde{T}(\sigma,\tet))\end{eqnarray}
$(T\bs{,}\tilde{T})$ contains in both terms a plain $\delta$-function
for the fermionic variables $\tet$, so that we can replace $\tet'$
by $\tet$. Integration by parts of $\partial_{\mu}'$ (where possible
boundary terms again do not contribute because of the vanishing of
the delta function and its derivative on the boundary) delivers the
desired result\begin{equation}
\left[\de T,\tilde{T}\right]_{(1)}^{\Delta}(\sigma,\tet)=-(-)^{t-t'}\int\de^{d_{\textrm{w}}}\sigma'\int\mu(\tet')\quad(T(\sigma',\tet')\bs{,}\tilde{T}(\sigma,\tet))=-(-)^{t-t'}A(\sigma,\tet)\end{equation}
This completes the proof of proposition 3b. $\qquad\square$

\chapter{Applications in string theory or 2d CFT}

\label{sec:Applications-in-string}

In the previous section the dimension of the worldvolume was arbitrary
or even dimensional. The appearance of derived brackets (including
e.g. the Dorfman bracket) is thus not a special feature of a 2-dimensional
sigma-model like string theory. There are, however, special features
in string theory. Currents in string theory (which have conformal
weight one) naturally are sums of 1-forms and vectors, if one takes
the identification $\partial_{1}x^{m}(\sigma)\leftrightarrow\de x^{m}$
and $p_{m}(\sigma)\leftrightarrow\pe_{m}$, as in \cite{Alekseev:2004np}
(see footnote \ref{foot:Strobl}), e.g. $\partial x^{m}=\partial_{1}x^{m}-\partial_{0}x^{m}\hat{=}\de x^{m}-\eta^{mn}\pe_{n}$
. This is closely related to the identification in our previous section
in the antifield formalism. In addition, only in two dimensions a
single $\tet$ can be interpreted as a worldsheet Weyl spinor (in
1 dimension it can be seen as a Dirac-spinor, but in higher dimensions
the interpretation of $\tet$ as worldvolume spinor breaks down).
As we ended the last section with the antifield formalism, which therefore
is perhaps still more present, let us start this section in the reversed
order, beginning with the application in the antifield formalism.

\section{Poisson sigma-model and Zucchini's {}``Hitchin sigma-model''}

\label{sub:Zucchini} \index{sigma-model!Poison $\sim$}\index{sigma-model!Hitchin $\sim$}\index{Poisson sigma model}\index{Hitchin sigma model}Remember
for a moment the Poisson-$\sigma$-model \cite{Schaller:1994es,Cattaneo:1999fm}.
It is a two-dimensional sigma-model ($d_{\textrm{w}}=2$) of the form
\begin{equation}
S_{0}=\int_{\Sigma}\,\bs{\eta}_{m}\dew x^{m}+\frac{1}{2}P^{mn}(x)\bs{\eta}_{m}\bs{\eta}_{n}\end{equation}
where $\bs{\eta}_{m}$ is a worldsheet one-form. This model is topological
if and only if the Poisson-structure $P^{mn}(x)$ is integrable, i.e.
the Schouten\index{Schouten bracket}-bracket of $P$ with itself
vanishes\begin{eqnarray}
S_{0}\textrm{ topological} & \iff & \left[P\bs{,}P\right]=0\end{eqnarray}
It gives on the one hand a field theoretic implementation of Kontsevich's
star\index{star product} product \cite{Cattaneo:1999fm} and is on
the other hand related to string theory via a topological limit (big
antisymmetric part in the open string metric), which leads to the
relation between string theory and noncommutative geometry.

The necessary ghost fields for the action can be introduced by extending
$x$ and $\eta$ to de Rham\index{de Rham superfield} superfield\index{superfield!de Rham $\sim$}s
as in (\ref{eq:deRhamSuperfieldPhi},\ref{eq:deRhamSuperfieldPhipl})\index{$\beta_m$@$\bs{\beta}_{m}$}\index{$\eta_{\mu m}$}\begin{eqnarray}
\Phi^{m}(\sigma,\tet) & \equiv & x^{m}(\sigma)+\underbrace{\bs{x}_{\mu}^{m}(\sigma)}_{\epsilon_{\mu\nu}\bs{\eta}^{+\nu n}}\tet^{\mu}+\underbrace{x_{\mu_{1}\mu_{2}}^{m}(\sigma)}_{-\frac{1}{2}\eps_{\mu_{1}\mu_{2}}\beta^{+\, m}}\tet^{\mu_{1}}\tet^{\mu_{2}}\\
\Ph_{m}(\sigma',\tet') & \equiv & \underbrace{\frac{1}{2!}\epsilon_{\mu_{1}\mu_{2}}\ix_{m}\hoch{\mu_{1}\mu_{2}}(\sigma')}_{\equiv\bs{\beta}_{m}(\sigma')}+\tet'^{\mu_{1}}\underbrace{\epsilon_{\mu_{1}\mu_{2}}x_{m}^{+}\hoch{\mu_{2}}(\sigma')}_{\eta_{\mu_{1}m}}+\frac{1}{2}\epsilon_{\mu_{1}\mu_{2}}\tet'^{\mu_{1}}\tet'^{\mu_{2}}\ix_{m}(\sigma')\end{eqnarray}
One can use Hodge-duality to rename some component fields as indicated.
$\bs{\beta}_{m}$ is then the ghost field related to the gauge symmetry.
The action including ghost fields and antifields simply reads\index{$P^{mn}$|itext{Poisson structure}}\begin{eqnarray}
S & = & \int d^{2}\sigma\,\int\mu(\tet)\quad\Ph_{m}\dew\Phi^{m}+\frac{1}{2}P^{mn}(\Phi)\Ph_{m}\Ph_{n}\end{eqnarray}
The expression under the integral corresponds to the tensor $-\delta_{m}\hoch{n}\de x^{m}\wedge\pe_{n}+\frac{1}{2}P^{mn}\pe_{m}\wedge\pe_{n}$
and the antibracket in the master-equation $(S,S)$ implements the
Schoutenbracket on $P$, which is a well known relation. Therefore
we will concentrate on a second example, which is very similar, but
less known. 

Zucchini suggested in \cite{Zucchini:2004ta} a 2-dimensional sigma-model
which is topological if a generalized complex structure in the target
space is integrable (see subsection \ref{sub:generalized-complex-structure}
on page \pageref{sub:generalized-complex-structure} and \ref{sub:Integrability-of-J}
on page \pageref{sub:Integrability-of-J} to learn more about generalized
complex structures). His model is of the form\begin{eqnarray}
S & = & \int d^{2}\sigma\,\int\mu(\tet)\quad\left(\Ph_{m}\dew\Phi^{m}\,+\,\right)\quad\frac{1}{2}P^{mn}(\Phi)\Ph_{m}\Ph_{n}-\frac{1}{2}Q_{mn}(\Phi)\dew\Phi^{m}\dew\Phi^{n}-J^{n}\tief{m}\dew\Phi^{m}\Ph_{n}\label{eq:Zucchini-Wirkung}\end{eqnarray}
where $P^{mn}$, $Q_{mn}$ and $J^{m}\tief{n}$ are the building blocks
of the generalized\index{generalized complex structure} complex\index{complex structure!generalized $\sim$}\index{almost complex structure|see{complex structure}}
structure (\ref{eq:J-matrix})\index{$J^M$@$\protect\mc{J}^M\tief{N}$|itext{generalized complex structure}}\index{$J^m\tief{n}$|itext{complex structure}}\index{$Q_{mn}$}\begin{eqnarray}
\mc{J}^{M}\tief{N} & = & \left(\begin{array}{cc}
J^{m}\tief{n} & P^{mn}\\
-Q_{mn} & -J^{n}\tief{m}\end{array}\right)\end{eqnarray}
The first term of (\ref{eq:Zucchini-Wirkung}) can be absorbed by
a field redefinition as already observed in \cite{Zucchini:2005rh}.
Ignoring thus the first term and using our notations of before, $S$
can be rewritten as\index{$J^Z$@$\protect\mc{J}(\Phi,\dew\Phi,\Ph)$}\begin{equation}
S=\int d^{2}\sigma\,\int\mu(\tet)\quad\frac{1}{2}\mc{J}(\Phi,\dew\Phi,\Ph)\end{equation}
Calculating the master equation explicitely and collecting the terms
which combine to the lengthy tensors for the integrability condition
(see (\ref{eq:integrability-tensor-I})-(\ref{eq:integrability-tensor-IV}))
is quite cumbersome, so we can enjoy using instead proposition 3b
on page \pageref{eq:PropositionIIIb}. For a worldsheet without boundary
its integrated version reads\begin{equation}
\left(\int d^{d_{\textrm{w}}}\sigma'\int\mu(\tet')K(\sigma',\tet'),\int\de^{d_{\textrm{w}}}\sigma\int\mu(\tet)L(\sigma,\tet)\right)=\int d^{d_{\textrm{w}}}\sigma\int\mu(\tet)\left[K,_{\de}L\right]_{(1)}^{\Delta}(\sigma,\tet)\label{eq:PropositionIIIb-integrated}\end{equation}
which leads to the relation\begin{eqnarray}
(S,S) & = & 0\qquad\iff\int d^{2}\sigma\int\mu(\tet)\left[\mc{J},_{\de}\mc{J}\right]_{(1)}^{\Delta}(\sigma,\tet)=0\label{eq:Zucchini-last-equation}\end{eqnarray}
\rem{no condition $J^2=-1$??}The derived bracket of the big bracket
of $\mc{J}$ with itself contains already the generalized\index{generalized Nijenhuis tensor}
Nijenhuis\index{Nijenhuis tensor!generalized $\sim$}\index{$N_{M}$@$\protect\mc{N}_{M_{1}M_{2}M_{3}}$|itext{generalized Nijenhuis tensor}}
tensor (see in the appendix in equation (\ref{eq:derived-bracket-for-J})
and in the discussion around)\index{$t^M$@$\protect\basis^{M}$} \begin{eqnarray}
\left[\mc{J},_{\de}\mc{J}\right]_{(1)}^{\Delta} & = & \mc{N}_{M_{1}M_{2}M_{3}}\basis^{M_{1}}\basis^{M_{2}}\basis^{M_{3}}-4\mc{J}^{JI}\mc{J}_{IM}\basis^{M}p_{J}=\\
 & \stackrel{\mc{J}^{2}=-1}{=} & \mc{N}_{M_{1}M_{2}M_{3}}\basis^{M_{1}}\basis^{M_{2}}\basis^{M_{3}}+4\oo\label{eq:relation-between-derived-and-Nij-Tens-in-main-part}\\
\basis^{M} & = & (\de x^{m},\pe_{m}),\qquad p_{J}=(p_{j},0)\\
\oo(\de x,p) & = & \de x^{m}p_{m}\end{eqnarray}
For $\mc{J}^{2}=-1$ the last term is proportional to the generator
$\oo$ (remember (\ref{eq:BRST-op})). In (\ref{eq:Zucchini-last-equation}),
however, it appears with $\de x$ and $p$ replaced by the superfields
as in (\ref{eq:o-of-sig-tet-II}) \begin{eqnarray}
\oo(\sigma,\tet) & = & \dew\Phi^{m}(\sigma,\tet)\dew\Ph_{m}(\sigma,\tet)=-\dew\left(\dew\Phi^{m}(\sigma,\tet)\Ph_{m}(\sigma,\tet)\right)\end{eqnarray}
which is a total worldsheet derivative and therefore drops during
the integration. We are left with the generalized Nijenhuis tensor
as a function of superfields\index{$N_{Z}$@$\protect\mc{N}(\sigma,\tet)$}\begin{eqnarray}
\mc{N}(\sigma,\tet) & = & \mc{N}_{M_{1}M_{2}M_{3}}(\Phi)\q{\basis}^{M_{1}}\q{\basis}^{M_{2}}\q{\basis}^{M_{3}}\\
\textrm{with }\q{\basis}^{M} & \equiv & (\dew\Phi^{m},\Ph_{m})\end{eqnarray}
Written in small indices\begin{eqnarray}
\mc{N}(\sigma,\tet) & = & \mc{N}_{m_{1}m_{2}m_{3}}(\Phi)\underbrace{\dew\Phi^{m_{1}}\dew\Phi^{m_{1}}\dew\Phi^{m_{1}}}_{=0}+3\mc{N}^{n}\tief{m_{1}m_{2}}(\Phi)\Ph_{n}\dew\Phi^{m_{1}}\dew\Phi^{m_{2}}+\nonumber \\
 &  & +3\mc{N}_{n}\hoch{m_{1}m_{2}}(\Phi)\dew\Phi^{n}\Ph_{m_{1}}\Ph_{m_{2}}+\mc{N}^{m_{1}m_{2}m_{3}}(\Phi)\Ph_{m}\Ph_{m}\Ph_{m}\end{eqnarray}
 One realizes that the first term vanishes identically (as mentioned
in \cite{Zucchini:2004ta}) and only the remaining three tensors are
required to vanish in order to satisfy (\ref{eq:Zucchini-last-equation}).
\rem{old\index{old and wrong} and wrong\index{wrong!old and $\sim$}:
The same model $S=\int d^{d_{\textrm{w}}}\sigma\,\int\mu(\tet)\quad\frac{1}{2}\mc{J}(\Phi,\dew\Phi,\Ph)$
in dimensions higher than 2 should however implement the complete
generalized Nijenhuis tensor. As the worldvolume has to be even-dimensional
for this setting (see proposition 3b on page \pageref{eq:PropositionIIIb}),
we need the worldvolume dimension $d_{\textrm{w }}$ to be at least
four.}\rem{twisted?\\
Quantum case \begin{eqnarray*}
W & = & S+\hbar M_{1}+\hbar^{2}M_{2}+\ldots\\
(W,W) & = & i\hbar\Delta W\\
\Delta & = & (-)^{A}\funktional{}{\Phi^{A}}\funktional{}{\Phi_{A}^{*}}\end{eqnarray*}
}

\section{Relation between a second worldsheet supercharge and generalized
complex geometry}

\label{sub:Zabzine}

In \cite{Lindstrom:2004iw} the relation between an extended worldsheet
supersymmetry in string theory and the presence of an integrable generalized
complex structure was explored. Zabzine clarified in \cite{Zabzine:2005qf}
the relation in an model independent way in a Hamiltonian description.
The structures appearing there are almost the same that we have discussed
before although we have to modify the procedure a little bit due to
the interpretation of $\tet$ as a worldsheet spinor.

Consider a sigma-model with 2-dimensional worldvolume (worldsheet\index{worldsheet})
with manifest $N=1$ supersymmetry on the worldsheet. In the phase
space there is only one $\sigma$-coordinate left. Let us denote the
corresponding superfields, following loosely \cite{Zabzine:2005qf},
by\index{$\lam^{m}(\sigma)$}\index{$\ro_{m}(\sigma)$}\begin{eqnarray}
\Phi^{m}(\sigma,\tet) & \equiv & x^{m}(\sigma)+\tet\lam^{m}(\sigma)\\
\Es_{m}(\sigma,\tet) & \equiv & \ro_{m}(\sigma)+\tet p_{m}(\sigma)\end{eqnarray}
In comparison to section \ref{sub:Natural-appearance}, there is a
change of notation from $\ce^{m}\To\lam^{m}$ \rem{so lassen?} and
$\be_{m}\To\ro_{m}$ as $\be$ and $\ce$ suggest the interpretation
as ghosts which is not true in this case, where $\lam$ and $\ro$
are worldsheet fermions. Introduce now, following Zabzine, the generator
$\qu$ \index{$Q_\theta$@$\qu$}of the \textbf{manifest SUSY}\index{SUSY!generator}
and the corresponding\index{SUSY!covariant derivative} \textbf{covariant
derivative} $\cov$\index{$D_\theta$@$\cov$}\begin{eqnarray}
\qu & \equiv & \partial_{\tet}+\tet\partial_{\sigma}\\
\cov & \equiv & \partial_{\tet}-\tet\partial_{\sigma}\end{eqnarray}
with the SUSY algebra\rem{%
\footnote{\index{footnote!\thefoot. further useful relations}Further useful
relations are \begin{eqnarray*}
\int\de\tet\mc{L} & = & \cov\mc{L}\mid=\partial_{\tet}\mc{L}=\qu\mc{L}\mid\\
\int\de\sigma\int\de\tet\mc{L} & = & \int\de\sigma\,\cov\mc{L}=\int\de\sigma\,\partial_{\tet}\mc{L}=\int\de\sigma\,\qu\mc{L}\qquad\fussend\end{eqnarray*}
}} \begin{eqnarray}
\left[\qu,\qu\right] & = & 2\partial_{\sigma}=-\left[\cov,\cov\right]\\
\left[\qu,\cov\right] & = & 0\end{eqnarray}
$\qu$ is the sum of two nilpotent differential operators, namely
$\partial_{\tet}$ and $\tet\partial_{\sigma}$. Acting on the Superfields
$\Phi^{m}$ and $\Es^{m}$, they induce the differentials $\es$ and
$\tilde{\es}$\index{$s$@$\tilde{\es}$} on the component fields,
which are in turn generated via the Poisson bracket by phase space
functions $\OO$ (the same as (\ref{eq:Omega})) and $\tilde{\OO}$.
\rem{(as a phase space function similar to (\ref{eq:tildOmega}),
but with a completely different action on multivector valued forms
in the present setting, as we will see below).}\index{$\Omega$@$\OO$}\index{$\Omega$@$\tilde{\OO}$}
\begin{eqnarray}
\OO & \equiv & \int d\sigma\:\lam^{k}p_{k}\label{eq:OmegaII}\\
\tilde{\OO} & = & -\int d\sigma\:\partial_{\sigma}x^{k}\ro_{k}\label{eq:tildOmegaII}\\
\es x^{m}\equiv\left\{ \OO,x^{m}\right\}  & = & \lam^{m}\leftrightarrow\de x^{m},\quad\es\ro_{m}\equiv\left\{ \OO,\ro_{m}\right\} =p_{m}\leftrightarrow\de(\pe_{m}),\\
\tilde{\es}\lam^{m}\equiv\left\{ \tilde{\OO},\lam^{m}\right\}  & = & -\partial_{\sigma}x^{m},\quad\tilde{\es}p_{k}=-\partial_{\sigma}\ro_{k}=\left\{ \tilde{\OO},p_{k}\right\} ,\\
\es\Phi^{m} & = & \partial_{\tet}\Phi^{m},\qquad\es\Es_{m}=\partial_{\tet}\Es_{m}\\
\tilde{\es}\Phi^{m} & = & \tet\partial_{\sigma}\Phi^{m},\qquad\tilde{\es}\Es_{m}=\tet\partial_{\sigma}\Es_{m}\end{eqnarray}
The Poisson-generator for the SUSY transformations of the component
fields induced by%
\footnote{\index{footnote!\thefoot. worldsheet SUSY transformations}We have
\begin{eqnarray*}
\qu\Phi^{m} & = & \lam^{m}+\tet\partial_{\sigma}x^{m},\qquad\qu\Es_{m}=p_{m}+\tet\partial_{\sigma}\ro_{m}\\
\cov\Phi^{m} & = & \lam^{m}(\sigma)-\tet\partial_{\sigma}x^{m},\qquad\cov\Es_{m}=p_{m}-\tet\partial_{\sigma}\ro_{m}\\
\delta_{\feps}x^{m} & = & \feps\lam^{m},\qquad\delta_{\feps}\lam^{m}=-\feps\partial_{\sigma}x^{m}\\
\delta_{\feps}\ro_{m} & = & \feps p_{m},\qquad\delta_{\feps}p_{m}=-\feps\partial_{\sigma}\ro_{m}\qquad\fussend\end{eqnarray*}
} $\qu$ is thus the sum of the generators of $\es$ and $\tilde{\es}$:
\begin{eqnarray}
\Q & = & \OO+\tilde{\OO}=\int d\sigma\:\lam^{k}p_{k}-\partial_{\sigma}x^{k}\ro_{k}=-\int d\sigma\int d\tet\,\qu\Phi^{k}\Es_{k}\label{eq:SUSY-generator}\end{eqnarray}
In (\ref{eq:superfield-definition}) superfields were defined via
$\partial_{\tet}Y=\es Y$ in order to implement the exterior derivative
directly with $\partial_{\tet}$. In that sense $\Phi$, $\Es$, $\de\Phi$,
$\de\Es$ and all analytic functions of them were superfields. In
the context of worldsheet supersymmetry, one prefers of course a supersymmetric
covariant formulation. Let us therefore define in this subsection
proper \textbf{superfield\index{superfield}s} via \begin{equation}
Y\textrm{ is a superfiled }\quad:\iff\quad\qu Y\stackrel{!}{=}\left\{ \Q,Y\right\} =(\es+\tilde{\es})Y\label{eq:superfield-def-susy}\end{equation}
 \rem{($\es+\tilde{\es}$ is invariant under the change $\es\leftrightarrow\tilde{\es}$
and therefore probably under the change of the two mechanisms (identifications))
}which holds for $\Phi$, $\Es$,$\cov\Phi$, $\cov\Es$, all analytic
functions of them (like our analytically continued multivector valued
forms) and worldsheet spatial derivatives $\partial_{\sigma}$ thereof
(but not for e.g. $\qu\Phi$. This means that although we have $\qu\Phi=(\es+\tilde{\es})\Phi$
this does not hold for a second action, i.e. $\qu^{2}\Phi\neq(\es+\tilde{\es})^{2}\Phi$,
which explains the somewhat confusing fact that the Poisson-generator
$\Q$ \index{$Q$@$\Q$|itext{SUSY generator}}has the opposite sign
in the algebra than $\qu$\rem{($\qu$ and ($\es+\tilde{\es}$) anticommute!)}\begin{eqnarray}
\left\{ \Q,\Q\right\}  & = & -2P\label{eq:SUSY-algebra}\end{eqnarray}
where we introduced the phase-space generator $P$ for the worldsheet
translation induced by $\partial_{\sigma}$\index{$P$} \begin{eqnarray}
P & \equiv & \int d\sigma\quad\partial_{\sigma}x^{k}p_{k}+\partial_{\sigma}\lam^{k}\ro_{k}=\int d\sigma\int d\tet\quad\partial_{\sigma}\Phi^{k}\Es_{k}\end{eqnarray}
The same phenomenon appears for the differentials $\es$ and $\tilde{\es}$.
The graded commutator of $\partial_{\tet}$ and $\tet\partial_{\sigma}$
is the worldsheet derivative $\left[\partial_{\tet},\tet\partial_{\sigma}\right]=\partial_{\sigma}$,
while the algebra for $\es$ and $\tilde{\es}$ has the opposite sign\begin{eqnarray}
\left[\es,\tilde{\es}\right]Y(\sigma,\tet) & = & -\partial_{\sigma}Y(\sigma,\tet)\\
\es\tilde{\OO}=\left\{ \OO,\tilde{\OO}\right\}  & = & -P=\tilde{\es}\OO\end{eqnarray}
One major statement in \cite{Zabzine:2005qf} is as follows: Making
a general ansatz for a generator of a second, non-manifest supersymmetry,
of the form (some signs are adopted to our conventions)\begin{eqnarray}
\Q_{2} & \equiv & \frac{1}{2}\int d\sigma\int d\tet\quad(P^{mn}(\Phi)\Es_{m}\Es_{n}-Q_{mn}(\Phi)\cov\Phi^{m}\cov\Phi^{n}+2J^{m}\tief{n}(\Phi)\Es_{m}\cov\Phi^{n})\label{eq:Qzwei}\end{eqnarray}
and requiring the same algebra as for $\Q$ in (\ref{eq:SUSY-algebra})\begin{eqnarray}
\left\{ \Q_{2},\Q_{2}\right\}  & = & -2P\label{eq:SUSY-alg-II}\\
\Big(\left\{ \Q,\Q_{2}\right\}  & = & 0\Big)\label{eq:useless-condition}\end{eqnarray}
is equivalent to \begin{eqnarray}
\mc{J}^{M}\tief{N} & \equiv & \left(\begin{array}{cc}
J^{m}\tief{n} & P^{mn}\\
-Q_{mn} & -J^{n}\tief{m}\end{array}\right)\end{eqnarray}
being an integrable generalized complex structure (see in the appendix
\ref{sub:generalized-complex-structure} on page \pageref{sub:generalized-complex-structure}
and \ref{sub:Integrability-of-J} on page \pageref{sub:Integrability-of-J}).
On a worldsheet without boundary, the second condition is actually
superfluous, because it is already implemented via the ansatz: The
expression in the integral is an analytic function of superfields
and therefore a superfield itself. According to (\ref{eq:superfield-def-susy})
we can replace at this point the commutator with $\Q$ with the action
of $\qu$ and get \begin{equation}
\left\{ \Q,\Q_{2}\right\} =\int d\sigma\int d\tet\quad\qu(\ldots)=\int d\sigma\quad\partial_{\sigma}(\ldots)=0\end{equation}
 For the other condition, the actual supersymmetry algebra (\ref{eq:SUSY-alg-II}),
the aim of the present considerations should now be clear. The generalized
complex structure $\mc{J}$ itself is a sum of multivector valued
forms\begin{eqnarray}
\mc{J} & \equiv & \mc{J}^{MN}(x)\basis_{M}\basis_{N}\equiv P^{mn}(x)\pe_{m}\wedge\pe_{n}-Q_{mn}(x)\de x^{m}\de x^{n}+2J^{m}\tief{n}(x)\pe_{m}\wedge\de x^{n}\end{eqnarray}
which can be seen as a function of $x$ and the basis elements \begin{equation}
\mc{J}=\mc{J}(x,\de x,\pe)\end{equation}
In \ref{sub:Natural-appearance} we replaced the arguments of functions
like this with {}``superfields'' $x^{m}\To\Phi^{m}$, $\de x^{m}\to\partial_{\tet}\Phi^{m}$
and $\pe_{m}\To\Es_{m}$. The name superfield might have been misleading,
as $\partial_{\tet}\Phi$ is only a superfield in the sense that it
implements the target-space exterior derivative via $\partial_{\tet}$,
but it is not a superfield in the sense of worldsheet supersymmetry.
In a supersymmetric theory one prefers a supersymmetric covariant
formulation. Working with $\partial_{\tet}\Phi$ as before is therefore
not desirable and we replace $\partial_{\tet}\Phi$ by $\cov\Phi$,
leading directly to $\Q_{2}$ (\ref{eq:Qzwei}) which now can be written
as\index{$Q_2$@$\Q_{2}$} \begin{eqnarray}
\Q_{2} & = & \frac{1}{2}\int d\sigma\int d\tet\,\mc{J}\left(\Phi(\sigma,\tet),\cov\Phi(\sigma,\tet),\Es(\sigma,\tet)\right)\label{eq:Qzwei-II}\end{eqnarray}
Apart from the change $\partial_{\tet}\Phi\To\cov\Phi$ we expect
from the previous section that the Poisson bracket of $\Q_{2}$ with
itself induces some algebraic and some derived bracket of $\mc{J}$
with itself which then corresponds to the integrability condition
for $\mc{J}$. This is indeed the case, but we first have to study
the changes coming from $\partial_{\tet}\Phi\To\cov\Phi$. In other
words, we need a new formulation of proposition 1 (\ref{eq:Proposition1})
in the case of two-dimensional supersymmetry (Proposition 1 is of
course still valid, but it is not formulated in a supersymmetric covariant
way. It should, however, be applicable to e.g. BRST symmetries ).
Let us redefine the meaning of $K(\sigma,\tet)$ in (\ref{eq:K-sig-tet-def})
for a multivector valued form $K^{(k,k')}$\index{$K^{(k,k')}(\sigma,\tet)$}
\begin{eqnarray}
K^{(k,k')}(\sigma,\tet) & \equiv & K^{(k,k')}\big(\Phi^{m}(\sigma,\tet),\cov\Phi^{m}(\sigma,\tet),\Es_{m}(\sigma,\tet)\big)=\label{eq:K-sig-tet-Zabz}\\
 &  & \hspace{-1.6cm}=K_{m_{1}\ldots m_{k}}\hoch{n_{1}\ldots n_{k'}}\left(\Phi(\sigma,\tet)\right)\,\cov\Phi^{m_{1}}(\sigma,\tet)\ldots\cov\Phi^{m_{k}}(\sigma,\tet)\Es_{n_{1}}(\sigma,\tet)\ldots\Es_{n_{k'}}(\sigma,\tet)\us{\stackrel{\tet=0}{=}}{(\ref{eq:K-of-sigma})}K^{(k,k')}(\sigma)\qquad\end{eqnarray}
Likewise for all the other examples in (\ref{eq:T-sig-tet})-(\ref{eq:db-sig-tet}):\index{$T^{(t,t',t'')}(\sigma,\tet)$}\index{$dK$@$\de K(\sigma,\tet)$}\index{$o$@$\oo(\sigma,\tet)$}\begin{eqnarray}
T^{(t,t',t'')}(\sigma,\tet) & \equiv & T^{(t,t',t'')}\left(\Phi(\sigma,\tet),\cov\Phi(\sigma,\tet),\Es(\sigma,\tet),\cov\Es(\sigma,\tet)\right)\stackrel{\tet=0}{=}T^{(t,t',t'')}(\sigma)\quad(\textrm{see }(\ref{eq:T-of-sigma}))\label{eq:T-sig-tet-III}\end{eqnarray}
\begin{eqnarray}
\textrm{e.g. }\de K(\sigma,\tet) & \equiv & \de K\left(\Phi(\sigma,\tet),\cov\Phi(\sigma,\tet),\Es(\sigma,\tet),\cov\Es(\sigma,\tet)\right)\label{eq:dK-of-sig-tet-III}\\
\textrm{or }\oo(\sigma,\tet) & \equiv & \oo\left(\cov\Phi(\sigma,\tet),\cov\Es(\sigma,\tet)\right)\stackrel{(\ref{eq:BRST-op})}{=}\cov\Phi^{m}(\sigma,\tet)\cov\Es_{m}(\sigma,\tet)\us{\stackrel{\tet=0}{=}}{(\ref{eq:o-of-sigma})}\oo(\sigma)\label{eq:o-of-sig-tet-III}\\
\hspace{-1.2cm}[K^{(k,k')},_{\de\,}L^{(l,l')}]_{(1)}^{\Delta}(\sigma,\tet) & \equiv & [K^{(k,k')}\bs{,}L^{(l,l')}]_{(1)}^{(\Delta)}\left(\Phi(\sigma,\tet),\cov\Phi(\sigma,\tet),\Es(\sigma,\tet),\cov\Es(\sigma,\tet)\right)\us{\stackrel{\tet=0}{=}}{(\ref{eq:bracket-of-sigma})}[K^{(k,k')}\bs{,}L^{(l,l')}]_{(1)}^{(\Delta)}(\sigma)\qquad\quad\\
\de x^{m}(\sigma,\tet) & \equiv & \cov\Phi^{m}(\sigma,\tet)=\lam^{m}(\sigma)-\tet\partial_{\sigma}x^{m}(\sigma)\\
\de\pe_{m}(\sigma,\tet) & \equiv & \cov\Es_{m}(\sigma,\tet)=p_{m}(\sigma)-\tet\partial_{\sigma}\ro_{m}(\sigma)\label{eq:db-sig-tet-III}\end{eqnarray}
Expanding $K$ in $\tet$ yields \begin{eqnarray}
K^{(k,k')}(\sigma,\tet) & = & K^{(k,k')}(\sigma)+\tet\left(\bei{\partial_{\tet'}K^{(k,k')}(\sigma,\tet')}{\tet'=0}\right)=\\
 & = & K^{(k,k')}(\sigma)+\tet\left(\bei{\qu\tief{'}K^{(k,k')}(\sigma,\tet')}{\tet'=0}\right)\end{eqnarray}
As $K$ is a superfield, we can replace $\qu$ by $\es+\tilde{\es}$\begin{eqnarray}
K^{(k,k')}(\sigma,\tet) & = & K^{(k,k')}(\sigma)+\tet(\es+\tilde{\es})K^{(k,k')}(\sigma)=\label{eq:wichtig-ZabzI}\\
 & = & K^{(k,k')}(\sigma)+\tet\bei{\left((\de+\ip_{v})K^{(k,k')}\right)(\sigma)}{v^{k}\To-\partial_{\sigma}x^{k}}\label{eq:wichtig-Zabz}\end{eqnarray}
This is the analogue to the non-supersymmetric (\ref{eq:wichtigII})
and delivers the exterior derivative which will lead to the appearance
of the derived bracket. The relation between $\tilde{\es}$ and the
inner product with a vector should perhaps be clarified. Remember
that all multivector forms at $\tet=0$, $K^{(k,k')}(\sigma)$, are
analytic functions of the component fields $x^{m},\lam^{m}$ and $\ro_{m}$
. But among those fields, $\tilde{\es}$ acts only on $\lam^{m}$
and we can express it with partial derivatives (instead of functional
ones) when acting on $K$:\begin{equation}
\tilde{\es}K(\sigma)=-\partial_{\sigma}x^{m}\partl{\lam^{m}}K(x,\lam,\ro)=\bei{\ip_{v}K(\sigma)}{v^{k}=-\partial_{\sigma}x^{k}}\label{eq:stilde-auf-K}\end{equation}
in the Poisson bracket of $\tilde{\es}K$ with another multivector
valued form $L$ at $\tet=0$, nothing acts on $v^{k}=-\partial_{\sigma}x^{k}$
(which would produce a derivative of a delta function), as $L$ does
not contain $p_{k}$. Therefore we have\begin{equation}
\left\{ \tilde{\es}K(\sigma'),L(\sigma)\right\} =\bei{[\ip_{v}K,L](\sigma)}{v^{k}=-\partial_{\sigma}x^{k}}\delta(\sigma-\sigma')\label{eq:stilde-in-bracket}\end{equation}
which we will need below. For superfields we have $Y(\sigma,\tet)=Y(\sigma)+\tet(\es+\tilde{\es})Y(\sigma)$.
Applying the same to $v$ yields\begin{eqnarray}
v^{k}(\sigma)+\tet(\es+\tilde{\es})v^{k}(\sigma) & = & -\partial_{\sigma}x^{k}-\tet(\es+\tilde{\es})\partial_{\sigma}x^{k}(\sigma)=\label{eq:v-sig-tet}\\
 & = & -\partial_{\sigma}x^{k}-\tet\partial_{\sigma}\lambda^{k}(\sigma)=-\partial_{\sigma}\Phi^{k}\label{eq:v-sig-tet-II}\end{eqnarray}
\rem{\begin{eqnarray*}
\tilde{\es}T & = & -\partial_{\sigma}x^{m}\partl{\lam^{m}}T-\partial_{\sigma}\ro_{k}\partl{p_{k}}\end{eqnarray*}
}

\paragraph{Proposition 1b }

\emph{\index{proposition!super Poisson bracket of multivector valued forms (1b)}For
all multivector valued forms $K^{(k,k')},L^{(l,l')}$ on the target
space manifold, in a local coordinate patch seen as functions of $x^{m}$,$\de x^{m}$
and $\pe_{m}$, the following equation holds for the corresponding
worldsheet-superfields (\ref{eq:K-sig-tet-Zabz}) }\\
\emph{\Ram{1}{\begin{eqnarray}
\{K^{(k,k')}(\sigma',\tet'),L^{(l,l')}(\sigma,\tet)\} & = & \cov\left(\delta(\tet-\tet')\delta(\sigma-\sigma')\right)\left[K,L\right]_{(1)}^{\Delta}(\sigma,\tet)+\nonumber \\
 &  & \hspace{-.6cm}+\delta(\tet'-\tet)\delta(\sigma-\sigma')\Big(\underbrace{[\de K,L]_{(1)}^{\Delta}(\sigma,\tet)}_{\lqn{-(-)^{k-k'}\left[K,_{\de}L\right]_{(1)}^{\Delta}}}+\underbrace{[\ip_{v}K,L]_{(1)}^{\Delta}(\sigma,\tet)}_{\quad-(-)^{k-k'}[K,_{\ip_{v}}L]}\Big|_{v^{k}=-\partial_{\sigma}\Phi^{k}}\Big)\qquad\label{eq:Proposition1b}\end{eqnarray}
}}\\
\emph{where  e.g. $[\de K,L]_{(1)}^{\Delta}(\sigma,\tet)\equiv[\de K,L]_{(1)}^{\Delta}\left(\Phi(\sigma,\tet),\cov\Phi(\sigma,\tet),\Es(\sigma,\tet),\cov\Es(\sigma,\tet)\right)$.
}\\
\emph{The integrated version for a worldsheet without boundary reads}\emph{\small \begin{equation}
\boxed{\Big\{\int d\sigma'\int d\tet'K^{(k,k')}(\sigma',\tet'),\int d\sigma\int d\tet\, L^{(l,l')}(\sigma,\tet)\Big\}=(\es+\tilde{\es})\int d\sigma\,\Big([K,_{\de}L]_{(1)}^{\Delta}-(-)^{k-k'}[\ip_{v}K,L]_{(1)}^{\Delta}\Big|_{v^{k}=-\partial_{\sigma}x^{k}}\Big)(\sigma)}\label{eq:Proposition1b-integrated}\end{equation}
\vspace{.2cm}}{\small \par}

\emph{Proof}$\quad$Let us use (\ref{eq:wichtig-ZabzI}) for both
multivector valued fields and plug into the lefthand side of (\ref{eq:Proposition1b})\begin{eqnarray}
\lqn{\left\{ K(\sigma',\tet'),L(\sigma,\tet)\right\} =}\nonumber \\
 & = & \left\{ K(\sigma')+\tet'(\es+\tilde{\es})K(\sigma')\,,\, L(\sigma)+\tet(\es+\tilde{\es})L(\sigma)\right\} =\\
 & = & \left\{ K(\sigma'),L(\sigma)\right\} +\tet'\left\{ (\es+\tilde{\es})K(\sigma'),L(\sigma)\right\} +(-)^{k-k'}\tet\left\{ K(\sigma'),(\es+\tilde{\es})L(\sigma)\right\} +\nonumber \\
 &  & +(-)^{k-k'}\tet\tet'\left\{ (\es+\tilde{\es})K(\sigma'),(\es+\tilde{\es})L(\sigma)\right\} =\\
 & = & \left\{ K(\sigma'),L(\sigma)\right\} +(\tet'-\tet)\left\{ (\es+\tilde{\es})K(\sigma'),L(\sigma)\right\} +\tet(\es+\tilde{\es})\left\{ K(\sigma'),L(\sigma)\right\} +\nonumber \\
 &  & +\tet'\tet(\es+\tilde{\es})\left\{ (\es+\tilde{\es})K(\sigma'),L(\sigma)\right\} -\tet'\tet\left\{ (\es+\tilde{\es})(\es+\tilde{\es})K(\sigma'),L(\sigma)\right\} =\\
 & = & \left(1+\tet(\es+\tilde{\es})\right)\left\{ K(\sigma'),L(\sigma)\right\} +(\tet'-\tet)\left(1+\tet(\es+\tilde{\es})\right)\left\{ (\es+\tilde{\es})K(\sigma'),L(\sigma)\right\} +\nonumber \\
 &  & -\tet'\tet\big\{\underbrace{[\es,\tilde{\es}]}_{-\partial_{\sigma'}}K(\sigma'),L(\sigma)\big\}=\\
 & = & \delta(\sigma-\sigma')\left(1+\tet(\es+\tilde{\es})\right)\left[K,L\right]_{(1)}^{\Delta}(\sigma)+(\tet'-\tet)\left(1+\tet(\es+\tilde{\es})\right)\left\{ (\es+\tilde{\es})K(\sigma'),L(\sigma)\right\} +\nonumber \\
 &  & -(\tet'-\tet)\tet\partial_{\sigma}\delta(\sigma-\sigma')\left[K,L\right]_{(1)}^{\Delta}(\sigma)\end{eqnarray}
Now let us make use of (\ref{eq:stilde-in-bracket}) and (\ref{eq:v-sig-tet-II})
to arrive at\begin{eqnarray}
\lqn{\left\{ K(\sigma',\tet'),L(\sigma,\tet)\right\} =}\nonumber \\
 & = & \cov\left(\delta(\tet-\tet')\delta(\sigma-\sigma')\right)\left[K,L\right]_{(1)}^{\Delta}(\sigma,\tet)+\delta(\tet'-\tet)\delta(\sigma-\sigma')\bei{\left[(\de+\ip_{v})K,L\right]_{(1)}^{\Delta}(\sigma,\tet)}{v^{k}=-\partial_{\sigma}\Phi^{k}}\qquad\end{eqnarray}
which is the first equation of the proposition. Integrating over $\tet'$
and $\sigma'$ results in\begin{eqnarray}
\int d\sigma'\int d\tet'\left\{ K(\sigma',\tet'),L(\sigma,\tet)\right\}  & = & \bei{\left[(\de+\ip_{v})K,L\right]_{(1)}^{\Delta}(\sigma,\tet)}{v^{k}=-\partial_{\sigma}\Phi^{k}}=\\
 & = & \bei{\left[(\de+\ip_{v})K,L\right]_{(1)}^{\Delta}(\sigma)}{v^{k}=-\partial_{\sigma}x^{k}}+\tet(\es+\tilde{\es})\bei{\left[(\de+\ip_{v})K,L\right]_{(1)}^{\Delta}(\sigma)}{v^{k}=-\partial_{\sigma}x^{k}}\end{eqnarray}
A second integration picks out the linear part in $\tet$ and adjusting
the order of the integrations gives the additional sign in (\ref{eq:Proposition1b-integrated}).$\qquad\qquad\qquad\qquad\qquad\qquad\qquad\qquad\qquad\qquad\qquad\qquad\qquad\qquad\qquad\qquad\square$

\subsubsection*{Application to the second supercharge $\Q_{2}$}

\index{SUSY!extended worldsheet $\sim$}\index{worldsheet SUSY!extended $\sim$}\index{extended worldsheet SUSY}We
are now ready to apply the proposition in the integrated form (\ref{eq:Proposition1b-integrated})
to the question of the existence of a second worldsheet supersymmetry
$\Q_{2}$. Remember, we want $\{\Q_{2},\Q_{2}\}=-2P$. Due to the
proposition, the lefthand side can be written as\begin{eqnarray}
\{\Q_{2},\Q_{2}\} & = & \frac{1}{4}(\es+\tilde{\es})\int d\sigma\,\Big([\mc{J},_{\de}\mc{J}]_{(1)}^{\Delta}-[\ip_{v}\mc{J},\mc{J}]_{(1)}^{\Delta}\Big|_{v=-\partial_{\sigma}x^{k}\ro_{k}}\Big)(\sigma)\label{eq:gruene-Pause}\end{eqnarray}
For $\mc{J}^{2}=-1$, the second term under the integral simplifies
significantly\begin{eqnarray}
-\frac{1}{4}\int d\sigma[\ip_{v}\mc{J},\mc{J}]_{(1)}^{\Delta}\Big|_{v=-\partial_{\sigma}x^{k}\ro_{k}} & = & -\int d\sigma\, v^{K}\mc{J}_{K}\hoch{L}\mc{J}_{L}\hoch{M}\basis_{M}\Big|_{v=-\partial_{\sigma}x^{k}\ro_{k}}=-\int d\sigma\,\partial_{\sigma}x^{k}\ro_{k}=\tilde{\OO}\end{eqnarray}
 Recalling that \begin{eqnarray}
(\es+\tilde{\es})\tilde{\OO} & = & \es\tilde{\OO}=\tilde{\es}\OO=(\es+\tilde{\es})\OO=-P\\
\textrm{and }\OO & = & \int d\sigma\,\oo(\sigma)\quad(\textrm{see }(\ref{eq:o-of-sigma}))\end{eqnarray}
we can rewrite (\ref{eq:gruene-Pause}) as\begin{eqnarray}
\dann\{\Q_{2},\Q_{2}\} & = & \frac{1}{4}(\es+\tilde{\es})\left(\int d\sigma\,[\mc{J},_{\de}\mc{J}]_{(1)}^{\Delta}+4\OO\right)=\\
 & = & \frac{1}{4}(\es+\tilde{\es})\left(\int d\sigma\,\left([\mc{J},_{\de}\mc{J}]_{(1)}^{\Delta}-4\oo\right)(\sigma)\right)+2\underbrace{\tilde{\es}\OO}_{-P}\end{eqnarray}
The righthand side clearly equals $-2P$ for \begin{eqnarray}
[\mc{J},_{\de}\mc{J}]_{(1)}^{\Delta}-4\oo & = & 0\end{eqnarray}
which is again (according to (\ref{eq:integrability-big-derived}))
just the integrability condition for the generalized almost complex
structure $\mc{J}$.

\rem{hier lungern explizite $(J,J)$-Rechnung und der Quanten-Fall!}

\section*{Conclusions to the Bracket Part}

\addcontentsline{toc}{chapter}{Conclusions to the Bracket Part}We
have seen two closely related mechanisms in sigma-models with a special
field content which lead to the derived bracket of the target space
algebraic bracket by the target space exterior derivative. This exterior
derivative is implemented in the sigma model in one case via the derivative
with respect to a (worldvolume-) Grassmann coordinate and in the other
case via the derivative with respect to the worldvolume coordinate
itself. In the latter case this derivative has to be contracted with
(worldvolume-) Grassmann coordinates in order to be an odd differential.
This leads to the problem that higher powers of the basis elements
vanish, as soon as the power exceeds the worldvolume dimension as
it happens in Zucchini's application. A big number of Grassmann-variables
is therefore advantageous in that approach. For the other mechanism
one rather prefers to have only one single Grassmann variable as there
is no need for any contraction. There is one worldvolume dimension
more in the Lagrangian formalism and for that reason it was preferable
to apply there the mechanism with worldvolume derivatives and use
the other one in the Hamiltonian formalism. \rem{Ist das wahr, dass beides bei beidem gehen wuerde?}

If one does not consider antisymmetric tensors of higher rank, but
only vectors or one-forms (or forms of worldvolume-dimension), the
partial worldvolume derivative without a Grassmann-coordinate is enough.
There is either no need for antisymmetrization or it can be performed
with the worldvolume epsilon tensor. The nature of the mechanism remains
the same and leads to the observations in \cite{Alekseev:2004np,Bonelli:2005ti}
that the Poisson bracket implements the Dorfman bracket for sums of
vectors and one-forms and the corresponding derived bracket for sums
of vectors and $p$-forms on a $p$-brane \cite{Bonelli:2005ti}.
In that sense, the present part of the thesis is a generalization
of those observations.

There remain a couple of things to do. It should be possible to implement
in the same manner by e.g. a BRST differential other target space
differentials which can depend on some extra-structure and repeat
the same analysis. Symmetric tensors then become more interesting
as well, because they need such an extra-structure anyway for a meaningful
differential. From the string theory point of view, the application
of extended worldsheet supersymmetry corresponds to applications in
the RNS string. But generalized complex geometry contains the tools
to allow RR-fluxes, which are hard to treat in RNS. It would therefore
be nice to find some topological limit in a string theory formalism
which is extendable to RR-fields, like the Berkovits-string \cite{Berkovits:2004px},
leading to a topological sigma model like Zucchini's, in order to
learn more about the correspondence between string theory and generalized
complex geometry.

\rem{noch was ueber Hull und ueber generalized Poisson? und ueber induced differential...}

\bibliographystyle{fullsort}
\bibliography{phd,Proposal}

\printindex{}
}

\part*{\hypertarget{Conclusion}{Conclusion}}

\index{conclusions}\label{par:Conclusion}\thispagestyle{plain}\addcontentsline{toc}{part}{\protect\hyperlink{Conclusion}{Conclusion}}{\inputTeil{0}\ifthenelse{\theinput=1}{}{}

\title{Conclusion}

\author{Sebastian Guttenberg}

\date{August 14, '07}

\maketitle
\begin{abstract}
Part of thesis, 
\end{abstract}
\tableofcontents{}

\rem{To do:

\begin{itemize}
\item Don't panic
\end{itemize}
}

After the conclusions on the bracket part, we would like to recall
the general idea of what we did. Apart from the presentation of the
explicit worldsheet BRST transformations, the result of the supergravity-constraint
calculations from Berkovits' pure spinor string in part \ref{par:PureSpinorString}
is not new in itself. It is, however, a very important result and
our contribution can be seen as an independent check. This is true
in particular, as we used different techniques at several points.
We established a covariant variation in this setting and derived everything
in the Lagrangian formalism, using {}``inverse Noether''. The argumentation
and calculation was done in detail, in order to allow checks by others,
and also some subtle points like the antighost gauge symmetry where
discussed carefully. Also our starting point was more general. Last
but not least, the insight from the first part about superspace conventions
served as a very powerful tool throughout. The aim of the calculation
in part \ref{par:PureSpinorString} was to make contact to generalized
geometry. The derivation of the generalized Calabi Yau condition has
been done so far from the supergravity point of view, and possible
quantum or string corrections to this geometry require a worldsheet
calculation. We have therefore derived the supergravity transformations
of the fermionic background fields which serve as the starting point
of these considerations. We did not yet calculate any string corrections,
but it could already be of big advantage to know the natural form
of the supergravity transformations as they come out from the string
and not from old supergravity considerations. In particular we expect
to obtain more insight about the geometric role of the RR-fields in
the super-geometrical setting. Non-commutativity considerations for
the open superstring (e.g. \cite{deBoer:2003dn,Berkovits:2003kj,Ooguri:2003qp}),
for example, assign a similar role to the RR-fields in superspace
as the $B$-field has in bosonic space. And the geometry of the latter
(with the field strength $H$ either seen as a twist or a torsion),
are understood much better. 

There are several directions ahead. One could try to establish the
tools of generalized (not necessarily complex) geometry already in
ten dimensions, before compactification. Having the superstring in
mind (embedded in superspace), it would be even more appealing to
consider some generalized supergeometry, i.e. structures on $T\oplus T^{*}$
of the supermanifold. String statements should simplify if one uses
a formulation where the structures of interest appear manifestly.
In this context it seems also reasonable to switch to a probably mixed
first-second order formalism of the pure spinor string in general
background. Topological limits of this formalism might lead to something
like the Hitchin sigma-model \cite{Zucchini:2004ta} or some supersymmetric
version of it. This again could shed light on the geometric role of
RR-fields. Similar to the last point would be the introduction of
doubled coordinates as suggested by Hull\cite{Hull:2004in,Hull:2006qs,Hull:2006va,Dabholkar:2005ve}.
Generalized complex geometry and this doubled geometry seem to be
very closely related. Deriving the first via supersymmetry conditions
in a formalism with doubled coordinates certainly could clarify this
relation. 

For all these considerations, our insight about brackets and sigma-models
and the relation to the integrability of generalized complex geometry
that we obtained in the last part of this thesis will be very useful.
What we learned about superspace conventions should even be useful
for everybody working with superspace.

\bibliographystyle{fullsort}
\bibliography{phd,Proposal}
\printindex{}
}

\appendix

\part*{\hypertarget{Appendix}{Appendix}}

\index{appendix}\renewcommand{\thefoot}{\Alph{chapter}.\twodigit{\value{footnote}}@\arabic{footnote}}\addcontentsline{toc}{part}{\protect\hyperlink{Appendix}{Appendix}}

\chapter{Notations and Conventions}

\index{conventions}\index{notations}\label{cha:Notations-and-Conventions}{\inputTeil{0}\renewcommand{\be}{\bs{b}}\renewcommand{\ce}{\bs{c}}\ifthenelse{\theinput=1}{}{}

\title{Conventions and Notations}

\author{Sebastian Guttenberg}

\date{August 09, 2007}

\maketitle
\begin{abstract}
Part of thesis-appendix...
\end{abstract}
\label{sec:Conventions}Within the thesis, a lot of different types
of tensors have to be denoted. The choices and sometimes some logic
behind, will be presented here.

The bracket part (\ref{part:Derived-Brackets-in}) (including appendices
\ref{cha:Generalized-Complex-Geometry} and \ref{cha:Derived-Brackets})
differs a bit in the notation from the rest, as it does not treat
a superspace. In any case we denote bosonic target space coordinates
via $x^{m}$. In the bracket part, however, world-volume-coordinates
are denoted by $\sigma^{\mu}$, while in the worldsheet coordinates
in the rest are most often chosen to be complex ($z$,$\bar{z}$).
At some places we write the real coordinates $\sigma^{\xi}$ with
an worldsheet index $\xi$ or $\zeta$, in order to distinguish it
from the curved spinorial indices $\mu,\nu,\ldots$. Our metric signature
is 'mostly plus':$\eta_{ab}=\diag(-1,1,\ldots,1)$\index{metric!signature}\index{signature of the metric}\index{$\eta_{ab}$}.

\paragraph{Superspace}

In the superspace parts we have $x^{M}\equiv(x^{m},\tet^{\mu},\hat{\tet}^{\hat{\mu}}$),
where $\tet$ and $\hat{\tet}$ are anticommuting coordinates with
the dimension 16 of a Majorana Weyl spinor in ten dimensions. The
hatted index should include both versions of superspace: IIA (with
$\hat{\tet}^{\hat{\mu}}=\hat{\tet}_{\mu})$ and $IIB$ (with $\hat{\tet}^{\hat{\mu}}=\hat{\tet}^{\mu})$.
The grading of the coordinate $x^{M}$ depends on the index. We therefore
prefer to write $x^{M}\equiv(x^{m},x^{\bs{\mu}},x^{\hat{\bs{\mu}}})$\index{$x^m$}\index{$x^{\hat{\bs{\mu}}}$}\index{$x^{\bs{\mu}}$}\index{$x^M$}.
Writing the fermionic indices boldface is just a reminder and will
not be substantial. A vielbein $E_{M}\hoch{A}$ will transform curved\index{curved index}\index{index!curved}
indices (from the middle of the alphabet) into flat indices\index{index!flat}\index{flat index}
(from the beginning of the alphabet) and vice verse, e.g. for the
pullbacks of the supersymmetric invariant form $\Pi_{z}^{A}=\partial x^{M}E_{M}\hoch{A}$.
The entries then have a corresponding index structure with letters
from the beginning of the alphabet: $\Pi_{z}^{A}=(\Pi_{z}^{a},\Pi_{z}^{\bs{\alpha}},\Pi_{z}^{\hat{\bs{\alpha}}})$\index{$\Pi_z^A$}\index{$\Pi_z^a$}\index{$\Pi_z^{\bs\alpha}$}\index{$\Pi_z^{\hat{\bs\alpha}}$}.
When we want to combine the spinorial indices only, we write $x^{\bs{\mc{M}}}\equiv(x^{\bs{\mu}},x^{\hat{\bs{\mu}}})$\index{$x^{\bs{\mc{M}}}$}
or $\tet^{\mc{M}}\equiv(\tet^{\mu},\hat{\tet}^{\hat{\mu}})$\index{$\tetb$@$\hat\tet^{\hat\mu}$}\index{$\tet^{\mu}$}\index{$\tet^{\mc{M}}$}
or $\Pi_{z}^{\bs{\mc{A}}}\equiv(\Pi_{z}^{\bs{\alpha}},\Pi_{z}^{\hat{\bs{\alpha}}})$.\index{$\Pi_z^{\bs{\mc{A}}}$}
If we want to omit the indices, (e.g. in functions of the coordinates)
we write $\xfull$\index{$x$@$\xfull$} for $x^{M}$, $\xboson$\index{$x$@$\xboson$}
for $x^{m}$, $\xbothtetas$\index{$\tet$@$\xbothtetas$} for $\tet^{\mc{M}}$,
$\tet$\index{$\tet$} for $\tet^{\bs{\mu}}$ and $\hat{\tet}$\index{$\tet$@$\hat\tet$}
for $\hat{\tet}^{\hat{\bs{\mu}}}$.

\paragraph{Notation for tensors in the bracket part}

In the bracket-part, we mainly denote target space vector-fields by
$a,b,\ldots$ or $v,w,\ldots$, 1-forms by small Greek letters $\alpha,\beta,\ldots$
and generalized $T\oplus T^{*}$-vectors by $\mf{a},\mf{b},\ldots$
or $\mf{v},\mf{w},\ldots$~. For an explicit split in vector and
1-form, the letters from the beginning of the alphabet are better
suited, as there is a better correspondence between Latin and Greek
symbols or one can visually better distinguish between Latin and Greek
symbols. Compare e.g. $\mf{a}=a+\alpha$ and $\mf{v}=v+(?\nu)$.\\
Higher order forms will be in general denoted by $\alpha^{(p)},\beta^{(q)},\ldots$
or $\omega^{(p)},\eta^{(q)},\rho^{(r)},\ldots$. There will be exceptions,
however , for specific forms like the $B$-field $B=B_{mn}\de x^{m}\wedge\de x^{n}$.
Following this logic, we will also denote multivectors (tensors with
antisymmetric upper indices) by small letters, indicating their multivector-degree
in brackets: $a^{(p)},b^{(q)},\ldots$ or $v^{(p)},w^{(q)},\ldots$.
There are again exceptions, e.g. a Poisson structure will often be
denoted by $P=P^{mn}\pe_{m}\wedge\pe_{n}$. The most horrible exception
is the one of the beta-transformation, which is denoted by a large
beta $\Beta^{mn}$ in (\ref{eq:SOnn}), in order to distinguish it
from forms.

Tensors of mixed type will be denoted by capital letters where we
denote in brackets first the number of lower indices and then the
number of upper indices, e.g. $T^{(p,q)}$. Most of the time, we treat
multivector valued forms, e.g. the lower indices as well as the upper
indices are antisymmetrized. The letters denoting form degree and
multivector degree will often be adapted to the letter of the tensor,
e.g.\index{$K^{(k,k')}$} $K^{(k,k')},L^{(l,l')},\ldots$\textbf{}\\
\textbf{Attention:} $k$ and $l$ are also used as dummy indices!
Sometimes (I'm sorry for that) the same letter appears with different
meanings. However, in those situations the dummy indices will carry
indices which might even be one of the degrees $k$ or $k'$, e.g.
$K_{\ldots}\hoch{k_{1}\ldots k_{k'}}L_{k_{k'}\ldots k_{1}\ldots}\hoch{\ldots}$.

Working all the time with graded algebras with a graded symmetric
product (the wedge product), everything in this thesis has to be understood
as \textbf{graded}. I.e. with commutator we mean the graded commutator
and with the Poisson bracket the graded Poisson bracket. They will
not be denoted differently than the non-graded operations. Relevant
for the sign rules is the \textbf{total\index{total degree} degree}\index{degree!total $\sim$}
which we define to be form degree minus the multivector degree. In
the field language, it corresponds to the total ghost number which
is the pure ghost number minus the antighost number. It will be denoted
in the bracket part by \begin{eqnarray}
\abs{K^{(k,k')}} & = & k-k'\end{eqnarray}
In the rest of the thesis, $\abs{\ldots}$ will only denote the parity,
i.e. $+1$ for commuting and $-1$ for anticommuting variables. As
only degrees or parities appear in the exponent of a minus sign, a
simplified notation is used there\index{$(-)^{AB}$}\begin{equation}
(-)^{A}\equiv(-1)^{\abs{A}},\quad(-)^{A+B}\equiv(-)^{\abs{A}+\abs{B}},\quad(-)^{AB}\equiv(-)^{\abs{A}\abs{B}}\quad\forall A,B\end{equation}

\paragraph{Poisson bracket and derivatives}

For the Poisson\index{Poisson bracket!sign convention} bracket\index{bracket!Poisson},
the following (less common) sign convention is chosen: \begin{eqnarray}
\left\{ p_{m},x^{n}\right\}  & = & \delta_{m}^{n}=-\left\{ x^{n},p_{m}\right\} \\
\left\{ b_{m},c^{n}\right\}  & = & \delta_{m}^{n}=-(-)^{bc}\left\{ c^{n},b_{m}\right\} \end{eqnarray}
Derivatives with respect to $x^{m}$ are denoted by $\partiell{}{x^{m}}f\equiv\partial_{m}f\equiv f_{,m}$.
For graded variables left\index{left derivative} and right\index{right derivative}
derivatives\index{derivative!right $\sim$}\index{derivative!left $\sim$}
are denoted respectively by\begin{eqnarray}
\partiell{f}{\ce}\equiv\partl{\ce}f(\ce)\equiv\frac{\vec{\partial}}{\partial\ce}f(\ce),\qquad\partial f(\ce)/\partial\ce & \equiv & f\partr{\ce}\end{eqnarray}
The corresponding notations are used for functional derivatives $\funktl{\ce(\sigma)}$.

\paragraph{Boldface philosophy and antisymmetrizations}

\index{boldface philosophy}With respect to the wedge product, the
basis element $\pe_{m}$ is an odd object ($\pe_{m}\wedge\pe_{n}=-\pe_{n}\wedge\pe_{m}$).
The partial derivative $\partial_{k}$ acting on some coefficient
function, however, is an even operator (it does not change the parity
as long as it is not contracted with a basis element $\de x^{k}$).
That is why we denote the odd basis element $\pe_{m}$\index{$\partial$@$\pe_{m}$}
and $\de x^{m}$\index{$dx$@$\de x^{m}$} as well as the odd exterior
derivative $\de\,$ with boldface symbols. The interior product itself
does not carry a grading in the sense that $\abs{\ip_{K}\rho}=\abs{K}+\abs{\rho}$,
while for the Lie derivative $\Lie_{K}=\left[\ip_{K},\de\,\right]$
the $\Lie$ carries a grading in the sense $\abs{\Lie_{K}\rho}=\abs{K}+\abs{\rho}+1$.
That is why the Lie derivative is denoted with a boldface $\Lie$
which is also very good to distinguish it from generalized multivectors
$\mc{K},\mc{L},\ldots$. The philosophy of writing odd objects in
boldface style is also extended to the combined\index{combined basis element $\basis_M$}
basis\index{basis element!combined $\sim$} element\index{$t_M$@$\basis_M$}

\begin{equation}
\basis_{M}\equiv(\pe_{m},\de x^{m}),\quad\basis^{M}\equiv(\de x^{m},\pe_{m})\end{equation}
and to the comma in the derived bracket $\left[\,\bs{,}\,\right]$
in contrast to the commutator $\left[\,,\,\right]$. This should be,
however, just a reminder. It will be obvious for other reasons, which
bracket is meant. But we do \textbf{not} extend this philosophy to
vectors and 1-forms, where it would be consistent (but too much effort)
to write the vectors and basis elements in boldface style and the
coefficients in standard style. We will instead write the vector in
the same style as the coefficient $a=a_{m}\de x^{m}$. 

A square bracket is used as usual to denote the antisymmetrization
of, say $p$, indices (including a normalization factor $\frac{1}{p!}$).
A vertical line is used to exclude some indices from antisymmetrization.
An extreme example would be\index{antisymmetrization}\index{$A^{[ab\mid cd\mid e\mid fg\mid hi]}$}\begin{equation}
A^{[ab\mid cd|e|fg|hi]}\end{equation}
where $A$ is antisymmetrized only in $a,b,e,h$ and $i$, but not
in $c,d,f$ and $g$. Normally we use only expressions like $A^{[ab\mid cd|efg]}$,
where $a,b,e,f$ and $g$ are antisymmetrized.

\paragraph{Wedge product }

\label{Wedge-product} A significant difference from usual conventions
is that for multivectors, forms and generalized multivectors we include
the normalization of the factor already in the definition of the wedge
product\index{$()$@$\wedge$}\index{wedge product}\begin{eqnarray}
\de x^{m_{1}}\cdots\de x^{m_{n}}\equiv\de x^{m_{1}}\wedge\ldots\wedge\de x^{m_{n}} & \equiv & \de x^{[m_{1}}\otimes\ldots\otimes\de x^{m_{n}]}\equiv\sum_{P}\frac{1}{n!}\de x^{m_{P(1)}}\otimes\ldots\otimes\de x^{m_{P(n)}}\\
\pe_{m_{1}}\cdots\pe_{m_{n}}\equiv\pe_{m_{1}}\wedge\cdots\wedge\pe_{m_{n}} & \equiv & \pe_{[m_{1}}\otimes\cdots\otimes\pe_{m_{n}]}\equiv\sum_{P}\frac{1}{n!}\pe_{m_{P(1)}}\otimes\cdots\otimes\pe_{m_{P(n)}}\\
\basis_{M_{1}}\ldots\basis_{M_{n}}\equiv\basis_{M_{1}}\wedge\ldots\wedge\basis_{M_{n}} & \equiv & \basis_{[M_{1}}\otimes\ldots\otimes\basis_{M_{n}]}\equiv\sum_{P}\frac{1}{n!}\basis_{M_{P(1)}}\otimes\ldots\otimes\basis_{M_{P(n)}}\end{eqnarray}
(where we sum over all permutations $P$), such that we omit the usual
factor of $\frac{1}{p!}$ in the coordinate expression of a $p$-form,
or a $p$-vector \begin{eqnarray}
\alpha^{(p)} & \equiv & \alpha_{m_{1}\ldots m_{p}}\de x^{m_{1}}\wedge\cdots\wedge\de x^{m_{p}}\equiv\alpha_{m_{1}\ldots m_{p}}\de x^{m_{1}}\cdots\de x^{m_{p}}\\
v^{(p)} & \equiv & v^{m_{1}\ldots m_{p}}\partial_{m_{1}}\wedge\ldots\wedge\partial_{m_{p}}\end{eqnarray}
Readers who prefer the $\frac{1}{p!}$, can easily reintroduce it
in every equation by replacing e.g. the coefficient functions $v^{m_{1}\ldots m_{p}}\To\frac{1}{p!}v^{m_{1}\ldots m_{p}}$.
The equation for the Schouten bracket ( \ref{eq:Schouten-bracketI}),
for example, would change as follows: \begin{eqnarray}
\left[v^{(p)}\bs{,}w^{(q)}\right]^{m_{1}\ldots m_{p+q-1}} & = & pv^{[m_{1}\ldots m_{p-1}|k}\partial_{k}w^{|m_{p}\ldots m_{p+q-1}]}-qv^{[m_{1}\ldots m_{p}\mid}\tief{,k}w^{k\,\mid m_{p+1}\ldots m_{p+q-1}]}\\
\hspace{-1cm}\To\frac{1}{(p+q-1)!}\left[v^{(p)}\bs{,}w^{(q)}\right]^{m_{1}\ldots m_{p+q-1}} & = & \frac{1}{(p-1)!}\frac{1}{q!}v^{[m_{1}\ldots m_{p-1}|k}\partial_{k}w^{|m_{p}\ldots m_{p+q-1}]}+\nonumber \\
 &  & -\frac{1}{p!}\frac{1}{(q-1)!}v^{[m_{1}\ldots m_{p}\mid}\tief{,k}w^{k\,\mid m_{p+1}\ldots m_{p+q-1}]}\quad\end{eqnarray}

\paragraph{Schematic index notation }

\label{par:Schematic-index-notation}\index{schematic index notation}\index{index!schematic $\sim$ notation}\index{notation!schematic index $\sim$}For
longer calculations in coordinate form it is useful to introduce the
following notation, where every boldface index is assumed to be contracted
with the corresponding basis element (at the same position of the
index), s.th. the indices are automatically antisymmetrized. \label{fat-index}\begin{eqnarray}
\omega^{(p)} & = & \omega_{m_{1}\ldots m_{p}}\de x^{m_{1}}\cdots\de x^{m_{p}}\equiv\omega_{\bs{m}\ldots\bs{m}}\\
a^{(p)} & = & a^{n_{1}\ldots n_{p}}\pe_{n_{1}}\wedge\ldots\pe_{n_{p}}\equiv a^{\nn}\\
\mc{K}^{(p)} & = & \mc{K}_{M_{1}\ldots M_{p}}\basis^{M_{1}}\ldots\basis^{M_{p}}\equiv\mc{K}_{\bs{M}\ldots\bs{M}}=\\
 & = & \mc{K}^{M_{1}\ldots M_{p}}\basis_{M_{1}}\ldots\basis_{M_{p}}\equiv\mc{K}^{\bs{M}\ldots\bs{M}}\end{eqnarray}
or for products of tensors e.g.\index{$\omega_{\bs{m}\ldots\bs{m}}\eta_{\bs{m}\ldots\bs{m}}$}\index{$Kb$@$\mc{K}_{\bs{M}\ldots\bs{M}}$}
\begin{eqnarray}
\omega_{\bs{m}\ldots\bs{m}}\eta_{\bs{m}\ldots\bs{m}} & \equiv & \omega_{[m_{1}\ldots m_{p}}\eta_{m_{p+1}\ldots m_{p+q}]}\de x^{m_{1}}\cdots\de x^{m_{p+q}}=\\
 & = & \omega_{m_{1}\ldots m_{p}}\eta_{m_{p+1}\ldots m_{p+q}}\de x^{m_{1}}\cdots\de x^{m_{p+q}}=(-)^{pq}\eta_{\bs{m}\ldots\bs{m}}\omega_{\bs{m}\ldots\bs{m}}\end{eqnarray}
A boldface index might be hard to distinguish from an ordinary one,
but this notation is nevertheless easy to recognize, as normally several
coinciding indices appear (which are not summed over as they are at
the same position). Similarly, for multivector valued forms we define%
\footnote{Upper and lower signs are thus treated independently. For calculational
reasons this is not the best way to do. We can interpret every boldface
index on the lefthand side of (\ref{eq:grosser-Vorzeichenkummer})
as a basis element sitting at the position of the index, so that the
order of the basis elements on the lefthand side is first $k\times\de x^{m}$,
$(k'-1)\pe_{m}$, $(l-1)\times\de x^{m}$ and $l'\times\pe_{m}$,
s.th., in order to get the order of the righthand side, we have to
interchange $(k'-1)\pe_{m}$ with $(l-1)\times\de x^{m}$, which gives
a sign factor of $(-)^{(k'-1)(l-1)}$. This is a natural sign factor
which appears all the way in the equations, which could be easily
absorbed into the definition. However, we wanted to keep the sign
factors explicitly in the equations in order to keep the notation
as self-explaining as possible and not confuse the reader too much.$\qquad\fussend$%
}\index{$K_{\bs{m}\ldots\bs{m}}\hoch{\bs{n}\ldots\bs{n}}$}\begin{eqnarray}
K_{\bs{m}\ldots\bs{m}}\hoch{\bs{n}\ldots\bs{n}} & \equiv & K_{m_{1}\ldots m_{k}}\hoch{n_{1}\ldots n_{k'}}\de x^{m_{1}}\wedge\ldots\wedge\de x^{m_{k}}\otimes\pe_{m_{1}}\wedge\ldots\wedge\pe_{m_{k'}}\\
\hspace{-1cm}K_{\bs{m}\ldots\bs{m}}\hoch{\bs{n}\ldots\bs{n}p}L_{p\bs{m}\ldots\bs{m}}\hoch{\bs{n}\ldots\bs{n}} & \equiv & \!\! K_{m_{1}\ldots m_{k}}\hoch{n_{1}\ldots n_{k'-1}p}L_{pm_{1}\ldots m_{l-1}}\hoch{n_{1}\ldots n_{l'}}\de x^{m_{1}}\cdots\de x^{m_{k+l-1}}\!\!\otimes\!\pe_{m_{1}}\cdots\pe_{m_{k'+l'-1}}\quad\label{eq:grosser-Vorzeichenkummer}\end{eqnarray}
\printindex{}
}

\chapter{Generalized Complex Geometry}

\label{cha:Generalized-Complex-Geometry}{\remch\inputTeil{0}\renewcommand{\be}{\bs{b}}\renewcommand{\ce}{\bs{c}}\ifthenelse{\theinput=1}{}{}\remch

\title{Some aspects of generalized (complex) geometry}

\author{Sebastian Guttenberg}

\date{August 08, 2008 }

\maketitle
\begin{abstract}
Part of thesis-appendix
\end{abstract}
\label{sec:Generalized-complex-geometry}

For introductions into Hitchin's \cite{Hitchin:2004ut} generalized
complex geometry (GCG) see e.g. Zabzine's review \cite{Zabzine:2006uz}
or Gualtieri's thesis \cite{Gualtieri:007}. In the appendix of \cite{Grana:2006kf}
there is another nice introduction with emphasis on the pure spinor
formulation of GCG. For a survey of compactification with fluxes and
its relation to GCG see Gra$\tilde{\textrm{n}}$a's review \cite{Grana:2005jc}.

\section{Basics}

In \textbf{generalized\index{generalized geometry} geometry\index{geometry!generalized $\sim$}}
one is looking at structures (e.g. a complex structure) on the direct
sum of tangent and cotangent bundle $T\oplus T^{*}$. Let us call
a section of this bundle a \textbf{generalized\index{generalized!vector field}
vector} (field) or synonymously \textbf{generalized\index{generalized!one-form}
1-form}, which is the sum of a vector field and a 1-form\index{$a$@$\mathfrak{a}=\protect\mf{a}^M\protect\basis_M$|itext{generalized vector field}}\begin{eqnarray}
\mf{a} & = & a+\alpha=\\
 & = & a^{m}\pe_{m}+\alpha_{m}\de x^{m}\end{eqnarray}
Using the \textbf{combined basis elements\index{combined basis element $\protect\basis_M$}\index{basis element!combined $\sim$ $\protect\basis_M$}}\index{$t_M$@$\protect\basis_M$}
\begin{eqnarray}
\basis_{M} & \equiv & (\pe_{m},\de x^{m})\label{eq:combined-basis}\end{eqnarray}
a generalized vector $\mf{a}$ can be written as \begin{eqnarray}
\mf{a} & = & \mf{a}^{M}\basis_{M}\label{eq:generalized-vector}\\
\mf{a}^{M} & = & (a^{m},\alpha_{m})\end{eqnarray}
There is a \textbf{canonical\index{canonical metric $\protect\mc{G}_{MN}$ of $T\oplus T^*$}
metric\index{metric!canonical $\sim$ $\protect\mc{G}_{MN}$ of $T\oplus T^*$}}
$\mc{G}$ on $T\oplus T^{*}$\index{$(<>)$@$\erw{\ldots,\ldots}$|itext{canonical inner product on $T\oplus T^*$}}\index{$G_{MN}$@$\mc{G}_{MN}$|itext{canonical metric on $T\oplus T^*$}}\begin{eqnarray}
\erw{\mf{a},\mf{b}} & \equiv & \mf{\alpha}(b)+\mf{\beta}(a)=\\
 & = & \alpha_{m}b^{m}+\beta_{m}a^{m}\equiv\\
 & \equiv & \mf{a}^{M}\mc{G}_{MN}\mf{b}^{N}\end{eqnarray}
with \begin{eqnarray}
\mc{G}_{MN} & \equiv & \left(\begin{array}{cc}
0 & \delta_{m}^{n}\\
\delta_{n}^{m} & 0\end{array}\right)\label{eq:canonical-metric}\end{eqnarray}
which has \textbf{signature}\index{signature!of the canonical metric on $T\oplus T^*$}
(d,-d) (if d is the dimension of the base manifold). The above definition
differs by a factor of 2 from the most common one. We prefer, however,
to have an inverse metric of the same form\begin{eqnarray}
\mc{G}^{MN} & \equiv & \left(\mc{G}^{-1}\right)^{MN}=\left(\begin{array}{cc}
0 & \delta_{n}^{m}\\
\delta_{m}^{n} & 0\end{array}\right)\label{eq:inverse-canonical-metric}\end{eqnarray}
As it is constant, we can always pull it through partial derivatives.
Using this metric to lower and raise indices just interchanges vector
and form component. We can equally rewrite $\mf{a}$ in (\ref{eq:generalized-vector})
with a basis with upper capital indices and the vector coefficients
with lower indices\index{$t^M$@$\protect\basis^{M}$}\begin{eqnarray}
\basis^{M} & \equiv & \left(\de x^{m},\pe_{m}\right)\label{eq:combined-basisII}\\
\mf{a} & = & \mf{a}_{M}\basis^{M}\\
\mf{a}_{M} & = & (\alpha_{m},a^{m})\end{eqnarray}
Note that in the present text there is no existence of any metric
on the tangent bundle assumed. Therefore we cannot raise or lower
small indices. In cases where 1-form and vector have a similar symbol,
the position of the small index therefore uniquely determines which
is which (e.g. $\omega_{m}$ and $w^{m}$). 

In addition to the canonical metric $\mc{G}_{MN}$ there is also a
\textbf{canonical\index{canonical antisymmetric 2-form} antisymmetric
2-form} $\mc{B}$, s.th. $\alpha(b)-\beta(a)=\mf{a}^{M}\mc{B}_{MN}\mf{b}^{N}$
with coordinate form\index{$B_{MN}$@$\mc{B}_{MN}$} \begin{eqnarray}
\mc{B}_{MN} & \equiv & \left(\begin{array}{cc}
0 & -\delta_{m}^{n}\\
\delta_{n}^{m} & 0\end{array}\right)\label{eq:canonical-B-tensor}\end{eqnarray}
Raising the indices with $\mc{G}^{MN}$ yields \begin{eqnarray}
\mc{B}^{M}\tief{N} & = & \left(\begin{array}{cc}
\delta_{n}^{m} & 0\\
0 & -\delta_{m}^{n}\end{array}\right)=-B_{N}\hoch{M}\\
\mc{B}^{MN} & = & \left(\begin{array}{cc}
0 & \delta_{n}^{m}\\
-\delta_{m}^{n} & 0\end{array}\right)\label{eq:upper-canonical-B-tensor}\end{eqnarray}
We can thus use $\mc{B}$ and $\mc{G}$ to construct \textbf{projection
operators} $\mc{P_{T}}$ and $\mc{P_{T^{\!*}}}$ to tangent and cotangent
space\begin{eqnarray}
\mc{P}_{\mc{T}}\hoch{M}\tief{N} & \equiv & \frac{1}{2}\left(\delta^{M}\tief{N}+B^{M}\tief{N}\right)=\left(\begin{array}{cc}
\delta_{n}^{m} & 0\\
0 & 0\end{array}\right)\\
\mc{P}_{\mc{T}^{*}}\hoch{M}\tief{N} & \equiv & \frac{1}{2}\left(\delta^{M}\tief{N}-B^{M}\tief{N}\right)=\left(\begin{array}{cc}
0 & 0\\
0 & \delta_{m}^{n}\end{array}\right)\\
\mc{P}_{\mc{T}}\mf{a} & = & a,\qquad\mc{P}_{\mc{T}^{*}}\mf{a}=\alpha\end{eqnarray}

\section{Generalized almost complex structure}

\label{sub:generalized-complex-structure}A \textbf{generalized\index{generalized!(almost) complex structure|fett}
almost complex\index{complex structure!generalized $\sim$|fett} structure}
is a linear map from $T\oplus T^{*}$ to itself which squares to minus
the identity-map, i.e. in components\index{$J$@$\mc{J}^M\tief{N}$}\begin{eqnarray}
\mc{J}^{M}\tief{K}\mc{J}^{K}\tief{N} & = & -\delta_{N}^{M}\label{eq:Jsquare-is-one}\end{eqnarray}
It is called a \textbf{generalized complex structure} if it is integrable
(see subsection \ref{sub:Integrability-of-J}). It should be \textbf{compatible}
with our canonical metric $\mc{G}$ which means that it should behave
like multiplication with $i$ in a Hermitian scalar product of a complex
vector space%
\footnote{\index{footnote!\thefoot. compatibility of GCS with canonical metric}\label{scalar-product}
In a complex vector space with Hermitian scalar product $\erw{a,b}=\overline{\erw{b,a}}$
we have $\erw{a,ib}=-\erw{ia,b}$.$\qquad\fussend$ %
}\begin{eqnarray}
\erw{\mf{v},\mc{J}\mf{w}} & = & -\erw{\mc{J}\mf{v},\mf{w}}\iff(\mc{G}\mc{J})^{T}=-\mc{G}\mc{J}\iff\mc{J}_{MN}=-\mc{J}_{NM}\label{eq:Jis-antisym}\end{eqnarray}
This property is also known as \textbf{antihermiticity\index{antihermiticity!of the generalized complex structure}}
of $\mc{J}$. Because of (\ref{eq:Jis-antisym}), $\mc{J}$ can be
written as \begin{eqnarray}
\mc{J}^{M}\tief{N} & = & \left(\begin{array}{cc}
J^{m}\tief{n} & P^{mn}\\
-Q_{mn} & -J^{n}\tief{m}\end{array}\right)\qquad\mc{J}_{MN}=\left(\begin{array}{cc}
-Q_{mn} & -J^{n}\tief{m}\\
J^{m}\tief{n} & P^{mn}\end{array}\right)\label{eq:J-matrix}\end{eqnarray}
where $P^{mn}$ and $Q_{mn}$ are antisymmetric matrices, and (\ref{eq:Jsquare-is-one})
translates into \begin{eqnarray}
J^{2}-PQ & = & -\one\label{eq:alg-PQJ-cond}\\
JP-PJ^{T} & = & 0\label{eq:alg-JP-cond}\\
-QJ+J^{T}Q & = & 0\label{eq:alg-JQ-cond}\end{eqnarray}
Here it becomes obvious that the generalized complex structure contains
the case of an ordinary almost complex structure $J$ with $J^{2}=-1$
for $Q=P=0$ as well as the case of an almost symplectic structure
of a non-degenerate 2-form $Q$ with existing inverse $PQ=\one$ for
$J=0$. In addition to those algebraic constraints, the integrability
of the generalized almost complex structure gives further differential
conditions (see subsection \ref{sub:Integrability-of-J}) which boil
down in the two special cases to the integrability of the ordinary
complex structure or to the integrability of the symplectic structure.

Because of $\mc{J}^{2}=-\one$, $\mc{J}$ has eigenvalues $\pm i$.
The corresponding eigenvectors span the space of \textbf{generalized\index{generalized!holomorphic}
holomorphic\index{holomorphic!generalized $\sim$} vectors} $L$\index{$L$|itext{generalized holomorphic bundle}}
or generalized\index{generalized!antiholomorphic} antiholomorphic\index{antiholomorphic!generalized $\sim$}
vectors $\bar{L}$\index{$L$@$\bar{L}$|itext{generalized antiholomorphic bundle}}
respectively. This provides a natural splitting of the complexified
bundle \begin{equation}
(T\oplus T^{*})\otimes\mathbb{C}=L\oplus\bar{L}\end{equation}
The \textbf{projector} $\Pi$ to the space of eigenvalue $+i$ (namely
$L$) can be be written as\index{$\Pi^M\tief{N}$}\index{$\Pi^M\tief{N}$@$\bar\Pi^M\tief{N}$}
\begin{eqnarray}
\Pi & \equiv & \frac{1}{2}\left(\one-i\mc{J}\right)\end{eqnarray}
while the projector to $\bar{L}$ is just the complex conjugate $\bar{\Pi}=\frac{1}{2}\left(\one+i\mc{J}\right)=G^{-1}\Pi^{T}G$.
Indeed, for any generalized vector field $\mf{v}$ we have \begin{eqnarray}
\mc{J}\Pi\mf{v} & = & i\Pi\mf{v}\end{eqnarray}
$L$ and $\bar{L}$ are what one calls \textbf{maximally\index{maximally isotropic subspace}
isotropic\index{isotropic!maximally $\sim$ subspace} subspaces},
i.e. spaces which are \emph{isotropic} \begin{eqnarray}
\erw{\mf{v},\mf{w}} & = & 0\quad\forall\mf{v},\mf{w}\in L\end{eqnarray}
(this is because $\Pi^{T}G\Pi=\mc{G}\bar{\Pi}\Pi=0$) and which have
half the dimension of the complete bundle. As the canonical metric
$\langle\cdots\rangle$ is nondegenerate, this is the maximal possible
dimension for isotropic subbundles.

\section{Dorfman and Courant bracket }

Something which seems to be a bit unnatural in this whole business
in the beginning is the introduction of the Courant bracket, which
is the antisymmetrization of the so-called Dorfman-bracket. The \textbf{Dorfman\index{Dorfman bracket}
bracket}\index{bracket!Dorfman $\sim$} in turn is the natural generalization
of the Lie bracket from the point of view of derived brackets (\ref{eq:derived-bracketI})%
\footnote{\index{footnote!\thefoot. twisted Dorfman bracket}\label{twisted-Dorfman}
The twisted\index{twisted!Dorfman bracket} Dorfman bracket is defined
similarly via \begin{eqnarray*}
\left[\left[\ip_{\mf{a}},\de+H\wedge\,\right],\ip_{\mf{b}}\right] & \equiv & \ip_{\left[\mf{a}\bs{,}\mf{b}\right]_{H}}\end{eqnarray*}
Remembering that $H\wedge=\ip_{H}$ and using $[\ip_{a},\ip_{H}]=\ip_{[a,H]^{\Delta}}=\ip_{\ip_{a}^{(1)}H}$,
we get\begin{eqnarray*}
\left[\mf{a}\bs{,}\mf{b}\right]_{H} & \equiv & \left[a\bs{,}b\right]-\ip_{b}\ip_{a}H\qquad\fussend\end{eqnarray*}
}\rem{Achtung! sind sicher irgendwelche Trivialitaeten in folgender
Fussnote!%
\footnote{\index{footnote!\thefoot. amputated bracket}If we take Buttin's generalization
in the sense that $\left[\Lie_{\mf{a}},\Lie_{\mf{b}}\right]=\Lie_{\left[\mf{a}\bs{,}\mf{b}\right]_{\textrm{B}}}$
, we get the same expression as in (\ref{eq:Dorfman-bracket}), but
without the total derivative at the end. In fact, this bracket obeys
the Jacobi identity and is antisymmetric at the same time which makes
it it quite attractive. It is not known to me whether there are objections
to use Buttin's bracket to define integrability in generalized geometry.
It would certainly lead to a different geometry.$\qquad\fussend$%
}}\begin{eqnarray}
\left[\left[\ip_{\mf{a}},\de\,\right],\ip_{\mf{b}}\right] & = & \ip_{\left[\mf{a}\bs{,}\mf{b}\right]}\\
\textrm{where }\left[\mf{a}\bs{,}\mf{b}\right] & \equiv & \left[a\bs{,}b\right]+\Lie_{a}\beta-\Lie_{b}\alpha+\de\,(\ip_{b}\alpha)=\label{eq:Dorfman-bracket}\\
 & = & \left[a\bs{,}b\right]+\Lie_{a}\beta-\ip_{b}(\de\alpha)=\label{eq:Dorfman-bracketII}\\
 & = & \Lie_{a}\mf{b}-\ip_{b}(\de\alpha)\label{eq:Dorfman-bracketIII}\end{eqnarray}
To get a homogeneous coordinate expression, we define\index{$\partial_{M}$ on $T\oplus T^*$}\index{$\partial^{M}$ on $T\oplus T^*$}\begin{eqnarray}
\partial_{M} & \equiv & \left(\partial_{m},0\right)\quad\dann\partial^{M}=\left(0,\partial_{m}\right)\end{eqnarray}
\newpage\noindent The Dorfman bracket can then be written as%
\footnote{\index{footnote!\thefoot. dual coordinate; relation to Hull's doubled geometry}\label{foot:dual-coord}It
is perhaps interesting to note that this notation of the partial derivative
with capital index suggests the extension to a derivative with respect
to some dual coordinate \[
\partial^{m}\equiv\partial_{\tilde{x}_{m}}\]
We could understand this as coordinates of a dual manifold whose tangent
space coincides in some sense with the cotangent space of the original
space and vice versa. This might be connected to Hull's doubled geometry
\cite{Dabholkar:2005ve,Hull:2006qs,Hull:2006va,Hull:2004in,Morris:2007ga}.

To see that such an ad-hoc extension of the Dorfman bracket is not
completely unfounded, note that there is a more general notion of
a Dorfman bracket (or Courant bracket) in the context of Lie-bialgebroids
(for a definition see e.g. \cite[p.32,20]{Gualtieri:007}). There
we have two Lie algebroids $L$ and $L^{*}$ which are dual with respect
to some inner product and which both carry some Lie bracket. (For
$T$ and $T^{*}$, only $T$ carries a Lie bracket in the beginning.
For a non-trivial Lie bracket of forms on $T^{*}$ we need some extra
structure like e.g. a Poisson structure which would lead to the Koszul
bracket on forms.) The Lie bracket on $L$ induces a differential
$\de$ on $L^{*}$ and the Lie bracket on $L^{*}$ induces a differential
$\de^{*}$ on $L$. The definition for the Dorfman bracket on the
Lie bialgebroid $L\oplus L^{*}$ is then \begin{eqnarray*}
\left[\mf{a}\bs{,}\mf{b}\right] & \equiv & \left[a\bs{,}b\right]+\Lie_{a}\beta-\Lie_{b}\alpha+\de\,(\ip_{b}\alpha)+\\
 &  & +\left[\alpha\bs{,}\beta\right]+\Lie_{\alpha}b-\Lie_{\beta}a+\de^{*}(\ip_{\beta}a)\end{eqnarray*}
The first line is the part we are used to from our usual Dorfman bracket
on $T\oplus T^{*}$, while second line is the corresponding part coming
from the nontrivial structure on $L^{*}$. Taking now $L=T$, $L^{*}=T^{*}$
and assuming that $[\alpha\bs{,}\beta]$ and $\Lie_{\alpha}$ and
$\de^{*}$ are a Lie bracket, Lie derivative and exterior derivative
built in the ordinary way, but with the new partial derivative w.r.t.
the dual coordinates $\partial^{m}$, the coordinate form of the Dorfman
bracket remains exactly the one of (\ref{eq:Dorfman-bracket-coord},\ref{eq:Dorfman-bracket-coordII}),
but with $\partial_{M}=(\partial_{m},0)$ replaced by $\partial_{M}=(\partial_{m},\partial^{m}).\qquad\fussend$\rem{relating $\pe^m$ and $\de x^m$: integrable structures: like GCS}%
}\begin{eqnarray}
\left[\mf{a}\bs{,}\mf{b}\right]^{M} & = & \mf{a}^{K}\partial_{K}\mf{b}^{M}+\left(\partial^{M}\mf{a}_{K}-\partial_{K}\mf{a}^{M}\right)\mf{b}^{K}\label{eq:Dorfman-bracket-coord}\\
\textrm{or }\left[\mf{a}\bs{,}\mf{b}\right]_{M} & = & \mf{a}^{K}\partial_{K}\mf{b}_{M}+2\partial_{[M}\mf{a}_{K]}\mf{b}^{K}\label{eq:Dorfman-bracket-coordII}\end{eqnarray}
Apart from the term in the middle $\partial^{M}\mf{a}_{K}$, (\ref{eq:Dorfman-bracket-coord})
looks formally the same as the Lie bracket of vector fields (\ref{eq:vector-Lie-bracket}).
The Dorfman bracket is in general not antisymmetric but it obeys a
\textbf{Jacobi\index{Jacobi-identity!for Dorfman bracket}-identity}
(Leibniz from the left) of the form\begin{eqnarray}
\left[\mf{a}\bs{,}\left[\mf{b}\bs{,}\mf{c}\right]\right] & = & \left[\left[\mf{a}\bs{,}\mf{b}\right]\bs{,}\mf{c}\right]+\left[\mf{b}\bs{,}\left[\mf{a}\bs{,}\mf{c}\right]\right]\label{eq:Jacobi-for-Dorfman}\end{eqnarray}
Although the Dorfman bracket is all we need, most of the literature
on generalized complex geometry so far works with its antisymmetrization,
which is called \textbf{Courant\index{Courant bracket} bracket\index{bracket!courant $\sim$}\index{$([])$@$\left[\ldots\bs{,}\ldots\right]_{-}$|itext{Courant bracket}}}
\begin{eqnarray}
\left[\mf{a}\bs{,}\mf{b}\right]_{-} & \equiv & \left[a\bs{,}b\right]+\Lie_{a}\beta-\Lie_{b}\alpha+\frac{1}{2}\de\,(\ip_{b}\alpha-\ip_{a}\beta)\label{eq:Courant-bracket}\\
\left[\mf{a}\bs{,}\mf{b}\right]_{-M} & = & \mf{a}^{K}\partial_{K}\mf{b}_{M}-\partial_{K}\mf{a}_{M}\mf{b}^{K}+\frac{1}{2}\left(\partial_{M}\mf{a}_{K}\mf{b}^{K}-\mf{a}^{K}\partial_{M}\mf{b}_{K}\right)\label{eq:Courant-bracket-coord}\end{eqnarray}
and which does not obey any Jacobi identity. As it is much simpler
to go from Dorfman to Courant, than the other way round, we will only
work with the Dorfman bracket. On any isotropic subspace ($\ip_{b}\alpha+\ip_{a}\beta=0$)
the two coincide anyway, i.e. they become a Lie bracket, obeying Jacobi
and being antisymmetric. 

We call a transformation a \textbf{symmetry\index{symmetry!of the Dorfman bracket}
of the bracket} when the bracket of two vectors transforms in the
same way as the vectors\begin{eqnarray}
\left[(\mf{b}+\delta\mf{b})\bs{,}(\mf{c}+\delta\mf{c})\right] & = & \left[\mf{b}\bs{,}\mf{c}\right]+\delta\left[\mf{b}\bs{,}\mf{c}\right]\\
\delta\left[\mf{b}\bs{,}\mf{c}\right] & = & \left[\delta\mf{b}\bs{,}\mf{c}\right]+\left[\mf{b}\bs{,}\delta\mf{c}\right]+\left[\delta\mf{b}\bs{,}\delta\mf{c}\right]\end{eqnarray}
I.e. infinitesimal symmetry transformations (where the last term drops)
have to obey a product rule. Similar as for the Lie-bracket of vector
fields, infinitesimal transformations are generated by the bracket
itself. Let us call the corresponding derivative, in analogy to the
Lie derivative, the \textbf{Dorfman\index{Dorfman derivative} derivative\index{derivative!Dorfman $\sim$}}
of a generalized vector with respect to a generalized vector.\index{$D_a$@$\Dorf_{\protect\mf{a}}$|itext{Dorfman derivative}}
\begin{equation}
\delta\mf{b}=\Dorf_{\mf{a}}\mf{b}\equiv\left[\mf{a},\mf{b}\right]\end{equation}
 These transformations are therefore, due to the Jacobi-identity (\ref{eq:Jacobi-for-Dorfman})
always symmetries of the bracket. From (\ref{eq:Dorfman-bracketIII})
we can see that the Dorfman derivative consists of a usual Lie derivative
and second part which acts only on the vector part of $\mf{b}$ by
contracting it with the exact 2-form $\de\alpha$\begin{eqnarray}
\Dorf_{a}\mf{b} & = & \Lie_{a}\mf{b}\label{eq:Dorfman-transformationI}\\
\Dorf_{\alpha}b & = & -\ip_{b}(\de\alpha)=b^{m}(\partial_{n}\alpha_{m}-\partial_{m}\alpha_{n})\de x^{n}\label{eq:Dorfman-transformationII}\end{eqnarray}
In fact, it is enough for the 2-form to be closed, in order to get
a symmetry. If we replace $-\de\alpha$ by a \emph{closed 2-form}
$B$, the transformation is known as $B$\textbf{\index{B-transform@$B$-transform}-transform\index{transform!$B$-$\sim$}\begin{eqnarray}
\delta_{B}b & = & \ip_{b}B\end{eqnarray}
}

Finally, we should note that the $B$-transform is part of the\index{$O(d,d)$}
$O(d,d)$-transformations, i.e. the transformations which leave the
canonical metric invariant. As usual for orthogonal groups the infinitesimal
generators are antisymmetric when the second index is pulled down
with the corresponding metric. The generators of an $O(d,d)$-transformation
can therefore be written as \cite[p.6]{Gualtieri:007}\begin{eqnarray}
\Omega_{MN} & = & \left(\begin{array}{cc}
B_{mn} & -A_{m}\hoch{n}\\
A_{n}\hoch{m} & \textrm{\Beta}^{mn}\end{array}\right)\\
\Omega^{M}\tief{N} & = & \left(\begin{array}{cc}
A_{n}\hoch{m} & \Beta^{mn}\\
B_{mn} & -A_{m}\hoch{n}\end{array}\right)\label{eq:SOnn}\end{eqnarray}
\index{$\beta^{mn}$@$\sBeta^{mn}$|itext{beta-transform}}In addition
to the $B$-transform, acting with $\Omega$ on a generalized vector
induces the so-called \textbf{beta\index{beta-transform}-transform}\index{transform!beta-$\sim$}
on the 1-form component%
\footnote{\index{footnote!\thefoot. letter for beta transform $\sBeta^{mn}$}The
letter $\beta$ for the beta-transformations does not really fit into
the philosophy of the present notations, where we use small Greek
letters for 1-forms (or sometimes p-forms) only, but not for multivectors.
As the transformation is, however, commonly known as beta-transformation,
we use a large $\sBeta$, in order to distinguish it from the one-forms
$\beta$, which are floating around.$\quad\fussend$%
} as well as $Gl(d)$-transformations of vector and 1-form component
via $A$. For constant tensors, the Lie-derivative is just a $Gl(d)$
transformation. Therefore both symmetries of the Dorfman bracket are
symmetries of the canonical metric $\mc{G}$ as well. For this reason
the canonical metric is invariant under the \textbf{Dorfman\index{Dorfman derivative}
derivative} $\Dorf_{\mf{v}}$with respect to a generalized vector
$\mf{v}$, which we define on generalized rank $p$ tensors using
(\ref{eq:Dorfman-bracket-coord}) in a way that it acts via Leibniz
on tensor products (like the Lie derivative) and as a directional
derivative on scalars\begin{eqnarray}
(\Dorf_{\mf{v}}\mc{T})^{M_{1}\ldots M_{p}} & \equiv & \mf{v}^{K}\partial_{K}\mc{T}^{M_{1}\ldots M_{p}}+\sum_{i}(\partial^{M_{i}}\mf{v}_{K}-\partial_{K}\mf{v}^{M_{i}})T^{M_{1}\ldots M_{i-1}KM_{i+1}\ldots M_{p}}\label{eq:Dorfman-derivative}\\
\Dorf_{\mf{v}}(\mc{A}\otimes\mc{B}) & = & \Dorf_{v}\mc{A}\otimes\mc{B}+\mc{A}\otimes\Dorf_{v}\mc{B}\\
\Dorf_{\mf{v}}(\phi) & = & \mf{v}^{K}\partial_{K}\phi=v^{k}\partial_{k}\phi\end{eqnarray}
Acting on the canonical metric, one recovers the fact, that the Dorfman
derivative contains the isometries of the metric \begin{eqnarray}
\Dorf_{\mf{v}}\mc{G} & = & 2(\partial^{M_{1}}\mf{v}_{K}-\partial_{K}\mf{v}^{M_{1}})\mc{G}^{KM_{2}}=0\end{eqnarray}
Comparing the role of Lie-derivative and Dorfman-derivative, the $B$-transform
should be understood as an extension of diffeomorphisms. In string
theory it shows up in the Buscher-rules for T-duality (\cite{Buscher:1987sk,Buscher:1987qj})
\rem{ist das wahr?} and can perhaps be better understood geometrically
via Hull's \rem{and Dhabolka's?}  doubled geometry \cite{Dabholkar:2005ve,Hull:2006qs,Hull:2006va}
(compare to footnote \ref{foot:dual-coord}). The beta-transform is
not a symmetry of the Dorfman bracket as it stands. However, if we
introduce dual coordinates as suggested in footnote \ref{foot:dual-coord},
the beta-transform would show up in the symmetry-transformations of
the extended Dorfman bracket generated by itself.%
\footnote{\index{footnote!\thefoot. contribution of beta transformation to extended Dorfman derivative}Taking
the Dorfman bracket of footnote \ref{foot:dual-coord}, we get as
Dorfman derivative of a generalized vector $\mf{c}$ instead of (\ref{eq:Dorfman-transformationI},\ref{eq:Dorfman-transformationII})
the extended transformation \begin{eqnarray*}
\Dorf_{a}\mf{c} & \equiv & \Lie_{a}\mf{c}-\ip_{\gamma}(\de^{*}a)\\
\Dorf_{\alpha}\mf{c} & \equiv & -(\ip_{c}\de\alpha)+\Lie_{\alpha}\mf{c}\end{eqnarray*}
I.e. the first line is extended by a beta-transformation of $\gamma$
with $\sBeta=-\de^{*}a$ and the $B$-transform of $\alpha$ ($B=-\de\alpha$)
in the second line is extended by a Lie derivative with respect to
$\alpha$.$\qquad\fussend$%
}

On an isotropic subspace $L$ (e.g. the generalized holomorphic subspace)
Courant- and Dorfman-bracket coincide and have the properties of a
Lie bracket. It is therefore possible to define a Schouten\index{Schouten bracket!on generalized multivectors}
bracket\index{bracket!Schouten $\sim$ on generalized multivectors}
on generalized\index{generalized multivector} multivector\index{multivector!generalized $\sim$}s
on $\bigwedge^{\bullet}L$\index{$()$@$\bigwedge^{\bullet}L$\addcontentsline{toc}{chapter}{\protect\hyperlink{Index}{Index}}\protect\hypertarget{Index}{}}
which have e.g. only generalized holomorphic indices (compare \cite[p.21]{Gualtieri:007}).
If we use again the notation with repeated boldface indices\index{$A^{(p)}$@$\mc{A}^{(p)}$|itext{generalized multivector}}\index{$A^{(p)}$@$\mc{A}_{\bs{M}\ldots\bs{M}}$}\begin{equation}
\mc{A}^{(p)}\equiv\mc{A}_{\bs{M}\ldots\bs{M}}\equiv\mc{A}_{M_{1}\ldots M_{p}}\basis^{M_{1}}\cdots\basis^{M_{2}}\end{equation}
we get as coordinate form for this \textbf{Dorfman\index{Dorfman-Schouten bracket}-Schouten}
\textbf{bracket}\index{bracket!Dorfman-Schouten $\sim$} \rem{Schouten-bracket on $\Lambda^\bullet L$}\begin{eqnarray}
\left[\mc{A}^{(p)}\bs{,}\mc{B}^{(q)}\right] & = & p\mc{A}^{\bs{M}\ldots\bs{M}K}\partial_{K}\mc{B}^{\bs{M}\ldots\bs{M}}+q\left(p\partial^{\bs{M}}\mc{A}_{K}\hoch{\bs{M}\ldots\bs{M}}-\partial_{K}\mc{A}^{\bs{M}\ldots\bs{M}}\right)\mc{B}^{K\bs{M}\ldots\bs{M}}\label{eq:Dorfman-Schouten-bracket}\end{eqnarray}
In the first term in the bracket on the righthand side, the $\partial^{\bs{M}}$
can as well be shifted with a minus sign to $\mc{B}$, because in
$\bigwedge^{\bullet}L$ we have only isotropic indices in the sense
that\begin{eqnarray}
\mc{A}^{\bs{M}\ldots\bs{M}}\tief{K}\mc{B}^{K\bs{M}\ldots\bs{M}} & = & 0\label{eq:isotropic-multivectors}\end{eqnarray}
For this reason, the Dorfman-Schouten bracket has really the required
skew-symmetry of a Schouten-bracket\begin{eqnarray}
\left[\mc{A}^{(p)}\bs{,}\mc{B}^{(q)}\right] & = & -(-)^{(q+1)(p+1)}\left[\mc{B}^{(q)}\bs{,}\mc{A}^{(p)}\right]\end{eqnarray}
On $\bigwedge^{\bullet}L$ this bracket coincides with the derived
bracket of the big bracket, as the extra term with $p_{M}$ in (\ref{eq:derived-of-big-generalized})
vanishes because of (\ref{eq:isotropic-multivectors}).

\section{Integrability}

\label{sub:Integrability-of-J} \index{integrability!of a generalized complex structure}Integrability
for an ordinary complex structure means that there exist in any chart
$\dim_{M}/2$ holomorphic vector fields (with respect to the almost
complex structure) which can be integrated to holomorphic coordinates
$z^{a}$ in this chart of the manifold and make it a complex manifold.
Those vector fields are then just $\partial/\partial z^{a}$. Those
coordinate differentials have vanishing Lie bracket among each other
(partial derivatives commute). In turn, every set of vectors with
vanishing Lie bracket can be integrated to coordinates. The existence
of such a set of integrable holomorphic vector fields is guaranteed
when the holomorphic subbundle is closed under the Lie bracket, i.e.
the Lie bracket of two holomorphic vector fields is again a holomorphic
vector field. \rem{Ist das wahr?} 

As the Dorfman bracket restricted to the generalized holomorphic subbundle
$L\subset(T\oplus T^{*})\otimes\mathbb{C}$ has the properties of
a Lie bracket, we can demand exactly the same for generalized holomorphic
vectors as above for holomorphic ones. \rem{The interpretation will
then be that there is some complex manifold \rem{die gleiche, oder eine andere??? Hull's geometry, oder doch ganz normale?} 
whose complex coordinates define the generalized holomorphic vector
fields. Anyway,} The condition for the generalized complex structure
to be integrable is thus that the generalized holomorphic subbundle
$L$ is closed under the Dorfman bracket, i.e. in terms of the projectors\begin{eqnarray}
\bar{\Pi}\left[\Pi\mf{v}\bs{,}\Pi\mf{w}\right] & = & 0\label{eq:generalized-integrability}\\
\iff\left[\mf{v}\bs{,}\mf{w}\right]-\left[\mc{J}\mf{v}\bs{,}\mc{J}\mf{w}\right]+\mc{J}\left[\mc{J}\mf{v}\bs{,}\mf{w}\right]+\mc{J}\left[\mf{v}\bs{,}\mc{J}\mf{w}\right] & = & 0\label{eq:generalized-integrabilityII}\end{eqnarray}
In the following two sub-subsections we will show that this is equivalent
to the vanishing of a \textbf{generalized\index{generalized!Nijenhuis tensor}
Nijenhuis\index{Nijenhuis tensor!generalized $\sim$}-tensor} \cite[p.25]{Gualtieri:007}
of the coordinate form%
\footnote{\label{foot:holomorphic-indices}\index{footnote!\thefoot. generalized Nijenhuis tensor versus generalized Schouten bracket}This
looks formally like the generalized Schouten bracket (e.g. \cite[p.21]{Gualtieri:007})
on $\bigwedge^{\bullet}L$ (with L being the generalized holomorphic
bundle)\rem{ (\ref{eq:Dorfman-Schouten-bracket})} of $\mc{J}$
with itself (see also the statement below (\ref{eq:derived-of-big-generalized})),
but it is not, as $\mc{J}$ has neither holomorphic nor antiholomorphic
indices\begin{eqnarray*}
\Pi\mc{J} & = & i\Pi\neq\mc{J}\\
\bar{\Pi}\mc{J} & = & -i\Pi\neq\mc{J}\end{eqnarray*}
In fact, we get zero if we contract both indices with the holomorphic
projector \rem{holomorphic upper indices are antiholomorphic lower indices!}\begin{eqnarray*}
\Pi^{N}\tief{L}\Pi^{M}\tief{K}\mc{J}^{KL} & = & \Pi\mc{J}\Pi^{T}=i\Pi\bar{\Pi}=0\end{eqnarray*}
The same happens for two antiholomorphic projectors. But we can project
one index with an holomorphic projector and the other one with an
antiholomorphic one. This yields\begin{eqnarray*}
\bar{\Pi}^{N}\tief{L}\Pi^{M}\tief{K}\mc{J}^{KL} & = & \Pi\mc{J}\Pi=i\Pi\end{eqnarray*}
Up to a constant prefactor the bracket of $\Pi$ with $\Pi$ coincides
with the bracket of $\mc{J}$ with $\mc{J}$. And like for the ordinary
complex structure, where we have the Nijenhuis bracket of the complex
structure with itself, which has one index in $T$ and the second
in $T^{*}$, we could here take $\Pi$ with one index in $L$ and
the other in $\bar{L}$ and regard the bracket as generalized Nijenhuis
bracket of $\Pi$ with itself.\rem{besser rausnehmen?}$\quad\fussend$%
}$^{,}$%
\footnote{\index{footnote!\thefoot. twisted generalized Nijenhuis tensor}If
instead the twisted Dorfman bracket (see footnote \ref{twisted-Dorfman})
is used, one gets the integrability condition for a twisted generalized
complex structure with a twisted generalized Nijenhuis tensor. Consider
the closed three form $H=H_{M_{1}M_{2}M_{3}}\basis^{M_{1}}\basis^{M_{2}}\basis^{M_{3}}$
with $H_{m_{1}m_{2}m_{3}}$ the only nonvanishing components. The
twisted\index{twisted generalized Nijenhuis tensor} generalized\index{generalized Nijenhuis tensor!twisted $\sim$}
Nijenhuis\index{Nijenhuis tensor!twisted generalized $\sim$} tensor
then reads\rem{siehe Notizblock 2006, S.73 und S.60} \begin{eqnarray*}
\mc{N}_{M_{1}M_{2}M_{3}}^{H} & = & \mc{N}_{M_{1}M_{2}M_{3}}+6H_{M_{1}M_{2}M_{3}}-18\mc{J}_{M_{1}}\hoch{K}H_{KM_{2}L}\mc{J}^{L}\tief{M_{3}}\end{eqnarray*}
Like (\ref{eq:integrability-tensor-I})-(\ref{eq:integrability-tensor-IV})
this twisted generalized Nijenhuis tensor as well matches with the
tensors given in \cite{Zucchini:2004ta} if one redefines $H_{mnk}\to\frac{1}{3!}H_{mnk}$.$\qquad\fussend$%
} \rem{Referenz zu Gualtieri! Nur Koordinatenform ist neu....}\index{$N^{M}$@$\mc{N}^{M_{1}M_{2}M_{3}}$|iText{generalized Nijenhuis tensor}}\begin{equation}
\boxed{\frac{1}{4}\mc{N}^{M_{1}M_{2}M_{3}}\equiv\mc{J}^{[M_{1}|K}\partial_{K}\mc{J}^{|M_{2}M_{3}]}+\mc{J}^{[M_{1}|K}\mc{J}_{K}\hoch{|M_{2},M_{3}]}\stackrel{!}{=}0}\label{eq:generalized-integrabilityIII}\end{equation}
Recalling that \begin{eqnarray}
\mc{J}^{MN} & = & \left(\begin{array}{cc}
P^{mn} & J^{m}\tief{n}\\
-J^{n}\tief{m} & -Q_{mn}\end{array}\right),\qquad\mc{J}_{M}\hoch{N}=\left(\begin{array}{cc}
-J^{n}\tief{m} & -Q_{mn}\\
P^{mn} & J^{m}\tief{n}\end{array}\right),\qquad\partial^{M}=(0,\partial_{m})\label{eq:J-tensor-II}\end{eqnarray}
we can rewrite this condition in ordinary tensor components, just
to compare it with the conditions given in literature (for the antisymmetrization
of the capital indices we take into account that in the last term
of (\ref{eq:generalized-integrabilityIII}) the indices $M_{1}$ and
$M_{2}$ are automatically antisymmetrized because of $\mc{J}^{2}=-1$):\begin{eqnarray}
\frac{1}{4}\mc{N}^{m_{1}m_{2}m_{3}} & = & P^{[m_{1}|k}\partial_{k}P^{|m_{2}m_{3}]}\stackrel{!}{=}0\label{eq:integrability-tensor-I}\\
\frac{1}{4}\mc{N}_{n}\hoch{m_{1}m_{2}} & = & \frac{1}{3}\left(-J^{k}\tief{n}\partial_{k}P^{[m_{1}m_{2}]}+2P^{[m_{1}|k}\partial_{k}J^{|m_{2}]}\tief{n}-P^{[m_{1}|k}J^{|m_{2}]}\tief{k,n}+J^{[m_{1}|}\tief{k}P^{k|m_{2}]}\tief{,n}\right)\stackrel{!}{=}0\qquad\quad\\
\frac{1}{4}\mc{N}^{n}\tief{m_{1}m_{2}} & = & \frac{1}{3}\left(-P^{nk}\partial_{k}Q_{[m_{1}m_{2}]}+2J^{k}\tief{[m_{1}|}\partial_{k}J^{n}\tief{|m_{2}]}+2J^{n}\tief{k}J^{k}\tief{[m_{1},m_{2}]}-2P^{nk}Q_{k[m_{1},m_{2}]}\right)\stackrel{!}{=}0\\
\frac{1}{4}\mc{N}_{m_{1}m_{2}m_{3}} & = & J^{k}\tief{[m_{1}|}\partial_{k}Q_{|m_{2}m_{3}]}+J^{k}\tief{[m_{1}|}Q_{k|m_{2},m_{3}]}-Q_{[m_{1}|k}J^{k}\tief{|m_{2},m_{3}]}\stackrel{!}{=}0\label{eq:integrability-tensor-IV}\end{eqnarray}
If we compare those expressions with the tensors $A,B,C$ and $D$
given in (2.16) of \cite[p.7]{Zucchini:2004ta}, we recognize (replacing
$Q$ by $-Q$) that our first line is just $\frac{1}{3}A$, the second
line is $-\frac{1}{3}B$ (using (\ref{eq:alg-JP-cond})), the third
$\frac{1}{3}C$ and the fourth line is $-\frac{1}{3}D$. There, in
turn, it is claimed that the expressions are equivalent to those originally
given in (3.16)-(3.19) of \cite[p.7]{Lindstrom:2004iw}.

\subsection{Coordinate based way to derive the generalized Nijenhuis-tensor}

\label{sub:Coordinate-based-way}\rem{dasselbe fuer twisted case machen! nur 3-Form, oder auch hoeher?}
In this sub-subsection we will see that calculations with capital-index
notation is rather convenient. So we simply calculate (\ref{eq:generalized-integrabilityII})
brute force by using the explicit coordinate formula for the Dorfman-bracket\bref{eq:Dorfman-bracket-coord}
\begin{equation}
\left[\mf{v}\bs{,}\mf{w}\right]^{M}=\mf{v}^{K}\partial_{K}\mf{w}^{M}+\left(\partial^{M}\mf{v}_{K}-\partial_{K}\mf{v}^{M}\right)\mf{w}^{K}\end{equation}
\eref  The brackets of interest are: \begin{eqnarray}
\left[\mf{v}\bs{,}\mc{J}\mf{w}\right]^{N} & = & \mf{v}^{K}\partial_{K}\mc{J}^{N}\tief{L}\mf{w}^{L}+\mc{J}^{N}\tief{L}\mf{v}^{K}\partial_{K}\mf{w}^{L}+\left(\partial^{N}\mf{v}_{K}-\partial_{K}\mf{v}^{N}\right)(\mc{J}\mf{w})^{K}\\
(\mc{J}\left[\mf{v}\bs{,}\mc{J}\mf{w}\right])^{M} & = & \underline{\mf{v}^{K}\mc{J}^{M}\tief{N}\partial_{K}\mc{J}^{N}\tief{L}\mf{w}^{L}}-\mf{v}^{K}\partial_{K}\mf{w}^{M}+\mc{J}^{M}\tief{N}\left(\partial^{N}\mf{v}_{K}-\partial_{K}\mf{v}^{N}\right)(\mc{J}\mf{w})^{K}\\
\left[\mc{J}\mf{v}\bs{,}\mf{w}\right]^{N} & = & \mc{J}^{K}\tief{L}\mf{v}^{L}\partial_{K}\mf{w}^{N}+\left(\partial^{N}\mc{J}_{KL}-\partial_{K}\mc{J}^{N}\tief{L}\right)\mf{v}^{L}\mf{w}^{K}+\left(\mc{J}_{K}\hoch{L}\partial^{N}\mf{v}_{L}-\mc{J}^{N}\tief{L}\partial_{K}\mf{v}^{L}\right)\mf{w}^{K}\\
(\mc{J}\left[\mc{J}\mf{v}\bs{,}\mf{w}\right])^{M} & = & \mc{J}^{M}\tief{N}(\mc{J}\mf{v})^{K}\partial_{K}\mf{w}^{N}+\underline{\mc{J}^{M}\tief{N}\left(\partial^{N}\mc{J}_{KL}-\partial_{K}\mc{J}^{N}\tief{L}\right)\mf{v}^{L}\mf{w}^{K}}+\nonumber \\
 &  & -(\mc{J}\mf{w})^{L}\mc{J}^{M}\tief{N}\partial^{N}\mf{v}_{L}+\partial_{K}\mf{v}^{M}\mf{w}^{K}\\
\left[\mc{J}\mf{v}\bs{,}\mc{J}\mf{w}\right]^{M} & = & \mc{J}^{K}\tief{N}\mf{v}^{N}\partial_{K}\mc{J}^{M}\tief{L}\mf{w}^{L}+\mc{J}^{K}\tief{N}\mf{v}^{N}\mc{J}^{M}\tief{L}\partial_{K}\mf{w}^{L}+\nonumber \\
 &  & \left(\partial^{M}\mc{J}_{KN}\mf{v}^{N}-\partial_{K}\mc{J}^{M}\tief{N}\mf{v}^{N}\right)\mc{J}^{K}\tief{L}\mf{w}^{L}+\left(\mc{J}_{KN}\partial^{M}\mf{v}^{N}-\mc{J}^{M}\tief{N}\partial_{K}\mf{v}^{N}\right)\mc{J}^{K}\tief{L}\mf{w}^{L}=\\
 & = & (\mc{J}\mf{v})^{K}\mc{J}^{M}\tief{L}\partial_{K}\mf{w}^{L}-\mc{J}^{M}\tief{N}\partial_{K}\mf{v}^{N}(\mc{J}\mf{w})^{K}+\nonumber \\
 &  & +\underline{\left(\mc{J}^{K}\tief{L}\partial^{M}\mc{J}_{KN}+2\mc{J}^{K}\tief{[N|}\partial_{K}\mc{J}^{M}\tief{|L]}\right)\mf{v}^{N}\mf{w}^{L}}+\partial^{M}\mf{v}_{L}\mf{w}^{L}\end{eqnarray}
The underlined terms sum up in the complete expression to the generalized
Nijenhuis tensor, while the rest cancels\begin{eqnarray}
0 & \stackrel{!}{=} & \left[\mf{v}\bs{,}\mf{w}\right]^{M}-\left[\mc{J}\mf{v}\bs{,}\mc{J}\mf{w}\right]^{M}+(\mc{J}\left[\mc{J}\mf{v}\bs{,}\mf{w}\right])^{M}+(\mc{J}\left[\mf{v}\bs{,}\mc{J}\mf{w}\right])^{M}=\\
 & = & \left(2\mc{J}^{M}\tief{K}\partial_{[N}\mc{J}^{K}\tief{L]}-\mc{J}^{K}\tief{L}\partial^{M}\mc{J}_{KN}+\mc{J}^{MK}\partial_{K}\mc{J}_{LN}-2\mc{J}^{K}\tief{[N|}\partial_{K}\mc{J}^{M}\tief{|L]}\right)\mf{v}^{N}\mf{w}^{L}=\\
 & = & \mf{v}_{N}\left(3\mc{J}^{[M|}\tief{K}\mc{J}^{K|L,N]}+3\mc{J}^{[N|K}\partial_{K}\mc{J}^{|ML]}\right)\mf{w}_{L}=\\
 & = & \frac{3}{4}\mf{v}_{N}\mc{N}^{NML}\mf{w}_{L}\end{eqnarray}

\subsection{Derivation via derived brackets}

\label{sub:Derivation-via-derived-bracket}Eventually we want to see
directly how the generalized Nijenhuis tensor is connected to derived
brackets. We will use our insight from the subsections \ref{sub:Algebraic-brackets}
and \ref{sub:Extended-exterior-derivative}. Remember, our basis $\basis^{M}=(\de x^{m},\pe_{m})$
was identified with the conjugate (ghost-)variables $\basis^{M}\equiv(\ce^{m},\be_{m})$.
One can define generalized multi-vector fields of the form\begin{eqnarray}
\mc{K}^{(\textsc{k})} & \equiv & \mc{K}_{\bs{M}\ldots\bs{M}}\equiv\mc{K}_{M_{1}\ldots M_{\textsc{k}}}\basis^{M_{1}}\cdots\basis^{M_{\textsc{k}}}\end{eqnarray}
They are in fact just sums of multivector valued forms:\begin{equation}
\mc{K}_{\bs{M}\ldots\bs{M}}=\sum_{k=0}^{\textsc{k}}\left(\zwek{\textsc{k}}{k}\right)\mc{K}_{\underbrace{{\scriptstyle \mm}}_{k}}\underbrace{\hoch{\nn}}_{\textsc{k}-k}\equiv\sum_{k=0}^{\textsc{k}}K^{(k,\textsc{k}-k)}\end{equation}
The big bracket, or Buttin's algebraic bracket is then just the canonical
Poisson bracket \begin{eqnarray}
\left[\mc{K},\mc{L}\right]_{(1)}^{\Delta} & \equiv & \textsc{kl}\mc{K}_{\bs{M}\ldots\bs{M}}\hoch{I}\mc{L}_{I\bs{M}\ldots\bs{M}}=\left\{ \mc{K},\mc{L}\right\} \label{eq:multvec-bigbrack}\\
\left\{ \basis_{M},\basis_{N}\right\}  & = & \mc{G}_{MN}\label{eq:canon-Poisson}\end{eqnarray}
The coordinate expression for its derived bracket (compare to (\ref{eq:bc-derived-of-bigbracket},\ref{eq:bc-derived-of-bigbracket-coord}))
reads \begin{eqnarray}
(-)^{\textsc{k}-1}\left[\de\mc{K}^{(\textsc{k})},\mc{L}^{(\textsc{L})}\right]_{(1)}^{\Delta} & = & \textsc{k}\cdot\mc{K}_{\bs{M}\ldots\bs{M}}\hoch{I}\partial_{I}\mc{L}_{\bs{M}\ldots\bs{M}}-(-)^{(\textsc{k}+1)(\textsc{l}+1)}\textsc{l}\cdot\mc{L}_{\bs{M}\ldots\bs{M}}\hoch{I}\partial_{I}\mc{K}_{\bs{M}\ldots\bs{M}}+\nonumber \\
 &  & +(-)^{\textsc{k}-1}\textsc{kl}\partial_{\bs{M}}\mc{K}_{\bs{M}\ldots\bs{M}}\hoch{I}\mc{L}_{I\bs{M}\ldots\bs{M}}+\textsc{k}\left(\textsc{k}-1\right)\textsc{l}\mc{K}_{\bs{M}\ldots\bs{M}}\hoch{IJ}\mc{L}_{I\bs{M}\ldots\bs{M}}p_{J}\label{eq:derived-of-big-generalized}\end{eqnarray}
with $p_{J}\equiv(p_{j},0)$ and $\partial_{I}\equiv(\partial_{i},0)$.
In the case were both $\mc{K}$ and $\mc{L}$ only have generalized
holomorphic indices, the $p$-term drops and this expression should
coincide with the Schouten-bracket on $\bigwedge^{\bullet}L$ for
the holomorphic Lie-algebroid $L$ (see e.g. \cite[p.21]{Gualtieri:007}
and footnote \ref{foot:holomorphic-indices}). For two rank-two objects,
like the generalized complex structure $\mc{J}$, this reduces to
\begin{eqnarray}
\left[\mc{K},_{\de\,}\mc{L}\right]_{(1)}^{\Delta} & = & 2\cdot\mc{K}_{\bs{M}}\hoch{I}\partial_{I}\mc{L}_{\bs{M}\bs{M}}+2\cdot\mc{L}_{\bs{M}}\hoch{I}\partial_{I}\mc{K}_{\bs{M}\bs{M}}-4\partial_{\bs{M}}\mc{K}_{\bs{M}}\hoch{I}\mc{L}_{I\bs{M}}+4\mc{K}^{IJ}\mc{L}_{I\bs{M}}p_{J}\end{eqnarray}
which reads for two coinciding tensors $\mc{J}$\begin{eqnarray}
\left[\mc{J},_{\de\,}\mc{J}\right]_{(1)}^{\Delta} & = & 4\cdot\mc{J}_{\bs{M}}\hoch{I}\partial_{I}\mc{J}_{\bs{M}\bs{M}}-4\partial_{\bs{M}}\mc{J}_{\bs{M}}\hoch{I}\mc{J}_{I\bs{M}}-4\mc{J}^{JI}\mc{J}_{I\bs{M}}p_{J}=\label{eq:derived-bracket-for-J}\\
 & \us{\stackrel{(\ref{eq:generalized-integrabilityIII})}{=}}{\mc{J}^{2}=-1} & \mc{N}_{\bs{M}\ldots\bs{M}}+4\underbrace{p_{M}\basis^{M}}_{\lqn{=\oo\textrm{ (\ref{eq:BRST-op})}}}\label{eq:relation-between-Nij-und-der-big}\end{eqnarray}
 where $\oo=\de x^{k}p_{k}=-\de(\de x^{k}\wedge\pe_{k})$. We will
verify this relation between the generalized Nijenhuis tensor and
the derived bracket in the following calculation, where we calculate
$\mc{N}$ using the big bracket (\ref{eq:multvec-bigbrack}) all the
time. This bracket is like a matrix multiplication if one of the objects
has only one index. We will use this fact frequently for the multiplication
of $\mc{J}$ with a vector\begin{eqnarray}
\mc{J}\mf{v} & \equiv & \mc{J}^{M}\tief{N}\mf{v}^{N}\basis_{M}=\frac{1}{2}\left\{ \mc{J},\mf{v}\right\} \\
\dann\left\{ \mc{J},\left\{ \mc{J},\mf{v}\right\} \right\}  & = & 4\mc{J}^{2}\mf{v}=-4\mf{v}=\left\{ \left\{ \mf{v},\mc{J}\right\} ,\mc{J}\right\} \\
\left\{ \left\{ \mf{v},\mc{J}\right\} ,\left\{ \mc{J},\mf{w}\right\} \right\}  & = & -4\mf{v}^{K}\mf{w}_{K}=-4\left\{ \mf{v},\mf{w}\right\} \end{eqnarray}
If both objects are of higher rank, however, antisymmetrization of
the remaining indices modifies the result. We thus have to be careful
with the following examples \rem{the first does not enter the calculations!}\begin{eqnarray}
\left\{ \mc{J},\mc{J}\right\}  & = & 4\mc{J}_{\bs{M}}\hoch{K}\mc{J}_{K\bs{M}}=-4\mc{G}_{\bs{M}\bs{M}}=0\quad(!\textrm{ because of antisymmetrization)}\label{eq:doesnt-enter}\\
\left\{ \mc{J},\left\{ \mc{J},\de\mf{v}\right\} \right\}  & = & \mc{J}_{\bs{M}}\hoch{K}\mc{J}_{[K|}\hoch{L}(\de\mf{v})_{L|\bs{M}]}\neq-4\de\mf{v}\end{eqnarray}
As mentioned earlier, the Dorfman bracket (\ref{eq:Dorfman-bracket})
used in our integrability condition is just the derived bracket of
the algebraic bracket. I.e. we have \begin{eqnarray}
\left[\mf{v}\bs{,}\mf{w}\right] & = & \left[\de\mf{v},\mf{w}\right]^{\Delta}=\label{eq:Dorf-ist-der}\\
 & = & \left[\de\mf{v},\mf{w}\right]_{(1)}^{\Delta}+\underbrace{\sum_{p\geq2}\left[\de\mf{v},\mf{w}\right]_{(p)}^{\Delta}}_{=0}=\\
 & = & \left\{ \de\mf{v},\mf{w}\right\} \label{eq:Dorf-ist-der-of-big}\end{eqnarray}
where the differential $\de$ has to be understood in the extended
sense of (\ref{eq:exterior-derivative-via-BRST},\ref{eq:dK}), namely
as Poisson-bracket with the BRST-like generator \begin{eqnarray}
\oo & = & \basis^{M}p_{M}=\ce^{m}p_{m}\stackrel{\textrm{locally }}{=}\de(x^{m}p_{m})=-\de(\ce^{m}\be_{m})\\
p_{M} & \equiv & (p_{m},0)\\
\de\mf{v} & \equiv & \left\{ \oo,\mf{v}\right\} =\partial_{\bs{M}}v_{\bs{M}}+\mf{v}^{K}p_{K}\end{eqnarray}
where $p_{m}$ is the conjugate variable to $x^{m}$. We can now rewrite
the integrability condition (\ref{eq:generalized-integrabilityII})
as \begin{eqnarray}
\left\{ \de\mf{v},\mf{w}\right\} -\frac{1}{4}\left\{ \de\left\{ \mc{J},\mf{v}\right\} ,\left\{ \mc{J},\mf{w}\right\} \right\} +\frac{1}{4}\left\{ \mc{J},\left\{ \de\left\{ \mc{J},\mf{v}\right\} ,\mf{w}\right\} \right\} +\frac{1}{4}\left\{ \mc{J},\left\{ \de\mf{v},\left\{ \mc{J},\mf{w}\right\} \right\} \right\}  & \stackrel{!}{=} & 0\label{eq:generalized-integrability-in-brackets}\end{eqnarray}
 Remember that the Poisson bracket is a graded one, and $\mf{v},\mf{w}$
and $\de$ are odd, while $\mc{J}$ is even. \rem{hier schlummert ein Einschub ueber das graded equal sign} 

Let us now start with applying Jacobi to the second term of (\ref{eq:generalized-integrability-in-brackets})
\begin{eqnarray}
-\frac{1}{4}\left\{ \de\left\{ \mc{J},\mf{v}\right\} ,\left\{ \mc{J},\mf{w}\right\} \right\}  & = & -\frac{1}{4}\left\{ \left\{ \de\left\{ \mc{J},\mf{v}\right\} ,\mc{J}\right\} ,\mf{w}\right\} -\frac{1}{4}\left\{ \mc{J},\left\{ \de\left\{ \mc{J},\mf{v}\right\} ,\mf{w}\right\} \right\} \end{eqnarray}
so that we get\begin{eqnarray}
0 & \stackrel{!}{=} & \left\{ \de\mf{v},\mf{w}\right\} -\frac{1}{4}\left\{ \left\{ \de\left\{ \mc{J},\mf{v}\right\} ,\mc{J}\right\} ,\mf{w}\right\} +\frac{1}{4}\left\{ \mc{J},\left\{ \de\mf{v},\left\{ \mc{J},\mf{w}\right\} \right\} \right\} =\\
 & = & \left\{ \de\mf{v},\mf{w}\right\} -\frac{1}{4}\left\{ \left\{ \left\{ \de\mc{J},\mf{v}\right\} ,\mc{J}\right\} ,\mf{w}\right\} -\frac{1}{4}\left\{ \left\{ \left\{ \mc{J},\de\mf{v}\right\} ,\mc{J}\right\} ,\mf{w}\right\} +\frac{1}{4}\left\{ \mc{J},\left\{ \de\mf{v},\left\{ \mc{J},\mf{w}\right\} \right\} \right\} =\\
 & = & \left\{ \de\mf{v},\mf{w}\right\} -\frac{1}{4}\left\{ \left\{ \left\{ \mf{v},\de\mc{J}\right\} ,\mc{J}\right\} ,\mf{w}\right\} +\frac{1}{4}\left\{ \left\{ \left\{ \de\mf{v},\mc{J}\right\} ,\mc{J}\right\} ,\mf{w}\right\} +\frac{1}{4}\left\{ \mc{J},\left\{ \de\mf{v},\left\{ \mc{J},\mf{w}\right\} \right\} \right\} \label{eq:kurzePause}\end{eqnarray}
It would be nice to separate $\mf{w}$ completely by moving it for
the last term into the last bracket like in the first three terms.
We thus consider only the last term for a moment and calculate it
in two different ways (first using Jacobi for second and third bracket
and after that using Jacobi for first and second bracket):\begin{eqnarray}
\frac{1}{4}\left\{ \mc{J},\left\{ \de\mf{v},\left\{ \mc{J},\mf{w}\right\} \right\} \right\}  & \stackrel{1.}{=} & \frac{1}{4}\left\{ \mc{J},\left\{ \left\{ \de\mf{v},\mc{J}\right\} ,\mf{w}\right\} \right\} +\frac{1}{4}\left\{ \mc{J},\left\{ \mc{J},\left\{ \de\mf{v},\mf{w}\right\} \right\} \right\} =\\
 & = & \frac{1}{4}\left\{ \mc{J},\left\{ \left\{ \de\mf{v},\mc{J}\right\} ,\mf{w}\right\} \right\} -\left\{ \de\mf{v},\mf{w}\right\} \\
 & \stackrel{2.}{=} & \frac{1}{4}\left\{ \left\{ \mc{J},\de\mf{v}\right\} ,\left\{ \mc{J},\mf{w}\right\} \right\} +\frac{1}{4}\left\{ \de\mf{v},\left\{ \mc{J},\left\{ \mc{J},\mf{w}\right\} \right\} \right\} =\\
 & = & \frac{1}{4}\left\{ \mc{J},\left\{ \left\{ \mc{J},\de\mf{v}\right\} ,\mf{w}\right\} \right\} +\frac{1}{4}\left\{ \left\{ \left\{ \mc{J},\de\mf{v}\right\} ,\mc{J}\right\} ,\mf{w}\right\} -\left\{ \de\mf{v},\mf{w}\right\} =\\
 & = & -\frac{1}{4}\left\{ \mc{J},\left\{ \left\{ \de\mf{v},\mc{J}\right\} ,\mf{w}\right\} \right\} +\left\{ \de\mf{v},\mf{w}\right\} -2\left\{ \de\mf{v},\mf{w}\right\} +\frac{1}{4}\left\{ \left\{ \left\{ \mc{J},\de\mf{v}\right\} ,\mc{J}\right\} ,\mf{w}\right\} \end{eqnarray}
Comparing both calculations yields\begin{eqnarray}
\frac{1}{4}\left\{ \mc{J},\left\{ \de\mf{v},\left\{ \mc{J},\mf{w}\right\} \right\} \right\}  & = & -\frac{1}{8}\left\{ \left\{ \mc{J},\left\{ \mc{J},\de\mf{v}\right\} \right\} ,\mf{w}\right\} -\left\{ \de\mf{v},\mf{w}\right\} \end{eqnarray}
We can plug this back in (\ref{eq:kurzePause}) and leave away the
outer bracket with $\mf{w}$:\begin{eqnarray}
0 & \stackrel{!}{=} & \de\mf{v}-\frac{1}{4}\left\{ \left\{ \mf{v},\de\mc{J}\right\} ,\mc{J}\right\} +\frac{1}{4}\left\{ \left\{ \de\mf{v},\mc{J}\right\} ,\mc{J}\right\} -\frac{1}{8}\left\{ \mc{J},\left\{ \mc{J},\de\mf{v}\right\} \right\} -\de\mf{v}=\\
 & = & -\frac{1}{4}\left\{ \left\{ \mf{v},\de\mc{J}\right\} ,\mc{J}\right\} +\frac{1}{8}\left\{ \left\{ \de\mf{v},\mc{J}\right\} ,\mc{J}\right\} =\\
 & = & -\frac{1}{8}\left\{ \left\{ \mf{v},\de\mc{J}\right\} ,\mc{J}\right\} +\frac{1}{8}\left\{ \de\left\{ \mf{v},\mc{J}\right\} ,\mc{J}\right\} =\\
 & = & -\frac{1}{8}\left\{ \left\{ \mf{v},\de\mc{J}\right\} ,\mc{J}\right\} +\frac{1}{8}\de\left\{ \left\{ \mf{v},\mc{J}\right\} ,\mc{J}\right\} +\frac{1}{8}\left\{ \left\{ \mf{v},\mc{J}\right\} ,\de\mc{J}\right\} =\\
 & = & -\frac{1}{8}\left\{ \mf{v},\left\{ \de\mc{J},\mc{J}\right\} \right\} -\frac{1}{2}\de\mf{v}=\\
 & = & \frac{1}{8}\Big(\big\{\left[\mc{J},_{\de}\mc{J}\right]_{(1)}^{\Delta},\mf{v}\big\}-4\de\mf{v}\Big)=\\
 & = & \frac{1}{8}\big\{\left[\mc{J},_{\de}\mc{J}\right]_{(1)}^{\Delta}-4\oo,\mf{v}\big\}\end{eqnarray}
where we used \begin{eqnarray}
\de\mf{v} & = & \left\{ \oo,\mf{v}\right\} \end{eqnarray}
The integrability condition is thus (explaining the normalization
of $\mc{N}$ of above) as promised in (\ref{eq:relation-between-Nij-und-der-big})\index{integrability!in terms of a derived bracket}\begin{equation}
\boxed{\mc{N}\equiv\left[\mc{J},_{\de}\mc{J}\right]_{(1)}^{\Delta}-4\oo\stackrel{!}{=}0}\label{eq:integrability-big-derived}\end{equation}
The derived bracket $\left[\mc{J},_{\de}\mc{J}\right]_{(1)}^{\Delta}$
indeed contains the term $4\oo=4\basis^{M}p_{M}$ \rem{(compare ...) Querverweis!!Achtung! Derived bracket ist nicht die volle, sondern nur die von big bracket abgeleitete!}
which therefore is exactly cancelled. 

Precisely the same calculation can be performed by calculating with
the complete algebraic bracket $\left[\,,\,\right]^{\Delta}$ instead
of the Poisson-bracket, its first order part. Similarly to above,
we have \begin{eqnarray}
\mc{J}\mf{v} & \equiv & \frac{1}{2}[\mc{J},\mf{v}]^{\Delta}\\
\dann[\mc{J},[\mc{J},\mf{v}]^{\Delta}]^{\Delta} & = & 4\mc{J}^{2}\mf{v}=-4\mf{v}\end{eqnarray}
 In combination with (\ref{eq:Dorf-ist-der}) this is enough to redo
the same calculation and get as integrability condition (using $\left[\mc{J}\bs{,}\mc{J}\right]\equiv-[\de\mc{J},\mc{J}]^{\Delta}$)\begin{equation}
\boxed{\mc{N}\equiv\left[\mc{J}\bs{,}\mc{J}\right]-4\oo\stackrel{!}{=}0}\label{eq:integrability-derived}\end{equation}
which also proves that the derived bracket bracket of the big bracket
(which is not necessarily geometrically well defined) coincides in
this case with the complete derived bracket\begin{eqnarray}
\left[\mc{J},_{\de}\mc{J}\right]_{(1)}^{\Delta} & = & \left[\mc{J}\bs{,}\mc{J}\right]\end{eqnarray}
As discussed in (\ref{eq:relation-Buttin-derived}) and (\ref{eq:relation-Buttin-derived-on-Tensor-level}),
throwing away the $\de$-closed part corresponds to taking Buttin's
bracket instead of the derived one. Remember that $\oo=\de x^{k}p_{k}=-\de(\de x^{k}\wedge\pe_{k})$,
s.th. $\de\oo=0$. We can thus equally write \begin{eqnarray}
\mc{N} & = & \left[\mc{J}\bs{,}\mc{J}\right]_{B}\label{eq:Nijenhuis-Buttin}\end{eqnarray}

\section{SO(d,d) pure spinors}

\label{sub:OddSpinors} There exists an alternative description of
a generalized complex structure and its integrability with the help
of pure spinors (see e.g. \cite[p.8]{Gualtieri:007} or in section
3 of \cite{Grana:2006kf}). {}``Spinor'' here refers to the special
orthonormal group $SO(d,d)$ ($d$ being the dimension of the manifold
$M$) of transformations on $T\oplus T^{*}$ which leave the canonical
metric $\erw{\ldots,\ldots}$ or $\mc{G}_{MN}$ (which has signature
$(d,d)$) invariant. It turns out that $T\oplus T^{*}$ itself, embedded
via \begin{eqnarray}
\ip_{X+\alpha}\rho & \equiv & \underbrace{\ip_{X}\rho}_{\equiv X\llcorner\rho}+\alpha\wedge\rho,\qquad X\in TM,\quad\alpha\in T^{*}M,\quad\rho\in\wedge^{\bullet}T^{*}M\end{eqnarray}
into the space of endomorphisms of $\wedge^{\bullet}T^{*}M$ (formal
sum of differential forms on $M$), forms a representation of the
Clifford algebra. The spinors\index{spinor!$SO(d,d)$ $\sim$}\label{T+T* spinors}
are thus differential forms $\rho\in\wedge^{\bullet}T^{*}M$ and the
gamma {}``matrices'' $\bs{\Gamma}^{M}$ are up to a normalization
factor just the interior products $\ip_{\basis^{M}}=\frac{i}{\hbar}\hat{\basis}_{M}$
with respect to the basis elements $\basis^{M}=(\de x^{m},\pe_{m})\equiv(\bs{c}^{m},\bs{b}_{m})$,
i.e. $\bs{\Gamma}^{M}=\{\sqrt{2}\ip_{\partial_{m}},\sqrt{2}\ip_{\de x^{n}}\equiv\de x^{n}\wedge\}.$
Indeed, the graded commutator (i.e. anticommutator) of the basis elements
reads $[\ip_{\pe_{m}},\de x^{n}\wedge]=\delta_{m}^{n}$ and therefore%
\footnote{\index{footnote!\thefoot. Poisson bracket of $T\oplus T^*$ basis forms a Clifford algebra}\frem{It
is very natural that $\ip_{\basis^{M}}=\frac{i}{\hbar}\hat{\basis}_{M}$
with $\basis^{M}=(\bs{c}^{m},\bs{b}_{m})$ forms a Clifford algebra,
since Clifford algebra and generator-annihilator algebra coincide
up to a basis change.}Note that one can think of $\ip_{\partial_{m}}$
as $\partl{\de x^{m}}$. Another observation is that the Poisson bracket
of the $T\oplus T^{*}$ basis elements also forms a Clifford-algebra\begin{eqnarray*}
\left\{ \basis^{M},\basis^{N}\right\}  & = & \mc{G}^{MN}\qquad\fussend\end{eqnarray*}
\frem{but \begin{eqnarray*}
\bs{\Gamma}^{M}\rho & \equiv & \sqrt{2}\left\{ \basis^{M},\rho\right\} \end{eqnarray*}
does not work...}%
}\begin{equation}
\left[\bs{\Gamma}^{M},\bs{\Gamma}^{N}\right]=2\mc{G}^{MN}\end{equation}
For general elements of the algebra (generalized vectors) $\mf{v}=\mf{v}_{M}\basis^{M},\,\mf{w}=\mf{w}_{N}\basis^{N}$,
the Clifford algebra becomes as usual $\left[\ip_{\mf{v}},\ip_{\mf{w}}\right]=2\erw{\mf{v},\mf{w}}$. 

One can further define a chirality matrix $\Gamma^{\#}$. It is characterized
by the properties that it squares to 1 and anticommutes with all other
$\bs{\Gamma}$-matrices.\rem{here really anticommute, because $\Gamma^{\#}$
is an even object} Usually it is proportional to the product of all
$\bs{\Gamma}$-matrices, but this is only true in a basis where $\mc{G}_{MN}$
is diagonal. In our basis $\basis_{M}$ it is off-diagonal. The definition
of the $\bs{\Gamma}$-matrices as $\tilde{\bs{\Gamma}}^{M}=\left\{ \ip_{(\de x^{m}-\pe_{m})},\ip_{(\de x^{m}+\pe_{m})}\right\} $
thus would be more appropriate in this context. The overall sign is
a matter of taste and we choose it such that the eigenvalues of rank
$r$ forms in (\ref{eq:GCG:chirality}) do not depend on the dimension.
The chirality\index{chirality!w.r.t. $SO(d,d)$} matrix is then given
by \begin{eqnarray}
\Gamma^{\#} & \equiv & (-)^{d}\prod_{k=0}^{d-1}\ip_{(\de x^{k}-\pe_{k})}\ip_{(\de x^{k}+\pe_{k})}=\\
 & = & (-)^{d}\left(\ip_{\de x^{0}}\ip_{\pe_{0}}-\ip_{\pe_{0}}\ip_{\de x^{0}}\right)\cdots\left(\ip_{\de x^{d-1}}\ip_{\pe_{d-1}}-\ip_{\pe_{d-1}}\ip_{\de x^{d-1}}\right)=\\
 & = & (-)^{d}\left(2\ip_{\de x^{0}}\ip_{\pe_{0}}-1\right)\cdots\left(2\ip_{\de x^{d-1}}\ip_{\pe_{d-1}}-1\right)=\\
 & = & (-)^{d}\prod_{k=0}^{d-1}\left(2\mf{n}_{\de x^{k}}-1\right)\end{eqnarray}
where $\mf{n}_{\de x^{k}}\equiv\not\!\!\!\sum\ip_{\de x^{k}}\ip_{\pe_{k}}$
counts the number of $\de x^{k}$ (with fixed $k$) of the differential
form $\rho^{(r)}$ on which $\Gamma^{\#}$ is acting. This number
can be (in each term of the expansion in basis elements) either zero
or one, because $(\de x^{k})^{2}=0$. The terms $(2\mf{n}_{\de x^{k}}-1)$
are therefore either $-1$ (if $\de x^{k}$ does not appear) or $1$
(if it appears). In a form $\rho^{(r)}$ of rank $r$, there are of
course in any term of the expansion $r$ basis elements $\de x^{k}$
which appear and $d-r$ which do not appear. We thus have\begin{eqnarray}
\Gamma^{\#}\rho^{(r)} & = & (-)^{d}(-1)^{d-r}\rho^{(r)}=(-1)^{r}\rho^{(r)}\label{eq:GCG:chirality}\end{eqnarray}
The chiral and antichiral spinors (those with eigenvalues $+1$ or
$-1$ ) therefore correspond to even and odd forms respectively.

A \emph{pure spinor\index{pure spinor!$SO(d,d)$ $\sim$}} is defined
to be a spinor which is annihilated by half of the gamma matrices.
(The same was true for the pure spinor in the Berkovits context, although
it is not obvious due to the formulation via the quadratic constraint
$\ce\gamma^{m}\ce=0$.) :\begin{equation}
\rho\text{ is pure :}\iff\zwek{L_{\rho}\equiv\{\mf{a}\in(T^{*}M\oplus TM)\otimes\mathbb{C}|\quad i_{\mf{a}}\rho=0\}}{\text{is of dimension }d=\dim M}\end{equation}
In other words, the Clifford action of $\left(T\oplus T^{\ast}\right)$
is maximally light-like. How is this related to an almost generalized
complex structure $\mc{J}$? The structure $\mc{J}$ induces a splitting
of $(T^{*}M\oplus TM)\otimes\mathbb{C}$ into a subbundle of eigenvalue
$i$ and another one of eigenvalue $-i$:\begin{eqnarray}
(T^{*}M\oplus TM)\otimes\mathbb{C} & = & L_{\mc{J}}\oplus L_{\mc{J}}^{\ast}\nonumber \\
L_{\mc{J}} & \equiv & \{\mf{a}\in(T^{*}M\oplus TM)\otimes\mathbb{C}|\,\mc{J}(\mf{a})=i\mf{a}\}\qquad\end{eqnarray}
Setting $L_{\mc{J}}\stackrel{!}{=}L_{\rho_{\mc{J}}}$ induces a map
from generalized complex structures to pure spinors and one can prove
that it is well-defined and one-to-one (up to a rescaling of the pure
spinor) \cite{Gualtieri:007}. The previosly discussed (twisted) integrability
condition can also be refomulated in the pure spinor language. Integrability
of $L_{\mc{J}}$ is closed under the action of the (twisted) Dorfman
bracket. $\mf{a},\mf{b}\in L_{\rho_{\mc{J}}}\dann[\mf{a}\bs{,}\mf{b}]=[[\ip_{\mf{a}},\de],\ip_{\mf{b}}]\in L_{\rho_{\mc{J}}}$.
In other words $[[\ip_{\mf{a}},\de+H\wedge],\ip_{\mf{b}}]\rho_{\mc{J}}=0\quad\forall\mf{a},\mf{b}$
with $\ip_{\mf{a}}\rho_{\mc{J}}=\ip_{\mf{b}}\rho_{\mc{J}}=0$. Writing
the graded commutator explicitely and using $\ip_{\mf{a}}\rho_{\mc{J}}=\ip_{\mf{b}}\rho_{\mc{J}}=0$,
this becomes \cite{Grana:2006kf} \begin{equation}
\mathcal{J}\quad\text{is twisted integrable}\quad:\iff\quad\ip_{\mf{b}}\ip_{\mf{a}}\de_{H}\rho_{\mc{J}}\equiv\ip_{\mf{b}}\ip_{\mf{a}}\left(\de+H\wedge\right)\rho_{\mc{J}}=0\quad\forall\mf{a},\mf{b}\in L_{\rho_{\mc{J}}}\end{equation}
One can think of $\rho_{\mc{J}}$ as a Clifford vacuum and of the
elements of $L_{\rho_{\mc{J}}}$ as annihilation operators. The creation
operators then lie in $L_{\mc{J}}^{\ast}$ and $\de_{H}\rho_{\mc{J}}$
must be at most at creator level two. However, as any creator changes
parity, and $\de\rho$ is of opposite parity than $\rho$ itself,
it can only be at odd creator-levels, i.e. level one. The above condition
is thus equivalent to\begin{equation}
\mathcal{J}\quad\text{is twisted integrable}\quad:\iff\de_{H}\rho_{\mc{J}}=\ip_{\mf{c}}\rho_{\mc{J}}\quad\mbox{for some }\mf{c}\in L_{\mc{J}}^{\ast}\end{equation}

\bibliographystyle{fullcream}
\addcontentsline{toc}{section}{\refname}\bibliography{phd,Proposal}
\printindex{}
}

\chapter{Derived Brackets}

\label{cha:Derived-Brackets}{\remch\inputTeil{0}\renewcommand{\be}{{\bs{b}}}\renewcommand{\ce}{{\bs{c}}}\ifthenelse{\theinput=1}{}{}

\title{Review of geometric brackets as derived brackets}

\author{Sebastian Guttenberg}

\date{August 08, 2007}

\maketitle
\begin{abstract}
Part of thesis-appendix
\end{abstract}
\tableofcontents{}\ifthenelse{\theinput=1}{}{\newpage}

\label{sec:bracket-review} Mathematics in this section is based on
the review article on derived brackets by Kosmann-Schwarzbach \cite{Kosmann-Schwarzbach:2003en}.
The presentation, however, will be somewhat different and in addition
to (or sometimes instead of) the abstract definitions coordinate expressions
will be given.

\section{Lie bracket of vector fields, Lie derivative and Schouten bracket}

\label{sub:Lie-and-Schouten} This first subsection is intended to
give a feeling, why the Schouten bracket is a very natural extension
of the Lie bracket of vector fields. It is a good example to become
more familiar with the subject, before we become more general in the
subsequent subsections, but it can be skipped without any harm (note
however the notation introduced before (\ref{eq:first-fat-notation})). 

Consider the ordinary \textbf{Lie-bracket\index{bracket!Lie $\sim$ of vector fields}\index{Lie-bracket!of vector fields}
of vector\index{vector field!Lie bracket} fields} which turns the
tangent space of a manifold into a Lie algebra or the tangent bundle
into a Lie algebroid\index{Lie algebroid} \rem{stimmt das?}  and
which takes in a local coordinate basis the familiar form\begin{eqnarray}
\left[v\bs{,}w\right]^{m} & = & v^{k}\partial_{k}w^{m}-w^{k}\partial_{k}v^{m}\label{eq:vector-Lie-bracket}\end{eqnarray}
We will convince ourselves in the following that numerous other common
differential brackets are just natural extensions of this bracket
and can be regarded as one and the same bracket. Such a generalized
bracket is e.g. useful to formulate integrability conditions and it
can serve via the Jacobi identity as a powerful tool in otherwise
lengthy calculations\rem{Hier waere das Beispiel mit Nijenhuis angebracht, wenn Leibniz funktionieren wuerde} .
In addition it shows up naturally in some sigma-models as is discussed
in section \ref{sec:sigma-model-induced}. 

Given the Lie-bracket of vector fields, it seems natural to extend
it to higher rank tensor fields by demanding a Leibniz rule on tensor
products of the form $\left[v\bs{,}w_{1}\otimes w_{2}\right]=\left[v\bs{,}w_{1}\right]\otimes w_{2}+w_{1}\otimes\left[v\bs{,}w_{2}\right]$.
Remembering that the Lie-bracket of two vector fields is just the
Lie derivative of one vector field with respect to the other\begin{eqnarray}
\left[v\bs{,}w\right] & = & \Lie_{v}w\end{eqnarray}
the \textbf{Lie\index{Lie derivative} derivative\index{derivative!Lie $\sim$}}
of a general tensor $T=T_{m_{1}\ldots m_{p}}^{n_{1}\ldots n_{q}}\de x^{m_{1}}\otimes\ldots\otimes\de x^{m_{p}}\otimes\pe_{n_{1}}\otimes\cdots\otimes\pe_{n_{q}}$with
respect to a vector field $v$ can be seen as a first extension of
the Lie bracket:\begin{eqnarray}
\left[v\bs{,}T\right] & \equiv & \Lie_{v}T\\
\left[v\bs{,}T\right]_{m_{1}\ldots m_{p}}^{n_{1}\ldots n_{q}} & = & v^{k}\partial_{k}T_{m_{1}\ldots m_{p}}^{n_{1}\ldots n_{q}}-\sum_{i}\partial_{k}v^{n_{i}}T_{m_{1}\ldots m_{p}}^{n_{1}\ldots n_{i-1}k\, n_{i+1}\ldots n_{q}}+\sum_{j}\partial_{m_{j}}v^{k}T_{m_{1}\ldots m_{j-1}k\, m_{j+1}\ldots m_{p}}^{n_{1}\ldots n_{q}}\quad\label{eq:Lie-derivative}\end{eqnarray}
The Lie derivative obeys (as a derivative should) the \textbf{Leibniz\index{Leibniz rule!for Lie derivative}
rule} \begin{eqnarray}
\left[v\bs{,}T_{1}\otimes T_{2}\right] & = & \left[v\bs{,}T_{1}\right]\otimes T_{2}+T_{1}\otimes\left[v\bs{,}T_{2}\right]\end{eqnarray}
In fact, giving as input only the Lie derivative of a scalar $\phi$,
namely the directional derivative $\left[v\bs{,}\phi\right]\equiv v^{k}\partial_{k}\phi$,
and the Lie bracket of vector fields (\ref{eq:vector-Lie-bracket}),
the Lie derivative of general tensors (\ref{eq:Lie-derivative}) is
determined by the Leibniz-rule. Insisting on antisymmetry of the bracket,
we have to define \begin{eqnarray}
\left[T\bs{,}v\right] & \equiv & -\left[v\bs{,}T\right]\label{eq:tensor-vector-bracket}\end{eqnarray}
Indeed, it can be checked that the above definitions lead to a valid
Jacobi-identity of the form \rem{Herleitung im 2006-Planer, S.23, beim 23.1., am 6.8.} \begin{eqnarray}
\left[v\bs{,}\left[w\bs{,}T\right]\right] & = & \left[\left[v\bs{,}w\right]\bs{,}T\right]+\left[w\bs{,}\left[v\bs{,}T\right]\right]\quad\textrm{for arbitrary tensors }T\label{eq:Jacobi-for-arbitraryT}\end{eqnarray}
which is perhaps better known in the form \begin{eqnarray}
\left[\Lie_{v},\Lie_{w}\right]T & = & \Lie_{\left[v\bs{,}w\right]}T\label{eq:Jacobi-for-arbitraryTII}\end{eqnarray}
 We have now vectors acting via the bracket on general tensors, but
tensors only acting on vectors via (\ref{eq:tensor-vector-bracket})
. It is thus natural to use Leibniz again to define the action of
tensors on tensors. To make a long story short, this is not possible
for general tensors. It is possible, however, for tensors with only
upper indices which are either antisymmetrized (\textbf{multivector\index{multivector}s})
or symmetrized (\textbf{symmetric\index{symmetric multivector} multivector\index{multivector!symmetric $\sim$}s}).
We will concentrate in this paper on tensors with antisymmetrized
indices (the reason being the natural given differential for forms
which also have antisymmetrized indices), but the symmetric case makes
perfect sense and at some points we will give short comments. (See
e.g. \cite{Dubois-Violette:1994gy}\rem{In  nur im Abstract. \cite{Michor:1987spd}
Trotzdem hilfreich...} for more information on the Schouten bracket
of symmetric tensor fields.)

Given two \textbf{multivector fields} (note that the prefactor $1/p!$
is intentionally missing (see page \pageref{Wedge-product}). \begin{eqnarray}
v^{(p)} & \equiv & v^{m_{1}\ldots m_{p}}\pe_{m_{1}}\wedge\ldots\wedge\pe_{m_{p}},\qquad w^{(q)}\equiv w^{m_{1}\ldots m_{q}}\pe_{m_{1}}\wedge\ldots\wedge\pe_{m_{q}}\end{eqnarray}
their Schouten(-Nijenhuis) bracket, or \textbf{Schouten\index{Schouten bracket|fett}
bracket}\index{bracket!Schouten $\sim$} for short, is given in a
local coordinate basis by\begin{eqnarray}
\left[v^{(p)}\bs{,}w^{(q)}\right]^{m_{1}\ldots m_{p+q-1}} & = & pv^{[m_{1}\ldots m_{p-1}|k}\partial_{k}w^{|m_{p}\ldots m_{p+q-1}]}-qv^{[m_{1}\ldots m_{p}\mid}\tief{,k}w^{k\,\mid m_{p+1}\ldots m_{p+q-1}]}\qquad\label{eq:Schouten-bracketI}\end{eqnarray}
Realizing that the Lie-derivative (\ref{eq:Lie-derivative}) of a
multivector field $w^{(q)}$ with respect to a vector $v^{(1)}$ is
\begin{eqnarray}
\left[v\bs{,}w^{(q)}\right]^{n_{1}\ldots n_{q}} & = & v^{k}\partial_{k}w^{n_{1}\ldots n_{q}}-q\partial_{k}v^{[n_{1}|}w^{k\,|n_{2}\ldots n_{q}]}\end{eqnarray}
one recognizes that (\ref{eq:Schouten-bracketI}) is a natural extension
of this, obeying a Leibniz rule, which we will write down below in
(\ref{eq:Leibniz-for-Schouten}). However, as the coordinate form
of generalized brackets will become very lengthy at some point, we
will first introduce some \textbf{notation} which is more schematic,
although still exact. Namely we imagine that every \textbf{boldface
index} $\bs{m}$ is an ordinary index $m$ contracted with the corresponding
basis vector $\pe_{m}$ at the position of the index:\index{notation!schematic index $\sim$}\index{index!schematic $\sim$ notation}\index{schematic index notation}\begin{equation}
v^{(p)}=v^{m_{1}\ldots m_{p}}\pe_{m_{1}}\wedge\ldots\wedge\pe_{m_{p}}\equiv v^{\mm}\end{equation}
This saves us the writing of the basis vectors as well as the enumeration
or manual antisymmetrization of the indices. As a boldface index might
be hard to distinguish from an ordinary one, we will use this notation
only for several indices, s.th. we get repeated indices $\mm$ which
are easily to recognize (and are not summed over, as they are at the
same vertical position). See in the appendix \ref{sec:Conventions}
on page \pageref{fat-index} for a more detailed explanation. The
Schouten bracket then reads\begin{eqnarray}
\left[v^{(p)}\bs{,}w^{(q)}\right] & = & pv^{\bs{m}\ldots\bs{m}k}\partial_{k}w^{\bs{m}\ldots\bs{m}}-qv^{\bs{m}\ldots\bs{m}}\tief{,k}w^{k\,\bs{m}\ldots\bs{m}}=\label{eq:first-fat-notation}\\
 & = & pv^{\bs{m}\ldots\bs{m}k}\partial_{k}w^{\bs{m}\ldots\bs{m}}-(-)^{p(q-1)}qw^{k\,\bs{m}\ldots\bs{m}}v^{\bs{m}\ldots\bs{m}}\tief{,k}=\\
 & = & pv^{\bs{m}\ldots\bs{m}k}\partial_{k}w^{\bs{m}\ldots\bs{m}}-(-)^{(p-1)(q-1)}qw^{\bs{m}\ldots\bs{m}k}\partial_{k}v^{\bs{m}\ldots\bs{m}}\label{eq:Schouten-bracket-condensed}\end{eqnarray}
In the last line it becomes obvious that the bracket is \textbf{skew\index{skew-symmetric}-symmetric\index{symmetric!skew-$\sim$}}
in the sense of a Lie algebra of degree%
\footnote{\index{footnote!\thefoot. Lie bracket of degree n}\label{Lie-bracket-of-degree}A
\textbf{Lie\index{Lie bracket!of degree n} bracket\index{bracket!Lie $\sim$ of degree n|itext{$[\ldots,_{(n)}\ldots]$}}\index{$([])$@$[\ldots,_{(n)}\ldots]$|itext{Lie bracket of degree n}}}
$\left[\:,_{(n)}\:\right]$ \textbf{of degree} $n$ in a graded algebra
increases the degree (which we denote by $\abs{\ldots}$) by $n$\[
\Abs{\left[A,_{(n)}B\right]}=\abs{A}+\abs{B}+n\]
It can be understood as an ordinary graded\index{graded!Lie algebra}\index{graded!Lie bracket}
Lie-bracket, when we redefine the grading $\norm{\ldots}\equiv\abs{\ldots}+n$,
such that the Lie bracket itself does not carry a grading any longer\begin{eqnarray*}
\Norm{\left[A,_{(n)}B\right]} & = & \Norm{A}+\Norm{B}\end{eqnarray*}
The symmetry properties are thus (\textbf{skew\index{skew symmetry of degree n}
symmetry of degree} $n$) \begin{eqnarray*}
\left[A,_{(n)}B\right] & = & -(-)^{(\abs{A}+n)(\abs{A}+n)}\left[B,_{(n)}A\right]\end{eqnarray*}
and it obeys the usual graded\index{graded Jacobi identity} Jacobi-identity
(with shifted degrees)\begin{eqnarray*}
\left[A,_{(n)}\left[B,_{(n)}C\right]\right] & = & \left[\left[A,_{(n)}B\right],_{(n)}C\right]+(-)^{(\abs{A}+n)(\abs{A}+n)}\left[B,_{(n)}\left[A,_{(n)}C\right]\right]\end{eqnarray*}
In addition there might be a Poisson-relation with respect to some
other product which respects the original grading. To be consistent
with both gradings, this relation has to read\begin{eqnarray*}
\left[A,_{(n)}B\cdot C\right] & = & \left[A,_{(n)}B\right]\cdot C+(-)^{(\abs{A}+n)\abs{B}}B\cdot\left[A,_{(n)}C\right]\end{eqnarray*}
This is consistent with $B\cdot C=(-)^{\abs{B}\abs{C}}C\cdot B$ on
the one hand and the skew symmetry of the bracket on the other hand.
One can imagine the grading of the bracket to sit at the position
of the comma.

For the bracket of multivectors we have as degree the vector degree.
Later, when we will have tensors of mixed type (vector and form),
we will use the form degree minus the vector degree as total degree.
Then the Schouten-bracket is of degree +1, which should not confuse
the reader.$\qquad\fussend$%
} \textbf{$-1$}:\begin{eqnarray}
\left[v^{(p)}\bs{,}w^{(q)}\right] & = & -(-)^{(p-1)(q-1)}\left[w^{(q)}\bs{,}v^{(p)}\right]\end{eqnarray}
It obeys the corresponding \textbf{Jacobi identity} \begin{eqnarray}
\left[v_{1}^{(p_{1})}\bs{,}\left[v_{2}^{(p_{2})}\bs{,}v_{3}^{(p_{3})}\right]\right] & = & \left[\left[v_{1}^{(p_{1})}\bs{,}v_{2}^{(p_{2})}\right]\bs{,}v_{3}^{(p_{3})}\right]+(-)^{(p_{1}-1)(p_{2}-1)}\left[v_{2}^{(p_{2})}\bs{,}\left[v_{1}^{(p_{1})}\bs{,}v_{3}^{(p_{3})}\right]\right]\end{eqnarray}
Our starting point was to extend the bracket in a way that it acts
via Leibniz on the wedge product. A Lie algebra which has a second
product on which the bracket acts via Leibniz is known as Poisson
algebra. However, here the bracket has degree $-1$ (it reduces the
multivector degree by one) while the wedge product has no degree (the
degree of the wedge product of multivectors is just the sum of the
degrees). According to footnote \ref{Lie-bracket-of-degree}, we have
to adjust the Leibniz rule. The resulting algebra for Lie brackets
of degree -1 is known as \textbf{Gerstenhaber\index{bracket!Gerstenhaber $\sim$}\index{Gerstenhaber algebra}
algebra\index{algebra!Gerstenhaber $\sim$}} or in this special case
\textbf{Schouten\index{Schouten algebra} algebra}\index{algebra!Schouten $\sim$}
(which is the standard example for a Gerstenhaber algebra). The \textbf{Leibniz
rule} is\begin{eqnarray}
\left[v_{1}^{(p_{1})}\bs{,}\, v_{2}^{(p_{2})}\wedge v_{3}^{(p_{3})}\right] & = & \left[v_{1}^{(p_{1})}\bs{,}\, v_{2}^{(p_{2})}\right]\wedge v_{3}^{(p_{3})}+(-)^{(p_{1}-1)p_{2}}v_{2}^{(p_{2})}\wedge\left[v_{1}^{(p_{1})}\bs{,}\, v_{3}^{(p_{3})}\right]\label{eq:Leibniz-for-Schouten}\end{eqnarray}
The standard example in field theory for a Poisson algebra is the
phase space equipped with the Poisson bracket or the commutator of
operators or matrices.%
\footnote{\index{footnote!\thefoot. Poisson algebra for symmetric multivectors}In
fact, working with totally symmetric\index{symmetric multivector}
multivector\index{multivector!symmetric $\sim$} fields would have
lead to a Poisson algebra instead of a Gerstenhaber algebra. \rem{ist das richtig?}$\qquad\fussend$%
} The Schouten algebra is naturally realized by the \textbf{antibracket}\index{bracket!anti-$\sim$}\index{antibracket}
of the BV antifield formalism (see subsection \ref{sub:antibracket}).

\section{Embedding of vectors into the space of differential operators}

\label{sub:Embedding-of-vectors} \index{embedding!of tensors into the space of differential operators}The
Leibniz rule is not the only concept to generalize the vector Lie
bracket to higher rank tensors. The major difficulty in the definition
of brackets between higher rank tensors is the Jacobi-identity, which
should hold for them. It is therefore extremely useful to have a mechanism
which automatically guarantees the Jacobi identity. A way to get such
a mechanism is to \textbf{embed} the tensors into some space of differential
operators, as for the operators we have the commutator as natural
Lie bracket which might in turn induce some bracket on the tensors
we started with. Vector fields e.g. naturally act on differential
forms via the \textbf{interior\index{interior product} product}\index{product!interior $\sim$}\begin{eqnarray}
\ip_{v}\omega^{(p)} & \equiv & p\cdot v^{k}\omega_{k\bs{m}\ldots\bs{m}}\label{eq:vector-interior-product}\end{eqnarray}
This can be seen as the embedding of vector fields in the space of
differential operators acting on forms, because the interior product
with respect to a vector is a graded derivative with the grading -1
of the vector (we take as total degree the form degree minus the multivector
degree, which for a vector is just -1)\begin{eqnarray}
\ip_{v}\left(\omega^{(p)}\wedge\eta^{(q)}\right) & = & \ip_{v}\omega^{(p)}\wedge\eta^{(q)}+(-)^{q}\omega^{(p)}\wedge\ip_{v}\eta^{(q)}\end{eqnarray}
Taking the idea of above we can take the commutator of two interior
products. We note, however, that it only induces a trivial (always
vanishing) bracket on the vectorfields\begin{eqnarray}
\left[\ip_{v},\ip_{w}\right] & = & 0=\ip_{0}\label{eq:trivial-alg-bracket}\end{eqnarray}
As the interior product (\ref{eq:vector-interior-product}) does not
include any partial derivative on the vector-coefficient, it was clear
from the beginning that this ansatz does not lead to the Lie bracket
of vector fields or any generalization of it. We have to bring the
exterior derivative into the game, in our notation \begin{equation}
\de\omega^{(p)}=\partial_{\bs{m}}\omega_{\bs{m}\ldots\bs{m}}\end{equation}
 There are two ways to do this

\begin{itemize}
\item \emph{Change the embedding:} Instead of embedding the vectors via
the interior product acting on forms, we can embed them via the Lie-derivative
acting on forms. When acting on forms, the Lie derivative can be written
as the (graded) commutator of interior product and exterior derivative\rem{spaeter
nicht immer injektiv!%
\footnote{This is perhaps a good point to demonstrate how to calculate with
our schematic notation\begin{eqnarray*}
\Lie_{v}\rho^{(r)} & = & \left[\ip_{v},\de\,\right]\rho^{(r)}=\\
 & = & \ip_{v}\de\rho^{(r)}+\de\ip_{v}\rho^{(r)}=\\
 & = & (p+1)\cdot v^{k}\partial_{[k}\omega_{\bs{m}\ldots\bs{m}]}+\partial_{\bs{m}}\left(p\cdot v^{k}\omega_{k\bs{m}\ldots\bs{m}}\right)=\\
 & = & v^{k}\partial_{k}\omega_{\bs{m}\ldots\bs{m}}-p\cdot v^{k}\partial_{\bs{m}}\omega_{k\bs{m}\ldots\bs{m}}+p\cdot\partial_{\bs{m}}v^{k}\omega_{k\bs{m}\ldots\bs{m}}+p\cdot v^{k}\partial_{\bs{m}}\omega_{k\bs{m}\ldots\bs{m}}=\\
 & = & v^{k}\partial_{k}\omega_{\bs{m}\ldots\bs{m}}+p\cdot\partial_{\bs{m}}v^{k}\omega_{k\bs{m}\ldots\bs{m}}\qquad\fussend\end{eqnarray*}
}} \begin{eqnarray}
\Lie_{v} & = & \left[\ip_{v},\de\,\right]\\
\Lie_{v}\omega^{(p)} & = & v^{k}\partial_{k}\omega_{\bs{m}\ldots\bs{m}}+p\cdot\partial_{\bs{m}}v^{k}\omega_{k\,\bs{m}\ldots\bs{m}}\label{eq:vector-Lie-derivative}\end{eqnarray}
Indeed, using the Lie derivative as embedding $v\mapsto\Lie_{v}$,
the commutator of Lie derivatives induces the Lie bracket of vector
fields (a special case of (\ref{eq:Jacobi-for-arbitraryTII}) \begin{eqnarray}
\left[\Lie_{v},\Lie_{w}\right] & = & \Lie_{[v\bs{,}w]}\label{eq:induced-bracket-via-Lie-derivative}\end{eqnarray}

\item \emph{Change the bracket:} In the space of differential operators
acting on forms, the commutator is the most natural Lie bracket. However,
the existence of a nilpotent odd operator acting on our algebra, namely
the commutator with the exterior derivative, enables the construction
of what is called a \textbf{derived bracket}%
\footnote{\index{footnote!\thefoot. derived bracket}\index{derived bracket|fett}\label{foot-derived-bracket}Given
a bracket $\left[\,,_{(n)}\,\right]$ of degree $n$ (not necessarily
a Lie bracket. It can be as well a \textbf{Loday\index{Loday bracket}
bracket}\index{bracket!Loday $\sim$} where the skew-symmetry property
as compared to footnote \ref{Lie-bracket-of-degree} is missing, but
the Jacobi identity still holds) and a differential $\De$ (derivation
of degree 1 and square 0), its \textbf{derived bracket}\index{bracket!derived $\sim$|fett}
\cite{Kosmann-Schwarzbach:1996a,Kosmann-Schwarzbach:1996b,Kosmann-Schwarzbach:2003en}
(which is of degree $n+1$) is defined by\index{$([])$@$[\ldots,_{(\De)}\ldots]$|itext{derived bracket by $\De$}}\[
\left[a,_{(\De)}b\right]=(-)^{n+a+1}\left[\De a,_{(n)}b\right]\]
We put the subscript $(\De\,)$ at the position of the comma, to indicate
that the grading of D is sitting there. The strange sign is just to
make the definition nicer for the most frequent case of an interior
derivation, where $\De a=\left[d,_{(n)}a\right]$ with $d$ some element
of the algebra with degree $\abs{d}=1-n$ and $\left[d,_{(n)}d\right]=0$,
s.th. we have\index{$([])$@$[\ldots,_{d}\ldots]$|itext{derived bracket by $\De=[d,\ldots]$}}\[
\left[a,_{d}b\right]=\left[\left[a,_{(n)}d\right],_{(n)}b\right]\]
The derived bracket is then again a Loday bracket (of degree $n+1$)
and obeys the corresponding Jacobi-identity (that is always the nontrivial
part). If $a,b$ are elements of a commuting subalgebra ($[a,_{(n)}b]=0$),
the derived bracket even is skew-symmetric and thus a Lie bracket
of degree $n+1$.

In the case at hand we start with a Lie bracket of degree 0 (the commutator)
and take as interior derivation the commutator with the exterior derivative
$\left[\de\,,\ldots\right]$. Note that the exterior derivative itself
is a derivative on forms, but not on the space of differential operators
on forms. Therefore we need the commutator.$\quad\fussend$%
}.\begin{eqnarray}
\left[\ip_{v},_{\de}\ip_{w}\right] & \equiv & \left[\left[\ip_{v},\de\,\right],\ip_{w}\right]\end{eqnarray}
This derived bracket (which is in this case a Lie bracket again, as
we are considering the abelian subalgebra of interior products of
vector fields) indeed induces the Lie bracket of vector fields when
we use the interior product as embedding\begin{eqnarray}
\left[\ip_{v},_{\de}\ip_{w}\right] & = & \ip_{\left[v\bs{,}w\right]}\label{eq:vector-derived-bracket}\end{eqnarray}

\end{itemize}
The above equations plus two additional ones are the well known \textbf{Cartan\index{Cartan formulae}
formulae} \begin{eqnarray}
\left[\ip_{v},\ip_{w}\right] & = & 0=\left[\de\,,\de\,\right]\\
\Lie_{v} & = & \left[\ip_{v},\de\,\right]\\
\left[\Lie_{v},\de\,\right] & = & 0\\
\left[\Lie_{v},\Lie_{w}\right] & = & \Lie_{\left[v,w\right]}\\
\big[\underbrace{\left[\ip_{v},\de\,\right]}_{\Lie_{v}},\ip_{w}\big]] & = & \ip_{\left[v,w\right]}\end{eqnarray}
(\ref{eq:induced-bracket-via-Lie-derivative}) can be rewritten, using
Jacobi's identity and $\left[\de\,,\de\,\right]=0$, as \begin{eqnarray}
\left[\left[\left[\ip_{v},\de\,\right],\ip_{w}\right],\de\,\right] & = & \left[\ip_{\left[v,w\right]},\de\,\right]\end{eqnarray}
Starting from (\ref{eq:vector-derived-bracket}), one thus arrives
at (\ref{eq:induced-bracket-via-Lie-derivative}) by simply taking
the commutator with $\de\,$. We will therefore concentrate in the
following on the second possibility, using the derived bracket, as
the first one can be deduced from it. Let us just mention that the
generalization in the spirit of the derived bracket (\ref{eq:vector-derived-bracket})
(or more precise its skew-symmetrization) is known as \textbf{Vinogradov\index{Vinogradov bracket}
bracket}\index{bracket!Vinogradov $\sim$} \cite{Vinogradov:1990,Vinogradov:1992}
(see footnote \ref{Vinogradov-bracket}), while the generalization
in the spirit of (\ref{eq:induced-bracket-via-Lie-derivative}) is
known as \textbf{Buttin\index{Buttin's bracket}'s bracket}\index{bracket!Buttin's $\sim$}
\cite{Buttin:1974}.\label{ite:Buttins-bracket}

\section{Derived bracket for multivector valued forms}

\label{sub:multivector-form-brackets} \rem{section similar zum Text!}\index{multivector valued form}\index{form!multivector valued $\sim$}
Let us now consider a much more general case, namely the space of
multivector valued forms, i.e. tensors which are antisymmetric in
the upper as well as in the lower indices. With the Schouten bracket
we have a bracket for multivectors, which are antisymmetric in all
(upper) indices. There exists as well a bracket for vector\index{vector valued form}
valued forms, namely tensors with one upper index and arbitrary many
antisymmetrized lower indices. This bracket (which we have not yet
discussed) is the (Fr\"ohlicher-) Nijenhuis bracket (see (\ref{eq:Nijenhuis-bracket-coord})),
which shows up in the integrability condition for almost complex structures.
Multivector valued forms have arbitrary many antisymmetrized upper
and arbitrary antisymmetrized lower indices and thus contain both
cases. The antisymmetrization appears quite naturally in field theory
(we give only a few remarks about completely symmetric indices, which
appear as well, but which will not be subject of this paper). It makes
also sense to define brackets on sums of tensors of different type
(e.g. the Dorfman bracket for generalized complex geometry). Those
brackets are then simply given by linearity.

So let us consider two multivector valued forms (we denote the number
of lower indices and the number of upper indices in this order via
superscripts)%
\footnote{\index{footnote!\thefoot. order of the indices of a multivector valued form}One
can certainly map a tensor $K_{m}\hoch{n}\de x^{m}\otimes\pe_{n}$
to one where the basis elements are antisymmetrized $K_{m}\hoch{n}\de x^{m}\wedge\pe_{n}\stackrel{\textrm{see page }\pageref{Wedge-product}}{\equiv}\frac{1}{2}K_{m}\hoch{n}\de x^{m}\otimes\pe_{n}-\frac{1}{2}K_{m}\hoch{n}\pe_{n}\otimes\de x^{m}$
and vice versa. In the field theory applications we will always get
a complete antisymmetrization. This mapping is the reason why we take
care for the horizontal positions of the indices. It should just indicate
the order of the basis elements which was chosen for the mapping.$\quad\fussend$%
}\index{$K^{(k,k')}$}\index{$K_{\bs{m}\ldots\bs{m}}\hoch{\bs{n}\ldots\bs{n}}$}\index{$K_{m_{1}\ldots m_{k}}\hoch{n_{1}\ldots n_{k'}}$}\begin{eqnarray}
K^{(k,k')} & \equiv & K_{\bs{m}\ldots\bs{m}}\hoch{\bs{n}\ldots\bs{n}}\equiv K_{m_{1}\ldots m_{k}}\hoch{n_{1}\ldots n_{k'}}\de x^{m_{1}}\cdots\de x^{m_{k}}\otimes\pe_{n_{1}}\cdots\pe_{n_{k'}}\\
L^{(l,l')} & \equiv & L_{\underbrace{{\scriptstyle \bs{m}\ldots\bs{m}}}_{l}}\hoch{\bs{n}\ldots\bs{n}}_{\underbrace{}_{l'}}\end{eqnarray}
Note the use of the schematic index notation, which we used for upper
indices already in subsection \ref{sub:Lie-and-Schouten} and which
is explained in the appendix \ref{sec:Conventions} on page \pageref{fat-index}.
Following the ideas of above, we want to embed those vector valued
forms in some space of differential operators. As we have upper as
well as lower indices now, it is less clear why we should choose the
space of operators acting on forms and not on some other tensors for
the embedding. However, the space of forms is the only one where we
have a natural exterior derivative without using any extra structure%
\footnote{\index{footnote!\thefoot. Lichnerowicz-Poisson differential $\de_P$}One
can define an exterior derivative -- the \textbf{Lichnerowicz\index{Lichnerowicz-Poisson differential $\de_P$}-Poisson\index{Poisson!Lichnerowicz $\sim$ differential $\de_P$}
differential\index{differential!Lichnerowicz-Poisson $\sim$}\index{$d_P$@$\de_P$|itext{Lichnerowicz-Poisson differential}}}
-- on the space of multivectors as well (via the Schouten bracket),
but for this we need an integrable Poisson structure: $\de_{P}N^{(q)}\equiv\left[P^{(2)}\bs{,}N^{(q)}\right]$,
with $\left[P^{(2)}\bs{,}P^{(2)}\right]=0\qquad\fussend$%
}. Therefore we will define again a natural embedding into the space
of differential operators acting on forms as a generalization of the
interior product. Namely, we will act with a multivector valued form
$K$ on a form $\rho$ by just contracting all upper indices with
form-indices and antisymmetrizing the remaining lower indices s.th.
we get again a form as result. The formal definition goes in two steps.
First one defines the interior product with multivectors. For a decomposable\index{decomposable!multivector}
multivector $v^{(p)}=v_{1}\wedge\ldots\wedge v_{p}$ set\begin{eqnarray}
\ip_{v_{1}\wedge\ldots\wedge v_{p}}\rho^{(r)} & \equiv & \ip_{v_{1}}\cdots\ip_{v_{p}}\rho^{(r)}\label{eq:multivec-inter-prod}\end{eqnarray}
This fixes the interior product for a generic multivector uniquely
(contracting all indices with form-indices). The next step is to define
for a multivector valued form $K^{(k,k')}=\eta^{(k)}\wedge v^{(k')}$
which is decomposable\index{decomposable!multivector valued form}
in a form and a multivector, that it acts on a form by first acting
with the multivector as above and then wedging the result with the
form \begin{eqnarray}
\ip_{\eta^{(k)}\wedge v^{(k')}}\rho & \equiv & \eta^{(k)}\wedge\ip_{v^{(k)}}\rho=(-)^{k'k}\ip_{v^{(k')}\wedge\eta^{(k)}}\rho\end{eqnarray}
 It is kind of a normal\index{normal ordering} ordering\index{ordering!normal $\sim$}
that $\ip_{v^{(k')}}$ acts first:\begin{equation}
\ip_{\eta}\ip_{v}=\ip_{\eta^{(k)}\wedge v^{(k')}}=(-)^{kk'}\ip_{v^{(k')}\wedge\eta^{(k)}}\neq\ip_{v}\ip_{\eta}\end{equation}
  For a generic multivector valued form, the above definitions fix
the following coordinate form of the \textbf{interior\index{interior product!with a multivector valued form|fett}
product}\index{product!interior $\sim$!with a multivector valued form}%
\footnote{\index{footnote!\thefoot. interior product (of maximal order)}\label{foot:interior-product}The
name 'interior product' is misleading in the sense that the operation
is (for decomposable tensors) a composition of interior and exterior
wedge product. It will, however, in the generalizations of Cartan's
formulae play the role of the interior product. We will therefore
stick to this name. We can also see it as a short name for 'interior
product of maximal order' in the sense that all upper indices are
contracted as opposed to an interior 'product of order $p$', where
we contract only $p$ upper indices. 'Order' is in the sense of the
order of a derivative. While $\ip_{v}$ is a derivative for any vector
$v$, the general interior product acts like a higher order derivative.$\qquad\fussend$%
} with a multivector valued form\index{$i_K$@$\ip_{K^{(k,k')}}\rho^{(r)}$}\begin{eqnarray}
\ip_{K^{(k,k')}}\rho^{(r)} & \equiv & (k')!\left(\zwek{r}{k'}\right)K_{\bs{m}\ldots\bs{m}}\hoch{l_{1}\ldots l_{k'}}\rho_{\underbrace{{\scriptstyle l_{k'}\ldots l_{1}\bs{m}\ldots\bs{m}}}_{r}}\label{eq:interior-productI}\end{eqnarray}
So we are just contracting all the upper indices of $K$ with an appropriate
number of indices of the form and are wedging the remaining lower
indices. The origin of the combinatorial prefactor is perhaps more
transparent in the phase space formulation (\ref{eq:bc-interior-product-II})
in subsection \ref{sub:bc-phase-space}. For multivectors $v^{(p)}$
and $w^{(q)}$ the operator product of $\ip_{v^{(p)}}$ and $\ip_{w^{(q)}}$
induces, due to (\ref{eq:multivec-inter-prod}) simply the wedge product
of the multivectors\begin{equation}
\ip_{v^{(p)}}\ip_{w^{(q)}}=\ip_{v^{(p)}\wedge w^{(q)}}\label{eq:multivector-product-of-int-pr}\end{equation}
But for general multivector-valued forms we have instead%
\footnote{\index{footnote!\thefoot. star product induced by composition of interior products}\label{foot:noncomm-prod}The
product of interior products in (\ref{eq:product-of-interior-products})
induces a noncommutative\index{noncommutative product} product\index{product!noncommutative $\sim$}
(star\index{star product} product\index{product!star $\sim$}) for
the multivector-valued forms, whose commutator is the algebraic bracket,
namely\begin{eqnarray*}
K*L & \equiv & \sum_{p\geq0}\ip_{K}^{(p)}L\\
\left[K,L\right]^{\Delta} & = & K*L-(-)^{(k-k')(l-l')}L*K\qquad\fussend\end{eqnarray*}
}\begin{eqnarray}
\ip_{K^{(k,k')}}\ip_{L^{(l,l')}} & = & \sum_{p=0}^{k'}\ip_{\ip_{K}^{(p)}L}=\ip_{K\wedge L}+\sum_{p=1}^{k'}\ip_{\ip_{K}^{(p)}L}\label{eq:product-of-interior-products}\end{eqnarray}
with\index{$i_K$@$\ip_{K^{(k,k')}}^{(p)}$} \begin{eqnarray}
\ip_{K^{(k,k')}}^{(p)}L^{(l,l')} & \equiv & (-)^{(k'-p)(l-p)}p!\left(\zwek{k'}{p}\right)\left(\zwek{l}{p}\right)K_{\bs{m}\ldots\bs{m}}\hoch{\bs{n}\ldots\bs{n}l_{1}\ldots l_{p}}L_{l_{p}\ldots l_{1}\bs{m}\ldots\bs{m}}\hoch{\bs{n}\ldots\bs{n}}\label{eq:interior-productIII}\end{eqnarray}
\rem{(see (\ref{eq:bc-product-of-interior-products}) and the comments
above). }For $p=k'$, $\ip_{K}^{(p)}$ reduces to the interior\index{interior product!of order p}
product (\ref{eq:interior-productI}). Both are in general not a derivative
any longer. $\ip^{(p)}$ is, however, a $p$-th order derivative,
as contracting $p$ indices means taking the $p$-th derivative with
respect to $p$ basis elements (see \ref{eq:bc-interior-pproduct-I}
in subsection \ref{sub:bc-phase-space}). Our embedding $\ip_{K^{(k,k')}}$
in (\ref{eq:interior-productI}) is therefore a $k'$-th order derivative.
For $p=0$ on the other hand, $\ip_{K}^{(p)}$ is just a wedge product
with $K$ \rem{For convenience, we also give a last further generalization,
which we will only rarely use\index{$i_K$@$\ip_{K^{(k,k')}}^{(p,q)}$}\begin{eqnarray}
\ip_{K^{(k,k')}}^{(p,q)}L^{(l,l')} & \equiv & (-)^{q(l+l'+k+k')+(k'-p)(l-p)}p!q!\left(\zwek{k}{q}\right)\left(\zwek{k'}{p}\right)\left(\zwek{l}{p}\right)\left(\zwek{l'}{q}\right)\times\nonumber \\
 &  & \times K_{l_{q}\ldots l_{1}\bs{m}\ldots\bs{m}}\hoch{\bs{n}\ldots\bs{n}l_{1}\ldots l_{p}}L_{l_{p}\ldots l_{1}\bs{m}\ldots\bs{m}}\hoch{\bs{n}\ldots\bs{n}l_{1}\ldots l_{q}}\label{eq:interior-productIV}\end{eqnarray}
$\ip_{K^{(k,k')}}^{(p,q)}$ in (\ref{eq:interior-productIV}) is a
$(p+q)$-th order derivative. Let us also define a generalized \textbf{wedge\index{wedge product!of multivector valued forms}
product}\index{product!wedge $\sim$ of multivector valued forms}\index{$K^{(k,k')}\wedge L^{(l,l')}$}
via \rem{wie genau definier ich jetzt die fetten Indizes?siehe auch 2. Gl oben}\begin{eqnarray*}
\hspace{-1cm}K^{(k,k')}\wedge L^{(l,l')} & \equiv & (-)^{k'l}K_{\bs{m}\ldots\bs{m}}\hoch{\bs{n}\ldots\bs{n}}L_{\bs{m}\ldots\bs{m}}\hoch{\bs{n}\ldots\bs{n}}\equiv\\
 & \equiv & (-)^{k'l}K_{m_{1}\ldots m_{k}}\hoch{n_{1}\ldots n_{k}}L_{m_{k+1}\ldots m_{k+l}}\hoch{n_{k'+1}\ldots n_{k'+l'}}\de x^{m_{1}}\cdots\de x^{m_{k+l}}\otimes\pe_{n_{1}}\cdots\pe_{n_{k'+l'}}=\qquad\\
 & = & \ip_{K}^{(0)}L\end{eqnarray*}
} 

While for vectors the commutator of two interior products (\ref{eq:trivial-alg-bracket})
did only induce a trivial bracket on vectors, which is the same for
multivectors due to (\ref{eq:multivector-product-of-int-pr}), this
is different for multivector-valued forms. \begin{eqnarray}
\hspace{-1cm}\left[\ip_{K^{(k,k')}},\ip_{L^{(l,,l')}}\right] & = & \ip_{\left[K,L\right]^{\Delta}}\label{eq:algebraic-bracket}\\
\left[K,L\right]^{\Delta} & \equiv & \sum_{p\geq1}\underbrace{\ip_{K}^{(p)}L-(-)^{(k-k')(l-l')}\ip_{L}^{(p)}K}_{\equiv[K,L]_{(p)}^{\Delta}}=\label{eq:algebraic-bracketI}\\
 & = & \sum_{p\geq1}\;(-)^{(k'-p)(l-p)}p!\left(\zwek{k'}{p}\right)\left(\zwek{l}{p}\right)K_{\bs{m}\ldots\bs{m}}\hoch{\bs{n}\ldots\bs{n}l_{1}\ldots l_{p}}L_{l_{p}\ldots l_{1}\bs{m}\ldots\bs{m}}\hoch{\bs{n}\ldots\bs{n}}+\nonumber \\
 &  & -(-)^{(k-k')(l-l')}(-)^{(l'-p)(k-p)}p!\left(\zwek{l'}{p}\right)\left(\zwek{k}{p}\right)L_{\bs{m}\ldots\bs{m}}\hoch{\bs{n}\ldots\bs{n}l_{1}\ldots l_{p}}K_{l_{p}\ldots l_{1}\bs{m}\ldots\bs{m}}\hoch{\bs{n}\ldots\bs{n}}\qquad\label{eq:algebraic-bracket-coord}\end{eqnarray}
where we introduced an \textbf{algebraic\index{algebraic bracket}
bracket}\index{bracket!algebraic $\sim$} $\left[K,L\right]^{\Delta}$\index{$([])$@$[\ldots,\ldots]^\Delta$}\index{$([])$@$[\ldots,\ldots]^\Delta$!$[K,L]^\Delta$}
in the second line, which is is due to Buttin\index{Buttin's algebraic bracket}
\cite{Buttin:1974}, and which is a generalization of the Nijenhuis-Richardson
bracket for vector-valued forms (\ref{eq:Richardson-Nijenhuis-bracket-coord}).
As it was induced via the embedding from the graded commutator, it
has the same properties, i.e. it is graded antisymmetric and obeys
the graded Jacobi identity. Actually, the term with lowest $p$, so
$[K,L]_{(p=1)}^{\Delta}$, is itself an algebraic bracket, which appears
in subsection \ref{sub:Algebraic-brackets} as canonical Poisson bracket\rem{Kosmann-Schwarzbach: Poisson-bracket on $T^*$?}.
It is known under the name \textbf{Buttin\index{Buttin's!algebraic bracket}'s
algebraic\index{algebraic bracket!Buttin's $\sim$} bracket\index{bracket!Buttin's algebraic $\sim$}}
(\cite{Buttin:1974}, denoted in \cite{Kosmann-Schwarzbach:2003en}
by $\left[\,,\,\right]_{B}^{0}$) or as \textbf{big\index{big bracket}
bracket\index{bracket!big $\sim$}}\index{$([])$@$[\ldots,\ldots]_{(1)}^\Delta$}\index{$([])$@$[\ldots,\ldots]_{(1)}^\Delta$!$[K,L]_{(1)}^\Delta$}\textbf{\begin{eqnarray}
[K,L]_{(1)}^{\Delta} & = & \ip_{K}^{(1)}L-(-)^{(k-k')(l-l')}\ip_{L}^{(1)}K=\label{eq:bigbracket}\\
 & = & (-)^{(k'-1)(l-1)}k'l\cdot K_{\bs{m}\ldots\bs{m}}\hoch{\bs{n}\ldots\bs{n}l_{1}}L_{l_{1}\bs{m}\ldots\bs{m}}\hoch{\bs{n}\ldots\bs{n}}+\nonumber \\
 &  & -(-)^{(k-k')(l-l')}(-)^{(l'-1)(k-1)}l'k\cdot L_{\bs{m}\ldots\bs{m}}\hoch{\bs{n}\ldots\bs{n}l_{1}}K_{l_{1}\bs{m}\ldots\bs{m}}\hoch{\bs{n}\ldots\bs{n}}\label{eq:bigbracket-coord}\end{eqnarray}
} But as for the vector fields in subsection \ref{sub:Embedding-of-vectors},
we are rather interested in the derived bracket of $\left[K,L\right]^{\Delta}$,
or at the bracket induced via an embedding based on the Lie derivative.
An obvious generalization of the Lie derivative is the commutator
$\left[\ip_{K},\de\,\right]$, which will be a derivative of the same
order as $\ip_{K}$ and therefore is not a derivative in the sense
that it obeys the Leibniz rule. Although it is common to use this
generalization, I am not aware of an appropriate name for it. Let
us just call it the \textbf{Lie\index{Lie derivative!with respect to a multivector valued form}
derivative} \textbf{with respect to $K$} (being a derivative of order
$k'$) \rem{Achtung, hier hab ich 'I' verwendet}\index{$L_K$@$\Lie_{K^{(k,k')}}$|itext{Lie derivative w.r.t. $K$}}
\begin{eqnarray}
\Lie_{K^{(k,k')}} & \equiv & \left[\ip_{K^{(k,k')}},\de\,\right]\label{eq:Lie-derivativeI}\\
\Lie_{K^{(k,k')}}\rho & = & (k')!\left(\zwek{r+1}{k'}\right)K_{\bs{m}\ldots\bs{m}}\hoch{l_{1}\ldots l_{k'}}\partial_{[l_{k'}}\rho_{l_{k'-1}\ldots l_{1}\bs{m}\ldots\bs{m}]}+\nonumber \\
 &  & -(-)^{k-k'}(k')!\left(\zwek{r}{k'}\right)\partial_{\bs{m}}\left(K_{\bs{m}\ldots\bs{m}}\hoch{l_{1}\ldots l_{k'}}\rho_{l_{k'}\ldots l_{1}\bs{m}\ldots\bs{m}}\right)=\\
 & = & (k')!\left(\zwek{r}{k'-1}\right)K_{\bs{m}\ldots\bs{m}}\hoch{l_{1}\ldots l_{k'}}\partial_{l_{k'}}\rho_{l_{k'-1}\ldots l_{1}\bs{m}\ldots\bs{m}}+\nonumber \\
 &  & -(-)^{k-k'}(k')!\left(\zwek{r}{k'}\right)\partial_{\bs{m}}K_{\bs{m}\ldots\bs{m}}\hoch{l_{1}\ldots l_{k'}}\rho_{l_{k'}\ldots l_{1}\bs{m}\ldots\bs{m}}\label{eq:Lie-derivativeII}\end{eqnarray}
The Lie derivative above is an ingredient to calculate the \textbf{derived\index{derived bracket}
bracket}\index{bracket!derived $\sim$} (remember footnote \ref{foot-derived-bracket}
on page \pageref{foot-derived-bracket}) which is given by%
\footnote{\label{Vinogradov-bracket} \index{footnote!\thefoot. Vinogradov bracket}The
\textbf{Vinogradov\index{Vinogradov bracket|fett} bracket}\index{bracket!Vinogradov $\sim$|fett}\index{$([])$@$[\ldots\bs{,}\ldots]_V$|itext{Vinogradov bracket}}
\cite{Vinogradov:1992,Vinogradov:1990} (see also \cite{Kosmann-Schwarzbach:2003en})
is a bracket in the space of all graded endomorphisms in the space
of differential forms $\Omega^{\bullet}(M)$\begin{eqnarray*}
\left[a\bs{,}b\right]_{V} & = & \frac{1}{2}\left(\left[\left[a,d\right],b\right]-(-)^{b}\left[a,\left[b,d\right]\right]\right)\quad\forall a,b\in\Omega^{\bullet}(M)\end{eqnarray*}
It is the skew symmetrization of a derived bracket. The embedding
of the multivector valued forms into the endomorphisms $\Omega^{\bullet}(M)$
via the interior product which we consider is neither closed under
the Vinogradov bracket nor under the derived bracket in the general
case.$\quad\fussend$%
}\index{$([])$@$[\ldots\bs{,}\ldots]$}\index{$([])$@$[\ldots,_{\de}\ldots]$}
\begin{eqnarray}
\left[\ip_{K},_{\de}\ip_{L}\right] & \equiv & \left[\left[\ip_{K},\de\,\right],\ip_{L}\right]\equiv\ip_{\left[K\bs{,}L\right]}\quad\textrm{if possible}\label{eq:derived-bracketI}\end{eqnarray}
One should distinguish the derived bracket on the level of operators
on the left from the derived bracket on the tensors $\left[K\bs{,}L\right]$
on the right. Only in special cases the result of the commutator on
the left can be written as the interior product of another tensorial
object which then can be considered as the derived bracket with respect
to the algebraic bracket $\left[\,,\,\right]^{\Delta}$. Therefore
one normally does not find an explicit general expression for this
derived bracket in literature. In \ref{sub:Extended-exterior-derivative},
however, the meaning of exterior derivative and interior product are
extended in order to be able to write down an explicit general coordinate
expression (\ref{eq:bc-derived-bracket-coord}) which reduces in the
mentioned special cases to the well known results (see e.g. \ref{sub:Nijenhuis-bracket}).
\rem{other derived brackets with other differentials!}

Closely related to the derived bracket in (\ref{eq:derived-bracketI})
of above is \textbf{Buttin\index{Buttin's!differential bracket}'s
differential\index{differential bracket|see{derived bracket}}\index{differential bracket!Buttin's $\sim$}
bracket}\index{bracket!Buttin's differential $\sim$}, given by\index{$([])$@$[\ldots\bs{,}\ldots]_B$|itext{Buttin's differential bracket}}
\begin{eqnarray}
\left[\Lie_{K},\Lie_{L}\right] & \equiv & \Lie_{\left[K\bs{,}L\right]_{B}}\quad\textrm{if possible}\label{eq:Buttins-bracket}\end{eqnarray}
Because of $\left[\de\,,\de\,\right]=0$ and $\Lie_{K}=\left[\ip_{K},\de\right]$
we have (using Jacobi)\begin{eqnarray}
\left[\Lie_{K},\Lie_{L}\right] & = & \left[\left[\ip_{K},_{\de}\ip_{L}\right],\,\de\,\right]=\left[\left[\ip_{K},_{\de}\ip_{L}\right],\,\de\,\right]\stackrel{!}{=}[\ip_{\left[K\bs{,}L\right]_{B}},\,\de\,]\label{eq:relation-Buttin-derived}\end{eqnarray}
Comparing with (\ref{eq:derived-bracketI}) s.th. in cases where $\left[K\bs{,}L\right]$
exists, the brackets have to coincide up to a closed term, or locally
a total derivative\begin{eqnarray}
\ip_{\left[K\bs{,}L\right]} & = & \ip_{\left[K\bs{,}L\right]_{B}}+\left[\de\,,\ldots\right]\end{eqnarray}
Using again the extended definition of exterior derivative and interior
product of \ref{sub:Extended-exterior-derivative}, this relation
can be rewritten as\begin{eqnarray}
\left[K\bs{,}L\right] & = & \left[K\bs{,}L\right]_{B}+\de\left(\ldots\right)\label{eq:relation-Buttin-derived-on-Tensor-level}\end{eqnarray}
The Nijenhuis bracket (\ref{eq:Nij-derived-coord}) is the major example
for this relation. \rem{Hier schlummert ein grosser ausgeschnittener Part mit erweiterter auesserer Ableitung und $L^(p)$, Koordinatenform der derived bracket usw...!} \rem{und hier schlummert gleich noch eine Subsection ueber Sums of multivector valued forms}

\section{Examples}

\label{sub:Examples}

\subsection{Schouten(-Nijenhuis) bracket}

\label{sub:Schouten-bracketII} Let us shortly review the Schouten
bracket under the new aspects. For multivector\index{multivector}s
$v^{(p)},w^{(q)}$ the algebraic bracket vanishes\begin{eqnarray}
\left[\ip_{v^{(p)}},\ip_{w^{(q)}}\right] & = & 0\end{eqnarray}
 The \textbf{Schouten\index{Schouten bracket} bracket}\index{bracket!Schouten-Nijenhuis $\sim$|see{Schouten $\sim$}}\index{bracket!Schouten $\sim$}\index{Nijenhuis!Schouten-Nijenhuis bracket|see{Schouten bracket}}
$\left[v^{(p)}\bs{,}w^{(q)}\right]$ coincides with the derived bracket
as well as with Buttin's differential bracket, i.e. we have \begin{eqnarray}
\left[\left[\ip_{v^{(p)}},\de\,\right],\ip_{w^{(q)}}\right] & = & \ip_{\left[v^{(p)}\bs{,}w^{(q)}\right]}\\
\left[\Lie_{v^{(p)}},\Lie_{w^{(q)}}\right] & = & \Lie_{\left[v^{(p)}\bs{,}w^{(q)}\right]}\end{eqnarray}
Its coordinate form -- given already before in (\ref{eq:Schouten-bracket-condensed})
-- is\begin{eqnarray}
\left[v^{(p)}\bs{,}w^{(q)}\right] & = & pv^{\bs{m}\ldots\bs{m}k}\partial_{k}w^{\bs{m}\ldots\bs{m}}-(-)^{(p-1)(q-1)}qw^{\bs{m}\ldots\bs{m}k}\partial_{k}v^{\bs{m}\ldots\bs{m}}\end{eqnarray}

The vector Lie bracket is a special case of the Schouten bracket as
well as of the Nijenhuis bracket.

\subsection{(Fr\"ohlicher-)Nijenhuis bracket and its relation to the Richardson-Nijenhuis
bracket}

\label{sub:Nijenhuis-bracket} Consider vector\index{vector valued form}
valued forms, i.e. tensors of the form\begin{eqnarray}
K^{(k,1)} & \equiv & K_{m_{1}\ldots m_{k}}\hoch{n}\de x^{m_{1}}\wedge\cdots\wedge\de x^{m_{k}}\wedge\pe_{n}\cong K_{m_{1}\ldots m_{k}}\hoch{n}\de x^{m_{1}}\wedge\cdots\wedge\de x^{m_{k}}\otimes\pe_{n}\end{eqnarray}
The algebraic bracket of two such tensors, defined via the graded
commutator (note that $\abs{\ip_{K}}=\abs{K}=k-1$)\begin{eqnarray}
\left[\ip_{K},\ip_{L}\right] & = & \ip_{\left[K,L\right]^{\Delta}}\end{eqnarray}
consists only of the first term in the expansion, because we have
only one upper index to contract.\begin{eqnarray}
\left[K^{(k,1)},L^{(l,1)}\right]^{\Delta} & = & \left[K^{(k,1)},L^{(l,1)}\right]_{(1)}^{\Delta}=\ip_{K}^{(1)}L-(-)^{(k-1)(l-1)}\ip_{L}^{(1)}K=\\
 & = & \hspace{-1cm}\stackrel{(\ref{eq:bigbracket-coord})}{=}l\, K_{\bs{m}\ldots\bs{m}}\hoch{j}L_{j\bs{m}\ldots\bs{m}}\hoch{\bs{n}}-(-)^{(k-1)(l-1)}k\, L_{\bs{m}\ldots\bs{m}}\hoch{j}K_{j\bs{m}\ldots\bs{m}}\hoch{\bs{n}}\label{eq:Richardson-Nijenhuis-bracket-coord}\end{eqnarray}
It is thus just the big bracket or Buttin's algebraic bracket but
in this case it is known as \textbf{Richardson\index{Richardson-Nijenhuis bracket}-Nijenhuis\index{Nijenhuis!Richardson-$\sim$ bracket}-bracket}\index{bracket!Richardso-Nijenhuis $\sim$}. 

The Lie derivative of a form with respect to $K$ (in the sense of
(\ref{eq:Lie-derivativeI})) is because of $k'=1$ really a (first
order) derivative and takes the form\begin{eqnarray}
\Lie_{K^{(k,1)}} & \equiv & \left[\ip_{K^{(k,1)}},\de\,\right]\\
\Lie_{K^{(k,1)}}\rho^{(r)} & = & K_{\bs{m}\ldots\bs{m}}\hoch{l}\partial_{l}\rho_{\bs{m}\ldots\bs{m}}+(-)^{k}r\partial_{\bs{m}}K_{\bs{m}\ldots\bs{m}}\hoch{l}\rho_{l\bs{m}\ldots\bs{m}}\end{eqnarray}
The \textbf{(Froehlicher\index{Froehlicher-Nijenhuis bracket|see{Nijenhuis bracket}}-)Nijenhuis}\index{Nijenhuis bracket}
\textbf{bracket}\index{bracket!(Froehlicher-)Nijenhuis $\sim$}\index{$([])$@$[\ldots\bs{,}\ldots]_N$|itext{Nijenhuis bracket}}
is defined as the unique tensor $\left[K\bs{,}L\right]_{N}$, s.th.\begin{equation}
[\Lie_{K}\bs{,}\Lie_{L}]=\Lie_{[K\bs{,}L]_{N}}\end{equation}
It is therefore an example of Buttin's differential bracket. Its explicit
coordinate form reads\begin{eqnarray}
\left[K\bs{,}L\right]_{N} & \equiv & K_{\bs{m}\ldots\bs{m}}\hoch{j}\partial_{j}L_{\bs{m}\ldots\bs{m}}\hoch{\bs{n}}+(-)^{k}l\partial_{\bs{m}}K_{\bs{m}\ldots\bs{m}}\hoch{j}L_{j\bs{m}\ldots\bs{m}}\hoch{\bs{n}}+\nonumber \\
 &  & -(-)^{kl}L_{\bs{m}\ldots\bs{m}}\hoch{j}\partial_{j}K_{\bs{m}\ldots\bs{m}}\hoch{\bs{n}}-(-)^{kl}(-)^{l}k\partial_{\bs{m}}L_{\bs{m}\ldots\bs{m}}\hoch{j}K_{j\bs{m}\ldots\bs{m}}\hoch{\bs{n}}\label{eq:Nijenhuis-bracket-coord}\\
 & = & "\Lie_{K}L-(-)^{kl}\Lie_{L}K"\end{eqnarray}

A different point of view on the Nijenhuis bracket is via the \textbf{derived
bracket} on the level of the differential operators acting on forms:
\begin{equation}
\left[\ip_{K},_{\de\,}\ip_{L}\right]\equiv\left[\left[\ip_{K},\de\right],\ip_{L}\right]\end{equation}
It induces the Nijenhuis-bracket only up to a total derivative (the
Lie-derivative-term) \begin{eqnarray}
\left[\ip_{K},_{\de\,}\ip_{L}\right] & \equiv & \ip_{\left[K\bs{,}L\right]_{N}}-(-)^{k(l-1)}\Lie_{\ip_{L}K}\label{eq:zwei-drueber}\end{eqnarray}
Using the extended definition of the exterior derivative in the sense
of (\ref{eq:d-auf-partial}) and of the interior product (\ref{eq:i-TII}),
one can write the Lie derivative as an interior product (see \ref{eq:dK-und-Lie})
$\Lie_{\ip_{L}K}=-(-)^{l+k}\ip_{\de\,(\ip_{L}K)}$ and $\left[\left[\ip_{K},\de\right],\ip_{L}\right]=(-)^{k}\left[\ip_{\de K},\ip_{L}\right]=(-)^{k}\ip_{\left[\de K,L\right]^{\Delta}}$,
so that we can rewrite (\ref{eq:zwei-drueber}) as\begin{eqnarray}
\left[K\bs{,}L\right] & \equiv & \left[K\bs{,}L\right]_{N}+(-)^{(k-1)l}\de\,(\ip_{L}K)\\
\textrm{with }\left[K\bs{,}L\right] & \equiv & (-)^{k}\left[\de K,L\right]^{\Delta}\end{eqnarray}
In that sense, $\left[K\bs{,}L\right]$ is the derived bracket of
the Richardson Nijenhuis bracket while the Nijenhuis bracket differs
by a total derivative. The explicit coordinate form can be read off
from (\ref{eq:bc-derived-bracketIII},\ref{eq:bc-derived-bracket-coord})
(with only the $p=1$ term surviving) \begin{eqnarray}
\left[K\bs{,}L\right] & = & (-)^{k}\ip_{\de K}^{(1)}L+(-)^{kl}(-)^{l}\ip_{\de L}^{(1)}K+(-)^{(k-1)l}\de(\ip_{L}^{(p)}K)=\\
 & = & K_{\bs{m}\ldots\bs{m}}\hoch{j}\partial_{j}L_{\bs{m}\ldots\bs{m}}\hoch{\bs{n}}+(-)^{k}l\partial_{\bs{m}}K_{\bs{m}\ldots\bs{m}}\hoch{j}L_{j\bs{m}\ldots\bs{m}}\hoch{\bs{n}}+\nonumber \\
 &  & -(-)^{kl}L_{\bs{m}\ldots\bs{m}}\hoch{j}\partial_{j}K_{\bs{m}\ldots\bs{m}}\hoch{\bs{n}}-(-)^{kl}(-)^{l}k\partial_{\bs{m}}L_{\bs{m}\ldots\bs{m}}\hoch{j}K_{j\bs{m}\ldots\bs{m}}\hoch{\bs{n}}+\nonumber \\
 &  & +(-)^{(k-1)l}\de\big(\underbrace{kL_{\bs{m}\ldots\bs{m}}\hoch{j}K_{j\bs{m}\ldots\bs{m}}\hoch{\bs{n}}}_{\ip_{L}K}\big)\label{eq:Nij-derived-coord}\end{eqnarray}
\rem{von der anderen Koordinatenform bekommt man direkt ganz genau dasselbe!}
where the last part is non-tensorial due to the appearance of the
basis element $p_{i}$ (see subsection \ref{sub:Extended-exterior-derivative}):\begin{eqnarray}
\de\left(\ip_{L}K\right)=\de\big(kL_{\bs{m}\ldots\bs{m}}\hoch{j}K_{j\bs{m}\ldots\bs{m}}\hoch{\bs{n}}\big) & = & k\partial_{\bs{m}}\left(L_{\bs{m}\ldots\bs{m}}\hoch{j}K_{j\bs{m}\ldots\bs{m}}\hoch{\bs{n}}\right)-(-)^{l+k}L_{\bs{m}\ldots\bs{m}}\hoch{j}K_{j\bs{m}\ldots\bs{m}}\hoch{i}p_{i}\end{eqnarray}
The remaining part coincides with the coordinate form of the \textbf{Nijenhuis
bracket} as given in (\ref{eq:Nijenhuis-bracket-coord}).

One can nicely summarize the algebra of graded derivations on forms
as \begin{eqnarray}
\lqn{\left[\Lie_{K_{1}^{(k_{1})}}+\ip_{L_{1}^{(l_{1})}}\,,\,\Lie_{K_{2}^{(k_{2})}}+\ip_{L_{2}^{(l_{2})}}\right]=}\nonumber \\
 & = & \Lie_{[K_{1}\bs{,}K_{2}]_{N}+\ip_{L_{1}}K_{2}-(-)^{(l_{2}-1)k_{1}}\,\ip_{L_{2}}K_{1}}+\ip_{\left[K_{1}\bs{,}L_{2}\right]_{N}-(-)^{(l_{1}-1)k_{2}}\left[K_{2}\bs{,}L_{1}\right]_{N}+\left[L_{1},L_{2}\right]^{\Delta}}\label{eq:Max-algebra}\end{eqnarray}
\rem{hier schlummern generalized complex examples. hauptsaechlich \begin{eqnarray*}
\lqn{(-)^{\textsc{k}-1}\left[\de\mc{K},\mc{L}\right]_{(p)}^{\Delta}=}\\
 & = & p!\left(\zwek{\textsc{k}}{p}\right)\left(\zwek{\textsc{l}}{p-1}\right)\cdot\mc{K}_{\bs{M}\ldots\bs{M}}\hoch{i_{1}\ldots i_{p-1}i_{p}}\partial_{i_{p}}\mc{L}_{i_{p-1}\ldots i_{1}\bs{M}\ldots\bs{M}}+\\
 &  & -(-)^{(\textsc{k}+1)(\textsc{l}+1)}p!\left(\zwek{\textsc{l}}{p}\right)\left(\zwek{\textsc{k}}{p-1}\right)\mc{L}_{\bs{M}\ldots\bs{M}}\hoch{i_{1}\ldots i_{p}}\partial_{i_{p}}\mc{K}_{i_{p-1}\ldots i_{1}\bs{M}\ldots\bs{M}}+\\
 &  & +(-)^{\textsc{k}-1}p!\left(\zwek{\textsc{k}}{p}\right)\left(\zwek{\textsc{l}}{p}\right)\left(\partial_{\bs{M}}\mc{K}_{\bs{M}\ldots\bs{M}}\hoch{i_{1}\ldots i_{p}}\mc{L}_{i_{p}\ldots i_{1}\bs{M}\ldots\bs{M}}-(-)^{p}\partial_{\bs{M}}\mc{K}_{\bs{M}\ldots\bs{M}i_{p}\ldots i_{1}}\mc{L}^{i_{1}\ldots i_{p}}\tief{\bs{M}\ldots\bs{M}}\right)+\\
 &  & +(p+1)!\left(\zwek{\textsc{k}}{p+1}\right)\left(\zwek{\textsc{l}}{p}\right)\left(\mc{K}_{\bs{M}\ldots\bs{M}}\hoch{i_{1}\ldots i_{p}L}\mc{L}_{i_{p}\ldots i_{1}\bs{M}\ldots\bs{M}}-(-)^{p}\mc{K}_{\bs{M}\ldots\bs{M}i_{p}\ldots i_{1}}\hoch{L}\mc{L}^{i_{1}\ldots i_{p}}\tief{\bs{M}\ldots\bs{M}}\right)p_{L}\end{eqnarray*}
For $p=1$, we get the derived bracket of the big bracket which is
most pleasant, because it can be written in terms of capital indices:\begin{eqnarray*}
(-)^{\textsc{k}-1}\left[\de\mc{K},\mc{L}\right]_{(1)}^{\Delta} & = & \textsc{k}\cdot\mc{K}_{\bs{M}\ldots\bs{M}}\hoch{i_{1}}\partial_{i_{1}}\mc{L}_{\bs{M}\ldots\bs{M}}-(-)^{(\textsc{k}+1)(\textsc{l}+1)}\textsc{l}\cdot\mc{L}_{\bs{M}\ldots\bs{M}}\hoch{i_{1}}\partial_{i_{1}}\mc{K}_{\bs{M}\ldots\bs{M}}+\\
 &  & +(-)^{\textsc{k}-1}\textsc{kl}\partial_{\bs{M}}\mc{K}_{\bs{M}\ldots\bs{M}}\hoch{I}\mc{L}_{I\bs{M}\ldots\bs{M}}+\textsc{k}\left(\textsc{k}-1\right)\textsc{l}\mc{K}_{\bs{M}\ldots\bs{M}}\hoch{IJ}\mc{L}_{I\bs{M}\ldots\bs{M}}p_{J}\end{eqnarray*}
}

\printindex{}

\bibliographystyle{fullsort}
\bibliography{Proposal,phd}

}

\chapter{Gamma-Matrices in 10 Dimensions}

\label{cha:Gamma-Matrices}{\inputTeil{0}\ifthenelse{\theinput=1}{}{}

\title{Gamma Matrices}

\author{Sebastian Guttenberg}

\date{Juli 29, '07}

\maketitle
\begin{abstract}
Part of the thesis. Formerly part of Berkovits\_in\_background.lyx
\end{abstract}
\tableofcontents{}

\label{app:gamma}

\section{Clifford algebra, Fierz identity and more for the Dirac matrices}

In the following we will collect some general relations for Dirac-$\Gamma$-matrices
in $d$ dimensions. In contrast to the rest of this document, we are
not using graded conventions in most of this appendix. In other words,
the spinorial indices are not understood to carry a grading and we
are thus using neither graded summation conventions nor the graded
equal sign. The reason is that a lot of people (me included) are used
to calculate with $\Gamma$-matrices in ordinary conventions, and
it therefore seemed to be simpler for me to translate only the results
into the graded conventions\rem{, which will be done in the last
section of this appendix}. This does not mean, however, that calculating
in the graded conventions would be more complicated. Let us give two
examples, how to translate the results. Remember first that in northwest-southeast
(NW) $\delta_{\alpha}^{\beta}=\delta_{\bs{\alpha}}\hoch{\bs{\beta}}=-\delta^{\bs{\beta}}\tief{\bs{\alpha}}$.
The equation $\delta_{\alpha}^{\alpha}=16$ therefore becomes $16=\delta_{\alpha}^{\alpha}=\sum_{\alpha}\delta_{\alpha}^{\alpha}=\sum_{\bs{\alpha}}\delta_{\bs{\alpha}}\hoch{\bs{\alpha}}=-\sum_{\bs{\alpha}}(-)^{\bs{\alpha}}\delta_{\bs{\alpha}}\hoch{\bs{\alpha}}=-\delta_{\bs{\alpha}}\hoch{\bs{\alpha}}$.
When there are naked indices, we also have to take into account the
graded equal sign, which compares the order of the indices in each
term: $\gamma_{\alpha\beta}^{c}=\gamma_{\beta\alpha}^{c}$ becomes
$\gamma_{\bs{\alpha}\bs{\beta}}^{c}=(-)^{\bs{\alpha\beta}}\gamma_{\bs{\beta\alpha}}^{c}=-\gamma_{\bs{\beta\alpha}}^{c}$.

Remember the form of the Clifford\index{Clifford algebra} algebra\index{algebra!Clifford $\sim$}\index{$\Gamma^a$|itext{gamma matrix}}\begin{eqnarray}
\{\Gamma^{a},\Gamma^{b}\} & = & 2\eta^{ab}\one\qquad\iff\Gamma^{(a}\Gamma^{b)}=\eta^{ab}\one\end{eqnarray}
\rem{Note that with graded matrix multiplication, we would get no
different sign. To see this, we have to write the spinorial indices.
The equation of above in the non-graded / graded version then looks
as follows\begin{equation}
\Gamma^{(a\,\q{\alpha}}\tief{\q{\gamma}}\Gamma^{b)\,\q{\gamma}}\tief{\q{\beta}}=\eta^{ab}\delta_{\q{\beta}}^{\q{\alpha}}\qquad\iff\qquad\Gamma^{(a|\,\q{\bs{\alpha}}}\tief{\q{\bs{\gamma}}}\Gamma^{|b)\,\q{\bs{\gamma}}}\tief{\q{\bs{\beta}}}=\eta^{ab}\delta^{\q{\bs{\alpha}}}\tief{\q{\bs{\beta}}}\end{equation}
On the righthand side we have the graded Kronecker delta and the graded
summation convention. For northeast-conventions both reduce to ordinary
Kronecker delta and ordinary summation convention. For northwest-convention
both get an additional minus sign which cancels. This reasoning also
works for mixed conventions, where we have northwest or northeast
for the chiral indices}Define as ususal $\Gamma^{a_{1}\ldots a_{p}}\equiv\Gamma^{[a_{1}}\cdots\Gamma^{a_{p}]}$.
The set $\{\Gamma^{I}\}\equiv\{\one,\Gamma^{a},\Gamma^{a_{1}a_{2}},\ldots,\Gamma^{a_{1}\ldots a_{10}}\}$
then builds a basis of $Gl(2^{[d/2]})$ where $2^{[d/2]}$ is the
dimension of the representation space.

\paragraph{Product\index{product!of antisymmetrized $\Gamma$-matrix-products}
of antisymmetrized\index{Antisymmetrized!product of $\Gamma$-matrices}
products of $\Gamma$-matrices}

One can in particular expand any product of antisymmetrized gamma
matrices in the basis $\{\Gamma^{I}\}$:\index{$\Gamma^{a_1\ldots a_p}$|itext{antisymmetrized product of gamma matrices}}
\begin{eqnarray}
\Gamma^{a_{1}\ldots a_{p}}\Gamma^{b_{1}\ldots b_{q}} & = & \sum_{k=0}^{min\{p,q\}}k!\left(\zwek{p}{k}\right)\left(\zwek{q}{k}\right)\eta^{\tief{[}a_{p}\tief{|}\hoch{[}b_{1}\hoch{|}}\eta^{\tief{|}a_{p-1}\tief{|}\hoch{|}b_{2}\hoch{|}}\cdots\eta^{\tief{|}a_{p+1-k}\tief{|}\hoch{|}b_{k}\hoch{|}}\Gamma^{\tief{|}a_{1}\ldots a_{p-k}\tief{]}\hoch{|}b_{k+1}\ldots b_{q}\hoch{]}}\label{eq:product-expansion}\end{eqnarray}
The antisymmetrization brackets on the righthand side shall indicate
that all the $a_{i}$'s and all the $b_{i}$'s are independently antisymmetrized.
The expressions become quite lengthy, if one spells out the antisymmetrization
explicitely. Let us write down the first terms only, using the notation
where a check $\check{}$ above an index means that this index is
omitted:%
\footnote{\index{footnote!\thefoot. product of antisymmetrized products of gamma-matrices}For
the proof of (\ref{eq:product-expansion}) one can simply study independently
the cases of how many indices $a_{i}$ and $b_{i}$ coincide. For
a nonvanishing lefthand side all the $a$'s are different and all
the $b$'s are different. If even none of the $a$'s coincides with
one of the $b$'s, we have simply $\Gamma^{a_{1}\ldots a_{k}}\Gamma^{b_{1}\ldots b_{l}}=\Gamma^{a_{1}\ldots a_{k}b_{1}\ldots b_{l}}$.
If $a_{1}=b_{1}$ and all others are different, we have $\Gamma^{a_{1}\ldots a_{k}}\Gamma^{b_{1}\ldots b_{l}}=(-)^{k-1}\eta^{a_{1}b_{1}}\Gamma^{a_{2}\ldots a_{k}b_{2}\ldots b_{l}}$.
If two indices coincide, e.g. $a_{1}=b_{1},a_{2}=b_{2}$, then we
have $\Gamma^{a_{1}\ldots a_{k}}\Gamma^{b_{1}\ldots b_{l}}=(-)^{k-1+k-2}\eta^{a_{1}b_{1}}\eta^{a_{2}b_{2}}\Gamma^{a_{3}\ldots a_{k}b_{3}\ldots b_{l}}$.
And so on...\frem{Perhaps (\ref{eq:product-expansion}) can also
be written as an exponential. To this end, introduce anticommuting
variables $\bs{x}_{a},\bs{y}_{a}$ and define $\Gamma(\bs{x})\equiv\Gamma^{a}\bs{x}_{a}$.
Then we have $\Gamma(\bs{x})^{p}=\Gamma^{a_{1}\ldots a_{p}}\bs{x}_{a_{1}}\cdots\bs{x}_{a_{p}}$...}
$\qquad\fussend$%
}\begin{eqnarray}
\Gamma^{a_{1}\ldots a_{k}}\Gamma^{b_{1}\ldots b_{l}} & = & \Gamma^{a_{1}\ldots a_{k}b_{1}\ldots b_{l}}+\sum_{i=1}^{k}\sum_{j=1}^{l}(-)^{k-i+j-1}\eta^{a_{i}b_{j}}\Gamma^{a_{1}\ldots\check{a}_{i}\ldots a_{k}b_{1}\ldots\check{b}_{j}\ldots b_{l}}+\nonumber \\
 &  & +\sum_{i_{1}=1}^{k}\sum_{j_{1}=1}^{l}\sum_{i_{2}=1}^{i_{1}-1}\Big(\sum_{j_{2}=1}^{j_{1}-1}\underbrace{(-)^{k-i_{1}+j_{1}-1+k-1-i_{2}+j_{2}-1}}_{-(-)^{2k+i_{1}+i_{2}+j_{1}+j_{2}}}\eta^{a_{i_{1}}b_{j_{1}}}\eta^{a_{i_{2}}b_{j_{2}}}\Gamma^{a_{1}\ldots\check{a}_{i_{2}}\ldots\check{a}_{i_{1}}\ldots a_{k}b_{1}\ldots\check{b}_{j_{2}}\ldots\check{b}_{j_{1}}\ldots b_{l}}+\nonumber \\
 &  & +\sum_{j_{2}=j_{1}+1}^{l}\underbrace{(-)^{k-i_{1}+j_{1}-1+k-1-i_{2}+j_{2}-2}}_{(-)^{2k+i_{1}+i_{2}+j_{1}+j_{2}}}\eta^{a_{i_{1}}b_{j_{1}}}\eta^{a_{i_{2}}b_{j_{2}}}\Gamma^{a_{1}\ldots\check{a}_{i_{2}}\ldots\check{a}_{i_{1}}\ldots a_{k}b_{1}\ldots\check{b}_{j_{1}}\ldots\check{b}_{j_{2}}\ldots b_{l}}\Big)+\ldots\end{eqnarray}
For some applications the precise coefficients are not important,
and a schematic version is enough. Let us denote $\Gamma^{a_{1}\ldots a_{k}}$
schematically simply by $\Gamma^{[k]}$\index{$\Gamma^{[k]}$|itext{schematic for  $\Gamma^{a_1\ldots a_k}$}}.
Neglecting all coefficients, we can write \begin{equation}
\boxed{\Gamma^{[k]}\Gamma^{[l]}\propto\Gamma^{[|k-l|]}+\Gamma^{[|k-l|+2]}+\ldots+\Gamma^{[k+l]}}\label{eq:product-expansion-schematic}\end{equation}
Some simpler cases are of particular interest for us:

\begin{eqnarray}
\Gamma^{a_{1}}\Gamma^{b_{1}\ldots b_{l}} & = & \Gamma^{a_{1}b_{1}\ldots b_{l}}+l\cdot\eta^{a_{1}[b_{1}}\Gamma^{b_{2}\ldots b_{l}]}\label{eq:einGammaMalGammas}\\
\Gamma^{a_{1}a_{2}}\Gamma^{b_{1}\ldots b_{l}} & = & \Gamma^{a_{1}a_{2}b_{1}\ldots b_{l}}-l\cdot\eta^{a_{1}[b_{1}|}\Gamma^{a_{2}|b_{2}\ldots b_{l}]}+l\cdot\eta^{a_{2}[b_{1}|}\Gamma^{a_{1}|b_{2}\ldots b_{l}]}-l(l-1)\eta^{a_{1}[b_{1}|}\eta^{a_{2}|b_{2}}\Gamma^{b_{3}\ldots b_{l}]}\qquad\\
\Gamma^{a_{1}a_{2}}\Gamma^{b_{1}b_{2}} & = & \Gamma^{a_{1}a_{2}b_{1}b_{2}}+\eta^{a_{2}b_{1}}\Gamma^{a_{1}b_{2}}+\eta^{a_{1}b_{2}}\Gamma^{a_{2}b_{1}}-\eta^{a_{1}b_{1}}\Gamma^{a_{2}b_{2}}-\eta^{a_{2}b_{2}}\Gamma^{a_{1}b_{1}}+\nonumber \\
 &  & +\eta^{a_{1}b_{2}}\eta^{a_{2}b_{1}}-\eta^{a_{1}b_{1}}\eta^{a_{2}b_{2}}\end{eqnarray}
Contracting (\ref{eq:einGammaMalGammas}) with $\Gamma_{a_{1}}$from
the left yields\begin{eqnarray}
(d-l)\Gamma^{b_{1}\ldots b_{l}} & = & \Gamma_{a_{1}}\Gamma^{a_{1}b_{1}\ldots b_{l}}\end{eqnarray}
Acting instead from the righthand side yields\begin{eqnarray}
\Gamma^{a}\Gamma^{b_{1}\ldots b_{l}}\Gamma_{a} & = & \Gamma^{ab_{1}\ldots b_{l}}\Gamma_{a}+l\eta^{a[b_{1}}\Gamma^{b_{2}\ldots b_{l}]}\Gamma_{a}=\nonumber \\
 & = & (-)^{l}(d-2l)\cdot\Gamma^{b_{1}\ldots b_{l}}\label{eq:d-2l}\end{eqnarray}
In particular for $l=0$ and $l=1$, we have\begin{eqnarray}
\Gamma^{a}\Gamma_{a} & = & d\\
\Gamma^{a}\Gamma^{b}\Gamma_{a} & = & -(d-2)\cdot\Gamma^{b}\end{eqnarray}
For even dimensions the righthand side of (\ref{eq:d-2l}) vanishes
for $l=d/2$. We will need this fact for ten dimensions: \begin{equation}
\boxed{\Gamma^{a}\Gamma^{b_{1}\ldots b_{5}}\Gamma_{a}=0}\mbox{ for }d=10\label{eq:Gamma5zwischenGammas}\end{equation}

\paragraph{Chirality\index{chirality matrix} matrix as a {}``Hodge\index{Hodge dual}
star''}

Remember the definition and the basic properties of the chirality
matrix in even dimensions:\index{$\Gamma^\#$|itext{chirality matrix}}\index{$\epsilon_{c_1\ldots c_d}$}\index{$\epsilon_{(d)}$}
\begin{eqnarray}
\Gamma^{\#} & \equiv & \sqrt{-\epsilon_{(d)}}\Gamma^{0}\cdots\Gamma^{d-1}=\frac{1}{d!}\sqrt{-\epsilon_{(d)}}\epsilon_{c_{1}\ldots c_{d}}\Gamma^{c_{1}\ldots c_{d}},\quad\mbox{with }\left\{ \zwek{\epsilon_{01\ldots(d-1)}\equiv1}{\epsilon_{(d)}\equiv(-)^{d(d-1)/2}=(-)^{[d/2]}}\right.\\
(\Gamma^{\#})^{2} & = & \one\\
\{\Gamma^{a},\Gamma^{\#}\} & = & 0\quad\forall a\in\{0,1,\ldots,d-1\},\qquad\mbox{for even }d,\qquad\Gamma^{\#}=\pm\one\quad\mbox{for odd }d\end{eqnarray}
The sign $\epsilon_{(d)}$ is the sign that one obtains when reversing
the order of $d$ indices of an antisymmetric object. Likewise if
we have an antisymmetric object with an arbitrary number $p$ of indices,
reversing the order yields the sign $\epsilon_{(p)}\equiv(-)^{\sum_{k=0}^{(p-1)}k}=(-)^{p(p-1)/2}=(-)^{[p/2]}$\index{$[d/2]$|itext{integer part of $d/2$}}.
It takes the explicit values\begin{equation}\mbox{ {\footnotesize }\begin{tabular}{|c|c|c|c|c|c|c|c|c|c|c|c|c|}
\hline 
{\footnotesize $d$} & {\footnotesize 0} & {\footnotesize 1} & {\footnotesize 2} & {\footnotesize 3} & {\footnotesize 4} & {\footnotesize 5} & {\footnotesize 6} & {\footnotesize 7} & {\footnotesize 8} & {\footnotesize 9} & {\footnotesize 10} & {\footnotesize 11}\tabularnewline
\hline
{\footnotesize $\epsilon_{(d)}=(-)^{[d/2]}$} & {\footnotesize 1} & {\footnotesize 1} & {\footnotesize -1} & {\footnotesize -1} & {\footnotesize 1} & {\footnotesize 1} & {\footnotesize -1} & {\footnotesize -1} & {\footnotesize 1} & {\footnotesize 1} & {\footnotesize -1} & {\footnotesize -1}\tabularnewline
\hline
\end{tabular}}\end{equation} and has the properties\begin{eqnarray}
\epsilon_{(p+q)} & = & (-)^{pq}\epsilon_{(p)}\epsilon_{(q)},\quad\epsilon_{(p)}^{2}=1,\quad\epsilon_{(2p)}=(-)^{p},\quad\epsilon_{(-p)}=\underbrace{\epsilon_{(p+1)}}_{-\epsilon_{(p-1)}}=(-)^{p}\epsilon_{(p)},\quad\epsilon_{(d-p)}=\epsilon_{(d)}\epsilon_{(p)}(-)^{p(d-p)}\qquad\label{eq:epsilon-p-multprop}\end{eqnarray}
The prefactor $\sqrt{-\epsilon_{(d)}}$ in the definition of the chirality
matrix guarantees the fact that it squares to the unity. For half
of the dimensions the square root is ill-defined, because $-\epsilon_{(d)}$
is negative. It should simply be understood via $\sqrt{-1}=i$, i.e.
$\sqrt{-\epsilon_{(d)}}\equiv i^{\frac{1}{2}(1+\epsilon_{(d)})}\stackrel{d=10}{=}1$.
Of course, a redefinition of $\Gamma^{\#}$ with an overall (perhaps
$d$-dependent) sign does not change its properties and might be useful
in certain situations. Because $\Gamma^{\#}$ squares to $\one$,
it can have eigenvalues $\pm1$. The corresponding eigenvectors are
chiral and antichiral spinors. For odd dimension, when $\Gamma^{\#}$
coincides with unity, there is only the eigenvalue $1$ and there
is no such split.

There is a natural isomorphism between the antisymmetrized product
of $\Gamma$-matrices $\Gamma^{a_{1}\ldots a_{p}}$ and the wedge
product of the cotangent basis elements (vielbeins) $e^{a_{1}}\wedge\ldots\wedge e^{a_{p}}$.
The multiplication with the chirality matrix on the one side then
corresponds to the application of the Hodge star on the other. It
maps $p$-forms to $(d-p)$-forms in the following sense: \begin{eqnarray}
\Gamma^{\#}\Gamma^{a_{1}\ldots a_{p}} & = & \frac{1}{d!}\sqrt{-\epsilon_{(d)}}\epsilon_{c_{d}\ldots c_{1}}\Gamma^{c_{d}\ldots c_{1}}\Gamma^{a_{1}\ldots a_{p}}=\nonumber \\
 & \stackrel{(\ref{eq:product-expansion})}{=} & \frac{1}{d!}\sqrt{-\epsilon_{(d)}}\epsilon_{c_{d}\ldots c_{1}}p!\left(\zwek{p}{p}\right)\left(\zwek{d}{p}\right)\eta^{c_{1}a_{1}}\cdots\eta^{c_{p}a_{p}}\Gamma^{c_{d}c_{d-1}\ldots c_{p+1}}=\nonumber \\
 & = & \frac{1}{(d-p)!}\sqrt{-\epsilon_{(d)}}\Gamma^{c_{d}\ldots c_{p+1}}\epsilon_{c_{d}\ldots c_{p+1}}\hoch{a_{p}\dots a_{1}}\label{eq:chirality-on-pform}\end{eqnarray}
Up to a sign $(-)^{p(d-p)}$ ($(-)^{p}$ for even $d$ and $1$ for
odd $d$) the same result is obtained when acting from the right,
s.t. we can summarize \begin{equation}
\boxed{\Gamma^{a_{1}\ldots a_{p}}\Gamma^{\#}=\epsilon_{(p)}\sqrt{-\epsilon_{(d)}}\frac{1}{(d-p)!}\epsilon^{a_{1}\ldots a_{p}}\tief{c_{1}\ldots c_{d-p}}\Gamma^{c_{1}\ldots c_{d-p}}=(-)^{(d-p)p}\Gamma^{\#}\Gamma^{a_{1}\ldots a_{p}}}\label{eq:chirality-on-pformII}\end{equation}
The above calculation is also true if we are in odd dimensions where
$\Gamma^{\#}$ is the unity. The antisymmetrized products $\Gamma^{a_{1}\ldots a_{p}}$
do then not correspond to $e^{a_{1}}\wedge\ldots\wedge e^{a_{p}}$,
but (at least in dimensions where $-\epsilon_{(d)}=1$, i.e. $d\in\{3,7,11,\ldots\}$)
to self dual forms $e^{a_{1}}\wedge\ldots\wedge e^{a_{p}}+\star(e^{a_{1}}\wedge\ldots\wedge e^{a_{p}})$
(see intermezzo below for the discussion of the Hodge star). The same
will be true in the even dimensions $d\in\{2,6,10\}$ for the chiral
blocks $\gamma^{a_{1}\ldots a_{p}}$ that will be discussed in particular
for $d=10$ later. In order to understand better the correspondence
between the multiplication with $\Gamma^{\#}$ and the Hodge star
operation, let us give a short review of the latter.\vspace{.5cm}

\lyxline{\normalsize}\vspace{-.25cm}\lyxline{\normalsize}

\subsubsection*{Intermezzo\index{intermezzo!Clifford map and Hodge star} on Clifford
map and Hodge star operator}

\label{intermezzo:Hodge} In order to avoid confusion about prefactors,
note first that we use a definition of the wedge product that absorbs
the normalization factor $\tfrac{1}{p!}$ which is therefore absent
at other places:\begin{eqnarray}
\omega^{(p)} & = & \omega_{m_{1}\ldots m_{p}}\de x^{m_{1}}\wedge\ldots\wedge\de x^{m_{p}}\end{eqnarray}
Replacing $\omega_{m_{1}\ldots m_{p}}\To\tfrac{1}{p!}\omega_{m_{1}\ldots m_{p}}$
everywhere leads to the equations in the standard convention. 

In even dimensions $d$ there is a natural isomorphism, the \textbf{Clifford\index{Clifford map}
map}\index{map!Clifford $\sim$}, from bispinors (which can be expanded
in the complete basis of antisymmetrized products of $\Gamma$-matrices)
and the formal sum of $p$-forms in $\bigwedge^{\bullet}T^{*}M\equiv\oplus_{p}\bigwedge^{p}T^{*}M$.
The basis elements map simply as \begin{equation}
\slash^{-1}:\quad\Gamma^{a_{1}\ldots a_{p}}\mapsto e^{a_{1}}\cdots e^{a_{p}}\equiv e^{a_{1}}\wedge\ldots\wedge e^{a_{p}}\label{eq:slashinverse}\end{equation}
where $e^{a}=\de x^{m}e_{m}\hoch{a}$ is an orthonormal vielbein-basis.
\rem{We consider the wedge product to be a graded commutative product
between the odd objects $e^{a}$ and allow to take formal derivatives
$\partl{e^{a}}$ with respect to them.} Its inverse map is often
denoted by a slash\index{slash}\index{$\slash$}\begin{eqnarray}
\slash:\quad e^{a_{1}}\cdots e^{a_{p}} & \mapsto & \Gamma^{a_{1}\ldots a_{p}}\label{eq:slash}\\
\rho=\sum_{p}\rho_{a_{1}\ldots a_{p}}e^{a_{1}}\cdots e^{a_{p}} & \mapsto & \not\!\rho\equiv\sum_{p}\rho_{a_{1}\ldots a_{p}}\Gamma^{a_{1}\ldots a_{p}}\end{eqnarray}
See in particular \cite{Grana:2004bg,Grana:2004??,Jeschek:2004wy,Witt:2005sk,Jeschek:2005ek}
for frequent use of this map in the context of generalized complex
geometry. Operations on the one side can then be translated to the
other. There is in particular the multiplication with the chirality
matrix on the bispinor side which corresponds more or less to the
Hodge star operator on the other side. The 'more or less' statement
depends on how exactly one defines the Hodge star, and we will simply
define it in such a way, that it corresponds exactly to the multiplication
with the chirality matrix, at least with the multiplication from the
righthand side.

The Hodge\index{Hodge star|ibold} star operation on a manifold with
metric maps $p$-forms to $(n-p)$-forms using the metric and the
$\eps$-tensor%
\footnote{\index{footnote!\thefoot. antisymmetrized Kronecker symbol}\label{fn:antisymKron}In
the following we will use some identities for the epsilon-symbol and
for the antisymmetrized Kronecker-delta, which we would like to recall.
Remember first the definition of the antisymmetrized\index{antisymmetrized!Kronecker delta}
Kronecker\index{Kronecker delta!antisymmetrized $\sim$} symbols\index{$\delta_{d_{1}\ldots d_{n}}^{c_{1}\ldots c_{n}}$|itext{antisymmetrized Kronecker delta}}
\begin{eqnarray*}
\delta_{d_{1}\ldots d_{n}}^{c_{1}\ldots c_{n}} & \equiv & \delta_{[d_{1}}^{c_{1}}\cdots\delta_{d_{n}]}^{c_{n}}\frem{=\frac{1}{n}\delta_{d_{1}\ldots d_{n-1}}^{c_{1}\ldots c_{n-1}}\delta_{d_{n}}^{c_{n}}-(-)^{n}\frac{n-1}{n}\delta_{d_{n}[d_{1}\ldots d_{n-2}}^{c_{1}\ldots c_{n-1}}\delta_{d_{n-1}]}^{c_{n}}}\end{eqnarray*}
If we contract one index pair, we arrive at \begin{eqnarray*}
\delta_{d_{1}\ldots d_{n-1}c_{n}}^{c_{1}\ldots c_{n-1}c_{n}} & = & \frac{d-(n-1)}{n}\delta_{d_{1}\ldots d_{n-1}}^{c_{1}\ldots c_{n-1}}\end{eqnarray*}
Contracting several indices leads to \[
\boxed{\delta_{d_{1}\ldots d_{n-p}a_{1}\ldots a_{p}}^{c_{1}\ldots c_{n-p}a_{1}\ldots a_{p}}=\frac{\tbinom{d-n+p}{p}}{\tbinom{n}{p}}\delta_{d_{1}\ldots d_{n-p}}^{c_{1}\ldots c_{n-p}}}\]
In particular, if all indices are contracted ($p=n$) or if the original
number of indices matches the dimension ($n=d$), we end up with\[
\delta_{a_{1}\ldots a_{p}}^{a_{1}\ldots a_{p}}=\left(\zwek{d}{p}\right),\qquad\delta_{d_{1}\ldots d_{d-p}a_{1}\ldots a_{p}}^{c_{1}\ldots c_{d-p}a_{1}\ldots a_{p}}=\left(\zwek{d}{p}\right)^{-1}\delta_{d_{1}\ldots d_{d-p}}^{c_{1}\ldots c_{d-p}}\]
(see also \cite[p.456]{Kugo:1997}). The last identities are important
to derive the identities for the Levi-Civita symbol $\epsilon$. The
first observation is that we have \begin{eqnarray*}
\epsilon_{a_{1}\ldots a_{d}}\epsilon^{b_{1}\ldots b_{d}} & = & -d!\delta_{a_{1}\ldots a_{d}}^{b_{1}\ldots b_{d}}\end{eqnarray*}
Both sides are completely antisymmetric in all $a$ and all $b$.
It is therefore enough to check the validity for $(a_{1},\ldots,a_{d})=(b_{1},\ldots,b_{d})=(0,\ldots,d-1)$.
The minus sign is coming from the different definition of the $\epsilon$-symbol
with upper sign, i.e. $\epsilon_{0\ldots d-1}=-\epsilon^{0\ldots d-1}=1$.
Using the above formula for contractions of the antisymmetrized Kronecker-delta,
we obtain \[
\boxed{\epsilon_{a_{1}\ldots a_{d-p}c_{1}\ldots c_{p}}\epsilon^{b_{1}\ldots b_{d-p}c_{1}\ldots c_{p}}=-p!(d-p)!\delta_{a_{1}\ldots a_{d-p}}^{b_{1}\ldots b_{d-p}}}\]
This equation remains the same if replace the Levi Civita symbol $\epsilon$
with the $\eps$-tensor (\ref{eq:eps-tensor-down}) and (\ref{eq:eps-tensor-up}),
as the normalization factors cancel.$\qquad\fussend$%
}\index{$\epsilon$@$\eps_{m_1\ldots m_d}$|itext{volume $\eps$ tensor}}
\begin{equation}
\eps_{m_{1}\ldots m_{d}}\equiv\sqrt{\abs{g}}\epsilon_{m_{1}\ldots m_{d}},\quad\epsilon_{0\ldots d-1}\equiv1\label{eq:eps-tensor-down}\end{equation}
where $\epsilon_{m_{1}\ldots m_{d}}$ is the totally antisymmetric
Levi Civita symbol\index{Levi Civita symbol}. Let us define the same
symbol with upper indices with a different sign, i.e. as $\epsilon^{0\ldots d-1}\equiv-1$
(corresponding to the $\eps$-tensor in flat Minkowski spacetime where
raising a zero-index yields the minus). Using that $\det g^{-1}=\epsilon_{m_{1}\ldots m_{d}}g^{m_{1}0}\cdots g^{m_{d}d-1}$\index{determinant!definition with Levi Civita symbol}
the $\eps$-tensor with upper indices takes the familiar form \begin{eqnarray}
\eps^{m_{1}\ldots m_{d}} & = & \tfrac{1}{\sqrt{\abs{g}}}\epsilon^{m_{1}\ldots m_{d}},\quad\epsilon^{0\ldots d-1}\equiv-1\label{eq:eps-tensor-up}\end{eqnarray}
\index{$\star$|ibold}The definition of the Hodge star on a manifold
with metric $\star:\quad\bigwedge^{p}T^{*}M\To\bigwedge^{d-p}T^{*}M$
has some ambiguity in the sign, depending on which behaviour one prefers
$\star$ to have. For us it will be most convenient to define it simply
in the way as $\Gamma^{\#}$ acts (at least for even dimensions).
One still has the freedom to decide whether it should correspond to
an action from the left or from the right, which differs by a factor
of $(-)^{(d-p)p}$ according to (\ref{eq:chirality-on-pformII}).
We choose the Hodge star corresponding to multiplication of $\Gamma^{\#}$
from the right as given in (\ref{eq:chirality-on-pformII}). The
dimension dependent prefactor $\sqrt{-\epsilon_{(d)}}$, however,
will not be included, because it is complex in some dimensions (but
fortunately equal one in 10 dimensions) and the definition of the
Hodge dual should make sense for real manifolds. We therefore define
\begin{eqnarray}
\star(\de x^{k_{1}}\wedge\ldots\wedge\de x^{k_{p}}) & \equiv & \frac{\epsilon_{(p)}}{(d-p)!}\eps\hoch{k_{1}\ldots k_{p}}\tief{m_{1}\ldots m_{d-p}}\de x^{m_{1}}\wedge\ldots\wedge\de x^{m_{d-p}}\label{eq:Hodge-DefI}\\
(\star\omega^{(p)})_{m_{1}\ldots m_{d-p}} & = & \frac{(-)^{p(d-p)}\epsilon_{(p)}}{(d-p)!}\eps\tief{m_{1}\ldots m_{d-p}}\hoch{k_{1}\ldots k_{p}}\omega_{k_{1}\ldots k_{p}}^{(p)}\label{eq:Hodge-DefII}\end{eqnarray}
The sign prefactor $\epsilon_{(p)}=(-)^{p(p-1)/2}$ is usually not
present in the old definitions in the literature. At some places (e.g.
in \cite{Jeschek:2005ek}) the Hodge star is defined such that it
coincides with multiplication of $\Gamma^{\#}$ from the left. This
corresponds to a redefinition of our Hodge star by $(-)^{p(d-p)}$.
Let us denote with \begin{equation}
\tilde{\omega}^{(p)}\equiv\omega^{m_{1}\ldots m_{p}}\pe_{m_{1}}\wedge\ldots\wedge\pe_{m_{p}}\end{equation}
 the multivector that arises when raising all the indices of the differential
form $\omega^{(p)}$ with the metric $g^{mn}$ and remember the definition
of the interior product (\ref{eq:interior-productI}) with respect
to multivector fields:\begin{eqnarray}
\ip_{\tilde{\omega}^{(p)}}\rho^{(r)} & \equiv & \tfrac{r!}{(r-p)!}\underbrace{\omega\hoch{l_{1}\ldots l_{p}}\rho_{{\scriptstyle l_{p}\ldots l_{1}m_{1}\ldots m_{r-p}}}}_{\epsilon_{(p)}\omega\hoch{l_{1}\ldots l_{p}}\rho_{{\scriptscriptstyle l_{1}\ldots l_{p}m_{1}\ldots m_{r-p}}}}\de x^{m_{1}}\wedge\ldots\wedge\de x^{m_{r-p}},\quad\ip_{\tilde{\omega}^{(p)}}\rho^{(r)}=0\mbox{ for }p>r\end{eqnarray}
Using (\ref{eq:epsilon-p-multprop}) and the identities for the $\eps$-tensor
given in footnote \vref{fn:antisymKron}, we obtain the following
relations for the Hodge star operator%
\footnote{\index{footnote!\thefoot. alternative Hodge-definition}Because of
the uncommon definition of the Hodge star, we'll provide here the
equations also for a redefined $\star$. Let us replace the sign factor
$(-)^{p(d-p)}\epsilon_{(p)}=(-)^{p(d-p)+p(p-1)/2}$ in the definition
(\ref{eq:Hodge-DefII}) of the Hodge star by some arbitrary $d$ and
$p$ dependent sign factor $\epsilon_{(d,p)}$\[
(\star\omega^{(p)})_{m_{1}\ldots m_{d-p}}\equiv\frac{\epsilon_{(d,p)}}{(d-p)!}\eps\tief{m_{1}\ldots m_{d-p}}\hoch{k_{1}\ldots k_{p}}\omega_{k_{1}\ldots k_{p}}^{(p)}\]
where some natural choices for $\epsilon_{(d,p)}$ are $1,(-)^{p(d-p)},\epsilon_{(p)}$
and $(-)^{p(d-p)}\epsilon_{(p)}$. The last one corresponds to our
definition, while the second is quite common in the literature. With
this more general ansatz we have\begin{eqnarray*}
(\star1)_{m_{1}\ldots m_{d}} & = & \frac{\epsilon_{(d,0)}}{d!}\eps\tief{m_{1}\ldots m_{d}}\\
\star^{2} & = & -(-)^{p(d-p)}\epsilon_{(d,p)}\epsilon_{(d,d-p)}\\
\star(\omega^{(p)}\wedge\eta^{(q)}) & = & (-)^{pq+p(d-p)}\epsilon_{(d,p+q)}\epsilon_{(d,q)}(-)^{p(p-1)/2}\ip_{\tilde{\omega}^{(p)}}\star\eta^{(q)}=(-)^{q(d-q)}\epsilon_{(d,p+q)}\epsilon_{(d,p)}(-)^{q(q-1)/2}\ip_{\tilde{\eta}^{(q)}}\omega^{(p)}\end{eqnarray*}
In particular for $\epsilon_{(d,p)}=(-)^{p(d-p)}$ one obtains the
more familiar equations\begin{eqnarray*}
(\star1)_{m_{1}\ldots m_{d}} & = & \frac{1}{d!}\eps\tief{m_{1}\ldots m_{d}}\\
\star^{2} & = & -(-)^{p(d-p)}\\
\omega^{(p)}\wedge\star\eta^{(p)} & = & -\ip_{\epsilon_{(p)}\tilde{\omega}^{(p)}}\eta^{(p)}\frac{1}{d!}\eps\tief{m_{1}\ldots m_{d}}\end{eqnarray*}
where the last equation follows from $\star(\omega^{(p)}\wedge\eta^{(q)})=(-)^{pq}\ip_{\epsilon_{(p)}\tilde{\omega}^{(p)}}\star\eta^{(q)}=\ip_{\epsilon_{(q)}\tilde{\eta}^{(q)}}\star\omega^{(p)}$
with $\eta^{(q)}$ replaced by $\star\eta^{(p)}$. The nice feature
of our present definition (with $\epsilon_{(d,p)}=(-)^{p(d-p)}\epsilon_{(p)}$)
is that the expression for $\star^{2}$ in (\ref{eq:sternquadrat})
does not depend on the form degree. $\qquad\fussend$ %
}\begin{eqnarray}
\star^{2} & = & -\epsilon_{(d)}\label{eq:sternquadrat}\\
(\star1)_{m_{1}\ldots m_{d}} & = & \frac{1}{d!}\eps\tief{m_{1}\ldots m_{d}}\\
\star(\omega^{(p)}\wedge\eta^{(q)}) & = & (\ip_{\tilde{\omega}^{(p)}}\star\eta^{(q)})=(-)^{pq}(\ip_{\tilde{\eta}^{(q)}}\star\omega^{(p)}),\quad\mbox{for }p+q\leq d\label{eq:star-omega-eta}\end{eqnarray}
This implies $\star(\omega^{(p)}\wedge\star\eta^{(q)})=-\epsilon_{(d)}(\ip_{\tilde{\omega}^{(p)}}\eta^{(q)})$
($p\leq q$) and $\star(\star\omega^{(p)}\wedge\eta^{(q)})=-\epsilon_{(d)}(-)^{(d-p)q}(\ip_{\tilde{\eta}^{(q)}}\omega^{(p)})$
($q\leq p$) and in particular for $p=q$ \begin{eqnarray}
(\omega^{(p)}\wedge\star\eta^{(p)})_{m_{1}\ldots m_{d}} & = & -\epsilon_{(d)}(\ip_{\tilde{\omega}^{(p)}}\eta^{(p)})\frac{1}{d!}\eps\tief{m_{1}\ldots m_{d}}\\
(\star\omega^{(p)}\wedge\eta^{(p)}) & = & -\epsilon_{(d)}(-)^{(d-p)p}(\ip_{\tilde{\eta}^{(q)}}\omega^{(p)})\frac{1}{d!}\eps\tief{m_{1}\ldots m_{d}}\end{eqnarray}
Note that wedge product and inner product play both the role as
an embedding $\ip$ of forms or vectors into the space of endomorphisms
acting on forms. Thus the equation (\ref{eq:star-omega-eta}) can
be written as $\star(\ip_{\omega^{(p)}}\eta^{(q)})=(\ip_{\tilde{\omega}^{(p)}}\star\eta^{(q)})$.
In turn, the same equation acted upon with an overall $\star$ and
in addition with $\eta$ replaced by $\star\eta$ and $\tilde{\omega}$
renamed as $v$ becomes $(\ip_{\tilde{v}^{(p)}}\star\eta^{(q)})=\star(\ip_{v^{(p)}}\eta^{(q)})$
(where $\tilde{v}$ is the $p$-form obtained from the $p$-vector
$v$ by lowering all indices). For decomposable multivector valued
forms $\omega^{(p)}\otimes v^{(k)}$ (with $\omega$ a $p$-form and
$v$ a $k$-multivector) the embedding is defined as $\ip_{\omega\otimes v}=\ip_{\omega}\ip_{v}=\omega\wedge\ip_{v}$
(see (\ref{eq:interior-productI}) on page \pageref{eq:interior-productI}).
We thus obtain \begin{eqnarray}
\star(\ip_{\omega^{(p)}\otimes v^{(k)}}\eta^{(q)}) & = & \ip_{\tilde{\omega}^{(p)}}\star(\ip_{v^{(k)}}\eta^{(q)})=\ip_{\tilde{\omega}^{(p)}}\ip_{\tilde{v}^{(k)}}\star\eta^{(q)}\end{eqnarray}
The order of the operators on the righthand side is not the {}``normal
order''. The wedge product acts before the interior product, while
the definition of the embedding of a multivector valued form is the
other way round. In order to write it as an embedding again, we need
to apply the commutator which yields the algebraic bracket $[\ip_{\tilde{\omega}^{(p)}},\ip_{\tilde{v}^{(k)}}]\equiv\ip_{[\tilde{\omega}^{(p)},\tilde{v}^{(k)}]}$
(see (\ref{eq:algebraic-bracket})). The above righthand side then
becomes $\ip_{((-)^{pk}\tilde{v}^{(k)}\otimes\tilde{\omega}^{(p)}+[\tilde{\omega}^{(p)},\tilde{v}^{(k)}]^{\Delta})}\star\eta^{(q)}$.
For general multivector valued forms $K^{(k,k')}$ of form-degree
$k$ and multivector degree $k'$ we therefore cannot set $\star(\ip_{K^{(k,k')}}\eta^{(q)})$
equal to $\ip_{\tilde{K}^{(k',k)}}\star\eta^{(q)}$, although this
would be tempting. Instead, we get in the schematic index notation
of page \ref{par:Schematic-index-notation} \begin{equation}
\star(\ip_{K^{(k,k')}}\eta^{(q)})=(k)!\left(\zwek{d-q+k'}{k}\right)K\hoch{l_{1}\ldots l_{k}}\tief{[l_{k}\ldots l_{1}\bs{m}\ldots\bs{m}}(\star\eta)_{\bs{m}\ldots\bs{m}]}^{(d-q)}\label{eq:stariKeta}\end{equation}
Only for multivector valued forms with vanishing contractions (e.g.
for a torsion which is completely antisymmetric after pulling down
one index) the righthand side reduces to $\ip_{\tilde{K}^{(k',k)}}\star\eta^{(q)}$,
where $\tilde{K}^{(k',k)}$ is obtained from $K^{(k,k')}$ by raising
all $k$ form indices and lowering all $k'$ multivector indices with
the metric.

Finally we can use (\ref{eq:star-omega-eta}) formally also to calculate
the action of $\star\de\star$, if we consider the exterior derivative
as wedge product $\de\wedge$. In flat space and Cartesian coordinates,
there is no contribution from the action of the derivative on the
metric and we arrive formally at $\star(\de\wedge\star\eta^{(q)})=-\epsilon_{(d)}(\ip_{\tilde{\de}}\eta^{(q)})$,
or explicitely $(\star\de\star\eta^{(q)})_{m_{1}\ldots m_{q-1}}=-q\epsilon_{(d)}\partial^{k}\eta_{km_{1}\ldots m_{q-1}}^{(q)}$.
In curved space this result gets covariantized to \begin{equation}
(\star\de\star\eta^{(q)})_{m_{1}\ldots m_{q-1}}=-q\epsilon_{(d)}\nabla^{(LC)\, k}\eta_{km_{1}\ldots m_{q-1}}^{(q)}\end{equation}
where the Levi-Civita connection arises from the action of the divergence
on the metric ($\frac{1}{\sqrt{\abs{g}}}\partial^{k}(\sqrt{\abs{g}}\rho_{k})=\nabla^{(LC)k}\rho_{k}$).
Note that for a Levi-Civita connection the covariant antisymmetrized
derivative $\nabla_{[m_{0}}^{(LC)}\omega_{m_{1}\ldots m_{p}]}$ reduces
to the exterior derivative $\partial_{[m_{0}}\omega_{m_{1}\ldots m_{p}]}$
because of the symmetry of the connection. This is not true any longer,
if a torsion is present. In that case it makes sense to define a different
exterior derivative via\index{$\protect\bs{\nabla}$} \begin{eqnarray}
(\bs{\nabla}\omega^{(p)})_{m_{0}\ldots m_{p}} & \equiv & \nabla_{[m_{0}}\omega_{m_{1}\ldots m_{p}]}=(\de\omega^{(p)})_{m_{0}\ldots m_{p}}-pT_{[m_{0}m_{1}|}\hoch{k}\omega_{k|m_{2}\ldots m_{p}]}\label{eq:fatnabla-comp}\\
\mbox{or }\bs{\nabla} & \equiv & \de-\ip_{T}\label{eq:fatnabla}\end{eqnarray}
The relation for $\star\de\star$ then turns into \begin{eqnarray}
(\star\bs{\nabla}\star\eta^{(q)})_{m_{1}\ldots m_{q-1}} & = & -q\epsilon_{(d)}\nabla^{k}\eta_{km_{1}\ldots m_{q-1}}^{(q)}\end{eqnarray}

Apart from the Hodge duality (induced by $\Gamma^{\#}$-multiplication)
there are other interesting operations on the bispinor side which
get translated to the form side via $\slash^{-1}$ (\ref{eq:slashinverse}).
E.g. the matrix multiplications with a $\Gamma$-matrix either from
the left or from the right translate due to (\ref{eq:einGammaMalGammas})
into \begin{eqnarray}
\Gamma^{a}\cdot\not\!\rho & \stackrel{\slash^{-1}}{\mapsto} & \quad e^{a}\wedge\rho+\eta^{ab}\underbrace{\partl{e^{b}}\rho}_{\ip_{e_{b}}\rho}=\\
 & \stackrel{(\ref{eq:star-omega-eta})}{=} & e^{a}\wedge\rho-\epsilon_{(d)}\star\left(e^{a}\wedge\star\rho\right)\\
\not\!\rho\cdot\Gamma^{a} & \stackrel{\slash^{-1}}{\mapsto} & \quad\rho\wedge e^{a}+\eta^{ab}\partial\rho/\partial e^{b}=\\
 &  & =(-)^{r}e^{a}\wedge\rho+(-)^{r-1}\eta^{ab}\underbrace{\partl{e^{b}}\rho}_{\ip_{e_{b}}\rho}=\\
 &  & =(-)^{r}\left(e^{a}\wedge\rho+\epsilon_{(d)}\star\left(e^{a}\wedge\star\rho\right)\right)\end{eqnarray}
The form degree $r$ in the last line makes strictly speaking only
sense if $\rho=\rho^{(r)}$ is a form of definite degree. If it is
instead a formal sum, $r$ should be understood as an operator (acting
on $\rho$) whose eigenvalues are the form degrees (i.e. $e^{a}\partl{e^{a}}$). 

In order to obtain the action of the Dirac\index{Dirac operator}
operator on the first or on the second index of a bispinor, the above
equations can be contracted with a covariant derivative $\nabla_{a}$
(whose connection is compatible with the metric $\eta^{ab}$, the
$\Gamma$-matrices and the vielbein-components, i.e. leaves each of
them invariant):\index{$\nabla$@$\bs{\nabla}$}\begin{eqnarray}
\underbrace{\Gamma^{a}\nabla_{a}}_{\not\nabla_{a}}\cdot\not\!\rho & \mapsto & \bs{\nabla}\rho-\epsilon_{(d)}\star\bs{\nabla}\star\rho\label{eq:actionOfDirac-op}\\
\nabla_{a}\not\!\rho\cdot\Gamma^{a} & \mapsto & \quad\sum_{r}(-)^{r}\left(\bs{\nabla}\rho^{(r)}+\epsilon_{(d)}\star\bs{\nabla}\star\rho^{(r)}\right)\label{eq:actionOfDirac-op-right}\end{eqnarray}
Vanishing of both expressions on the bispinor side yields (because
of the different relative signs in the brackets of both results) $\bs{\nabla}\rho=\star\bs{\nabla}\star\rho=0$,
which for vanishing torsion corresponds to $\de\rho=\star\de\star\rho=0$.
Let us try to recover $\de$ and $\star\de\star$ also in the case
with torsion. According to (\ref{eq:OmegaInTermsOfSymAndTorsion})
or (\ref{eq:GammaInTermsOfGAndTandM}) any connection which is compatible
with the metric can be written as \begin{eqnarray}
\Gamma_{mn}\hoch{k} & = & \Gamma_{mn}^{(LC)\, k}+T_{mn}\hoch{k}+T^{k}\tief{m|n}-T_{n}\hoch{k}\tief{|m}\\
\omega_{ca}\hoch{b} & = & \omega_{ca}^{(LC)\, b}+T_{ca}\hoch{b}+2T^{b}\tief{(c|a)}\end{eqnarray}
so that\begin{eqnarray}
\underbrace{r\nabla^{c}\rho_{ca_{1}\ldots a_{r-1}}^{(r)}}_{-\epsilon_{(d)}(\star\bs{\nabla}\star\rho^{(r)})} & = & \underbrace{r\nabla^{(LC)\, c}\rho_{ca_{1}\ldots a_{r-1}}^{(r)}}_{-\epsilon_{(d)}(\star\de\star\rho^{(r)})_{a_{1}\ldots a_{r-1}}}+\underbrace{2rT^{cd}\tief{c}\rho_{da_{1}\ldots a_{r-1}}^{(r)}-r(r-1)T^{bc}\tief{[a_{1}|}\rho_{cb|a_{2}\ldots a_{r-1}]}^{(r)}}_{\epsilon_{(d)}(\star\ip_{T}\star\rho^{(r)})_{a_{1}\ldots a_{r-1}}}\end{eqnarray}
As indicated below the brackets, the same result is obtained via $\star\bs{\nabla}\star\rho=\star(\de\star\rho-\ip_{T}\star\rho)$
and then using (\ref{eq:stariKeta}) for $\star(\ip_{T}\star\rho)$,
considering $T$ as a vector valued 2-form. 

As a next step we should study the effect of multiplying the bispinor
with another bispinor which again can be expanded in antisymmetrized
products $\Gamma^{b_{1}\ldots b_{p}}$ of $\Gamma$-matrices. Using
(\ref{eq:product-expansion}), we obtain \begin{eqnarray}
\not\!\omega^{(p)}\not\!\rho^{(r)} & = & \sum_{k=0}^{min\{p,r\}}k!\left(\zwek{p}{k}\right)\left(\zwek{r}{k}\right)\omega_{a_{1}\ldots a_{p-k}}\hoch{c_{k}\ldots c_{1}}\rho_{c_{1}\ldots c_{k}a_{p-k+1}\ldots a_{p+r-2k}}^{(r)}\Gamma^{a_{1}\ldots a_{p+r-2k}}\label{eq:Clifford-multI}\end{eqnarray}
The $\Gamma^{a_{1}\ldots a_{p+r-2k}}$'s get mapped to $e^{a_{1}}\cdots e^{a_{p+r-2q}}$
by $\slash^{-1}$. For forms which are not of definite degree, the
result can then be written as\begin{eqnarray}
\not\!\omega\not\!\rho & \stackrel{\slash^{-1}}{\mapsto} & \sum_{k\geq0}\frac{1}{k!}\omega\partr{e^{a_{1}}}\cdots\partr{e^{a_{k}}}\eta^{a_{1}b_{1}}\cdots\eta^{a_{k}b_{k}}\partl{e^{b_{k}}}\cdots\partl{e^{b_{1}}}\rho\label{eq:Clifford-multII}\end{eqnarray}
which defines the \textbf{Clifford\index{Clifford multiplication}
multiplication\index{multiplication!Clifford $\sim$} }between forms.
The Clifford multiplication of two self dual forms is either 0 or
another self-dual form:\begin{equation}
\not\!\omega^{(p)}\tfrac{1}{2}(\one+\Gamma^{\#})\not\!\rho^{(r)}\tfrac{1}{2}(\one+\Gamma^{\#})=\left\{ \zwek{\not\!\omega^{(p)}\not\!\rho^{(r)}\tfrac{1}{2}(\one+\Gamma^{\#})\quad\mbox{for }r\mbox{ even}}{0\quad\mbox{for }r\mbox{ }\mbox{odd}}\right.\label{eq:Clifford-mult-self-dual}\end{equation}
 Note finally that the matrix-commutator on the bispinor side naturally
defines an (algebraic\index{algebraic bracket!between forms}) bracket\index{bracket!some algebraic $\sim$ between forms}
on the form-side \begin{equation}
[\not\!\omega,\not\!\rho]\stackrel{\slash^{-1}}{\mapsto}\sum_{k\geq0}\frac{1}{k!}\Bigl(\omega\partr{e^{a_{1}}}\cdots\partr{e^{a_{k}}}\eta^{a_{1}b_{1}}\cdots\eta^{a_{k}b_{k}}\partl{e^{b_{k}}}\cdots\partl{e^{b_{1}}}\rho-\rho\partr{e^{a_{1}}}\cdots\partr{e^{a_{k}}}\eta^{a_{1}b_{1}}\cdots\eta^{a_{k}b_{k}}\partl{e^{b_{k}}}\cdots\partl{e^{b_{1}}}\omega\Bigr)\qquad\end{equation}
Although this is a valid and consistent map, it is not the most natural
object from the form point of view. On the lefthand side we have the
possibility to think of the gamma matrices as fermionic supermatrices
as suggested in section \vref{sec:gradedGamma} and consider the graded
commutator which would include an additional sign $(-)^{pr}$ in front
of the second term for forms $\omega^{(p)}$ and $\rho^{(r)}$ of
definite degree. Then one can use that $\omega^{(p)}\partr{e^{a}}=-(-)^{p}\partl{e^{a}}\omega^{(p)}$
and therefore $\omega^{(p)}\partr{e^{a_{1}}}\cdots\partr{e^{a_{k}}}=(-)^{kp+k}\epsilon_{(k)}\partl{e^{a_{k}}}\cdots\partl{e^{a_{1}}}\omega^{(p)}$
in order to interchange the position of $\omega$ and $\rho$ and
arrives at \begin{eqnarray}
\underbrace{[\not\!\omega,\not\!\rho]}_{{\rm with\, odd\,\bs{\Gamma}'s}} & \stackrel{\slash^{-1}}{\mapsto} & \quad\sum_{k\geq0}\left(1-(-)^{k}\right)\frac{1}{k!}\omega\partr{e^{a_{1}}}\cdots\partr{e^{a_{k}}}\eta^{a_{1}b_{1}}\cdots\eta^{a_{k}b_{k}}\partl{e^{b_{k}}}\cdots\partl{e^{b_{1}}}\rho=\\
 &  & =\sum_{k\geq0}\frac{2}{(2k+1)!}\omega\partr{e^{a_{1}}}\cdots\partr{e^{a_{2k+1}}}\eta^{a_{1}b_{1}}\cdots\eta^{a_{2k+1}b_{2k+1}}\partl{e^{b_{2k+1}}}\cdots\partl{e^{b_{1}}}\rho\end{eqnarray}
This contains as a special case the anticommutator of the gamma-matrices
themselves\begin{equation}
\left\{ \Gamma^{a},\Gamma^{b}\right\} \mapsto2e^{a}\partr{e^{a_{1}}}\eta^{a_{1}b_{1}}\partl{e^{b_{1}}}e^{b}=2\eta^{ab}\end{equation}

\lyxline{\normalsize}\vspace{-.25cm}\lyxline{\normalsize}\vspace{.5cm} The
Hodge star as defined in the previous intermezzo corresponds to a
multiplication with $\sqrt{-\epsilon_{(d)}}\Gamma^{\#}$ from the
right. It would of course be possible to absorb the prefactor in the
definition of $\Gamma^{\#}$. This, however, would spoil $(\Gamma^{\#})^{2}=1$
in general dimensions. Let us now continue with the discussion of
the properties of the chirality matrix. From (\ref{eq:chirality-on-pformII})
we obtain in particular \begin{eqnarray}
\lqn{\Gamma^{\#}\Gamma^{a_{1}\ldots a_{p}}\otimes\Gamma_{a_{p}\ldots a_{1}}\Gamma^{\#}=(-)^{p(d-p)}\Gamma^{\#}\Gamma^{a_{1}\ldots a_{p}}\otimes\Gamma^{\#}\Gamma_{a_{p}\ldots a_{1}}=}\nonumber \\
 & = & (-)^{p(d-p)}\left(\frac{\sqrt{-\epsilon_{(d)}}}{(d-p)!}\right)^{2}\epsilon_{c_{d}\ldots c_{p+1}}\hoch{a_{p}\dots a_{1}}\Gamma^{c_{d}\ldots c_{p+1}}\otimes\epsilon_{b_{d}\ldots b_{p+1}a_{1}\ldots a_{p}}\Gamma^{b_{d}\ldots b_{p+1}}\end{eqnarray}
Using $\epsilon\hoch{c_{d}\ldots c_{p+1}a_{p}\dots a_{1}}\epsilon_{b_{d}\ldots b_{p+1}}\tief{a_{1}\ldots a_{p}}=-\epsilon_{(p)}p!(d-p)!\eta_{c_{d}\ldots c_{p+1},b_{d}\ldots b_{p+1}}$
(see footnote \ref{fn:antisymKron}) we get\begin{eqnarray}
\Gamma^{\#}\Gamma^{a_{1}\ldots a_{p}}\otimes\Gamma_{a_{p}\ldots a_{1}}\Gamma^{\#} & = & (-)^{p(d-p)}\epsilon_{(d)}\epsilon_{(p)}\frac{p!}{(d-p)!}\Gamma_{b_{d}\ldots d_{p+1}}\otimes\Gamma^{b_{d}\ldots d_{p+1}}\end{eqnarray}
Reversing the order of the indices of one of the $\Gamma$'s on the
righthand side of the equation (contributing a factor $\epsilon_{(d-p)}=\epsilon_{(d)}\epsilon_{(p)}(-)^{p(d-p)}$),
we arrive at \begin{equation}
\boxed{\Gamma^{\#}\Gamma^{a_{1}\ldots a_{p}}\otimes\Gamma_{a_{p}\ldots a_{1}}\Gamma^{\#}=\frac{p!}{(d-p)!}\Gamma^{b_{1}\ldots b_{d-p}}\otimes\Gamma_{b_{d-p}\ldots b_{1}}}\label{eq:dual-pform-sum}\end{equation}
In particular in ten dimensions and for $p=5$, we obtain \begin{equation}
\Gamma^{\#}\Gamma^{a_{1}\ldots a_{5}}\otimes\Gamma_{a_{5}\ldots a_{1}}\Gamma^{\#}=\Gamma_{b_{1}\ldots b_{5}}\otimes\Gamma^{b_{1}\ldots b_{5}}\qquad\mbox{for }d=10\label{eq:dual5form}\end{equation}

\paragraph{Trace }

The trace\index{trace!of gamma matrices} of all antisymmetrized products
of Gamma-matrices vanishes in even dimensions: \begin{eqnarray*}
\tr\Gamma^{a_{1}\ldots a_{2k+1}}=\tr\Gamma^{a_{1}\ldots a_{2k+1}}\Gamma^{\#}\Gamma^{\#}\stackrel{{\rm even}\, d}{=}\pm\tr\Gamma^{\#}\Gamma^{a_{1}\ldots a_{2k+1}}\Gamma^{\#} & \dann & \tr\Gamma^{a_{1}\ldots a_{2k+1}}=0\\
\tr\Gamma^{a_{1}\ldots a_{2k}}=\pm\tr\Gamma^{a_{2k}a_{1}\ldots a_{2k-1}} & \dann & \tr\Gamma^{a_{1}\ldots a_{2k}}=0\end{eqnarray*}
\begin{equation}
\boxed{\tr\Gamma^{a_{1}\ldots a_{p}}=0}\quad\forall p\geq1\quad\mbox{for even }d\label{eq:spurlos}\end{equation}

\paragraph{Fierz identity}

(see e.g. \cite{Kreuzer:2001gm}) The Fierz identity is simply a completeness
relation. Given a basis $\left\{ \ket{e^{k}}\right\} $ of a vector
space, define its dual basis via $\bra{e_{k}}\ket{e^{l}}=\delta_{k}^{l}$.
The completeness relation then reads \begin{equation}
\sum_{k}\ket{e^{k}}\bra{e_{k}}=\one\end{equation}
In our case the vector space is the space of all $2^{[d/2]}\times2^{[d/2]}$-matrices
and in \textbf{even dimensions} the antisymmetrized products of $\Gamma$-matrices
form a basis of it: $\{\one,\Gamma^{a},\Gamma^{a_{1}a_{2}},\ldots,\Gamma^{a_{1}\ldots a_{d}}\}\equiv\{\Gamma^{I}\}$.
In odd dimensions this is still a generating set, but not linearly
independent. The dual basis to $\{\Gamma^{I}\}$ in even dimensions
is simply given by $2^{-d/2}\cdot\{\one,\Gamma_{a},\Gamma_{a_{2}a_{1}},\ldots,\Gamma_{a_{d}\cdots a_{1}}\}\equiv\{\Gamma_{I}\}$
(acting on the original basis by contracting all spinor indices).
One can convince oneself that we have indeed (using $\tr\Gamma^{a_{1}\ldots a_{p}}=0$)\begin{eqnarray}
2^{-d/2}\delta_{\q{\beta}}^{\q{\alpha}}\delta_{\q{\alpha}}^{\q{\beta}} & = & 1\\
\frac{2^{-d/2}}{p!}\Gamma_{a_{p}\ldots a_{1}}\hoch{\q{\alpha}}\tief{\q{\beta}}\Gamma^{b_{1}\ldots b_{q}\,\q{\beta}}\tief{\q{\alpha}} & = & \delta_{p}^{q}\delta_{a_{1}\ldots a_{p}}^{b_{1}\ldots b_{p}}\equiv\delta_{p}^{q}\delta_{[a_{1}}^{b_{1}}\cdots\delta_{a_{p}]}^{b_{p}}\end{eqnarray}
The completeness relation or \textbf{Fierz}\index{Fierz identity}
\textbf{identity}\index{identity!Fierz} thus reads\begin{equation}
\boxed{\sum_{p=0}^{d}\frac{2^{-d/2}}{p!}\Gamma^{a_{1}\ldots a_{p}\,\q{\alpha}}\tief{\q{\beta}}\Gamma_{a_{p}\ldots a_{1}}\hoch{\q{\gamma}}\tief{\q{\delta}}=\delta_{\q{\delta}}^{\q{\alpha}}\delta_{\q{\beta}}^{\q{\gamma}}}\label{eq:Fierz}\end{equation}
Using (\ref{eq:dual-pform-sum}) it can be rewritten as\begin{equation}
\boxed{\sum_{p=0}^{d/2-1}\frac{2^{-d/2}}{p!}\left(\Gamma^{a_{1}\ldots a_{p}\,\q{\alpha}}\tief{\q{\beta}}\Gamma_{a_{p}\ldots a_{1}}\hoch{\q{\gamma}}\tief{\q{\delta}}+(-)^{p}(\Gamma^{a_{1}\ldots a_{p}}\Gamma^{\#})^{\q{\alpha}}\tief{\q{\beta}}(\Gamma_{a_{p}\ldots a_{1}}\Gamma^{\#})\hoch{\q{\gamma}}\tief{\q{\delta}}\right)+\frac{2^{-d/2}}{(d/2)!}\Gamma^{a_{1}\ldots a_{d/2}\,\q{\alpha}}\tief{\q{\beta}}\Gamma_{a_{d/2}\ldots a_{1}}\hoch{\q{\gamma}}\tief{\q{\delta}}=\delta_{\q{\delta}}^{\q{\alpha}}\delta_{\q{\beta}}^{\q{\gamma}}}\label{eq:FierzII}\end{equation}
which further simplifies when contracted with chiral spinors for which
$\Gamma^{\#}\To\one$. The identities (\ref{eq:Fierz}) and equivalently
(\ref{eq:FierzII}) can be rewritten in various ways. One appearance
of the Fierz identity which is of particular interest, is to contract
the identity (\ref{eq:Fierz}) with $\Gamma^{c\,\tilde{\q{\alpha}}}\tief{\q{\alpha}}\Gamma_{c}\hoch{\q{\tilde{\gamma}}}\tief{\q{\gamma}}$
which yields (after relabeling in the result $\q{\tilde{\alpha}}\To\q{\alpha}$,
$\tilde{\q{\gamma}}\To\q{\gamma}$) \begin{eqnarray}
\sum_{p=0}^{d}\frac{2^{-d/2}}{p!}(\underbrace{\Gamma^{c}\Gamma^{a_{1}\ldots a_{p}}}_{\Gamma^{ca_{1}\ldots a_{p}}+p\eta^{c[a_{1}}\Gamma^{a_{2}\ldots a_{p}]}})\hoch{\q{\alpha}}\tief{\q{\beta}}(\underbrace{\Gamma_{c}\Gamma_{a_{p}\ldots a_{1}}}_{\Gamma_{ca_{p}\ldots a_{1}}+p\eta_{c[a_{p}}\Gamma_{a_{p-1}\ldots a_{1}]}})\hoch{\q{\gamma}}\tief{\q{\delta}} & = & \Gamma^{c\,\q{\alpha}}\tief{\q{\delta}}\Gamma_{c}\hoch{\q{\gamma}}\tief{\q{\beta}}\end{eqnarray}
Some relabeling yields\begin{eqnarray}
\sum_{p=0}^{d}\frac{(-)^{p}}{2^{d/2}p!}\left(d-2p\right)(\Gamma^{a_{1}\ldots a_{p}})\hoch{\q{\alpha}}\tief{\q{\beta}}(\Gamma_{a_{p}\ldots a_{1}})\hoch{\q{\gamma}}\tief{\q{\delta}} & = & \Gamma^{c\,\q{\alpha}}\tief{\q{\delta}}\Gamma_{c}\hoch{\q{\gamma}}\tief{\q{\beta}}\end{eqnarray}
Finally we can use again (\ref{eq:dual-pform-sum}), in order to arrive
at \begin{eqnarray}
\sum_{p=0}^{d/2-1}\frac{(-)^{p}}{2^{d/2}p!}\left(d-2p\right)\left((\Gamma^{a_{1}\ldots a_{p}})\hoch{\q{\alpha}}\tief{\q{\beta}}(\Gamma_{a_{p}\ldots a_{1}})\hoch{\q{\gamma}}\tief{\q{\delta}}-(-)^{p}(\Gamma^{a_{1}\ldots a_{p}}\Gamma^{\#})^{\q{\alpha}}\tief{\q{\beta}}(\Gamma_{a_{p}\ldots a_{1}}\Gamma^{\#})\hoch{\q{\gamma}}\tief{\q{\delta}}\right) & = & \Gamma^{c\,\q{\alpha}}\tief{\q{\delta}}\Gamma_{c}\hoch{\q{\gamma}}\tief{\q{\beta}}\qquad\end{eqnarray}
Contracting the identity \textbf{with chiral spinors} $\Psi^{\q{\beta}}=(\psi^{\beta},0)$
and $\Phi^{\q{\delta}}=(\phi^{\delta},0)$ leads to \begin{equation}
\boxed{\sum_{p=1,\mbox{ odd}}^{d/2-1}\frac{2\left(d-2p\right)}{2^{d/2}p!}(\Gamma^{a_{1}\ldots a_{p}}\Psi)\hoch{\q{\alpha}}(\Gamma_{a_{p}\ldots a_{1}}\Phi)\hoch{\q{\gamma}}=-(-)^{\Phi\Psi}(\Gamma^{c}\Phi)\hoch{\q{\alpha}}(\Gamma_{c}\Psi)\hoch{\q{\gamma}}}\end{equation}
$d=4,6:$\begin{eqnarray}
(\Gamma^{c}\Psi)\hoch{\q{\alpha}}(\Gamma_{c}\Phi)\hoch{\q{\gamma}} & = & -(-)^{\Phi\Psi}(\Gamma^{c}\Phi)\hoch{\q{\alpha}}(\Gamma_{c}\Psi)\hoch{\q{\gamma}}\end{eqnarray}
$d=10$:\begin{eqnarray}
\frac{1}{2}(\Gamma^{a_{1}}\Psi)\hoch{\q{\alpha}}(\Gamma_{a_{p}}\Phi)\hoch{\q{\gamma}}+\frac{1}{24}(\Gamma^{a_{1}a_{2}a_{3}}\Psi)\hoch{\q{\alpha}}(\Gamma_{a_{3}a_{2}a_{1}}\Phi)\hoch{\q{\gamma}} & = & -(-)^{\Phi\Psi}(\Gamma^{c}\Phi)\hoch{\q{\alpha}}(\Gamma_{c}\Psi)\hoch{\q{\gamma}}\label{eq:FierzIIIten}\end{eqnarray}
In 10 dimensions this can be further rewritten, using the symmetry
properties of the gamma matrices in their fermionic indices. We will
come back to that in subsection \ref{sub:chiralFierz}.

\section{Explicit 10d-representation}

In the following we will give an explicit representation\index{representation!of gamma matrices}
of the Dirac\index{Dirac!gamma matrices!representation}-$\Gamma$-matrices
in 10 dimensions which we are using throughout this document. The
presentation is based on the one given in the appendix of \cite{Nh:2003cm}.

\subsection{D=(2,0): Pauli-matrices (2x2)}

We start with the 3 Pauli matrices\begin{equation}
\tau^{1}\equiv\left(\begin{array}{cc}
0 & 1\\
1 & 0\end{array}\right),\quad\tau^{2}\equiv\left(\begin{array}{cc}
0 & -i\\
i & 0\end{array}\right),\quad\tau^{3}\equiv\left(\begin{array}{cc}
1 & 0\\
0 & -1\end{array}\right)\end{equation}
\begin{eqnarray}
\tau^{i}\tau^{j} & = & i\epsilon^{ijk}\tau^{k}+\delta^{ij}\one\\
{}[\tau^{i},\tau^{j}] & = & 2i\epsilon^{ijk}\tau^{k}\\
\{\tau^{i},\tau^{j}\} & = & 2\delta^{ij}\one\\
\tr\tau^{i} & = & 0,\quad\det(\sigma^{i})=-1\\
(\tau^{i})^{\dagger} & = & \tau^{i}\end{eqnarray}

\subsection{D=(3,1), 4x4}

Define $\gamma^{k}\equiv\tau^{k}\otimes\tau^{2}$, $\gamma^{4}\equiv\one\otimes\tau^{1}$,$\gamma^{5}\equiv\one\otimes\tau^{3}$.
The tensor product can be understood in different ways when writing
down the resulting matrices. We understand it as plugging the lefthand
matrix into the righthand one:\begin{equation}
\gamma^{k}\equiv\left(\begin{array}{cc}
0 & -i\tau^{k}\\
i\tau^{k} & 0\end{array}\right),\quad\gamma^{4}\equiv\left(\begin{array}{cc}
0 & \one\\
\one & 0\end{array}\right)\equiv i\gamma^{0},\quad\gamma^{5}\equiv\left(\begin{array}{cc}
\one & 0\\
0 & -\one\end{array}\right)\end{equation}
\begin{eqnarray}
\left\{ \gamma^{\mu},\gamma^{\nu}\right\}  & = & 2\delta^{\mu\nu}\one\\
\tr(\gamma^{\mu}) & = & 0\\
(\gamma^{\mu})^{\dagger} & = & \gamma^{\mu}\\
\gamma^{1}\gamma^{2}\gamma^{3}\gamma^{4} & = & \left(\begin{array}{cc}
0 & -i\tau^{1}\tau^{2}\tau^{3}\\
i\tau^{1}\tau^{2}\tau^{3} & 0\end{array}\right)\left(\begin{array}{cc}
0 & \one\\
\one & 0\end{array}\right)=\left(\begin{array}{cc}
\one & 0\\
0 & -\one\end{array}\right)=\gamma^{5}\end{eqnarray}
$\gamma^{2},\gamma^{4}$ and $\gamma^{5}$ are real and symmetric,
while $\gamma^{1}$ and $\gamma^{3}$ are imaginary and antisymmetric.

\subsection{D=(7,0), 8x8}

We can define seven purely imaginary $8\times8$ matrices $\lambda^{i}$
as follows:\begin{eqnarray}
\lambda^{i} & = & \left\{ \gamma^{2}\otimes\tau^{2},\gamma^{4}\otimes\tau^{2},\gamma^{5}\otimes\tau^{2},\gamma^{1}\otimes\one,\gamma^{3}\otimes\one,i\gamma^{2}\gamma^{4}\gamma^{5}\otimes\tau^{1},i\gamma^{2}\gamma^{4}\gamma^{5}\otimes\tau^{3}\right\} \\
 &  & \mbox{with }i\gamma^{2}\gamma^{4}\gamma^{5}=i\tau^{2}\otimes\tau^{2}\tau^{1}\tau^{3}=\tau^{2}\otimes\one=\left(\begin{array}{cc}
\tau_{2} & 0\\
0 & \tau_{2}\end{array}\right)\nonumber \end{eqnarray}
\begin{eqnarray}
\{\lambda^{i},\lambda^{j}\} & = & 2\delta^{ij}\one\\
\tr(\lambda^{i}) & = & 0\\
(\lambda^{i})^{\dagger} & = & \lambda^{i}\\
\lambda^{1}\cdots\lambda^{6} & = & (\gamma^{2}\gamma^{4}\gamma^{5}\gamma^{1}\gamma^{3}i\gamma^{2}\gamma^{4}\gamma^{5})\otimes\tau^{2}\tau^{1}=-(\gamma^{1}\gamma^{3})\otimes\tau^{3}=(i\tau^{2}\otimes\one)\otimes\tau^{3}=ii\gamma^{2}\gamma^{4}\gamma^{5}\otimes\tau^{3}=i\lambda^{7}\end{eqnarray}

\subsection{D=(8,0), 16x16}

Now we can define 8 real symmetric $16\times16$ matrices $\sigma^{\mu}\equiv\{\lambda^{i}\otimes\tau^{2},\one\otimes\tau^{1}\}$\begin{eqnarray}
\sigma^{i} & \equiv & \left(\begin{array}{cc}
0 & -i\lambda^{i}\\
i\lambda^{i} & 0\end{array}\right),\quad\sigma^{8}\equiv\left(\begin{array}{cc}
0 & \one\\
\one & 0\end{array}\right)\end{eqnarray}
\begin{eqnarray}
\left\{ \sigma^{\mu},\sigma^{\nu}\right\}  & = & 2\delta^{\mu\nu}\one\\
(\sigma^{\mu})^{\dagger} & = & \sigma^{\mu}\\
\tr(\sigma^{\mu}) & = & 0\\
\chi\equiv\sigma^{1}\cdots\sigma^{8} & = & \lambda^{1}\cdots\lambda^{7}\otimes\tau^{2}\tau^{1}=\one\otimes\tau^{3}=\left(\begin{array}{cc}
\one & 0\\
0 & -\one\end{array}\right)\end{eqnarray}

\subsection{D=(9,1),32x32}

Finally we define the real Dirac-matrices for 10-dimensional Minkowski-space
as $\Gamma^{a}\equiv\left\{ \one\otimes i\tau^{2},\sigma^{\mu}\otimes\tau_{1},\chi\otimes\tau_{1}\right\} $\begin{eqnarray}
\Gamma^{0} & \equiv & \left(\begin{array}{cc}
0 & \one\\
-\one & 0\end{array}\right)\equiv-i\Gamma^{10},\quad\Gamma^{\mu}\equiv\left(\begin{array}{cc}
0 & \sigma^{\mu}\\
\sigma^{\mu} & 0\end{array}\right),\Gamma^{9}\equiv\left(\begin{array}{cc}
0 & \chi\\
\chi & 0\end{array}\right)\end{eqnarray}
\begin{eqnarray}
\Gamma^{a\,\q{\alpha}}\tief{\q{\beta}} & \equiv & \left(\begin{array}{cc}
0 & \gamma^{a\,\alpha\beta}\\
\gamma_{\alpha\beta}^{a} & 0\end{array}\right),\quad\mbox{with }\gamma^{a\,\alpha\beta}\equiv\{\delta^{\alpha\beta},\sigma^{\mu\,\alpha}\tief{\beta},\chi^{\alpha}\tief{\beta}\},\quad\gamma_{\alpha\beta}^{a}\equiv\left\{ -\delta_{\alpha\beta},\sigma^{\mu\,\alpha}\tief{\beta},\chi^{\alpha}\tief{\beta}\right\} \end{eqnarray}
\index{$\gamma_{\bs{\alpha\beta}}^a$|itext{chiral gamma matrix}}The
small $\gamma^{a}$ (chiral\index{chiral gamma matrices} gamma\index{gamma matrix!chiral $\sim$}
matrices) are thus all real and symmetric! The Dirac matrices obey\begin{eqnarray}
\{\Gamma^{a},\Gamma^{b}\} & = & 2\eta^{ab}\one\\
\Gamma^{\#} & \equiv & \Gamma^{0}\cdots\Gamma^{9}=i\Gamma^{1}\cdots\Gamma^{10}=\sigma^{1}\cdots\sigma^{8}\chi\otimes i\tau^{2}(\tau^{1})^{9}=\one\otimes\tau^{3}=\left(\begin{array}{cc}
\one & 0\\
0 & -\one\end{array}\right)\\
(\Gamma^{\#})^{2} & = & \one,\quad\Gamma^{\#}\Gamma^{a}=-\Gamma^{a}\Gamma^{\#}\\
(\Gamma^{a})^{\dagger} & = & \Gamma^{a},\quad(\Gamma^{\#})^{\dagger}=\Gamma^{\#}\\
\tr\Gamma^{a} & = & 0,\quad\tr\Gamma^{\#}=0\end{eqnarray}

\paragraph{Intertwiner\index{intertwiner}s}

The unitary intertwiners $A$, $B$ and $C$ are defined via\begin{eqnarray}
(\Gamma^{a})^{\dagger} & = & A\Gamma^{a}A^{\dagger},\qquad-(\Gamma^{a})^{*}=B^{\dagger}\Gamma^{a}B,\qquad-(\Gamma^{a})^{T}=C^{\dagger}\Gamma^{a}C\end{eqnarray}
We can choose\begin{eqnarray}
A_{\q{\alpha}\q{\beta}} & = & -\Gamma^{0}\Gamma^{\#}=\left(\begin{array}{cc}
0 & \delta_{\alpha}^{\beta}\\
\delta_{\beta}^{\alpha} & 0\end{array}\right)\\
B & = & \Gamma^{\#}\\
C & = & BA^{\dagger}=-\Gamma^{\#}\Gamma^{0}\Gamma^{\#}=\Gamma^{0}\end{eqnarray}
The Dirac\index{Dirac!conjugate} conjugate is $\bar{\psi}\equiv\psi^{\dagger}A$.
In the Lorentz-covariant expression $\bar{\psi}\Gamma^{m}\phi$, there
appears therefore the combination

\begin{eqnarray}
(A\Gamma^{a})_{\q{\alpha}\q{\beta}} & = & \left(\begin{array}{cc}
\gamma_{\alpha\beta}^{a} & 0\\
0 & \gamma^{a\,\alpha\beta}\end{array}\right),\quad\gamma_{\alpha\beta}^{a}\mbox{ sym and real}\end{eqnarray}
The other conjugate is the charge\index{charge conjugate} conjugate
spinor $\psi^{c}\equiv C\bar{\psi}^{T}=CA^{T}\psi^{*}=B\psi^{*}=\Gamma^{\#}\psi^{*}$.

\section{Clifford algebra, Fierz identity and more for the chiral blocks in
10 dimensions}

Above we have defined \begin{equation}
\Gamma^{a\,\q{\alpha}}\tief{\q{\beta}}=\left(\begin{array}{cc}
0 & \gamma^{a\,\alpha\beta}\\
\gamma_{\alpha\beta}^{a} & 0\end{array}\right)\end{equation}
The Clifford algebra for the $\Gamma'$s reads in terms of the smallo\index{Clifford algebra!chiral $\sim$}\index{chiral!Clifford algebra}
$\gamma'$s: \begin{eqnarray}
\gamma^{(a|\,\alpha\gamma}\gamma_{\gamma\beta}^{|b)} & = & \eta^{ab}\delta_{\beta}^{\alpha}\label{eq:smallClifford}\\
\gamma^{(a|\,\alpha\beta}\gamma_{\beta\alpha}^{|b)} & = & 16\eta^{ab}\label{eq:smallCliffordContracted}\end{eqnarray}

\subsection{Product of antisymmetrized products of gamma-matrices}

Antisymmetrized products of $\Gamma'$s are block-diagonal for even
number of factors and block-offdiagonal for odd number of factors%
\footnote{\index{footnote!\thefoot. explicit form of antisymmetrized product of two Gamma-matrices}For
example, the product of two gamma-matrices reads\begin{eqnarray*}
\Gamma^{a_{1}a_{2}\,\q{\alpha}}\tief{\q{\beta}} & \equiv & \Gamma^{[a_{1}|\,\q{\alpha}}\tief{\q{\gamma}}\Gamma^{|a_{1}]\,\q{\gamma}}\tief{\q{\beta}}=\\
 & = & \left(\begin{array}{cc}
\gamma^{[a_{1}|\,\alpha\gamma}\gamma_{\gamma\beta}^{|a_{2}]} & 0\\
0 & \gamma_{\alpha\gamma}^{[a_{1}}\gamma^{a_{2}]\,\gamma\beta}=-\gamma^{[a_{1}|\,\beta\gamma}\gamma_{\gamma\alpha}^{|a_{2}]}\end{array}\right)\equiv\left(\begin{array}{cc}
\gamma^{a_{1}a_{2}\,\alpha}\tief{\beta} & 0\\
0 & \gamma^{a_{1}a_{2}}\tief{\alpha}\hoch{\beta}\end{array}\right)\\
\gamma^{a_{1}a_{2}\,\alpha}\tief{\beta} & = & -\gamma^{a_{1}a_{2}}\tief{\beta}\hoch{\alpha}\\
\gamma^{[0]\,\alpha}\tief{\beta} & \equiv & \delta_{\beta}^{\alpha}\quad(\mbox{no index-grading here!)}\qquad\fussend\end{eqnarray*}
}. The chiral blocks read: \begin{eqnarray}
\gamma^{a_{1}\ldots a_{2k}\,\alpha}\tief{\beta} & \equiv & \gamma^{[a_{1}|\,\alpha\gamma_{1}}\gamma_{\gamma_{1}\gamma_{2}}^{|a_{2}|}\cdots\gamma_{\gamma_{2k-1}\beta}^{|a_{2k}]}=(-)^{k}\gamma^{a_{1}\ldots a_{2k}}\tief{\beta}\hoch{\alpha}\label{eq:antisymSmallGammasEven}\\
\gamma_{\alpha\beta}^{a_{1}\ldots a_{2k+1}} & = & (-)^{k}\gamma_{\beta\alpha}^{a_{1}\ldots a_{2k+1}},\qquad\gamma^{a_{1}\ldots a_{2k+1}\,\alpha\beta}=(-)^{k}\gamma^{a_{1}\ldots a_{2k+1}\,\beta\alpha}\label{eq:antisymSmallGammasOdd}\end{eqnarray}
\index{$\gamma^{a_{1}\ldots a_{2k}\,\bs\alpha}\tief{\bs\beta}$}The
schematic expansion of antisymmetrized products of $\Gamma$-matrices
given in (\ref{eq:product-expansion-schematic}) has the same form
for the chiral blocks, if we suppress the index structure:\index{$\gamma^{[k]}$|itext{schematic for $\gamma^{a_1 \ldots a_k}$}}
\begin{equation}
\boxed{\gamma^{[k]}\gamma^{[l]}\propto\gamma^{[|k-l|]}+\gamma^{[|k-l|+2]}+\ldots+\gamma^{[k+l]}}\label{eq:product-exp-schem-chiral}\end{equation}
Indeed, without the spinorial indices, even the exact equations (including
the correct prefactors) look identically for the small $\gamma'$s:\begin{eqnarray}
\gamma^{a_{1}\ldots a_{p}}\gamma^{b_{1}\ldots b_{q}} & = & \sum_{k=0}^{min\{p,q\}}k!\left(\zwek{p}{k}\right)\left(\zwek{q}{k}\right)\eta^{\tief{[}a_{p}\tief{|}\hoch{[}b_{1}\hoch{|}}\eta^{\tief{|}a_{p-1}\tief{|}\hoch{|}b_{2}\hoch{|}}\cdots\eta^{\tief{|}a_{p+1-k}\tief{|}\hoch{|}b_{k}\hoch{|}}\gamma^{\tief{|}a_{1}\ldots a_{p-k}\tief{]}\hoch{|}b_{k+1}\ldots b_{q}\hoch{]}}\label{eq:product-expansion-chiral}\end{eqnarray}
 In particular we have \begin{eqnarray}
\gamma^{a_{1}}\gamma^{b_{1}\ldots b_{l}} & = & \gamma^{a_{1}b_{1}\ldots b_{l}}+l\cdot\eta^{a_{1}[b_{1}}\gamma^{b_{2}\ldots b_{l}]},\qquad\gamma^{b_{1}\ldots b_{l}}\gamma^{a_{1}}=\gamma^{b_{1}\ldots b_{l}a_{1}}+l\cdot\gamma^{[b_{1}\ldots b_{l-1}}\eta^{b_{l}]a_{1}}\label{eq:gammaIgammal}\\
\gamma^{a_{1}a_{2}}\gamma^{b_{1}\ldots b_{l}} & = & \gamma^{a_{1}a_{2}b_{1}\ldots b_{l}}-l\cdot\eta^{a_{1}[b_{1}|}\gamma^{a_{2}|b_{2}\ldots b_{l}]}+l\cdot\eta^{a_{2}[b_{1}|}\gamma^{a_{1}|b_{2}\ldots b_{l}]}+\nonumber \\
 &  & -l(l-1)\eta^{a_{1}[b_{1}|}\eta^{a_{2}|b_{2}}\gamma^{b_{3}\ldots b_{l}]}\label{eq:gammaIIgammalohneIndex}\\
\gamma^{a_{1}a_{2}}\gamma^{b_{1}b_{2}} & = & \gamma^{a_{1}a_{2}b_{1}b_{2}}-2\eta^{a_{1}[b_{1}|}\gamma^{a_{2}|b_{2}]}+2\eta^{a_{2}[b_{1}|}\gamma^{a_{1}|b_{2}]}-2\eta^{a_{1}[b_{1}|}\eta^{a_{2}|b_{2}]}=\nonumber \\
 & = & \gamma^{a_{1}a_{2}b_{1}b_{2}}+\eta^{a_{2}b_{1}}\gamma^{a_{1}b_{2}}+\eta^{a_{1}b_{2}}\gamma^{a_{2}b_{1}}-\eta^{a_{1}b_{1}}\gamma^{a_{2}b_{2}}-\eta^{a_{2}b_{2}}\gamma^{a_{1}b_{1}}+\nonumber \\
 &  & +\eta^{a_{1}b_{2}}\eta^{a_{2}b_{1}}-\eta^{a_{1}b_{1}}\eta^{a_{2}b_{2}}\label{eq:gammaIIgammaIIohneIndex}\end{eqnarray}
Reintroducing the spinorial indices for the last line yields (remember
that we do not use our graded conventions in this part of the appendix):\begin{eqnarray}
\gamma^{a_{1}a_{2}}\tief{\alpha}\hoch{\gamma}\gamma^{b_{1}b_{2}}\tief{\gamma}\hoch{\beta} & = & \gamma^{a_{1}a_{2}b_{1}b_{2}}\tief{\alpha}\hoch{\beta}+\eta^{a_{2}b_{1}}\gamma^{a_{1}b_{2}}\tief{\alpha}\hoch{\beta}+\eta^{a_{1}b_{2}}\gamma^{a_{2}b_{1}}\tief{\alpha}\hoch{\beta}-\eta^{a_{1}b_{1}}\gamma^{a_{2}b_{2}}\tief{\alpha}\hoch{\beta}-\eta^{a_{2}b_{2}}\gamma^{a_{1}b_{1}}\tief{\alpha}\hoch{\beta}+\nonumber \\
 &  & +\eta^{a_{1}b_{2}}\eta^{a_{2}b_{1}}\delta_{\alpha}^{\beta}-\eta^{a_{1}b_{1}}\eta^{a_{2}b_{2}}\delta_{\alpha}^{\beta}\label{eq:gammaIIgammaII}\end{eqnarray}
If we regard $\gamma^{a_{1}a_{2}}\tief{\alpha}\hoch{\gamma}$ as a
matrix with collected indices $(a_{1},\alpha)$ and $(a_{2},\gamma$),
we can use the above equation also to construct an inverse to this
matrix: Contracting $a_{2}$ and $b_{1}$, we obtain\begin{eqnarray}
\gamma^{a_{1}}\tief{c\alpha}\hoch{\gamma}\gamma^{cb_{2}}\tief{\gamma}\hoch{\beta} & = & 8\gamma^{a_{1}b_{2}}\tief{\alpha}\hoch{\beta}+9\eta^{a_{1}b_{2}}\delta_{\alpha}^{\beta}\label{eq:gammaIIgammaIIContr}\end{eqnarray}
and therefore\begin{eqnarray}
\frac{1}{9}\gamma^{a_{1}}\tief{c\alpha}\hoch{\gamma}\left(\gamma^{cb_{2}}\tief{\gamma}\hoch{\beta}-8\eta^{cb_{2}}\delta_{\gamma}^{\beta}\right) & = & \eta^{a_{1}b_{2}}\delta_{\alpha}^{\beta}\label{eq:inverseOfGammatwo}\end{eqnarray}
If two indices in (\ref{eq:gammaIIgammaII}) are contracted, it turns
into \begin{eqnarray}
\gamma^{ab}\tief{\alpha}\hoch{\gamma}\gamma_{ba}\tief{\gamma}\hoch{\beta} & = & 90\delta_{\alpha}^{\beta}\label{eq:gammaIIgammaIIContrII}\end{eqnarray}
The equations (\ref{eq:gammaIIgammaIIContr}) and (\ref{eq:gammaIIgammaIIContrII})
are special cases of the following equations (which are in turn a
direct consequence of (\ref{eq:gammaIgammal}) and (\ref{eq:gammaIIgammalohneIndex})):\begin{eqnarray}
\gamma_{b_{1}}\gamma^{b_{1}\ldots b_{l}} & = & l\cdot\delta_{b_{1}}^{[b_{1}}\gamma^{b_{2}\ldots b_{l}]}=(11-l)\gamma^{b_{2}\ldots b_{l}},\qquad\gamma^{b_{1}\ldots b_{l}}\gamma_{b_{l}}=(11-l)\gamma^{b_{1}\ldots b_{l-1}}\label{eq:gammaIgammalContr}\\
\gamma^{a_{1}}\tief{b_{1}}\gamma^{b_{1}\ldots b_{l}} & = & (10-l)\cdot\gamma^{a_{1}b_{2}\ldots b_{l}}+\left(11-l\right)(l-1)\eta^{a_{1}[b_{2}}\gamma^{b_{3}\ldots b_{l}]}\label{eq:gammaIIgammalContrI}\\
\gamma\tief{b_{2}b_{1}}\gamma^{b_{1}\ldots b_{l}} & = & \left(11-l\right)\left(12-l\right)\gamma^{b_{3}\ldots b_{l}}\label{eq:gammaIIgammalContrII}\end{eqnarray}

\subsection{Hodge duality}

\index{Hodge duality!for chiral gamma matrices}In the intermezzo
on page \pageref{intermezzo:Hodge}, we had defined the Hodge star
operator such that it coincides with the multiplication of $\Gamma^{\#}$
from the right. Remember\begin{eqnarray}
\Gamma^{\#\,\q{\alpha}}\tief{\q{\beta}} & \equiv & \Gamma^{0\ldots9\,\q{\alpha}}\tief{\q{\beta}}=\left(\begin{array}{cc}
\one & 0\\
0 & -\one\end{array}\right)\end{eqnarray}
\begin{equation}
\Gamma^{\#}\Gamma^{a_{1}\ldots a_{p}}=\frac{1}{(10-p)!}(-)^{p(p+1)/2}\epsilon^{a_{1}\ldots a_{p}}\tief{c_{1}\ldots c_{10-p}}\Gamma^{c_{1}\ldots c_{10-p}}=\frac{1}{(10-p)!}\Gamma^{c_{10}\ldots c_{p+1}}\epsilon_{c_{10}\ldots c_{p+1}}\hoch{a_{p}\dots a_{1}}\end{equation}
The chiral blocks of $\Gamma^{\#}$coincide either with plus or minus
the unit matrix:\begin{eqnarray}
\gamma^{\#\,\alpha}\tief{\beta}\equiv\gamma^{0\ldots9\,\alpha}\tief{\beta} & = & \delta_{\beta}^{\alpha}=\frac{1}{10!}\epsilon_{c_{1}\ldots c_{10}}\gamma^{c_{1}\ldots c_{10}\,\alpha}\tief{\beta}\quad\mbox{with }\epsilon_{01\ldots9}\equiv1\\
\gamma_{\alpha}^{\#}\hoch{\beta}\equiv\gamma^{0\ldots9}\tief{\alpha}\hoch{\beta} & = & -\delta_{\alpha}^{\beta}=\frac{1}{10!}\epsilon_{c_{1}\ldots c_{10}}\gamma^{c_{1}\ldots c_{10}}\tief{\alpha}\hoch{\beta}\end{eqnarray}
Any chiral block $\gamma^{[p]}$ of $\Gamma^{[p]}$ is therefore always
equal (not only {}``Hodge-dual'') to a $\gamma^{[10-p]}$:{\footnotesize \begin{eqnarray}
\gamma^{a_{1}\ldots a_{2k}\,\alpha}\tief{\beta} & = & \frac{1}{(10-2k)!}(-)^{k}\epsilon^{a_{1}\ldots a_{2k}}\tief{c_{1}\ldots c_{10-2k}}\gamma^{c_{1}\ldots c_{10-2k}\,\alpha}\tief{\beta}=\frac{1}{(10-2k)!}\gamma^{c_{10}\ldots c_{2k+1}\,\alpha}\tief{\beta}\epsilon_{c_{10}\ldots c_{2k+1}}\hoch{a_{2k}\ldots a_{1}}\label{eq:chiralHodgeI}\\
-\gamma^{a_{1}\ldots a_{2k}}\tief{\alpha}\hoch{\beta} & = & \frac{1}{(10-2k)!}(-)^{k}\epsilon^{a_{1}\ldots a_{2k}}\tief{c_{1}\ldots c_{10-2k}}\gamma^{c_{1}\ldots c_{10-2k}}\tief{\alpha}\hoch{\beta}=\frac{1}{(10-2k)!}\gamma^{c_{10}\ldots c_{2k+1}}\tief{\alpha}\hoch{\beta}\epsilon_{c_{10}\ldots c_{2k+1}}\hoch{a_{2k}\ldots a_{1}}\\
\gamma^{a_{1}\ldots a_{2k+1}\,\alpha\beta} & = & \frac{1}{(9-2k)!}(-)^{(k+1)}\epsilon^{a_{1}\ldots a_{2k+1}}\tief{c_{1}\ldots c_{9-2k}}\gamma^{c_{1}\ldots c_{9-2k}\,\alpha\beta}=\frac{1}{(9-2k)!}\gamma^{c_{10}\ldots c_{2k+2}\,\alpha\beta}\epsilon_{c_{10}\ldots c_{2k+2}}\hoch{a_{2k+1}\dots a_{1}}\qquad\\
-\gamma_{\alpha\beta}^{a_{1}\ldots a_{2k+1}} & = & \frac{1}{(9-2k)!}(-)^{(k+1)}\epsilon^{a_{1}\ldots a_{2k+1}}\tief{c_{1}\ldots c_{9-2k}}\gamma_{\alpha\beta}^{c_{1}\ldots c_{9-2k}}=\frac{1}{(9-2k)!}\gamma_{\alpha\beta}^{c_{10}\ldots c_{2k+2}}\epsilon_{c_{10}\ldots c_{2k+2}}\hoch{a_{2k+1}\dots a_{1}}\label{eq:chiralHodgeIV}\end{eqnarray}
}In particular this leads to a self\index{self duality of $\gamma^{[5]}$}
duality constraint for $\gamma^{[5]}$:\begin{eqnarray}
\gamma^{a_{1}\ldots a_{5}\,\alpha\beta} & = & -\frac{1}{5!}\epsilon^{a_{1}\ldots a_{5}}\tief{c_{1}\ldots c_{5}}\gamma^{c_{1}\ldots c_{5}\,\alpha\beta}\label{eq:selfDualityGammaFiveI}\\
\gamma_{\alpha\beta}^{a_{1}\ldots a_{5}} & = & \frac{1}{5!}\epsilon^{a_{1}\ldots a_{5}}\tief{c_{1}\ldots c_{5}}\gamma_{\alpha\beta}^{c_{1}\ldots c_{5}}\label{eq:selfDualityGammaFiveII}\end{eqnarray}
This is the same behaviour as for the $\Gamma^{[p]}$'s themselves
in odd dimensions, where $\Gamma^{\#}$ coincides with the unit matrix.
This means that a bispinor with two chiral indices cannot just be
seen as a sum of odd (same chirality) or even (opposite chirality)
forms, but as a self-dual sum of odd an even forms. This is also further
discussed in the intermezzo on RR-fields on page \pageref{intermezzo:RR}. 

For the five-form we had $\Gamma^{\#}\Gamma^{a_{1}\ldots a_{5}}\otimes\Gamma_{a_{5}\ldots a_{1}}\Gamma^{\#}=\Gamma_{d_{1}\ldots d_{5}}\otimes\Gamma^{d_{1}\ldots d_{5}}$,
which turns into $-\gamma^{a_{1}\ldots a_{5}\,\alpha\beta}\gamma_{a_{5}\ldots a_{1}}^{\gamma\delta}=\gamma_{d_{1}\ldots d_{5}}^{\alpha\beta}\gamma^{d_{1}\ldots d_{5}\,\gamma\delta}$
and $-\gamma_{\alpha\beta}^{a_{1}\ldots a_{5}}\gamma_{a_{5}\ldots a_{1}\,\gamma\delta}=\gamma_{d_{1}\ldots d_{5}\,\alpha\beta}\gamma_{\gamma\delta}^{d_{1}\ldots d_{5}}$
and thus \begin{equation}
\boxed{\gamma^{a_{1}\ldots a_{5}\,\alpha\beta}\gamma_{a_{5}\ldots a_{1}}^{\gamma\delta}=\gamma_{\alpha\beta}^{a_{1}\ldots a_{5}}\gamma_{a_{5}\ldots a_{1}\,\gamma\delta}=0}\label{eq:5formcontractionvanishes}\end{equation}

\subsection{Vanishing of gamma\index{trace!of chiral gamma matrices}-traces
and projector\index{projector!for gamma matrix expansion}s for the
gamma-matrix expansion }

\label{sub:Vanishing-of-gamma-traces}For any even $p$ ($2\leq p\leq8$)
we have \begin{eqnarray}
\gamma^{a_{1}\ldots a_{p}\,\alpha}\tief{\alpha} & = & 0,\quad2\leq p\leq8,\, p\mbox{ even}\label{eq:gammatracelessness}\end{eqnarray}
The reason is that there is no invariant constant tensor with $p$
antisymmetrized indices apart from the $\epsilon$-tensor for $p=10$
and the Kronecker delta for $p=0$:\begin{equation}
\gamma^{a_{1}\ldots a_{10}}\tief{\alpha}\hoch{\alpha}=-\gamma^{a_{1}\ldots a_{10}\,\alpha}\tief{\alpha}=16\epsilon^{a_{1}\ldots a_{10}},\qquad\gamma^{[0]}\tief{\alpha}\hoch{\alpha}\equiv\gamma^{[0]\,\alpha}\tief{\alpha}\equiv\delta_{\alpha}^{\alpha}=16\qquad\label{eq:gammaTenGammaZeroTrace}\end{equation}
With the same argument we get $\gamma_{\alpha\beta}^{a}\gamma_{b}^{\alpha\beta}\propto\delta_{b}^{a}$
and fixing the proportionality by taking the trace yields\begin{equation}
\gamma_{\alpha\beta}^{a}\gamma_{b}^{\beta\alpha}=16\delta_{b}^{a}\label{eq:gammagammaSpur}\end{equation}
Alternatively this can be derived from $\gamma_{\alpha\beta}^{a}\gamma^{b\,\beta\gamma}=\eta^{ab}\delta_{\alpha}^{\gamma}+\gamma^{ab}\tief{\alpha}\hoch{\gamma}$
(the Clifford algebra for the chiral blocks and thus a special case
of (\ref{eq:product-expansion-chiral})) together with (\ref{eq:gammatracelessness}).
In the same manner we get for all other forms (using (\ref{eq:product-expansion-chiral})
and (\ref{eq:gammatracelessness}))\begin{eqnarray}
\gamma_{\alpha\beta}^{a_{1}\ldots a_{p}}\gamma_{b_{p}\ldots b_{1}}^{\beta\alpha} & = & 16p!\delta_{b_{1}\ldots b_{p}}^{a_{1}\ldots a_{p}}\quad\mbox{for }p\in\{1,3\}\label{eq:gammapgammapSpurOdd}\\
\gamma_{\alpha\beta}^{a_{1}\ldots a_{5}}\gamma_{b_{5}\ldots b_{1}}^{\beta\alpha} & = & 16\epsilon^{a_{1}\ldots a_{5}}\tief{b_{5}\ldots b_{1}}+16\cdot5!\delta_{b_{1}\ldots b_{5}}^{a_{1}\ldots a_{5}}\label{eq:gammaFiveGammaFiveSpur}\\
\gamma^{a_{1}\ldots a_{p}\,\alpha}\tief{\beta}\gamma_{b_{p}\ldots b_{1}}\hoch{\beta}\tief{\alpha} & = & 16p!\delta_{b_{1}\ldots b_{p}}^{a_{1}\ldots a_{p}}\quad\mbox{for }p\in\{2,4\}\label{eq:gammapgammapSpurEven}\end{eqnarray}
The extra term in the $\gamma^{[5]}\gamma_{[5]}$ contraction on
the righthand side of the second line is due to the fact that the
trace of $\gamma^{[10]}$ does not vanish according to (\ref{eq:gammaTenGammaZeroTrace}).
Any other contraction, where the number of bosonic indices does not
match, vanishes\begin{eqnarray}
\tr(\gamma^{[p]}\gamma^{[q]}) & = & 0\quad\mbox{for }p\neq q,\mbox{ and }p,q\leq5\end{eqnarray}
The results of above can be used to project to the coefficients of
$\gamma$-matrix expansions:\begin{eqnarray}
A^{\alpha\beta} & = & A_{a}\gamma^{a\,\alpha\beta}+A_{a_{1}a_{2}a_{3}}\gamma^{a_{1}a_{2}a_{3}\,\alpha\beta}+A_{a_{1}\ldots a_{5}}\gamma^{a_{1}\ldots a_{5}\,\alpha\beta},\nonumber \\
 &  & \mbox{with }A_{a_{1}\ldots a_{p}}=\frac{1}{16p!}\gamma_{a_{p}\ldots a_{1}\,\beta\alpha}A^{\alpha\beta}\mbox{ for }p\in\{1,3\}\mbox{ and }A_{a_{1}\ldots a_{5}}=\frac{1}{32\cdot5!}\gamma_{a_{5}\ldots a_{1}\,\beta\alpha}A^{\alpha\beta}\qquad\label{eq:gamma:gammaexpansionOddI}\\
D_{\alpha\beta} & = & D_{a}\gamma_{\alpha\beta}^{a}+D_{a_{1}a_{2}a_{3}}\gamma_{\alpha\beta}^{a_{1}a_{2}a_{3}}+D_{a_{1}\ldots a_{5}}\gamma_{\alpha\beta}^{a_{1}\ldots a_{5}},\nonumber \\
 &  & \mbox{with }D_{a_{1}\ldots a_{p}}=\frac{1}{16p!}\gamma_{a_{p}\ldots a_{1}}^{\beta\alpha}D_{\alpha\beta}\mbox{ for }p\in\{1,3\}\mbox{ and }D_{a_{1}\ldots a_{5}}=\frac{1}{32\cdot5!}\gamma_{a_{5}\ldots a_{1}}^{\beta\alpha}D_{\alpha\beta}\qquad\label{eq:gamma:gammaexpansionOddII}\\
B^{\alpha}\tief{\beta} & = & B_{[0]}\delta_{\beta}^{\alpha}+B_{a_{1}a_{2}}\gamma^{a_{1}a_{2}\,\alpha}\tief{\beta}+B_{a_{1}a_{2}a_{3}a_{4}}\gamma^{a_{1}a_{2}a_{3}a_{4}\,\alpha}\tief{\beta},\qquad B_{a_{1}\ldots a_{p}}=\frac{1}{16p!}\gamma_{a_{p}\ldots a_{1}}\hoch{\beta}\tief{\alpha}B^{\alpha}\tief{\beta}\label{eq:gamma:gammaexpansionEven}\\
C_{\alpha}\hoch{\beta} & = & C_{[0]}\delta_{\alpha}^{\beta}+C_{a_{1}a_{2}}\gamma^{a_{1}a_{2}}\tief{\alpha}\hoch{\beta}+C_{a_{1}a_{2}a_{3}a_{4}}\gamma^{a_{1}a_{2}a_{3}a_{4}}\tief{\alpha}\hoch{\beta},\qquad C_{a_{1}\ldots a_{p}}=\frac{1}{16p!}\gamma_{a_{p}\ldots a_{1}\beta}\hoch{\alpha}C_{\alpha}\hoch{\beta}\label{eq:gamma:gammaexpansionEvenAltern}\end{eqnarray}
For the first two expansions it was used that due to the restrictions
(\ref{eq:selfDualityGammaFiveI}) and (\ref{eq:selfDualityGammaFiveII})
on $\gamma^{[5]}$, the corresponding expansion coefficients can always
be chosen to obey (anti) self-duality constraints of the form \begin{eqnarray}
A_{a_{1}\ldots a_{5}} & = & -\frac{1}{5!}A_{c_{1}\ldots c_{5}}\epsilon^{c_{1}\ldots c_{5}}\tief{a_{1}\ldots a_{5}}\\
D_{a_{1}\ldots a_{5}} & = & \frac{1}{5!}D_{c_{1}\ldots c_{5}}\epsilon^{c_{1}\ldots c_{5}}\tief{a_{1}\ldots a_{5}}\end{eqnarray}
which lead together with (\ref{eq:gammaFiveGammaFiveSpur}) to an
extra factor of two and thus to a normalization factor $\tfrac{1}{32}$
instead of $\frac{1}{16}$ for $p=5$.

\subsection{Chiral Fierz}

\label{sub:chiralFierz}Remember \begin{equation}
\sum_{p=0}^{10}\frac{1}{32p!}\Gamma^{a_{1}\ldots a_{p}\,\q{\alpha}}\tief{\q{\beta}}\Gamma_{a_{p}\ldots a_{1}}\hoch{\q{\gamma}}\tief{\q{\delta}}=\delta_{\q{\delta}}^{\q{\alpha}}\delta_{\q{\beta}}^{\q{\gamma}}\end{equation}
or\begin{equation}
\sum_{p=0}^{4}\frac{1}{32p!}\left(\Gamma^{a_{1}\ldots a_{p}\,\q{\alpha}}\tief{\q{\beta}}\Gamma_{a_{p}\ldots a_{1}}\hoch{\q{\gamma}}\tief{\q{\delta}}+(\Gamma^{\#}\Gamma^{a_{1}\ldots a_{p}})^{\q{\alpha}}\tief{\q{\beta}}(\Gamma_{a_{p}\ldots a_{1}}\Gamma^{\#})\hoch{\q{\gamma}}\tief{\q{\delta}}\right)+\frac{1}{32\cdot5!}\Gamma^{a_{1}\ldots a_{5}\,\q{\alpha}}\tief{\q{\beta}}\Gamma_{a_{5}\ldots a_{1}}\hoch{\q{\gamma}}\tief{\q{\delta}}=\delta_{\q{\delta}}^{\q{\alpha}}\delta_{\q{\beta}}^{\q{\gamma}}\end{equation}
We want to make a distinction of the different cases corresponding
to the chiral indices:\begin{eqnarray}
\sum_{p\in\{0,2,4\}}\frac{1}{16p!}\left(\gamma^{a_{1}\ldots a_{p}\,\alpha}\tief{\beta}\gamma_{a_{p}\ldots a_{1}}\hoch{\gamma}\tief{\delta}\right) & = & \delta_{\delta}^{\alpha}\delta_{\beta}^{\gamma}\\
0\cdot+\frac{-4}{4}\sum_{p\in\{1,3\}}\frac{1}{16p!}\gamma^{a_{1}\ldots a_{p}\,\alpha\beta}\gamma_{a_{p}\ldots a_{1}}\hoch{\gamma\delta}+\frac{1}{32\cdot5!}\underbrace{\gamma^{a_{1}\ldots a_{5}\,\alpha\beta}\gamma_{a_{5}\ldots a_{1}}\hoch{\gamma\delta}}_{=0} & = & 0\\
0\cdot\sum_{p\in\{1,3\}}\frac{1}{16p!}\gamma^{a_{1}\ldots a_{p}}\tief{\alpha\beta}\gamma_{a_{p}\ldots a_{1}\,\gamma\delta}+\frac{1}{32\cdot5!}\underbrace{\gamma^{a_{1}\ldots a_{5}}\tief{\alpha\beta}\gamma_{a_{5}\ldots a_{1}\,\gamma\delta}}_{=0} & = & 0\\
\sum_{p\in\{1,3\}}\frac{1}{16p!}\gamma^{a_{1}\ldots a_{p}\,\alpha\beta}\gamma_{a_{p}\ldots a_{1}\,\gamma\delta}+\frac{1}{32\cdot5!}\gamma^{a_{1}\ldots a_{5}\,\alpha\beta}\gamma_{a_{5}\ldots a_{1}\,\gamma\delta} & = & \delta_{\delta}^{\alpha}\delta_{\gamma}^{\beta}\end{eqnarray}
Only the first and the last give nontrivial information. \begin{eqnarray}
\delta_{\beta}^{\alpha}\delta_{\delta}^{\gamma}+\frac{1}{2}\gamma^{a_{1}a_{2}\,\alpha}\tief{\beta}\gamma_{a_{2}a_{1}}\hoch{\gamma}\tief{\delta}+\frac{1}{4!}\gamma^{a_{1}a_{2}a_{3}a_{4}\,\alpha}\tief{\beta}\gamma_{a_{4}a_{3}a_{2}a_{1}}\hoch{\gamma}\tief{\delta} & = & 16\delta_{\delta}^{\alpha}\delta_{\beta}^{\gamma}\label{eq:chiralFierzI}\\
\gamma^{a\,\alpha\beta}\gamma_{a\,\gamma\delta}+\frac{1}{3!}\gamma^{a_{1}a_{2}a_{3}\,\alpha\beta}\gamma_{a_{3}a_{2}a_{1}\,\gamma\delta}+\frac{1}{2\cdot5!}\gamma^{a_{1}\ldots a_{5}\,\alpha\beta}\gamma_{a_{5}\ldots a_{1}\,\gamma\delta} & = & 16\delta_{\delta}^{\alpha}\delta_{\gamma}^{\beta}\label{eq:chiralFierzII}\end{eqnarray}
Contracting $\gamma,\delta$ in (\ref{eq:chiralFierzI}) yields $16\delta_{\beta}^{\alpha}=16\delta_{\beta}^{\alpha}$,
contracting $\gamma,\beta$ instead, yields%
\footnote{\index{footnote!\thefoot. combinatorical consistency check}As a consitency
check we can in addition contract $\alpha,\delta$ and get for the
first Fierz \begin{eqnarray*}
16+16\frac{1}{2}2!\delta_{a_{1}a_{2}}^{a_{1}a_{2}}+16\frac{1}{4!}4!\delta_{a_{1}\ldots a_{4}}^{a_{1}\ldots a_{4}} & = & (16)^{3}\\
1+\underbrace{\left(\zwek{10}{2}\right)}_{45}+\underbrace{\left(\zwek{10}{4}\right)}_{210} & = & (16)^{2}=256\end{eqnarray*}
and for the second one\[
10+\underbrace{\left(\zwek{10}{3}\right)}_{120}+\frac{1}{2}\underbrace{\left(\zwek{10}{5}\right)}_{252}=256\qquad\fussend\]
}\begin{eqnarray}
\delta_{\delta}^{\alpha}+\frac{1}{2}\gamma^{a_{1}a_{2}\,\alpha}\tief{\gamma}\gamma_{a_{2}a_{1}}\hoch{\gamma}\tief{\delta}+\frac{1}{4!}\gamma^{a_{1}a_{2}a_{3}a_{4}\,\alpha}\tief{\gamma}\gamma_{a_{4}a_{3}a_{2}a_{1}}\hoch{\gamma}\tief{\delta} & = & (16)^{2}\delta_{\delta}^{\alpha}\\
\underbrace{\gamma^{a\,\alpha\beta}\gamma_{a\,\beta\delta}}_{10\delta_{\delta}^{\alpha}}+\frac{1}{3!}\gamma^{a_{1}a_{2}a_{3}\,\alpha\beta}\gamma_{a_{3}a_{2}a_{1}\,\beta\delta}+\frac{1}{2\cdot5!}\gamma^{a_{1}\ldots a_{5}\,\alpha\beta}\gamma_{a_{5}\ldots a_{1}\,\beta\delta} & = & (16)^{2}\delta_{\delta}^{\alpha}\end{eqnarray}
We can also contract (\ref{eq:chiralFierzI}) with $\gamma_{\alpha\rho}^{b}\gamma_{b\,\gamma\sigma}$
to arrive at\begin{eqnarray}
0 & = & \gamma_{\beta\rho}^{b}\gamma_{b\,\delta\sigma}+\frac{1}{2}\underbrace{\gamma^{a_{1}a_{2}\,\alpha}\tief{\beta}\gamma_{\alpha\rho}^{b}}_{\gamma^{[3]}+\gamma^{[1]}}\underbrace{\gamma_{b\,\gamma\sigma}\gamma_{a_{2}a_{1}}\hoch{\gamma}\tief{\delta}}_{\gamma_{[3]}+\gamma_{[1]}}+\frac{1}{4!}\underbrace{\gamma^{a_{1}a_{2}a_{3}a_{4}\,\alpha}\tief{\beta}\gamma_{\alpha\rho}^{b}}_{\gamma^{[5]}+\gamma^{[3]}}\underbrace{\gamma_{b\,\gamma\sigma}\gamma_{a_{4}a_{3}a_{2}a_{1}}\hoch{\gamma}\tief{\delta}}_{\gamma_{[5]}+\gamma_{[3]}}-16\gamma_{\delta\rho}^{b}\gamma_{b\,\beta\sigma}\end{eqnarray}
Now we use that $\gamma^{[3]}$is antisymmetric in $\beta\rho$ and
that $\gamma^{[5]}\gamma_{[5]}=0$ (mixed terms like $\gamma^{[5]}\gamma_{[3]}$
also vanish, because some $\eta$ are contracted with antisymmetric
indices of $\gamma^{[5]}$). Symmetrizing the above equation in $\beta\rho$
yields\begin{eqnarray}
0 & = & \gamma_{\beta\rho}^{b}\gamma_{b\,\delta\sigma}+2\eta^{b[a_{1}}\gamma_{\rho\beta}^{a_{2}]}\eta_{b[a_{2}}\gamma_{a_{1}]\,\sigma\delta}-16\gamma_{\delta(\rho|}^{b}\gamma_{b\,|\beta)\sigma}=\nonumber \\
 & = & \gamma_{\beta\rho}^{b}\gamma_{b\,\delta\sigma}+2\delta_{a_{2}}^{[a_{1}}\gamma_{\rho\beta}^{a_{2}]}\gamma_{a_{1}\,\sigma\delta}-16\gamma_{\delta(\rho|}^{b}\gamma_{b\,|\beta)\sigma}=\nonumber \\
 & = & \gamma_{\beta\rho}^{b}\gamma_{b\,\delta\sigma}+\delta_{a_{2}}^{a_{1}}\gamma_{\rho\beta}^{a_{2}}\gamma_{a_{1}\,\sigma\delta}-\delta_{a_{2}}^{a_{2}}\gamma_{\rho\beta}^{a_{1}}\gamma_{a_{1}\,\sigma\delta}-16\gamma_{\delta(\rho|}^{b}\gamma_{b\,|\beta)\sigma}=\nonumber \\
 & = & \gamma_{\beta\rho}^{b}\gamma_{b\,\delta\sigma}+\gamma_{\rho\beta}^{a}\gamma_{a\,\sigma\delta}-10\gamma_{\rho\beta}^{a_{1}}\gamma_{a_{1}\,\sigma\delta}-16\gamma_{\delta(\rho|}^{b}\gamma_{b\,|\beta)\sigma}=\nonumber \\
 & = & -8\gamma_{\beta\rho}^{b}\gamma_{b\,\delta\sigma}-16\gamma_{\delta(\rho|}^{b}\gamma_{b\,|\beta)\sigma}\end{eqnarray}
\index{Fierz identity!chiral $\sim$}\index{little Fierz}\index{chiral Fierz identity}\begin{equation}
\boxed{\gamma_{(\beta\rho|}^{b}\gamma_{b\,|\delta)\sigma}=0}\label{eq:LittleFierz}\end{equation}
We could have used directly equation (\ref{eq:FierzIIIten}) to derive
this result. This is a very important identity because it is so simple
and can be used to derive many other identities. One example will
be useful for us in the main part. Consider the contraction of the
bosonic indices of two $\gamma^{[2]}$'s: \begin{eqnarray}
\gamma^{ab\,\alpha}\tief{\beta}\gamma_{ab}\hoch{\gamma}\tief{\delta} & = & \left(\gamma^{a\,\alpha\rho}\gamma_{\rho\beta}^{b}-\eta^{ab}\delta_{\beta}^{\alpha}\right)\left(\gamma_{a}^{\gamma\sigma}\gamma_{b\,\sigma\delta\,}-\eta_{ab}\delta_{\delta}^{\gamma}\right)=\\
 & = & \gamma^{a\,\alpha\rho}\gamma_{a}^{\gamma\sigma}\gamma_{\rho\beta}^{b}\gamma_{b\,\sigma\delta\,}-\gamma^{a\,\alpha\rho}\gamma_{a\,\rho\beta}\delta_{\delta}^{\gamma}-\gamma^{b\,\gamma\sigma}\gamma_{b\,\sigma\delta\,}\delta_{\beta}^{\alpha}+10\delta_{\beta}^{\alpha}\delta_{\delta}^{\gamma}\end{eqnarray}
In order to make use of $(\ref{eq:LittleFierz})$ we symmetrize the
lower spinorial indices and obtain \begin{eqnarray}
\gamma^{ab\,\alpha}\tief{(\beta|}\gamma_{ab}\hoch{\gamma}\tief{|\delta)} & = & \left(\gamma^{a\,\alpha\rho}\gamma_{\rho\beta}^{b}-\eta^{ab}\delta_{\beta}^{\alpha}\right)\left(\gamma_{a}^{\gamma\sigma}\gamma_{b\,\sigma\delta\,}-\eta_{ab}\delta_{\delta}^{\gamma}\right)=\\
 & = & \gamma^{a\,\alpha\rho}\gamma_{a}^{\gamma\sigma}\underbrace{\gamma_{\rho(\beta|}^{b}\gamma_{b\,|\delta)\sigma\,}}_{-\tfrac{1}{2}\gamma_{\beta\delta}^{b}\gamma_{b\,\rho\sigma}\,(\ref{eq:LittleFierz})}-\underbrace{\gamma^{a\,\alpha\rho}\gamma_{a\,\rho(\beta}}_{10\delta_{(\beta}^{\alpha}}\delta_{\delta)}^{\gamma}-\underbrace{\gamma^{b\,\gamma\sigma}\gamma_{b\,\sigma(\delta\,}}_{10\delta_{(\delta}^{\gamma}}\delta_{\beta)}^{\alpha}+10\delta_{(\beta}^{\alpha}\delta_{\delta)}^{\gamma}=\\
 & = & -\tfrac{1}{2}\underbrace{\gamma^{a\,\alpha\rho}\gamma_{b\,\rho\sigma}\gamma_{a}^{\sigma\gamma}}_{-8\gamma_{b}^{\alpha\gamma}\,(\ref{eq:d-2l})}\gamma_{\beta\delta}^{b}-10\delta_{(\beta}^{\alpha}\delta_{\delta)}^{\gamma}\end{eqnarray}
We can thus express $\gamma^{[2]}\gamma_{[2]}$ by $\gamma^{[1]}\gamma_{[1]}$
and Kronecker deltas\begin{equation}
\boxed{\gamma^{ab\,\alpha}\tief{(\beta|}\gamma_{ab}\hoch{\gamma}\tief{|\delta)}=4\gamma_{a}^{\alpha\gamma}\gamma_{\beta\delta}^{a}-10\delta_{(\beta}^{\alpha}\delta_{\delta)}^{\gamma}}\label{eq:FromLittleFierz}\end{equation}
\rem{

\section{with graded summation convention}

In the \textbf{graded conventions}, we understand $\bs{\alpha}$ to
be graded. The definition of matrix multiplication and the Kronecker
delta changes and also the notation of symmetric and antisymmetric
switches. Let us study NW-conventions\begin{eqnarray}
A_{\q{\bs{\alpha}}\q{\bs{\beta}}} & = & \left(\begin{array}{cc}
0 & \delta_{\bs{\alpha}}^{\:\bs{\beta}}\\
-(-)^{\bs{\alpha\beta}}\delta_{\bs{\beta}}^{\:\bs{\alpha}} & 0\end{array}\right)=\left(\begin{array}{cc}
0 & \delta_{\bs{\alpha}}^{\:\bs{\beta}}\\
-\delta_{\:\bs{\beta}}^{\bs{\alpha}} & 0\end{array}\right)\\
\gamma^{(a|\,\bs{\alpha\gamma}}\gamma_{\bs{\gamma\beta}}^{|b)} & = & -\eta^{ab}\delta^{\bs{\alpha}}\tief{\bs{\beta}},\quad\gamma^{(a|\,\bs{\alpha\beta}}\gamma_{\bs{\beta\alpha}}^{|b)}=-\eta^{ab}\delta^{\bs{\alpha}}\tief{\bs{\alpha}}=16\eta^{ab}\\
\gamma_{\bs{\alpha\gamma}}^{(b}\gamma^{a)\,\bs{\gamma\beta}} & = & -\eta^{ab}\delta_{\bs{\alpha}}\hoch{\bs{\beta}},\quad\gamma_{\bs{\alpha\beta}}^{(b}\gamma^{a)\,\bs{\beta\alpha}}=-\eta^{ab}\delta_{\bs{\alpha}}\hoch{\bs{\alpha}}=16\eta^{ab}\\
\dann\Gamma^{(a|\,\q{\bs{\alpha}}}\tief{\q{\bs{\gamma}}}\Gamma^{|b)\,\q{\bs{\gamma}}}\tief{\q{\bs{\beta}}} & = & -\eta^{ab}\delta^{\q{\bs{\alpha}}}\tief{\q{\bs{\beta}}}\\
(\gamma^{[a_{1}}\cdots\gamma^{a_{2k}]})^{\bs{\alpha}}\tief{\bs{\beta}} & = & (-)^{k-1}\gamma^{a_{1}\ldots a_{k}\,\bs{\alpha}}\tief{\bs{\beta}}\\
(\gamma^{[a_{1}}\cdots\gamma^{a_{2k+1}]})^{\bs{\alpha}\bs{\beta}} & = & (-)^{k}\gamma^{a_{1}\ldots a_{k}\,\bs{\alpha\beta}}\\
(\gamma^{[a_{1}}\cdots\gamma^{a_{2k}]})_{\bs{\beta}}\hoch{\bs{\alpha}} & = & (-)^{k}\gamma^{a_{1}\ldots a_{2k}}\tief{\bs{\beta}}\hoch{\bs{\alpha}}\\
(\gamma^{[a_{1}}\cdots\gamma^{a_{2k+1}]})_{\bs{\beta\alpha}} & = & (-)^{k}\gamma^{a_{1}\ldots a_{2k+1}}\tief{\bs{\beta\alpha}}\\
\gamma^{a_{1}\ldots a_{2k}\,\bs{\alpha}}\tief{\bs{\beta}} & \greq & -(-)^{k}\gamma^{a_{1}\ldots a_{2k}}\tief{\bs{\beta}}\hoch{\bs{\alpha}}\quad\left(\gamma^{a_{1}a_{2}\,\bs{\alpha}}\tief{\bs{\beta}}\greq\gamma^{a_{1}a_{2}}\tief{\bs{\beta}}\hoch{\bs{\alpha}},\gamma^{[0]\,\bs{\alpha}}\tief{\bs{\beta}}=-\delta^{\bs{\alpha}}\tief{\bs{\beta}}\greq-\gamma^{[0]}\tief{\bs{\beta}}\hoch{\bs{\alpha}}\right)\\
(\gamma^{[a_{1}}\cdots\gamma^{a_{2k}]})^{\bs{\alpha}}\tief{\bs{\beta}} & \greq & (-)^{k}(\gamma^{[a_{1}}\cdots\gamma^{a_{2k}]})_{\bs{\beta}}\hoch{\bs{\alpha}}\\
\gamma_{\bs{\alpha\beta}}^{a_{1}\ldots a_{2k+1}} & \greq & -(-)^{k}\gamma_{\bs{\beta\alpha}}^{a_{1}\ldots a_{2k+1}},\qquad\gamma^{a_{1}\ldots a_{2k+1}\,\bs{\alpha\beta}}=-(-)^{k}\gamma^{a_{1}\ldots a_{2k+1}\,\bs{\beta\alpha}}\\
(\gamma^{[a_{1}}\cdots\gamma^{a_{2k+1}]})_{\bs{\beta\alpha}} & \greq & -(-)^{k}(\gamma^{[a_{1}}\cdots\gamma^{a_{2k+1}]})_{\bs{\alpha\beta}},\quad(\gamma^{[a_{1}}\cdots\gamma^{a_{2k+1}]})^{\bs{\beta\alpha}}=-(-)^{k}(\gamma^{[a_{1}}\cdots\gamma^{a_{2k+1}]})^{\bs{\alpha\beta}}\end{eqnarray}
}

\printindex{}

\bibliographystyle{fullsort}
\bibliography{phd,Proposal}

}

\chapter{Noether}

\label{cha:Noether}\rem{see WZNW-NW.lyx (and WZNW.lyx), almost nothing in noether.lyx}{\inputTeil{0}
\ifthenelse{\theinput=1}{}{}

\title{Noether}

\author{Sebastian Guttenberg}

\date{last modified January 30, 2009}

\maketitle
\begin{abstract}
Part of thesis
\end{abstract}
\tableofcontents{}

\rem{To do:

\begin{itemize}
\item keep cool\newpage
\end{itemize}
}\index{Noether|fett}

\section{Noether's theorem and the inverse Noether method}

\label{sec:NoetherLagrange}Most of the following presentation is
based on \cite[p.67f, p.95]{Henneaux:1992ig}, although somewhat modified.
Consider an action of the quite general form \begin{eqnarray}
S[\allfields{I}] & \equiv & \int d^{n}\sigma\quad\mc{L}(\allfields{I},\partial_{\mu}\allfields{I},\partial_{\mu_{1}}\partial_{\mu_{2}}\allfields{I},\ldots)\end{eqnarray}
\index{$\Phi$@$\allfields{I}$}In most of the applications there appear
no higher derivatives than $\partial_{\mu}\allfields{I}$. Let us
treat global and local symmetries at the same time and consider a
symmetry transformation with infinitesimal transformation parameter
$\rho(\sigma)$.\rem{ \begin{eqnarray}
\rho^{a}(\sigma) & \equiv & \rho_{0}^{a}+\sigma^{\mu}\rho_{\mu}^{a}+\frac{1}{2}\sigma^{\mu_{1}}\sigma^{\mu_{2}}\rho_{\mu_{1}\mu_{2}}^{a}+\ldots\end{eqnarray}
} The transformation can be expanded in derivatives of the transformation
parameter:\begin{eqnarray}
\delta_{(\rho)}\allfields{I} & \equiv & \underbrace{\rho^{a}\delta_{a}\allfields{I}}_{\delta_{(\rho)}^{0}\allfields{I}}+\underbrace{\partial_{\mu}\rho^{a}\delta_{a}^{\mu}\allfields{I}}_{\delta_{(\rho)}^{1}\allfields{I}}+\underbrace{\partial_{\mu_{1}}\partial_{\mu_{2}}\rho^{a}\delta_{a}^{\mu_{1}\mu_{2}}\allfields{I}}_{\delta_{(\rho)}^{2}\allfields{I}}+\ldots\label{eq:noet:deltaPhiExp}\end{eqnarray}
\rem{\begin{eqnarray*}
 & = & \rho^{a}\delta_{a}\allfields{I}+\left(\rho_{\mu}^{a}+\sigma^{\mu_{2}}\rho_{\mu\mu_{2}}^{a}+\ldots\right)\delta_{a}^{\mu}\allfields{I}+\left(\rho_{\mu_{1}\mu_{2}}^{a}+\sigma^{\mu_{3}}\rho_{\mu_{1}\mu_{2}\mu_{3}}^{a}+\ldots\right)\delta_{a}^{\mu_{1}\mu_{2}}\allfields{I}+\ldots=\\
 & = & \rho^{a}\delta_{a}\allfields{I}+\rho_{\mu}^{a}\delta_{a}^{\mu}\allfields{I}+\rho_{\mu_{1}\mu_{2}}^{a}\left(\sigma^{\mu_{2}}\delta_{a}^{\mu_{1}}\allfields{I}+\delta_{a}^{\mu_{1}\mu_{2}}\allfields{I}\right)+\rho_{\mu_{1}\mu_{2}\mu_{3}}^{a}\left(\frac{1}{2}\sigma^{\mu_{2}}\sigma^{\mu_{3}}\delta_{a}^{\mu_{1}}\allfields{I}+\sigma^{\mu_{3}}\rho_{\mu_{1}\mu_{2}\mu_{3}}^{a}\delta_{a}^{\mu_{1}\mu_{2}}\allfields{I}+\rho_{\mu_{1}\mu_{2}\mu_{3}}^{a}\delta_{a}^{\mu_{1}\mu_{2}\mu_{3}}\allfields{I}\right)+\ldots=\end{eqnarray*}
}In order to define properly the variational derivatives for this
more general case, consider first the variation of the Lagrangian%
\footnote{\label{foot:iteratedPartialIntegration} \index{footnote!\thefoot. iterated partial integration }In
(\ref{eq:noet:LagrangianVariation}) we have reformulated the variations
containing derivatives of the fields $\allfields{I}$ using schematically
the following iterated 'partial integration':\begin{eqnarray*}
\partial^{k}a\cdot b & = & \partial\left(\partial^{k-1}a\cdot b\right)-\partial^{k-1}a\cdot\partial b=\\
 & = & \partial\left(\partial^{k-1}a\cdot b\right)-\partial\left(\partial^{k-2}a\cdot\partial b\right)+\partial^{k-2}a\cdot\partial^{2}b=\\
 & = & \partial\left[\partial^{k-1}a\cdot b-\partial^{k-2}a\cdot\partial b+\ldots+(-)^{k-1}a\cdot\partial^{k-1}b\right]+(-)^{k}a\cdot\partial^{k}b=\\
 & = & \partial\left[\sum_{i=0}^{k-1}(-)^{i}\partial^{k-1-i}a\cdot\partial^{i}b\right]+(-)^{k}a\cdot\partial^{k}b\end{eqnarray*}
 \frem{\begin{eqnarray*}
\partial^{n}\left(a\cdot b\right) & = & \sum_{k=0}^{n}\left(\zwek{n}{k}\right)\partial^{k}a\cdot\partial^{n-k}b=\\
 & = & a\cdot\partial^{n}b+\sum_{k=1}^{n-1}\left(\zwek{n}{k}\right)\partial^{k}a\cdot\partial^{n-k}b+\partial^{n}a\cdot b\end{eqnarray*}
}This equation is applicable in (\ref{eq:noet:LagrangianVariation}),
because the indices of the partial derivatives are all contracted
and symmetrized and therefore behave like one-dimensional derivatives.
In our case the above formula takes the explicit form \begin{eqnarray*}
\lqn{\delta(\partial_{\mu_{1}}\ldots\partial_{\mu_{k}}\allfields{I})\cdot\partiell{\mc{L}}{(\partial_{\mu_{1}}\ldots\partial_{\mu_{k}}\allfields{I})}=}\\
 & = & \partial_{\mu}\Big[\sum_{i=0}^{k-1}(-)^{i}\partial_{\nu_{1}}\ldots\partial_{\nu_{k-i-1}}\delta\allfields{I}\cdot\partial_{\nu_{k-i}}\ldots\partial_{\nu_{k-1}}\partiell{\mc{L}}{(\partial_{\mu}\partial_{\nu_{1}}\ldots\partial_{\nu_{k-1}}\allfields{I})}\Big]+(-)^{k}\delta\allfields{I}\cdot\partial_{\mu_{1}}\ldots\partial_{\mu_{k}}\partiell{\mc{L}}{(\partial_{\mu_{1}}\ldots\partial_{\mu_{k}}\allfields{I})}\qquad\fussend\end{eqnarray*}
\frem{Boundary of boundary is zero $\dann$$(-)^{k-1}k\delta\allfields{I}\cdot\partial_{\mu_{2}}\ldots\partial_{\mu_{k}}\partiell{\mc{L}}{(\partial_{\mu_{1}}\ldots\partial_{\mu_{k}}\allfields{I})}$}%
} 

\begin{eqnarray}
\delta\mc{L} & = & \delta\allfields{I}\biggl(\partiell{\mc{L}}{\allfields{I}}-\partial_{\mu}\partiell{\mc{L}}{(\partial_{\mu}\allfields{I})}+\partial_{\mu_{1}}\partial_{\mu_{2}}\partiell{\mc{L}}{(\partial_{\mu_{1}}\partial_{\mu_{2}}\allfields{I})}-\ldots\biggr)+\nonumber \\
 &  & +\partial_{\mu}\biggl(\delta\allfields{I}\cdot\partiell{\mc{L}}{(\partial_{\mu}\allfields{I})}+\sum_{k\geq2}\sum_{i=0}^{k-1}(-)^{i}\partial_{\nu_{1}}\ldots\partial_{\nu_{k-1-i}}\delta\allfields{I}\cdot\partial_{\nu_{k-i}}\ldots\partial_{\nu_{k-1}}\partiell{\mc{L}}{(\partial_{\mu}\partial_{\nu_{1}}\ldots\partial_{\nu_{k-1}}\allfields{I})}\biggr)\qquad\label{eq:noet:LagrangianVariation}\end{eqnarray}
The total derivative term reduces to a boundary term in the variation
of the action, while the remaining term defines the variational derivative.
As the boundary of a boundary vanishes, one can further partially
integrate the boundary term in order to obtain a convenient form that
determines the boundary conditions:%
\footnote{\label{foot:Stokes-theorem}\index{Stokes' theorem}\index{theorem!Stokes}\index{footnote!\thefoot. Stokes' theorem}Stokes'
theorem reads \begin{eqnarray*}
\int_{\Sigma^{(n)}}\de\omega & = & \int_{\partial\Sigma}\omega^{(n-1)}\end{eqnarray*}
For any $\Sigma$ that can be covered by one single coordinate patch,
we can write \begin{eqnarray*}
\int_{\Sigma}\de\sigma^{\mu_{1}}\wedge\ldots\wedge\de\sigma^{\mu_{n}}\partial_{[\mu_{1}}\omega_{\mu_{2}\ldots\mu_{n}]} & = & \int_{\partial\Sigma}\de\sigma^{\mu_{1}}\wedge\ldots\wedge\de\sigma^{\mu_{n-1}}\omega_{\mu_{1}\ldots\mu_{n-1}}\end{eqnarray*}
where on the righthand side the coordinate differentials $\de\sigma^{\mu}$
have to be understood as pullbacks $\de\tau^{i}\partial_{i}\sigma^{\mu}(\tau)$
on the boundary. 

For the integral of a divergence term like \begin{eqnarray*}
\int_{\Sigma}d^{n}\sigma\,\partial_{\mu}v^{\mu} & \equiv & \int_{\Sigma}\de\sigma^{1}\wedge\ldots\wedge\de\sigma^{n}\,\partial_{\mu}v^{\mu}\frem{=\int_{\Sigma}\frac{1}{n!}\,\epsilon_{\mu_{1}\ldots\mu_{n}}\de\sigma^{\mu_{1}}\wedge\ldots\wedge\de\sigma^{\mu_{n}}\,\partial_{\nu}v^{\nu}}\end{eqnarray*}
we can use the fact that \begin{eqnarray*}
\de\sigma^{1}\wedge\ldots\wedge\de\sigma^{n}\,\partial_{\mu}v^{\mu} & = & \de\omega\end{eqnarray*}
with \[
\omega\equiv\frac{1}{(n-1)!}v^{\mu}\epsilon_{\mu\mu_{1}\ldots\mu_{n-1}}\de\sigma^{\mu_{1}}\wedge\ldots\wedge\de\sigma^{\mu_{n-1}}\]
\frem{\begin{eqnarray*}
\de\omega & \equiv & \frac{1}{(n-1)!}\partial_{[\mu_{1}|}v^{\mu}\epsilon_{\mu|\mu_{2}\ldots\mu_{n}]}\de\sigma^{\mu_{1}}\wedge\ldots\wedge\de\sigma^{\mu_{n}}=\\
 & = & \frac{1}{(n-1)!}n!\partial_{[1|}v^{\mu}\epsilon_{\mu|2\ldots n]}\de\sigma^{1}\wedge\ldots\wedge\de\sigma^{n}=\\
 & = & \left(\partial_{1}v^{1}\epsilon_{12\ldots n}+(-)^{n-1}\partial_{2}v^{2}\epsilon_{23\ldots n1}+\ldots\right)\de\sigma^{1}\wedge\ldots\wedge\de\sigma^{n}=\\
 & = & \partial_{\nu}v^{\nu}\de\sigma^{1}\wedge\ldots\wedge\de\sigma^{n}\end{eqnarray*}
\begin{eqnarray*}
v^{\mu} & = & \omega_{\mu_{2}\ldots\mu_{n}}\epsilon^{\mu\mu_{2}\ldots\mu_{n}},\qquad\epsilon_{\nu\mu_{2}\ldots\mu_{n}}\epsilon^{\mu\mu_{2}\ldots\mu_{n}}=(n-1)!\delta_{\nu}^{\mu}\\
\omega_{\mu_{1}\ldots\mu_{n-1}} & = & \frac{1}{(n-1)!}v^{\mu}\epsilon_{\mu\mu_{1}\ldots\mu_{n-1}},\quad\epsilon^{\mu\nu_{1}\ldots\nu_{n-1}}\epsilon_{\mu\mu_{1}\ldots\mu_{n-1}}=(n-1)!\delta_{\mu_{1}\ldots\mu_{n-1}}^{\nu_{1}\ldots\nu_{n-1}}\end{eqnarray*}
}Applying Stokes then leads to \[
\int_{\Sigma}d^{n}\sigma\,\partial_{\mu}v^{\mu}=\int_{\partial\Sigma}\frac{1}{(n-1)!}v^{\mu}\epsilon_{\mu\mu_{1}\ldots\mu_{n-1}}\de\sigma^{\mu_{1}}\wedge\ldots\wedge\de\sigma^{\mu_{n-1}}\qquad\fussend\]
}\begin{eqnarray}
\delta S & = & \int_{\Sigma}d^{n}\sigma\qquad\delta\allfields{I}\underbrace{\left(\partiell{\mc{L}}{\allfields{I}}-\partial_{\mu}\partiell{\mc{L}}{(\partial_{\mu}\allfields{I})}+\partial_{\mu_{1}}\partial_{\mu_{2}}\partiell{\mc{L}}{(\partial_{\mu_{1}}\partial_{\mu_{2}}\allfields{I})}-\ldots\right)}_{\equiv\funktional{S}{\allfields{I}}}+\nonumber \\
 &  & +\int_{\partial\Sigma}\quad\delta\allfields{I}\underbrace{\left(\partiell{\mc{L}}{(\partial_{\mu}\allfields{I})}-2\partial_{\mu_{2}}\partiell{\mc{L}}{(\partial_{\mu}\partial_{\mu_{2}}\allfields{I})}+3\partial_{\mu_{2}}\partial_{\mu_{3}}\partiell{\mc{L}}{(\partial_{\mu}\partial_{\mu_{2}}\partial_{\mu_{3}}\allfields{I})}-\ldots\right)}_{\bdry{\mu}{\mc{I}}}\times\nonumber \\
 &  & \qquad\qquad\qquad\qquad\qquad\qquad\qquad\qquad\qquad\qquad\times\frac{1}{(n-1)!}\epsilon_{\mu\nu_{1}\ldots\nu_{n-1}}\de\sigma^{\nu_{1}}\wedge\cdots\wedge\de\sigma^{\nu_{n-1}}\end{eqnarray}
A general variation $\delta\allfields{I}$ determines via $\delta S=0$
the equations of motion $\funktional{S}{\allfields{I}(\sigma)}=0$
(and the boundary conditions $n_{\mu}\bdry{\mu}{\mc{I}}=0$ with $n_{\mu}$
the normal one form), while for a symmetry transformation $\delta_{(\rho)}\allfields{I}$
the variation of the action has to vanish off-shell. Then the variation
of the Lagrangian has to be a divergence independent from the equations
of motion:\index{$K^\mu_{(\rho)}$@$\mathcal{K}^\mu_{(\rho)}$|itext{divergence term of symmetry trafo}}\begin{equation}
\delta_{(\rho)}\mc{L}\stackrel{!}{=}\partial_{\mu}\mc{K}_{(\rho)}^{\mu}\qquad\textrm{with }\bei{n_{\mu}\mc{K}_{(\rho)}^{\mu}}{\partial\Sigma}=0\end{equation}
The symmetry variation of the Lagrangian is thus on the one hand equal
to a divergence and on the other hand (according to (\ref{eq:noet:LagrangianVariation}))
equal to the equations of motion plus another divergence. One can
therefore define an object whose divergence is proportional to the
equations of motion. So let us define the current \begin{eqnarray}
j_{(\rho)}^{\mu} & \equiv & \delta\allfields{I}\cdot\partiell{\mc{L}}{(\partial_{\mu}\allfields{I})}+\sum_{k\geq1}\sum_{i=0}^{k}(-)^{i}\partial_{\nu_{1}}\ldots\partial_{\nu_{k-i}}\delta\allfields{I}\cdot\partial_{\nu_{k-i+1}}\ldots\partial_{\nu_{k}}\partiell{\mc{L}}{(\partial_{\mu}\partial_{\nu_{1}}\ldots\partial_{\nu_{k}}\allfields{I})}-\mc{K}_{(\rho)}^{\mu}\qquad\label{eq:noet:Noether-current}\end{eqnarray}
\index{$j^\mu_{(\rho)}$|itext{Noether current}}\index{Noether current}Note
that $\mc{K}_{(\rho)}^{\mu}$ is determined only up to off-shell divergence
free terms. The same is of course true for the current. Using this
definition, we can deduce from the above (\ref{eq:noet:LagrangianVariation})
that\begin{equation}
\boxed{\partial_{\mu}j_{(\rho)}^{\mu}=-\delta_{(\rho)}\allfields{I}\funktional{S}{\allfields{I}}}\label{eq:noet:currentdivergence}\end{equation}
This equation shows one direction of Noether's theorem:

\begin{thm}[Noether]\index{theorem!Noether's|fett}\index{Noether's theorem|fett}To
every transformation $\delta_{(\rho)}\allfields{I}$ which leaves
the action $S$ invariant, there is an on-shell divergence-free current
$j_{(\rho)}^{\mu}$ whose explicit form is given in (\ref{eq:noet:Noether-current}).
Its off-shell divergence is given in (\ref{eq:noet:currentdivergence}).
The such defined Noether current is unique up to trivially conserved\index{trivially conserved|itext{$\partial_\nu S^{[\nu\mu]}$ }}
terms of the form $\partial_{\nu}S^{[\nu\mu]}$.

In turn, for any given on-shell divergence-free current $\tilde{j}^{\mu}$
(see (\ref{eq:noet:generalCurrentDivergence})), which is furthermore
itself on-shell neither vanishing nor trivial, there is a corresponding
nonzero symmetry transformation $\delta\allfields{I}$ of the form
given in (\ref{eq:noet:trafoForCurrent}) .\end{thm}

\paragraph{Remark:}

The equation (\ref{eq:noet:currentdivergence}) for the off-shell
divergence can serve for reconstructing the symmetry transformations
for a given current. In the Hamiltonian formalism , the current (or
better the charge) generates the transformations via the Poisson bracket.
In the Lagrangian formalism one can simply calculate all functional
derivatives $\funktional{S}{\allfields{I}}$ (i.e. the equations of
motion) and try to express the divergence of the current as a linear
combination of them. This method -- let's call it \textbf{inverse
Noether} \index{Noether!inverse $\sim$|fett}\index{inverse Noether|fett}--
determines the transformations up to trivial gauge transformations
(see e.g. \cite[p.69]{Henneaux:1992ig}) and we are using it frequently
in the main part, in particular to derive the BRST transformations.

\paragraph{Proof}

of the theorem: We have already shown the first part (every symmetry
transformation induces a conserved current) by deriving (\ref{eq:noet:currentdivergence}).
The uniqueness up to trivial terms follows from the algebraic Poincaré
lemma. This does not yet show the inverse. For a given on-shell divergence-free
current $\tilde{j}^{\mu}$ we do not necessarily have the form (\ref{eq:noet:currentdivergence}),
but its off-shell divergence can also depend on derivatives of the
equations of motion:\begin{eqnarray}
\partial_{\mu}\tilde{j}^{\mu} & = & -y_{(0)}^{\mc{I}}\funktional{S}{\allfields{I}}-y_{(1)}^{\mc{I}\mu_{1}}\partial_{\mu_{1}}\funktional{S}{\allfields{I}}-\ldots-y_{(N)}^{\mc{I}\mu_{N}\ldots\mu_{1}}\partial_{\mu_{1}}\ldots\partial_{\mu_{N}}\funktional{S}{\allfields{I}}\label{eq:noet:generalCurrentDivergence}\end{eqnarray}
However, one can always redefine the current such that we get the
form (\ref{eq:noet:currentdivergence}). This is achieved by performing
the iterated 'partial integration' of footnote \vref{foot:iteratedPartialIntegration}.
We have schematically\begin{eqnarray}
y_{(k)}^{\mc{I}}\partial^{k}\funktional{S}{\allfields{I}} & = & \partial\left[\sum_{i=0}^{k-1}(-)^{i}\partial^{i}y_{(k)}^{\mc{I}}\cdot\partial^{k-1-i}\funktional{S}{\allfields{I}}\right]+(-)^{k}\partial^{k}y_{(k)}^{\mc{I}}\cdot\funktional{S}{\allfields{I}}\end{eqnarray}
We can then rewrite schematically the divergence of the current as
follows\begin{eqnarray}
\partial_{\mu}\tilde{j}^{\mu} & = & -\sum_{k=0}^{N}y_{(k)}^{\mc{I}}\partial^{k}\funktional{S}{\allfields{I}}=\nonumber \\
 & = & -\partial\left[\sum_{k=1}^{N}\sum_{i=0}^{k-1}(-)^{i}\partial^{i}y_{(k)}^{\mc{I}}\cdot\partial^{k-1-i}\funktional{S}{\allfields{I}}\right]-\sum_{k=0}^{N}(-)^{k}\partial^{k}y_{(k)}^{\mc{I}}\cdot\funktional{S}{\allfields{I}}\end{eqnarray}
 To summarize, if we define\begin{eqnarray}
j^{\mu} & \equiv & \tilde{j}^{\mu}+\sum_{k=1}^{N}\sum_{i=0}^{k-1}(-)^{i}\partial_{\mu_{1}}\ldots\partial_{\mu_{i}}y_{(k)}^{\mc{I}\mu\,\mu_{1}\ldots\mu_{k-1}}\cdot\partial_{\mu_{i+1}}\ldots\partial_{\mu_{k-1}}\funktional{S}{\allfields{I}}\\
\delta\allfields{I} & \equiv & \sum_{k=0}^{N}(-)^{k}\partial_{\mu_{1}}\ldots\partial_{\mu_{k}}y_{(k)}^{\mc{I}\,\mu_{1}\ldots\mu_{k}}\label{eq:noet:trafoForCurrent}\end{eqnarray}
we get $\partial j^{\mu}=-\delta\allfields{I}\funktional{S}{\allfields{I}}$
and thus discover that the above defined $\delta\allfields{I}$ is
a symmetry transformation.\rem{boundary?} We assumed that the current
was on-shell neither vanishing nor trivial, while we redefined it
with on-shell zero terms only. Therefore the new current will not
be trivial and its divergence is off-shell non-zero. The symmetry
transformations constructed above are therefore (at least off-shell)
non-zero as well. This completes the proof. $\hfill\square$

We should add that an on-shell vanishing current does not in general
imply vanishing transformations. In fact all Noether currents of gauge
transformations are vanishing on-shell. The gauge transformations
will be discussed in the following, where one discovers that the equations
of motion are not independent but are related via the Noether identities.
Going back to our construction of the transformations from an arbitrarily
conserved current one can make use of these dependencies instead of
only redefining the current. This avoids ending up with an identically
vanishing current after the redefinitions.

\section{Noether identities and on-shell vanishing gauge currents}

\index{gauge transformation!Noether identities and vanishing currents}Equation
(\ref{eq:noet:currentdivergence}) is valid for any symmetry transformation,
global as well as local ones. For local ones, however, the relation
has to hold for any local parameter $\rho^{a}$ which is much more
restrictive and allows to extract additional information. Let us assume
that there is some highest component $j_{a}^{\mu_{N}\,\mu_{N-1}\ldots\mu_{1}}$,
or in other words $\exists N$, s.t. $j_{a}^{\mu_{k}\,\mu_{k-1}\ldots\mu_{1}}=0\qquad\forall k>N$.
The expansion of $j_{(\rho)}^{\mu}$ in derivatives of the transformation
parameter $\rho$ takes the form\begin{eqnarray}
j_{(\rho)}^{\mu} & \equiv & \rho^{a}j_{a}^{\mu}+\partial_{\mu_{1}}\rho^{a}j_{a}^{\mu{\,\mu}_{1}}+\ldots+\partial_{\mu_{1}}\ldots\partial_{\mu_{N-1}}\rho^{a}j_{a}^{\mu\,\mu_{1}\ldots\mu_{N-1}}\label{eq:noet:currentExpansion}\end{eqnarray}
Now we plug this expansion and the one of $\delta_{(\rho)}\allfields{I}$
(\ref{eq:noet:deltaPhiExp}) into the equation for the current-divergence
(\ref{eq:noet:currentdivergence}):\begin{eqnarray}
\lqn{\rho^{a}\partial_{\mu}j_{a}^{\mu}+\partial_{\mu_{1}}\rho^{a}\left(j_{a}^{\mu_{1}}+\partial_{\mu}j_{a}^{\mu\mu_{1}}\right)+\partial_{\mu_{1}}\partial_{\mu_{2}}\rho^{a}\left(j_{a}^{(\mu_{1}\mu_{2})}+\partial_{\mu}j_{a}^{\mu\mu_{1}\mu_{2}}\right)+\ldots=}\nonumber \\
 & = & -\rho^{a}\delta_{a}\allfields{I}\funktional{S}{\allfields{I}}-\partial_{\mu_{1}}\rho^{a}\delta_{a}^{\mu_{1}}\allfields{I}\funktional{S}{\allfields{I}}-\partial_{\mu_{1}}\partial_{\mu_{2}}\rho^{a}\delta_{a}^{\mu_{1}\mu_{2}}\allfields{I}\funktional{S}{\allfields{I}}-\ldots\end{eqnarray}
Depending on whether we have a local or global symmetry, we get a
number of recursive relations:\begin{eqnarray}
\lqn{\Ramm{.4}{\drek{}{}{}}}\quad\partial_{\mu_{1}}j_{a}^{\mu_{1}} & = & -\delta_{a}\allfields{I}\funktional{S}{\allfields{I}}\qquad\textrm{if }\rho^{a}\neq0\label{eq:noet:NoethersTheorem}\\
\partial_{\mu_{2}}j_{a}^{\mu_{2}\mu_{1}} & = & -j_{a}^{\mu_{1}}-\delta_{a}^{\mu_{1}}\allfields{I}\funktional{S}{\allfields{I}}\qquad\textrm{if }\partial_{\mu_{1}}\rho^{a}\neq0\label{eq:rekursiveNoetherII}\\
\partial_{\mu_{3}}j_{a}^{\mu_{3}\mu_{2}\mu_{1}} & = & -j_{a}^{(\mu_{2}\mu_{1})}-\delta_{a}^{\mu_{2}\mu_{1}}\allfields{I}\funktional{S}{\allfields{I}}\qquad\textrm{if }\partial_{\mu_{1}}\partial_{\mu_{2}}\rho^{a}\neq0\label{eq:rekursiveNoetherIII}\\
 & \ddots\nonumber \\
\partial_{\mu_{N}}j_{a}^{\mu_{N}\,\mu_{N-1}\ldots\mu_{1}} & = & -j_{a}^{(\mu_{N-1}\,\mu_{N-2}\ldots\mu_{1})}-\delta_{a}^{\mu_{N-1}\ldots\mu_{1}}\allfields{I}\funktional{S}{\allfields{I}}\qquad\textrm{if }\partial_{\mu_{1}}\ldots\partial_{\mu_{N-1}}\rho^{a}\neq0\label{eq:rekursiveNoetherk}\\
0 & = & -j_{a}^{(\mu_{N}\,\mu_{N-1}\ldots\mu_{1})}-\delta_{a}^{\mu_{N}\ldots\mu_{1}}\allfields{I}\funktional{S}{\allfields{I}}\qquad\textrm{if }\partial_{\mu_{1}}\ldots\partial_{\mu_{N}}\rho^{a}\neq0\label{eq:rekursiveNoetherLast}\end{eqnarray}
The first equation (\ref{eq:noet:NoethersTheorem}) is present already
for a global symmetry and corresponds to the Noether's theorem for
global symmetries\index{theorem!Noether's}\index{Noether's theorem}.\rem{\begin{eqnarray*}
Q_{a}(\sigma^{0}) & \equiv & \int d\sigma^{1}\cdots d\sigma^{n-1}\: j_{a}^{0}\\
\partial_{0}Q_{a}(\sigma^{0}) & \stackrel{\textrm{on-shell}}{=} & -\int d\sigma^{1}\cdots d\sigma^{n-1}\,\partial_{i}j_{a}^{i}=-\int_{\partial\Sigma|_{\sigma^{0}}}10d\sigma^{1}\cdots d\sigma^{n-2}\, n_{i}j_{a}^{i}\end{eqnarray*}
}  If the transformation parameters are instead local and arbitrary,
the complete set of equations is forced. Taking then the divergence
of the second equation, the double divergence of the third and so
on, and adding them with appropriate signs, we can remove all currents
from the equations and arrive at a version of the \textbf{Noether's
identities}\index{Noether identities}\index{identities!Noether $\sim$}:\rem{\begin{eqnarray*}
\partial_{\mu_{1}}j_{a}^{\mu_{1}} & = & -\delta_{a}\allfields{I}\funktional{S}{\allfields{I}}\qquad\textrm{if }\rho^{a}\neq0\\
\partial_{\mu_{1}}\left(j_{a}^{\mu_{1}}+\partial_{\mu_{2}}j_{a}^{\mu_{2}\mu_{1}}\right) & = & -\partial_{\mu_{1}}\left(\delta_{a}^{\mu_{1}}\allfields{I}\funktional{S}{\allfields{I}}\right)\qquad\textrm{if }\partial_{\mu_{1}}\rho^{a}\neq0\\
\partial_{\mu_{1}}\partial_{\mu_{2}}\left(j_{a}^{(\mu_{2}\mu_{1})}+\partial_{\mu_{3}}j_{a}^{\mu_{3}\mu_{2}\mu_{1}}\right) & = & -\partial_{\mu_{1}}\partial_{\mu_{2}}\left(\delta_{a}^{\mu_{2}\mu_{1}}\allfields{I}\funktional{S}{\allfields{I}}\right)\qquad\textrm{if }\partial_{\mu_{2}}\partial_{\mu_{1}}\rho^{a}\neq0\\
 & \ddots\\
\partial_{\mu_{1}}\ldots\partial_{\mu_{k}}\left(j_{a}^{(\mu_{k}\,\mu_{k-1}\ldots\mu_{1})}+\partial_{\mu_{k+1}}j_{a}^{\mu_{k+1}\,\mu_{k}\ldots\mu_{1}}\right) & = & -\partial_{\mu_{1}}\ldots\partial_{\mu_{k}}\left(\delta_{a}^{\mu_{k}\ldots\mu_{1}}\allfields{I}\funktional{S}{\allfields{I}}\right)\qquad\textrm{if }\partial_{\mu_{k}}\ldots\partial_{\mu_{1}}\rho^{a}\neq0\end{eqnarray*}
If chain breaks at $j_{a}^{\mu_{N+2}\,\mu_{N+1}\ldots\mu_{1}}=0$,
we get (I)-(II)+(III)-...=0.} \begin{eqnarray}
\delta_{a}\allfields{I}\funktional{S}{\allfields{I}}-\partial_{\mu_{1}}\left(\delta_{a}^{\mu_{1}}\allfields{I}\funktional{S}{\allfields{I}}\right)+\ldots+(-)^{N+1}\partial_{\mu_{1}}\ldots\partial_{\mu_{N+1}}\left(\delta_{a}^{\mu_{N+1}\ldots\mu_{1}}\allfields{I}\funktional{S}{\allfields{I}}\right) & = & 0\label{eq:noeth:NoetherIdentities}\end{eqnarray}
From the recursive equations above, one can also obtain an interesting
statement about the current of a gauge symmetry (compare \cite[p.95]{Henneaux:1992ig}): 

\begin{prop}\label{prop:currentOnshellZero}\index{vanishing current}\index{proposition!on-shell vanishing current}\index{on-shell!vanishing current}:
The Noether current of a gauge symmetry vanishes on-shell up to trivially
conserved\index{trivially conserved} terms (see (\ref{eq:noet:claim})).
In turn, if a given global symmetry transformation has an on-shell
vanishing current (see (\ref{eq:noet:globalOnShellZeroCurrent})),
then one can extend the transformation to a local one (see (\ref{eq:localSymmetryToOnShellVanishingCurrent})).\rem{Or,
every on-shell vanishing current corresponds to a gauge symmetry or
to a trivial symmetry. }\end{prop}

\paragraph{Proof}

Start with a given gauge symmetry $\delta_{(\rho)}\allfields{I}$
and its corresponding current $j_{(\rho)}^{\mu}$ with the expansion
given in (\ref{eq:noet:currentExpansion}), which defines the number
$N$ of the highest derivative on $\rho$. We want to show that the
current of a local symmetry is of the form \begin{eqnarray}
j_{(\rho)}^{\mu} & = & \sum_{k=0}^{N}\lambda_{(\rho)}^{\mu\mc{I}\mu_{1}\ldots\mu_{k}}\partial_{\mu_{1}}\ldots\partial_{\mu_{k}}\funktional{S}{\allfields{I}}+t_{(\rho)}^{\mu}\label{eq:noet:claim}\end{eqnarray}
for some coefficients $\lambda_{(\rho)}^{\mu\mc{I}\mu_{1}\ldots\mu_{k}}$
and with a term $t^{\mu}$ whose divergence vanishes off-shell: $\partial_{\mu}t_{(\rho)}^{\mu}\equiv0$.
(Due to the algebraic Poincaré lemma, this means that there is some
antisymmetric tensor $S_{(\rho)}^{[\mu\nu]}$ such that $t_{(\rho)}^{\mu}=\partial_{\nu}S_{(\rho)}^{[\mu\nu]}$.
)

In order to reduce the length of the equations, define first%
\footnote{\index{footnote!\thefoot. symmetrized current components}Note that
from \begin{eqnarray*}
k\cdot j_{a}^{(\mu_{k}\,\mu_{k-1}\ldots\mu_{1})} & = & j_{a}^{\mu_{k}\,\mu_{k-1}\ldots\mu_{1}}+(k-1)j_{a}^{(\mu_{k-1}\,\mu_{k-2}\ldots\mu_{1})\mu_{k}}\end{eqnarray*}
one can deduce\frem{\begin{eqnarray*}
j_{a}^{\mu_{k}\,\mu_{k-1}\ldots\mu_{1}} & = & j_{a}^{(\mu_{k}\,\mu_{k-1}\ldots\mu_{1})}+(k-1)\left(j_{a}^{(\mu_{k}\,\mu_{k-1}\ldots\mu_{1})}-j_{a}^{(\mu_{k-1}\,\mu_{k-2}\ldots\mu_{1})\mu_{k}}\right)=\\
 & = & j_{a}^{(\mu_{k}\,\mu_{k-1}\ldots\mu_{1})}+(k-1)\left(\frac{1}{k}j_{a}^{\mu_{k}\,\mu_{k-1}\ldots\mu_{1}}+\frac{k-1}{k}j_{a}^{(\mu_{k-1}\ldots\mu_{1})\mu_{k}}-\frac{k}{k}j_{a}^{(\mu_{k-1}\,\mu_{k-2}\ldots\mu_{1})\mu_{k}}\right)=\\
 & = & j_{a}^{(\mu_{k}\,\mu_{k-1}\ldots\mu_{1})}+\frac{(k-1)}{k}\left(j_{a}^{\mu_{k}\,\mu_{k-1}\ldots\mu_{1}}-j_{a}^{(\mu_{k-1}\,\mu_{k-2}\ldots\mu_{1})\mu_{k}}\right)\end{eqnarray*}
}\begin{eqnarray*}
j_{a}^{\mu_{k}\,\mu_{k-1}\ldots\mu_{1}}-j_{a}^{(\mu_{k}\,\mu_{k-1}\ldots\mu_{1})} & = & \frac{2}{k}\sum_{i=1}^{k-1}j_{a}^{[\mu_{k}|\,\mu_{k-1}\ldots|\mu_{i}]\ldots\mu_{1}}\qquad\fussend\end{eqnarray*}
} \begin{eqnarray}
E_{a}^{\mu_{k}\ldots\mu_{1}} & \equiv & \delta_{a}^{\mu_{k}\ldots\mu_{1}}\allfields{I}\funktional{S}{\allfields{I}},\qquad E_{a}^{\mu_{k}\ldots\mu_{1}}=E_{a}^{(\mu_{k}\ldots\mu_{1})}\\
A_{a}^{\mu_{k+1}\,\mu_{k}\ldots\mu_{1}} & \equiv & j_{a}^{\mu_{k+1}\,\mu_{k}\ldots\mu_{1}}-j_{a}^{(\mu_{k+1}\,\mu_{k}\ldots\mu_{1})},\quad A_{a}^{\mu_{k+1}\,\mu_{k}\ldots\mu_{1}}=A_{a}^{\mu_{k+1}\,(\mu_{k}\ldots\mu_{1})},\quad A_{a}^{(\mu_{k+1}\,\mu_{k}\ldots\mu_{1})}=0\qquad\end{eqnarray}
The first object is symmetric in all indices and the second is symmetric
in the last k indices and vanishes when symmetrized in all indices.
Using this notation, we can rewrite the recursive equations (\ref{eq:rekursiveNoetherII})-(\ref{eq:rekursiveNoetherLast})
in the following form\begin{eqnarray}
j_{a}^{\mu_{1}} & = & -E_{a}^{\mu_{1}}-\partial_{\mu_{2}}j_{a}^{\mu_{2}\mu_{1}}\\
j_{a}^{\mu_{2}\,\mu_{1}} & = & A_{a}^{\mu_{2}\,\mu_{1}}-E_{a}^{\mu_{2}\mu_{1}}-\partial_{\mu_{3}}j_{a}^{\mu_{3}\mu_{2}\mu_{1}}\\
 & \ddots\nonumber \\
j_{a}^{\mu_{N-1}\,\mu_{N-2}\ldots\mu_{1}} & = & A_{a}^{\mu_{N-1}\,\mu_{N-2}\ldots\mu_{1}}-E_{a}^{\mu_{N-1}\ldots\mu_{1}}-\partial_{\mu_{N}}j_{a}^{\mu_{N}\,\mu_{N-1}\ldots\mu_{1}}\\
j_{a}^{\mu_{N}\,\mu_{N-1}\ldots\mu_{1}} & = & A_{a}^{\mu_{N}\,\mu_{N-1}\ldots\mu_{1}}-E_{a}^{\mu_{N}\ldots\mu_{1}}\end{eqnarray}
This set of equations can now formally be solved for all components
of the current, starting from the $N$-th equation. We end up with
\begin{eqnarray}
j_{a}^{\mu_{1}} & = & -\partial_{\mu_{2}}A_{a}^{\mu_{2}\,\mu_{1}}+\partial_{\mu_{2}}\partial_{\mu_{3}}A_{a}^{\mu_{3}\,\mu_{2}\mu_{1}}-\partial_{\mu_{2}}\partial_{\mu_{3}}\partial_{\mu_{4}}A_{a}^{\mu_{4}\,\mu_{3}\mu_{2}\mu_{1}}+\ldots+\nonumber \\
 &  & -E_{a}^{\mu_{1}}+\partial_{\mu_{2}}E_{a}^{\mu_{2}\mu_{1}}-\partial_{\mu_{2}}\partial_{\mu_{3}}E_{a}^{\mu_{3}\,\mu_{2}\mu_{1}}+\partial_{\mu_{2}}\partial_{\mu_{3}}\partial_{\mu_{4}}E_{a}^{\mu_{4}\,\mu_{3}\mu_{2}\mu_{1}}-\ldots\\
j_{a}^{\mu_{2}\,\mu_{1}} & = & A_{a}^{\mu_{2}\,\mu_{1}}-\partial_{\mu_{3}}A_{a}^{\mu_{3}\,\mu_{2}\mu_{1}}+\partial_{\mu_{3}}\partial_{\mu_{4}}A_{a}^{\mu_{4}\,\mu_{3}\mu_{2}\mu_{1}}-\ldots+\nonumber \\
 &  & -E_{a}^{\mu_{2}\mu_{1}}+\partial_{\mu_{3}}E_{a}^{\mu_{3}\,\mu_{2}\mu_{1}}-\partial_{\mu_{3}}\partial_{\mu_{4}}E_{a}^{\mu_{4}\,\mu_{3}\mu_{2}\mu_{1}}+\ldots\\
 & \ddots\nonumber \\
j_{a}^{\mu_{k}\,\mu_{k-1}\ldots\mu_{1}} & = & A_{a}^{\mu_{k}\,\mu_{k-1}\ldots\mu_{1}}-\partial_{\mu_{k+1}}A_{a}^{\mu_{k+1}\,\mu_{k}\ldots\mu_{1}}+\ldots+(-)^{N-k}\partial_{\mu_{k+1}}\ldots\partial_{\mu_{N}}A_{a}^{\mu_{N}\,\mu_{N-1}\ldots\mu_{1}}+\nonumber \\
 &  & -E_{a}^{\mu_{k}\ldots\mu_{1}}+\partial_{\mu_{k+1}}E_{a}^{\mu_{k+1}\ldots\mu_{1}}-\ldots-(-)^{N-k}\partial_{\mu_{k+1}}\ldots\partial_{\mu_{N}}E_{a}^{\mu_{N}\ldots\mu_{1}}\\
\frem{j_{a}^{\mu_{N-2}\,\mu_{N-3}\ldots\mu_{1}}} & \ddots & \frem{A_{a}^{\mu_{N-2}\,\mu_{N-3}\ldots\mu_{1}}-\partial_{\mu_{N-1}}A_{a}^{\mu_{N-1}\,\mu_{N-2}\ldots\mu_{1}}+\partial_{\mu_{N-1}}\partial_{\mu_{N}}A_{a}^{\mu_{N}\,\mu_{N-1}\ldots\mu_{1}}-E_{a}^{\mu_{N-2}\ldots\mu_{1}}+\partial_{\mu_{N-1}}E_{a}^{\mu_{N-1}\ldots\mu_{1}}-\partial_{\mu_{N-1}}\partial_{\mu_{N}}E_{a}^{\mu_{N}\ldots\mu_{1}}}\nonumber \\
j_{a}^{\mu_{N-1}\,\mu_{N-2}\ldots\mu_{1}} & = & A_{a}^{\mu_{N-1}\,\mu_{N-2}\ldots\mu_{1}}-\partial_{\mu_{N}}A_{a}^{\mu_{N}\,\mu_{N-1}\ldots\mu_{1}}-E_{a}^{\mu_{N-1}\ldots\mu_{1}}+\partial_{\mu_{N}}E_{a}^{\mu_{N}\ldots\mu_{1}}\\
j_{a}^{\mu_{N}\,\mu_{N-1}\ldots\mu_{1}} & = & A_{a}^{\mu_{N}\,\mu_{N-1}\ldots\mu_{1}}-E_{a}^{\mu_{N}\ldots\mu_{1}}\end{eqnarray}
In order to obtain the complete current $j_{(\rho)}^{\mu_{1}}$ we
have to contract the $k$-th term $j_{a}^{\mu_{1}\,\mu_{k}\ldots\mu_{2}}$
(with interchanged $\mu_{1}\leftrightarrow\mu_{k}$!) with $\partial_{\mu_{2}}\ldots\partial_{\mu_{k}}\rho^{a}$
and then add all the terms.\rem{\begin{eqnarray*}
j_{a}^{\mu_{1}} & = & \qquad\quad-\partial_{\mu_{2}}A_{a}^{\mu_{2}\,\mu_{1}}+\partial_{\mu_{2}}\partial_{\mu_{3}}A_{a}^{\mu_{3}\,\mu_{2}\mu_{1}}-\partial_{\mu_{2}}\partial_{\mu_{3}}\partial_{\mu_{4}}A_{a}^{\mu_{4}\,\mu_{3}\mu_{2}\mu_{1}}+\ldots-E_{a}^{\mu_{1}}+\partial_{\mu_{2}}E_{a}^{\mu_{2}\mu_{1}}-\partial_{\mu_{2}}\partial_{\mu_{3}}E_{a}^{\mu_{3}\mu_{2}\mu_{1}}+\partial_{\mu_{2}}\partial_{\mu_{3}}\partial_{\mu_{4}}E_{a}^{\mu_{4}\mu_{3}\mu_{2}\mu_{1}}-\ldots\qquad(1)\\
j_{a}^{\mu_{1}\,\mu_{2}} & = & A_{a}^{\mu_{1}\,\mu_{2}}-\partial_{\mu_{3}}A_{a}^{\mu_{3}\,\mu_{2}\mu_{1}}+\partial_{\mu_{3}}\partial_{\mu_{4}}A_{a}^{\mu_{4}\,\mu_{3}\mu_{2}\mu_{1}}-\ldots-E_{a}^{\mu_{2}\mu_{1}}+\partial_{\mu_{3}}E_{a}^{\mu_{3}\mu_{2}\mu_{1}}-\partial_{\mu_{3}}\partial_{\mu_{4}}E_{a}^{\mu_{4}\mu_{3}\mu_{2}\mu_{1}}+\ldots\qquad(2)\\
j_{a}^{\mu_{1}\,\mu_{3}\mu_{2}} & = & A_{a}^{\mu_{1}\,\mu_{3}\mu_{2}}-\partial_{\mu_{4}}A_{a}^{\mu_{4}\,\mu_{3}\mu_{2}\mu_{1}}+\partial_{\mu_{4}}\partial_{\mu_{5}}A_{a}^{\mu_{5}\,\mu_{4}\mu_{3}\mu_{2}\mu_{1}}-\ldots-E_{a}^{\mu_{3}\mu_{2}\mu_{1}}+\partial_{\mu_{4}}E_{a}^{\mu_{4}\mu_{3}\mu_{2}\mu_{1}}-\partial_{\mu_{4}}\partial_{\mu_{5}}E_{a}^{\mu_{5}\mu_{4}\mu_{3}\mu_{2}\mu_{1}}+\ldots\qquad(3)\\
 & \ddots\\
j_{a}^{\mu_{1}\,\mu_{k}\ldots\mu_{2}} & = & A_{a}^{\mu_{1}\,\mu_{k}\ldots\mu_{2}}-\partial_{\mu_{k+1}}A_{a}^{\mu_{k+1}\,\mu_{k}\ldots\mu_{1}}+\ldots+(-)^{N-k}\partial_{\mu_{k+1}}\ldots\partial_{\mu_{N}}A_{a}^{\mu_{N}\,\mu_{N-1}\ldots\mu_{1}}+\\
 &  & -E_{a}^{\mu_{k}\ldots\mu_{1}}+\partial_{\mu_{k+1}}E_{a}^{\mu_{k+1}\ldots\mu_{1}}-\ldots-(-)^{N-k}\partial_{\mu_{k+1}}\ldots\partial_{\mu_{N}}E_{a}^{\mu_{N}\ldots\mu_{1}}\qquad(k)\\
 & \ddots\\
j_{a}^{\mu_{1}\,\mu_{N-2}\ldots\mu_{2}} & = & A_{a}^{\mu_{1}\,\mu_{N-2}\ldots\mu_{2}}-\partial_{\mu_{N-1}}A_{a}^{\mu_{N-1}\,\mu_{N-2}\ldots\mu_{1}}+\partial_{\mu_{N-1}}\partial_{\mu_{N}}A_{a}^{\mu_{N}\,\mu_{N-1}\ldots\mu_{1}}-E_{a}^{\mu_{N-2}\ldots\mu_{1}}+\partial_{\mu_{N-1}}E_{a}^{\mu_{N-1}\ldots\mu_{1}}-\partial_{\mu_{N-1}}\partial_{\mu_{N}}E_{a}^{\mu_{N}\ldots\mu_{1}}\qquad(N-2)\\
j_{a}^{\mu_{1}\,\mu_{N-1}\ldots\mu_{2}} & = & A_{a}^{\mu_{1}\,\mu_{N-1}\ldots\mu_{2}}-\partial_{\mu_{N}}A_{a}^{\mu_{N}\,\mu_{N-1}\ldots\mu_{1}}-E_{a}^{\mu_{N-1}\ldots\mu_{1}}+\partial_{\mu_{N}}E_{a}^{\mu_{N}\ldots\mu_{1}}\qquad(N-1)\\
j_{a}^{\mu_{1}\,\mu_{N}\ldots\mu_{2}} & = & A_{a}^{\mu_{1}\,\mu_{N}\ldots\mu_{2}}-E_{a}^{\mu_{N}\ldots\mu_{1}}\qquad(N)\end{eqnarray*}
\begin{eqnarray*}
j_{(\rho)}^{a} & = & -\rho^{a}\partial_{\mu_{2}}A_{a}^{\mu_{2}\,\mu_{1}}+\partial_{\mu_{2}}\rho^{a}A_{a}^{\mu_{1}\,\mu_{2}}+\\
 &  & +\rho^{a}\partial_{\mu_{2}}\partial_{\mu_{3}}A_{a}^{\mu_{3}\,\mu_{2}\mu_{1}}-\partial_{\mu_{2}}\rho^{a}\partial_{\mu_{3}}A_{a}^{\mu_{3}\,\mu_{2}\mu_{1}}+\partial_{\mu_{2}}\partial_{\mu_{3}}\rho^{a}A_{a}^{\mu_{1}\,\mu_{3}\mu_{2}}+\\
 &  & -\rho^{a}\partial_{\mu_{2}}\partial_{\mu_{3}}\partial_{\mu_{4}}A_{a}^{\mu_{4}\,\mu_{3}\mu_{2}\mu_{1}}+\partial_{\mu_{2}}\rho^{a}\partial_{\mu_{3}}\partial_{\mu_{4}}A_{a}^{\mu_{4}\,\mu_{3}\mu_{2}\mu_{1}}-\partial_{\mu_{2}}\partial_{\mu_{3}}\rho^{a}\partial_{\mu_{4}}A_{a}^{\mu_{4}\,\mu_{3}\mu_{2}\mu_{1}}+\partial_{\mu_{2}}\partial_{\mu_{3}}\partial_{\mu_{4}}\rho^{a}A_{a}^{\mu_{1}\,\mu_{4}\mu_{3}\mu_{2}}+\\
 &  & \ldots-(-)^{k}\rho^{a}\partial_{\mu_{2}}\ldots\partial_{\mu_{k}}A_{a}^{\mu_{k}\,\mu_{k-1}\ldots\mu_{1}}+(-)^{k}\partial_{\mu_{2}}\rho^{a}\partial_{\mu_{3}}\ldots\partial_{\mu_{k}}A_{a}^{\mu_{k}\,\mu_{k-1}\ldots\mu_{1}}-\ldots-\partial_{\mu_{2}}\ldots\partial_{\mu_{k-1}}\rho^{a}\partial_{\mu_{k}}A_{a}^{\mu_{k}\,\mu_{k-1}\ldots\mu_{1}}+\partial_{\mu_{2}}\ldots\partial_{\mu_{k}}\rho^{a}A_{a}^{\mu_{1}\,\mu_{k}\ldots\mu_{2}}+\\
 &  & +\ldots-(-)^{N}\rho^{a}\partial_{\mu_{2}}\ldots\partial_{\mu_{N}}A_{a}^{\mu_{N}\,\mu_{N-1}\ldots\mu_{1}}+(-)^{N}\partial_{\mu_{2}}\rho^{a}\partial_{\mu_{3}}\ldots\partial_{\mu_{N}}A_{a}^{\mu_{N}\,\mu_{N-1}\ldots\mu_{1}}-\ldots-\partial_{\mu_{2}}\ldots\partial_{\mu_{N-1}}\rho^{a}\partial_{\mu_{N}}A_{a}^{\mu_{N}\,\mu_{N-1}\ldots\mu_{1}}+\partial_{\mu_{2}}\ldots\partial_{\mu_{N}}\rho^{a}A_{a}^{\mu_{1}\,\mu_{N}\ldots\mu_{2}}+\\
 &  & -\rho^{a}E_{a}^{\mu_{1}}+\rho^{a}\partial_{\mu_{2}}E_{a}^{\mu_{2}\,\mu_{1}}-\partial_{\mu_{2}}\rho^{a}E_{a}^{\mu_{1}\,\mu_{2}}+\\
 &  & -\rho^{a}\partial_{\mu_{2}}\partial_{\mu_{3}}E_{a}^{\mu_{3}\,\mu_{2}\mu_{1}}+\partial_{\mu_{2}}\rho^{a}\partial_{\mu_{3}}E_{a}^{\mu_{3}\,\mu_{2}\mu_{1}}-\partial_{\mu_{2}}\partial_{\mu_{3}}\rho^{a}E_{a}^{\mu_{1}\,\mu_{3}\mu_{2}}+\ldots\end{eqnarray*}
} Interchanging $\mu_{k}$ and $\mu_{1}$ for the $k$-th equation
affects (because of the symmetries) only the term $A_{a}^{\mu_{k}\,\mu_{k-1}\ldots\mu_{1}}\mapsto A_{a}^{\mu_{1}\,\mu_{k}\ldots\mu_{2}}$.
We will sort the $A_{a}$-terms with respect to the number of indices
on $A_{a}$ and the $E_{a}$-terms with respect to the number of derivatives
on $\rho^{a}$: \begin{eqnarray}
j_{(\rho)}^{a} & = & \sum_{k=2}^{N}\underbrace{\left(\sum_{i=0}^{k-2}-(-)^{k-i}\partial_{\mu_{2}}\ldots\partial_{\mu_{2+i-1}}\rho^{a}\partial_{\mu_{2+i}}\ldots\partial_{\mu_{k}}A_{a}^{\mu_{k}\,\mu_{k-1}\ldots\mu_{1}}+\partial_{\mu_{2}}\ldots\partial_{\mu_{k}}\rho^{a}A_{a}^{\mu_{1}\,\mu_{k}\ldots\mu_{2}}\right)}_{\equiv t_{(\rho,k)}^{\mu_{1}}}+\nonumber \\
 &  & -\sum_{k=1}^{N}\partial_{\mu_{2}}\ldots\partial_{\mu_{k}}\rho^{a}\sum_{i=0}^{N-k}(-)^{i}\partial_{\mu_{k+1}}\ldots\partial_{\mu_{k+i}}E_{a}^{\mu_{k+i}\ldots\mu_{k+1}\mu_{k}\ldots\mu_{1}}\end{eqnarray}
The second line vanishes on-shell, but it remains to show that the
first line $t_{(\rho)}^{\mu_{1}}\equiv\sum_{k=2}^{N}t_{(\rho)}^{\mu_{1}}$
has trivially vanishing divergence. The second term in the first line
is written separately (not in the sum over $i$), because in contrast
to the other terms it has the $\mu_{1}$ index at the first position
(which is not symmetrized like the other positions). This difference
in treatment disappears in the divergence with contracted $\mu_{1}$.
We use this fact to show the trivial vanishing (without the use of
equations of motion) of the divergence of for every single $t_{(\rho,k)}^{\mu_{1}}$:\begin{eqnarray}
\lqn{\partial_{\mu_{1}}t_{(\rho,k)}^{\mu_{1}}=}\nonumber \\
 & = & \sum_{i=0}^{k-1}(-)^{k-i+1}\partial_{\mu_{1}}\ldots\partial_{\mu_{i+1}}\rho^{a}\partial_{\mu_{i+2}}\ldots\partial_{\mu_{k}}A_{a}^{\mu_{k}\,\mu_{k-1}\ldots\mu_{1}}-\sum_{i=0}^{k-1}(-)^{k-i}\partial_{\mu_{2}}\ldots\partial_{\mu_{2+i-1}}\rho^{a}\partial_{\mu_{2+i}}\ldots\partial_{\mu_{k}}\lqn{\partial_{\mu_{1}}A_{a}^{\mu_{1}\,\mu_{k}\ldots\mu_{2}}}\nonumber \\
 & = & \sum_{i=1}^{k-1}-(-)^{k-i+1}\partial_{\mu_{1}}\ldots\partial_{\mu_{i}}\rho^{a}\partial_{\mu_{i+1}}\ldots\partial_{\mu_{k}}A_{a}^{\mu_{k}\,\mu_{k-1}\ldots\mu_{1}}-\sum_{i=1}^{k-1}(-)^{k-i}\partial_{\mu_{1}}\ldots\partial_{\mu_{i}}\rho^{a}\partial_{\mu_{i+1}}\ldots\partial_{\mu_{k}}A_{a}^{\mu_{k}\,\mu_{k-1}\ldots\mu_{1}}\lqn{+}\nonumber \\
 &  & -(-)^{k-i}\partial_{\mu_{1}}\ldots\partial_{\mu_{k}}\rho^{a}\underbrace{A_{a}^{(\mu_{k}\,\mu_{k-1}\ldots\mu_{1})}}_{=0}-(-)^{k}\rho^{a}\partial_{\mu_{1}}\ldots\partial_{\mu_{k}}\underbrace{A_{a}^{(\mu_{k}\,\mu_{k-1}\ldots\mu_{1})}}_{=0}\qquad=0\end{eqnarray}
This completes the proof of (\ref{eq:noet:claim}) or of one direction
of the proposition. 

Now consider that we have a global transformation (constant parameter
$\rho_{c}$) $\delta_{(\rho_{c})}^{0}\allfields{I}=\rho_{c}^{a}\delta_{a}\allfields{I}$
with Noether current $j_{(\rho_{c})}^{\mu}=\rho_{c}^{a}j_{a}^{\mu}$,
which itself vanishes on-shell\begin{eqnarray}
j_{a}^{\mu} & = & \sum_{k=0}^{N}\lambda_{a}^{\mu\mc{I}\mu_{1}\ldots\mu_{k}}\partial_{\mu_{1}}\ldots\partial_{\mu_{k}}\funktional{S}{\allfields{I}}\label{eq:noet:globalOnShellZeroCurrent}\\
\partial_{\mu}j_{a}^{\mu} & = & -\delta_{a}\allfields{I}\funktional{S}{\allfields{I}}\label{eq:noet:divergenceOfGOSZC}\end{eqnarray}
If we plug (\ref{eq:noet:globalOnShellZeroCurrent}) into (\ref{eq:noet:divergenceOfGOSZC})
we already discover relations between the equations of motion, which
look like the Noether identities for local symmetries. Indeed, if
$j_{a}^{\mu}$ vanishes on-shell, also $\rho^{a}j_{a}^{\mu}$ vanishes
on-shell, even for local $\rho^{a}$. For consistent equations of
motion (some which have solutions at all) certainly also its derivative
vanishes on-shell. The combination $j_{(\rho)}^{0}\equiv\rho^{a}j_{a}^{\mu}$
therefore corresponds to a symmetry transformation with a local parameter,
i.e. a gauge symmetry, although this current is in general not yet
in the standard form of a Noether current (where its divergence does
not contain derivatives of $\funktional{S}{\allfields{I}}$, but only
the plain equations of motion):\begin{eqnarray}
\partial_{\mu}(\rho^{a}j_{a}^{\mu}) & = & \partial_{\mu}\rho^{a}\cdot j_{a}^{\mu}+\rho^{a}\partial_{\mu}j_{a}^{\mu}=\\
 & = & \sum_{k=1}^{N}\partial_{\mu}\rho^{a}\lambda_{a}^{\mu\mc{I}\mu_{1}\ldots\mu_{k}}\partial_{\mu_{1}}\ldots\partial_{\mu_{k}}\funktional{S}{\allfields{I}}-\left(\rho^{a}\delta_{a}\allfields{I}-\partial_{\mu}\rho^{a}\lambda_{a}^{\mu\mc{I}}\right)\funktional{S}{\allfields{I}}\end{eqnarray}
In order to get a proper Noether current (where the righthand side
does not contain any derivatives of the equations of motion) we can
use our insights from the proof of Noether's theorem, i.e. equations
(\ref{eq:noet:generalCurrentDivergence})-(\ref{eq:noet:trafoForCurrent}).
We learn that if we define the whole current to be\begin{eqnarray}
j_{(\rho)}^{\mu} & \equiv & \rho^{a}j_{a}^{\mu}-\sum_{k=1}^{N}\sum_{i=0}^{k-1}(-)^{i}\partial_{\mu_{1}}\ldots\partial_{\mu_{i}}\partial_{\nu}\rho^{a}\lambda_{a}^{\nu\mc{I}\mu\mu_{1}\ldots\mu_{k-1}}\cdot\partial_{\mu_{i+1}}\ldots\partial_{\mu_{k-1}}\funktional{S}{\allfields{I}}\end{eqnarray}
we get a proper Noether current with corresponding symmetry transformations\begin{eqnarray}
\delta_{(\rho)}\allfields{I} & \equiv & \rho^{a}\delta_{a}\allfields{I}-\partial_{\mu}\rho^{a}\lambda_{a}^{\mu\mc{I}}+\sum_{k=1}^{N}(-)^{k+1}\partial_{\mu_{1}}\ldots\partial_{\mu_{k}}\left(\partial_{\nu}\rho^{a}\lambda_{a}^{\nu\mc{I}\,\mu_{1}\ldots\mu_{k}}\right)\label{eq:localSymmetryToOnShellVanishingCurrent}\end{eqnarray}
\rem{The divergence of an on-shell vanishing current $\tilde{\jmath}^{\mu}=\sum_{k=0}^{N}\lambda^{\mu\mc{I}\mu_{1}\ldots\mu_{k}}\partial_{\mu_{1}}\ldots\partial_{\mu_{k}}\funktional{S}{\allfields{I}}$contains
always derivatives on $\funktional{S}{\allfields{I}}$. Redefining
it in the way it is done in (\ref{eq:noet:generalCurrentDivergence})-(\ref{eq:noet:trafoForCurrent})
leads to identically vanishing symmetry transformations. One can avoid
this only when there is a gauge transformation which induces Noether
identities and makes it possible to express derivatives of $\funktional{S}{\allfields{I}}$
by $\funktional{S}{\allfields{I}}$ itself. } The transformation
(\ref{eq:localSymmetryToOnShellVanishingCurrent}) is a local symmetry
transformation which completes the proof of the $\mbox{proposition. }\hfill\square$

\begin{thm}\index{theorem!on-shell vanishing symmetry transformation}\index{on-shell!vanishing transformation}\index{vanishing transformation}\label{thm:trafoOnshellZero}Every
on-shell vanishing symmetry transformation is a \textbf{trivial gauge
transformation}\index{gauge transformation!trivial $\sim$}\index{trivial gauge transformation}
as defined below:\begin{eqnarray}
\delta\allfields{I}\stackrel{{\rm on-shell}}{=}0,\quad\delta S=0 & \dann & \delta\allfields{I}=\int d^{d}\sigma\quad\mc{A}^{\mc{IJ}}(\sigma,\sigma')\funktional{S}{\allfields{I}(\sigma')}\quad\mbox{with}{\:\mc{A}}^{\mc{IJ}}(\sigma,\sigma')=-\mc{A}^{\mc{JI}}(\sigma',\sigma)\qquad\end{eqnarray}
\end{thm}

See in \cite{Henneaux:1992ig} (theorem 17.3 on page 414 or theorem
3.1 on page 17) for a proof of this theorem. See \cite[p.69]{Henneaux:1992ig}
for a discussion of trivial gauge transformations.

\section{Shortcut to calculate the Noether current}

\index{Noether current!trick to calculate the $\sim$}\index{trick!to calculate the Noether current}\index{shortcut!to calculate the Noether current}There
is a nice shortcut to calculate the current: multiply both sides of
(\ref{eq:noet:currentdivergence}) with some local parameter $\eta(\sigma)$,
integrate over the world-volume $\Sigma$ and perform a partial integration
to arrive at\begin{equation}
\boxed{\int_{\Sigma}d^{n}\sigma\:\partial_{\mu}\eta\cdot j_{(\rho)}^{\mu}+\int_{\partial\Sigma}(\ldots)=\delta_{(\eta,\rho)}S}\label{eq:noet:trick}\end{equation}
where $\delta_{(\eta,\rho)}\allfields{I}\equiv\eta\cdot\delta_{(\rho)}\allfields{I}$.
One thus obtains the current by multiplying the variation with an
independent local parameter $\eta$ and reading off the coefficient
of $\partial_{\mu}\eta$. This trick is better known for global symmetries%
\footnote{\label{foot:noet:trick}\index{footnote!\thefoot. trick for Noether current}If
one is just interested in $j_{a}^{\mu}$ one can consider a variation
not with the full variation $\delta_{(\rho)}\allfields{I}$, but only
with its derivative free part $\delta_{(\rho)}^{0}\allfields{I}\equiv\rho^{a}\delta_{a}\allfields{I}$
(see (\ref{eq:noet:deltaPhiExp})) and allow local $\rho^{a}$ even
in the case of a global symmetry. Multiplying both sides of (\ref{eq:noet:NoethersTheorem})
with $\rho^{a}$ we get $\rho^{a}\partial_{\mu}j_{a}^{\mu}=-\delta_{(\rho)}^{0}\allfields{I}\funktional{S}{\allfields{I}}$.
Integrating over $\Sigma$ and partially integrating finally yields\[
\delta_{(\rho)}^{0}S=\int_{\Sigma}d^{n}\sigma\quad\partial_{\mu}\rho^{a}\, j_{a}^{\mu}+\int_{\partial\Sigma}(\ldots)\]
The (conserved) Noether current thus can be read off from the derivative-free
variation of the action as the coefficient of $\partial_{\mu}\rho^{a}$.
We could then proceed with a variation $\delta_{(\rho)}^{1}\allfields{I}\equiv\partial_{\mu}\rho^{a}\delta_{a}^{\mu}\allfields{I}$
to derive $j_{a}^{\mu\mu_{1}}$ from the coefficient of $\partial_{\mu}\partial_{\mu_{1}}\rho^{a}$,
and so on. All this is done at the same time in (\ref{eq:noet:trick}).$\qquad\fussend$%
} calculating just $j_{a}^{\mu}$. 

\rem{Savvidy!?:\begin{eqnarray}
\delta A_{\mu}^{a} & = & \partial_{\mu}\rho^{a}\\
\delta A_{\mu\mu_{1}}^{a} & = & \partial_{\mu}(\partial_{\mu_{1}}\rho^{a})\\
\delta A_{\mu\mu_{1}\mu_{2}}^{a} & = & \partial_{\mu}(\partial_{\mu_{1}}\partial_{\mu_{2}}\rho^{a})\end{eqnarray}
} \rem{

\section{Hamiltonian formalism}

\section{Noether procedure}

}

\section{Noether current for the commutator of two symmetries}

Determining the Noether charge for the commutator of two symmetries
is a very simple task in the Hamiltonian formalism. As the charges
generate the symmetries via the Poisson bracket, we have $\delta_{1}\allfields{I}=\left\{ Q_{1},\allfields{I}\right\} $
and $\delta_{2}\allfields{I}=\left\{ Q_{2},\allfields{I}\right\} $.
The Jacobi identity for the Poisson bracket then implies for the commutator
of the symmetry transformations that $\left[\delta_{1},\delta_{2}\right]\allfields{I}=\left\{ \left\{ Q_{1},Q_{2}\right\} ,\allfields{I}\right\} $.
In other words $\left\{ Q_{1},Q_{2}\right\} =\delta_{1}Q_{2}=-\delta_{2}Q_{1}$
is the charge corresponding to the symmetry transformation $\left[\delta_{1},\delta_{2}\right]$.
After dropping the integration over space, this relation also holds
for the currents, i.e. $\delta_{1}j_{2}^{\mu}=-\delta_{2}j_{1}^{\mu}$
is the divergence-free (on-shell) current corresponding to the transformation
$\left[\delta_{1},\delta_{2}\right]$.

Of course one expects to obtain the same result within the Lagrangian
formalism. And on-shell this indeed has to be the case. Off-shell,
however, there might be a difference to the Hamiltonian formalism.
In order to capture all the subtleties, we will therefore derive in
the following the off-shell Noether current corresponding to $\left[\delta_{1},\delta_{2}\right]$
within in the Lagrangian formalism. As it turns out, the derivation
is a bit more involved than one might expect.

The current corresponding to the symmetry transformation $[\delta_{1},\delta_{2}]$
can in principle easily be computed if we know $\mc{K}_{1}^{\mu}$
and $\mc{K}_{2}^{\mu}$ with $\delta\mc{L}=\partial_{\mu}\mc{K}^{\mu}$
for the symmetries $\delta_{1}$ and $\delta_{2}$. By acting with
the commutator symmetry on the Lagrangian, we get a simple expression
for the total derivative term for this symmetry:\begin{eqnarray}
[\delta_{1},\delta_{2}]\mc{L} & = & \delta_{1}\partial_{\mu}\mc{K}_{2}^{\mu}-\delta_{2}\partial_{\mu}\mc{K}_{1}^{\mu}=\partial_{\mu}\left(2\delta_{[1}\mc{K}_{2]}^{\mu}\right)\end{eqnarray}
Knowing the total derivative term (up to trivially conserved terms),
the corresponding current is simply (according to (\ref{eq:noet:Noether-current}))
\begin{eqnarray}
j_{[\delta_{1},\delta_{2}]}^{\mu} & = & [\delta_{1},\delta_{2}]\allfields{I}\partl{(\partial_{\mu}\allfields{I})}\mc{L}+\nonumber \\
 &  & +\sum_{k\geq1}\sum_{i=0}^{k}(-)^{i}\partial_{\nu_{1}}\ldots\partial_{\nu_{k-i}}[\delta_{1},\delta_{2}]\allfields{I}\cdot\partial_{\nu_{k-i+1}}\ldots\partial_{\nu_{k}}\partiell{\mc{L}}{(\partial_{\mu}\partial_{\nu_{1}}\ldots\partial_{\nu_{k}}\allfields{I})}-2\delta_{[1}\mc{K}_{2]}^{\mu}\qquad\label{eq:jforcomm}\end{eqnarray}
The nontrivial part is now to show that this current is (at least
on-shell) equal to $\delta_{1}j_{2}^{\mu}$ or $-\delta_{2}j_{1}^{\mu}$,
which was suggested by the Hamiltonian formalism. We start with two
currents corresponding to two symmetry transformations \begin{eqnarray}
\partial_{\mu}j_{1}^{\mu} & = & -\delta_{1}\allfields{I}\funktional{S}{\allfields{I}},\quad\partial_{\mu}j_{2}^{\mu}=-\delta_{2}\allfields{I}\funktional{S}{\allfields{I}}\label{eq:joneandjtwo}\end{eqnarray}

\paragraph{How not to do it. }

The way to derive the desired result presented in the main part of
the original version of this thesis was unfortunately wrong (although
luckily without bad consequences). Let me shortly sketch it and point
out the trap. Acting in (\ref{eq:joneandjtwo}) with $\delta_{1}$
on $\partial_{\mu}j_{2}^{\mu}$ and subtracting $\delta_{2}$ of $\partial_{\mu}j_{1}^{\mu}$,
one obtains \begin{eqnarray}
\partial_{\mu}\left(\delta_{1}j_{2}^{\mu}-\delta_{2}j_{1}^{\mu}\right) & = & -[\delta_{1},\delta_{2}]\allfields{I}\funktional{S}{\allfields{I}}+2\delta_{[1}\allfields{I}\delta_{2]}\funktional{S}{\allfields{I}}\end{eqnarray}
So far everything is correct, and it is tempting to argue that the
last term is vanishing. The reasoning would be $\delta_{[1}\allfields{I}\delta_{2]}\funktional{S}{\allfields{I}}\stackrel{?}{=}\delta_{[1}\allfields{I}\delta_{2]}\allfields{J}\funktional{^{2}S}{\allfields{J}\delta\allfields{I}}=0$.
The last step is true for symmetry reasons, but the step before is
simply wrong, because it misses an integration of the form $\delta_{[1}\allfields{I}\delta_{2]}\funktional{S}{\allfields{I}}=\int d\tilde{\sigma}\quad\delta_{[1}\allfields{I}\delta_{2]}\allfields{J}(\tilde{\sigma})\funktional{^{2}S}{\allfields{J}(\tilde{\sigma})\delta\allfields{I}}$.
This integration, however, destroys the symmetry argument. Moreover,
not only the derivation is wrong, but also the result (by a factor
of two). Following the wrong argument of above, $\delta_{1}j_{2}^{\mu}-\delta_{2}j_{1}^{\mu}$
would be the current of $[\delta_{1},\delta_{2}]$ instead of $\delta_{1}j_{2}=-\delta_{2}j_{1}=\tfrac{1}{2}\left(\delta_{1}j_{2}^{\mu}-\delta_{2}j_{1}^{\mu}\right)$
(the result from the Hamiltonian reasoning).

\paragraph{Correct derivation in the Lagrangian formalism.}

It will be very useful in the following to use a shorthand notation
in which repeated indices which are at the same vertical position
are simply symmetrized, like for example in $(\partial_{\nu})^{2}A_{\nu}\equiv\partial_{\nu}\partial_{\nu}A_{\nu}\tilde{\equiv}\partial_{(\nu_{1}}\partial_{\nu_{2}}A_{\nu{}_{3})}$.
Only if they are at opposite vertical position they are summed over.
In this context one should also be aware that lower index positions
in the denominator correspond to upper index positions in the nominator.
This notation is similar to the one introduced on page \pageref{par:Schematic-index-notation}
for antisymmetrized indices.

Let us now once more act in (\ref{eq:joneandjtwo}) with $\delta_{1}$
on $\partial_{\mu}j_{2}^{\mu}$ (without subtracting $\delta_{2}\partial_{\mu}j_{1}^{\mu}$)
and reformulate the righthand side such that we obtain the desired
result plus some rest: \begin{eqnarray}
\partial_{\mu}\left(\delta_{1}j_{2}^{\mu}\right) & = & -\delta_{1}\delta_{2}\allfields{I}\funktional{S}{\allfields{I}}-\delta_{2}\allfields{I}\delta_{1}\funktional{S}{\allfields{I}}=\\
 & = & -[\delta_{1},\delta_{2}]\allfields{I}\funktional{S}{\allfields{I}}+\nonumber \\
 &  & -\delta_{2}\delta_{1}\allfields{I}\funktional{S}{\allfields{I}}-\delta_{2}\allfields{I}\delta_{1}\funktional{S}{\allfields{I}}\label{eq:noet:intermediate}\end{eqnarray}
This time we should be more careful about the variation $\delta_{1}$
of the variational derivative $\funktional{S}{\allfields{I}}$ and
we assume that we study a point $\sigma^{\mu}$ which is not at the
boundary of the manifold $\Sigma$ (which means that the variational
derivative of boundary terms with respect to $\allfields{I}(\sigma)$
vanishes):\begin{eqnarray}
\delta_{1}\funktional{S}{\allfields{I}(\sigma)} & = & \funktional{(\overbrace{\delta_{1}S}^{0})}{\allfields{I}(\sigma)}+[\delta_{1},\funktional{}{\allfields{I}(\sigma)}]S=\\
 & = & \int d^{n}\tilde{\sigma}\quad\Biggl(\delta_{1}\allfields{J}(\tilde{\sigma})\funktional{^{2}S}{\allfields{J}(\tilde{\sigma})\delta\allfields{I}(\sigma)}-\funktional{}{\allfields{I}(\sigma)}\bigl(\delta_{1}\allfields{J}(\tilde{\sigma})\funktional{S}{\allfields{J}(\tilde{\sigma})}\bigr)\Biggr)=\\
 & = & -\int d^{n}\tilde{\sigma}\quad\underbrace{\Biggl(\funktional{}{\allfields{I}(\sigma)}\delta_{1}\allfields{J}(\tilde{\sigma})\Biggr)}_{\delta(\sigma-\tilde{\sigma})\partiell{(\delta_{1}\allfields{J})}{\allfields{I}}(\sigma)\lqn{{\scriptstyle +\tilde{\partial}_{\mu}\delta(\sigma-\tilde{\sigma})\partiell{(\delta_{1}\allfields{J})}{(\partial_{\mu}\allfields{I})}(\tilde{\sigma})+\tilde{\partial}_{\mu}\tilde{\partial}_{\mu}\delta(\sigma-\tilde{\sigma})\partiell{(\delta_{1}\allfields{J})}{(\partial_{\mu}\partial_{\mu}\allfields{I})}(\tilde{\sigma})+\ldots}}}\funktional{S}{\allfields{J}(\tilde{\sigma})}=\\
 & = & -\partiell{(\delta_{1}\allfields{J})}{\allfields{I}}\funktional{S}{\allfields{J}}-\sum_{k\geq1}(-)^{k}(\partial_{\mu})^{k}\Biggl(\partiell{(\delta_{1}\allfields{J})}{((\partial_{\mu})^{k}\allfields{I})}\funktional{S}{\allfields{J}}\Biggr)\end{eqnarray}
The righthand side vanishes on-shell which shows that the symmetry
transformation of an equation of motion is always another valid equation
of motion. Likewise we can expand $\delta_{2}\delta_{1}\allfields{I}$
as \begin{eqnarray}
\delta_{2}\delta_{1}\allfields{I} & = & \delta_{2}\allfields{K}\partiell{(\delta_{1}\allfields{I})}{\allfields{K}}+\sum_{k\geq1}(\partial_{\mu})^{k}\delta_{2}\allfields{K}\cdot\partiell{(\delta_{1}\allfields{I})}{(\partial_{\mu})^{k}\allfields{K}}\end{eqnarray}
Plugging the above two expansions into the variation (\ref{eq:noet:intermediate})
of the current-divergence yields

\begin{eqnarray}
\partial_{\mu}\left(\delta_{1}j_{2}^{\mu}\right) & = & -[\delta_{1},\delta_{2}]\allfields{I}\funktional{S}{\allfields{I}}+\nonumber \\
 &  & -\sum_{k\geq1}(\partial_{\mu})^{k}\delta_{2}\allfields{K}\cdot\partiell{(\delta_{1}\allfields{I})}{(\partial_{\mu})^{k}\allfields{K}}\funktional{S}{\allfields{I}}+\sum_{k\geq1}(-)^{k}\delta_{2}\allfields{I}(\partial_{\mu})^{k}\Biggl(\partiell{(\delta_{1}\allfields{J})}{((\partial_{\mu})^{k}\allfields{I})}\funktional{S}{\allfields{J}}\Biggr)\end{eqnarray}
Now we can use the schematic formula $-\partial^{k}a\cdot b+(-)^{k}a\partial^{k}b=-\partial\left(\sum_{l=0}^{k-1}(-)^{l}\partial^{k-1-l}a\cdot\partial^{l}b\right)$
from footnote \ref{foot:iteratedPartialIntegration}. The total derivative
can then be added to $\delta_{1}j_{2}^{\mu}$ on the lefthand side.
Therefore the current defined by\index{Noether current!$\sim$ for commutator of symmetries}\begin{equation}
\boxed{j_{[\delta_{1},\delta_{2}]}^{\mu}\equiv\delta_{1}j_{2}^{\mu}+\sum_{k\geq1}\sum_{l=0}^{k-1}(-)^{l}(\partial_{\mu})^{l}\Bigl(\funktional{S}{\allfields{I}}\partiell{(\delta_{1}\allfields{I})}{(\partial_{\mu})^{k}\allfields{K}}\Bigr)\cdot(\partial_{\mu})^{k-1-l}\delta_{2}\allfields{K}}\label{eq:noet:jcommisdeltaj}\end{equation}
obeys \begin{equation}
\partial_{\mu}j_{[\delta_{1},\delta_{2}]}^{\mu}=-[\delta_{1},\delta_{2}]\allfields{I}\funktional{S}{\allfields{I}}\end{equation}
and is thus the off-shell Noether current corresponding to the commutator
symmetry $[\delta_{1},\delta_{2}]$. Remember that this Noether current
is defined only up to trivially conserved terms. The fact that the
current $j_{[\delta_{1},\delta_{2}]}^{\mu}$ is antisymmetric in $1$
and $2$ also implies that\begin{eqnarray}
\delta_{1}j_{2}^{\mu} & = & -\delta_{2}j_{1}^{\mu}-2\sum_{k\geq1}\sum_{l=0}^{k-1}(-)^{l}(\partial_{\mu})^{k-1-l}\delta_{(1|}\allfields{K}\cdot(\partial_{\mu})^{l}\Bigl(\partiell{(\delta_{|2)}\allfields{I})}{(\partial_{\mu})^{k}\allfields{K}}\funktional{S}{\allfields{I}}\Bigr)\end{eqnarray}
Only on-shell these results coincide with the ones from the Hamiltonian
formalism.

Note that one could also start with equation (\ref{eq:noet:Noether-current})
for the current $j_{2}^{\mu}$ and act on it with $\delta_{1}$ (instead
of acting on the divergence of this equation). In order to turn the
result into something resembling (\ref{eq:jforcomm}), one needs to
make use of several commutators like $[\partl{((\partial_{\nu})^{k}\allfields{I})},\partial_{\mu}]=\delta_{\mu}^{\nu}\partl{((\partial_{\nu})^{k-1}\allfields{I})}\quad\forall k\geq1\quad(0$
for $k=0)$, which imply by induction $[\partl{(\partial_{\nu}^{k}\allfields{I})},\partial_{\rho}^{l}]=\sum_{c=1}^{k}\binom{l}{c}\delta_{\rho\ldots\rho}^{\nu\ldots\nu}\partial_{\rho}^{l-c}\partl{(\partial_{\nu}^{k-c}\allfields{I})}\quad\forall k\geq1$.
The derivation of the latter commutators involves the formula\index{combinatorical formula}
$\sum_{i=0}^{r}\binom{i}{c}=\binom{r+1}{c+1}$. Following this path
becomes extremely clumsy and I managed to follow it to the end only
if the Lagrangian depends maximally on first order derivatives.

\bibliographystyle{/home/basti/fullsort}
\bibliography{/home/basti/phd,/home/basti/Proposal}
\printindex{}
}

\chapter{Torsion, Curvature H-field and their Bianchi identities}

\label{cha:BIs}{\inputTeil{0} \ifthenelse{\theinput=1}{}{}

\title{Torsion, Curvature, $H$-field and their Bianchi Identities}

\author{Sebastian Guttenberg}

\date{January 12, '08}

\maketitle
\begin{abstract}
Part of thesis, originally part of Berkovits\_in\_background.lyx
\end{abstract}
\tableofcontents{}

\rem{To do:

\begin{itemize}
\item BI's: Was ist bekannt, was will man wissen?
\end{itemize}
}

In the following we are frequently making use of the (super)vielbein\index{supervielbein|see{vielbein}}\index{vielbein}
and its inverse\index{vielbein!inverse $\sim$}\index{inverse vielbein},
i.e. a local frame in (co)tangent space different from the coordinate
basis. We denote it via\index{$E^A$|itext{vielbein 1-form}}\index{$E_M\hoch{A}$|itext{vielbein components}}\index{$E_A\hoch{M}$|itext{inverse vielbein components}}\index{$E_A$|itext{vielbein basis vector}}\begin{eqnarray}
E^{A} & \equiv & \de x^{M}E_{M}\hoch{A}\\
E_{A}\hoch{K}E_{K}\hoch{B} & \equiv & \delta_{A}\hoch{B}\\
E_{A} & \equiv & E_{A}\hoch{K}\pe_{K}\end{eqnarray}
The one forms $E^{A}$ are chosen in such a way that they obey nice
properties, i.e. in a Riemannian space it is natural to choose an
orthonormal\index{orthonormal frame} frame, while if no metric is
present, it can be replaced by other requirements like e.g. invariance
under supersymmetry for flat superspace. The structure group is then
the set of transformations of the vielbein which do not change these
properties.

To be a useful concept, the frame should be invariant under the covariant
derivative.\begin{eqnarray}
0 & \stackrel{!}{=} & \nabla_{M}E_{N}\hoch{A}\equiv\partial_{M}E_{N}\hoch{A}+\Omega_{MB}\hoch{A}E_{N}\hoch{B}-\Gamma_{MN}\hoch{K}E_{K}\hoch{A}\end{eqnarray}
This relates the spacetime connection to the structure group connection.

\section{Definition of torsion and curvature and $H$-field}

\subsection{Torsion}

\label{sub:Torsion}\index{torsion|fett}There are at least three
ways to define the torsion. Let us start with the component based
one and derive from this the more geometric (coordinate independent)
definintion. So at first we define the (super) torsion components
simply as the antisymmetric part of the connection coefficients\index{$T_{MN}\hoch{K}$|itext{torsion components}}\begin{equation}
\boxed{T_{MN}\hoch{K}\equiv\Gamma_{[MN]}\hoch{K}}\label{eq:TorsionDefI}\end{equation}
The structure group connection $\Omega_{MA}\hoch{B}$ is given by
demanding that the covariant derivative of the vielbein vanishes\begin{eqnarray}
0\stackrel{!}{=}\nabla_{M}E_{N}\hoch{A} & = & \partial_{M}E_{N}\hoch{A}-\Gamma_{MN}\hoch{K}E_{K}\hoch{A}+\Omega_{MB}\hoch{A}E_{N}\hoch{B}\label{eq:OmegaGammaRelation}\end{eqnarray}
Antisymmetrizing in $(M,N)$ and comparing with (\ref{eq:TorsionDefI})
yields%
\footnote{\label{foot:wedge}\index{footnote!\thefoot. missing factor in wedge product}Note
that in the present text form components are defined as e.g. $T^{A}=T_{MN}\hoch{A}\de x^{M}\wedge\de x^{N}$
with no (!) factor $\frac{1}{2}$ in front which corresponds to a
definition of the wedge product as $\de x^{M}\de x^{N}\equiv\de x^{M}\wedge\de x^{N}\equiv\de x^{[M}\otimes\de x^{N]}\equiv\frac{1}{2}\left(\de x^{M}\otimes\de x^{N}-\de x^{M}\otimes\de x^{N}\right)$.
You will thus usually find in literature a factor of 2 on the righthand
side of (\ref{eq:TorsionDefI}) and a factor $\frac{1}{2}$ in (\ref{eq:TorsionDefIII}).
To go from one convention to the other, simply replace $T_{MN}\hoch{K}$
by $2T_{MN}\hoch{K}$ in all equations in component form. (For a p-form
the factor is of course $p!$). Coordinate independent equations like
(\ref{eq:TorsionDefII}) remain untouched because of the compensating
redefinition of the wedge product and the resulting redefinition of
the exterior product.$\qquad\fussend$%
}\index{$T^A$|itext{torsion 2-form}}\begin{equation}
\boxed{T^{A}=\de E^{A}-E^{B}\wedge\Omega_{B}\hoch{A}}\label{eq:TorsionDefII}\end{equation}
This can be used as an alternative definition to (\ref{eq:TorsionDefI}).
Consider now the commutator of two covariant derivatives on a scalar
(super) field (with $\nabla_{K}\varphi=\partial_{K}\varphi$)\begin{eqnarray}
\left[\nabla_{M},\nabla_{N}\right]\varphi & = & 2\nabla_{[M}\partial_{N]}\varphi=\\
 & = & -2\Gamma_{[MN]}\hoch{K}\partial_{K}\varphi\end{eqnarray}
or simply \begin{equation}
\boxed{\nabla_{[M}\nabla_{N]}\varphi=-T_{MN}\hoch{K}\nabla_{K}\varphi}\label{eq:TorsionDefIII}\end{equation}
which is yet an alternative and equivalent definition of the torsion.

\subsection{Curvature}

\label{sub:Curvature}\index{curvature|fett}For the curvature, let
us start with the definition via the commutator of covariant derivatives
acting on vector fields\begin{equation}
\boxed{\nabla_{[M}\nabla_{N]}v^{A}=-T_{MN}\hoch{K}\nabla_{K}v^{A}+R_{MNB}\hoch{A}v^{B}}\label{eq:curvatureDefIII}\end{equation}
This is not only a definition, but also a proposition that the commutator
takes this form. Let us check this and by doing this get a definition
of the curvature in component form\begin{eqnarray}
\lqn{\nabla_{[M}\nabla_{N]}v^{A}=}\nonumber \\
 & = & \partial_{[M}(\partial_{N]}v^{A}+\Omega_{N]B}\hoch{A}v^{B})+\Omega_{[M|C}\hoch{A}(\partial_{|N]}v^{C}+\Omega_{|N]B}\hoch{C}v^{B})-\Gamma_{[MN]}\hoch{K}(\partial_{K}v^{A}+\Omega_{KB}\hoch{A}v^{B})=\\
 & = & \partial_{[M}\Omega_{N]B}\hoch{A}v^{B}+\Omega_{[N|B}\hoch{A}\partial_{|M]}v^{B}+\Omega_{[M|C}\hoch{A}(\partial_{|N]}v^{C}+\Omega_{|N]B}\hoch{C}v^{B})-T_{[MN]}\hoch{K}\nabla_{K}v^{A}=\\
 & = & -T_{[MN]}\hoch{K}\nabla_{K}v^{A}+\left(\partial_{[M}\Omega_{N]B}\hoch{A}+\Omega_{[M|C}\hoch{A}\Omega_{|N]B}\hoch{C}\right)v^{B}\end{eqnarray}
We can thus read off \index{$R_{MNA}\hoch{B}$|itext{curvature components}}\begin{eqnarray}
R_{MNB}\hoch{A} & = & \partial_{[M}\Omega_{N]B}\hoch{A}-\Omega_{[M|B}\hoch{C}\Omega_{|N]C}\hoch{A}\end{eqnarray}
which in form language reads\index{$R_A\hoch{B}$|itext{curvature 2-form}}\begin{equation}
\boxed{R_{A}\hoch{B}=\de\Omega_{A}\hoch{B}-\Omega_{A}\hoch{C}\wedge\Omega_{C}\hoch{B}}\label{eq:curvatureDefII}\end{equation}
We finally can rewrite this in terms of $\Gamma$ by using (\ref{eq:OmegaGammaRelation})
in the simplified form\begin{equation}
\Omega_{MB}\hoch{A}=\Gamma_{MB}\hoch{A}-E_{B}\hoch{R}\partial_{M}E_{R}\hoch{A}\label{eq:OmegaGammaRelationb}\end{equation}
$\dann$\begin{eqnarray}
R_{MNB}\hoch{A} & = & \partial_{[M|}\left(\Gamma_{|N]B}\hoch{A}-E_{B}\hoch{R}\partial_{|N]}E_{R}\hoch{A}\right)-\left(\Gamma_{[M|B}\hoch{C}-E_{B}\hoch{R}\partial_{[M|}E_{R}\hoch{C}\right)\left(\Gamma_{|N]C}\hoch{A}-E_{C}\hoch{S}\partial_{|N]}E_{S}\hoch{A}\right)\\
R_{MNK}\hoch{L} & = & \partial_{[M|}\Gamma_{|N]K}\hoch{L}+E_{K}\hoch{B}\partial_{[M|}E_{B}\hoch{R}\Gamma_{|N]R}\hoch{L}+E_{A}\hoch{L}\partial_{[M|}E_{S}\hoch{A}\Gamma_{|N]K}\hoch{S}-E_{K}\hoch{B}E_{A}\hoch{L}\partial_{[M|}E_{B}\hoch{R}\partial_{|N]}E_{R}\hoch{A}+\nonumber \\
 &  & -\left(\Gamma_{[M|K}\hoch{C}-\partial_{[M|}E_{K}\hoch{C}\right)\left(\Gamma_{|N]C}\hoch{L}-E_{C}\hoch{S}\partial_{|N]}E_{S}\hoch{A}E_{A}\hoch{L}\right)=\\
 & = & \partial_{[M|}\Gamma_{|N]K}\hoch{L}-\Gamma_{[M|K}\hoch{P}\Gamma_{|N]P}\hoch{L}\end{eqnarray}
\begin{equation}
\boxed{R_{MNK}\hoch{L}=\partial_{[M|}\Gamma_{|N]K}\hoch{L}-\Gamma_{[M|K}\hoch{P}\Gamma_{|N]P}\hoch{L}}\label{eq:curvatureDefI}\end{equation}
The same expression can be derived (even simpler) by acting with the
commutator of covariant deriavtives on a vector $v^{M}$ with a curved
index instead of the flat index.

\subsection{Summary, including $H$-field-strength}

Let us add the field strength $H$\index{H@$H$-field}\index{$H$|itext{$3$-form field strength of $B$}}
of the antisymmetric\index{antisymmetric!rank 2 tensor field $B$}
tensor field $B$\index{$B$|itext{B-field 2-form}}\index{B@$B$-field}
to our considerations. We then have \begin{eqnarray}
H & \equiv & \de B\label{eq:HfieldDef}\\
T^{A} & \equiv & \de E^{A}-E^{C}\wedge\Omega_{C}\hoch{A}\\
R_{A}\hoch{B} & \equiv & \de\Omega_{A}\hoch{B}-\Omega_{A}\hoch{C}\wedge\Omega_{C}\hoch{B}\end{eqnarray}
In coordinate basis ('curved indices') we have\index{$H_{MNK}$|itext{$H$-field components}}\index{$B_{MN}$|itext{$B$-field components}}
\begin{eqnarray}
H_{MNK} & \equiv & \partial_{[M}B_{NK]}\\
T_{MN}\hoch{K} & \equiv & \Gamma_{[MN]}\hoch{K}\\
R_{MNK}\hoch{L} & \equiv & \partial_{[M|}\Gamma_{|N]K}\hoch{L}-\Gamma_{[M|K}\hoch{C}\Gamma_{|N]C}\hoch{L}\end{eqnarray}
The commutator of covariant derivatives on an arbitrary rank (p,q)-tensor
fields (as a generalization of (\ref{eq:TorsionDefIII}) and (\ref{eq:curvatureDefIII}))
reads\index{commutator!of covariant derivatives}\index{general!commutator of covariant derivatives}\begin{eqnarray}
\lqn{\nabla_{[M}\nabla_{N]}t_{B_{1}\ldots B_{p}}^{A_{1}\ldots A_{q}}=}\nonumber \\
 & = & -T_{MN}\hoch{K}\nabla_{K}t_{B_{1}\ldots B_{p}}^{A_{1}\ldots A_{q}}+\sum_{i=1}^{q}R_{MNC}\hoch{A_{i}}t_{B_{1}\ldots B_{p}}^{A_{1}\ldots A_{i-1}CA_{i+1}\ldots A_{q}}-\sum_{i=1}^{q}R_{MNB_{i}}\hoch{C}t_{B_{1}\ldots B_{i-1}CB_{i+1}\ldots B_{p}}^{A_{1}\ldots A_{q}}\label{eq:generalCommutatorOfCovDer}\end{eqnarray}
This can be generalized yet a bit more, if we want to include fields
that do not transform tensorial, like e.g. the compensator field.
If we denote the representation of the structure group transformation,
or better the representation of an Lie algebra element, by $\group{L_{\,\cdot}\hoch{\cdot}}$\index{$R$@$\mc{R}(L_{\cdot}\hoch{\cdot})$}\index{structure group!representation $\mc{R}$}\index{representation!of the structure group}
(where $L_{A}\hoch{B}$ is the matrix of the fundamental representation),
the covariant derivative can be written as\begin{eqnarray}
\nabla_{M} & = & \partial_{M}+\mc{R}(\Omega_{M\,\cdot}\hoch{\cdot})\label{eq:covDerWithRepresentation}\end{eqnarray}
The commutator takes the general form\begin{eqnarray}
\nabla_{[M}\nabla_{N]} & = & -T_{MN}\hoch{K}\nabla_{K}+\mc{R}(R_{MN\,\cdot}\hoch{\cdot})\label{eq:mostgeneralCommutatorOfCovDer}\end{eqnarray}
This is in particular interesting for the compensator field, where
we have a negative shift as representations and therefore%
\footnote{\index{footnote!\thefoot. covariant derivative of a connection}\label{foot:covDerOfConn}It
is even possible now to define a covariant derivative of a connection
(see (\ref{eq:connectionTransformsLikeConnection}) on page \pageref{eq:connectionTransformsLikeConnection}
or footnote \vref{foot:structuregroupTrafoConnection} for the representation
of the structure group and its algebra on the connection) \begin{eqnarray*}
\nabla_{M}\tilde{\Omega}_{NA}\hoch{B} & = & \partial_{M}\tilde{\Omega}_{NA}\hoch{B}\underbrace{-\partial_{N}\Omega_{MA}\hoch{B}-[\Omega_{M},\tilde{\Omega}_{N}]_{A}\hoch{B}}_{\mc{R}(\Omega_{M\,\cdot}\hoch{\cdot})\tilde{\Omega}_{NA}\hoch{B}}\end{eqnarray*}
If the two connections coincide, we obtain\begin{eqnarray*}
\nabla_{M}\Omega_{NA}\hoch{B} & = & \partial_{M}\Omega_{NA}\hoch{B}-\partial_{N}\Omega_{MA}\hoch{B}-[\Omega_{M},\Omega_{N}]_{A}\hoch{B}=2R_{MNA}\hoch{B}\qquad\fussend\end{eqnarray*}
} \begin{eqnarray}
\nabla_{M}\Phi & = & \partial_{M}\Phi-\Omega_{M}^{(D)}\label{eq:covDerCompensator}\\
\nabla_{[M}\nabla_{N]}\Phi & = & -T_{MN}\hoch{K}\nabla_{K}\Phi-F_{MN}^{(D)}\label{eq:CommOfCovDerCompensator}\end{eqnarray}

Using the definition of the torsion, exterior derivatives of p-forms
$\eta^{(p)}$ can be rewritten with covariant derivatives, thus allowing
to switch to flat coordinates\index{exterior covariant derivative}\index{covariant derivative!exterior $\sim$}\begin{eqnarray}
\partial_{[M_{1}}\eta_{M_{2}\ldots M_{p+1}]} & = & \nabla_{[M_{1}}\eta_{M_{2}\ldots M_{p+1}]}+pT_{[M_{1}M_{2}|}\hoch{K}\eta_{K|M_{3}\ldots M_{p+1}]}\label{eq:extDerInTermsOfCovDer}\end{eqnarray}
In particular \begin{equation}
H=\partial_{\bs{M}}B_{\bs{MM}}=\nabla_{\bs{A}}B_{\bs{AA}}+2T_{\bs{AA}}\hoch{C}B_{C\bs{A}}\label{eq:HinFlatCoord}\end{equation}

\section{The Bianchi identities}

\label{sec:The-Bianchi-identities}Bianchi identities all base on
the nilpotency of the exterior derivative $\de^{2}=0$. The objects
$H$, $T^{A}$ and $R_{A}\hoch{B}$ are all defined using the exterior
derivative. Acting a second time with the exterior derivative (using
$\de^{2}=0$) yields consitency conditions (the Bianchi identities)
which have to be fulfilled by any valid $H$, $T^{A}$or $R_{A}\hoch{B}$.
While these identities are trivially fulfilled, if the original definitions
for these objects are used, the imposure of constraints on them makes
a check necessary.%
\footnote{\index{footnote!\thefoot. example for use of BI's}Let us look at
an example to make this point clear: one of the supergravity constraints
that we get is $H_{\bs{\alpha\beta\gamma}}=0$. As $H$ was defined
via $H=\de B$ in the beginning, this is actually a differential equation
for $B$ of the form $E_{\bs{\alpha}}\hoch{M}E_{\bs{\beta}}\hoch{N}E_{\bs{\gamma}}\hoch{K}\left(\partial_{[M}B_{NK]}\right)=0$.
One could try to calculate the general solution for this equation
(which might be quite hard) and then calculate the $H$-field via
$H=\de B$ which will of course trivially obey the Bianchi identities.
However, one prefers not to solve for $B$, but to calculate additional
constraints on $H$ using the Bianchi identities. The idea is to get
the full information about $H$ without solving for $B$. The same
story holds for the other objects.$\qquad\fussend$%
}

\subsection{BI for $H_{ABC}$}

The most simple Bianchi identity is the one of the $H$-field $H=\de B$
(\ref{eq:HfieldDef}). It just reads

\begin{eqnarray}
\de H & \stackrel{!}{=} & 0\label{eq:BIforH}\end{eqnarray}
The supergravity constraints on $H$ that we will obtain, however,
are all in flat coordinates, so that it is convenient to rewrite the
Bianchi identity (using (\ref{eq:extDerInTermsOfCovDer})) with covariant
derivatives\index{Bianchi identity!$H$-field $\sim$}\index{H@$H$-field!Bianchi identity}
and then contract with vielbeins in order to turn the curved indices
into flat ones:\begin{equation}
\boxed{\nabla_{\bs{A}}H_{\bs{AAA}}\stackrel{!}{=}-3T_{\bs{AA}}\hoch{C}H_{C\bs{AA}}}\label{eq:BIforHcov}\end{equation}
Regarding the torsion as a vector valued 2-form and using the generalized
definition of the interior product\rem{given in...?}, this can also
be written as \begin{eqnarray}
\bs{\nabla}H & \equiv & \de H-\ip_{T}H\stackrel{!}{=}-\ip_{T}H\label{eq:BIforHcovCondensed}\end{eqnarray}

\rem{

\paragraph{Remark}

Using the connection with torsion to define $\tilde{H}$, we get \begin{eqnarray}
\tilde{H}_{\bs{MMM}} & \equiv & \nabla_{\bs{M}}B_{\bs{MM}}=\\
 & = & \partial_{\bs{M}}B_{\bs{MM}}-2T_{\bs{MM}}\hoch{K}B_{K\bs{M}}\\
\tilde{H}=\bs{\nabla}B & = & \de B-\ip_{T}B\\
\bs{\nabla}\tilde{H} & = & \de\tilde{H}-\ip_{T}\tilde{H}=\\
 & = & \de\left(\de B-\ip_{T}B\right)-\ip_{T}\left(\de B-\ip_{T}B\right)=\\
 & = & -\ip_{\de T}B+\ip_{T}\de B-\ip_{T}\de B+\frac{1}{2}[\ip_{T},\ip_{T}]B=\\
 & = & -\underbrace{\ip_{\de T}B}_{\mc{L}_{T}B=[T\bs{,}B]}+\frac{1}{2}\ip_{[T,T]^{\Delta}}B=\\
 & = & -\ip_{\de T-\frac{1}{2}[T,T]^{\Delta}}B\\
\textrm{with }{}[T,T]^{\Delta} & = & 2\ip_{T}T\end{eqnarray}
where \begin{eqnarray}
\de T & \equiv & \de T^{A}\cdot E_{A}+T^{A}\de(E_{A}\hoch{K}\pe_{K})=\\
 & = & \de T^{A}\cdot E_{A}+T^{A}\underbrace{\partial_{M}E_{A}\hoch{K}}_{-E_{A}\hoch{L}\partial_{M}E_{L}\hoch{B}E_{B}\hoch{K}}\de x^{M}\pe_{K}+T^{K}p_{K}=\\
 & \stackrel{T-\textrm{BI}}{=} & (E^{C}\wedge R_{C}\hoch{A}-T^{C}\wedge\Omega_{C}\hoch{A})\cdot E_{A}-T^{L}\partial_{M}E_{L}\hoch{B}\de x^{M}E_{B}+T^{K}p_{K}=\\
 & = & \left(E^{C}\wedge R_{C}\hoch{A}-T^{L}\Gamma_{ML}\hoch{A}\de x^{M}\right)\cdot E_{A}+T^{K}p_{K}=\\
 & = & E^{C}\wedge R_{C}\hoch{A}E_{A}+T^{K}p_{K}-T^{L}\Gamma_{ML}\hoch{K}\de x^{M}\pe_{K}\end{eqnarray}
\begin{eqnarray}
\ip_{\de T}\rho & = & \ip_{R}\rho+(\ip_{T}\nabla)\rho+\frac{1}{2}\ip_{\ip_{T}T}\rho\end{eqnarray}

}

\subsection{BI for $T^{A}$}

Remember $T^{A}=\de E^{A}-E^{C}\wedge\Omega_{C}\hoch{A}$ (\ref{eq:TorsionDefII}).
Acting on this equation with the exterior derivative yields\begin{eqnarray}
\de T^{A} & = & -\de E^{C}\wedge\Omega_{C}\hoch{A}+E^{C}\wedge\de\Omega_{C}\hoch{A}=\\
 & \os{(\ref{eq:curvatureDefII})}{=} & -T^{C}\wedge\Omega_{C}\hoch{A}-E^{D}\wedge\Omega_{D}\hoch{C}\wedge\Omega_{C}\hoch{A}+E^{C}\wedge R_{C}\hoch{A}+E^{C}\wedge\Omega_{C}\hoch{D}\wedge\Omega_{D}\hoch{A}=\\
 & = & -T^{C}\wedge\Omega_{C}\hoch{A}+E^{C}\wedge R_{C}\hoch{A}\end{eqnarray}
The Bianchi identity for the torsion (sometimes also called the first
Bianchi identity\index{first Bianchi identity|itext{see torsion BI}}\index{Bianchi identity!first $\sim$|see{torsion $\sim$}})
thus reads \begin{eqnarray}
\de T^{A}+T^{C}\wedge\Omega_{C}\hoch{A} & \stackrel{!}{=} & E^{C}\wedge R_{C}\hoch{A}\label{eq:BIforT}\end{eqnarray}
Again we want to rewrite it in terms of the covariant derivative.
The {}``exterior\index{exterior covariant derivative}'' covariant
derivative\index{covariant derivative!exterior $\sim$} of $\, T$
reads\begin{eqnarray}
\nabla_{\bs{M}}T_{\bs{MM}}\hoch{A} & = & \partial_{\bs{M}}T_{\bs{MM}}\hoch{A}-2T_{\bs{MM}}\hoch{K}T_{K\bs{M}}\hoch{A}+\Omega_{\bs{M}B}\hoch{A}T_{\bs{MM}}\hoch{B}\label{eq:covExteriorDerOnT}\\
\bs{\nabla}T^{A} & = & \de T^{A}+T^{B}\wedge\Omega_{B}\hoch{A}-\ip_{T}T^{A}\label{eq:covExteriorDerivativeOnTcondensed}\end{eqnarray}
The above Bianchi-identity can thus be rewritten as\index{torsion!Bianchi identity}\index{Bianchi identity!torsion $\sim$ }
\begin{eqnarray}
\lqn{\Ramm{.4}{\zwek{\Big.}{}}}\quad\nabla_{\bs{A}}T_{\bs{AA}}\hoch{D}+2T_{\bs{AA}}\hoch{C}T_{C\bs{A}}\hoch{D} & \stackrel{!}{=} & R_{\bs{AAA}}\hoch{D}\label{eq:BIforTcov}\\
\bs{\nabla}T^{D}+\ip_{T}T^{D} & \stackrel{!}{=} & R^{D}\equiv E^{C}\wedge R_{C}\hoch{D}\label{eq:BIforTcovCondensed}\end{eqnarray}

\subsection{BI for $R_{A}\hoch{B}$}

Remember $R_{A}\hoch{B}=\de\Omega_{A}\hoch{B}-\Omega_{A}\hoch{C}\wedge\Omega_{C}\hoch{B}$
(\ref{eq:curvatureDefII}). Acting on it with the exterior derivative
yields \begin{eqnarray}
\de R_{A}\hoch{B} & = & -\de\Omega_{A}\hoch{C}\wedge\Omega_{C}\hoch{B}+\Omega_{A}\hoch{C}\wedge\de\Omega_{C}\hoch{B}=\\
 & = & -R_{A}\hoch{C}\wedge\Omega_{C}\hoch{B}-\Omega_{A}\hoch{D}\wedge\Omega_{D}\hoch{C}\wedge\Omega_{C}\hoch{B}+\Omega_{A}\hoch{C}\wedge R_{C}\hoch{B}+\Omega_{A}\hoch{C}\wedge\Omega_{C}\hoch{D}\wedge\Omega_{D}\hoch{B}=\\
 & = & -R_{A}\hoch{C}\wedge\Omega_{C}\hoch{B}+\Omega_{A}\hoch{C}\wedge R_{C}\hoch{B}\end{eqnarray}
The Bianchi identity for the curvature (also called second Bianchi
identity) thus reads\begin{eqnarray}
\de R_{A}\hoch{B}+\underbrace{R_{A}\hoch{C}\wedge\Omega_{C}\hoch{B}-\Omega_{A}\hoch{C}\wedge R_{C}\hoch{B}}_{[R,\Omega]_{A}\hoch{C}} & \stackrel{!}{=} & 0\label{eq:BIforR}\end{eqnarray}
Again we want to rewrite this in terms of covariant derivatives and
flat indices and therefore consider the antisymmetrized covariant
derivative\index{exterior covariant derivative}\index{covariant derivative!exterior $\sim$}\begin{eqnarray}
\nabla_{\bs{M}}R_{\bs{MM}A}\hoch{B} & = & \partial_{\bs{M}}R_{\bs{MM}A}\hoch{B}-2T_{\bs{MM}}\hoch{K}R_{K\bs{M}A}\hoch{B}-\Omega_{\bs{M}A}\hoch{C}R_{\bs{MM}C}\hoch{B}+\Omega_{\bs{M}C}\hoch{B}R_{\bs{MM}A}\hoch{C}\label{eq:covExteriorDerOnR}\\
\bs{\nabla}R_{A}\hoch{B} & = & \de R_{A}\hoch{B}-\Omega_{A}\hoch{C}\wedge R_{C}\hoch{B}+R_{A}\hoch{C}\wedge\Omega_{C}\hoch{B}-\ip_{T}R_{A}\hoch{B}\label{eq:covExteriorDerOnRcondensed}\end{eqnarray}
We thus can rewrite the above Bianchi-identity as\index{Bianchi identity!curvature $\sim$}\index{Bianchi identity!second $\sim$|see{curvature $\sim$}}\index{second Bianchi identity|see{curvature $\sim$}}\index{curvature!Bianchi identity}
\begin{eqnarray}
\lqn{\Ramm{.41}{\big.}}\quad\nabla_{\bs{M}}R_{\bs{MM}A}\hoch{B}+2T_{\bs{MM}}\hoch{K}R_{K\bs{M}A}\hoch{B} & = & 0\label{eq:BIforRcov}\\
\bs{\nabla}R_{A}\hoch{B}+\ip_{T}R_{A}\hoch{B} & = & 0\label{eq:BIforRcovCondensed}\end{eqnarray}
\rem{Folgendes zu den speziellen Eigenschaften?}If the structure
group is restricted to e.g. Lorentz plus scale transformations (see
section \vref{sec:Restricted-structure-group}), we get\index{$R^{(L)}_{MNa}\hoch{b}$|itext{Lorentz curvature}}\index{Lorentz curvature}\index{field strength!scale $\sim$}\index{$F^{(D)}$|itext{scale curvature 2-form}}\index{$F_{MN}^{(D)}$|itext{scale curvature components}}\index{curvature!Lorentz $\sim$}\index{curvature!scale $\sim$}
\begin{eqnarray}
R_{\bs{MM}a}\hoch{b} & = & F_{\bs{MM}}^{(D)}\delta_{a}^{b}+R_{\bs{MM}a}^{(L)}\hoch{b}\\
\mbox{and }R_{\bs{MM}\bs{\alpha}}\hoch{\bs{\beta}} & = & \frac{1}{2}F_{\bs{MM}}^{(D)}\delta_{\bs{\alpha}}\hoch{\bs{\beta}}+\frac{1}{4}R_{\bs{MM}ab}^{(L)}\gamma^{ab}\tief{\bs{\alpha}}\hoch{\bs{\beta}}\end{eqnarray}
The above Bianchi identity then has to hold seperately for Lorentz
and Dilatation part. In particular we have\index{Bianchi identity!scale curvature}\index{scale curvature!Bianchi identity}
\begin{equation}
\boxed{\nabla_{\bs{M}}F_{\bs{MM}}^{(D)}+2T_{\bs{MM}}\hoch{K}F_{K\bs{M}}^{(D)}=0}\label{eq:BIforFcov}\end{equation}

\subsection{Alternative derivation from the Jacobi identity}

The above derivations of the Bianchi identities were based on the
nilpotency $\de²=0$ of the exterior derivative. The Bianchi identities
for curvature and torsion are equivalently obtained from the Jacobi
identity for commutators:\begin{eqnarray}
\left[A,[B,C]\right]+\left[C,[A,B]\right]+\left[B,[C,A]\right] & = & 0\label{eq:Jacobi-for-comm}\end{eqnarray}
Applying this to covariant derivatives, using (\ref{eq:mostgeneralCommutatorOfCovDer})
yields\begin{eqnarray}
0 & = & \left[\nabla_{\bs{M}},[\nabla_{\bs{M}},\nabla_{\bs{M}}]\right]=\\
 & = & -2\left[\nabla_{\bs{M}},\, T_{\bs{MM}}\hoch{K}\nabla_{K}\right]+2\left[\nabla_{\bs{M}},\,\mc{R}(R_{\bs{MM}\,\cdot}\hoch{\cdot})\right]=\\
 & = & -2\nabla_{\bs{M}}T_{\bs{MM}}\hoch{K}\nabla_{K}-2T_{\bs{MM}}\hoch{K}[\nabla_{\bs{M}},\nabla_{K}]+2\mc{R}(\nabla_{\bs{M}}R_{\bs{MM}\,\cdot}\hoch{\cdot})+2R_{\bs{MMM}}\hoch{K}\nabla_{K}=\\
 & = & 2\left(R_{\bs{MMM}}\hoch{K}-\nabla_{\bs{M}}T_{\bs{MM}}\hoch{K}\right)\nabla_{K}-2T_{\bs{MM}}\hoch{K}\left(-2T_{\bs{M}K}\hoch{L}\nabla_{L}+2\mc{R}(R_{\bs{M}K\,\cdot}\hoch{\cdot})\right)+2\mc{R}(\nabla_{\bs{M}}R_{\bs{MM}\,\cdot}\hoch{\cdot})=\qquad\\
 & = & 2\left(R_{\bs{MMM}}\hoch{K}-\nabla_{\bs{M}}T_{\bs{MM}}\hoch{K}-2T_{\bs{MM}}\hoch{L}T_{L\bs{M}}\hoch{K}\right)\nabla_{K}+2\mc{R}\left(\nabla_{\bs{M}}R_{\bs{MM}\,\cdot}\hoch{\cdot}+2T_{\bs{MM}}\hoch{K}R_{K\bs{M}\,\cdot}\hoch{\cdot}\right)\end{eqnarray}
Both brackets have to vanish separately, which correctly reproduces
the identities (\ref{eq:BIforTcov}) and (\ref{eq:BIforRcov}). \rem{

Let us try to find a similar derivation for the Bianchi identity for
$H$\begin{eqnarray*}
\nabla_{M}B_{NK} & \equiv & \partial_{M}B_{NK}-\Gamma_{MN}\hoch{L}B_{LK}-\Gamma_{MK}\hoch{L}B_{NL}-2\partial_{[N|}\Sigma_{M\,|K]}\end{eqnarray*}
}

\section{Shifting the connection}

\index{connection!shift in $\sim$}\index{shift in connection}Some
expressions might look simpler if one changes the connection $\Omega_{MA}\hoch{B}$
to some new connection $\tilde{\Omega}_{MA}\hoch{B}$. As usual, the
difference \begin{equation}
\Delta_{MA}\hoch{B}\equiv\tilde{\Omega}_{MA}\hoch{B}-\Omega_{MA}\hoch{B}\end{equation}
transforms as a tensor (the inhomogenous term in the transformation
cancels). The new torsion looks as follows: \begin{eqnarray}
\tilde{T}^{A} & = & \de E^{A}-E^{C}\wedge\tilde{\Omega}_{C}\hoch{A}=\\
 & = & T^{A}-E^{C}\wedge\Delta_{C}\hoch{A}=\end{eqnarray}
Or simply\index{torsion!with shifted connection} \begin{equation}
\boxed{\tilde{T}_{\bs{MM}}\hoch{A}=T_{\bs{MM}}\hoch{A}+\Delta_{\bs{MM}}\hoch{A}}\label{eq:TwithNewConn}\end{equation}
The expression for the new curvature is a bit more involved and reads%
\footnote{\label{foot:rotatedVielbein}\index{footnote!\thefoot. rotated vielbein}Of
similar interest is a change in the definition of the vielbein. Note
that local structure group transformations of the vielbein which go
along with a structure group transformation of torsion and curvature
also include a corresponding transformation of the connection. Instead
we want to look at an independent transformation of the vielbein and
consider general local $Gl(n)$ transformations. \[
\tilde{E}^{A}=E^{B}J_{B}\hoch{A}\]
 with $\tilde{\nabla}_{M}\tilde{E}^{A}=0$. For the new torsion, we
get\begin{eqnarray*}
\tilde{T}^{A} & = & \de\tilde{E}^{A}-\tilde{E}^{C}\wedge\Omega_{C}\hoch{A}=\\
 & = & \de E^{B}J_{B}\hoch{A}-E^{B}\wedge\de J_{B}\hoch{A}-E^{B}J_{B}\hoch{C}\wedge\Omega_{C}\hoch{A}=\\
 & = & T^{B}J_{B}\hoch{A}-E^{B}\wedge\bs{\nabla}J_{B}\hoch{A}\end{eqnarray*}
or \[
\boxed{\tilde{T}_{\bs{MM}}\hoch{B}=T_{\bs{MM}}\hoch{B}J_{B}\hoch{A}+\nabla_{\bs{M}}J_{\bs{M}}\hoch{A}}\]
The curvature remains untouched\[
\boxed{\tilde{R}_{A}\hoch{B}=R_{A}\hoch{B}}\]
Alternatively one might be interested in shifts of the vielbein (resulting
in $\tilde{T}=T+\de(\Delta E)^{A}-(\Delta E)^{C}\wedge\Omega_{C}\hoch{A}$)
or linear transformations of the connection of the form $\tilde{\Omega}=J\Omega J^{-1}\qquad\fussend$%
}\index{curvature!with shifted connection}\begin{eqnarray}
\hat{R}_{A}\hoch{B} & = & \de\tilde{\Omega}_{A}\hoch{B}-\tilde{\Omega}_{A}\hoch{C}\wedge\tilde{\Omega}_{C}\hoch{B}=\\
 & = & R_{A}\hoch{B}+\de\Delta_{A}\hoch{B}-\Delta_{A}\hoch{C}\wedge\Omega_{C}\hoch{B}-\Omega_{A}\hoch{C}\wedge\Delta_{C}\hoch{B}-\Delta_{A}\hoch{C}\wedge\Delta_{C}\hoch{B}=\\
 & = & R_{A}\hoch{B}+\bs{\nabla}\Delta_{A}\hoch{B}+T^{K}\Delta_{KA}\hoch{B}-\Delta_{A}\hoch{C}\wedge\Delta_{C}\hoch{B}\end{eqnarray}
\begin{equation}
\boxed{\tilde{R}_{\bs{MM}A}\hoch{B}=R_{\bs{MM}A}\hoch{B}+\nabla_{\bs{M}}\Delta_{\bs{M}A}\hoch{B}+T_{\bs{MM}}\hoch{K}\Delta_{KA}\hoch{B}-\Delta_{\bs{M}A}\hoch{C}\Delta_{\bs{M}C}\hoch{B}}\label{eq:RwithNewConn}\end{equation}
or equivalently \begin{equation}
\boxed{\tilde{R}_{\bs{MM}A}\hoch{B}=R_{\bs{MM}A}\hoch{B}+\tilde{\nabla}_{\bs{M}}\Delta_{\bs{M}A}\hoch{B}+\tilde{T}_{\bs{MM}}\hoch{K}\Delta_{KA}\hoch{B}+\Delta_{\bs{M}A}\hoch{C}\Delta_{\bs{M}C}\hoch{B}}\label{eq:RwithNewConnII}\end{equation}
\rem{\begin{eqnarray*}
\hat{R}_{\bs{CC}a}\hoch{b} & \stackrel{\check{\Omega}=\Omega}{=} & R_{\bs{CC}a}\hoch{b}+\gemnabla_{\bs{C}}\Delta_{\bs{C}a}\hoch{b}+\gemT_{\bs{CC}}\hoch{E}\Delta_{Ea}\hoch{b}-\Delta_{\bs{C}a}\hoch{e}\Delta_{\bs{C}e}\hoch{b}\\
\hat{R}_{\bs{CC}\bs{\alpha}}\hoch{\bs{\beta}} & = & R_{\bs{CC}\bs{\alpha}}\hoch{\bs{\beta}}+\nabla_{\bs{C}}\Delta_{\bs{C}\bs{\alpha}}\hoch{\bs{\beta}}+T_{\bs{CC}}\hoch{E}\Delta_{E\bs{\alpha}}\hoch{\bs{\beta}}-\Delta_{\bs{C}\bs{\alpha}}\hoch{\bs{\eps}}\Delta_{\bs{C}\bs{\eps}}\hoch{\bs{\beta}}\end{eqnarray*}
}\begin{prop}\label{prop:BI-with-shifted-connection}The Bianchi
identities\index{proposition!Bianchi identities for shifted connection}
for $T^{A}$ and $R_{A}\hoch{B}$ on the one hand and $\tilde{T}^{A}$
and $\tilde{R}_{A}\hoch{B}$ on the other hand are equivalent if the
objects are related via (\ref{eq:TwithNewConn}) and (\ref{eq:RwithNewConn}).\end{prop}

\subparagraph{Proof}

In fact this is a rather trivial statement. The Bianchi identities
do not put restrictions on the elementary objects (the connection
and the vielbein), but on the derived objects (torsion and curvature).
In the same way they do not put restrictions on the difference tensor.
Let us make this statement more precise. If the Bianchi identity for
$T^{A}$and $R_{A}\hoch{B}$ is fulfilled, then these objects can
locally be written as $T^{A}=\de E^{A}-E^{C}\wedge\Omega_{C}\hoch{A}$
and $R_{A}\hoch{B}=\de\Omega_{A}\hoch{B}-\Omega_{A}\hoch{C}\wedge\Omega_{C}\hoch{B}$
for some $E^{A}$ and some $\Omega_{A}\hoch{B}$. If we revert the
derivation of (\ref{eq:TwithNewConn}) and (\ref{eq:RwithNewConn}),
these equations then simply imply that $\tilde{T}^{A}$ and $\tilde{R}_{A}\hoch{B}$
can locally be written as $\tilde{T}^{A}=\de E^{A}-E^{C}\wedge\tilde{\Omega}_{C}\hoch{A}$
and $\tilde{R}_{A}\hoch{B}=\de\tilde{\Omega}_{A}\hoch{B}-\tilde{\Omega}_{A}\hoch{C}\wedge\tilde{\Omega}_{C}\hoch{B}$
with $\tilde{\Omega}_{MA}\hoch{B}=\Omega_{MA}\hoch{B}+\Delta_{MA}\hoch{B}$
and therefore necessarily obey the Bianchi identities. This proves
the proposition.$\qquad\square$

For the first Bianchi identity, we will also provide a brute force
proof: Remember the first Bianchi identity (\ref{eq:BIforTcov}) for
which we temporarily introduce the symbol $J$:\begin{equation}
J_{\bs{AAA}}\hoch{D}\equiv\nabla_{\bs{A}}T_{\bs{AA}}\hoch{D}+2T_{\bs{AA}}\hoch{C}T_{C\bs{A}}\hoch{D}-R_{\bs{AAA}}\hoch{D}\stackrel{!}{=}0\label{eq:JasBIforTcov}\end{equation}
The transformed $J$ reads\begin{eqnarray}
\tilde{J}_{\bs{AAA}}\hoch{D} & \stackrel{(\ref{eq:TwithNewConn})(\ref{eq:RwithNewConn})(\ref{eq:JasBIforTcov})}{=} & J_{\bs{AAA}}\hoch{D}+\nabla_{\bs{A}}\Delta_{\bs{AA}}\hoch{D}+\Delta_{\bs{A}C}\hoch{D}(T_{\bs{AA}}\hoch{C}+\Delta_{\bs{AA}}\hoch{C})-2\Delta_{\bs{AA}}\hoch{C}(T_{C\bs{A}}\hoch{D}+\Delta_{[C\bs{A}]}\hoch{D})+\nonumber \\
 &  & +2\Delta_{\bs{AA}}\hoch{C}(T_{C\bs{A}}\hoch{D}+\Delta_{[C\bs{A}]}\hoch{D})+2T_{\bs{AA}}\hoch{C}\Delta_{[C\bs{A}]}\hoch{D}+\nonumber \\
 &  & -\nabla_{\bs{A}}\Delta_{\bs{AA}}\hoch{D}-T_{\bs{AA}}\hoch{C}\Delta_{C\bs{A}}\hoch{D}+\Delta_{\bs{AA}}\hoch{C}\Delta_{\bs{A}C}\hoch{D}=\\
 & = & J_{\bs{AAA}}\hoch{D}\end{eqnarray}
This proves the proposition again for the first Bianchi identity.
The brute force proof for the second is left to the reader as an exercise
;-)

\section{Restricted structure group}

\label{sec:Restricted-structure-group}\index{restricted structure group}\index{structure group}\index{group!structure $\sim$}As
we discussed earlier\rem{Referenz!}, the (infinitesimal) local structure
group transformations in the type II supergravity context are block-diagonal
$\Lambda_{A}\hoch{B}=\diag(\Lambda_{a}\hoch{b},\Lambda_{\bs{\alpha}}\hoch{\bs{\beta}},\Lambda_{\hat{\bs{\alpha}}}\hoch{\hat{\bs{\beta}}})$
and are in addition restricted to Lorentz transformations and scale
transformations in order to leave invariant the supersymmetry structure
constants $\gamma_{\bs{\alpha\beta}}^{c}$:\begin{eqnarray}
\Lambda_{a}\hoch{b} & = & \Lambda^{(D)}\delta_{a}^{b}+\Lambda_{\: a_{1}}^{(L)}\hoch{a_{2}}\\
\Lambda_{\bs{\alpha}}\hoch{\bs{\beta}} & = & \frac{1}{2}\Lambda^{(D)}\delta_{\bs{\alpha}}\hoch{\bs{\beta}}+\frac{1}{4}\Lambda_{a_{1}a_{2}}^{(L)}\gamma^{a_{1}a_{2}}\tief{\bs{\alpha}}\hoch{\bs{\beta}}\\
\Lambda_{\hat{\bs{\alpha}}}\hoch{\hat{\bs{\beta}}} & = & \frac{1}{2}\Lambda^{(D)}\delta_{\hat{\bs{\alpha}}}\hoch{\hat{\bs{\beta}}}+\frac{1}{4}\Lambda_{a_{1}a_{2}}^{(L)}\gamma^{a_{1}a_{2}}\tief{\hat{\bs{\alpha}}}\hoch{\hat{\bs{\beta}}}\end{eqnarray}
Also the connection is a sum of a scaling connection and a Lorentz
connection which makes perfect sense as it is supposed to be a Lie
algebra valued one form: \begin{eqnarray}
\Omega_{Ma}\hoch{b} & = & \Omega_{M}^{(D)}\delta_{a}^{b}+\Omega_{Ma_{1}}^{(L)}\hoch{a_{2}}\\
\Omega_{M\bs{\alpha}}\hoch{\bs{\beta}} & = & \frac{1}{2}\Omega_{M}^{(D)}\delta_{\bs{\alpha}}\hoch{\bs{\beta}}+\frac{1}{4}\Omega_{M\, a_{1}a_{2}}^{(L)}\gamma^{a_{1}a_{2}}\tief{\bs{\alpha}}\hoch{\bs{\beta}}\\
\Omega_{M\hat{\bs{\alpha}}}\hoch{\hat{\bs{\beta}}} & = & \frac{1}{2}\Omega_{M}^{(D)}\delta_{\hat{\bs{\alpha}}}\hoch{\hat{\bs{\beta}}}+\frac{1}{4}\Omega_{M\, a_{1}a_{2}}^{(L)}\gamma^{a_{1}a_{2}}\tief{\hat{\bs{\alpha}}}\hoch{\hat{\bs{\beta}}}\end{eqnarray}
with \begin{equation}
\Omega_{M\, a_{1}a_{2}}^{(L)}\equiv\Omega_{Ma_{1}}^{(L)}\hoch{c}\eta_{ca_{2}}=-\Omega_{M\, a_{2}a_{1}}^{(L)}\end{equation}

\subsection{Curvature}

\index{curvature!form of $\sim$ for restricted structure group}It
is well known that the curvature is a Lie algebra valued two form.
Let us quickly recall the reason. The curvature is defined to be\begin{eqnarray}
R_{A}\hoch{B} & = & \de\Omega_{A}\hoch{B}-\Omega_{A}\hoch{C}\wedge\Omega_{C}\hoch{B}\label{eq:curvatureImBIteil}\end{eqnarray}
If $\Omega_{A}\hoch{B}$ is Lie algebra valued, $\de\Omega_{A}\hoch{B}$
is still Lie algebra valued, as the exterior derivative acts only
on the coefficient functions and not on the Lie algebra generator.
In addition, the term $\Omega_{A}\hoch{C}\wedge\Omega_{C}\hoch{B}$
can be written as $\frac{1}{2}[\Omega,\Omega]_{A}\hoch{B}$, and the
commutator of two Lie algebra elements is again a Lie algebra element. 

Let us now see how the structure group reduces into irreducible parts
or in particular how the curvature decays into the Lorentz part and
the scaling part (if the latter is present). First of all, the result
is clearly block diagonal if the connection is of this type\begin{eqnarray}
R_{A}\hoch{B} & = & \diag(R_{a}\hoch{b},R_{\bs{\alpha}}\hoch{\bs{\beta}},R_{\hat{\bs{\alpha}}}\hoch{\hat{\bs{\beta}}})\end{eqnarray}
such that the curvature definition (\ref{eq:curvatureImBIteil}) decays
into the three blocks\begin{eqnarray}
R_{a}\hoch{b} & = & \de\Omega_{a}\hoch{b}-\Omega_{a}\hoch{c}\wedge\Omega_{c}\hoch{b}\\
R_{\bs{\alpha}}\hoch{\bs{\beta}} & = & \de\Omega_{\bs{\alpha}}\hoch{\bs{\beta}}-\Omega_{\bs{\alpha}}\hoch{\bs{\gamma}}\wedge\Omega_{\bs{\gamma}}\hoch{\bs{\beta}}\\
R_{\hat{\bs{\alpha}}}\hoch{\hat{\bs{\beta}}} & = & \de\Omega_{\hat{\bs{\alpha}}}\hoch{\hat{\bs{\beta}}}-\Omega_{\hat{\bs{\alpha}}}\hoch{\hat{\bs{\gamma}}}\wedge\Omega_{\hat{\bs{\gamma}}}\hoch{\hat{\bs{\beta}}}\end{eqnarray}
 For the bosonic part of the curvature the seperation of scaling part
and Lorentz part is quite obvious\begin{eqnarray}
R_{a}\hoch{b} & = & \de\left(\Omega^{(D)}\delta_{a}^{b}+\Omega_{\: a}^{(L)}\hoch{b}\right)-\left(\Omega^{(D)}\delta_{a}^{c}+\Omega_{\: a}^{(L)}\hoch{c}\right)\wedge\left(\Omega^{(D)}\delta_{c}^{b}+\Omega_{\: c}^{(L)}\hoch{b}\right)=\\
 & = & \underbrace{\de\Omega^{(D)}}_{\equiv F^{(D)}}\delta_{a}^{b}+\underbrace{\left(\de\Omega_{\, a}^{(L)}\hoch{b}-\Omega_{\: a}^{(L)}\hoch{c}\wedge\Omega_{\: c}^{(L)}\hoch{b}\right)}_{R_{\: a}^{(L)}\hoch{b}}\label{eq:R-Zerfall-bosonic}\end{eqnarray}
Where the Lorentz curvature $R_{\: a}^{(L)}\hoch{b}$ is antisymmetric
if we pull down the index $b$ with the Minkowski metric. We can thus
extract from the complete curvature the scale part and the Lorentz
part (here for 10 spacetime dimensions)\begin{eqnarray}
F^{(D)} & = & \frac{1}{10}R_{a}\hoch{a}\end{eqnarray}
For the fermionic parts we get similarly ($\delta_{\bs{\alpha}}\hoch{\bs{\alpha}}=-16$
in our conventions)%
\footnote{\index{footnote!\thefoot. curvature decays in scale and Lorentz part}In
order to see how the curvature decays into Lorentz and scale part,
let us first consider the building blocks seperately:\begin{eqnarray*}
\partial_{\bs{M}}\Omega_{\bs{M}\bs{\alpha}}\hoch{\bs{\beta}} & = & \frac{1}{2}\partial_{\bs{M}}\Omega_{\bs{M}}\delta_{\bs{\alpha}}\hoch{\bs{\beta}}+\frac{1}{4}\partial_{\bs{M}}\Omega_{\bs{M}a_{1}a_{2}}\gamma^{a_{1}a_{2}}\tief{\bs{\alpha}}\hoch{\bs{\beta}}\end{eqnarray*}
\frem{\[
\big(=\frac{1}{2}\partial_{\bs{M}}\Omega_{\bs{M}}\delta_{\bs{\alpha}}\hoch{\bs{\beta}}+\frac{1}{4}\partial_{\bs{M}}\tilde{\Omega}_{\bs{M}a_{1}a_{2}}\tilde{\gamma}^{a_{1}a_{2}}\tief{\bs{\alpha}}\hoch{\bs{\beta}}-\frac{1}{2}\partial_{\bs{M}}\Phi\tilde{\Omega}_{\bs{M}a_{1}a_{2}}\tilde{\gamma}^{a_{1}a_{2}}\tief{\bs{\alpha}}\hoch{\bs{\beta}}\big)\]
}\begin{eqnarray*}
\Omega_{\bs{M}\bs{\alpha}}\hoch{\bs{\gamma}}\Omega_{\bs{M}\bs{\gamma}}\hoch{\bs{\beta}} & = & \left(\frac{1}{2}\Omega_{\bs{M}}\delta_{\bs{\alpha}}\hoch{\bs{\gamma}}+\frac{1}{4}\Omega_{\bs{M}a_{1}a_{2}}\gamma^{a_{1}a_{2}}\tief{\bs{\alpha}}\hoch{\bs{\gamma}}\right)\left(\frac{1}{2}\Omega_{\bs{M}}\delta_{\bs{\gamma}}\hoch{\bs{\beta}}+\frac{1}{4}\Omega_{\bs{M}b_{1}b_{2}}\gamma^{b_{1}b_{2}}\tief{\bs{\gamma}}\hoch{\bs{\beta}}\right)=\\
 & = & \frac{1}{16}\underbrace{\Omega_{\bs{M}a_{1}a_{2}}\Omega_{\bs{M}b_{1}b_{2}}}_{\mbox{antisym in }(a_{1}a_{2})\leftrightarrow(b_{1}b_{2})}\gamma^{a_{1}a_{2}}\tief{\bs{\alpha}}\hoch{\bs{\gamma}}\gamma^{b_{1}b_{2}}\tief{\bs{\gamma}}\hoch{\bs{\beta}}=\\
 & \stackrel{(\ref{eq:gammaIIgammaII})}{=} & \frac{1}{4}\Omega_{\bs{M}a_{1}c}\eta^{cd}\Omega_{\bs{M}da_{2}}\gamma^{a_{1}a_{2}}\tief{\bs{\alpha}}\hoch{\bs{\beta}}\end{eqnarray*}
The curvature thus takes the form\begin{eqnarray*}
\dann R_{\bs{MM}\bs{\alpha}}\hoch{\bs{\beta}} & = & \frac{1}{2}\partial_{\bs{M}}\Omega_{\bs{M}}^{(Dil)}\delta_{\bs{\alpha}}\hoch{\bs{\beta}}+\frac{1}{4}\left(\partial_{\bs{M}}\Omega_{\bs{M}a_{1}a_{2}}^{(Lor)}-\Omega_{\bs{M}a_{1}c}^{(Lor)}\eta^{cd}\Omega_{\bs{M}da_{2}}^{(Lor)}\right)\gamma^{a_{1}a_{2}}\tief{\bs{\alpha}}\hoch{\bs{\beta}}\equiv\\
 & \equiv & \frac{1}{2}F^{(Dil)}\delta_{\bs{\alpha}}\hoch{\bs{\beta}}+\frac{1}{4}R^{(Lor)}\tief{a_{1}}\hoch{b}\eta_{ba_{2}}\gamma^{a_{1}a_{2}}\tief{\bs{\alpha}}\hoch{\bs{\beta}}\qquad\fussend\end{eqnarray*}
}\rem{extract Lorentz part! Und erlaube unterschiedl Konnexionen!Und Fussnote ueberarbeiten!}\begin{eqnarray}
R_{\bs{\alpha}}\hoch{\bs{\beta}} & = & \frac{1}{2}F^{(D)}\delta_{\bs{\alpha}}\hoch{\bs{\beta}}+\frac{1}{4}R^{(L)}\tief{a_{1}}\hoch{b}\eta_{ba_{2}}\gamma^{a_{1}a_{2}}\tief{\bs{\alpha}}\hoch{\bs{\beta}}\label{eq:R-Zerfall-ferm}\\
F^{(D)} & = & -\frac{1}{8}R_{\bs{\alpha}}\hoch{\bs{\alpha}}\label{eq:R-Zerfall-Ferm-Dil}\end{eqnarray}
and \begin{eqnarray}
R_{\hat{\bs{\alpha}}}\hoch{\hat{\bs{\beta}}} & = & \frac{1}{2}F^{(D)}\delta_{\hat{\bs{\alpha}}}\hoch{\hat{\bs{\beta}}}+\frac{1}{4}R^{(L)}\tief{a_{1}}\hoch{b}\eta_{ba_{2}}\gamma^{a_{1}a_{2}}\tief{\hat{\bs{\alpha}}}\hoch{\hat{\bs{\beta}}}\label{eq:R-Zerfall-ferm-hut}\\
F^{(D)} & = & -\frac{1}{8}R_{\hat{\bs{\alpha}}}\hoch{\hat{\bs{\alpha}}}\label{eq:R-Zerfall-Ferm-hut-Dil}\end{eqnarray}

\subsection{Alternative version of the first Bianchi identity}

\rem{Ueberarbeiten!}The ordinary Riemannian curvature (without torsion)
obeys $R_{abcd}=-R_{bacd}=-R_{abdc}$, $R_{[abc]d}=0$ and $R_{abcd}=R_{cdab}$
(The last is a consequence of the others). For the bosonic components
of our curvature we have (using $G_{ab}=e^{2\Phi}\eta_{ab}$ with
$\nabla_{M}G_{ab}=2(\partial_{M}\Phi-\Omega_{M}^{(Dil)})G_{ab}$ to
pull down bosonic indices)\begin{eqnarray}
R_{abcd} & = & -R_{bacd},\qquad R_{(ab)cd}=0\\
R_{abcd} & = & -R_{abdc}+2F_{ab}^{(Dil)}G_{cd},\qquad R_{ab(cd)}=F_{ab}^{(Dil)}G_{cd}\\
R_{[abc]d} & = & \nabla_{[a}T_{bc]|d}-2(\partial_{[a}\Phi-\Omega_{[a}^{(Dil)})T_{bc|d}+2T_{[ab|}\hoch{E}T_{E|c]|d}\end{eqnarray}
Let us write down the antisymmetrization of the indices in $R_{[abc]d}$
explicitely and several times, with permuted indices: \begin{eqnarray}
R_{[abc]d} & = & R_{abcd}+R_{cabd}+R_{bcad}\label{eq:ReqI}\\
R_{[dab]c} & = & R_{dabc}+R_{bdac}+R_{abdc}\label{eq:ReqII}\\
R_{[cda]b} & = & R_{cdab}+R_{acdb}+R_{dacb}\label{eq:ReqIII}\\
R_{[bcd]a} & = & R_{bcda}+R_{dbca}+R_{cdba}\label{eq:ReqIV}\end{eqnarray}
From this we learn, how we can express the difference $R_{abcd}-R_{cdab}$
(which vanishes in the Riemannian case), in terms of antisymmetrized
and symmetrized terms. Consider the sum (\ref{eq:ReqI})-(\ref{eq:ReqII})-(\ref{eq:ReqIII})+(\ref{eq:ReqIV}):\begin{eqnarray}
\lqn{R_{[abc]d}-R_{[dab]c}-R_{[cda]b}+R_{[bcd]a}=}\nonumber \\
 & = & 2R_{abcd}-2R_{ab(cd)}-2R_{cdab}+2R_{cd(ab)}+2R_{(ca)bd}-2R_{ac(db)}+2R_{bc(da)}-2R_{da(bc)}-2R_{bd(ac)}+2R_{(db)ca}=\nonumber \\
 & = & 2\left(R_{abcd}-R_{cdab}\right)+2\left(-F_{ab}G_{cd}+F_{cd}G_{ab}-F_{ac}G_{db}+F_{bc}G_{da}-F_{da}G_{bc}-F_{bd}G_{ac}\right)\end{eqnarray}
The identity corresponding to $R_{abcd}=R_{cdab}$ in the Riemannian
case thus reads \begin{eqnarray}
\lqn{2\left(R_{abcd}-R_{cdab}\right)=}\\
 & = & 2\left(F_{ab}G_{cd}-F_{cd}G_{ab}+F_{ac}G_{db}-F_{bc}G_{da}+F_{da}G_{bc}+F_{bd}G_{ac}\right)+R_{[abc]d}-R_{[dab]c}-R_{[cda]b}+R_{[bcd]a}\nonumber \end{eqnarray}
with $R_{[abc]d}=\nabla_{[a}T_{bc]|d}-2(\partial_{[a}\Phi-\Omega_{[a}^{(Dil)})T_{bc|d}+2T_{[ab|}\hoch{E}T_{E|c]|d}$.

\subsection{Scaling-curvature}

\index{scaling field strength}A covariant way to calculate the scaling
field strength $F_{MN}^{(D)}$ is as follows: Consider the covariant
derivative $\covPhi{M}=\partial_{M}\Phi-\Omega_{M}^{(D)}$ of a compensator
field $\Phi$ (a field transforming with a shift under scaling transformations
$\delta\Phi=-\Lambda^{(D)}$). We can calculate $F_{MN}^{(D)}$ via
the ususal commutator of covariant derivatives%
\footnote{\index{footnote!\thefoot. commutator of covariant derivatives on compensator field}\index{commutator!of covariant derivatives on compensator}\index{covariant derivative!commutator of $\sim$ on compensator field}\index{compensator field!commutator of covariant derivatives}Let
us check explicitely the validity of (\ref{eq:covDerAlgOnCompensator}):
\begin{eqnarray*}
\nabla_{[M}\covPhi{N]} & = & \partial_{[M}\covPhi{N]}-\Gamma_{[MN]}\hoch{K}\covPhi{K}=\\
 & = & \partial_{[M}(\partial_{N]}\Phi-\Omega_{N]}^{(D)})-T_{[MN]}\hoch{K}\covPhi{K}=\\
 & = & -F_{MN}^{(D)}-T_{[MN]}\hoch{K}\covPhi{K}\qquad\fussend\end{eqnarray*}
}\begin{eqnarray}
\nabla_{[M}\covPhi{N]} & = & -T_{MN}\hoch{K}\nabla_{K}\Phi\underbrace{-F_{MN}^{(D)}}_{\group{F_{MN}^{(D)}}\Phi}\label{eq:covDerAlgOnCompensator}\end{eqnarray}
Note that the curvature (or field strength) appears {}``naked''
in difference to any action on tensor fields. The above equation will
be particularly useful when we have constraints on $\covPhi{M}$ which
then determine the scaling curvature via\rem{%
\footnote{Remember the difference tensor $\Delta_{M}^{(D)}=\hat{\Omega}_{M}-\Omega_{M}$.
Using it, we can seperate the connection in a tensorial part and a
total derivative.\frem{nur lokal?}\\
\begin{tabular}{ccc}
$\Omega_{M}^{(D)}=\partial_{M}\Phi-E_{M}\hoch{\bs{\alpha}}\Delta_{\bs{\alpha}}^{(D)}$,  & $\hat{\Omega}_{M}^{(D)}=\partial_{M}\Phi+E_{M}\hoch{\hat{\bs{\alpha}}}\Delta_{\hat{\bs{\alpha}}}^{(D)}$,  & $\avOm_{M}^{(D)}=\partial_{M}\Phi-\frac{1}{2}E_{M}\hoch{\bs{\alpha}}\Delta_{\bs{\alpha}}^{(D)}+\frac{1}{2}E_{M}\hoch{\hat{\bs{\alpha}}}\Delta_{\hat{\bs{\alpha}}}^{(D)}$\tabularnewline
\end{tabular}\\
or equivalently \\
\begin{tabular}{ccc}
$\Delta_{\bs{\alpha}}^{(D)}=\covPhi{\bs{\alpha}}$,  & $\Delta_{\hat{\bs{\alpha}}}^{(D)}=-\hatcovPhi{\hat{\bs{\alpha}}}$,  & $\avcovPhi{\bs{\alpha}}=\frac{1}{2}\Delta_{\bs{\alpha}}^{(D)}\quad\avcovPhi{\hat{\bs{\alpha}}}=-\frac{1}{2}\Delta_{\hat{\bs{\alpha}}}^{(D)}$\tabularnewline
\end{tabular}\\
Only the mixed connection has a different dilatation for each block:\begin{eqnarray*}
\gemOm_{MA}^{(D)}\hoch{B} & = & \left(\begin{array}{ccc}
\check{\Omega}_{M}^{(D)}\delta_{a}^{b} & 0 & 0\\
0 & \frac{1}{2}\Omega_{M}^{(D)}\delta_{\bs{\alpha}}\hoch{\bs{\beta}} & 0\\
0 & 0 & \frac{1}{2}\hat{\Omega}_{M}^{(D)}\delta_{\hat{\bs{\alpha}}}\hoch{\hat{\bs{\beta}}}\end{array}\right)\end{eqnarray*}
where $\check{\Omega}_{M}^{(D)}$ can be either $\Omega_{M}^{(D)}$,
$\hat{\Omega}_{M}^{(D)}$ or $\avOm_{M}^{(D)}$. The scaling curvatures
(field strengths) built from these scaling connections read\index{scale field strength}\\
\begin{tabular}{cc}
$F_{MN}^{(D)}=E_{[M}\hoch{\bs{\alpha}}\nabla_{N]}\Delta_{\bs{\alpha}}^{(D)}-T_{MN}\hoch{\bs{\alpha}}\Delta_{\bs{\alpha}}^{(D)}$,  & $\hat{F}_{MN}^{(D)}=-E_{[M}\hoch{\hat{\bs{\alpha}}}\hat{\nabla}_{N]}\Delta_{\hat{\bs{\alpha}}}^{(D)}+\hat{T}_{MN}\hoch{\hat{\bs{\alpha}}}\Delta_{\hat{\bs{\alpha}}}^{(D)}$, \tabularnewline
$\av{F}_{MN}^{(D)}=\frac{1}{2}\Big(E_{[M}\hoch{\bs{\alpha}}\nabla_{N]}\Delta_{\bs{\alpha}}^{(D)}-T_{MN}\hoch{\bs{\alpha}}\Delta_{\bs{\alpha}}^{(D)}\lqn{-E_{[M}\hoch{\hat{\bs{\alpha}}}\hat{\nabla}_{N]}\Delta_{\hat{\bs{\alpha}}}^{(D)}+\hat{T}_{MN}\hoch{\hat{\bs{\alpha}}}\Delta_{\hat{\bs{\alpha}}}^{(D)}\Big)}$ & \tabularnewline
$\av{F}_{MN}^{(D)}=\frac{1}{2}\Big(E_{[M}\hoch{\bs{\alpha}}\av{\nabla}_{N]}\Delta_{\bs{\alpha}}^{(D)}-\av{T}_{MN}\hoch{\bs{\alpha}}\Delta_{\bs{\alpha}}^{(D)}\lqn{-E_{[M}\hoch{\hat{\bs{\alpha}}}\av{\nabla}_{N]}\Delta_{\hat{\bs{\alpha}}}^{(D)}+\av{T}_{MN}\hoch{\hat{\bs{\alpha}}}\Delta_{\hat{\bs{\alpha}}}^{(D)}\Big)\qquad\fussend}$ & \tabularnewline
\end{tabular}%
}}\begin{equation}
\boxed{F_{MN}^{(D)}=-\nabla_{[M}\covPhi{N]}-T_{MN}\hoch{K}\nabla_{K}\Phi}\label{eq:scalingCurvAsComm}\end{equation}

\section{Dragon's theorem}

In the following we will need the commutator of two covariant derivatives
acting on the torsion with afterwards all lower indices antisymmetrized.
Due to (\ref{eq:generalCommutatorOfCovDer}), it is given by%
\footnote{\index{footnote!\thefoot. weakest possible condition for Dragon's theorem}Of
course (\ref{eq:generalCommutatorOfCovDer}) implies a more general
relation than (\ref{eq:commOfCovDerOnT}), namely one of the form
$[\nabla_{M},\nabla_{N}]T_{KL}\hoch{A}=\ldots$. However, the lower
indices are intentionally antisymmetrized in (\ref{eq:commOfCovDerOnT}),
in order to get the weakest possible condition that we need to proof
the theorem later on. You'll see... $\qquad\fussend$%
}\begin{eqnarray}
\nabla_{\bs{M}}\nabla_{\bs{M}}T_{\bs{MM}}\hoch{A} & = & -T_{\bs{MM}}\hoch{K}\nabla_{K}T_{\bs{MM}}\hoch{A}-2R_{\bs{MMM}}\hoch{K}T_{K\bs{M}}\hoch{A}+R_{\bs{MM}B}\hoch{A}T_{\bs{MM}}\hoch{B}\label{eq:commOfCovDerOnT}\end{eqnarray}
and can, using the first Bianchi identity (\ref{eq:BIforTcov}), be
rewritten as\begin{eqnarray}
\lqn{R_{\bs{MM}B}\hoch{A}T_{\bs{MM}}\hoch{B}=}\nonumber \\
 & = & \nabla_{\bs{M}}\nabla_{\bs{M}}T_{\bs{MM}}\hoch{A}+T_{\bs{MM}}\hoch{K}\nabla_{K}T_{\bs{MM}}\hoch{A}+2\left(\nabla_{\bs{M}}T_{\bs{MM}}\hoch{K}+2T_{\bs{MM}}\hoch{L}T_{L\bs{M}}\hoch{K}\right)T_{K\bs{M}}\hoch{A}\label{eq:RinTermsOfT}\end{eqnarray}
It is convenient to introduce a new symbol for the terms of the curvature
Bianchi identity\index{$I_{\bs{CCC}A}\hoch{B}$}\index{$I_A\hoch{B}$}
\begin{equation}
I_{A}\hoch{B}\equiv I_{\bs{CCC}A}\hoch{B}\equiv\nabla_{\bs{C}}R_{\bs{CC}A}\hoch{B}+2T_{\bs{CC}}\hoch{D}R_{D\bs{C}A}\hoch{B}\label{eq:Iab2ndBI}\end{equation}
so that the Bianchi identity (\ref{eq:BIforRcov}) simply reads $I_{A}\hoch{B}\stackrel{!}{=}0$.
Then the following theorem holds (originally due to Dragon in \cite{Dragon:1978nf};
slightly modified in order to include dilatations):

\begin{thm}[Dragon]\index{theorem!Dragon's $\sim$}\index{Dragon's theorem}\label{thm:Dragon}Given
a block diagonal structure group consisting of Lorentz transformation
and dilatation in a type II superspace, the torsion Bianchi identity
(\ref{eq:BIforTcov}) together with the algebra (\ref{eq:commOfCovDerOnT})
or equivalently (\ref{eq:RinTermsOfT}) imply the curvature Bianchi
identities (\ref{eq:BIforRcov}) $I_{A}\hoch{B}=0$ up to one remaining
equation for the scale part, namely $I_{\bs{\gamma}\hat{\bs{\gamma}}c}^{(D)}\stackrel{!}{=}0$
or equivalently \begin{equation}
\nabla_{[\bs{\gamma}}F_{\hat{\bs{\gamma}}c]}^{(D)}+2T_{[\bs{\gamma}\hat{\bs{\gamma}}|}\hoch{D}F_{D|c]}^{(D)}\stackrel{!}{=}0\label{eq:remaining2ndBI}\end{equation}
where $F_{MN}^{(D)}$ is the field strength of the scale connection
$\Omega_{M}^{(D)}$.\end{thm}

It is natural to proof this theorem in two steps, the first being
useful enough to write it as a seperate proposition. Let us include
one more index into the antisymmetrization of $I_{A}\hoch{B}$ and
define\index{$I^B$}\index{$I_{\bs{CCCC}}\hoch{B}$}\begin{equation}
I^{B}\equiv I_{\bs{CCCC}}\hoch{B}\equiv\nabla_{\bs{C}}R_{\bs{CCC}}\hoch{B}+2T_{\bs{CC}}\hoch{D}R_{D\bs{CC}}\hoch{B}\label{eq:Ib2ndBI}\end{equation}
so that we can make direct use of the torsion-Bianchi-identity (\ref{eq:BIforTcov})
due to the appearance of $R_{\bs{CCC}}\hoch{B}$. Clearly $I^{B}\stackrel{!}{=}0$
is a consequence of $I_{A}\hoch{B}\stackrel{!}{=}0$ and is in general
a weaker condition. The following proposition treats this weaker condition:

\begin{prop}\label{prop:weakDragon}\index{proposition!weak Dragon}In
any dimension and for any structure group, the equation $I^{B}\stackrel{!}{=}0$
(with $I^{B}$ given by (\ref{eq:Ib2ndBI})) is implied by the first
Bianchi identity (\ref{eq:BIforTcov}) and the algebra (\ref{eq:commOfCovDerOnT})
or equivalently (\ref{eq:RinTermsOfT}).\end{prop}

\subparagraph{Proof}

of the proposition\rem{%
\footnote{\index{footnote!\thefoot. weak Dragon in condensed notation}An alternative
way of writing this is via the more condensed notation (using the
torsion Bianchi-identity of the form $R^{B}=\bs{\nabla}T^{B}+\ip_{T}T^{B}$
(\ref{eq:BIforTcovCondensed}))\begin{eqnarray*}
0 & \stackrel{!}{=} & \bs{\nabla}R^{B}+\ip_{T}R^{B}-T^{A}R_{A}\hoch{B}=\\
 & \stackrel{(\ref{eq:BIforTcovCondensed})}{=} & \bs{\nabla}\left(\bs{\nabla}T^{B}+\ip_{T}T^{B}\right)+\ip_{T}\left(\bs{\nabla}T^{B}+\ip_{T}T^{B}\right)-T^{A}R_{A}\hoch{B}=\\
 & \stackrel{(\ref{eq:generalCommutatorOfCovDer})}{=} & -T^{C}\nabla_{C}T^{B}+R_{A}\hoch{B}T^{A}-\ip_{R}T^{B}+\bs{\nabla}(-2T^{C}T_{\, C}\hoch{B})+T^{C}\nabla_{C}T^{B}+2T^{C}\bs{\nabla}T_{\, C}\hoch{B}+\ip_{T}(\ip_{T}T^{B})-T^{A}R_{A}\hoch{B}=\\
 & = & 2\left(R^{C}-\bs{\nabla}T^{C}\right)T_{\, C}\hoch{B}-2T^{C}\bs{\nabla}T_{\, C}\hoch{B}+2T^{C}\bs{\nabla}T_{\, C}\hoch{B}+\underbrace{\ip_{T}(\ip_{T}T^{B})}_{\ip_{\ip_{T}T}T^{B}+\ip_{T\wedge T}T^{B}}=\\
 & \stackrel{(\ref{eq:BIforTcovCondensed})}{=} & 2(\ip_{T}T^{C})T_{\, C}\hoch{B}+\ip_{\ip_{T}T}T^{B}+\ip_{\underbrace{T\wedge T}_{=0}}T^{B}=0\end{eqnarray*}
It would be nice, however to use some product rule $\bs{\nabla}(\ip_{K}L)\stackrel{?}{=}\ip_{\bs{\nabla}K}L+\ip_{K}\bs{\nabla}L$
or similar...\begin{eqnarray*}
\bs{\nabla}K & = & \de K-k\Gamma_{\bs{mm}}\hoch{p}K_{p\mm}\hoch{\nn}+(-)^{k}k'\Gamma_{\bs{m}p}\hoch{\bs{n}}K_{\mm}\hoch{p\nn}=\\
 & = & \de K-\ip_{T}K-(-)^{k-k'}k'K_{\mm}\hoch{\nn p}\Gamma_{\bs{m}p}\hoch{\bs{n}}=\\
 & = & \partial_{\bs{m}}K_{\mm}\hoch{\nn}+(-)^{(k-k')}K_{\mm}\hoch{\nn p}\underbrace{(\partial_{\bs{m}}(\pe_{p})-\Gamma_{p\bs{m}}\hoch{n}\pe_{n})}_{\bs{\nabla}(\pe_{p})}-\ip_{T}K\end{eqnarray*}
Hmm... Another question: How is $\ip_{\nabla E_{A}}T^{B}$ related
to $\nabla_{A}T^{B}$? $\qquad\fussend$%
}}:

\begin{eqnarray}
I^{B} & = & \nabla_{\bs{M}}R_{\bs{MMM}}\hoch{B}+2T_{\bs{MM}}\hoch{K}R_{K\bs{MM}}\hoch{B}=\\
 & \stackrel{(\ref{eq:BIforTcov})}{=} & \nabla_{\bs{M}}\left(\nabla_{\bs{M}}T_{\bs{MM}}\hoch{B}+2T_{\bs{MM}}\hoch{C}T_{C\bs{M}}\hoch{B}\right)+2T_{\bs{MM}}\hoch{K}R_{K\bs{MM}}\hoch{B}=\\
 & \stackrel{(\ref{eq:commOfCovDerOnT})}{=} & -T_{\bs{MM}}\hoch{C}\nabla_{C}T_{\bs{M}\bs{M}}\hoch{B}-2R_{\bs{MMM}}\hoch{C}T_{C\bs{M}}\hoch{B}+R_{\bs{MM}C}\hoch{B}T_{\bs{MM}}\hoch{C}+\\
 &  & +2\nabla_{\bs{M}}T_{\bs{MM}}\hoch{C}T_{C\bs{M}}\hoch{B}+2T_{\bs{MM}}\hoch{C}\nabla_{\bs{M}}T_{C\bs{M}}\hoch{B}+2T_{\bs{MM}}\hoch{K}R_{K\bs{MM}}\hoch{B}=\\
 & = & 3T_{\bs{MM}}\hoch{C}\left(R_{[C\bs{MM}]}\hoch{B}-\nabla_{[C}T_{\bs{M}\bs{M}]}\hoch{B}\right)-2\left(R_{\bs{MMM}}\hoch{C}-\nabla_{\bs{M}}T_{\bs{MM}}\hoch{C}\right)T_{C\bs{M}}\hoch{B}=\\
 & \stackrel{(\ref{eq:BIforTcov})}{=} & 6T_{\bs{MM}}\hoch{C}T_{[C\bs{M}|}\hoch{D}T_{D|\bs{M}]}\hoch{B}-4T_{\bs{MM}}\hoch{D}T_{D\bs{M}}\hoch{C}T_{C\bs{M}}\hoch{B}=\\
 & = & 2T_{\bs{MM}}\hoch{C}T_{\bs{MM}}\hoch{D}T_{DC}\hoch{B}=0\end{eqnarray}
Indeed $I^{B}=0$ is a consequence of the torsion Bianchi identity
(\ref{eq:BIforTcov}) $R_{\bs{MMM}}\hoch{B}=\nabla_{\bs{M}}T_{\bs{MM}}\hoch{B}+2T_{\bs{MM}}\hoch{C}T_{C\bs{M}}\hoch{B}$
and (\ref{eq:commOfCovDerOnT}).$\quad\square$

\subparagraph{Proof}

of the theorem: Let us now show that in the case of the type II superspace
the antisymmetrized version already implies (up to one term) the complete
one. Remember the object $I_{\bs{CCC}A}\hoch{B}\equiv\nabla_{\bs{C}}R_{\bs{CC}A}\hoch{B}+2T_{\bs{CC}}\hoch{D}R_{D\bs{M}A}\hoch{B}$
introduced in (\ref{eq:Iab2ndBI}). It is Lie algebra valued and thus
has (for our block diagonal structure group) no mixed components in
$A,B$:\begin{equation}
I_{\bs{CCC}A}\hoch{B}=\diag(I_{\bs{CCC}a}\hoch{b},I_{\bs{CCC}\bs{\alpha}}\hoch{\bs{\beta}},I_{\bs{CCC}\hat{\bs{\alpha}}}\hoch{\hat{\bs{\beta}}})\end{equation}
In addition it splits into dilatation and Lorentz part\begin{equation}
I_{\bs{CCC}A}\hoch{B}=I_{\bs{CCC}}^{(D)}\delta_{A}\hoch{B}+I_{\bs{CCC}A}^{(L)}\hoch{B}\end{equation}
with the latter term being antisymmetric in $A,B$ for bosonic $a,b$.
The complete object is fixed by determing%
\footnote{\index{footnote!\thefoot. remark about connection w.r.t. proof of Dragon's theorem}The
following proof is based on a block-diagonal connection of the form
$\Omega_{MA}\hoch{B}=\diag(\Omega_{Ma}\hoch{b},\Omega_{M\bs{\alpha}}\hoch{\bs{\beta}},\Omega_{M\hat{\bs{\alpha}}}\hoch{\hat{\bs{\beta}}})$
where the three entries are related by $\nabla_{M}\gamma_{\bs{\alpha}\bs{\beta}}^{a}=\nabla_{M}\gamma_{\hat{\bs{\alpha}}\hat{\bs{\beta}}}^{a}=0$
which in turn is equivalent to $\Omega_{M\bs{\alpha}}\hoch{\bs{\beta}}=\frac{1}{4}\Omega_{Ma}\hoch{b}\gamma^{a}\tief{b\,\bs{\alpha}}\hoch{\bs{\beta}}$
and $\Omega_{M\hat{\bs{\alpha}}}\hoch{\hat{\bs{\beta}}}=\frac{1}{4}\Omega_{Ma}\hoch{b}\gamma^{a}\tief{b\,\hat{\bs{\alpha}}}\hoch{\hat{\bs{\beta}}}$.
The Bianchi identity for its torsion $T^{A}=(T^{a},T^{\bs{\alpha}},T^{\hat{\bs{\alpha}}})$
is equivalent to the one for the Torsion $\gem{T}^{A}=(\check{T}^{a},T^{\bs{\alpha}},\hat{T}^{\hat{\bs{\alpha}}})$
when information about the connection-difference $\Delta_{MA}\hoch{B}$
is available.$\qquad\fussend$%
} $I_{\bs{CCC}a}\hoch{b}$. Given the equation $I_{\bs{CCCC}}\hoch{B}=0$,
we want to show that $I_{\bs{CCC}A}\hoch{B}=0$. Consider first $B=b$:\begin{eqnarray}
0 & = & 4I_{[\bs{\mc{C}\mc{C}\mc{C}}a]}\hoch{b}=I_{\bs{\mc{C}\mc{C}\mc{C}}a}\hoch{b}\end{eqnarray}
Similarly, for $B=\bs{\beta}$:\begin{eqnarray}
0 & = & 4I_{[\hat{\bs{\gamma}}\hat{\bs{\gamma}}\hat{\bs{\gamma}}\bs{\alpha}]}\hoch{\bs{\beta}}=I_{\hat{\bs{\gamma}}\hat{\bs{\gamma}}\hat{\bs{\gamma}}\bs{\alpha}}\hoch{\bs{\beta}}=0\\
0 & = & 4I_{[c\hat{\bs{\gamma}}\hat{\bs{\gamma}}\bs{\alpha}]}\hoch{\bs{\beta}}=I_{c\hat{\bs{\gamma}}\hat{\bs{\gamma}}\bs{\alpha}}\hoch{\bs{\beta}}=0\\
0 & = & 4I_{cc\hat{\bs{\gamma}}\bs{\alpha}]}\hoch{\bs{\beta}}=I_{cc\hat{\bs{\gamma}}\bs{\alpha}}\hoch{\bs{\beta}}=0\\
0 & = & 4I_{ccc\bs{\alpha}]}\hoch{\bs{\beta}}=I_{ccc\bs{\alpha}}\hoch{\bs{\beta}}=0\end{eqnarray}
This implies\begin{eqnarray}
I_{c\hat{\bs{\gamma}}\hat{\bs{\gamma}}a}\hoch{b} & = & 0\\
I_{cc\hat{\bs{\gamma}}a}\hoch{b} & = & 0\\
I_{ccca}\hoch{b} & = & 0\end{eqnarray}
Equivalently we get from the equations for $B=\hat{\bs{\beta}}$:\begin{eqnarray}
I_{c\bs{\gamma}\bs{\gamma}a}\hoch{b} & = & 0\\
I_{cc\bs{\gamma}a}\hoch{b} & = & 0\end{eqnarray}
There is thus only one component of $I_{\bs{\gamma}\hat{\bs{\gamma}}ca}\hoch{b}$
left to determine. For this we get\begin{eqnarray}
0 & = & I_{\bs{\gamma}\hat{\bs{\gamma}}[ca]}\hoch{b}=\\
 & = & I_{\bs{\gamma}\hat{\bs{\gamma}}[c}^{(D)}\delta_{a]}^{b}+I_{\bs{\gamma}\hat{\bs{\gamma}}[ca]}^{(L)}\hoch{b}\end{eqnarray}
Taking the trace in (a,b) yields\begin{eqnarray}
0 & = & 9I_{\bs{\gamma}\hat{\bs{\gamma}}c}^{(D)}+I_{\bs{\gamma}\hat{\bs{\gamma}}ac}^{(L)}\hoch{a}\end{eqnarray}
In order that they vanish independently, it is thus enough to check
only one equation, namely $I_{\bs{\gamma}\hat{\bs{\gamma}}c}^{(D)}\stackrel{!}{=}0$
which reads explicitely \begin{equation}
\boxed{\nabla_{[\bs{\gamma}}F_{\hat{\bs{\gamma}}c]}^{(D)}+2T_{[\bs{\gamma}\hat{\bs{\gamma}}|}\hoch{D}F_{D|c]}^{(D)}\stackrel{!}{=}0}\end{equation}
\rem{\begin{eqnarray}
0 & \stackrel{!}{=} & \nabla_{[\bs{\gamma}}F_{\hat{\bs{\gamma}}c]}^{(D)}+2T_{[\bs{\gamma}\hat{\bs{\gamma}}|}\hoch{D}F_{D|c]}^{(D)}=\\
 & = & \nabla_{[\bs{\gamma}}\left(\nabla_{\hat{\bs{\gamma}}}\Omega_{c]}+T_{\hat{\bs{\gamma}}c]}\hoch{D}\Omega_{D}\right)+2T_{[\bs{\gamma}\hat{\bs{\gamma}}|}\hoch{D}F_{D|c]}^{(D)}=\\
 & = & -T_{[\bs{\gamma}\hat{\bs{\gamma}}|}\hoch{D}\nabla_{D}\Omega_{|c]}-R_{[\bs{\gamma}\hat{\bs{\gamma}}c]}\hoch{D}\Omega_{D}+\nabla_{[\bs{\gamma}}T_{\hat{\bs{\gamma}}c]}\hoch{D}\Omega_{D}+T_{[\hat{\bs{\gamma}}c|}\hoch{D}\nabla_{|\bs{\gamma}]}\Omega_{D}+\nonumber \\
 &  & +2T_{[\bs{\gamma}\hat{\bs{\gamma}}|}\hoch{D}\left(\frac{1}{2}\nabla_{D}\Omega_{|c]}-\frac{1}{2}\nabla_{|c]}\Omega_{D}\right)=\\
 & = & -R_{[\bs{\gamma}\hat{\bs{\gamma}}c]}\hoch{D}\Omega_{D}+\nabla_{[\bs{\gamma}}T_{\hat{\bs{\gamma}}c]}\hoch{D}\Omega_{D}+\frac{1}{3}(-R_{\hat{\bs{\gamma}}c\bs{\gamma}}\hoch{\bs{\delta}}+\nabla_{\bs{\gamma}}T_{\hat{\bs{\gamma}}c}\hoch{\bs{\delta}})\Omega_{\bs{\delta}}=\\
 & = & \frac{1}{3}(-\gamma_{c\,\hat{\bs{\gamma}}\hat{\bs{\delta}}}\nabla_{\bs{\gamma}}\RR^{\bs{\delta}\hat{\bs{\delta}}}+\nabla_{\bs{\gamma}}(\gamma_{c\,\hat{\bs{\gamma}}\hat{\bs{\delta}}}\RR^{\bs{\delta}\hat{\bs{\delta}}}))\Omega_{\bs{\delta}}=0\end{eqnarray}
}\rem{Unsinn! natuerlich ist BI erfuellt, wenn man elementare Def von F verwendet!}\rem{

\subsection{Alternative proof of proposition \ref{prop:weakDragon}}

Let us give a more direct proof of proposition \ref{prop:weakDragon}
by giving a somewhat unconventional derivation of the Bianchi identities.
This will be based on the superspace connection instead of the structure
group connection. Remember the definitons (\ref{eq:TorsionDefI})
and (\ref{eq:curvatureDefI}) for torsion and curvature. Defining
\begin{eqnarray}
\Gamma_{N}\hoch{K} & \equiv & \de x^{M}\Gamma_{MN}\hoch{K}\end{eqnarray}
they can -- at least formally -- be written as\begin{eqnarray}
T^{K} & = & -\de x^{N}\wedge\Gamma_{N}\hoch{K}\\
R_{K}\hoch{L} & = & \de\Gamma_{K}\hoch{L}-\Gamma_{K}\hoch{P}\wedge\Gamma_{P}\hoch{L}\end{eqnarray}
The first Bianchi identity is obtained by acting on $T^{K}$ with
the exterior derivative\begin{eqnarray}
\de T^{K} & = & \de x^{N}\wedge\de\Gamma_{N}\hoch{K}=\\
 & = & \de x^{N}\wedge\left(R_{N}\hoch{K}+\Gamma_{N}\hoch{P}\wedge\Gamma_{P}\hoch{K}\right)\end{eqnarray}
Define now \begin{eqnarray}
I_{K}\hoch{L} & \equiv & \de\left(R_{K}\hoch{L}+\Gamma_{K}\hoch{P}\wedge\Gamma_{P}\hoch{L}\right)=\\
 & = & \de R_{K}\hoch{L}+R_{K}\hoch{P}\wedge\Gamma_{P}\hoch{L}-\Gamma_{K}\hoch{P}\wedge R_{P}\hoch{L}\end{eqnarray}
\rem{Es macht einen Unterschied, ob man erste oder zweite Zeile als Definition fuer $I$ sieht!}With
this definition the second Bianchi identity (obtained by acting with
the exterior derivative on the definition of $R_{K}\hoch{L}$) simply
reads \begin{equation}
I_{K}\hoch{L}=0\end{equation}
 It is now obvious that acting a second time with the exterior derivative
on the first Bianchi identity leads to\begin{eqnarray}
0 & = & -\de x^{N}\wedge I_{N}\hoch{K}\end{eqnarray}
}

\bibliographystyle{fullsort}
\bibliography{phd,Proposal}
\printindex{}
}

\chapter{About the Connection}

\label{cha:ConnectionAppend}{\inputTeil{0} \ifthenelse{\theinput=1}{}{}

\title{Form of the connection}

\author{Sebastian Guttenberg}

\date{January 3, '08}

\maketitle
\begin{abstract}
Part of thesis-appendix. Please check for typos and other errors ;-) 
\end{abstract}
\tableofcontents{}

\index{connection|fett}\rem{To do:...} Let us refer to both, spacetime
and structure group connection, simply as {}``the connection''.
Properties of the one are translated to the other via the condition
of covariantly constant vielbeins $\nabla_{M}E_{N}\hoch{A}=0$:\begin{eqnarray}
\Gamma_{MN}\hoch{A} & = & \partial_{M}E_{N}\hoch{A}+\Omega_{MN}\hoch{A}\label{eq:GammaOmegaRelationInConnectionPart}\end{eqnarray}
We will use symbols without any decoration (like hats or whatever)
to describe a general connection and objects derived from it. In our
application to the Berkovits string, however, we use the undecorated
symbol $\Omega_{M\bs{\alpha}}\hoch{\bs{\beta}}$ for the leftmoving
connection only, which hopefully does not lead to confusions. To be
more explicit, in the application we work with several different connections
which are all blockdiagonal. In the action there appear only $\Omega_{M\bs{\alpha}}\hoch{\bs{\beta}}$
and $\hat{\Omega}_{M\hat{\bs{\alpha}}}\hoch{\hat{\bs{\beta}}}$. The
spinorial $\Omega_{M\bs{\alpha}}\hoch{\bs{\beta}}$ induces via $\nabla_{M}\gamma_{\bs{\alpha\beta}}^{c}$
a connection $\Omega_{Ma}\hoch{b}$ for the bosonic subspace which
in turn induces a connection $\Omega_{M\hat{\bs{\alpha}}}\hoch{\hat{\bs{\beta}}}$
via $\nabla_{M}\gamma_{\hat{\bs{\alpha}}\hat{\bs{\beta}}}^{c}=0$.
The collection of those will be denoted by $\Omega_{MA}\hoch{B}$
(\textbf{left-mover connection}\index{connection!left mover $\sim$}\index{left mover connection}\index{$\Omega_{MA}\hoch{B}$|itext{left mover connection}}).
The same can be done for $\hat{\Omega}_{M\hat{\bs{\alpha}}}\hoch{\hat{\bs{\beta}}}$
leading to a connection $\hat{\Omega}_{MA}\hoch{B}$ which we call
the \textbf{right-mover connection}\index{connection!right mover $\sim$}\index{right mover connection}\index{$\Omega$@$\hat{\Omega}_{MA}\hoch{B}$|itext{right mover connection}}.\begin{eqnarray}
\Omega_{MA}\hoch{B} & = & \left(\begin{array}{ccc}
\Omega_{Ma}\hoch{b} & 0 & 0\\
0 & \Omega_{M\bs{\alpha}}\hoch{\bs{\beta}} & 0\\
0 & 0 & \Omega_{M\hat{\bs{\alpha}}}\hoch{\hat{\bs{\beta}}}\end{array}\right),\qquad\hat{\Omega}_{MA}\hoch{B}=\left(\begin{array}{ccc}
\hat{\Omega}_{Ma}\hoch{b} & 0 & 0\\
0 & \hat{\Omega}_{M\bs{\alpha}}\hoch{\bs{\beta}} & 0\\
0 & 0 & \hat{\Omega}_{M\hat{\bs{\alpha}}}\hoch{\hat{\bs{\beta}}}\end{array}\right)\end{eqnarray}
The supergravity constraints are derived from the Berkovits string
using a \textbf{mixed connection}\index{connection!mixed $\sim$}\index{mixed connection}\index{$\Omega$@$\gemOm_{MA}\hoch{B}$|itext{mixed connection}}\begin{eqnarray}
\gemOm_{MA}\hoch{B} & \equiv & \left(\begin{array}{ccc}
\check{\Omega}_{Ma}\hoch{b} & 0 & 0\\
0 & \Omega_{M\bs{\alpha}}\hoch{\bs{\beta}} & 0\\
0 & 0 & \hat{\Omega}_{M\hat{\bs{\alpha}}}\hoch{\hat{\bs{\beta}}}\end{array}\right)\end{eqnarray}
where $\check{\Omega}_{Ma}\hoch{b}$ is an a priori independent connection
for the bosonic part which is only at some parts of the calculation
set to either the right or the left mover connection. In order to
have covariantly constant structure constants ($\gamma_{\bs{\alpha}\bs{\beta}}^{c},\gamma_{\hat{\bs{\alpha}}\hat{\bs{\beta}}}^{c}$)
the latter connection is inadequate and we need to use either one
of the first two or s.th. inbetween, an \textbf{average connection},
which we denote by\index{$\Omega$@$\avOm_{MA}\hoch{B}$|itext{average connection}}\index{average connection $\avOm_{MA}\hoch{B}$}\index{connection!average $\sim$ $\avOm_{MA}\hoch{B}$}\begin{eqnarray}
\avOm_{MA}\hoch{B} & \equiv & \frac{1}{2}\left(\Omega_{MA}\hoch{B}+\hat{\Omega}_{MA}\hoch{B}\right)\end{eqnarray}
By definition the connections $\Omega_{MA}\hoch{B}$, $\hat{\Omega}_{MA}\hoch{B}$
and $\avOm_{MA}\hoch{B}$ (but not $\gemOm_{MA}\hoch{B}$) obey \begin{eqnarray}
\nabla_{M}\gamma_{\bs{\alpha}\bs{\beta}}^{c} & = & \hat{\nabla}_{M}\gamma_{\bs{\alpha}\bs{\beta}}^{c}=\av{\nabla}_{M}\gamma_{\bs{\alpha}\bs{\beta}}^{c}=0\\
\nabla_{M}\gamma_{\hat{\bs{\alpha}}\hat{\bs{\beta}}}^{c} & = & \hat{\nabla}_{M}\gamma_{\hat{\bs{\alpha}}\hat{\bs{\beta}}}^{c}=\av{\nabla}_{M}\gamma_{\hat{\bs{\alpha}}\hat{\bs{\beta}}}^{c}=0\end{eqnarray}
This relates the three matrix-blocks of the connection components.
E.g. for the left-mover connection the spinorial connection $\Omega_{M\bs{\alpha}}\hoch{\bs{\beta}}$(being
a sum of scale and Lorentz connection) determines the remaining two
blocks (see footnote \vref{foot:LorentzScaleReason} for a derivation):\begin{eqnarray}
\Omega_{Ma}\hoch{b} & = & \Omega_{M}^{(D)}\delta_{a}^{b}+\Omega_{Ma}^{(L)}\hoch{b},\qquad\mbox{with }\Omega_{Mab}^{(L)}=-\Omega_{Mba}^{(L)}\\
\Omega_{M\bs{\alpha}}\hoch{\bs{\beta}} & = & \frac{1}{2}\Omega_{M}^{(D)}\delta_{\bs{\alpha}}\hoch{\bs{\beta}}+\frac{1}{4}\Omega_{Mab}^{(L)}\gamma^{ab}\tief{\bs{\alpha}}\hoch{\bs{\beta}}\\
\Omega_{M\hat{\bs{\alpha}}}\hoch{\hat{\bs{\beta}}} & = & \frac{1}{2}\Omega_{M}^{(D)}\delta_{\hat{\bs{\alpha}}}\hoch{\hat{\bs{\beta}}}+\frac{1}{4}\Omega_{Mab}^{(L)}\gamma^{ab}\tief{\hat{\bs{\alpha}}}\hoch{\hat{\bs{\beta}}}\end{eqnarray}

Please note again that the considerations in the following sections
are for a general connection $\Omega_{MA}\hoch{B}$ and not specific
to the leftmoving one. In particular the block diagonality and also
$\nabla_{M}\gamma_{\bs{\alpha}\bs{\beta}}^{c}=\nabla_{M}\gamma_{\hat{\bs{\alpha}}\hat{\bs{\beta}}}^{c}=0$
are only used if this is explicitely mentioned.

\section{Connection in terms of torsion and vielbein (or metric)}

A given torsion and vielbein do not determine yet the connection completely.
It can be determined by having additional structures (like metric
or some group structure constants) that one wants to be covariantly
constant. In the case where a metric is present, the connection is
uniquely determined by the torsion and the (non)metricity of the metric.
Remember the form of the torsion:\begin{eqnarray}
T^{A} & = & \de E^{A}-E^{C}\wedge\Omega_{C}\hoch{A}\\
T_{[MN]}\hoch{A} & = & \partial_{[M}E_{N]}\hoch{A}+\Omega_{[MN]}\hoch{A}\label{eq:torsionForBelow}\end{eqnarray}
Assume that there is some given symmetric tensor field $G_{AB}$ (call
it metric, although it might be degenerate). In flat indices, (non)metricity\index{$M_{ABC}$|itext{nonmetricity}}\index{nonmetricity $M_{ABC}$}
(metricity\index{metricity} for $M_{ABC}=0$) reads\begin{eqnarray}
M_{ABC} & \equiv & \nabla_{A}G_{BC}=\\
 & = & E_{A}\hoch{M}\left(\partial_{M}G_{BC}-2\Omega_{M(B|}\hoch{D}G_{D|C)}\right)=\\
 & \equiv & E_{A}\hoch{M}\left(\partial_{M}G_{BC}-2\Omega_{M(B|C)}\right)\label{eq:nonMetricity}\end{eqnarray}
Here we used $G_{AB}$ to pull down indices, although there might
be no inverse to pull indices up. It is quite common that the metric
in the comoving frame (i.e. in flat indices) is constant, like the
Minkowski metric, and then the derivative part above vanishes. This
is, however, not obligatory. In any case, nonmetricity is part of
the symmetric part (in the last two indices) of $\Omega_{MB|C}$ only.
Let us directly compare (\ref{eq:nonMetricity}) (solved for the connection
term) with (\ref{eq:torsionForBelow}) (rewritten in terms of flat
indices and with one index pulled down via $G_{AB}$ \rem{%
\footnote{\index{footnote!\thefoot. (Nonmetricity mit B statt G)}One might
ask oneself, if the same procedure is possible with the antisymmetric
tensor field $B_{AB}$ instead of $G_{AB}$. Use now $B_{AB}$ (from
the right) to pull down indices and consider its covariant derivative\begin{eqnarray*}
N_{ABC} & \equiv & \nabla_{A}B_{BC}=\\
 & = & E_{A}\hoch{M}(\partial_{M}B_{BC}-2\Omega_{M[B|}\hoch{D}B_{D|C]})\\
\dann\Omega_{A[B|C]} & = & \frac{1}{2}\left(E_{A}\hoch{M}\partial_{M}B_{BC}-N_{ABC}\right)\end{eqnarray*}
The relation to the torsion now reads\begin{eqnarray*}
\Omega_{[AB]|C} & = & T_{AB|C}-(\de E^{D})_{AB}B_{DC}\end{eqnarray*}
$\Omega_{AB|C}$ cannot be expressed in terms of $\Omega_{A[B|C]}$and
$\Omega_{[AB]|C}$ only. (The number of independent degrees does not
match).$\qquad\fussend$ %
} }\begin{eqnarray}
\Omega_{A(B|C)} & = & \frac{1}{2}\left(E_{A}\hoch{M}\partial_{M}G_{BC}-M_{ABC}\right)\label{eq:nonMetricityInOmega}\\
\Omega_{[AB]|C} & = & T_{AB|C}-\underbrace{E_{A}\hoch{M}E_{B}\hoch{N}\partial_{[M}E_{N]}\hoch{D}G_{DC}}_{(\de E^{D})_{AB}G_{DC}}\label{eq:torsionInOmega}\end{eqnarray}
From those two equations we can derive the $\Omega_{AB|C}$ without
any symmetrization. To this end, write down the antisymmetrized connection
three times with permuted indices\begin{eqnarray}
\Omega_{AB|C}-\Omega_{BA|C} & = & 2\Omega_{[AB]|C}\label{eq:OmegaAntisymI}\\
\Omega_{BC|A}-\Omega_{CB|A} & = & 2\Omega_{[BC]|A}\label{eq:OmegaAntisymII}\\
\Omega_{CA|B}-\Omega_{AC|B} & = & 2\Omega_{[CA]|B}\label{eq:OmegaAntisymIII}\end{eqnarray}
Note that \begin{eqnarray}
\Omega_{AB|C} & = & -\Omega_{AC|B}+2\Omega_{A(B|C)}\end{eqnarray}
and consider $\frac{1}{2}\left((\ref{eq:OmegaAntisymI})+(\ref{eq:OmegaAntisymIII})-(\ref{eq:OmegaAntisymII})\right)$:\begin{eqnarray}
\Omega_{AB|C}-\Omega_{A(C|B)}+\Omega_{C(B|A)}-\Omega_{B(C|A)} & = & \Omega_{[AB]|C}+\Omega_{[CA]|B}-\Omega_{[BC]|A}\end{eqnarray}
or \begin{equation}
\boxed{\Omega_{AB|C}=\Omega_{[AB]|C}+\Omega_{[CA]|B}-\Omega_{[BC]|A}+\Omega_{A(C|B)}+\Omega_{B(C|A)}-\Omega_{C(B|A)}}\label{eq:OmegaInTermsOfSymAndAsym}\end{equation}
$\textrm{with }\Omega_{AB|C}\equiv E_{A}\hoch{M}\Omega_{MB}\hoch{D}G_{DC}$.
Now one can plug in (\ref{eq:nonMetricityInOmega}) and (\ref{eq:torsionInOmega}),
in order to get the relation to non-metricity and torsion. For our
purpose it is, however, more convenient to use only the torsion (\ref{eq:torsionInOmega})
and leave $\Omega_{A(B|C)}$ instead of replacing it by nonmetricity.\vRam{.85}{\begin{eqnarray}
\Omega_{AB|C} & = & T_{AB|C}+T_{CA|B}-T_{BC|A}-(\de E^{D})_{AB}G_{DC}-(\de E^{D})_{CA}G_{DB}+(\de E^{D})_{BC}G_{DA}+\nonumber \\
 &  & +\Omega_{A(C|B)}+\Omega_{B(C|A)}-\Omega_{C(B|A)}\label{eq:OmegaInTermsOfSymAndTorsion}\end{eqnarray}
} Some readers might be more familiar with the derivation in curved
indices (defining $\Gamma_{MN|K}\equiv\Gamma_{MN}\hoch{L}G_{LK}$):
\begin{eqnarray}
\Gamma_{[MN]|K} & = & T_{MN|K}\label{eq:torsionInGamma}\\
\Gamma_{K(M|N)} & = & \frac{1}{2}\big(\partial_{K}G_{MN}-\underbrace{\nabla_{K}G_{MN}}_{\equiv M_{KMN}}\big)\label{eq:nonMetricityInGamma}\end{eqnarray}
Equation (\ref{eq:OmegaInTermsOfSymAndAsym}) of course holds likewise
for the spacetime connection\begin{equation}
\boxed{\Gamma_{MN|K}=\Gamma_{[MN]|K}+\Gamma_{[KM]|N}-\Gamma_{[NK]|M}+\Gamma_{M(N|K)}+\Gamma_{N(K|M)}-\Gamma_{K(M|N)}}\label{eq:GammaInTermsOfSymAndAntisym}\end{equation}
This time we replace not only the terms antisymmetrized in the first
two indices with the torsion (\ref{eq:torsionInGamma}) but also the
terms symmetrized in the last two indices with the (non)metricity
(\ref{eq:nonMetricityInGamma}): \begin{equation}
\boxed{\Gamma_{MN|K}=\frac{1}{2}\left(\partial_{M}G_{NK}+\partial_{N}G_{KM}-\partial_{K}G_{MN}\right)+T_{MN|K}+T_{KM|N}-T_{NK|M}-\frac{1}{2}\left(M_{MNK}+M_{NKM}-M_{KMN}\right)}\label{eq:GammaInTermsOfGAndTandM}\end{equation}

If the metric $G_{MN}$ is nondegenerate, one can raise the index
and the connection is completely determined. In ten-dimensional superspace,
however, the situation is different as we have a nondegenerate metric
only in the bosonic subspace.

Consider finally a second connection \begin{equation}
\tilde{\Omega}_{MA}\hoch{B}\equiv\Omega_{MA}\hoch{B}+\Delta_{MA}\hoch{B}\end{equation}
Due to (\ref{eq:GammaOmegaRelationInConnectionPart}), we also have
\begin{eqnarray}
\tilde{\Gamma}_{MK}\hoch{L} & = & \Gamma_{MK}\hoch{L}+\Delta_{MK}\hoch{L}\\
\dann\tilde{T}_{MK}\hoch{L} & = & T_{MK}\hoch{L}+\Delta_{[MK]}\hoch{L}\label{eq:torsiondifference}\end{eqnarray}
The equations (\ref{eq:OmegaInTermsOfSymAndAsym}) and (\ref{eq:GammaInTermsOfSymAndAntisym})
certainly also hold for $\Delta$:\begin{equation}
\boxed{\Delta_{AB|C}=\Delta_{[AB]|C}+\Delta_{[CA]|B}-\Delta_{[BC]|A}+\Delta_{A(C|B)}+\Delta_{B(C|A)}-\Delta_{C(B|A)}}\label{eq:diffTensorInTermsOfSymAndAsym}\end{equation}
The vielbein part of (\ref{eq:OmegaInTermsOfSymAndTorsion}) drops
out in the difference of two connections and we get with (\ref{eq:torsiondifference})%
\footnote{\label{foot:formOfDiffTensor}\index{footnote!\thefoot. form of difference tensor}Some
of our supergravity constraints \frem{Referenz!}will determine $\Delta_{[ab]|c}=-3H_{abc}$,
$\Delta_{[\bs{\alpha}b]|c}=-T_{\bs{\alpha}b|c}$, $\Delta_{[\hat{\bs{\alpha}}b]|c}=\hat{T}_{\hat{\bs{\alpha}}b|c}$,
$\Delta_{a(b|c)}=0$, $\Delta_{\bs{\alpha}(b|c)}=\covPhi{\bs{\alpha}}G_{bc}$
and $\Delta_{\hat{\bs{\alpha}}(b|c)}=-\hatcovPhi{\hat{\bs{\alpha}}}G_{bc}$,
so that the difference tensor reads\begin{eqnarray*}
\Delta_{ab|c} & = & -3H_{abc}\qquad(=-2T_{ab|c}=2\hat{T}_{ab|c})\\
\Delta_{\bs{\alpha}b|c} & = & -2T_{\bs{\alpha}[b|c]}+\covPhi{\bs{\alpha}}G_{bc}=-2T_{\bs{\alpha}b|c}\\
\Delta_{\hat{\bs{\alpha}}b|c} & = & 2\hat{T}_{\hat{\bs{\alpha}}[b|c]}-\hatcovPhi{\hat{\bs{\alpha}}}G_{bc}=2\hat{T}_{\hat{\bs{\alpha}}b|c}\qquad\fussend\end{eqnarray*}
}\begin{equation}
\boxed{\Delta_{AB|C}=(\tilde{T}-T)_{AB|C}+(\tilde{T}-T)_{CA|B}-(\tilde{T}-T)_{BC|A}+\Delta_{A(C|B)}+\Delta_{B(C|A)}-\Delta_{C(B|A)}}\end{equation}

\section{Connection in Superspace}

At least in the ten dimensional type II superspace, there is no natural
nondegenerate superspace metric. Only the bosonic part $G_{MN}$ can
be inverted and the remaining undetermined connection coefficients
have to be fixed by additional conditions. The expression (\ref{eq:OmegaInTermsOfSymAndTorsion})
for the structure group connection in flat indices is more appropriate
than (\ref{eq:GammaInTermsOfGAndTandM}), because in flat indeces
we have a clear split of the bosonic and fermionic subspace of the
tangent space and the only nonvanishing components of the metric $G_{AB}$
is the bosonic (and invertible) metric $G_{ab}$. The connection is
from now on block diagonal of the form $\Omega_{MA}\hoch{B}=\diag(\Omega_{Ma}\hoch{b},\Omega_{m\bs{\alpha}}\hoch{\bs{\beta}},\Omega_{m\hat{\bs{\alpha}}}\hoch{\hat{\bs{\beta}}})$.
Due to the degeneracy of $G_{AB}$, equation (\ref{eq:OmegaInTermsOfSymAndTorsion})
determines only the components $\Omega_{Ab}\hoch{c}$ or equivalently
$\Omega_{Mb}\hoch{c}$ of the structure group connection, i.e. those
with bosonic Lie algebra indices. 

In order to determine the remaining components $\Omega_{M\bs{\alpha}}\hoch{\bs{\beta}}$
and $\Omega_{M\hat{\bs{\alpha}}}\hoch{\hat{\bs{\beta}}}$, we have
to give additional information on what properties we want our connection
to have. In supergravity it is a reasonable demand that the structure
constants of the supersymmetry algebra, i.e. the gamma matrices, are
covariantly constant:\begin{eqnarray}
\nabla_{M}\gamma_{\bs{\alpha}\bs{\beta}}^{a} & \stackrel{!}{=} & 0\\
\nabla_{M}\gamma_{\hat{\bs{\alpha}}\hat{\bs{\beta}}}^{a} & \stackrel{!}{=} & 0\end{eqnarray}
This does not only fix uniquely the form of $\Omega_{M\bs{\alpha}}\hoch{\bs{\beta}}$
and $\Omega_{M\hat{\bs{\alpha}}}\hoch{\hat{\bs{\beta}}}$ in terms
of $\Omega_{Ma}\hoch{b}$, but it also restricts the latter to be
the sum of a Lorentz connection and a scale (or dilatation) connection:%
\footnote{\index{footnote!\thefoot. argument for Lorentz plus scale connection}Let
us give at this point only a short argument for this. According to
(\ref{eq:product-expansion})-(\ref{eq:product-expansion-schematic})
we have schematically $\Gamma^{[k]}\Gamma^{[1]}\propto\Gamma^{[|k-1|]}+\Gamma^{[k+1]}\quad\forall k$,
if $\Gamma^{[k]}$ denotes a term proportional to a completely antisymmetrized
product of $k$ gamma matrices. Let us restrict now to ten dimensions.
The same schematic equation then holds for the chiral submatrices
$\gamma^{[k]}$. The connection can due to its index structure be
expanded in even antisymmetrized products:\begin{eqnarray*}
\Omega_{M\bs{\alpha}}\hoch{\bs{\beta}} & \propto & \gamma^{[0]}+\gamma^{[2]}+\gamma^{[4]}\end{eqnarray*}
When this connection acts on another gamma matrix, we get schematically\begin{eqnarray*}
\Omega_{M[\bs{\alpha}|}\hoch{\bs{\gamma}}\gamma_{\bs{\gamma}|\bs{\beta}]}^{c} & \propto & (\gamma^{[0]}+\gamma^{[2]}+\gamma^{[4]})\gamma^{[1]}\propto\gamma^{[1]}+(\gamma^{[1]}+\underbrace{\gamma^{[3]}}_{0})+(\underbrace{\gamma^{[3]}}_{0}+\gamma^{[5]})\end{eqnarray*}
 The $\gamma^{[3]}$-parts vanish due to the graded antisymmetrization
of the indices. The $\gamma^{[1]}$ parts are fine because they can
be absorbed by acting with the bosonic connection on the bosonic index.
Only the $\gamma^{[5]}$ part remains and cannot be removed. As it
stems from the $\gamma^{[4]}$-part in $\Omega_{M\bs{\alpha}}\hoch{\bs{\beta}}$,
we conclude that the corresponding coefficient has to vanish and only
scale and Lorentz connection remain. The sketched argumentation can
be done rigorously which leads to the stated results for the relation
between bosonic and fermionic connection.\frem{noch irgendwo explizit? -- antighost gauge sym.}$\qquad\fussend$%
} \begin{eqnarray}
\Omega_{M\bs{\alpha}}\hoch{\bs{\beta}} & = & \frac{1}{4}\Omega_{Ma}\hoch{b}\gamma^{a}\tief{b\,\bs{\alpha}}\hoch{\bs{\beta}}+\frac{1}{2}\Omega_{M}^{(D)}\delta_{\bs{\alpha}}\hoch{\bs{\beta}}\label{eq:conn:relateBosonicAndFermConn}\\
\Omega_{M\hat{\bs{\alpha}}}\hoch{\hat{\bs{\beta}}} & = & \frac{1}{4}\Omega_{Ma}\hoch{b}\gamma^{a}\tief{b\,\hat{\bs{\alpha}}}\hoch{\hat{\bs{\beta}}}+\frac{1}{2}\Omega_{M}^{(D)}\delta_{\hat{\bs{\alpha}}}\hoch{\hat{\bs{\beta}}}\label{eq:conn:relateBosonicAndFermConnHat}\end{eqnarray}
with \begin{eqnarray}
\Omega_{Ma}\hoch{b} & \equiv & \underbrace{\Omega_{M[ac]}G^{cb}}_{\equiv\Omega_{\, Ma}^{(L)}\hoch{b}}+\Omega_{M}^{(D)}\delta_{a}^{b}\end{eqnarray}
Because of the split in Lorentz and scale connection, the block-diagonality
of the structure group and the degeneracy of the superspace metric,
equation (\ref{eq:OmegaInTermsOfSymAndTorsion}) can be rewritten
as\begin{equation}
\Omega_{Ab|c}=T_{Ab|c}+T_{cA|b}-T_{bc|A}-(\de E^{d})_{Ab}G_{dc}-(\de E^{d})_{cA}G_{db}+(\de E^{d})_{bc}G_{dA}+\Omega_{A}^{(D)}G_{cb}+\Omega_{b}^{(D)}G_{cA}-\Omega_{c}^{(D)}G_{bA}\label{eq:ConnectionInTermsOfTorsionAndVielbein}\end{equation}
or \begin{eqnarray}
\Omega_{ab|c} & = & T_{ab|c}+T_{ca|b}-T_{bc|a}-(\de E^{d})_{ab}G_{dc}-(\de E^{d})_{ca}G_{db}+(\de E^{d})_{bc}G_{da}+\Omega_{a}^{(D)}G_{cb}+\Omega_{b}^{(D)}G_{ca}-\Omega_{c}^{(D)}G_{ba}\qquad\label{eq:CITOTAV-I}\\
\Omega_{\bs{\alpha}b|c} & = & T_{\bs{\alpha}b|c}+T_{c\bs{\alpha}|b}-(\de E^{d})_{\bs{\alpha}b}G_{dc}-(\de E^{d})_{c\bs{\alpha}}G_{db}+\Omega_{\bs{\alpha}}^{(D)}G_{cb}\label{eq:CITOTAV-II}\\
\Omega_{\hat{\bs{\alpha}}b|c} & = & T_{\hat{\bs{\alpha}}b|c}+T_{c\hat{\bs{\alpha}}|b}-(\de E^{d})_{\hat{\bs{\alpha}}b}G_{dc}-(\de E^{d})_{c\hat{\bs{\alpha}}}G_{db}+\Omega_{\hat{\bs{\alpha}}}^{(D)}G_{cb}\label{eq:CITOTAV-III}\end{eqnarray}
which determines $\Omega_{Ma}\hoch{b}$ via \begin{equation}
\Omega_{Ma}\hoch{b}=E_{M}\hoch{C}\Omega_{Ca|d}G^{db}\quad\textrm{with }G_{ac}G^{cb}\equiv\delta_{a}^{b}\label{eq:OmegaM}\end{equation}
The remaining components $\Omega_{M\bs{\alpha}}\hoch{\bs{\beta}}$
and $\Omega_{M\hat{\bs{\alpha}}}\hoch{\hat{\bs{\beta}}}$ are then
fixed via (\ref{eq:conn:relateBosonicAndFermConn}) and (\ref{eq:conn:relateBosonicAndFermConnHat}).

Let us in the following calculate $\Omega_{Ma}\hoch{b}$ more explicitely
in the WZ gauge in order to extract the Levi Civita connection of
the bosonic subspace.

\section{Extracting Levi Civita from whole superspace connection (in WZ-gauge)}

\index{Levi Civita!extracting $\sim$ from superspace connection}Remember
our definition $G_{MN}=E_{M}\hoch{a}\underbrace{e^{2\Phi}\eta_{ab}}_{G_{ab}}E_{N}\hoch{b}$
in the application to the Berkovits string and the Wess Zumino gauge
(\ref{eq:WZ-gauge},\ref{eq:WZ-gauge-inverse},\ref{eq:WZ-gauge-connection}):\begin{eqnarray}
\bei{E_{M}\hoch{A}}{\xbothtetas=0} & = & \left(\begin{array}{ccc}
e_{m}\hoch{a} & \psi_{m}\hoch{\bs{\alpha}} & \hat{\psi}_{m}\hoch{\hat{\bs{\alpha}}}\\
0 & \delta_{\bs{\mu}}\hoch{\bs{\alpha}} & 0\\
0 & 0 & \delta_{\hat{\bs{\mu}}}\hoch{\hat{\bs{\alpha}}}\end{array}\right),\quad\bei{E_{A}\hoch{M}}{}=\left(\begin{array}{ccc}
e_{a}\hoch{m} & -\psi_{a}\hoch{\bs{\mu}} & -\hat{\psi}_{a}\hoch{\hat{\bs{\mu}}}\\
0 & \delta_{\bs{\alpha}}\hoch{\bs{\mu}} & 0\\
0 & 0 & \delta_{\hat{\bs{\alpha}}}\hoch{\hat{\bs{\mu}}}\end{array}\right),\quad\bei{\Omega_{\bs{\mc{M}}A}\hoch{B}}{}=0\label{eq:WZgauge-in-connectionappend}\\
\mbox{with } &  & e_{m}\hoch{a}e_{a}\hoch{n}=\delta_{m}^{n},\quad\psi_{a}\hoch{\bs{\mu}}\equiv e_{a}\hoch{m}\psi_{m}\hoch{\bs{\alpha}}\delta_{\bs{\alpha}}\hoch{\bs{\mu}},\quad\hat{\psi}_{a}\hoch{\hat{\bs{\mu}}}\equiv e_{a}\hoch{m}\psi_{m}\hoch{\hat{\bs{\alpha}}}\delta_{\hat{\bs{\alpha}}}\hoch{\hat{\bs{\mu}}}\nonumber \end{eqnarray}
As bosonic metric, we could either take just the leading component
in the $\xbothtetas$-expansion of $G_{mn}$, or the one given by
the bosonic vielbein $e_{m}\hoch{a}$ and the Minkowski metric:\begin{equation}
\tilde{g}_{mn}\equiv\bei{G_{mn}}{}=e_{m}\hoch{a}\underbrace{e^{2\phi}\eta_{ab}}_{\tilde{g}_{ab}}e_{n}\hoch{b},\qquad g_{mn}\equiv e_{m}\hoch{a}\eta_{ab}e_{n}\hoch{b}=e^{-2\phi}\tilde{g}_{mn}\end{equation}
 The first is naturally induced by the superspace 'metric', while
the second is by construction covariantly conserved with respect to
the connection $\omega_{ma}\hoch{b}\equiv\bei{\Omega_{ma}\hoch{b}}{}$
(in contrast to $\tilde{g}_{mn}$ because of the scaling compensator
field $\phi$). We want to write the superspace connection at $\xbothtetas=0$
as the Levi Civita connection w.r.t. $\tilde{g}_{mn}$ or $g_{mn}$
plus additional terms. 

The superspace connection was derived above starting from (\ref{eq:OmegaInTermsOfSymAndAsym})
or (\ref{eq:OmegaInTermsOfSymAndTorsion}), arriving at the equations
(\ref{eq:CITOTAV-I}-\ref{eq:CITOTAV-III}) for $\Omega_{ab|c},\Omega_{\bs{\alpha}b|c}$
and $\Omega_{\hat{\bs{\alpha}}b|c}$ in terms of the torsion and the
exterior derivative of the supervielbein $\de E^{d}$. We can also
use the general equation (\ref{eq:OmegaInTermsOfSymAndTorsion}),
in order to determine the form of the Levi Civita connection for $g_{mn}$
in terms of the bosonic vielbein. We just have to set the torsion
and the symmetric part to zero. However, as we already use the supervielbein
in order to switch from flat to curved indices and vice versa, we
better should write the bosonic vielbeins explicitely in the resulting
equation: \begin{eqnarray}
\lqn{\Ramm{.8}{\zwek{}{}}}\quad e_{a}\hoch{m}\omega_{mb}^{LC}\hoch{d}[g]\cdot\eta_{dc} & = & -e_{a}\hoch{m}e_{b}\hoch{n}(\de e^{d})_{mn}\eta_{dc}-e_{c}\hoch{m}e_{a}\hoch{n}(\de e^{d})_{mn}\eta_{db}+e_{b}\hoch{m}e_{c}\hoch{n}(\de e^{d})_{mn}\eta_{da}\label{eq:LCconnOfeta}\end{eqnarray}
For the metric $\tilde{g}_{mn}$ instead, the symmetric part of the
Levi Civita connection is no longer zero. We still have torsionlessness
and metric compatibility as characterizing properties. The latter
condition implies via (\ref{eq:nonMetricity}) that \begin{equation}
\omega_{m(b|c)}^{(LC)}[\tilde{g}]=\frac{1}{2}\partial_{m}\tilde{g}_{bc}=\partial_{m}\phi\cdot\tilde{g}_{bc}\end{equation}
Using again (\ref{eq:OmegaInTermsOfSymAndTorsion}) with vanishing
torsion, we arrive at \vRam{.8}{\begin{eqnarray}
e_{a}\hoch{m}\omega_{mb}^{LC}\hoch{d}[\tilde{g}]\cdot\tilde{g}_{dc} & = & -e_{a}\hoch{m}e_{b}\hoch{n}(\de e^{d})_{mn}\tilde{g}_{dc}-e_{c}\hoch{m}e_{a}\hoch{n}(\de e^{d})_{mn}\tilde{g}_{db}+e_{b}\hoch{m}e_{c}\hoch{n}(\de e^{d})_{mn}\tilde{g}_{da}+\nonumber \\
 &  & +e_{a}\hoch{m}\partial_{m}\phi\cdot\tilde{g}_{bc}+e_{b}\hoch{m}\partial_{m}\phi\cdot\tilde{g}_{ca}-e_{c}\hoch{m}\partial_{m}\phi\cdot\tilde{g}_{ab}\label{eq:LC-conn-of-g}\end{eqnarray}
} In both cases (for $\tilde{g}$ and $g$) the corresponding Levi
Civita connection is certainly sitting in the superspace connection
in the terms with $\de E^{d}$ in (\ref{eq:CITOTAV-I}-\ref{eq:CITOTAV-III})
at $\xbothtetas=0$. Indeed one can write%
\footnote{\index{footnote!\thefoot. exterior derivative of supervielbein and vielbein}In
the Wess Zumino gauge we can express $\bei{\de E^{a}}{}$ by $\de e^{a}$
plus torsion terms as we will demonstrate now. First we have \begin{eqnarray*}
\bei{\de E^{a}}{} & = & \overbrace{\bei{\partial_{[m}E_{n]}\hoch{a}}{}\de x^{m}\de x^{n}}^{\de e^{a}}+2\bei{\partial_{[m}E_{\bs{\mc{N}}]}\hoch{a}}{}\de x^{m}\de x^{\bs{\mc{N}}}+\bei{\partial_{[\bs{\mc{M}}}E_{\bs{\mc{N}}]}}{}\de x^{\bs{\mc{M}}}\de x^{\bs{\mc{N}}}\end{eqnarray*}
As $\bei{E_{m}\hoch{a}}{}=e_{m}\hoch{a}$, we have \[
\bei{(\de E^{a})_{mn}}{}=(\de e^{a})_{mn}\]
Now remember the definition of the torsion $T^{A}=\de E^{A}-E^{B}\wedge\Omega_{B}\hoch{A}$
which reads for fermionic form indices at $\xbothtetas=0$ in the
Wess-Zumino gauge (\ref{eq:WZI},\ref{eq:WZII}):\begin{eqnarray*}
\bei{\partial_{[\bs{\mc{M}}}E_{\bs{\mc{N}}]}\hoch{A}}{} & = & \bei{T_{\bs{\mc{MN}}}\hoch{A}}{}-\bei{\Omega_{[\bs{\mc{MN}}]}\hoch{A}}{}\stackrel{(\ref{eq:WZII})}{=}\bei{T_{\bs{\mc{MN}}}\hoch{A}}{}\end{eqnarray*}
Similarly we have \begin{eqnarray*}
\bei{\partial_{[\bs{\mc{M}}}E_{n]}\hoch{A}}{} & = & \bei{T_{\bs{\mc{M}}n}\hoch{A}}{}-\bei{\Omega_{[\bs{\mc{M}}n]}\hoch{A}}{}\stackrel{(\ref{eq:WZII})}{=}\bei{T_{\bs{\mc{M}}n}\hoch{A}}{}+\frac{1}{2}\delta_{\bs{\mc{M}}}\hoch{\bs{\mc{B}}}\bei{\Omega_{n\bs{\mc{B}}}\hoch{A}}{}\end{eqnarray*}
For $A=a$, we can thus write in summary\begin{eqnarray*}
\bei{(\de E^{a})_{\bs{\mc{M}}N}}{} & = & \bei{T_{\bs{\mc{M}}N}\hoch{a}}{}\qquad\fussend\end{eqnarray*}
} \begin{eqnarray}
\bei{(\de E^{a})_{mn}}{} & = & (\de e^{a})_{mn}\\
\bei{(\de E^{a})_{\bs{\mc{M}}N}}{} & = & \bei{T_{\bs{\mc{M}}N}\hoch{a}}{}\end{eqnarray}
This is consistent with the fact that $\Omega_{\bs{\alpha}b|c}$ and
$\Omega_{\hat{\bs{\alpha}}b|c}$ as given in (\ref{eq:CITOTAV-II})
and (\ref{eq:CITOTAV-III}) vanish at $\xbothtetas=0$ in the WZ-gauge
(where $\bei{E_{\bs{\alpha}}\hoch{M}}{}=\delta_{\bs{\alpha}}\hoch{M}$
and $\bei{\Omega_{\bs{\mu}A}\hoch{B}}{}=0$. In order to calculate
$\bei{\Omega_{ab|c}}{}$ as given in (\ref{eq:CITOTAV-I}), we need
the the exterior derivative of the vielbein (as given above) with
flat bosonic indices. As the constraints on the torsion components
will also be given in flat indices, we will express everything in
terms of torsion components with flat indices: \begin{eqnarray}
\bei{(\de E^{d})_{ab}}{} & = & e_{a}\hoch{m}e_{b}\hoch{n}(\de e^{d})_{mn}-2\psi_{[a}\hoch{\bs{\mc{M}}}e_{b]}\hoch{n}\bei{T_{\bs{\mc{M}}n}\hoch{d}}{}+\psi_{a}\hoch{\bs{\mc{M}}}\psi_{b}\hoch{\bs{\mc{N}}}\bei{T_{\bs{\mc{M}}\bs{\mc{N}}}\hoch{d}}{}=\label{eq:intermediate-dE}\\
 & = & e_{a}\hoch{m}e_{b}\hoch{n}\left((\de e^{d})_{mn}+\psi_{m}\hoch{\bs{\mc{A}}}\psi_{n}\hoch{\bs{\mc{B}}}\bei{T_{\bs{\mc{A}}\bs{\mc{B}}}\hoch{d}}{}\right)-2e_{[a|}\hoch{m}\psi_{m}\hoch{\bs{\mc{A}}}e_{b]}\hoch{n}\left(e_{n}\hoch{c}\bei{T_{\bs{\mc{A}}c}\hoch{d}}{}+\psi_{n}\hoch{\bs{\mc{B}}}\bei{T_{\bs{\mc{A}}\bs{\mc{B}}}\hoch{d}}{}\right)=\\
 & = & e_{a}\hoch{m}e_{b}\hoch{n}\left((\de e^{d})_{mn}-\psi_{m}\hoch{\bs{\mc{A}}}\psi_{n}\hoch{\bs{\mc{B}}}\bei{T_{\bs{\mc{A}}\bs{\mc{B}}}\hoch{d}}{}\right)-2e_{[a|}\hoch{m}\psi_{m}\hoch{\bs{\mc{A}}}\bei{T_{\bs{\mc{A}}|b]}\hoch{d}}{}\end{eqnarray}
Plugging this result into (\ref{eq:CITOTAV-I}) yields $\Omega_{ab|c}$
at $\xbothtetas=0$ in terms of torsion components with flat indices
and derivatives of the bosonic vielbein only:\begin{eqnarray}
\bei{\Omega_{ab|c}}{} & = & \bei{T_{ab|c}}{}+\bei{T_{ca|b}}{}-\bei{T_{bc|a}}{}-\left(e_{a}\hoch{m}e_{b}\hoch{n}\left((\de e^{d})_{mn}\tilde{g}_{dc}-\psi_{m}\hoch{\bs{\mc{A}}}\psi_{n}\hoch{\bs{\mc{B}}}\bei{T_{\bs{\mc{A}}\bs{\mc{B}}|c}}{}\right)-2e_{[a|}\hoch{m}\psi_{m}\hoch{\bs{\mc{A}}}\bei{T_{\bs{\mc{A}}|b]c}}{}\right)+\nonumber \\
 &  & -\left(e_{c}\hoch{m}e_{a}\hoch{n}\left((\de e^{d})_{mn}\tilde{g}_{db}-\psi_{m}\hoch{\bs{\mc{A}}}\psi_{n}\hoch{\bs{\mc{B}}}\bei{T_{\bs{\mc{A}}\bs{\mc{B}}|b}}{}\right)-2e_{[c|}\hoch{m}\psi_{m}\hoch{\bs{\mc{A}}}\bei{T_{\bs{\mc{A}}|a]b}}{}\right)+\nonumber \\
 &  & +\left(e_{b}\hoch{m}e_{c}\hoch{n}\left((\de e^{d})_{mn}\tilde{g}_{da}-\psi_{m}\hoch{\bs{\mc{A}}}\psi_{n}\hoch{\bs{\mc{B}}}\bei{T_{\bs{\mc{A}}\bs{\mc{B}}|a}}{}\right)-2e_{[b|}\hoch{m}\psi_{m}\hoch{\bs{\mc{A}}}\bei{T_{\bs{\mc{A}}|c]a}}{}\right)+\nonumber \\
 &  & +\bei{\Omega_{a}}{}\tilde{g}_{cb}+\bei{\Omega_{b}}{}\tilde{g}_{ca}-\bei{\Omega_{c}}{}\tilde{g}_{ba}\label{eq:intermediate-connection}\end{eqnarray}
Now we can express everything in terms of the Levi Civita connection
w.r.t. $\tilde{g}$ (\ref{eq:LC-conn-of-g}), torsion terms with flat
indices and covariant derivatives of the compensator field:

\begin{eqnarray}
\bei{\Omega_{ab|c}}{} & = & e_{a}\hoch{m}\omega_{mb}^{LC\, d}[\tilde{g}]\tilde{g}_{dc}+\bei{T_{ab|c}}{}+\bei{T_{ca|b}}{}-\bei{T_{bc|a}}{}+\nonumber \\
 &  & -\underbrace{\left(e_{a}\hoch{m}\partial_{m}\phi-\bei{\Omega_{a}}{}\right)}_{\bei{\covPhi{a}}{}+\psi_{a}\hoch{\bs{\mc{M}}}\underbrace{\bei{(\partial_{\bs{\mc{M}}}\Phi)}{}}_{\delta_{\bs{\mc{M}}}\hoch{\bs{\mc{A}}}\bei{(\nabla_{\bs{\mc{A}}}\Phi)}{}}}\tilde{g}_{cb}-\left(e_{b}\hoch{m}\partial_{m}\phi-\bei{\Omega_{b}}{}\right)\tilde{g}_{ca}+\left(e_{c}\hoch{m}\partial_{m}\phi-\bei{\Omega_{c}}{}\right)\tilde{g}_{ba}+\nonumber \\
 &  & +\left(e_{a}\hoch{m}e_{b}\hoch{n}\tilde{g}_{cd}+e_{c}\hoch{m}e_{a}\hoch{n}\tilde{g}_{bd}-e_{b}\hoch{m}e_{c}\hoch{n}\tilde{g}_{ad}\right)\psi_{m}\hoch{\bs{\mc{A}}}\psi_{n}\hoch{\bs{\mc{B}}}\bei{T_{\bs{\mc{A}}\bs{\mc{B}}}\hoch{d}}{}\nonumber \\
 &  & +2e_{[a|}\hoch{m}\psi_{m}\hoch{\bs{\mc{A}}}\bei{T_{\bs{\mc{A}}|b]c}}{}+2e_{[c|}\hoch{m}\psi_{m}\hoch{\bs{\mc{A}}}\bei{T_{\bs{\mc{A}}|a]b}}{}-2e_{[b|}\hoch{m}\psi_{m}\hoch{\bs{\mc{A}}}\bei{T_{\bs{\mc{A}}|c]a}}{}\end{eqnarray}
While for the use of $\omega_{mb}^{LC\, d}[\tilde{g}]$ above the
partial derivatives of the compensator $\compcomp$ combine with the
scale connections to covariant derivatives, either the scale connections
or the partial derivatives remain explicitely for the use of $\omega_{mb}^{LC\, d}[g]$
(\ref{eq:LCconnOfeta}). In summary we have for the two cases \vRam{1.02}{\begin{eqnarray}
\underbrace{\bei{\Omega_{ab|c}}{}}_{e_{a}\hoch{m}\omega_{mb}\hoch{d}e^{2\compcomp}\eta_{dc}} & = & e_{a}\hoch{m}\omega_{mb}^{LC\, d}[\tilde{g}]\tilde{g}_{dc}+2\bei{T_{a[b|c]}}{}-\bei{T_{bc|a}}{}+\nonumber \\
 &  & -2\bei{\covPhi{[b}}{}\tilde{g}_{c]a}-\bei{\covPhi{a}}{}\tilde{g}_{bc}-\left(2e_{[b}\hoch{m}\tilde{g}_{c]a}+e_{a}\hoch{m}\tilde{g}_{bc}\right)\psi_{m}\hoch{\bs{\mc{A}}}\bei{(\nabla_{\bs{\mc{A}}}\Phi)}{}+\nonumber \\
 &  & +\left(2e_{a}\hoch{m}e_{[b}\hoch{n}\tilde{g}_{c]d}-e_{[b}\hoch{m}e_{c]}\hoch{n}\tilde{g}_{ad}\right)\psi_{m}\hoch{\bs{\mc{A}}}\psi_{n}\hoch{\bs{\mc{B}}}\bei{T_{\bs{\mc{A}}\bs{\mc{B}}}\hoch{d}}{}\nonumber \\
 &  & +2e_{a}\hoch{m}\psi_{m}\hoch{\bs{\mc{A}}}\bei{T_{\bs{\mc{A}}[b|c]}}{}-2e_{[b|}\hoch{m}\psi_{m}\hoch{\bs{\mc{A}}}\bei{T_{\bs{\mc{A}}a|c]}}{}-2e_{[b|}\hoch{m}\psi_{m}\hoch{\bs{\mc{A}}}\bei{T_{\bs{\mc{A}}|c]a}}{}=\\
 & = & e_{a}\hoch{m}\omega_{mb}^{LC\, d}[g]\eta_{dc}e^{2\compcomp}+2\bei{T_{a[b|c]}}{}-\bei{T_{bc|a}}{}+\nonumber \\
 &  & -2(\bei{\covPhi{[b|}}{}-e_{[b|}\hoch{n}\partial_{n}\phi)\eta_{c]a}-(\bei{\covPhi{a}}{}-e_{a}\hoch{n}\partial_{n}\phi)\eta_{bc}-\left(2e_{[b}\hoch{m}\tilde{g}_{c]a}+e_{a}\hoch{m}\tilde{g}_{bc}\right)\psi_{m}\hoch{\bs{\mc{A}}}\bei{(\nabla_{\bs{\mc{A}}}\Phi)}{}+\nonumber \\
 &  & +e^{2\compcomp}\left(2e_{a}\hoch{m}e_{[b}\hoch{n}\eta_{c]d}-e_{[b}\hoch{m}e_{c]}\hoch{n}\eta_{ad}\right)\psi_{m}\hoch{\bs{\mc{A}}}\psi_{n}\hoch{\bs{\mc{B}}}\bei{T_{\bs{\mc{A}}\bs{\mc{B}}}\hoch{d}}{}+\nonumber \\
 &  & +2e_{a}\hoch{m}\psi_{m}\hoch{\bs{\mc{A}}}\bei{T_{\bs{\mc{A}}[b|c]}}{}-2e_{[b|}\hoch{m}\psi_{m}\hoch{\bs{\mc{A}}}\bei{T_{\bs{\mc{A}}a|c]}}{}-2e_{[b|}\hoch{m}\psi_{m}\hoch{\bs{\mc{A}}}\bei{T_{\bs{\mc{A}}|c]a}}{}\label{eq:OmegaInTermsOfLCandFlatTorsion:bosonic}\end{eqnarray}
} We have written the terms in a way that one can clearly distinguish
between terms anti-symmetric in $b,c$ (Lorentz-part) and terms symmetric
in $b,c$ (scale-part). In the second version (\ref{eq:OmegaInTermsOfLCandFlatTorsion:bosonic}),
the whole second line could be written as $+2\bei{\Omega_{[b}^{(D)}}{}e^{2\compcomp}\eta_{c]a}+\bei{\Omega_{a}^{(D)}}{}e^{2\compcomp}\eta_{bc}$
which is, however, less convenient for plugging the constraints into
it. The Levi Civita connection $\omega_{mb}^{LC\, d}[g]$ does not
transform under scale transformations in the way it should, which
is repaired by the non-covariantly transforming partial derivatives
$\partial_{k}\phi$. They are thus the minimal extension of the Levi-Civita
connection to make it transforming properly under the whole structure
group. Combining these terms with $\omega_{mb}^{LC\, d}[g]$ just
leads back to $\omega_{mb}^{LC\, d}[\tilde{g}]$ which apparently
has a scale part. This seems strange for a Levi Civita connection,
but is only true in the frame $e_{m}\hoch{a}$ where the flat metric
is not Minkowski.

Assuming that $\nabla_{M}\gamma_{\bs{\alpha\beta}}^{a}=\nabla_{M}\gamma_{\hat{\bs{\alpha}}\hat{\bs{\beta}}}^{a}=0$,
we can finally (according to (\ref{eq:conn:relateBosonicAndFermConn})
and (\ref{eq:conn:relateBosonicAndFermConnHat})) write down the connection
when acting on fermionic indices. We restrict to the version with
the Levi Civita action for $g_{mn}=e_{m}\hoch{a}\eta_{ab}e_{n}\hoch{b}$:
\begin{eqnarray}
\bei{\Omega_{m\bs{\gamma}}\hoch{\bs{\alpha}}}{} & = & \frac{1}{4}\omega_{m[b|c]}\tilde{\gamma}^{bc}\tief{\bs{\gamma}}\hoch{\bs{\alpha}}+\frac{1}{2}\omega_{m}^{(D)}\delta_{\bs{\gamma}}\hoch{\bs{\alpha}}=\nonumber \\
 & = & \frac{1}{4}e_{m}\hoch{a}\biggl\{ e_{a}\hoch{n}\omega_{n[b|}^{LC\, d}[g]\eta_{d|c]}+2e^{-2\compcomp}\bei{T_{a[b|c]}}{}-e^{-2\compcomp}\bei{T_{bc|a}}{}+\nonumber \\
 &  & -2(\bei{\covPhi{[b|}}{}-e_{[b|}\hoch{n}\partial_{n}\phi)\eta_{c]a}-2e_{[b}\hoch{m}\eta_{c]a}\psi_{m}\hoch{\bs{\mc{A}}}\bei{(\nabla_{\bs{\mc{A}}}\Phi)}{}+\nonumber \\
 &  & +\left(2e_{a}\hoch{k}e_{[b}\hoch{n}\eta_{c]d}-e_{b}\hoch{k}e_{c}\hoch{n}\eta_{ad}\right)\psi_{k}\hoch{\bs{\mc{A}}}\psi_{n}\hoch{\bs{\mc{B}}}\bei{T_{\bs{\mc{A}}\bs{\mc{B}}}\hoch{d}}{}+\nonumber \\
 &  & +e^{-2\compcomp}\Bigl(2e_{a}\hoch{n}\psi_{n}\hoch{\bs{\mc{A}}}\bei{T_{\bs{\mc{A}}[b|c]}}{}\underbrace{-2e_{b}\hoch{n}\psi_{n}\hoch{\bs{\mc{A}}}\bei{T_{\bs{\mc{A}}(a|c)}}{}+2e_{c}\hoch{n}\psi_{n}\hoch{\bs{\mc{A}}}\bei{T_{\bs{\mc{A}}(a|b)}}{}}_{-2e_{[b|}\hoch{n}\psi_{n}\hoch{\bs{\mc{A}}}\bei{T_{\bs{\mc{A}}a|c]}}{}-2e_{[b|}\hoch{n}\psi_{n}\hoch{\bs{\mc{A}}}\bei{T_{\bs{\mc{A}}|c]a}}{}}\Bigr)\biggr\}\gamma^{bc}\tief{\bs{\gamma}}\hoch{\bs{\alpha}}\nonumber \\
 &  & -\frac{1}{2}\underbrace{\Bigl(\bei{\covPhi{a}}{}-e_{a}\hoch{n}\partial_{n}\phi+e_{a}\hoch{m}\psi_{m}\hoch{\bs{\mc{A}}}\bei{(\nabla_{\bs{\mc{A}}}\Phi)}{}\Bigr)}_{\equiv-\omega_{m}^{(D)}=-\bei{\Omega_{m}^{(D)}}{}}\delta_{\bs{\gamma}}\hoch{\bs{\alpha}}\label{eq:LCinSuperspaceConnectionFermionic}\end{eqnarray}
An equivalent expression with $\gamma^{bc}\tief{\bs{\gamma}}\hoch{\bs{\alpha}}$
and $\delta_{\bs{\gamma}}\hoch{\bs{\alpha}}$ replaced by $\gamma^{bc}\tief{\hat{\bs{\gamma}}}\hoch{\hat{\bs{\alpha}}}$
and $\delta_{\hat{\bs{\gamma}}}\hoch{\hat{\bs{\alpha}}}$ is obtained
for $\Omega_{m\hat{\bs{\gamma}}}\hoch{\hat{\bs{\alpha}}}$. 

A second useful way to write the connection $\bei{\Omega_{ab|c}}{}$
is to bring it to a form which is the bosonic version of (\ref{eq:OmegaInTermsOfSymAndTorsion})
and from which we can read off the bosonic torsion and nonmetricity.
To this end, we rewrite (\ref{eq:intermediate-dE}) as \begin{eqnarray}
\bei{(\de E^{d})_{ab}}{} & = & e_{a}\hoch{m}e_{b}\hoch{n}\left((\de e^{d})_{mn}-T_{mn}\hoch{d}\right)+\bei{T_{ab}\hoch{d}}{}\end{eqnarray}
Plugging this into (\ref{eq:CITOTAV-I}) yields \vRam{0.75}{\begin{eqnarray}
\quad\underbrace{\bei{\Omega_{ab|c}}{}}_{e_{a}\hoch{m}\omega_{mb|c}} & = & -e_{a}\hoch{m}e_{b}\hoch{n}(\de e^{d})_{mn}\tilde{g}_{dc}-e_{c}\hoch{m}e_{a}\hoch{n}(\de e^{d})_{mn}\tilde{g}_{db}+e_{b}\hoch{m}e_{c}\hoch{n}(\de e^{d})_{mn}\tilde{g}_{da}+\nonumber \\
 &  & +e_{a}\hoch{m}e_{b}\hoch{n}\bei{T_{mn}\hoch{d}}{}\tilde{g}_{dc}+e_{c}\hoch{m}e_{a}\hoch{n}\bei{T_{mn}\hoch{d}}{}\tilde{g}_{db}-e_{b}\hoch{m}e_{c}\hoch{n}\bei{T_{mn}\hoch{d}}{}\tilde{g}_{da}\nonumber \\
 &  & +\bei{\Omega_{a}^{(D)}}{}\tilde{g}_{cb}+\bei{\Omega_{b}^{(D)}}{}\tilde{g}_{ca}-\bei{\Omega_{c}^{(D)}}{}\tilde{g}_{ba}\label{eq:OmegaInTermsOfvielbeinAndTorsion:bosonic}\end{eqnarray}
} As we have in the Wess-Zumino gauge $\bei{\Omega_{mb}\hoch{e}}{}=e_{m}\hoch{a}\omega_{ab}\hoch{e}$,
the obtained equation is simply the bosonic version of (\ref{eq:OmegaInTermsOfSymAndTorsion})
with $\omega_{ma}^{(D)}\hoch{b}=\bei{\Omega_{m}^{(D)}}{}\delta_{a}^{b}$.
The bosonic torsion coincides with $\bei{T_{mn}\hoch{d}}{}$. \index{induced bosonic torsion}\rem{antisymmetrisch in allen Indizes?} \begin{eqnarray}
\bei{T_{mn}\hoch{d}}{} & = & e_{m}\hoch{a}e_{n}\hoch{b}\bei{T_{ab}\hoch{d}}{}+2e_{[m}\hoch{a}\psi_{n]}\hoch{\bs{\mc{B}}}\bei{T_{a\bs{\mc{B}}}\hoch{d}}{}+\psi_{m}\hoch{\bs{\mc{A}}}\psi_{n}\hoch{\bs{\mc{B}}}\bei{T_{\bs{\mc{AB}}}\hoch{d}}{}\label{eq:inducedBosonicTorsion}\end{eqnarray}
 \rem{hier noch ein paar alte LC-Rechnungen verborgen} \rem{ und hier noch ein paar alte explizite $\Omega_{ab|c}$s}

\bibliographystyle{/home/basti/fullsort}
\bibliography{/home/basti/phd,/home/basti/Proposal}

\printindex{}
}

\chapter[Supergauge Transformations, their Algebra and the WZ Gauge]{Supergauge Transformations, their Algebra and the Wess Zumino Gauge}

\label{cha:Supergauge-Transformations}{\inputTeil{0} \ifthenelse{\theinput=1}{}{} 

\title{SUSY Transformation in Supergravity}

\author{Sebastian Guttenberg}

\date{January 27, '08}

\maketitle
\begin{abstract}
Part of thesis-appendix
\end{abstract}
\tableofcontents{}\ifthenelse{\theinput=1}{}{\newpage}

This appendix contains, like most of the others, considerations which
are valid not only for our application to the Berkovits string in
ten dimensions, but as well for other dimensions and for different
supergravity theories. The curved indices ${\scriptstyle M}$ as well
as the flat indices ${\scriptstyle A}$ contain bosonic indices ${\scriptstyle m}$
or ${\scriptstyle a}$ as well as fermionic indices ${\scriptstyle \bs{\mc{M}}}$
or ${\scriptstyle \bs{\mc{A}}}$. For extended supersymmetry the latter
are further split into several irreducible fermionic indices. E.g.
for type II in ten dimensions (our application) we have ${\scriptstyle \bs{\mc{M}}=(\bs{\mu},\hat{\bs{\mu}})}$
and ${\scriptstyle \bs{\mc{A}}=(\bs{\alpha},\hat{\bs{\alpha}})}$
where ${\scriptstyle \hat{\bs{\alpha}}}$ is either of the same or
of opposite chirality as ${\scriptstyle \bs{\alpha}}$. We only assume
the presence of a (super)vielbein $E_{M}\hoch{A}$ and of a (super)connection
$\Omega_{MA}\hoch{B}$ in the supergravity theory. Discussions of
other fields (like the $B$-field) are of course only relevant for
theories containing these fields.

\label{chapter:susy}The supergravity\index{supergravity!transformation}
transformation (local supersymmetry\index{supersymmetry!transformation})
is in some sense a special class of superdiffeomorphism transformations.
If the general superdiffeomorphisms are parametrized by a vector field
$\xi^{A}(\xfull)\equiv\xi^{A}(x,\xbothtetas)$\index{$\xi^{A}(\xfull)$},
the local supersymmetry will be parametrized by only $\xi^{\bs{\mc{A}}}(x,0)$.
Likewise, general coordinate transformations in the bosonic submanifold
are parametrized by $\xi^{a}(x,0)$, while all the higher $\xbothtetas$-components
of $\xi^{A}$ correspond to additional auxiliary gauge degrees of
freedom\index{auxiliary!gauge degrees of freedom}. Similarly, the
local structure group transformations $L_{ab}(\xfull)$\index{$L_{ab}(x,\tet)$}
(e.g. Lorentz-transformations or in our application also scale transformations)
have auxiliary gauge degrees in the higher $\xbothtetas$-parts. Following
roughly \cite[p.127-144]{Wess:1992cp}, we want to bring e.g. the
vielbein into a particular form, using (and thereby fixing) some of
those shift\index{shift!symmetries} symmetries\index{symmetries!shift$\sim$},
and to identify the bosonic spacetime diffeomorphisms and the local
supersymmetry transformations with the bosonic and fermionic stabilizers
of this (Wess-Zumino-like) gauge respectively. But let us at first
have a look at the general transformation properties of the superfields.

\section{Supergauge transformations of the superfields}

\subsection{Infinitesimal form}

In the following, we make frequent use of some structure group connection
$\Omega_{MA}\hoch{B}$ and the corresponding covariant derivative
$\nabla_{M}$. As long as nothing else is announced, the equations
are valid for any connection (in particular, it is not meant to be
the left-moving connection only).

\paragraph{Transformation of a general tensor field}

We are interested in a combination of an infinitesimal superdiffeomorphism
transformation (or better the corresponding Lie derivative) and a
local structure group transformation. For an object with only curved
indices, the transformation reduces to the Lie derivative. The Lie
derivative of a vector field $\vecfull{v}\equiv v^{M}\pe_{M}$ e.g.
reads as usual \begin{eqnarray}
\Lie_{\vecfull{\xi}}v^{M} & \equiv & (\Lie_{\vecfull{\xi}}\vecfull{v})^{M}=\label{eq:importantPointComponents}\\
 & = & \xi^{K}\partial_{K}v^{M}-\partial_{K}\xi^{M}v^{K}\label{eq:LiederivativeVectorNoncov}\end{eqnarray}
It can be rewritten in terms of covariant derivatives as\index{Lie derivative!in terms of covariant derivatives}
\begin{eqnarray}
\Lie_{\vecfull{\xi}}v^{M} & = & \xi^{K}\nabla_{K}v^{M}-\nabla_{K}\xi^{M}v^{K}-2\xi^{K}T_{KL}\hoch{M}v^{L}\label{eq:LiederivativeVector}\end{eqnarray}
For one-forms the covariant expression of the Lie derivative contains
a torsion term with opposite sign:\begin{eqnarray}
\Lie_{\vecfull{\xi}}\omega_{M} & \equiv & \big(\Lie_{\vecfull{\xi}}(\omega_{N}\de x^{N})\big)_{M}\\
 & = & \xi^{K}\partial_{K}\omega_{M}+\partial_{M}\xi^{K}\omega_{K}=\label{eq:LiederivativeOneformNoncov}\\
 & = & \xi^{K}\nabla_{K}\omega_{M}+\nabla_{M}\xi^{K}\omega_{K}+2\xi^{K}T_{KM}\hoch{L}\omega_{L}\label{eq:LiederivativeOneform}\end{eqnarray}
In contrast to the above, it is convenient for objects with flat indices,
not to consider them as being contracted with basis elements, when
acting with the Lie derivative, but to really only act on the component
functions, which transform like scalars under diffeomorphisms%
\footnote{\label{foot:componentsOfLiederivative}\index{footnote!\thefoot. components of Lie derivative}Note
the (common) convention used in (\ref{eq:importantPointComponents})
to define $\Lie_{\vecfull{\xi}}v^{M}$ as the $M$-th component of
the Lie derivative of $\vecfull{v}$ and not the Lie derivative of
the $M$-th component function! This convention is extended to objects
with an arbitrary number of curved indices, i.e.\[
\Lie_{\vecfull{\xi}}t_{M_{1}\ldots M_{p}}^{N_{1}\ldots N_{q}}\equiv\left(\Lie_{\vecfull{\xi}}\big(t_{K_{1}\ldots K_{p}}^{L_{1}\ldots L_{q}}\de x^{K_{1}}\otimes\ldots\otimes\de x^{K_{p}}\otimes\pe_{L_{1}}\otimes\ldots\otimes\pe_{L_{q}}\big)\right)_{M_{1}\ldots M_{p}}^{N_{1}\ldots N_{q}}\]
In cases where we want to act explicitely on e.g. the component functions,
we can denote it with e.g. $\Lie_{\vecfull{\xi}}(v^{M})=\xi^{K}\partial_{K}v^{M}$.
This is of course not the component of a tensor, but it makes sense
in calculations like $\Lie_{\vecfull{\xi}}(v^{M}\pe_{M})=\Lie_{\vecfull{\xi}}(v^{M})\cdot\pe_{M}+v^{M}\Lie_{\vecfull{\xi}}(\pe_{M})$\frem{Formulierung!}.
From the Lie derivatives for general vectors (\ref{eq:LiederivativeVectorNoncov})
and one forms (\ref{eq:LiederivativeOneformNoncov}) we can in turn
read off the transformation of the basis elements\begin{eqnarray*}
\Lie_{\vecfull{\xi}}(\pe_{M}) & = & -\partial_{M}\xi^{N}\,\pe_{N}\\
\Lie_{\vecfull{\xi}}(\de x^{M}) & = & \partial_{N}\xi^{M}\,\de x^{N}\end{eqnarray*}
For flat indices, however, we use just the opposite convention, i.e.
we do not regard the flat index to be contracted with any basis element
when acting with the Lie derivative. The action on an object with
both, flat and curved indices will thus be defined as follows\begin{eqnarray*}
\Lie_{\vecfull{\xi}}t_{MA}^{NB} & \equiv & \left(\Lie_{\vecfull{\xi}}\big(t_{KA}^{LB}\de x^{K}\otimes\pe_{L}\big)\right)_{M}^{N}\end{eqnarray*}
In cases where we want to calculate something different we will use
a more explicit notation like on the righthand side in the above equation.
The reason for this convention is the following. Starting in a coordinate
basis, it is natural to express the transformed tensor in the coordinate
basis again, while if one starts in a non-coordinate frame $e_{A}$,
it is more natural to express the result in the transformed basis:\begin{eqnarray*}
\tilde{v} & \equiv & v+\Lie_{\vecfull{\xi}}v=v+\Lie_{\vecfull{\xi}}v^{A}\cdot e_{A}+v^{A}\Lie_{\vecfull{\xi}}e_{A}=\left(v^{A}+\Lie_{\vecfull{\xi}}v^{A}\right)\underbrace{\left(e_{A}+\Lie_{\vecfull{\xi}}e_{A}\right)}_{\tilde{e}_{A}}\end{eqnarray*}
\frem{\begin{eqnarray*}
\Lie_{\vecfull{\xi}}(v^{M}\pe_{M}) & = & \Lie_{\vecfull{\xi}}v^{M}\cdot\pe_{M}=\Lie_{\vecfull{\xi}}(v^{M})\cdot\pe_{M}+v^{M}\cdot\Lie_{\vecfull{\xi}}(\pe_{M})\\
\Lie_{\vecfull{\xi}}(v^{A}e_{A}) & = & \Lie_{\vecfull{\xi}}v^{A}\cdot e_{A}+v^{A}\cdot\underbrace{\Lie_{\vecfull{\xi}}e_{A}}_{\Lie_{\vecfull{\xi}}e_{A}\hoch{M}\cdot\pe_{M}}\stackrel{!}{=}\Lie_{\vecfull{\xi}}v^{M}\cdot\pe_{M}\\
\Lie_{\vecfull{\xi}}v^{A} & = & \left(\Lie_{\vecfull{\xi}}v^{M}-v^{B}\cdot\Lie_{\vecfull{\xi}}e_{B}\hoch{M}\right)e_{M}\hoch{A}\end{eqnarray*}
}

Let us finally give the Lie derivative of the local vielbein and its
inverse (using (\ref{eq:LiederivativeVector}) and (\ref{eq:LiederivativeOneform}))
which will also be discussed in the equations (\ref{eq:vielbeinTrafoNoncov})
and following:\begin{eqnarray*}
\Lie_{\vecfull{\xi}}(E_{A}) & = & \left(\xi^{K}\Omega_{KA}\hoch{B}-\nabla_{A}\xi^{B}-2\xi^{K}T_{KA}\hoch{B}\right)E_{B}\\
\Lie_{\vecfull{\xi}}(E^{A}) & = & \left(-\xi^{K}\Omega_{KB}\hoch{A}+\nabla_{B}\xi^{A}+2\xi^{K}T_{KB}\hoch{A}\right)E^{B}\qquad\fussend\end{eqnarray*}
}.\begin{eqnarray}
\Lie_{\vecfull{\xi}}v^{A} & = & \xi^{K}\partial_{K}v^{A}=\label{eq:LieVonVnoncov}\\
 & = & \xi^{K}\nabla_{K}v^{A}-\xi^{K}\Omega_{KB}\hoch{A}v^{B}\label{eq:LieVonVnoncovII}\end{eqnarray}
This is a covariant object from the diffeomorphism point of view,
but the connection transforms inhomogenously under the structure group
transformations. The entire gauge transformation of $v^{A}$, however,
contains also a local structure group transformation:\index{$L$@$\tilde{L}_{B}\hoch{A}$}
\begin{equation}
\delta v^{A}=\Lie_{\vecfull{\xi}}v^{A}+\tilde{L}_{B}\hoch{A}v^{B}\end{equation}
 As the structure group connection itself is Lie algebra valued, the
second term in (\ref{eq:LieVonVnoncovII}) can be absorbed in the
structure group transformation:\index{$L_{B}\hoch{A}$} \begin{equation}
L_{B}\hoch{A}\equiv\tilde{L}_{B}\hoch{A}-\xi^{K}\Omega_{KB}\hoch{A}\label{eq:tildeL}\end{equation}
 The combined diffeomorphism and local structure group transformation
can thus be written as \begin{eqnarray}
\delta v^{A} & = & \xi^{K}\nabla_{K}v^{A}+L_{B}\hoch{A}v^{B}\end{eqnarray}
The first term is a covariantized (w.r.t. the structure group) version
of the Lie derivative\index{covariantized Lie derivative|seep{supergauge transformation}}\index{Lie derivative!covariantized $\sim$|seep{supergauge transformation}}\index{$Liecov$@$\Liecov_{\vecfull{\xi}}$}
(\ref{eq:LieVonVnoncov}), and we will therefore denote it by \begin{eqnarray}
\Liecov_{\vecfull{\xi}}v^{A} & \equiv & \xi^{K}\nabla_{K}v^{A}\end{eqnarray}
In general $\Liecov_{\vecfull{\xi}}$ will be defined as the $L_{A}\hoch{B}=0$
part of the complete transformation, i.e. a Lie derivative w.r.t.
$\vecfull{\xi}$, accompanied by a structure group transformation
with $\tilde{L}_{A}\hoch{B}=\xi^{K}\Omega_{KA}\hoch{B}$ whose representation\index{representation!of the structure group: $\mc{R}$}
we denote with $\group{\tilde{L}_{\,\cdot}\hoch{\cdot}}$\index{$R(L)$@$\mc{R}(L_\cdot\hoch{\cdot})$}
(see also before (\ref{eq:covDerWithRepresentation}) on page \pageref{eq:covDerWithRepresentation}):\begin{equation}
\boxed{\Liecov_{\vecfull{\xi}}\equiv\Lie_{\vecfull{\xi}}+\group{\xi^{K}\Omega_{K\cdot}\hoch{\cdot}}}\end{equation}
On one-forms we thus have $\Liecov_{\vecfull{\xi}}\omega_{A}\equiv\xi^{K}\nabla_{K}\omega_{A}$,
while on objects with curved index the structure group transformation
has no effect and the covariantized Lie derivative reduces to the
ordinary Lie derivative. When acting on a more general tensor with
curved and flat indices, $\Liecov_{\vecfull{\xi}}$ thus takes the
following form:\\
\Ram{0.85}{ \begin{eqnarray}
\Liecov_{\vecfull{\xi}}t_{MA}^{NB} & = & \xi^{K}\partial_{K}t_{MA}^{NB}-\partial_{K}\xi^{N}t_{MA}^{KB}+\partial_{M}\xi^{K}t_{MA}^{NB}+\xi^{K}\Omega_{KC}\hoch{B}t_{MA}^{NC}-\xi^{K}\Omega_{KA}\hoch{C}t_{MC}^{NB}=\label{eq:covLiederOnGenTensorNoncov}\\
 & = & \xi^{K}\nabla_{K}t_{MA}^{NB}-\left(\nabla_{L}\xi^{N}+2\xi^{K}T_{KL}\hoch{N}\right)t_{MA}^{LB}+\left(\nabla_{M}\xi^{L}+2\xi^{K}T_{KM}\hoch{L}\right)t_{LA}^{NB}\label{eq:covLiederOnGenTensor}\end{eqnarray}
}\\
This transformation is usually called a \textbf{supergauge transformation}\index{supergauge transformation|fett}\index{supergauge transformation!supervielbein}
\cite[chapter XVI]{Wess:1992cp}. As it reduces for curved indices
to the ordinary Lie derivative, its action on tensor components (given
above) is determined by the Lie derivative, the Leibniz rule and the
transformation of the supervielbein. In addition the transformation
of the structure group connection will be of interest, as it transforms
inhomogenously under the structure group transformation. For completeness
(even if the given information will be a bit redundant), let us write
down explicitely the transformations (supergauge + structure group)
for all the type II supergravity superfields of our interest:

\paragraph{Supervielbein\index{supervielbein!supergauge transformation}}

A general infinitesimal gauge transformation (a Lie derivative corresponding
to a superdiffeomorphism plus a local structure group transformation)
of the supervielbein $E_{M}\hoch{A}$ looks as follows:\begin{eqnarray}
\delta E_{M}\hoch{A} & = & \xi^{K}\partial_{K}E_{M}\hoch{A}+\partial_{M}\xi^{K}E_{K}\hoch{A}+E_{M}\hoch{B}\tilde{L}_{B}\hoch{A}\label{eq:vielbeinTrafoNoncov}\end{eqnarray}
Redefining the local structure group transformation parameter, this
can be written in terms of covariant derivatives\begin{eqnarray}
\delta E_{M}\hoch{A} & = & \xi^{K}\underbrace{\nabla_{K}E_{M}\hoch{A}}_{0}+\nabla_{M}\xi^{K}E_{K}\hoch{A}+\xi^{K}\underbrace{\left(\Gamma_{KM}\hoch{L}-\Gamma_{MK}\hoch{L}\right)E_{L}\hoch{A}}_{2T_{KM}\hoch{A}}+E_{M}\hoch{B}\underbrace{\left(\tilde{L}_{B}\hoch{A}-\xi^{K}\Omega_{KB}\hoch{A}\right)}_{L_{B}\hoch{A}}=\label{eq:tildeLuL}\\
 & = & \underbrace{\nabla_{M}\xi^{A}+2\xi^{C}T_{CM}\hoch{A}}_{\equiv\Liecov_{\vecfull{\xi}}E_{M}\hoch{A}}+L_{B}\hoch{A}E_{M}\hoch{B}\label{eq:supergauge-vielbein}\end{eqnarray}
For some purposes, also the explicit form with partial derivatives
(but in the new parametrization) will be useful:\begin{equation}
\boxed{\delta E_{M}\hoch{A}=\underbrace{\overbrace{\partial_{M}\xi^{A}+\Omega_{MC}\hoch{A}\xi^{C}}^{\nabla_{M}\xi^{A}}+2\xi^{C}T_{CM}\hoch{A}}_{\Liecov_{\vecfull{\xi}}E_{M}\hoch{A}}+\underbrace{L_{B}\hoch{A}E_{M}\hoch{B}}_{\group{L}E_{M}\hoch{A}}}\label{eq:vielbeinTrafo}\end{equation}
For the inverse vielbein we get likewise (or via $\delta E^{-1}=-E^{-1}\delta E\cdot E^{-1}$)\begin{eqnarray}
\delta E_{A}\hoch{M} & = & \xi^{K}\partial_{K}E_{A}\hoch{M}-\partial_{K}\xi^{M}E_{A}\hoch{K}-\tilde{L}_{A}\hoch{B}E_{B}\hoch{M}\label{eq:inversevielbeintrafoNoncov}\\
\textrm{or }\delta E_{A}\hoch{M} & = & \underbrace{-\nabla_{A}\xi^{M}-2\xi^{C}T_{CA}\hoch{M}}_{\Liecov_{\vecfull{\xi}}E_{A}\hoch{M}}-L_{B}\hoch{A}E_{A}\hoch{N}\label{eq:inversevielbeintrafo}\end{eqnarray}

\paragraph{The structure group connection\index{connection!supergauge transformation} }

transforms tensorial with respect to the superdiffeomorphisms but
of course not like a tensor (but inhomogenous) with respect to the
structure group transformation.%
\footnote{\label{foot:structuregroupTrafoConnection}\index{footnote!\thefoot. transformation of $\Omega_{MA}\hoch{B}$}\index{connection!structure group transformation}\index{transformation!of the connection under the structure group}Let
us quickly rederive the correct structure group transformation of
the connection via the transformation property of the covariant derivative:\begin{eqnarray*}
\delta_{(L)}v^{A} & = & v^{B}L_{B}\hoch{A}\\
\delta_{(L)}\nabla_{M}v^{A} & = & \delta_{(L)}\left(\partial_{M}v^{A}+\Omega_{MB}\hoch{A}v^{B}\right)=\\
 & = & \partial_{M}\left(v^{B}L_{B}\hoch{A}\right)+\delta_{L}\Omega_{MB}\hoch{A}v^{B}+\Omega_{MB}\hoch{A}\delta_{L}v^{B}=\\
 & = & \partial_{M}v^{B}\cdot L_{B}\hoch{A}+v^{B}\partial_{M}L_{B}\hoch{A}+\delta_{L}\Omega_{MB}\hoch{A}v^{B}+\Omega_{MB}\hoch{A}v^{C}L_{C}\hoch{B}=\\
 & = & \left(\partial_{M}v^{B}+\Omega_{MC}\hoch{B}v^{C}\right)\cdot L_{B}\hoch{A}+v^{C}\left(\partial_{M}L_{C}\hoch{A}+\delta_{L}\Omega_{MC}\hoch{A}+L_{C}\hoch{B}\Omega_{MB}\hoch{A}-\Omega_{MC}\hoch{B}L_{B}\hoch{A}\right)\end{eqnarray*}
For $\nabla_{M}v^{A}$ to transform covariantly, we need to have \begin{eqnarray*}
\delta_{(L)}\Omega_{MC}\hoch{A} & = & -\partial_{M}L_{C}\hoch{A}\underbrace{-L_{C}\hoch{B}\Omega_{MB}\hoch{A}+\Omega_{MC}\hoch{B}L_{B}\hoch{A}}_{\equiv-[L,\Omega_{M}]_{C}\hoch{A}}=\\
 & = & -\nabla_{M}L_{C}\hoch{A}\qquad\fussend\end{eqnarray*}
}\begin{eqnarray}
\delta\Omega_{MA}\hoch{B} & = & \xi^{K}\partial_{K}\Omega_{MA}\hoch{B}+\partial_{M}\xi^{K}\Omega_{KA}\hoch{B}-\partial_{M}\underbrace{\tilde{L}_{A}\hoch{B}}_{\lqn{{\scriptstyle L_{A}\hoch{B}+\xi^{K}\Omega_{KA}\hoch{B}}}}-[\tilde{L},\Omega_{M}]_{A}\hoch{B}=\\
 & = & \xi^{K}\partial_{K}\Omega_{MA}\hoch{B}+\partial_{M}\xi^{K}\Omega_{KA}\hoch{B}-\partial_{M}L_{A}\hoch{B}-\partial_{M}\xi^{K}\Omega_{KA}\hoch{B}-\xi^{K}\partial_{M}\Omega_{KA}\hoch{B}+\nonumber \\
 &  & -[L+\xi^{K}\Omega_{K},\Omega_{M}]_{A}\hoch{B}=\\
 & = & 2\xi^{K}\partial_{[K}\Omega_{M]A}\hoch{B}-\xi^{K}[\Omega_{K},\Omega_{M}]_{A}\hoch{B}-\partial_{M}L_{A}\hoch{B}-[L,\Omega_{M}]_{A}\hoch{B}\end{eqnarray}
$\dann$\index{supergauge transformation!connection}\begin{equation}
\boxed{\delta\Omega_{MA}\hoch{B}=\underbrace{2\xi^{K}R_{KMA}\hoch{B}}_{\Liecov_{\vecfull{\xi}}\Omega_{MA}\hoch{B}}\underbrace{-\partial_{M}L_{A}\hoch{B}-[L,\Omega_{M}]_{A}\hoch{B}}_{\quad-\nabla_{M}L_{A}\hoch{B}=\group{L}\Omega_{MA}\hoch{B}}}\label{eq:connectionTrafo}\end{equation}

\paragraph{The scale connection\index{scale connection!supergauge transformation}}

The above transformation of the connection is valid for a general
one. In our application to the Berkovits-string, however, the structure
group on the supermanifold is restricted as follows. Firstly, the
connection is block-diagonal. Secondly, each block decays into Lorentz-
plus scale transformation. Finally, the blocks are not independent
in the end, but let us assume for the moment, that they are. Then
we have three scale connections, namely the trace of each block respectively.
In detail we have for the {}``mixed connection'' (see appendix \ref{cha:ConnectionAppend})
\begin{eqnarray}
\gemOm_{MA}\hoch{B} & = & \left(\begin{array}{ccc}
\check{\Omega}_{Ma}\hoch{b} & 0 & 0\\
0 & \Omega_{M\bs{\alpha}}\hoch{\bs{\beta}} & 0\\
0 & 0 & \hat{\Omega}_{M\hat{\bs{\alpha}}}\hoch{\hat{\bs{\beta}}}\end{array}\right)=\\
 & = & \left(\begin{array}{ccc}
\check{\Omega}_{M}^{(D)}\delta_{a}^{b} & 0 & 0\\
0 & \frac{1}{2}\Omega_{M}^{(D)}\delta_{\bs{\alpha}}\hoch{\bs{\beta}} & 0\\
0 & 0 & \frac{1}{2}\hat{\Omega}_{M}^{(D)}\delta_{\hat{\bs{\alpha}}}\hoch{\hat{\bs{\beta}}}\end{array}\right)+\left(\begin{array}{ccc}
\check{\Omega}_{Ma}^{(L)}\hoch{b} & 0 & 0\\
0 & \frac{1}{4}\Omega_{Mab}^{(L)}\gamma^{ab}\tief{\bs{\alpha}}\hoch{\bs{\beta}} & 0\\
0 & 0 & \frac{1}{4}\hat{\Omega}_{Mab}^{(L)}\gamma^{ab}\tief{\hat{\bs{\alpha}}}\hoch{\hat{\bs{\beta}}}\end{array}\right)\label{eq:gemOm}\\
\gemR_{MNA}\hoch{B} & = & \left(\begin{array}{ccc}
\check{F}_{MN}^{(D)}\delta_{a}^{b} & 0 & 0\\
0 & \frac{1}{2}F_{MN}^{(D)}\delta_{\bs{\alpha}}\hoch{\bs{\beta}} & 0\\
0 & 0 & \frac{1}{2}\hat{F}_{MN}^{(D)}\delta_{\hat{\bs{\alpha}}}\hoch{\hat{\bs{\beta}}}\end{array}\right)+\left(\begin{array}{ccc}
\check{R}_{MNa}^{(L)}\hoch{b} & 0 & 0\\
0 & \frac{1}{4}R_{MNab}^{(L)}\gamma^{ab}\tief{\bs{\alpha}}\hoch{\bs{\beta}} & 0\\
0 & 0 & \frac{1}{4}\hat{R}_{MNab}^{(L)}\gamma^{ab}\tief{\hat{\bs{\alpha}}}\hoch{\hat{\bs{\beta}}}\end{array}\right)\qquad\end{eqnarray}
 The scale connection (or dilatation\index{dilatation connection|see{trace connection}}
connection) simply transforms as \index{supergauge transformation!scale connection}\begin{eqnarray}
\delta\Omega_{M}^{(D)} & = & \xi^{K}\partial_{K}\Omega_{M}^{(D)}+\partial_{M}\xi^{K}\Omega_{K}^{(D)}-\partial_{M}\tilde{L}^{(D)},\qquad\delta\hat{\Omega}_{M}^{(D)}=\xi^{K}\partial_{K}\hat{\Omega}_{M}^{(D)}+\partial_{M}\xi^{K}\hat{\Omega}_{K}^{(D)}-\partial_{M}\tilde{\hat{L}}^{(D)}\quad\label{eq:scaleConnectionTrafoTilde}\\
\lqn{\Ramm{.32}{\zwek{}{}}}\quad\delta\Omega_{M}^{(D)} & = & 2\xi^{K}F_{KM}^{(D)}-\partial_{M}L^{(D)}\quad,\qquad\qquad\qquad\qquad\;\;\delta\hat{\Omega}_{M}^{(D)}=2\xi^{K}\hat{F}_{KM}^{(D)}-\partial_{M}\hat{L}^{(D)}\label{eq:scaleConnectionTrafo}\\
\textrm{with }F_{KM}^{(D)} & = & \partial_{[K}\Omega_{M]},\qquad\qquad\qquad\qquad\qquad\qquad\quad\;\;\;\hat{F}_{KM}^{(D)}\;=\partial_{[K}\hat{\Omega}_{M]}\end{eqnarray}
We also could have started with the pure left-mover connection $\Omega_{MA}\hoch{B}=\diag(\Omega_{Ma}\hoch{b},\Omega_{M\bs{\alpha}}\hoch{\bs{\beta}},\Omega_{M\hat{\bs{\alpha}}}\hoch{\hat{\bs{\beta}}})$
to derive $\delta\Omega_{M}^{(D)}$ or the pure right-mover connection
$\hat{\Omega}_{MA}\hoch{B}$ to derive $\delta\hat{\Omega}_{M}^{(D)}$.
We will now return to the notation of this appendix, where $\Omega_{MA}\hoch{B}$
is just a general connection, and not necessarily the left-mover one.

\paragraph{The superspace connection \index{connection!Lie derivative of superspace $\sim$}\index{Lie derivative!of superspace connection}}

We will not need the superspace connection $\Gamma_{MN}\hoch{K}$
as frequently as the structure group connection, but let us discuss
its transformation for completeness. As it is inert under structure
group transformations, the supergauge transformation reduces to the
Lie derivative. Remember the relation \begin{eqnarray}
\Gamma_{MN}\hoch{K} & = & \Omega_{MN}\hoch{K}+\partial_{M}E_{N}\hoch{A}\cdot E_{A}\hoch{K}\label{eq:superspaceConnInTermsOfSGC}\end{eqnarray}
which is a direct consequence of $\nabla_{M}E_{M}\hoch{A}=0$. The
Lie derivative of $\Gamma_{MN}\hoch{K}$ can thus be derived from
the Lie derivative (or alternatively from the supergauge transformation)
of the structure group transformation and the vielbein. Both, vielbein
and structure group transformation are tensorial with respect to diffeomorphisms
and thus the inhomogenity in the transformation of $\Gamma_{MN}\hoch{K}$
can only result from the inhomogenity of the Lie derivative of $\partial_{M}E_{N}\hoch{A}$,
which is (using commutativity of partial and Lie derivative%
\footnote{\index{footnote!\thefoot. commutation of Lie derivative and partial derivative}\label{foot:commutationOfLieAndPartial}For
a scalar field $\dil$, whose partial derivative becomes the component
of a vector field, it is quite obvious that partial and Lie derivative
commute:\begin{eqnarray*}
\Lie_{\vecfull{\xi}}\partial_{M}\dil & = & \xi^{K}\partial_{K}\partial_{M}\dil+\partial_{M}\xi^{K}\partial_{K}\dil=\partial_{M}(\xi^{K}\partial_{K}\dil)=\partial_{M}\Lie_{\vecfull{\xi}}\dil\end{eqnarray*}
For a nontensorial object like $\partial_{M}t_{M_{1}\ldots M_{p}}^{N_{1}\ldots N_{q}}$
(or also the connection) it is less clear whether it makes sense to
define a Lie derivative on it. However, it will be very convenient
to do so, and we will simply take the definition coming from infinitesimal
diffeomorphisms (with $x'=x+\xi$). Note that $\bei{\partial'_{M}{t'}_{M_{1}\ldots M_{p}}^{N_{1}\ldots N_{q}}(x')}{x'=x}=\partial_{M}{t'}_{M_{1}\ldots M_{p}}^{N_{1}\ldots N_{q}}(x)$,
which leads to\begin{eqnarray*}
\Lie_{\vecfull{\xi}}\partial_{M}t_{M_{1}\ldots M_{p}}^{N_{1}\ldots N_{q}}(x) & \equiv & \partial_{M}t_{M_{1}\ldots M_{p}}^{N_{1}\ldots N_{q}}(x)-\bei{\partial'_{M}{t'}_{M_{1}\ldots M_{p}}^{N_{1}\ldots N_{q}}(x')}{x'=x}=\partial_{M}(\Lie_{\vecfull{\xi}}t_{M_{1}\ldots M_{p}}^{N_{1}\ldots N_{q}}(x))\end{eqnarray*}

We can likewise extend the definition of $\Liecov_{\vecfull{\xi}}=\Lie_{\vecfull{\xi}}+\group{\xi^{K}\Omega_{K\,\cdot}\hoch{\cdot}}$
to nontensorial objects by defining e.g.\[
\group{L}\,\partial_{P}t_{MA}^{NB}\equiv\partial_{P}\left(\group{L}t_{MA}^{NB}\right)\]
The structure group transformation $\group{L}$ thus commutes with
the partial derivative by definition and we thus have the same property
for the covariantized Lie derivative \[
\Liecov_{\vecfull{\xi}}\partial_{P}t_{MA}^{NB}=\partial_{P}(\Liecov_{\vecfull{\xi}}t_{MA}^{NB})\]
Note that this is also consistent with a proper transformation property
of the covariant derivative\label{foot:transformationOfCovDer}: \begin{eqnarray*}
\Liecov_{\vecfull{\xi}}\nabla_{P}t_{MA}^{NB} & = & \Liecov_{\vecfull{\xi}}\left(\partial_{P}t_{MA}^{NB}+\Gamma_{PK}\hoch{N}t_{MA}^{KB}-\Gamma_{PM}\hoch{K}t_{KA}^{NB}+\group{\Omega_{P\,\cdot}\hoch{\cdot}}t_{MA}^{NB}\right)=\\
 & = & \partial_{P}\left(\Liecov_{\vecfull{\xi}}t_{MA}^{NB}\right)+\left(\Liecov_{\vecfull{\xi}}\Gamma_{PK}\hoch{N}\right)t_{MA}^{KB}+\Gamma_{PK}\hoch{N}\Liecov_{\vecfull{\xi}}t_{MA}^{KB}-\left(\Liecov_{\vecfull{\xi}}\Gamma_{PM}\hoch{K}\right)t_{KA}^{NB}-\Gamma_{PM}\hoch{K}\Liecov_{\vecfull{\xi}}t_{KA}^{NB}+\\
 &  & +\group{\Liecov_{\vecfull{\xi}}\Omega_{P\,\cdot}\hoch{\cdot}}t_{MA}^{NB}+\group{\Omega_{P\,\cdot}\hoch{\cdot}}\Liecov_{\vecfull{\xi}}t_{MA}^{NB}=\\
 & = & \nabla_{P}\left(\Liecov_{\vecfull{\xi}}t_{MA}^{NB}\right)+\left(\Lie_{\vecfull{\xi}}\Gamma_{PK}\hoch{N}\right)t_{MA}^{KB}-\left(\Lie_{\vecfull{\xi}}\Gamma_{PM}\hoch{K}\right)t_{KA}^{NB}+\group{\Liecov_{\vecfull{\xi}}\Omega_{P\,\cdot}\hoch{\cdot}}t_{MA}^{NB}=\\
 & = & \nabla_{P}\left(\xi^{K}\nabla_{K}t_{MA}^{NB}+\left(\nabla_{M}\xi^{K}+2\xi^{L}T_{LM}\hoch{K}\right)t_{KA}^{NB}-\left(\nabla_{K}\xi^{N}+2\xi^{L}T_{LK}\hoch{N}\right)t_{MA}^{KB}\right)+\\
 &  & +\left(2\xi^{L}R_{LPK}\hoch{N}+\nabla_{P}(\nabla_{K}\xi^{N}+2\xi^{L}T_{LK}\hoch{N})\right)t_{MA}^{KB}-\left(2\xi^{L}R_{LPM}\hoch{K}+\nabla_{P}(\nabla_{M}\xi^{K}+2\xi^{L}T_{LM}\hoch{K})\right)t_{KA}^{NB}+\\
 &  & +\group{2\xi^{L}R_{LP\,\cdot}\hoch{\cdot}}t_{MA}^{NB}=\\
 & = & \xi^{K}\underbrace{\nabla_{P}\nabla_{K}t_{MA}^{NB}}_{\nabla_{K}\nabla_{P}t_{MA}^{NB}-2T_{PK}\hoch{L}\nabla_{L}t_{MA}^{NB}\lqn{+2R_{PKL}\hoch{N}t_{MA}^{LB}-2R_{PKM}\hoch{L}t_{LA}^{NB}+\group{2R_{PK\,\cdot}\hoch{\cdot}}t_{MA}^{NB}}}+\left(\nabla_{M}\xi^{K}+2\xi^{L}T_{LM}\hoch{K}\right)\nabla_{P}t_{KA}^{NB}-\left(\nabla_{K}\xi^{N}+2\xi^{L}T_{LK}\hoch{N}\right)\nabla_{P}t_{MA}^{KB}+\\
 &  & +\nabla_{P}\xi^{K}\nabla_{K}t_{MA}^{NB}+\nabla_{P}\left(\nabla_{M}\xi^{K}+2\xi^{L}T_{LM}\hoch{K}\right)t_{KA}^{NB}-\nabla_{P}\left(\nabla_{K}\xi^{N}+2\xi^{L}T_{LK}\hoch{N}\right)t_{MA}^{KB}\\
 &  & +\left(2\xi^{L}R_{LPK}\hoch{N}+\nabla_{P}(\nabla_{K}\xi^{N}+2\xi^{L}T_{LK}\hoch{N})\right)t_{MA}^{KB}-\left(2\xi^{L}R_{LPM}\hoch{K}+\nabla_{P}(\nabla_{M}\xi^{K}+2\xi^{L}T_{LM}\hoch{K})\right)t_{KA}^{NB}+\\
 &  & +\group{2\xi^{L}R_{LP\,\cdot}\hoch{\cdot}}t_{MA}^{NB}=\\
 & = & \xi^{K}\nabla_{K}\nabla_{P}t_{MA}^{NB}+\left(\nabla_{P}\xi^{K}+2\xi^{L}T_{LP}\hoch{K}\right)\nabla_{K}t_{MA}^{NB}+\left(\nabla_{M}\xi^{K}+2\xi^{L}T_{LM}\hoch{K}\right)\nabla_{P}t_{KA}^{NB}-\left(\nabla_{K}\xi^{N}+2\xi^{L}T_{LK}\hoch{N}\right)\nabla_{P}t_{MA}^{KB}\qquad\fussend\end{eqnarray*}
}) $\partial_{M}\partial_{N}\xi^{L}\, E_{L}\hoch{A}$. The Lie derivative
of the connection thus reads\begin{eqnarray}
\Lie_{\vecfull{\xi}}\Gamma_{MN}\hoch{K} & = & \xi^{L}\partial_{L}\Gamma_{MN}\hoch{K}+\partial_{M}\xi^{L}\Gamma_{LN}\hoch{K}+\underbrace{\partial_{N}\xi^{L}\Gamma_{ML}\hoch{K}-\partial_{L}\xi^{K}\Gamma_{MN}\hoch{L}+\partial_{M}\partial_{N}\xi^{K}}_{[\partial\xi,\Gamma_{M}]_{N}\hoch{L}+\partial_{M}(\partial\xi)_{N}\hoch{K}}\label{eq:LietrafosuperspaceconnNoncov}\end{eqnarray}
The first two terms are just the Lie derivative of a matrix valued
one form $\de x^{M}\Gamma_{MN}\hoch{K}$, while the last three terms
are the usual inhomogenous transformation of a structure group connection
(compare (\ref{eq:connectionTrafo})), here with the Gl(n)-matrix
$\tilde{M}_{N}\hoch{K}\equiv-\partial_{N}\xi^{K}$. The same transformation
can be derived by comparing e.g. the tensorial transformation of $\Lie_{\vecfull{\xi}}\nabla_{M}v^{K}$
on the one side with $\partial_{M}(\Lie_{\vecfull{\xi}}v^{K})+\Lie_{\vecfull{\xi}}\Gamma_{MN}\hoch{K}\cdot v^{N}+\Gamma_{MN}\hoch{K}\Lie_{\vecfull{\xi}}v^{N}$
on the other side (using again that Lie and partial derivative commute).
The Lie derivative of the connection is in some sense the difference
of two connections and is therefore a tensor. This can be seen by
expressing the partial derivatives on $\xi^{M}$ in terms of covariant
ones and discover that the remaining connection terms combine to curvature
and torsion.\rem{calculation hidden in Note}%
\footnote{\index{footnote!\thefoot. Lie derivative of connection}Alternatively
we can derive the same result, starting from (\ref{eq:superspaceConnInTermsOfSGC})
\begin{eqnarray*}
\Lie_{\vecfull{\xi}}\Gamma_{MN}\hoch{K} & = & \Liecov_{\vecfull{\xi}}(\Omega_{MA}\hoch{B}E_{N}\hoch{A}E_{B}\hoch{K})+\partial_{M}(\Liecov_{\vecfull{\xi}}E_{N}\hoch{A})\cdot E_{A}\hoch{K}+\partial_{M}E_{N}\hoch{A}\cdot\Liecov_{\vecfull{\xi}}E_{A}\hoch{K}\end{eqnarray*}
Using the covariant expressions of the supergauge transformation of
$\Omega_{MA}\hoch{B}$ and $E_{M}\hoch{A}$ then leads to (\ref{eq:Lietrafosuperspaceconn}).$\qquad\fussend$%
}\begin{equation}
\boxed{\Lie_{\vecfull{\xi}}\Gamma_{MN}\hoch{K}=2\xi^{L}R_{LMN}\hoch{K}+\nabla_{M}\underbrace{\left(\nabla_{N}\xi^{K}+2\xi^{L}T_{LN}\hoch{K}\right)}_{\equiv-M_{N}\hoch{K}}}\label{eq:Lietrafosuperspaceconn}\end{equation}
Remember that above we have seen the Lie derivative of the superspace
connection as a combination of a Lie derivative on its form index
(the first lower index) plus a Gl(n) structure group transformation
with transformation matrix $\tilde{M}_{N}\hoch{K}\equiv-\partial_{N}\xi^{K}$.
Equivalently it can be seen as a combination of a supergauge transformation
(regarding only the first index as curved one) plus a modified Gl(n)
transformation with the matrix (compare (\ref{eq:tildeL}))\begin{eqnarray}
M_{N}\hoch{K} & \equiv & -\partial_{N}\xi^{K}-\xi^{P}\Gamma_{PN}\hoch{K}=\\
 & = & -\nabla_{N}\xi^{K}-2\xi^{P}T_{PN}\hoch{K}\end{eqnarray}
Indeed the above Lie transformation can be written as \begin{eqnarray}
\Lie_{\vecfull{\xi}}\Gamma_{MN}\hoch{K} & = & 2\xi^{L}R_{LMN}\hoch{K}\underbrace{-\partial_{M}M_{N}\hoch{K}-[M,\Gamma_{M}]_{N}\hoch{K}}_{=-\nabla_{M}M_{N}\hoch{K}}\end{eqnarray}
which perfectly agrees with the form of a gauge transformation of
a structure group connection given in (\ref{eq:connectionTrafo}). 

Let us finally note that \begin{eqnarray}
[\Lie_{\vecfull{\xi}},\nabla_{M}]v^{K} & = & (\Lie_{\vecfull{\xi}}\Gamma_{MN}\hoch{K})v^{N}\end{eqnarray}
which provides another way to calculate the Lie derivative of the
connection. For the Levi Civita connection this equation implies that
the Lie derivative commutes with the covariant derivative, if $\vecfull{\xi}$
is a killing vector.\rem{more carefully in next section?} \rem{Hier in Note sind Trafos der einzelnen Felder (u.a. Compensator field), die in den Hauptteil gehoeren}

\paragraph{Tensorial superfields}

Usually, all additional fields present in a supergravity theory (like
$B$-field, RR-fields or dilaton) are contained in superfields that
transform homogenously (tensorial) under supergauge transformations
and structure group transformations. The gauge transformation of a
tensor field with index structure $t_{MA}^{NB}$ transforms as \begin{equation}
\delta t_{MA}^{NB}=\Liecov_{\vecfull{\xi}}t_{MA}^{NB}+\underbrace{L_{C}\hoch{B}t_{MA}^{NC}-L_{A}\hoch{C}t_{MC}^{NB}}_{\group{L_{\,\cdot}\hoch{\cdot}}t_{MA}^{NB}}\label{eq:generalTensorGaugeTrafo}\end{equation}
where $\Liecov_{\vecfull{\xi}}$ was given in (\ref{eq:covLiederOnGenTensor}).
The above transformation is of course also valid for scalar fields
where simply the structure group transformation vanishes. If a $B$-field
(a two form, i.e. an antisymmetric rank two tensor) is present, its
general gauge transformation contains in addition the one-form gauge
transformation $B\To B+\de\Lambda$ which will briefly be discussed
in a separate section at a later point\rem{cross-reference}. Another
example of a tensorial superfield in our application to the Berkovits
string is the bispinor-superfield $\RR^{\bs{\alpha}\hat{\bs{\beta}}}$
which contains the RR-fields in the leading component in the $\xbothtetas$-expansion.
In order to act with the structure group transformation $L_{A}\hoch{B}$
(appearing in the general transformation (\ref{eq:generalTensorGaugeTrafo}))
on the bispinor indices, we need $L_{A}\hoch{B}$ to be block diagonal.
This is described in the main part (see (\ref{eq:BlockDiagLambda})).
A final remark about our application in the main part is about the
appearance of a compensator field $\Phi$ which does not transform
homogenously under the structure group, but via a shift (see discussion
below (\ref{eq:metricWithCompensator})).

\subsection{Algebra of Lie derivatives and supergauge transformations}

\subsubsection{Commutator of Lie derivatives}

The SUSY algebra on scalar fields and tensors with curved indices
should be entirely implemented in the superdiffeomorphisms (independent
from any accompanying local structure group transformation which appeared
above). The commutator of two diffeomorphisms yields the vector Lie
bracket of the transformation parameters\begin{eqnarray}
\left[\mc{L}_{\vecfull{\xi}_{1}},\mc{L}_{\vecfull{\xi}_{2}}\right] & = & \mc{L}_{[\vecfull{\xi}_{1},\vecfull{\xi}_{2}]}\end{eqnarray}
where the vector Lie bracket reads \begin{eqnarray}
\left[\vecfull{\xi}_{1},\vecfull{\xi}_{2}\right]^{M} & = & \xi_{1}^{K}\partial_{K}\xi_{2}^{M}-\xi_{2}^{K}\partial_{K}\xi_{1}^{M}=\\
 & = & \xi_{1}^{K}\nabla_{K}\xi_{2}^{M}-\xi_{2}^{K}\nabla_{K}\xi_{1}^{M}-2\xi_{1}^{K}T_{KL}\hoch{M}\xi_{2}^{L}\end{eqnarray}
If we plug in the local basis elements $\vecfull{E_{A}}\equiv E_{A}\hoch{M}\pe_{M}$
in place of $\xi_{1/2}$, the above equation only holds, if the covariant
derivative acts only on the curved index. The covariant derivatives
do not vanish when we act on the curved index of $E_{A}\hoch{M}$
only. We thus do not only get the torsion term, as one would naively
expect, but instead\begin{eqnarray}
[\vecfull{E_{A}},\vecfull{E_{B}}] & = & \left(2\Omega_{[AB]}\hoch{C}-2T_{AB}\hoch{C}\right)\vecfull{E_{C}}=\\
 & = & -2(\de E^{C})_{AB}\vecfull{E_{C}}\end{eqnarray}
  For objects with flat indices it is thus convenient to extend the
Lie derivative to the supergauge transformation, which is covariantized
with respect to the structure group.

\subsubsection{Algebra of covariant Lie derivative and structure group action}

\label{sub:algOfLieUsg}Let us restrict our considerations for a moment
to a structure group vector $v^{A}$. We first want to study the commutator
of two covariantized Lie derivatives. 

\begin{eqnarray}
[\Liecov_{\vecfull{\xi}},\Liecov_{\vecfull{\eta}}]v^{A} & = & \xi^{L}\nabla_{L}\left(\eta^{K}\nabla_{K}v^{A}\right)-(\xi\leftrightarrow\eta)=\\
 & = & \left(\xi^{L}\nabla_{L}\eta^{K}-\eta^{L}\nabla_{L}\xi^{K}\right)\nabla_{K}v^{A}+\xi^{L}\eta^{K}\left[\nabla_{L},\nabla_{K}\right]v^{A}=\\
 & = & \left(\xi^{L}\nabla_{L}\eta^{K}-\eta^{L}\nabla_{L}\xi^{K}-2\xi^{L}T_{LP}\hoch{K}\eta^{P}\right)\nabla_{K}v^{A}+2\xi^{L}\eta^{K}R_{LKB}\hoch{A}v^{B}=\\
 & = & \Liecov_{[\vecfull{\xi},\vecfull{\eta}]}v^{A}+2\xi^{L}\eta^{K}R_{LKB}\hoch{A}v^{B}\end{eqnarray}
For a one form we arrive likewise at \begin{eqnarray}
[\Liecov_{\vecfull{\xi}},\Liecov_{\vecfull{\eta}}]\omega_{A} & = & \Liecov_{[\vecfull{\xi},\vecfull{\eta}]}\omega_{A}-2\xi^{L}\eta^{K}R_{LKA}\hoch{B}\omega_{B}\end{eqnarray}
On curved indices, however, the super gauge transformation reduces
to the Lie derivative\begin{eqnarray}
\left[\Liecov_{\vecfull{\xi}},\Liecov_{\vecfull{\eta}}\right]v^{M} & = & \left[\Lie_{\vecfull{\xi}},\Lie_{\vecfull{\eta}}\right]v^{M}=\Lie_{[\vecfull{\xi},\vecfull{\eta}]}v^{M}=\Liecov_{[\vecfull{\xi},\vecfull{\eta}]}v^{M}\\
\left[\Liecov_{\vecfull{\xi}},\Liecov_{\vecfull{\eta}}\right]\omega_{M} & = & \Liecov_{[\vecfull{\xi},\vecfull{\eta}]}\omega_{M}\end{eqnarray}
On a more general tensor $t_{MA}^{NB}$ we therefore have the following
commutator of supergauge transformations (remember footnote \ref{foot:componentsOfLiederivative})\begin{equation}
\boxed{\left[\Liecov_{\vecfull{\xi}},\Liecov_{\vecfull{\eta}}\right]t_{MA}^{NB}=\Liecov_{[\vecfull{\xi},\vecfull{\eta}]}t_{MA}^{NB}+\underbrace{2\xi^{K}\eta^{L}R_{KLC}\hoch{B}t_{MA}^{NC}-2\xi^{K}\eta^{L}R_{KLA}\hoch{C}t_{MC}^{NB}}_{\group{-\ip_{\vecfull{\xi}}\ip_{\vecfull{\eta}}(R_{\cdot}\hoch{\cdot})}t_{MA}^{NB}}}\end{equation}
In particular we have for supergauge transformations along the coordinate
basis\begin{equation}
[\Liecov_{\pe_{K}},\Liecov_{\pe_{L}}]t_{MA}^{NB}=2R_{KLC}\hoch{B}t_{MA}^{NC}-2R_{KLA}\hoch{C}t_{MC}^{NB}=\group{-\ip_{\pe_{K}}\ip_{\pe_{L}}(R_{C}\hoch{D})}t_{MA}^{NB}\end{equation}
The algebra of two infinitesimal structure group transformations is
rather simple%
\footnote{\index{footnote!\thefoot. minus sign in structure group algebra}The
minus sign comes from our definition how the structure group matrix
acts on vectors and forms. E.g. on a vector we have $\group{L_{1}}\group{L_{2}}v^{A}=\group{L_{1}}(L_{2\, B}\hoch{A}v^{B})=L_{1\, C}\hoch{A}L_{2\, B}\hoch{C}v^{B}=(L_{2}L_{1})_{B}\hoch{A}v^{B}=\group{L_{2}L_{1}}v^{A}$$\dann[\group{L_{1}},\group{L_{2}}]v^{A}=-\group{[L_{1},L_{2}]}v^{A}$.
Similarly for one forms $\group{L_{1}}\group{L_{2}}\omega_{A}=\group{L_{1}}(-L_{2\, A}\hoch{B}\omega_{B})=L_{1\, A}\hoch{C}L_{2\, C}\hoch{B}\omega_{B}=(L_{1}L_{2})_{A}\hoch{B}\omega_{B}=-\group{L_{1}L_{2}}\omega_{A}$$\dann[\group{L_{1}},\group{L_{2}}]\omega_{A}=-\group{[L_{1},L_{2}]}\omega_{A}$.
If one prefers, one can get rid of the minus sign by either redefining
the action of $\group{L}$ with a minus sign or with a transposed
$L$ (not only for antisymmetric $L$). This is because $[L_{1}^{T},L_{2}^{T}]^{T}=-[L_{1},L_{2}]$
and $-[-L_{1},-L_{2}]=-[L_{1},L_{2}].\qquad\fussend$%
}\begin{equation}
\boxed{\left[\group{L_{1}},\group{L_{2}}\right]=-\group{[L_{1},L_{2}]}}\end{equation}
The commutator between supergauge transformation and structure group
transformation finally reads\begin{equation}
\boxed{\left[\Liecov_{\vecfull{\xi}},\group{L}\right]=\group{(\Liecov_{\vecfull{\xi}}L)}}\end{equation}
which is easily checked by acting e.g. on a vector $v^{A}$. The complete
algebra can be written in one single equation as \begin{equation}
\hspace{-.5cm}\boxed{\left[\Liecov_{\vecfull{\xi}}+\group{L_{1}}\,,\,\Liecov_{\vecfull{\eta}}+\group{L_{2}}\right]=\Liecov_{[\vecfull{\xi},\vecfull{\eta}]}+\group{2\xi^{K}\eta^{L}R_{KL\cdot}\hoch{\cdot}+\Liecov_{\vecfull{\xi}}L_{2\,\cdot}\hoch{\cdot}-\Liecov_{\vecfull{\eta}}L_{1\,\cdot}\hoch{\cdot}-[L_{1},L_{2}]_{\cdot}\hoch{\cdot}\Big)}}\label{eq:LiecovAndStructureComm}\end{equation}

\subsubsection{Commutator of covariantized Lie derivative (supergauge) and covariant
derivative}

In Riemannian geometry the commutator of Lie derivative and covariant
derivative vanishes, if the vector along which the Lie derivative
is taken is a killing vector. We want to see what relation there is
for a more general connection. Let us first consider the commutator
of the Lie derivative and the covariant derivative with curved index
on a superspace vector\begin{eqnarray}
\left[\Lie_{\vecfull{\xi}},\nabla_{M}\right]v^{K} & = & \underbrace{[\Lie_{\vecfull{\xi}},\partial_{M}]}_{=0}v^{K}+\Lie_{\vecfull{\xi}}\Gamma_{MN}\hoch{K}\cdot v^{N}\end{eqnarray}
According to footnote \ref{foot:commutationOfLieAndPartial}, the
first term vanishes and we have \begin{equation}
\left[\Lie_{\vecfull{\xi}},\nabla_{M}\right]=0\quad\iff\quad0=\Lie_{\vecfull{\xi}}\Gamma_{MN}\hoch{K}\quad\left(\stackrel{(\ref{eq:Lietrafosuperspaceconn})}{=}2\xi^{L}R_{LMN}\hoch{K}+\nabla_{M}\left(\nabla_{N}\xi^{K}+2\xi^{L}T_{LN}\hoch{K}\right)\right)\end{equation}
 In the case of a Levi Civita connection, the Lie derivative of the
connection vanishes, if the Lie derivative of the metric vanishes,
i.e. if $\vecfull{\xi}$ is a killing vector%
\footnote{\index{footnote!\thefoot. killing vectors and Lie derivative of the connection}This
is quite natural, as the Levi Civita connection is built only out
of the metric. Nevertheless, let us check this statement explicitly
with the derived formula, in order to see whether it is consistent.
In the Riemannian case we have \[
\Lie_{\vecboson{\xi}}\Gamma_{mn}\hoch{k}=2\xi^{l}R_{lmn}\hoch{k}+\nabla_{m}\nabla_{n}\xi^{k}\]
and the killing vector\index{killing vector} condition reads (pulling
down the indices with the covariantly conserved metric $g_{mn}$)\[
\nabla_{(m}\xi_{n)}=0\]
We can rewrite the above Lie derivative as\begin{eqnarray*}
\Lie_{\vecboson{\xi}}\Gamma_{mn|k} & = & 2\xi^{l}R_{lmnk}+\nabla_{m}\nabla_{n}\xi_{k}=\\
 & = & 2\xi^{l}R_{lmnk}+\frac{1}{2}\nabla_{m}\nabla_{n}\xi_{k}+\frac{1}{2}\nabla_{n}\nabla_{m}\xi_{k}-R_{mnk}\hoch{l}\xi_{l}=\\
 & \stackrel{\mbox{killing}}{=} & 2\xi^{l}R_{lmnk}-\frac{1}{2}\nabla_{m}\nabla_{k}\xi_{n}-\frac{1}{2}\nabla_{n}\nabla_{k}\xi_{m}-R_{mnk}\hoch{l}\xi_{l}=\\
 & = & 2\xi^{l}R_{lmnk}-\frac{1}{2}\nabla_{k}\nabla_{m}\xi_{n}+R_{mkn}\hoch{l}\xi_{l}-\frac{1}{2}\nabla_{k}\nabla_{n}\xi_{m}+R_{nkm}\hoch{l}\xi_{l}-R_{mnk}\hoch{l}\xi_{l}\\
 & = & 2\xi^{l}\underbrace{R_{lmnk}}_{-R_{nkml}}-R_{kmn}\hoch{l}\xi_{l}+R_{nkm}\hoch{l}\xi_{l}-R_{mnk}\hoch{l}\xi_{l}=\\
 & = & -\left(R_{nkm}\hoch{l}+R_{kmn}\hoch{l}+R_{mnk}\hoch{l}\right)\xi_{l}=0\qquad\fussend\end{eqnarray*}
}\rem{Was sind die Bedingungen an Kruemmung und Torsion fuer Existenz von Killing vector? Wie verallgemeinern im Superfall? Evtl.$\Lie_{\vecfull{\xi}}G_{\bs{\mu}N}=0$?}.
In general, however, we have the condition that the Lie derivative
of the connection has to vanish. 

Let us introduce just for the moment the symbol $\tilde{\mc{R}}$
to denote the action of a Gl(n) matrix (like the superspace connection
$\Gamma_{M\,\cdot}\hoch{\cdot}$) on the curved indices. Acting on
an arbitrary tensor, the commutator of above becomes\begin{eqnarray}
\left[\Liecov_{\vecfull{\xi}},\nabla_{M}\right] & = & \tilde{\mc{R}}\left(\Lie_{\vecfull{\xi}}\Gamma_{M\cdot}\hoch{\cdot}\right)+\group{\Liecov_{\vecfull{\xi}}\Omega_{M\cdot}\hoch{\cdot}}\end{eqnarray}
How does this commutator modify, if we choose the covariant derivative
with flat index?\begin{eqnarray}
\left[\Liecov_{\vecfull{\xi}},\nabla_{A}\right] & = & \left[\Liecov_{\vecfull{\xi}},E_{A}\hoch{M}\nabla_{M}\right]=\\
 & \stackrel{(\ref{eq:inversevielbeintrafo})}{=} & -\left(\nabla_{A}\xi^{M}+2\xi^{C}T_{CA}\hoch{M}\right)\nabla_{M}+E_{A}\hoch{M}\tilde{\mc{R}}\left(\Lie_{\vecfull{\xi}}\Gamma_{M\cdot}\hoch{\cdot}\right)+E_{A}\hoch{M}\group{\Liecov_{\vecfull{\xi}}\Omega_{M\cdot}\hoch{\cdot}}\qquad\end{eqnarray}
Finally we allow for an additional structure group transformation,
in order to see the commutator of a general gauge transformation with
the covariant derivative:\vRam{.88}{\begin{eqnarray}
\quad\left[\Liecov_{\vecfull{\xi}}+\group{L_{\cdot}\hoch{\cdot}}\:,\:\nabla_{A}\right] & = & \Bigl(\underbrace{-(\nabla_{A}\xi^{D}+2\xi^{C}T_{CA}\hoch{D})}_{(\Liecov_{\vecfull{\xi}}E_{A}\hoch{M})E_{M}\hoch{D}}-L_{A}\hoch{D}\Bigr)\nabla_{D}+\nonumber \\
 &  & +\tilde{\mc{R}}\underbrace{\left(2\xi^{L}R_{LA\,\cdot}\hoch{\cdot}+\nabla_{A}\left(\nabla_{\cdot}\xi^{\cdot}+2\xi^{L}T_{L\,\cdot}\hoch{\cdot}\right)\right)}_{E_{A}\hoch{M}\Lie_{\vecfull{\xi}}\Gamma_{M\,\cdot}\hoch{\cdot}}+\quad\underbrace{\!\!\!\!\!\!\group{2\xi^{C}R_{CA\,\cdot}\hoch{\cdot}-\nabla_{A}L_{\cdot}\hoch{\cdot}}\!\!\!\!\!\!\!\!\!\!\!\!\!\!\!\!\!\!\!\!\!\!\!\!}_{E_{A}\hoch{M}\Liecov_{\vecfull{\xi}}\Omega_{M\,\cdot}\hoch{\cdot}}\quad\qquad\qquad\label{eq:commutatorOfCovDerFlatAndLie}\end{eqnarray}
}\vorRam\\  When acting on scalar fields, only the first term remains. 

The idea of the above considerations was of course that part of the
gauge transformations become just the local supersymmetry transformations,
while the fermionic components of the covariant derivative should
contain the supersymmetric covariant derivative. We therefore expect,
at least for the flat case, a vanishing result for the fermionic components
of this commutator. We will come back to this question after having
established the WZ-gauge.

\subsubsection{Algebra of the gauge transformations}

The algebra in subsection \ref{sub:algOfLieUsg} was assuming that
the variation acts on all objects, including the transformation parameter
of the first transformation. This is not true for field-independent
transformation parameters. If $\vecfull{\xi}$ is just the transformation
parameter of the symmetry, then this parameter does not transform
itself. On the other hand, there is no need for the transformation
parameter to coincide with $\vecfull{\xi}$. Instead, $\vecfull{\xi}$
can be a functional of transformation parameter and of the the fields.
We thus have to treat its variation seperately. A general gauge variation
has the form  $\delta t_{MA}^{NB}=\Liecov_{\vecfull{\xi}}t_{MA}^{NB}+\group{L\cdot\hoch{\cdot}}t_{MA}^{NB}$,
where $\vecfull{\xi}$ and the structure group matrix $L$ are local
and may or may not depend on the fields of the theory. Acting a second
time with such a variation yields\begin{eqnarray}
\lqn{\delta_{1}\delta_{2}(\ldots)=}\nonumber \\
 & = & \delta_{1}\left(\Liecov_{\vecfull{\xi_{2}}}+\group{L_{2\,\cdot}\hoch{\cdot}}\right)=\\
 & = & \delta_{1}\left(\Lie_{\vecfull{\xi_{2}}}+\group{\xi_{2}^{K}\Omega_{K\cdot}\hoch{\cdot}+L_{2\,\cdot}\hoch{\cdot}}\right)(\ldots)=\\
 & = & \left(\Lie_{\delta_{1}\vecfull{\xi_{2}}}+\group{\delta_{1}\xi_{2}^{K}\Omega_{K\cdot}\hoch{\cdot}+\xi_{2}^{K}\delta_{1}\Omega_{K\cdot}\hoch{\cdot}+\delta_{1}L_{2\,\cdot}\hoch{\cdot}}\right)(\ldots)+\left(\Lie_{\vecfull{\xi_{2}}}+\group{\xi_{2}^{K}\Omega_{K\cdot}\hoch{\cdot}+L_{2\,\cdot}\hoch{\cdot}}\right)\delta_{1}(\ldots)=\qquad\\
 & = & \left(\Liecov_{\delta_{1}\vecfull{\xi_{2}}}+\group{\xi_{2}^{K}\left(\Liecov_{\vecfull{\xi_{1}}}\Omega_{K\cdot}\hoch{\cdot}-\nabla_{K}L_{1\,\cdot}\hoch{\cdot}\right)+\delta_{1}L_{2\,\cdot}\hoch{\cdot}}\right)(\ldots)+\nonumber \\
 &  & +\left(\Liecov_{\vecfull{\xi_{2}}}+\group{L_{2\,\cdot}\hoch{\cdot}}\right)\left(\Liecov_{\vecfull{\xi_{1}}}+\group{L_{1}}\right)(\ldots)=\\
 & = & \Big[\Liecov_{\delta_{1}\vecfull{\xi_{2}}}+\group{2\xi_{2}^{K}\xi_{1}^{L}R_{LK\cdot}\hoch{\cdot}-\xi_{2}^{K}\nabla_{K}L_{1}+\delta_{1}L_{2\,\cdot}\hoch{\cdot}}+\left(\Liecov_{\vecfull{\xi_{2}}}+\group{L_{2\,\cdot}\hoch{\cdot}}\right)\left(\Liecov_{\vecfull{\xi_{1}}}+\group{L_{1\,\cdot}\hoch{\cdot}}\right)\Big](\ldots)\qquad\end{eqnarray}
Finally we take the commutator and use the commutation relation (\ref{eq:LiecovAndStructureComm})
of above\begin{eqnarray}
[\delta_{1},\delta_{2}] & = & \group{4\xi_{2}^{K}\xi_{1}^{L}R_{LK\cdot}\hoch{\cdot}+\xi_{1}^{K}\nabla_{K}L_{2}-\xi_{2}^{K}\nabla_{K}L_{1}+\delta_{1}L_{2}\cdot\hoch{\cdot}-\delta_{2}L_{1}}+\nonumber \\
 &  & +\Liecov_{[\vecfull{\xi_{2}},\vecfull{\xi_{1}}]+\delta_{1}\vecfull{\xi_{2}}-\delta_{2}\vecfull{\xi_{1}}}+\group{2\xi_{2}^{K}\xi_{1}^{L}R_{KL\cdot}\hoch{\cdot}+\Liecov_{\vecfull{\xi_{2}}}L_{1}-\Liecov_{\vecfull{\xi_{1}}}L_{2}-[L_{2},L_{1}]}\end{eqnarray}
\begin{equation}
\boxed{[\delta_{1},\delta_{2}]=\Liecov_{[\vecfull{\xi_{2}},\vecfull{\xi_{1}}]+\delta_{1}\vecfull{\xi_{2}}-\delta_{2}\vecfull{\xi_{1}}}+\group{2\xi_{1}^{K}\xi_{2}^{L}R_{KL\,\cdot}\hoch{\cdot}+[L_{1},L_{2}]_{\cdot}\hoch{\cdot}+\delta_{1}L_{2\,\cdot}\hoch{\cdot}-\delta_{2}L_{1\,\cdot}\hoch{\cdot}}}\label{eq:supergauge-algebra}\end{equation}
If $\vecfull{\xi}$ and $L$ are field dependent and transform like
all the other fields, we have $\delta_{1}\vecfull{\xi_{2}}=[\vecfull{\xi_{1}},\vecfull{\xi_{2}}]$
and $\delta_{1}L_{2}=\Liecov_{\vecfull{\xi_{1}}}L_{2}-[L_{1},L_{2}]$
and the above equation is the same as (\ref{eq:LiecovAndStructureComm}),
while if both parameters do not transform at all, we have a similar,
but still different algebra with some different signs and some terms
missing. The above important equation will help us to find the SUSY-algebra
in this huge algebra. By going to the WZ-gauge, we will fix part of
the superdiffeomorphisms and local structure group transformations.
The remaining transformations, which stabilize this gauge will then
have a field-dependent $\vecfull{\xi}$, which we can plug into the
above equation.

\rem{Let us now consider transformation vector fields of the form
$\vecfull{\xi_{1/2}}=\eps_{1/2}^{A}q_{A}\hoch{M}\pe_{M}$, with $\eps^{A}$
being inert under variations, while $q_{A}\hoch{M}$ is built from
the fields and transforms in the way its indices indicate. The transformation
of $\vecfull{\xi}$ then reads\begin{eqnarray}
\delta_{1}\xi_{2}^{M} & = & \eps_{2}^{B}\left(\Liecov_{\vecfull{\xi_{1}}}q_{B}\hoch{M}-L_{1\, B}\hoch{C}q_{C}\hoch{M}\right)=\\
 & = & \eps_{2}^{B}\left\{ \eps_{1}^{A}q_{A}\hoch{K}\nabla_{K}q_{B}\hoch{M}-\left(\nabla_{L}(\eps_{1}^{A}q_{A}\hoch{M})+2\eps_{1}^{A}q_{A}\hoch{K}T_{KL}\hoch{M}\right)q_{B}\hoch{L}-L_{1\, B}\hoch{C}q_{C}\hoch{M}\right\} \\
\delta_{1}\xi_{2}^{M}-\delta_{2}\xi^{M} & = & 2\eps_{1}^{A}\eps_{2}^{B}\left(q_{A}\hoch{K}\nabla_{K}q_{B}\hoch{M}-q_{B}\hoch{L}\nabla_{L}q_{A}\hoch{M}-2q_{A}\hoch{K}T_{KL}\hoch{M}q_{B}\hoch{L}\right)-2\eps_{[2}^{B}q_{B}\hoch{L}\nabla_{L}\eps_{1]}^{A}q_{A}\hoch{M}+\nonumber \\
 &  & -\eps_{2}^{B}L_{1\, B}\hoch{C}q_{C}\hoch{M}+\eps_{1}^{B}L_{2\, B}\hoch{C}q_{C}\hoch{M}\end{eqnarray}
On the other hand we have \begin{eqnarray}
\left[\vecfull{\xi_{1}},\vecfull{\xi_{2}}\right]^{M} & = & \xi_{1}^{K}\nabla_{K}\xi_{2}^{M}-\xi_{2}^{K}\nabla_{K}\xi_{1}^{M}-2\xi_{1}^{K}T_{KL}\hoch{M}\xi_{2}^{L}=\\
 & = & \eps_{1}^{A}q_{A}\hoch{K}\nabla_{K}(\eps_{2}^{B}q_{B}\hoch{M})-\eps_{2}^{B}q_{B}\hoch{K}\nabla_{K}(\eps_{1}^{A}q_{A}\hoch{M})-2\eps_{1}^{A}\eps_{2}^{B}q_{A}\hoch{K}T_{KL}\hoch{M}q_{B}\hoch{L}=\\
 & = & \eps_{1}^{A}\eps_{2}^{B}\left(q_{A}\hoch{K}\nabla_{K}q_{B}\hoch{M}-q_{B}\hoch{K}\nabla_{K}q_{A}\hoch{M}-2q_{A}\hoch{K}T_{KL}\hoch{M}q_{B}\hoch{L}\right)+2\eps_{[1}^{A}q_{A}\hoch{K}\nabla_{K}\eps_{2]}^{B}q_{B}\hoch{M}\end{eqnarray}
which means that \begin{eqnarray}
\lqn{\delta_{1}\xi_{2}^{M}-\delta_{2}\xi^{M}=}\nonumber \\
 & = & 2\left[\vecfull{\xi_{1}},\vecfull{\xi_{2}}\right]^{M}-2\eps_{[1}^{A}q_{A}\hoch{K}\nabla_{K}\eps_{2]}^{B}q_{B}\hoch{M}-\eps_{2}^{B}L_{1\, B}\hoch{C}q_{C}\hoch{M}+\eps_{1}^{B}L_{2\, B}\hoch{C}q_{C}\hoch{M}=\\
 & = & \left[\vecfull{\xi_{1}},\vecfull{\xi_{2}}\right]^{M}+\eps_{1}^{A}\eps_{2}^{B}\left(q_{A}\hoch{K}\nabla_{K}q_{B}\hoch{M}-q_{B}\hoch{K}\nabla_{K}q_{A}\hoch{M}-2q_{A}\hoch{K}T_{KL}\hoch{M}q_{B}\hoch{L}\right)+2\eps_{[1}^{B}L_{2]\, B}\hoch{C}q_{C}\hoch{M}\end{eqnarray}
The gauge algebra thus becomes\vRam{.75}{\begin{eqnarray}
[\delta_{1},\delta_{2}] & = & \Liecov_{\eps_{1}^{A}\eps_{2}^{B}\left(q_{A}\hoch{K}\nabla_{K}q_{B}\hoch{M}-q_{B}\hoch{K}\nabla_{K}q_{A}\hoch{M}-2q_{A}\hoch{K}T_{KL}\hoch{M}q_{B}\hoch{L}\right)\pe_{M}+2\eps_{[1}^{B}L_{2]\, B}\hoch{C}\vecfull{q_{C}}}+\nonumber \\
 &  & +\group{2\eps_{1}^{A}\eps_{2}^{B}q_{A}\hoch{L}q_{B}\hoch{K}R_{LK\cdot}\hoch{\cdot}+[L_{1},L_{2}]_{\cdot}\hoch{\cdot}+\delta_{1}L_{2\,\cdot}\hoch{\cdot}-\delta_{2}L_{1\,\cdot}\hoch{\cdot}}\label{eq:supergaugeAlgebra-q}\end{eqnarray}
}
In particular for $\eps_{1}^{C}=\delta_{A}^{C}$ and $\eps_{2}^{D}=\delta_{B}^{D}$
and $q_{A}\hoch{M}=E_{A}\hoch{M}$ (corresponding to $\vecfull{\xi_{1}}=\vecfull{E_{A}}$,
$\vecfull{\xi_{2}}=\vecfull{E_{B}}$) we get\begin{eqnarray}
[\delta_{A},\delta_{B}] & = & \Liecov_{\left(-2T_{AB}\hoch{C}+L_{2\, A}\hoch{C}-L_{1\, B}\hoch{C}\right)\vecfull{E_{C}}}+\group{2R_{AB\cdot}\hoch{\cdot}+[L_{1},L_{2}]_{\cdot}\hoch{\cdot}+\delta_{1}L_{2\,\cdot}\hoch{\cdot}-\delta_{2}L_{1\,\cdot}\hoch{\cdot}}\qquad\end{eqnarray}
When acting on flat indices, $\delta_{A}$ and $\delta_{B}$ coincide
for $L_{1}=L_{2}=0$ with the covariant derivative. This is consistent
with the above algebra.

In the case of supersymmetry transformations, we will have some $q_{\bs{\alpha}}\hoch{M}\neq E_{\bs{\alpha}}\hoch{M}$
and in addition an accompanying structure group transformation $L_{A}\hoch{B}=\eps^{\bs{\alpha}}L_{\bs{\alpha}A}\hoch{B}$...
(see \ref{eq:SUSY-xi-L-cond}). The result will be the supersymmetry-algebra.}

\subsection{Finite gauge transformations}

In order to choose an explicit gauge it is useful to know the finite
form of the gauge transformations (only then you can decide whether
a particular gauge is accessible or not). For superdiffeomorphisms
and local structure group transformations (i.e. Lorentz transformations
and perhaps dilatations), we know the finite form anyway. Let us denote
the transformed fields by a prime (for superdiffeomorphisms) and by
a tilde (for structure group transformations). The vielbein transforms
homogenously under both transformations, i.e. $E'_{M}\hoch{A}(\xfull')=\partiell{x^{N}}{x'^{M}}E_{N}\hoch{A}(\xfull')$
under superdiffeomorphisms and $\tilde{E}_{M}\hoch{A}(\xfull)=E_{M}\hoch{B}(\xfull)\Lambda_{B}\hoch{A}(\xfull)$
under structure group transformations. Altogether this reads\begin{eqnarray}
\tilde{E}'_{M}\hoch{A}(\xfull') & = & \partiell{x^{N}}{x'^{M}}\left(E_{N}\hoch{B}(\xfull)\Lambda_{B}\hoch{A}(\xfull)\right)=\left(\partiell{x^{N}}{x'^{M}}E_{N}\hoch{B}(\xfull)\right)\Lambda'_{B}\hoch{A}(\xfull')\end{eqnarray}
Likewise a more general tensor field with index structure $t_{MA}^{NB}$
transfoms as\begin{eqnarray}
\tilde{t}'{}_{MA}^{NB}(\xfull') & = & \partiell{x^{K}}{x'^{M}}\partiell{x'^{N}}{x^{L}}t_{KC}^{LD}(\xfull)\Lambda_{A}\hoch{C}(\xfull)(\Lambda^{-1})_{D}\hoch{B}(\xfull)\end{eqnarray}
Other examples for such homogenous transformations (apart from the
vielbein) are a RR-superfield with $\tilde{\RR}'^{\bs{\delta}\hat{\bs{\delta}}}(\xfull')=\RR^{\bs{\gamma}\hat{\bs{\gamma}}}(\xfull)\Lambda_{\bs{\gamma}}\hoch{\bs{\delta}}\Lambda_{\hat{\bs{\gamma}}}\hoch{\hat{\bs{\delta}}}$
(where the structure group transformation $\Lambda_{A}\hoch{B}$ is
supposed to be a blockdiagonal one), or a dilaton scalar superfield
with simply $\widetilde{\dil}'(\xfull')=\dil(\xfull)$. 

The finite inhomogenous transformation of the connection superfield
reads%
\footnote{\index{footnote!\thefoot. finite transformation of scale connection}Defining
$\Omega_{M}^{(D)}\equiv\frac{1}{\dim}\Omega_{Ma}\hoch{a}$ and $\Lambda^{(D)}\equiv\frac{1}{\dim}\Lambda_{a}\hoch{a}$
yields the transformation (\ref{eq:finitesupergaugeOmegaScale}) in
the second line. However, having in mind the definition of the mixed
connection (\ref{eq:gemOm}) yields the same transformation for each
of the scale connections $\Omega_{M}^{(D)}$ (with $\Lambda^{(D)}$),
$\hat{\Omega}_{M}^{(D)}$ (with $\hat{\Lambda}^{(D)}$) and $\check{\Omega}_{M}^{(D)}$
(with $\check{\Lambda}^{(D)}$) respectively.

In our application to the Berkovits string, we have introduced a compensator
field $\Phi$ via $G_{ab}=e^{2\Phi}\eta_{ab}$ which transforms under
the bosonic scale transformations $\check{\Lambda}$. The distinction,
however, is not important, as $\Lambda$, $\hat{\Lambda}$ and $\check{\Lambda}$
get coupled by the gauge fixing of $T_{\bs{\alpha\beta}}\hoch{c}=\gamma_{\bs{\alpha}\bs{\beta}}^{c}$
and $T_{\hat{\bs{\alpha}}\hat{\bs{\beta}}}\hoch{c}=\gamma_{\hat{\bs{\alpha}}\hat{\bs{\beta}}}^{c}$
anyway.$\quad\fussend$%
}\begin{eqnarray}
\tilde{\Omega}'_{MA}\hoch{B}(\xfull') & = & \partiell{x^{N}}{x'^{M}}\left(-\partial_{N}\Lambda_{A}\hoch{B}+(\Lambda^{-1})_{A}\hoch{D}\Omega_{ND}\hoch{C}(\xfull)\Lambda_{C}\hoch{B}\right)\label{eq:finitesupergaugeOmega}\\
\tilde{\Omega}^{(D)}{}'_{M}(\xfull') & = & \partiell{x^{N}}{x'^{M}}\left(\Omega_{N}^{(D)}(\xfull)-\partial_{N}\Lambda^{(D)}(\xfull)\right)\label{eq:finitesupergaugeOmegaScale}\end{eqnarray}
In the main part of this thesis we have also introduced a compensator
field $\Phi$, which transforms by a shift under scale transformations,
i.e. $\tilde{\Phi}'(\xfull')=\Phi(\xfull)-\check{\Lambda}^{(D)}(\xfull)$
(where $\check{\Lambda}^{(D)}$ denotes the dilatation or scale part
of the bosonic block).

\rem{In the main text, we would denote it via $\gem{\Lambda}_{A}\hoch{B}=\diag(\check{\Lambda}_{a}\hoch{b},\Lambda_{\bs{\alpha}}\hoch{\bs{\beta}},\hat{\Lambda}_{\hat{\bs{\alpha}}}\hoch{\hat{\bs{\beta}}})$,
consisting of Lorentz and scale transformations: \begin{eqnarray}
\gem{\Lambda}_{A}\hoch{B} & \equiv & \left(\begin{array}{ccc}
\check{\Lambda}_{a}\hoch{b} & 0 & 0\\
0 & \Lambda_{\bs{\alpha}}\hoch{\bs{\beta}} & 0\\
0 & 0 & \hat{\Lambda}_{\hat{\bs{\alpha}}}\hoch{\hat{\bs{\beta}}}\end{array}\right)=(\exp\{\gem{L}\})_{A}\hoch{B}\label{eq:gemLam}\\
\gem{L}_{A}\hoch{B} & = & \left(\begin{array}{ccc}
\check{L}^{(D)}\delta_{a}^{b} & 0 & 0\\
0 & \frac{1}{2}L^{(D)}\delta_{\bs{\alpha}}\hoch{\bs{\beta}} & 0\\
0 & 0 & \frac{1}{2}\hat{L}^{(D)}\delta_{\hat{\bs{\alpha}}}\hoch{\hat{\bs{\beta}}}\end{array}\right)+\left(\begin{array}{ccc}
\check{L}_{a}^{(L)}\hoch{b} & 0 & 0\\
0 & \frac{1}{4}L_{ab}^{(L)}\gamma^{ab}\tief{\bs{\alpha}}\hoch{\bs{\beta}} & 0\\
0 & 0 & \frac{1}{4}\hat{L}_{ab}^{(L)}\gamma^{ab}\tief{\hat{\bs{\alpha}}}\hoch{\hat{\bs{\beta}}}\end{array}\right)\end{eqnarray}
How are $L$ and $\hat{L}$ connected? They should respect the gaugings
$T_{\bs{\alpha\beta}}\hoch{c}=\gamma_{\bs{\alpha\beta}}^{c}$ and
$\hat{T}_{\hat{\bs{\alpha}}\hat{\bs{\beta}}}\hoch{c}=\gamma_{\hat{\bs{\alpha}}\hat{\bs{\beta}}}^{c}$.\begin{equation}
\delta T_{\bs{\alpha\beta}}\hoch{c}=0=\delta\hat{T}_{\bs{\alpha\beta}}\hoch{c}\end{equation}
which means that $\Lambda=\hat{\Lambda}=\check{\Lambda}$. That does
not mean the same for the corresponding connections (they are not
equal), but in fact -- if the gauge fixings should remain the same
under parallel transport -- it tells us that one should take only
one of the connections as the one which defines parallel transport
and rewrite the other in terms of this one plus a difference tensor.
The equations written in terms of the mixed connection are still valid,
but should be taken as an abbreviation for the interpretation that
we just have given.}

\section{Wess-Zumino gauge}

\subsection{WZ gauge for the vielbein}

\index{Wess-Zumino gauge|fett}Superdiffeomorphisms $x'^{M}=F^{M}(\xfull)\stackrel{inf}{=}x^{M}+\xi^{M}(\xfull)$
with $\xfull=(\xboson,\xbothtetas)$ parametrise many more gauge degrees
of freedom than just the bosonic diffeomorphisms $x'^{m}=f^{m}(\xboson)\stackrel{inf}{=}x^{m}+\xi^{m}(\xboson,\xbothtetas=0)$.
Let us write $\xfull'$ as \begin{eqnarray}
x'^{M} & = & {x'}_{0}^{M}(\xboson)+\underbrace{x^{\bs{\mc{N}}}}_{\tet^{\mc{N}}}{x'}_{\bs{\mc{N}}}^{M}(\xboson)+\mc{O}(\xbothtetas^{2})\end{eqnarray}
We have \begin{eqnarray}
\partiell{x'^{M}}{x^{N}} & = & \left(\begin{array}{cc}
\partiell{x'^{m}}{x^{n}} & \partiell{x'^{m}}{x^{\bs{\mc{N}}}}\\
\partiell{x'^{\bs{\mc{M}}}}{x^{n}} & \partiell{x'^{\bs{\mc{M}}}}{x^{\bs{\mc{N}}}}\end{array}\right)\stackrel{\xbothtetas=0}{=}\left(\begin{array}{cc}
\partiell{{x'}_{0}^{m}}{x^{n}} & {x'}_{\bs{\mc{N}}}^{m}\\
\partiell{{x'}_{0}^{\bs{\mc{M}}}}{x^{n}} & {x'}_{\bs{\mc{N}}}^{\bs{\mc{M}}}\end{array}\right)\end{eqnarray}
In the following we will see that it is possible to fix the vielbein
for vanishing $\xbothtetas$ to\index{Wess-Zumino gauge!for the vielbein}
\\
\Ram{.4}{\begin{eqnarray}
\bei{E_{M}\hoch{A}}{} & = & \left(\begin{array}{cc}
e_{m}\hoch{a} & \psi_{m}\hoch{\bs{\mc{A}}}\\
0 & \delta_{\bs{\mc{M}}}\hoch{\bs{\mc{A}}}\end{array}\right)\label{eq:WZ-gauge}\end{eqnarray}
 } \\
with inverse\begin{eqnarray}
\lqn{{\Ramm{.4}{\drek{}{\drek{}{}{}}{}}}}\qquad\bei{E_{A}\hoch{M}}{} & = & \left(\begin{array}{cc}
e_{a}\hoch{m} & -\psi_{a}\hoch{\bs{\mc{M}}}\\
0 & \delta_{\bs{\mc{A}}}\hoch{\bs{\mc{M}}}\end{array}\right)\qquad\quad\qquad\label{eq:WZ-gauge-inverse}\\
\mbox{where }\psi_{a}\hoch{\bs{\mc{M}}} & \equiv & e_{a}\hoch{m}\psi_{m}\hoch{\bs{\mc{A}}}\delta_{\bs{\mc{A}}}\hoch{\bs{\mc{M}}}\\
e_{m}\hoch{a}e_{a}\hoch{n} & = & \delta_{m}^{n}\end{eqnarray}
We want to show that the above gauge can always be reached if the
original supervielbein had full rank. To this end, let us call the
supervielbein in the above gauge $E'_{M}\hoch{A}(x')$ and only the
original general one $E_{M}\hoch{A}$. We should have the relation
$\partiell{x'^{M}}{x^{N}}E'_{M}\hoch{A}(x')=E_{N}\hoch{A}(x)$. Indeed,
multiplying $E'_{M}\hoch{A}(x')$ from the left with the transposed
$(\xbothtetas=0)$-Jacobian without ordinary diffeos ($\partiell{{x'}_{0}^{m}}{x^{n}}=\delta_{n}^{m}$)
yields\begin{equation}
\left(\begin{array}{cc}
\delta_{n}^{m} & \partiell{{x'}_{0}^{\bs{\mc{M}}}}{x^{n}}\\
{x'}_{\bs{\mc{N}}}^{m} & {x'}_{\bs{\mc{N}}}^{\bs{\mc{M}}}\end{array}\right)\left(\begin{array}{cc}
e_{m}\hoch{a} & \psi_{m}\hoch{\bs{\mc{A}}}\\
0 & \delta_{\bs{\mc{M}}}\hoch{\bs{\mc{A}}}\end{array}\right)=\left(\begin{array}{cc}
e_{n}\hoch{a} & \left(\psi_{n}\hoch{\bs{\mc{A}}}+\partiell{{x'}_{0}^{\bs{\mc{M}}}}{x^{n}}\delta_{\bs{\mc{M}}}\hoch{\bs{\mc{A}}}\right)\\
{x'}_{\bs{\mc{N}}}^{m}e_{m}\hoch{a} & \left({x'}_{\bs{\mc{N}}}^{m}\psi_{m}\hoch{\bs{\mc{A}}}+{x'}_{\bs{\mc{N}}}^{\bs{\mc{M}}}\delta_{\bs{\mc{M}}}\hoch{\bs{\mc{A}}}\right)\end{array}\right)\stackrel{!}{=}\bei{E_{N}\hoch{A}}{}\qquad\end{equation}
This fixes some of the auxiliary gauge parameters: \begin{eqnarray}
{x'}_{\bs{\mc{N}}}^{m} & = & e_{a}\hoch{m}\bei{E_{\bs{\mc{N}}}\hoch{a}}{},\qquad{x'}_{\bs{\mc{N}}}^{\bs{\mc{M}}}=\left(E_{\bs{\mc{N}}}\hoch{\bs{\mc{A}}}-{x'}_{\bs{\mc{N}}}^{m}\psi_{m}\hoch{\bs{\mc{A}}}\right)\delta_{\bs{\mc{A}}}\hoch{\bs{\mc{M}}}\end{eqnarray}
So all the ${x'}_{\bs{\mc{N}}}^{M}$ are fixed. In contrast, ${x'}_{0}^{M}(\xboson)$
are still free and they parametrize bosonic diffeomorphisms and local
supersymmmetry. We still have many more unfixed auxiliary gauge parameters
(the higher $\tet$-derivatives of $x'$) whose fixing we will discuss
in subsection \ref{sub:fix-aux}.

\subsection{Calculus with the gauge fixed vielbein}

Before we proceed with the gauge fixing of the connection, let us
have a look at some consequences of the special vielbein gauge. The
new bosonic vielbein $e_{m}\hoch{a}(\xboson)=E_{m}\hoch{a}(\xboson,0)$
offers a second possibility to switch from curved to flat indices
and one has to be careful, in order not to mix up things. The inverse
of the supervielbein behaves differently than the inverse of the bosonic
vielbein. While in superspace the inverse is with respect to a sum
over all superspace indices, the sum for the bosonic inverse runs
only over the bosonic indices\begin{eqnarray}
E_{M}\hoch{A}E_{B}\hoch{M} & = & \delta^{A}\tief{B}\quad\dann\bei{E_{M}\hoch{a}}{}\bei{E_{b}\hoch{M}}{}=\delta_{b}^{a}\\
\bei{E_{m}\hoch{a}}{}e_{b}\hoch{m} & = & \delta_{b}^{a}\end{eqnarray}
It therefore makes a difference which vielbein is used to change from
flat to curved indices and vice verse. Consider an arbitrary supervector
$V_{M}$:\begin{eqnarray}
\bei{V_{m}}{}e_{a}\hoch{m} & = & \bei{V_{C}E_{m}\hoch{C}}{}e_{a}\hoch{m}=\\
 & = & \bei{V_{c}E_{m}\hoch{c}}{}e_{a}\hoch{m}+\bei{V_{\bs{\mc{C}}}E_{m}\hoch{\bs{\mc{C}}}}{}e_{a}\hoch{m}\end{eqnarray}
or in summary \begin{equation}
\boxed{\bei{V_{m}}{}e_{a}\hoch{m}=\bei{V_{a}}{}+\bei{V_{\bs{\mc{C}}}}{}\psi_{m}\hoch{\bs{\mc{C}}}e_{a}\hoch{m}}\label{eq:bosvielb-calcI}\end{equation}
For upper bosonic indices the situation is better because the WZ-gauge
removes the disturbing additional term: \begin{eqnarray}
\bei{V^{a}}{}e_{a}\hoch{m} & = & \bei{V^{N}E_{N}\hoch{a}}{}e_{a}\hoch{m}=\\
 & = & \bei{V^{n}E_{n}\hoch{a}}{}e_{a}\hoch{m}+V^{\bs{\mc{N}}}\underbrace{\bei{E_{\bs{\mc{N}}}\hoch{a}}{}}_{=0}e_{a}\hoch{m}\end{eqnarray}
so that we get the nice relation\begin{equation}
\boxed{\bei{V^{a}}{}e_{a}\hoch{m}=\bei{V^{m}}{}}\label{eq:bosvielb-calcII}\end{equation}
We can do the same considerations for fermionic indices and arrive
at the opposite situation\vRam{.4}{ \begin{eqnarray}
\bei{\Psi_{\bs{\mc{M}}}}{}\delta_{\bs{\mc{A}}}\hoch{\bs{\mc{M}}} & = & \bei{\Psi_{\bs{\mc{A}}}}{}\\
\bei{\Xi^{\bs{\mc{M}}}}{}\delta_{\bs{\mc{M}}}\hoch{\bs{\mc{A}}} & = & \bei{\Xi^{\bs{\mc{A}}}}{}-\bei{\Xi^{b}}{}\psi_{b}\hoch{\bs{\mc{M}}}\delta_{\bs{\mc{M}}}\hoch{\bs{\mc{A}}}\end{eqnarray}
}

\subsection{WZ gauge for the connection}

Similar to the supervielbein-case it is likewise possible to reach
a special gauge at $\xbothtetas=0$ for the connection componets with
fermionic form-index:\index{Wess-Zumino gauge!for the connection}\begin{equation}
\boxed{\bei{\Omega_{\bs{\mc{M}}\, A}\hoch{B}}{}=0}\label{eq:WZ-gauge-connection}\end{equation}
Let us show that this gauge fixing is really accessible. We would
like to reach the gauge (\ref{eq:WZ-gauge-connection}) using the
local structure group transformations of higher order in $\xbothtetas$
(i.e. with $\bei{\Lambda_{A}\hoch{B}}{}=\delta_{A}\hoch{B}$). Remember
the structure group transformation of the connection \begin{eqnarray}
\tilde{\Omega}_{MA}\hoch{B}(x) & = & -\partial_{M}\Lambda_{A}\hoch{B}+(\Lambda^{-1})_{A}\hoch{D}\Omega_{MD}\hoch{C}(x)\Lambda_{C}\hoch{B}\end{eqnarray}
Reaching the gauge fixing condition (\ref{eq:WZ-gauge-connection})
is thus possible by simply choosing \begin{eqnarray}
\Lambda_{\bs{\mc{M}}A}\hoch{B} & \equiv & \bei{\partial_{\bs{\mc{M}}}\Lambda_{A}\hoch{B}}{}\stackrel{!}{=}\bei{\Omega_{\bs{\mc{M}}A}\hoch{B}(x)}{}\end{eqnarray}

\subsection{Gauge fixing the remaining auxiliary gauge freedom}

\label{sub:fix-aux}In addition to the ordinary Wess Zumino\index{Wess-Zumino gauge!extension to $\sim$}
gauge\begin{eqnarray}
\bei{E_{\bs{\mc{M}}}\hoch{A}}{} & = & \delta_{\bs{\mc{M}}}\hoch{A}\label{eq:WZI}\\
\bei{\Omega_{\bs{\mc{M}}A}\hoch{B}}{} & = & 0\label{eq:WZII}\end{eqnarray}
 we can demand the gauge fixing condition $\bei{\partial_{(\bs{\mc{M}}}E_{\bs{\mc{N}})}\hoch{A}}{}\stackrel{!}{=}0$
using the gauge parameter $\bei{\partial_{\bs{\mc{M}}}\partial_{\bs{\mc{N}}}\xi^{A}}{}$.
Indeed all the other higher components of $\xi^{A}$ and $L_{A}\hoch{B}$
can be fixed by imposing%
\footnote{\index{footnote!\thefoot. accessibility of extended WZ gauge}Looking
at the infinitesimal transformations\begin{eqnarray*}
\delta\bei{\left(\partial_{\bs{\mc{M}}_{1}}\ldots\partial_{\bs{\mc{M}}_{n}}E_{\bs{\mc{M}}_{n+1}}\hoch{A}\right)}{} & = & \partial_{\bs{\mc{M}}_{1}}\ldots\partial_{\bs{\mc{M}}_{n}}\bei{\left(\partial_{\bs{\mc{M}}_{n+1}}\xi^{A}+\Omega_{\bs{\mc{M}}_{n+1}B}\hoch{A}\xi^{B}+2\xi^{C}T_{CM}\hoch{A}\right)}{}=\\
\delta\bei{\left(\partial_{\bs{\mc{M}}_{1}}\ldots\partial_{\bs{\mc{M}}_{n}}\Omega_{\bs{\mc{M}}_{n+1}A}\hoch{B}\right)}{} & = & -\partial_{\bs{\mc{M}}_{1}}\ldots\partial_{\bs{\mc{M}}_{n}}\bei{\left(\partial_{\bs{\mc{M}}_{n+1}}L_{A}\hoch{B}+[L,\Omega_{\bs{\mc{M}}_{n+1}}]\right)}{}\end{eqnarray*}
it seems quite obvious that the parameters $\bei{\partial_{\bs{\mc{M}}_{1}}\ldots\partial_{\bs{\mc{M}}_{n+1}}\xi^{A}}{}$
and $\bei{\partial_{\bs{\mc{M}}_{1}}\ldots\partial_{\bs{\mc{M}}_{n+1}}L_{A}\hoch{B}}{}$
can be used to shift $\bei{\partial_{(\bs{\mc{M}}_{1}}\ldots\partial_{\bs{\mc{M}}_{n}}E_{\bs{\mc{M}}_{n+1})}\hoch{A}}{}$
and $\bei{\partial_{(\bs{\mc{M}}_{1}}\ldots\partial_{\bs{\mc{M}}_{n}}\Omega_{\bs{\mc{M}}_{n+1})A}\hoch{B}}{}$
to whatever value one likes. A rigorous proof that (\ref{eq:gauge-fix-sym-vielbein})
and (\ref{eq:gauge-fix-sym-Omega}) are accessible, however, should
consider the finite transformations.$\quad\fussend$%
} (see e.g. \cite{Tsimpis:2004gq})\begin{eqnarray}
\bei{\partial_{(\bs{\mc{M}}_{1}}\ldots\partial_{\bs{\mc{M}}_{n}}E_{\bs{\mc{M}}_{n+1})}\hoch{A}}{} & \stackrel{!}{=} & 0\label{eq:gauge-fix-sym-vielbein}\\
\bei{\partial_{(\bs{\mc{M}}_{1}}\ldots\partial_{\bs{\mc{M}}_{n}}\Omega_{\bs{\mc{M}}_{n+1})A}\hoch{B}}{} & \stackrel{!}{=} & 0\qquad\forall n\in\{1,\ldots,\dim({\scriptstyle \bs{\mc{M}}})-1\}\label{eq:gauge-fix-sym-Omega}\end{eqnarray}
where $\dim({\scriptstyle \bs{\mc{M}}})$ shall denote the number
of fermionic dimensions, e.g. 32 for type II in ten dimensions. Actually
the above equations even hold for $n=\dim({\scriptstyle \bs{\mc{M}}})$
(the highest components of $E$ and $\Omega$), but then trivially,
as the total graded symmetrization of $n+1$ fermionic indices (which
is an antisymmetrization in fact) in $n$ dimensions always vanishes.
For $n>\dim({\scriptstyle \bs{\mc{M}}})$ even the derivative without
graded symmetrization vanishes trivially as usual. The second equation
is even true for $n=0$ (due to (\ref{eq:WZII})) while the first
is modified for $n=0$ to $\bei{E_{\bs{\mc{M}}}\hoch{A}}{}=\delta_{\bs{\mc{M}}}\hoch{A}$
(\ref{eq:WZI}).

This gauge is useful to calculate explicitely higher orders in the
$\xbothtetas$-expansion of the vielbein or the connection in terms
of torsion and curvature. Let us consider at first the connection.
For the $n$-th partial derivative of the component with fermionic
form index we can write\begin{eqnarray}
\lqn{\bei{\partial_{\bs{\mc{M}}_{1}}\ldots\partial_{\bs{\mc{M}}_{n}}\Omega_{\bs{\mc{M}}_{n+1}A}\hoch{B}}{}=}\nonumber \\
 & = & \underbrace{\bei{\partial_{(\bs{\mc{M}}_{1}}\ldots\partial_{\bs{\mc{M}}_{n}}\Omega_{\bs{\mc{M}}_{n+1})A}\hoch{B}}{}}_{=0\:(\ref{eq:gauge-fix-sym-Omega})}+\frac{2}{n+1}\sum_{i=1}^{n}\bei{\partial_{\bs{\mc{M}}_{1}}\ldots\partial_{[\bs{\mc{M}}_{i}|}\ldots\partial_{\bs{\mc{M}}_{n}}\Omega_{|\bs{\mc{M}}_{n+1}]A}\hoch{B}}{}=\\
 & = & \frac{2}{n+1}\sum_{i=1}^{n}\bei{\partial_{\bs{\mc{M}}_{1}}\ldots\partial_{\bs{\mc{M}}_{i-1}}\partial_{\bs{\mc{M}}_{i+1}}\ldots\partial_{\bs{\mc{M}}_{n}}\left(R_{\bs{\mc{M}}_{i}\bs{\mc{M}}_{n+1}A}\hoch{B}+\Omega_{[\bs{\mc{M}}_{i}|A}\hoch{C}\Omega_{|\bs{\mc{M}}_{n+1}]C}\hoch{B}\right)}{}=\\
 & = & \frac{n}{n+1}\bei{\partial_{(\bs{\mc{M}}_{1}}\ldots\partial_{\bs{\mc{M}}_{n-1}|}\left(2R_{|\bs{\mc{M}}_{n})\bs{\mc{M}}_{n+1}A}\hoch{B}+\Omega_{|\bs{\mc{M}}_{n})A}\hoch{C}\cdot\Omega_{\bs{\mc{M}}_{n+1}C}\hoch{B}-\Omega_{\bs{\mc{M}}_{n+1}A}\hoch{C}\cdot\Omega_{|\bs{\mc{M}}_{n})C}\hoch{B}\right)}{}\qquad\end{eqnarray}
\begin{equation}
\stackrel{(\ref{eq:gauge-fix-sym-Omega})}{\dann}\boxed{\bei{\partial_{\bs{\mc{M}}_{1}}\ldots\partial_{\bs{\mc{M}}_{n}}\Omega_{\bs{\mc{M}}_{n+1}A}\hoch{B}}{}=\frac{2n}{n+1}\bei{\partial_{(\bs{\mc{M}}_{1}}\ldots\partial_{\bs{\mc{M}}_{n-1}}R_{\bs{\mc{M}}_{n})\bs{\mc{M}}_{n+1}A}\hoch{B}}{}}\quad\forall n\geq1\label{eq:connectionInTermsOfR}\end{equation}
Unfortunately, due to the $n$-dependent factor $\frac{2n}{n+1}$,
this relation cannot easily be integrated. In particular, although
the above equation implies $\bei{\partial_{\bs{\mc{M}}}\Omega_{\bs{\mc{N}}A}\hoch{B}}{}=\bei{R_{\bs{\mc{M}}\bs{\mc{N}}A}\hoch{B}}{}$,
we have in general $\partial_{\bs{\mc{M}}}\Omega_{\bs{\mc{N}}A}\hoch{B}\neq R_{\bs{\mc{M}}\bs{\mc{N}}A}\hoch{B}$.
Also $\Omega_{\bs{\mc{N}}A}\hoch{B}\neq x^{\bs{\mc{M}}}R_{\bs{\mc{M}}\bs{\mc{N}}A}\hoch{B}$.\rem{%
\footnote{\index{footnote!\thefoot. Taylor expansion of the connection}The
Taylor expansion of $\Omega_{\bs{\mc{M}}A}\hoch{B}$ reads\begin{eqnarray*}
\Omega_{\bs{\mc{M}}A}\hoch{B}(\xboson,\xbothtetas) & = & \Omega_{\bs{\mc{M}}A}\hoch{B}(\xboson,0)+\sum_{n\geq1}\frac{1}{n!}x^{\bs{\mc{M}}_{1}}\cdots x^{\bs{\mc{M}}_{n}}\partial_{\bs{\mc{M}}_{1}}\ldots\partial_{\bs{\mc{M}}_{n}}\bei{\Omega_{\bs{\mc{M}}A}\hoch{B}}{}=\\
 & = & \Omega_{\bs{\mc{M}}A}\hoch{B}(\xboson,0)+\sum_{n\geq1}\frac{1}{n!}\frac{2n}{n+1}x^{\bs{\mc{M}}_{1}}\cdots x^{\bs{\mc{M}}_{n}}\bei{\partial_{\bs{\mc{M}}_{1}}\ldots\partial_{\bs{\mc{M}}_{n-1}}R_{\bs{\mc{M}}_{n}\bs{\mc{M}}A}\hoch{B}}{}=\\
 & = & \Omega_{\bs{\mc{M}}A}\hoch{B}(\xboson,0)+2\sum_{n\geq1}\frac{1}{(n+1)!}x^{\bs{\mc{M}}_{1}}\cdots x^{\bs{\mc{M}}_{n}}\bei{\partial_{\bs{\mc{M}}_{1}}\ldots\partial_{\bs{\mc{M}}_{n}}(x^{\bs{\mc{N}}}R_{\bs{\mc{N}}\bs{\mc{M}}A}\hoch{B})}{}\qquad\fussend\end{eqnarray*}
}} 

The calculation for the components of the vielbein is very similar\begin{eqnarray}
\lqn{\bei{\partial_{\bs{\mc{M}}_{1}}\ldots\partial_{\bs{\mc{M}}_{n}}E_{\bs{\mc{M}}_{n+1}}\hoch{A}}{}=}\nonumber \\
 & = & \underbrace{\bei{\partial_{(\bs{\mc{M}}_{1}}\ldots\partial_{\bs{\mc{M}}_{n}}E_{\bs{\mc{M}}_{n+1})}\hoch{A}}{}}_{=0\:(\ref{eq:gauge-fix-sym-vielbein})}+\frac{2}{n+1}\sum_{i=1}^{n}\bei{\partial_{\bs{\mc{M}}_{1}}\ldots\partial_{[\bs{\mc{M}}_{i}|}\ldots\partial_{\bs{\mc{M}}_{n}}E_{|\bs{\mc{M}}_{n+1}]}\hoch{A}}{}=\\
 & = & \frac{2}{n+1}\sum_{i=1}^{n}\bei{\partial_{\bs{\mc{M}}_{1}}\ldots\partial_{\bs{\mc{M}}_{i-1}}\partial_{\bs{\mc{M}}_{i+1}}\ldots\partial_{\bs{\mc{M}}_{n}}\left(T_{\bs{\mc{M}}_{i}\bs{\mc{M}}_{n+1}}\hoch{A}+E_{[\bs{\mc{M}}_{i}}\hoch{B}\Omega_{\bs{\mc{M}}_{n+1}]B}\hoch{A}\right)}{}=\\
 & = & \frac{n}{n+1}\bei{\partial_{(\bs{\mc{M}}_{1}}\ldots\partial_{\bs{\mc{M}}_{n-1}|}\left(2T_{|\bs{\mc{M}}_{n})\bs{\mc{M}}_{n+1}}\hoch{A}+E_{|\bs{\mc{M}}_{n})}\hoch{B}\Omega_{\bs{\mc{M}}_{n+1}B}\hoch{A}-E_{\bs{\mc{M}}_{n+1}}\hoch{B}\Omega_{|\bs{\mc{M}}_{n})B}\hoch{A}\right)}{}\end{eqnarray}
For the second and third term in the bracket we can use (\ref{eq:gauge-fix-sym-vielbein})
and (\ref{eq:gauge-fix-sym-Omega}) again, so that the third term
will vanish while from the second term we get a contribution only
when all derivatives act on the connection, because $\bei{E_{\bs{\mc{M}}_{n}}\hoch{B}}{}=\delta_{\bs{\mc{M}}_{n}}\hoch{B}$.
Using (\ref{eq:connectionInTermsOfR}), we arrive at \vRam{.84}{\begin{eqnarray}
\lqn{\:\bei{\partial_{\bs{\mc{M}}_{1}}\ldots\partial_{\bs{\mc{M}}_{n}}E_{\bs{\mc{M}}_{n+1}}\hoch{A}}{}=\qquad\qquad\qquad\qquad\qquad\qquad\qquad\qquad\forall n\geq1}\nonumber \\
 & = & \frac{2n}{n+1}\bei{\partial_{(\bs{\mc{M}}_{1}}\ldots\partial_{\bs{\mc{M}}_{n-1}}T_{\bs{\mc{M}}_{n})\bs{\mc{M}}_{n+1}}\hoch{A}}{}+\frac{2(n-1)}{n+1}\delta_{(\bs{\mc{M}}_{1}}\hoch{\bs{\mc{B}}}\bei{\partial_{\bs{\mc{M}}_{2}}\ldots\partial_{\bs{\mc{M}}_{n-1}}R_{\bs{\mc{M}}_{n})\bs{\mc{M}}_{n+1}\bs{\mc{B}}}\hoch{A}}{}\label{eq:vielbeinInTermsOfTuR}\end{eqnarray}
 }\\
In particular we get for $n=1$\begin{eqnarray}
\bei{\partial_{\bs{\mc{M}}}E_{\bs{\mc{N}}}\hoch{A}}{} & = & \bei{T_{\bs{\mc{MN}}}\hoch{A}}{},\qquad\bei{\partial_{\bs{\mc{M}}}\Omega_{\bs{\mc{N}}A}\hoch{B}}{}=\bei{R_{\bs{\mc{MN}}A}\hoch{B}}{}\end{eqnarray}

The higher $\xbothtetas$-components of the vielbein and connection
parts with bosonic form index ($E_{m}\hoch{A}$ and $\Omega_{mA}\hoch{B}$)
can likewise be expressed in terms of torsion and curvature:\begin{eqnarray}
\bei{\partial_{\bs{\mc{M}}_{1}}\ldots\partial_{\bs{\mc{M}}_{n}}\Omega_{mA}\hoch{B}}{} & = & \frac{2}{n}\sum_{i=1}^{n}\bei{\partial_{\bs{\mc{M}}_{1}}\ldots\partial_{[\bs{\mc{M}}_{i}|}\ldots\partial_{\bs{\mc{M}}_{n}}\Omega_{|m]A}\hoch{B}}{}+\partial_{m}\underbrace{\bei{\partial_{(\bs{\mc{M}}_{1}}\ldots\partial_{\bs{\mc{M}}_{n-1}}\Omega_{\bs{\mc{M}}_{n})A}\hoch{B}}{}}_{=0\:(\ref{eq:gauge-fix-sym-Omega})}=\\
 & = & 2\bei{\partial_{(\bs{\mc{M}}_{1}}\ldots\partial_{\bs{\mc{M}}_{n-1}|}\left(R_{|\bs{\mc{M}}_{n})mA}\hoch{B}+\frac{1}{2}\Omega_{|\bs{\mc{M}}_{n})A}\hoch{C}\Omega_{mC}\hoch{B}-\frac{1}{2}\Omega_{mA}\hoch{C}\Omega_{|\bs{\mc{M}}_{n})C}\hoch{B}\right)}{}\qquad\quad\end{eqnarray}
\begin{equation}
\stackrel{(\ref{eq:gauge-fix-sym-Omega})}{\dann}\boxed{\bei{\partial_{\bs{\mc{M}}_{1}}\ldots\partial_{\bs{\mc{M}}_{n}}\Omega_{mA}\hoch{B}}{}=2\bei{\partial_{(\bs{\mc{M}}_{1}}\ldots\partial_{\bs{\mc{M}}_{n-1}|}R_{|\bs{\mc{M}}_{n})mA}\hoch{B}}{}}\:\forall n\geq1\label{eq:bosConnectionInTermsOfR}\end{equation}
Although in contrast to (\ref{eq:connectionInTermsOfR}) we do not
have an $n$-dependent factor, we have in general $\partial_{\bs{\mc{M}}}\Omega_{mA}\hoch{B}\neq2R_{\bs{\mc{M}}mA}\hoch{B}$
away from $\xbothtetas=0$. The reason for this fact is the symmetrization
on the righthand side. Also we have $\Omega_{mA}\hoch{B}\neq2x^{\bs{\mc{M}}}R_{\bs{\mc{M}}mA}\hoch{B}$
for $\xbothtetas\neq0$. 

For the vielbein the situation is again similar:\begin{eqnarray}
\bei{\partial_{\bs{\mc{M}}_{1}}\ldots\partial_{\bs{\mc{M}}_{n}}E_{n}\hoch{A}}{} & = & \frac{2}{n}\sum_{i=1}^{n}\bei{\partial_{\bs{\mc{M}}_{1}}\ldots\partial_{[\bs{\mc{M}}_{i}|}\ldots\partial_{\bs{\mc{M}}_{n}}E_{|m]}\hoch{A}}{}+\underbrace{\partial_{m}\bei{\partial_{(\bs{\mc{M}}_{1}}\ldots\partial_{\bs{\mc{M}}_{n-1}}E_{\bs{\mc{M}}_{n})}\hoch{A}}{}}_{=0\:(\ref{eq:gauge-fix-sym-vielbein}),(\ref{eq:WZI})}=\\
 & = & 2\bei{\partial_{(\bs{\mc{M}}_{1}}\ldots\partial_{\bs{\mc{M}}_{n-1}|}\left(T_{|\bs{\mc{M}}_{n})m}\hoch{A}+\frac{1}{2}E_{|\bs{\mc{M}}_{n})}\hoch{B}\Omega_{mB}\hoch{A}-\frac{1}{2}E_{m}\hoch{B}\Omega_{|\bs{\mc{M}}_{n})B}\hoch{A}\right)}{}=\qquad\\
 & \ous{(\ref{eq:gauge-fix-sym-vielbein}),(\ref{eq:gauge-fix-sym-Omega})\qquad\quad}{=}{(\ref{eq:WZI})} & 2\bei{\partial_{(\bs{\mc{M}}_{1}}\ldots\partial_{\bs{\mc{M}}_{n-1}}T_{\bs{\mc{M}}_{n})m}\hoch{A}}{}+\delta_{(\bs{\mc{M}}_{n}}\hoch{B}\bei{\partial_{\bs{\mc{M}}_{1}}\ldots\partial_{\bs{\mc{M}}_{n-1})}\Omega_{mB}\hoch{A}}{}\end{eqnarray}
In particular for $n=1$ we get\begin{equation}
\boxed{\partial_{\bs{\mc{M}}}\bei{E_{m}\hoch{A}}{}=2\bei{T_{\bs{\mc{M}}m}\hoch{A}}{}+\delta_{\bs{\mc{M}}}\hoch{B}\bei{\Omega_{mB}\hoch{A}}{}}\label{eq:bosVielbeinInTermsOfTandOmega}\end{equation}
while for $n>1$ we can use (\ref{eq:bosConnectionInTermsOfR}) to
arrive at\begin{equation}
\boxed{\bei{\partial_{\bs{\mc{M}}_{1}}\ldots\partial_{\bs{\mc{M}}_{n}}E_{n}\hoch{A}}{}=2\bei{\partial_{(\bs{\mc{M}}_{1}}\ldots\partial_{\bs{\mc{M}}_{n-1}}T_{\bs{\mc{M}}_{n})m}\hoch{A}}{}+2\delta_{(\bs{\mc{M}}_{1}}\hoch{\bs{\mc{B}}}\bei{\partial_{\bs{\mc{M}}_{2}}\ldots\partial_{\bs{\mc{M}}_{n-1}}R_{\bs{\mc{M}}_{n})m\bs{\mc{B}}}\hoch{A}}{}}\:\forall n\geq2\label{eq:bosVielbeinInTermsOfTandR}\end{equation}
In practice we are given constraints on torsion and curvature components
with only flat indices. Rewriting the equations (\ref{eq:connectionInTermsOfR}),(\ref{eq:vielbeinInTermsOfTuR}),(\ref{eq:bosConnectionInTermsOfR}),(\ref{eq:bosVielbeinInTermsOfTandOmega})
and (\ref{eq:bosVielbeinInTermsOfTandR}) with flat components yields
the following rekursion realtions\index{rekursion realtions for vielbein and connection components}\index{components!of $\xbothtetas$-expansion}\vRam{1.01}{\begin{eqnarray}
\bei{\partial_{\bs{\mc{M}}_{1}}\ldots\partial_{\bs{\mc{M}}_{n}}\Omega_{\bs{\mc{M}}_{n+1}A}\hoch{B}}{} & = & \frac{2n}{n+1}\bei{\delta_{(\bs{\mc{M}}_{n}}\hoch{\bs{\mc{C}}}\partial_{\bs{\mc{M}}_{1}}\ldots\partial_{\bs{\mc{M}}_{n-1})}(E_{\bs{\mc{M}}_{n+1}}\hoch{D}R_{\bs{\mc{C}}DA}\hoch{B})}{}\quad\forall n\geq1\label{eq:connectionInTermsOfRII}\\
\bei{\partial_{\bs{\mc{M}}_{1}}\ldots\partial_{\bs{\mc{M}}_{n}}E_{\bs{\mc{M}}_{n+1}}\hoch{A}}{} & = & \frac{2n}{n+1}\bei{\delta_{(\bs{\mc{M}}_{n}}\hoch{\bs{\mc{C}}}\partial_{\bs{\mc{M}}_{1}}\ldots\partial_{\bs{\mc{M}}_{n-1})}(E_{\bs{\mc{M}}_{n+1}}\hoch{D}T_{\bs{\mc{C}}D}\hoch{A})}{}+\qquad(\forall n\geq1)\nonumber \\
 &  & +\frac{2(n-1)}{n+1}\delta_{(\bs{\mc{M}}_{n-1}}\hoch{\bs{\mc{C}}}\delta_{\bs{\mc{M}}_{n}}\hoch{\bs{\mc{B}}}\bei{\partial_{\bs{\mc{M}}_{1}}\ldots\partial_{\bs{\mc{M}}_{n-2})}(E_{\bs{\mc{M}}_{n+1}}\hoch{D}R_{\bs{\mc{C}}D\bs{\mc{B}}}\hoch{A})}{}\label{eq:vielbeinInTermsOfTandRII}\\
\bei{\partial_{\bs{\mc{M}}_{1}}\ldots\partial_{\bs{\mc{M}}_{n}}\Omega_{mA}\hoch{B}}{} & = & 2\delta_{(\bs{\mc{M}}_{n}}\hoch{\bs{\mc{C}}}\bei{\partial_{\bs{\mc{M}}_{1}}\ldots\partial_{\bs{\mc{M}}_{n-1})}(E_{m}\hoch{D}R_{\bs{\mc{C}}DA}\hoch{B})}{}\:\forall n\geq1\label{eq:bosConnectionInTermsOfRII}\\
\partial_{\bs{\mc{M}}}\bei{E_{m}\hoch{A}}{} & = & 2\delta_{\bs{\mc{M}}}\hoch{\bs{\mc{C}}}\bei{E_{m}\hoch{D}T_{\bs{\mc{C}}D}\hoch{A}}{}+\delta_{\bs{\mc{M}}}\hoch{\bs{\mc{B}}}\bei{\Omega_{m\bs{\mc{B}}}\hoch{A}}{}\label{eq:bosVielbeinInTermsOfTandOmegaII}\\
\bei{\partial_{\bs{\mc{M}}_{1}}\ldots\partial_{\bs{\mc{M}}_{n}}E_{n}\hoch{A}}{} & = & 2\delta_{(\bs{\mc{M}}_{n}}\hoch{\bs{\mc{C}}}\bei{\partial_{\bs{\mc{M}}_{1}}\ldots\partial_{\bs{\mc{M}}_{n-1})}(E_{m}\hoch{D}T_{\bs{\mc{C}}D}\hoch{A})}{}+\nonumber \\
 &  & +2\delta_{(\bs{\mc{M}}_{n}}\hoch{\bs{\mc{B}}}\delta_{\bs{\mc{M}}_{n-1}}\hoch{\bs{\mc{C}}}\bei{\partial_{\bs{\mc{M}}_{1}}\ldots\partial_{\bs{\mc{M}}_{n-2})}(E_{m}\hoch{D}R_{\bs{\mc{C}}D\bs{\mc{B}}}\hoch{A})}{}\:\forall n\geq2\label{eq:bosVielbeinInTermsOfTandRII}\end{eqnarray}
} \\
Let us do the first steps of the iteration, in order to see what is
happening:\begin{eqnarray}
\hspace{-.8cm}n=0:\quad\bei{\Omega_{\bs{\mc{M}}A}\hoch{B}}{} & = & 0,\quad\bei{\Omega_{mA}\hoch{B}}{}\equiv\omega_{mA}\hoch{B}\frem{,\quad\bei{\Omega_{m\bs{\alpha}}\hoch{\bs{\beta}}}{}\equiv\frac{1}{4}\omega_{mab}\gamma^{ab}\tief{\bs{\alpha}}\hoch{\bs{\beta}},\quad\bei{\Omega_{m\hat{\bs{\alpha}}}\hoch{\hat{\bs{\beta}}}}{}\equiv\frac{1}{4}\omega_{mab}\gamma^{ab}\tief{\hat{\bs{\alpha}}}\hoch{\hat{\bs{\beta}}}}\qquad\\
\bei{E_{\bs{\mc{M}}}\hoch{A}}{} & = & \delta_{\bs{\mc{M}}}\hoch{A},\quad\bei{E_{m}\hoch{a}}{}\equiv e_{m}\hoch{a},\quad\bei{E_{m}\hoch{\bs{\mc{A}}}}{}\equiv\psi_{m}\hoch{\bs{\mc{A}}}\\
\hspace{-.8cm}n=1:\quad\bei{\partial_{\bs{\mc{M}}_{1}}\Omega_{\bs{\mc{M}}_{2}A}\hoch{B}}{} & = & \delta_{\bs{\mc{M}}_{1}}\hoch{\bs{\mc{C}}}\delta_{\bs{\mc{M}}_{2}}\hoch{\bs{\mc{D}}}\bei{R_{\bs{\mc{C}}\bs{\mc{D}}A}\hoch{B}}{},\quad\bei{\partial_{\bs{\mc{M}}}\Omega_{nA}\hoch{B}}{}=2\delta_{\bs{\mc{M}}}\hoch{\bs{\mc{C}}}e_{n}\hoch{d}\bei{R_{\bs{\mc{C}}dA}\hoch{B}}{}+2\delta_{\bs{\mc{M}}}\hoch{\bs{\mc{C}}}\psi_{n}\hoch{\bs{\mc{D}}}\bei{R_{\bs{\mc{C}}\bs{\mc{D}}A}\hoch{B}}{}\qquad\\
\bei{\partial_{\bs{\mc{M}}_{1}}E_{\bs{\mc{M}}_{2}}\hoch{A}}{} & = & \delta_{\bs{\mc{M}}_{1}}\hoch{\bs{\mc{C}}}\delta_{\bs{\mc{M}}_{2}}\hoch{\bs{\mc{D}}}\bei{T_{\bs{\mc{C}}\bs{\mc{D}}}\hoch{A}}{},\quad\bei{\partial_{\bs{\mc{M}}}E_{n}\hoch{a}}{}=2\delta_{\bs{\mc{M}}}\hoch{\bs{\mc{C}}}e_{n}\hoch{d}\bei{T_{\bs{\mc{C}}d}\hoch{a}}{}+2\delta_{\bs{\mc{M}}}\hoch{\bs{\mc{C}}}\psi_{n}\hoch{\bs{\mc{D}}}\bei{T_{\bs{\mc{C}}\bs{\mc{D}}}\hoch{a}}{}\nonumber \\
 &  & \qquad\bei{\partial_{\bs{\mc{M}}}E_{n}\hoch{\bs{\mc{A}}}}{}=2\delta_{\bs{\mc{M}}}\hoch{\bs{\mc{C}}}e_{n}\hoch{d}\bei{T_{\bs{\mc{C}}d}\hoch{\bs{\mc{A}}}}{}+2\delta_{\bs{\mc{M}}}\hoch{\bs{\mc{C}}}\psi_{n}\hoch{\bs{\mc{D}}}\bei{T_{\bs{\mc{C}}\bs{\mc{D}}}\hoch{\bs{\mc{A}}}}{}+\delta_{\bs{\mc{M}}}\hoch{\bs{\mc{B}}}\omega_{n\bs{\mc{B}}}\hoch{\bs{\mc{A}}}\\
\hspace{-.8cm}n=2:\bei{\partial_{\bs{\mc{M}}_{1}}\partial_{\bs{\mc{M}}_{2}}\Omega_{\bs{\mc{M}}_{3}A}\hoch{B}}{} & = & \frac{4}{3}\delta_{(\bs{\mc{M}}_{2}|}\hoch{\bs{\mc{C}}}\delta_{|\bs{\mc{M}}_{1})}\hoch{\bs{\mc{E}}}\delta_{\bs{\mc{M}}_{3}}\hoch{\bs{\mc{F}}}\bei{T_{\bs{\mc{E}}\bs{\mc{F}}}\hoch{D}}{}\bei{R_{\bs{\mc{C}}DA}\hoch{B}}{}+\frac{4}{3}\delta_{(\bs{\mc{M}}_{2}|}\hoch{\bs{\mc{C}}}\delta_{\bs{\mc{M}}_{3}}\hoch{\bs{\mc{D}}}\partial_{|\bs{\mc{M}}_{1})}R_{\bs{\mc{C}}\bs{\mc{D}}A}\hoch{B}\\
\bei{\partial_{\bs{\mc{M}}_{1}}\partial_{\bs{\mc{M}}_{2}}\Omega_{mA}\hoch{B}}{} & = & 2\delta_{(\bs{\mc{M}}_{2}}\hoch{\bs{\mc{C}}}\left(2\delta_{\bs{\mc{M}}_{1})}\hoch{\bs{\mc{E}}}e_{n}\hoch{f}\bei{T_{\bs{\mc{E}}f}\hoch{D}}{}+2\delta_{\bs{\mc{M}}_{1})}\hoch{\bs{\mc{E}}}\psi_{m}\hoch{\bs{\mc{F}}}\bei{T_{\bs{\mc{E}}\bs{\mc{F}}}\hoch{D}}{}+\delta_{\bs{\mc{M}}_{1})}\hoch{\bs{\mc{E}}}\omega_{m\bs{\mc{E}}}\hoch{D}\right)\bei{R_{\bs{\mc{C}}DA}\hoch{B}}{}+\nonumber \\
 &  & +2\delta_{(\bs{\mc{M}}_{2}}\hoch{\bs{\mc{C}}}e_{m}\hoch{d}\bei{\partial_{\bs{\mc{M}}_{1})}R_{\bs{\mc{C}}dA}\hoch{B}}{}+2\delta_{(\bs{\mc{M}}_{2}}\hoch{\bs{\mc{C}}}\psi_{m}\hoch{\bs{\mc{D}}}\bei{\partial_{\bs{\mc{M}}_{1})}R_{\bs{\mc{C}}\bs{\mc{D}}A}\hoch{B}}{}\\
\bei{\partial_{\bs{\mc{M}}_{1}}\partial_{\bs{\mc{M}}_{2}}E_{\bs{\mc{M}}_{3}}\hoch{A}}{} & = & \frac{4}{3}\delta_{(\bs{\mc{M}}_{2}}\hoch{\bs{\mc{C}}}\delta_{\bs{\mc{M}}_{1})}\hoch{\bs{\mc{E}}}\delta_{\bs{\mc{M}}_{3}}\hoch{\bs{\mc{F}}}\bei{T_{\bs{\mc{E}}\bs{\mc{F}}}\hoch{D}}{}\bei{T_{\bs{\mc{C}}D}\hoch{A}}{}+\frac{4}{3}\delta_{(\bs{\mc{M}}_{2}}\hoch{\bs{\mc{C}}}\delta_{\bs{\mc{M}}_{3}}\hoch{\bs{\mc{D}}}\bei{\partial_{\bs{\mc{M}}_{1})}T_{\bs{\mc{C}}\bs{\mc{D}}}\hoch{A}}{}+\nonumber \\
 &  & +\frac{2}{3}\delta_{(\bs{\mc{M}}_{1}}\hoch{\bs{\mc{C}}}\delta_{\bs{\mc{M}}_{2})}\hoch{\bs{\mc{B}}}\delta_{\bs{\mc{M}}_{3}}\hoch{\bs{\mc{D}}}\bei{R_{\bs{\mc{C}}\bs{\mc{D}}\bs{\mc{B}}}\hoch{A}}{}\\
\bei{\partial_{\bs{\mc{M}}_{1}}\partial_{\bs{\mc{M}}_{2}}E_{n}\hoch{A}}{} & = & 2\delta_{(\bs{\mc{M}}_{2}}\hoch{\bs{\mc{C}}}\left(2\delta_{\bs{\mc{M}}_{1})}\hoch{\bs{\mc{E}}}e_{m}\hoch{f}\bei{T_{\bs{\mc{E}}f}\hoch{D}}{}+2\delta_{\bs{\mc{M}}_{1})}\hoch{\bs{\mc{E}}}\psi_{m}\hoch{\bs{\mc{F}}}\bei{T_{\bs{\mc{E}}\bs{\mc{F}}}\hoch{D}}{}+\delta_{\bs{\mc{M}}_{1})}\hoch{\bs{\mc{E}}}\omega_{m\bs{\mc{E}}}\hoch{D}\right)\bei{T_{\bs{\mc{C}}D}\hoch{A}}{}+\nonumber \\
 &  & +2\delta_{(\bs{\mc{M}}_{2}}\hoch{\bs{\mc{C}}}e_{m}\hoch{d}\bei{\partial_{\bs{\mc{M}}_{1})}T_{\bs{\mc{C}}d}\hoch{A}}{}+2\delta_{(\bs{\mc{M}}_{2}}\hoch{\bs{\mc{C}}}\psi_{m}\hoch{\bs{\mc{D}}}\bei{\partial_{\bs{\mc{M}}_{1})}T_{\bs{\mc{C}}\bs{\mc{D}}}\hoch{A}}{}+\nonumber \\
 &  & +2\delta_{(\bs{\mc{M}}_{2}}\hoch{\bs{\mc{B}}}\delta_{\bs{\mc{M}}_{1})}\hoch{\bs{\mc{C}}}e_{m}\hoch{d}\bei{R_{\bs{\mc{C}}d\bs{\mc{B}}}\hoch{A}}{}+2\delta_{(\bs{\mc{M}}_{2}}\hoch{\bs{\mc{B}}}\delta_{\bs{\mc{M}}_{1})}\hoch{\bs{\mc{C}}}\psi_{m}\hoch{\bs{\mc{D}}}\bei{R_{\bs{\mc{C}}\bs{\mc{D}}\bs{\mc{B}}}\hoch{A}}{}\\
 & \ddots\nonumber \end{eqnarray}
Apparently this iteration gets very involved for higher orders, but
in principle we can express every supervielbein component and superconnection
component in terms of the bosonic vielbein, the gravitinos, the bosonic
connection and the torsion and curvature components. Note finally
that the components $\Gamma_{\bs{\mc{M}}N}\hoch{K}$ of the superspace
connection do not vanish at leading order like the structure group
connection. Instead we find because of $\Gamma_{MN}\hoch{K}=\left(\partial_{M}E_{N}\hoch{C}+E_{N}\hoch{B}\Omega_{MB}\hoch{C}\right)E_{C}\hoch{K}$
for the leading order that \begin{eqnarray}
\bei{\Gamma_{\bs{\mc{M}}N}\hoch{K}}{} & \stackrel{(\ref{eq:WZII})}{=} & \bei{\partial_{\bs{\mc{M}}}E_{N}\hoch{C}}{}\bei{E_{C}\hoch{K}}{}\end{eqnarray}
Using some of the equations above, this implies in particular \begin{eqnarray}
\bei{\Gamma_{\bs{\mc{M}}n}\hoch{K}}{} & = & 2\bei{T_{\bs{\mc{M}}n}\hoch{a}}{}\bei{E_{a}\hoch{K}}{}+2\bei{T_{\bs{\mc{M}}n}\hoch{\bs{\mc{A}}}}{}\delta_{\bs{\mc{A}}}\hoch{K}+\delta_{\bs{\mc{M}}}\hoch{\bs{\mc{B}}}\omega_{n\bs{\mc{B}}}\hoch{\bs{\mc{A}}}\delta_{\bs{\mc{A}}}\hoch{K}=\label{eq:GammaWZgaugeI}\\
 & = & 2\delta_{\bs{\mc{M}}}\hoch{\bs{\mc{C}}}e_{n}\hoch{d}\bei{T_{\bs{\mc{C}}d}\hoch{a}}{}\bei{E_{a}\hoch{K}}{}+2\delta_{\bs{\mc{M}}}\hoch{\bs{\mc{C}}}\psi_{n}\hoch{\bs{\mc{D}}}\bei{T_{\bs{\mc{C}}\bs{\mc{D}}}\hoch{a}}{}\bei{E_{a}\hoch{K}}{}+\nonumber \\
 &  & +2\delta_{\bs{\mc{M}}}\hoch{\bs{\mc{C}}}e_{n}\hoch{d}\bei{T_{\bs{\mc{C}}d}\hoch{\bs{\mc{A}}}}{}\delta_{\bs{\mc{A}}}\hoch{K}+2\delta_{\bs{\mc{M}}}\hoch{\bs{\mc{C}}}\psi_{n}\hoch{\bs{\mc{D}}}\bei{T_{\bs{\mc{CD}}}\hoch{\bs{\mc{A}}}}{}\delta_{\bs{\mc{A}}}\hoch{K}+\delta_{\bs{\mc{M}}}\hoch{\bs{\mc{B}}}\omega_{n\bs{\mc{B}}}\hoch{\bs{\mc{A}}}\delta_{\bs{\mc{A}}}\hoch{K}\\
\bei{\Gamma_{\bs{\mc{MN}}}\hoch{K}}{} & = & \delta_{\bs{\mc{M}}}\hoch{\bs{\mc{C}}}\delta_{\bs{\mc{N}}}\hoch{\bs{\mc{D}}}\bei{T_{\bs{\mc{C}}\bs{\mc{D}}}\hoch{a}}{}\bei{E_{a}\hoch{K}}{}+\delta_{\bs{\mc{M}}}\hoch{\bs{\mc{C}}}\delta_{\bs{\mc{N}}}\hoch{\bs{\mc{D}}}\bei{T_{\bs{\mc{C}}\bs{\mc{D}}}\hoch{\bs{\mc{A}}}}{}\delta_{\bs{\mc{A}}}\hoch{K}\label{eq:GammaWZgaugeII}\end{eqnarray}
 \rem{Let us finally add what this means for the superspace connection
$\Gamma_{MN}\hoch{K}=\left(\partial_{M}E_{N}\hoch{C}+E_{N}\hoch{B}\Omega_{MB}\hoch{C}\right)E_{C}\hoch{K}$:\begin{eqnarray}
\bei{\Gamma_{\bs{\mc{M}}N}\hoch{K}}{} & = & \bei{\partial_{\bs{\mc{M}}}E_{N}\hoch{C}}{}\bei{E_{C}\hoch{K}}{}+\bei{\Omega_{\bs{\mc{M}}N}\hoch{K}}{}=\\
 & \stackrel{(\ref{eq:WZII})}{=} & \bei{\partial_{\bs{\mc{M}}}E_{N}\hoch{C}}{}\bei{E_{C}\hoch{K}}{}\\
\bei{\Gamma_{\bs{\mc{M}}n}\hoch{K}}{} & = & 2\bei{\partial_{[\bs{\mc{M}}}E_{n]}\hoch{C}}{}\bei{E_{C}\hoch{K}}{}=\\
 & = & 2\bei{T_{\bs{\mc{M}}n}\hoch{K}}{}+\bei{\Omega_{n\bs{\mc{M}}}\hoch{K}}{}\\
\bei{\Gamma_{\bs{\mc{MN}}}\hoch{K}}{} & = & \bei{\partial_{\bs{\mc{M}}}E_{\bs{\mc{N}}}\hoch{C}}{}\bei{E_{C}\hoch{K}}{}=\\
 & \stackrel{(\ref{eq:gauge-fix-sym-vielbein})}{=} & \bei{T_{\bs{\mc{MN}}}\hoch{K}}{}\end{eqnarray}
\begin{eqnarray}
\bei{\Lie_{\vecfull{\xi}}\Gamma_{\bs{\mc{M}}N}\hoch{K}}{} & = & \left(\bei{\partial_{\bs{\mc{M}}}\Liecov_{\vecfull{\xi}}E_{N}\hoch{C}}{}+\bei{E_{N}\hoch{B}}{}\bei{\Lie_{\vecfull{\xi}}\Omega_{\bs{\mc{M}}B}\hoch{C}}{}\right)\bei{E_{C}\hoch{K}}{}+\nonumber \\
 &  & +\bei{\partial_{\bs{\mc{M}}}E_{N}\hoch{C}}{}\bei{\Liecov_{\vecfull{\xi}}E_{C}\hoch{K}}{}\end{eqnarray}
For a stabilizer-transformation (see later), the first line vanishes.
Let us distinguish in the second\begin{eqnarray}
\bei{\Lie_{\vecfull{\xi}}\Gamma_{\bs{\mc{M}}n}\hoch{K}}{} & = & \bei{\partial_{\bs{\mc{M}}}E_{n}\hoch{c}}{}\bei{\Liecov_{\vecfull{\xi}}E_{c}\hoch{K}}{}=\\
 & = & 2\bei{E_{n}\hoch{B}}{}\bei{T_{\bs{\mc{M}}B}\hoch{c}}{}\bei{\Liecov_{\vecfull{\xi}}E_{c}\hoch{K}}{}\\
\bei{\Lie_{\vecfull{\xi}}\Gamma_{\bs{\mc{M}}\bs{\mc{N}}}\hoch{K}}{} & = & \bei{T_{\bs{\mc{MN}}}\hoch{c}}{}\bei{\Liecov_{\vecfull{\xi}}E_{c}\hoch{K}}{}\end{eqnarray}
}

\section{Partial Gauge Fixing of the B-superfield}

Although the gauge fixing of the $B$-field is not necessary in order
to obtain the supergravity transformations, we will discuss it at
this place, as it is very similar to the gauge fixings of connection
and vielbein. Again we want to fix only the auxiliary gauge degrees
but leave the gauge freedom of the bosonic two-form. The B-field gauge
symmetry is of the form $B\To B+\de\Lambda$, with some one-form $\Lambda$.
Let us split the gauge transformation into three cases with different
index structures:\begin{eqnarray}
B_{\bs{\mc{M}}\bs{\mc{N}}} & \To & B_{\bs{\mc{M}}\bs{\mc{N}}}+\partial_{[\bs{\mc{M}}}\Lambda_{\bs{\mc{N}}]}\\
B_{\bs{\mc{M}}n} & \To & B_{\bs{\mc{M}}n}+\partial_{[\bs{\mc{M}}}\Lambda_{n]}\\
B_{mn} & \To & B_{mn}+\partial_{[m}\Lambda_{n]}\end{eqnarray}
In the $\xbothtetas$-expansion, we thus have \begin{eqnarray}
\bei{\partial_{\bs{\mc{K}}_{1}}\ldots\partial_{\bs{\mc{K}}_{p}}B_{\bs{\mc{M}}\bs{\mc{N}}}}{} & \To & \bei{\partial_{\bs{\mc{K}}_{1}}\ldots\partial_{\bs{\mc{K}}_{p}}B_{\bs{\mc{M}}\bs{\mc{N}}}}{}+\frac{1}{2}\bei{\partial_{\bs{\mc{K}}_{1}}\ldots\partial_{\bs{\mc{K}}_{p}}\partial_{\bs{\mc{M}}}\Lambda_{\bs{\mc{N}}}}{}-\frac{1}{2}\bei{\partial_{\bs{\mc{K}}_{1}}\ldots\partial_{\bs{\mc{K}}_{p}}\partial_{\bs{\mc{N}}}\Lambda_{\bs{\mc{M}}}}{}\\
\bei{\partial_{\bs{\mc{K}}_{1}}\ldots\partial_{\bs{\mc{K}}_{p}}B_{\bs{\mc{M}}n}}{} & \To & \bei{\partial_{\bs{\mc{K}}_{1}}\ldots\partial_{\bs{\mc{K}}_{p}}B_{\bs{\mc{M}}n}}{}+\frac{1}{2}\bei{\partial_{\bs{\mc{K}}_{1}}\ldots\partial_{\bs{\mc{K}}_{p}}\partial_{\bs{\mc{M}}}\Lambda_{n}}{}-\frac{1}{2}\partial_{n}\bei{\partial_{\bs{\mc{K}}_{1}}\ldots\partial_{\bs{\mc{K}}_{p}}\Lambda_{\bs{\mc{M}}}}{}\\
\bei{\partial_{\bs{\mc{K}}_{1}}\ldots\partial_{\bs{\mc{K}}_{p}}B_{mn}}{} & \To & \bei{\partial_{\bs{\mc{K}}_{1}}\ldots\partial_{\bs{\mc{K}}_{p}}B_{mn}}{}+\frac{1}{2}\partial_{m}\bei{\partial_{\bs{\mc{K}}_{1}}\ldots\partial_{\bs{\mc{K}}_{p}}\Lambda_{n}}{}-\frac{1}{2}\partial_{n}\bei{\partial_{\bs{\mc{K}}_{1}}\ldots\partial_{\bs{\mc{K}}_{p}}\Lambda_{m}}{}\end{eqnarray}
The gauge symmetries of the first two lines can be used to set $\bei{\partial_{(\bs{\mc{K}}_{1}}\ldots\partial_{\bs{\mc{K}}_{p}}B_{\bs{\mc{M}})\bs{\mc{N}}}}{}-\bei{\partial_{(\bs{\mc{K}}_{1}}\ldots\partial_{\bs{\mc{K}}_{p}}B_{\bs{\mc{N}})\bs{\mc{M}}}}{}$
and $\bei{\partial_{(\bs{\mc{K}}_{1}}\ldots\partial_{\bs{\mc{K}}_{p}}B_{\bs{\mc{M}})n}}{}$
to any value one likes. This fixes $\Lambda_{M}$ up to a de-Rham
closed term (as usual) and up to the bosonic gauge parameter $\bei{\Lambda_{m}}{}$.
We want to choose a gauge in such a way that for $p\geq1$, the higher
orders in the $\xbothtetas$-expansion can be expressed in a simple
way in terms of the $H$-flux $H_{MNK}\equiv\partial_{[M}B_{NK]}$.
To this end consider \begin{eqnarray}
\lqn{3p\cdot\partial_{(\bs{\mc{K}}_{1}}\ldots\partial_{\bs{\mc{K}}_{p-1}}H_{\bs{\mc{K}}_{p})\bs{\mc{M}}\bs{\mc{N}}}=}\nonumber \\
 & = & 3\sum_{i=1}^{p}\partial_{\bs{\mc{K}}_{1}}\ldots\partial_{[\bs{\mc{K}}_{i}|}\ldots\partial_{\bs{\mc{K}}_{p}}B_{|\bs{\mc{M}}\bs{\mc{N}}]}=\\
 & = & p\partial_{\bs{\mc{K}}_{1}}\ldots\partial_{\bs{\mc{K}}_{p}}B_{\bs{\mc{M}}\bs{\mc{N}}}-\sum_{i=1}^{p}\left(\partial_{\bs{\mc{K}}_{1}}\ldots\partial_{\bs{\mc{M}}}\ldots\partial_{\bs{\mc{K}}_{p}}B_{\bs{\mc{K}}_{i}\bs{\mc{N}}}-\partial_{\bs{\mc{K}}_{1}}\ldots\partial_{\bs{\mc{N}}}\ldots\partial_{\bs{\mc{K}}_{p}}B_{\bs{\mc{K}}_{i}\bs{\mc{M}}}\right)=\\
 & = & (p+2)\partial_{\bs{\mc{K}}_{1}}\ldots\partial_{\bs{\mc{K}}_{p}}B_{\bs{\mc{M}}\bs{\mc{N}}}-(p+1)\left(\partial_{(\bs{\mc{K}}_{1}}\ldots\partial_{\bs{\mc{K}}_{p}}B_{\bs{\mc{M}})\bs{\mc{N}}}-\partial_{(\bs{\mc{K}}_{1}}\ldots\partial_{\bs{\mc{K}}_{p}}B_{\bs{\mc{N}})\bs{\mc{M}}}\right)\label{eq:3pHI}\end{eqnarray}
This suggests to choose the gauge\begin{eqnarray}
\bei{\partial_{(\bs{\mc{K}}_{1}}\ldots\partial_{\bs{\mc{K}}_{p}}B_{\bs{\mc{M}})\bs{\mc{N}}}}{}-\bei{\partial_{(\bs{\mc{K}}_{1}}\ldots\partial_{\bs{\mc{K}}_{p}}B_{\bs{\mc{N}})\bs{\mc{M}}}}{} & \stackrel{!}{=} & 0\quad\forall p\label{eq:BgaugefixI}\end{eqnarray}
which fixes $\bei{\partial_{\bs{\mc{K}}_{1}}\ldots\partial_{\bs{\mc{K}}_{p}}\partial_{[\bs{\mc{M}}}\Lambda_{\bs{\mc{N}}]}}{}$
. The above equation is a trivial statement for $p$ equal or bigger
as the fermionic dimensions (i.e. 32 for a ten-dimensional spacetime
and type II), because the graded symmetrization of fermionic indices
(i.e. their antisymmetrization) vanishes when the number of indices
exceeds the dimension. On the other hand the statement is a very strong
one for $p=0$, where we simply get $\bei{B_{\bs{\mc{MN}}}}{}=0$.

The choice for the gauge in the case with mixed index structure is
not as obvious as above:

\begin{eqnarray}
\lqn{3p\cdot\partial_{(\bs{\mc{K}}_{1}}\ldots\partial_{\bs{\mc{K}}_{p-1}}H_{\bs{\mc{K}}_{p})\bs{\mc{M}}n}=}\nonumber \\
 & = & 3\sum_{i=1}^{p}\partial_{\bs{\mc{K}}_{1}}\ldots\partial_{[\bs{\mc{K}}_{i}|}\ldots\partial_{\bs{\mc{K}}_{p}}B_{|\bs{\mc{M}}n]}=\\
 & = & p\partial_{\bs{\mc{K}}_{1}}\ldots\partial_{\bs{\mc{K}}_{p}}B_{\bs{\mc{M}}n}-\sum_{i=1}^{p}\left(\partial_{\bs{\mc{K}}_{1}}\ldots\partial_{\bs{\mc{M}}}\ldots\partial_{\bs{\mc{K}}_{p}}B_{\bs{\mc{K}}_{i}n}-\partial_{\bs{\mc{K}}_{1}}\ldots\partial_{n}\ldots\partial_{\bs{\mc{K}}_{p}}B_{\bs{\mc{K}}_{i}\bs{\mc{M}}}\right)=\\
 & = & (p+1)\partial_{\bs{\mc{K}}_{1}}\ldots\partial_{\bs{\mc{K}}_{p}}B_{\bs{\mc{M}}n}-\left((p+1)\partial_{(\bs{\mc{K}}_{1}}\ldots\partial_{\bs{\mc{K}}_{p}}B_{\bs{\mc{M}})n}-p\partial_{n}\partial_{(\bs{\mc{K}}_{1}}\ldots\partial_{\bs{\mc{K}}_{p-1}}B_{\bs{\mc{K}}_{p})\bs{\mc{M}}}\right)\label{eq:3pHII}\end{eqnarray}
Instead of setting $\bei{\partial_{(\bs{\mc{K}}_{1}}\ldots\partial_{\bs{\mc{K}}_{p}}B_{\bs{\mc{M}})n}}{}$
to zero (which is of course a valid choice, too), it seems more convenient
here to choose\begin{eqnarray}
\bei{\partial_{(\bs{\mc{K}}_{1}}\ldots\partial_{\bs{\mc{K}}_{p}}B_{\bs{\mc{M}})n}}{} & = & \frac{p}{p+1}\partial_{n}\bei{\partial_{(\bs{\mc{K}}_{1}}\ldots\partial_{\bs{\mc{K}}_{p-1}}B_{\bs{\mc{K}}_{p})\bs{\mc{M}}}}{}\quad\forall p\label{eq:BgaugefixII}\end{eqnarray}
which fixes $\bei{\partial_{\bs{\mc{K}}_{1}}\ldots\partial_{\bs{\mc{K}}_{p}}\partial_{\bs{\mc{M}}}\Lambda_{n}}{}$.
Now we have fixed as much as we can and hope that the remaining components
behave in a nice way: \begin{eqnarray}
\lqn{3p\cdot\partial_{(\bs{\mc{K}}_{1}}\ldots\partial_{\bs{\mc{K}}_{p-1}}H_{\bs{\mc{K}}_{p})mn}=}\nonumber \\
 & = & 3\sum_{i=1}^{p}\partial_{\bs{\mc{K}}_{1}}\ldots\partial_{[\bs{\mc{K}}_{i}|}\ldots\partial_{\bs{\mc{K}}_{p}}B_{|mn]}=\\
 & = & p\partial_{\bs{\mc{K}}_{1}}\ldots\partial_{\bs{\mc{K}}_{p}}B_{mn}-\sum_{i=1}^{p}\left(\partial_{\bs{\mc{K}}_{1}}\ldots\partial_{m}\ldots\partial_{\bs{\mc{K}}_{p}}B_{\bs{\mc{K}}_{i}n}-\partial_{\bs{\mc{K}}_{1}}\ldots\partial_{n}\ldots\partial_{\bs{\mc{K}}_{p}}B_{\bs{\mc{K}}_{i}m}\right)=\\
 & = & p\partial_{\bs{\mc{K}}_{1}}\ldots\partial_{\bs{\mc{K}}_{p}}B_{mn}-p\left(\partial_{m}\partial_{(\bs{\mc{K}}_{1}}\ldots\partial_{\bs{\mc{K}}_{p-1}}B_{\bs{\mc{K}}_{p})n}-\partial_{n}\partial_{(\bs{\mc{K}}_{1}}\ldots\partial_{\bs{\mc{K}}_{p-1}}B_{\bs{\mc{K}}_{p})m}\right)\label{eq:3pHIII}\end{eqnarray}
Indeed, the gauge fixing condition (\ref{eq:BgaugefixII}) is fine
to remove the last terms for $\xbothtetas=0$. Plugging (\ref{eq:BgaugefixI})
in (\ref{eq:3pHI}) and (\ref{eq:BgaugefixII}) in (\ref{eq:3pHII})
and (\ref{eq:3pHIII}), we can express all auxiliary components of
$B$ in terms of some $H$-field components:\vRam{.85}{\begin{eqnarray}
\bei{\partial_{\bs{\mc{K}}_{1}}\ldots\partial_{\bs{\mc{K}}_{p}}B_{\bs{\mc{M}}\bs{\mc{N}}}}{} & = & \frac{3p}{p+2}\cdot\bei{\partial_{(\bs{\mc{K}}_{1}}\ldots\partial_{\bs{\mc{K}}_{p-1}}H_{\bs{\mc{K}}_{p})\bs{\mc{M}}\bs{\mc{N}}}}{}\quad\forall p\geq1,\qquad\bei{B_{\bs{\mc{MN}}}}{}=0\quad(p=0)\\
\bei{\partial_{\bs{\mc{K}}_{1}}\ldots\partial_{\bs{\mc{K}}_{p}}B_{\bs{\mc{M}}n}}{} & = & \frac{3p}{p+1}\cdot\bei{\partial_{(\bs{\mc{K}}_{1}}\ldots\partial_{\bs{\mc{K}}_{p-1}}H_{\bs{\mc{K}}_{p})\bs{\mc{M}}n}}{}\quad\forall p\geq1,\qquad\bei{B_{\bs{\mc{M}}n}}{}=0\quad(p=0)\\
\bei{\partial_{\bs{\mc{K}}_{1}}\ldots\partial_{\bs{\mc{K}}_{p}}B_{mn}}{} & = & \underbrace{3}_{\frac{3p}{p}}\bei{\partial_{(\bs{\mc{K}}_{1}}\ldots\partial_{\bs{\mc{K}}_{p-1}}H_{\bs{\mc{K}}_{p})mn}}{}\quad\forall p\geq1\end{eqnarray}
} Again, the constraints on the components of $H$ wil be given in
flat coordinates. Rewriting the above set of equations correspondingly,
produces derivatives acting on the vielbein. We thus get again a recursion
relation which is coupled to the recursion relation for the vielbein.

\section{Stabilizer}

In order to recover the supergravity transformations, we need to determine
those supergauge transformations which leave the Wess-Zumino-gauge
and the additional gauge fixing conditions untouched.

\subsection{Stabilizer\index{stabilizer!of the WZ gauge} of the Wess Zumino
gauge}

Let us start with the vielbein which was fixed to $\bei{E_{\bs{\mc{M}}}\hoch{A}}{}=\delta_{\bs{\mc{M}}}\hoch{A}$
(\ref{eq:WZ-gauge}), and remember the general transformation (\ref{eq:vielbeinTrafo})
\begin{eqnarray}
\delta E_{M}\hoch{A} & = & \underbrace{\partial_{M}\xi^{A}+\Omega_{MC}\hoch{A}\xi^{C}}_{\nabla_{M}\xi^{A}}+2\xi^{C}T_{CM}\hoch{A}+L_{B}\hoch{A}E_{M}\hoch{B}\end{eqnarray}
Let us denote the first components in the $\xbothtetas$-expansion
of the transformation parameters as follows\begin{eqnarray}
\xi^{A} & \equiv & \xi_{0}^{A}+x^{\bs{\mc{M}}}\xi_{\bs{\mc{M}}}^{A}+\ldots\\
L_{A}\hoch{B} & \equiv & L_{0\, A}\hoch{B}+x^{\bs{\mc{M}}}L_{\bs{\mc{M}}\, A}\hoch{B}+\ldots\end{eqnarray}
The $\xbothtetas=0$ component of $E_{\bs{\mc{M}}}\hoch{A}$ in the
WZ gauge then transforms as \begin{eqnarray}
\delta\bei{E_{\bs{\mc{M}}}\hoch{A}}{} & = & \xi_{\bs{\mc{M}}}^{A}+\underbrace{\bei{\Omega_{\bs{\mc{M}}C}\hoch{A}}{}}_{=0\:(\ref{eq:WZ-gauge-connection})}\xi_{0}^{C}+2\xi_{0}^{C}\bei{T_{C\bs{\mc{M}}}\hoch{A}}{}+L_{0\, B}\hoch{A}\underbrace{\bei{E_{\bs{\mc{M}}}\hoch{B}}{}}_{\delta_{\bs{\mc{M}}}\hoch{B}\:(\ref{eq:WZ-gauge})}=\\
 & = & \xi_{\bs{\mc{M}}}^{A}+2\xi_{0}^{C}\bei{T_{C\bs{\mc{M}}}\hoch{A}}{}+L_{0\,\bs{\mc{B}}}\hoch{A}\delta_{\bs{\mc{M}}}\hoch{\bs{\mc{B}}}\label{eq:deltaVielbeinBeiTetNull}\end{eqnarray}
In order to preserve the gauge of the vielbein, we thus need that
the above variation vanishes\index{stabilizer!of vielbein WZ gauge}\begin{equation}
\boxed{\xi_{\bs{\mc{M}}}^{A}=-\delta_{\bs{\mc{M}}}\hoch{\bs{\mc{B}}}\left(2\xi_{0}^{C}\bei{T_{C\bs{\mc{B}}}\hoch{A}}{}+L_{0\,\bs{\mc{B}}}\hoch{A}\right)}\label{eq:stabilizingCondWZxi}\end{equation}
This result is very general, without any restriction on the structure
group. In order to become more explicit, let us now assume that the
structure group is \textbf{block-diagonal} and split the index $A$
into $(a,\bs{\mc{A}})$. (Remember, the fermionic index might further
decay, e.g. for type II in ten dimensions into $\bs{\mc{A}}=(\bs{\alpha},\hat{\bs{\alpha}})$.)
The vector $\xi^{A}$ can then be written as \begin{eqnarray}
\xi^{a} & = & \xi_{0}^{a}-2x^{\bs{\mc{M}}}\delta_{\bs{\mc{M}}}\hoch{\bs{\mc{B}}}\xi_{0}^{C}\bei{T_{C\bs{\mc{B}}}\hoch{a}}{}+\mc{O}(\xbothtetas^{2})\label{eq:stabilizer-xia}\\
\xi^{\bs{\mc{A}}} & = & \xi_{0}^{\bs{\mc{A}}}-x^{\bs{\mc{M}}}\delta_{\bs{\mc{M}}}\hoch{\bs{\mc{B}}}\left(2\xi_{0}^{C}\bei{T_{C\bs{\mc{B}}}\hoch{\bs{\mc{A}}}}{}+L_{0\,\bs{\mc{B}}}\hoch{\bs{\mc{A}}}\right)+\mc{O}(\xbothtetas^{2})\label{eq:stabilizer-xialpha}\end{eqnarray}
In this appendix, we will not make use of any torsion constraints.
This will be done in the main part. 

The gauge fixing condition of the connection was $\bei{\Omega_{\bs{\mc{M}}A}\hoch{B}}{}=0$,
while its general gauge transformation reads (\ref{eq:connectionTrafo})\begin{equation}
\delta\Omega_{MA}\hoch{B}=2\xi^{K}R_{KMA}\hoch{B}-\partial_{M}L_{A}\hoch{B}-[L,\Omega_{M}]_{A}\hoch{B}\end{equation}
 The gauge is thus preserved if \index{stabilizer!of connection WZ gauge}\begin{equation}
\boxed{L_{\bs{\mc{M}}A}\hoch{B}\stackrel{!}{=}2\delta_{\bs{\mc{M}}}\hoch{\bs{\mc{D}}}\xi_{0}^{C}\bei{R_{C\bs{\mc{D}}A}\hoch{B}}{}}\label{eq:stabilizingCondWZL}\end{equation}
or \begin{eqnarray}
L_{A}\hoch{B}(\xboson,\xbothtetas) & = & L_{0\, A}\hoch{B}(\xboson)+2x^{\bs{\mc{M}}}\delta_{\bs{\mc{M}}}\hoch{\bs{\mc{D}}}\xi_{0}^{C}\bei{R_{C\bs{\mc{D}}A}\hoch{B}}{}+\mc{O}(\xbothtetas^{2})\label{eq:stabilizer-L}\end{eqnarray}

\subsection{Stabilizer of the additional gauge fixing conditions}

Remember the additional gauge fixing conditions (\ref{eq:gauge-fix-sym-vielbein})
and (\ref{eq:gauge-fix-sym-Omega})\begin{equation}
\bei{\partial_{(\bs{\mc{M}}_{1}}\ldots\partial_{\bs{\mc{M}}_{n}}E_{\bs{\mc{M}}_{n+1})}\hoch{A}}{}\stackrel{!}{=}0,\qquad\bei{\partial_{(\bs{\mc{M}}_{1}}\ldots\partial_{\bs{\mc{M}}_{n}}\Omega_{\bs{\mc{M}}_{n+1})A}\hoch{B}}{}\stackrel{!}{=}0\qquad\forall n\geq1\end{equation}
Stabilizing the first condition\begin{eqnarray}
\lqn{\delta\bei{\partial_{(\bs{\mc{M}}_{1}}\ldots\partial_{\bs{\mc{M}}_{n}}E_{\bs{\mc{M}}_{n+1})}\hoch{A}}{}=}\nonumber \\
 & = & \partial_{(\bs{\mc{M}}_{1}}\ldots\partial_{\bs{\mc{M}}_{n}|}\bei{\left(\partial_{|\bs{\mc{M}}_{n+1})}\xi^{A}+\Omega_{|\bs{\mc{M}}_{n+1})C}\hoch{A}\xi^{C}+2\xi^{C}T_{C|\bs{\mc{M}}_{n+1})}\hoch{A}+L_{\bs{\mc{B}}}\hoch{A}E_{|\bs{\mc{M}}_{n+1})}\hoch{\bs{\mc{B}}}\right)}{}=\\
 & = & \partial_{(\bs{\mc{M}}_{1}}\ldots\partial_{\bs{\mc{M}}_{n}|}\bei{\left(\partial_{|\bs{\mc{M}}_{n+1})}\xi^{A}+\delta_{|\bs{\mc{M}}_{n+1})}\hoch{\bs{\mc{B}}}\left(2\xi^{C}T_{C\bs{\mc{B}}}\hoch{A}+L_{\bs{\mc{B}}}\hoch{A}\right)\right)}{}\end{eqnarray}
implies \index{stabilizer!of additional vielbein gauge}\begin{equation}
\boxed{\bei{\partial_{\bs{\mc{M}}_{1}}\ldots\partial_{\bs{\mc{M}}_{n+1}}\xi^{A}}{}=-\partial_{(\bs{\mc{M}}_{1}}\ldots\partial_{\bs{\mc{M}}_{n}|}\bei{\left(2\xi^{C}T_{C\bs{\mc{B}}}\hoch{A}+L_{\bs{\mc{B}}}\hoch{A}\right)}{}\delta_{|\bs{\mc{M}}_{n+1})}\hoch{\bs{\mc{B}}}}\qquad\forall n\geq1\label{eq:stabilizingCondAdd-xi}\end{equation}
This is actually recursion relation again. For the second fermionic
derivative of the transformation parameter e.g., we get\begin{eqnarray}
\bei{\partial_{\bs{\mc{M}}_{1}}\partial_{\bs{\mc{M}}_{2}}\xi^{A}}{} & = & -2\xi_{(\bs{\mc{M}}_{1}|}^{C}\bei{T_{C|\bs{\mc{M}}_{2})}\hoch{A}}{}-2\xi_{0}^{C}\bei{\partial_{(\bs{\mc{M}}_{1}|}T_{C|\bs{\mc{M}}_{2})}\hoch{A}}{}-L_{(\bs{\mc{M}}_{1}\bs{\mc{M}}_{2})}\hoch{A}=\\
 & = & 2\xi_{0}^{C}\left(2\bei{T_{C(\bs{\mc{M}}_{1}|}\hoch{E}}{}\bei{T_{E|\bs{\mc{M}}_{2})}\hoch{A}}{}-\bei{\partial_{(\bs{\mc{M}}_{1}|}T_{C|\bs{\mc{M}}_{2})}\hoch{A}}{}-\bei{R_{C(\bs{\mc{M}}_{1}\bs{\mc{M}}_{2})}\hoch{A}}{}\right)+\nonumber \\
 &  & +2L_{0\,(\bs{\mc{M}}_{1}|}\hoch{C}\bei{T_{C|\bs{\mc{M}}_{2})}\hoch{A}}{}\qquad\end{eqnarray}
\rem{Term in brackets is similar to Bianchi-identity }Stabilizing
finally the second additional condition (the one on the connection)
\begin{eqnarray}
\lqn{\delta\bei{\partial_{(\bs{\mc{M}}_{1}}\ldots\partial_{\bs{\mc{M}}_{n}}\Omega_{\bs{\mc{M}}_{n+1})A}\hoch{B}}{}=}\nonumber \\
 & = & \bei{\partial_{(\bs{\mc{M}}_{1}}\ldots\partial_{\bs{\mc{M}}_{n}|}\left(2\xi^{K}R_{K|\bs{\mc{M}}_{n+1})A}\hoch{B}-\partial_{|\bs{\mc{M}}_{n+1})}L_{A}\hoch{B}-[L,\Omega_{|\bs{\mc{M}}_{n+1})}]_{A}\hoch{B}\right)}{}=\\
 & = & \bei{\partial_{(\bs{\mc{M}}_{1}}\ldots\partial_{\bs{\mc{M}}_{n}|}\left(2\xi^{K}R_{K|\bs{\mc{M}}_{n+1})A}\hoch{B}-\partial_{|\bs{\mc{M}}_{n+1})}L_{A}\hoch{B}\right)}{}\end{eqnarray}
implies \index{stabilizer!of additional connection gauge}\begin{equation}
\boxed{\bei{\partial_{\bs{\mc{M}}_{1}}\ldots\partial_{\bs{\mc{M}}_{n+1}}L_{A}\hoch{B}}{}=2\bei{\partial_{(\bs{\mc{M}}_{1}}\ldots\partial_{\bs{\mc{M}}_{n}|}\left(\xi^{C}R_{C\bs{\mc{D}}A}\hoch{B}\right)}{}\delta_{|\bs{\mc{M}}_{n+1})}\hoch{\bs{\mc{D}}}}\qquad\forall n\geq1\label{eq:stabilizingCondAddL}\end{equation}
Like above, this is a recursion relation, starting with the second
fermionic derivative\begin{eqnarray*}
\bei{\partial_{\bs{\mc{M}}_{1}}\partial_{\bs{\mc{M}}_{2}}L_{A}\hoch{B}}{} & = & 2\xi_{(\bs{\mc{M}}_{1}|}^{C}\bei{R_{C|\bs{\mc{M}}_{2})A}\hoch{B}}{}+2\xi_{0}^{C}\bei{\partial_{(\bs{\mc{M}}_{1}|}R_{C|\bs{\mc{M}}_{2})A}\hoch{B}}{}=\\
 & = & 2\xi_{0}^{C}\left(-2\bei{T_{C(\bs{\mc{M}}_{1}|}\hoch{E}}{}\bei{R_{E|\bs{\mc{M}}_{2})A}\hoch{B}}{}+\bei{\partial_{(\bs{\mc{M}}_{1}|}R_{C|\bs{\mc{M}}_{2})A}\hoch{B}}{}\right)-2L_{0\,(\bs{\mc{M}}_{1}|}\hoch{C}\bei{R_{C|\bs{\mc{M}}_{2})A}\hoch{B}}{}\end{eqnarray*}
\rem{Again: the term in brackets is similar to Bianchi-identity }The
two conditions (\ref{eq:stabilizingCondAdd-xi}) and (\ref{eq:stabilizingCondAddL})
are restricting only terms of order 2 and higher in $\xbothtetas$
of the transformation parameters $\xi^{A}$ and $L_{A}\hoch{B}$ and
therefore do not affect our earlier result (\ref{eq:stabilizer-xia})-(\ref{eq:stabilizer-xialpha})
and (\ref{eq:stabilizer-L}) for the stabilizer of the WZ gauge.\rem{

\subsection{Restricted structure group}

We have seen above that the stablizer of the gauge fixing conditions
contains only the free parameters $\xi_{0}^{A}$ and $L_{0\, A}\hoch{B}$.
All higher orders in the $\xbothtetas$-expansion are fixed by the
conditions. Apart from some remarks, we did not assume any restrictions
of the structure group where $L_{0\, A}\hoch{B}$ takes its values.
In most of the applications, the structure group will be block diagonal
and won't mix between bosonic and fermionic and between the two different
fermionic indices. This is also the case in our application to the
Berkovits string in the main part. In addition, each of the three
blocks is restricted to be a sum of Lorentz and scale transformations
(being a stabilizer of the pure spinor constraint). At the same time,
the connection (there it is denoted by $\gemOm_{MA}\hoch{B}\equiv\diag(\check{\Omega}_{Ma}\hoch{b},\Omega_{M\bs{\alpha}}\hoch{\bs{\beta}},\hat{\Omega}_{M\hat{\bs{\alpha}}}\hoch{\hat{\bs{\beta}}})$
is Lie algebra valued, i.e. it is also block-diagonal and a sum of
Lorentz and scale transformations. There is, however, a subtle point:
the transformations of each block get at some point fixed by manually
demanding $\check{T}_{\bs{\alpha\beta}}\hoch{c}=\gamma_{\bs{\alpha\beta}}^{c},\quad\check{T}_{\hat{\bs{\alpha}}\hat{\bs{\beta}}}\hoch{c}=\gamma_{\hat{\bs{\alpha}}\hat{\bs{\beta}}}^{c}$.
The stabilizer of this gauge fixing has only one independent block.
The gamma-matrices $\gamma_{\bs{\alpha\beta}}^{c}$ and $\gamma_{\hat{\bs{\alpha}}\hat{\bs{\beta}}}^{c}$
should then be covariantly constant w.r.t. a corresponding Lie algebra
valued connection. The original connection $\gemOm_{MA}\hoch{B}$
is not any longer Lie algebra valued and should be split into a Lie-algebra
valued part (e.g. $\Omega_{MA}\hoch{B}$, $\hat{\Omega}_{MA}\hoch{B}$
or $\avOm_{MA}\hoch{B}$) and a difference tensor. The same problem
occurs for every restriction of the structure group where the corresponding
components of the connection do not vanish automatically. If, for
example, we decide to fix the dilatation, then the dilatation part
of the connection has to be seen as independent tensor. All the equations
in this appendix hold for the Lie algebra valued connection. In particular,
the definition of the supergauge transformation has to be modified,
if the structure group gets restricted.}

\subsection{Local Lorentz\index{Lorentz transformations!local $\sim$} transformations
as part of the stabilizer}

For a reasonable gauge fixing we should still have local\index{local Lorentz transformation}
Lorentz invariance and the bosonic diffeomorphism as part of the stabilizer
group. We recover the local structure group transformations, if we
set \begin{eqnarray}
\xi_{0}^{C} & = & 0\end{eqnarray}
which leads to \begin{eqnarray}
L_{A}\hoch{B}(\xboson,\xbothtetas) & = & L_{0\, A}\hoch{B}(\xboson)+\mc{O}(\xbothtetas^{2})\\
\xi^{a} & = & \mc{O}(\xbothtetas^{2})\\
\xi^{\bs{\mc{A}}} & = & -x^{\bs{\mc{M}}}\delta_{\bs{\mc{M}}}\hoch{\bs{\mc{B}}}L_{0\,\bs{\mc{B}}}\hoch{\bs{\mc{A}}}+\mc{O}(\xbothtetas^{2})\end{eqnarray}
The leading components of all superfields with flat indices obviously
then transform only under the local structure group transformation
$L_{0\, A}\hoch{B}$, because the coupled superdiffeomorphism affects
only higher orders in $\xbothtetas$. When acting on a more general
tensor of e.g. the form $t_{MA}^{NB}$, the coupled diffeomorphism
contributes via the matrix $(\nabla_{L}\xi^{K}+2\xi^{C}T_{CL}\hoch{K})$
acting on the curved indices (compare (\ref{eq:covLiederOnGenTensor})).
For the leading component, i.e. $\xbothtetas=0$, the nonvanishing
part of this matrix is just \begin{eqnarray}
\bei{-(\nabla_{\bs{\mc{K}}}\xi^{P}+2\xi^{D}T_{CK}\hoch{P})}{} & = & \delta_{\bs{\mc{K}}}\hoch{\bs{\mc{B}}}L_{0\,\bs{\mc{B}}}\hoch{\bs{\mc{A}}}\delta_{\bs{\mc{A}}}\hoch{\bs{\mc{P}}}\end{eqnarray}
In other words, the bosonic curved indices $m,n,\ldots$ do not transform,
while the fermionic curved indices $\bs{\mc{M}},\bs{\mc{N}},\ldots$
transform under the structure group. 

For the behaviour on first order in $\xbothtetas$, it is already
instructive to consider the action of the above transformation on
a scalar superfield like a dilaton superfield $\dil$: \begin{eqnarray}
\delta\dil & = & \xi^{C}\nabla_{C}\dil=-x^{\bs{\mc{M}}}\delta_{\bs{\mc{M}}}\hoch{\bs{\mc{B}}}L_{0\,\bs{\mc{B}}}\hoch{\bs{\mc{C}}}\nabla_{\bs{\mc{C}}}\dil+\mc{O}(\xbothtetas^{2})\end{eqnarray}
That means for the $\xbothtetas$-component $\dilo_{\bs{\mc{M}}}\equiv\bei{\nabla_{\bs{\mc{M}}}\dil}{}$,
that it transforms, as if ${\scriptstyle \bs{\mc{M}}}$ was a spinor
index.\begin{eqnarray}
\delta\lambda_{\bs{\mc{M}}} & = & \bei{\partial_{\bs{\mc{M}}}\delta(\dil)}{}=\\
 & = & -\delta_{\bs{\mc{M}}}\hoch{\bs{\mc{B}}}L_{0\,\bs{\mc{B}}}\hoch{\bs{\mc{C}}}\bei{\nabla_{\bs{\mc{C}}}\dil}{}=\\
 & = & -\delta_{\bs{\mc{M}}}\hoch{\bs{\mc{B}}}L_{0\,\bs{\mc{B}}}\hoch{\bs{\mc{C}}}\delta_{\bs{\mc{C}}}\hoch{\bs{\mc{N}}}\dilo_{\bs{\mc{N}}}\end{eqnarray}
Although it might seem intuitive that (curved) fermionic indices transform
under the structure group, it is important to note that this is only
due to the WZ-gauge, which couples part of the superdiffeomorphisms
to the local structure group transformations. Originally, the curved
fermionic index ${\scriptstyle \bs{\mc{M}}}$ does not transform under
structure group transformations.

\subsection{Bosonic diffeomorphism\index{diffeomorphism!bosonic $\sim$ as part of WZ-stabilizer}s
as part of the stabilizer}

The equations for the stabilizer are given in flat indices $\xi^{A}$.
We will need this to extract the local supersymmetry transformations.
But in order to see whether the transformation with parameters $\xi^{M}(\xfull)=(\xi_{0}^{m}(\xboson),0,0)$
and $\tilde{L}_{A}\hoch{B}=0$ (not $L_{A}\hoch{B}$, which has absorbed
part of the diffeomorphism), corresponding to bosonic diffeomorphisms,
is contained in the stabilizer, a change to curved indices is preferable.
Instead of using the vielbein to switch from flat to curved index,
we check this directly. The transformation of the vielbein components
with this parameter is \begin{eqnarray}
\delta\bei{E_{\bs{\mc{M}}}\hoch{A}}{} & = & \xi_{0}^{k}\partial_{k}\underbrace{\bei{E_{\bs{\mc{M}}}\hoch{A}}{}}_{\delta_{\bs{\mc{M}}}\hoch{A}}+\underbrace{\bei{\partial_{\bs{\mc{M}}}\xi^{K}}{}}_{=0}\bei{E_{K}\hoch{A}}{}=0\\
\bei{\delta\partial_{(\bs{\mc{M}}_{1}}\ldots\partial_{\bs{\mc{M}}_{n}}E_{\bs{\mc{M}}_{n+1})}\hoch{A}}{} & = & \bei{\partial_{(\bs{\mc{M}}_{1}}\ldots\partial_{\bs{\mc{M}}_{n}|}\left(\xi^{k}\partial_{k}E_{|\bs{\mc{M}}_{n+1})}\hoch{A}+\partial_{|\bs{\mc{M}}_{n+1})}\xi^{k}E_{k}\hoch{A}\right)}{}=\\
 & = & \bei{\xi^{k}\partial_{k}\partial_{(\bs{\mc{M}}_{1}}\ldots\partial_{\bs{\mc{M}}_{n}|}E_{|\bs{\mc{M}}_{n+1})}\hoch{A}}{}=0\end{eqnarray}
The same is true for the connection\begin{eqnarray}
\delta\bei{\Omega_{\bs{\mc{M}}A}\hoch{B}}{} & = & \xi_{0}^{k}\partial_{k}\bei{\Omega_{\bs{\mc{M}}A}\hoch{B}}{}+\underbrace{\bei{\partial_{\bs{\mc{M}}}\xi^{K}}{}}_{=0}\bei{\Omega_{KA}\hoch{B}}{}=0\\
\bei{\delta\partial_{(\bs{\mc{M}}_{1}}\ldots\partial_{\bs{\mc{M}}_{n}}\Omega_{\bs{\mc{M}}_{n+1})A}\hoch{B}}{} & = & \ldots=0\end{eqnarray}

\section{Local\index{local SUSY} SUSY\index{SUSY!local $\sim$}-transformation}

This section could actually be another subsection of the {}``stabilizer''
section. But as we have special interest in the local SUSY transformations,
we make it a seperate section.

\subsection{The transformation parameter}

The supersymmetry transformations are defined to be the set of transformations
within the stabilizer with \begin{equation}
\textrm{SUSY:}\quad\xi_{0}^{c}=L_{0\, A}\hoch{B}=0,\qquad0\neq\xi_{0}^{\bs{\mc{C}}}\equiv\eps^{\bs{\mc{C}}}\label{eq:SUSY-def}\end{equation}
From (\ref{eq:stabilizingCondWZxi}) and (\ref{eq:stabilizingCondWZL})
we thus get\begin{equation}
\xi_{\bs{\mc{M}}}\hoch{A}=-2\eps^{\bs{\mc{C}}}\bei{T_{\bs{\mc{C}}\bs{\mc{M}}}\hoch{A}}{},\qquad L_{\bs{\mc{M}}A}\hoch{B}=2\eps^{\bs{\mc{C}}}\bei{R_{\bs{\mc{CM}}A}\hoch{B}}{}\label{eq:SUSY-xi-L-cond}\end{equation}
 Or more explicitely (compare (\ref{eq:stabilizer-xia}),(\ref{eq:stabilizer-xialpha})
and (\ref{eq:stabilizer-L})):\begin{eqnarray}
\xi^{a}(\eps) & = & -2x^{\bs{\mc{M}}}\delta_{\bs{\mc{M}}}\hoch{\bs{\mc{D}}}\eps^{\bs{\mc{C}}}\bei{T_{\bs{\mc{C}}\bs{\mc{D}}}\hoch{a}}{}+\mc{O}(\xbothtetas^{2})\label{eq:SUSY-xia}\\
\xi^{\bs{\mc{A}}}(\eps) & = & \eps^{\bs{\mc{A}}}-2x^{\bs{\mc{M}}}\delta_{\bs{\mc{M}}}\hoch{\bs{\mc{D}}}\eps^{\bs{\mc{C}}}\bei{T_{\bs{\mc{C}}\bs{\mc{D}}}\hoch{\bs{\mc{A}}}}{}+\mc{O}(\xbothtetas^{2})\label{eq:SUSY-xialpha}\\
L_{A}\hoch{B}(\eps) & = & 2x^{\bs{\mc{M}}}\delta_{\bs{\mc{M}}}\hoch{\bs{\mc{D}}}\eps^{\bs{\mc{C}}}\bei{R_{\bs{\mc{C}}\bs{\mc{D}}A}\hoch{B}}{}+\mc{O}(\xbothtetas^{2})\label{eq:SUSY-L}\end{eqnarray}
Remember that the gauge transformation corresponding to these parameters
is of the form\begin{eqnarray}
\delta_{\eps} & = & \Liecov_{\vecfull{\xi}(\eps)}+\group{L(\eps)_{\,\cdot}\hoch{\cdot}}\end{eqnarray}
We should finally note that the separation of the gauge transformations
into local structure group transformations, local bosonic diffeomorphisms
and local supersymmetry contains some arbitraryness. In particular
when the structure group contains an abelian subgroup (e.g. dilatations),
a redefinition of local supersymmetry with such an abelian structure
group transformation does not change the supersymmetry algebra. In
fact the choice $L_{0\, A}\hoch{B}=0$ as part of the stabilizer of
the gauge fixing is not possible any longer if such a subgroup (e.g.
the local scale transformation) is fixed. In the case where we fix
for example (in our application in the main part) the leading component
of the (bosonic) compensator field $\Phi$ to $\bei{\Phi}{}\stackrel{!}{=}0$
or $\bei{\Phi}{}\stackrel{!}{=}\bei{\dil}{}$, we get the additional
stabilizer condition $\bei{\left(\xi^{C}"\nabla_{C}\Phi"-L^{(D)}\right)}{}\stackrel{!}{=}0$
or $\bei{\left(\xi^{C}"\nabla_{C}\Phi"-L^{(D)}\right)}{}\stackrel{!}{=}\xi^{C}\nabla_{C}\bei{\dil}{}$
or equivalently \begin{eqnarray}
L_{0}^{(D)}(\eps)\stackrel{!}{=}\eps^{\bs{\mc{C}}}\bei{\covPhi{\bs{\mc{C}}}}{} & \& & \xi^{\bs{\mc{A}}}(\eps)\To\xi^{\bs{\mc{A}}}(\eps)-\frac{1}{2}x^{\bs{\mc{M}}}\delta_{\bs{\mc{M}}}\hoch{\bs{\mc{A}}}\eps^{\bs{\mc{C}}}\bei{\covPhi{\bs{\mc{C}}}}{}\label{eq:additionalScaleStabilizer}\\
\mbox{or }L_{0}^{(D)}(\eps)\stackrel{!}{=}\eps^{\bs{\mc{C}}}\left(\bei{"\nabla_{\bs{\mc{C}}}\Phi"}{}-\nabla_{\bs{\mc{C}}}\bei{\dil}{}\right) & \& & \xi^{\bs{\mc{A}}}(\eps)\To\xi^{\bs{\mc{A}}}(\eps)-\frac{1}{2}x^{\bs{\mc{M}}}\delta_{\bs{\mc{M}}}\hoch{\bs{\mc{A}}}\eps^{\bs{\mc{C}}}\left(\bei{"\nabla_{\bs{\mc{C}}}\Phi"}{}-\nabla_{\bs{\mc{C}}}\bei{\dil}{}\right)\qquad\label{eq:additionalScaleStabilizerII}\end{eqnarray}
 Alternatively\index{scale invariance!two ways of fixing the $\sim$}\index{two ways of fixing the scale invariance},
we could have fixed the complete superfield $\Phi$ to zero (before
going to WZ-gauge). Then the scale part of the connection is not structure
group valued and therefore has to be treated as a difference tensor.
Only the Lorentz part can then be used for the implementation of the
WZ-gauge.\rem{Here are the two ways of fixing the scale invariance!}

\subsection{The supersymmetry\index{SUSY algebra} algebra\index{algebra!SUSY $\sim$}}

\rem{

\subsubsection{In terms of Lie derivatives}

}In order to read off the algebra of the local supersymmetry transformations
from (\ref{eq:supergauge-algebra}), we need the transformation of
$\vecfull{\xi}$ itself under a second supersymmetry transformation\begin{eqnarray}
\delta_{\eps_{1}}\xi^{A}(\eps_{2}) & = & -2x^{\bs{\mc{M}}}\eps_{2}^{\bs{\mc{C}}}\bei{\delta_{\eps_{1}}T_{\bs{\mc{CM}}}\hoch{A}}{}+\mc{O}(\xbothtetas^{2})=\\
 & = & -2x^{\bs{\mc{M}}}\eps_{2}^{\bs{\mc{C}}}\delta_{\bs{\mc{M}}}\hoch{\bs{\mc{B}}}\bei{\Liecov_{\vecfull{\xi}(\eps_{1})}T_{\bs{\mc{C}}\bs{\mc{B}}}\hoch{A}}{}+\mc{O}(\xbothtetas^{2})=\\
 & = & -2x^{\bs{\mc{M}}}\delta_{\bs{\mc{M}}}\hoch{\bs{\mc{D}}}\eps_{2}^{\bs{\mc{C}}}\eps_{1}^{\bs{\mc{B}}}\bei{\nabla_{\bs{\mc{B}}}T_{\bs{\mc{CD}}}\hoch{A}}{}+\mc{O}(\xbothtetas^{2})\end{eqnarray}
\rem{\begin{eqnarray}
\bei{\delta_{\eps_{1}}\xi^{M}(\eps_{2})}{} & = & \bei{\delta_{\eps_{1}}\xi^{A}(\eps_{2})}{}\cdot\bei{E_{A}\hoch{M}}{}+\underbrace{\bei{(\xi^{a}\delta E_{a}\hoch{M})}{}}_{0}=\\
\dann\bei{\delta_{\eps_{1}}\xi^{m}(\eps_{2})}{} & = & \bei{\delta_{\eps_{1}}\xi^{a}(\eps_{2})}{}\cdot e_{a}^{m}\\
\bei{\delta_{\eps_{1}}\xi^{\bs{\mc{M}}}(\eps_{2})}{} & = & -\bei{\delta_{\eps_{1}}\xi^{a}(\eps_{2})}{}\cdot\psi_{a}^{\bs{\mc{M}}}+\bei{\delta_{\eps_{1}}\xi^{\bs{\mc{A}}}(\eps_{2})}{}\cdot\delta_{\bs{\mc{A}}}^{\bs{\mc{M}}}\end{eqnarray}
}and also the transformation of $L_{A}\hoch{B}$ under supersymmetry:\begin{eqnarray}
\delta_{\eps_{1}}L_{A}\hoch{B}(\eps_{2}) & = & 2x^{\bs{\mc{M}}}\eps_{2}^{\bs{\mc{C}}}\bei{\delta_{\eps_{1}}R_{\bs{\mc{CM}}A}\hoch{B}}{}+\mc{O}(\xbothtetas^{2})=\\
 & = & 2x^{\bs{\mc{M}}}\eps_{2}^{\bs{\mc{C}}}\delta_{\bs{\mc{M}}}\hoch{\bs{\mc{D}}}\bei{\Liecov_{\vecfull{\xi}(\eps_{1})}R_{\bs{\mc{C}}\bs{\mc{D}}A}\hoch{B}}{}+\mc{O}(\xbothtetas^{2})=\\
 & = & 2x^{\bs{\mc{M}}}\eps_{2}^{\bs{\mc{C}}}\delta_{\bs{\mc{M}}}\hoch{\bs{\mc{D}}}\eps_{1}^{\bs{\mc{E}}}\bei{\nabla_{\bs{\mc{E}}}R_{\bs{\mc{C}}\bs{\mc{D}}A}\hoch{B}}{}+\mc{O}(\xbothtetas^{2})\end{eqnarray}
For the algebra (\ref{eq:supergauge-algebra}), we still need the
Lie bracket of the vector field:\begin{eqnarray}
[\vecfull{\xi}_{1},\vecfull{\xi}_{2}]^{A} & = & \xi_{1}^{C}\nabla_{C}\xi_{2}^{A}-\xi_{2}^{C}\nabla_{C}\xi_{1}^{A}-2\xi_{1}^{C}T_{CB}\hoch{A}\xi_{2}^{B}\end{eqnarray}
For simplicity, let us restrict to the leading component, although
we would have enough information to calculate higher orders as well:\begin{eqnarray}
\bei{[\vecfull{\xi}(\eps_{1}),\vecfull{\xi}(\eps_{2})]^{A}}{} & = & \eps_{1}^{\bs{\mc{C}}}\delta_{\bs{\mc{C}}}\hoch{\bs{\mc{M}}}\xi_{\bs{\mc{M}}}\hoch{A}(\eps_{2})-\eps_{2}^{\bs{\mc{B}}}\delta_{\bs{\mc{B}}}\hoch{\bs{\mc{M}}}\xi_{\bs{\mc{M}}}\hoch{A}(\eps_{1})-2\eps_{1}^{\bs{\mc{C}}}\bei{T_{\bs{\mc{CB}}}\hoch{A}}{}\eps_{2}^{\bs{\mc{B}}}=\\
 & = & -2\eps_{1}^{\bs{\mc{C}}}\eps_{2}^{\bs{\mc{B}}}\bei{T_{\bs{\mc{B}}\bs{\mc{C}}}\hoch{A}}{}+2\eps_{2}^{\bs{\mc{B}}}\eps_{1}^{\bs{\mc{C}}}\bei{T_{\bs{\mc{C}}\bs{\mc{B}}}\hoch{A}}{}-2\eps_{1}^{\bs{\mc{C}}}\bei{T_{\bs{\mc{CB}}}\hoch{A}}{}\eps_{2}^{\bs{\mc{B}}}=\\
 & = & 2\eps_{1}^{\bs{\mc{C}}}\bei{T_{\bs{\mc{C}}\bs{\mc{B}}}\hoch{A}}{}\eps_{2}^{\bs{\mc{B}}}\end{eqnarray}
\rem{we would have enough information to calculate any order. And at least the second $\theta$ derivative of $\xi$ was formerly explicitely given... Would be nice thus to calculate the first order of the vector Lie bracket.}Having
derived only the leading component of the vector-Lie bracket, we should
restrict to the leading component for the rest as well. The algebra
(\ref{eq:supergauge-algebra}) then becomes\begin{equation}
\boxed{[\delta_{\eps_{1}},\delta_{\eps_{2}}]=\Liecov_{\left(-2\eps_{1}^{\bs{\mc{C}}}\bei{T_{\bs{\mc{C}}\bs{\mc{D}}}\hoch{A}}{}\eps_{2}^{\bs{\mc{D}}}+\mc{O}(\xbothtetas)\right)\vecfull{E}_{A}}+\group{2\eps_{1}^{\bs{\mc{C}}}\eps_{2}^{\bs{\mc{D}}}R_{\bs{\mc{CD}}\,\cdot}\hoch{\cdot}+\mc{O}(\xbothtetas)}}\end{equation}
\rem{

\subsubsection{In terms of the superspace differential operator}

When acting on scalar fields, the Lie derivative corresponds to the
differential operator $\xi^{A}E_{A}\hoch{M}\partial_{M}=\xi_{0}^{\bs{\mc{A}}}\delta_{\bs{\mc{A}}}\hoch{\bs{\mc{M}}}\partial_{\bs{\mc{M}}}-2x^{\bs{\mc{M}}}\xi_{0}^{\bs{\mc{C}}}\bei{T_{\bs{\mc{C}}\bs{\mc{M}}}\hoch{A}}{}\bei{E_{A}\hoch{M}}{}\partial_{M}+\mc{O}(\vecfull{\tet}^{2})$,
which corresponds to \begin{eqnarray}
\Qsp_{\bs{\mc{A}}} & \equiv & \delta_{\bs{\mc{A}}}\hoch{\bs{\mc{M}}}\partial_{\bs{\mc{M}}}-2x^{\bs{\mc{M}}}\bei{T_{\bs{\mc{A}}\bs{\mc{M}}}\hoch{C}}{}\bei{E_{C}\hoch{M}}{}\partial_{M}+\mc{O}(\vecfull{\tet}^{2})\\
\textrm{or }\Qsp_{\bs{\mc{A}}} & \equiv & \left(\partl{\xi_{0}^{\bs{\mc{A}}}}\xi^{M}\right)\partial_{M}\end{eqnarray}
\begin{eqnarray}
\left[\Qsp_{\bs{\mc{A}}},E_{\bs{\mc{B}}}\hoch{N}\partial_{N}\right] & = & \delta_{\bs{\mc{A}}}\hoch{\bs{\mc{M}}}\partial_{\bs{\mc{M}}}\left(E_{\bs{\mc{B}}}\hoch{N}\partial_{N}\left(\ldots\right)\right)-2x^{\bs{\mc{M}}}\bei{T_{\bs{\mc{A}}\bs{\mc{M}}}\hoch{C}}{}\bei{E_{C}\hoch{M}}{}\partial_{M}\left(E_{\bs{\mc{B}}}\hoch{N}\partial_{N}\left(\ldots\right)\right)-\nonumber \\
 &  & -E_{\bs{\mc{B}}}\hoch{N}\partial_{N}\left(\delta_{\bs{\mc{A}}}\hoch{\bs{\mc{M}}}\partial_{\bs{\mc{M}}}\left(\ldots\right)-2x^{\bs{\mc{M}}}\bei{T_{\bs{\mc{A}}\bs{\mc{M}}}\hoch{C}}{}\bei{E_{C}\hoch{M}}{}\partial_{M}\left(\ldots\right)\right)+\mc{O}(\vecfull{\tet})=\\
 & = & \delta_{\bs{\mc{A}}}\hoch{\bs{\mc{M}}}\partial_{\bs{\mc{M}}}E_{\bs{\mc{B}}}\hoch{N}\partial_{N}\left(\ldots\right)-2x^{\bs{\mc{M}}}\bei{T_{\bs{\mc{A}}\bs{\mc{M}}}\hoch{C}}{}\bei{E_{C}\hoch{M}}{}\partial_{M}\left(E_{\bs{\mc{B}}}\hoch{N}\right)\partial_{N}\left(\ldots\right)-\nonumber \\
 &  & +2E_{\bs{\mc{B}}}\hoch{M}\partial_{M}\left(x^{\bs{\mc{M}}}\bei{T_{\bs{\mc{A}}\bs{\mc{M}}}\hoch{C}}{}\bei{E_{C}\hoch{N}}{}\right)\partial_{N}\left(\ldots\right)+\mc{O}(\vecfull{\tet})=\\
 & = & \delta_{\bs{\mc{A}}}\hoch{\bs{\mc{M}}}\partial_{\bs{\mc{M}}}E_{\bs{\mc{B}}}\hoch{N}\partial_{N}\left(\ldots\right)-2x^{\bs{\mc{M}}}\bei{T_{\bs{\mc{A}}\bs{\mc{M}}}\hoch{C}}{}\bei{E_{C}\hoch{M}}{}\partial_{M}\left(E_{\bs{\mc{B}}}\hoch{N}\right)\partial_{N}\left(\ldots\right)-\nonumber \\
 &  & +2E_{\bs{\mc{B}}}\hoch{\bs{\mc{M}}}\bei{T_{\bs{\mc{A}}\bs{\mc{M}}}\hoch{C}}{}\bei{E_{C}\hoch{N}}{}\partial_{N}\left(\ldots\right)+\nonumber \\
 &  & +2E_{\bs{\mc{B}}}\hoch{m}x^{\bs{\mc{M}}}\partial_{m}\left(\bei{T_{\bs{\mc{A}}\bs{\mc{M}}}\hoch{C}}{}\bei{E_{C}\hoch{N}}{}\right)\partial_{N}\left(\ldots\right)+\mc{O}(\vecfull{\tet})\end{eqnarray}
\begin{equation}
\left[\Qsp_{\bs{\mc{A}}},\Qsp_{\bs{\mc{B}}}\right]=\ldots\end{equation}
}\rem{Hier gibt's noch was zu tun!}

\subsection{Transformation of the fields}

\label{sub:The-supersymmetry-transformation}The supersymmetry\index{SUSY!trafo of the fields}
transformation of the fields is simply given by \begin{equation}
\delta_{\eps}=\Liecov_{\vecfull{\xi}(\eps)}+\group{L(\eps)_{\cdot}\hoch{\cdot}}\label{eq:versteckteSUSYtrafo}\end{equation}
where $\xi^{A}(\eps)$ and $L_{A}\hoch{B}(\eps)$ are of the special
form given in (\ref{eq:SUSY-def})-(\ref{eq:SUSY-L}). Let us derive
the transformations of all the fields that we will need. In order
to extract the transformation of the (leading) components, we will
again make frequent use of the Wess Zumino gauge (\ref{eq:WZ-gauge})
and (\ref{eq:WZ-gauge-connection}) (using $\bei{E_{m}\hoch{a}}{}\equiv e_{m}\hoch{a}$,
$\bei{E_{m}\hoch{\bs{\mc{A}}}}{}\equiv\psi_{m}\hoch{\bs{\mc{A}}}$).
In any supergravity theory we have a vielbein and a structure group
connection which we will consider first.

\subsubsection{Vielbein (bosonic vielbein and gravitino)}

Remember, the vielbein transforms according to (\ref{eq:vielbeinTrafo})
as\begin{eqnarray}
\delta E_{M}\hoch{A} & = & \underbrace{\partial_{M}\xi^{A}+\Omega_{MC}\hoch{A}\xi^{C}}_{\nabla_{M}\xi^{A}}+2\xi^{C}T_{CM}\hoch{A}+L_{B}\hoch{A}E_{M}\hoch{B}\end{eqnarray}
In practice, we will be given constraints on torsion components with
flat indices, s.t. it is useful to rewrite the equations in those
components. In addition, we plug in the explicit form of $\xi^{A}(\eps)$
and $L_{B}\hoch{A}(\eps)$ given in (\ref{eq:SUSY-xia})-(\ref{eq:SUSY-L})
to obtain the local supersymmetry transformation of the nonvanishing
leading vielbein components (the bosonic vielbein and the gravitino(s))::\vRam{0.7}{\begin{eqnarray}
\delta_{\eps}e_{m}\hoch{a} & = & 2\eps^{\bs{\mc{C}}}e_{m}\hoch{b}\bei{T_{\bs{\mc{C}}b}\hoch{a}}{}+2\eps^{\bs{\mc{C}}}\psi_{m}\hoch{\bs{\mc{B}}}\bei{T_{\bs{\mc{C}}\bs{\mc{B}}}\hoch{a}}{}\\
\delta_{\eps}\psi_{m}\hoch{\bs{\mc{A}}} & = & \underbrace{\partial_{m}\eps^{\bs{\mc{A}}}+\omega_{m\bs{\mc{C}}}\hoch{\bs{\mc{A}}}\eps^{\bs{\mc{C}}}}_{\nabla_{m}\eps^{\bs{\mc{A}}}}+2\eps^{\bs{\mc{C}}}e_{m}\hoch{b}\bei{T_{\bs{\mc{C}}b}\hoch{\bs{\mc{A}}}}{}+2\eps^{\bs{\mc{C}}}\psi_{m}\hoch{\bs{\mc{B}}}\bei{T_{\bs{\mc{C}}\bs{\mc{B}}}\hoch{\bs{\mc{A}}}}{}\label{eq:generalGravitinoSusyTrafo}\end{eqnarray}
}

\subsubsection{Connection}

Remember the general gauge transformation of the structure group connection
(\ref{eq:connectionTrafo})\begin{eqnarray}
\delta\Omega_{MA}\hoch{B} & = & 2\xi^{K}R_{KMA}\hoch{B}-\partial_{M}L_{A}\hoch{B}-[L,\Omega_{M}]_{A}\hoch{B}\end{eqnarray}
In the case where a scale part of the connection is present, this
transforms accordingly as (see (\ref{eq:scaleConnectionTrafo})) \begin{equation}
\delta\Omega_{M}^{(D)}=2\xi^{C}F_{CM}^{(D)}-\partial_{M}L^{(D)}\end{equation}
For the stabilizer of WZ-gauge with $\bei{\Omega_{\bs{\mc{M}}A}\hoch{B}}{}=0$
and $\delta\bei{\Omega_{\bs{\mc{M}}A}\hoch{B}}{}=0$ and for the choice
$\xi_{0}^{c}=L_{0\, A}\hoch{B}$ (corresponding to local supersymmetry
(\ref{eq:SUSY-def}) and (\ref{eq:SUSY-xi-L-cond})) the nontrivial
part of the above equations becomes (for $\xbothtetas=0$): \begin{eqnarray}
\delta\bei{\Omega_{mA}\hoch{B}}{} & = & 2\xi_{0}^{\bs{\mc{C}}}\bei{R_{\bs{\mc{C}}mA}\hoch{B}}{}\\
\delta\bei{\Omega_{m}^{(D)}}{} & = & 2\xi_{0}^{\bs{\mc{C}}}\bei{F_{\bs{\mc{C}}m}^{(D)}}{}\end{eqnarray}
More explicitely (replacing $\eps^{\bs{\gamma}}\equiv\xi_{0}^{\bs{\gamma}},\quad\hat{\eps}^{\hat{\bs{\gamma}}}\equiv\xi_{0}^{\hat{\bs{\gamma}}})$
this reads \vRam{0.7}{\begin{eqnarray}
\delta\bei{\Omega_{ma}\hoch{b}}{} & = & 2\eps^{\bs{\gamma}}\left(e_{m}\hoch{d}\bei{R_{\bs{\gamma}da}\hoch{b}}{}+\psi_{m}\hoch{\bs{\delta}}\bei{R_{\bs{\gamma}\bs{\delta}a}\hoch{b}}{}+\hat{\psi}_{m}\hoch{\hat{\bs{\delta}}}\bei{R_{\bs{\gamma}\hat{\bs{\delta}}a}\hoch{b}}{}\right)+\nonumber \\
 &  & +2\eps^{\hat{\bs{\gamma}}}\left(e_{m}\hoch{d}\bei{R_{\hat{\bs{\gamma}}da}\hoch{b}}{}+\psi_{m}\hoch{\bs{\delta}}\bei{R_{\hat{\bs{\gamma}}\bs{\delta}a}\hoch{b}}{}+\hat{\psi}_{m}\hoch{\hat{\bs{\delta}}}\bei{R_{\hat{\bs{\gamma}}\hat{\bs{\delta}}a}\hoch{b}}{}\right)\\
\delta\bei{\Omega_{m}^{(D)}}{} & = & 2\eps^{\bs{\gamma}}\left(e_{m}\hoch{d}\bei{F_{\bs{\gamma}d}^{(D)}}{}+\psi_{m}\hoch{\bs{\delta}}\bei{F_{\bs{\gamma}\bs{\delta}}^{(D)}}{}+\hat{\psi}_{m}\hoch{\hat{\bs{\delta}}}\bei{F_{\bs{\gamma}\hat{\bs{\delta}}}^{(D)}}{}\right)+\nonumber \\
 &  & +2\eps^{\hat{\bs{\gamma}}}\left(e_{m}\hoch{d}\bei{F_{\hat{\bs{\gamma}}d}^{(D)}}{}+\psi_{m}\hoch{\bs{\delta}}\bei{F_{\hat{\bs{\gamma}}\bs{\delta}}^{(D)}}{}+\hat{\psi}_{m}\hoch{\hat{\bs{\delta}}}\bei{F_{\hat{\bs{\gamma}}\hat{\bs{\delta}}}^{(D)}}{}\right)\end{eqnarray}
}

\subsubsection{Compensator field}

A compensator field is not necessarily present in a supergravity theory.
In our context such a field $\Phi$ is used to allow a scale transformation
of the metric in flat indices:\begin{eqnarray}
G_{AB} & \equiv & e^{2\Phi}\eta_{AB}\end{eqnarray}
Where $\eta_{AB}$ is some constant metric which is invariant under
the orthogonal transformations. In our case, its bosonic part is just
the Minkowski metric and the rest is zero. There is no way, how a
constant metric can scale. Therefore the compensator field $\Phi$
takes over the scaling of $G_{AB}$ under scale transformation by
simply getting shifted with the scale parameter\begin{equation}
\group{L}\Phi=\Phi-L^{(D)}\end{equation}
Similarly, the covariant derivative will be defined to act only on
$\Phi$ (and not on $\eta_{AB}$) in such a way that the covariant
derivative of $G_{AB}$ has the form that is indicated by its indices.
\begin{eqnarray}
\nabla_{M}G_{AB} & = & 2(\partial_{M}\Phi-\Omega_{M}^{(D)})G_{AB}\\
\dann"\nabla_{M}\Phi" & = & 2(\partial_{M}\Phi-\Omega_{M}^{(D)})\end{eqnarray}
The general gauge transformation of the compensator field thus reads\begin{eqnarray}
\delta\Phi & = & \xi^{K}\underbrace{\left(\partial_{K}\Phi-\Omega_{K}^{(D)}\right)}_{"\nabla_{K}\Phi"}-L^{(D)}\end{eqnarray}
Define \begin{eqnarray}
\phi & \equiv & \bei{\Phi}{}\\
\phi_{\bs{\mc{M}}} & \equiv & \bei{\partial_{\bs{\mc{M}}}\Phi}{}\end{eqnarray}
For the lowest component , this implies the following local SUSY transformation
in the WZ gauge\begin{equation}
\boxed{\delta_{\eps}\phi=\eps^{\bs{\gamma}}\phi_{\bs{\mc{\gamma}}}+\hat{\eps}^{\hat{\bs{\gamma}}}\phi_{\hat{\bs{\gamma}}}}\label{eq:compensator-field-SUSY:leading}\end{equation}
The transformation is zero, if we combine it with an additional scale
stabilizer transformation (\ref{eq:additionalScaleStabilizer})\begin{equation}
L^{(D)}=\xi_{0}^{\bs{\mc{C}}}\phi_{\bs{\mc{C}}}\end{equation}
Note that the transformation of the connection is such that the covariant
derivative of the compensator field transforms like a vector\begin{eqnarray}
\delta\nabla_{A}\Phi & = & \xi^{B}\nabla_{B}\nabla_{A}\Phi-L_{A}\hoch{B}\nabla_{B}\Phi\end{eqnarray}
In particular we have for the SUSY transformation of the first theta-components\begin{equation}
\boxed{\delta_{\eps}\bei{\nabla_{\bs{\mc{A}}}\Phi}{}=\eps^{\bs{\mc{B}}}\bei{\nabla_{\bs{\mc{B}}}\nabla_{\bs{\mc{A}}}\Phi}{}}\label{eq:generalDilatinoSusyTrafo}\end{equation}

\subsubsection{Scalar super field (e.g. dilaton and dilatino)}

The Dilaton field is a scalar and thus has the simple transformation\begin{eqnarray}
\delta\dil & = & \xi^{C}\underbrace{\nabla_{C}\dil}_{E_{C}\hoch{M}\partial_{M}\dil}=\Lie_{\vecfull{\xi}}\dil\label{eq:dilatonsf-trafo}\end{eqnarray}
Define now the \textbf{dilatino\index{dilatino}} to be \begin{eqnarray}
\dilo_{\bs{\mc{A}}} & \equiv & \bei{\nabla_{\bs{\mc{A}}}\dil}{}=\delta_{\bs{\mc{A}}}\hoch{\bs{\mc{M}}}\bei{\partial_{\bs{\mc{M}}}\dil}{}\\
\dilo_{\bs{\mc{M}}} & = & \bei{\partial_{\bs{\mc{M}}}\dil}{}\\
\dann\dil & = & \phi_{(ph)}+x^{\bs{\mu}}\dilo_{\bs{\mu}}+x^{\hat{\bs{\mu}}}\hat{\lambda}_{\hat{\bs{\mu}}}+\frac{1}{2}x^{\bs{\mc{M}}}x^{\bs{\mc{N}}}\bei{\partial_{\bs{\mc{M}}}\partial_{\bs{\mc{N}}}\Phi}{}+\ldots\end{eqnarray}
This definition of the dilatino implies according to (\ref{eq:dilatonsf-trafo})
for the dilaton\index{dilaton} $\dilcomp$ the transformation \begin{equation}
\boxed{\delta\dil=\eps^{\bs{\mc{C}}}\dilo_{\bs{\mc{C}}}}\end{equation}
For the transformation of the dilatino we use the fact that the variation
of a covariant derivative is simply the covariantized Lie derivative
(supergauge transformation) plus the structure group transformation
of the new tensor according to the new index structure (see footnote
\ref{foot:transformationOfCovDer} on page \pageref{foot:transformationOfCovDer}
and (\ref{eq:covLiederOnGenTensor})). We thus have\rem{hier ist ein Note ueber Stabilizator des inv. Vielbeins}\begin{eqnarray}
\delta(\nabla_{A}\dil) & = & \xi^{C}\nabla_{C}\nabla_{A}\dil-L_{A}\hoch{B}\nabla_{B}\dil\end{eqnarray}
with $\xi^{C}$ and $L_{A}\hoch{B}$ given in (\ref{eq:SUSY-def})-(\ref{eq:SUSY-L}).
For the fermionic components at $\xbothtetas=0$, this reads simply\begin{equation}
\boxed{\delta\dilo_{\bs{\mc{A}}}=\eps^{\bs{\mc{C}}}\bei{\nabla_{\bs{\mc{C}}}\nabla_{\bs{\mc{A}}}\dil}{}}\label{eq:generalDilatinoSusyTrafo:old}\end{equation}
 Apparently, we need some equations of motion at this point, in order
to say more. We can, however, relate this expression explicitely to
the $\xbothtetas^{2}$ component $\bei{\partial_{\bs{\mc{M}}}\partial_{\bs{\mc{N}}}\dil}{}$
of the dilaton:\begin{eqnarray}
\delta\dilo_{\bs{\mc{A}}} & = & \eps^{\bs{\mc{C}}}\delta_{\bs{\mc{C}}}\hoch{\bs{\mc{M}}}\bei{\partial_{\bs{\mc{M}}}(E_{\bs{\mc{A}}}\hoch{K}\partial_{K}\dil)}{}=\\
 & = & \eps^{\bs{\mc{C}}}\delta_{\bs{\mc{C}}}\hoch{\bs{\mc{M}}}\left(\bei{\partial_{\bs{\mc{M}}}E_{\bs{\mc{A}}}\hoch{K}}{}\bei{\partial_{K}\dil}{}+\delta_{\bs{\mc{A}}}\hoch{\bs{\mc{K}}}\bei{\partial_{\bs{\mc{M}}}\partial_{\bs{\mc{K}}}\dil}{}\right)\end{eqnarray}
Now we can use that\begin{eqnarray}
\bei{\partial_{\bs{\mc{M}}}E_{\bs{\mc{A}}}\hoch{K}}{} & = & -\bei{E_{\bs{\mc{A}}}\hoch{L}}{}\bei{\partial_{\bs{\mc{M}}}E_{L}\hoch{B}}{}\bei{E_{B}\hoch{K}}{}=\\
 & = & -\bei{E_{\bs{\mc{A}}}\hoch{\bs{\mc{L}}}}{}\underbrace{\bei{\partial_{\bs{\mc{M}}}E_{\bs{\mc{L}}}\hoch{B}}{}}_{\bei{\partial_{[\bs{\mc{M}}}E_{\bs{\mc{L}}]}\hoch{B}}{}}\bei{E_{B}\hoch{K}}{}=\\
 & = & -\delta_{\bs{\mc{A}}}\hoch{\bs{\mc{L}}}\bei{T_{\bs{\mc{M}}\bs{\mc{L}}}\hoch{B}}{}\bei{E_{B}\hoch{K}}{}\end{eqnarray}
The transformation of before can then be rewritten as \vRam{.7}{\begin{eqnarray}
\delta\dilo_{\bs{\mc{A}}} & = & -\eps^{\bs{\mc{C}}}\bei{T_{\bs{\mc{C}}\bs{\mc{A}}}\hoch{b}}{}e_{b}\hoch{k}\partial_{k}\dilcomp+\eps^{\bs{\mc{C}}}\bei{T_{\bs{\mc{C}}\bs{\mc{A}}}\hoch{b}}{}\psi_{b}\hoch{\bs{\mc{K}}}\dilo_{\bs{\mc{K}}}-\eps^{\bs{\mc{C}}}\bei{T_{\bs{\mc{C}}\bs{\mc{A}}}\hoch{\bs{\mc{B}}}}{}\dilo_{\bs{\mc{B}}}+\nonumber \\
 &  & +\eps^{\bs{\mc{C}}}\delta_{\bs{\mc{C}}}\hoch{\bs{\mc{M}}}\delta_{\bs{\mc{A}}}\hoch{\bs{\mc{K}}}\bei{\partial_{\bs{\mc{M}}}\partial_{\bs{\mc{K}}}\dil}{}\label{eq:generalDilatinoSusyTrafo:long}\end{eqnarray}
}

\subsubsection{Bispinor fields (RR-fields)}

Apart from that we will be interested in the transformation of RR-fields\begin{equation}
\delta\RR^{\bs{\alpha}\hat{\bs{\beta}}}=\xi^{C}\nabla_{C}\RR^{\bs{\alpha}\hat{\bs{\beta}}}+L_{\bs{\gamma}}\hoch{\bs{\alpha}}\RR^{\bs{\gamma}\hat{\bs{\beta}}}+L_{\hat{\bs{\gamma}}}\hoch{\hat{\bs{\beta}}}\RR^{\bs{\alpha}\hat{\bs{\gamma}}}\end{equation}
The leading component, that we defined in the main text as $\rr^{\bs{\alpha}\hat{\bs{\beta}}}=e^{-8\dilcomp}\bei{\RR^{\bs{\alpha}\hat{\bs{\beta}}}}{}$,
then transforms as\begin{equation}
\boxed{\delta\rr^{\alpha\hat{\bs{\beta}}}=-8\eps^{\bs{\mc{C}}}\dilo_{\bs{\mc{C}}}\rr^{\alpha\hat{\bs{\beta}}}+e^{-8\dilcomp}\eps^{\bs{\mc{C}}}\bei{\nabla_{\bs{\mc{C}}}\RR^{\bs{\alpha}\hat{\bs{\beta}}}}{}}\end{equation}

\subsubsection{Two or three form (e.g. $B$-field and $H$-field)}

Finally we consider the transformation of a two form (e.g. the B-field)
and of a three form (e.g the $H$-field):\begin{eqnarray}
\delta B_{AB} & = & \xi^{D}\nabla_{D}B_{AB}-2L_{[A|}\hoch{D}B_{D|B]}\\
\delta B_{MN} & = & \xi^{D}\nabla_{D}B_{MN}+2(\nabla_{[M|}\xi^{L}+2\xi^{P}T_{P[M|}\hoch{L})B_{L|N]}=\xi^{K}\partial_{K}B_{MN}+2\partial_{[M|}\xi^{L}B_{L|N]}\\
\delta H_{ABC} & = & \xi^{D}\nabla_{D}H_{ABC}-3L_{[A|}\hoch{D}H_{D|BC]}\\
\delta H_{MNK} & = & \xi^{D}\nabla_{D}H_{MNK}+3(\nabla_{[M|}\xi^{L}+2\xi^{P}T_{P[M|}\hoch{L})H_{L|NK]}=\xi^{L}\partial_{L}H_{MNK}+3\partial_{[M|}\xi^{L}H_{L|NK]}\qquad\end{eqnarray}
It makes some difference whether we consider the fields with flat
or with curved coordinates. The difference lies in the transformation
of the vielbeins. Physically, we are interested in the transformation
of the bosonic $B$-field $\bei{B_{mn}}{}$ and $H$-field $\bei{H_{mnk}}{}$
with curved indices. If we assume that $H=\de B$ and $B$ thus is
a gauge field, we can make use of the WZ-like gauge $\bei{B_{\bs{\mc{MN}}}}{}=\bei{B_{m\bs{\mc{N}}}}{}=0$
and $\bei{\partial_{\bs{\mc{K}}}B_{mn}}{}=3\bei{H_{\bs{\mc{K}}mn}}{}$,
in order to become more explicit for the transformation of $\bei{B_{mn}}{}$.
For the B-field transformation it thus makes sense to take the version
in terms of partial derivatives instead of covariant ones.\begin{eqnarray}
\delta\bei{B_{mn}}{} & = & \eps^{\bs{\mc{D}}}\delta_{\bs{\mc{D}}}\hoch{\bs{\mc{K}}}\bei{\partial_{\bs{\mc{K}}}B_{mn}}{}+2\partial_{[m|}\eps^{\bs{\mc{D}}}\bei{B_{\bs{\mc{D}}|n]}}{}=\\
 & = & 3\eps^{\bs{\mc{D}}}\bei{H_{\bs{\mc{D}}mn}}{}\end{eqnarray}
Rewritten in flat coordinates, the result becomes \vRam{.8}{\begin{eqnarray}
\delta_{\eps}\bei{B_{mn}}{} & = & 3\eps^{\bs{\mc{D}}}e_{m}\hoch{a}e_{n}\hoch{b}\bei{H_{\bs{\mc{D}}ab}}{}+6\eps^{\bs{\mc{D}}}\psi_{[m}\hoch{\bs{\mc{A}}}e_{n]}\hoch{b}\bei{H_{\bs{\mc{DA}}b}}{}+3\eps^{\bs{\mc{D}}}\psi_{[m}\hoch{\bs{\mc{A}}}\psi_{n]}\hoch{\bs{\mc{B}}}\bei{H_{\bs{\mc{DAB}}}}{}\end{eqnarray}
} So far we have only used simplifications coming from the WZ-like
gauge but no supergravity constraints yet.\rem{

\section{Notation for the component fields}

Notation for the component fields: In order not to get too confused
in what follows, we should introduce some systematics in the notation
of the components (with respect to the $\xbothtetas$-expansion) of
the superfields. For some components we will reserve special names
and for the rest we will apply some standard notation. Given a scalar
superfield $A(\xfull)=A(\xboson,\xbothtetas)$, its leading component
will be called $A_{0}$ or in some cases $a$ (if it does not lead
to confusions with existing variables). The next component will be
called $A_{\bs{\mc{M}}}$ or $a_{\bs{\mc{M}}}$. \begin{eqnarray}
\bei{A}{} & \equiv & A_{0}\equiv a\\
\bei{\partial_{\bs{\mc{M}}}A}{} & \equiv & A_{\bs{\mc{M}}}\equiv a_{\bs{\mc{M}}}\end{eqnarray}
In particular we define the components of the dilaton superfield (although
it is not a scalar with respect to dilatations...) as \begin{eqnarray}
\Phi(\xboson,\xbothtetas) & = & \phi(\xboson)+x^{\bs{\mc{M}}}\phi_{\bs{\mc{M}}}+\mc{O}(\xbothtetas^{2})\end{eqnarray}
The components of the other superfields read:\begin{eqnarray}
E_{M}\hoch{a}(\xboson,\xbothtetas) & \equiv & e_{M}\hoch{a}(\xboson)+\underbrace{x^{\bs{\mc{M}}}}_{\xbothtetas^{\mc{M}}}e_{\bs{\mc{M}}M}\hoch{a}(\xboson)+\ldots\\
E_{M}\hoch{\bs{\mc{A}}}(\xboson,\xbothtetas) & \equiv & \psi_{M}\hoch{\bs{\mc{A}}}(\xboson)+x^{\bs{\mc{M}}}\psi_{\bs{\mc{M}}M}\hoch{\bs{\mc{A}}}(\xboson)+\ldots\\
\Omega_{MA}\hoch{B}(\xboson,\xbothtetas) & \equiv & \underbrace{\left(\omega_{MA}\hoch{B}(\xboson)+\omega_{MA}^{(D)}\hoch{B}(\xboson)\right)}_{\Omega_{0\, MA}\hoch{B}(\xboson)}+x^{\bs{\mc{M}}}\Omega_{\bs{\mc{M}}MA}\hoch{B}(\xboson)+\ldots\\
\RR^{\bs{\alpha}\hat{\bs{\alpha}}}(\xboson,\xbothtetas) & \equiv & \rr^{\bs{\alpha}\hat{\bs{\alpha}}}(\xboson)+x^{\bs{\mc{M}}}\RR_{\bs{\mc{M}}}^{\quad\bs{\alpha}\hat{\bs{\alpha}}}(\xboson)+\ldots\\
H_{abc}(\xboson,\xbothtetas) & \equiv & h_{abc}(\xboson)\\
G_{ab}(\xboson,\xbothtetas) & = & e^{2\Phi(\xboson,\xbothtetas)}\eta_{ab}=e^{2\phi(\xboson)}\eta_{ab}+\ldots\\
G_{mn}(\xboson,\xbothtetas) & \equiv & g_{mn}(\xboson)+\ldots\end{eqnarray}
}\rem{Baustelle}

\printindex{}\bibliographystyle{/home/basti/fullsort}
\bibliography{/home/basti/phd,/home/basti/Proposal}

\begin{thebibliography}{10%
0}

\addcontentsline{toc}{chapter}{Bibliography}

\bibitem{D'Auria:2008ny}
R.~D'Auria, P.~Fre', P.~A. Grassi, and M.~Trigiante, ``{Pure Spinor
  Superstrings on Generic type IIA Supergravity Backgrounds},''
\href{http://www.arXiv.org/abs/arXiv:0803.1703 [hep-th]}{{\tt arXiv:0803.1703
  [hep-th]}}.

\bibitem{Kluson:2008as}
J.~Kluson, ``{Note About Redefinition of BRST Operator for Pure Spinor String
  in General Background},''
\href{http://www.arXiv.org/abs/0803.4390}{{\tt 0803.4390}}.

\bibitem{Candelas:1985en}
P.~Candelas, G.~T. Horowitz, A.~Strominger, and E.~Witten, ``Vacuum
  configurations for superstrings\protect\hypertarget{Bib}{},'' {\em Nucl.
  Phys.} {\bf B258} (1985)
46--74.

\bibitem{Strominger:1986uh}
A.~Strominger, ``Superstrings with torsion,'' {\em Nucl. Phys.} {\bf B274}
  (1986)
253.

\bibitem{Grana:2004??}
M.~Grana, R.~Minasian, M.~Petrini, and A.~Tomasiello, ``Generalized structures
  of n=1 vacua,'' {\em JHEP} {\bf 11} (2005) 020,
\href{http://www.arXiv.org/abs/hep-th/0505212}{{\tt hep-th/0505212}}.

\bibitem{Grana:2004bg}
M.~Grana, R.~Minasian, M.~Petrini, and A.~Tomasiello, ``Supersymmetric
  backgrounds from generalized {Calabi-Yau} manifolds,'' {\em JHEP} {\bf 08}
  (2004) 046,
\href{http://www.arXiv.org/abs/hep-th/0406137}{{\tt hep-th/0406137}}.

\bibitem{Berkovits:2000fe}
N.~Berkovits, ``Super-{Poincare} covariant quantization of the superstring,''
  {\em JHEP} {\bf 04} (2000) 018,
\href{http://www.arXiv.org/abs/hep-th/0001035}{{\tt hep-th/0001035}}.

\bibitem{Nh:2001ug}
P.~A. Grassi, G.~Policastro, M.~Porrati, and P.~van Nieuwenhuizen, ``Covariant
  quantization of superstrings without pure spinor constraints,'' {\em JHEP}
  {\bf 10} (2002) 054,
\href{http://www.arXiv.org/abs/hep-th/0112162}{{\tt hep-th/0112162}}.

\bibitem{Nh:2003cm}
P.~A. Grassi, G.~Policastro, and P.~van Nieuwenhuizen, ``An introduction to the
  covariant quantization of superstrings,'' {\em Class. Quant. Grav.} {\bf 20}
  (2003) S395--S410,
\href{http://www.arXiv.org/abs/hep-th/0302147}{{\tt hep-th/0302147}}.

\bibitem{Nh:2003kq}
P.~A. Grassi, G.~Policastro, and P.~van Nieuwenhuizen, ``The quantum
  superstring as a {WZNW} model,'' {\em Nucl. Phys.} {\bf B676} (2004) 43--63,
\href{http://www.arXiv.org/abs/hep-th/0307056}{{\tt hep-th/0307056}}.

\bibitem{Guttenberg:2004ht}
S.~Guttenberg, J.~Knapp, and M.~Kreuzer, ``On the covariant quantization of
  type {II} superstrings,'' {\em JHEP} {\bf 06} (2004) 030,
\href{http://www.arXiv.org/abs/hep-th/0405007}{{\tt hep-th/0405007}}.

\bibitem{Berkovits:2004px}
N.~Berkovits, ``Multiloop amplitudes and vanishing theorems using the pure
  spinor formalism for the superstring,'' {\em JHEP} {\bf 09} (2004) 047,
\href{http://www.arXiv.org/abs/hep-th/0406055}{{\tt hep-th/0406055}}.

\bibitem{Berkovits:2001ue}
N.~Berkovits and P.~S. Howe, ``Ten-dimensional supergravity constraints from
  the pure spinor formalism for the superstring,'' {\em Nucl. Phys.} {\bf B635}
  (2002) 75--105,
\href{http://www.arXiv.org/abs/hep-th/0112160}{{\tt hep-th/0112160}}.

\bibitem{Chandia:2006ix}
O.~Chandia, ``A note on the classical {BRST} symmetry of the pure spinor string
  in a curved background,'' {\em JHEP} {\bf 07} (2006) 019,
\href{http://www.arXiv.org/abs/hep-th/0604115}{{\tt hep-th/0604115}}.

\bibitem{Dragon:1978nf}
N.~Dragon, ``Torsion and curvature in extended supergravity,'' {\em Z. Phys.}
  {\bf C2} (1979)
29--32.

\bibitem{Guttenberg:2006zi}
S.~Guttenberg, ``Brackets, sigma models and integrability of generalized
  complex structures,''
\href{http://www.arXiv.org/abs/hep-th/0609015}{{\tt hep-th/0609015}}.

\bibitem{Wess:1992cp}
J.~Wess and J.~Bagger, ``Supersymmetry and supergravity,''. Princeton, USA:
  Univ. Pr. (1992) 259 p.

\bibitem{VanNieuwenhuizen:1981ae}
P.~Van~Nieuwenhuizen, ``Supergravity,'' {\em Phys. Rept.} {\bf 68} (1981)
189--398.

\bibitem{DeWitt:1992cy}
B.~S. DeWitt, ``Supermanifolds,''. Cambridge, UK: Univ. Pr. (1992) 407 p.
  (Cambridge monographs on mathematical physics). (2nd ed.),.

\bibitem{Frydryszak:2006wk}
A.~Frydryszak, ``{Nilpotent classical mechanics},'' {\em Int. J. Mod. Phys.}
  {\bf A22} (2007) 2513--2534,
\href{http://www.arXiv.org/abs/hep-th/0609072}{{\tt hep-th/0609072}}.

\bibitem{Cartier:2002zp}
P.~Cartier, C.~DeWitt-Morette, M.~Ihl, and C.~Saemann, ``{Supermanifolds -
  Application to Supersymmetry},''
\href{http://www.arXiv.org/abs/math-ph/0202026}{{\tt math-ph/0202026}}.

\bibitem{Schmitt:1996hp}
T.~Schmitt, ``{Supergeometry and quantum field theory, or: What is a classical
  configuration?},'' {\em Rev. Math. Phys.} {\bf 9} (1997) 993--1052,
\href{http://www.arXiv.org/abs/hep-th/9607132}{{\tt hep-th/9607132}}.

\bibitem{Cvitanovic:1979qz}
P.~Cvitanovic, ``Supersymmetry, negative dimensions and the emergence of {E7}
  symmetry,''. Print-79-1010 (NORDITA).

\bibitem{Cv2007:bt}
P.~Cvitanovic, {\em Group Theory}.
\newblock Princeton University Press, 2007.

\bibitem{Frappat:1996pb}
L.~Frappat, P.~Sorba, and A.~Sciarrino, ``Dictionary on lie superalgebras,''
\href{http://www.arXiv.org/abs/hep-th/9607161}{{\tt hep-th/9607161}}.

\bibitem{Siegel:1986xj}
W.~Siegel, ``Classical superstring mechanics,'' {\em Nucl. Phys.} {\bf B263}
  (1986)
93.

\bibitem{Berkovits:2005bt}
N.~Berkovits, ``Pure spinor formalism as an n = 2 topological string,'' {\em
  JHEP} {\bf 10} (2005) 089,
\href{http://www.arXiv.org/abs/hep-th/0509120}{{\tt hep-th/0509120}}.

\bibitem{Berkovits:2005ng}
N.~Berkovits and C.~R. Mafra, ``Equivalence of two-loop superstring amplitudes
  in the pure spinor and rns formalisms,'' {\em Phys. Rev. Lett.} {\bf 96}
  (2006) 011602,
\href{http://www.arXiv.org/abs/hep-th/0509234}{{\tt hep-th/0509234}}.

\bibitem{Berkovits:2006bk}
N.~Berkovits and C.~R. Mafra, ``Some superstring amplitude computations with
  the non- minimal pure spinor formalism,'' {\em JHEP} {\bf 11} (2006) 079,
\href{http://www.arXiv.org/abs/hep-th/0607187}{{\tt hep-th/0607187}}.

\bibitem{Berkovits:2006vi}
N.~Berkovits and N.~Nekrasov, ``Multiloop superstring amplitudes from
  non-minimal pure spinor formalism,'' {\em JHEP} {\bf 12} (2006) 029,
\href{http://www.arXiv.org/abs/hep-th/0609012}{{\tt hep-th/0609012}}.

\bibitem{Stahn:2007uw}
C.~Stahn, ``Fermionic superstring loop amplitudes in the pure spinor
  formalism,'' {\em JHEP} {\bf 05} (2007) 034,
\href{http://www.arXiv.org/abs/arXiv:0704.0015 [hep-th]}{{\tt arXiv:0704.0015
  [hep-th]}}.

\bibitem{Berkovits:2004tw}
N.~Berkovits and D.~Z. Marchioro, ``Relating the {Green-Schwarz} and pure
  spinor formalisms for the superstring,'' {\em JHEP} {\bf 01} (2005) 018,
\href{http://www.arXiv.org/abs/hep-th/0412198}{{\tt hep-th/0412198}}.

\bibitem{Nekrasov:2005wg}
N.~A. Nekrasov, ``Lectures on curved beta-gamma systems, pure spinors, and
  anomalies,''
\href{http://www.arXiv.org/abs/hep-th/0511008}{{\tt hep-th/0511008}}.

\bibitem{Berkovits:2006ik}
N.~Berkovits, ``Explaining pure spinor superspace,''
\href{http://www.arXiv.org/abs/hep-th/0612021}{{\tt hep-th/0612021}}.

\bibitem{Berkovits:1994wr}
N.~Berkovits, ``Covariant quantization of the {Green-Schwarz} superstring in a
  {Calabi-Yau} background,'' {\em Nucl. Phys.} {\bf B431} (1994) 258--272,
\href{http://www.arXiv.org/abs/hep-th/9404162}{{\tt hep-th/9404162}}.

\bibitem{Kappeli:2006fj}
J.~Kappeli, S.~Theisen, and P.~Vanhove, ``Hybrid formalism and topological
  amplitudes,''
\href{http://www.arXiv.org/abs/hep-th/0607021}{{\tt hep-th/0607021}}.

\bibitem{Linch:2006ig}
I.~Linch, William~D. and B.~C. Vallilo, ``Hybrid formalism, supersymmetry
  reduction, and ramond- ramond fluxes,''
\href{http://www.arXiv.org/abs/hep-th/0607122}{{\tt hep-th/0607122}}.

\bibitem{Chesterman:2002ey}
M.~Chesterman, ``Ghost constraints and the covariant quantization of the
  superparticle in ten dimensions,'' {\em JHEP} {\bf 02} (2004) 011,
\href{http://www.arXiv.org/abs/hep-th/0212261}{{\tt hep-th/0212261}}.

\bibitem{Chesterman:2004xt}
M.~Chesterman, ``On the cohomology and inner products of the {Berkovits}
  superparticle and superstring,''
\href{http://www.arXiv.org/abs/hep-th/0404021}{{\tt hep-th/0404021}}.

\bibitem{Aisaka:2002sd}
Y.~Aisaka and Y.~Kazama, ``A new first class algebra, homological perturbation
  and extension of pure spinor formalism for superstring,'' {\em JHEP} {\bf 02}
  (2003) 017,
\href{http://www.arXiv.org/abs/hep-th/0212316}{{\tt hep-th/0212316}}.

\bibitem{Aisaka:2003mw}
Y.~Aisaka and Y.~Kazama, ``Operator mapping between {RNS} and extended pure
  spinor formalisms for superstring,'' {\em JHEP} {\bf 08} (2003) 047,
\href{http://www.arXiv.org/abs/hep-th/0305221}{{\tt hep-th/0305221}}.

\bibitem{Aisaka:2004ga}
Y.~Aisaka and Y.~Kazama, ``Relating {Green-Schwarz} and extended pure spinor
  formalisms by similarity transformation,'' {\em JHEP} {\bf 04} (2004) 070,
\href{http://www.arXiv.org/abs/hep-th/0404141}{{\tt hep-th/0404141}}.

\bibitem{Aisaka:2005vn}
Y.~Aisaka and Y.~Kazama, ``Origin of pure spinor superstring,'' {\em JHEP} {\bf
  05} (2005) 046,
\href{http://www.arXiv.org/abs/hep-th/0502208}{{\tt hep-th/0502208}}.

\bibitem{Aisaka:2006by}
Y.~Aisaka and Y.~Kazama, ``Towards pure spinor type covariant description of
  supermembrane: An approach from the double spinor formalism,'' {\em JHEP}
  {\bf 05} (2006) 041,
\href{http://www.arXiv.org/abs/hep-th/0603004}{{\tt hep-th/0603004}}.

\bibitem{SorokinMatone:2002ft}
M.~Matone, L.~Mazzucato, I.~Oda, D.~Sorokin, and M.~Tonin, ``The superembedding
  origin of the {Berkovits} pure spinor covariant quantization of
  superstrings,'' {\em Nucl. Phys.} {\bf B639} (2002) 182--202,
\href{http://www.arXiv.org/abs/hep-th/0206104}{{\tt hep-th/0206104}}.

\bibitem{Oda:2004bg}
I.~Oda and M.~Tonin, ``On the b-antighost in the pure spinor quantization of
  superstrings,'' {\em Phys. Lett.} {\bf B606} (2005) 218--222,
\href{http://www.arXiv.org/abs/hep-th/0409052}{{\tt hep-th/0409052}}.

\bibitem{Oda:2005sd}
I.~Oda and M.~Tonin, ``Y-formalism in pure spinor quantization of
  superstrings,''
\href{http://www.arXiv.org/abs/hep-th/0505277}{{\tt hep-th/0505277}}.

\bibitem{Oda:2005wu}
I.~Oda and M.~Tonin, ``The b-field in pure spinor quantization of
  superstrings,''
\href{http://www.arXiv.org/abs/hep-th/0510223}{{\tt hep-th/0510223}}.

\bibitem{Oda:2007ak}
I.~Oda and M.~Tonin, ``Y-formalism and $b$ ghost in the non-minimal pure spinor
  formalism of superstrings,'' {\em Nucl. Phys.} {\bf B779} (2007) 63--100,
\href{http://www.arXiv.org/abs/arXiv:0704.1219 [hep-th]}{{\tt arXiv:0704.1219
  [hep-th]}}.

\bibitem{Nh:2002tz}
P.~A. Grassi, G.~Policastro, and P.~van Nieuwenhuizen, ``The massless spectrum
  of covariant superstrings,'' {\em JHEP} {\bf 11} (2002) 001,
\href{http://www.arXiv.org/abs/hep-th/0202123}{{\tt hep-th/0202123}}.

\bibitem{Nh:2002xf}
P.~A. Grassi, G.~Policastro, and P.~van Nieuwenhuizen, ``The covariant quantum
  superstring and superparticle from their classical actions,'' {\em Phys.
  Lett.} {\bf B553} (2003) 96--104,
\href{http://www.arXiv.org/abs/hep-th/0209026}{{\tt hep-th/0209026}}.

\bibitem{Nh:2004nz}
P.~A. Grassi, G.~Policastro, and P.~van Nieuwenhuizen, ``Superstrings and
  {WZNW} models,''
\href{http://www.arXiv.org/abs/hep-th/0402122}{{\tt hep-th/0402122}}.

\bibitem{Nh:2004cz}
P.~A. Grassi and P.~van Nieuwenhuizen, ``Gauging cosets,''
\href{http://www.arXiv.org/abs/hep-th/0403209}{{\tt hep-th/0403209}}.

\bibitem{Grassi:2004tv}
P.~A. Grassi and G.~Policastro, ``Super-chern-simons theory as superstring
  theory,''
\href{http://www.arXiv.org/abs/hep-th/0412272}{{\tt hep-th/0412272}}.

\bibitem{Knapp:2004Dipl}
J.~Knapp, ``Covariant quantization of the superstring,'' Master's thesis, TU
  Wien, 2004.
\newblock Diploma Thesis.

\bibitem{Nh:2004we}
P.~A. Grassi and P.~van Nieuwenhuizen, ``{N = 4} superconformal symmetry for
  the covariant quantum superstring,''
\href{http://www.arXiv.org/abs/hep-th/0408007}{{\tt hep-th/0408007}}.

\bibitem{Gotz:2006qp}
G.~Gotz, T.~Quella, and V.~Schomerus, ``The {WZNW} model on {PSU(1,1|2)},''
  {\em JHEP} {\bf 03} (2007) 003,
\href{http://www.arXiv.org/abs/hep-th/0610070}{{\tt hep-th/0610070}}.

\bibitem{Chandia:2003hn}
O.~Chandia and B.~C. Vallilo, ``Conformal invariance of the pure spinor
  superstring in a curved background,'' {\em JHEP} {\bf 04} (2004) 041,
\href{http://www.arXiv.org/abs/hep-th/0401226}{{\tt hep-th/0401226}}.

\bibitem{Bedoya:2006ic}
O.~A. Bedoya and O.~Chandia, ``One-loop conformal invariance of the type {II}
  pure spinor superstring in a curved background,'' {\em JHEP} {\bf 01} (2007)
  042,
\href{http://www.arXiv.org/abs/hep-th/0609161}{{\tt hep-th/0609161}}.

\bibitem{Kluson:2006wq}
J.~Kluson, ``Note about classical dynamics of pure spinor string on {AdS(5) x
  S**5} background,'' {\em Eur. Phys. J.} {\bf C50} (2007) 1019--1030,
\href{http://www.arXiv.org/abs/hep-th/0603228}{{\tt hep-th/0603228}}.

\bibitem{Bianchi:2006im}
M.~Bianchi and J.~Kluson, ``{Current algebra of the pure spinor superstring in
  AdS(5) x S(5)},'' {\em JHEP} {\bf 08} (2006) 030,
\href{http://www.arXiv.org/abs/hep-th/0606188}{{\tt hep-th/0606188}}.

\bibitem{Grassi:2004ih}
P.~A. Grassi and L.~Tamassia, ``Vertex operators for closed superstrings,''
  {\em JHEP} {\bf 07} (2004) 071,
\href{http://www.arXiv.org/abs/hep-th/0405072}{{\tt hep-th/0405072}}.

\bibitem{Minkevich:1982a}
A.~V. Minkevich and F.~Karakura, ``On the relativistic dynamics of spinning
  matter in space-time with curvature and torsion,'' {\em J. Phys. A: Math.
  Gen.} (1983) 1409--1418.

\bibitem{Luckock:1989jr}
H.~Luckock and I.~Moss, ``The quantum geometry of random surfaces and spinning
  membranes,'' {\em Class. Quant. Grav.} {\bf 6} (1989)
1993.

\bibitem{Minkevich:1968a}
A.~Minkevich and F.~I. Fedorov {\em Izv. Akad. Nauk BSSR, Ser. Fiz.-Mat.} {\bf
  5} (1968) 35.

\bibitem{Minkevich:1975a}
A.~Minkevich and A.~A. Sokolski {\em Izv. Akad. Nauk BSSR, Ser. Fiz.-Mat.} {\bf
  4} (1975) 72.

\bibitem{Bergshoeff:2001pv}
E.~Bergshoeff, R.~Kallosh, T.~Ortin, D.~Roest, and A.~Van~Proeyen, ``New
  formulations of {D = 10} supersymmetry and {D8 - O8} domain walls,'' {\em
  Class. Quant. Grav.} {\bf 18} (2001) 3359--3382,
\href{http://www.arXiv.org/abs/hep-th/0103233}{{\tt hep-th/0103233}}.

\bibitem{Guttenberg:2007ha}
S.~Guttenberg, ``Derived brackets from super-{P}oisson brackets,''
\href{http://www.arXiv.org/abs/hep-th/0703085}{{\tt hep-th/0703085}}.

\bibitem{Bering:2006eb}
K.~Bering, ``{On non-commutative Batalin-Vilkovisky algebras, strongly homotopy
  Lie algebras and the Courant bracket},'' {\em Commun. Math. Phys.} {\bf 274}
  (2007) 297--341,
\href{http://www.arXiv.org/abs/hep-th/0603116}{{\tt hep-th/0603116}}.

\bibitem{Kosmann-Schwarzbach:2003en}
Y.~Kosmann-Schwarzbach, ``Derived brackets,'' {\em Lett. Math. Phys.} {\bf 69}
  (2004) 61--87,
\href{http://www.arXiv.org/abs/math.dg/0312524}{{\tt math.dg/0312524}}.

\bibitem{Alekseev:2004np}
A.~Alekseev and T.~Strobl, ``Current algebra and differential geometry,'' {\em
  JHEP} {\bf 03} (2005) 035,
\href{http://www.arXiv.org/abs/hep-th/0410183}{{\tt hep-th/0410183}}.

\bibitem{Gualtieri:007}
M.~Gualtieri, ``Generalized complex geometry,'' {\em Oxford University DPhil
  thesis} (2003) 107, \href{http://www.arXiv.org/abs/math.DG/0401221}{{\tt
  math.DG/0401221}}.

\bibitem{Bonelli:2005ti}
G.~Bonelli and M.~Zabzine, ``From current algebras for p-branes to topological
  m- theory,'' {\em JHEP} {\bf 09} (2005) 015,
\href{http://www.arXiv.org/abs/hep-th/0507051}{{\tt hep-th/0507051}}.

\bibitem{Hitchin:2004ut}
N.~Hitchin, ``Generalized {Calabi-Yau} manifolds,'' {\em Quart. J. Math. Oxford
  Ser.} {\bf 54} (2003) 281--308,
\href{http://www.arXiv.org/abs/math.dg/0209099}{{\tt math.dg/0209099}}.

\bibitem{Grana:2005ny}
M.~Grana, J.~Louis, and D.~Waldram, ``Hitchin functionals in {N=2}
  supergravity,''
\href{http://www.arXiv.org/abs/hep-th/0505264}{{\tt hep-th/0505264}}.

\bibitem{Grana:2005jc}
M.~Grana, ``Flux compactifications in string theory: A comprehensive review,''
  {\em Phys. Rept.} {\bf 423} (2006) 91--158,
\href{http://www.arXiv.org/abs/hep-th/0509003}{{\tt hep-th/0509003}}.

\bibitem{Kapustin:2004gv}
A.~Kapustin and Y.~Li, ``Topological sigma-models with {H-flux} and twisted
  generalized complex manifolds,''
\href{http://www.arXiv.org/abs/hep-th/0407249}{{\tt hep-th/0407249}}.

\bibitem{Pestun:2005rp}
V.~Pestun and E.~Witten, ``The {Hitchin} functionals and the topological
  {B-model} at one loop,''
\href{http://www.arXiv.org/abs/hep-th/0503083}{{\tt hep-th/0503083}}.

\bibitem{Pestun:2006rj}
V.~Pestun, ``Topological strings in generalized complex space,''
\href{http://www.arXiv.org/abs/hep-th/0603145}{{\tt hep-th/0603145}}.

\bibitem{Jeschek:2004je}
C.~Jeschek, ``Generalized {Calabi-Yau} structures and mirror symmetry,''
\href{http://www.arXiv.org/abs/hep-th/0406046}{{\tt hep-th/0406046}}.

\bibitem{Jeschek:2005ek}
C.~Jeschek and F.~Witt, ``Generalised geometries, constrained critical points
  and ramond-ramond fields,''
\href{http://www.arXiv.org/abs/math.dg/0510131}{{\tt math.dg/0510131}}.

\bibitem{Cassani:2007pq}
D.~Cassani and A.~Bilal, ``Effective actions and n=1 vacuum conditions from
  su(3) x su(3) compactifications,''
\href{http://www.arXiv.org/abs/arXiv:0707.3125 [hep-th]}{{\tt arXiv:0707.3125
  [hep-th]}}.

\bibitem{Grange:2004ah}
P.~Grange and R.~Minasian, ``Modified pure spinors and mirror symmetry,'' {\em
  Nucl. Phys.} {\bf B732} (2006) 366--378,
\href{http://www.arXiv.org/abs/hep-th/0412086}{{\tt hep-th/0412086}}.

\bibitem{Tomasiello:2007zq}
A.~Tomasiello, ``Reformulating supersymmetry with a generalized dolbeault
  operator,''
\href{http://www.arXiv.org/abs/arXiv:0704.2613 [hep-th]}{{\tt arXiv:0704.2613
  [hep-th]}}.

\bibitem{Ikeda:2006pd}
N.~Ikeda and T.~Tokunaga, ``Topological membranes with 3-form h flux on
  generalized geometries,''
\href{http://www.arXiv.org/abs/hep-th/0609098}{{\tt hep-th/0609098}}.

\bibitem{Ikeda:2007rn}
N.~Ikeda and T.~Tokunaga, ``An alternative topological field theory of
  generalized complex geometry,''
\href{http://www.arXiv.org/abs/arXiv:0704.1015 [hep-th]}{{\tt arXiv:0704.1015
  [hep-th]}}.

\bibitem{Lindstrom:2004iw}
U.~Lindstrom, R.~Minasian, A.~Tomasiello, and M.~Zabzine, ``Generalized complex
  manifolds and supersymmetry,'' {\em Commun. Math. Phys.} {\bf 257} (2005)
  235--256,
\href{http://www.arXiv.org/abs/hep-th/0405085}{{\tt hep-th/0405085}}.

\bibitem{Zabzine:2006uz}
M.~Zabzine, ``Lectures on generalized complex geometry and supersymmetry,''
\href{http://www.arXiv.org/abs/hep-th/0605148}{{\tt hep-th/0605148}}.

\bibitem{Lindstrom:2006ee}
U.~Lindstrom, ``A brief review of supersymmetric non-linear sigma models and
  generalized complex geometry,''
\href{http://www.arXiv.org/abs/hep-th/0603240}{{\tt hep-th/0603240}}.

\bibitem{Zabzine:2005qf}
M.~Zabzine, ``Hamiltonian perspective on generalized complex structure,'' {\em
  Commun. Math. Phys.} {\bf 263} (2006) 711--722,
\href{http://www.arXiv.org/abs/hep-th/0502137}{{\tt hep-th/0502137}}.

\bibitem{Zucchini:2004ta}
R.~Zucchini, ``A sigma model field theoretic realization of {Hitchin's}
  generalized complex geometry,'' {\em JHEP} {\bf 11} (2004) 045,
\href{http://www.arXiv.org/abs/hep-th/0409181}{{\tt hep-th/0409181}}.

\bibitem{Zucchini:2005rh}
R.~Zucchini, ``Generalized complex geometry, generalized branes and the
  {Hitchin} sigma model,'' {\em JHEP} {\bf 03} (2005) 022,
\href{http://www.arXiv.org/abs/hep-th/0501062}{{\tt hep-th/0501062}}.

\bibitem{Zucchini:2005cq}
R.~Zucchini, ``A topological sigma model of {biKaehler} geometry,'' {\em JHEP}
  {\bf 01} (2006) 041,
\href{http://www.arXiv.org/abs/hep-th/0511144}{{\tt hep-th/0511144}}.

\bibitem{Zucchini:2007ie}
R.~Zucchini, ``The {Hitchin} model, {Poisson}-quasi-{Nijenhuis} geometry and
  symmetry reduction,''
\href{http://www.arXiv.org/abs/arXiv:0706.1289 [hep-th]}{{\tt arXiv:0706.1289
  [hep-th]}}.

\bibitem{Henneaux:1992ig}
M.~Henneaux and C.~Teitelboim, {\em Quantization of gauge systems}.
\newblock Princeton, USA: Univ. Pr. (1992) 520 p.

\bibitem{Buttin:1974}
C.~Buttin, ``Th\'eorie des op\'erateurs diff\'erentiels gradu\'es sur les
  formes diff\'erentielles,'' {\em Bull. Soc. Math. Fr.} {\bf 102} (1974)
  49--73.

\bibitem{Cattaneo:1999fm}
A.~S. Cattaneo and G.~Felder, ``A path integral approach to the {Kontsevich}
  quantization formula,'' {\em Commun. Math. Phys.} {\bf 212} (2000) 591--611,
\href{http://www.arXiv.org/abs/math.qa/9902090}{{\tt math.qa/9902090}}.

\bibitem{Schaller:1994es}
P.~Schaller and T.~Strobl, ``Poisson structure induced (topological) field
  theories,'' {\em Mod. Phys. Lett.} {\bf A9} (1994) 3129--3136,
\href{http://www.arXiv.org/abs/hep-th/9405110}{{\tt hep-th/9405110}}.

\bibitem{deBoer:2003dn}
J.~de~Boer, P.~A. Grassi, and P.~van Nieuwenhuizen, ``Non-commutative
  superspace from string theory,'' {\em Phys. Lett.} {\bf B574} (2003) 98--104,
\href{http://www.arXiv.org/abs/hep-th/0302078}{{\tt hep-th/0302078}}.

\bibitem{Berkovits:2003kj}
N.~Berkovits and N.~Seiberg, ``Superstrings in graviphoton background and {N =
  1/2 + 3/2} supersymmetry,'' {\em JHEP} {\bf 07} (2003) 010,
\href{http://www.arXiv.org/abs/hep-th/0306226}{{\tt hep-th/0306226}}.

\bibitem{Ooguri:2003qp}
H.~Ooguri and C.~Vafa, ``The {C-deformation} of gluino and non-planar
  diagrams,'' {\em Adv. Theor. Math. Phys.} {\bf 7} (2003) 53--85,
\href{http://www.arXiv.org/abs/hep-th/0302109}{{\tt hep-th/0302109}}.

\bibitem{Hull:2004in}
C.~M. Hull, ``A geometry for non-geometric string backgrounds,'' {\em JHEP}
  {\bf 10} (2005) 065,
\href{http://www.arXiv.org/abs/hep-th/0406102}{{\tt hep-th/0406102}}.

\bibitem{Hull:2006qs}
C.~M. Hull, ``Global aspects of {T}-duality, gauged sigma models and {T}-
  folds,''
\href{http://www.arXiv.org/abs/hep-th/0604178}{{\tt hep-th/0604178}}.

\bibitem{Hull:2006va}
C.~M. Hull, ``Doubled geometry and {T}-folds,''
\href{http://www.arXiv.org/abs/hep-th/0605149}{{\tt hep-th/0605149}}.

\bibitem{Dabholkar:2005ve}
A.~Dabholkar and C.~Hull, ``Generalised {T}-duality and non-geometric
  backgrounds,'' {\em JHEP} {\bf 05} (2006) 009,
\href{http://www.arXiv.org/abs/hep-th/0512005}{{\tt hep-th/0512005}}.

\bibitem{Grana:2006kf}
M.~Grana, R.~Minasian, M.~Petrini, and A.~Tomasiello, ``A scan for new n=1
  vacua on twisted tori,'' {\em JHEP} {\bf 05} (2007) 031,
\href{http://www.arXiv.org/abs/hep-th/0609124}{{\tt hep-th/0609124}}.

\bibitem{Morris:2007ga}
S.~Morris, ``Doubled geometry versus generalized geometry,'' {\em Class. Quant.
  Grav.} {\bf 24} (2007)
2879--2900.

\bibitem{Buscher:1987sk}
T.~H. Buscher, ``A symmetry of the string background field equations,'' {\em
  Phys. Lett.} {\bf B194} (1987)
59.

\bibitem{Buscher:1987qj}
T.~H. Buscher, ``Path integral derivation of quantum duality in nonlinear sigma
  models,'' {\em Phys. Lett.} {\bf B201} (1988)
466.

\bibitem{Dubois-Violette:1994gy}
M.~Dubois-Violette and P.~W. Michor, ``A common generalization of the
  {F}r{\"o}hlicher-{N}ijenhuis bracket and the {S}chouten bracket for symmetric
  multivector fields,''
\href{http://www.arXiv.org/abs/alg-geom/9401006}{{\tt alg-geom/9401006}}.

\bibitem{Kosmann-Schwarzbach:1996a}
Y.~Kosmann-Schwarzbach, ``From {P}oisson algebras to {G}erstenhaber algebras,''
  {\em Ann. Inst. Fourier (Grenoble)} {\bf 46} (1996) 1241--1272.

\bibitem{Kosmann-Schwarzbach:1996b}
Y.~Kosmann-Schwarzbach, ``Derived brackets and the gauge algebra of closed
  string field theory,'' {\em Quantum Group Symposium at GROUP 21
  (Goslar,1996), H.-D. Doebner and V. K. Dobrev, eds., Heron Press, Sofia}
  (1997) 53--61.

\bibitem{Vinogradov:1990}
A.~M. Vinogradov, ``Unication of the {S}chouten and {N}ijenhuis brackets,
  cohomology, and superdifferential operators,'' {\em Mat. Zametki} {\bf 47
  (6)} (1990) 138--140. not translated in Math. Notes.

\bibitem{Vinogradov:1992}
A.~Cabras and A.~M. Vinogradov, ``Extensions of the {P}oisson bracket to
  differential forms and multi-vector fields,'' {\em J. Geom. Phys.} {\bf 9}
  (1992) 75--100.

\bibitem{Jeschek:2004wy}
C.~Jeschek and F.~Witt, ``{Generalised G(2)-structures and type IIB
  superstrings},'' {\em JHEP} {\bf 03} (2005) 053,
\href{http://www.arXiv.org/abs/hep-th/0412280}{{\tt hep-th/0412280}}.

\bibitem{Witt:2005sk}
F.~Witt, ``{Special metric structures and closed forms},''
\href{http://www.arXiv.org/abs/math/0502443}{{\tt math/0502443}}.

\bibitem{Kugo:1997}
T.~Kugo, {\em Eichtheorie}.
\newblock Berlin/Heidelberg, Germany: Springer (1997) 522 P.

\bibitem{Kreuzer:2001gm}
M.~Kreuzer, {\em Geometrische Methoden der Theoretischen Physik}.
\newblock 2001.
\newblock http://hep.itp.tuwien.ac.at/$\sim$kreuzer/inc/gmtp.ps.gz.

\bibitem{Tsimpis:2004gq}
D.~Tsimpis, ``Curved 11d supergeometry,'' {\em JHEP} {\bf 11} (2004) 087,
\href{http://www.arXiv.org/abs/hep-th/0407244}{{\tt hep-th/0407244}}.

\bibitem{Mafra:2009wq}
C.~R. Mafra, ``{Superstring Scattering Amplitudes with the Pure Spinor
  Formalism},''
\href{http://www.arXiv.org/abs/0902.1552}{{\tt 0902.1552}}.

\bibitem{Bedoya:2008yw}
O.~A. Bedoya, ``{Superstring Sigma Model Computations Using the Pure Spinor
  Formalism},''
\href{http://www.arXiv.org/abs/0808.1755}{{\tt 0808.1755}}.

\end{thebibliography}

}\rem{

\chapter{Solving the Pure Spinor Constraint}

\label{cha:Solving-the-PureSpinor}see conventions.lyx!}

\rem{

\chapter{Relation between operator products and commutators}

\label{cha:OPE}see OPE.lyx and conventions.lyx}

\rem{

\chapter{WZNW Model}

\label{cha:WZNW-Model}see WZNW-NW.lyx} \providecommand{\href}[2]{#2}\begingroup\raggedright\endgroup

\renewcommand{\be}{\bs{b}}\renewcommand{\ce}{\bs{c}}\label{index}\printindex{}

\chapter*{Curriculum Vitae}

\addcontentsline{toc}{chapter}{Curriculum Vitae}{\inputTeil{0}\ifthenelse{\theinput=1}{}{}

\title{Curriculum Vitae}

\author{Sebastian Guttenberg}

\date{{}}

\maketitle

\rem{To do:

\begin{itemize}
\item Don't panic
\end{itemize}
}

\subsection*{Personal data}

\begin{tabular}{ll}
Name: & Sebastian Guttenberg\tabularnewline
Date of birth: & May 13, 1977\tabularnewline
Place of birth: & Rosenheim, Germany\tabularnewline
Nationality: & German\tabularnewline
Parents / brother: & Sibylle Guttenberg, Andreas Guttenberg$\qquad/\qquad$Philipp Guttenberg\tabularnewline
\end{tabular}

\subsection*{Education / research experience}

\begin{itemize}
\item 1996 Abitur (final exams) at the {}``Gymnasium Miesbach'' (highschool)
in Germany
\item 10/1997 - 10/2002: physics studies at the Munich University of Technology
(TUM)
\item 1999/2000: three and a half months studies as a guest at the {}``State
University of New York, Stony Brook'' 
\item 10/2001 - 10/2002 extramural diploma thesis with the title {}``effektive
Wirkungen in der Stringtheorie'' (effective actions in string theory)
at the {}``Ludwig Maximilians Universität'' in Munich with advisor
Ivo Sachs. Part of this work was done from May '02 until July '02
at the Trinity College in Dublin.
\item 10/1998 - fall 1999: additional studies of mathematics at the TUM,
terminated after successfully having passed the first part ({}``Vordiplom'')
\item 03/2003-09/2007 PhD studies at the Vienna University of Technology
(TU Wien) with supervisor Maximilian Kreuzer. 
\item 11/2005 - 09/2006: ten months visit in Paris, Saclay (CEA/SPhT), working
in the string group with Ruben Minasian, Mariana Gra\~na, Pierre
Vanhove et al.
\item 10/2007 planned start of a postdoctoral position at the 'Demokritos
Nuclear Research Centre' in Athens/Greece in the group of George Savvidy
\end{itemize}

\subsection*{Awards and Fellowships}

\begin{itemize}
\item 1995 Second prize in the first round of the {}``Bundeswettbewerb
Mathematik~'95'' (German high school math competition)
\item 1995 reaching the third round of the German selection procedure for
the International Physics Olympiade '96 (One of top 50 German high
school students) 
\item 1996 First prize in the first round and second prize in the second
round of the {}``Bundeswettbewerb Mathematik '96''
\item From Nov.'98 until the end of the physics studies: fellowship of the
{}``Studienstiftung des deutschen Volkes''.
\item May '05 to July '05 {}``Junior Research Fellowship in Mathematics
and Mathematical Physics'' at the Erwin Schrödinger Institute, Vienna
\item Mobility fellowship ''Mobilitätsstipendium der Akademisch-sozialen
Arbeitsgemeinschaft Österreichs (ASAG)'' and fellowship to go abroad
{}``Auslands\-Stipendium der TU Wien'' in order to allow a three
months visit in Saclay which was then extended to a ten months stay
financed by EGIDE in France
\end{itemize}

}

\chapter*{Lebenslauf}

\addcontentsline{toc}{chapter}{Lebenslauf}{\inputTeil{0} \ifthenelse{\theinput=1}{}{\pagestyle{empty}}

\title{Lebenslauf}

\author{Sebastian Guttenberg}

\date{{}}

\maketitle

\rem{To do:

\begin{itemize}
\item Don't panic
\end{itemize}
}

\subsection*{Persönliche Daten}

\begin{tabular}{ll}
Name: & Sebastian Guttenberg\tabularnewline
Geburtsdatum: & 13. Mai 1977\tabularnewline
Geburtsort: & Rosenheim, Deutschland\tabularnewline
Nationalität: & deutsch\tabularnewline
Eltern / Bruder: & Sibylle Guttenberg, Andreas Guttenberg$\qquad/\qquad$Philipp Guttenberg\tabularnewline
\end{tabular}

\subsection*{Ausbildung / Forschung}

\begin{itemize}
\item 1996 Abitur am Gymnasium Miesbach
\item 10/1997 - 10/2002: Physikstudium an der Technischen Universität München
(TUM)
\item 1999/2000: dreieinhalb-monatiges Gaststudium an der {}``State University
of New York, Stony Brook'' 
\item 10/2001 - 10/2002 Externe Diplomarbeit mit dem Titel {}``effektive
Wirkungen in der Stringtheorie'' an der Ludwig Maximilians Universität
in München unter der Betreuung von Ivo Sachs. Ein Teil dieser Arbeit
wurde zwischen Mai '02 und Juli '02 am 'Trinity College' in Dublin
angefertigt.
\item 10/1998 - Herbst 1999: zusätzliches Studium der Mathematik an der
TUM, beendet nach erfolgreichem Ablegen der Vordiplomsprüfungen.
\item 03/2003-09/2007 Doktoratsstudium an der Technischen Universität Wien
unter der Leitung von \\
Maxi\-milian Kreuzer. 
\item 11/2005 - 09/2006: zehnmonatiger Aufenthalt in Paris, Saclay (CEA/SPhT);
Gast der dortigen Stringtheorie-Gruppe bestehend aus Ruben Minasian,
Mariana Gra\~na, Pierre Vanhove et al.
\item 10/2007 geplanter Beginn einer Postdoc-Stelle am 'Demokritos Nuclear
Research Centre' in Athen/Griechenland bei George Savvidy
\end{itemize}

\subsection*{Auszeichnungen und Stipendien}

\begin{itemize}
\item 1995 Zweiter Preis in der ersten Runde des Bundeswettbewerbs Mathematik~1995
\item 1995 Erreichen der dritten Runde des deutschen Auswahlverfahrens zur
Internationalen Physik~Olympiade~'96
\item 1996 Erster Preis in der ersten Runde und zweiter Preis in der zweiten
Runde des Bundeswettbewerbs Mathematik 1996
\item Von Nov.'98 bis zum Ende des Physikstudiums: Stipendium der {}``Studienstiftung
des deutschen Volkes''.
\item Mai '05 bis Juli '05 {}``Junior Research Fellowship in Mathematics
and Mathematical Physics'' am Erwin Schrödinger Institut in Wien
\item ''Mobilitätsstipendium der Akademisch-sozialen Arbeitsgemeinschaft
Österreichs (ASAG)'' und {}``Auslands\-Stipendium der TU Wien''
zur Ermöglichung eines dreimonatigen Aufenthaltes in Saclay, der dann
\\
(finanziert durch EGIDE von französischer Seite) auf zehn Monate verlängert
wurde.
\end{itemize}

}
\end{document}